# Business cycles in economics

**Viktor O. Ledenyov and Dimitri O. Ledenyov**

James Cook University, Townsville, Queensland, Australia

V. N. Karazin Kharkiv National University, Kharkiv, Ukraine

March, 2018

*To our lovely parents Oleg P. Ledenyov and Tamara V. Ledenyova.*

# Contents

# Introduction

The business cycles are generated by the oscillating macro-/micro-/nano- economic output variables in the economy of the scale and the scope in the amplitude/frequency/phase/time domains in the economics. The accurate forward looking assumptions on the business cycles oscillation dynamics can optimize the financial capital investing and/or borrowing by the economic agents in the capital markets. The book's main objective is to study the business cycles in the economy of the scale and the scope, formulating the Ledenyov unified business cycles theory in the Ledenyov classic and quantum econodynamics. Chapter 1 makes a historical overview on the development of the economies of the scales and the scopes in the world over the centuries. Chapter 2 adds some insights on the continuous-time economic output waves in the economy of the scale and the scope in the Ledenyov classic econodynamics. Chapter 3 considers the Ledenyov discrete-time digital economic output waves in the economy of the scale and the scope in the Ledenyov classic econodynamics. Chapter 4 researches the Ledenyov discrete-time digital economic output waves in the form of the vector-modulated discrete-time digital direct sequence spread spectrum signal's bursts in the economy of the scale and the scope in the Ledenyov classic econodynamics. Chapter 5 deals with the Ledenyov discrete-time digital economic output waves in the form of the vector-modulated discrete-time digital direct sequence spread spectrum signal's short/wide/ultra wide band pulses in the economy of the scale and the scope in the Ledenyov classic econodynamics. Chapter 6 describes the Ledenyov discrete-time digital economic output waves in the form of the vector-modulated discrete-time digital direct sequence spread spectrum signal's short/wide/ultra wide band pulses generated by the quantum fluctuations in the economy of the scale and the scope in the Ledenyov quantum econodynamics. Chapter 7 discusses a problem on the precise measurement of the econodynamic variables in the economy of the scale and the scope in the Ledenyov classic and quantum econodynamics. Chapter 8 writes on the accurate forecast of the economic and financial trends with the business cycles oscillation dynamics analysis in the economy of the scale and the scope in the Ledenyov classic and quantum econodynamics. Conclusion summarizes all the important research findings, highlighting the original contributions in a plain language format. The book presents a wonderful opportunity for the thinking readers to learn more on the business cycles in the economics.

# Chapter 1

## Economies of scales and scopes from ancient days to modern time

The modern human civilization is created as a result of evolutionary development in the increasingly-interconnected closely-integrated fields of the economics, finances, science, culture, religion, politics, defense and sport in the highly organized human societies over the thousands of years. In an ideal case, the modern human civilization's evolutionary development goal is to reach the economical prosperity, peace and stability for all the members of the highly organized human society at the national / global levels in the short /long time perspectives. However, in the reality, the human civilization experiences the numerous altering phases of the progressive / regressive social/economic developments as a result of the growing / declining progress in the human activities in the fields of the economics, finances, science, culture, religion, politics, defense and sport in the national / global economies of the scales and the scopes in the short /long historical periods.

Let us permit that the human civilization aims to reach a state of prosperity, which is uniquely predetermined by the produced economic output magnitude in the global economy of the scale and the scope over the certain time period among other factors. Presumably, in order to reach the increasing economic output magnitude, we can expect that the new wealth creation process by the economic agents must be established, applying the three main objects:

*1.* The physical matter / the land;

*2.* The human labour;

*3.* The financial capital;

in the economy of the scale and the scope in accordance with the fundamental principles of the political economy in Joseph Penso de la Vega (1668, 1996), Mortimer (1765), Smith (1776, 2008), Ricardo (1817, 1821), Bentham (1839), Mill (1862), Hirsch (1896). Perhaps, at this point, we can agree to the fact that the modern human civilization's progressive prosperous development ultimately depends on the degree of increase of the economic output magnitude over the selected time period.

Fig. 1 shows a simple scheme of the new wealth synthesis process with the use of the matter, labour, capital in the economy of the scale and the scope in the time scale ($t_1 < t_2 < t_3$).

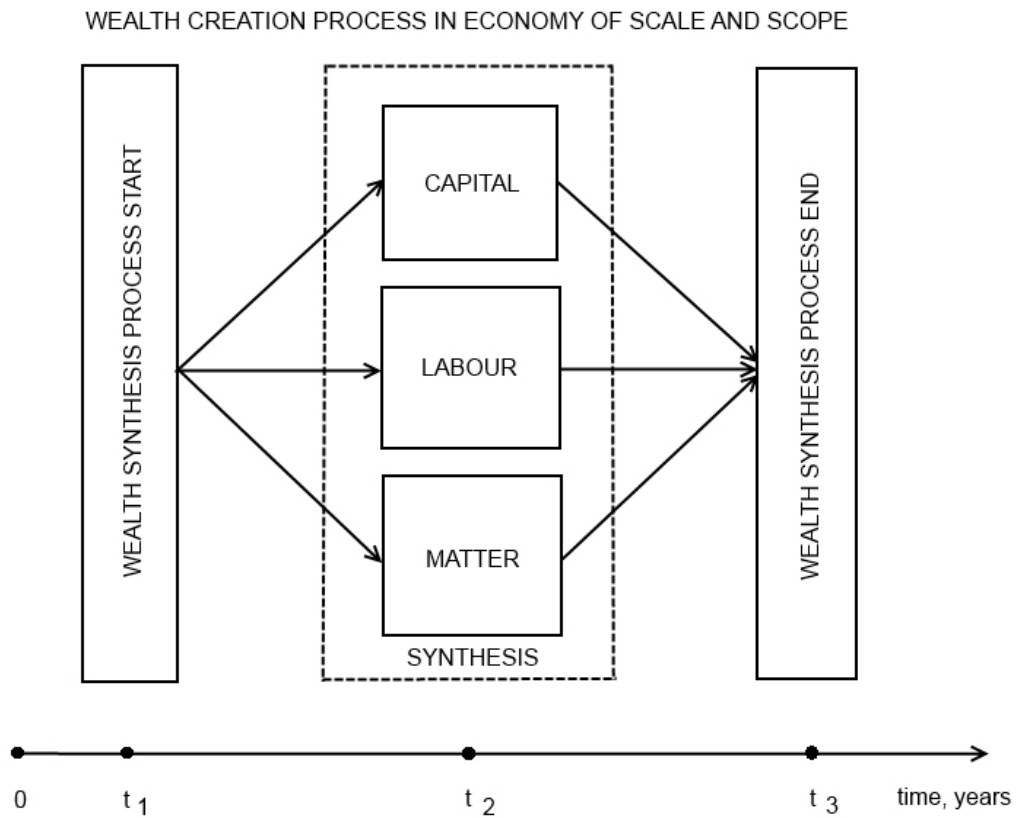

**Fig. 1.** Scheme of new wealth creation process with application of matter, labour, capital in economy of scale and scope in time scale ($t_1 < t_2 < t_3$).

In the recent years, there has been a significant increase in a number of scientific publications with the expressed research opinions that the human civilization's evolution is uniquely predetermined by the frequently changing economic growth / decline phases in the national/global economies of the scales and the scopes over the time. Thus, we can assume that an every economy of the scale and the scope experiences the rapidly changing phases of the economic growth / decline with the market resources optimization and their possible subsequent re-allocation in the other newly created industries or even the other developing economies of the scales and scopes over the time. Therefore, let us research these business cycles, which can be characterized by the growth / recession / depression / stagnation phases in the economies of the scales and scopes in the corresponding historical time periods.

A quite interesting fact is that the new wealth synthesis process can be considered as a nonlinear function of the multiple nonlinearly changing macro-/micro-/nano- economic variables, resulting in the strong/weak nonlinear fluctuations of the economic output magnitudes in the nonlinear economies of the scales and the scopes in the macroeconomics.

Fig. 2 displays a scheme of the new wealth synthesis process with an application of the matter, labour, capital in the nonlinear economy of the scale and the scope in the time scale ($t_1 < t_2 < t_3$). There is a nonlinear connection in the form of a positive/negative feedback loop between the wealth synthesis process start and the wealth synthesis process end in the nonlinear economy of the scale and the scope.

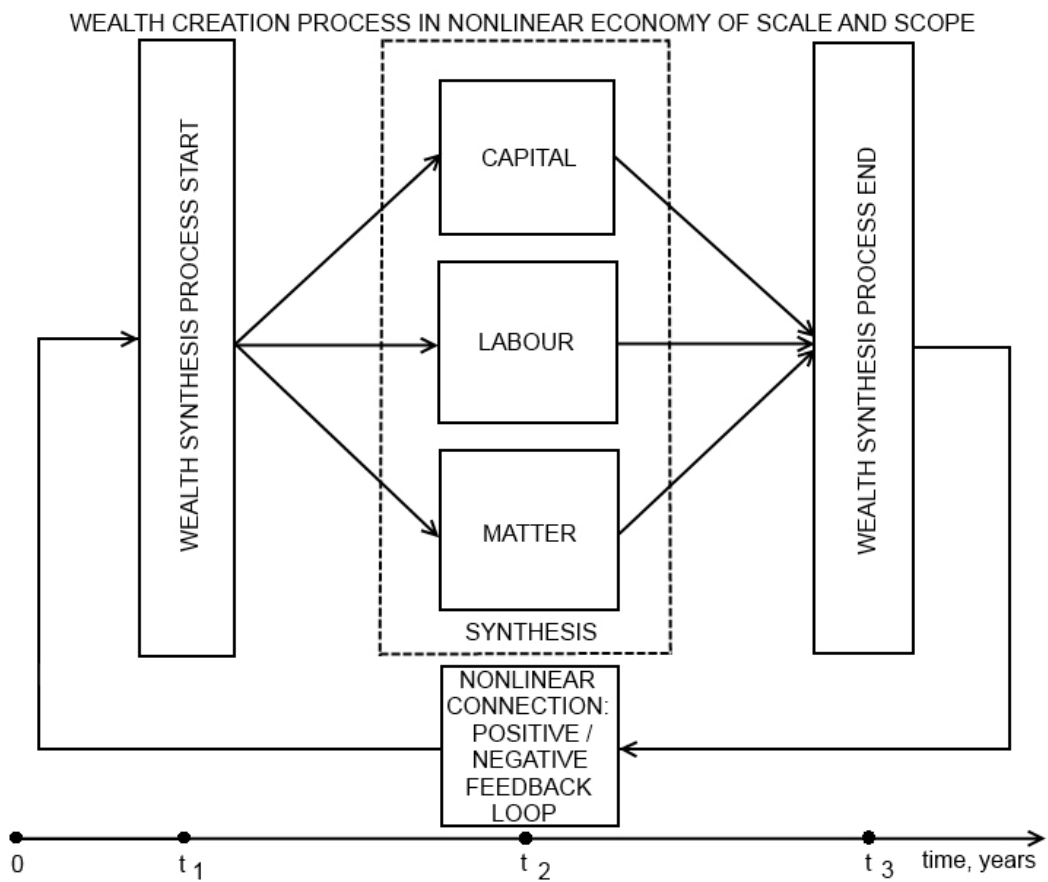

**Fig. 2.** Scheme of new wealth synthesis process with application of matter, labour, capital in nonlinear economy of scale and scope in time domain ($t_1 < t_2 < t_3$). Positive/negative feedback loops between wealth synthesis process start and end points can adjust economic output magnitude in nonlinear economy of scale and scope over time.

We assume that, in the course of the human civilization evolution, the first economies of the scales and scopes have been established in the Asia, Middle East and Europe in different time periods. We can see that the changing periods of the economic output convergence and/or divergence during the economic progressive development / regressive decline processes had place, because of introduction of the disruptive technological / managerial / social / cultural / political innovation(s) by the economic agent(s) into the corresponding economy of the scale and the scope.

So let us review some of the historical milestones during the modern human civilization development through the processes of the constant technological modernization in the economies of the scales and the scopes in many regions around the World.

Researching the old history of economic development in Asia, we can evidently write that one of the first known economies of the scales and the scopes has been created by the Chinese people in the mainland China more than the four thousand years ago in Chen Shou (280s-290s; 1977), Landes (1998), Liu Guanglin (2015), Ma Debin, Yuan Weipeng (2016), Xu Yi, Ni Yuping, Van Leeuwen (2016), Xu Yi, Shi Zhihong, Van Leeuwen, Ni Yuping, Zhang Zipeng, Ma Ye (2016), Mitchener, Ma Debin (2017). In the beginning of the historical evolutionary process, the Chinese people used the land, labour and capital to produce various crops, including the rice, millet, sorghum, barley, wheat and other cereals. Indeed, the Han people accumulated a formidable scientific and technological knowledge in the fields of the agriculture development, stone processing, ore extraction, metal melting in the mainland China over the centuries. They were able to successfully complete the complex engineering projects towards the stone-made building construction, subsequently reaching a considerable progress in the stone-made fortifications design such as the Great Wall in the Northern China. In addition, the created/invented/derived the Chinese language with the 3000 characters, paper to write letters/books, mathematical formulas for astronomic/physical/chemical/engineering calculations, and other things.

Focusing on the historical records of economic development in Mesopotamia, we can point out to the fact that there was a relatively highly organized economy of the scale and the scope centered around the City of Babylon in the Akkadian Empire and the Neo-Babylonian Empire since around c. 2300 B.C. Babylon was a well established place to live with the numerous stone/bricks made buildings, big fortress walls and other architectural sites such as the Temple of Babel, the Ishtar Gate, the Walls of Babylon. It was also known as a center of the commerce, education and science in Mignan (1829).

Discussing a state of economic development in Europe from the ancient history perspective, we can comment that the first economy of the scale and the scope was organized in the Kingdom of Macedonia with the capital cities of Aigai and Pella since around 808 B.C. until 168 B.C. The first economy of the scale and the scope included the faming, livestock herding, minerals mining, homes construction, ships building and some other industries. A minting of the precious metals coins was introduced in Macedonia under Alexander I sometime after 479 B.C. in Kremydi (2011), Treister (1996). Since that time, the minting of the different golden and silver coins at a number of the officially permitted mints in the Macedonian commonwealth at various historical periods was conducted. One of the main tasks of the financial system was to facilitate the exchange by the goods and services in the Macedonian economy of the scale and the scope.

Considering the subsequent economic developments in Europe, we can point out to a well known historical fact that the City of Rome was a capital of the Roman Empire since c. 30 B.C. until approximately 1461 in Abbott (1901). The Roman economy of the scale and the scope was established in Roman Empire on that time. The Roman economy of the scale and the scope included a big number of the smaller economies of the scales and scopes within the Roman Empire. We can certainly say that there were the agriculture, minerals mining, building construction, commerce and transportation industries in the Roman economy of the scale and the scope. For example, it may be interesting to note that a big number of the renowned architectural sites, including the Coliseum, Flavian amphitheatre, numerous villas, dams, roads, were constructed in the Roman Empire. Most interestingly, the financial system was characterized by a quite advanced level of the capital market in terms of the monetization, banking and taxation in the Roman economy of the scale and the scope on that time period in Duncan-Jones (1994), Harl (19 June 1996), Andreau (1999), Scheidel (2009), Harris (2010), Kessler, Temin (2010).

Analyzing the economic developments in America in the modern history context during the recent two hundred years, we can higlight the well described historical facts that the American economy of the scale and the scope was mainly based on the farming, cattle production and mining in the initial phase of its development. It was relatively isolated, because of natural geographical factors such as the remote location from the Asia and Europe regions. The transformation of the American economy of the scale and the scope took place as a result of the multiple industrial revolutions, which were facilitated by the steam engine invention, the steam engine locomotive development, the East-West coasts railroad construction by the Chinese people, the steam/combustion engines automobiles design, the Ford automobiles mass production, the steam/diesel/gas turbine engine cargo ships production, the Morse cable/wireless information transmission code invention, the Marconi trans-ocean wireless/communication links design, etc. The American financial system was established in an analogy with the European financial system on that time. It was strongly influenced by the scientific ideas by the Austrian school of the financial and economic thinking, namely by the Austrian scientists-immigrants at University of Chicago in the State of Illinois in the USA in Menger (1871), von Böhm-Bawerk (1884, 1889, 1921), Von Mises (1912), Hayek (1931, 1935, 1948, 1980, 2008). The Austrian school of the financial and economic thinking used the financial and monetary interventions as a possible remedy to solve the existing misbalances in the economy of the scale and the scope, presuming that the sustainable economic development can be achieved by adding an excessive financial liquidity to the capital market. The arising problem on the high inflation expectations can be effectively solved by the new financial products / treasure papers introduction in a speculative sector of the economy of the scale and the scope. This line of modern economic thinking prevails among a big number of the American influential economists in the USA in our days.

An ancient history of the economies of the scales and scopes shows their paramount importance in the human civilization development in various time perspectives. There were the multiple periods of the economic output expansion and the economic output contraction in all the economies of the scales and the scopes in the World. The altering periods of the growth and decrease of the economic output magnitude were somehow dependant on a state of the matters in the following two economic sectors:

1.  The real economic industrial sector (the agriculture, minerals mining, minerals/metals/stones processing, chemicals production, machine tools development, homes/offices/fortifications construction, electronics development, ships building, aircrafts development, automobiles/motorcycles/bicycles production industries),

2.  The speculative economic industrial sector (the domestic/international trade, stock exchange trade, financial services, insurance services, juridical services, medical services, leisure/tourism services, games services, video/music services industries).

Fig. 3 presents a scheme of the economy of the scale and the scope with the real- and the imaginary- (speculative) economic industrial sectors.

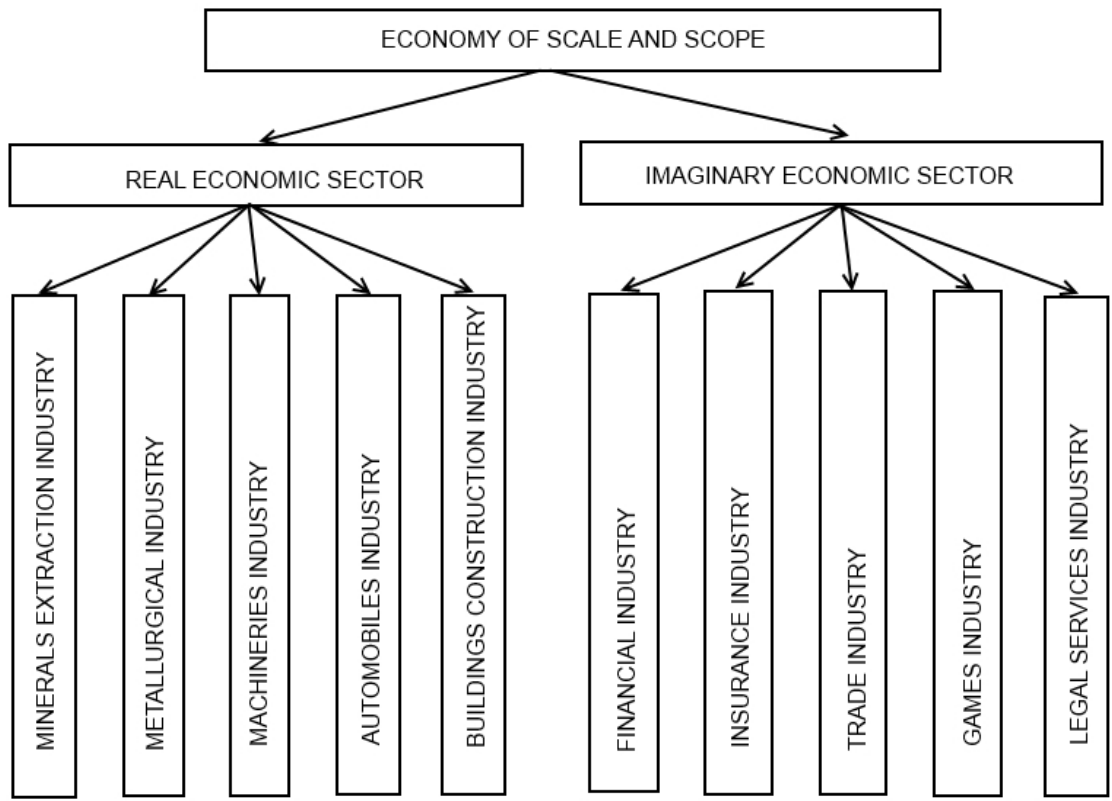

**Fig. 3.** Scheme of economy of scale and scope with real- and imaginary- (speculative) economic sectors with different possible industries.

At this point in our discussion, one practical question may arise: How can we evaluate the economy of the scale and the scope? Indeed, a measurement of the economic output of the economy of the scale and the scope over the selected time period is one of the central research problems in the macroeconomics. Presently, the economists came to a certain consensus among the expressed research opinions on a most appropriate approach to the total economic output measurement by introducing a special economic variable such as the Gross Domestic Product (GDP) in the macroeconomics in Kuznets (1934). At the first consideration, the GDP(t) can be interpreted as a monetary measure of the total economic output value created by all the wealth synthesis processes in the economy of the scale and the scope over the selected time period.

Fig. 4 presents a measurement scheme of the Gross Domestic Product GDP(t), which is a sum of the economic outputs values by the numerous wealth synthesis processes in the economy of the scale and the scope over the selected time period (t=$t_2$-$t_1$).

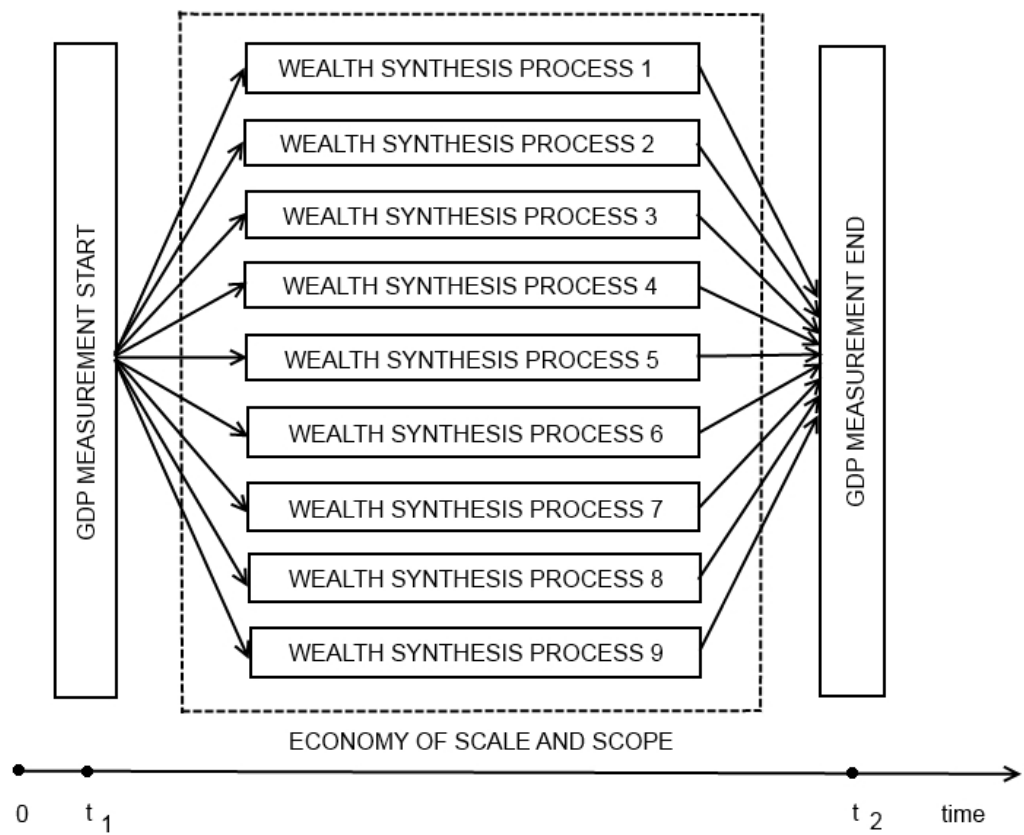

**Fig. 4.** General scheme of Gross Domestic Product GDP(t) measurement in economy of scale and scope over selected time period (t=$t_2$-$t_1$).

However, in a real life situation, it makes more sense to associate an every selected wealth synthesis process with the corresponding industry in the economy of the scale and the scope. Then, we can say that the GDP is equal to a sum of the economic outputs of wealth synthesis processes by the numerous industries in the real and imaginary economic sectors in the economy of the scale and the scope over the selected time period ($t=t_2-t_1$).

Fig. 5 demonstrates a scheme of the Gross Domestic Product (GDP) measurement in the economy of the scale and the scope over the selected time period ($t=t_2-t_1$). GDP is equal to a sum of the economic outputs values of the numerous operating industries in the real and imaginary economic sectors in the economy of the scale and the scope over the selected time period ($t=t_2-t_1$).

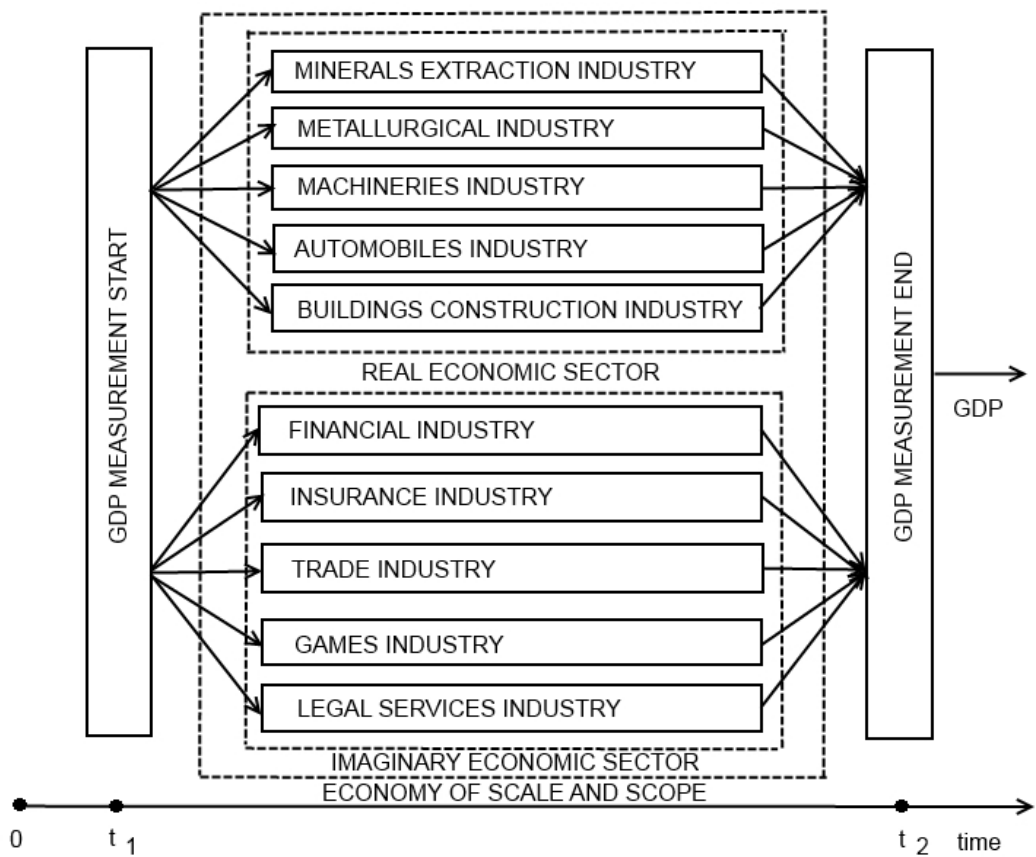

**Fig. 5.** Detailed scheme of Gross Domestic Product (GDP) measurement in economy of scale and scope over selected time period ($t=t_2-t_1$). GDP is equal to sum of economic outputs values of numerous industries in real and imaginary economic sectors in economy of scale and scope over selected time period ($t=t_2-t_1$).

At this point in our research discussion, let us write a set of the formulas for the GDP(t) magnitude measurement in the economy of the scale and the scope over the selected time period in the three cases of possible measurement approaches in Kuznets (1934), Coyle (2014):

1. Production measurement approach:

    GDP(t) = Gross Value of Domestic Economic Output – Intermediate Production Expenses value;

2. Income measurement approach:

    GDP(t) = Gross Value of Domestic Economic Income + Income Taxes Value – Production Subsidies Value;

3. Expenditure measurement approach:

    GDP(t) = Gross Value of Domestic Economic Private/Public Expenditures – Imports Value.

Let us sum up some expressed ideas on the GDP(t) measurement in the economies of the scales and the scopes in macroeconomics science. Going from the fundamental macroeconomic analysis, we came up to an understanding that the GDP(t) magnitude can be precisely measured at a certain time moment, and it may change over the selected measurement time period.

In fact, we will see that the GDP(t) magnitude may change nonlinearly over the time, depending on the gross values of the domestic economic output/income/private-public expenditures created in the course of the wealth synthesis processes in the economy of the scale and the scope at any selected time frame. Therefore, we realized that it may be quite difficult to perform the precise measurement of the GDP(t) magnitude due to various (non)objective factors such as the very soft definitions of the numerous economic variables, which are used for the GDP(t) magnitude calculation.

In Chapter 2, we will comprehensively discuss the problem on the continuous-time economic output waves in economy of scale and scope in Ledenyov classic econodynamics with particular focus on the dependence of the Ledenyov General Information Product on the time GIP(t), the dependence of the General (Gross) Domestic Product on the time GDP(t), the dependence of the General National Product on the time GNP(t), the dependence of the Purchase Power Parity on the time PPP(t).

# Chapter 2

# Continuous-time economic output waves in economy of scale and scope in classic econodynamics

In the course of evolution of the fundamental and applied economics sciences, the new empirical theories, sophisticated mathematical models and original practical solutions to understand the nature of the complex economic phenomena in the economy of the scale and the scope have been proposed. A quite significant attention by eminent scientists was paid to an accurate characterization of the fundamental and applied properties of the complex economic phenomena in the economy of the scale and the scope over the time as discussed in the academic literature in Joseph Penso de la Vega (1668, 1996), Mortimer (1765), Bagehot (1873, 1897), von Böhm-Bawerk (1884, 1889, 1921), Hirsch (1896), Bachelier (1900), Schumpeter (1906, 1911, 1933, 1939, 1961, 1939, 1947), Slutsky (1910, 1915 1923), von Mises (1912), Hayek (1945), Ellis, Metzler (1949), Friedman (1953), Baumol (1957), Debreu (1959), Dodd (2014).

According to the modern economic theories, the economy of the scale and the scope is uniquely predefined by an existing state of the philosophical thinking in an every society of research interest. Reaching a high level of the philosophical thinking on the scientific problems in the fundamental and applied economics sciences, the researchers become capable to formulate, comprehensively discuss and then practically implement a set of progressive economic policies toward the new limitless economic opportunities development on the way to the social and economical prosperity building in any innovative country in Joseph Penso de la Vega (1668, 1996), Mortimer (1765), Smith (1776, 2008), Menger (1871), Bagehot (1873, 1897), von Böhm-Bawerk (1884, 1889, 1921), Hirsch (1896), Bachelier (1900), Schumpeter (1906, 1911, 1933, 1939, 1961, 1939, 1947), Slutsky (1910, 1915 1923), von Mises (1912), Hayek (1931, 1935, 2008; 1948, 1980), Keynes (1936, 1992), Ellis, Metzler (1949), Friedman (1953), Baumol (1957), Debreu (1959), Krugman, Wells (2005), Stiglitz (2005, 2015).

A present state of philosophical thinking in the modern economics science is predetermined, to a certain degree, by an existing level of research advancements in the scientific fields of the macroeconomics, microeconomics and nanoeconomics, constituting the integral parts of the fundamental and applied economics sciences.

In this book, we mainly intend to focus our research discussion on the macroeconomics by making an innovative research on the fluctuations of the economy output in the form of the oscillating quantity of the produced goods and provided services in the economy of the scale and the scope over the finite time period in Juglar (1862), George (1881, 2009), Kondratieff (1922, 1925, 1926, 1928, 1935, 1984, 2002), Kitchin (1923), Kuznets (1930a, b, 1973a, b), Schumpeter (1939), Burns, Mitchell (1946), Dupriez (1947), Samuelson (1947), Hicks (1950), Akamatsu (March-August 1962), Inada, Uzawa (1972), Bernanke (1979), Marchetti (1980), Kleinknecht (1981), Dickson (1983), Hodrick, Prescott (1980, 1997), Baxter, King (1999), Kim, Nelson (1999), McConnell, Pérez-Quirós (2000), Devezas, Corredine (2001, 2002), Devezas (editor) (2006), Arnord (2002), Stock, Watson (2002), Helfat, Peteraf (2003), Sussmuth (2003), Hirooka (2006), Kleinknecht, Van der Panne (2006), Jourdon (2008), Taniguchi, Bando, Nakayama (2008), Drehmann, Borio, Tsatsaronis (2011), Iyetomi, Nakayama, Yoshikawa, Aoyama, Fujiwara, Ikeda, Souma (2011), Ikeda, Aoyama, Fujiwara, Iyetomi, Ogimoto, Souma, Yoshikawa (2012), Swiss National Bank (2012, 2013), Uechi, Akutsu (2012), Central Banking Newsdesk (2013), Ledenyov D O, Ledenyov V O (2013c, 2015d, 2015e, 2016r), Union Bank of Switzerland (2013), Wikipedia (2015a, b, c).

It would be interesting to explain that a present state of philosophical thinking in the modern economics science was shaped, to some degree, by the important research findings in the fluid mechanics science (the fluid dynamics), namely: the theoretical concept on the continuous-time economic output waves in the classic macroeconomics in Juglar (1862) was probably derived in analogy with the existing theoretical concepts on the Newtonian / non Newtonian fluids as well as the Newton's laws of motion in the fluid dynamics, including both: *1)* the continuous-time air – phase waves in the aerodynamics and the continuous-time liquid-phase waves in the hydrodynamics in Newton (5 July 1687, 1999).

At later date, the Maxwell electrodynamics with a number of the new theoretical representations was created Maxwell (1890). We would like to formulate the new Ledenyov classic and quantum econodynamics sciences, based on the Maxwell electrodynamics mainly. So, let us draw a chart with various known types of the electromagnetic signals such as *1)* the continuous-time analog signals, 2) the discrete-time digital signals, *3)* the discrete-time quantum signals, in the electrodynamics, highlighting a fact that all the early discovered economic output waves belong to a class of the continuous-time economic output waves in the classic macroeconomics.

Fig. 6 shows the electromagnetic signals, including *1)* the continuous-time analog signals, *2)* the discrete-time digital signals, *3)* the discrete-time quantum signals and their possible technical applications such as the analog / discrete / quantum signals generation / processing / computing.

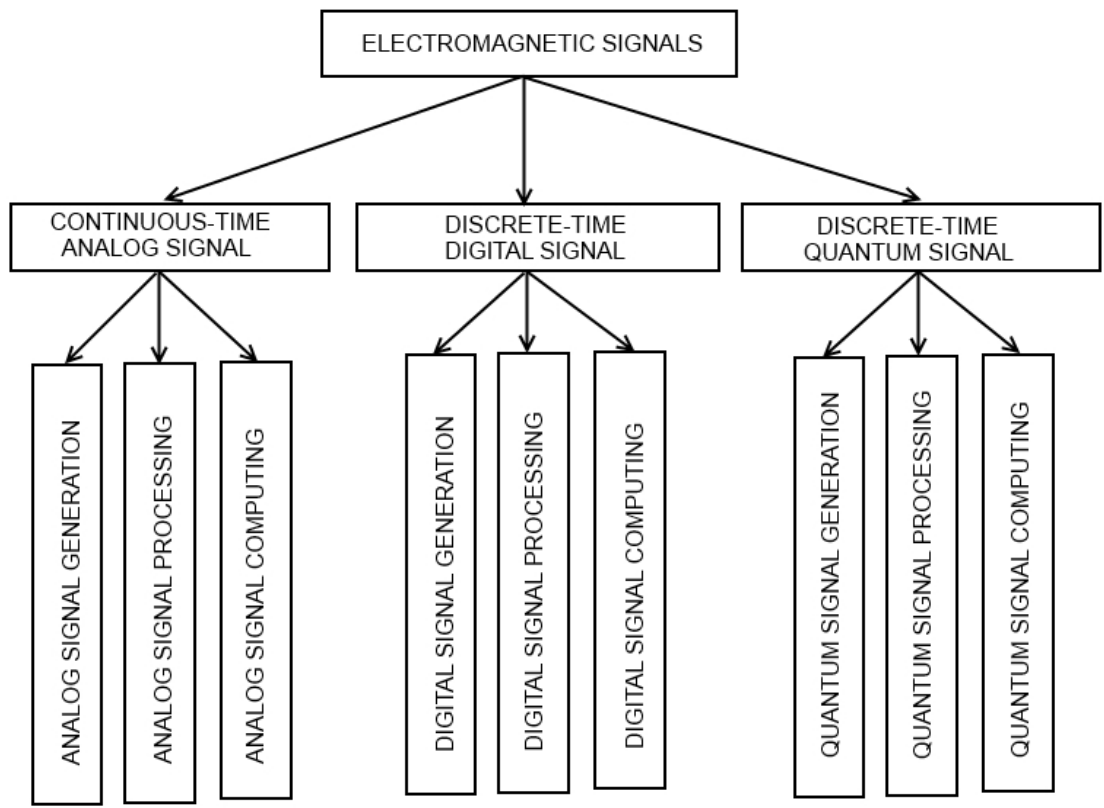

**Fig. 6.** Electromagnetic signals in electrodynamics: *1)* continuous-time analog signals, *2)* discrete-time digital signals, *3)* discrete-time quantum signals, and their possible technical applications.

Now, let us make the continuous-time analog electromagnetic signal(s) as a central topic of our ongoing research discussion, particularly focusing on the various aspects of the continuous-time analog electromagnetic signal(s) processing and filtering with an application of the low/band/high pass/reject electromagnetic signal filters in the Ledenyov classic econodynamics science.

Fig. 7 shows the continuous-time analog electromagnetic signal, which is normally regarded as the continuous wave (CW): *1)* The CW dependence of the amplitude on the time in an ideal case without the harmonics generation; *2)* The CW dependence of the amplitude on the frequency in an ideal case without the harmonics generation; *3)* The CW dependence of the amplitude on the frequency in a real case scenario with the signal's harmonics generation.

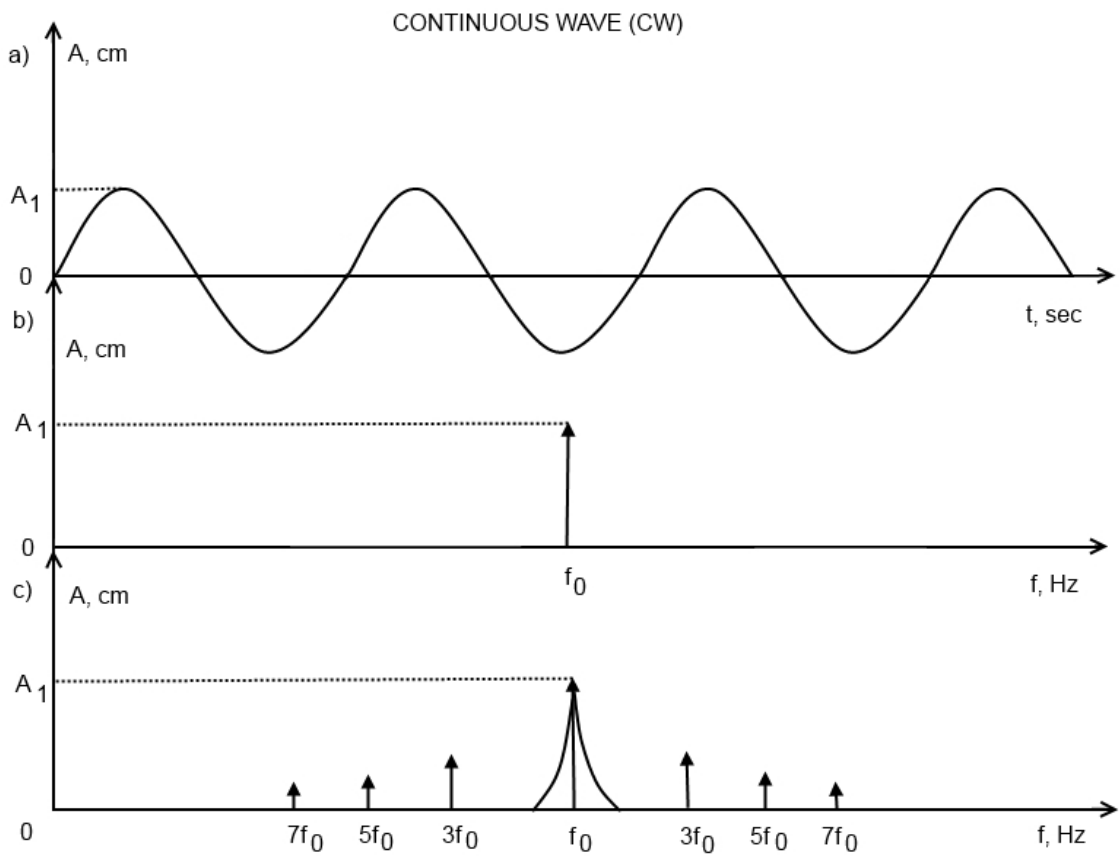

**Fig. 7.** Continuous-time analog electromagnetic signal (Continuous Wave (CW)): *1)* CW dependence of amplitude on time in ideal case without harmonics generation; *2)* CW dependence of amplitude on frequency in ideal case without harmonics generation; *3)* CW dependence of amplitude on frequency in real case scenario with harmonics generation.

Let us describe the main technical parameters of the electromagnetic signal filter, which can be used to filter out the continuous-time analog electromagnetic signal(s) in the frequency domain over the time.

Fig. 8 displays graphically the main technical parameters of the electromagnetic signal filter for the filtering of the continuous-time analog electromagnetic signal(s) (the Continuous Wave(s) (CW)). The main technical parameters include: the central frequency, passband, stopband, signal transfer function, amplitude/frequency/shape of ripples in passband, flatness of passband, amplitude/frequency/shape of ripples in stopband, flatness of stopband, curvature of slopes.

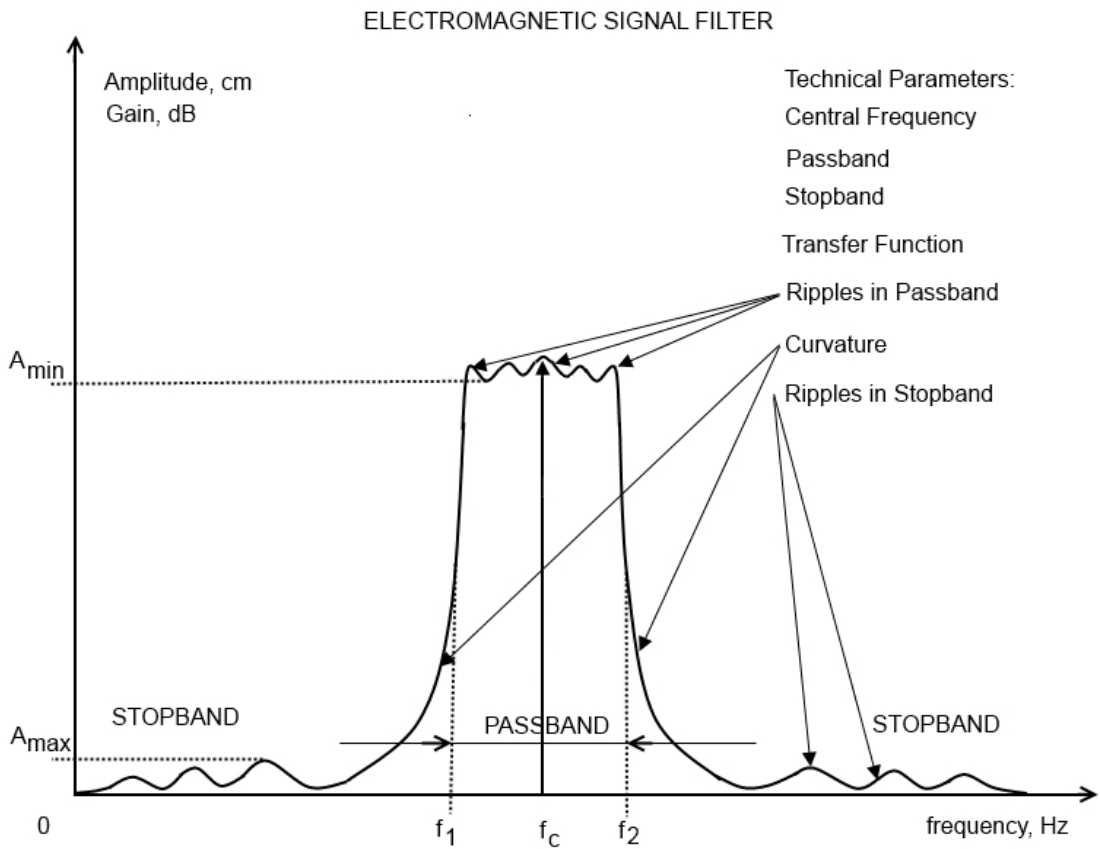

**Fig. 8.** Electromagnetic signal filter in application to continuous-time analog electromagnetic signal (Continuous Wave (CW)). Main technical parameters: *1)* central frequency, *2)* passband, *3)* stopband, *4)* signal transfer function, *5)* amplitude/frequency/shape of ripples in passband, *6)* flatness of passband, *7)* amplitude/frequency/shape of ripples in stopband, *8)* flatness of stopband, *9)* curvature of slopes.

We would like to discuss the lowpass electromagnetic signal filters, because all other types of the electromagnetic signal filters, including the highpass, bandpass, stopband electromagnetic signal filters can be designed by using the lowpass electromagnetic signal filter as explained in Wanhammar (February 24 1999, June 2 2009).

Fig. 9 displays the analog lowpass RF electromagnetic signal filters transfer functions approximations, including: the Butterworth lowpass filter in Butterworth (1930); Chebyshev Type I lowpass filter in Chebyshev (1854); Chebyshev Type II lowpass filter in Chebyshev (1854); Cauer elliptic lowpass filter Cauer (1927); Bessel lowpass filter in Bessel (1832).

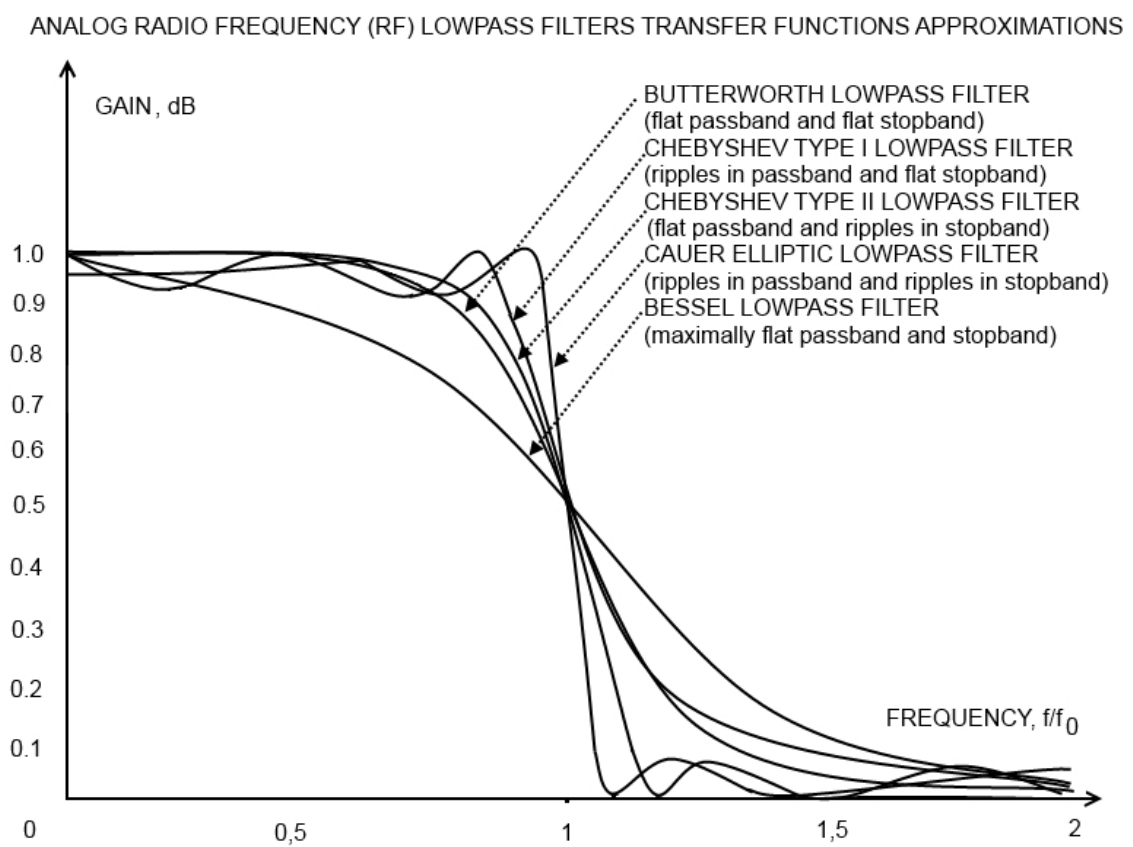

**Fig. 9.** Analog lowpass RF electromagnetic signal filters transfer functions approximations: *1)* Butterworth lowpass filter in Butterworth (1930); *2)* Chebyshev Type I lowpass filter in Chebyshev (1854); *3)* Chebyshev Type II lowpass filter in Chebyshev (1854); *4)* Cauer elliptic lowpass filter Cauer (1927); *5)* Bessel lowpass filter in Bessel (1832).

Considering the classic macroeconomics research findings from historical perspective, we can say that a number of the continuous-time cyclic oscillations of the economic output variables with the different amplitudes, frequencies and phases were discovered. Most interestingly, all the continuous time waves with the trends can be filtered out by the differential data filters in the classic macroeconomics similarly to the analog electromagnetic signal filters in the Maxwell electrodynamics, and analyzed with the mathematical statistical analysis techniques in Pollock (November 2007), Ledenyov D O, Ledenyov V O (2013c).

Juglar (1862) discovered the 7 –11 years Juglar fixed investment cycle, which is still in the scope of research interest by many scientists in Schumpeter (1939), Grinin, Korotayev, Malkov (2010), Korotayev, Tsirel (2010), Ledenyov V O, Ledenyov D O (2012), Ledenyov D O, Ledenyov V O (2013). Here, it makes sense to explain that Schumpeter (1939) showed that there are the four stages in the Juglar cycle: 1) the expansion; 2) the crisis; 3) the recession; 4) the recovery.

Kitchin (1923) proposed that there is the 3 – 7 years Kitchin inventory cycle. This proposition was investigated in Schumpeter (1939), Korotayev, Tsirel (2010), Ledenyov V O, Ledenyov D O (2012), Ledenyov D O, Ledenyov V O (2013).

Kondratieff (1922, 1925, 1926, 1928, 1935, 1984, 2002) made a significant contribution to the science of economics. The Kondratieff's early research was focused on the big cycles of conjuncture in the World economy in Kondratieff (1922, 1925, 1926, 1928). The discovery of the 45 – 60 years Kondratieff long wave cycle in Kondratieff, Stolper (1935) had a considerable impact on the science of economics. The Kondratieff's research achievements are comprehensively analyzed in Kondratieff (1984, 2002). Since that time, the Kondratieff long wave cycle has been a subject of intensive research by many scientists in Schumpeter (1939), Garvy (1943), Silberling (1943), Rostow (1975), Kuczynski (1978, 1982), Forrester (1978, 1981, 1985), Barr (1979), Van Duijn (1979, 1981, 1983), Eklund (1980), Mandel (1980), Van der Zwan (1980), Tinbergen (1981), Van Ewijk (1982), Cleary, Hobbs (1983), Glismann, Rodemer, Wolter (1983), Wallerstein (1984), Bieshaar, Kleinknecht (1984), Zarnowitz (1985), Summers (1986), Freeman (1987), Goldstein (1988), Solomou (1989), Berry (1991), Tylecote (1992), Metz (1992, 1998, 2006), Cooley (1995), Freeman, Louçã (2001),

Modelski (2001, 2006), Perez (2002), Rennstich (2002), Rumyantseva (2003), Diebolt, Doliger (2006), Linstone (2006), Thompson (2007), Papenhausen (2008), Taniguchi, Bando, Nakayama (2008), Korotayev, Tsirel (2010), Ikeda, Aoyama, Fujiwara, Iyetomi, Ogimoto, Souma, Yoshikawa (2012), Ledenyov V O, Ledenyov D O (2012), Ledenyov D O, Ledenyov V O (2013).

Akamatsu conducted his advanced research and formulated idea on a new possible origin of the 45 – 60 years Kondratieff long wave cycle, using the representations from Asian cultural-scientific heritage in Akamatsu (March-August 1962).

Kuznets (1930a, b, 1973a, b) introduced the 15 – 25 years Kuznets infrastructural investment cycle in Kuznets (1973), based on his research on the cyclical fluctuations of the production and prices in Kuznets (1930). The researches on the nature of the Kuznets cycles were conducted by Abramovitz (1961), Rostow (1975), Solomou (1989); Diebolt, Doliger (2006, 2008), Korotayev, Tsirel (2010), Ledenyov V O, Ledenyov D O (2012), Ledenyov D O, Ledenyov V O (2013). Most recently, Korotayev, Tsirel (2010) conducted the spectral analysis and proposed that there is a tight connection between the Kondratieff long wave cycle and the Kuznets infrastructural investment cycle, suggesting that the Kuznets swings represent a third frequency harmonic of the main frequency oscillation, which is generated by the Kondratieff long wave cycle, hence the Kuznets cycle is not an independent oscillation in Korotayev, Tsirel (2010).

Let us present graphically the following continuous-time economic output business cycles in the economy of the scale and the scope in the classical macroeconomics.

1. *3 – 7 years Kitchin continuous-time inventory cycle in Kitchin (1923);*
2. *7 –11 years Juglar continuous-time fixed investment cycle in Juglar (1862);*
3. *15 – 25 years Kuznets continuous-time infrastructural investment cycle in Kuznets (1930a, b, 1973a, b);*
4. *45 – 60 years Kondratieff continuous-time long wave cycle in Kondratieff, Stolper (1935);*

**5.** 45 – 60 years Akamatsu continuous-time long wave cycle in Akamatsu (March-August 1962);

**6.** 70+ years Grand continuous-time super-cycle.

Let us draw a graphical representation of the 3 – 7 years Kitchin, 7 –11 years Juglar, 15 – 25 years Kuznets, 45 – 60 years Kondratieff, 45 – 60 years Akamatsu, 70+ years Grand continuous-time economic output business cycles in economy of scale and scope in the classical macroeconomics.

Fig. 10 shows the main parameters of *1)* 3 – 7 years Kitchin, *2)* 7 –11 years Juglar, *3)* 15 – 25 years Kuznets, *4)* 45 – 60 years Kondratieff, and 45 – 60 years Akamatsu, *5)* 70+ years Grand continuous-time economic output business cycles in the economy of the scale and the scope in the time domain in Kitchin (1923), Juglar (1862), Kuznets (1930a, b, 1973a, b), Kondratieff, Stolper (1935), Akamatsu (March-August 1962).

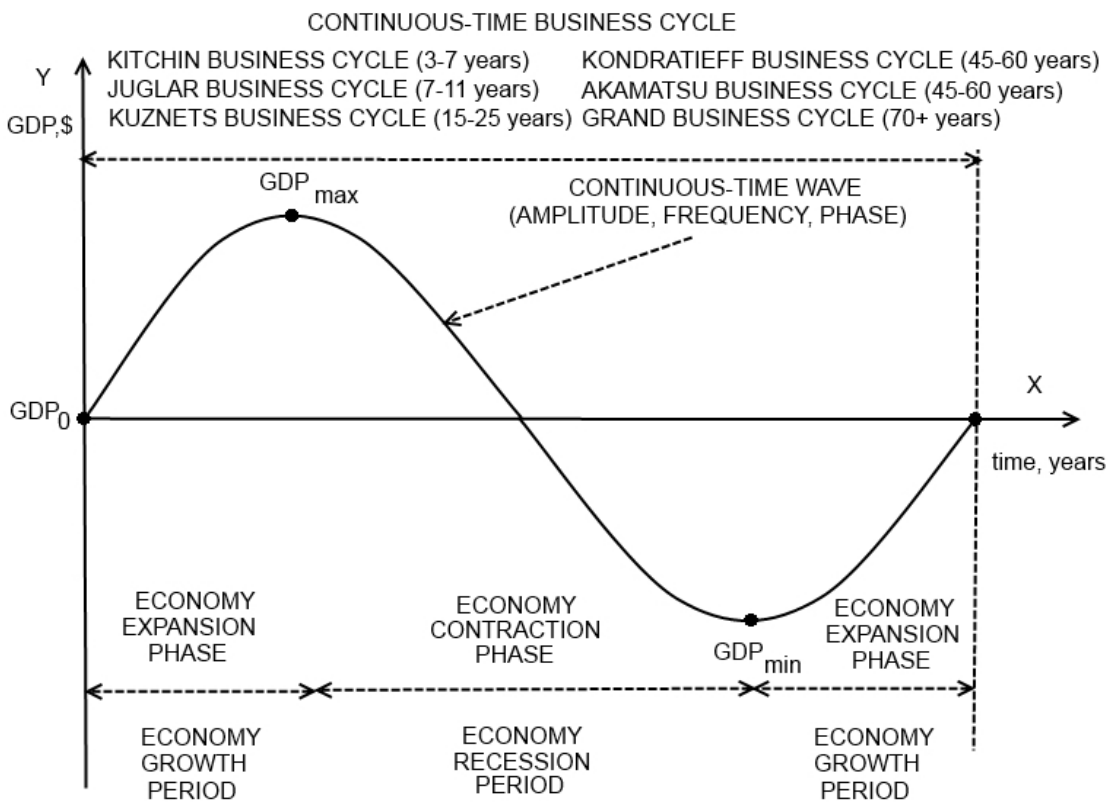

**Fig. 10.** Parameters of continuous-time business cycles in economy of scale and scope: *1)* 3 – 7 years Kitchin in Kitchin (1923), *2)* 7 –11 years Juglar in Juglar (1862), *3)* 15 – 25 years Kuznets in Kuznets (1930a, b, 1973a, b), *4)* 45 – 60 years Kondratieff in Kondratieff, Stolper (1935), and 45 – 60 years Akamatsu in Akamatsu (March-August 1962), *5)* 70+ years Grand.

Many economic theories on the nature of the continuous-time economic output waves were proposed in Ledenyov D O, Ledenyov V O (2013c, 2015d, e, f, g, h, 2016r), Ledenyov V O, Ledenyov D O (2016, 2017):

1. The classical efficient free market theory in Smith (1776, 2008), Say (1803, 1834), Mill (1862);

2. The new classical efficient free market theory with the rational expectation and utility maximization hypotheses in Lucas (1977, 1980);

3. The Marx theory with the hypothesis on the firm's profitability change due to the fluctuation in the firm's capital accumulation/consumption levels in Marx (1867, July 1893, October 1894);

4. The George theory with the land value oscillation hypothesis in George (1881, 2009);

5. The Austrian business cycle theory with the hypothesis on the monetary base (financial credit) expansion/contraction oscillations in von Mises (1912), Hayek (1931, 1935, 2008; 1948, 1980; 2012), Minsky (1974, May 1992, 2015);

6. The Keynes theory with the hypothesis on the dependence of the economic output on the aggregate demand fluctuations (the marginal efficiency of invested capital) in Keynes (1936);

7. The new Keynes theory with the hypothesis that the asymmetry in the information flows between the market agents results in the goods/services prices adjustment delays in conditions of the imperfect market competition, causing the economic output fluctuations in Mankiw (March 1989);

8. The Real Business Cycle (RBC) theory with an assumption on the real exogenous technological shocks in Schumpeter (1939), Lucas (1977, 1980), Mankiw (1989), Kydland, Prescott (Spring 1990);

9. The globalization vs. protectionism theory with a hypothesis on the globalization vs. protectionism state policies implementation oscillation dynamics in Wolf (2004).

10. The goods markets supply and demand fluctuations theory in Inada, Uzawa (1972), Iyetomi, Nakayama, Yoshikawa, Aoyama, Fujiwara, Ikeda, Souma (2011), Ikeda, Aoyama, Fujiwara, Iyetomi, Ogimoto, Souma, Yoshikawa (2012).

Fig. 11 graphically demonstrates the waveforms of *1)* 3 – 7 years Kitchin, *2)* 7 –11 years Juglar, *3)* 15 – 25 years Kuznets, *4)* 45 – 60 years Kondratieff/Akamatsu, *5)* 70+ years Grand continuous-time economic output waves in the economy of the scale and the scope in the classical macroeconomics in Kitchin (1923), Juglar (1862), Kuznets (1930a, b, 1973a, b), Kondratieff, Stolper (1935), Akamatsu (March-August 1962).

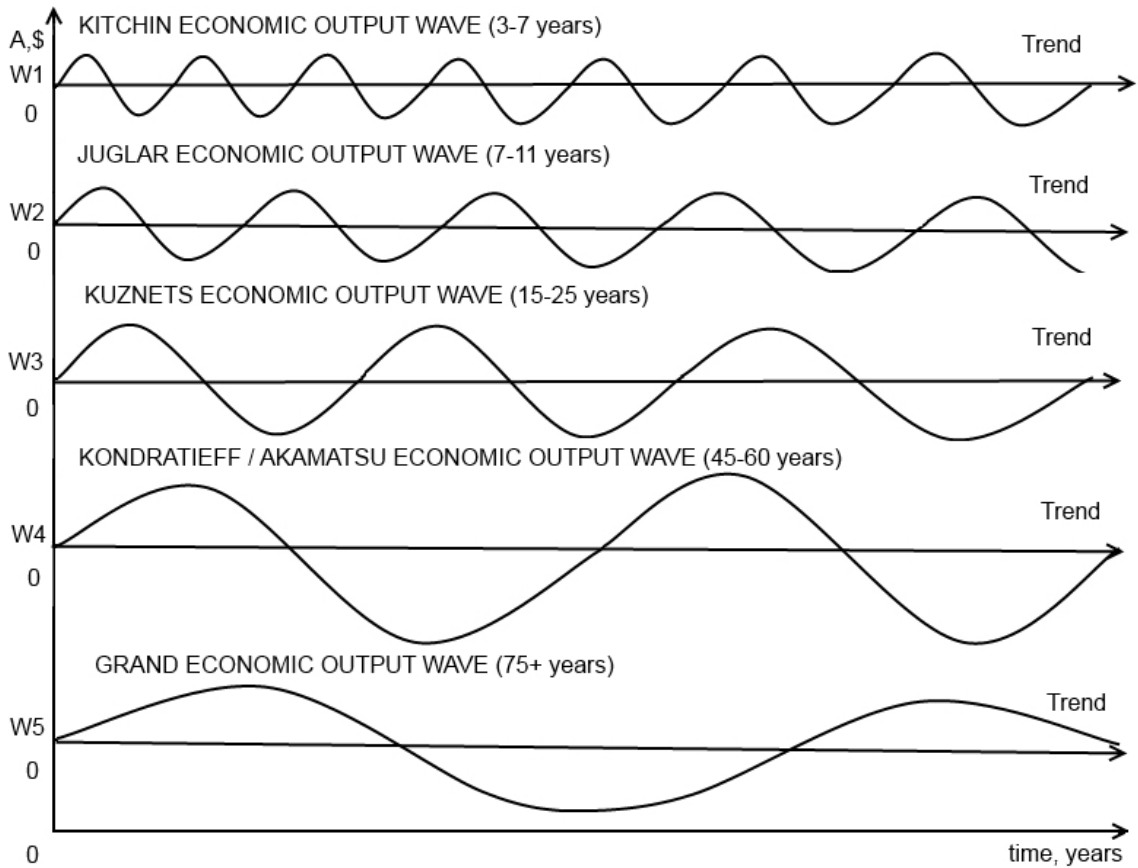

**Fig. 11.** Waveforms of continuous-time business cycles in economy of scale and scope: *1)* 3 – 7 years Kitchin in Kitchin (1923), *2)* 7 –11 years Juglar in Juglar (1862), *3)* 15 – 25 years Kuznets in Kuznets (1930a, b, 1973a, b), *4)* 45 – 60 years Kondratieff in Kondratieff, Stolper (1935), and 45 – 60 years Akamatsu in Akamatsu (March-August 1962), *5)* 70+ years Grand.

As we can see, there is one big drawback with the above representation of GDP(t), namely an uncontrolled expansion of the monetary base by the central bank/treasure/government can lead to a so called "incorrect" increase of GDP(t) in the economy of the scale and scope over the time.

Therefore, in the Ledenyov classic econodynamics science, we propose to use the Ledenyov graphic scheme for a precise characterization of the continuous-time economic output waves of GIP(t, monetary base), GDP(t, monetary base), GNP(t, monetary base), PPP(t, monetary base) with the changing amplitude, frequency, period, phase parameters in the economy of the scale and the scope over the time in the XYZ coordinates space; because the economic output can only be precisely measured, knowing the exact increased values of the monetary base in the economy of the scale and the scope at the measurement time moments.

Fig. 12 provides the Ledenyov graphic representation scheme of the continuous-time economic output wave of GIP(t, monetary base), GDP(t, monetary base), GNP(t, monetary base), PPP(t, monetary base) with the changing amplitude, frequency, period, phase parameters in the economy of the scale and the scope at the certain monetary base over the selected time period in the XYZ coordinates space.

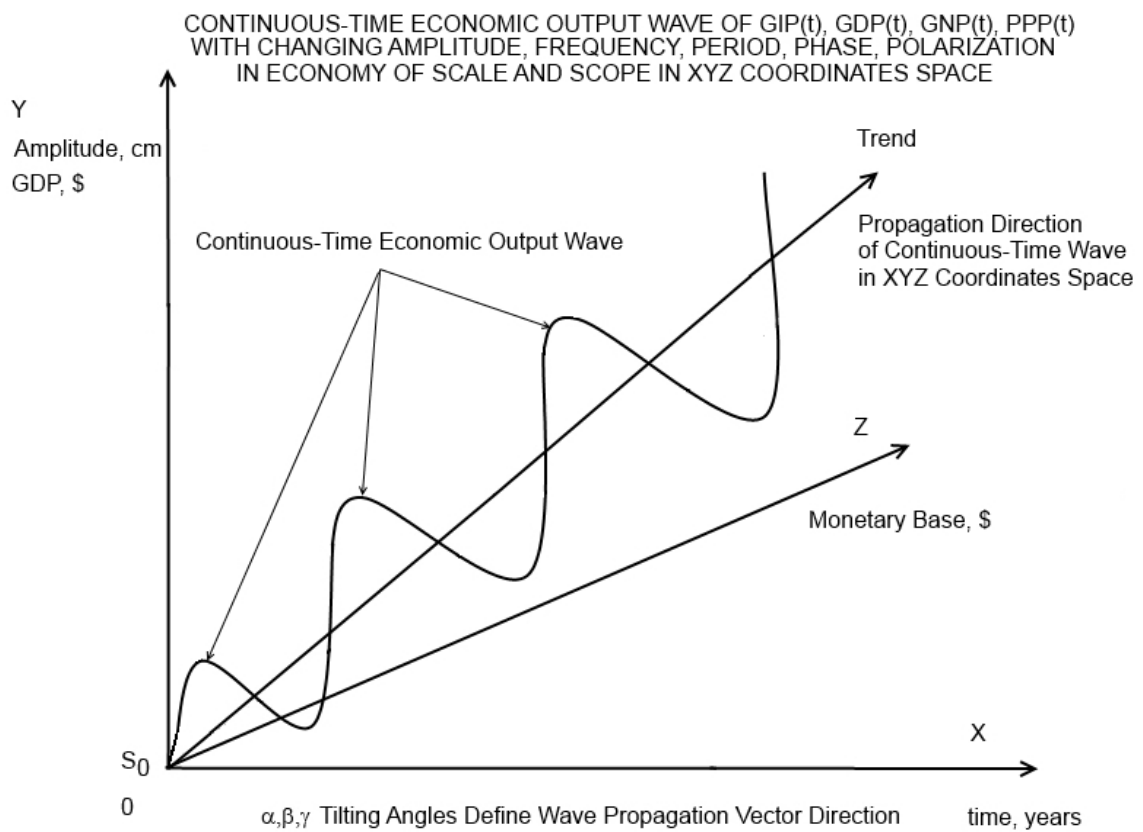

**Fig. 12.** Ledenyov graphic scheme of continuous-time economic output wave of GIP(t, monetary base), GDP(t, monetary base), GNP(t, monetary base), PPP(t, monetary base) with changing amplitude, frequency, period, phase parameters in economy of scale and scope at certain monetary base over selected time period in XYZ coordinates space.

Fig. 13 displays the Ledenyov graphic scheme of *1)* 3 – 7 years Kitchin, *2)* 7 –11 years Juglar, *3)* 15 – 25 years Kuznets, *4)* 45 – 60 years Kondratieff/Akamatsu, *5)* 70+ years Grand continuous-time economic output waves in the economy of the scale and the scope at the certain monetary bases over the selected time periods in the XYZ coordinates space in Kitchin (1923), Juglar (1862), Kuznets (1930a, b, 1973a, b), Kondratieff, Stolper (1935), Akamatsu (March-August 1962). The trends vectors directions and the tilting angles for every considered wave with the different amplitudes, frequencies, periods, and phases in the economy of the scale and the scope at the certain monetary bases over the selected time periods are shown.

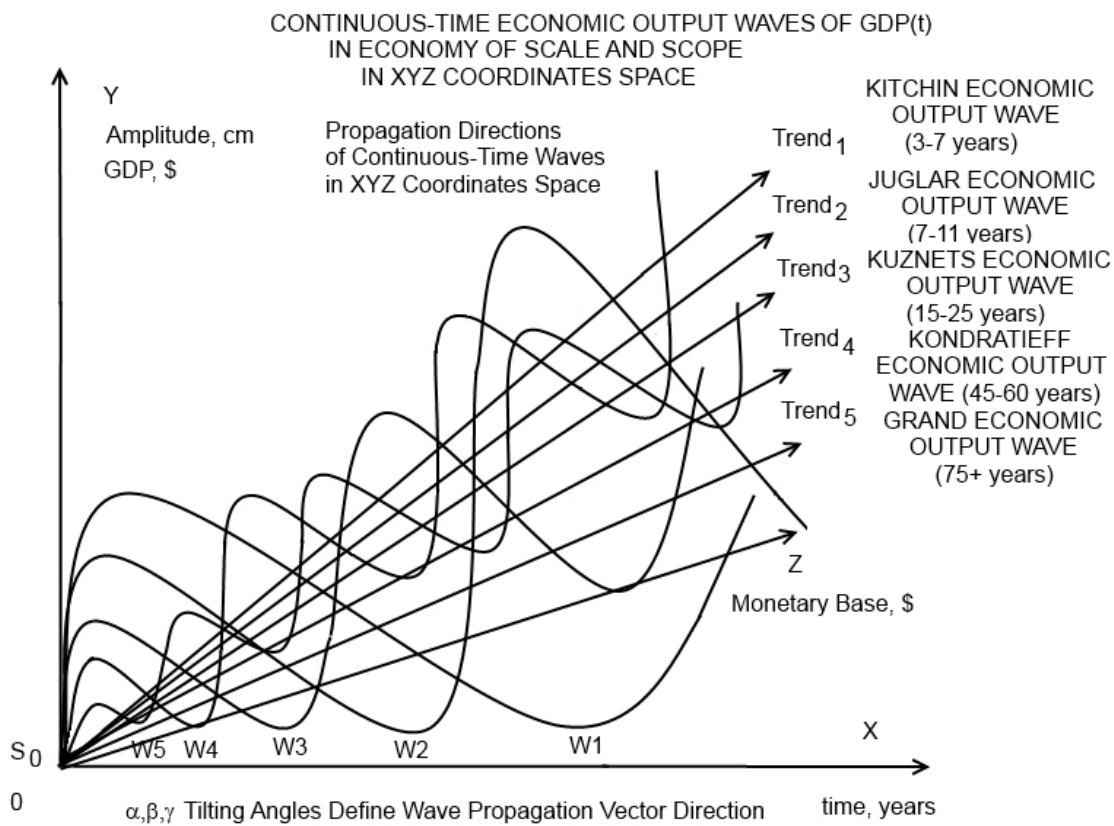

**Fig. 13.** Ledenyov graphic scheme of continuous-time economic output waves in economy of scale and scope at certain monetary base over selected time period in XYZ coordinates space: *1)* 3 – 7 years Kitchin in Kitchin (1923), *2)* 7 –11 years Juglar in Juglar (1862), *3)* 15 – 25 years Kuznets in Kuznets (1930a, b, 1973a, b), *4)* 45 – 60 years Kondratieff in Kondratieff, Stolper (1935), and 45 – 60 years Akamatsu in Akamatsu (March-August 1962), *5)* 70+ years Grand.

In the macroeconomics, the continuous-time economic wave(s) can be filtered out from the statistical data, creating the so called business cycles: the periodic oscillations of the macro-/micro-/nano- economic output variables in the economy of the scale and the scope over the selected time period. These continuous-time economic output waves can be described by the certain mathematical variables: the amplitude, frequency, period, phase, and polarization, changing continuously in the time scale.

Once again, the General Domestic Product GDP(t) was introduced in Kuznets (1930a, b, 1973a, b), and it is frequently viewed as a continuous-time fluctuation of the economy output in the form of the oscillating quantity of the produced goods and provided services in the economy of the scale and the scope over the specified time period in Kuznets (1930a, b, 1973a, b).

Presently, in the Ledenyov classic econodynamics science, we can say that it better to write the continuous-time economic waves of the GIP(t, monetary base), GDP(t, monetary base), GNP(t, monetary base), PPP(t, monetary base) oscillations in the economy of the scale and the scope over the selected time period.

Up to this date, a number of the different sophisticated research techniques to analyse the economic time series were proposed, aiming to detect, filter, precisely measure the continuous-time economic output waves in the economy of the scale and scope over the time. The statistical data analysis methods in the macroeconomics are based on the statistical data analysis methods in the natural sciences:

| | |
|---|---|
| *1.* Statistical mathematics; | *4.* Electrodynamics physics; |
| *2.* Statistical physics; | *5.* Analog signals processing; |
| *3.* Hydrodynamics physics; | *6.* Differential/Integral Equations. |

**Tab.1.** Natural sciences with statistical data analysis methods.

The appropriate modifications to the statistical data analysis methods are usually made, having the final goal to better analyze and accurately characterize the economic output time series in the macroeconomics.

Let us demonstrate the spectrum analysis of the oscillating economic output variables with the detected *1)* 3 – 7 years Kitchin, *2)* 7 –11 years Juglar, *3)* 15 – 25 years Kuznets, *4)* 45 – 60 years Kondratieff/Akamatsu, *5)* 70+ years Grand continuous-time economic output waves in the economy of the scale and scope over the time in Kitchin (1923), Juglar (1862), Kuznets (1930a, b, 1973a, b), Kondratieff, Stolper (1935), Akamatsu (March-August 1962).

Fig. 14 shows the spectrum analysis of *1)* 3 – 7 years Kitchin, *2)* 7 –11 years Juglar, *3)* 15 – 25 years Kuznets, *4)* 45 – 60 years Kondratieff/Akamatsu, *5)* 70+ years Grand continuous-time economic output waves in the economy of the scale and the scope at the certain monetary base over the selected time period in Kitchin (1923), Juglar (1862), Kuznets (1930a, b, 1973a, b), Kondratieff, Stolper (1935), Akamatsu (March-August 1962).

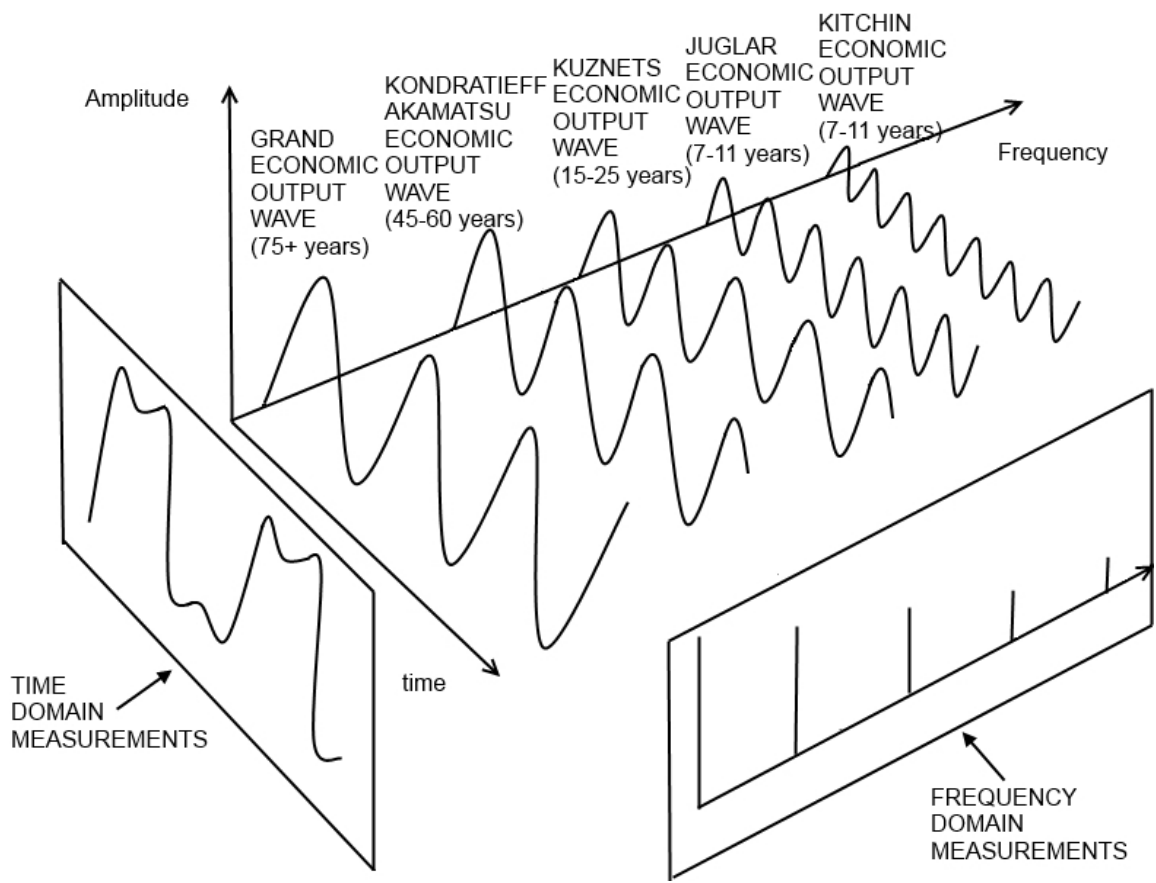

**Fig. 14.** Spectrum analysis of continuous-time economic output waves in economy of scale and scope at certain monetary base over selected time period in XYZ coordinates space: *1)* 3 – 7 years Kitchin in Kitchin (1923), *2)* 7 –11 years Juglar in Juglar (1862), *3)* 15 – 25 years Kuznets in Kuznets (1930a, b, 1973a, b), *4)* 45 – 60 years Kondratieff in Kondratieff, Stolper (1935), *5)* 45 – 60 years Akamatsu in Akamatsu (March-August 1962), *6)* 70+ years Grand.

In the modern macroeconomics, a full list of most frequently applied econometric methods with the statistical time-series estimation to analyse the economic output data in the selected economy of the scale and the scope over the selected time period includes the following econodynamic tests:

1. The unit root test to detect the trends in the economic output time series in Dickey, Fuller (1979), Bhargava (1986), Stock, Watson (1988a, b), Warne (1993), Patterson (2011, 2012);

2. The Wiener–Kolmogorov (W-K) filter to filter out the continuous-time economic output waves/cycles/trends in Tintner (1940), Wiener (1941), Kolmogorov (1941), Pollock (1999, 2000, 2003a,b, 2006, November 2007);

3. The Hodrick-Prescott filter to filter out the continuous-time economic output waves/cycles/trends in Hodrick, Prescott (1980, 1997), Cogley, Nason (1995), Pedregal, Young (2001), Ravn, Uhlig (2002), Harvey, Trimbur (2003), Pollock (1999, 2000, 2003a,b, 2006, November 2007);

4. The Butterworth band-pass filters to filter out the continuous-time economic output waves/cycles/trends in Baxter, King (1999), Gómez (2001), Christiano, Fitzgerald (2003), Valle e Azevedo, Koopman, Rua (2004), Pelagatti (2004) Pollock (1999, 2000, 2003a,b, November 2007);

5. The Bayesian filters to filter out the continuous-time economic output waves/cycles/trends propagation dynamics in Harvey (1989), West, Harrison (1989);

6. The cross correlation/synchronization comparative analysis between the continuous-time economic output waves in Koopman, Valle e Azevedo (2004), Ikeda, Aoyama, Fujiwara, Iyetomi, Ogimoto, Souma, Yoshikawa (2012), Ikeda, Aoyama, Yoshikawa (2013a, b), Ikeda (2013);

7. The Vector Autoregression Model (VAR) to analyse the dynamic behaviour of the continuous-time economic output waves in Sims (1980), Krolzig (1997, 1999), Yao (2001), Asteriou, Hall (2011), Qin Duo (2011);

8. The impulse response function (IRF) techniques to analyse the propagation of the continuous-time economic output waves in the conditions of the economic shocks in Frisch (1933), Koop, Pesaran, Potter (1996), Ehrmann, Elison, Valla (2001, 2003);

9. The Granger causality test to predict the change dynamics of the continuous-time economic output waves in Granger (1969, 1980, 2004), Achilleas (March 2016);

10. The Vector Error Correction Model (VECM) test to detect an existence of the long term interconnections between the economic output time series in Yule (1926), Granger, Newbold (1978), Engle, Granger (1987), Peel, Davidson (1998), Psaradakis, Sola, Spagnolo (2001), Krolzig, Marcellino, Mizon (2002);

11. The coupled oscillator model test to describe the nature of the oscillating dynamics of business cycles in Ikeda, Aoyama, Fujiwara, Iyetomi, Ogimoto, Souma, Yoshikawa (2012), etc.

Presently, the economic output of the economy of the scale and the scope is usually evaluated by its monetary value in the major currency unit(s) (US\$, EU€, UK£) over the selected time period (month, quarters, years) in National Bureau of Economic Research (2018), US Department of Commerce (2018). In some cases, the economic output of the economy of the scale and the scope is measured by its monetary value in the certain currency unit(s) (US\$, EU€, UK£) per a head of population over the selected time period. In other cases, the observed data of $\Delta G(i) = GDP(i) - GPD(i-1)$ over the time, calculated from the GDP per capita, are reported in Taniguchi, Bando, Nakayama (2008). In addition, the economic output of the economy of the scale and the scope can be characterized as the dependence of the grow rate of GDP(t), which is defined as $x_i(t) = \frac{(GDP_i(t) - GDP_i(t-1))}{GDP_i(t-1)}$ in Ikeda, Aoyama, Yoshikawa (2013). The statistical data on the Greece economy of the scale and the scope were analyzed with an application of the above econometric methods in Achilleas (March 2016). Therefore, it is noteworthy to say that the above listed econometric methods are still quite popular among the macroeconomics researchers.

Considering the economy of the scale and the scope, let us highlight an interesting fact that all the interactions between the continuous-time economic output waves with the certain amplitudes, frequencies and phases in the nonlinear dynamic diffusion-type economy of the scale and the scope can be divided into the three main types in the Ledenyov classic and quantum econodynamics science in Ledenyov D O, Ledenyov V O (2013c):

*1.* The linear interactions;

*2.* The non-linear interactions;

*3.* The mixed linear/non-linear interactions.

Fig. 15 shows the linear vs. nonlinear continuous-time economic output signals' waveforms in the nonlinear dynamic diffusion-type economy of the scale and the scope in the amplitude, frequency, phase, time domains over the time.

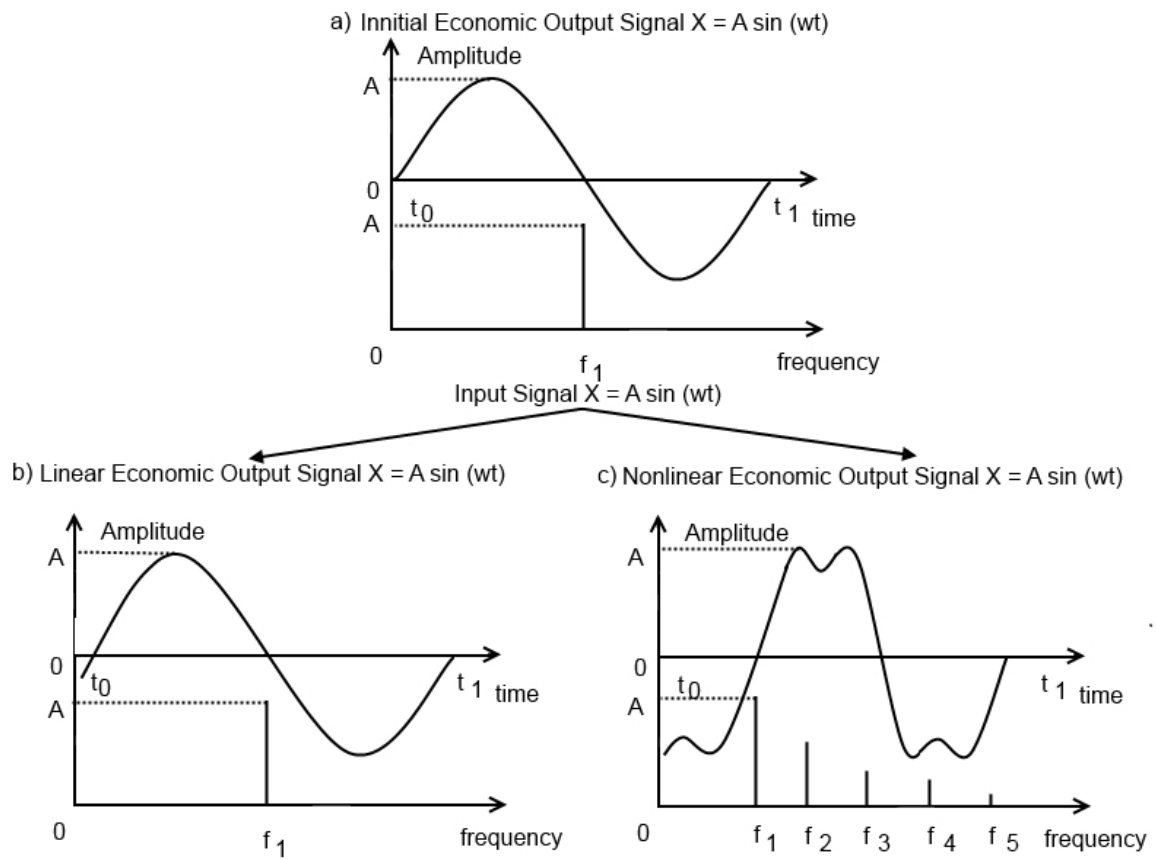

**Fig. 15.** Linear vs. nonlinear continuous-time economic output signals waveforms in nonlinear dynamic diffusion-type economy of scale and scope in amplitude, frequency, phase and time measurement domains over time.

Speaking scientifically, in the physics, we know that the micro-systems can exhibit the physical properties of macro-systems, and the macro-systems can exhibit the physical characteristics of the micro-systems. In application to the economics, it means that, hypothetically, we can assume that an accurate characterization of the economic output waves can be done much more better by drawing parallels between *1)* the economic output waves in the economy of the scale and the scope in Ledenyov classic and quantum econodynamics as well as *2)* the electromagnetic waves in the continuous media in the Maxwell electrodynamics. Therefore, we think that the characteristic properties of the economic output waves in the economy of the scale and the scope over the time can be studied by applying the scientific theories and methods, derived in the research on the electromagnetic waves in the continuous media. Obviously, the proposed scientific approach implies many various benefits, including the effective use of an accumulated scientific knowledge base in the Maxwell electrodynamics science to solve the challenging problem on the precise characterization of the economic output waves in the economy of the scale and the scope in the in Ledenyov classic and quantum econodynamics. Thus, let us draw some comparative parallels between *1)* the Ledenyov classic and quantum econodynamics science and *2)* the Maxwell electrodynamics science, by considering the new theoretical mechanisms on possible realization of a mixing of the continuous-time economic output waves in the economy of the scale and the scope in Ledenyov classic and quantum econodynamics.

Korotayev, Tsirel (2010) made an interesting research proposition that the Kuznets continuous-time economic output wave in Kuznets (1930a, b, 1973a, b) may, in fact, be one of the harmonics, which to be generated in the process of the high power economic output waves mixing in the nonlinear dynamic diffusion-type economic system in the economy of the scale and the scope. We think that this research proposition Korotayev, Tsirel (2010) deserves a focused scientific attention, because the continuous-time economic output signal's harmonics can be generated in the following possible cases:

*1.* An interaction of the one continuous-time economic output waves with the nonlinear dynamic diffusion-type economic system in the economy of the scale and the scope over the time;

*2.* A mixing of the two continuous-time economic output waves in the nonlinear dynamic diffusion-type economic system in the economy of the scale and the scope over the time;

*3.* A mixing of a number of the continuous-time economic output waves in the nonlinear dynamic diffusion-type economic system in the economy of the scale and the scope over the time.

Let us illustrate graphically the possible theoretical mechanisms of the economic output signal's harmonics generation in the nonlinear dynamic diffusion-type economy of the scale and the scope over the time in Ledenyov classic and quantum econodynamics by using an optional analogy with the theoretical mechanisms on the electromagnetic signal's harmonic generation in the nonlinear medium in Maxwell electrodynamics.

Fig. 16 presents a scheme of the Radio Frequency (RF) mixer for the mixing of the two electro-magnetic signals in accordance with the analog signal processing theory in the electrodynamics.

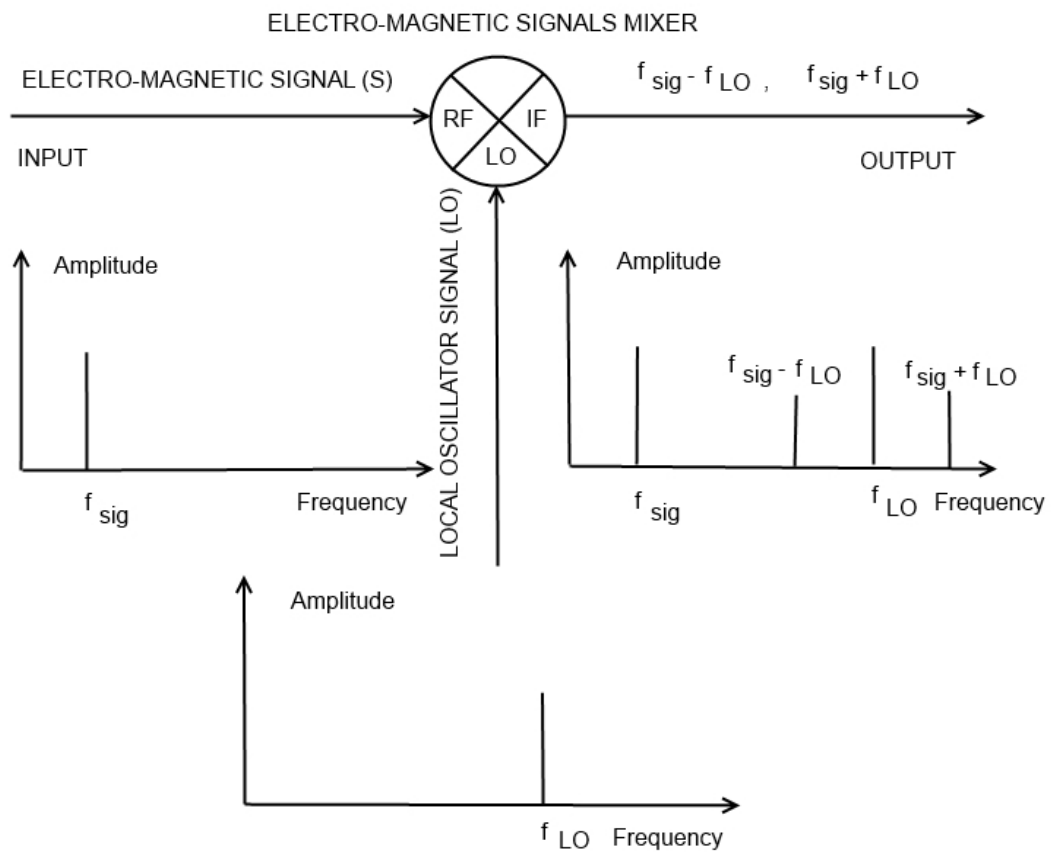

**Fig. 16.** Scheme of Radio Frequency (RF) mixer for mixing of two electro-magnetic signals in accordance with analog signal processing theory in electrodynamics.

Now, let us imagine that the single continuous-time economic output wave interacts with the nonlinear dynamic diffusion-type economic system in the economy of the scale and the scope over the time in the Ledenyov classic and quantum econodynamics similarly to the case, when the one high power electro-magnetic signal interacts with the nonlinear physical medium over the time in the Maxwell electrodynamics. We can expect that the second, third and other high order harmonics can be generated.

Fig. 17 depicts the harmonics generation in the case of the single continuous-time economic output wave interaction with the nonlinear dynamic diffusion-type economic system in the economy of the scale and the scope over time in the Ledenyov classic and quantum econodynamics. It is similar to the single high power electro-magnetic signal interaction with the nonlinear medium over the time in the Maxwell electrodynamics.

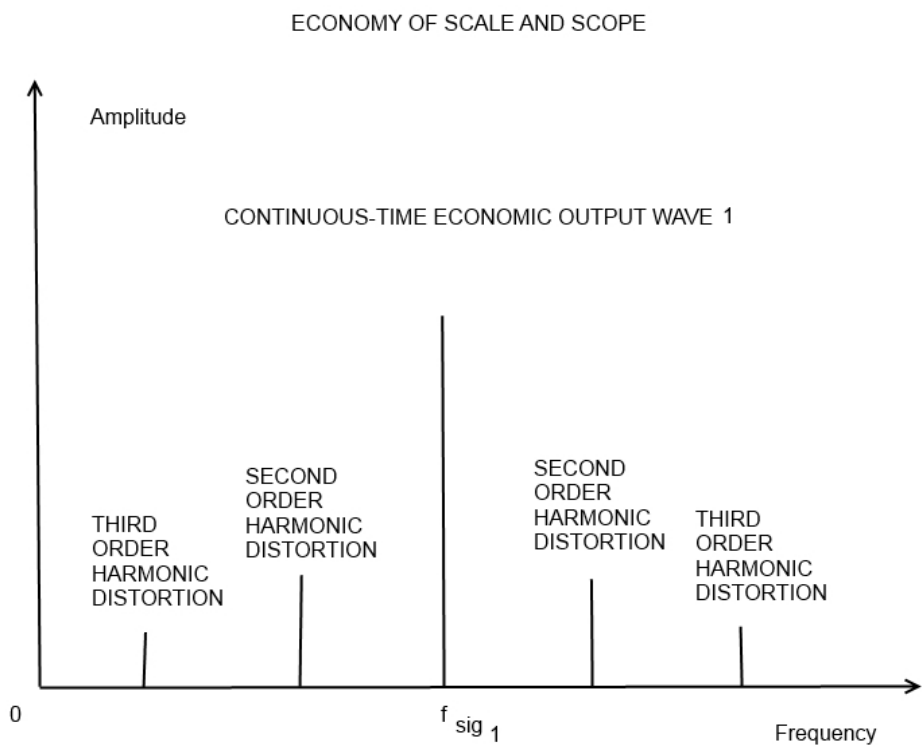

**Fig. 17.** Ledenyov One Wave Interaction (LOWI): Harmonics generation in case of interaction between single continuous-time economic output wave and nonlinear dynamic diffusion-type economy of scale and scope over time in Ledenyov classic and quantum econodynamics. It is similar to single electro-magnetic signal interaction with nonlinear medium over time in Maxwell electrodynamics.

Going to the next case scenario, let us imagine that the two continuous-time economic output waves interact in the nonlinear dynamic diffusion-type economic system in the economy of the scale and the scope over the time in the Ledenyov classic and quantum econodynamics similarly to the case, when the two high power electro-magnetic signals interact in the nonlinear physical medium over the time in the Maxwell electrodynamics. In this case, the second, third and high order harmonics can be generated.

Fig. 18 shows the harmonics generation in the case of two continuous-time economic output waves mixing in the nonlinear dynamic diffusion-type economic system in the economy of the scale and the scope over the time in the Ledenyov classic and quantum econodynamics. It is similar to the numerous electro-magnetic signals mixing in the nonlinear medium over the time in the Maxwell electrodynamics.

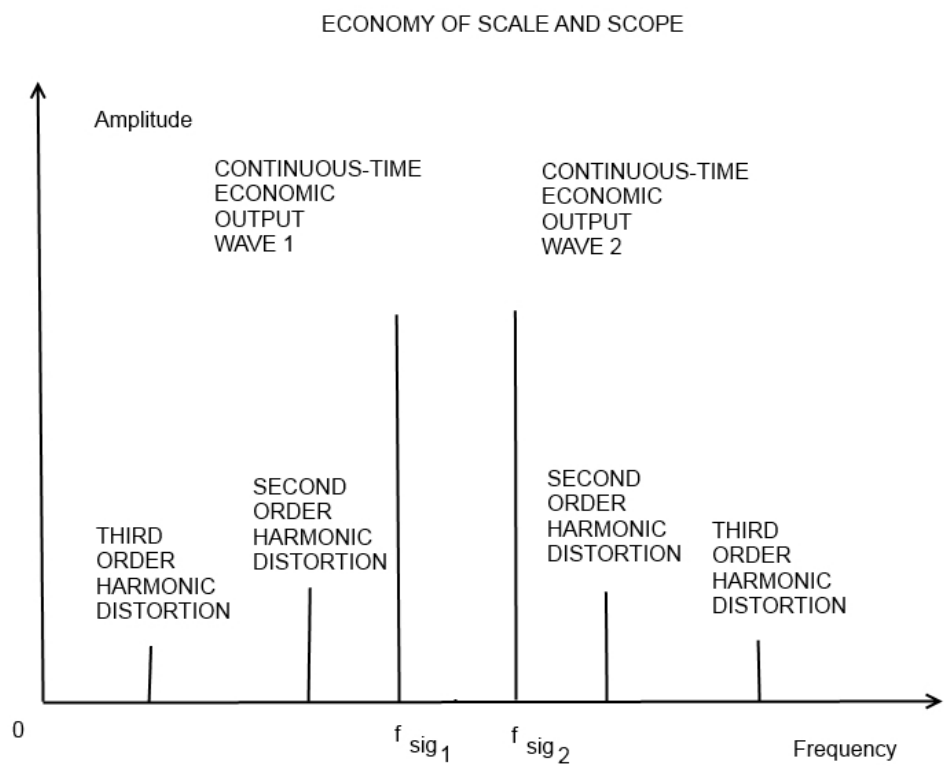

**Fig. 18.** Ledenyov Two Waves Mixing (LTWM): Harmonics generation in case of mixing of two continuous-time economic output waves in nonlinear dynamic diffusion-type economy of scale and scope over time in Ledenyov classic and quantum econodynamics. It is similar to numerous electro-magnetic signals mixing in nonlinear medium over time in Maxwell electrodynamics.

Of course, we cannot limit our research discussion by the only consideration on a possible realization of the harmonics generation effects during a mixing of the continuous-time economic output waves in the nonlinear dynamic diffusion-type economy of the scale and the scope over the time in the Ledenyov classic and quantum econodynamics. There is a big number of the researched nonlinear electromagnetic effects at various frequencies bands in the Maxwell electrodynamics. Therefore, we think that the four extra nonlinear high power effects can originate in the nonlinear dynamic diffusion-type economy of the scale and the scope over the time in the Ledenyov classic and quantum econodynamics in Ledenyov D O, Ledenyov V O (2013c):

*1.* The Ledenyov Four Waves Mixing (LFWM) effect in the nonlinear dynamic diffusion-type economy of the scale and the scope over the time in the econodynamics in analogy with the Four Waves Mixing (FWM) nonlinear high power effect in the nonlinear optical media such as the optical crystal/fiber over the time in the electrodynamics;

*2.* The Stimulated Ledenyov Scattering (SLS1) effect in the nonlinear dynamic diffusion-type economy of the scale and the scope over the time in the econodynamics in analogy with the Stimulated Brillouin Scattering (SBS) nonlinear high power effect in the nonlinear optical media such as the optical crystal/fiber over the time in the electrodynamics;

*3.* The Stimulated Ledenyov Scattering (SLS2) effect in the nonlinear dynamic diffusion-type economy of the scale and the scope over the time in the econodynamics in analogy with the Stimulated Raman Scattering (SRS) nonlinear high power effect in the nonlinear optical media such as the optical crystal/fiber over the time in the electrodynamics;

*4.* The Ledenyov Carrier-Induced Phase Modulation effect (LCIPM) in the nonlinear dynamic diffusion-type economy of the scale and the scope over the time in the econodynamics can exhibit itself and be tracked down in analogy with the Carrier-Induced Phase Modulation nonlinear high power effect in the nonlinear optical media such as the optical crystal/fiber over the time in the electrodynamics.

According to the above thinking approach, we propose that the Ledenyov Four Waves Mixing (LFWM) nonlinear effect in the nonlinear dynamic diffusion-type economy of the scale and the scope over the time in the Ledenyov classic and quantum econodynamics can originate in analogy with the Four Waves Mixing (FWM) nonlinear high power effect in the optical media over the time in the Maxwell electrodynamics, when the optical signals at frequency of $\omega_1$ and $\omega_2$ can generate the two new optical signals at the frequencies of $\omega_3 = 2\omega_1 - \omega_2$ and $\omega_4 = 2\omega_2 - \omega_1$, propagating in the initial signals direction in the optical crystal/fiber over the time in Dutton (1998).

Fig. 19 shows the Ledenyov Four Waves Mixing (LFWM) nonlinear high power effect in the nonlinear dynamic diffusion-type economy of the scale and the scope over the time in the in the Ledenyov classic and quantum econodynamics in analogy with the Four Waves Mixing (FWM) nonlinear high power effect in the nonlinear optical media (the optical crystal/fiber) over the time in the Maxwell electrodynamics.

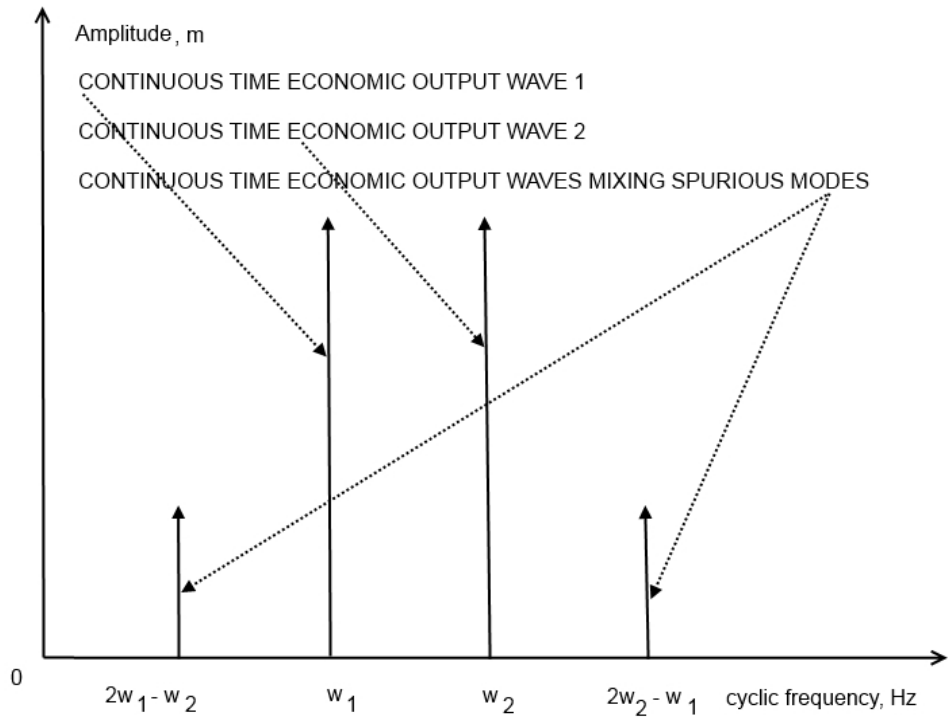

**Fig. 19.** Ledenyov Four Waves Mixing (LFWM) nonlinear high power effect in nonlinear dynamic diffusion-type economy of scale and scope over time in in the Ledenyov classic and quantum econodynamics.

We suggest that the Stimulated Ledenyov Scattering (SLS1) effect in the nonlinear dynamic diffusion-type economy of the scale and the scope over the time in the in the Ledenyov classic and quantum econodynamics can appear in analogy with the Stimulated Brillouin Scattering (SBS) nonlinear high power effect in the nonlinear optical media over the time in the Maxwell electrodynamics, when the optical signal at frequency $\omega_1$ (the photons signal) propagates, experiencing the backward reflection and subsequent attenuation as a result of its interaction with the crystal grating's periodic correlated acoustic mechanical vibrations (the phonons signal, causing the refraction index change) in the optical crystal/fiber. The Stimulated Brillouin Scattering (SBS) was discovered in Brillouin (1922), Mandelstam (1926).

Fig. 20 displays the Stimulated Ledenyov Scattering (SLS1) nonlinear high power effect in the nonlinear dynamic diffusion-type economy of the scale and the scope over the time in the Ledenyov classic and quantum econodynamics in analogy with the Stimulated Brillouin Scattering (SBS) nonlinear high power effect in the nonlinear optical media (the optical crystal/fiber) in the electrodynamics.

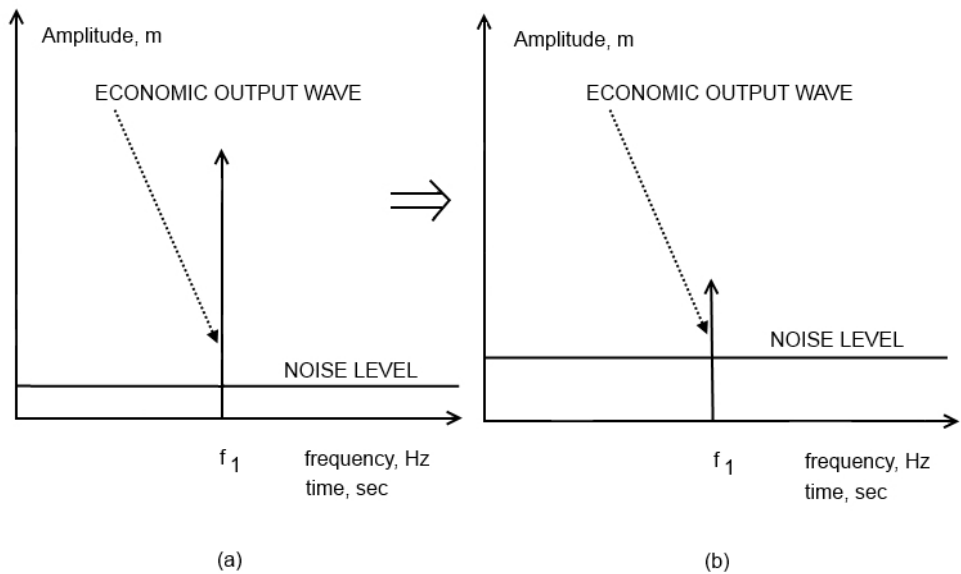

**Fig. 20.** Stimulated Ledenyov Scattering (SLS1) nonlinear high power effect in nonlinear dynamic diffusion-type economy of scale and scope over time in the Ledenyov classic and quantum econodynamics.

We think that the Stimulated Ledenyov Scattering (SLS2) effect in the nonlinear dynamic diffusion-type economy of the scale and the scope over the time in the in the Ledenyov classic and quantum econodynamics can be measured in analogy with the Stimulated Raman Scattering (SRS) nonlinear high power effect in the nonlinear optical media over the time in the Maxwell electrodynamics, when the optical signal at frequency $\omega_1$ (the photons signal) propagates, experiencing the backward reflection and subsequent attenuation as a result of its interaction with the molecules/atoms bonds vibrations/rotations (the infrared signal, causing the refraction index change) in the optical crystal/fiber. The Stimulated Raman Scattering (SRS) was discovered in Raman (1928), Landsberg, Mandelstam (1928).

Fig. 21 pictures the Stimulated Ledenyov Scattering (SLS2) nonlinear high power effect in the nonlinear dynamic diffusion-type economy of the scale and the scope over the time in the Ledenyov classic and quantum econodynamics in analogy with the Stimulated Raman Scattering (SRS) nonlinear high power effect in the nonlinear optical media (the optical crystal/fiber) in the Maxwell electrodynamics.

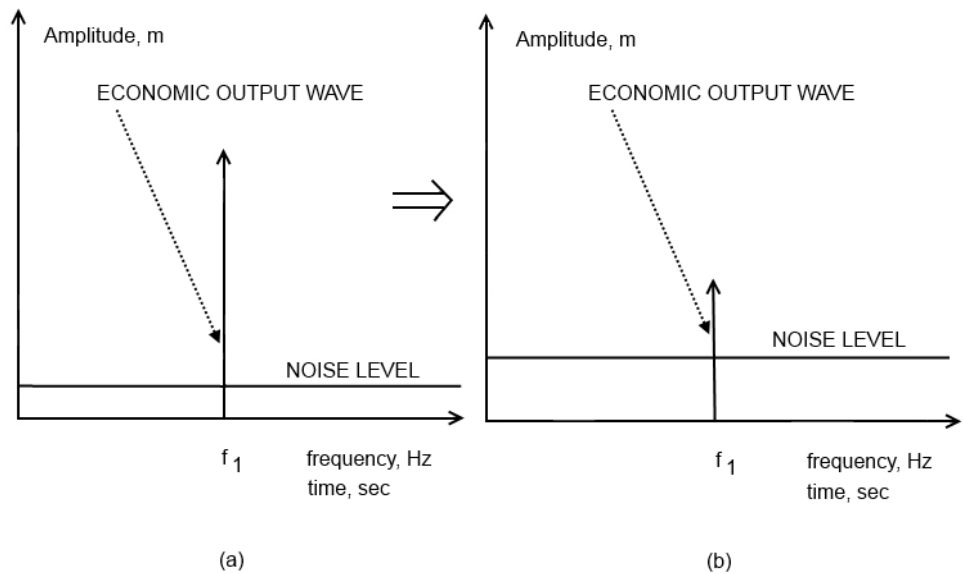

**Fig. 21.** Stimulated Ledenyov Scattering (SLS2) nonlinear high power effect in nonlinear dynamic diffusion-type economy of scale and scope over time in the Ledenyov classic and quantum econodynamics.

We believe that the Ledenyov Carrier-Induced Phase Modulation effect (LCIPM) in the nonlinear dynamic diffusion-type economy of the scale and the scope over the time in the in the Ledenyov classic and quantum econodynamics can exhibit itself in analogy with the Carrier-Induced Phase Modulation (the Self-Phase Modulation and/or the Cross-Phase Modulation) nonlinear high power effect in the nonlinear optical media in the Maxwell electrodynamics, when the optical signal at frequency $\omega_1$ (the photons signal) propagates in the nonlinear optical crystal/fiber, experiencing the optical signal phase widening due to the Kerr effect, i.e. the refraction index change due to the optical signal's electromagnetic field action on the atoms/molecules in Kerr (1875a, b), Stolen, Lin (April 1978).

Fig. 22 demonstrates the Ledenyov Carrier-Induced Phase Modulation effect (LCIPM) in the nonlinear dynamic diffusion-type economy of the scale and the scope over the time in the Ledenyov classic and quantum econodynamics in analogy with the Carrier-Induced Phase Modulation nonlinear high power effect in the nonlinear optical media (the optical crystal/fiber) in the Maxwell electrodynamics.

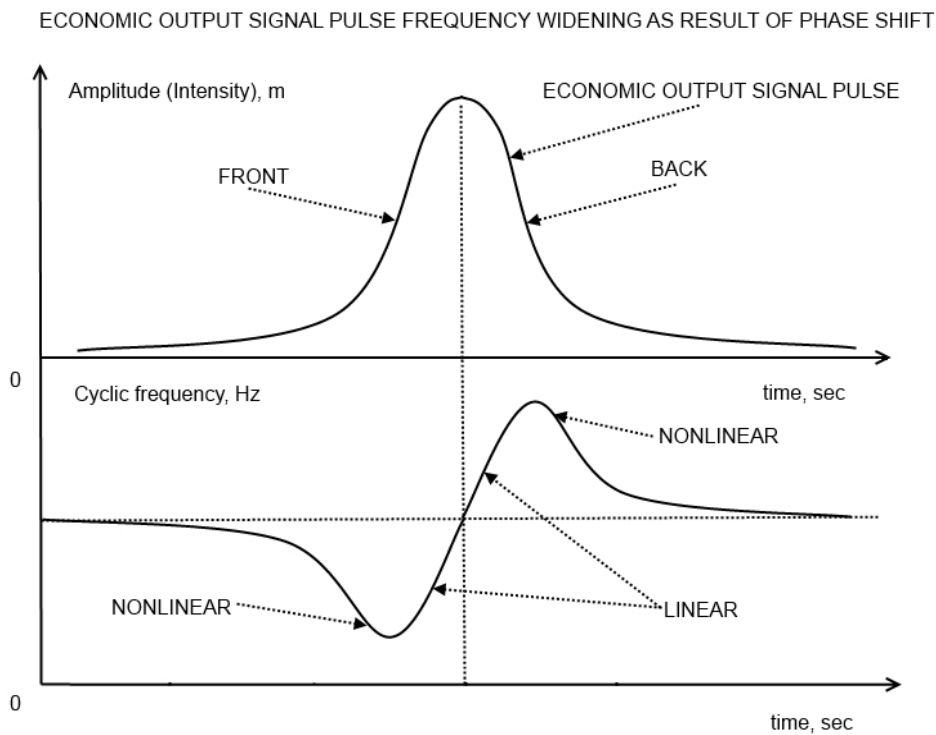

**Fig. 22.** Ledenyov Carrier-Induced Phase Modulation effect (LCIPM) in nonlinear dynamic diffusion-type economy of scale and scope over time in the Ledenyov classic and quantum econodynamics.

Concluding our research discussion on the nonlinearities, we would like to note that the nonlinear effects in the economy of the scale and scope over the time were researched comprehensively in Purica, Caraiani (2009).

Taking our research discussion to a next qualitative level, we would like to emphasis that, in the classic macroeconomics, a considerable scientific attention is paid to an accurate characterization of the continuous-time economic output waves, including the Juglar continuous-time economic output wave in Juglar (1862), Kitchin continuous-time economic output wave in Kitchin (1923), Kondratieff continuous-time economic output wave in Kondratieff (1922, 1925, 1926, 1928, 1935, 1984, 2002), Kuznets continuous-time economic output wave in Kuznets (1930a, b, 1973a, b).

In addition, it makes sense to remind that, in the beginning of $19^{th}$ century, Prof. Joseph Alois Schumpeter started to think on the accurate characterization of the business cycles in the economics at University of Czernowitz in the City of Czernowitz in the State of Ukraine in 1909 – 1911, completing the writing of his well known book on the business cycles in Schumpeter (1939). Since that time, the groundbreaking ideas on an accurate characterization of the cyclic oscillations of the economic output variables have been further discussed in Burns, Mitchell (1946), Hicks (1950), Bernanke (1979), Sussmuth (2003), Devezas (editor) (2006) among others.

Thus, at present time, we can speak in the terms of technical parameters such as the amplitude, frequency and phase during an accurate characterization of the continuous-time economic output waves in the space-time domain in agreement with the theoretical representations in the analog signal processing theory in the electronics and the Maxwell electrodynamics theory in the physics.

Well, in this context, we would like to highlight an interesting observation that the scientific notion of the mathematical field is not used in application to parameters measurement of the continuous-time economic output waves in the space-time domains in the fundamental economics. Here, it is necessary to explain that, for the first time, an abstract meaning on the mathematical field was in the mathematics introduced by the Euclid as described in Ledenyov D O, Ledenyov V O (2015a). In the mathematics, it is a well known fact that the mathematical field can be considered as one of the most important geometrical characteristics of the abstract mathematical space-time domain in Ledenyov D O, Ledenyov V O (2015a).

In the Maxwell electrodynamics within the physics, the electro-magnetic signal can radiate by the physical fields in Maxwell (1890), Ledenyov D O, Ledenyov V O (2015a). Presently, a list of the possible physical fields includes the gravitation field in space physics, the calibrating field in the quantum physics, the information field in the mathematical physics, etc.

Fig. 23 shows the directions of the electric field vector (E) and the magnetic field vector (H) in case of continuous-time electromagnetic wave propagation in medium/vacuum in Maxwell electrodynamics.

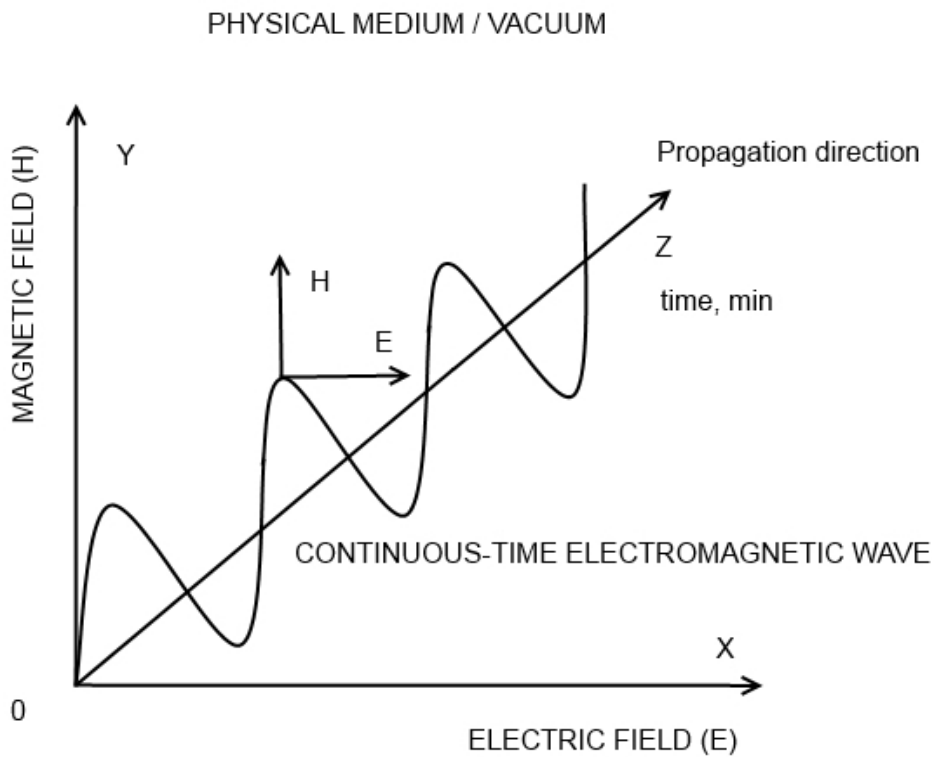

**Fig. 23.** Directions of electric field vector (**E**) and magnetic field vector (**H**) in case of continuous-time electromagnetic wave propagation in medium/vacuum in Maxwell electrodynamics.

Let us formulate the Ledenyov information money fields theory of the continuous-time economic output waves in the nonlinear dynamic diffusion-type economic system in the economy of the scale and the scope.

The Ledenyov information money fields theory of the continuous-time economic output waves in the nonlinear dynamic diffusion-type economic system in the economy of the scale and the scope postulates that the continuous-time economic output waves may have the Ledenyov information money fields, transmitting the financial/ economic information in the nonlinear dynamic diffusion-type economic system in the economy of the scale and the scope.

In other words, we make a theoretical proposition that the continuous-time economic output waves in the nonlinear dynamic diffusion-type economy of the scale and the scope, including:

*1.* The Juglar fixed investment continuous-time economic output wave;

*2.* The Kitchin inventory continuous-time economic output wave;

*3.* The Kondratieff long continuous-time economic output wave;

*4.* The Kuznets infrastructural investment continuous-time economic output wave;

may be characterized by the Ledenyov information money fields.

Speaking clearly, our theoretical proposition on the Ledenyov information money fields in the Ledenyov classic econodynamics is made in an analogy with the case of the electromagnetic continuous-time waves, which can be characterized by the electric and magnetic fields in the electrodynamics theory in the physics in Ledenyov D O, Ledenyov V O (2015a).

In our opinion, the Ledenyov information money fields of the cyclic oscillations of economic variables in the nonlinear dynamic economic system play an important role of the information transmission about a characteristic state of the radiating source of the information money fields in the economy of the scale and the scope.

We think that a econodynamic description of the direction, structure and density distribution of the Ledenyov information money fields in the Ledenyov classic econodynamics can be done in analogy with the physical description of the electromagnetic fields by the Maxwell equations in the Maxwell electromagnetism theory in the electrodynamics in Maxwell (1890), Ledenyov D O, Ledenyov V O (2015a).

Fig. 24 presents the directions of the real and imaginary components of the Ledenyov information money field vector in the case of the continuous-time economic output wave of GIP(t, monetary base), GDP(t, monetary base), GNP(t, monetary base), PPP(t, monetary base) with the changing amplitude, frequency, period, phase parameters in the nonlinear dynamic diffusion-type economy of scale and scope at the certain monetary bases over the selected time period in the XYZ coordinates space in the Ledenyov classic econodynamics.

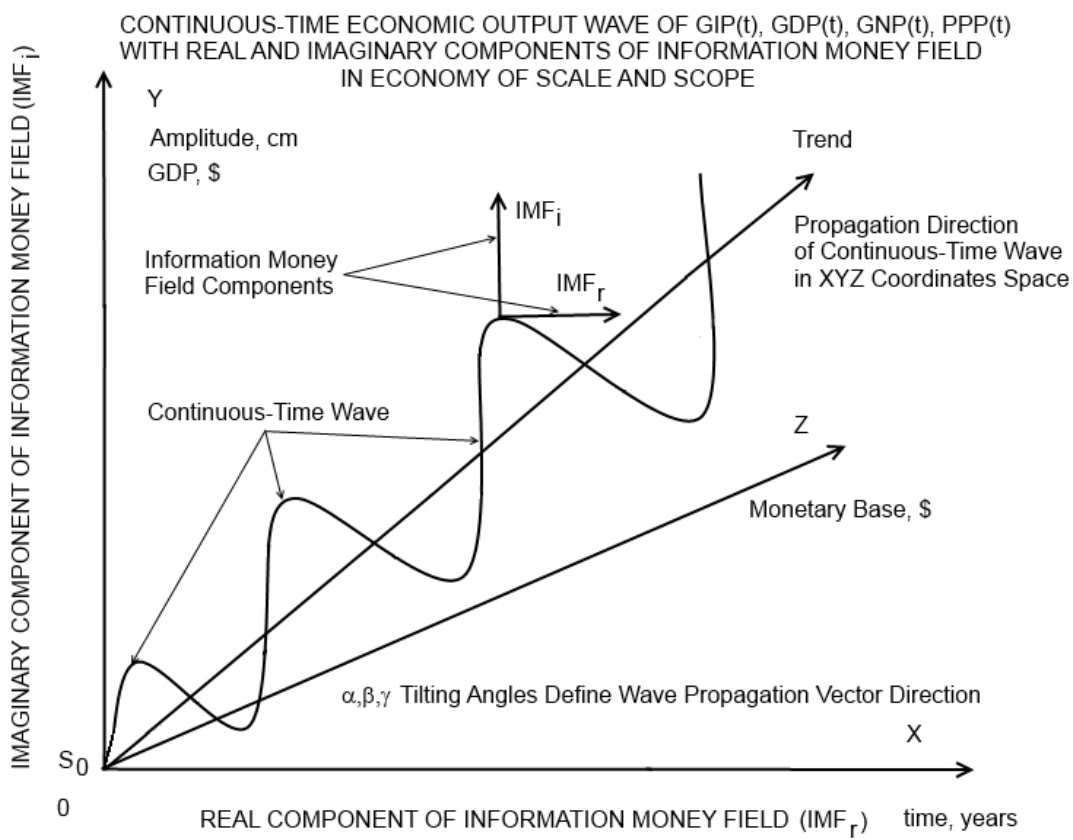

**Fig. 24.** Directions of real and imaginary components of Ledenyov information money field vector in case of continuous-time economic output wave of GIP(t, monetary base), GDP(t, monetary base), GNP(t, monetary base), PPP(t, monetary base) with changing amplitude, frequency, period, phase parameters in nonlinear dynamic diffusion-type economy of scale and scope at certain monetary bases over selected time period in XYZ coordinates space in Ledenyov classic econodynamics.

We assume that the following interactions can have place in the frames of the Ledenyov information money fields theory of the continuous-time economic output waves in the nonlinear dynamic diffusion-type economic system in the economy of the scale and the scope over the time in the Ledenyov classic econodynamics:

1. The Ledenyov information money fields of one continuous-time economic output wave can interact with other information money fields of other continuous-time economic output waves in the nonlinear dynamic diffusion-type economic system in the economy of the scale and the scope in the space-time domain;

2. The Ledenyov information money fields of one continuous-time economic output wave can interact with the nonlinear dynamic diffusion-type economic system in the economy of the scale and the scope by means of the weak and strong interactions with other information money fields in the space-time domain.

3. The magnitude of the Ledenyov information money fields' density distribution in the case of the continuous-time economic output waves in the nonlinear dynamic economic system in the economy of the scale and the scope can change in the space-time domain.

Therefore, in our opinion, the Ledenyov information money fields, which are associated with the Juglar fixed investment continuous-time economic output wave, Kitchin inventory continuous-time economic output wave, Kondratieff long continuous-time economic output wave, Kuznets infrastructural investment continuous-time economic output wave, can interact with each other and/or with the medium such as the nonlinear dynamic diffusion-type economic system in the economy of the scale and the scope, resulting in the change of the magnitude of the Ledenyov information money fields' density distribution in the case of the continuous-time economic output waves in the nonlinear dynamic economic system in the economy of the scale and the scope can change in the space-time domain.

In general, we believe that the Ledenyov information money fields theory of the continuous-time economic output waves in the nonlinear dynamic economic system in the economy of the scale and the scope can be applied in the forecasting process with the application of the complex algorithms at the supercomputers by the central banks with the purpose to make the decisions on the monetary policies, the financial stability policies creation/implementation as well as by other financial institutions/banks/funds with an aim to perform the minimum capital allocation, the countercyclical capital buffer creation, and the capital investments.

To the present day, the nature of the continuous-time economic output waves in the economy of the scale and the scope at the selected observation time periods as well as the scientific methods for their accurate characterization were investigated (in a chronological order) in Ricardo (1817, 1821), Juglar (1862), Benner (1875), George (1881, 2009), Marshall (1890), Pareto (1890), Walras (1898), Wicksell (1898), Parvus (1901, 1999), Marx (1867, July 1893, October 1894), Tugan-Baranovsky (1901, 1923), Spiethoff (1902), Schumpeter (1908; 1911, 1934, 1955; 1917, 1956; 1934; 1936; 1939, 1989; 1945; 1952; 1954; 1961), Aftalion (1909, 1913), Mitchell (1913; 1923, 1951; 1923; 1927; 1928; 1951), Mitchell, King (1923), Burns, Mitchell (1946), Van Gelderen (1913, 1996), Moore (1914, 1967), Persons (1914, 1919a, b, c, d, e, f, g, h, i, j, k, l, 1921, 1922a, b, c, 1924a, b, 1926, 1927, 1928), Persons, Coyle (1921), Kondratieff (1922; 1925, 1984; December 1926, Spring 1979; 1928; 1979; 1984; 1998a, b; 2002), Kondratieff, Stolper (1935, 1996), Polanyi (1922), Kitchin (1923), De Wolff (1924; 1924, 1999), Kuznets (1924, 1929, 1930a, b, c, d, 1934, 1937, 1940, 1941, 1954, March 1955, October 1962, 1963, 1965, 1966, 1968, 1971, 1973a, b), Yule (1926), Ashby (1927), Mitchell (1927; 1941, 1971; 1951), Copeland (1929), Dopsch (1930), Heckscher (1930), Souter (1930), Wagemann (1930), Wilson (May 1930, 22 October 2007), Hayek (1931, 1935, 2008; 1933; 1948, 1980; 1966; 1974; 2012), Funk (1932), Hansen (1932; 1941; 1951a; b, 1991; 1997), Mortara (1932), Fisher (1933), Frisch (1933), Working (1934), Keynes (1919, 1930, 1934, 1936, 1939, 1998), Leontief (1936, 1941, 1973, 1977), Haberler (1937), Slutsky (1937), Varga (1937), Trakhtenberg (1939), Kolmogorov (1941), Metzler (1941), Northrop (1941), Rose (March 1941), Fromm (1942), Garvy (November 1943), Samuelson (1943, 1947), Silberling (1943), Ayres (1944, 1962), Åkerman (1947), Dupriez (1947, September 1978), Williams (16 Apr 1947), Duesenberry (1949), Von Mises (1949, 1962), Dahmén (1950, 1988), Hicks

(1950), Goodwin (1951, 1991), Tintner (1953), Wold (1954, 1955, 1960), Levi-Strauss (1955, 1961), Solow (1956, 1970), Wolfson (1958, 1991), Mendelson (1959-1964), Oldak (1959), Maddison (June 1960, 1981, 1995, 2003), Olivera (1960), Abramovitz (1961, 1986a, March 1986b), Date (1961, 1991), Akamatsu (March-August 1962), Friedman, Schwartz (1963), Friedman (1993), Hickman (1963), Korenjak (1963), Mandel (1964; 1975; 1980, 1995; 1981), Adelman (June 1965), Perroux (1965), Granger (1966, 1969, 1980, 2004), Granger, Newbold (1978), Granger, Teräsvirta (1993), Granger, Inoue, Morin (1997), Schmookler (1966), Easterlin (1968), Eckstein (1968), Frey (1968, 1974, 1976a, b, 1978), Frey, Ramser (1976), Jenkins, Watts (1968), Mansfield (1968, 1983), Rezneck (1968), De Cecco (1969), Hutchings (1969), Matthews (1969), Palmer, Colton (1969), Goodhart, Bhansali (1970), Goodhart (2003), Link (1970), Bajt (1971), Bry, Boschan (1971), Harley (1971), Luce, Krantz, Suppes, Tversky (1971, 1989, 1990), Lundberg (1971), Arrow (1972), Attali (1972), Bernholz (1972, Inada, Uzawa (1972), Meadows (1972), Shuman, Rosenau (1972), Kindleberger (1973, 1978, 1989, 1996), Kindleberger, Laffargue (1979), Namenwirth (1973), Ryder, Heal (1973), Feiwel (1974), Tufte (1974), Ben-Porath (1975), Bloom, Price (1975), Boddy, Crotty (1975), Buchanan (1975), Lindbeck (1975) MacRae (1975), Mass (1975, 1980), Mass, Senge (1981), McClelland (1975), Mensch (1975a, b, 1979; 1981), Mensch et al (1987), Nordhaus (1975), Pruden (1975), Rostow (1975, 1978a, b, 1980), Rostow, Kennedy (1979), Day (1976), Forrester (1976, 1977, 1999; 1978, 1979, 1981a, b, 1982, 1984, 1985), Forrester, Mass, Ryan (July 1976), Forrester, Graham, Senge, Sterman (1983), Gottlieb (1976), Keran (4 June 1976), Minsky (1976, 2008; 1986, 2008), Georgescu-Roegen (1977), Lucas (1977; November 1980, 1981; 1987; 1988; 1990; 1993; 2000; 2002), Anikine, Entov (1978), Glismann, Rodemer, Wolter (Juni 1978), Gordon (1978, 1980, October 1982), Gordon, Weisskopf, Bowles (1983, 1996), Gordon, Edwards, Reich (1994), Kuczynski (1978, 1982), Lewis (1978), Modelski (April 1978, 1981, 1987, 2001, 2006, 2008), Modelski, Johnston, Wu (March 1979), Modelski, Thompson (1988, 1996), Senge (1978, 1980, 1982), Shaikh (1978, 1992), Aglietta (1979, 1997), Aglietta, Rebérioux (27 April 2005), Barr (Spring 1979), Bernanke (1979, 1995, 2000), Bernanke, Lown (1991), Bernanke,

Gartler (1989), Dickey, Fuller (1979), Freeman (1979, 1982, 1983, 1984, 1987, 1993, 1996, 1998, 2001, 2007), Clark, Freeman, Soete (1981), Freeman, Clark, Soete (1982), Freeman, Pérez (1988, 1996), Freeman, Louca (2001), Nerlove, Grether, Carvalho (1979), Van Duijn (1979, 1981, 1983a, b), Wallerstein (1979, 1980, 1982, 1984a, b), Bousquet (1980), Choudri, Kochin (1980), Doran, Parsons (1980), Doran (2003), Eklund (1980), Graham, Senge (1980), Homer (1980), Low (1980), Marchetti (1980, 1988, 1998), Frank (1980), Hodrick, Prescott (1980, 1997), Kydland, Prescott (November 1982, Spring 1990), Prescott (1986, 1998a, b), Parente, Prescott (1993, 1994, 2000), Sterman (1980, 1981, 1982a, b, 18 March 1983a, b, 1984, 1985, June 17 2012, October 2 2013), Sterman, Mosekilde (February 28 2014, February 14 2015, August 25 2017), Van der Zwan (1980), Weber (1980), Beveridge, Nelson (1981), Brunner (1981), Delbeke (1981), Hall (26 March 1981, 1997), Kleinknecht (1981a b, 1982, 1986 1987, 1990, 1993), Bieshaar, Kleinknecht (1984), Kleinknecht, Mandel, Wallerstein (1992), Kleinknecht, Van der Panne (2006, 30 July 2008), Pasinetti (1981, 1995, 1998, December 2000), Tinbergen (1981, 1984), Priestley (1981), Warren (1981), Boltho (1982), Chitre (1982), Haustein, Neuwirth (1982), Moore (1982), Moore, Klein (1989), Nelson, Winter (1982), Nelson, Plosser (1982), Senge (1982, 1983), Stewart (1982), Thompson, Zuk (1 December 1982), Thompson (2007), Van Ewijk (1982, 1994), Beckman (1983), Blatt (1983), Cass, Shell (April 1983), Cleary, Hobbs (1983), Dickson (25 February 1983), Glismann, Rodemer, Wolter (1983), Goldstein (1983, 1988, 1991), Hill (1983), Hozelitz (January 1983), Long, Plosser (1983), Pérez (1983, 1996; 1985; 2002; 2015), Rosenberg, Frischtak (1983), Rosier, Dockes (1983), Tobin (1983), Bell (1984), Bergstrom (1984, 1988), Bergstrom, Nowman (2007), King, Plosser (1984), King, Rebelo (1988, 1993, 1999), King, Plosser, Rebelo (1988a, b), King, Plosser, Stock, Watson (1991), Screpanti (1984, 1999), Rothbard (1984), Foders, Glismann (1985), Grandmont (1985), Hansen (1985), Hansen, Wright (1992), Harvey (1985, 1989), Harvey, Jaeger (1993), Harvey, Koopman (1997), Harvey, Trimbur (2003), Harold (1985, 1986), Elliot (1985, 1991), Zarnowitz (1985, 1989, June 1992), Zarnowitz, Moore (1986), Zarnowitz, Stock (1992), Bhargava (1986), Farmer (1986), Kitwood (1 January 1986), Mosekilde, Rasmussen (April 1986), Rasmussen,

Mosekilde, Holst (16 October 1989), Mosekilde (1996-1997), Reichlin (1986), Romer (1986, 1990, February 1991, September 1994, Spring 1999), Semmler (1986), Solomou (1986a, b, 1987, 1989, 1998), Staley (1986), Summers (1986, 2005, 2014), Engle, Granger (1987), Kennedy (1987), Kirk (1987), Klimenko, Menshikov (1987), Mager (1987), Nakicenovic (1987), Puu (1987, 2001), Ritzen (1987), Tremblay (28 December 1987), Volland (1987), Alesina (1988, 1989), Alesina, Tabellini (1990), Alesina, Ozler, Roubini, Swagel (1992, June 1996), Alesina, Perotti (1994), Alesina, Rodrik (1994), Alesina, Roubini, Cohen (1997), Ben-Porath (April 1988), Cochrane (1988), Cuadrado-Roura, Rubalcaba-Bermejo (1998), Greenwald (1988, 1993), Greenwood, Hercowitz, Huffman (June 1988), Greenwood, Hercowitz, Krusell (2000), Schön (1988, 1989, 1991, 1994, 1998, 2000, 2004, 2009), Stock, Watson (1988a, b, 1989, 1993, September 2002, 2004), Street (June 1988), Blanchard, Fischer (1989), Hamilton (1989, 1994), İmrohoroglu (1989), Mankiw (March 1989), Ploser (1989a, b), Shiller (1989), Stewart (1989), Tarascio (1989) West, Harrison (1989), Harrison, Berman (2016), Ayres (1990a, b, 2006), Boldrin, Woodford (1990), Boldrin, Christiano, Fisher (1995), Boldrin (September 2000), Boschan, Banerji (1990), Dua, Banerji (1999, 2004a, b, August 2006), Layton, Banerji (2004), Danthine, Donaldson (1990), Danthine, Neftci (1990), Danthine, Donaldson (1991, 1993, 1995), Escudier (1990), Grubler, Nowotny (1990), Hedtke (1990), Jovanovic, Rob (1990), Kontorovich (1990), Reijnders (1990, 2009), Rogoff (1990), Baxter, King (1991, 1999), Berry (1991, 2000), Berry, Kim (1993, 1994), Diebold, Rudebusch (1991), Bangia, Diebold, Kronimus, Schlagen, Schuermann (2002), Diebold (2004), Aruoba, Diebold, Kose, Terrones (2011), French, Sichel (1991, 1993), Grossman, Helpman (1991), Jang-Ok Cho, Cooley (1991, 1995), Jang-Ok Cho, Cooley, Hyung Seok Kim (2015), Levy (1991), Madaràsz (1991), Schubert (1991), Sterken (1991), Thio (1991), Pope (1991), Aghion, Howitt (1998), Backus, Kehoe, Kydland (1992), Backus, Kehoe (1992), Backus, Kehce, Kydland (1995), Christiano, Eichenbaum (1992), Christiano, Fitzgerald (1998), Christiano, Eichenbaum, Evans (2002), Christiano, Fitzgerald (2003), Christiano, Eichenbaum, Vigfusson (2004), Christiano, Ilut, Motto, Rostagno (2007), Gilles (October 1992), Kenwood, Lougheed (1992), Laidler (1992), Mellander, Vredin,

Warne (1992), Merrill, Szidarovszky (1992), Metz (1992, 1998, 2006), Phelan (1992), Reati (1992, 1998), Reati, Toporowski (December 2004), Sowell (1992), Tylecote (1992, 1994), Atkeson, Kehoe (1993), Atkeson, Phelan (1994), Bowles, Edwards (1993), Bruno (1993), Bruno, Easterly (1998) Chari, Christiano, Kehce (1993), Dolado, Sebastián, Vallés (Septiembre 1993), Fischer (1993), Goodwin (1993), Labini (1993), Saint-Paul (1993), Silverberg, Lehnert (1993, 1996), Silverberg, Verspagen (1996, 2003a, b), Silverberg (2003, 2005, 2006), Stiglitz (1993, 1994a, b, 1999, 2002, June 28 2011), Warne (1993), Bils, Cho (1994), Bils, Klenow (1998), Crafts (1994), Kashyap, Stein, Lamont (August 1994), Kim (1994), Kim, Yoo (1995), Kim, Nelson (1998, 1999a, b), Kim, Piger (2002), Kim, Morley, Piger (2002), McGrattan (1994a, b), Barro, Sala-I-Martin (1995), Barro (1996, 1997), Burnside, Eichenbaum, Rebelo (1995, 1996), Burnside, Eichenbaum (1996), Calomiris, Himmelberg, Wachtel (1995), Cogley, Nason (1995), Cooley (1995), Cooley, Prescott (1995), Kapuria-Foreman, Perlman (November 1995), Pedregal (1995, 2001), Pedregal, Young (1996, 2001), Uhlig, Xu (1995), Witt (1995), Akerlof, Dickens, Perry (1996), Arena, Festré (1996), Bernard, Gerlach (1996), Cheng, Dinopoulos (1996), Crafts, Mills (1996), Crafts, Toniolo (1996), De la Croix, Deneulin (1996), De la Croix (29 February 2000), Galor (1996), Haken (1996), Justman (1996, 1997), Koop, Pesaran, Potter (1996), Lee (1996), McDonald (1996), Quah (1996), Silverberg, Lehnert (1996), Silverberg, Verspagen (2000), Silverberg (2002), Zimmermann (1996), Argandoña, Gamez, Mochón (1997), Artis, Zhang (1997), Artis, Krolzig, Toro (2004), Artis, Marcellino, Proietti (2004), Balke, Fomby (1997), Bierens (1997), Budnevich, Le Fort Varela (1997), Canton (1997), Gandolfo (1997), Harrod (1997), Kiyotaki, Moore (1997), Gertler, Kiyotaki (2009), Klenow, Rodriguez-Clare (1997), Krolzig (1997, 1999), Krolzig, Toro (1999), Krolzig, Marcellino, Mizon (2002), Louçã (1997, 1999), Louçã, Reijnders (1999), Neumann (1997), Pagan (1997), Stein (1997), Barnett (1998a, b), Brenner (May-June 1998), Canova (1998), Chauvet (1998, 1999), Edison et al (March 1998), Estrella, Mishkin (1998), Gowdy, Mesner (Summer 1998), Helpman, Trajtenberg (1998), Israel (November 1998), Kouparitsas (1998, 1999, 2000, December 2001), Lee (1998), Peel, Davidson (1998), Thomas, Nefiodow (1998), Volkmann (1998), Anderson, Ramsey (1999), Basu, Taylor (1999), Gali (1999), Gordon (1999), Jones (1999), Jones, Klenow (2011), Koopman, Shephard, Doornik (1999), Koopman, Valle e Azevedo (2004), Lettau, Uhlig (1999), Ljungqvist, Uhlig (1999), Pollock (1999, April 2008), Quigley (1999), Xiangkang Yin, Zuscovitch (October 1999), Zeira (1999), Alt, Lassen (2000), Breschi, Malerba, Orsenigo (2000), Chin, Geweke, Miller (2000), Collard, de la Croix (2000), Den Haan (2000), Devezas (2000, 2006), Devezas, Corredine (2001, 2002), Devezas, Modelski (2003, 2008), Devezas, Linstone, Santos (2005), Devezas, Grinin, Korotayev (2012), Drazen (2000), Fogel R W (2000), Fogel R W, Fogel E M, Guglielmo, Grotte (2013), Hossein-Zadeh, Gabb (1 September 2000), Imbs (2000), McConnell, Pérez-Quirós (2000), Persson, Tabellini (2000), Blanchard, Simon (2001), Brillet (2001), Camacho (2001, 2003), Chistilin (September 2001, 2008), Chol-Won Li (2001), Ehrmann, Elison, Valla (2001, 2003), Gómez (2001), Gonzalo, Ng (2001), Inklaar, de Haan (2001), Kongsamut, Rebelo, Danyang Xie (2001), Kouparitsas (2001), Psaradakis, Sola, Spagnolo (2001), Racorean (2001), Rothman, Van Dijk, Franses (2001), Trimbur (2001), Vošvrda (January 2001), Weder (2001), Wen (2001, 2002, 2004, March 2006), Yao (2001), Agénor (2002), Arnord (2002), Beaudry, Portier (2002, September 2004, September 2006, July 2007), Beaudry, Dupaigne, Portier (2011), Beaudry, Fve, Guay, Portier (2015), Festré (2002), Fisher, Hornstein (2002), Harding, Pagan (2002, 2003), Jordan, Rosengren (2002), Kim, Burnie (2002), Livio (2002), O' Hara (2002, 2003), Pedregal (2002), Ravn, Uhlig (2002), Rennstich (2002), Ritschl (2002a, b, 2003, 2005), Ritschl, Straumann (January 2009), Albu, Nicolae-Balan, Iordan, Caraiani (2003), Álvarez Vázquez (2003), Brian (2003), Duecker, Wesche (2003), Helenius (2003, 2009, 2010, 2012), Helenius, Pagni (2011), Helfat, Peteraf (2003), Hirooka (2003, 2006), Mariano, Murasawa (2003), Mills (2003), Nakajima (2003), Ogawa (2003), Rumyantseva (2003), Selover, Jensen, Kroll (2003), Sussmuth (2003), Tsoulfidis (2003, October 2006), Zhang, Zhang, Lee (2003), Benhabib, Wen (2004), Caraiani (2004, 2007a, b), Jaeger, Schucknecht (2004), Kaminsky, Reinhart, Vegh (2004), Kobayashi, Inaba (20 March 2004), Manfredi, Fanti (2004), McCauley (2004), Ngai (2004), Ngai, Pissarides (2007), Ohn, Taylor, Pagan (2004) Pelagatti (2004), Schnabel (2004), Sergienko (2004),

Syed, Mohammad (2004), Verspagen (2004), Valle e Azevedo, Koopman, Rua (2004), Andergassen, Nardini (2005), Banerjee, Duflo (2005), Cover, Pecorino (2005), Darvas, Rose, Szapary (2005), De Groot, Franses (2005, 27 March 2006, 2008, 2012), Francis, Ramey (2005), Jonung, Schucknecht, Tujula (2005), Miles, Scott (2005), Neumeyer, Perri (2005), Ozawa (2005), Peaucelle (2005), Rebelo (2005, 2010), Shimer (March 2005), Steehouwer (2005), Woo (2005), Andersen (2006), Chian, Rempel, Borotto, Rogers (2006), Comin, Gertler (2006), Congcong Dong (27 February 2006), Diebolt, Doliger (2006, 2008), Lee, Sang-Hyop Lee, Mason (July 2006), Linstone (2006), McMinn (2006, 2007), Monteiro (2006), Papageorgiou, Tsoulfidis (20 July 2012), Yamashiro, Uesugi (2006), Aguiar, Gopinath (2007), Fernández-Villaverde, Rubio-Ramirez, Sargent, Watson (June 2007), Fernández-Villaverde, Rubio-Ramirez (2010), Fernández-Villaverde, Guerrn, Kuester, Rubio-Ramrez (2012), Fernández-Villaverde, Rubio-Ramírez, Schorfheide (2016), Flodén (2007),Honjo (1 August 2007), Knotek (2007), Krantz, Schön (2007), Rodriguez Mora, Schulstad (November 2007), Adrian, Estrella (2008), Adrian, Estrella, Hyun Song Shin (January 2010), Battilossi, Foreman-Peck, Kling (September 2008), Kling, Foreman-Peck, Battilossi (2008), Hagedorn, Manovskii (September 2008), Haltmaier (April 2008), Hori (2008), Ilzetzki, Vegh (2008), Ilzetzki (2011), Iyetomi, Aoyama, Ikeda, Souma, Fujiwara (2008), Iyetomi, Nakayama, Yoshikawa, Aoyama, Fujiwara, Ikeda, Souma (2011), Iyetomi, Aoyama, Fujiwara, Sato (2012), Jourdon (2008), Papenhausen (2008), Roa, Jose, Dulce (2008), Taniguchi, Bando, Nakayama (2008), Angeletos, La'O (May 2009), Boone, Johnson (2009), Boretos (2009), Coccia (2009a, b, 2010), Den Haan, Kaltenbrunner (April 2009), Jaimovich, Rebelo (September 2009), McCauley (2009), OECD (2009), Purica, Caraiani (2009), Schmitt-Grohé, Uribe (2009), Sims (2009), Walentin (April 2009), Akaev, Sadovnichiy, Korotaev (December 2010, 15 May 2011, 2012), Sadovnichiy, Akaev, Korotaev, Malkov (2012), Crafts, Fearon (2010), Fanti, Gori (15 January 2010), Grigor'ev (2010), Grinin, Korotayev, Malkov (2010), Korotayev, Tsirel (2010), Grinin, Korotayev, Tausch (2016), Ritschl, Straumann (2010), Krusell, McKay (Fourth Quarter 2010), Rossiter (2010), Xu Feiqiong (2010), Asteriou, Hall (2011), Barseghyan, Battaglini, Coate (2011), Bazzi, Blattman (2011), Buera,

Monge-Naranjo, Primiceri (2011), Canes-Wrone, Park (2011), Claessens, Kose, Terrones (2011), Drehmann, Borio, Tsatsaronis (2011), Gokcekus, Suzuki (2011), Jacks, Novy, Meissner (2011), Jenkins, Brandolini, Micklewright, Nolan (2011), Jenkins, Taylor (2012), Jin Guantao, Liu Jinfeng (2011), Jordá, Schularick, Taylor (2011), Lopes (2011), Lostun (2011), Lucchese, Pianta (February 2011), Lucchese, Pianta (June 2012), Patterson (2011, 2012), Qin Duo (2011), Restuccia (3$^{rd}$ Quarter 2011), Roper (2011), Shiyan (2011), Aimar, Bismans, Diebolt (2012), Albers S, Albers A L (31 March 2012), Ales, Maziero, Yared (2012, 2013, 2014), Bandura (2012), Borio (December 2012), Borio, Karroubi, Upper, Zampoli (14 April 2016), Branca, Pina, Catalão-Lopes (2012), Camacho, Perez Quiros, Poncela (2012, 2015), Camacho, Martinez-Martin (2014, February 2015), Dementiev (2012, 2013, December 2014), Ikeda, Aoyama, Fujiwara, Iyetomi, Ogimoto, Souma, Yoshikawa (2012), Ikeda, Aoyama, Yoshikawa (2013a, b), Ikeda (2013), Pérez Quirós (December 2012), Podlesnaya (2012a, b), Pustovoit (2012), Swiss National Bank (2012, 2013), Uechi, Akutsu (2012), Azzimonti, Talbert (November 2013), Bachmann, Bai (2013), Beber, Brandt, Luisi (2013), Central Banking Newsdesk (2013), Duran (2013), Klepach, Kuranov (2013), Maziero, Yared, Ales (2013), Raicu, Stanca, Raicu (2013), Ternyik (1 January 2013a, b), Union Bank of Switzerland (2013), Coyle (2014), Estrada (2014), Ferrara, Marsilli (2014), Golinelli, Parigi (2014), Grossman, Mack, Martinez-Garcia (2014), Lucas, Gevorkyan, Palley (2014), Quast, Gonzalez (2014), Da Costa (2015), Desai, King, Goodhart (2015), Federal Reserve Bank of St Louis (2015), Galli (2015), Ledenyov D O, Ledenyov V O (2013c, 2015d, e, f, g, h, 2016r), Ledenyov V O, Ledenyov D O (2016, 2017), Lucarelli, Paulré (2015), Maloney, Pickering (2015), Yaguang Zhang, Guo Fan, Whalley (October 2015), Wikipedia (2015a b, 2017a, b, c), Achilleas (March 2016), Bazillier, Magris, Mirza (February 2016), Gîlcă (2016), Kronick (July 2016), Legrand, Assous, Hagemann (2016), Nopphawan Photphisutthiphong, Weder (2016), Olkhov (October 2016, 2017), Yaguang Zhang, Guo Fan, Whalley (January 2016), Xu, Ni, Van Leeuwen (2016), Bazillier, Magris, Mirza (2017), Daianu (2017), McGregor, Wills (March 2017), Siena (15 February 2017), Villarreal, Bielma (2017).

# Chapter 3

## Discrete-time digital economic output waves in form of vector-modulated direct sequence spread spectrum signal in economy of scale and scope in classic econodynamics

Let us begin our advanced research on the discrete-time digital economic output waves in the economy of the scale and scope in the Ledenyov classic econodynamics in the social science by applying an accumulated scientific knowledge base in the Maxwell electrodynamics in the natural science in Maxwell (1890). Speaking concisely, we prefer to think philosophically on a highlighted subject by expressing our new original research ideas toward the formulation of the fundamental theoretical principles on the Ledenyov discrete-time digital economic output waves in the economy of the scale and the scope in the Ledenyov classic econodynamics.

In this context, we believe that a short introduction to the nature of the electromagnetic signals in the electrodynamics can be a good starting point for our research discussion. Thus, let us begin by making a quite trivial research statement that the electromagnetic signals generation, amplification, attenuation, propagation and processing is a central research subject in the electromagnetic signals theory in the Maxwell electrodynamics in Maxwell (1890).

At this point, let us explain that there may be the following electromagnetic signals in the Maxwell electrodynamics in Maxwell (1890):

1. The electromagnetic signals with the continuously changing physical parameters:
    a) The continuous-amplitude signals;
    b) The continuous-frequency signals;
    c) The continuous-phase signals;
    d) The continuous-time signals.
2. The electromagnetic signals with the discretely changing physical parameters.

*a)* The discrete-amplitude signals;

*b)* The discrete -frequency signals;

*c)* The discrete -phase signals;

*d)* The discrete -time signals.

All the existing electromagnetic signals in our nature can be accurately characterized by the physical parameters in the amplitude, frequency, phase and time physical characterization domains in the Maxwell electrodynamics in Maxwell (1890).

Fig. 25 displays a modern scientific approach to the accurate characterization of the electromagnetic signals in the Maxwell electrodynamics in Maxwell (1890).

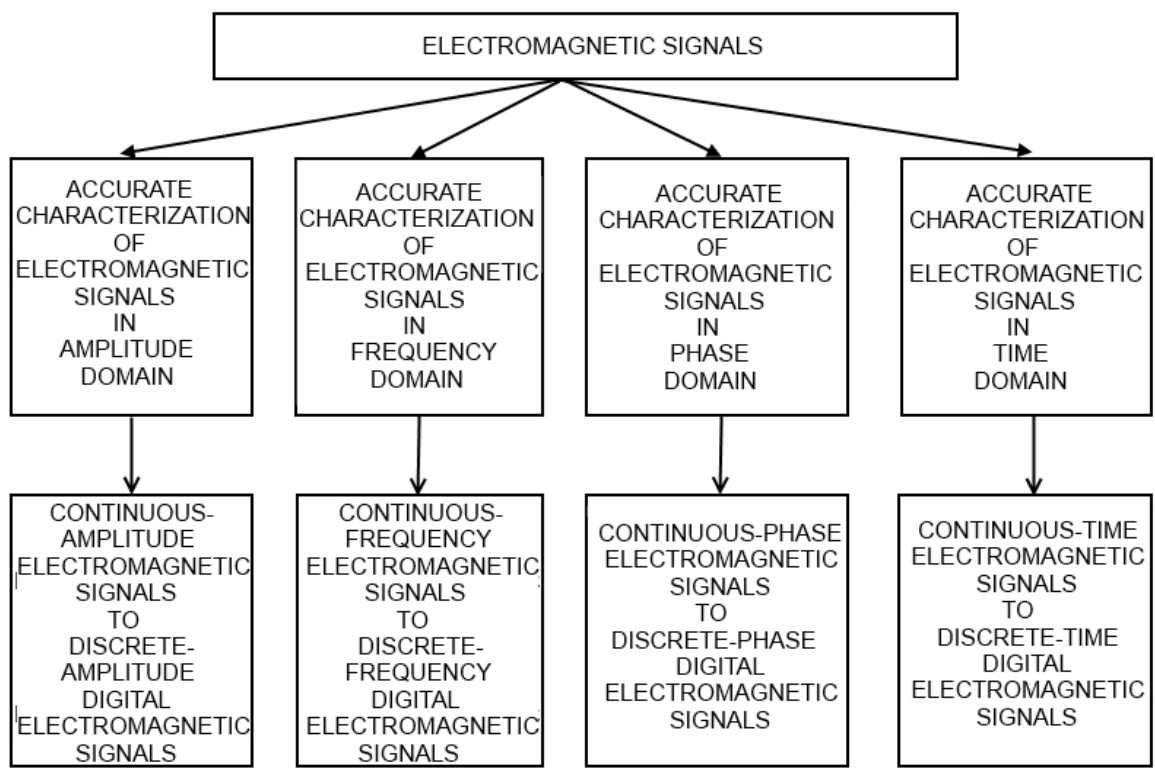

**Fig. 25.** Modern scientific approach to accurate characterization of continuous-/discrete-time electromagnetic signals in amplitude, frequency, phase and time measurement domains in Maxwell electrodynamics.

Giving enough research emphasis to a general classification of the electromagnetic signals into the two main types: *1)* The electromagnetic signals with the continuously changing physical parameters, *2)* The electromagnetic signals with the discretely changing physical parameters;

we would like to describe these electromagnetic signals more clearly.

The **continuous-time signals** (the continuous-wave signals (CW)) are characterized in the continuous-time (CW) signals generation/processing/filtering theory in the Maxwell electrodynamics in Maxwell (1890), Ledenyov D O, Ledenyov V O (2015a).

Fig. 26 displays the continuous-time signal's waveform as a dependence of the amplitude on the time, $S_1(t)$, in the Maxwell electrodynamics in Maxwell (1890), Ledenyov D O, Ledenyov V O (2015a).

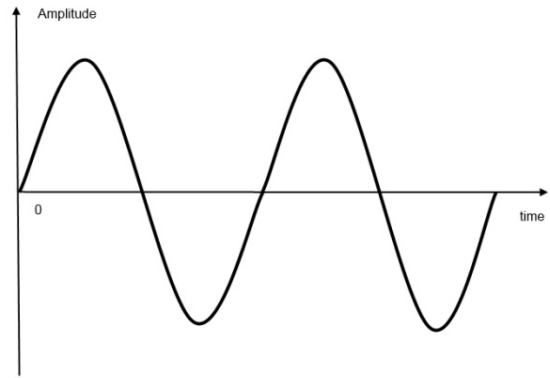

$$y = f(t), y \in C, t \in C;$$

$$y_i = A_i \sin\left(2\pi f_i(t) + \phi_i(t)\right),$$

$$y_i = A_i e^{j\pi\left(2\pi f_i(t) + \phi_i(t)\right)},$$

**Fig. 26.** Continuous-time signal in form of dependence of amplitude on time in Maxwell electrodynamics, $S_1(t)$.

We can also display the continuous-time signal in the form of a dependence of the amplitude on the frequency, $S_1(f)$, in the Maxwell electrodynamics in Maxwell (1890), Ledenyov D O, Ledenyov V O (2015a). The main parameters of the continuous-time signal are:

*1.* The amplitude of signal;

*2.* The frequency of signal;

*3.* The period of signal;

*4.* The phase of signal;

*5.* The amplitude/frequency/period/phase generation accuracies;

*6.* The phase noise of signal;

*7.* The non/sub harmonic/spurious emissions of signal.

Fig. 27 pictures the continuous-time signal in the form of a dependence of the amplitude on the frequency in the Maxwell electrodynamics, $S_1(f)$.

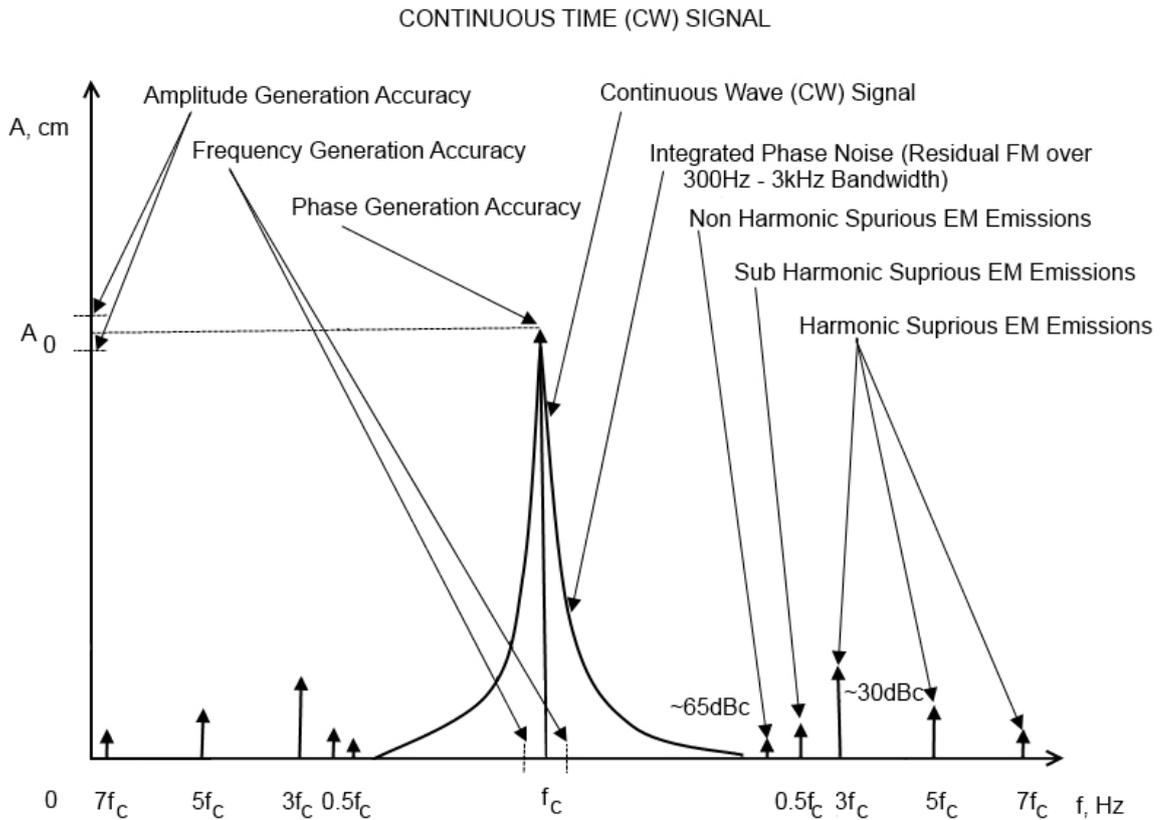

**Fig. 27.** Dependence of amplitude on frequency for continuous-time electromagnetic signal in Maxwell electrodynamics, $S_1(f)$.

In practice, the continuous-time signal can be synthesized by radiating an electromagnetic wave with the continuous- amplitude, frequency, phase modifications over the certain time period. Most importantly, the information can be coded into the CW signal carrier by applying the analog modulations. Let us write a set of formulas to represent the analog modulations by amplitude/frequency/phase modifications of the continuous-time signal carrier over the certain time period:

*1.* The analog modulation by the amplitude modification;

*2.* The analog modulation by the frequency modification;

*3.* The analog modulation by the phase modification;

*4.* The analog modulation by the amplitude switching on/off (the pulse(s) generation);

However, let us note that the analog modulation by the on/off switching of the continuous-time signal carrier will result in a generation of the discrete-time pulse modulated signal.

$$S_{1,2,3,4}(t) = A(t)\sin(2\pi f(t) + \phi(t)),$$

*where*

$A(t)$ *is the amplitude*, $\left[\textbf{Amplitude Modulation } (AM), \textbf{Pulse Modulation } (PM)\right]$;

$f(t)$ *is the frequency*, $\left[\textbf{Frequency Modulation } (FM), \textbf{Angle Modulation } (AM)\right]$;

$\phi(t)$ *is the phase*, $\left[\textbf{Phase Modulation } (PM), \textbf{Angle Modulation } (AM)\right]$.

Fig. 28 pictures the analog signal modulations by the amplitude, frequency, phase modifications over the selected time period in the analog electronics in the Maxwell electrodynamics.

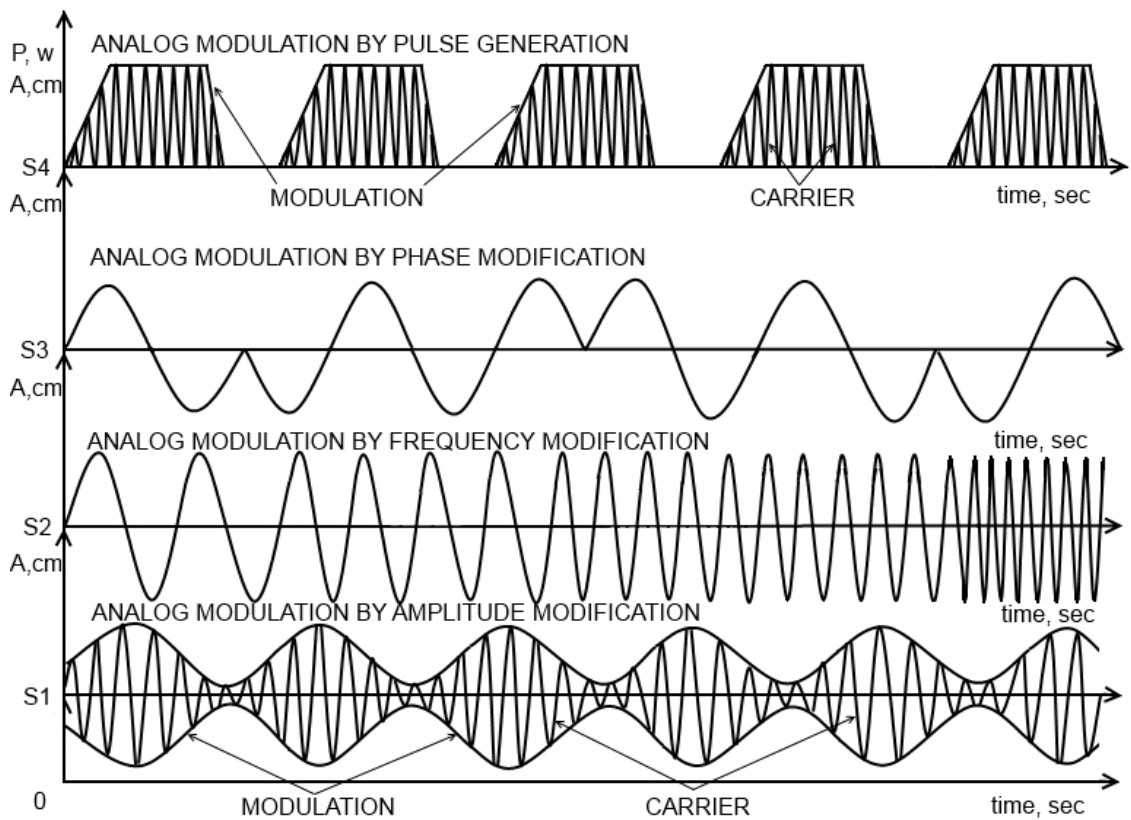

**Fig. 28.** Analog signal modulations by amplitude, frequency, phase modifications over selected time period in Maxwell electrodynamics: *1)* Analog modulation by amplitude modification, $S_1$; *2)* Analog modulation by ferquency modification, $S_2$; *3)* Analog modulation by phase modification, $S_3$; *4)* Analog modulation by pulse generation (amplitude switching on/off), $S_4$.

The **discrete-time signals** (the discrete-wave signals (DW)) are characterized in the discrete-time digital signals generation/processing/ filtering theory in the Walsh discrete-time digital signal processing science in Walsh (1923a, b), Wiener (1923, 1930, 1949), Ito (1944, 1951a, b, 2000), Pugachev (1944, 1956a, b, 1960, 1961, 1962, 1971, 1973, 1974, 1975, 1974, 1978, 1979a, b, 1980a, b, 1981, 1982a, b, 1984, 1985, 1986), Pugachev, Sinitsyn (1986, 1989, 1990, 1999, 2004), Pugachev, Sinitsyn, Shin (1986a, b, 1987a, b, c), Bartlett (1954), Tukey (1957), Stratonovich (1958, 1959a, b, 1960a, b, 1961, 1964, 1965, 1966, 1967a, b, 1968, 1975), Stratonovich, Kuznetsov, Tikhonov (1965), Kalman, Koepcke (1958, 1959), Kalman, Bertram (1958, 1959), Kalman (1960a, b, 1963), Kalman, Bucy (1961), Oppenheim, Schafer (1989), Simon, Hinedi, Lindsey (1995), Proakis, Manolakis (1996), Prisch (1998), Wanhammar (February 24 1999), Sklar (2001), Rice (2008), Ledenyov D O, Ledenyov V O (2015a).

Fig. 29 shows the discrete-time digital signal's waveform in the Walsh discrete-time digital signal processing science in Walsh (1923a, b), Wanhammar (February 24 1999), Ledenyov D O, Ledenyov V O (2015a).

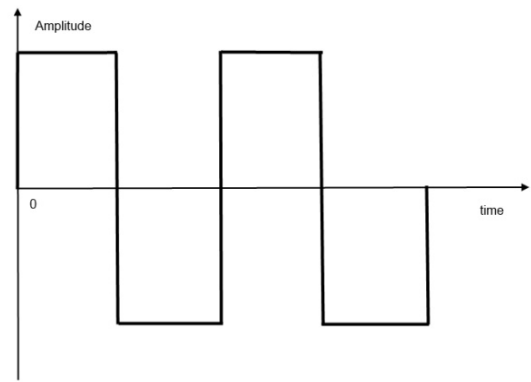

$$y = f(nT), y \in C, n \in Z, T > 0;$$
$$y = f(nT), y \in Z, n \in Z, I > 0.$$

**Fig. 29.** Discrete-time digital signal in form of dependence of amplitude on time in Walsh discrete-time digital signal processing science, $S_2(t)$.

We can draw the discrete-time digital signal in the form of a dependence of the amplitude on the frequency, $S_2(f)$, in the Walsh discrete-time digital signal processing science in Ledenyov D O, Ledenyov V O (2015a). The main parameters of the discrete-time digital signal are:

1. The amplitude of signal;
2. The frequency of signal;
3. The period of signal;
4. The phase of signal;
5. The amplitude/frequency/period/phase generation accuracies;
6. The phase noise of signal;
7. The non harmonic / harmonic / sub harmonic spurious emissions.

Fig. 30 pictures the discrete-time signal in the form of a dependence of the amplitude on the frequency in the electrodynamics, $S_2(f)$.

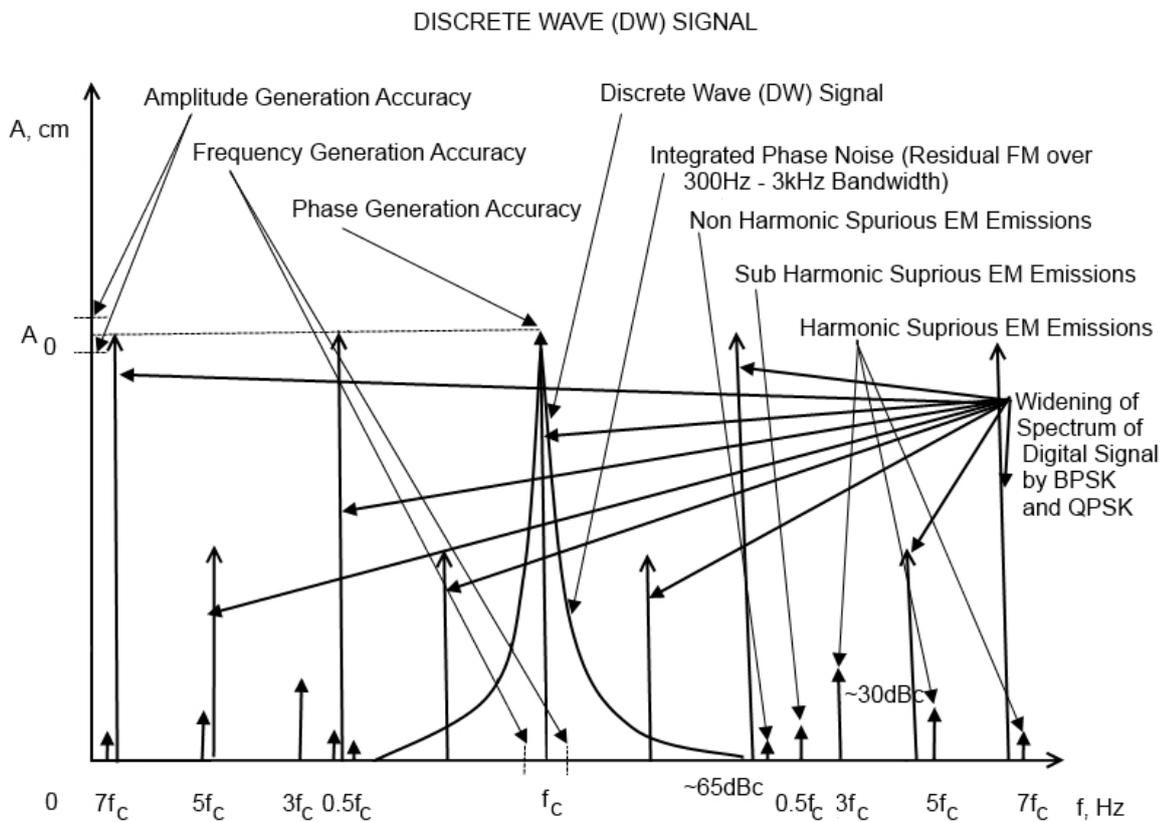

**Fig. 30.** Dependence of amplitude on frequency for discrete-time digital signal in Walsh discrete-time digital signal processing science, $S_2(f)$.

Discrete-time digital signal can be generated by applying BPSK, QPSK, and other high order digital modulations (Square Wave) to continuous-time signal carrier (Continuous Wave). Digital modulations make sharp abrupt instant transitions between states of 1 and 0, resulting in discrete-time digital signal spectrum widening. Digital filters can be used to reduce occupied discrete-time digital signal spectrum, but phase transitions will become less sharp/abrupt in this case.

At this point, we brought to the detailed consideration the two main types of the electromagnetic signals: *1)* The continuous-time signals; *2)* The discrete-time digital signals. Here, let us highlight the fact that the discrete-time changes of the electromagnetic signal's parameters, including:

| | |
|---|---|
| *1.* Amplitude; | *3.* Phase; |
| *2.* Frequency; | *4.* Time; |

can be considered as a main characteristic distinction between both *1)* the continuous-time signals and *2)* the discrete-time digital signals in the Walsh discrete-time digital signal processing science in Walsh (1923a, b).

In the theory, we can say that a primary purpose of the discrete-time digital signals generation is to synthesis the digital signal states, which correspond to the digits: 1 and 0 in Walsh (1923a, b).

In practice, the discrete-time digital signals can be generated, using many technical methods by making the discrete- amplitude, frequency, phase modifications to the continuous-time carrier signal over the time period:

   *1.* The digital modulation by the amplitude modification, when the change in the amplitude corresponds to the discrete states: 1 and 0;

   *2.* The digital modulation by the frequency modification, when the change in the frequency corresponds to the discrete states: 1 and 0;

   *3.* The digital modulation by the phase modification, when the change in the phase corresponds to the discrete states: 1 and 0;

   *4.* The digital modulation by the amplitude/phase modifications, when the change in the amplitude/phase refers to the discrete states: 1 and 0.

$$S_{1,2,3,4}(t) = A(t)\sin\left(2\pi f(t) + \phi(t)\right),$$

*where*

$A(t)$ *is the amplitude*, $\left[\text{*Amplitude Modulation* }(AM)\right]$;

$f(t)$ *is the frequency*, $\left[\text{*Frequency Modulation* }(FM)\right]$;

$\phi(t)$ *is the phase*, $\begin{bmatrix} \text{*Phase Modulation* }(PM), \text{*Binary Phase Shift Keying* }(BPSK), \\ \text{*Quadrature Phase Shift Keying* }(QPSK) \end{bmatrix}$;

$A(t), \phi(t)$ *is the amplitude and the phase*, $\left[\text{*Qudrature Amplitude Modulation* }(QAM)\right]$.

Let us graphically demonstrate the possible discrete-time digital signal's modulations by making the discrete- amplitude, frequency, phase and pulse modifications over the selected time period as it is usually done in the digital electronics.

Fig. 31 shows the possible discrete-time digital signal's modulations by making the discrete- amplitude, frequency, phase modifications over the selected time period in the in Walsh discrete-time digital signal processing science: *1)* The digital modulation by amplitude modification, $S_1$; *2)* The digital modulation by frequency modification, $S_2$; *3)* The digital modulation by phase modification, $S_3$; *4)* The digital modulation by amplitude and phase modification, $S_4$.

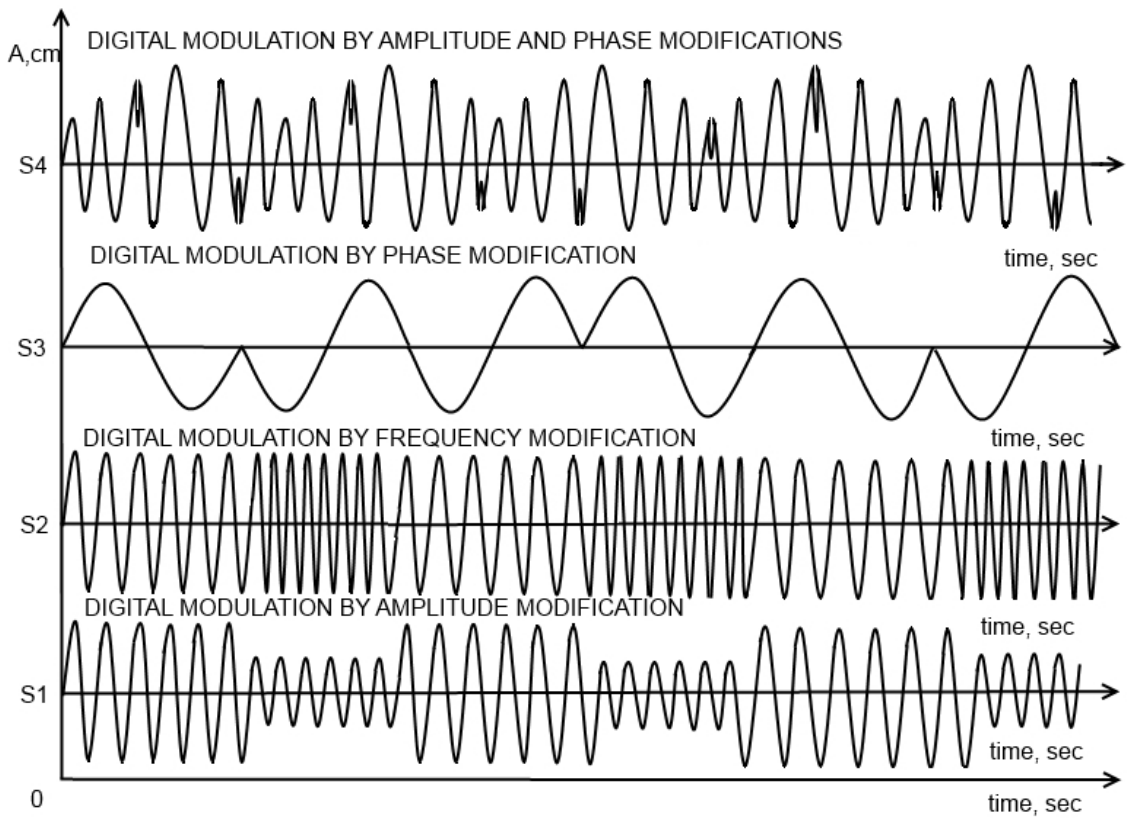

**Fig. 31.** Digital signal modulations by discrete- amplitude, frequency, phase modifications over selected time period in Walsh discrete-time digital signal processing science: *1)* The digital modulation by amplitude modification, $S_1$; *2)* The digital modulation by frequency modification, $S_2$; *3)* The digital modulation by phase modification, $S_3$; *4)* The digital modulation by amplitude and phase modification, $S_4$.

As we know, the pioneering groundbreaking research on the orthogonal functions played an important role in the formulation of the Walsh discrete-time digital signals science in Walsh (1923a, b). Since that time, the Walsh discrete-time digital signals theory is extensively used to design/develop/produce various electronics components, devices, modules and systems for the information processing in the electronics engineering:

1. ***Information Signals Conversion***: Analog-to-Digital Converter(s) (ADC) and Digital-to-Analog Converter(s) (DAC);

2. ***Information Computing***: The mathematical computation, applying the digital signals in the digital microprocessors in the form of the Integrated Circuit (IC) or the Very Large Scale Integrated (VLSI) circuit, which is fabricated on the crystal made of Silicon ($Si_{111}$);

3. ***Information Communication***: The digital signals generation, filtering, modulation, computing, coding and spreading over the wireline, wireless, and optical communication channels in the information communication technologies (ICT) in the telecommunications. The audio/video information streams can be coded in the digital signal processors (DSP) and the communication protocols stacks can be created in the microprocessors with the Very Large Scale Integrated (VLSI) circuits;

4. ***Information Coding***: The digital signals generation, filtering, computing, coding during the audio/still picture/video signals processing in the digital signal processors (DSP). The digital signal processors can be designed as the Reduced Instruction Set Computers (RISC) or the Complex Instruction Set Computers (CISC) and fabricated on the crystal made of Silicon ($Si_{111}$). The Field Programmable Gate Arrays (FPGA) chipsets can be used instead of the DSPs for the custom designs of electronic circuits in the small series productions. The Application Specific Integrated Circuit (ASIC) for the specific digital signal processing tasks can also be employed;

5. ***Information Storing***: The digital signals memorizing in the different types of the stand alone electronic/photonic/magnetic memory chips (RAM, DRAM, ROM, EEPROM, etc) and/or embedded memory chips inside the VLSI microchips in the central processors units (CPU) in the digital computers.

6. ***Information Displaying***: The Organic Light Emitting Diod (OLED) / Thin Film Transistor (TFT) displays in TV sets, desktop & laptop computers, avionics, etc.

Firstly, let us make a brief review on the problem on the analog-to-digital and the digital-to-analog ***signals conversions*** with an application of the Analog-to-Digital Converter(s) (ADC) and the Digital-to-Analog Converter(s) (DAC). The information can be transmitted in the space over the time, using the two possible types of the electromagnetic signal-carriers:

*a)* The analog continuous-time signal carrier, which can be modulated by the analog modulation;

*b)* The digital discrete-time signal carrier, which can be modulated by the digital modulation, and in some complex cases, by the additional analog modulation (see the next chapters).

In many practical cases, there is a specific need to covert the analog-to-digital signals and/or the digital-to-analog signals, which can be solved with an application of the ADC and/or the DAC. In the case of the ADC conversion, the main aim is to keep the information in the digital signal from the analog signal, satisfying the certain technical conditions:

1. The analog signal has to be in the form of the finite band-limited signal in Wanhammar (1999);

2. The Whittaker-Nyquist- Küpfmüller-Kotelnikov-Shannon theorem on the sample frequency must be satisfied in Whittaker (1915, 1935), Nyquist (1928), Küpfmüller (1928), Kotelnikov (1933), Shannon (July 1948, 1949). It means that the Nyquist sample rate must be at least in the two times bigger than the analog signal's bandwidth;

3. The interpolators and/or the decimators in the form of the digital filters can be used as the sample rate converters to increase and/or decrease the sample rates in the multivariate systems, preserving the information in the modulated signal carrier in Wanhammar (1999).

In the another opposite case of the DAC conversion, the analog signal can be reconstructed by applying the linear signal filtering to the digital signal in Wanhammar (1999).

Fig. 32 displays the analog-to-digital signal conversion process in electronics: ***a)*** The analog continuous-time electromagnetic signal waveform in ideal case with no distortions, $S_0$; ***b)*** The discrete-time sampled signal waveform in ideal case with no distortions, $S_1$; ***c)*** The discrete-time digital signal waveform in ideal case with no distortions, $S_2$; ***d)*** The discrete-time digital signal waveform in practical case with the severe distortions, $S_3$.

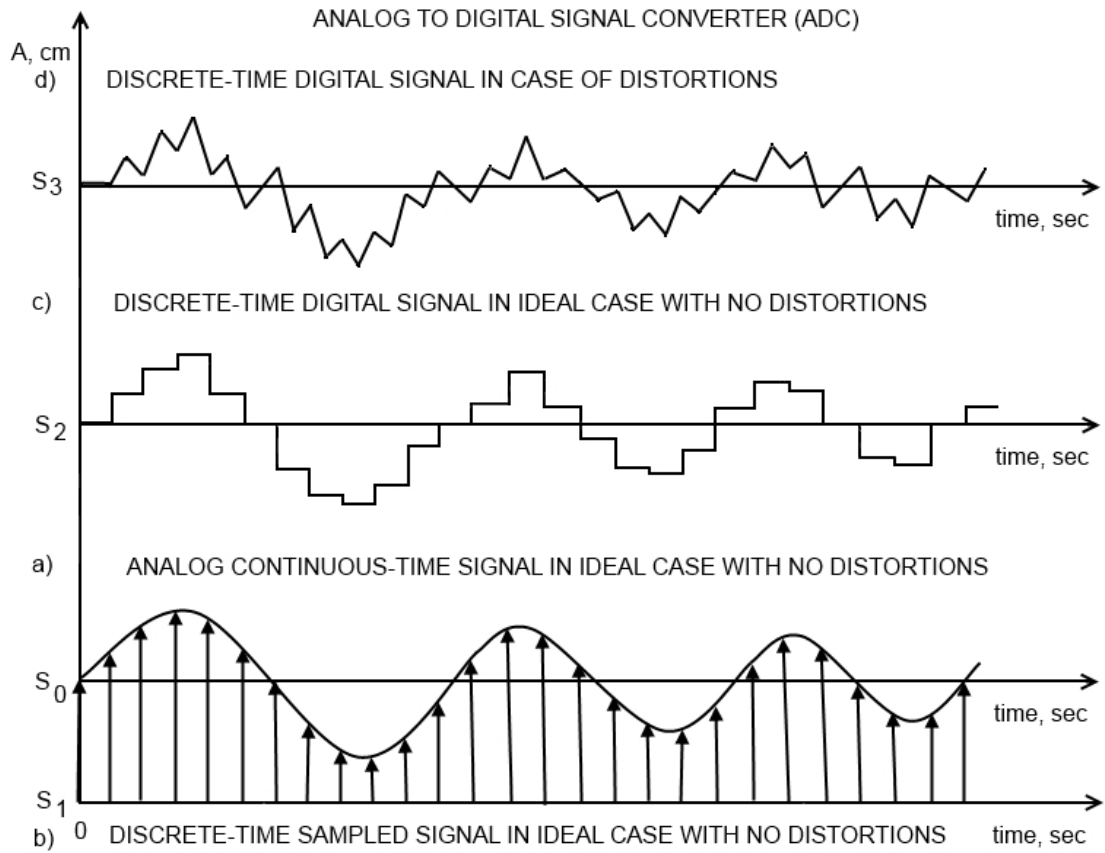

**Fig. 32.** Analog-to-digital signal conversion process in Walsh discrete-time digital signal processing science: ***a)*** Analog continuous-time electromagnetic signal waveform in ideal case with no distortions, $S_0$; ***b)*** Discrete-time sampled signal waveform in ideal case with no distortions, $S_1$; ***c)*** Discrete-time digital signal waveform in ideal case with no distortions, $S_2$; ***d)*** Discrete-time digital signal waveform in case with severe distortions, $S_3$.

Speaking about the problem on the ***information computing*** with the microprocessors to perform the computation on the mathematical binary logic with the two digits: 1 and 0. A binary number can be written as 10011010010, that is $(0x2^0) + (1x2^1) + (0x2^2) + (0x2^3) + (1x2^4) + (0x2^5) + (1x2^6) + (1x2^7) + (0x2^8) + (0x2^9) + (1x2^{10}) = 1234_{10}$. The computers operate with the binary numbers of 2, 4, 8, 32, 64 length. The computers values are represented in the hexadecimal format $1234_{10}= 10011010010_2 = 4D2_{16}$ . The computers use the floating point arithmetic with floating point numbers to make the computations, based on the IEEE standard of the floating point numbers representation. In the computers, the digital logic gates made of the Low Voltage Complementary Metal Oxyde Semiconductor (LVCMOS) transistors are used to perform the real computing, including:

| *a)* inverter; | *b)* AND gate; | *c)* OR gate; | *d)* NAND gate; |
|---|---|---|---|
| *e)* NOR gate; | *f)* negative logic AND gate; | *g)* negative logic OR gate; | *h)* multiplexer. |

The typical microprocessor in the form of the LVCMOS VLSI circuit can contain numerous design elements/parts in Wanhammar (1999), Chandrakasan, Bowhill, Fox (October 2000), Harris, Harris (August 7 2012), Hennessy, Patterson (December 17 2017):

*1.* The processing element(s) (PE) with the computing core(s):

| *a)* Arihmetic Logic Unit(s) (ALU); | *g)* Address multiplexer (Addr MUX); |
|---|---|
| *b)* Register bank(s); | *h)* Data multiplexer (Data MUX); |
| *c)* Accumulator register; | *k)* Program counter; |
| *d)* Instruction register; | *l)* Timing logic; |
| *e)* Indirect address register; | *m)* Control store; |
| *f)* Data out register; | *n)* ALU result internal bus; |

*2.* The shared cache in the form of the dynamic random access memory (RAM/DRAM) of different levels;

*3.* The random number generator (RNG) implemented in the hardware;

*4.* The memory controller;

*5.* The internal data path with the data queue;

*6.* The input/output (I/O) data interfaces.

The microprocessors can have the two possible memory architectures in Wanhammar (1999):

*1.* The Von Neumann microprocessor memory architecture, which has one memory space for the software program code and data;

*2.* The Harvard microprocessor memory architecture, which has the two separate memory spaces: one memory space for the software program code and another memory space for the data.

Fig. 33 pictures the Von Neumann microprocessor memory architecture with the microprocessor, the one external memory, and the data/address busses.

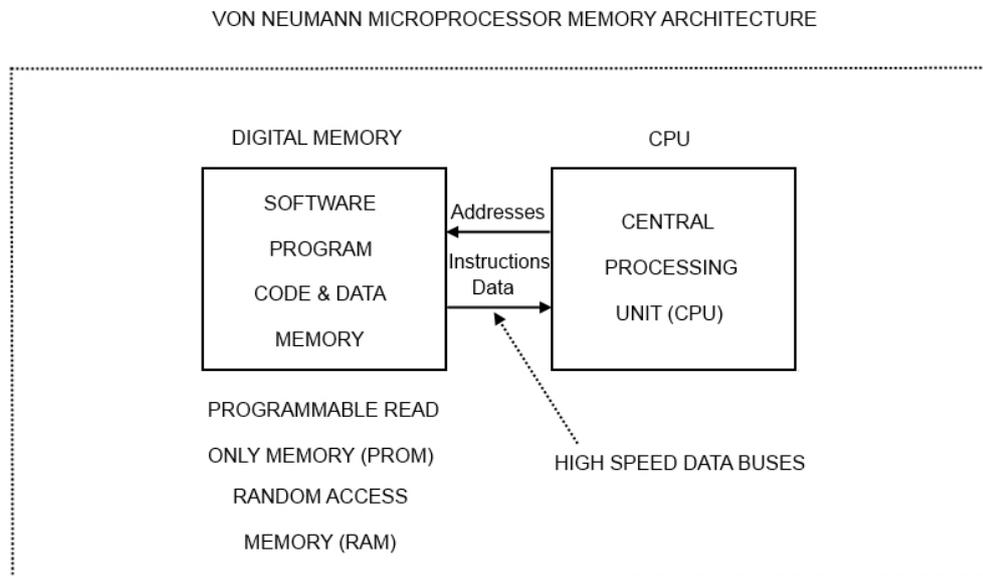

**Fig. 33.** Von Neumann microprocessor memory architecture.

Fig. 34 depicts the Harvard microprocessor memory architecture with the microprocessor, the two external memory, and the data/address busses.

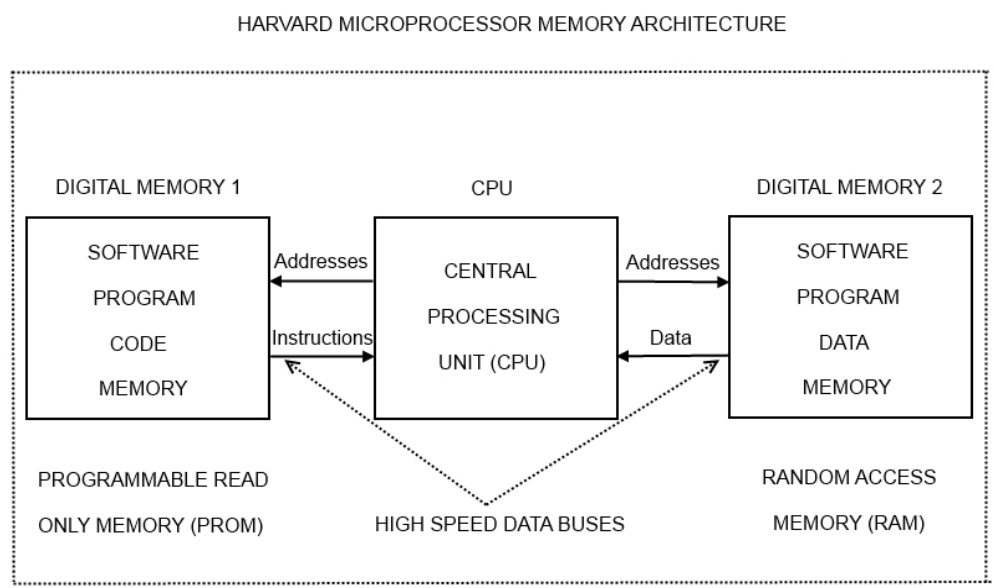

**Fig. 34.** Harvard microprocessor memory architecture.

Fig. 35 shows a general scheme of the microprocessor design layout (the floor plan), including the processing element(s) (PE), the shared cache in the form of the random access memory (RAM), the random number generator (RNG) in the hardware implementation, the memory controller, the data path with the data queue, the input/output (I/O) data interfaces.

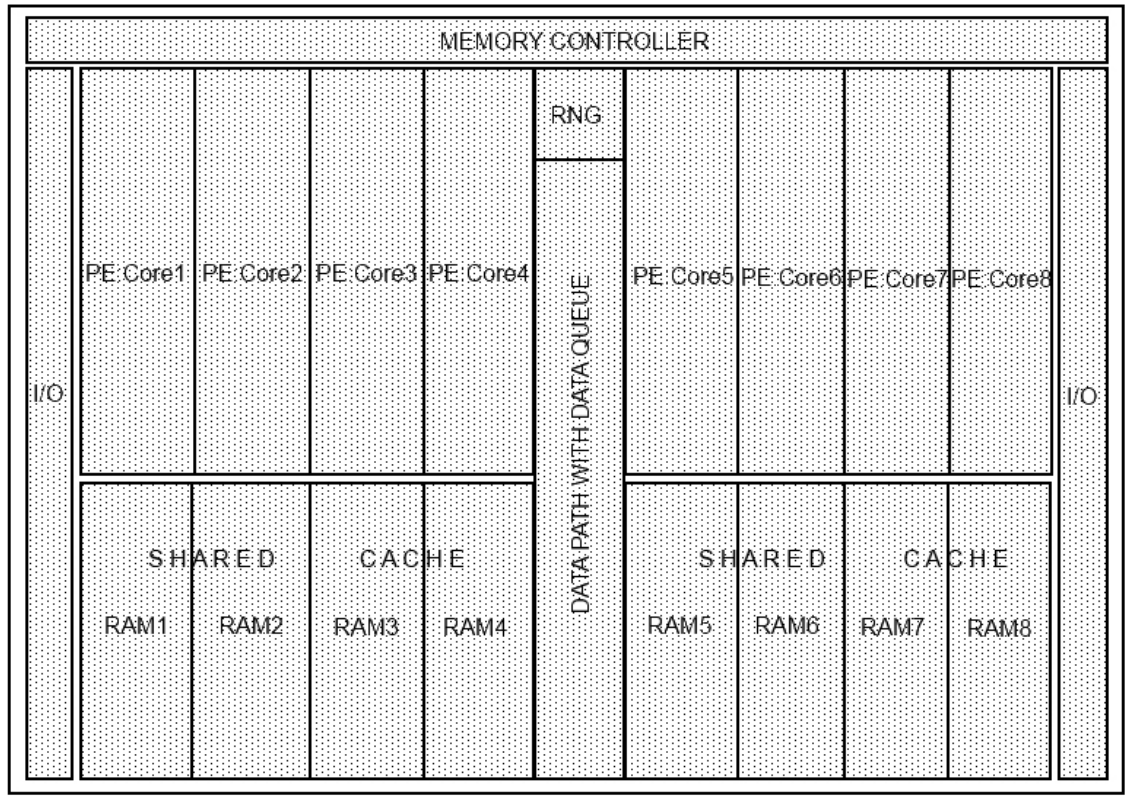

**Fig. 35.** Scheme of microprocessor design layout in microelectronics.

Speaking about the ***information communication***, we would like to comment that the discrete-time digital signals are frequently employed for the digital information communication purposes in microelectronics:

*1.* to create the communication links between the processing element with the computing kernel and the memory over the high speed data path with the digital communication protocols inside/outside the digital microprocessor(s);

*2.* to provide the communication links between the multiple digital microprocessor(s) at the same multilayered motherboard;

*3.* to establish the communication links between the microprocessor and the test equipment over the I$^2$C/JTAG/RS232 protocols converters for the testing/tuning purposes;

*4.* to make the optical communication links between the microprocessors at the same/different multilayered motherboards.

Discussing the information communication technologies more comprehensively, let us focus precisely on the vector-modulated discrete-time digital direct sequence spread spectrum signal, which is usually employed to transmit the information in the modern electronic devices over the wireline, wireless, and optical communication channels in the telecommunications.

Fig. 36 displays the vector-modulated discrete-time digital direct sequence spread spectrum signal in the form of a dependence of the amplitude on the frequency in the electrodynamics/the telecommunications.

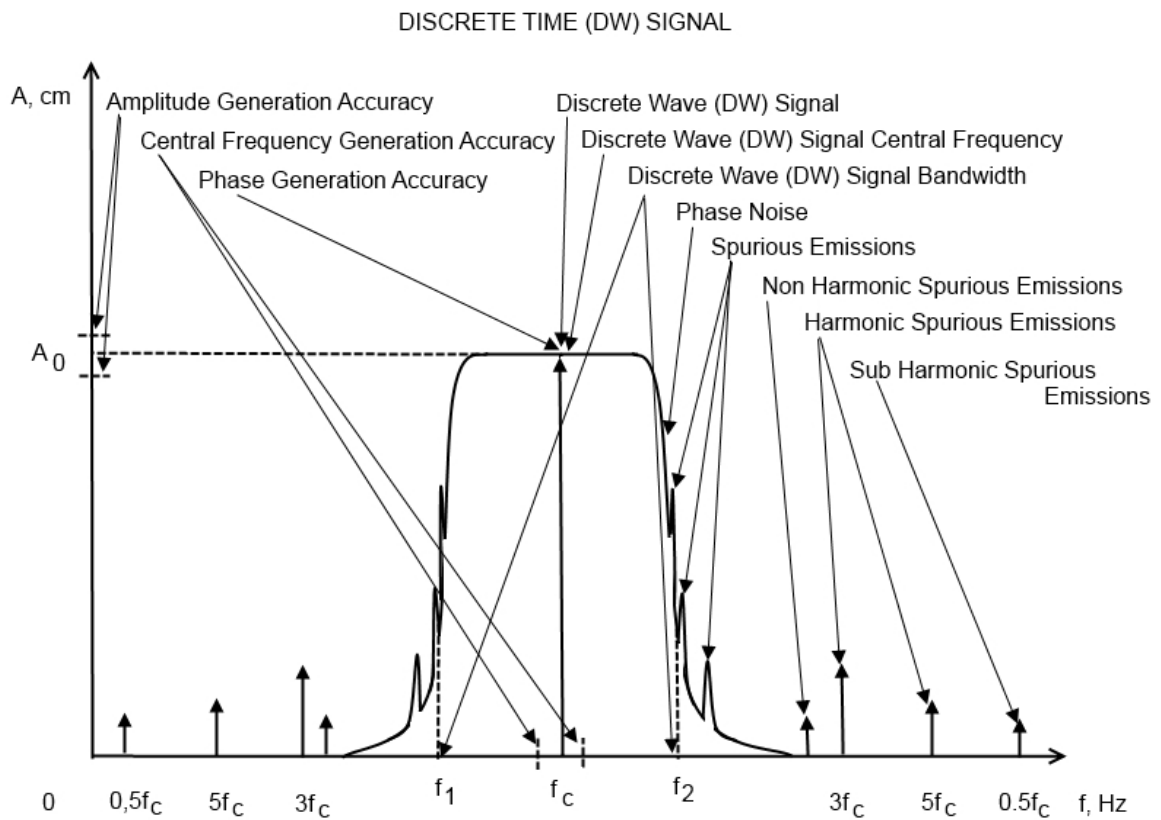

**Fig. 36.** Vector-modulated discrete-time digital direct sequence spread spectrum signal in form of dependence of amplitude on frequency in Maxwell electrodynamics and in Walsh discrete-time digital signal processing science. Note: Constant signal power level is radiated.

We would like to explain that, in the high speed wireline/wireless/optical data communication links, the communicated information (data) can be encoded into the discrete-time digital signal by applying the high order digital modulation techniques. The digital modulation by the phase modification is a most frequently used method for the information encoding, because the electromagnetic signal's phase change can be precisely controlled in the state-of-the-art digital electronic circuits:

$$y_i = A_i \sin\left(2\pi f_i(t) + \phi_i(t)\right),$$

$$\textbf{\textit{where}} \quad \textbf{\textit{BPSK}} : \phi(t) = 1, 2$$

$$\textbf{\textit{QPSK}} : \phi(t) = 1, 2, 3, 4$$

$$\textbf{\textit{MPSK}} \, \phi(t) = 1, 2, 3, 4, ..., \textbf{\textit{i}}.$$

There is a number of very efficient high-order digital modulation techniques with the amplitude/phase modifications in Matlab (2012, 2014):

1. The Bipolar (Binary) Phase-Shift Keying phase shift keying (BPSK);
2. The Differentially Encoded Binary Phase-Shift Keying (DE-BPSK);
3. The Quadrature (Quaternary) Phase-Shift Keying (QPSK);
4. The Differentially Encoded Quaternary Phase-Shift Keying (DE-QPSK);
5. The Offset Quaternary Phase-Shift Keying (OQPSK);
6. The Differentially Encoded Offset Quaternary Phase-Shift Keying (DE-OQPSK);
7. The M-ary Phase-Shift keying (M-PSK);
8. The M-ary Differential Phase-Shift Keying (M-DPSK);
9. The M-ary Pulse Amplitude Modulation (M-PAM);
10. The M-ary Quadrature Amplitude Modulation (M-QAM);
11. The M-ary Frequency-Shift Keying (M-FSK);
12. The Differentially Encoded M-ary Phase-Shift Keying (DE-M-PSK)
13. The M-ary Continuous-Phase Frequency-Shift Keying (M-CPFSK)
14. The Minimum Shift Keying (MSK).

Discussing the information encoding with the digital modulation by the phase modification in details, let us introduce a scientific notion of the phasor in the phase space of the discrete-time digital signal. We can assume that the phasor is a vector of certain magnitude, which is tilted on the alpha angle (the phase) in the phase space of the discrete-time digital signal in the XY coordinates system. The phasor notion is frequently used to explain a basic idea on the information encoding by the digital modulations with the phase/amplitude modification to the input discrete-time digital signal in the digital electronics. The Imaginary-Quadrature (IQ) modulation vector diagram displays the discretely changing position of the digital modulation phasor vector in Wikipedia (2016k, l), Matlab (2012, 2014).

Fig. 37 pictures the phasor vector for the digital signal modulation by the phase modification over the time at IQ constellation diagram in the digital electronics.

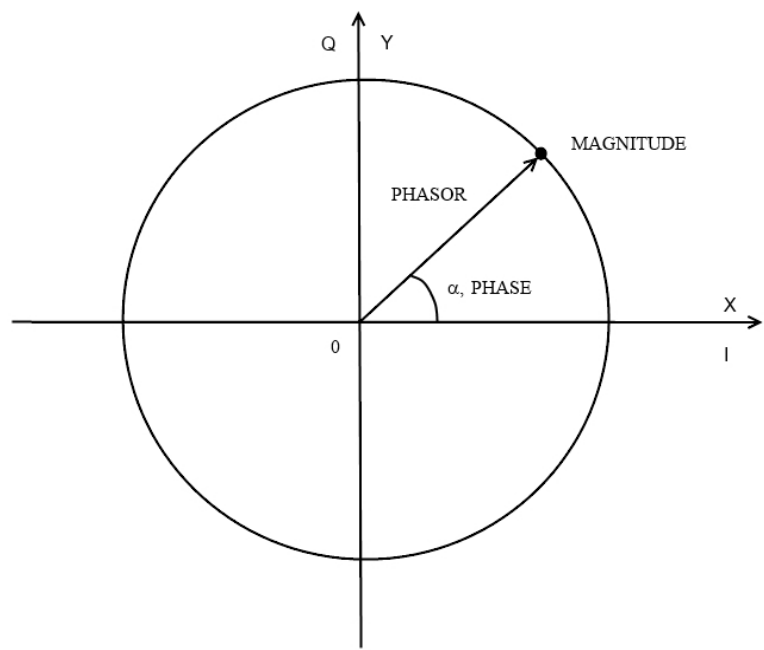

**Fig. 37.** Phasor vector to demonstrate digital signal modulation by phase modification over time at IQ constellation diagram in digital electronics.

We can graphically demonstrate the high order digital modulation techniques in the digital transceivers in the digital electronics by using the phasor notion. For example, let us show the following high order digital modulation techniques at the IQ constellation diagrams:

*1.* The Bipolar Phase Shift Keying (BPSK) digital modulation technique;

*2.* The Quadrature Phase Shift Keying (QPSK) digital modulation technique;

*3.* The 64 Quadrature Amplitude Modulation (64 QAM) digital modulation technique.

Fig. 38 displays the information encoding with the Bipolar Phase Shift Keying (BPSK) technique at the IQ constellation diagram in the digital electronics.

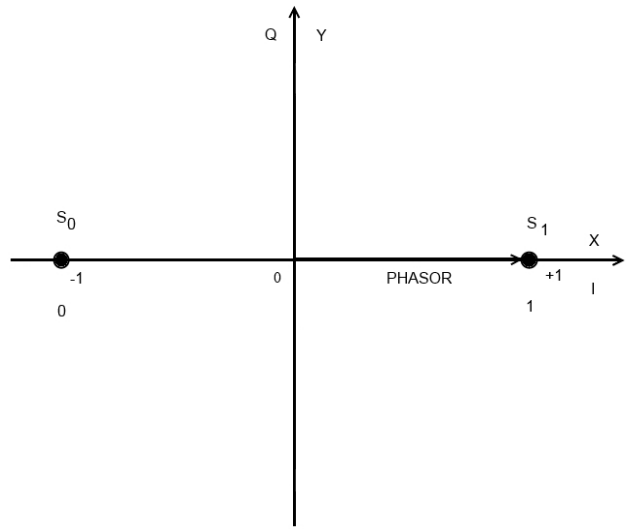

**Fig. 38.** Information encoding with bipolar phase shift keying (BPSK) at IQ constellation diagram in digital electronics.

Fig. 39 demonstrates the information encoding with the Quadrature Phase Shift Keying (QPSK) technique at the IQ constellation diagram in the digital electronics.

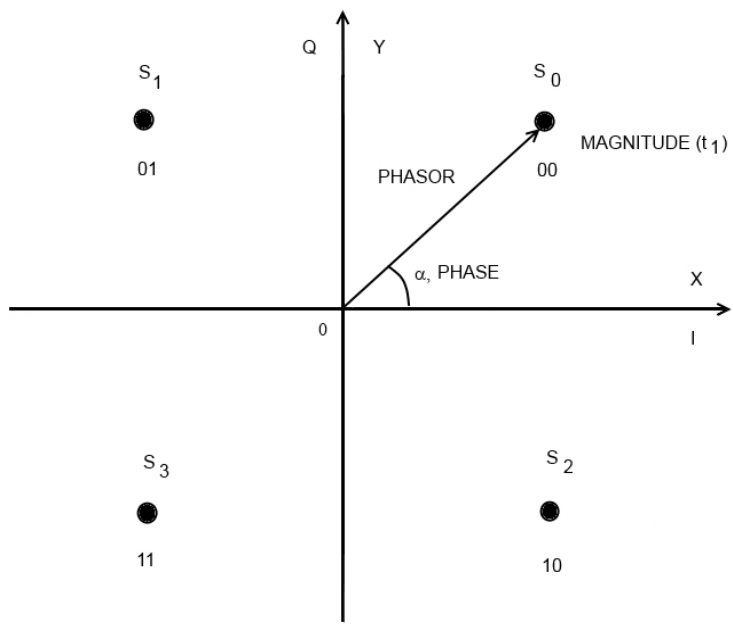

**Fig. 39.** Information encoding with Quadrature Phase Shift Keying (QPSK) technique at IQ constellation diagram in digital electronics.

Fig. 40 shows the information encoding with the 64 Quadrature Amplitude Modulation (64 QAM) technique at the IQ constellation diagram, using the amplitude and the phase simultaneous modifications over the time in the digital electronics.

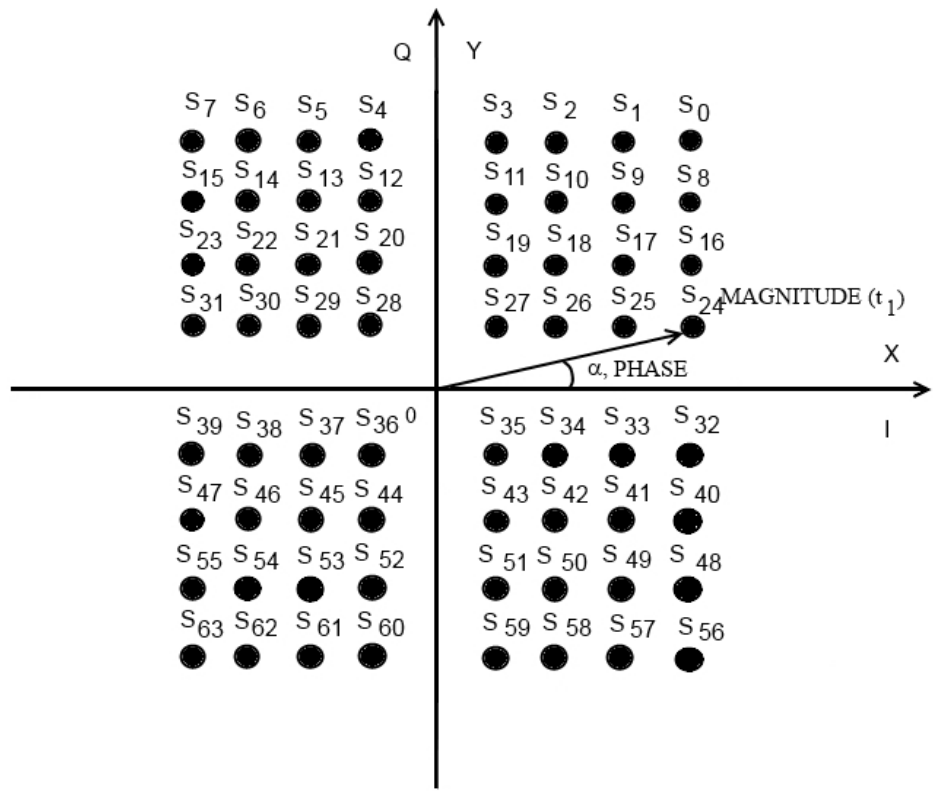

**Fig. 40.** Information encoding with 64 Quadrature Amplitude Modulation (64 QAM) at IQ constellation diagram in digital electronics.

The high order digital modulation techniques quality can be measured, using the Error Vector Magnitude (EVM) theoretical concept. Therefore, we can say that the Error Vector Magnitude theoretical concept has been formulated by the researchers to analyze the quality of the information encoding with the high order digital modulation techniques (BPSK, DE-BPSK, QPSK, DE-QPSK, OQPSK, DE-OQPSK, M-PSK, M-DPSK, M-PAM, M-QAM, M-FSK, DE-M-PSK, M-CPFSK, MSK) in the digital transceivers in the digital electronics. The EVM can be measured by making a comparison between the comparing the measured digitally modulated signal and the reference digitally modulated signal. The EVM can be displayed on the IQ constellation diagram.

Fig. 41 presents the modulation accuracy conception demonstration, based on the error vector magnitude (EVM) representation at the constellation diagram in the digital electronics.

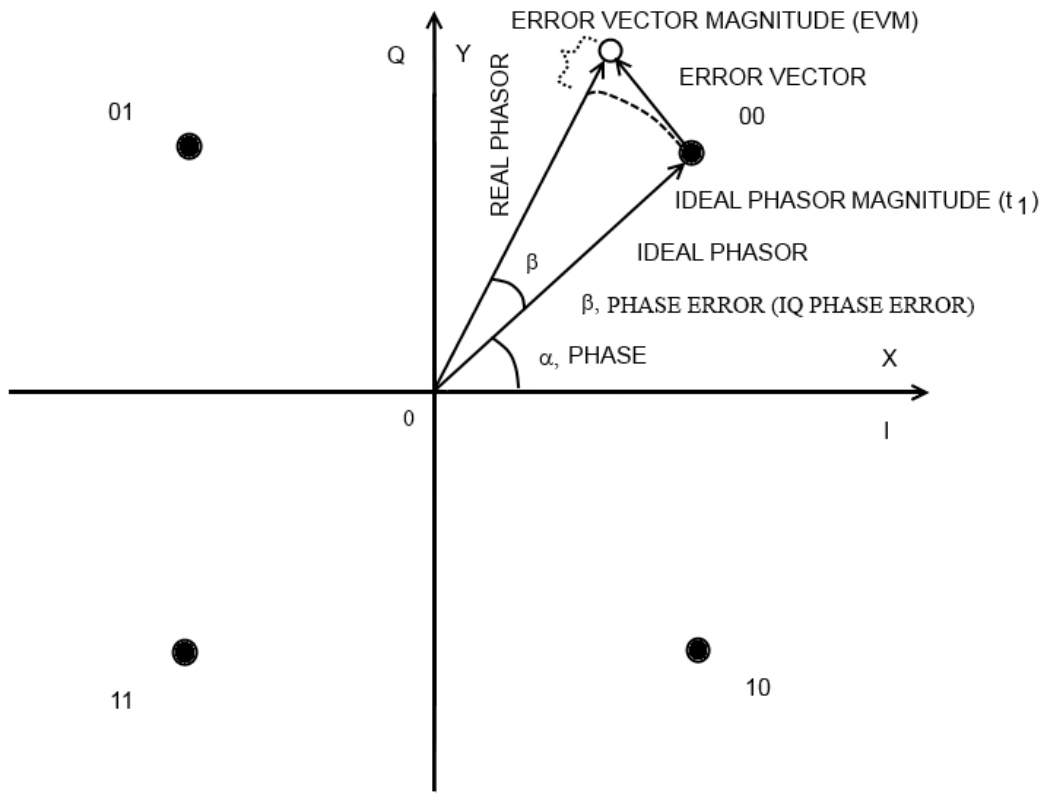

**Fig. 41.** Modulation accuracy measurement conception demonstration, based on error vector magnitude (EVM) representation at IQ constellation diagram in digital electronics.

Now, let us draw the RF/digital electronic elements scheme of the In-phase and Quadrature (IQ) (de)modulator of the analog/digital signals in electronics. The main technical purpose of the In-phase and Quadrature (IQ) (de)modulator of the analog/digital signals is to encode the information (the data streams) into the discrete-time digital signal by applying the high order digital modulation techniques (BPSK, DE-BPSK, QPSK, DE-QPSK, OQPSK, DE-OQPSK, M-PSK, M-DPSK, M-PAM, M-QAM, M-FSK, DE-M-PSK, M-CPFSK, MSK) in transceivers in the digital electronics.

Fig. 42 provides a scheme of the In-phase and Quadrature (IQ) (de)modulator of the analog/digital signals in the digital electronics.

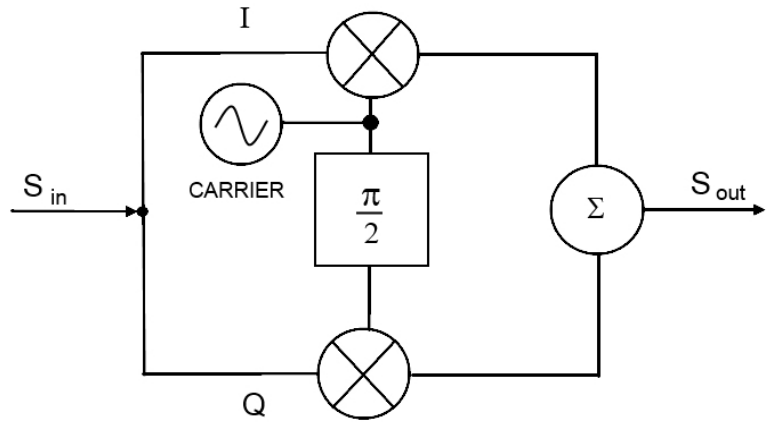

**Fig. 42.** Scheme of In-phase and Quadrature (IQ) (de)modulator of analog/digital signals in digital electronics.

As we discussed above, the communicated information (data) can be encoded into the discrete-time digital signal by applying the high order digital modulation techniques (BPSK, DE-BPSK, QPSK, DE-QPSK, OQPSK, DE-OQPSK, M-PSK, M-DPSK, M-PAM, M-QAM, M-FSK, DE-M-PSK, M-CPFSK, MSK) in transceivers in the digital electronics.

In practice, it is done by applying the In-phase and Quadrature (IQ) (de)modulator of the digital signals in the digital electronics. More specifically, the electromagnetic continuous-time carrier signal, is modulated by shifting the phase and by changing the amplitude simultaneously and independently in the process of operation of the In-phase and Quadrature (IQ) (de)modulator of digital signals in the digital electronics.

The main idea on the digital modulation is demonstrated, using the phasor-vector conception in the above I/Q constellation diagrams.

However, it is necessary to explain that the generated discrete-time digital vector-modulated signals can be increasingly complex, because of the digital modulations, coding and spreading techniques applied. Therefore, the following parameters of the generated discrete-time digital vector-modulated signals must be measured, analyzed and controlled precisely:

| | |
|---|---|
| *1.* Amplitude; | *8.* Gain; |
| *2.* Frequency; | *9.* Group delay; |
| *3.* Phase; | *10.* Modulation errors; |
| *4.* Period; | *11.* Coding errors; |
| *5.* Time; | *12.* Spreading errors; |
| *6.* Power; | *13.* Bits/Symbols/Chips timing; |
| *7.* Waveform; | *14.* Bits/Symbols/Chips offset. |

**Tab. 2.** Parameters of vector-modulated discrete-time digital signal.

In the software defined radio in the digital electronics, the generated discrete-time digital vector-modulated signals can be accurately characterized by using the following measurement diagrams:

*1.* Spectral diagram (Signal spectrum parameters);

*2.* I/Q Constellation diagram (Modulation parameters);

*3.* Error Vector Magnitude (EVM) diagram (Modulation parameters);

*4.* Bit Error Rate (BER) diagram (Information transmission parameters);

*5.* Eye diagram (Data rates parameters).

In principle, the software defined radio includes the RF transmitter and the RF receiver, which can be regarded as the RF transceiver in many cases.

The RF transmitter in the software defined radio has both parts:

*1.* The digital electronics part;

*2.* The analog electronics part.

Going from the RF engineering point of view, it may be interesting to explain that:

*1.* The digital electronics part of RF transmitter in the software defined radio is made in the form of the digital signal processor (DSP);

*2.* The analog electronics part of RF transmitter in the software defined radio is made with application of Au/Ag/Cu microstrip line technology at the multilayered electronic elements board, and the RF high-power wide-band solid-state amplifier, and the RF high power dielectric resonator-filter(s).

The transmitter in the software defined radio is usually designed, using the following digital- and analog- electronics parts block scheme.

Fig. 43 shows the digital transmitter scheme in software defined radio in the digital electronics.

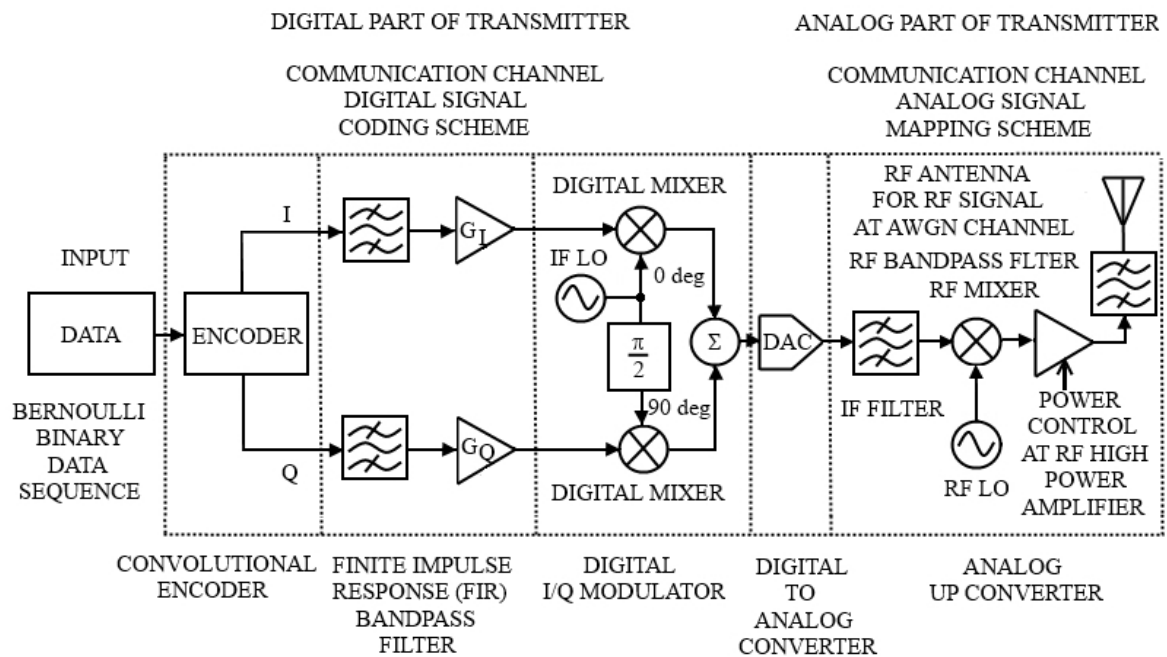

**Fig. 43.** Digital transmitter scheme in software defined radio in digital electronics.

The RF receiver in the software defined radio has also 1) the analog electronics part and 2) the digital electronics part.

Once again, it is necessary to highlight an interesting fact that:

*1.* The analog electronics part of RF receiver in the software defined radio is made with application of Au/Ag/Cu microstrip line technology at the multilayered electronic elements board, and the RF high power wide band solid state amplifier, and the RF high power dielectric resonator filter(s).

*2.* The digital electronics part of RF receiver in the software defined radio is produced in the form of the digital signal processor (DSP) in the case of big series production or the Field Programmable Gate Array (FPGA) in the case of small series production;

The RF receiver in the software defined radio is normally designed, utilizing the following analog- and digital- electronics parts block scheme.

Fig. 44 shows the digital receiver scheme in software defined radio in the digital electronics.

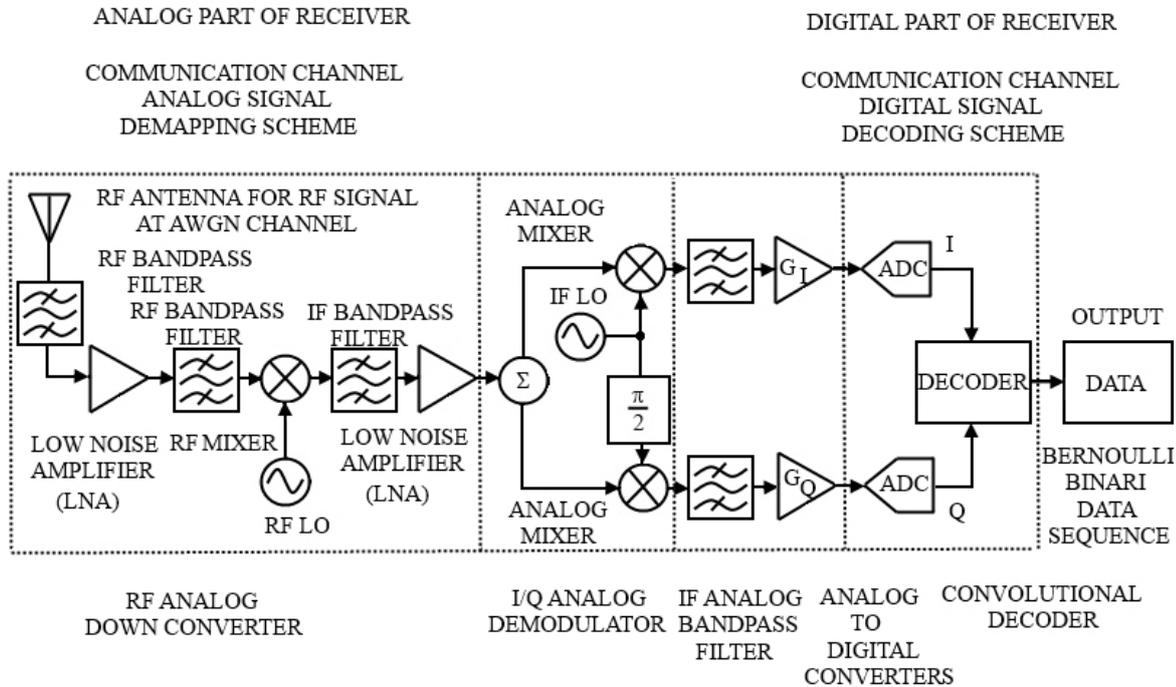

**Fig. 44.** Digital receiver scheme in software defined radio in digital electronics.

Now, let us return back to the discussion on the economics related problems, explaining that all the economic analysis methods, which attempt to collect the statistical time-series data, then to filter out the continuous-time economic output wave, and subsequently to accurately characterize the continuous-time economic output wave in the selected economy of the scale and the scope over the predetermined time period, are restricted to a slightly outdated questionable scientific representation that all the economic output parameters must change continuously in the selected economy of the scale and the scope in the time domain.

In reality, the economic output parameters in the modern economy of the scale and the scope have the discrete-time nature instead of the continuous-time nature. More clearly, the economic output parameters in the modern economy of the scale and the scope change discretely in the amplitude, frequency, period, phase and time domains. Therefore, in our opinion, all the continuous–time signal theories and testing approaches cannot be blindly applied to analyze the discretely changing economic output parameters in the modern economy of the scale and the scope.

Well, switching our research attention to the Ledenyov discrete-time digital signals to accurately characterize the business cycles in the Ledenyov classic econodynamics, let us explain that the Ledenyov discrete-time digital signals theory was created on the knowledge base in the Walsh discrete-time signals processing theory science in Ledenyov V O, Ledenyov D O (2016s, 2017). One of the most important discrete-time signals properties is that the discrete-time digital signals can sharply/abruptly/instantly change the amplitude, frequency and phase over the time, because of their digital nature. In other words, the Ledenyov discrete-time digital waves, which characterize the dependence of GDP(t, monetary base), can change abruptly in the time domain. This fact is further demonstrated/confirmed during the spectral analysis of the existing GDP(t, monetary base) dependences with the complex discrete-time digital signal waveforms in various economies of the scales and the scopes over the time.

Accurately characterizing the Ledenyov discrete-time digital waves, it is necessary to highlight a fact that the nature of the oscillations by the economic variables in the macroeconomics is discrete, because these fluctuations are mainly caused by the discrete-time events in Ledenyov D O, Ledenyov V O (2015e). The possible examples of the discrete-time events are: *1)* the Schumpeter's creative disruptive innovations in Schumpeter (1939, 1942, 1947), Isaacson (6 October 2015), *2)* the unexpected changes in the supply/demand chain goods/services delivery in various markets, *3)* the instant sharp change in the financial stability / the monetary stability policies by the central bank, *4)* the unpredicted change in the economic state course by the government; the sharp change of political state course by the government in Ledenyov D O, Ledenyov V O (2015e).

The discrete nature of the innovation breakthrough processes, which are known as the Schumpeter's creative innovative disruptions during the capitalism evolution was researched in Schumpeter (1911, 1939, 1947), Christensen (June 16, 1977; Fall, 1992a, b; 1997; 1998; December, 1998; April, 1999a, b, c; 1999a, b; Summer, 2001; June, 2002; 2003; March, April, 2003; January, 2006), Bower, Christensen (January, February, 1995; 1997; 1999), Christensen, Armstrong (Spring, 1998), Christensen, Cape (December, 1998), Christensen, Dann (June, 1999), Christensen, Tedlow (January, February, 2000), Christensen, Donovan (March, 2000; May, 2010), Christensen, Overdorf (March, April, 2000), Christensen, Bohmer, Kenagy (September, October, 2000), Christensen, Craig, Hart (March, April, 2001), Christensen, Milunovich (March, 2002), Bass, Christensen (April, 2002), Anthony, Roth, Christensen (April, 2002), Kenagy, Christensen (May, 2002; 2002), Christensen, Johnson, Rigby (Spring, 2002), Hart, Christensen (Fall, 2002), Christensen, Verlinden, Westerman (November, 2002), Shah, Brennan, Christensen (April, 2003), Christensen, Raynor (2003), Burgelman, Christensen, Wheelwright (2003), Christensen, Anthony (January, February, 2004), Christensen, Anthony, Roth (2004), Christensen, Baumann, Ruggles, Sadtler (December, 2006), Christensen, Horn, Johnson (2008), Christensen, Grossman, Hwang (2009), Dyer, Gregersen, Christensen (December, 2009; 2011), Christensen, Talukdar, Alton, Horn (Spring, 2011), Christensen, Wang, van Bever (October, 2013)).

Thus, let us draw the Ledenyov discrete-time digital economic output signal waveforms of GIP(t, monetary base), GDP(t, monetary base), GNP(t, monetary base), PPP(t, monetary base): $S_{1digital}(t)$, $S_{2digital}(t)$, $S_{3digital}(t)$, in the three special considered cases in the economy of the scale and the scope at the certain monetary bases in the seleced time periods in the XY coordinates space. Analyzing the below presented waveforms, we can see that the Ledenyov discrete-time digital economic output signal waveform in the case of strong distortions, $S_{3digital}(t)$ represents the fluctuations of the magnitudes of the statistical data of GDP(t), which are usually observed in the real economies of the scales and the scopes in the real life conditions. The distortions would certainly be able to distort the Ledenyov discrete-time digital economic output signal during its propagation in the real economies of the scales and the scopes over the time in the real life case scenario. They can be caused by the different regulatory/economic/ financial/political cultural impact factors, which can be present in the real economies of the scales and the scopes at certain time periods.

Fig. 45 shows the Ledenyov discrete-time digital economic output signal waveforms of GIP(t, monetary base), GDP(t, monetary base), GNP(t, monetary base), PPP(t, monetary base) in XY coordinates in the Ledenyov classic econodynamics science: *a)* The Ledenyov discrete-time digital economic output signal waveform in the case of no distortions, $S_{1digital}(t)$; *b)* The Ledenyov discrete-time digital economic output signal waveform in the case of weak distortions, $S_{2digital}(t)$; *c)* The Ledenyov discrete-time digital economic output signal waveform in the case of strong distortions, $S_{3digital}(t)$.

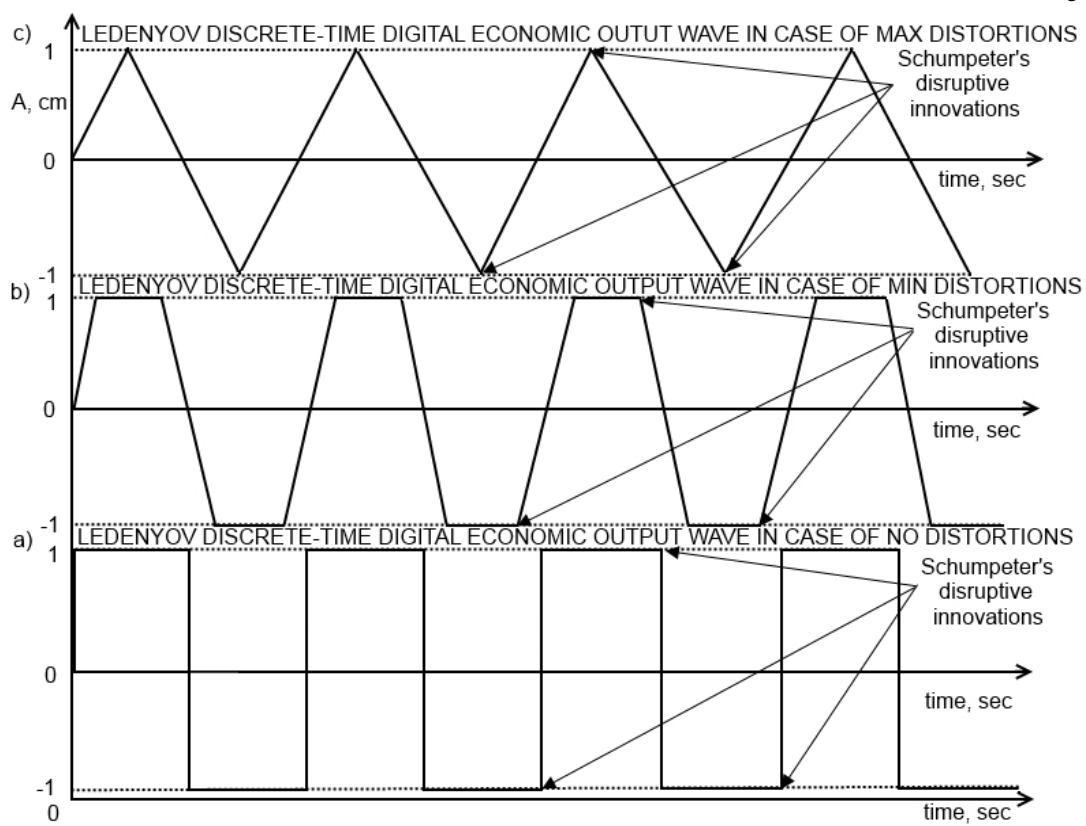

**Fig. 45.** Ledenyov discrete-time digital economic output signal waveforms of GIP(t, monetary base), GDP(t, monetary base), GNP(t, monetary base), PPP(t, monetary base) in XY coordinates space: *a)* Ledenyov discrete-time digital economic output signal waveform in case of no distortions, $S_{1digital}(t)$; *b)* Ledenyov discrete-time digital economic output signal waveform in case of weak distortions, $S_{2digital}(t)$; *c)* Ledenyov discrete-time digital economic output signal waveform in case of strong distortions, $S_{3digital}(t)$.

Now, let us add the digital signals generated/encoded/spread by the high order modulation/encoding/spreading techniques under the action by the Schumpeter disruptive innovations of various origins and other disruptions, to our initial consideration, W3 and W4.

Fig. 46 shows the Ledenyov discrete-time digital economic output signal waveforms of GIP(t, monetary base), GDP(t, monetary base), GNP(t, monetary base), PPP(t, monetary base) in the XY coordinates space in the Ledenyov classic econodynamics science: *a)* The Ledenyov discrete-time digital wave in the ideal case of no distortions; *b)* The Ledenyov discrete-time digital wave in the real case of present distortions; *c)* The Ledenyov vector-modulated discrete-time digital wave in the ideal case of no distortions, when the high order modulation is applied; *d)* The Ledenyov vector-modulated discrete-time digital wave in the real case of present distortions, when the high order modulation is applied.

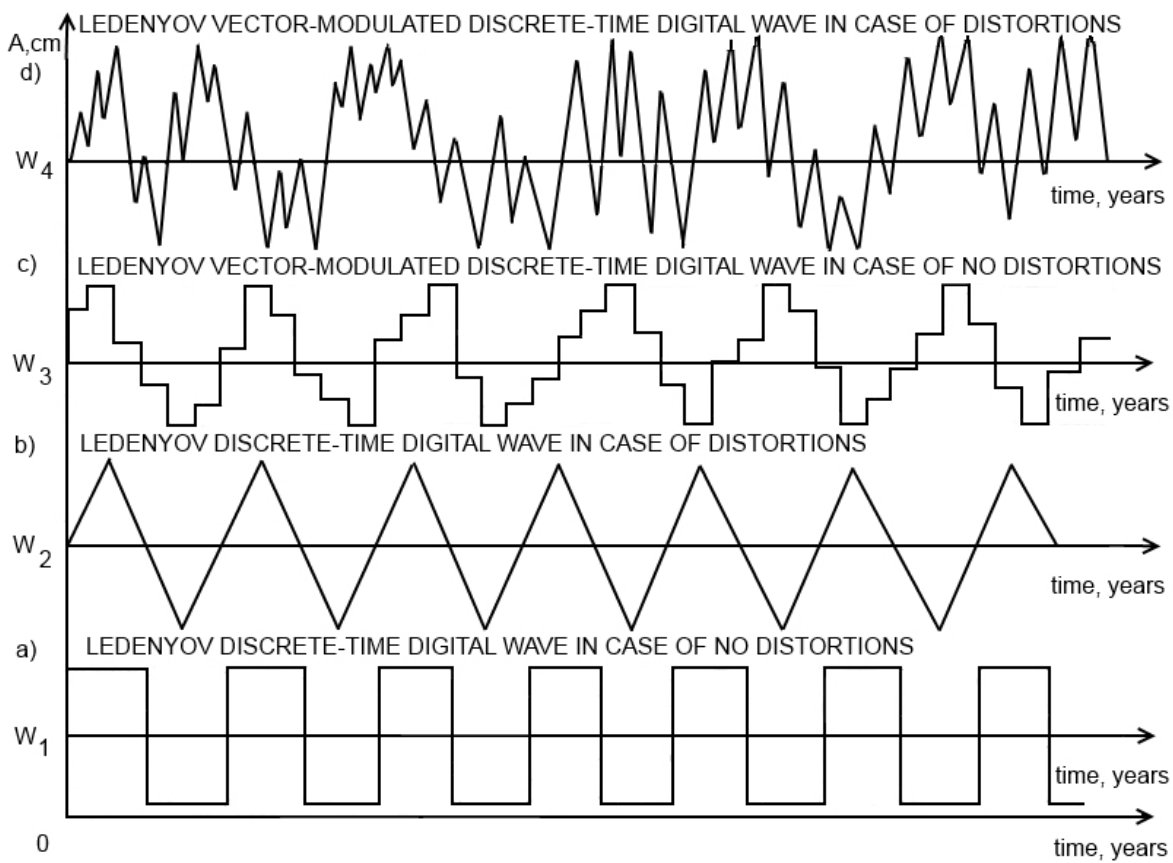

**Fig. 46.** Ledenyov discrete-time digital economic output signal waveforms of GIP(t, monetary base), GDP(t, monetary base), GNP(t, monetary base), PPP(t, monetary base) in XY coordinates space in Ledenyov classic econodynamics science: *a)* Ledenyov discrete-time digital wave in cases of no distortions; *b)* Ledenyov discrete-time digital wave in case of present distortions; *c)* Ledenyov vector-modulated discrete-time digital wave in case

of no distortions; *d)* Ledenyov vector-modulated discrete-time digital wave in case of present distortions.

Now, we are going to consider the phase diagram of the Ledenyov discrete-time digital economic output wave of GIP(t, monetary base), GDP(t, monetary base), GNP(t, monetary base), PPP(t, monetary base) on the Ledenyov IQ constellation diagram in the Ledenyov classic econodynamics.

Fig. 47 presents a phase diagram of the economic output data stream encoding by using the econodynamic high-order Ledenyov Quadrature Amplitude Modulation (QAM) into the Ledenyov discrete-time digital economic output wave of GIP(t, monetary base), GDP(t, monetary base), GNP(t, monetary base), PPP(t, monetary base) at the Ledenyov IQ constellation diagram. The amplitude / phase simultaneous modifications in the economy of the scale and the scope at the certain monetary base in the selected time periods in the Ledenyov classic econodynamics are applied.

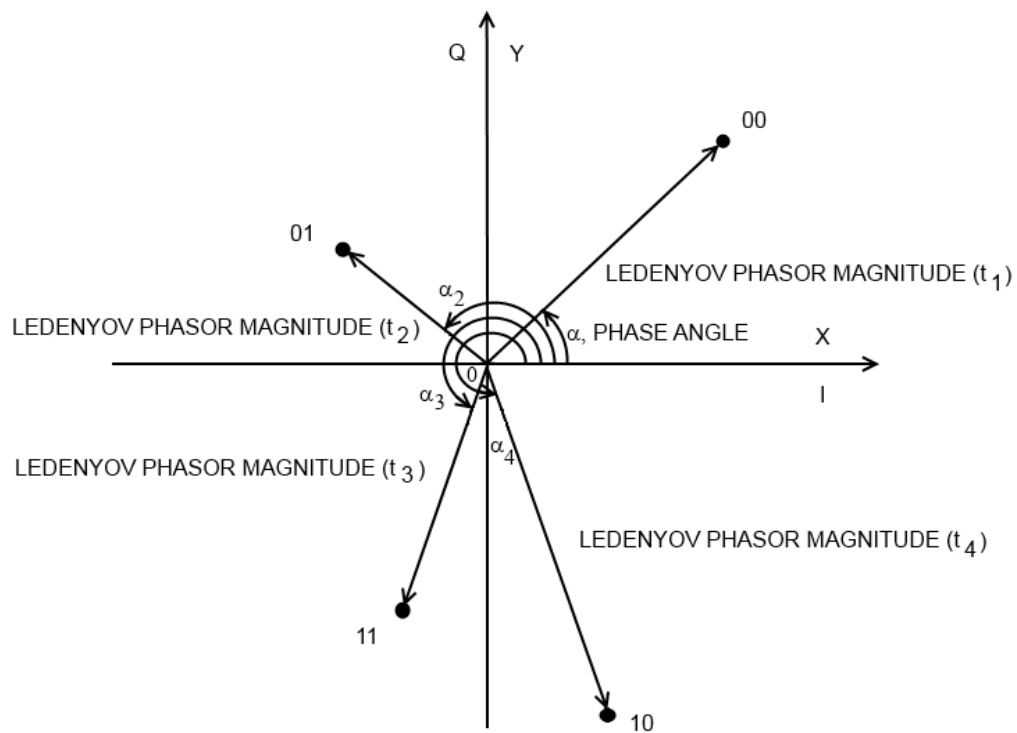

**Fig. 47.** Information encoding with econodynamic high order Ledenyov Quadrature Amplitude Modulation (QAM) into Ledenyov discrete-time digital economic output wave of GIP(t, monetary base), GDP(t, monetary base), GNP(t, monetary base), PPP(t, monetary base) at Ledenyov IQ

constellation diagram, using amplitude and phase simultaneous modifications in time in economy of scale and scope in Ledenyov classic econodynamics.

Now, let us draw the Ledenyov discrete-time economic output wave as the Ledenyov vector-modulated discrete-time digital direct sequence spread spectrum signal of GIP(t, monetary base), GDP(t, monetary base), GNP(t, monetary base), PPP(t, monetary base) in the economy of the scale and the scope at the certain monetary base over the time in the XYZ coordinates in the ideal case without the distortions in Ledenyov classic econodynamics.

Fig. 48 pictures Ledenyov vector-modulated discrete-time digital direct sequence spread spectrum signal of GIP(t, monetary base), GDP(t, monetary base), GNP(t, monetary base), PPP(t, monetary base) in the economy of the scale and the scope at the certain monetary base in the selected time period in the XYZ coordinates space in an ideal case without the distortions in Ledenyov classic econodynamics.

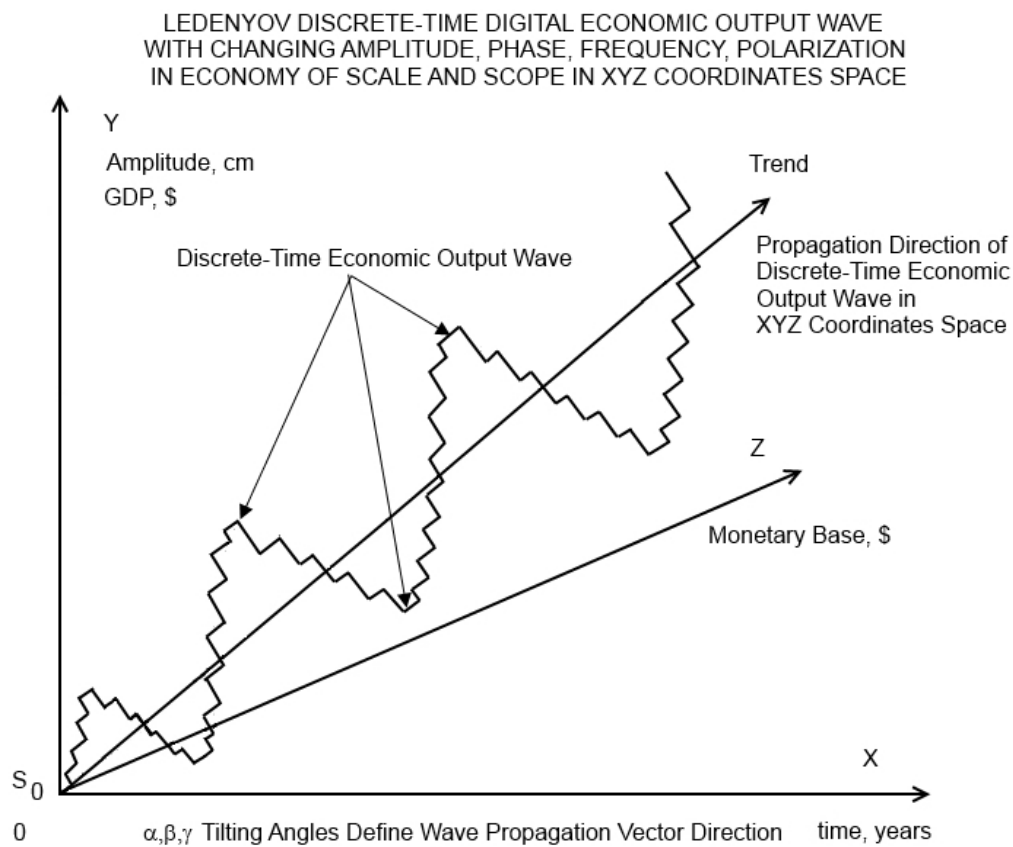

**Fig. 48.** Ledenyov graphic scheme of Ledenyov discrete-time economic output wave in form of vector-modulated discrete-time digital direct sequence spread spectrum signal of GIP(t, monetary base), GDP(t, monetary base), GNP(t, monetary base), PPP(t, monetary base) with changing

amplitude, frequency, period, phase in economy of scale and scope at certain monetary bases in selected time period in XYZ coordinates space in the ideal case without the distortions in Ledenyov classic econodynamics.

Let us draw the Ledenyov discrete-time economic output wave as the vector-modulated discrete-time digital direct sequence spread spectrum signal of GIP(t, monetary base), GDP(t, monetary base), GNP(t, monetary base), PPP(t, monetary base) in the economy of the scale and the scope at the certain monetary base in the selected time period in the XYZ coordinates in the case of the strong distortions in Ledenyov classic econodynamics.

Fig. 49 pictures the Ledenyov discrete-time economic output wave in form of the vector-modulated discrete-time digital direct sequence spread spectrum signal of GIP(t, monetary base), GDP(t, monetary base), GNP(t, monetary base), PPP(t, monetary base) in the economy of the scale and the scope at the certain monetary base in the selected time period in the XYZ coordinates space in the case of the strong distortions in Ledenyov classic econodynamics.

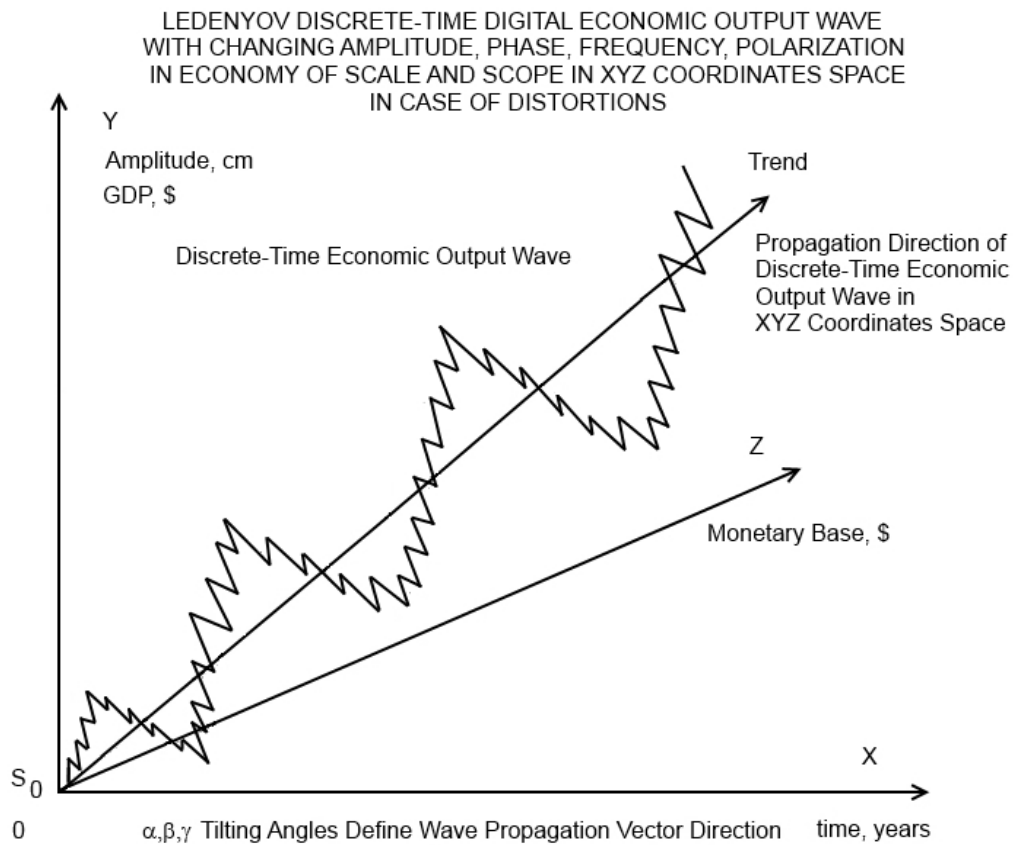

**Fig. 49.** Ledenyov graphic scheme of Ledenyov discrete-time economic output wave in form of vector-modulated discrete-time digital direct

sequence spread spectrum signal of GIP(t, monetary base), GDP(t, monetary base), GNP(t, monetary base), PPP(t, monetary base) with changing amplitude, frequency, period, phase in economy of scale and scope at certain monetary bases in selected time period in XYZ coordinates space in the case of the strong distortions in Ledenyov classic econodynamics.

In the Chapter 3, we investigated some aspects on the Ledenyov discrete-time digital economic output wave in the form of the Ledenyov vector-modulated direct sequence spread spectrum signal of GIP(t, monetary base), GDP(t, monetary base), GNP(t, monetary base), PPP(t, monetary base), which can be synthesized/encoded/spread/propagated due to the Schumpeter disruptive innovations introduction into the economy of the scale and the scope at the certain monitary base over the selected time period in an agreement with the Ledenyov classic econodynamics.

However, we may assume that there may be many other more complex types of the Ledenyov discrete-time digital waves of GIP(t, monetary base), GDP(t, monetary base), GNP(t, monetary base), PPP(t, monetary base) with the quite complex waveforms, generated by an application of the high order digital modulation, encoding and spreading techniques in the economy of the scale and the scope at the certain monetary bases over the selected time periods as permitted by the Ledenyov classic econodynamics.

Therefore, in the next Chapter 4, we want to study the Ledenyov discrete-time digital economic output wave in the form of the vector-modulated direct sequence spread spectrum signal bursts of GIP(t, monetary base), GDP(t, monetary base), GNP(t, monetary base), PPP(t, monetary base), which can be generated in the economy of the scale and the scope at the certain monetary base over the selected time in the frames of the Ledenyov classic econodynamics.

# Chapter 4

## Discrete-time digital economic output waves in form of vector-modulated discrete-time digital direct sequence spread spectrum signal's bursts in economy of scale and scope in classic econodynamics

In the Ledenyov classic econodynamics science, we would like to consider the Ledenyov discrete-time digital economic output waves in the form of the Ledenyov discrete-time digital vector-modulated direct sequence spread spectrum signal bursts of GIP(t, monetary base), GDP(t, monetary base), GNP(t, monetary base), PPP(t, monetary base), which, in our opinion, can originate, propagate and be observed in the economy of the scale and the scope at the certain monetary base over the selected observation time period in agreement with the in Ledenyov classic econodynamics.

We prefer to begin by providing some technical information on the Frequency Division Duplex (FDD) / Time Division Duplex (TDD) electromagnetic spectrum access technologies for the discrete-time digital wave(s), which can be generated/transmitted/propagated in form of vector-modulated discrete-time digital Direct Sequence Spread Spectrum (DSSS) signal bursts in the electromagnetic spectrum (the multidimensional space) over the time in agreement with the well known principles and theories in the Maxwell electrodynamics science. The Frequency Division Duplex (FDD) / Time Division Duplex (TDD) vector-modulated discrete-time digital Direct Sequence Spread Spectrum (DSSS) signal bursts in the multidimensional space over the time, were researched in Dixon (1976), Viterbi (May 1979), Pickholtz, Schilling, Milstein (May 1982), Simon, Omura, Scholtz, Levitt (1985, 1994), Simon, Moher (2008), Rappaport (January 2010), Miao, Zander, Sung, Slimane (2016).

Presently, the achieved technical progress in the advanced research on the (FDD) / (TDD) discrete-time digital waves in the form of the vector-modulated Direct Sequence Spread Spectrum (DSSS) signal bursts resulted in an appearance of the electronic/photonic devices with the innovative designs of the wireless, wireless, and optical digital high-speed data communication channels/links for the information communication purposes over the short/long distances in the space at the selected time periods.

Fig. 50 displays *a)* The Frequency Division Duplex (FDD) / Time Division Duplex (TDD) vector-modulated discrete-time digital Direct Sequence Spread Spectrum (DSSS) signal bursts in the space over the time in the Maxwell electrodynamics; *b)* The I/Q constellation diagram with the 64 Quadrature Amplitude Modulation (64 QAM) in the Maxwell electrodynamics; *c)* The Spectrum analysis diagram with a dependence of the amplitude on the frequency for the Frequency Division Duplex (FDD) / Time Division Duplex (TDD) vector-modulated discrete-time digital Direct Sequence Spread Spectrum (DSSS) signal bursts in Maxwell electrodynamics.

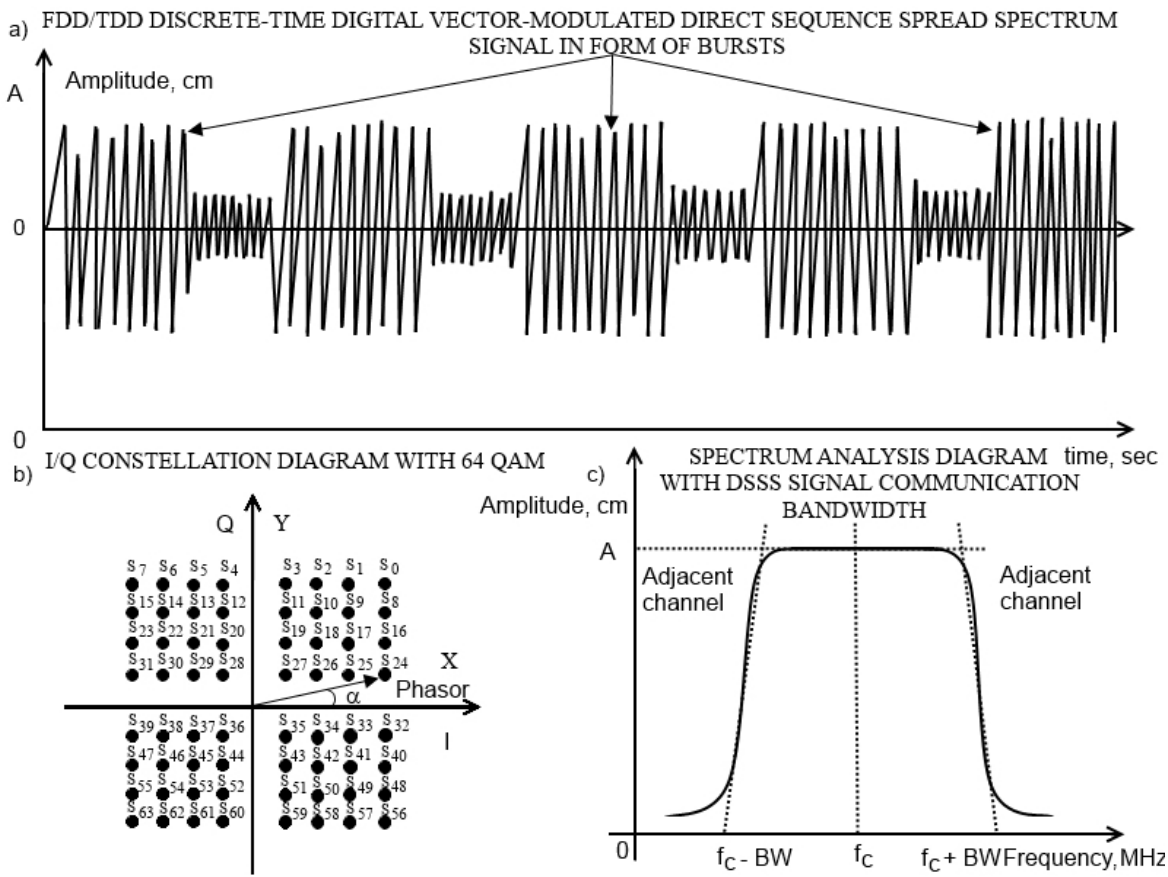

**Fig. 50.** *a)* FDD) / TDD discrete-time digital wave in form of FDD/TDD vector-modulated discrete-time digital DSSS signal bursts in space over time in Maxwell electrodynamics; *b)* I/Q constellation diagram with 64 Quadrature Amplitude Modulation for FDD/TDD vector-modulated discrete-time digital DSSS signal bursts over time in Maxwell

electrodynamics; *c)* Spectrum analysis diagram with dependence of amplitude on frequency for FDD / TDD vector-modulated discrete-time digital Direct Sequence Spread Spectrum (DSSS) signal bursts over time in Maxwell electrodynamics.

At this point in our discussion, we would like to switch our research attention to the Ledenyov classic econodynamics science. Going from the foundational principles of the Ledenyov classic econodynamics science, we tend to think that the Ledenyov discrete-time digital economic output waves in the form of the vector-modulated discrete-time digital direct sequence spread spectrum DSSS signal bursts could be generated at very same moments of time, when the Schumpeter disruptive innovations would be introduced into the economy of the scale and the scope.

In this case, we assume that the economic output carrier signal of the certain amplitude, frequency, phase can be digitally modulated by the high order digital modulation techniques, and then, it can be encoded by the economic specific encoding codes, and after that, it can be digitally spread with the unique economic spreading codes, resulting in a generation of the Ledenyov discrete-time digital economic output wave in the form of the vector-modulated discrete-time digital direct sequence spread spectrum DSSS signal bursts, propagating in the economy of the scale and the scope over the time in the Ledenyov classic econodynamics. In other words, the Ledenyov discrete-time digital economic output waves in the form of the vector-modulated discrete-time digital direct sequence spread spectrum DSSS signal bursts represents the modulated, encoded and spread carrier signals of the certain amplitude, frequency, phase, propagating in the economy of the scale and the scope over the time in the Ledenyov classic econodynamics. The modulation/encoding/spreading takes place, because of the Schumpeter disruptive innovations introduction and can also be defined by some other factors (the economic regulatory law base).

Generally, we can say that the Ledenyov discrete-time digital economic output waves in the form of the vector-modulated discrete-time digital direct sequence spread spectrum DSSS signal bursts represent the dependencies of GIP(t, monetary base), GDP(t, monetary base), GNP(t, monetary base), PPP(t, monetary base), which can be measured in the economy of the scale and the scope at the certain monetary base over the selected observation time periods in Ledenyov D O, Ledenyov V O (2015d, e, f), Ledenyov D O, Ledenyov V O (2016r), Ledenyov V O, Ledenyov D O (2016s, 2017).

Fig. 51 pictures *a)* The Ledenyov discrete-time digital economic output waves in the form of the vector-modulated discrete-time digital direct sequence spread spectrum DSSS signal bursts of GIP(t, monetary base), GDP(t, monetary base), GNP(t, monetary base), PPP(t, monetary base), propagating in the economy of the scale and the scope over the time in the Ledenyov classic econodynamics; *b)* The Ledenyov I/Q constellation diagram with the econodynamic 64 Qudrature Amplitude Modulation (QAM) in the Ledenyov classic econodynamics; *c)* The spectrum analysis diagram with the Ledenyov transmission bandwidth of the Ledenyov discrete-time digital economic output waves in the form of the vector-modulated discrete-time digital direct sequence spread spectrum DSSS signal bursts of GIP(t, monetary base), GDP(t, monetary base), GNP(t, monetary base), PPP(t, monetary base) in the economy of the scale and the scope over the time in the Ledenyov classic econodynamics.

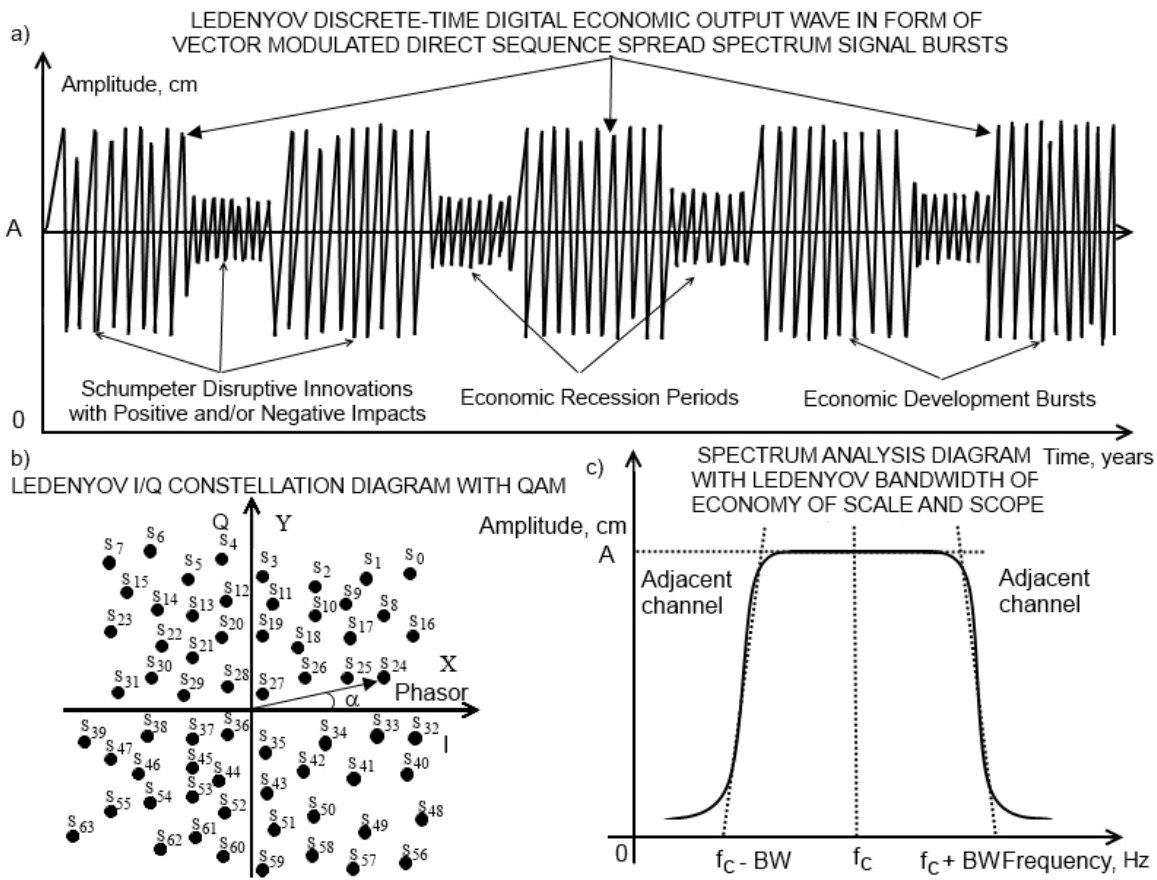

**Fig. 51.** *a)* Ledenyov discrete-time digital economic output waves in the form of the vector-modulated discrete-time digital direct sequence spread spectrum DSSS signal bursts of GIP(t, monetary base), GDP(t, monetary base), GNP(t, monetary base), PPP(t, monetary base), propagating in economy of scale and scope over time in Ledenyov classic econodynamics; *b)* Ledenyov I/Q constellation diagram with econodynamic 64 Qudrature Amplitude Modulation (QAM) in Ledenyov classic econodynamics; *c)* Spectrum analysis diagram with Ledenyov transmission bandwidth of Ledenyov discrete-time digital economic output waves in the form of the vector-modulated discrete-time digital direct sequence spread spectrum DSSS signal bursts of GIP(t, monetary base), GDP(t, monetary base), GNP(t, monetary base), PPP(t, monetary base) in economy of scale and scope in Ledenyov classic econodynamics.

Thus, in Ledenyov classic econodynamics, we tend to believe that the Schumpeter disruptive innovations have a natural ability to modulate, encode and spread the economic output carrier signal digitally by discretely changing its magnitude, phase, frequency parameters over the selected time period during the high order digital modulations, encoding and spreading processes with the unique economy-specific modulation, encoding and spreading codes, which are defined by the Schumpeter's disruptive innovations origination/introduction/ adaptation/application origins. Therefore, let us consider comprehensively the waveforms of the Ledenyov discrete-time digital economic output waves in the form of the vector-modulated discrete-time digital direct sequence spread spectrum DSSS signal bursts of GIP(t, monetary base), GDP(t, monetary base), GNP(t, monetary base), PPP(t, monetary base), propagating in economy of scale and scope over time in Ledenyov classic econodynamics.

Fig. 52 presents the Ledenyov discrete-time waves of of GIP(t, monetary base), GDP(t, monetary base), GNP(t, monetary base), PPP(t, monetary base), in the economy of the scale and the scope at certain monetary base over time. The Ledenyov simple discrete-time digital wave, Wave 1 (W1); the Ledenyov complex discrete-time digital wave, Wave 2 (W2); the Ledenyov complex discrete-time digital wave with the multiple distortions, Wave 3 (W3); the Ledenyov discrete-time digital economic output wave in the form of the vector-modulated discrete-time digital direct sequence spread spectrum DSSS signal bursts in the case of multiple distortions, Wave 4 (W4) in Ledenyov V O, Ledenyov D O (2017).

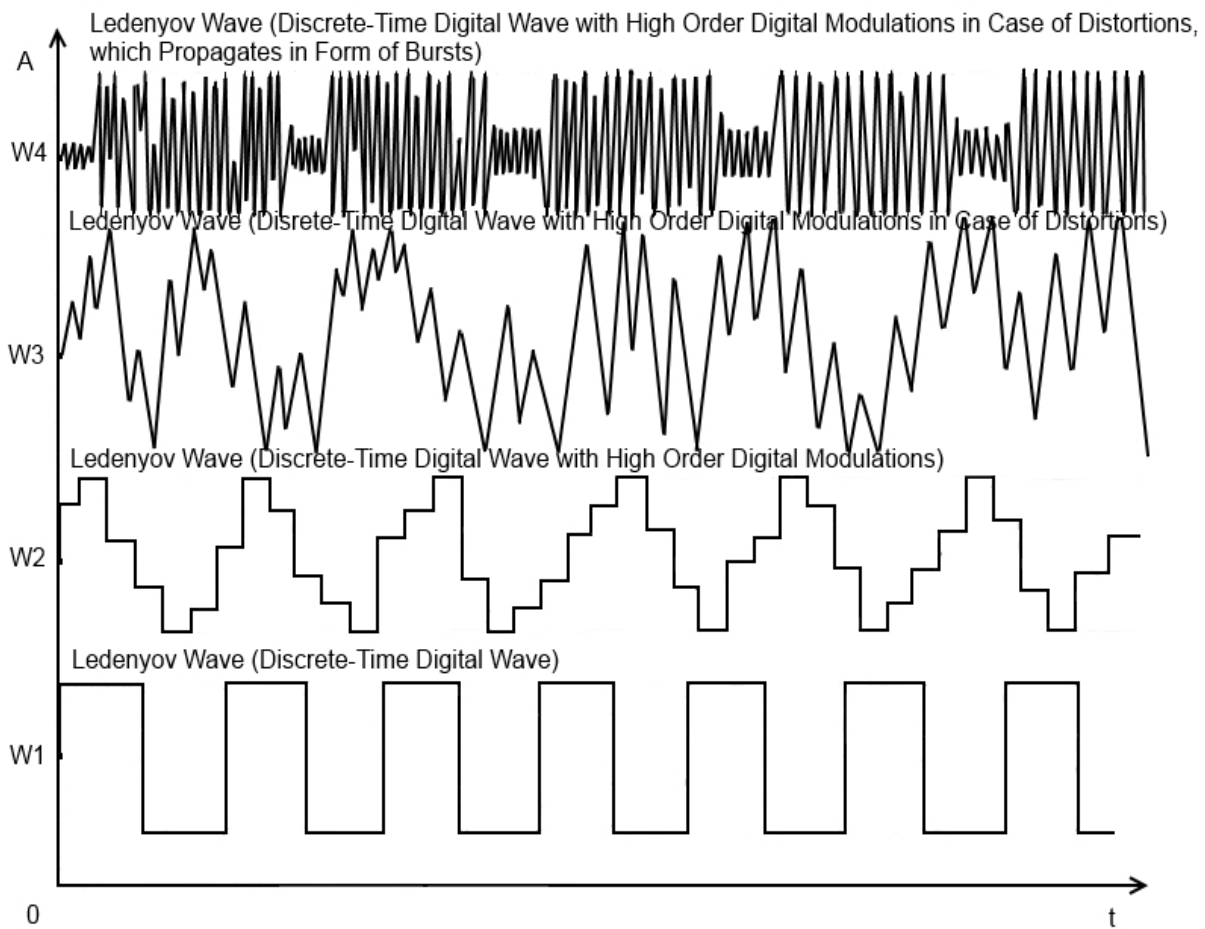

**Fig. 52.** Ledenyov discrete-time digital waves of GIP(t, monetary base), GDP(t, monetary base), GNP(t, monetary base), PPP(t, monetary base) in economy of scale and scope at certain monetary base over time in Ledenyov classic econodynamics: ***a)*** Ledenyov simple discrete-time digital wave, Wave 1 (W1); ***b)*** Ledenyov complex discrete-time digital wave, Wave 2 (W2); ***c)*** Ledenyov complex discrete-time digital wave with multiple distortions, Wave 3 (W3); ***d)*** Ledenyov discrete-time digital economic output wave in the form of the vector-modulated discrete-time digital direct sequence spread spectrum DSSS signal bursts in the case of multiple distortions, Wave 4 (W4).

Now, let us take a close look on the Ledenyov discrete-time digital economic output wave in the form of the vector-modulated discrete-time digital direct sequence spread spectrum DSSS signal bursts of GIP(t, monetary base), GDP(t, monetary base), GNP(t, monetary base), PPP(t, monetary base) in the economy of the scale and the scope at the certain monetary base in the selected observation time period in the case of multiple distortions the Ledenyov classic econodynamics science in Ledenyov V O, Ledenyov D O (2017).

Fig. 53 shows the Ledenyov discrete-time digital economic output wave in the form of the vector-modulated discrete-time digital direct sequence spread spectrum DSSS signal bursts of GIP(t, monetary base), GDP(t, monetary base), GNP(t, monetary base), PPP(t, monetary base) in the economy of the scale and the scope at the certain monetary base in the selected observation time period in the case of multiple distortions the Ledenyov classic econodynamics science.

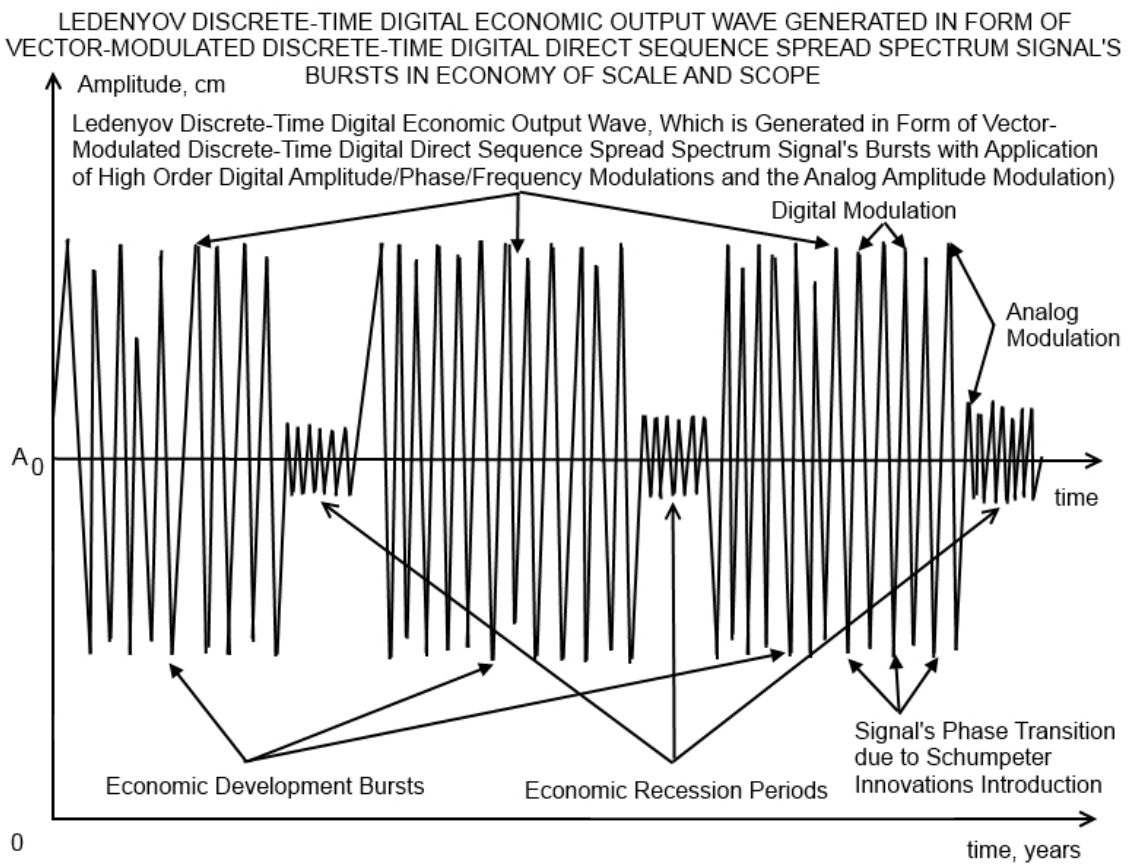

**Fig. 53.** Ledenyov discrete-time digital economic output wave in the form of the vector-modulated discrete-time digital direct sequence spread spectrum DSSS signal bursts of GIP(t, monetary base), GDP(t, monetary base), GNP(t, monetary base), PPP(t, monetary base) in the economy of the scale and the scope at the certain monetary base in the selected observation time period in the case of multiple distortions the Ledenyov classic econodynamics science.

Fig. 54 shows a Ledenyov graphic scheme of the Ledenyov discrete-time digital economic output wave in the form of the vector-modulated discrete-time digital direct sequence spread spectrum signal's bursts of GIP(t, monetary base), GDP(t, monetary base), GNP(t, monetary base), PPP(t, monetary base) with the changing amplitude, frequency, period, phase in the economy of the scale and the scope at the certain monetary base in the selected time period in the XYZ coordinates space in Ledenyov classic econodynamics.

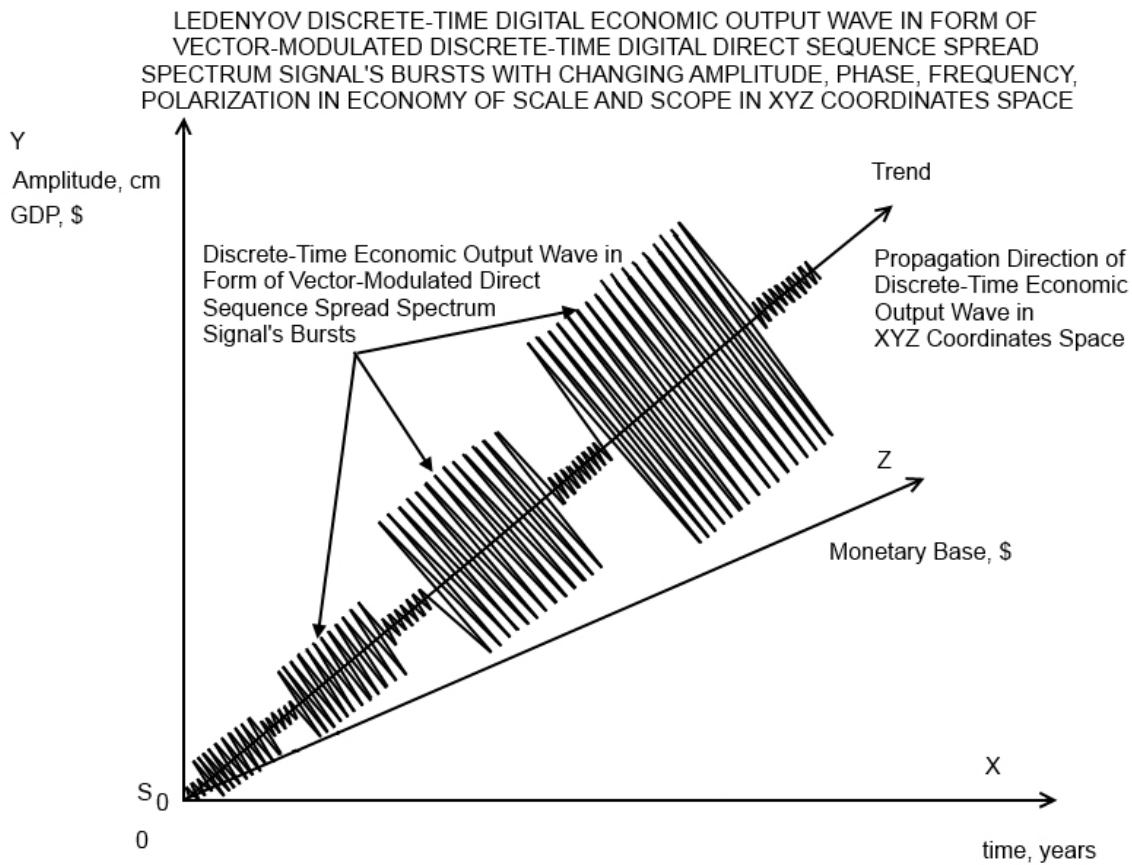

**Fig. 54.** Ledenyov graphic scheme of Ledenyov discrete-time digital economic output wave in form of vector-modulated discrete-time digital direct sequence spread spectrum signal's bursts of GIP(t, monetary base), GDP(t, monetary base), GNP(t, monetary base), PPP(t, monetary base) with changing amplitude, frequency, period, phase in economy of scale and scope at certain monetary base in selected time period in XYZ coordinates space in Ledenyov classic econodynamics.

Presently, we live in the time of great transition from the continuous-time finances/economics in Merton (November 3 1992) to the discrete-time digital finances in the economies of the scales and scopes in Ledenyov V O, Ledenyov D O (2017). The formulation of the Ledenyov classic econodynamics science founding principles requires us to re-think, re-consider and re-conceptualize the completely outdated theoretical ideas on the continuous-time business cycles, and to focus our research mainly on the innovative theoretical ideas on the discrete-time digital business cycles in the economy of the scale and the scope over the selected time period in Ledenyov V O, Ledenyov D O (2017).

Completing the information gathering, thoughtful thinking and rigorous analysis, we make a scientific proposition that there are the five main types of the discrete-time digital business cycles in the economy of the scale and the scope in agreement with the Ledenyov classic econodynamics:

*1.* 3 – 7 years Ledenyov discrete-time inventory cycle in Ledenyov V O, Ledenyov D O (2017);

*2.* 7 – 11 years Ledenyov discrete-time fixed investment cycle in Ledenyov V O, Ledenyov D O (2017);

*3.* 15 – 25 years Ledenyov discrete-time infrastructural investment cycle in Ledenyov V O, Ledenyov D O (2017);

*4.* 45 – 60 years Ledenyov discrete-time long wave cycle in Ledenyov V O, Ledenyov D O (2017);

*5.* 70+ years Ledenyov grand discrete-time super-cycle in Ledenyov V O, Ledenyov D O (2017);

instead of the five outdated types of the continuous-time business cycles in the economy of the scale and the scope in the classic macroeconomics:

*1.* 3 – 7 years Kitchin continuous-time inventory cycle in Kitchin (1923);

*2.* 7 –11 years Juglar continuous -time fixed investment cycle in Juglar (1862);

*3.* 15 – 25 years Kuznets continuous -time infrastructural investment cycle in Kuznets (1973a, b);

*4.* 45 – 60 years Kondratieff continuous-time long wave cycle in Kondratieff, Stolper (1935);

*5.* 70+ years Grand continuous-time super-cycle.

Fig. 55 shows the Ledenyov discrete-time digital economic output wave in the form of the vector-modulated direct sequence spread spectrum signals' bursts in the economy of the scale and the scope over the time in Ledenyov classic econodynamics. The 3 – 7 years Ledenyov discrete-time inventory cycle, Wave 5 (W5); the 7 –11 years Ledenyov discrete-time fixed investment cycle, Wave 4 (W4); the 15 – 25 years Ledenyov discrete-time infrastructural investment cycle, Wave 3 (W3); the 45 – 60 years Ledenyov discrete-time long wave cycle, Wave 2 (W2); the 70+ years Ledenyov grand discrete-time super-cycle, Wave 1 (W1).

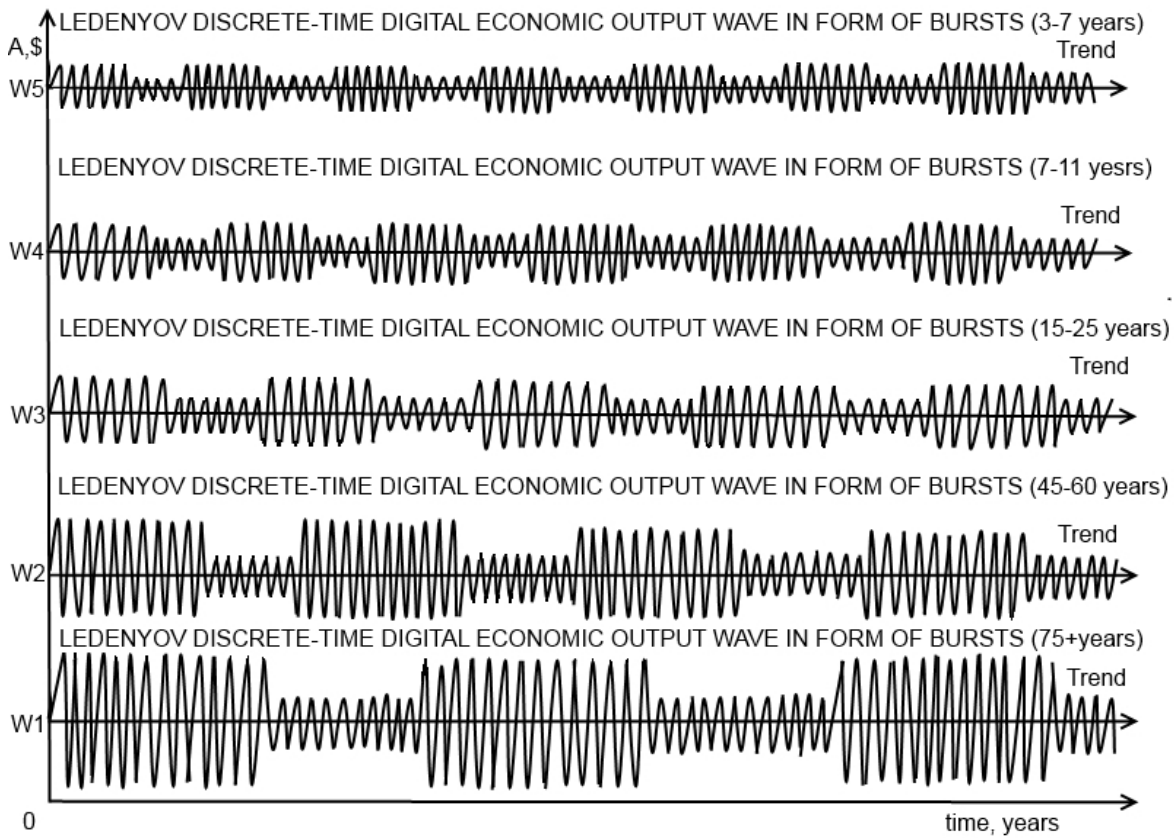

**Fig. 55.** Ledenyov discrete-time digital economic output waves in form of vector-modulated direct sequence spread spectrum signals' bursts in economy of scale and scope at certain monetary base over time in Ledenyov classic econodynamics: *1)* 3 – 7 years Ledenyov discrete-time inventory cycle, Wave 5 (W5); *2)* 7 –11 years Ledenyov discrete-time fixed investment cycle, Wave 4 (W4); *3)* 15 – 25 years Ledenyov discrete-time infrastructural investment cycle, Wave 3 (W3); *4)* 45 – 60 years Ledenyov discrete-time long wave cycle, Wave 2 (W2); *5)* 70+ years Ledenyov grand discrete-time super-cycle, Wave 1 (W1).

Fig. 56 shows the Ledenyov discrete-time digital economic output waves in the form of the vector-modulated discrete-time digital direct sequence spread spectrum signal's bursts of GIP(t, monetary base), GDP(t, monetary base), GNP(t, monetary base), PPP(t, monetary base) with the changing amplitude, frequency, period, phase in the economy of the scale and the scope at the certain monetary base in the selected time period in the XYZ coordinates space in Ledenyov classic econodynamics: *a)* 3 – 7 years Ledenyov discrete-time inventory cycle, Wave 5 (W5); *b)* 7 –11 years Ledenyov discrete-time fixed investment cycle, Wave 4 (W4); *c)* 15 – 25 years Ledenyov discrete-time infrastructural investment cycle, Wave 3 (W3); *d)* 45 – 60 years Ledenyov discrete-time long wave cycle, Wave 2 (W2); *e)* 70+ years Ledenyov grand discrete-time super-cycle, Wave 1 (W1). The wave propagation vector direction is defined by the α, β, γ angles in the XYZ coordinates space on the base of the statistical data.

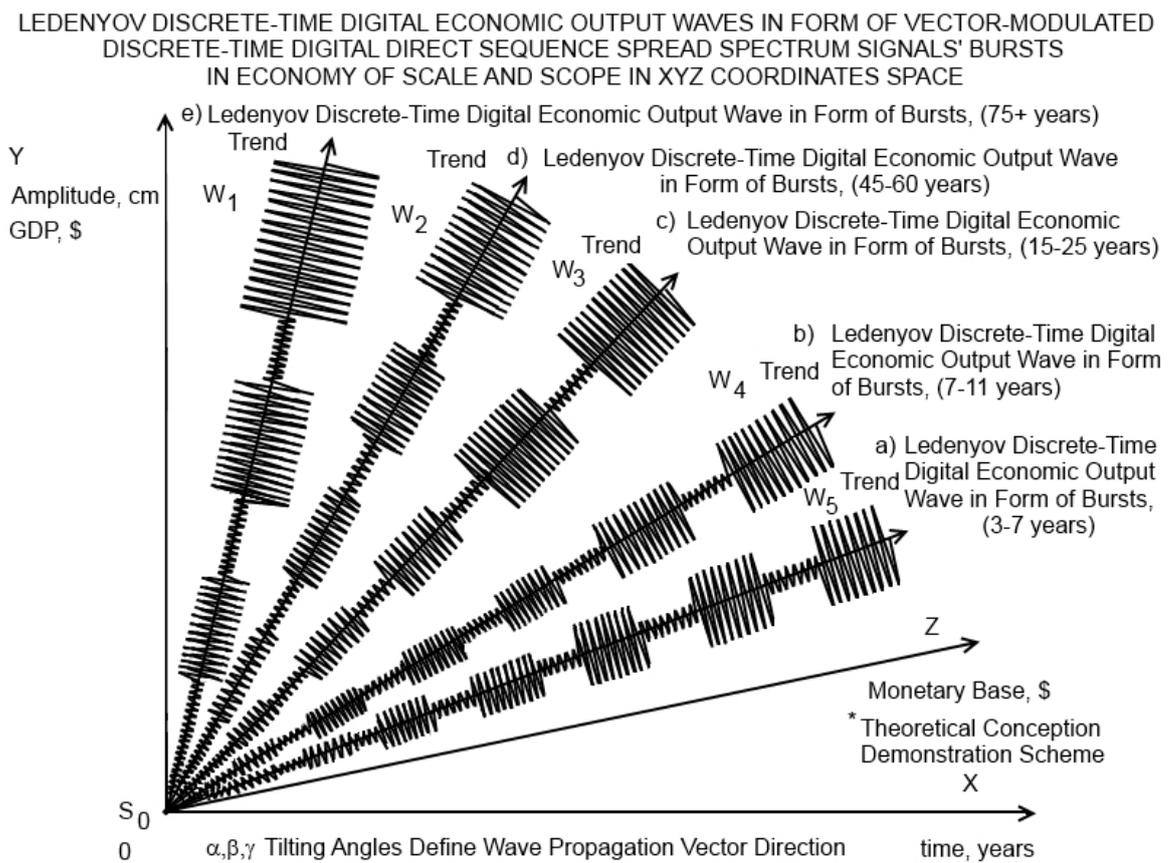

**Fig. 56.** Ledenyov graphic scheme of Ledenyov discrete-time digital economic output wave in form of vector-modulated discrete-time digital direct sequence spread spectrum signal's bursts of GIP(t, monetary base), GDP(t, monetary base), GNP(t, monetary base), PPP(t, monetary base) with changing amplitude, frequency, period, phase in economy of scale and scope at certain monetary base in selected time periods in XYZ coordinates space in Ledenyov classic econodynamics: *a)* 3 – 7 years Ledenyov discrete-time inventory cycle, Wave 5 (W5); *b)* 7 –11 years Ledenyov discrete-time fixed investment cycle, Wave 4 (W4); *c)* 15 – 25 years Ledenyov discrete-time infrastructural investment cycle, Wave 3 (W3); *d)* 45 – 60 years Ledenyov discrete-time long wave cycle, Wave 2 (W2); *e)* 70+ years Ledenyov grand discrete-time super-cycle, Wave 1 (W1).

In the Chapter 4, we researched some aspects on the Ledenyov discrete-time digital economic output wave in the form of the vector-modulated discrete-time digital direct sequence spread spectrum DSSS signal bursts of GIP(t, monetary base), GDP(t, monetary base), GNP(t, monetary base), PPP(t, monetary base) with the changing amplitude, frequency, period, phase, which can be generated and measured in the economy of the scale and the scope at the certain monetary base in the selected time periods in the XYZ coordinates space in Ledenyov classic econodynamics.

At the same time, we think that there may be many other discrete-time digital waves with the even more complex waveforms, which can be originated by an application of the high order digital modulation, encoding and spreading techniques in the economy of the scale and the scope at the certain monetary base in the selected time periods in the XYZ coordinates space in Ledenyov classic econodynamics.

Therefore, in the Chapter 5, we intend to investigate the Ledenyov discrete-time digital economic output wave in the form of the vector-modulated discrete-time digital direct sequence spread spectrum DSSS signal short/wide/ultra-wide band pulses in the economy of the scale and the scope at the certain monetary base in the selected time periods in the XYZ coordinates space in Ledenyov classic econodynamics.

# Chapter 5

## Discrete-time digital economic output waves in form of vector-modulated discrete-time digital direct sequence spread spectrum signals' short/long/ultra long pulses in economy of scale and scope in classic econodynamics

Let us continue our research discussion with consideration on the generation, propagation and possible absorption/attenuation/amplification of the following economic output waves:

1. The Ledenyov discrete-time digital economic output waves in the form of the continuous-time modulated signal's short/long/ultra long time duration pulses of GIP(t, monetary base), GDP(t, monetary base), GNP(t, monetary base), PPP(t, monetary base) in the economy of the scale and the scope in the Ledenyov classic econodynamics;

2. The Ledenyov discrete-time digital economic output waves in the form of the discrete-time digital vector-modulated direct sequence spread spectrum signal's short/long/ultra long time duration pulses of GIP(t, monetary base), GDP(t, monetary base), GNP(t, monetary base), PPP(t, monetary base) in the economy of the scale and the scope in the Ledenyov classic econodynamics.

Following our early adopted discussion style, we prefer to begin the research by making a focus precisely on the following broad research topics in the Maxwell electrodynamics:

1. The generation of the discrete-time digital electromagnetic waves by the analog pulse modulation of the continuous-time signal carrier, analyzing the short/long/ultra long time duration pulses;

2. The generation of the discrete-time digital electromagnetic waves by the analog pulse modulation of the discrete-time digital signal carrier, considering the short/ long/ultra long time duration pulses.

At this point, we would like to explain that the pulse/pulse code modulations of the continuous-time electromagnetic signals and the discrete-time electromagnetic signals are intensively used in the modern electromagnetic signal communication systems/radars/jammers, including the data communication transceivers, electronically-steered electronically-scanned phased array radars, the synthetic aperture radars, the noise radars in a big number of different industrial applications.

Fig. 57 demonstrates the discrete-time electromagnetic pulse signal generation with an application of the analog pulse modulation of the continuous-time signal carrier in Maxwell electrodynamics: *a)* The power spectrum of the discrete-time pulse signal with the continuous-time signal inside the pulse; *b)* The power spectrum of the discrete-time pulse signal with the continuous-time signal inside the pulse.

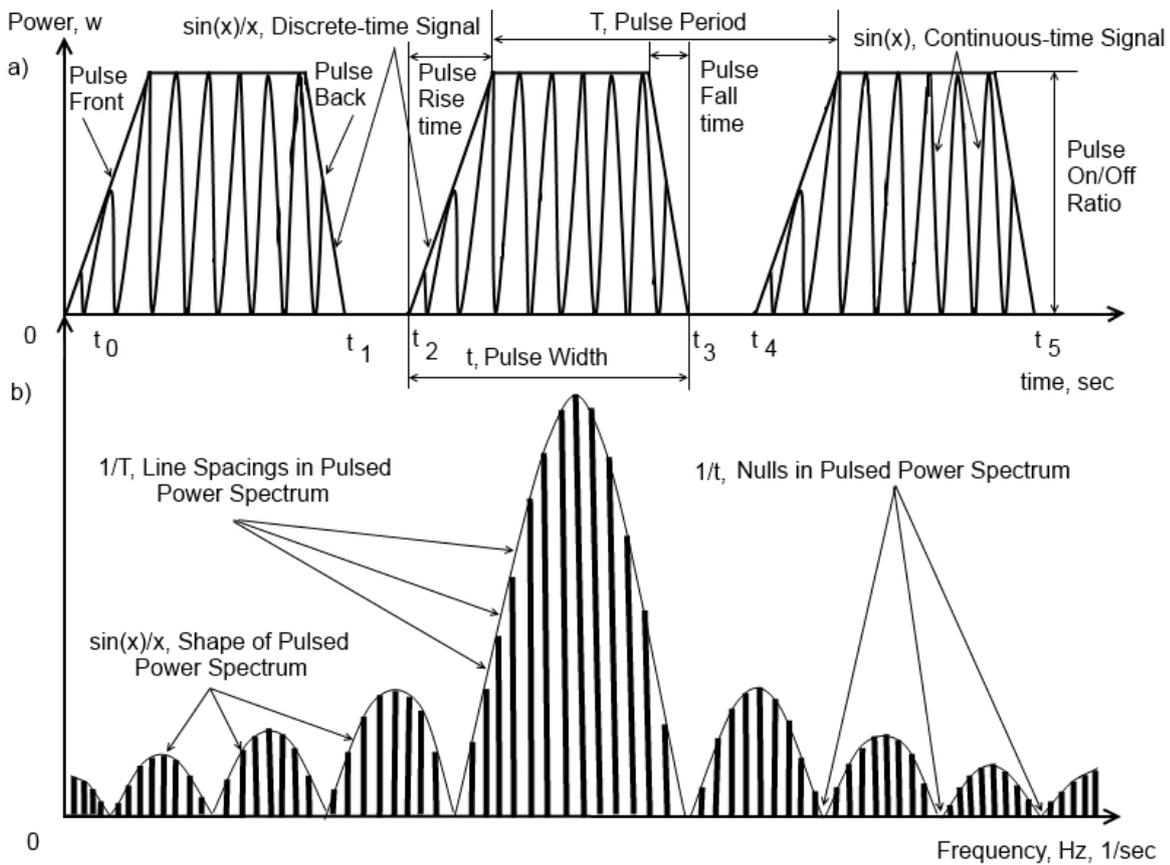

**Fig. 57.** Discrete-time electromagnetic pulse signal generation with application of analog pulse modulation of continuous-time signal carrier in Maxwell electrodynamics: *a)* Power spectrum of discrete-time pulse signal with continuous-time signal inside pulse; *b)* Power spectrum of discrete-time pulse signal with continuous-time signal inside pulse.

Let us add some complexity to our representation and consider the discrete-time pulse signal generation with an application of the analog pulse modulation of the discrete-time digital signal carrier in the Maxwell electrodynamics. These signals found their application in the noise radars.

Fig. 58 The discrete-time electromagnetic pulse signal generation with an application of the analog pulse modulation of the discrete-time digital signal carrier in the Maxwell electrodynamics and in the Walsh discrete-time digital signal processing science: *a)* The power spectrum of the discrete-time pulse signal with the discrete-time digital signal inside the pulse; *b)* The power spectrum of the discrete-time pulse signal with the discrete-time digital signal inside the pulse.

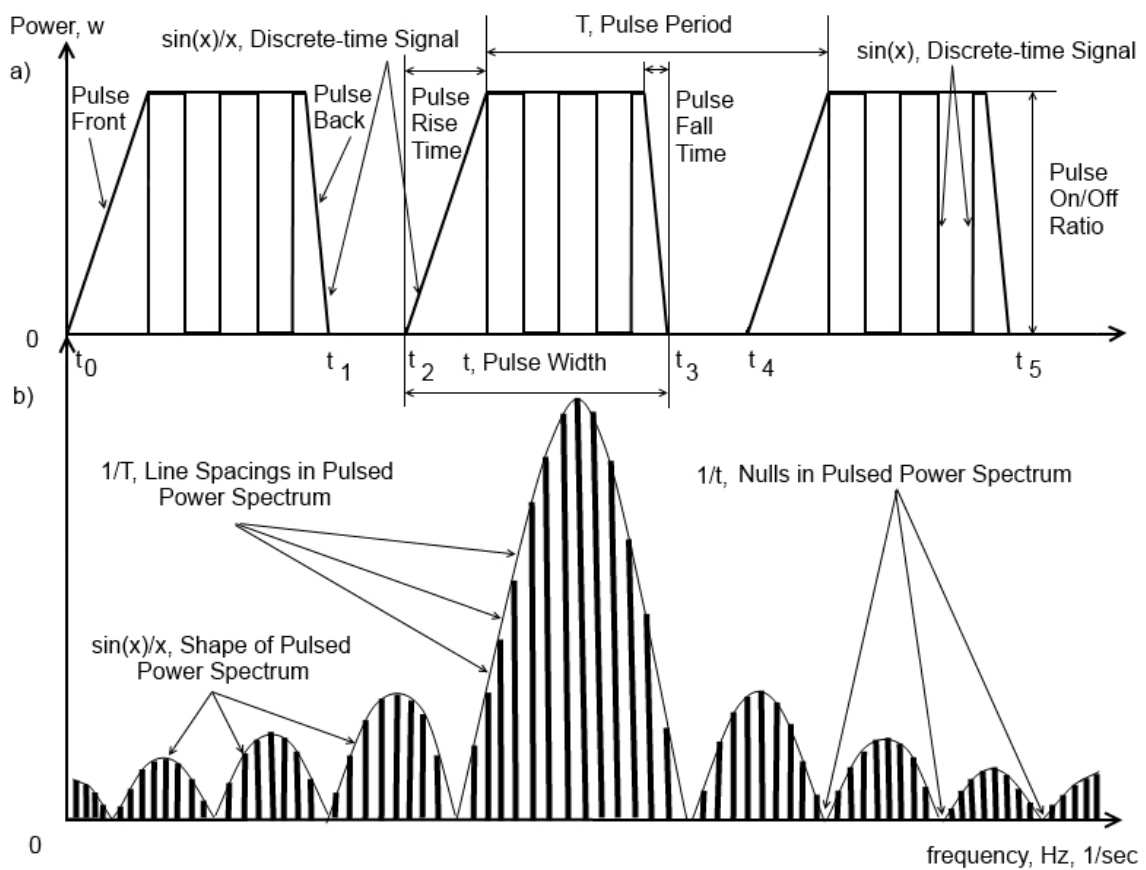

**Fig. 58.** Discrete-time electromagnetic pulse signal generation with application of analog pulse modulation of discrete-time digital signal carrier in Maxwell electrodynamics and in Walsh discrete-time digital signal processing science: *a)* Power spectrum of discrete-time pulse signal with discrete-time digital signal inside pulse; *b)* Power spectrum of discrete-time pulse signal with discrete-time digital signal inside pulse.

Now, let us make a first possible abstract analogy between:

1. The discrete-time electromagnetic waves in the form of the short/long/ultra long time duration electromagnetic pulses, generated by the analog pulse modulation of the continuous-time electromagnetic signal carrier in the space over the time in the Maxwell electrodynamics; and

2. The Ledenyov discrete-time digital economic output waves in the form of the short/wide/ultra wide band pulses of GIP(t, monetary base), GDP(t, monetary base), GNP(t, monetary base, PPP(t, monetary base)], generated by the analog pulse modulation of the continuous-time signal carrier in the economy of the scale and the scope over the time in the Ledenyov classic econodynamics.

Fig. 59 displays the Ledenyov discrete-time digital economic output waves as the short/wide/ultra wide band pulses of GIP(t, monetary base), GDP(t, monetary base), GNP(t, monetary base), PPP(t, monetary base) ), generated by the analog pulse modulation of the continuous-time signal carrier in the economy of the scale and the scope at the certain monetary base over the selected time period in the Ledenyov classic econodynamics.

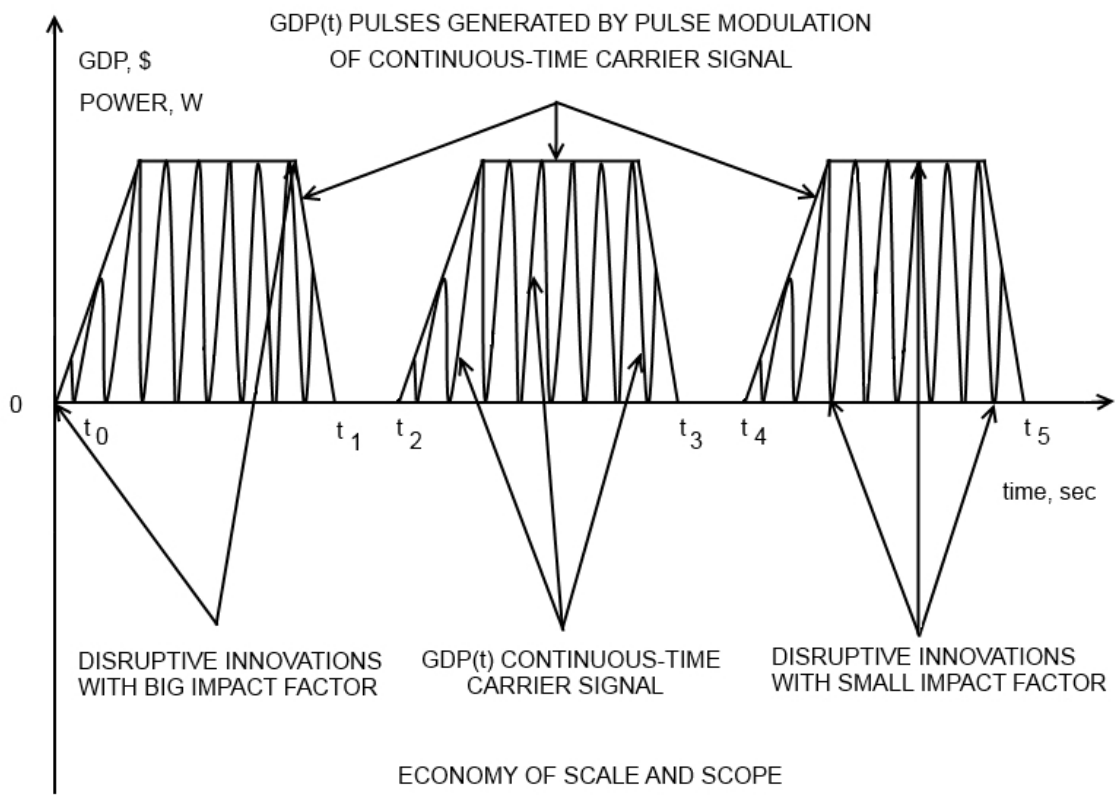

**Fig. 59.** Ledenyov discrete-time digital economic output wave in form of short/wide/ultra wide band pulses of GIP(t, monetary base), GDP(t, monetary base), GNP(t, monetary base), PPP(t, monetary base), generated by analog pulse modulation of continuous-time signal carrier in economy of scale and scope at certain monetary base over selected time period in Ledenyov classic econodynamics.

Then, let us make a second abstract analogy between:

1. The discrete-time digital electromagnetic wave in the form of the short/long/ultra long time duration electromagnetic pulses, created by the analog pulse modulation of the vector-modulated discrete-time digital direct sequence spread spectrum signal carrier in the space over the time in the Maxwell electrodynamics; and

2. The Ledenyov discrete-time digital economic output wave in the form of the short/long/ultra long time duration pulses of GIP(t, monetary base), GDP(t, monetary base), GNP(t, monetary base), PPP(t, monetary base), generated by the analog pulse modulation of the vector-modulated discrete-time digital direct sequence spread spectrum signal carrier in the economy of the scale and the scope over the time in the Ledenyov classic econodynamics.

Now, let us show the Ledenyov discrete-time digital economic output wave in the form of the short/long/ultra long time duration pulses of GIP(t, monetary base), GDP(t, monetary base), GNP(t, monetary base), PPP(t, monetary base), generated by the analog pulse modulation of the vector-modulated discrete-time digital direct sequence spread spectrum signal carrier in the economy of the scale and the scope over the time in the Ledenyov classic econodynamics.

Fig. 60 displays Ledenyov discrete-time digital economic output wave in the form of the short/long/ultra long time duration pulses of GIP(t, monetary base), GDP(t, monetary base), GNP(t, monetary base), PPP(t, monetary base), generated by the analog pulse modulation of the vector-modulated discrete-time digital direct sequence spread spectrum signal carrier in the economy of the scale and the scope over the time in the Ledenyov classic econodynamics.

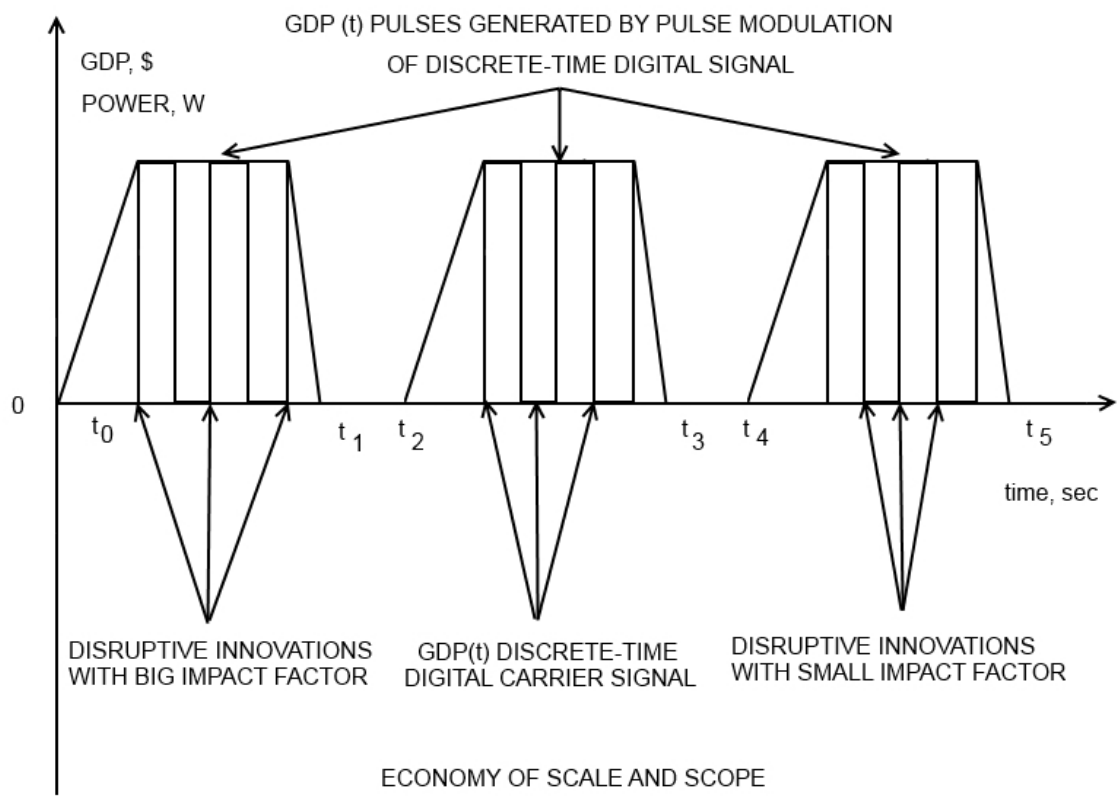

**Fig. 60.** Ledenyov discrete-time digital economic output wave in form of short/long/ultra long time duration pulses of GIP(t, monetary base), GDP(t, monetary base), GNP(t, monetary base), PPP(t, monetary base), generated by analog pulse modulation of vector-modulated discrete-time digital direct sequence spread spectrum signal carrier in economy of scale and scope over time in Ledenyov classic econodynamics.

Thus, we think that the Ledenyov discrete-time digital economic output waves in the form of the vector-modulated discrete-time digital direct sequence spread spectrum signals' short/long/ultra long pulses of GIP(t, monetary base), GDP(t, monetary base), GNP(t, monetary base, PPP(t, monetary base)] in the economy of the scale and the scope in the Ledenyov classic econodynamics can be generated.

Going to the next point, it is logical to assume that the Ledenyov discrete-time digital economic output waves in the form of the vector-modulated discrete-time digital direct sequence spread spectrum signals' short/long/ultra long pulses of GIP(t, monetary base), GDP(t, monetary base), GNP(t, monetary base, PPP(t, monetary base)] in the economy of the scale and the scope in the Ledenyov classic econodynamics, may have the different econodynamic spectral parameters in terms of the amplitude/power/frequency/time scales, including:

*1.* The pulse econodynamic spectral parameters:

  *a)* the pulse/average/envelope powers;

  *b)* the pulse rise/fall time;

  *c)* the pulse repetition frequencies;

  *d)* the pulse period;

  *e)* the pulse width.

*2.* The vector-modulated direct sequence spread spectrum signal carrier econodynamic spectral parameters:

  *a)* the amplitude;

  *b)* the frequency;

  *c)* the phase;

  *d)* the modulation accuracy;

  *e)* the encoding sequence;

  *f)* the spreading sequence.

Thus, let us present graphically the Ledenyov discrete-time digital economic output waves in the form of the vector-modulated discrete-time digital direct sequence spread spectrum signals' short/long/ultra long pulses of GIP(t, monetary base), GDP(t, monetary base), GNP(t, monetary base, PPP(t, monetary base)] in the economy of the scale and the scope in the Ledenyov classic econodynamics.

Fig. 61 displays the Ledenyov discrete-time digital economic output waves in the form of the vector-modulated discrete-time digital direct sequence spread spectrum signals' short/long/ultra long pulses of GIP(t, monetary base), GDP(t, monetary base), GNP(t, monetary base, PPP(t, monetary base)] in the economy of the scale and the scope in the Ledenyov classic econodynamics: *a)* Ledenyov discrete-time digital economic output wave in form of vector-modulated discrete-time digital direct sequence spread spectrum signal's short pulses; *b)* Ledenyov discrete-time digital economic output wave in form of vector-modulated discrete-time digital direct sequence spread spectrum signal's long pulses; *c)* Ledenyov discrete-time digital economic output wave in form of vector-modulated discrete-time digital direct sequence spread spectrum signal's ultra long pulses.

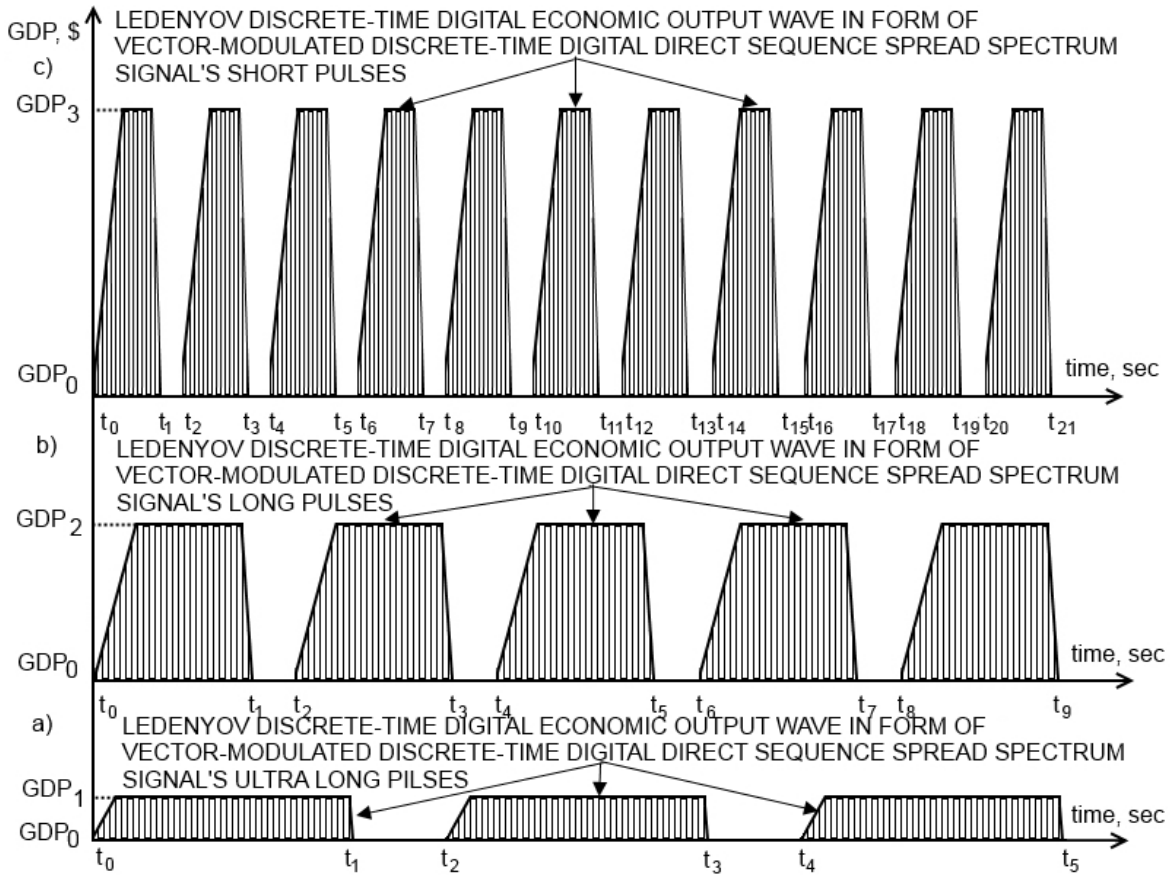

**Fig. 61.** Ledenyov discrete-time digital economic output waves in the form of the vector-modulated discrete-time digital direct sequence spread spectrum signals' short/long/ultra long pulses of GIP(t, monetary base), GDP(t, monetary base), GNP(t, monetary base, PPP(t, monetary base)] in the economy of the scale and the scope in the Ledenyov classic econodynamics: *a)* Ledenyov discrete-time digital economic output wave in form of vector-modulated discrete-time digital direct sequence spread spectrum signal's short pulses; *b)* Ledenyov discrete-time digital economic output wave in form of vector-modulated discrete-time digital direct sequence spread spectrum signal's long pulses; *c)* Ledenyov discrete-time digital economic output wave in form of vector-modulated discrete-time digital direct sequence spread spectrum signal's ultra long pulses.

Fig. 62 displays a Ledenyov graphic representation scheme of the Ledenyov discrete-time digital economic output wave in the form of the vector-modulated discrete-time digital direct sequence spread spectrum signals' short/long/ultra long time duration pulses of GIP(t, monetary base), GDP(t, monetary base), GNP(t, monetary base), PPP(t, monetary base) with the changing amplitude, frequency, period, phase in the economy of the scale and the scope at the certain monetary base in the selected time periods in the XYZ coordinates space in Ledenyov classic econodynamics. The wave propagation vector direction is defined by the α, β, γ tilting angles in the XYZ coordinates space on the base of the statistical data on the economic output.

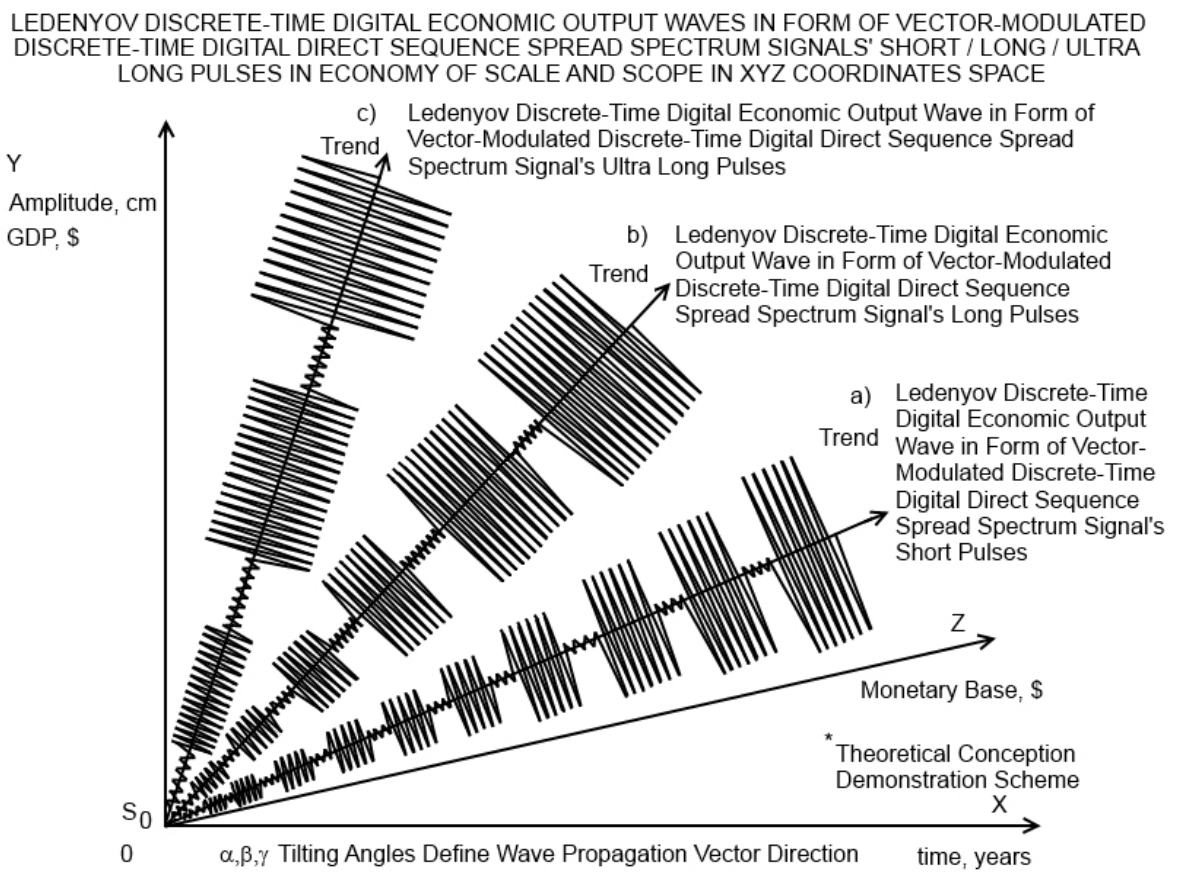

**Fig. 62.** Ledenyov graphic scheme of Ledenyov discrete-time digital economic output wave in form of vector-modulated discrete-time digital direct sequence spread spectrum signals' short/long/ultra long time duration pulses of GIP(t, monetary base), GDP(t, monetary base), GNP(t, monetary base), PPP(t, monetary base) with changing amplitude, frequency, period, phase in the economy of the scale and the scope at the certain monetary base in the selected time periods in the XYZ coordinates space in Ledenyov

classic econodynamics: *a)* Ledenyov discrete-time digital economic output wave in form of vector-modulated discrete-time digital direct sequence spread spectrum signal's short pulses; *b)* Ledenyov discrete-time digital economic output wave in form of vector-modulated discrete-time digital direct sequence spread spectrum signal's long pulses; *c)* Ledenyov discrete-time digital economic output wave in form of vector-modulated discrete-time digital direct sequence spread spectrum signal's ultra long pulses.

Making some explanatory, summarizing and concluding scientific remarks, we would like to highlight the following important research findings and proposals:

*1.* We think that the Ledenyov discrete-time digital economic output waves in the form of the vector-modulated discrete-time digital direct sequence spread spectrum signals' short/long/ultra long pulses of GIP(t, monetary base), GDP(t, monetary base), GNP(t, monetary base, PPP(t, monetary base)] in the economy of the scale and the scope in the Ledenyov classic econodynamics can be generated;

*2.* We believe that the Schumpeter numerous disruptive scientific/technological/economical/financial/social/political/cultural innovations have the discrete-time nature in agreement with the Schumpeter disruptive innovations theory. It means that the Schumpeter numerous disruptive innovations can generate/switch/trigger the streams of the bits/chips, making it possible to perform the processes of the modulation, encoding and spreading of GIP(t, monetary base), GDP(t, monetary base), GNP(t, monetary base), PPP(t, monetary base) signals. Therefore, in our opinion, the Schumpeter numerous disruptive innovations have a natural ability to digitally generated, modulate, encode and spread the Ledenyov discrete-time digital economic output waves in the form of the vector-modulated discrete-time digital direct sequence spread spectrum signals' short/long/ultra long pulses of GIP(t, monetary base), GDP(t, monetary base), GNP(t, monetary base, PPP(t, monetary base)] in the economy of the scale and the scope in the Ledenyov classic econodynamics;

*3.* We assume that the Schumpeter numerous disruptive innovations may have the different impacts/influences on the resulting changes of the total magnitude of GIP(t, monetary base), GDP(t, monetary base), GNP(t, monetary base), PPP(t, monetary base) in the economy of the scale and the scope over the selected time period. For instance, some Schumpeter's disruptive innovations may have a big impact and/or other Schumpeter's disruptive innovations may exhibit a small impact as far as the resulting changes to the total magnitude of GIP(t, monetary base), GDP(t, monetary base), GNP(t, monetary base), PPP(t, monetary base) in the economy of the scale and the scope at the selected time period is concerned. For example, in the transportation industry, we can say that the steam locomotive/automobile/aircraft inventions can be classified as a disruptive innovation with the big impact factor, whereas the numerous steam/petrol/diesel/electric locomotive/automobile/aircraft models developments can be considered as the multiple disruptive innovations with the small impact factor. Of course, we can continue to make the multiple examples, considering many other inventions in various industries in the economies of the scales and the scopes over the centuries.

In Chapter 5, we studied the Ledenyov discrete-time digital economic output waves in the form of the vector-modulated discrete-time digital direct sequence spread spectrum signals' short/long/ultra long pulses of GIP(t, monetary base), GDP(t, monetary base), GNP(t, monetary base, PPP(t, monetary base)] in the economy of the scale and the scope in the Ledenyov classic econodynamics.

Now, the main problem to understand is: Can the phase transitions in the Ledenyov discrete-time digital economic output waves in the form of the vector-modulated discrete-time digital direct sequence spread spectrum signals' short/long/ultra long pulses of GIP(t, monetary base), GDP(t, monetary base), GNP(t, monetary base, PPP(t, monetary base)] in the economy of the scale and the scope in the Ledenyov classic econodynamics occur as a result of the quantum jumps of GIP(t, monetary base), GDP(t, monetary base), GNP(t, monetary base), PPP(t, monetary base), because of the Schumpeter's disruptive innovations introduction into in the economy of the scale and the scope over the selected time period?

In next Chapter 6, we intend to research the Ledenyov discrete-time digital economic output waves in the form of the vector-modulated discrete-time digital direct sequence spread spectrum signals' short/long/ultra long pulses of GIP(t, monetary base), GDP(t, monetary base), GNP(t, monetary base, PPP(t, monetary base)] in the economy of the scale and the scope in the Ledenyov classic econodynamics, which can be generated by the quantum leaps of GIP(t, monetary base), GDP(t, monetary base), GNP(t, monetary base), PPP(t, monetary base) due to the Schumpeter's disruptive innovations introduction into the economy of the scale and the scope at the certain monetary base in the selected time moments/periods in the XYZ coordinates space in Ledenyov quantum econodynamics.

# Chapter 6

## Discrete-time digital economic output waves in form of vector-modulated discrete-time digital direct sequence spread spectrum signals' short/long/ultra long pulses generated by quantum leaps in economy of scale and scope in quantum econodynamics

In the quantum technology driven century, we would like to formulate the Ledenyov quantum econodynamics science's foundational scientific principles. More clearly, we intend to advance our research by discussing the nature of the Ledenyov discrete-time digital economic output waves in the form of the vector-modulated discrete-time digital direct sequence spread spectrum signal's short/long/ultra ultra long time duration bursts/pulses, generated by the quantum jumps/leaps/fluctuations in the economy of the scale and the scope in the frames of the Ledenyov quantum econodynamics.

We prefer to begin our discussion with a short introductory historical overview on a number of important theoretical propositions by the famous eminent physicists, which were made in the quantum mechanics in the beginning of XX century. The creation of the quantum mechanics can certainly be regarded as one of the key milestones, which tremendously shaped the science in XX century.

Let us reflect the main scientific contributions to the quantum mechanics in a historical order:

1. The Planck constant in Planck (1900 a, b, c, d);

   **$E=hw$**

   *where: E is the energy;*

          *h is the Plank constant;*

          *w is the circular frequency.*

2. The Bohr atom model: *1)* the electrons at the different discrete-space electrons' orbits and *2)* the quantum discrete transitions by the orbiting electrons between the orbits as a result of the photon

absorption/radiation in Thomson (1897a, b, 30 April 1897, 1904, 1912, 1913, 1923), Bohr (1913a, b, c, 1921, 1922);

3. The de Broglie matter wave by electron in atom in de Broglie (1924, 1925, 1926, 1929);

4. The Schrödinger wave function in Schrödinger quantum mechanical wave equation in Schrödinger (1926a, b)

$$i\hbar\frac{\partial}{\partial t}\big|\psi(r,t)\big\rangle = \hat{H}\big|\psi(r,t)\big\rangle,$$

*where* : *$\hbar$ is the reduced Plank constant,*
  *$\psi$ is the wave function,*
  *H is the Hamiltonian.*

5. The Heisenberg uncertainty principle in Heisenberg (1927), Kennard (1927).

$$\sigma_x\sigma_p \geq \frac{\hbar}{2},$$

*where* : *$\sigma_x$ is standard deviation of position,*
  *$\sigma_p$ is the standard deviation of momentum,*
  *$\hbar$ is the reduced Plank constant.*

Let us consider the Bohr atom model in details, highlighting the fact that Bohr (1922) came up with the so called Bohr's atom model, suggesting that *1)* the electrons at the different discrete-space electrons' orbits and *2)* the quantum discrete transitions by the orbiting electrons between the orbits as a result of the photon absorption/radiation in Bohr (1913a, b, c, 1921, 1922).

As we know, sometime later, the Bohr's atom model was complemented by the theoretical proposition that the electron's orbiting movement with the changing electron's impulse around the nucleus in the atom can be precisely described by the de Broglie continuous-time wave in de Broglie (1924, 1925, 1926).

Let us draw a scheme of the Bohr's atom model, based on the Niels Bohr's theoretical proposal on the electrons discrete-time transitions from/to the electrons discrete-space orbits during the electrons rotational movement at the orbits with the de Broglie matter waves trajectories in de Broglie (1924, 1925, 1926) around the nucleus in the atom in agreement with principles in the condensed matter physics and in the quantum physics.

Fig. 63 shows a scheme of the Bohr atom model in Thomson (1897a, b, 30 April 1897, 1904, 1912, 1913, 1923), Bohr (1913a, b, c, 1921, 1922), de Broglie (1924, 1925, 1926): *1)* the electrons rotational movement around the nucleus at the different discrete-space elliptic orbits in Bohr (1913a, b, c, 1921, 1922); *2)* the quantum discrete-space discrete-time transitions by the orbiting electrons between the elliptic orbits as a result of the photon(s) absorption/radiation in Bohr (1913a, b, c, 1921, 1922); *3)* the electrons rotational movement around nucleus at different discrete-space elliptic orbits in the atom is described by the de Broglie matter waves in de Broglie (1924, 1925, 1926). The discrete-space discrete-time quantum leaps by the electrons have place in the atom at the nano-scale.

NIELS BOHR ATOM MODEL IN QUANTUM MECHANICS

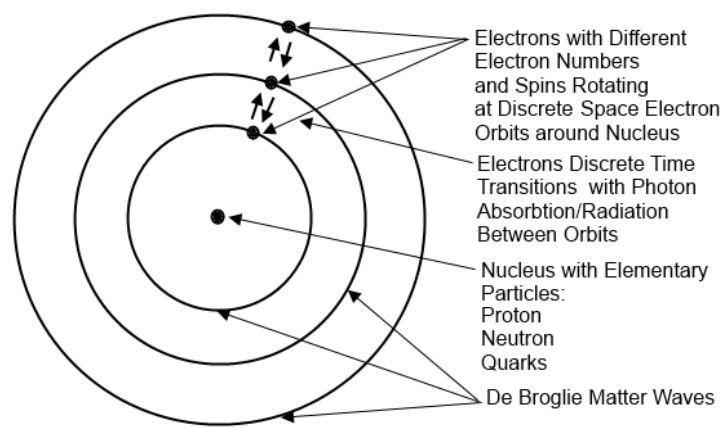

**Fig. 63.** Scheme of Bohr atom model in Thomson (1897a, b, 30 April 1897, 1904, 1912, 1913, 1923), Bohr (1913a, b, c, 1921, 1922), de Broglie (1924, 1925, 1926): *1)* electrons rotation movement around nucleus at different discrete-space elliptic orbits in atom in Bohr (1913a, b, c, 1921, 1922); *2)* quantum discrete-space discrete-time transitions by orbiting electrons between elliptic orbits as result of photon absorption/radiation in Bohr (1913a, b, c, 1921, 1922); *3)* electrons rotational movement around nucleus at different discrete-space elliptic orbits in form of de Broglie matter waves in de Broglie (1924, 1925, 1926) in atom.

The discrete-space discrete-time quantum leaps can also play an important role in the stimulated emission in Einstein (1916, 1917) with the population inversion mechanism in an application to the laser and/or maser devices in the quantum photonics science.

Fig. 64 shows a scheme of the stimulated emission with the population inversion mechanism in an application to the laser in quantum optics.

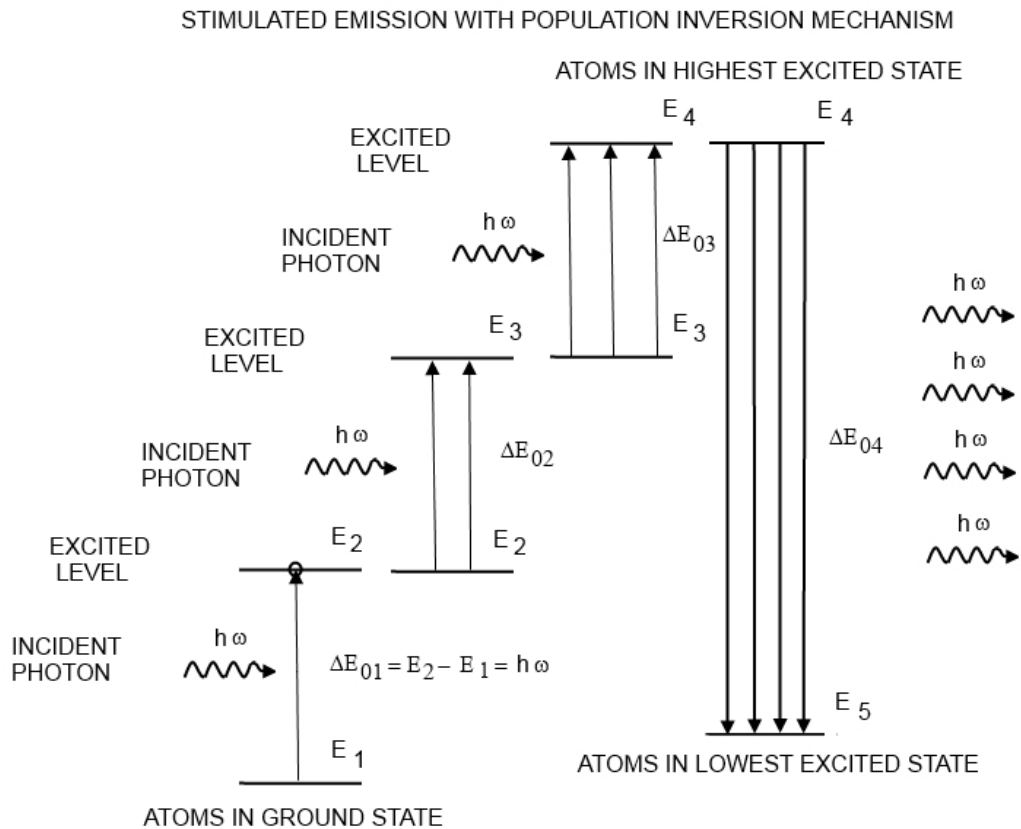

STIMULATED EMISSION WITH POPULATION INVERSION MECHANISM

**Fig. 64.** Scheme of stimulated emission with population inversion mechanism in laser in quantum optics.

Taking presented opportunity, let us note that the numerous R&D programs in the quantum physics/nanoelectronics, resulted in the design of quantum fusing / amplifying / sensing / computing devices: *1)* The nuclear/thermonuclear reactor(s) in Blokhintsev (1954), Fortov (2016); *2)* The tunneling semiconductor diode / light emitting semiconductor diode / laser semiconductor diode in Alferov (1996), Bimberg, Grundmann, Ledentsov (1999); *3)* The single electron transistor (SET); *4)* The Light Amplification by Stimulated Emission of Radiation (LASER) of different designs in in Prokhorov, Basov (1955), Gordon, Zeiger, Townes (15 August 1955), Gould (1959); *5)* The RF/DC Superconducting Quantum Interference

Device (SQUID); *6)* The Quantum Random Number Generator on Magnetic Flux Qubits chipset (1024 QRNG_MFQ chipset) in Ledenyov V O, Ledenyov D O, Ledenyov O P (2001); *7)* The quantum memory on knots of magnetic vortices chipset in Ledenyov V O, Ledenyov D O, Ledenyov O P (1999).

In the Ledenyov quantum econodynamics, we think that the statistical oscillations in the form of the dependences of GIP(t, monetary base), GDP(t, monetary base), GNP(t, monetary base), PPP(t, monetary base) can actually be viewed as the Ledenyov discrete-time digital economic output wave in the form of the vector-modulated discrete-time digital direct sequence spread spectrum signal's short/long/ultra long time duration bursts/pulses in the case of present distortions in the economy of the scale and the scope. In addition, we believe that the two types of the modulations of GIP(t, monetary base), GDP(t, monetary base), GNP(t, monetary base), PPP(t, monetary base) signals can occur as the results of the Type I and II phase transitions:

1. ***Discrete-Time Digital Modulation with Type I Phase Transition***: In this case, the high order amplitude/phase modulation together with the encoding/spreading by the Schumpeter disruptive innovations take place in the economy of the scale and the scope over the selected time period. Therefore, we may assume that the Schumpeter disruptive innovations introduction can lead to the quantum leaps of GIP(t, monetary base), GDP(t, monetary base), GNP(t, monetary base), PPP(t, monetary base), which can cause the discrete-time digital modulation of the Ledenyov discrete-time digital economic output wave in the form of the vector-modulated discrete-time digital direct sequence spread spectrum signal's short/wide/ultra wide bursts/pulses.

2. ***Discrete-Time Analog Modulation with Type II Phase Transition:*** In this case, the analog amplitude/phase modulation by the Schumpeter disruptive innovations takes place in the economy of the scale and the scope over the selected time period. Therefore, we may assume that the Schumpeter disruptive innovations introduction can lead to the quantum leaps of GIP(t, monetary base), GDP(t, monetary base), GNP(t, monetary base), PPP(t, monetary base), which can cause the discrete-time analog modulation of the Ledenyov discrete-time digital economic output wave in the form of the vector-modulated discrete-time digital direct sequence spread spectrum signal's short/wide/ultra wide bursts/pulses.

Fig. 65 displays: *a)* The Ledenyov discrete-time digital economic output wave in the form of the vector-modulated discrete-time digital direct sequence spread spectrum signal's short/ long/ultra long time duration pulses, generated by the quantum leaps in an ideal theoretical case with the no distortions in the economy of the scale and the scope; *b)* The Ledenyov discrete-time digital economic output wave in the form of the vector-modulated discrete-time digital direct sequence spread spectrum signal's short/long/ultra long time duration pulses, generated by the quantum leaps in a real practical case with the different distortions in the economy of the scale and the scope. The Ledenyov Type I and Type II phase transitions are shown in both considered cases in the Ledenyov quantum econodynamics.

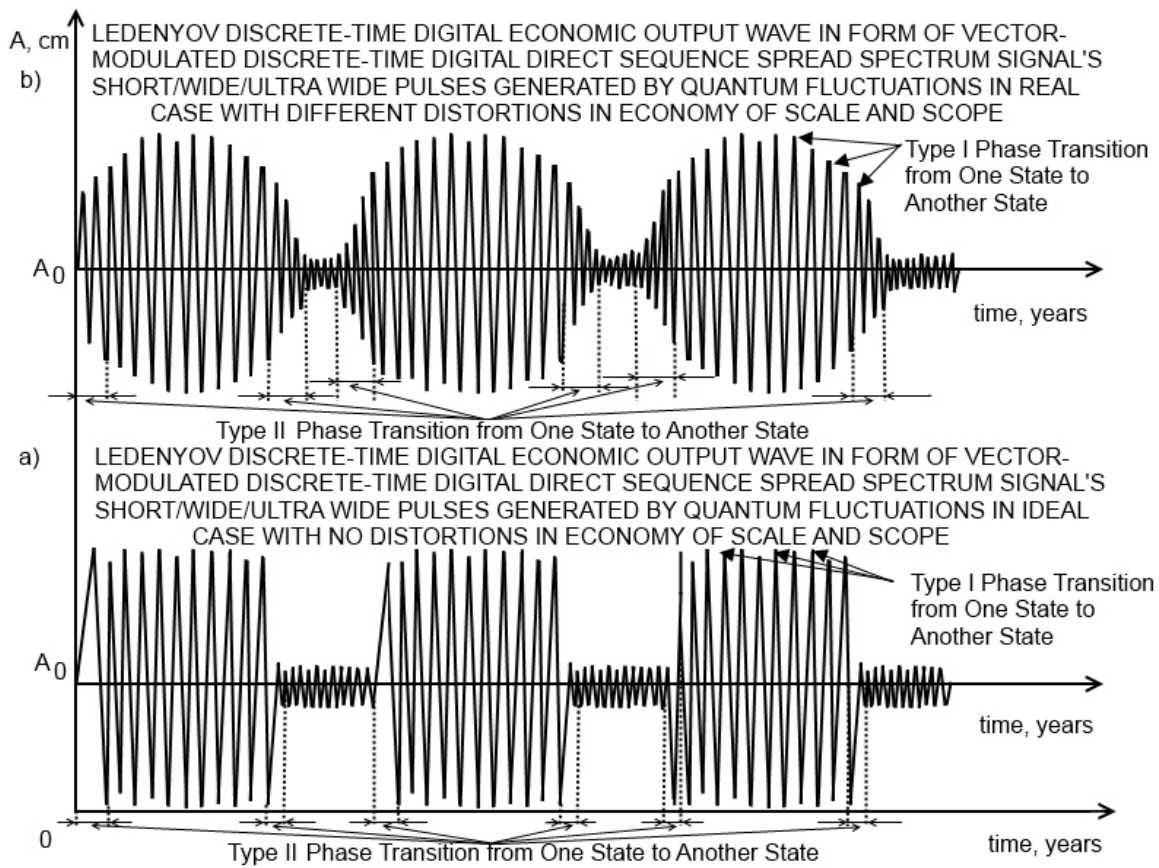

**Fig. 65. *a)*** Ledenyov discrete-time digital economic output wave in form of vector-modulated discrete-time digital direct sequence spread spectrum signal's short/long/ultra long time duration pulses, generated by quantum leaps in ideal theoretical case with no distortions in economy of scale and scope; *b)* Ledenyov discrete-time digital economic output wave in form of vector-modulated discrete-time digital direct sequence spread spectrum signal's short/ long/ultra long time duration pulses generated by quantum leaps in real practical case with different distortions in economy of scale and scope. The Ledenyov Type I and Type II phase transitions are shown in both studied cases in Ledenyov quantum econodynamics.

Now, let us discuss some possible reasons to believe that there may be the quantum leaps of GIP(t, monetary base), GDP(t, monetary base), GNP(t, monetary base), PPP(t, monetary base) in the economy of the scale and the scaope over the selected time period in the frames of the Ledenyov quantum econodynamics. Before outlining our theoretical concept on the possible quantum leaps of GIP(t, monetary base), GDP(t, monetary base), GNP(t, monetary base), PPP(t, monetary base), we would like to assume the Ledenyov quantum econodynamics integrates in itself all the important research findings in both:

1. The Ledenyov quantum microeconomics science in Ledenyov D O, Ledenyov V O (2015j), and

2. The Ledenyov quantum macroeconomics science in Ledenyov D O, Ledenyov V O (2015h), Jakimowicz (2016).

In the **Ledenyov quantum microeconomics** in Ledenyov D O, Ledenyov V O (2015j), we would like to introduce the new terminology such as the Ledenyov firm of the scale and the scope. In other words, we think that the firm has its certain operational boundaries, functioning in the nonlinear media such as the nonlinear dynamic economy of the scale and the scope.

We permit that the Schumpeter disruptive innovation introduction can discretely change the EBITDA of the firm in the dynamic nonlinear economy of the scale and the scope over the selected time period.

Then, in the Ledenyov quantum microeconomics, we can assume that the discrete changes of the EBITDA of the firm in the dynamic nonlinear economy of the scale and the scope over the selected time period may occur in the form of the quantum leaps.

These quantum leaps in the EBITDA of the Ledenyov firms of the scales and the scopes in the economy of the scale and scope over the selected time period in the Ledenyov quantum microeconomics may in turn result into a presence of the quantum leaps in the dependencies of GIP(t, monetary base), GDP(t, monetary base), GNP(t, monetary base), PPP(t, monetary base) in the nonlinear dynamic diffusion type economy of the scale and the scope at the certain monetary bases in the selected time periods in Ledenyov Ledenyov quantum macroeconomics in agreement with a general framework of the Ledenyov quantum econodynamics.

Fig. 66 provides a scheme of the Ledenyov stimulated emission with the population inversion mechanism in economy of scale and scope in the Ledenyov quantum microeconomics.

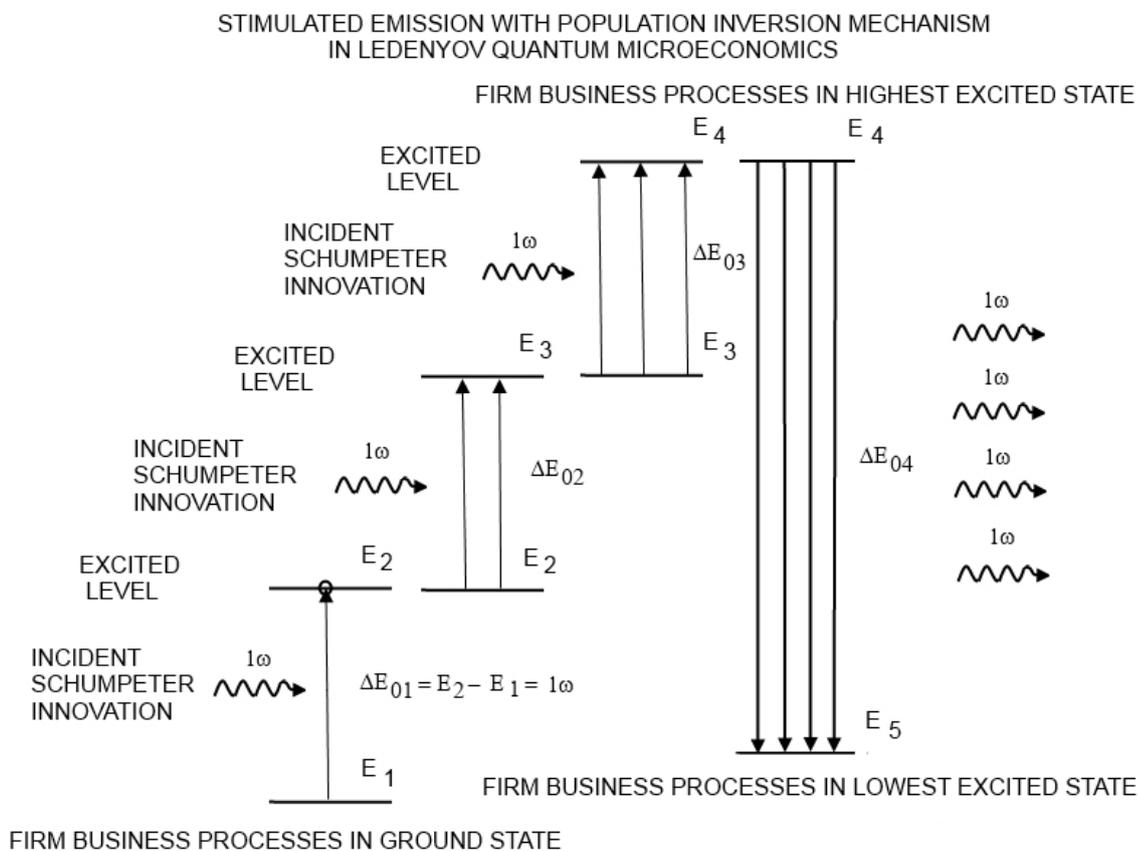

**Fig. 66.** Scheme of Ledenyov stimulated emission with population inversion mechanism in economy of scale and scope in Ledenyov quantum microeconomics.

In the Ledenyov quantum microeconomics, the quantum theory of the firm postulates that the discrete-time transitions from one level of the firm's economic performance to another level of the firm's economic performance will occur in the nonlinear dynamic economic system at the time moment, when in Ledenyov D O, Ledenyov V O (2015j):

*1.* The present land, labour and capital resources are (added and absorbed) / (released and radiated) in the form of quanta, decreasing or increasing the general energy entropy in the nonlinear dynamic economic system (the nonlinear medium);

*2.* The disruptive scientific/technological/financial/social/political innovation(s) is/are introduced into or withdrawn from the nonlinear dynamic economic system (the nonlinear medium), creating the resonance conditions to amplify/attenuate the value of the firm's economic performance, during the evolution process of the economy of the scale and the scope in the time domain (Note: the resonance can result in the increase/decrease of the energy of the electromagnetic wave in the electrodynamics science);

*3.* The firm's business processes population inversion mechanism is present, when a) the every business process in the firm can be conditionally compared to the electron in the atom, b) the discrete increase of business process value in the firm can be conditionally associated with the discrete increase of electron's energy in the atom during the energy pumping process in the laser, c) the land, labour and capital resources release at the population inversion mechanism realization in the firm can be conditionally regarded as the light radiation at the population inversion mechanism action in the laser;

*4.* The derived formula to describe the discrete-time EBITDA changes during the firm's economic performance variations in terms of the quantum theory of the firm is

$$\hbar_{micro}\omega_{m,n} = \triangle \textbf{\textit{EBITDA}}(t) = \textbf{\textit{EBITDA}}\left(t\right)_m - \textbf{\textit{EBITDA}}\left(t\right)_n$$

$$\hbar_{micro}\omega_{m,n} = \triangle \textbf{\textit{firm's value}}(t) = \textbf{\textit{firm's value}}\left(t\right)_m - \textbf{\textit{firm's value}}\left(t\right)_n$$

where: $\hbar_{micro}$ – the Ledenyov constant,

$\omega$ – the cyclic velocity,

t – the time,

EBITDA – the Earnings Before Interest Tax Depreciation Amortization,

Firm's value – the firm's market capitalization minus the firm's long term investments and debt.

*5.* The Ledenyov distribution of a number of excited firms' business processes of certain value at a selected level (state) in the economy of scale and the scope in terms of the quantum microeconomics theory is

$$\frac{N_m}{N_n} = \exp\frac{-\left(EBIDTA(t)_m - EBIDTA(t)_n\right)}{\lambda_{micro}T},$$

$$\frac{N_m}{N_n} = \exp\frac{-\left(firm's\ value(t)_m - firm's\ value(t)_n\right)}{\lambda_{micro}T},$$

where: $\lambda_{micro}$ – the Ledenyov constant,

$N_m$ – the number of firms' processes of certain value at the state (m),

$N_n$ – the number of firms' business processes of certain value at the state (n),

$N = N_m + N_n$ – the general number of firms' processes of certain value in the economy of the scale and the scope,

t – the time,

T – the temperature of the economy of scale and scope, which corresponds to the level of entropy of the economy of scale and scope (the level of information/business activities by the firms),

EBITDA – the Earnings Before Interest Tax Depreciation and Amortization,

Firm's value – the firm's market capitalization minus the firm's long term investments and debt.

In other words, let us emphasis the fact that the quantum theory of the firm states that there may be the discrete-time induced transition(s) between the different levels of the firm's EBITDAs (the firm's values) in the nonlinear dynamic economic system at the time, when the following things are present in Ledenyov D O, Ledenyov V O (2015j):

*1.* The land, labour and capital, which can be added and absorbed / released and radiated in the form of quanta in the nonlinear dynamic economic system (the nonlinear medium);

*2.* The discrete-time fluctuational processes, which can appear in the form of the disruptive scientific/technological/financial/social/political innovation(s) that absorb or release the available land, labour and capital resources, creating the resonance, in the nonlinear dynamic economic system (the nonlinear medium) during the evolution process of the firm in the economy of the scale and the scope in the time domain;

**3.** The firm's business processes population inversion mechanism, which occurs at the following condition: $N_2/N_1 > 1$.

In the ***Ledenyov quantum macroeconomics*** in Ledenyov D O, Ledenyov V O (2015h), let us highlight a research observation that the general information product GIP(t, monetary base), the general domestic product GDP(t, monetary base), and the general national product GNP(t, monetary base), purchasing power parity PPP(t, monetary base) usually change abruptly in the discrete values in the economy of the scale and the scope at the certain monetary base over the selected time period, which we prefer to call the quanta.

Fig. 67 shows a scheme of Ledenyov stimulated emission with population inversion mechanism in economy of scale and scope in Ledenyov quantum macroeconomics.

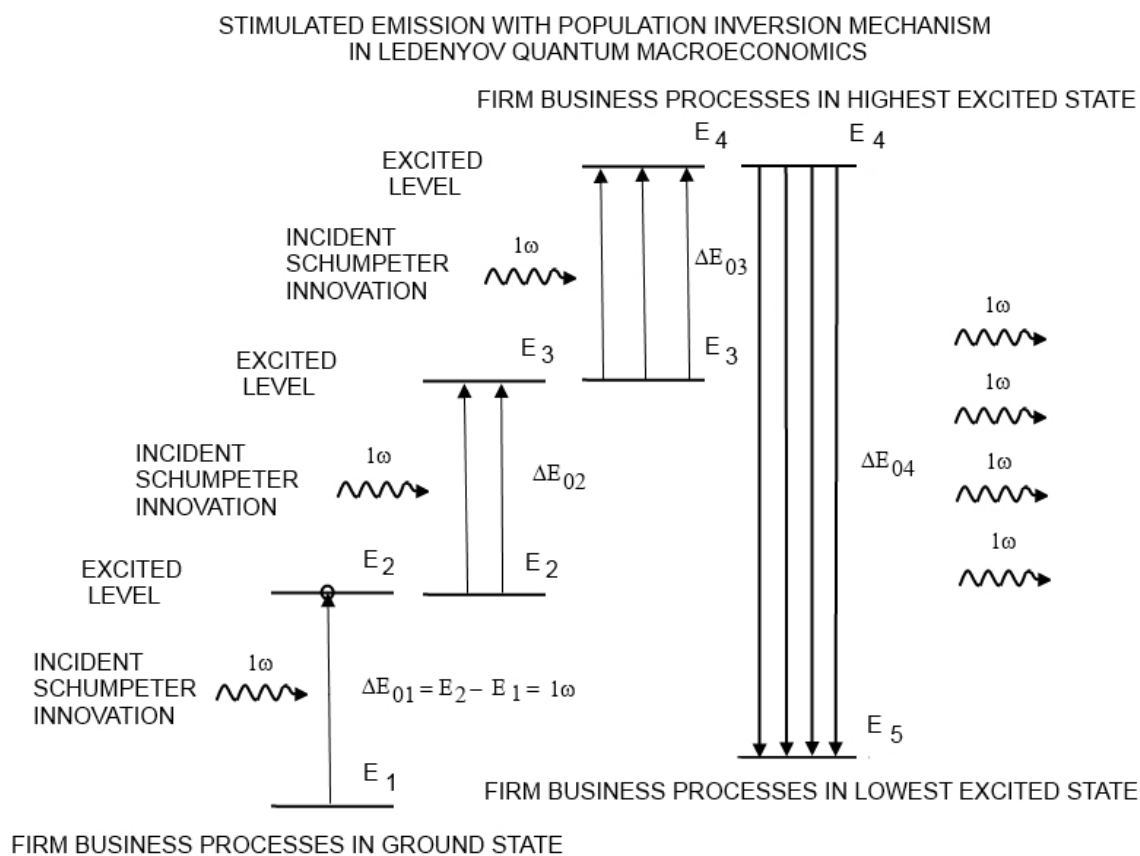

**Fig. 67.** Scheme of Ledenyov stimulated emission with population inversion mechanism in economy of scale and scope in Ledenyov quantum macroeconomics.

In our opinion, we have to focus our research attention on the two manifestations of a quantum nature of the GIP(t, monetary base), GDP(t, monetary base), GNP(t, monetary base), PPP(t, monetary base) dependences:

*1.* The presence of the discrete-output spectrum in GIP(t, monetary base), GDP(t, monetary base), GNP(t, monetary base), PPP(t, monetary base) dependences, which can be described by the discretely increasing/decreasing levels of GIP(t, monetary base), GDP(t, monetary base), GNP(t, monetary base), PPP(t, monetary base) in the nonlinear dynamic economies of the scale and the scope at the certain monetary base over the selected time period;

*2.* The presence of the discrete-time digital signals (the Ledenyov discrete-time digital waves with the Markov information in Ledenyov D O, Ledenyov V O (2015 e, f, g)), which describe the business cycle envelope waveform of GIP(t, monetary base), GDP(t, monetary base), GNP(t, monetary base), PPP(t, monetary base) in the nonlinear dynamic economies of scale and scope in the time domain.

Let us formulate the Ledenyov quantum macroeconomics theory, using the quantum econophysics principles and assuming that the characteristic dependences of GIP(t, monetary base), GDP(t, monetary base), GNP(t, monetary base), PPP(t, monetary base) are the discrete-time digital signals (the Ledenyov discrete-time digital waves with the Markov information) in distinction from the early researched continuous-time signals (the Kitchin, Juglar, Kuznets, Kondratieff continuous-time waves). The discrete-time digital nature of the disruptive scientific/technological/financial/social/political innovation(s) introduction and implementation generate the GIP(t, monetary base), GDP(t, monetary base), GNP(t, monetary base), PPP(t, monetary base) oscillations in the nonlinear dynamic economies of the scopes and scales in the time domain in Ledenyov D O, Ledenyov V O (2013c, 2015d, 2015e, 2015f).

The Ledenyov quantum macroeconomics theory postulates that the discrete-time transitions from one level of GIP(t, monetary base), GDP(t, monetary base), GNP(t, monetary base), PPP(t, monetary base) to another discrete level of GIP(t, monetary base), GDP(t, monetary base), GNP(t, monetary base), PPP(t, monetary base) will occur in the nonlinear dynamic economic systems at the time moment, when:

**1.** The land, labour and capital resources are added and absorbed / released and radiated in the form of quanta, decreasing or increasing the general energy entropy in the nonlinear dynamic economic system (the nonlinear medium);

**2.** The disruptive scientific/technological/financial/social/political innovation(s) is/are introduced into or withdrawn from the nonlinear dynamic economic system (the nonlinear medium), creating the resonance conditions to amplify/attenuate the value of GIP(t, monetary base), GDP(t, monetary base), GNP(t, monetary base), PPP(t, monetary base) during the evolution process of the nonlinear dynamic economy of the scale and the scope in the time domain (Note: the resonance can result in the increase/decrease of energy of the electromagnetic wave in the electrodynamics science);

**3.** The derived formula to describe the discrete-time output change of the nonlinear dynamic economy of the scale and the scope in terms of the Ledenyov quantum macroeconomics theory is

$$\hbar\omega_{m,n} = \triangle GIP(t, \ monetary \ base) = GIP\left(t, \ monetary \ base\right)_m - GIP\left(t, \ monetary \ base\right)_n$$

$$\hbar\omega_{m,n} = \triangle GDP(t, \ monetary \ base) = GDP\left(t, \ monetary \ base\right)_m - GDP\left(t, \ monetary \ base\right)_n$$

$$\hbar\omega_{m,n} = \triangle GNP(t, \ monetary \ base) = GNP\left(t, \ monetary \ base\right)_m - GNP\left(t, \ monetary \ base\right)_n$$

$$\hbar\omega_{m,n} = \triangle PPP(t, \ monetary \ base) = PPP\left(t, \ monetary \ base\right)_m - PPP\left(t, \ monetary \ base\right)_n$$

where: $\hbar$ - Ledenyov constant, $\omega$ - cyclic velocity, t – time, GIP(t, monetary base) - the general information product on the time GIP(t, monetary base), GDP(t, monetary base) - the general domestic product on the time, GNP(t, monetary base) - the general national product on the time, PPP(t, monetary base) – the purchasing power parity on the time.

**4.** The Ledenyov distribution of a number of excited business processes of certain value at the selected level (state) in the nonlinear dynamic economy of the scale and the scope in terms of the Ledenyov quantum macroeconomics theory is

$$\frac{N_m}{N_n} = \exp \frac{-\left(GIP\left(t, \ monetary \ base\right)_m - GIP\left(t, \ monetary \ base\right)_n\right)}{\lambda T},$$

$$\frac{N_m}{N_n} = \exp \frac{-\left(GDP\left(t, \ monetary \ base\right)_m - GDP\left(t, \ monetary \ base\right)_n\right)}{\lambda T},$$

$$\frac{N_m}{N_n} = \exp \frac{-\left(GNP\left(t, \ monetary \ base\right)_m - GNP\left(t, \ monetary \ base\right)_n\right)}{\lambda T},$$

$$\frac{N_m}{N_n} = \exp \frac{-\left(PPP\left(t, \ monetary \ base\right)_m - PPP\left(t, \ monetary \ base\right)_n\right)}{\lambda T}$$

where: $\lambda$ - Ledenyov constant, $N_m$ - number of information/business processes of certain value at the state (m), $N_n$ - number of information/business processes of certain value at the state (n), $N = N_m + N_n$ - general number of information/business processes of certain value in the economy of scale and scope, t - time, T - temperature of the economy of scale and scope, which corresponds to the level of entropy of the economy of scale and scope (the level of information/business activities by the economic agents), GIP(t, monetary base) - the general information product on the time, GDP(t, monetary base) - the general domestic product on the time, GNP(t, monetary base) - the general national product on the time, PPP(t, monetary base) – the purchasing power parity on the time.

In other words, let us emphasis that the Ledenyov quantum macroeconomics theory states that there may be the discrete-time transition between the different levels of GIP(t, monetary base), GDP(t, monetary base), GNP(t, monetary base), PPP(t, monetary base) in the nonlinear dynamic economic system at the time, when the following things are present:

*1.* The land, labour and capital, which can be added and absorbed / released and radiated in the form of quanta in the nonlinear dynamic economic system (the nonlinear medium);

*2.* The discrete-time fluctuational processes, which can appear in the form of the disruptive scientific/technological/financial/social/political innovation(s) that absorb or release the available land, labour and capital resources, creating the resonance, in the nonlinear dynamic economic system (the nonlinear medium) during the evolution process of the economy of scale and scope in the time domain;

*3.* The business processes population inversion mechanism, which occurs at the following condition: $N_2/N_1 > 1$.

Let us, once again, give the possible examples of the above discussed disruptive scientific/technological/financial/social/political innovation(s):

*1)* Scientific innovation: the discovery of new scientific phenomena or the derivation of the new formula in the physics in Landes (1998);

*2)* Technological innovation: the creation of new materials / devices such as the new metals / steam engines, new metals / combustion engines, semiconductors / transistors, semiconductors / lasers, superconductors / electric motors, superconductors / single electron transistors, superconductors / Josephson junctions, superconductors / quantum random number generators, superconductors / quantum processors in Ledenyov D O, Ledenyov V O (2015a);

*3)* Financial innovation: the creation of new financial products and services such as the cryptocurrencies, derivatives and mobile banking;

*4)* Social innovation: the introduction of new socioeconomic models, for instance: the shared-value initiative, which can be defined as: "the policies and operating practices that enhance the competitiveness of a company while simultaneously advancing the economic and social conditions in the communities in which it operates" in Porter, Kramer (2006, 2011);

*5)* Political innovation: the establishment of the new effective governmental system.

We can provide the illustrations of the Ledenyov quantum macroeconomics theory by making a comparative analogy and by finding the possible parallels between the Ledenyov quantum macroeconomics theory and the quantum physics theory:

*1.* The discrete nature of the value change of GIP(t, monetary base), GDP(t, monetary base), GNP(t, monetary base), PPP(t, monetary base) in the quantum macroeconomics theory can be analogous to the discrete nature of the electrical charge change (the single electron charge is $1.6 \times 10^{-19}$ Coulombs) in the atom in the quantum physics theory in Ledenyov D O, Ledenyov V O (2015a);

**2.** The discrete nature of the value change of GIP(t, monetary base), GDP(t, monetary base), GNP(t, monetary base), PPP(t, monetary base) in the quantum macroeconomics theory can be similar to the discrete nature of the electromagnetic energy change ($\hbar\omega$ - the photon energy, $\hbar$ - the Planck constant, $\omega$ - the cyclic frequency) in the physical world as described in the quantum physics theory in Loudon (2001), Ledenyov D O, Ledenyov V O (2015a);

**3.** The discrete nature of the value change of GIP(t, monetary base), GDP(t, monetary base), GNP(t, monetary base), PPP(t, monetary base) in the Ledenyov quantum macroeconomics theory can also be collated with the discrete nature of the magnetic flux change ($\Phi_0$ - the flux quantum) in the superconducting circuits in the physical world as described in the quantum physics theory. (For example: In the superconducting ring, the product of the magnetic field times the area of the closed loop superconducting circuit has to be equal to the multiple of a ratio of the fundamental physical constants $\frac{\hbar}{2e}$, where $\hbar$ - the Planck constant, 2e – the charge of an electron pair in Tesche, Clarke (1977), Clarke (1989), Muck (1998), Ledenyov D O, Ledenyov V O (2015a));

**4.** The discrete-time transitions of GIP(t, monetary base), GDP(t, monetary base), GNP(t, monetary base), PPP(t, monetary base) in the Ledenyov quantum  macroeconomics theory can be compared with the discrete-time transitions of the electronic excitations of different energies between the possible orbits in the atom. (The Bohr's atom model in the condensed matter physics in Bohr (1922), when the multiple electrons orbit an atomic nucleus and can transit from one orbit to another orbit, making the absorption or radiation of the energy quanta);

**5.** The discrete-time transitions of GIP(t, monetary base), GDP(t, monetary base), GNP(t, monetary base), PPP(t, monetary base) in the Ledenyov quantum macroeconomics theory can also be compared with the discrete-time transitions of the electronic excitations between the energy levels in the laser (the light amplification by stimulated emission of radiation) - a quantum electronic device that generates the coherent electromagnetic wave radiation of high energy by converting and amplifying the incident non-coherent electromagnetic waves radiation of low energy in the nonlinear medium such as the electron/ion plasma, which is created in:

*1)* The special cesium/nitrogen/carbonic gas in a tube terminated by the optically flat reflecting parallel mirrors like in Fabry-Perot interferometer; or

*2)* The semiconductor-hetero-structures diode with the different energy band gaps with the Brag reflectors to select the mode) at the resonance, created by various types of resonators,

in Townes (1939, 1964, 1966, 1969, 1995, 1999), Townes, Schawlow (1955), Gordon, Zeiger, Townes (1955), Shimoda, Wang, Townes (1956), Prokhorov, Basov (1955), Prokhorov, Fedorov (1963), Prokhorov (1964, 1965, 1979), Karlov, Prokhorov (1976), Prokhorov, Buzzi, Sprangle, Wille (1992), Schawlow, Townes (1958), Schawlow (1963, 1964), Gould (1959), Basov (1964, 1965), Yokoyama, Ujihara (1995), Alferov (1996), Milonni, Eberly (1998), Bimberg, Grundmann, Ledentsov (1999).

As we know, during the laser operation process, the charge carriers undertake the discrete-time radiative transitions between the multiple energy levels, which occur with the absorption or radiation of the energy quanta, as characterized by the electronic excitations population inversion mechanism, achieving the resonant optical photons emission in Townes (1939, 1964, 1966, 1969, 1995, 1999), Townes, Schawlow (1955), Gordon, Zeiger, Townes (1955), Shimoda, Wang, Townes (1956), Prokhorov, Basov (1955), Prokhorov, Fedorov (1963), Prokhorov (1964, 1965, 1979), Karlov, Prokhorov (1976), Prokhorov, Buzzi, Sprangle, Wille (1992), Schawlow, Townes (1958), Schawlow (1963, 1964), Gould (1959), Basov (1964, 1965).

In Chapter 6, we researched the Ledenyov discrete-time digital economic output waves in form of vector-modulated discrete-time digital direct sequence spread spectrum signals' short/long/ultra long pulses of GIP(t, monetary base), GDP(t, monetary base), GNP(t, monetary base), PPP(t, monetary base), generated by the quantum leaps in the economy of the scale and the scope at the certain monetary base over the selected time period in the Ledenyov quantum econodynamics.

Now, let us think: Is it possible to accurately characterize the econodynamic variables of the GIP(t, monetary base), GDP(t, monetary base), GNP(t, monetary base), PPP(t, monetary base) in the economy of the scale and the scope at the certain monetary base over the selected time period in the Ledenyov classic and quantum econodynamics?

In next Chapter 7, we will discuss some aspects of the problem on the precise measurement of the econodynamic variables of the GIP(t, monetary base), GDP(t, monetary base), GNP(t, monetary base), PPP(t, monetary base) in the economy of the scale and the scope at the certain monetary base over the selected time period in the Ledenyov classic and quantum econodynamics.

# Chapter 7

## Precise measurement of econodynamic variables in economy of scale and scope in classic and quantum econodynamics

The classic macroeconomics has the scientific knowledge, including the macroeconomics theories, the statistical data collection tools, the statistical data approximation techniques, the statistical data rigorous analysis methodologies, to describe the economy of the scale and the scope with a certain degree of accuracy over the time in Joseph Penso de la Vega (1668, 1996), Mortimer (1765), Smith (1776, 2008), Menger (1871), Bagehot (1873, 1897), von Böhm-Bawerk (1884, 1889, 1921), Hirsch (1896), Bachelier (1900), Schumpeter (1906, 1911, 1933, 1939, 1961, 1939, 1947), Slutsky (1910, 1915 1923), von Mises (1912), Hayek (1931, 1935, 2008; 1948, 1980), Keynes (1936, 1992), Ellis, Metzler (1949), Friedman (1953), Baumol (1957), Debreu (1959), Krugman, Wells (2005), Stiglitz (2005, 2015).

Naturally, for some time period, the classic macroeconomics was considered as an empirical science, which uses the purely empirical methods to solve economic problems in Krugman, Wells (2005), Stiglitz (2005, 2015), Desai, King, Goodhart (2015). However, in the multi-petabit digital information age, the classic macroeconomics successfully transformed into the multidisciplinary science with a particular focus on the formulation of the new macroeconomics theories by applying the mathematics, econometrics, econophysics sciences in Jakimowicz (2016). As a result, presently, a big number of the well established macroeconomic thinking schools with the classical empirical approaches to the macroeconomics research continue to disappear quickly. At the same time, the new innovative macroeconomics schools of thinking with a growing interest into the multidisciplinary scientific approach to understand the macroeconomic problems by applying the research findings in the mathematics, econometrics, econophysics sciences are created, bringing the numerous unbounded opportunities for the scientific innovation in Jakimowicz (2016).

Well, we have discussed some approaches to the precise measurement of the GIP(t, monetary base), GDP(t, monetary base), GNP(t, monetary base), PPP(t, monetary base) in line with the classic macroeconomics science in the previous chapter, realizing that the GIP(t, monetary base), GDP(t, monetary base), GNP(t, monetary base), PPP(t, monetary base) can be used with the purpose to analyze the macroeconomic processes in the economies of the scales and the scopes in Kuznets (1973a, b). Making the next step forward, we would like to consider a research problem on the precise measurement of the macroeconomic variables at the economies of the scales and the scopes in the amplitude, frequency, phase and time domains.

More specifically, we would like apply a classical socioeconomic approach, which is based on the universal fundamental knowledge in the macroeconomics, complementing it by the innovative econophysical theories in the econophysics with the aim to precisely measure the economic output in the economy of the scale and the scope. In other words, we pretend to use the econometrical and econophysical principles, theories and perspectives in our advanced research in the Ledenyov classic and quantum econodynamics in Schumpeter (1906, 1933), Bowley (1924), Fogel (1964), Box, Jenkins (1970), Grangel, Newbold (1977), Van Horne (1984), Taylor S (1986), Tong (1986, 1990), Judge, Hill, Griffiths, Lee, Lutkepol (1988), Hardle (1990), Grangel, Teräsvirta (1993), Pesaran, Potter (1993), Banerjee, Dolado, Galbraith, Hendry (1993), Hamilton (1994), Karatzas, Shreve (1995), Campbell, Lo, MacKinlay (1997), Rogers, Talay (1997), Hayashi (2000), Durbin, Koopman (2000, 2002, 2012), Ilinski (2001), Greene (2003), Koop (2003), Davidson, MacKinnon (2004), Cameron, Trivedi (2005), Iyetomi, Aoyama, Ikeda, Souma, Fujiwara (2008), Iyetomi, Aoyama, Fujiwara, Sato (editors) (2012), Vialar, Goergen (2009), Jakimowicz (2016).

In the Schumpeterian creative-destruction digital-information century, the information in the form of the knowledge in the science, business, culture and society has been generated, transmitted, propagated, received, processed and analyzed by the economic agents at the economies of the scales and the scopes in various countries at the different continents. The information in the form of a numerical measure of the knowledge has been researched in the frames of the information processing theory the information communication science, which is concerned with the scientific ideas on the generation, transmission, gathering, classification, storage, retrieval and analysis of the acquired "bits" of the information in the in Maxwell (1890), Gabor (1946), Shannon (1948). The mathematical analysis of the information is normally performed with an application of the mathematical statistics and the probability sciences in De Laplace (1812), Bunyakovsky (1846), Chebyshev (1846, 1867, 1891), Markov (1890, 1899, 1900, 1906, 1907, 1908, 1910, 1911, 1912, 1913), Kolmogorov (1938, 1985, 1986), Wiener (1949), Brush (1968, 1977), Shiryaev (1995).

In the Ledenyov classic and quantum econodynamics, we propose the Ledenyov theory of the general information product GIP(t, monetary base) in the economies of scales and scopes for the first time. It worth to say that all the generated economic information can be structured, coded, stored, retrieved and analyzed, representing a most valuable asset in possession by the economic agent(s) in the modern economies of the scales and the scopes in the information societies in an information age. In our theory, we introduce a notion on the general information product GIP(t, monetary base), which represents a dependence of the general information product on the time. The dependence of the general information product on the time GIP(t, monetary base) can be interpreted as the ratio of the measured total information data stream by the economic agents to the finite time period (the bits per month/quarter/year) in accordance with the digital signal processing science in Hwang, Briggs (1984), Anceau (1986), Fountain (1987), Chen (editor) (1988), Van de Goor (1989), Priemer (1991), Hsu (1995), Lathi (1998), Prisch (1998), Wanhammar (February 24 1999), McMahon (2007), Ledenyov D O, Ledenyov V O (2015a). Speaking clearly, the measured information has to include all the meaningful data at the multiple information layers, which are generated by the economic agents within the economy of the scale and scope over the finite time period. The exact formula for the GIP(t, monetary base) is Ledenyov D O, Ledenyov V O (2015g)

$$GIP(t, \ monetary \ base) = \frac{Total \ Generated \ Information}{Time} \left[ \frac{Bits}{Month, \ Quarter, Year} \right].$$

In other words, we would like to state that the GIP(t, monetary base) is a main parameter, evaluating a performance of the economies of the scales and the scopes from the macroeconomics point of view. Hence, we think that the Ledenyov economic indicator such as the general information product per the time GIP(t, monetary base), can complement the Kuznets economic indicator such as the general (gross) domestic product per the time GDP(t) in Kuznets (1973a, b), at the accurate measurement of the economic performance of any economy of the scale and the scope in the time domain in agreement with the Ledenyov theory on the GIP(t, monetary base) in the Ledenyov classic and quantum econodynamics.

It worth saying that, going from the macroeconomic point of view, the five main possible origins of the discrete-time fluctuations of the dependence of the general information product on the monetary base on the time GIP(t, monetary base) in the economies of the scales and the scopes can include:

*1.* The discrete-time fluctuations in the technical innovation origination;

*2.* The discrete-time fluctuations in the financial capital availability;

*3.* The discrete-time fluctuations in the qualified labour presence;

*4.* The discrete-time fluctuations in the material/technical resources access;

*5.* The discrete-time fluctuations in the economic/political/social regimes establishment.

In general, we know that the information data streams with the discrete-time nature are constantly generated by the various economic agents in all the existing economic industrial sectors in the modern economies of the scales and the scopes in the information societies in 21st century. Therefore, taking to the consideration the oscillating nature of GIP(t,monetary base), we conclude that the GIP(t, monetary base) represents the discrete-time digital signal (the so called Ledenyov discrete-time digital waves with Markov information) rather than the continuous-time signals (the continuous waves);

because of the discrete-time digital nature of the information generation process by the economic agents in the economies of the scales and the scopes as early researched in the theories on the disruptive innovation in Schumpeter (1911, 1939, 1947), Christensen (June 16, 1977; Fall, 1992a, b; 1997; 1998; December, 1998; April, 1999a, b, c; 1999a, b; Summer, 2001; June, 2002; 2003; March, April, 2003; January, 2006), Bower, Christensen (January, February, 1995; 1997; 1999), Christensen, Armstrong (Spring, 1998), Christensen, Cape (December, 1998), Christensen, Dann (June, 1999), Christensen, Tedlow (January, February, 2000), Christensen, Donovan (March, 2000; May, 2010), Christensen, Overdorf (March, April, 2000), Christensen, Bohmer, Kenagy (September, October, 2000), Christensen, Craig, Hart (March, April, 2001), Christensen, Milunovich (March, 2002), Bass, Christensen (April, 2002), Anthony, Roth, Christensen (April, 2002), Kenagy, Christensen (May, 2002; 2002), Christensen, Johnson, Rigby (Spring, 2002), Hart, Christensen (Fall, 2002), Christensen, Verlinden, Westerman (November, 2002), Shah, Brennan, Christensen (April, 2003), Christensen, Raynor (2003), Burgelman, Christensen, Wheelwright (2003), Christensen, Anthony (January, February, 2004), Christensen, Anthony, Roth (2004), Christensen, Baumann, Ruggles, Sadtler (December, 2006), Christensen, Horn, Johnson (2008), Christensen, Grossman, Hwang (2009), Dyer, Gregersen, Christensen (December, 2009; 2011), Christensen, Talukdar, Alton, Horn (Spring, 2011), Christensen, Wang, van Bever (October, 2013)), Bhattacharya, Ritter (1983), Scherer (1984).

Here, let us explain that the continuous-time signals (the continuous waves, CW) empirical / experimental models of GDP(t, monetary base) cannot be used to finely analyze/approximate/forecast the real-life dependences of GDP(t, monetary base), because of the existing limitations, connected with the nature of the discrete-time statistical data of GIP(t, monetary base), GDP(t, monetary base), GNP(t, monetary base), PPP(t, monetary base). The following problems must be highlighted:

*1.* The discrete-time digital signals are different from the continuous-time signals, hence they cannot be processed with the mathematical formulas derived for the continuous-time signal of GIP(t, monetary base), GDP(t, monetary base), GNP(t, monetary base), PPP(t, monetary base);

2. The discrete-time digital signals are generated by the discrete-time statistical data, hence it makes no sense to apply the continuous-time wave differential filtering technique in Hodrick, Prescott (1980, 1997) with the purpose to filter out the continuous-time wave of GIP(t, monetary base), GDP(t, monetary base), GNP(t, monetary base), PPP(t, monetary base), considering their propagation directions as the so called "trend";

3. The discrete-time digital signals are generated by the discrete-time statistical data, hence the discrete-time statistical data smoothing/approximation techniques with the sinusoid/ co-sinusoid like waveforms in Hodrick, Prescott (1980, 1997) cannot be used in the forecast of GIP(t, monetary base), GDP(t, monetary base), GNP(t, monetary base), PPP(t, monetary base) trends. The problem is that the Ledenyov discrete-time digital waves can change sharply/abruptly/instantly, whereas the continuous-time waves can change slowly/smoothly.

Therefore, we think that all the continuous-time signals (the continuous waves) empirical / experimental outdated models of GDP(t) have to be considered as inaccurate in some sense, including the models discussed in Juglar (1862), George (1881, 2009), Kondratieff (1922, 1925, 1926, 1928, 1935, 1984, 2002), Kitchin (1923), Schumpeter (1939), Burns, Mitchell (1946), Dupriez (1947), Samuelson (1947), Hicks (1950), Goodwin (1951), Inada, Uzawa (1972), Kuznets (1973a, b), Bernanke (1979), Marchetti (1980), Kleinknecht (1981), Dickson (1983), Hodrick, Prescott (1980, 1997), Anderson, Ramsey (1999), Baxter, King (1999), Kim, Nelson (1999), McConnell, Pérez-Quirós (2000), Devezas, Corredine (2001, 2002), Devezas (editor) (2006), Arnord (2002), Stock, Watson (2002), Helfat, Peteraf (2003), Selover, Jensen, Kroll (2003), Sussmuth (2003), Hirooka (2006), Kleinknecht, Van der Panne (2006), Jourdon (2008), Taniguchi, Bando, Nakayama (2008), Drehmann, Borio, Tsatsaronis (2011), Iyetomi, Nakayama, Yoshikawa, Aoyama, Fujiwara, Ikeda, Souma (2011), Ikeda,

Aoyama, Fujiwara, Iyetomi, Ogimoto, Souma, Yoshikawa (2012), Ikeda, Aoyama, Yoshikawa (2013a, b), Uechi, Akutsu (2012).

At this point in our discussion, we can summarize all the acquired knowledge by clearly stating that the dependence of the GIP(t, monetary base), GDP(t, monetary base), GNP(t, monetary base), PPP(t, monetary base) are usually used to characterize the economy of the scale and the scope in Ledenyov D O, Ledenyov V O (2016r). However, looking at the dependences of GIP(t, monetary base), GDP(t, monetary base), GNP(t, monetary base), PPP(t, monetary base), it is not possible to answer the question: What are the separate contributions by both:

*1.* The economic output created by the real economic industrial sector of the economy of the scale and the scope, as well as

*2.* The economic output created by the speculative economic industrial sector of the economy of the scale and the scope,

to the total resulting magnitudes of GIP(t, monetary base), GDP(t, monetary base), GNP(t, monetary base), PPP(t, monetary base)?

The clear separation of the economic output contributions made by:

*1.* The real economic industrial sector of the economy of the scale and the scope, as well as

*2.* The speculative economic industrial sector of the economy of the scale and the scope,

to the total resulting magnitudes of GIP(t, monetary base), GDP(t, monetary base), GNP(t, monetary base), PPP(t, monetary base) is quite important, because of a number of the theoretical and practical reasons.

Here, let us explain that a huge increase in the monetary base by means of the money supply inflow during the quantitative easing program(s) implementation by the central banks / the federal reserve / the treasures in the various financial systems within the economies of the scales and the scopes in the developed(ing) countries may lead to the situation, when:

*1.* A financial sector of the economy of the scale and the scope, including the investment banks/funds with a lot of printed money by the central banks/treasures, contribute to the "fictional growth" of GIP(t, monetary base), GDP(t, monetary base), GNP(t, monetary base), PPP(t, monetary base) mainly;

*2.* A real estate sector of the economy of the scale and the scope, including the building construction/leasing companies, which become hugely overvalued and absorb a lot of money from the home owners/leasers/renters/buyers, which get the mortgages/loans from the investment/commercial banks, which in turn obtain a lot of money from the central banks, contribute to the "fictional growth" of GIP(t, monetary base), GDP(t, monetary base), GNP(t, monetary base), PPP(t, monetary base) mostly;

*3.* An Information Communication Technology (ICT) sector of the economy of the scale and the scope, including the software/hardware development/service providing companies, which become hugely overvalued by obtaining the big money from the institutional/private investors such as the investment/ commercial banks, investment boutiques, pension funds, which in turn get a lot of money from the central banks, contribute to the "fictional growth" of GIP(t, monetary base), GDP(t, monetary base), GNP(t, monetary base), PPP(t, monetary base)mainly.

In other words, we can state that the GIP(t, monetary base), GDP(t, monetary base), GNP(t, monetary base), PPP(t, monetary base) may grow disproportionally, because of the increasing contributions by the speculative sectors of the economies of the scales and the scopes as a result of huge increase of the monetary bases due to the quantitative easing programs implementations by the central banks in the economies of the scales and the scopes in the developed(ing) countries. At the same time, when the GIP(t, monetary base), GDP(t, monetary base), GNP(t, monetary base), PPP(t, monetary base) magnitudes increase disproportionally, the contributions by the real economic industrial sectors of the economies of the scales and the scopes to the changes of GIP(t, monetary base), GDP(t, monetary base), GNP(t, monetary base), PPP(t, monetary base) may strongly decrease/not change/slightly increase, depending on the various (non)objective financial/economic/technological/political factors.

Therefore, we can conclude that, in many practical cases, the reported values of GIP(t, monetary base), GDP(t, monetary base), GNP(t, monetary base), PPP(t, monetary base) can be considered as meaningless by the economists in view of the fact that it is almost not possible to distinguish the contributions made by the real sector of the economy of the scale and the scope as well as by the speculative sector of the economy of the scale and the scope to the total resulting magnitude of GIP(t, monetary base), GDP(t, monetary base), GNP(t, monetary base), PPP(t, monetary base).

We propose to solve the above described problem by introducing the Ledenyov three dimensional (3D) economic output wave diagram to accurately characterize and to clearly visualize all the contributions by real- and speculative- industrial sectors to the final resulting magnitude of GIP(t, monetary base), GDP(t, monetary base), GNP(t, monetary base), PPP(t, monetary base).

Let us take a minute and explain that the Ledenyov 3D wave diagram in the Ledenyov classic and quantum econodynamics has been created, using the theory on the continuous-time electromagnetic waves with the rotating circular polarization vector in the Maxwell electrodynamics in Wikipedia (2016i, j). In Maxwell electrodynamics, the circular polarization of an electromagnetic wave is a polarization state, when the electric field of the electromagnetic wave has a constant magnitude at each point, but its direction rotates with the time at constantly in a plane perpendicular to the direction of the wave.

$$\boldsymbol{E}(\boldsymbol{r},t) = |\boldsymbol{E}| \operatorname{Re}\left\{\boldsymbol{Q}|\psi\rangle \exp\left[i\left(kz - \omega t\right)\right]\right\},$$
$$\boldsymbol{B}(\boldsymbol{r},t) = \hat{\boldsymbol{z}} \times \boldsymbol{E}(\boldsymbol{r},t),$$

*where* $\boldsymbol{E}(\boldsymbol{r},t)$ *is the electric field*,

$\boldsymbol{B}(\boldsymbol{r},t)$ *is the magnetic field*,

$\boldsymbol{Q} = \begin{bmatrix} \hat{\boldsymbol{x}}, \hat{\boldsymbol{y}} \end{bmatrix}$ *is the orthogonal matrix* $2 \times 2$,

$|\psi\rangle \overset{def}{=} \begin{pmatrix} \psi_x \\ \psi_y \end{pmatrix} = \begin{pmatrix} \cos\theta\exp(i\alpha_x) \\ \sin\theta\exp(i\alpha_y) \end{pmatrix}$ *is the Jones vector in* $x - y$ *plane*,

$\omega = ck$ *is the angular frequency*,

$c$ *is the light speed*,

$k$ *is the wave number*.

Fig. 68 displays a scheme to illustrate the circular polarization of the continuous-time electromagnetic wave (the continuous wave (CW)), propagating in the Z direction in the XYZ coordinates space over the time in the Maxwell electrodynamics. The circular polarization vector of the electric field of the continuous-time electromagnetic wave is shown in XY coordinates space. The continuous-time electromagnetic wave propagates in the Z direction in the XYZ coordinates space over the time.

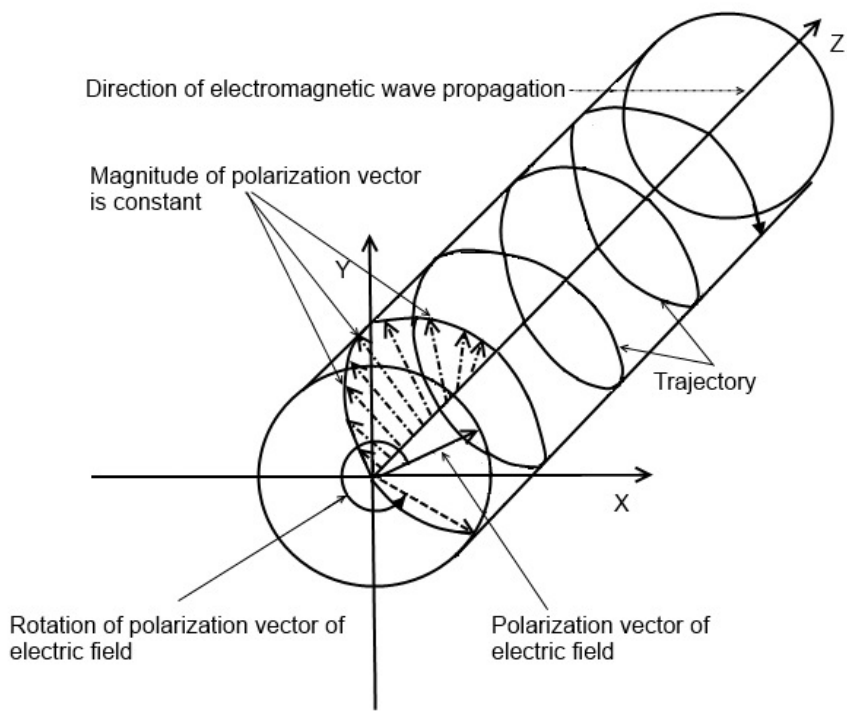

**Fig. 68.** Scheme to illustrate circular polarization of continuous-time electromagnetic wave (continuous wave (CW)), propagating in Z direction in XYZ coordinates space over time in Maxwell electrodynamics. Circular polarization vector of electric field of continuous-time electromagnetic wave is shown in XY coordinates space. Continuous-time electromagnetic wave propagates in Z direction in XYZ coordinates space over the time.

Usually, the continuous-time electromagnetic waves with the rotating circular polarization vector are normally used in the process of the information transmission over the different wireless/optical communication links/channels in the space and the terrestrial telecommunications:

*1.* The ground-to-space wireless communication links;

*2.* The space-to-ground satellite communication links;

*3.* The fiber optics communication channels.

In the Ledenyov classic and quantum econodynamics, let us introduce a notion on the Ledenyov economic activity vector and write a set of the mathematical expressions for the GIP(t, monetary base), GDP(t, monetary base), GNP(t, monetary base), PPP(t, monetary base), aiming to make the definition of GDP(t, monetary base), in this particular case, as in the formula:

$$GDP_{Total}\left(t, \ monetary \ base\right) = \sum GDP_n\left(t, \ monetary \ base\right)$$

$$GDP_{total} = EAV^2 = \left(EAV_{real}\right)^2 + \left(EAV_{speculative}\right)^2$$

$where:\ GDP\ is\ the\ Gross\ Domestic\ Product,\ which\ is\ a\ scalar\ value;$

$t\ is\ the\ time;$

$monetary\ base\ is\ the\ monetary\ base\ issued\ by\ Treasure\ /\ Central\ Bank$

$EAV\ is\ the\ economic\ activity\ vector,\ which\ is\ a\ vector\ magnitude;$

$EAV_{real}\ is\ the\ real\ economic\ activity\ vector\ component;$

$EAV_{speculative}\ is\ the\ speculative\ economic\ activity\ vector\ component.$

The Ledenyov 3D wave diagram shows the total GIP(t, monetary base), GDP(t, monetary base), GNP(t, monetary base), PPP(t, monetary base) as a sum of the two components, including the real- and the speculative-magnitudes of GIP(t, monetary base), GDP(t, monetary base), GNP(t, monetary base), PPP(t, monetary base). It can change as 1) the continuous-time wave in Kitchin (1923), Juglar (1862), Kuznets (1973a, b), Kondratieff, Stolper (1935) or as 2) the discrete-time wave in Ledenyov D O, Ledenyov V O (2013c, 2015d, 2015e).The phase angle φ defines the tilt of GIP(t, monetary base), GDP(t, monetary base), GNP(t, monetary base), PPP(t, monetary base), depending on the real- and the speculative- magnitudes.

Fig. 69 shows graphically the proposed Ledenyov three dimensional (3D) wave diagram in the macroeconomics science, which can be used to accurately characterize and finely display the GIP(t, monetary base), GDP(t, monetary base), GNP(t, monetary base), PPP(t, monetary base) dependences changes dynamics in the time domain in the two possible cases: the continuous-time waves of GIP(t, monetary base), GDP(t, monetary base), GNP(t, monetary base), PPP(t, monetary base) and the discrete-time waves of GIP(t, monetary base), GDP(t, monetary base), GNP(t, monetary base), PPP(t, monetary base) in the Ledenyov classic and quantum econodynamics.

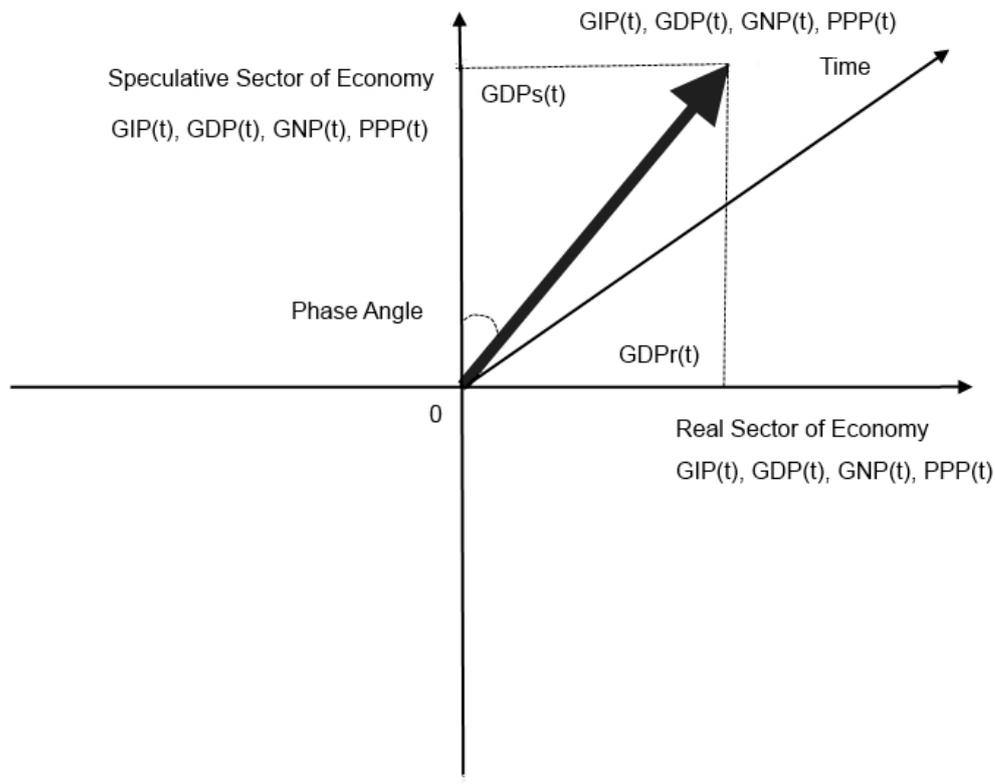

**Fig. 69.** Economic activity vector (EAV), defined by phase angle α in XY coordinates space, for dependences of GIP(t, monetary base), GDP(t, monetary base), GNP(t, monetary base), PPP(t, monetary base) in economy of scale and scope. GDP$_r$(t, monetary base) is positive projection of EAV on scale of real economic sector in XY coordinates space. GDP$_s$(t, monetary base) is positive projection of EAV on scale of speculative economic sector in XY coordinates space. EAV changes in time scale similar to continuous-/discrete- time wave with rotating polarization in Ledenyov classic and quantum econodynamics.

Fig. 70 displays graphically the proposed Ledenyov three dimensional (3D) wave diagram in the macroeconomics science, which can be used to accurately characterize and finely display the GIP(t, monetary base), GDP(t, monetary base), GNP(t, monetary base), PPP(t, monetary base) dependences changes dynamics in the time domain in the two possible cases: the continuous-time waves of GIP(t, monetary base), GDP(t, monetary base), GNP(t, monetary base), PPP(t, monetary base) and the discrete-time waves of GIP(t, monetary base), GDP(t, monetary base), GNP(t, monetary base), PPP(t, monetary base) in the Ledenyov classic and quantum econodynamics.

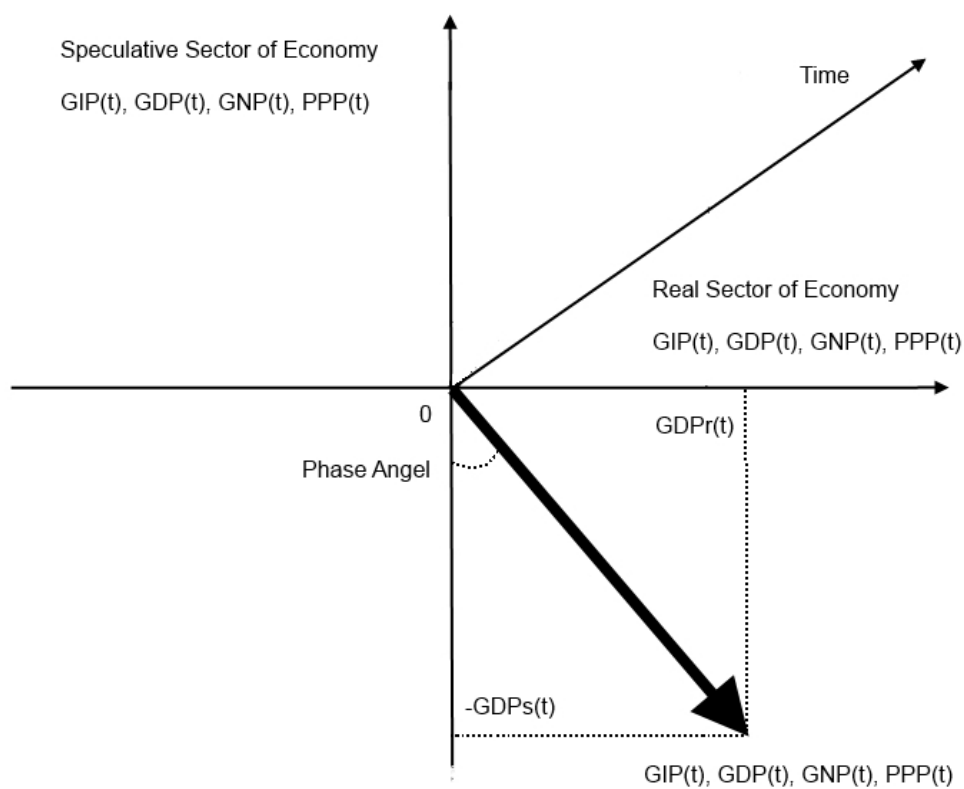

**Fig. 70.** Economic activity vector (EAV), defined by phase angle α in XY coordinates space, for dependences of GIP(t, monetary base), GDP(t, monetary base), GNP(t, monetary base), PPP(t, monetary base) in economy of scale and scope. $GDP_r$(t, monetary base) is positive projection of EAV on scale of real economic sector in XY coordinates space. $GDP_s$(t, monetary base) is negative projection of EAV on scale of speculative economic sector in XY coordinates space. EAV changes in time scale similar to continuous-/discrete- time wave with rotating polarization in Ledenyov classic and quantum econodynamics.

Fig. 71 demonstrates graphically the proposed Ledenyov three dimensional (3D) wave diagram in the macroeconomics science, which can be used to accurately characterize and finely display the GIP(t, monetary base), GDP(t, monetary base), GNP(t, monetary base), PPP(t, monetary base) dependences changes dynamics in the time domain in the two possible cases: the continuous-time waves of GIP(t, monetary base), GDP(t, monetary base), GNP(t, monetary base), PPP(t, monetary base) and the discrete-time waves of GIP(t, monetary base), GDP(t, monetary base), GNP(t, monetary base), PPP(t, monetary base) in the Ledenyov classic and quantum econodynamics.

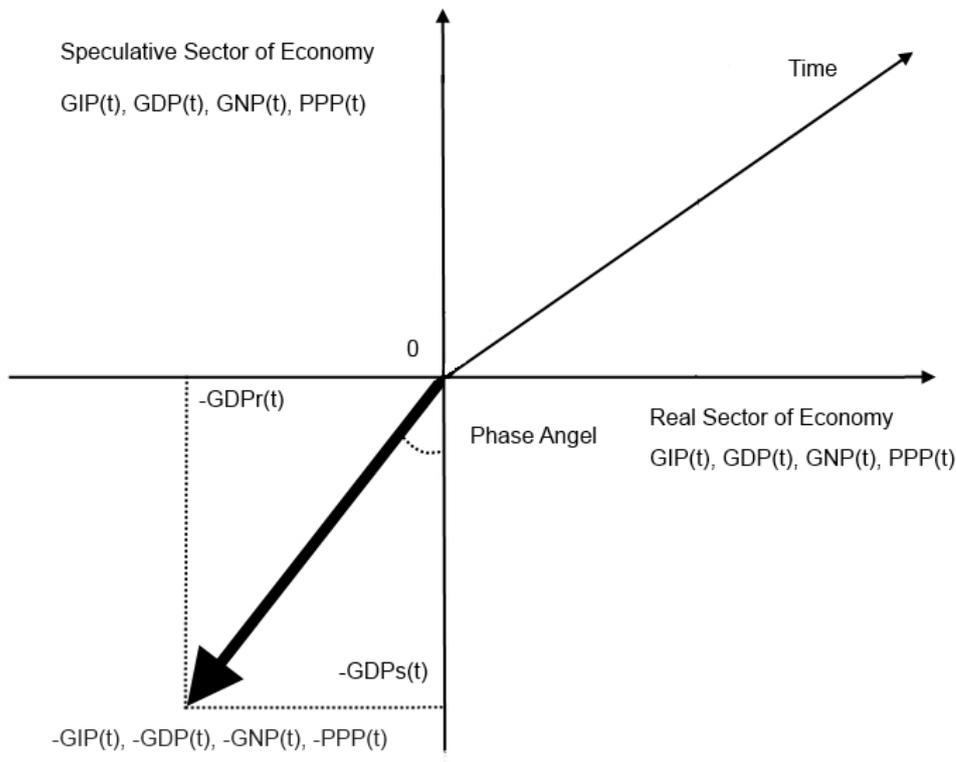

**Fig. 71.** Economic activity vector (EAV), defined by phase angle α in XY coordinates space, for dependences of GIP(t, monetary base), GDP(t, monetary base), GNP(t, monetary base), PPP(t, monetary base) in economy of scale and scope. $GDP_r$(t, monetary base) is negative projection of EAV on scale of real economic sector in XY coordinates space. $GDP_s$(t, monetary base) is negative projection of EAV on scale of speculative economic sector in XY coordinates space. EAV changes in time scale similar to continuous-/discrete- time wave with rotating polarization in Ledenyov classic and quantum econodynamics.

Fig. 72 depicts graphically the proposed Ledenyov three dimensional (3D) wave diagram in the macroeconomics science, which can be used to accurately characterize and finely display the GIP(t, monetary base), GDP(t, monetary base), GNP(t, monetary base), PPP(t, monetary base) dependences changes dynamics in the time domain in the two possible cases: the continuous-time waves of GIP(t, monetary base), GDP(t, monetary base), GNP(t, monetary base), PPP(t, monetary base) and the discrete-time waves of GIP(t, monetary base), GDP(t, monetary base), GNP(t, monetary base), PPP(t, monetary base) in the Ledenyov classic and quantum econodynamics.

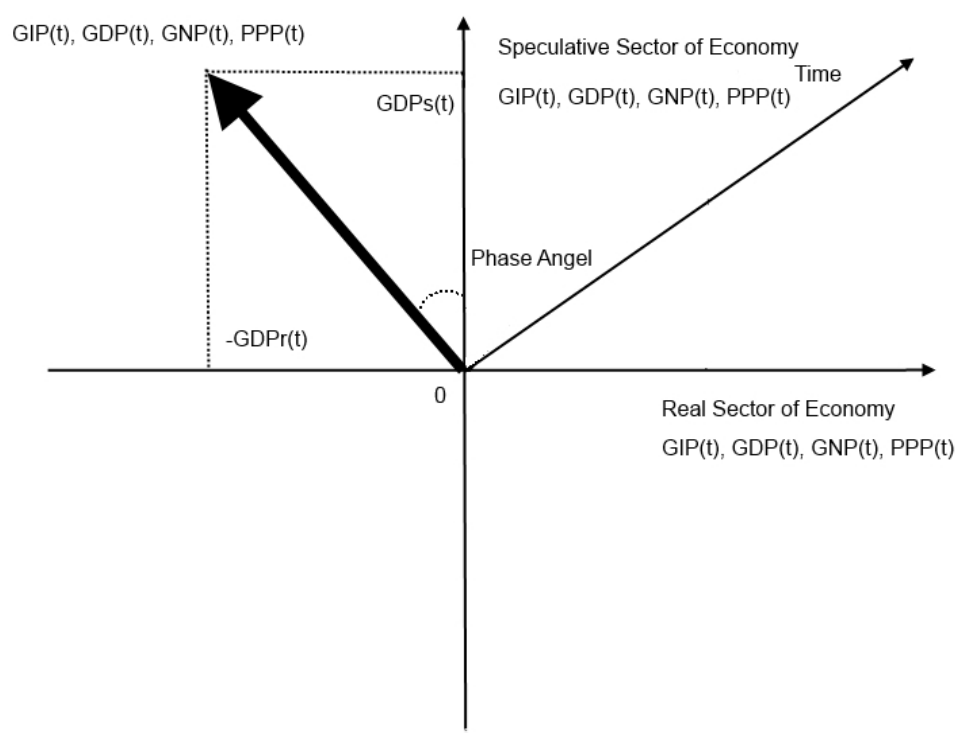

**Fig. 72.** Economic activity vector (EAV), defined by phase angle α in XY coordinates space, for dependences of GIP(t, monetary base), GDP(t, monetary base), GNP(t, monetary base), PPP(t, monetary base) in economy of scale and scope. $GDP_r$(t, monetary base) is negative projection of EAV on scale of real economic sector in XY coordinates space. $GDP_s$(t, monetary base) is positive projection of EAV on scale of speculative economic sector in XY coordinates space. EAV changes in time scale similar to continuous-/discrete- time wave with rotating polarization in Ledenyov classic and quantum econodynamics.

Going to the next point, let us make a few important scientific comments on the special features and the main differences between the continuous-time waves of GIP(t, monetary base), GDP(t, monetary base), GNP(t, monetary base), PPP(t, monetary base) and the discrete-time waves of GIP(t, monetary base), GDP(t, monetary base), GNP(t, monetary base), PPP(t, monetary base) at the Ledenyov 3D wave diagrams:

*1.* The continuous-time wave of GIP(t, monetary base), GDP(t, monetary base), GNP(t, monetary base), PPP(t, monetary base) changes its amplitude / frequency / wavelength / period / phase / polarization continuously in the time domain. Therefore, in this case, we assume that the continuous-time wave of GIP(t, monetary base), GDP(t, monetary base), GNP(t, monetary base), PPP(t, monetary base) can be modulated by the continuous-time economical, financial, political and social events in agreement with the old philosophical scientific views in the macroeconomics.

*2.* The discrete-time wave of GIP(t, monetary base), GDP(t, monetary base), GNP(t, monetary base), PPP(t, monetary base) changes its amplitude / frequency / wavelength / period / phase / polarization discretely in the time domain. Therefore, in this case we permit that the discrete-time wave of GIP(t, monetary base), GDP(t, monetary base), GNP(t, monetary base), PPP(t, monetary base) can be modulated by the discrete-time economical, financial, political and social events such as the disruptive innovations in the economies of scales and scopes in the time domain in Ledenyov D O, Ledenyov V O (2013c, 2015d, 2015e), Christensen, Denning (December 2015).

Let us summarize all the ideas by emphasizing that we considered a research problem on the precise measurement of the macroeconomic variables changes in the time domain in the macroeconomics science. We propose to use the Ledenyov three dimensional (3D) wave diagram in the macroeconomics science for the first time, aiming to accurately characterize and to clearly visualize the GIP(t, monetary base), GDP(t, monetary base), GNP(t, monetary base), PPP(t, monetary base) dependences changes in the time domain. We explain that the Ledenyov three dimensional (3D) wave diagram in the modern macroeconomics science has been created, using some limited analogy with the theory on the continuous-time waves with the rotating polarization vector in the electrodynamics science. We show that the Ledenyov three dimensional (3D) wave diagram in the macroeconomics science can be used to accurately characterize and finely display the GIP(t, monetary base), GDP(t, monetary base), GNP(t, monetary base), PPP(t, monetary base) dependences changes in the time domain in the two possible cases: the continuous-time waves of GIP(t, monetary base), GDP(t, monetary base), GNP(t, monetary base), PPP(t, monetary base) and the discrete-time waves of GIP(t, monetary base), GDP(t, monetary base), GNP(t, monetary base), PPP(t, monetary base). Thus, an introduction of the Ledenyov three dimensional (3D) wave diagram in the Ledenyov econodynamics and the modern macroeconomics sciences can help to solve a challenging research problem on the precise measurement of the macroeconomic variables in the time domain.

Well, at this point in our research, we can definitely say that the GIP(t, monetary base), GDP(t, monetary base), GNP(t, monetary base), PPP(t, monetary base) periodic oscillations in the amplitude, frequency, phase and time domains result in an origination of the business cycles in the economy of scale and the scope. Also, we can assume that, having an exact information about the business cycles fluctuation dynamics, it would be quite possible for us to make the accurate predictions on the GIP(t, monetary base), GDP(t, monetary base), GNP(t, monetary base), PPP(t, monetary base) changes dynamics in various time perspectives. However, the GIP(t, monetary base), GDP(t, monetary base), GNP(t, monetary base), PPP(t, monetary base) changes dynamics forecast is not a simple scientific problem to solve, and most frequently, it is a real challenge for the researchers to be able to accurately characterize the business cycles in the economy of scale and the scope in the short/long time periods. Having the technical capabilities to obtain the Ledenyov 3D wave diagram for the economy of the scale and scope in the real-time domain, we can evaluate precisely the state of matters in the economy of scale and scope in the time domain and to forecast accurately the possible developments in the economy of scale and scope in the time domain. In other words, once again, an introduction of the Ledenyov three dimensional (3D) wave diagram in the Ledenyov classic and quantum econodynamics and the modern macroeconomics sciences makes it possible to solve a challenging research problem on the precise measurement of the macroeconomic variables changes in the time domain.

In next Chapter 8, we will discuss the problem on the accurate forecast of the economic and financial trends with the business cycles oscillation dynamics analysis in the economy of the scale and the scope at the certain monetary base over the selected time period in the Ledenyov classic and quantum econodynamics.

# CHAPTER 8

## Accurate forecast of economic and financial trends with business cycles oscillation dynamics analysis in economy of scale and scope in classic and quantum econodynamics

The economic output magnitude in the economy of the scale and the scope at the certain monetary base over the time has an oscillating nature, because the economic macro-, micro-, nano- variables have the tendencies to change their values dependently/ periodically / chaotically in the economy of the scale and the scope at the certain monetary base over the time. It means that the economic indicators such as the General Information Product GIP(t, monetary base), Gross Domestic Product GDP(t, monetary base), Gross National Product GNP(t, monetary base), Purchasing Power Parity PPP(t, monetary base) may change sharply/abruptly/instantly in the economy of the scale and the scope at the certain monetary base over the selected time period. In addition, we can evidently say that the General Information Product GIP(t, monetary base), Gross Domestic Product GDP(t, monetary base), Gross National Product GNP(t, monetary base), Purchasing Power Parity PPP(t, monetary base) dependencies can only be measured with a certain degree of accuracy in the economy of the scale and the scope at the certain monetary base in the amplitude/frequency/phase/time domains in agreement with the Ledenyov classic / quantum econodynamics sciences.

As we know, in the political economy, there are multiple economic agents, which operate and create the new wealth during the wealth synthesis process in the real and/or speculative sectors in the economy of the scale and the scope at the certain monetary bases over the time. Basically, the wealth creation process includes the three essential components and takes some time period for its realization: *1)* The matter; *2)* The labor; *3)*The capital.

As we can see the financial capital is an essential part in the wealth synthesis process in the economy of the scale and the scope at the certain monetary bases over the time. The financial capital can be created, collected, accumulated, stored, used, and most importantly, increased by means of the investment process. In other words, the financial capital must be invested by the financial capital investors with an ultimate goal to gain an increased return premium on the invested financial capital during the certain period of time toward the financial/economic private/public prosperity building.

Applying the computer science terminology, let us define the investment process in the form of the Ledenyov hypothetical three layers investment process protocols stack in finances:

*1.* The investment products, which utilize the financial capital;

*2.* The investment vehicles, which invest the financial capital;

*3.* The investment mediums, which make the financial capital ecosystem.

Fig. 73 shows the Ledenyov hypothetical investment process protocols stack for the financial capital investment in the capital markets in the economies of the scales and the scopes over the short / long time periods.

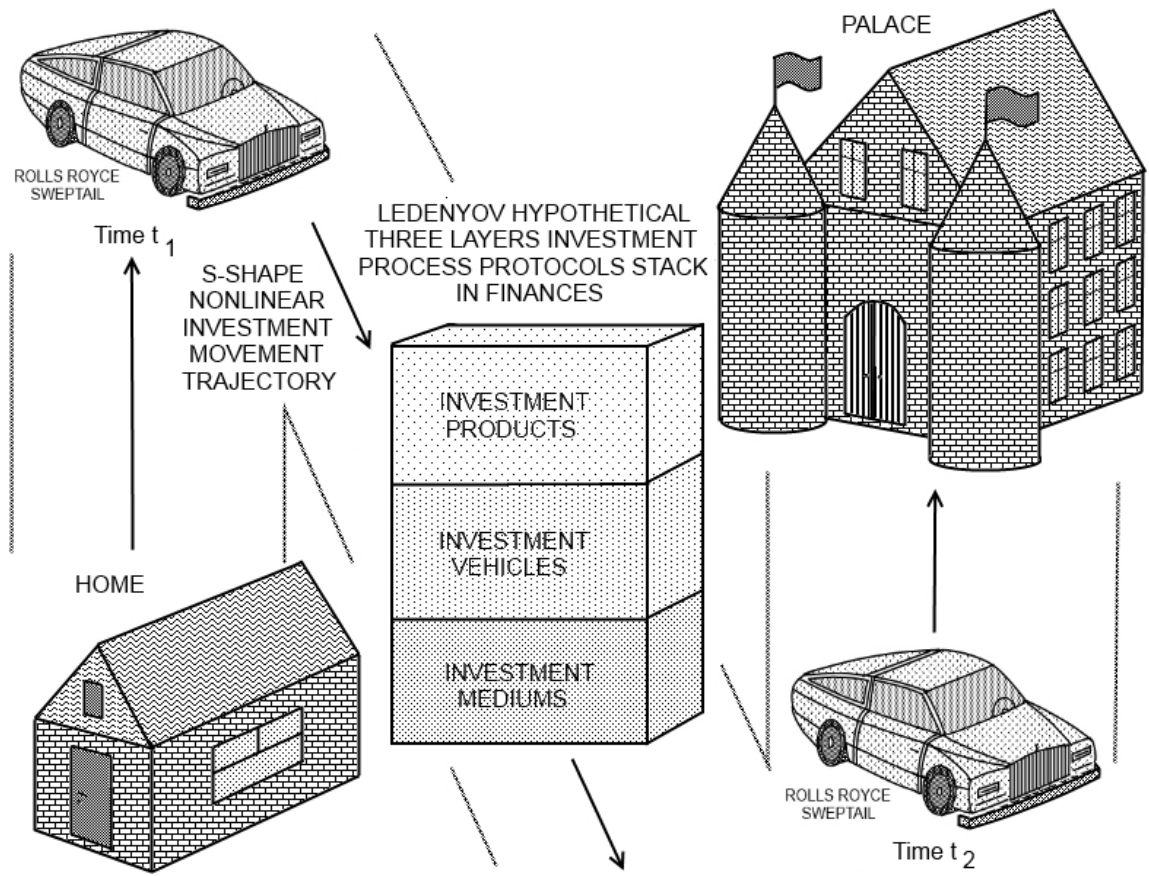

**Fig. 73.** Investment process essentials: *1)* Investment products, *2)* investment vehicles, *3)* investment mediums in Ledenyov hypothetical three layers investment process protocols stack in finances.

At this point in our discussion, we can notice that the two most frequently solved economic problems by the investors / the businessmen in the capital markets in the economies of the scales and the scopes are:

1. The problem on the financial capital investment opportunities finding by the private/institutional/corporate investors, aiming to derive a high return premium on the invested financial capital in the capital markets in the short / long time periods;

2. The problem on the financial capital borrowing opportunities search by the private/institutional/corporate borrowers, aiming to get a low interest rate on the borrowed financial capital in the capital markets in the short / long time periods.

Let us discuss the above two fundamental problems, keeping in mind the fact that any reasonably accurate prediction towards the investment opportunities appearance in the course of the economic and financial trends evolutions can be made by using the business cycles oscillation dynamics analysis in the economies of the scales and the scopes in the short and long time periods in agreement with in the Ledenyov classic and quantum econodynamics sciences.

We would like to continue with a challenging problem on the financial capital investment opportunities finding by the private/institutional investors, aiming to derive a high return premium on the invested capital in the capital markets in the short / long time periods with the purpose to make the extra money and increase the accumulated wealth. In this connection, let us imagine the Ledenyov hypothetical investment portfolio with the randomly diversified uncorrelated investment products, the investment vehicles, and the investment mediums in Ledenyov V O, Ledenyov D O (2017):

1. The investment products: the land, real estate, commodity, bond, company stock, company common/executive/preferred stock option, financial security, foreign currency, intellectual property, antiquaries, ancient/modern collectable art pieces, Swiss valuable time pieces;

2. The investment vehicles: the investment bank, investment fund, hedge fund, mutual fund, venture capital fund, angel investor, investment boutique, auction house;

***3.*** The investment mediums: the land exchange, real estate exchange, stock exchange, foreign currencies exchange, financial securities exchange, commodities exchange, precious metals exchange, intellectual property exchange, auction exchange.

Let us explain that a main characteristic distinction by the Ledenyov hypothetical investment portfolio from many other investment portfolios is in concluded in the fact that the Ledenyov hypothetical investment portfolio includes the randomly diversified uncorrelated selected investment products, the investment vehicles, and the investment mediums.

Fig. 74 displays a Ledenyov hypothetical investment capital portfolio with the diversified uncorrelated investment products, the investment vehicles and the investment mediums for the financial capital investment in the capital markets in the economies of the scales and the scopes in the short and long time periods.

INVESTMENT CAPITAL PORTFOLIO

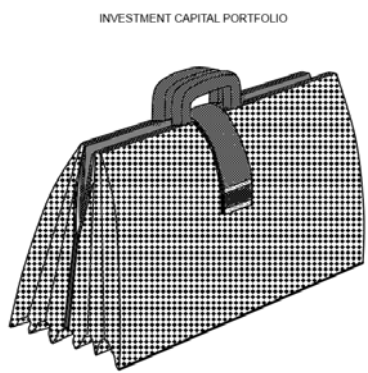

**Fig. 74.** Ledenyov hypothetical investment capital portfolio with diversified uncorrelated investment products, investment vehicles and investment mediums for financial capital investment in capital markets in economies of scales and scopes in short and long time periods.

In the real life investment scenarious in the finances, a Ledenyov modern investment capital portfolio is usually made of a big number of the investment capital portfolios with the diversified uncorrelated investment products, investment vehicles and investment mediums for the financial capital investment in the capital markets in the economies of the scales and the scopes in the short / long time periods. A main reason for a rising complexity in the structure of the Ledenyov modern investment capital portfolio is associated with a natural interest by the portfolio managers/the seasoned investors to minimize the total risk by decreasing the level of possible statistical distributions correlations between the investment products, the investment vehicles and the investment mediums in the case of the financial capital investment in the capital markets in the economies of the scales and the scopes in the short and/or long time periods.

Fig. 75 shows a modern investment capital portfolio made of a big number of the investment capital portfolios with the diversified investment products, investment vehicles and investment mediums for the financial capital investment in the capital markets in the economies of the scales and the scopes in the short and long time periods.

INVESTMENT CAPITAL PORTFOLIO MADE OF NUMBER OF INVESTMENT CAPITAL PORTFOLIOS

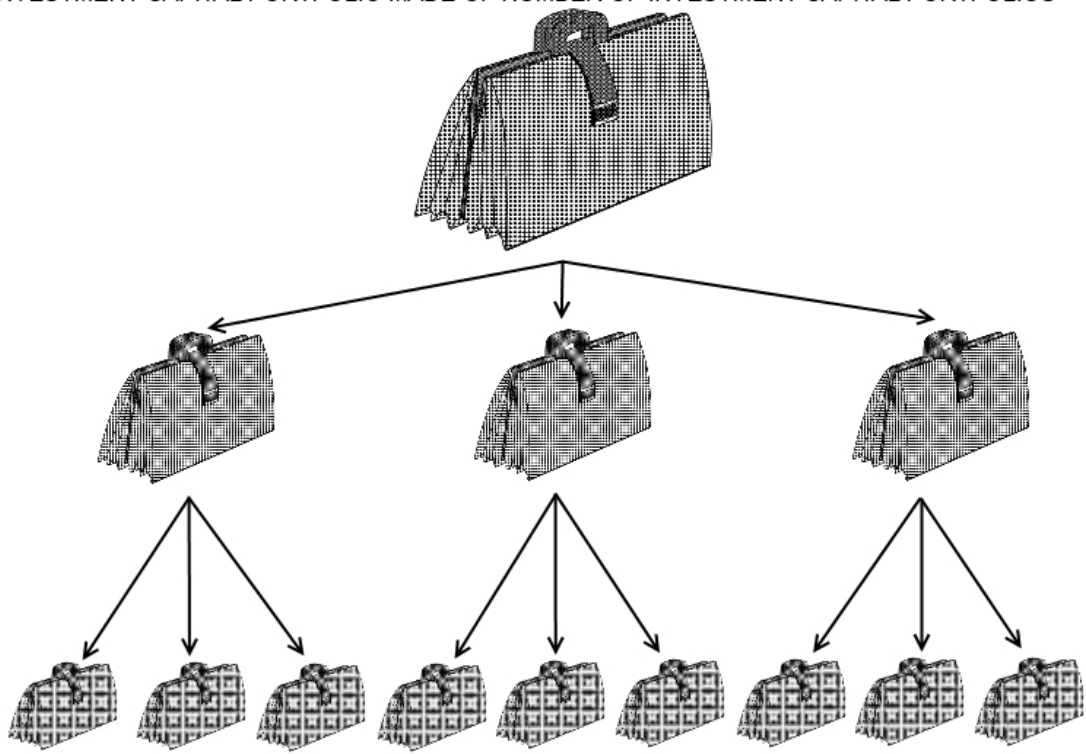

**Fig. 75.** Ledenyov modern investment capital portfolio made of big number of investment capital portfolios with diversified uncorrelated investment products, investment vehicles and investment mediums for financial capital investment in capital markets in economies of scales and scopes in short and long time periods.

Now, considering the modern investment theories and practices, let us highlight the main characteristics of the Ledenyov modern investment portfolio, advancing our initial representation to a qualitatively new level:

1. *1.* The Ledenyov modern investment portfolio has the layered structure, including the three basic layers: the investment products, the investment vehicles, the investment mediums;

2. *2.* The Ledenyov modern investment portfolio can be made of a big number of the investment portfolios with the diversified investment products, the investment vehicles and the investment mediums;

3. *3.* The Ledenyov modern investment portfolio with the diversified investment products, the investment vehicles and the investment mediums is created by having in mind an idea on the financial and economic risks optimization, minimization, and possible exception;

4. *4.* The Ledenyov quantum winning virtuous investment strategies with the Ledenyov quantum winning virtuous strategy search (LQWVSS) algorithm must be created and executed to get an increased return premium during the financial capital investment by means of the Ledenyov modern investment portfolios with the optimized designs in the capital markets in the economies of the scales and the scopes in the short and long time periods in Ledenyov V O, Ledenyov D O (2017);

5. *5.* The Ledenyov quantum winning virtuous investment strategies in the capital markets within the economies of the scales and the scopes can be multidimensional:

   *a)* in terms of the investment products, vehicles, mediums dimensions,

   *b)* in terms of the investment scales, spaces, dimensions,

   *c)* in terms of the investment strategies execution vectors direction, including the sequential, parallel, recursive, non-recursive, mixed execution vector directions dimensions,

   *d)* in terms of the investment decision making inductive, deductive, abductive, quantum (probabilistic) logics dimensions,

   *e)* in terms of the investment agility, stability, predictability factors dimensions.

6. *6.* The roboadviser(s) with the artificial intelligence (AI) can be used to make the near-real time investment decisions during the algorithmic investment/trading in the capital markets in the economies of the scales and the scopes in the short and long time periods;

*7.* The roboadviser(s) with the algorithmic investment/trading technical capabilities may have the unique technical characteristics to create and execute the different versatile non-uniform quantum winning virtuous investment strategies instead of the simple primitive investment strategies in the capital markets in the economies of the scales and the scopes in the short and long time periods;

*8.* The roboadviser(s) can create and execute the quantum winning virtuous investment strategies during the algorithmic investment/trading in the capital markets in the economies of the scales and the scopes in the short and long time at the ultra high frequencies (UHF).

*9.* The roboadviser(s) can get an access to the big data streams with the information on the capital markets during the algorithmic trading in the capital markets in the economies of the scales and the scopes in the short and long time at the ultra high frequencies (UHF).

Providing some illustrations on the Ledenyov modern investment capital portfolio management practices, let us say that a computer roboadviser normally makes the automated trading decisions on the financial capital investment in the capital markets at the ultra high frequencies, using the following things:

*1.* A set of the recursive trading decision making algorithms;

*2.* A number of the multiple relational databases with the accumulated trading patterns;

*3.* A number of the real-time high-speed filtered financial data streams.

Therefore, we my assume that a modern computer roboadviser with the artificial intelligence (AI) capabilities can efficiently manage the modern investment portfolio, which is made of a big number of the investment portfolios with the diversified investment products, the investment vehicles and the investment mediums for the financial capital investment in the capital markets in the economies of the scales and the scopes in the short and long time periods. The artificial intelligence (AI) principles and techniques were comprehensively researched in Turing (October 1950), Rich (1983), Winston (1984), Haugeland (1985), Edelson (1991), Jang (July 1991), Sun, Bookman (1994), John, Langley (1995), Dowe, Hajek (1997), Kohavi (1997), Mitchell (1997), Nilsson (1998, 2010), Poole, Mackworth, Goebel (1998), Calmet, Benhamou, Caprotti, Henocque, Sorge (2002), Russell, Norvig (2003), Luger, Stubblefield (2004), McNelis (2005), Bach (2008), Neapolitan, Jiang (2012).

Fig. 76 depicts a block scheme of a computer roboadviser with the artificial intelligence (AI) capabilities for the modern investment portfolio management in the capital markets in the economies of the scales and the scopes at certain monetary base in the short / long time periods.

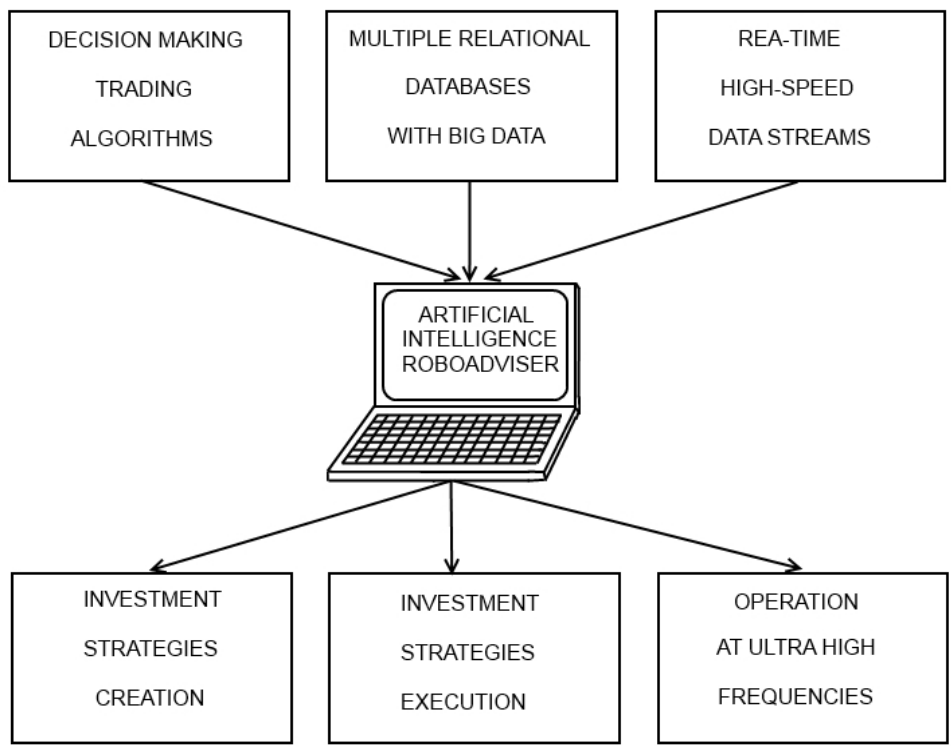

**Fig. 76.** Block scheme of computer roboadviser with artificial intelligence (AI) capabilities for modern investment portfolio management in capital markets in nonlinear diffusion type economy of scale and scope at short / long time periods in finances.

The Ledenyov modern investment portfolio's total risk calculation with an account for the macro-, micro- and nano- economic risks, generated by the investment products, investment vehicles and investment mediums during a process of the modern investment portfolio management in the capital markets in the economies of the scales and the scopes at the certain monetary base in the short / long time periods in the Ledenyov classic and quantum econodynamics sciences can certainly be considered as one of the most important tasks by a computer roboadviser. Therefore, let us draw an approximate block scheme to compute the modern investment portfolio's total risk magnitude below.

Fig. 77 shows a block scheme for the investment portfolio's total risk calculation with an account for the macro-, micro- and nano- economic risks by the investment products, investment vehicles and investment mediums in the capital markets in the nonlinear diffusion type economy of the scale and the scope in the finances.

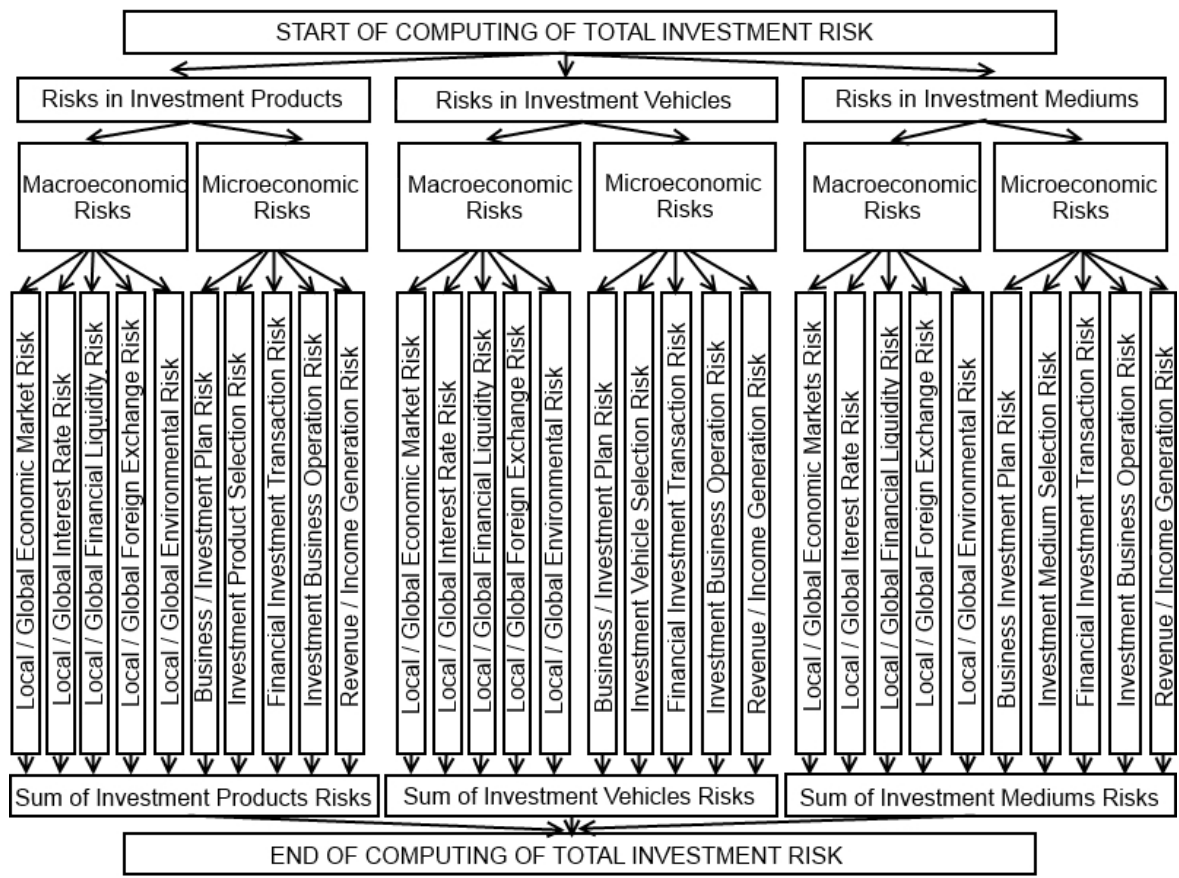

**Fig. 77.** Algorithm in form of block scheme for investment portfolio's total risk calculation with account for macro-, micro- and nano- economic risks by investment products, investment vehicles and investment mediums in capital markets in nonlinear diffusion type economy of scale and scope in finances.

Let us slightly modify the above block scheme, designing the Ledenyov high complexity watch's dial for the investment portfolio's total risk calculation with an account for the macro-, micro- and nano- economic risks by the investment products, investment vehicles and investment mediums in the diffusion type nonlinear financial system in the economy of the scale and the scope. This modification can help us to better memorize all the important research findings in the course of our research discussion.

Fig. 78 shows the Swiss high complexity watch's dial for the investment portfolio's total risk calculation with an account for the macro-, micro- and nano- economic risks by the investment products, investment vehicles and investment mediums in the capital markets in the nonlinear diffusion type economy of the scale and the scope in the finances.

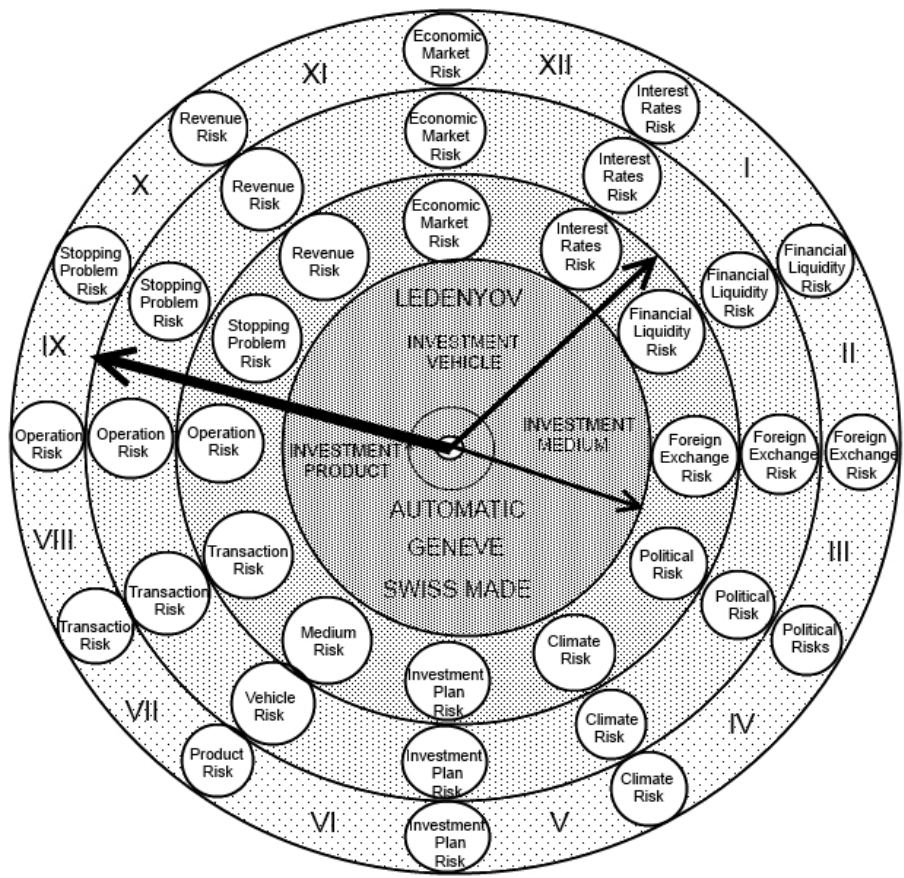

**Fig. 78.** Ledenyov high complexity automatic radial watch's dial for investment portfolio's total risk calculation with account for macro-, micro- and nano- economic risks by investment products, investment vehicles and investment mediums in capital markets in nonlinear diffusion type economy of scale and scope in finances.

The modern borrowing portfolio's total risk magnitude with an account for the macro-, micro- and nano- economic risks, generated by the borrowing products, borrowing vehicles and borrowing mediums during a process of the modern borrowing portfolio management in the capital markets in the economies of the scales and the scopes at the certain monetary base in the short / long time periods in the Ledenyov classic and quantum econodynamics sciences can also be calculated by taking to an account the appropriate risk factors. The block scheme to compute the modern borrowing portfolio's total risk magnitude is shown below.

Fig. 79 shows a block scheme for the borrowing portfolio's total risk calculation with an account for the macro-, micro- and nano- economic risks by the borrowing products, borrowing vehicles and borrowing mediums in the capital markets in the nonlinear diffusion type economy of the scale and the scope in the finances.

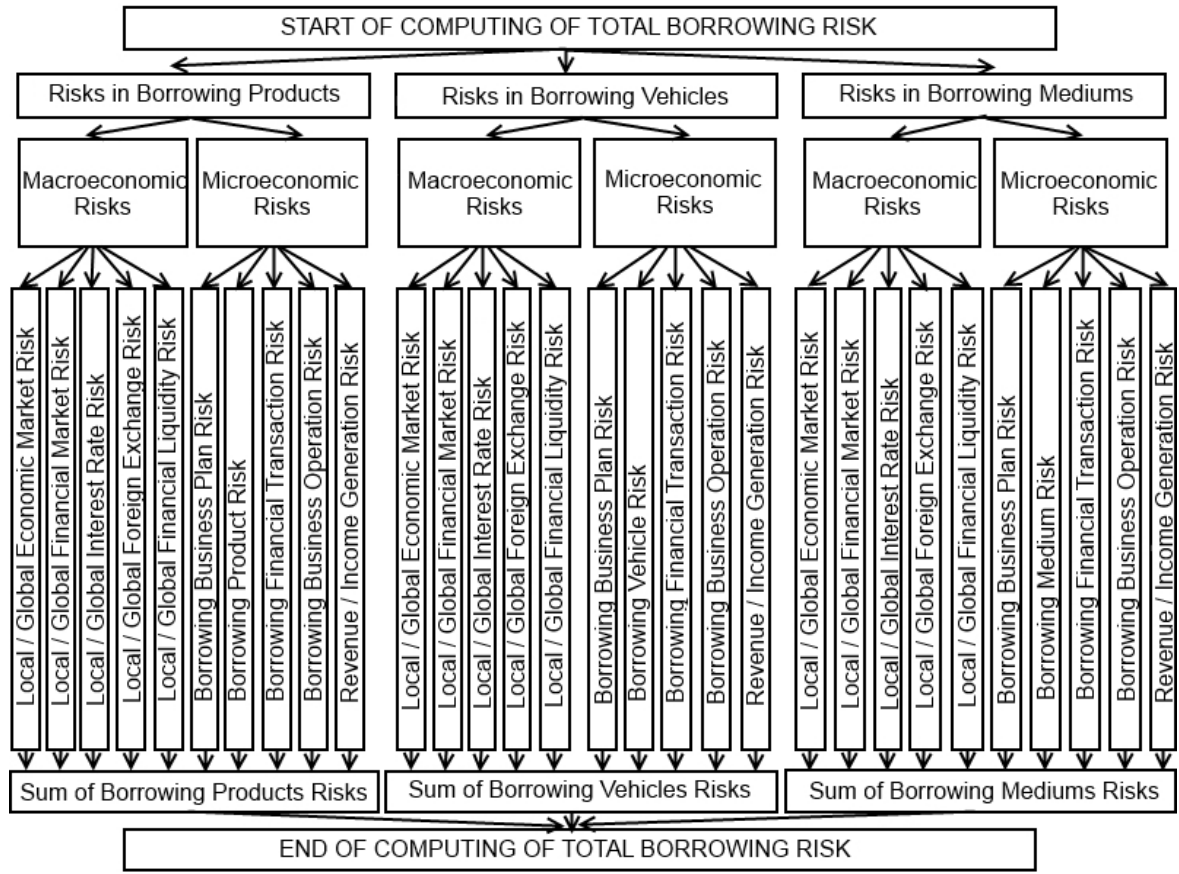

**Fig. 79.** Algorithm in form of block scheme for borrowing portfolio's total risk calculation with account for macro-, micro- and nano- economic risks by borrowing products, borrowing vehicles and borrowing mediums in capital markets in nonlinear diffusion type economy of scale and scope in finances.

We can design the Ledenyov high complexity watch's dial for the borrowing portfolio's total risk calculation with an account for the macro-, micro- and nano- economic risks by the borrowing products, borrowing vehicles and borrowing mediums during a process of the modern borrowing portfolio management in the capital markets in the economies of the scales and the scopes at the certain monetary base in the short / long time periods in the Ledenyov classic and quantum econodynamics sciences.

Fig. 80 displays the Ledenyov high complexity automatic watch's dial for the borrowing portfolio's total risk calculation with an account for the macro-, micro- and nano- economic risks by the borrowing products, borrowing vehicles and borrowing mediums in the capital markets in the nonlinear diffusion type economy of the scale and the scope in the finances.

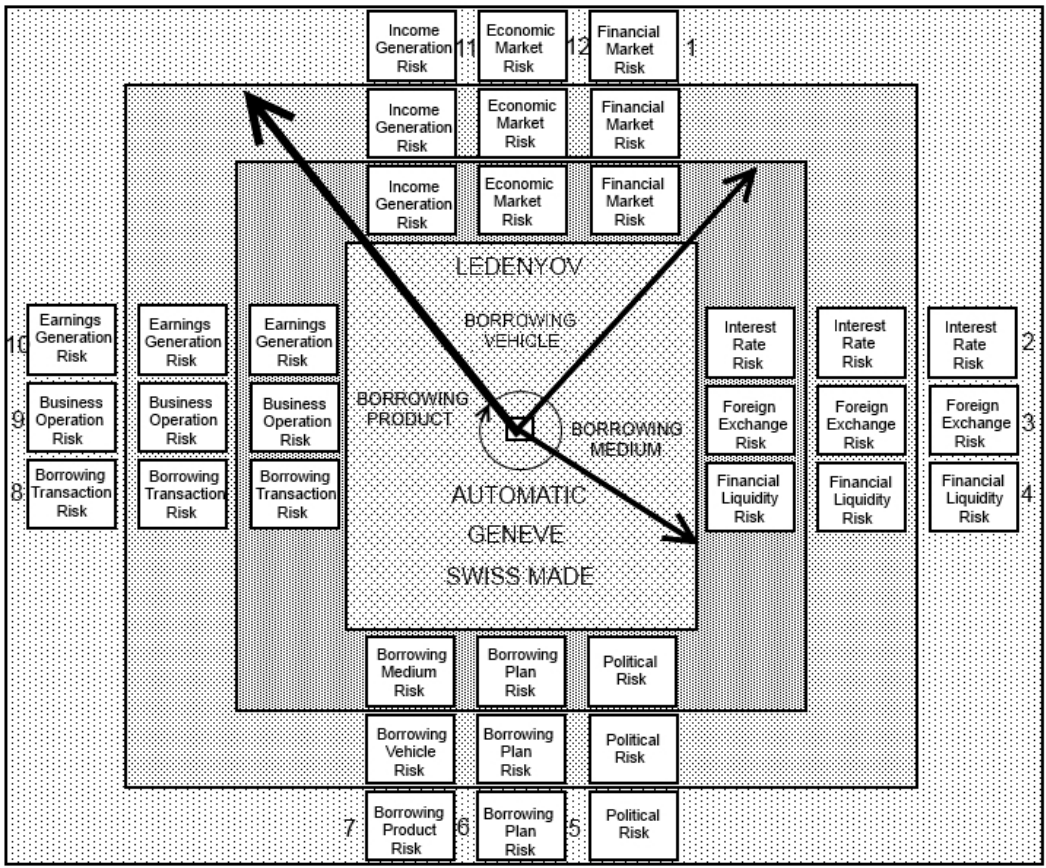

**Fig. 80.** Ledenyov high complexity automatic rectangular watch's dial for borrowing portfolio's total risk calculation with account for macro-, micro- and nano- economic risks by borrowing products, borrowing vehicles and borrowing mediums in capital markets in nonlinear diffusion type economy of scale and scope in finances.

Now let us explain that the investors intent to increase the wealth during the wealth management process with an application of the modern investment portfolio in the capital markets in the economies of the scales and the scopes in the short / long time periods. Therefore, the investors prefer to use the sophisticated theories, including the modern portfolio theory (MPT) to reach their investment goals, namely to select the diversified uncorrelated tangible / intangible assets for a modern investment portfolio in the capital markets in the nonlinear diffusion type economy of scale and scope over the short / long time periods in finances.

Let us define the weighted expected return of a portfolio Rp as in Mitra (2009)

$$R_p = \sum_{i=1}^{N} w_i \mu_i,$$

then the portfolio's variance $\sigma_p^2$ can be written as

$$\sigma_p^2 = \sum_{i=1}^{N} \sum_{j=1}^{N} \sigma_{ij} w_i w_j,$$

where

- N is the number of assets in a portfolio;
- i, j are the asset indices and $i, j \in \{1, ..., N\}$ ;
- $w_i$ is the asset weight, subject to the constraints:

$$0 \leq w_i \leq 1,$$

$$\sum_{i=1}^{N} w_i = 1;$$

- $\sigma_{ij}$ is the covariance of asset i with asset j;
- $\mu_i$ is the expected return for asset i.

Let us explain that the expected return premium in agreement with the Capital Asset Pricing Model (CAPM) theory can be written as in Sharpe (1965, 1966, 1968, 1992, 1994), Sharpe, Alexander, Bailey (1999), Lintner (1965), Mossin (1966), Engle (2003, 2006), Fama, French (2004), Capocci, Hubner (2004), Mitra (2009), Kantarelis (2017):

$$R_a = R_f + \beta(R_m - R_f) + \epsilon,$$

where:

- $R_a$ is the expected return of an asset;
- $R_f$ is the risk-free rate of return;

- $R_m$ is the expected market return;
- $\epsilon$ is the error term;
- $\beta = \dfrac{\sigma_{am}}{\sigma_{mm}}$ is the systematic risk with respect to the tangency portfolio ;
- $\sigma_{am}$ is the market and asset's covariance;
- $\sigma_{mm}$ is the market's variance.

The Sharpe ratio attempts to provide the portfolio risk measure in terms of the quality of the portfolio's return at its given level of risk. The Sharpe ratio is a return-to-risk measure in the frames of the Capital Asset Pricing Model (CAPM) theory in Sharpe (1966), Mitra (2009), Kantarelis (2017):

$$S = \frac{R_p - R_f}{\sigma_p}$$

where $\sigma_p$ is the portfolio return's standard deviation.

Let us illustrate the Markowitz efficient frontier theoretical conception in the wealth management process with the modern investment/borrowing portfolio in the capital markets in the economies of the scales and the scopes at the certain monetary base in the short / long time periods, which was proposed with the aim to show how to minimize the known risks by selecting the diversified uncorrelated tangible / intangible investment assets in Markowitz (1952, 1956, 1959, 1987) and it was further researched in Shiryaev (1998a, b), Hull (2005-2006, 2010, 2012), Demidova-Menzel; Heidorn (August 2007), Mitra (2009), Hassine, Roncalli (2013), Ledenyov D O, Ledenyov V O (2013a), Kantarelis (2017).

Fig. 81 shows the Markowitz efficient frontier, which is constructed graphically to select the diversified uncorrelated tangible / intangible assets for the modern investment portfolio in the capital markets in the nonlinear diffusion type economy of scale and scope over the short / long time periods in the frames of the Capital Asset Pricing Model (CAPM) theory within the Modern Portfolio Theory (MPT) in the finances. The financial capital investment at the wealth management process includes the following stages:

1. The Markowitz efficient frontier graphical construction;
2. The highest Sharpe ratio portfolio finding;

***3.*** The wealth investment between the highest Sharpe ratio portfolio point and the risk free lending/borrowing point with a certain acceptable risk tolerance level in mind.

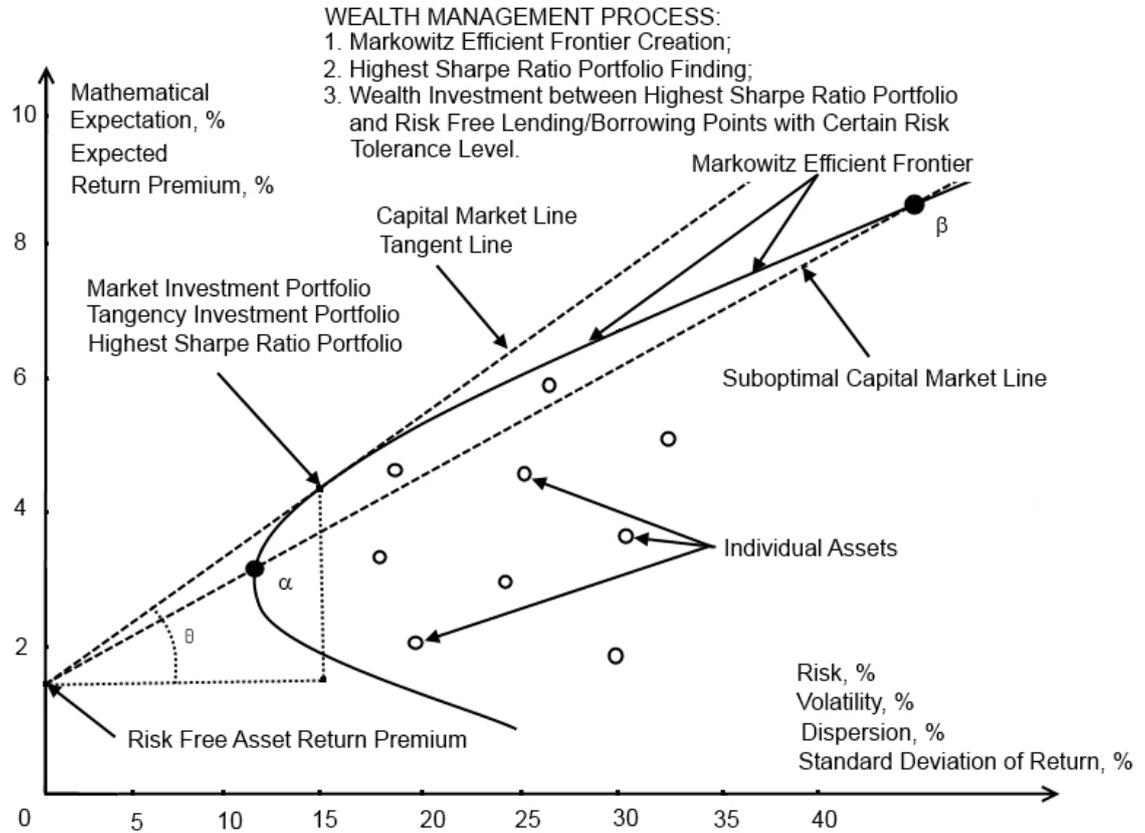

**Fig. 81.** Markowitz efficient frontier to select diversified uncorrelated tangible / intangible assets for modern investment portfolio in capital markets in nonlinear diffusion type economy of scale and scope over short / long time periods in agreement with Capital Asset Pricing Model (CAPM) theory within Modern Portfolio Theory (MPT) in finances.

Therefore, we can come to a logical thought that the seasoned investors must attempt to predict the new investment risks in the capital markets in the economies of the scales and the scopes in the short / long time periods in order to make the optimal investment decisions and to get an increased return premium on the invested financial capital in the capital markets. An analysis of the business cycle oscillation dynamics by the experienced investors is a good starting point in making of the future forecasts on the financial / the economic market behaviour.

In this connection, let us consider a challenging problem on the accurate forecast of the economic / the financial trends with the business cycles oscillation dynamics analysis, aiming to derive the most optimal solutions for the wealth management problem in the case, when the wealth is managed with an application of the modern investment portfolio in the capital markets in the economies of the scales and the scopes in the short / long time periods.

One of the well known forecast techniques is based on an application of the Hodrick-Preskot (HP) differential filter in Hodrick, Prescott (1980, 1997) to a massive of the statistical data, describing the change of the economic output in the economy of the scale and the scope over the selected time period time. The filtered CW signal can be used with the purpose to make the short/long time term economic output change forecasts, which have limited accuracy in Hodrick, Prescott (1980, 1997)

$$y_t = \tau_t + c_t + \in_t,$$

*where* $\tau_t$ *is the trend component*;

$c_t$ *is the cyclical component*;

$\in_t$ *is the error component*;

$$\min_\tau \left[ \sum_{t=1}^{T} \left( y_t - \tau_t \right)^2 + \lambda \sum_{t=2}^{T-1} \left[ \left( \tau_t - \tau_{t-1} \right) - \left( \tau_{t-1} - \tau_{t-2} \right) \right]^2 \right].$$

The slightly outdated prediction technique, which is based on the Hodrick-Prescott (HP) differential filter in Hodrick, Prescott (1980, 1997), includes the following main research stages:

*1.* to filter out the continuous-time economic output wave in the economy of the scale and the scope over the past and present time periods in Hodrick, Prescott (1980, 1997);

*2.* to analyze the continuous-time economic output wave oscillation dynamics in the economy of the scale and the scope over the past and present time periods;

*3.* to model the forecast horizon by shifting the window with the business cycle of the continuous-time economic output wave toward its propagation trend in the economy of the scale and the scope in the future time periods.

Let us demonstrate graphically the filtering of the continuous-time economic output wave in economy of scale and scope in past and present time periods by applying the Hodrick-Prescott (HP) differential filter in Hodrick, Prescott (1980, 1997), Federal Reserve Bank of St Louis (2012), Matlab (2012).

Fig. 82 depicts the filtering of the continuous-time economic output wave in economy of scale and scope in past and present time periods by applying Hodrick-Preskot (HP) differential filter in Hodrick, Prescott (1980, 1997), Federal Reserve Bank of St Louis (2012), Matlab (2012).

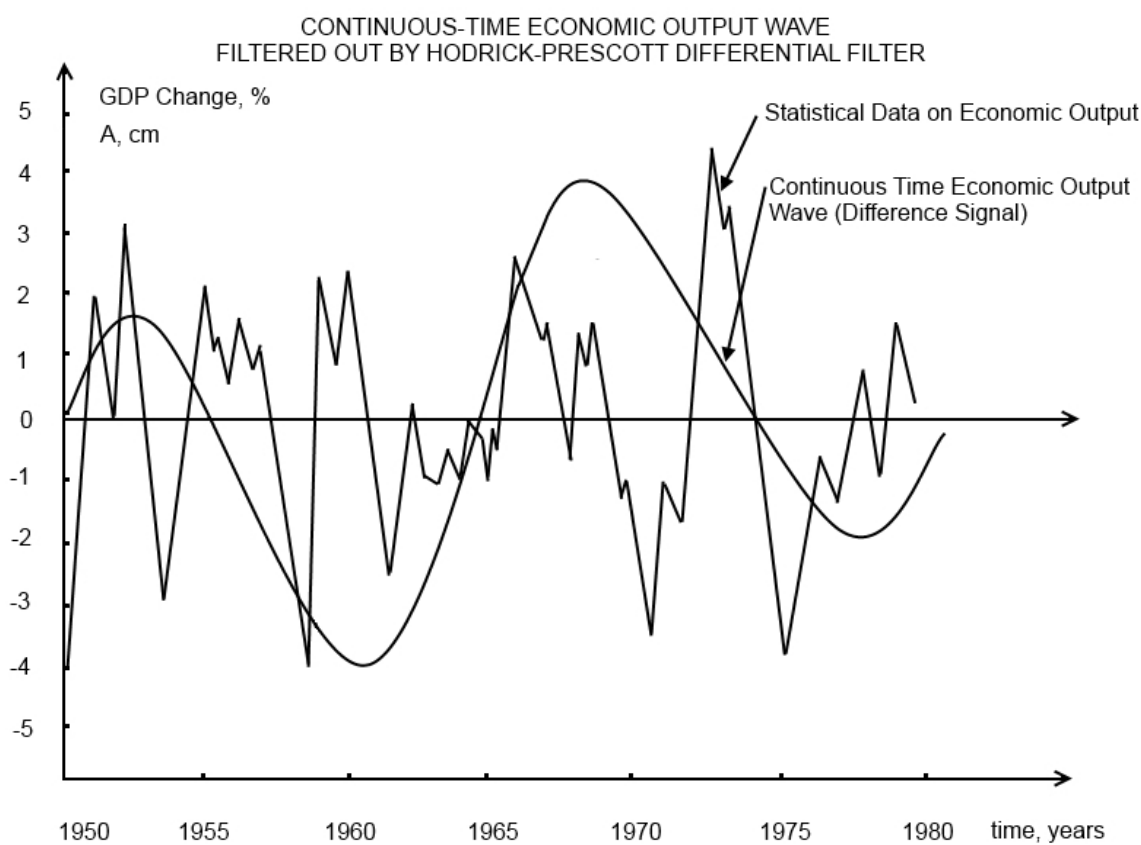

**Fig. 82.** Filtering of continuous-time economic output wave by applying Hodrick-Prescott (HP) differential filter in economy of scale and scope in past and present time periods.

Let us make a modeling of the forecast horizon for the continuous-time economic output wave oscillation dynamics in the economy of the scale and the scope in the forthcoming time periods. The detailed description of the used techniques to make assumptions on the continuous-time economic output wave oscillation dynamics in economy of the scale and scope in the forthcoming time periods can be found in Purica, Caraiani (2009).

Fig. 83 shows the long time term economic output change forecast technique with the forecast horizon modeling for the continuous-time economic output wave in the economy of the scale and the scope in the future time periods.

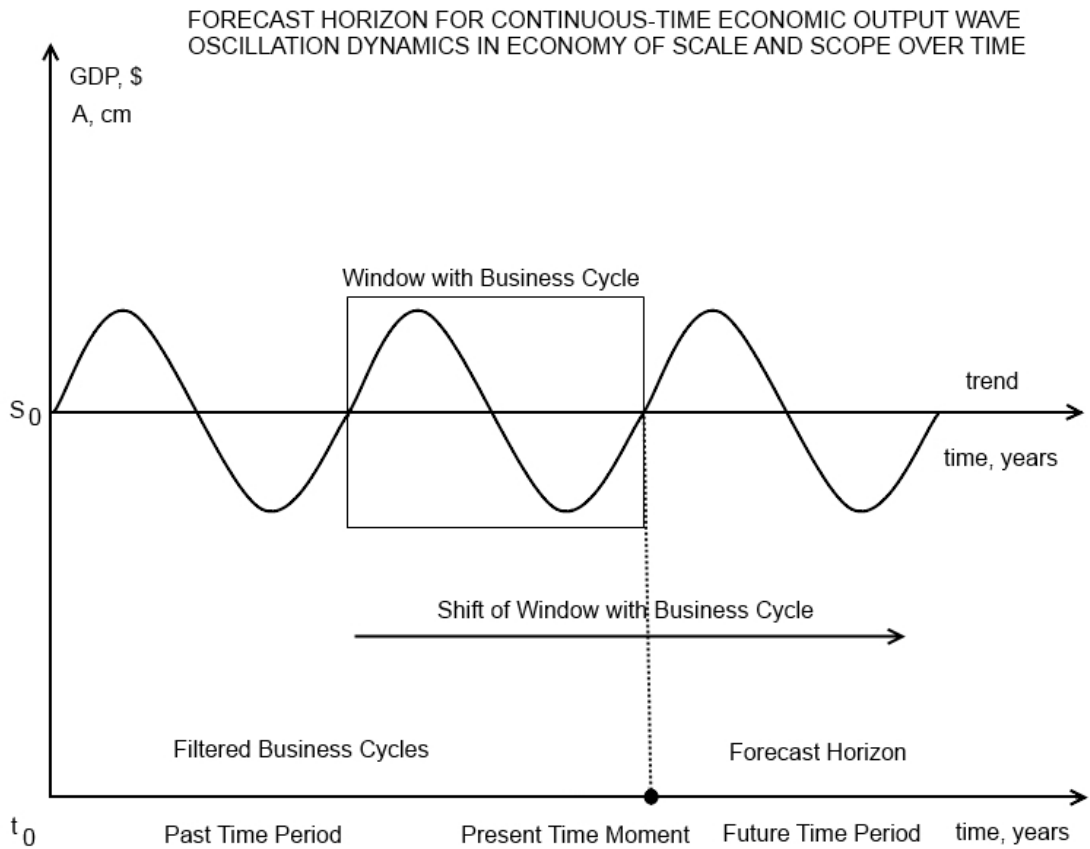

**Fig. 83.** Long time term economic output change forecast technique with forecast horizon modeling for continuous-time economic output wave in economy of scale and scope in future time periods.

Now, we would like to explain that our innovative approach to solve a an accurate forecast of the economic and financial trends with the business cycles oscillation dynamics analysis allows a seasoned investor to find the most optimal solutions for the wealth management problem in the case, when the wealth is managed with an application of the Ledenyov modern investment portfolio in the capital markets in the economies of the scales and the scopes at the certain monetary base in the short / long time periods in agreement with the Ledenyov classic and quantum econodynamics sciences.

Let us make our abstract theoretical conception tangible by explaining an essence of our research approach to solve the forecast problem by applying the Bayesian filtering such as the Stratanovich-Kalman-Bucy filter with the Stratanovich-Kalman-Bucy filtering algorithm to make the assumptions on the discrete-time digital economic output wave oscillation dynamics in the economies of the scales and the scopes at the certain monetary base in the short time periods in agreement with the Ledenyov classic and quantum econodynamics sciences. As we know the Stratonovich – Kalman - Bucy filters with the embedded Stratonovich – Kalman - Bucy filtering algorithm in the frames of the Stratonovich optimal non-linear filtering theory in the electronics engineering were researched in Mandel'shtam (1948-1955), Wiener (1949), Bode, Shannon (1950), Zadeh, Ragazzini (1950), Booton (1952), Davis (1952), Bartlett (1954), Doob (1955), Franklin (1955), Pugachev (1956a, b), Solodovnikov, Batkov (1956), Laning, Battin (1956), Lees (1956), Newton, Gould, Kaiser (1957), Tukey (1957), Rytov (1957), Bellman, Glicksberg, Gross (1958), Blum (1958), Darlington (1958), Davenport, Root (1958), Sherman (1958), Shinbrot (1958), Smith (1958), Merriam (1959), Stratonovich (1959a, b, 1960a, b), Kalman, Koepcke (1958, 1959), Kalman, Bertram (1958, 1959), Kalman (1960a, b, 1963), Kalman, Bucy (1961), US Air Forces Office of Scientific Research (1960 – 2013), Friedman (1962), Kushner (1967, 2000), Bryson, Ho (1969), Bucy, Joseph (1970), Jazwinski (1970), Sorenson (1970), Wright-Patterson Air Forces Base (1970 – 2013), Chow, Lin (1971, 1976), Maybeck (1972, 1974, 1990), Willner (1973), Leondes, Pearson (1973), Akaike (1974), Dempster, Laird, Rubin (1977), Griffiths (1977), Schwarz (1978), Falconer, Ljung (1978), Anderson, Moore (1979), Bozic (1979), Priestley (1981), Lewis (1986), Proakis, Manolakis (1988), Caines (1988), de Jong (1988, 1989, 1991), de Jong, Chu-Chun-Lin (1994), Bar-Shalom, Maybeck (1990), Franklin, Powell, Workman (1990), Brockwell, Davis (1991), Jang (1991), Brown, Hwang (1992, 1997), Xiao-Rong Li (1993), Gordon, Salmond, Smith (1993), Farhmeir, Tutz (1994), Grimble (1994), Lee, Ricker (1994), Ricker, Lee (1995), Fuller (1996), Hayes (1996), Haykin (1996), Golub, van Loan (1996), Schwaller, Parnisari (1997), Julier, Uhlmann (1997), Babbs, Nowman (1999), Ljung (1999), Wanhammar (1999), Ito,

Xiong (2000), Kushner, Budhiraja (2000), Welch, Bishop (2001), Haykin (editor) (2001), Arulampalam, Maskell, Gordon, Clapp (2002), Doucet, Tadic (2003), Litvin, Konrad, Karl (2003), de Jong, Penzer (2004), Ristic, Arulampalam, Gordon (2004), van Willigenburg, De Koning (2004), Voss, Timmer, Kurths (2004), Cappé, Moulines (2005), Capp´e, Moulines, Ryd´en (2005), Poyiadjis, Doucet, Singh (2005a, b), Misra, Enge (2006), Frühwirth-Schnatter (2006), Gamerman, Lopes (2006), Rajamani (2007), Olsson, Cappé, Douc, Moulines (2008), Rajamani, Rawlings (2009), Francke, Koopman, de Vos (2010), Xia, Tong (2011), Matisko, Havlena (2012), Durbin, Koopman (2012), Ledenyov D O, Ledenyov V O (2013g).   It   may also be interesting to note that the Stratonovich – Kalman - Bucy filters with the embedded Stratonovich – Kalman - Bucy filtering algorithm within the Stratonovich optimal non-linear filtering theory in the finances were researched in Athans (1974), Fernandez (1981), Geweke, Singleton (1981), Litterman (1983), Meinhold, Singpurwalla (1983), Engle, Watson (1983), Harvey, Pierse (1984), Harvey (1989), Engle, Lilien, Watson (1985), de Jong (1991), Doran (1992), Tanizaki (1993), Bomhoff (1994), Venegas, de Alba, Ordorica (1995), Roncalli (1996), Roncalli, Weisang (2008), Wells (1996), Hodrick, Prescott (1997), Krelle (1997), Pitt, Shephard (1999), Cuche, Hess (2000), Durbin, Koopman (2000), Javaheri, Lautier, Galli (2002), Morley, Nelson, Zivot (2002), Bahmani, Brown (2004), Broto, Ruiz (2004), Fernàndez-Villaverde, Primiceri (2005), Fernàndez-Villaverde, Rubio-Ramirez (2005, 2007), Fernàndez-Villaverde (2010), Ozbek, Ozale (2005), Proietti (2006, 2008), Luati, Proietti (2010), Proietti, Luati (2012a, b), Ochoa (2006), Horváth (2006), Cardamone (2006), Pasricha (2006), Bignasca, Rossi (2007), Dramani, Laye (2007), Paschke, Prokopczuk (2007), Andreasen (2008), Osman, Louis, Balli (2008), Gonzalez-Astudillo (2009), Bationo, Hounkpodote (2009), Mapa, Sandoval, Yap (2009), Chang, Miller, Park (2009), Theoret, Racicot (2010), Winschel, Kratzig (2010), Lai, Te (2011), Jungbacker, Koopman, van der Wel (2011), Moghaddam, Haleh, Ebrahimijam (2011), Creal (2012), Darvas, Varga (2012), Hang Qian (2012), Ledenyov D O, Ledenyov V O (2013g).

We developed the Ledenyov short / long time period economic output change forecast techniques to make the assumptions on the discrete-time digital economic output wave oscillation dynamics in the economies of the scales and the scopes at the certain monetary base in the short time periods in agreement with the Ledenyov classic and quantum econodynamics sciences:

*1.* The Ledenyov short time term economic output change forecast technique has the following stages (see also Chunyang Bai (2016)):

*1)* to analyze an immediate past state of the discrete-time digital economic output wave oscillation dynamics in the economy of the scale and the scope before the present time moment, using the Stratanovich-Kalman-Bucy filter;

*2)* to analyze a present state of the discrete-time digital economic output wave oscillation dynamics in the economy of the scale and the scope at the present time moment, applying the Stratanovich-Kalman-Bucy filter;

*3)* to model a future state of the Ledenyov discrete-time digital economic output wave in the economy of the scale and the scope in the forthcoming short time period, employing the Stratanovich-Kalman-Bucy filter.

*2.* The Ledenyov long time term economic output change forecast technique includes the following stages:

*1)* to filter out the Ledenyov discrete-time digital economic output wave in the economy of the scale and the scope over the past and present time periods, using the complex digital filters;

*2)* to analyze the Ledenyov discrete-time digital economic output wave oscillation dynamics in the economy of the scale and the scope over the past and present time periods, applying the complex digital filters;

*3)* to model the forecast horizon by shifting the window with the business cycle of the Ledenyov discrete-time digital economic output wave toward its propagation trend in the economy of the scale and the scope in the future long time periods.

Now, let us graphically demonstrate the Ledenyov long time term economic output change forecast technique with the forecast horizon modeling for the Ledenyov discrete-time digital economic output wave in the economy of the scale and the scope in the future time periods.

Fig. 84 displays the Ledenyov long time term economic output change forecast technique with the forecast horizon modeling for the Ledenyov discrete-time digital economic output wave in the economy of the scale and the scope in the future long time periods.

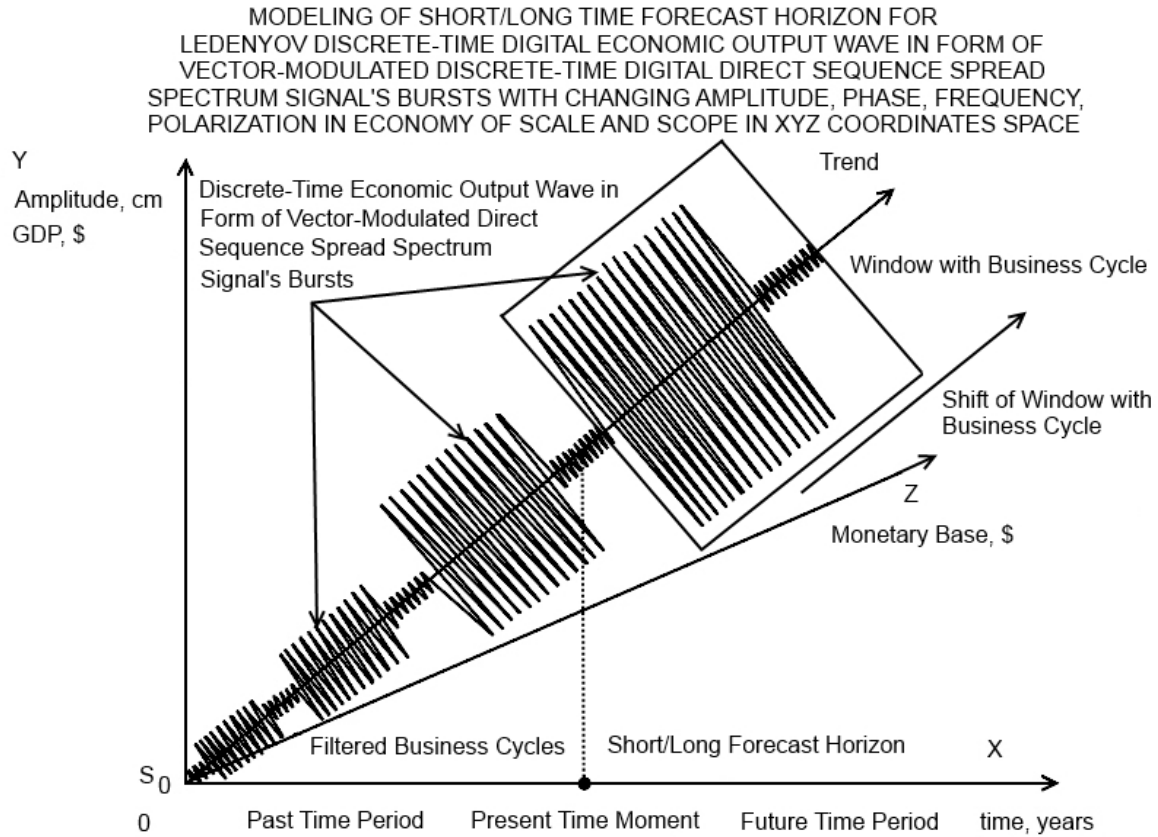

**Fig. 84.** Ledenyov long time term economic output change forecast technique with forecast horizon modeling for Ledenyov discrete-time digital economic output wave in economy of scale and scope in future long time periods.

Finally, let us note that the investment portfolio manager / the seasoned investor can design the Ledenyov modern investment portfolio with 1) the diversified uncorrelated tangible/intangible **investment products**, 2) the **investment vehicles** and 3) the **investment mediums** for the financial capital investment in the capital markets in the economies of the scales and the scopes in the short and long time periods, using the Ledenyov Quantum Winning Virtuous Strategy Search (LQWVSS) algorithm with the quantum logic theoretical conception in Ledenyov V O, Ledenyov D O (2017). One main thing to remember is that the Ledenyov modern investment portfolio must be designed keeping in mind the Ledenyov investment portfolio stability theorem in Ledenyov V O, Ledenyov D O (2017).

Now, let us summarize the most significant research contributions on the **investment products** for the financial capital investment in the capital markets made by talented scientists up to this date.

The **land** as an investment product was researched in Ricardo (1817, 1821), Denman (1956), Silberberg (1975), Veenman, Zonen (1961), Elvin, Ervin (1982), Mills, Hoover (1982), McFarlane (1983), Eaton (December 1984), Phipps (1984), Kaplan (Winter 1985), Fiske (October 7-8 1986), Wiltshaw (1986), Guzhvin (1987), Eaton (1988), Mercier (1988), Reeve (1988), Asako, Kuninori, Inoue, Murase (1989), Asako, Kuninori, Inoue, Murase (1997), Feder, Onchan (1989), Phillips, Bauer, Mercier, Mumey (1989), Phillips, Bauer, Akabua (1993), Roth, Barrows, Carter, Kanel (1989), Schnitkey, Taylor, Barry (1989), Barrett (1991), Kanemoto (1991), Lins, Hoffman, Kowalski (September 23-24 1991), Yoshida (1991), Butler, DeBraal (1993), Miller, Andrews (February 9 - 11 1993), Capozza, Yuming Li (1994), Werner (1994), Nieuwoudt (1995), Ogawa, Suzuki (1995), Ogawa Kazuo, Shin-Ichi Kitasaka, Hiroshi Yamaoka, Yasuharu Iwata (1996), Grepperud (February 1996), Echevarria (1997), Deininger, Feder (2001), Kiyotaki, West (2004), Sekine, Tachibana (March 2004), Feinerman, Peerlings (2005), Hennings, Sherrick, Barry (2005), Ray (2005), Turnbull (2005), Kiyotaki, West (2006), Holden, Deininger, Ghebru (2007), Sekine, Tachibana (2007), Głuszak (2008), Von Braun, Meinzen-Dick (April 2009), Cotula, Vermeulen, Leonard, Keeley (2009), Cotula (2012), Fakton, van der Post (2010), Fischer, Shah (2010), Zoomers (2010), Abdulai, Owusu, Goetz (2011), Arezki, Deininger, Selod (December 2011), De Schutter (2011), Hui-Wen Koo (2011), Palyanychko (2011), Zagema (September 22 2011), Di Corato, Hess (September 26-28 2012), Domeher, Abdulai (2012), Van der Kr Pasmans, Van der Heijden (2012), Todorovic, Vasiljevic, Rajic (2012), Oya (2013), Xianlei Ma, Heerink, van Ierland, van den Berg, Xiaoping Shi (2013), Zolin, Luzi (2013), Boehlje, Baker, Langemeier (January 3-5 2014),

Malashchuk (2014), Palenychak (2014), Arezki, Deininger, Selod (2015), Busha (2015), Szreder (2015), Bochco (2016), Borghesi, Giovannetti, Iannucci, Russu (2016).

The **real estate** as an investment product was researched in Grebler (1954), Taylor G S, Bailey (1963), Wendt, Wong (December 1965), Friedman (December 1970, March 1971), Roulac (1976, 1996), Smith, Shulman (September–October 1976), Lusht (1978), Pellatt (1972), Chapman Findlay III, Hamilton, Messner, Yormark (September 1979), Ibbotson (Fall 1979), Ibbotson, Sinquefield (1982), Ibbotson, Siegel (1983, 1984), Ibbotson, Siegel, Love (Fall 1985), Penny (1980), Rystrom (1980), Burns, Epley (Winter 1982), Miles, Estey (Winter 1982), Miles, McCue (Summer 1982, 1984), Brueggeman, Chen, Thibodeau (1984), Fogler (1984), Webb (1984), Webb, McIntosh (1986), Gau (1985), Gau, Wang (1990), Hartzell, Mengden (August 27 1986), Hartzell, Hekman, Miles (Spring 1987), Hartzell, Webb (1988), Hartzell, Liu, Kallberg (2004), Kuhle, Walther, Wurtzebach (1986), Kuhle (1987), Titman, Warga (Fall 1986), Haight, Fort (1987), Chen, Tzang (1988), Chinloy (1988), Firstenberg, Ross, Zisler (1988), Gyourko, Linneman (1988), Gyourko, Keim (1992), Hines (March 1988), Lusht (1988), Miller, Sklarz, Ordway (1988), Mooney S, Mooney K (1988), Wittner (Fall 1988), Geltner (1989), Goebel, Kim (1989), Rubens, Bond, Webb (1989), Sweeney (1989, 1993), Chan, Hendershott, Sanders (Winter 1990), Colwell, Park (September 1990), Giliberto (Summer 1990), Goetzmann, Ibbotson (1990), Howe, Shilling (1990), Liu, Hartzell, Grissom, Wylie (1990), McMahan (1990), Sagalyn (1990), Wheaton, Torto (1990), Asabere, Kleiman, McGowan (1991), Martin, Cook (1991), McIntosh, Liang, Tompkins (Spring 1991), Ross, Zisler (1991), Ziobrowski A J, Boyd (1991), Ziobrowski A J, Curcio (1991), Ziobrowski A J, Ziobrowski B J (1993), Ziobrowski B J, Ziobrowski A J (1995), Ziobrowski A J, McAlum, Ziobrowski B J (1996), Ambrose, Ancel, Griffiths (Spring 1992),

DiPascuale, Wheaton (1992), Kleiman, Farragher (1992), Liu, Mei (1992), Brueggeman, Fisher (1993), Newell, MacFarlane (March 1994), Worzala (1994), Worzala, Newel (1997), Zumpano, Elder (1994), Baum (1995, 2009), Corgel, McIntosh, Ott (1995), Jun Han, Youguo Liang (1995), Mueller (Spring 1995), Barkham, Ward, Henry (1996), Geurts, Jaffe (1996), Han (1996), Newell (1996), Newell, Webb (1996), Seck (1996), Solnik (1996), Bers, Springer (1997), Brueggeman, Fisher (1997), Farrell (1997), Hoesli, MacGregor, Matysiak, Nanthakumaran (1997), Mei, Saunders (1997), Redman, Manakyan, Liano (1997), Ball, Lizieri, MacGregor (1998), Chun, Shilling (1998), D'Arcy, Keogh (1998), Martens (1998), Svensson (1998), Adair, Berry, McGreal, Syacutekora, Ghanbari Parsa, Redding (Spring 1999), Keogh, D'Arcy (1999), Liao, Mei (1999), Moshirian, Pham (2000), Brounen, Cools, Schweizer (Summer 2001), Lee St (2001, 2005, 2006a, b), McGreal, Parsa, Keivani (2001), Thrall (2002), Blazenko, Pavlov (2004), Deqing Diane Li, Yung (2004), Engelberts, Suarez (2004), Hoskin, Higgins, Cardew (Spring 2004), Loo Lee Sim, Xing Quan Zhang, Jieming Zhu (2004), Pi-Ying, Lai Peddy (2004), Hardin, Liano, Gow-Cheng Huang (2005), Laposa, Lizieri (April 13 – 16 2005), Holsapple, Ozawa, Olienyk (2006), Johnson (2006), Lim, McGreal, Webb (2006), Adlington, Grover, Heywood, Keith, Munro-Faure, Perrotta (December 6-8 2008), Chan, Hardin, III, Liano, Zheng (2008), Lambie-Hanson (Fall 2008), Minye Zhang, Yongheng Deng (July 9 2008), Falkenbach (2009), Kurzrock, Rottke, Schiereck (2009), Edelstein, Qian, Tsang (2010), Lieser, Groh (2010, July 2011, 2014), Peralta-Alva (2011), Chaney, Sraer, Thesmar (2012), Baker, Chinloy (editors) (September 4 2014, September 8 2014), Cunat, Cvijanovic, Yuan (2014), Gauder, Houssard, Orsmond (2014), Anderson, Rottke, Krautz (2015), Hazama, Uesugi (October 2015), Maksimov, Bachurinskaya (2015), Papastamos, Matysiak, Stevenson (2015).

The **commodity** as an investment product was researched in Keynes (1923, September 1938, 1942, 1980, 1943), Hotelling (1931), Graham (1937), Kaldor (1939), Working (February 1948, 1949, 1960), Brennan (1958, 1991), Brennan, Schwartz (1985), Sharpe (1964), Sharpe, Alexander, Bailey (1995), Telser (1968), Dusak (1973), Lovell, Vogel (1973), Merton (1973), Stoll (1979), Stoll, Whaley (2009, 2010), Bodie, Rosansky (May June 1980), Breeden (1980), Spraos (March 1980), Newbery, Stiglitz (1981), Wright, Williams (1982), Abel (1983), Abel, Blanchard (1986), Abel, Mankiw, Summers, Zeckhauser (1989), Abel, Eberly (1994), Bernanke (1983), Carter, Rausser, Schmitz (April 1983), Garbade, Silber (1983), Edwards, Freebairn (1984), Kanbur (1984), Cox, Rubinstein (1985), Gilbert (September 1985, 1996), Jagannathan (1985), Taylor (1985), Fama, French (January 1987, December 1988), Hartzmark (1987), Chinloy (1988), Gyourko, Linneman (1988), Hirshleifer (1988, 1989, 1990), Lichtenberg, Ujihara (1988), Paddock, Siegel, Smith (1988), Baumol, Benhabib (1989), Frank, Stengos (1989), Jaffee (1989), Rubens, Bond, Webb (1989), De Gorter, Zilberman (1990), De Gorter, Tsur (1991), De Gorter, Nielson, Rausser (1992), De Gorter, Swinnen (1998), Gibson, Schwartz (1990), Kaminsky, Kumar (1990), Pindyck, Rotemberg (1990), Shleifer, Summers (1990), Siegel D R, Siegel D F (1990), Blank (1991), Brennan (1991), Chan K, Chan K C, Karolyi (1991), Subrahmanyam (1991), Williams, Wright (1991), Bessembinder (1992), Bessembinder, Chan K (1992), Bessembinder, Seguin (1992, 1993), Bessembinder, Coughenour, Seguin, Smoller (1995, Winter 1996), Deaton, Laroque (January 1992, 1996), DeCoster, Labys, Mitchell (1992), Edwards, Ma (1992), Kolb (1992, 1996, 1997), Maizels (1992), Ankrim, Hensel (May/June 1993), Bleaney, Geenaway (July 1993), Brueggeman, Fisher (1993), Lummer, Siegel (Summer 1993), Yang, Brorsen (1993), Dixit, Pindyck (1994), Satyanarayan, Varangis (March 1994), Litzenberger, Rabinowitz (December 1995), O'Hara (1995), Sachs, Warner

(1995a, b, 1997, 1999, 2001, Chambers, Bailey (1996), Duffie (1996), Hamilton (1996, 2005, Spring 2009a, b), Hamilton, Jing Wu (2013a b, 2014), Sanders, Irwin, Leuthold (1996), Sanders, Irwin (2010a b, 2011a, b, 2013), Sanders, Irwin, Merrin (2010), Hoesli, MacGregor, Matysiak, Nanthakumaran (1997), Kocagil (1997), Schneeweis, Spurgin (1997), Schwartz (July 1997), Schwartz, Smith (July 2000), Mabro (1998), Pilipovic (1998), Cashin, McDermott, Scott (1999), Greenaway, Morgan (1999), Hielscher (1999), Irwin, Yoshimaru (1999), Irwin, Holt (2004), Irwin, Sanders, Merrin (April 20-21 2009a, August 2009b), Irwin, Sanders (February 2011 2012a, b), Labys, Achouch, Terraza (1999), Becker, Finnerty (2000), Borenstein, Bushnell, Stoft (2000), Cashin, Liang, McDermott (2000), Cashin, McDermott (2002), Christie-David, Chaudry, Koch (2000), De Roon, Nijman, Veld (2000), De Roon, Szymanowska (2010), Greer (2000), Jensen, Johnson, Mercer (2000, Summer 2002), Routledge, Seppi, Spatt (2000), Swinnena, De Gorterc, Rausserd, Banerjeea (2000), Till (Fall 2000a, September 1 2000b, April 2003, 2006), Till, Eagleeye (May 2003, Fall 2005), Cochrane (2001), Fabozzi (2001), Kogan (November 2001, 2004), Kogan, Livdan, Yaron (2005), Richards, Padilla (August 5 - 8 2001), Barsky, Kilian (May 2002, 2004), Chatrath, Adrangi, Dhanda (2002), Cremer, Laffont (2002), Cremer, Gasmi, Laffont (2003), Ederington, Lee J H (2002), Weiner (2002), Bower, Kamel (June 2003), Sorenson (2003), Vrugt (2003), Wang (2003), Weiser (2003), Greer (Summer 2004, April 2005), Gorton, Rouwenhorst (2004, 2006), Gorton, Hayashi, Rouwenhorst (2013), Longstaff, Wang (2004), Beenen (2005), Casassus, Collin-Dufresne (2005), Casassus, Collin-Dufresne, Routledge (December 2005), Erb, Harvey (May 2005, 2006), Haigh, Hranaiova, Oswald (2005), Lautier (Summer 2005), Lewis (April 2005), Micu (December 2005), O'Connell (December 2005), Taylor (2005), Brown (January February 2006), Bryant, Bessler, Haigh (2006), Campbell, Orskaug, Williams (Spring 2006), Feldman, Till (2006),

Frankel (2006), Fusaro, Vasey (September 2006), Holmes (2006), McNee (July 2006), Upperman (2006), Blanchard, Gali (2007), Demidova-Menzel, Heidorn (August 2007), Domanski, Heath (March 2007), Kat, Oomen (2007a, b), Miffre, Rallis (2007), Chong, Miffre (2010), Basu, Miffre (2013), Miffre, Brooks (2013), Röthig, Chiarella (2007), Worthington, Pahlavani (2007), Adams, Füss, Kaiser (2008), Bachmeier, Qi Li, Liu (2008), Bhardwaj, Gorton, Rouwenhorst (2008), Büyükşahin, Haigh, Harris, Overdahl, Robe (2008), Büyükşahin, Haigh, Robe (2010), Büyükşahin, Robe (2010, 2011, 2014), Büyükşahin, Brunetti, Harris (2010), Büyükşahin, Harris (2011), Caballero, Farhi, Gourinchas (2008), Carney (2008), Fabozzi, Füss, Kaiser (2008), Johnson Matthey Plc (2008), Khan, Khoker, Simin (2008), Kilian (2008a, b, 2009, 2010, 2014), Kilian, Park (2009), Kilian, Vega (2011), Kilian, Vigfusson (2011), Kilian, Murphy (2013, 2014), Kilian, Hicks (2013), Kolos, Ronn (March 2008), Reisen (July 18 2008), Roache (September 2008), Roache, Rossi (July 2009), Scherer, He (2008), Woodward (2008), Attié, Roache (April 2009), Bose (2009), Du, Yu, Hayes (2009), Frankel, Rose (2009), Kaufmann, Ullman (2009), Korniotis (2009), Mayer (2009, 2012), Reitz, Slopek (2009), Smith (2009), Smith, Thompson, Lee (2013), Yamori (September 7 2009, May 20 2011, 2011), Yung, Liu (2009), Alquist, Kilian (2010), Alquist, Kilian, Vigfusson (2013), Anzuini, Lombardi, Pagano (2010), Aulerich, Irwin, Garcia (2010), Baffes, Haniotis (July 2010), Baker (2012), Baker, Routledge (2012), Basak, Pavlova (2012), Blanchard, Gali (2010), Calvo-Gonzalez, Shankar, Trezzi (2010), Cifarelli , Paladino (2010), Fuertes, Miffre, Rallis (2010), Elder, Serletis (2010), Etula (2010), Gilbert (March 2010, 2010), Hailu, Weersink (2010), Hernandez, Torero (2010), Nissanke (December 2010), Silvennoinen, Thorp (2010), Tang, Wei Xiong (2010, 2012), Tang, Zhu (2015), Wong, Smith (July 25 - 27 2010), Basu, Gavin (2011), Cárdenas, Ramírez, Tuzemen (December 2011), Connolly, Orsmond (2011), Daskalakis, Skiadopoulos (2011), Dwyer,

Gardner, Williams (2011), Dwyer, Holloway, Wright (2012), Fattouh (2011), Ghosh (2011), G20 Study Group on Commodities (November 2011), Hong, Yogo (2011), Inamura, Kimata, Kimura, Muto (March 2011), Lombardi, Robays (2011), Mou (2011), Peñaranda, Micola (2011), Reichsfeld, Roache (2011), Tilton, Humphreys, Radetzki (2011), Yiuman Tse, Williams (December 20 2011), UNCTAD (2011), Arbatli, Vasishtha (2012), Cespedes, Velasco (2012), Fattouh, Kilian, Mahadeva (2012), Gasmi, Oviedo (October 2012), Girardi (January 2 2012), Girardi (October 9 2012), Gruber, Vigfusson (2012), Henderson, Pearson, Li Wang (2012), Hong, Yogo (2012), Juvenal, Petrella (2012), Nissanke (2012), Rouwenhorst, Ke Tang (2012), Sockin, Wei Xiong (2012, 2013), Varadi (March 2012), Acharya, Lochstoer, Ramadorai (2013), Anghelache, Dincă, Sacală, Lixandru (2013), Aulerich, Irwin, Garcia (2013), Baumeister, Peersman (2013), Baumeister, Kilian (2014, 2016), Chevallier, Ielpo, Ling-Ni Boon (2013), Creti, Joëts, Mignon (2013), Fattouh, Kilian, Mahadeva (2013), Heumesser, Staritz (October 2013), Hurduzeu G, Hurduzeu R (2013), Ing-Haw Cheng, Wei Xiong (2013, November 2013, 2014), Ing-Haw Cheng, Kirilenko, Jovanovic (2013), Kang, Rouwenhorst, Ke Tang (2013), Le Pen, Sévi (2013), Marshall, Nguyen, Visaltanachoti (2013), Grimes, Hyland (2013), Santos (2013), Van Robays (2013), Venditti (2013), Adams, Glück (June 2014, August 2015), Basak, Pavlova (2014), Daskalakis, Kostakis, Skiadopoulos (2014), Fornero, Kirchner, Yany (2014), Gruss (2014), Jo (2014), Mykhailovska (2014), Sadorsky (2014), Bonato, Taschini (November 2015), Henderson, Pearson, Wang (2015), Novotný (2015), Novotný, Polach (2016), Joëts, Mignon, Razafindrabe (November 2016).

The **bond** as an investment product was researched in Macaulay (1938), Dougall (1953), Burrell (1953), Hickman (1957), Fisher (June 1959), Francis, Archer (1971), Barro (l974), Merton (1974), Ederlngton (December 1974), Fischer (June 1975), Black, Cox (1976), Black (1980), Jackson (1976), Khang (December 1979), Ibbotson, Sinquefield (1982), Bodie, Kane, McDonald (March 1983), Brennan, Schwartz (1983), Kaufman, Bierwag, Toevs (editors) (1983), Blume (1987), Blume, Keim, Patel (March 1991), Fons (March 1987, 1990), Levy, Lerman (1988), Altman (September 1989, Summer 1993), Asquith, Mullins, Wolff (September 1989), Tucker (Fall 1989), Varma (1989), Ambarish, Subrahmanyam (1990), Chance (June 1990), Rosengren (May 1990), Litterman, Iben (March-April 1991), Shreve, Soner, Xu (1991), Leland (November 1994), Merrill (January 1994), Morgan (1994), Babbel, Merrill, Panning (September 1995), Whittingham (Winter 1997), Zatti (1998), Narayan (August 1999), Bolton, Freixas (2000), Brennan, Xia (2000), Canterbery (2000), Hong, Warga (2000), Landen (2000), Campbell, Viceira (2001), Campbell, Taksler (2003), Elliott, Van Der Hoek (2001), Elliott, Nishide (2013), Schultz (2001), Livingston, Zhou (2002, 2010), Jacoby (2003), Hunter, Simon (2004), Cochrane, Piazzesi (2005), Koijen, Nijman, Werker (2005), Pu Shen (2005), Alderson, Betker, Stock (2006), Anghelache (2006), Harris, Piwowar (2006), Pilotte, Sterbenz (2006), Bierman (2007, 2010), Chen, Lesmond, Wei (2007), Connolly (2007), Edwards, Harris, Piwowar (2007), Goldstein, Hotchkiss (2007), Green, Hollifield, Schurhoff (2007a, b), Bessembinder, Maxwell (2008), Hanson, Liljeblom, Löflund (2008), Rosengren (2008), Santos, Winton (2008), Beber, Brandt, Kavajecz (2009), Dejun Xie (2009), Drut (March 2009), Webb (2009), Banko, Lei Zhou (2010), Chulho Jung, Shambora, Kyongwook Choi (2010), Grasso, Linciano, Pierantoni, Siciliano (2010), Güntay, Hackbarth (2010), Răscolean, Szabo (2010), Schneider, Ciobanu (2010), Chang, Krueger (2011), Ibarra-Ramirez (June 2011), Puzyrewicz (2011), Fabozzi (January 16 2012), Kovner, Chenyang Wei (March 2012), Prewysz-Kwinto (2012), Prewysz-Kwinto, Voss (2014), Zhiguo He, Milbradt (January 13 2012), Compton, Kunkel, Kuhlemeyer (2013), Hung-Gay Fung, Derrick Tzau, Jot Yau (2013), Janska (2013), Mazurek (2013), Booth,

Gounopoulos, Skinner (2014), Morellec, Valta, Zhdanov (2015), Coletta, Santioni (October 2016), Ranosz (2016), Yu-Sheng Lai (2016).

The **company stock** as an investment product was researched in Fama, Fisher, Jensen, Roll (1969), Fama (1970, 1998), Fama, French (1992, 1993, 1996), Fama, Hansen, French (2013), Akerlof (1970), Stoll, Curley (1970), Logue (1973), Logue, Rogalski, Seward, Foster-Johnson (2001), Reilly (1973), McDonald, Jacquillat (1974), Ibbotson, Jaffe (1975), Ibbotson (1975), Ibbotson, Sindeler, Ritter (1988, 1994), Benston, Smith (1976), Jensen, Meckling (1976), Jensen (1986), Miller (1977), Leland, Pyle (1977), Weinstein (1978), Kahneman, Tversky (1979), Wilson (1979), Brown, Warner (1980), Buckland, Herbert, Yeomans (1981), Milgrom (1981), Milgrom, Weber (1982), Myerson (1981), Baron (1982), Dretske (1983), Myers, Majluf (1984), Ritter (1984, 1987, 1991, 1998a, b, 2002, 2003a, b, 2005), Beatty, Ritter (1986), Loughran, Ritter (1995, 2002), Hamao, Packer, Ritter (1998), Kim, Ritter (1999), Chen, Ritter (2000), Ritter, Welch (2002), Ritter, Warr (2002), Kyle (1985), Amihud, Mendelson (1986), Amihud, Mendelson, Uno (1999), Amihud, Hauser, Kirsh (2001, 2003), Beatty, Ritter (1986), Beatty, Zajac (1994), Beatty, Welch (1996), Booth J, Smith (1986), Ridder (1986), Rock (1986), Shleifer, Vishny (1986), Shleifer, Wolfenzon (2002), Titman, Trueman (1986), Bernheim, Peleg, Whinston (1987), McAfee, McMillan (1987), Miller, Reilly (1987), Balvers, McDonald, Miller (1988), Johnson, Miller (1988), Tinic (1988), Allen, Faulhaber (1989), Barry (1989), Barry, Muscarella, Peavy, Vetsuypens (1990), Benveniste, Spindt (1989), Benveniste, Wilhelm (1990, 1997, 1998), Benveniste, Busaba, Wilhelm (1996), Benveniste, Busaba (1997), Benveniste, Erdal, Benveniste, Ljungqvist, Wilhelm, Yu (2003), Grinblatt, Hwang (1989), Koh, Walter (1989), Maskin, Riley (1989), Muscarella, Vetsuypens (1989a, b, 1990), Uhlir (1989), Welch (1989, 1992, 1996), Welch, Ritter (2002), Carter, Manaster (1990), Carter, Dark, Singh (1998), Clarkson, Thompson (1990),

Husson, Jacquillat (1990), Levis (1990), Lucas, McDonald (1990), Sahlman (1990), Allen (1991), Hasbrouck (1991), Lee Ch M C, Shleifer, Thaler (1991), Megginson, Weiss (1991), Megginson, Smart (2009), Menon, Williams (1991), Spatt, Srivastava (1991), Cotter (1992), Hughes, Thakor (1992), Mauer, Senbet (1992), Aggarwal, Leal, Hernandez (1993), Aggarwal (2000), Aggarwal, Conway (2000), Aggarwal, Prabhala, Puri (2002), Aggarwal, Krigman, Womack (2002), Aggarwal (2003), Affleck-Graves, Hegde, Miller, Reilly (1993), Choe, Masulis, Nanda (1993), Back, Zender (1993), Back, Zender (2001), Chemmanur (1993), Chemmanur, Fulghieri (1997, 1999), Chemmanur, Liu (2003), Chemmanur, Hu (2007), Chemmanur, Yan (2009), Chemmanur, He, Hu (2009), Chemmanur, He, Nandy (2010), Chemmanur, Krishnan (2012), Chemmanur, He (2012), Conrad, Kaul (1993), Dhatt, Kim, Lim (1993), Drake, Vetsuypens (1993), Figlewski, Webb (1993), Hanley (1993), Hanley, Wilhelm (1995), Hebner, Hiraki (1993), Jegadeesh, Weinstein, Titman (1993a), Jegadeesh, Weinstein, Welch (1993b), Keloharju (1993), Keloharju, Kulp (1996), Keloharju, Nyborg, Rydqvist (2004), Leleux (1993), Leleux, Muzyka (1997), Levis (1993, 2004), Loughran (1993), Loughran, Ritter, Rydqvist (1994), Loughran, Ritter (1995, 1997, 2000, 2002, 2003, 2004), Ruud (1993), Rydqvist (1993), Rydqvist, Högholm (1995), Vos, Cheung (1993), Friedlan (1994), Jain, Kini (1994), Jog, Srivastava (1994), Jog, McConomy (1999), Kunz, Aggarwal (1994), Lerner (1994), Michaely, Shaw (1994), Michaely, Womack (1999), Schultz, Zaman (1994), Schultz (2003), Degeorge (1995), Degeorge, Derrien, Womack (2005), Gerstein (1995, 1996), Gompers (1995), Gompers, Lerner (1997, 2001, 2003a, b), Kim, Krinsky, Lee (1995), Spiess, Affleck-Graves (1995, 1999), Spiess, Pettway (1997), Zingales (1995), Barber, Lyon (1996a, b, 1997), Barber, Odean (2008), Booth J R, Chua (1996), Booth J R, Booth L (2003), Borggreve, Dobrikat (1996), Brealey, Myers (1996), Brennan, Subrahmanyam (1996), Chowdhry, Sherman (1996),

Chowdhry, Nanda (1996), Easley, Kiefer, O'Hara, Paperman (1996), Hazen (1996), Houston, James (1996), Houston, James, Karceski (2004), Kogut, Zander (1996), Kothari, Warner (1996, 1997), Kothari (2001), Lee P J, Taylor, Walter (1996a, b, 1999), Nyborg, Sundaresan (1996), Pettway, Kaneko (1996), Périer (1996), Aussenegg (1997), Brav, Gompers (1997, 2002, 2003), Brav (2000), Brav, Geczy, Gompers (2000), Brennan, Franks (1997), Cai, Wei (1997), Carhart (1997), Datta, Iskandar-Datta, Patel (1997), Dechow, Sloan (1997), Dechow, Hutton, Sloan (1999, 2000), Ehrhardt (1997), Firth (1997), Gande, Puri, Saunders, Walter (1997), Gande, Puri, Saunders (1999), Gregg (1997), Huang, Stoll (1997), Kooli (2000), La Porta, Lopez-de-Silanes, Shleifer, Vishny (1997, 1998), La Porta, Lopez-de-Silanes, Shleifer (2002, 2006), Lee I (1997), Ljungqvist (1997, 2006), Ljungqvist, Nanda, Singh (2001), Ljungqvist, Wilhelm (2002, 2003), Ljungqvist, Jenkinson, Wilhelm (2003), Ljungqvist, Marston, Wilhelm (2003), Ljungqvist, Nanda, Singh (2003, 2006), Mikkelson, Partch, Shah (1997), Nanda, Youngkeol Yun (1997), Page, Reyneke (1997), Rajan, Servaes (1997a, b), Stehle (1997), Stehle, Ehrhardt (1999), Stehle, Ehrhardt, Przyborowsky (2000), Steib, Mohan (1997), Su, Fleisher (1997), Arkebauer (1998), Asquith, Jones, Kieschnick (1998), Ausubel, Cramton (1998a, b), Black, Gilson (1998), Daniel, Hirshleifer, Subrahmanyam (1998), Goergen (1998), Goergen, Renneboog (2002), Helwege, Kleiman (1998), Helwege, Liang (2001, 2004), Kahn, Winton (1998), Malvey, Archibald (1998), Mello, Parsons (1998), Mok, Hui (1998), Pagano, Panetta, Zingales (1998), Pagano, Röell (1998), Paudyal, Saadouni, Briston (1998), Poon, Firth, Fung (1998), Rangan (1998), Reese (1998), Sapusek (1998), Stoughton, Zechner (1998), Taylor, Whittred (1998), Theoh, Welch, Wong (1998a, b), Ansotegui, Fabregat (1999), Arcas, Ruiz (1999), Baker, Gompers (1999, 2001), Baker, Wurgler (2000), Baker, Nofsinger, Weaver (2002), Brown (1999), Cornelli, Goldreich (1999, 2001, 2002, 2003), Cornelli, Goldreich, Ljungqvist (2006),

Field (1999), Field, Hanka (2001), Field, Sheehan (2001, 2002), Field, Karpoff (2002), Field, Lowry (2009), Kandel, Sarig, Wohl (1999), Khurshed, Mudambi, Goergen (1999), Khurshed, Mudambi (2002), Krigman, Shaw, Womack (1999), Krigman, Shaw, Womack (2001), Lyon, Barber, Tsai (1999), Olson, Nelson (1999), Short, Keasey (1999), Stulz (1999, 2005, 2009), Subramanyam, Titman (1999), Thomas, Zhang (1999), Arosio, Giudici, Paleari (2000, 2001), Aussenegg (2000), Berkman, Bradbury, Ferguson (2000), Binmore, Swierzbinski (2000), Boehmer, Fishe (2000, 2001, 2005), Brailsford, Heaney, Powell, Shi (2000), Brailsford, Heaney, Shi (2004), Chen, Firth, Kim (2000), D'Mello, Ferris (2000), Draho (2000), Dunbar (2000), Dunbar, Foerster (2008), Duque, Almeida (2000), Eckbo, Masulis, Norli (2000), Eckbo, Norli (2001, 2002, 2005), Eckbo (2008), Ellis, Michaely, O'Hara (2000, 2002), Fabrizio (2000), Foerster (2000), Gilbert, Klemperer (2000), Jain, Kini (2000), Kiymaz (2000), Koskie, Michaely (2000), Lewis, Seward, Foster-Johnson (2000), Löffler (2000), Reuschenbach (2000), Sapusek (2000), Schultz (2000, 2001), Schultz, Zaman (2001), Sinclair (2000), Sherman (2000, 2001, 2003), Sherman, Titman (2002), Smart, Zutter (2000), Stehle, Ehrhardt, Przyborowsky (2000), Westerholm (2000), Von Eije, De Witte, Van der Zwaan (2000), Bernardo, Welch (2001), Bradley, Jordan, Ha-Chin Yi, Roten (2001), Bradley, Jordan (2002), Bradley, Jordan, Ritter (2003, 2008a), Bradley, Chan, Kim, Singh (2008b), Busaba, Benveniste, Guo (2001), Certo, Covin, Daily, Dalton (2001), Chan, Wang, Wei (2001), Cooney, Singh, Carter, Dark (2001), Daines, Klausner (2001), Danielsen, Sorescu (2001), Degeorge, Derrien (2001a, b), Derrien, Womack (2002), Derrien (2005, 2007), Derrien, Kecskés (2006), DuCharme, Rajgopal, Sefcik (2001), Francis, Hasan (2001), Gerke, Fleischer (2001), Habib, Ljungqvist (2001), Hahn, Ligon (2004), Hansen (2001), Heaton (2001), Hoffmann-Burchardi (2001), Holmén, Högfeldt (2001), Houge, Loughran, Suchanek, Yan (2001), Jakobsen, Sørensen

(2001), Jenkinson, Ljungqvist (2001), Jenkinson, Jones (2004, 2007), Jenkinson, Morrison, Wilhelm (2006), Killian, Smith, Smith (2001), Lowry, Schwert (2001, 2002), Lowry, Shu (2002), Lowry (2003), Mager (2001), Maksimovic, Pichler (2001), Purnanandam, Swaminathan (2001), Rehkugler, Schenek (2001), Schatt, Roy (2001), Schatt, Broye (2003), Sentis (2001, 2002, 2004), Severin (2001), Stoughton, Wong, Zechner (2001), Torstila (2001, 2003), Van Bommel, Vermaelen (2001), Van Frederikslust, Van der Geest (2001), Vayanos (2001), Zhang (2004), Biais, Bossaerts, Rochet (2002), Biais, Faugeron-Crouzet (2002), Blondell, Hoang, Powell, Shi (2002), Brau, Francis, Kohers (2002), Brounen, Eichholtz (2002), Bulow, Klemperer (2002), Cheng, Mak, Chan (2002), Deloof, De Maeseneire, Inghelbrecht (2002), Easton, Taylor, Shroff, Sougiannis (2002), Easton (2004, 2006), Easton, Sommers (2007), Faugeron-Crouzet, Ginglinger (2002), Filatotchev, Bishop (2002), Fishe (2002), Gao, Mao, Zhong (2002), Giudici, Roosenboom (2002, 2005), Houge, Loughran, Suchanek, Xuemin Yan (2002), Kim, Kitsabunnarat, Nofsinger (2002), Kiss, Stehle (2002), Kutsuna, Okamura, Cowling (2002), Logue, Rogalski, Seward, Foster-Johnson (2002), Martimort (2002), Moerland (2002), Schiereck, Wagner (2002), Schuster (2002, 2003), Wang, Zender (2002), Xie (2002), Baginski, Wahlen (2003), Barondes, Nyce, Sanger (2003), Bartlett, Shulman (2003), Binay, Pirinsky (2003), Bourjade (2003, 2008), Clarke, Dunbar, Kahle (2003), Derrien, Womack (2003), Doeswijk, Hemmes, Venekamp (2005), Ellul, Pagano (2003), Goergen, Khurshed, McCahery, Renneboog (2003), Gounopoulos (2003), Gulati, Higgins (2003), Higgins, Gulati (2003), Hoberg (2003), Hong, Kubik (2003), Huyghebaert, Van Hulle (2003), Jelic, Briston (2003), Kaneko, Pettway (2003), Karolyi, Stulz (2003), Kraus, Burghof (2003), Lemmens (2003, 2004, 2007), Manigart, De Maeseneire (2003), Neuhaus, Schremper (2003), Nounis (2003), Ofek, Richardson (2003), Peristiani (2003), Pham, Kalev, Steen (2003), Roosenboom, Van der Goot (2003, 2005), Roosenboom, Van der Goot, Mertens (2003), Roosenboom (2007), Smart, Zutter (2003), Van Bommel, Vermaelen (2003), Van der Goot (2003), Weber, Willenborg (2003), Arugaslan, Cook, Kieschnick (2004), Bodnaruk, Kandel, Massa, Simonov (2004), Burrowes, Jones (2004), Cassia, Giudici, Paleari, Redondi (2004), Cassia, Paleari, Vismara (2004), Cassia, Vismara (2009), Chahine (2004a, b), Chiang, Harikumar (2004), Cliff, Denis (2004), Corwin, Harris, Lipson (2004), Durnev, Morck, Yeung (2004), Fernando, Gatchev, Spindt (2004), Foerster (2004), Griffith (2004), Ganor (2004), Hahn, Ligon (2004), Hao (2004), Hoberg (2004), Kooli, Suret (2004), Kremer, Nyborg (2004a, b), Kutsuma, Smith (2004), Lamont (2004), Lee M, Wahal (2004), Levy (2004), Lubig (2004), Mayhew, Mihov (2004, 2005), Mira (2004), Peggy, Wahal (2004), Pollock, Porac, Wade (2004), Pollock, Chen, Jackson, Hambrick (2005), Pritsker (2004, 2005, 2006), Purnanandam, Swaminathan (2004), Rath, Tebroke, Tietze (2004), Reuter (2004), Rice (2004, 2006), Rindermann (2004), Sanders, Boivie (2004), Schenone (2004), Serve (2004), Alti (2005, 2006), Alvarez, Gonzalez (2005), Benninga, Helmantel, Sarig (20050, Berg, Neumann, Rietz (2005), Butler, Grullon, Weston (2005), Choo (2005), Corwin, Schultz (2005), Dolvin (2005), Drobetz, Kammermann, Wälchli (2005), Forestieri (2005), Hess (2005), Hurt (2005, 2006), Jagannathan, Gao (2005), Jain, Kini (2005), Jaskiewicz, Gonzàlez, Menéndez, Schiereck (2005), Khanna, Noe, Sonti (2005), Khanna, Noe, Sonti (2008), Li, McInish, Wongchoti (2005a), Li, Zheng, Melancon (2005b), LiCalzi, Pavan (2005), Malloy (2005), Nounis (2005), Pandey (2005), Parlour, Rajan (2005), Pastror, Veronesi (2005), Pastor, Taylor, Veronesi (2009), Pons-Sanz (2005), Sherman (2005), Yan (2005), Anand (2006), Aussenegg (2006), Aussenegg, Pichler, Stomper (2006), Boot A W A, Gopalan, Thakor (2006), Das, Guo, Zhang (2006), Damodaran (2006), Ellul, Pagano (2006), Gajewski, Gresse (2006), Goergen, Renneboog, Khurshed (2006), Hong (2006), Jagannathan, Sherman (2006),

James, Karceski (2006), Pastor-Llorca, Poveda-Fuentes (2006), Tirole (2006), Trauten, Schulz (2006), Yung, Colak, Wang (2006, 2008), Zhang (2006), Arnold, Fishe, North (2007), Berkeley (2007), Doran, Jiang, Peterson (2007, 2009), Hopp, Dreherdo (2007), Jog, Sun (2007), Kerins, Kutsuna, Smith (2007), Leite (2007), Paleari, Vismara (2007), Paleari, Pellizzoni, Vismara (2008), Paleari, Ritter, Vismara (2010), Penman (2007), Thomas (2007), Toniato (2007), Zheng, Stangeland (2007), An, Chan (2008), Casotti, Motta (2008), Farina (2008), Hale, Santos (2008), Hild (2008), Kaustia, Knupfer (2008), Khurshed, Pande, Singh (2008), Khurshed, Paleari, Pande, Vismara (2011), Rossetto (2008), Kim, Weisbach (2008), Poudyal (2008), Yongyuan Qiao (2008), Colak, Gunay (2011), Bouis (2009), Coakley, Hadass, Wood (2009), Deloof, De Maeseneire, Inghelbrecht (2009), Jiang, Leger (2009), Zhang (2009), Arikawa, Imad'eddine (2010), Bonardo, Paleari, Vismara (2010), Caglio, Weiss-Hanley, Marietta-Westberg (2010), Cogliati, Paleari, Vismara (2010), Chod, Lyandres (2010), Deb, Marisetty (2010), Elston, Yang (2010), Guo, Brooks, Shami (2010), Hsu, Reed, Rocholl (2010), Hussinger (2010, 2012), Jagannathan, Jirnyi, Sherman (2010), Pennacchio, Del Monte, Acconcia (2010), Acconcia, Del Monte, Pennacchio (2011), Pennacchio (2013), Sahoo, Rajib (2010), Shao, Wu, Qin, Wang (2010), Yao-Min Chiang, Hirshleifer, Yiming Qian, Sherman (2010), Adesoye, Atanda (2012), Boissin (2012), Cumming, Hass, Schweizer (2012), Datar, Emm, Ince (2012), Jacob (2012), Rodrigues, Stegemoller (2012), Saturnino, Saturnino, Lucena, Caetano, Dos Santos (2012), Chang-Yi Hsu, Jean Yu, Shiow-Ying Wen (2013), Zhiqiang Hu, Yizhu Wang (2013), Lakicevic, Shachmurove, Vulanovic (2013).

The **stock option** as an investment product was researched in Weinberg, Patton (1963), James Boness (April 1964), Bierman (September 1967), Lewellen (1968), Hirshleifer (1970), Black, Scholes (1973), Black (1975), Merton (1973, 1997), Fisher (March 1978), Klemkosky (1978),

Margrabe (March 1978), Cox, Ross, Rubinstein (1979), MacBeth, Merville (1979), Bookstaber (1981), Brown, Rainbow (1981), Hite, Long (1982), Whaley (1982), Geske, Roll (1984), Leland (1985), Rubinstein (1985), Barone-Adesi, Whaley (1987), Hull, White (1987, January-February 2004), Hull (1997, 2005 – 2006), Blomeyer, Johnson (1988), Detemple, Jorion (1989), Jorion, Stoughton (1989), Lambert, Lanen, Larker (1989), Defusco, Johnson, Zorn (1990), Vijh (1990), Zivney (1991), Boyle, Vorst (1992), French, Maberly (1992), Harvey, Whaley (1992), Bizjak, Brickley, Coles (1993), George, Longstaff (1993), Heston (1993), Dawson (1994, 2000), Diz, Finucane (1994), Fleming, Whaley (March 1994), Huddart (1994), Huddart, Lang (February 1996), Saly (1994), Sheikh, Ronn (1994), Kamara, Miller (1995), Mas-Colell, Whinston, Green (1995), Rubinstien (Fall 1995), Sung (1995), Abken, Madan, Buddhavarapu Sailesh Ramamurtie (1996), Aboody (1996), Bergman, Grundy, Weiner (1996), Chicago Board Options Exchange (1996), Corrado, Miller (1996), Ikenberry, Vermaelen (1996), Scott (1997), Easley, O'Hara, Subrahmanya Srinivas (1998), Alexander, Veronesi (1999), Bates (1999), Campbell, Canlin Li (1999), Heath, Huddart, Lang (May 1999), Chance, Kumar, Todd (2000), Etling, Miller (November 2000), Hall, Murphy (May 2000a, b, 2002, 2003), Johnson, Tian (2000), Lee, Nayar (2000), McMurray, Yadav (2000), Core, Guay (August 2001), Coval, Shumway (2001), Dueker, Miller (2002), Kalok Chan, Peter Chung, Wai-Ming Fong (2002), Poteshman, Serbin (2002), Sahlman (December 2002), Taylor (2002), Utsunomiya (2002), Bliss (April 2003), Bodie, Kaplan, Merton (March 2003), Campbell (June 2003), Cassano (2003), Craig, Glatzer, Keller, Scheicher (2003), Guay, Kothari, Sloan (May 2003), Haubrich (November 2003), Raupach (2003), Barenbaum, Schubert, O'Rourke (December 16-20 2004), Battalio, Hatch, Jennings (2004), Calomiris, Hubbard (January 2004), Arnold, Gillenkirch (September 2005), Bulow, Shoven (Fall 2005), Ciccotello, Grant, Wilder (September 2005),

Dahlgren, Korn (2005), Eberhart (2005), Lie (May 2005), Manzon (2005), Oyer, Schaefer (April 2005), Ross, Westerfield, Jaffe (2005), Sundaram (2005), Sautner, Weber (2005), Cetin, Jarrow, Protter, Warachka (2006), Chongwoo Choe, Xiangkang Yin (2006), Deshmukh, Howe, Luft (Spring 2006), Macminn, Page (January 2006), Pan, Poteshman (2006), Rhee (2006), Sharma (2006), Tirole (2006), Uchida (2006), Campbell (Winter 2007), Lakonishok, Inmoo Lee, Pearson, Poteshman (2007), Taylor (December 2007), Hurtt (2008), Ni, Pan, Poteshman (2008), Ye (2008), Cuny, Martin, Puthenpurackal (2009), Gârleanu, Pedersen, Poteshman (2009), Chen, Bong Soo Lee (2010), Engle, Neri (2010), Hallock, Olson (2010), Babenko, Lemmon, Tserlukevich (2011), Børsum (2011), Schürhoff, Ziegler (2011), Yang, Carleton (2011), Zhongjin Yang, Cassidy Yang (2011), Korn, Paschke, Uhrig-Homburg (2012), Morikawa (2012), Perobelli, De Souza Lopes, Da Silveira (2012), Timraz, Al-Shubiri (2012), Baule, Tallau (2013), Cao, Bing Han (2013), Aldatmaz, Ouimet, Van Wesep (2014), An, Ang, Bali, Cakici (2014), Boyer, Vorkink (2014), Câmara, Popova, Simkins (2014), Karakaya (2014), Sonika, Carline, Shackleton (2014), Christoffersen, Goyenko, Jacobs, Karoui (2015), Goyenko, Ornthanalai, Shengzhe Tang (2015), Choy, Wei (2016), Investopedia (2016), Kanne, Korn, Uhrig-Homburg (2016).

The **financial security** as an investment product was researched in Knight (1921), Fisher, Tippett (1928), Working (1948), Modigliani, Miller (1958), Borch (1960, 1961, 1962), Sharpe (1964), Akerlof (1970), Cox (1972), Black, Scholes (1973), Black, Cox (1976), Fama, MacBeth (1973), Merton (1974a, b), Jaffee, Russel (1976), Leland, Pyle (1977), Leland (1994), Stiglitz, Weiss (1981, 1983), Diamond (1984, 1991), Diamond, Rajan (2000), Fama (1985), Blazenko (1986), James (1987), Rudolph (1987, 1994), Berger, Udell (1990), Hellwig (1991), Berlin (1992), Heath, Jarrow, Morton (1992), Dewatripont, Tirole (1993), Froot, Scharfstein, Stein (1993),

Froot, Stein (1998), Cocheo (1994), Figlewski (Summer 1994), Hull, White (1994a, b, 1996, 2000, 2001, 2004, 2005a, b, 2008), Hull, Predescu, White (2004), Hull (2013, January 25 2014), Jarrow, Lando, Turnbull (1994, 1997), Jarrow, Turnbull (1995, 2000a, b), Jarrow (2001), Jarrow, Yu (2001), Jarrow, Yildirim (2002), McAllister, Mingo (May 1994), Neuberger (1994), Das (Spring 1995, 1998a, b, 2000), Das, Hanouna, Sarin (2009), Edwards (August 1995), Gorton, Pennacchi (1995), Gorton, Winton (1998), Gorton, Souleles (2005), Hasbrouck (1995), Longstaff, Schwartz (June 1995a, b), Longstaff, Mithal, Neis (August 2003, 2005), Longstaff, Rajan (2008), Whittaker, Kumar (1995), Whittaker, Frost (May 1997), Altman (September 1996), Duffee (1996), Duffee, Zhou (February 1997, November 1999, 2001), Duffie (1996a b, 1999, July 2008), Duffie, Singleton (1999, 2002), Duffie, Pan, Singleton (November 2000), Duffie, Gârleanu (2001, 2003), Duffie, Lando (2001), Duffie, Filipovic, Schachermayer (2003), Hattori (1996), Hyder, Bolger, Leung (July 1996), Neal (May 1996), Neal, Rolph (1999), Reoch, Masters (March 1996), Schönbucher (August 1996, 1997, Fall 1999, 2000a, b, 2003, 2006), Smithson, Holappa, Rai (June 1996), Drzik, Kuritzkes (July 1997), Freixas, Rochet (1997), Géczy, Minton, Schrand (1997), Hill (1997), Hüttemann (1997, 1998), Joe (1997), Kaufman (1997), Mahtani (August 23 1997), Ogden (1997), Wilson (September 1997a, October 1997b), Burghof, Henschel (1998), Burghof, Henke, Rudolph, Schönbucher, Sommer (2005), Henke, Burghof, Rudolph (1998), Fabozzi (January 1998, January 16 2012), Fabozzi, Davis, Choudhry (October 20 2006), Fabozzi, Kothari (2007), Lancaster, Schultz, Fabozzi (April 25 2008), Lando (1998, 2004), Scott–Quinn, Walmsley (1998), Steinherr (1998), Tavakoli (1998, 2001, 2003), Arvanitis, Gregory, Laurent (Spring 1999), DeMarzo, Duffie (1999), DeMarzo (2005), Gauvin (September 8 1999), Heidorn (1999), Embrechts, McNeil, Straumann (1999), Embrechts, Lindskog, McNeil (2001), Master, Bryson (1999), Nelken (1999), Rappoport (1999), Staehle,

Cumming (1999), Allen, Gale (2000), Allen , Carletti (2006), British Bankers' Association (2000), Brewer, Minton, Moser (2000), Kiff, Morrow (Autumn 2000), Kiff, Michaud, Mitchell (2002, June 2003a, b), Li (2000), Müller (2000), Saunders (2000), Wahl, Broll (2000), Arvantis, Gregory (Jon 2001), Bomfim (July 11 2001), Cossin, Hricko (May 2001), Collin-Dufresne, Goldstein, Martin (2001), Craddock, Platen (June 28 2001), Delianedis, Geske (2001), Elton, Gruber, Agrawal, Mann (2001), Giesecke (2001, 2002, 2003, 2004, 2006), Giesecke, Weber (2003), Giesecke, Goldber (2004), Goodman (2001), Lopez (2001), Lucas, Klaassen, Spreij, Staetmans (2001), Mashal, Naldi (2001), Ranciere (2001), Rehm (2001), Rule (June 2001), Schmidt (März 2001), Amihud (2002), Angelini (2002), Arping (May 2002), Aunon-Nerin, Cossin, Hricko, Huang (2002), Batten, Hogan (2002), Bielecki, Rutkowski (2002, 2004), Broll, Welzel (2002), Broll, Pausch, Welzel (July 2002), Broll, Schweimayer, Welzel (2003), Di Graziano, Rogers (2002), Dunne, Moore, Portes (2002), Grill, Perczynski, Int-Veen, Muthig, Platz (2002), Jobst (2002), Läger (2002), Prato (November 2002), Ranciere (April 2002), Skinner, Townend (2002), Albanese, Campolieti, Chen, Zavidonov (2003), Albanese, Chen (2005), Albanese, Vidler (October 9 2007), Amato, Remolona (December 2003), Amato (December 2005), BIS (2003), Blanco, Brennan, Marsch (2003, 2004), Codogno, Favero, Missale (2003), Fage (November 24 2003), Francis, Kakodkar, Martin (April 2003), Galiani (2003), Grundke (2003), Haas (June 2003), Li Chen, Filipović (July 4 2003), MacKenzie, Millo (2003), Packer, Suthiphongchai (December 2003), Schmidt (2003), Xu, Wilder (May 2003), Blanco, Brennan, March (2004), Cebenoyan, Strahan (2004), Chan-Lau, Kim (2004), Edwards, Morrison (2004), Eller, Markus (2004, 2005), Ericsson, Jacobs, Oviedo (2004), Ericsson, Reneby, Wang (2006), Favero, Von Thadden (2004), Favero, Pagano, Von Thadden (2009), Felsenheimer (June 2 2004, September 2004), Felsenheimer, Gisdakis, Zaiser (2005), Garcia, Maghakian,

Sharma (2004), Gibson (2004 , May 22 2007), Madan, Konikov, Marinescu (2004), Meneguzzo, Vecchiato (2004), Norden, Weber (2004a, b, March 26 2007), Olléon-Assouan (June 2004), Pagano, Von Thadden (2004), Pausch, Schweimayer (March 2004), Puffer (2004), Shelton (August 2004), Shimko (2004), Tavares, Nguyen, Chapovsky, Vaysburd (2004), Verdier (2004), Zhu (2004), Abid, Naifar (2005, 2006a, b), Acharya, Pedersen (2005), Acharya, Schaefer (2006), Acharya, Johnson (2007), Acharya, Schaefer, Zhang (May 2007), Albrecht (2005), Bielecki, Crepey, Jeanblanc, Rutkowski (March 2005), Bielecki, Vidozzi A, Vidozzi L (May 2006), Blanco, Brennan, Marsh (2005), Bomfim (2005), Brigo (January 2005), Brigo, Alfonsi (2005), Brigo, Cousot (2006), Brigo, El-Bachir (2006), Brigo, Pallavicini, Torresetti (June 2007), Burtschell, Gregory, Laurent (2005, 2008), Chiesa (2005), Cossin, Lu (June 2005), Cousseran, Rahmouni (June 2005), Franke, Krahnen (2005), Franke (2005), Greenspan (May 5 2005), Heinrich (2005), Houweling, Vorst (2005), Instefjord (2005), Jorion (2005), Jortzik (2005), Joshi, Stacey (2005), Kalemanova, Schnid, Werner (2005), Kim, Lucey, Wu (2005), Kim, Moshirian, Wu (2006), Lucas, Goodman, Fabozzi (2006, 2007), Lucas, Goodman, Fabozzi, Manning (2007), Minton, Stulz, Williamson (August 2005, June 2006), Morrison (2005), Nési (March 2005), Neske (2005), Nicolò, Pelizzon (October 17 2005), Parlour, Plantin (2005), Posthaus (2005), Pelizzon, Schaefer (2005), Skinner (2005), Stamicar, Finger (2005), Allen, Carletti (2006), Beitel, Dürr, Pritsch, Stegemann (2006), Beck, Lesko, Schlottmann, Wimmer (Ausgabe 14 2006), Chilcote (2006), Cremers (2006), Cremers, Walzner (June 2007, 2009a, b), De Wit (2006), Felsenheimer, Gisdakis, Zaiser (2006a, b), Geithner (May 16 2006), Gikhman (July 2006, February 4 2008, July 2008, 2008, November 2011), Goderis, Marsh, Castello, Wagner (2006), Gruber J, Gruber W, Braun (2005), Huault, Rainelli-Le, Montagner (2006, 2008), Joshi, Stacey (July 2006), Juselius (2006), Ludovici (November December 2006), Marsh (2006), Martin, Reitz,

Wehn (2006), Minton, Stulz, Williamson (June 2006), Mortensen (2006), Nicolò, Pelizzon (November 2006), Papenbrock (2006), Partnoy, Skeel (2006), Predescu (2006), Pugachevsky (2006), Sircar, Zariphopoulou (2006), Zhu (2006), Wagner, Marsh (2006), Ammer, Cai (2007), Ashcraft, Santos (July 2007), Baba, Inada (2007), Boulier, Brière, Viala (2007), Byström (2007, 2008, 2010), Chen, Lesmond, Wei (2007), Davydenko, Strebulaev (2007), Dötz (2007), Drucker, Puri (2007), Fathi, Nader (2007), Fulop, Lescourret (2007), Henrard (2007, 2009), Hirtle (February 2007, March 2008), In, Kang, Kim (2007), Mengle (2007), Nashikkar, Subrahmanyam (2007), Nashikkar, Subrahmanyam, Mahanti (2009), Pan, Singleton (2007), Papageorgiou, Sircar (2007), Pausch (2007), Pykhtin, Zhu (July August 2007), Rajan, McDermott (2007), Schulz, Wolf (2007), Summer (2007), Tang, Yan (2007), Thompson (2007), Truslow (March 22 2007), Weithers (Fourth Quarter 2007), Yan, Zivot (2007), Abid, Naifar (2008), Alexander, Kaeck (2008), Arnsdorf, Halperin (2008), Chen, Cheng, Liu (2008), Claes, De Ceuster (2008), Cont, Minca (2008), Coudert, Gex (2008), Cserna, Imbierowicz (2008), Curto, Nunes, Oliveira (2008), Jankowitsch, Pullirsch, Veža (2008), Le Roux (October 2008), Li, Zou (2008), Morgan (May 2008), O'Kane (2008), Peng, Kou (2008), Sougné, Heuchenne, Hübner (2008), Ametrano, Bianchetti (2009), Ametrano (January 18 2011), Bayraktar, Yang (2009), Elizalde, Doctor (2009), Elizalde, Doctor, Saltuk (2009), Forte, Peña (2009), Fujii, Shimada, Takahashi (2009a, b, 2010a, b), Fujii, Takahashi (2011), Mayordomo, Peña, Romo (September 2009, 2012), Rey (2009, November 20 2009), Panchenko, Wu (2009), Varga (May 2009), Buraschi, Porchia P, Porchia F (2010), Ciolpan, Adam (2010), Coudert, Gex (2010), Courtney (2008, May 19 2010), De Wit (2006), Gómez (2010), Gregory (2010), Fries (2010), Tsui (2010, May 18 2010), Piterbarg (2010a, b), Bianchetti (October 31 2011), Bianchetti, Mattia (March 28 2012), Calice, Ioannidis, Williams (2011), Coşkun (2011), Lahusen, Speyer (2011), Lijun

Bo, Ying Jiao, Xuewei Yang (December 13 2011), Ojo (May 24 2011, April 30 2012), Rennie, Lipton (2011), Avino, Lazar (2012), Heller, Vause (2012), Roman, Şargu (2012), Leung, Peng Liu (October 2 2012), Yildirim, Coskun, Caglar, Yildirak (2012).

The **foreign currency** as an investment product was researched in Ellis, Metzler (editors) (1949), Machlup (1949), Robinson (1949), Friedman (1953), Baumol (1957), Debreu (1959), Fama (1965, 1970, 1984, 1998), Fama, Blume (1966), Fama, French (1988, 1996), Fama, Hansen, Shiller (2013), Demsetz (1968), Radner (1968), Bates, Granger (1969), Akerlof (1970), Arrow (1970), Black (1971, 1986), Black, Scholes (1973), Merton (1973), Newbold, Granger(1974), Fleming (1975), Shapiro (1975), Dooley, Shafer (1976), Dornbusch (1976, 1987), Frankel (1976, 1979, 1982a, b, 1983, 1992, (editor) 1993), Frankel, Froot (1987, 1990a, b, c), Frankel, Goldstein, Mason (1991), Frankel, Rose (1994, 1995), Frankel, Galli, Giovannini (editors) (1996), Frankel, Poonawala (2004), Garman (1976), Grossman (1976), Grossman, Stiglitz (1980), Grossman, Miller (1988), Kouri (1976), McKinnon (1976), Mussa (1976, 1979, 1981, 1984), Williamson (1976), Branson (1977), Branson, Halttunen, Masson (1977), Branson, Henderson (1985), Clark, Logue, Sweeney (editors) (1977), Girton, Henderson (1977), Cornell, Dietrich (1978), Cornell (1982), Stoll (1978, 1985, 1989, 1995, 1998, 2006), Huang, Stoll (1996, 1997), Stoll, Schenzler (2005), Blanchard (1979), Brunner, Meltzer (editors) (1979), Deardorff (1979), Goodman (1979), Aliber (1980, 2002), Allen, Kenen (1980), Amihud, Mendelson (1980), Amihud, Ho, Schwartz (editors) (1985), Amihud (1994a, b, c) Amihud, Levich (editors) (1994), Hansen, Hodrick (1980), Hellwig (1980, 1982), Krugman (1980, 1984, 1991, 1999), Krugman, Miller (1993), Callier (1981), Cohen, Maier, Schwartz, Whitcomb (1981), Cox, Ingersoll, Ross (1981), Diamond, Verrecchia (1981), Diamond (1982), Fieleke (1981), Ho, Stoll (1981, 1983), Loosignian (1981), Stigum (1981, 1990), Dooley, Isard (1982), Hansen (1982), Hodder (1982), Milgrom, Stokey (1982), Taylor D (1982), Bigman, Taya (editors) (1983), Copeland, Galai (1983), Dooley, Shafer (1983), Edwards (1983), French (1983), Garman, Kohlhagen (1983), Meese , Rogoff (1983a, b, 1988) Rogoff (1984, 1985, 1996), Meese (1986, 1990), Obstfeld, Rogoff (1995, 1998), Robinson (1983), Adler, Dumas (1984), Backus (1984), Bilson, Marston (editors) (1984), Booth (1984), Engel, Frankel (1984a, b), Engel, Hamilton (1990), Engel (1992, 1995, 1996, 1999), Devereux, Engel (1999, 2002), Devereux, Shi (2005), Engel, West (2004a b, 2005, 2006), Engel, Mark, West (2007), Garner, Shapiro (1984), Loopesko (1984), Roll (1984, 1988), French, Roll (1986), White, Domowitz (1984), Bahmani-Oskooee, Das (1985), Cohen, Conroy, Maier (1985), Glosten, Milgrom (1985), Glosten, Harris (1988), Glosten (1989, 1994), Hakkio, Pearce (1985), Hardouvelis (1985), Jones, Kenen (editors) (1985), Kearney, Macdonald (1985), Kyle (1985, 1989), Kyle, Xiong (2001), Levich (1985), McInish, Wood (1985), Dominguez (1986, 1990, 1992, 1993, 1998, 2003a, b), Dominguez, Frankel (1993a, b, c), Bollerslev (1986, 1990), Baillie, Bollerslev (1989 1990, 1991), Bollerslev, Chou, Jayaraman, Kroner (1990), Bollerslev, Domowitz (1991, 1993), Bollerslev, Melvin (1994), Andersen, Bollerslev (1994, 1998), Bollerslev, Engle, Nelson (1995), Bollerslev, Cai, Song (2000), Andersen, Bollerslev, Diebold, Labys (2000, 2001, 2003), Andersen, Bollerslev, Diebold, Vega (2001, 2003), Andersen, Bollerslev, Diebold (2007), Engle (1982), Engle, Bollerslev (1986), Engle, Granger (1987), Engle, Rodriguez (1989), Engle, Ito, Lin Wen-Ling (1990), Engle, Russell (1995), Engle, Gallo (2006), Evans (1986), Flood, Lessard (1986), Grammatikos, Saunders, Swary (1986), Harris (1986, 1990), Hart, Kreps (1986), Lyons (1986, 1988, 1990, 1991, 1992, 1993a, b, c, 1994, 1995, 1996a, b, 1997a, b, c, 1998a, b, 2001, 2002a, b, 2003, 2006), Baldwin, Lyons (1994), Lyons, Rose (1995), Fan, Lyons (2001, 2003), Killeen, Lyons, Moore (2001, 2006), Killeen, Hau, Moore

(2001), O'Hara, Oldfield (1986) , Burdett, O'Hara (1987), O'Hara (1995, 1998), Shleifer (1986), Shleifer, Summers (1990), Sweeney (1986), DeLong, Shleifer, Summers, Waldmann (1990), Bilson, Hsieh (1987), Glassman (1987), Gerlach (1987), Hasbrouck, Ho (1987), Hasbrouck (1988, 1991), Hasbrouck, Sofianos (1993), Hasbrouck, Seppi (2001), Hodrick (1987), Ito, Roley (1987, 1990), Canova, Ito (1991), Ito, Engle, Lin (1992), Ito, Lin (1992), Ito, Isard, Symansky, Bayoumi (1996), Ito, Lyons, Melvin (1998), Ito (2002, 2005a, b), Ito, Hashimoto (2006), Mendelson (1987), Newey, West (1987), Rubinstein, Wolinsky (1987), Taylor M P (1987, 1989, 1995, 2005), Allen, Taylor (1989), Taylor M P, Allen (1992), Sarno, Taylor M P (2000, 2001a, b), Sager, Taylor M P (2005, 2006, 2008), Reitz, Taylor M P (2006), Schulmeister (1987), Melvin, Taylor M P (2009), Newey, West (1987), Wolff (1987), Admati, Pfleiderer (1988, 1989), Boothe (1988), Choi, Salandro, Shastri, Clinton (1988), Goodhart (1988, 1989, 1992), Goodhart, Demos (1990, 1991a, b), Goodhart, Curcio (1991), Goodhart, Figliuoli (1991), Goodhart, Hall, Henry, Pesaran (1993), Goodhart, Hesse (1993), Goodhart, Ito, Payne (1995, 1996), Goodhart, O'Hara (1995, 1997), Goodhart, Love, Payne, Rime (2002), Hardouvelis (1988), Lewis (1988, 1995), Baldwin, Krugman (1989), Baxter, Stockman (1989), Dooley, Lizondo, Mathieson (1989), Giovannini (1989), Golub (1989), Humpage (1989), Leach, Madhavan (1989), Leahy (1989), Miller, Eichengreen, Portes (editors) (1989), Van Hagen (1989), Allen, Taylor (1990), Allen, Karjalainen (1999), Courakis, Taylor (1990), Diebold, Nason (1990), Flood, Hodrick (1990), Flood, Rose (1995), Flood, Taylor M P(1996), Flood, Marion (2001), Foster, Viswanathan (1990, 1993), Holthausen, Leftwich, Mayers (1990), Domowitz (1990, 1993), Domowitz, Steil (1999), Johansen, Juselius (1990), Johansen (1991, 1992), Jorion (1990, 1991, 1996), Lo, MacKinley (1990), Melino, Turnbull (1990, 1995), Mishkin (1990), Müller, Dacorogna, Olsen, Pictet, Schwarz, Morgenegg (1990), Müller, Dacorogna, Dave, Pictet, Olsen,

Ward (1993), Müller, Dacorogna, Dave, Olsen, Pictet, von Weizsäcker (1995), Roell (1990), Seppi (1990), Bali (1991), Bhattacharya, Spiegel (1991), Black (1991), Bossaerts, Hillion (1991), Burnham (1991), Campbell, LaMaster, Smith, Van Boening (1991), Campbell, Lo, MacKinlay (1997), Chinn (1991), Chinn, Meese (1995), Chowdhry, Nanda (1991), Edwards (1991), Froot, Obstfeld (1991), Froot, Rogoff (1995), Froot, Ramadorai (2002, 2005), Froot, Donohue (2004), Georg, Kaul, Nimalendran (1991), Grabbe (1991), Harvey, Huang (1991), Khonry (editor) (1991), Kim, Verrecchia (1991, 1994, 1997), Klein (1991), Klein, Rosengren (1991), Lease, Masulis, Page (1991), LeBaron (1991), Lee Ch M C, Ready (1991), Messe, Rose (1991), Subrahmanyam (1991), Spiegel, Subrahmanyam (1992, 1995), Williamson (1991), Bekaert, Hodrick (1992), Choi, Elyasiani, Kopecky (1992), Choi, Elyasiani (1997), Curcio, Goodhart (1992), Curcio, Goodhart, Guillaume, Payne (1997), De Grauwe, Decupere (1992), De Grauwe, Grimaldi (2006a, b), Edison (1992, 1993, 2003), Edison, Liang (1999), Flood (1992, 1994), Flood, Rose (1995), Flood, Huisman, Koedijk, Mahieu (1996, 1998), Gosh (1992), Guillaume, Dacorogna, Dave, Müller, Olsen, Hamon, Jacquillat (1992), Guillaume, Pictet, Dacorogna (1995), Guillaume, Dacorogna, Dave, Müller, Olsen, Pictet (1997), Hansen (1992), Holden, Subrahmanyam (1992), Neal (1992), Pesaran, Samiei (1992), Rhee, Chang, Svensson (1992, 1993), Bertola, Svensson (1993), Rose, Svensson (1994), Taylor S J (1992), Zhou (1992a, b, 1997), Bank for International Settlements (1993, 1999a, b, 2001, 2002, 2004 2005, 2007, 2010), Bertola, Svensson (1993), Biais (1993), Chan, Weinstein (1993), Cheung (1993), Cheung, Ng (1996), Cheung, Chinn (1998, 2001), Cheung, Wong (1999, 2000), Cheung, Chinn, Marsh (2004), Cheung, Chinn, Pascual (2004, 2005), Dacorogna, Müller, Nagrel, Olsen, Pictet (1993), Dacorogna, Müller, Pictet, de Vries (1995), Dominguez, Frankel (1993), Dominguez (1998, 2006), Dominguez, Panthaki (2006), Ederington, Lee J (1993), Edin, Vredin (1993),

Goldstein, Folkerts-Landau, Garber, Rojas-Suarez, Spencer (1993), Griffiths, White (1993), Grimes (1993), Harris, Raviv (1993), Klein (1993), Levich, Thomas (1993), Matsuyama, Kiyotaki, Matsui (1993), Romer (1993), Schmidt, Iversen, Treske (1993), Schmidt, Iversen (1993), Schmidt, Oesterhelweg, Treske (1996), Wolinsky (1990), Ammer, Brunner (1994), Andrew, Broadbent (1994), Bakker, Boot, Sleijpen, Vanthoor (editors) (1994), Bartov, Bodnar (1994, 1995), Berry, Howe (1994), Bessembinder (1994), Ball, Roma (1994), Brousseau, Czarnecki (1994), De Jong (1994), De Jong, Nijman, Röell (1995, 1996), De Jong, Mahieu, Schotman (1998), De Jong, Ligterink, Macrae (2006), De Jong, Verschoor, Zwinkels (2010), Degryse, De Jong, van Kervel (2011), Dini (1994), Fialkowski, Petersen (1994), Glass (1994), Grünbichler, Longstaff, Schwartz (1994), Hansch, Naik, Viswanathan (1994), Hirschleifer, Subrahmanyam, Titman (1994), Hogan, Melvin (1994), Jones, Kaul, Lipson (1994), Jones, Lipson (1999), Kraus, Smith (1994), Massib, Phelps (1994), Mendelson, Peake (1994), Naidu, Rozeff (1994), Nieuwland, Verschoor, Wolff (1994), Pictet, Dacorogna, Müller, De Vries (1994), Sharpe (1994), Silber (1994), Slezak (1994), Szpiro (1994), Yadav, Pope, Paudyal (1994), Walsh (1994), Wei (1994), Watanabe (1992), Watanabe, Harada (2004), Watanabe, Yabu (2007), Almekinders (1995), Chiang, Jiang (1995), Dumas, Solnik (1995), Ederington, Lee J (1995), Evertsz (1995), Faruqee (1995), Frino, McCorry (1995), Frino, McInish, Toner (1998), Ghysels, Jasiak (1995), Grossman, Rogoff (1995), Havrilesky (1995), Hong, Wang (1995), Isard (1995), Kandel, Pearson (1995), Lewis (1995), Lin, Sanger, Booth (1995), Mantegna, Stanley (1995), Mark (1995, 2001, 2009), Mark, Wu (1998), Obstfeld, Rogoff (1995, 1998), Osler (1995, 1998, 2000, 2003, 2005, 2006, 2008, 2009, 2012), Carlson, Osler (1999, 2005), Kevin, Osler (1999), Osler, Vandrovych (2009), Osler, Yusim (2009), Osler, Mende, Menkhoff (2011), Osler, Savaser (2011), Dahl, Carlson, Osler (2011), Peiers (1995), Prasad,

Rajan (1995), Schnidrig, Würtz (1995), Schwartz (editor) (1995), Shyy, Lee J (1995), Shyy, Vijayraghavan, Scott-Quinn (1996), Vivex (1995), Zaheer A, Zaheer S(1995), Bonser – Neal, Tanner (1996), Claassen (1996), Danker, Haas, Henderson, Symanski, Tryon (1996), Dukas, Fatemi, Tavakkol (1996), Dwyer, Locke, Yu (1996), Easley, Kiefer, O'Hara, Paperman (1996, 1997a, b), Easley, O'Hara, Srinivas (1998), Flemming, Ostdiek, Whaley (1996), Gagnon (1996), Ghashghaie, Breymann, Peinke, Talkner, Dodge (1996), Hsieh, Kleidon (1996), Ingersoll (1996), Kaminsky, Lewis (1996), LeBaron (1996), MacDonald, Marsh (1996), Madrigal (1996), Mosekilde (1996), Pirrong (1996), Rosenberg (1996), Tsang (1996, 1998, 1999a, b), Tsang, Sin, Cheng (1999), Tsang, Yue (2002), Vermeiren, Ková (1996), Balke, Fomby (1997), Balke, Wohar (1998), Bhattacharya, Weller (1997), Campbell, Lo, MacKinlay (1997), Campbell, Viceira (2002), Chamberlain, Howe, Popper (1997), Clarida, Taylor M P (1997), Clarida, Sarno, Taylor M P, Valente (2003), Copejans, Domowitz (1997), DeGennaro, Shrieves (1997), Dewachter (1997, 2001), Dewachter, Lyrio (2005), Embrechts, Klueppelberg, Mikosch (1997), Evans (1997, 2001, 2002, 2005, 2009, 2010, 2011), Evans, Lyons (1999, 2001a, b, c, 2002a, b, c, d, 2003, 2004a, b, 2005a, b, c, d, 2006, 2007, 2008, 2009), Cao, Evans, Lyons (2003), Evans, Hnatkovska (2005), Fleming, Remolona (1997, 1999), Fleming (2002, 2003), Franke, Hess (1997), Goldberg, Tenorio (1997), Gosh, Ostry, Gulde, Wolf (1997), Harris, Schultz (1997), Hartmann (1997, 1998a, b, 1999), Hung (1997), Kirilenko (1997), Lamoureux, Schnitzlein (1997), Madhavan, Smidt (1991, 1993), Leach, Madhavan (1993), Keim, Madhavan (1996), Madhavan, Cheng (1997), Madhavan, Richardson, Roomans (1997), Madhavan, Sofianos (1997), Madhavan (2000a, b, c), Martens (1997), Montiel (1997), Pagano, Roell (1997), Peiers (1997), Reiss, Werner (1997), Sweeney (1997, 2000), Szakmary, Mathur (1997), Vogler (1997), Wei, Kim (1997), Werner (1997), Wren-Lewis (1997), Abhyankar (1998), Abrams, Beato (1998),

Anthony, MacDonald (1998, 1999), Bjønnes, Rime (1998, 2001, 2005), Bjønnes, Rime, Solheim (2005), Bjønnes, Osler, Rime (2011), Blennerhasset, Bowman (1998), Bodnar, Hayt, Marston (1998), Caramazza, Aziz (1998), Chang, Taylor S (1998), Choi, Hiraki, Takezawa (1998), Chow, Chen (1998), Clark, Macdonald (1998), Covrig, Melvin (1998), Eddelbuttel, McCurdy (1998), Edison (1998), Fleming, Kirby, Ostdiek (1998), Garfinkel, Nimalendran (1998), George (1998), Hansch, Naik, Viswanathan (1998), Hau (1998), Hau, Killeen, Moore (2000, 2002a, b), Hau, Rey (2002, 2003), He, Ng (1998), Helpman, Sadka (1998), Hong Kong Monetary Authority (1998), Isard, Faruqee (1998), Isard, Faruqee, Kincaid, Fetherston (2001), Kanas (1998), Lee R (1998), Litterman, Winkelmann (1998), Lui, Mole (1998), Menkhoff (1998, 2010), Gehrig, Menkhoff (2000, 2004), Mende, Menkhoff (2003, 2006), Menkhoff, Taylor M P (2007), Frömmel, Mende, Menkhoff (2008), Menkhoff, Schmeling (2008, 2010), Miller, Reuer (1998), Miville, DiMillo (1998), Nagayasu (1998), Neely (1998, 2000a, b, 2004, 2005), Pesaran, Hasem, Smith (1998), Portes, Rey (1998), Rey (2001), Reiss, Werner (1998), Sarkar, Tozzi (1998), Viswanathan, Wang (1998, 2000), Vitale (1998, 1999, 2000, 2003, 2004, 2006), Yao (1998), Alberola, Cervero, Lopez, Ubide (1999), Bos, Fetherstone (1999), Carrera (1999), Chaboud, LeBaron (1999, 2001), Chaboud, Humpage (2005), Chaboud, Chernenko, Wright (2008), Chaboud, Chiquoine, Hjalmarsson, Vega (2009), Chaboud, Chiquoine, Hjalmarsson, Loretan (2009), Fiess, MacDonald (1999, 2002), Fleming, Lopez (1999), Freihube, Kehr, Krahnen, Theissen (1999), Grammig, Schiereck, Theissen (1999), Isard, Razin, Rose (1999), Jeanne, Rose (1999), Kandel, Marx (1999), LeBaron (1999), Marks (1999), Macey, O'Hara (1999), Naik, Neuberger, Viswanathan (1999), Naik, Yadav (1999), Payne (1999, 2003), Payne, Vitale (2003), Moore, Payne (2011), Love, Payne (2004, 2008), Rigobon (1999), Saar (1999), Scalia, Vacca (1999), Scalia (2004, 2008), Shapiro, Varian (1999), Theissen (1999), Vayanos (1999, 2001), Wang (1999), Aliber, Chowdhry, Yan (2000), Ausloos (2000), Baillie, Humpage, Osterberg (2000), Carlson, Osler (2000), Carlson (2002), Ebrahim (2000), Eichengreen, Mathieson (2000), Greenspan (2000), Hüfner (2000), Franke, Hess (2000), Fujiwara (2000), Kanas (2000), Kaul, Mehrotra, Morck (2000), Kim, Kortian, Sheen (2000), Kim, Sheen (2002), Kim (2003), Lane, Milesi-Ferretti (2000), Lo (2000), Lee C, Swaminathan (2000), Ma, Kanas (2000), Ma, Tsang, Yiu, Wai-Yip Alex Ho (2010), Martin (2000), Martin, Mauer (2003, 2005), McCallum (2000), Melvin M, Yin (2000), Melvin M, Melvin B P (2003), Melvin M, Taylor M P (2009), Naranjo, Nimalendran (2000), Ng (2000), Ramaswamy, Samiei (2000), Rime (2000, 2001, 2003), Akram, Rime, Sarno (2005), Rime, Sarno, Sojli (2006, 2007, 2010), Schwartz (2000), US General Accounting Office (2000), Allayannis, Ofek (2001), Anderson, Vahid (2001), Brandt, Edelen, Kavajecz (2001), Brown (2001), Cai, Cheung, Lee R S K, Melvin (2001), Claessens, Forbes (2001), Clark, McCraken (2001), Collins, Rodrik (2001), Corsetti, Pesenti, Roubini (2001), Coval, Shumway (2001), Croushore, Stark (2001), Dacorogna, Gencay, Mueller, Olsen, Pictet (2001), D'Souza (2001), Duarte, Stockman (2001), Fischer (2001), Galati (2001), Griffin, Stulz (2001), Guembel, Sussman (2001), Hong (2001), Lane (2001), Montgomery, Popper (2001), Moore, Roche (2001, 2002), Rey (2001), Sato, Hawkins (2001), Sinn, Westermann (2001), Tse, Zabotina (2001), Williamson (2001), Yamaguchi (2001), Aguiar (2002), Beine (2002), Cavallo, Perri, Roubini, Kisselev (2002), Chari, Kehoe, McGrattan (2002), Chari (2006), Chordia, Roll, Subrahmanyam (2002), Covrig, Melvin (2002), Daníelsson, Payne (2002, 2011), Danielsson, Payne, Luo (2002), Daníelsson, Love (2006), Deutsche Bundesbank (2002), Doyne, Joshi (2002), Fatum, Hutchison (2002, 2003, 2006), Fatum, King (2005), King, Sarno, Sojli (2010), King, Rime (2010), King, Mallo (2010), King, Osler, Rime (2011a, b, 2012), Kantelhardt, Zschiegner, Koscielny-Bunde, Havlin, Bunde, Stanley (2002),

Galati (2002), Girardin, Horsewood (2002), Huang, Cai, Jeanne, Rose (2002), Kaul, Mehrotra (2002), Obadan (2002), Ryan, Worthington (2002), Abreu, Brunnermeier (2003), Aliber, Chowdry, Yan (2003), Bacchetta, van Wincoop (2003, 2004, 2006), Bergsten, Williamson (2003), Bodnar, Wong (2003), Burstein, Neves, Rebelo (2003), Carpenter, Wang (2003), Derviz (2003), Dominguez (2003), Dominguez, Panthaki (2006), Doukas, Hall, Lang (2003), Faust, Rogers, Wright (2003), Gordon (2003), Humpage (2003), Koutmos, Martin (2003), Laurenceson, Chai (2003), Mathisen (2003), Okunev, White (2003), Peng, Shu, Chow (2003), Rogers, Siklos (2003), Spiegel (2003), Westerhoff (2003), Wright (2003), Aitken, Frino, Hill, Jarnecic (2004), Anwar (2004), Bartram (2004), Bartram, Bodnar (2004), Bartram, Brown, Minton (2005), Bartram, Karolyi (2006), Bhanumurthy (2004), Brandt, Kavajecz (2004), Breedon, Vitale (2004), Cashin, Cespedes, Sahay (2004), Choi, Baek (2004), De Wet, Gebreselasie (2004), Dunne, Hau, Moore (2004), Fratzscher (2004), Hahm (2004), Hui, Neely, Higbee (2004, 2007), Hui, Yeung, Fung, Lo (2007), Hui, Fong (2007), Hui, Genberg, Chung (2009), Kim, Yoon (2004), Nagayasu (2004), National Bank of Poland (2004, 2007), Reinhart, Rogoff (2004), Rigobon, Sack (2004), Simatele (2004), Akram, Rime, Sarno (2005), Ates, Wang (2005), Bauwens, Omrane, Giot (2005), Campa, Goldberg (2005, 2006a, b), Chui, Gerlach, Yu (2005), DeGrauwe (editor) (2005), Dueker, Neely (2005), Eichengreen (2005), El-Shagi, Rübel (editors) (2005), Fung, Lien, Tse Y M, Tse Y K (2005), Hau, Rey (2005), Inoue, Kilian (2005), Marsh, O'Rourke (2005), Newsome (2006), Vaubel (2005), Yu, Fung, Hongyi (2005), Alexander, Barbosa (2006), Bacchetta, van Wincoop (2006), Bayoumi, Lee J, Jayanthi (2006), Boyen, Van Norden (2006), Cai, Howorka, Wongswan (2006, 2008), Cao, Evans, Lyons (2006), Carlson, Lo (2006), Charlebois, Sapp (2006), Chu, Mo, Wong, Lim (2006), Gilbert, Rijken (2006), Jeon, Oh, Yang (2006), Escribano, Pascual (2006), Kaul, Sapp (2006), Killeen, Lyons,

Moore (2006), Kim, Lee J W, Shin (2006), Kočenda, Valachy (2006), Kočenda, Kutan, Yigit (2008), Kočenda, Poghosyan (2009), LeBaron (2006), Mende (2006), Mende, Menkhoff (2006), Müller, Verschoor (2006), Norouzzadeh, Rahmani (2006), Pelham (2006), Rodrik (2006), Sager, Taylor M P(2006), Starks, Wei (2006), Tabak, Cajueiro (2006), Taylor A, Farstrup (2006) Taylor J B (2006), Tesfatsion, Judd (editors) (2006), Wong (2006), Adebiyi (2007), Barker (2007), Bhansali (2007), Broz, Frieden, Weymouth (2007), Burnside, Eichenbaum, Rebelo (2007, 2009), Burnside (2012), Canto, Kräussl (2007), Chi, Tripe, Young (2007), Christodoulou, O'Connor (2007), Dreher, Vaubel (2007), DuCharme (2007), Egstrup, Fischer (2007), Fleming, Mizrach (2007), Fung, Yu (2007), Genberg, He, Leung (2007a, b), Genberg, Hui (2009), Hong Kong Monetary Authority (2007), Jiang, Ma, Cai (2007), Leung, Ng (2007, 2008), Mitchell, Pedersen, Pulvino (2007), Pasquariello (2007), Sahminan (2007), Scarlat, Stan, Cristescu (2007), Van Wincoop, Tille (2007), Wong J, Wong E, Fong, Choi (2007), Wong E, Wong J, Leung (2008), Yu, Fung, Tam (2007), Acemoglu, Rogoff, Woodford (editors) (2008), Baglioni, Monticini (2008), Barndorff-Nielsen, Hansen, Lunde, Shephard (2008), Bartram (2008), Beaupain, Durré (2008), Berger, Chaboud, Chernenko, Howorka, Wright (2008), Brunnermeier, Nagel, Pedersen (2008), Burnside (2008), Burnside, Eichenbaum, Kleshchelski, Rebelo, Hall L, Hall H (2008), Chinn, Moore (2008, 2011), Gagnon, Chaboud (2008), Lam, Fung, Yu (2008), Lien (2008), Lindley (2008), Liu, Tsang (2008), Liu, Fung, Tse (2008), Lo, Sapp (2008, 2010), Ramadorai (2008), Sebastião (2008), Terada, Higashio, Iwasaki (2008), Adrian, Etula, Shin (2009), Bacchetta, Mertens, Van Wincoop (2009), Baba, Packer (2009), Brunnermeier, Nagel, Pedersen (2009), Brunnermeier, Crockett, Goodhart, Persaud, Shin (2009), Bubák, Zikes (2009), Bubák, Kočenda, Žikeš (2010), De Zwart, Markwat, Swinkels, van Dijk (2009), Ding (2009), Gallardo, Heath (2009), Gençay, Gradojevic (2009), Jiang, Zhou (2009), Hattori, Shin (2009), He, Zhang, Wang (2009), Heath, Whitelaw (2011), McGuire, von Peter (2009), Meyers (2009), Müller, Verschoor (2009), Nolte I, Nolte S (2009, 2011), Serban (2009), Simwaka, Mkandawire (2009), Breedon, Vitale (2010), Breedon, Rime, Vitale (2011), Dunne, Hau, Moore (2010), Fukuda, Kon (2010), Liu, Qian, Lu (2010), Maurer, Schäfer (2010), Nightingale, Ossolinski, Zurawski (2010), Pasquariello (2010), Yiu, Ho, Ma, Tsang (2010), Diamond (2011), Durčáková (2011), Heimer, Simon (2011), Marzo, Zagaglia (2011), Moore, Payne (2011), Plantin, Shin (2011), Rafferty (2011), Wang, Wu, Pan (2011), Banti, Phylaktis, Sarno (2012), James, Marsh, Sarno (editors) (2012), Mancini, Ranaldo, Wrampelmeyer (2012), Sheng (2012a, b, 2014), Trenca, Plesoianu, Căpusan (2012), Wang, Yu, Suo (2012), Lassmann (2013), Ledenyov V O, Ledenyov D O (2016s).

The **intellectual property** as an investment product was researched in Plant (1934a, b), Callmann (1947), Penrose (1951), Prager (1952), Arrow (1962), Scherer (1965, 1984), Baxter (1966), Hurt, Schuchman (1966), Barzel (1968), Nordhaus (1969), Bowman (1973, 1977), Taylor C, Silberston (1973), Roffe (1974), Adelman (1977), Loury (1979), Cheung (1982), Gilbert, Newbery (1982, 1984), Gilbert, Shapiro (1990), Mackaay (1982, 1985a, b, 1986, 1989, 1990a, b, 1991a, b, c, 1992a, b, 1994), Sieghart (1982), Ashford, Heaton (1983), Ashford et al (1985), Baird (1983), Beck (1983), Fudenberg, Gilbert, Stiglitz, Tirole (1983), Wright (1983), Mossinghoff (1984), Adelstein, Peretz (1985), Cave (1985), David (1985, 1993), Farrell, Saloner (1985, 1989), Gallini, Winter (1985), Gallini (1992), Judd (1985), Lehmann (1985, 1989, 1990), Pendleton (1985), Samuelson (1985), Department of Trade and Industry (1986), Hay (1986), Mansfield (1986), Priest (1986), Evenson, Putnam (1987), Menell (1987, 1989, 1994, 1998), Rozek (1987), Sirilli (1987), Tullock (1987), Feinberg (1988), Feinberg, Rousslang (1990), Hughes (1988a, b), Merges (1988, 1992, 1994a, b, 1995a, b, 1996a, b), Merges, Menell, Lemley, Jorde (1997), Von Hippel (1988),

Walker, Bloomfield (editors) (1988), Beier, Shricker (1989), Besen, Kirby (1989a, b), Besen, Raskind (1991), Besen, Kirby, Salop (1992), Braga (1989), Centner (1989), Centner, White (1989), Davis (1989), Krauss (1989), Palmer (1989, 1990), Brenner (1990), Chin, Grossman (1990), Easterbrook (1990), Gilbert, Shapiro (1990), Klemperer (1990), Rushing, Brown (editors) (1990), Caves, Whinston, Hurwitz (1991), Coombe (1991), Heald (1991), Scotchmer (1991), Segerstrom (1991), Teijl, Holzhauer (1991), Deardorff (1992), Gallini (1992), Gordon (1992a, b, 1993), Grady, Alexander (1992), Quaedvlieg (1992), Aoki (1993-1994, 1996a, b), Becker (1993), Brennan (1993), Carter (1993), Chou Chien-Fu, Shy (1993), Helpman (1993), Kay (1993), Lanjouw (1993), Lanjouw, Pakes, Putnam (1998), Lanjouw, Schankerman (2001a, b), Nelson (1993, 1994), Barlow (1994), Dam (1994, 1995), Japan Institute of Intellectual Property (1994), Karjala (1994), Lerner (1994, 1995, 2002), Chang (1995), Lemley (1995), Gould, Gruben (1996), Matutes, Regibeau, Rockett (1996), Brousseau, Bessy (September 19 - 21 1997), Ginarte, Park (1997), Grindley, Teece (1997), Park, Ginarte (1997), Besen (1998), Maskus (1998), Schankerman (1998), Templeman (1998), Reilly, Schweihs (1999), Reilly (2013), Gallini, Scotchmer (2001), Hall, Zeidonis (2001), McCalman (2001), Sakakibara, Branstetter (2001), Scotchmer (2001), Shapiro (2001), Boldrin, Levine (2002, 2004a, b, 2005, 2006, February 6 2007), Grossman, Lai (2002, 2004), Lasinski (2002), Maskus (2000a, b), Deli Yang (2003), Menell (2003), Anson, Suchy, Ahya (2005), Anson, Noble, Samala (2014), Blair, Cotter (June 2005), Ramello (2005), Smith, Parr (2005), Andersen (editor) (January 1 2006), Hisamitsu Arai (2006), Kamiyama, Sheehan, Martínez (2006), Kanwar (2006), Kumar (2006), Lakdawalla, Philipson, Wang (October 2006), Moerman, Laan (2006), Aoki, Schiff (2007), Bittelmeyer (2007), Holland, III, Reed, Lee S H, Kimmel, Peterson (2007), Holland, Benedikt (2014), Malackowski, Cardoza, Gray, Conroy (2007), Menell, Scotchmer (2007), Parr (2007), Siegel, Wright (2007), Van Caenegem (May 2007), Ruder (2008), Kite (2009), Blakeney, Ullrich, Stauder, Llewelyn, MacQueen, Jacob, Laddie, Chisum, Benyamini, Straus, Llewellyn, McCarthy, Dworkin, Soltysinski, Lahore, Dufty, Ricketson, Ginsburg, Christie, Goldstein, Tapper, Kamina (December 2010), Flanagan, Montagnani (editors) (January 2010), Baker, Pak Yee Lee, Mezzetti (2011), Bryer, Lebson, Asbell (2011), Cottier, Veron (2011), McCoy, Barton, McDermott (2011), Palfrey (October 2011), Bouchoux (2012), George (2012), Rüther (2012), Boldrin, Levine (2013), Frey (2013), Howe, Griffiths, Sherman, Pottage, Gangjee, Bently, Hudson, Dreier, Breakey, Balganesh, Carrier, Burrell, Hudson, Lametti, Dussollier (September 2013), Buchanan, Wilson (2014), Fawcett (2014), Gervais (editor) (2014), Guellec, Ménière (2014), Sople (2014), Schmitt (2016), Searle, Brassell (August 2 2016).

Next, let us summarize the most interesting research contributions on the **investment vehicles** for the financial capital investment in the capital markets made by distinguished scientists.

The **investment bank** as an investment vehicle was researched in in Howell (1953), O'Donnell (1957), Pontecorvo (1958), Mandelker, Raviv (1977), Beatty, Ritter (1986), Smith (1986), Keeley, Pozdena (June 19 1987), McDonough (1987), Walter, Smith (1989), Carter, Dark (1992), Chemmanur, Fulghieri (1994), Sussman (1994), Clark (April 1995), Brockman (1996), Ferguson (1996), Grant (1999), Parkan, Ming-Lu Wu (1999), Fleuriet (2000), Anand, Galetovic (July 01 2001), Smith (2001), Perez (2002), Benveniste, Ljungqvist, Wilhelm, Xiaoyun Yu (2003), Ritter (2003), Sirri (2004), Abor (Summer 2005), Morrison, Wilhelm (2007, November 15 2008a, b) Brambilla, Piluso (2008), Chindris-Vasioiu (2008), Jaffee Dwight, Perlow (2008), Bodnaruk, Massa, Simonov (2009), Bao, Edmans (2011), Blundell-Wignall (2011), Rubinton (June 17 2011), Ziman (2011), Berzins, Liu, Trzcinka (2013), Clare, Gulamhussen, Pinheiro (2013),

Erkan Celik, Hacioglu, Dincer (2013), Bonin (2014), Grullon, Underwood, Weston (2014), Balluck (2015), Corovei (2015), Mishra (2016).

The **investment fund** in the form of the **hedge fund** as an investment vehicle was researched in Brown, Harlow, Starks (1996), Brown, Goetzmann, Park (1997), Brown, Goetzmann, Ibbotson (1998), Brown, Goetzmann, Ibbotson (1999), Brown (2001), Brown, Goetzmann, Park (2000, 2001), Brown, Goetzmann (2001), Brown, Goetzmann (2003), Brown, Fraser, Liang (2008), Brown, Goetzmann, Liang, Schwarz (2008), Brown, Goetzmann, Liang, Schwarz (2010), Fung, Hsieh (1997a, b, 1999a, b, 2000a, b, 2001, 2002a, b, c, 2003, 2004a, b, 2006a, b, 2007), Fung, Hsieh, Naik, Ramadorai (2006, 2008), Ackermann, Ravenscraft (1998), Ackermann, McEnally, Ravenscraft (1999), Eichengreen, Mathieson, Chadha, Jansen, Kodres, Sharma (1998), Mathieson, Chadha, Jansen, Kodres, Eichengreen, Sharma (1998), Edwards (1999, 2000a, b, 2003, 2004a, b, 2006), Edwards, Caglayan (2001), Edwards, Gaon (2003), Liang (1999, 2000, 2003, 2004), US President's Working Group on Financial Markets (1999), Stonham (1999a, b), Tatsaronis (2000), Agarwal, Naik (2000, 2004), Aggarwal, Jorion (2010), Asness, Krail, Liew (2001), Braga (2001), Brealy, Kaplanis (2001), Brooks, Kat (2001), Amin, Kat (2001, 2003a, b), Kat (2003), Kat, Menexe (2003), Kat, Palaro (2005, 2006), Kat (2007, 2010), Capocci, Hübner (2001), Capocci, Corhay, Hübner (2003), Capocci, Hübner (2004), Kramer (2001), Goetzmann, Ingersoll, Ross (2001), Anson (2002), Favre, Galeano (2002), Gimbel, Gupta, Pines (2002), Ineichen (2002), Kao (2002), Locho (2002), Weismann (2002), Schneeweis, Kazemi, Martin (2002, 2003), Bacmann, Scholz (2003), Bares, Gibson, Gyger (2003), Geman, Kharoubi (2003), Gregoriou (2003), Gregoriou, Gueyie (2003), Gregoriou, Sedzro, Zhu (2005), Gregoriou, Kooli, Rouah (2008), Goetzmann, Ingersoll, Ross (2003), Gulko (2003), Ennis, Sebastian (2003), Popova I, Morton, Popova E (2003, 2006), Morton, Popova E, Popova I (2006), Amenc, El Bied, Martellini (2003),

Amenc, Géhin, Martellini, Meyfredi (2007), Amenc, Géhin, Martellini, Meyfredi, Ziemann (2008), Bacmann, Gawron (2004), Baquero, ter Horst, Verbeek (2004a, b), ter Horst, Verbeek (2007), Boido, Riente (2004), Brunnermeier, Nagel (2004), Feiger, Botteron (2004), Hedges (2004), Posthuma, van der Sluis (2004), Getmansky, Lo, Mei (2004), Getmansky, Lo, Makarov (2004), Lhabitant (2004), Nguyen-Thi-Thanh Huyen (2004, 2006), Huber, Kaiser (2004), Al-Sharkas (2005), Alexander, Dimitriu (2005), Carretta, Mattarocci (2005), Chan, Getmansky, Haas, Lo (2005, 2006), Chan, Getmansky, Lo, Haas (2007), Cremers, Kritzman, Page (2005), Danielsson, Taylor, Zigrand (2005), Do, Faff, Wickramanayake (2005), Eling, Schuhmacher (2005), Garbaravičius (2005), Garbaravičius, Dierick (2005), Gilroy, Lukas (2005), Gupta, Lang (2005), Kaiser, Kisling (2005), Malkiel, Saha (2005), Hodder, Jackwerth (2005), Jaeger, Wagner (2005), Azman-Saini (2006), Baba, Goko (2006), Boyson, Stahel, Stulz (2006, 2008), Ding, Shawky (2006), Izzo (2006), Jackwerth, Hodder (2006), Jagannathan, Malakhov, Novikov (2006), Heidorn, Hoppe, Kaiser (2006a, b), Sadka (2006), Adrian (2007), Becker, Clifton (2007), Goltz, Martellini, Vaissié (2007), Kambhu, Schuermann, Stiroh (2007), King, Maier (2007), Kosowski, Naik, Teo (2007), Li Sh, Linton O (2007), Hakamada, Takahashi, Yamamoto (2007), Hasanhodzic, Lo (2007), Papademos (2007), Smedts K, Smedts J (2007), Stulz (2007), Weber (2007), Carlson, Steinman (2008), Billio, Getmansky, Pelizzon (2008), de los Rios, Garcia (2008), Lo (2008), McGuire, Tsatsaronis (2008), Takahashi, Yamamoto (2008), Jackwerth, Kolokolova, Hodder (2008), Gray (2008), Gray, Kern (2008), Gupta, Szado, Spurgin (2008), Kazemi, Tu, Li (2008), Nahum, Aldrich (2008), Roncalli, Teiletche (2008), Roncalli, Weisang (2008), Hedge Fund Working Group & Hedge Fund Standards Board (2008), Bollen, Pool (2009), Brophy, Ouimet, Sialm (2009), Füss, Kaiser, Strittmatter (2009), Heidorn, Kaiser, Roder (2009), Jaeger (2009), Khanniche (2009), Minsky, Obradovic, Tang, Thapar (2009), Mitra (2009), Xiong, Idzorek, Chen, Ibbotson (2009), Gibson, Wang (2010), Heidorn, Kaiser, Voinea (2010), Maillard, Roncalli, Teiletche (2010), Ramadorai (2010), Titman (2010), Sadka (2010), Wallerstein, Tuchschmid, Zaker (2010), Ang, Gorovyy, van Inwegen (2011), Cao, Ogden, Tiu (2011, 2012), Freed, McMillan (2011), Eychenne, Martinetti, Roncalli (2011), Piluso, Amerise (2011), Chakravarty, Deb (2012), Chen, Tindall (2012), Roncalli, Weisang (2012), Bruder, Roncalli (2012), Hassine, Roncalli (2013), Agarwal, Vikram, Sugata (2013).

The **investment fund** in the form of the **pension fund** as an investment vehicle was researched in MacIntosh (1976), Sharpe (June 1976), Treynor (May 1977), Winklevass (1977), Bulow (1979, August 1982), Bulow, Morck, Summers (1987), Bodie (Fall 1980, October 1988), Bodie, Light, Morck, Taggart (1987), Bodie, Shoven, Wise (editors) (1987), Bodie, Kane, Marcus (1989), Black (September—October 1980), Black, Dewhurst (Summer 1981), Feldstein, Seligman (1981), Feldstein (1982), Feldstein, Morck (1983), Frankfurter, Hill (1981), Tepper (March 1981), Harrison, Sharpe (1983), Friedman (1983), Kotlikoff, Smith (1983), Warshawsky (1987, November 1988, 1989, 1990), Bernheim, Shoven (1985), Ang, Tsong-Yue Lai (1988), Bernheim, Shoven (1988), Coggin, Fabozzi, Rahman (1993), Haberman, Joo-Ho Sung (1994), Mitchell, Smith (1994), Davis (1995), Dyson, Exley (1995), Wahal (1996), Brown, Draper, McKenzie (March 1997), Cairns, Parker (1997), Blake (1998), Blake, Lehmann, Timmermann (1999), Cangiano, Cottarelli, Cubeddu (1998), Hemming (1998), Mitchell (1998), Clark (July 13 2000, March 27 2003, 2008), Sinn (2000), Srinivas, Whitehouse, Yermo (2000), Whitehouse (2000), Chapman, Gordon, Speed (2001), Head, Adkins, Cairns, Corvesor, Cule, Exley, Johnson, Spain, Wise (2001), Thomas, Tonks (2001), Tonks (2002), Besley, Prat (2003), Kakabadse N, Kakabadse A, Kouzmin (2003), Bateman, Mitchell (2004), Bergstresser, Desai, Rauh (2004), Eaton, Nofsinger (2004),

Munnell, Sundén (2004), Owadally, Haberman (2004), Sundén (2004), Cowling, Gordon, Speed (2005), Diamond (2005), Dobronogov, Murthi (2005), Franzoni, Marin (January 25 2005), McCarthy, Neuberger (2005), Greco (2006), Miao Jerry, Wang (2006), Sweeting (2006), Perotti, Schwienbacher (2007), Antolin (2008), Evans, Orszag, Piggott (editors) (January 1 2008), Impavido (2008), Adam, Moutos (2009), Inderst (January 1 2009), Kleinow (2011), Munnell, Aubry, Quinby (2011), Bovenberg, Mehlkopf (2014), Rossi, Blake, Timmermann, Tonks, Wermers (2015), Garon (2016), Dahlquist, Setty, Vestman (2016), Mitsel, Rekundal (2016).

The **investment fund** in the form of the **mutual fund** as an investment vehicle was researched in Sharpe (January 1966), Treynor , Mazuy (1966, July-August 1996), Allerdice, Farrar (1967), Jensen (June 1968), Arditti (1971), Fama (June 1972), Maurice Joy, Burr Porter (1974), Scott, Klemkosky (1975), Fabozzi, Francis (March 1978, December 1979, 1980), Kon, Jen (May 1978, April 1979), Kon (1983), Kim (1978), Gatto, Geske, Litzenberger, Sosin (1980), Miller, Gressis (1980), Eckardt, Bagamery (1983), Cowen, Kroszner (1990), Cook, Hebner (1992), Grinblatt, Titman (1992, 1993), Grinblatt, Titman, Wermers (1995), Hendricks, Patel, Zeckhauser (1993), Mack (November 1993), Wohlever (1993), Dickson, Shoven (1995), Dickson, Shoven, Sialm (2000), Malkiel (1995), Neely (1995), Chordia (1996), Droms, Walker (1996), Elton, Gruber, Blake (1996), Livingston, O'Neal (1996, 1998), Livingston, Lei Zhou (2015), Lockwood (1996), Brown, Goetzmann (1997), Carhart (1997), Carhart, Carpenter, Lynch, Musto (2002), Cortez, Armada (1997), Cortez, Paxson, Armada (1999), Daniel, Grinblatt, Titman, Wermers (1997), Goetzmann, Peles (1997), Goetzmann, Brown (1998), Goetzmann, Ivkovic, Rouwenhorst (2001), Malhotra, McLeod (1997), Tufano, Sevick (1997), Barclay, Pearson, Weisbach (1998), Blake, Timmermann (1998, 2003), Fortune (1998), Sirri, Tufano (1998), Bogle (1999), Chevalier, Ellison (1999), James, Ferrier,

Smalhout, Vittas (1999), Khorana, Servaes (1999), Kulldorff, Khanna (1999), Latzko (1999), Blake, Morey (2000), Chen, Jegadeesh, Wermers (2000), Kim, Shukla, Tomas (2000), Wermers (2000, 2003), Bollen (2001), Kothari (2001), Teo, Woo Sung-Jun (2001), Bergstresser, Poterba (2002), Chan, Hsiu-Lang Chen, Lakonishok (2002), Otten, Bams (2002), Pastor, Stamburgh (2002), Sengupta (2003), Berk, Green (2004), Kosowski, Timmermann, White, Wermers (April 2004), Mahoney (2004), Mamaysky, Spiegel, Zhang (2004), Nanda, Wang, Zheng (2004), Prather, Bertin, Henker (2004), Anderson, Ahmed (2005), Barber, Odean, Zheng (2005), Bollen, Busse (2005), Cuthbertson, Nitzsche, O'Sullivan (February 2005), Kacperczyk, Sialm, Zheng (2005, 2008), Nitzsche, Cuthbertson, O'Sullivan (2005), Raychaudhuri (2005), Busse, Irvine (2006), Gaspar, Massa, Matos (2006), Morey, O'Neal (2006), Alves, Mendes (2007), Bollen (2007), Cohen, Frazzini, Malloy (2007), Chen, Goldstein, Jiang (2008), Chen, Goldstein, Jiang (2009), Ferris, Yan (2007), Khorana, Tufano, Wedge (2007), Khorana, Servaes, Tufano (2008), Meschke (2007), Haslem, Baker, Smith (2008), Jans, Otten (2008), Jin-Li Hu, Tzu-Pu Chang (2008), Kacperczyk, Sialm, Zheng (2008), Kacperczyk, van Nieuwerburgh, Veldkamp (2014), Kempf, Ruenzi (2008), Kong, Tang (2008), Palmiter, Taha (2008), Sophie Xiaofei Kong, Dragon Yongjun Tang (2008), Bergstresser, Chalmers, Tufano (2009), Coggins, Beaulieu, Gendron (2009), Cremers, Driessen, Maenhout, Weinbaum (2009), Karoui, Meier (2009), Khorana, Servaes, Tufano (2009), Koehler, Mercer (2009), Kuhnen (2009), Lilian Ng, Qinghai Wang, Zaiats (2009), Qiang Bu, Lacey (2009), Sialm, Starks (2009), Evans (2010), Hsuan-Chi Chen, Lai (2010), Jin Zhang, Maringer (2010), Soongswang, Sanohdontree (2011), English, Demiralp, Dukes (2011), Adams, Mansi, Nishikawa (2012), Bhojraj, Young Jun Cho, Yehuda (2012), Cashman (2012), Gottesman, Morey (2012), Sialm, Starks (2012), Christensen (2013), Eisele, Nefedova, Parise (2013), Vidal-García (2013), Agarwal, Gay, Ling (2014), Jingjing Yang, Jing Chi, Martin Young (2014), Simutin (2014), Berk, van Binsbergen (2015), Chuprinin, Massa, Schumacher (2015), Gallaher, Kaniel, Starks (2015), Kopsch, Han-Suck Song, Wilhelmsson (2015), Meifen Qian, Bin Yu (2015), Ortiz, Ramírez, Vicente (2015), Panda, Mahapatra, Moharana (2015), Sisli-Ciamarra, Hornstein (2015), Grinblatt, Ikäheimo, Keloharju, Knüpfer (2016), Matallín-Sáez, Soler-Domínguez, Tortosa-Ausina (2016).

The **investment fund** in the form of the **venture capital fund** as an investment vehicle was researched in Rind (1981); Tyebjee, Bruno (1981), Bruno, Tyebjee (1983, 1986), Tyebjee, Bruno (1984), Chan (1983), Felda, DeNino, Salter (1983), Wilson (1983); Merkle (1984); Hutt, Thomas (1985), MacMillan, Siegel, Narasimha (1985), MacMillan, Zemann, Narasimha (1987); Beatty, Ritter (1986), Nevermann, Falk (1986), Timmons, Bygrave (1986); Block, Ornati (1987), Bygrave (1987), Bygrave, Timmons (1992), Robinson (1987), Ruhnka, Young (1987, 1991), Ruhnka, Felman, Dean (1992), Sandberg, Hofer (1987), Stedler (1987); Brophy, Guthner (1988), Clark (1988), Eisinger (1988, 1993), Florida, Kenney (1988), Florida, Smith (1993), Gladstone (1988), Harris, Raviv (1988), Hofer (1988), MacMillan, Kulow, Khoylian (1988), Sandberg, Schweiger, Schmidt (1988), Siegel R, Siegel E, MacMillan (1988), Tirole (1988); Benveniste, Spindt (1989), Gorman, Sahlman (1989), Holmstrom, Tirole (1989), Poterba (1989a, b); Amit, Glosten, Muller (1990a, b), Amit, Brander, Zott (1998), Barry, Muscarella, Peavy, Vetsuypens (1990), Barry (1994), Chan, Siegel, Thakor (1990), Hisrich, Jankowitz (1990), Sahlman (1990, 1993), Sykes (1990); Dixon (1991), Megginson, Weiss (1991); Sapienza (1992), Sapienza, Gupta (1994), Sapienza, Manigart, Vermeir (1996); Hall, Hofer (1993), Rosenstein, Bruno, Bygrave, Taylor (1993), Sahlman (1993); Admati, Pfleiderer (1994), Aghion, Tirole (1994), Anton, Yao (1994), Berglöf (1994), Bhidé (1994), Fried, Hisrich (1994), Gompers (1994, 1995, 1996, 1998, 2002, 2007),

Gompers, Lerner (1996, 1997, 1998a, b, c, 1999a, b, c, d, 2000a, b, 2001a, b), Brav, Gompers (1997), Baker, Gompers (2003), Brav, Gompers (2003), Gompers, Lerner, Scharfstein (2005), Gompers, Kovner, Lerner, Scharfstein (2006, 2008), Gompers, Kovner, Lerner (2009), Gompers, Lerner, Scharfstein, Kovner (2010), Knight (1994), Kroszner, Rajan (1994), Lerner (1994a, b, 1995a, b, 1998, 1999, 2002, 2008, 2009), Kortum, Lerner (1998, 2000), Lerner, Shane, Tsai (2003), Lerner, Schoar (2004, 2005), Lerner, Moore, Shepherd (2005), Lerner, Schoar, Wongsunwai (2007), Lerner, Sorensen, Strömberg (2009), Chen, Gompers, Kovner, Lerner (2009), Lerner, Tåg (2012), Puri (1994, 1996), Puri, Robinson (2011); Anton, Yao (1995), Elango, Fried, Hisrich, Polonchek (1995), Hart (1995), Jain, Kini (1995), Loughran, Ritter (1995), Willner (1995); Muzyka, Birley, Leleux (1996), Packer (1996), Pettway, Kaneko (1996); Amit, Brander, Zott (1997), Cai, Wei (1997), Chevalier, Ellison (1997), Gilford (1997), Karsai, Wright, Filatotchev (1997), Wright, Robbie, Ennew (1997), Manigart, Wright, Robbie, Desbrieres, De Waele (1997), Manigart, Baeyens, Hyfte (2002), Manigart, De Waele, Wright, Robbie, Desbrieres, Sapienza, Beekman (2000, 2002); Bergemann, Hege (1998), Berger, Udell (1998), Berger, Schaek (2011), Black, Gilson (1998), Cornelius , Isaksson (1998), Fried, Bruton, Hisrich (1998), Gerke (1998), Hellmann (1998, 2000, 2002, 2006, 2007), Hellmann, Puri (2000, 2002), Becker, Hellmann (2005), Hellmann, Lindsey, Puri (2004, 2008), Hyde (1998), Jacobs, Scheffler (1998), Karsai (1998, 2003, 2004), Karsai, Wright, Dudzinski, Morovic (1999), Lin, Smith (1998), Marx (1998), Marx, Strumsky, Fleming (2009), Murray, Marriott (1998), Prowse (1998), Rajan, Zingales (1998), Trester (1998), Wright, Robbie (1998), Zider (1998); Aernoudt (1999), Bliss (1999), Bygrave, Hay, Peeters (1999), Gilson (1999), Gilson, Schizer (2002, 2003), Gulati, Gargiulo (1999), Hamao, Packer, Ritter (1999), Leopold (1999), Neher (1999), Shepherd (1999), Shepherd, Zacharakis (1999), Stillman, Sunderland, Heyl, Swart

(1999); Baygan, Freudenberg (2000), Baygan (2003), Bharat, Galetovic (2000), Cumming (2000, 2001, 2008), Cumming, MacIntosh (2000, 2001, 2002a, 2002b, 2002c, 2002d, 2003a, b, 2006), Cumming, Fleming (2002), Cumming, Fleming, Schwienbacher (2005, 2006, 2009), Cumming, Fleming, Suchard (2005), Cumming, Johan (2008), Cumming, Walz (2010), Gans, Stern (2000, 2003), Gans, Hsu, Stern (2002), Jain, Kini (2000), Jeng, Wells (2000), Kaplan, Strömberg (2000, 2001, 2002, 2003, 2004, 2009), Kaplan, Schoar (2005), Kaplan, Martel, Strömberg (2007), Kaplan, Sensoy, Strömberg (2009), Kaplan, Lerner (2010), Karaömerlioğlu, Jacobsson (2000), Koski (2000), Lee (2000), Lehtonen (2000), Quindlen (2000), Schefczyk (2000), Schertler (2000); Bascha, Walz (2001), Engel (2001a, b, 2002); Francis, Hasan (2001), Fredriksen, Klofsten (2001), Hyytinen, Pajarinen (2001), Keuschnigg, Nielsen (2001, 2003a, b, 2004a, b), Keuschnigg (2003, 2004a, b), Kanniainen, Keuschnigg (2004), Kirilenko (2001), Lockett, Wright (2001), Maula, Murray (2001), Peng (2001), Seppä, Laamanen (2001), Seppä (2003), Shachmurove Y (2001, 2007a, b), Shachmurove A, Shachmurove Y (2004), Shachmurove E, Shachmurove Y (2004), Sorenson, Stuart (2001); Allen, Song (2002), Audretsch, Lehmann (2002), Bottazzi, Da Rin (2002a, b, 2004), Bottazzi, Da Rin, Giavazzi (2003), Bottazzi, Da Rin, Hellmann (2004a, b, 2008, 2009), Brander, Amit, Antweiler (2002), Brander, De Bettignies (2009), Brander, Egan, Hellmann (2008), Brander, Du, Hellmann (2010), Chesbrough (2002), Cestone (2002), Davis, Schachermayer, Tompkins (2002), Dossani, Kenney (2002), Kenney, Han, Tanaka (2002), Eisele, Habermann, Oesterle (2002), Everts (2002), Koh F C C, Koh W T H (2002), McGlue (2002), Moskowitz, Vissing-Jørgensen (2002), Shane, Cable (2002), Shane, Stuart (2002), Zook (2002); Becker, Hellman (2003), Bergemann, Hege (2003), Hege, Palomino, Schwienbacher (2009), Casamatta (2003), Casamatta, Haritchabalet (2007), Cornelli, Yosha (2003), Davila, Foster, Gupta (2003), Franzke, Grohs, Laux (2003), Gawlik,

Teczke (2003), Gilson, Schizer (2003), Hirukawa, Ueda (2003), Inderst, Muller (2003), Keilbach, Engel (2003), Leleux, Surlemont (2003), Rindermann (2003), Schertler (2003), Schmidt (2003), Schmidt, Wahrenburg (2003), Stuart, Sorenson (2003), Wang C K, Wang K, Lu (2003), Wasserman (2003, 2006), Woodward, Hall (2003); Aghion, Bolton, Tirole (2004), Avnimelech, Kenney, Teubal (2004), Avnimelech, Teubal (2004), Baum, Silverman (2004), Berk, Green, Naik (2004), Da Rin, Nicodano, Dittmann, Maug, Kemper (2004), Hsu (2004), Inderst, Müller (2004, 2009), Inderst, Müller, Muennich (2007), Jones, Rhodes-Kropf (2004), Lee, Wahal (2004), Megginson (2004), Michelacci, Suarez (2004), Mishra (2004), Peggy, Wahal (2004), Repullo, Suarez (2004), Roman, van Pottelsberghe de la Potterie (2004), Sembenelli (2004), Sternberg (2004), Ueda (2004); Bergemann, Hege (2005), Cochrane (2005), Da Rin, Hege, Llobet, Walz (2005), Da Rin, Nicodano, Sembenelli (2005, 2006), Da Rin, Hellmann, Puri (2011), De Carvalho, Calomiris, De Matos (2005), Dessein (2005), Dessí (2005), Dimov, Shepherd (2005), Dushnitsky, Lenox (2005, 2006), Dushnitsky, Lavie (2008), Ernst, Witt, Brachtendorf (2005), Ge, Mahoney J M, Mahoney J T (2005), Hsu, Kenney (2005), Hsu (2006, 2007), Klepper, Sleeper (2005), Klepper, Thompson (2010), Kõomägi (2005a, b, c), Kõomägi, Sander (2006), Lai (2005, 2007), Mäkelä, Maula (2005), Mayer, Schoors, Yafeh (2005), Mayer, Schoors, Yafeh (2005), Neus, Walz (2005), Wong (2005), Zook (2005); Antonelli, Teubal (2006), Cassiman, Ueda (2006), Colombo, Dimov, De Clercq (2006), Eckhardt, Shane, Delmar (2006), Ellul, Pagano (2006), Gebhardt, Schmidt (2006), Grilli, Piva (2006), Fallick, Fleischman, Rebitzer (2006), Franco, Filson (2006), Franco, Mitchell (2008), Isaksson (2006), Mathews (2006), Motohashi (2006, 2010), Nielsen, Keuschnigg (2006), Proimos, Murray (2006), Riyanto, Schwienbacher (2006), Wadhwa, Kotha (2006), Zhang (2006, 2007a, b); Bernile, Cumming, Lyandres (2007), Campbell, Kraeussl (2007), Colombo, Dawid, Kabus (2007), de Bettignies,

Brander (2007), de Bettignies, Chemla (2008), de Bettignies (2008), Engel, Keilbach (2007), Hochberg, Ljungqvist, Lu (2007, 2010), Hsu (2007), Jovanovic, Szentes (2007), Lai (2007), Li, Prabhala (2007), Luukkonen (2007, 2008), Mann, Sager (2007), Pintado, De Lema, Van Auken (2007), Robinson, Stuart (2007), Sau (2007), Schwienbacher (2007, 2008), Sørensen (2007), Tykvová (2007); Aizenman, Kendall (2008), Broughman (2008), Broughman, Fried (2010), Davidsson, Steffens, Gordon, Senyard (2008), Geronikolaou, Papachristou (2008), Hand (2008), Hirukawa, Ueda (2008a, b), Katila, Rosenberger, Eisenhardt (2008), Lindsey (2008), McMillan, Roberts, Livada, Wang (2008), Orman (2008), Rossetto (2008), Schwienbacher (2008, 2009), Sorenson, Stuart (2008), Winton, Yerramilli (2008), Nahata (2008), Phalippou (2008), Phalippou, Gottschalg (2009), Puri, Zarutskie (2008), Van Deventer, Mlambo (2008, 2009); Aberman (2009), Bengtsson, Ravid (2009), Bengtsson, Hand (2011), Bengtsson, Sensoy (2011), Block, Sandner (2009), Clarysse, Knockaert, Wright (2009), Cockburn, MacGarvie (2009), Duffner, Schmid, Zimmermann (2009), Fitza, Matusik, Mosakowski (2009), Fulghieri, Sevilir (2009a, b), Jones, Mlambo (2009), Krohmer, Lauterbach, Calanog (2009), Lingelbach, Murray, Gilbert (2009), Litvak (2009a, b), Masulis, Nahata (2009, 2011), Norbäck, Persson (2009), Samuelsson, Davidsson (2009), Van de Vrande, Vanhaverbeke, Duysters (2009); Arikawa, Imad'eddine (2010), Benson, Ziedonis (2010), Bienz, Walz (2010), Cantner, Stützer (2010), Cowling, Murray, Liu (2010), Dushnitsky, Shapira (2010), Elston, Yang (2010), Groh, Liechtenstein (2010), Hall, Woodward (2010), Inci, Barlo (2010), Ivanov, Xie (2010), Jegadeesh, Kräussl, Pollet (2010), Korteweg, Sørensen (2010), Metrick, Yasuda (2010, 2011), Obschonka, Silbereisen, Schmitt-Rodermund, StuetzerNascent (2010), Samila, Sorenson (2010, 2011), Sevilir (2010), Zarutskie (2010), Zacharakis, Erikson, George (2010); Ball, Chiu, Smith (2011), Cherif, Gazdar (2011), Das, Jo, Kim (2011), Ferretti, Meles (2011),

Kandel, Leshchinskii, Kraeussl, Krause (2011), Kerr, Nanda (2011), Li, Abrahamsson (2011), Samila, Sorenson (2011), Tian (2011), Yuklea (2011); Diaconu (2012), Gvazdaitytė (2012), Lazarevski, Mrsik, Smokvarski (2012), Lim, Cu (2012), Pommet (2012), Rosenbusch, Brinckmann, Müller (2012), Yitshaki (2012); Alqatawni (2013), Brettel, Mauer, Appelhoff (2013), Pennacchio (2013), Stuetzer, Obschonka, Davidsson, Schmitt-Rodermund (2013), Stuetzer, Obschonka, Schmitt-Rodermund (2013), Ledenyov D O, Ledenyov V O (2013i), Hsu, Haynie, Simmons, McKelvie (2014), Lahr, Mina (2014), Woike, Hoffrage, Petty (2015), Kvist, Rosengren (May 29 2016).

The **angel investor** as an investment vehicle was researched in Rubenstein (1958), Wetzel (1981, 1983, 1986, 1987), Landström (1992, 1993, 1995, 1998, 2007), Landström, Manigart, Mason, Sapienza (1998), Landström, Mason (editors) (2012), Landström , Mason (2016a, October 29 2016b), Mason, Harrison (1994, 1995, 2000a, b, 2002, 2003, 2008, 2015), Mason, Rogers (1997), Mason, Stark (2004), Mason (2006, 2007, 2009, 2011a, b, 2016), Mason, Botelho (2016), Mason, Harrison, Botelho (2016), Riding, Duxbury, Haines (1994), Coveney, Moore (1998), Lerner (1998 , 2009), Lerner, Schoar, Sokolinski, Wilson (August 2015), Prowse (1998), Aernoudt (1999), Tashiro (1999), Farrell (2000), Freear, Sohl, Wetzel (2000, 2002), Freear, Sohl (2012), Kelly, Hay (2000, 2003), Kelly (2007), Prasad, Bruton, Vozikis (2000), Van Osnabrugge (2000), Amis, Stevenson (2001), Amis, Stevenson, Liechtenstein (2003), Benjamin, Margulis (2001), Sørheim, Landström (2001), Sørheim (2003), Brettel (2002, 2003), Jensen (2002), Payne, Macarty (2002), Stadler, Peters (2003), Sohl (2003, 2006, 2012), Sohl, Hill (2007), San Jose, Roure, Aernoudt (2004), Jossi (July 2005), Månsson, Landström (2005), Maula, Autio, Arenius (2005), EBAN (European Business Angel Network) (2005), Shane (October 1 2005, January 1 2008, 2009), Vance (2005), Wiltbank (October 2005, 2009), Wiltbank,

Boeker (November 2007a, b), Wiltbank, Read, Dew, Sarasvathy (2009), Amatucci, Sohl (2006), Heukamp, Liechtenstein, Wakeling (October 2006), Mercil (2006), Sudek (2006), Chahine, Filatochev, Wright (2007), Harrison, Mason (2007), Harrison, Mason, Robson (2010), Knyphausen-Aufseß, Westphal (2007), Preston (2007), Clark (2008), Goldfarb, Hoberg, Kirsch, Triantis (2008), Ibrahim (2008), Ibrahim (2011), Riding (2008), Collewaert, Manigart (2009), Collewaert, Manigart, Aernoudt (2010), Collewaert (2016), DeGennaro, Dwyer (2009, 2010), DeGennaro (2010, 2012), Goldfarb, Hoberg, Kirsch, Triantis (2009), Goldfarb, Hoberg, Kirsch, Triantis (2012), Mehrotra (2009), Roach (2009), Wallisch (2009), Wong, Bhatia, Freeman (2009), Markova, Petkovska-Mirčevska (2010), Lahti (2011a, b), Lahti, Keinonen (2016), Maxwell, Jeffrey, Lévesque (2011), Maxwell, Levesque, Jeffrey (2014), Maxwell (2016), OECD (2011), Johnson, Sohl (2012a, b), Mitteness, Sudek, Cardon (2012), Scheela, Jittrapanun (2012), Wu Zhenyu, Yuan Wenlong, Wei Xueqi (2012), Festel, De Cleyn (2013), Florin, Dino, Huvaj (2013), Gregson, Mann, Harrison (2013), Hoyos, Santos (2013), Zachary, Mishra (2013), Carpentier, Suret (2014), Carpentier, Suret (2016), Parhankangas, Ehrlich (2014), Zhujun Ding, Sunny Sun, Kevin Au (2014), Altuntas (2015), Anselmo, Amati (2015), Ashish, Agarwal (2015), Banga, Lewis (2015), Baumann (2015), Dollinger, Rhodes (2015), Finlay, Witkin (2015), Gluntz (2015), Gray (2015), Green (2015), Gunther (2015), Hellmann, Thiele (2015), Henyon (2015), Hudson (2015), Litzka (2015), May, Liu (editors) (2015), May, Manhong Mannie Liu (2015), Mullett (2015), Neira (2015), Oker-Blom (2015), Payne (2015), Poh-Kam Wong (2015), Protopopov, Fokin (2015), Reijtenbagh (20150, Roure, San Jose (2015), Roza, Banha (2015), Sidman (2015), Silby, Nicholas (2015), Tooth (2015), Vossen, Munck (2015), Wallace, Conti (2015), Wang Jiani, Chen Su (2015), Wang Jiani, Yi Tan, Manhong Liu (2016), Yeung (2015), Zhujun Ding, Au, Chiang (2015), Avdeitchikova, Landström (2016), Amatucci (2016), Estapé-Dubreuil, Ashta, Hédou (2016), Hornuf, Schwienbacher (October 29 2016), Lingelbach (2016), Murnieks, Cardon, Sudek, White, Brooks (2016), Politis (2016), Qoqiauri (2016), Romaní, Atienza (2016), Scheela (2016), Sørheim, Botelho (2016).

The **investment boutique firm** as an investment vehicle was researched in Thrift (1994), Luenberger (1997), Hall (2007), Morrison, Wilhelm (Winter 2007), Office of Career Services August (2012), Weihong Song, Wei Jie, Lei Zhou (2013), Goodhart, Schoenmaker (March 2016), Thomson Reuters (2016), Wikipedia (2016).

Finally let us shortly summarize the priceless research contributions on the **investment mediums** for the financial capital investment in the capital markets made by world renowned scientists.

The **land exchange** as an investment medium was researched in (see a list of literature on the land as an investment product).

The **real estate exchange** as an investment medium was researched in (see a list of literature on the real estate as an investment product).

The **stock exchange** as an investment medium was researched in (see a list of literature on the company stock and the company stock options as the investment products).

The **foreign currencies exchange** as an investment medium was researched in (see a list of literature on the foreign currency as an investment product).

The **financial securities exchange** as an investment medium was researched in (see a list of literature on the financial security as an investment product).

The **commodities exchange** as an investment medium was researched in (see a list of literature on the commodity as an investment product).

The **precious metal exchange** as an investment medium was researched in Hourwich (1902, 1903), Goodman (1956), Tschoegl (1980),

Solt, Swanson (1981), Burke (June 4 1982), Mate (1984), Ho (1985), Aggarwal, Sundararaghavan (1987), Aggarwal, Soenen (1988), Aggarwal, Mohanty, Song (1995), Aggarwal (2004), Aggarwal, Lucey (2007), Aggarwal, Zong (2008), Aggarwal, Lucey, O'Connor (November 2014, 2015), Crowson (1987), Anikin (1988), Fama, French (1988), Luke, Chan, Mountain (1988), Frank, Stengos (1989), Jaffee (1989), Kaufmann, Winters (1989), Radetzki (1989), Sephton, Cochrane (1990), Vandeloise, Wael (1990), Akgiray, Booth, Hatem (August 1991), Agbeyegbe (1992), Cheung, Lai (May 1993), Chaudhuri (1994), Brunetti, Gilbert (1995), Moore, Cullen (1995), Qiang, Weber (1995), Qiang (1998), Wahab (1995), Sjaastad, Scacciavillani (December 1996), Sjaastad (June 2008), Escribano, Granger (1998), Taylor (1998), Labys, Achouch, Terraza (1999), Rockerbie (1999), Christie-David, Chaudhry, Koch (2000), Cai, Cheung, Wong (2001), Ciner (2001), Ciner, Gurdgiev, Lucey (2013), Hammoudeh, Malik, McAleer (2001, March 2011), Hammoudeh, Yuan (2008), Hammoudeh, Yuan, McAleer, Thompson (2009, 2010), Hammoudeh, Santos, Al-Hassan (2013), Mackenzie, Mitchell, Brooks, Faff (2001), Adrangi, Chatrath (April 2002), Smith (2002), Cavaletti, Factor, All (2004), Baron (2005), Capie, Mills, Wood (2005), Conover, Jensen, Johnson, Mercer (2005, 2009), Drelichman (2005), Papyrakis, Gerlagh (2005), Pulvermacher (March 2005a, b), Xiaoqing Eleanor Xu, Hung-Gay Fung (2005), Banken (2006), Draper, Faff, Hillier (2006), Hillier, Draper, Robert (2006), Batten, Lucey (2007, 2010), Batten, Ciner, Lucey (June 2008, 2010, 2013, 2015), Demidova-Menzel, Heidorn (August 2007a, b), Kyrtsou, Labys (2007), Sari, Hammoudeh, Ewing (2007), Sari, Hammoudeh, Soytas (2010), Tully, Lucey (2007), Worthington, Pahlavani (2007), Jerrett, Cuddington (2008), London Bullion Market Association (LBMA) (2008), Watkins, McAleer (2008), Roberts (2009), Soytas, Sari, Hammoudeh, Hacihasanoglu (2009), Baur, Lucey (2010), Baur, Mcdermott (2010), Baur (2012), Baur, Tran (2014), Chen (September 2010), Humphreys (2010), Kovalenko (2010), Lucey (2010), Lucey, Larkin, O'Connor (2013), Lucey, O'Connor (2014), Lucey, Sile Li (2015), Riley (Summer 2010), Roache, Rossi (2010), Shafiee, Topal (2010), Tsuchiya (2010), Zhang, Wei (September 2010), Khalifa, Miao, Ramchander (2011), Morales, Andreosso-O'Callaghan (2011, 2014), Pukthuanthong, Roll (2011), Arouri, Hammoudeh, Lahiani, Nguyen (2012), Arouri, Hammoudeh, Nguyen, Lahiani (2013), Cochran, Mansur, Odusami (2012), Elder, Miao, Ramchander (2012), Krezolek (2012), Mutafoglu, Tokat E, Tokat H A (2012), Papież, Śmiech (2012), Śmiech, Papież (2012), Yermilova (2012), Caporin, Ranaldo, Velo (2013, 2015), Emmirich, McGroarty (2013), Ewing, Malik (2013), Hood, Malik (2013), Jain, Ghosh (2013), Öztek, Ocal (2013), Reboredo (2013a, b), Reboredo, Ugolini (2015), Revenda (2013, 2016), Rizea Raluca, Sârbu, Condrea (2013), Sensoy (2013), Smales (2013), Westerlund (2013), Agyei-Ampomah, Gounopoulos, Mazouz (2014), Apergis, Christou, Payne (2014), Charles, Darné, Kim (June 19 2014), Charlot, Vêlayoudom Marimoutou (2014), Demiralay, Ulusoy (January 27 2014a, b), Giles, Qinlu Chen (2014), Golosnoy, Rossen (2014), Issler, Rodrigues, Burjack (2014), Papadamou, Markopoulos (2014), Tsolas (2014), Walczak (2014), Wanat, Papież, Śmiech (June 15 2014), Antonakakis, Kizys (2015), Auer (2015), Balcilar, Katzke, Gupta (2015), Balcilar, Hammoudeh, Nwin-Anefo Fru Asaba (2015), Bildirici, Türkmen (2015), Bosch, Pradkhan (2015), Gil-Alana, Chang, Balcilar, Aye, Gupta (2015), Mensi, Hammoudeh, Sang Hoon Kang (2015), Figuerola-Ferretti, McCrorie (2016), Novotný, Polach (2016), Pierdzioch, Risse, Rohloff (2016), Pradkhan (2016), Rand Kwong Yew Low, Yiran Yao, Faff (2016).

The **intellectual property exchange** as an investment medium was researched in (see a list of literature on the intellectual property as an investment product).

# Conclusion

The Ledenyov unified theory on the business cycles in the economy of the scale and the scope was formulated, accurately characterizing the business cycles spectral parameters. A spectroscopy analysis of the business cycles was conducted in the amplitude, frequency, phase, polarization and time domains by applying the theoretical findings in the Ledenyov classic and quantum econodynamics sciences.

More specifically, in the course of research, we assumed that the business cycles can be generated by the oscillating macro-/micro-/nano-economic output variables in the amplitude/frequency/phase and time domains in the economy of the scale and the scope at the certain monetary bases over the selected time periods. We also explained that the GIP (t, monetary base), GDP (t, monetary base), GNP (t, monetary base), PPP (t, monetary base) dependencies can be theoretically treated as both 1) the continuous-time economic output waves, and 2) the Ledenyov discrete-time digital economic output waves. We highlighted the fact that the accurate forward looking assumptions on the GIP(t, monetary base), GDP(t, monetary base), GNP(t, monetary base), PPP(t, monetary base) dependencies can be made, using the business cycles oscillation dynamics analysis in the economy of the scale and the scope at the certain monetary bases over the selected time periods. We noted that the accurate forward looking assumptions on the GIP(t, monetary base), GDP(t, monetary base), GNP(t, monetary base), PPP(t, monetary base) dependencies can help us to find the optimal solutions for the problems on 1) the financial capital investing by increasing the return-on-investment for the investors on one side, and 2) the financial capital borrowing by decreasing the cost-of-capital at various businesses financing schemes for the business founders/owners/managers on other side.

Considering the obtained research results in details, we can make the following general explanatory comments. Chapter 1 made a historical overview on the development of the national/global economies of the scales and the scopes in the world over the centuries. Chapter 2 discussed the continuous-time economic output waves in the economy of the scale and the scope in the Ledenyov classic econodynamics. Chapter 3 considered the Ledenyov discrete-time digital economic output waves in the economy of the scale and the scope in the Ledenyov classic econodynamics. Chapter 4 researched the Ledenyov discrete-time digital economic output waves in the form of the vector-modulated discrete-time digital direct sequence spread spectrum signal's bursts in the economy of the scale and the scope in the Ledenyov classic econodynamics. Chapter 5 dealt with the Ledenyov discrete-time digital economic output waves in the form of the vector-modulated discrete-time digital direct sequence spread spectrum signal's short/long/ultra long time duration pulses in the economy of the scale and the scope in the Ledenyov classic econodynamics. Chapter 6 described the Ledenyov discrete-time digital economic output waves in the form of the vector-modulated discrete-time digital direct sequence spread spectrum signal's short/long/ultra long time duration pulses generated by the quantum fluctuations in the economy of the scale and the scope in the Ledenyov quantum econodynamics. Chapter 7 highlighted some issues in the precise measurement of the econodynamic variables in the economy of the scale and the scope in the Ledenyov classic and quantum econodynamics. Chapter 8 focused on the accurate forecast of the economic and financial trends with the business cycles oscillation dynamics analysis in the economy of the scale and the scope in the Ledenyov classic and quantum econodynamics. Conclusion summarized all the important research findings, highlighting the original scientific contributions in the general-audience plain language format. As a result of completed research work, we formulated the Ledenyov unified theory on the nature of the business cycles in the economy of the scale and the scope in the Ledenyov classic and quantum econodynamics for the first time.

Summarizing the innovative research and making concluding remarks, we can definitely say that the written book presents a wonderful opportunity for the thinking readers with various professional backgrounds to learn more on the advanced theories, techniques and practices to precisely filter out, to accurately characterize and to thoughtfully analyze the business cycles in the Ledenyov unified theory on the nature of the business cycles in the economy of the scale and the scope in the Ledenyov classic and quantum econodynamics.

We believe that the obtained knowledge could be applied by the investors, financiers, economists, businessmen, academicians and other researchers in the process of decision making on the optimal financial capital investment / borrowing in the national / global capital markets to accelerate the business processes toward the sustainable development of the economies of the scales and the scopes on the way to the economic and social prosperity building in the World.

# Acknowledgement

Let us express our words of sincere gratitude for the valuable research discussions to a big number of the great scientists from various universities and institutions around the World during our scientific research over 30 years.

The first author began to be interested in the sound waves' nature, generation techniques and modulation principles during his education at the Spanish classic guitar class at the music school in Kharkiv, Ukraine in 1983. After the successful graduation from the music school and the high school in Kharkiv, Ukraine in 1988, he continued his university education, focusing on the microwave signal generators with the microstrip resonators, and then, shifting his research interest to the resonant structures for the charged particles accelerators at the Department of Radiophysics and Electronics at V. N. Karazin Kharkiv National University, and the National Scientific Centre Kharkiv Institute of Physics and Technology in Kharkiv, Ukraine in 1988-1993.

The second author conducted his research on the ultra high frequency electromagnetic signal filtering with application of the microwave resonators and filters at the Department of Radiophysics and Electronics at V. N. Karazin Kharkiv National University; and National Scientific Centre Kharkiv Institute of Physics and Technology in Kharkiv, Ukraine in 1994-2000. He continues his innovative research on the electromagnetic signal filtering and processing, using the microstrip / dielectric resonators and filters with the high temperature superconductors (HTS) at the microwaves at Department of Electrical and Computer Engineering at James Cook University in Townsville in Australia since 2000 and until present time.

The first author developed his strong research interest into the economic output waves generation and filtering in the macroeconomics in Kharkiv, Ukraine in 1988, when he read the book titled: "Progress and Poverty" by Henry George. His research interest was stimulated by the fact that Prof. Tugan-Baranovsky and Prof. Simon Kuznets conducted the research on the macroeconomics in Kharkiv, Ukraine in the beginning of 19 century. Also, the fact that Prof. Joseph Alois Schumpeter made his macroeconomics research in Chernovtsy, Ukraine in the beginning of 19 century played an important role on that time. Therefore, the first author made his first well known invited talk on the Ledenyov discrete-time digital economic output waves in the market economy at the International Union for the Free Trade and Land Value Taxation Conference in Roskilde, Denmark in 1995, advancing the scientific understanding on the nature of the business cycles in the free market economy, and formulating his new theory on the generation of the Ledenyov discrete-time digital economic output waves in the economies of the scales and scopes over the time.

In 1997-1998, the first author was a visiting researcher at Technical University of Denmark in Lyngby, Denmark, continuing his university-level research on the quantum computing in the quantum physics. It was the time of new knowledge accumulation on the Danish school of scientific thinking in the analog signals processing as well as the Swedish school of scientific thinking on the digital signals processing in GHz frequencies range. The Scandinavian period in the first author's research work was quite unique, because it allowed to make the innovative researches in the electronics and in the economics sciences in parallel in 1995, 1996, and 1997.

Speaking about the economics, we think that the Schumpeter disruptive innovations discovery by Prof. Joseph Alois Schumpeter at the University of Vienna in Austria in 1905 – 1908, University of Czernowitz in Ukraine in 1909 – 1911, University of Graz in Austria in 1912 – 1914, University of Bonn in Germany in 1925 – 1932, Harvard University in the USA in 1932 – 1950, had a considerable impact on the economics as a science. The analysis of business cycles by by Prof. Joseph Alois Schumpeter in his well known book can certainly be considered as one of the first successful attempts to understand a true nature of the periodic oscillations of the economic output in the economies of the scales and scopes over the time.

In this connection, the authors think that the discrete-time digital signal processing theory must be used to accurately characterize the economic output waves in the economy of the scale and the scope in the past, present and future time periods, because the Schumpeter disruptive innovations, which modulate the economic output waves, have the discrete-time nature. In turn, it means that the forecast techniques based on the continuous-time economic output waves cannot be considered as accurate and meaningful.

We used the digital signal processing theory outcomes, which were obtained in both 1) the modern research programs on the spread spectrum communications systems and the advanced multimode noise radars in Kharkiv, Ukraine since 1990s; and 2) the research consulting practices on the spread spectrum communications at the startups and multinational corporations over the recent decades. The accumulated knowledge base on the spread spectrum systems helped us to formulate the theory on the Ledenyov discrete-time digital economic output waves, which can propagate in the form of the vector-modulated direct sequence spread spectrum (DSSS) signal bursts/pulses in the economy of the scale ans the scope at the certain monetary base over the time.

The authors acknowledge the multiple scientific discussions on the econophysics and the quantum mechanics with our father, Oleg P. Ledenyov, in Kharkiv, Ukraine over the recent decades. Our farther, Oleg P. Ledenyov, researched the absorption phenomena, including the absorption of the electromagnetic signals in the high pure metals / superconductors at the ultrasonic frequencies at the ultra low temperatures; the absorption of the electromagnetic signals in the superconductor crystals/thin films at the ultra high frequencies at the low temperatures; the absorption of the electromagnetic signals in the quantum liquids such as the liquid Helium at the ultra high frequencies at the ultra low temperatures; and the absorption of the chemical elements and their isotopes in the soft condensed matter in the nuclear physics in Kharkiv, Ukraine for the five decades. Therefore, it was interesting for us to make the advanced research on the absorption of information by the investors during the financial capital investment in the capital markets in the economies of the scales and scopes in the economics and in the finances.

Most interestingly, our innovative research in the social sciences is much better known to the vivid readers around the World, because of research books publications. However, it may be interesting to explain that we also made some significant contributions to the natural sciences over the years. We are still quite active in the research in the electronics engineering, quantum physics/chemistry and the condensed matter physics, hence the second author thanks for a wonderful opportunity to deliver an invited research seminar, answer the multiple research questions, and make an exchange by the innovative research opinions on the nonlinear analog/digital signals processing at Electrical and Computer Engineering Department, James Cook University, Townsville, Australia in April, 2016. Also, he is grateful for an opportunity to continue his intensive high complexity level research program in the space electronics under Prof. Janina E. Mazierska supervision at Electrical and Computer Engineering Department, James Cook University, Townsville, Australia at present time.

Discussing the finances, the economics and the business administration sciences, the authors would like to explain that we formulated the quantum strategy theory, which represents a new research subject in the finances/the economics/the business administration for a big number of the leading research institutions and the research driven universities. Basically, we decided to extend the scientific knowledge base in the finances, the economics and the business administration sciences by applying the quantum physics principles, based on the Prof. Albert Einstein, Prof. Niels Bohr, Prof. Lev D. Landau theories in the quantum electrodynamics/physics. As a result we came up with an idea of the quantum logic in the decision making process. Presently, the quantum logic is a subject of intensive research in the artificial intelligence (AI), the strategy, the business administration, the jurisprudence, the finances, the economics and many other social/natural sciences. In this connection, we know that Prof. Michael E. Porter, Harvard University is a "guiding star" in the strategy science galaxy and his valuable research contributions to the strategy science resulted in a number of the positive social changes toward the shared value principles adaptation by the business leaders at the corporations, the governments and the universities on the global scale. Therefore, we sincerely acknowledge an enormous interest to our innovative hi-tech research on the quantum strategy creation, implementation and its possible application in the finances/the economics/the business administration sciences/the jurisprudence and other sciences. In addition, we can say that Prof. Niels Bohr, Copenhagen University, Denmark visit to Kharkov, Ukraine in 1933 led to the creation of the econophysics science, and the second author's visits to Roskilde, Lyngby, Denmark and Copenhagen, Denmark in 1995, 1996-1997 resulted in the new quantum theories formulation in the econophysics science.

In addition, let us explain that our Swiss automatic mechanic watches collecting lifestyle, encouraged the second author to visit the Patek Philippe museum in Genève, Switzerland in 2016. We also researched the time-pieces expositions at the Beyer Watch and Clock collecting museum in Zurich, Switzerland. The second author learned a lot about the clocks during his visit to an interesting exhibition: "In course of the time" organized by the Kungahuset in Stockholm, Sweden in 2017. Also, the historical clocks were examined during his visit to the Nobel museum in Stockholm, Sweden in June, 2017. As a result we decided to present the financial risks, the economic risks and the quantum strategy search algorithm in the form of the imagined dials in the Swiss mechanical time-pieces, aiming to make it easy to memorize the complex theoretical conceptions for the interested readers. Amazingly, we discovered that the social sciences students, researchers, professors and business executives can memorize the presented information on the financial risks, the economic risks and the Ledenyov Quantum Winning Virtuous Strategy Search (LQWVSS) algorithm in the form of an imagined dial in the Swiss mechanical time-pieces, much more better, comparing to the generally accepted engineering block-schemes, which may better suit to the scientific interests by the computer scientists, the electronics engineering professors, and the natural sciences professors.

One of the most important lessons, which we learned in the processes of our education and research at the universities over the years, is that the innovative groundbreaking research ideas matter a lot in the modern society. The innovative research ideas lead to the "quantum leaps" in the social-scientific-economic-financial progress in the developed/developing countries. Fortunately, we obtained the multidisciplinary knowledge, completing the university degrees in the Radio-Physics and Electronics at V.N. Karazin Kharkiv National University in Kharkiv, Ukraine in 1993 and 1999. Therefore, we would like to share an opinion that all our scientific discoveries were made due to the multi-disciplinary knowledge application, which is considered by the authors as a key factor on the way toward the modern science development in our increasingly-interconnected global-integrated World society.

Looking forward, we can see an existence of a serious contradiction in the minds of the reputable economists in the economics science. The problem is that all the classic theories in the economics and finances sciences are based on an assumption that the financial and economic processes in the economies of the scales and the scopes can be classified as the continuous-time processes. Indeed, the fact is that a big number of the economists created the outdated economic theories dealing with the continuous-time financial and economic processes in the economies of the scales and the scopes. However, presently, we came to an understanding that all the financial and economic processes in the economies of the scales and the scopes are the discrete-time digital processes. Therefore, it is necessary to recognize a profoundly important fact that a new way of thinking and a new mentality must be applied to understand and accurately characterize the discrete-time digital processes in the economies of the scale and the scope in the amplitude/power/frequency/phase/polarization/scale/time domains, that is why, we formulated our new innovative theories in the economics and finances sciences.

Most importantly, we know that the economy of the scale and the scope supposedly has its "DNA" with the "gene's sequence" in the form of the generated information sequence, which can uniquely characterize and identify the economy of the scale and the scope in Ledenyov V O, Ledenyov D O (2017). Therefore, the main question to understand is: Can we edit the "gene's sequence" of the economy of the scale and the scope with the aim to increase the economic output magnitude in the economy of the scale and scope at the certain monetary base over the time? We think that it is quite possible to edit the gene's sequence of the economy of the scale and the scope with the ultimate goal to increase the economic output magnitude of the economy of the scale and scope at the certain monetary base over the time. We simply need to create and implement a set of the new general regulatory policies in the finances, the new taxation policies in the economics, the new management policies in the business administration, encouraging the Schumpeter disruptive innovations introduction and adaptation in the economy of the scale and the scope in the modern high-tech society at the present and future time.

We thank Prof. Vernon L. Smith, Chapman University, California, USA; Prof. Dr. Jorg Franke, Karlsruhe Institute of Technology/Berliner Effektengesellschaft AG, Berlin, Germany; Prof. Mats Larsson, Stockholm University, Sweden; Prof. Charles K. Whitehead, Cornell University, New York, USA; Prof. Stacy McCaskill, Rock Valley College, Rockford, IL, USA; Prof. Petr Sáha, Tomas Bata University, Zlin, Czech Republic; Prof. Massimo Capacciolli, Iniversity of Naples, Italy; and many other world renowned distinguished scientists, who decided to meet with the first author and discuss our multidisciplinary research in the econodynamics during their visits to V.N. Karazin Kharkiv National University in Kharkiv, Ukraine in 2017.

In general, it worth to comment that the scientific thinking school in Bunyakovsky (1825a, b, c, 1846), who was born in Town of Bar, Region of Vinnytsia, Ukraine; influenced the authors' strategic scientific vision creation and helped to develop the authors' tactical approaches to the scientific solutions finding as far as the problem on the business cycles in the economics is concerned. Of course, it is not conceivable to write this book without the multiple useful research inputs from and encouragements by many brilliant people, who are not listed in the acknowledgement, because of some reasons. Indeed, playing the tennis at the tennis courts or the golf at the golf play grounds with our respected research collaborators, business partners, family friends in various countries around the World, we have already conducted many thousands of thoughtful discussions on the topics of our research interest, hence we would like to thank all our global Friends for their brilliant ideas, interesting opinions, numerous comments, wise suggestions and shared experiences on the subject of our research interest in the economics, finances, and econophysics.

# References:

*Economics science history:*

*1.* Chen Shou 280s-290s; 1977 Sanguozhi, Records of the Three Kingdoms: Book of Wei (Wei Shu), Book of Shu (Shu Shu), Book of Wu (Wu Shu) Nanchong Sichuan China; Pei Songzhi (editor) *Dingwen Printing* Taipei Taiwan.

*2.* Joseph Penso de la Vega 1668, 1996 Confusión de Confusiones *John Wiley and Sons Inc* (re-published) USA.

*3.* Mortimer Th 1765 Every man his own broker *4th edition* London UK.

*4.* Smith A 1776, 2008 An inquiry into the nature and causes of the wealth of nations *W Strahan and T Cadell* London UK, A Selected Edition Sutherland K (editor) *Oxford Paperbacks* Oxford UK.

*5.* Say J-B 1803, 1821, 1834 Traité d'economie politique Paris France, A treatise on political economy or the production, distribution and consumption of wealth 6[th] edition *Grigg & Elliott* Philadelphia USA
https://archive.org/details/ATreatiseOnThePoliticalEconomy .

*6.* Ricardo D 1817, 1821 On the principles of political economy and taxation 3[rd] edition *John Murray* Albemarle Street London UK.

*7.* Mignan R 1829 Travels in Chaldea, including a journey from Bussorah to Bagdad, Hillah, and Babylon, performed on foot in 1827 *H Colburn and R Bentley* ISBN 1-4021-6013-5 .

*8.* Bentham J 1839 A manual of political economy *G P Putnam* New York USA.

*9.* Mill J St 1862 Principles of political economy 5[th] edition *Appleton* New York USA.

*10.* Marx K 1867 Capital, Volume **1**.

*11.* Marx K July 1893 Capital, Volume 2: The process of circulation of capital *Engels publisher*.

*12.* Marx K October 1894 Capital, Volume 3: The process of capitalist production as a whole *Engels publisher*, *International Publishers* New York USA.

*13.* Menger C 1871 Principles of economics (Grundsätze der Volkswirtschaftslehre) *Ludwig von Mises Institute* Auburn Alabama USA
http://www.mises.org/etexts/menger/Mengerprinciples.pdf .

*14.* Bagehot W 1873, 1897; 1924 Lombard Street: A description of the money market *Charles Scribner's Sons* New York USA; *Clowes* London UK
http://www.gutenberg.org/ebooks/4359 .

*15.* George H 1879, 1881, 2009 Progress and poverty: An inquiry into the cause of industrial depressions and of increase of want with increase of wealth; The remedy *Kegan Paul* USA, *Cambridge University Press* UK ISBN 978-1-108-00361-2.

*16.* Von Böhm-Bawerk E 1884, 1889, 1921 Capital and interest: History and critique of interest theories, positive theory of capital, further essays on capital and interest Austria; 1890 *Macmillan and Co* Smart W A (translator) London UK
http://files.libertyfund.org/files/284/0188_Bk.pdf .

*17.* Marshall A 1890, 1920 Principles of economics 1[st] edition, 8[th] edition *MacMillan and Company* London UK.

*18.* Hirsch M 1896 Economic principles: A manual of political economy *The Russkin Press Pty Ltd* 123 Latrobe Street Melbourne Australia.

19. Bachelier L 1900 Théorie de la spéculation *Annales scientifiques de l'Ecole Normale Supérieure* 3e série tome **17** *Gauthier-Villars* Paris France pp 21 – 86.

20. Bachelier L 1914 Le jeu, la chance et le hazard *Bibliothèque de Philosophie scientifique Ernest Flammarion* Paris France

http://gallica.bnf.fr/ark:/12148/bpt6k61926m .

21. Bachelier L 1937 Les lois des grands nombres du calcul des probabilitiés *Gauthier-Villars* Paris France.

22. Bachelier L 19 May 1941 Probabilitiés des oscillations maxima *Comptes-rendus des Séances de l'Académie des Sciences* pp 836 – 838.

23. Courtault J-M, Kabanov Yu, Bru B, Crépel P, Lebon I, Le Marchand A 2000 Louis Bachelier on the centenary of théorie de la spéculation *Mathematical Finance* **10** (3) pp 339 – 353 doi:10.1111/1467-9965.00098 .

24. Bachelier L, Samuelson P A, Davis M, Etheridge A 2006 Louis Bachelier's theory of speculation: The origins of modern finance *Princeton University Press* Princeton NJ USA ISBN 978-0-691-11752-2.

25. Abbott F F 1901 A history and description of Roman political institutions *Elibron Classics* ISBN 0-543-92749-0 .

26. Schumpeter J A 1906 Über die mathematische methode der theoretischen ökonomie *ZfVSV* Austria.

27. Schumpeter J A 1933 The common sense of econometrics *Econometrica*.

28. Schumpeter J A 1911; 1939, 1961 Theorie der wirtschaftlichen entwicklung; The theory of economic development: An inquiry into profits, capital, credit, interest and the business cycle Redvers Opie (translator) *OUP* New York USA, *Harvard University Press* Cambridge MA USA.

29. Schumpeter J A 1939 Business cycle *McGraw-Hill* New York USA.

30. Schumpeter J A 1942, 1950, 1976 Capitalism, socialism and democracy *Harper & Row* New York USA.

31. Schumpeter J A 1947 The creative response in economic history *Journal of Economic History* vol **7** pp 149 – 159.

32. Slutsky E E 1910 Theory of marginal utility *M Sc Thesis* Vernadsky National Library Kiev Ukraine.

33. Slutsky E E 1915 Sulla teoria sel bilancio del consumatore *Giornale degli economisti e rivista di statistica* **51** no 1 pp 1 – 26 Italy.

34. Slutsky E E 1923 On calculation of state revenue from emission of paper money *Local Economy* **2** pp 39 – 62 Kiev Ukraine.

35. Slutsky E E 1937. The summation of random causes as the source of cyclic processes *Econometrica* **5** (April) pp 105 – 146.

36. Von Mises L 1912 The theory of money and credit *Ludwig von Mises Institute* Auburn Alabama USA

https://mises.org/library/theory-money-and-credit .

37. Keynes J M 1919 The economic consequences of the peace *Macmillan* London UK.

38. Keynes J M 1930 The applied theory of money: A treatise on money vol **2** *Macmillan* London UK.

39. Keynes J M 1934 A treatise on money *Macmillan* London UK.

**40.** Keynes J M 1936 The general theory of employment, interest and money *Macmillan Cambridge University Press* Cambridge UK.

**41.** Keynes J M 1998 The collected writings of John Maynard Keynes *Cambridge University Press* Cambridge UK ISBN 978-0-521-30766-6.

**42.** Pigou A C 1924 The economics of welfare 2$^{nd}$ edition *Macmillan* London UK.

**43.** Clark J B 1927 Essential of economic theory *Macmillan* New York USA.

**44.** Hayek F A 1931, 1935, 2008 Prices and production 1$^{st}$ edition *Routledge and Sons* London UK, 2$^{nd}$ edition *Routledge and Kegan Paul* London UK, 2008 edition *Ludwig von Mises Institute* Auburn Alabama USA.

**45.** Hayek F A 1948, 1980 Individualism and economic order *London School of Economics and Political Science* London UK, *University of Chicago Press* Chicago USA.

**46.** Hayek F A 2012 The Collected Works of F. A. Hayek: Business cycles: Part 1 and 2 *The University of Chicago Press* Chicago USA ISBN 9780226320441.

**47.** Hicks J R 1939 Value and capital *Clarendon Press Oxford University Press* Cambridge UK.

**48.** Kaldor N October 1939 Speculation and economic stability *Review of Economic Studies* **7** pp 1 – 27.

**49.** Marshall A 1947 Principles of economics. An introductory volume *Macmillan* London UK.

**50.** Ellis H, Metzler L (editors) 1949 Readings in the theory of international trade *Blakiston* Philadelphia USA.

**51.** Friedman M (editor) 1953 Essays in positive economics *Chicago University Press* Chicago USA.

**52.** Baumol W 1957 Speculation, profitability, and stability *Review of Economics and Statistics* **39** pp 263 – 271.

**53.** Debreu G 1959 Theory of value *Cowles Foundation Monograph* vol **17** *John Wiley & Sons Inc* New York USA.

**54.** Olson M 1965 The logic of collective action *Harvard University Press* Cambridge Massachusetts USA.

**55.** Olson M 1982 The rise and decline of nations: Economic growth, stagflation, and social rigidities *Yale University Press* New Haven Connecticut USA.

**56.** Landes D S 1969 The unbound Prometheus: Technological change and industrial development in Western Europe from 1750 to the present *Cambridge University Press* Cambridge UK.

**57.** Landes D S May 1990 Why are we so rich and they so poor *American Econ Review* **80** (2) pp 1 – 13.

**58.** Landes D S 1998; 1998; 1999 The wealth and poverty of nations: Why some are so rich and some are so poor *W W Norton & Company Inc*; *Little, Brown and Company*; *Abacus* London UK ISBN 0 34911166 9 pp 1 – 650.

**59.** Pareto V 1971 Manual of political economy *Macmillan* London UK.

**60.** Minsky H P 1974 The modeling of financial instability: An introduction *Modeling and Simulation* Proceedings of the Fifth Annual Pittsburgh Conference **5**.

**61.** Minsky H P May 1992 The financial instability hypothesis *Working Paper no 74*: 6–8

http://www.levy.org/pubs/wp74.pdf .

**62.** Minsky H P 2015 Minsky archive *The Levy Economics Institute of Bard College* Blithewood Bard College Annandale-on-Hudson New York USA

http://www.bard.edu/library/archive/minsky/ .

**63.** Stigler G J 1982 Nobel Prize in Economics *Nobel Foundation* Stockholm Sweden.

**64.** Stigler G J 1988 Chicago studies in political economy *University of Chicago Press* Chicago USA ISBN 0-226-77437-6.

**65.** Coase 1991 Nobel Prize in Economics *Nobel Foundation* Stockholm Sweden.

**66.** Roseveare H 1991 The financial revolution 1660-1760 *Longman* UK.

**67.** Becker 1992 Nobel Prize in Economics *Nobel Foundation* Stockholm Sweden.

**68.** Fogel 1993 Nobel Prize in Economics *Nobel Foundation* Stockholm Sweden.

**69.** Foray D, Freeman Ch (editors) 1993 Technology and the wealth of nations *Stanford University Press* Stanford California USA.

**70.** Lucas 1995 Nobel Prize in Economics *Nobel Foundation* Stockholm Sweden.

**71.** Obstfeld M, Rogoff K 1996 Foundations of international macroeconomics *MIT Press* USA.

**72.** Scornick-Gerstein F May, 1996 Private communications on land value taxation theory by Henry George *Royal Automobile Club* London UK.

**73.** Scornick-Gerstein F 1999 The future of taxation: The failure of the poll tax in the UK ISBN: 978-1-9022-5000-7.

**74.** Wolf M 2004 Why globalization works Yale Nota Bene *Yale University Press* New Haven, USA and London, UK ISBN 0-300-10777-3 pp 1 – 398.

**75.** Krugman P, Wells R 2005 Economics *Worth Publishers* 1[st] edition ISBN-10: 1572591501 ISBN-13: 978-1572591509 pp 1 – 1200.

**76.** Stiglitz J E 2005 Principles of macroeconomics *W W Norton* 4[th] edition ISBN-10: 0393926249 ISBN-13: 978-0393926248 pp 1 – 526.

**77.** Stiglitz J E 2015 The great divide *Public Lecture on 19.05.2015* London School of Economics and Political Science London UK

http://media.rawvoice.com/lse_publiclecturesandevents/richmedia.lse.ac.uk/publiclectures andevents/20150519_1830_greatDivide.mp4 .

**78.** Stiglitz J E 2016 The Euro: How a common currency threatens the future of Europe *W W Norton and Company Inc* ISBN 978-0-393-25402-0 pp 1 – 448.

**79.** Morris D, Heathcote J 2007 The price and quantity of residential land in the United States *Journal of Monetary Economics* **54** (8) pp 2595 – 2620.

**80.** Poitras G 2000 The early history of financial economics, 1478–1776 From commercial arithmetic to life annuities and joint stocks *Edward Elgar Publishing* ISBN: 978 1 84064 455 5 pp 1 – 544

http://www.e-elgar.com/shop/the-early-history-of-financial-economics-1478-1776.

**81.** Poitras G (editor) 2006 Pioneers of financial economics: Volume 1 Contributions prior to Irving Fisher *Edward Elgar Publishing* ISBN: 978 1 84542 381 0 pp 1 – 288

http://www.e-elgar.com/shop/pioneers-of-financial-economics-volume-1 .

**82.** Poitras G, Jovanovic F (editors) 2007 Pioneers of financial economics: Volume 2 Twentieth-century contributions *Edward Elgar Publishing* ISBN: 978 1 84542 382 7 pp 1 – 256

http://www.e-elgar.com/shop/pioneers-of-financial-economics-volume-2 .

*83.* Fama 2013 Nobel Prize in Economics Stockholm Sweden.

*84.* Hansen 2013 Nobel Prize in Economics Stockholm Sweden.

*85.* Piketty Th August 2013, August 15 2014 Le Capital au XXIe siècle, Capital in the twenty-first century Goldhammer A (translator) *Éditions du Seuil, Harvard University Press* France, USA ISBN 978-0674430006 pp 1 – 696

http://piketty.pse.ens.fr/en/capital21c2 ,

https://en.wikipedia.org/wiki/Capital_in_the_Twenty-First_Century .

*86.* Liu Guanglin 2015 The Chinese market economy, 1000–1500 *The State University of New York Press* Albany NY USA.

*87.* Ma Debin, Yuan Weipeng 2016 Discovering economic history in footnotes: The story of Tŏng Tàishēng Merchant archive (1790–1850) and the historiography of modern China *Modern China* **42** (5) pp 483 – 504.

*88.* Xu Yi, Ni Y, Van Leeuwen B 2016 Calculating China's historical economic aggregate: A GDP-centered overview *Soc Sci China* **37** (2) pp 56 – 75.

*89.* Xu Yi, Shi Zhihong, Van Leeuwen B, Ni Yuping, Zhang Zipeng, Ma Ye 2016 Chinese national income, ca. 1661–1933 *Aus Economic History Review*.

*90.* Mitchener K J, Debin Ma 2017 Introduction to the special issue: A new economic history of China *Explorations in Economic History* **63** pp 1 – 7.

***Finance science history:***

*91.* Marx K 1867 Capital, Volume **1**.

*92.* Marx K July 1893 Capital, Volume 2: The process of circulation of capital *Engels publisher*.

*93.* Marx K October 1894 Capital, Volume 3: The process of capitalist production as a whole *Engels publisher*.

*94.* Bagehot W 1873, 1897 Lombard Street: A description of the money market *Charles Scribner's Sons* New York USA

http://www.gutenberg.org/ebooks/4359 .

*95.* Von Böhm-Bawerk E 1884, 1889, 1921 Capital and interest: History and critique of interest theories, positive theory of capital, further essays on capital and interest Austria; 1890 *Macmillan and Co* Smart W A (translator) London UK

http://files.libertyfund.org/files/284/0188_Bk.pdf .

*96.* Fisher I 1911 The purchasing power of money New Haven CT USA.

*97.* Fisher I 1925 Mathematical investigations in the theory of value and prices *Yale University Press* New Haven USA.

*98.* Fisher I 1930 The theory of interest *Macmillan Company* New York USA.

*99.* Fisher I October 1933 The dept-deflation theory of great depression *Econometrica*.

*100.* Von Mises L 1912 The theory of money and credit *Ludwig von Mises Institute* Auburn Alabama USA

https://mises.org/library/theory-money-and-credit .

*101.* Friedman M, Schwartz A J 1960, 1963 A monetary history of the United States, 1867-1960 *Princeton University Press* Princeton USA.

*102.* Kindleberger C P 1984 A financial history of Western Europe *Allen & Unwin* London UK.

**103.** Merton R C November 3 1992 Continuous-time finance *Wiley-Blackwell* USA ISBN-13 978-0631185086 pp 1 – 754.

**104.** Duncan-Jones R 1994 Money and government in the Roman empire *Cambridge University Press* UK ISBN-13: 978-0-52-164829-5 pp 1 – 324.

**105.** Harl K W 19 June 1996 Coinage in the Roman economy, 300 B.C. to A.D. 700 *Johns Hopkins University Press* ISBN 978-0-8018-5291-6 pp 1 – 472.

**106.** Andreau J 1999 Banking and business in the Roman world *Cambridge University Press* UK ISBN-13: 978-0-52-138932-7 pp 1 – 200.

**107.** Harris W V 2010 The nature of Roman money *in* The monetary systems of the Greeks and Romans *Oxford University Press* UK ISBN-13: 978-0-1995-8671-4 pp 1 – 344.

**108.** Scheidel W 2009 The monetary systems of the Han and Roman empires in Rome and China. Comparative perspectives on ancient world empires *Oxford University Press* ISBN 978-0-19-533690-0 pp 1 – 258.

**109.** Kessler D, Temin P 2010 Money and prices in the early Roman empire *in* The monetary systems of the Greeks and Romans *Oxford University Press* UK ISBN-13: 978-0-1995-8671-4 pp 1 – 344.

**110.** Antonopoulos A M December 20 2014 Mastering bitcoin: Unlocking digital cryptocurrencies *O'Reilly Media* ISBN-13: 978-1-4493-7404-4.

**111.** Antonopoulos A M September 5 2016 The Internet of money *Merkle Bloom LLC* ISBN-13: 978-1-5370-0045-9.

**112.** Dodd N 2014 The social life of money *Princeton University Press* NJ USA ISBN: 978-0-6911-4142-8 pp 1 – 456.

**113.** Ledenyov V O, Ledenyov D O 2016s Forecast in capital markets *Lambert Academic Publishing* Saarbrücken Germany ISBN 978-3-659-91698-4; *MPRA Paper no 72286* Munich University Munich Germany; *SSRN Paper no SSRN-id2802085 Social Sciences Research Network* New York USA pp 1 – 260

www.lap-publishing.com ,

www.forecastcapitalmarkets.com ,

http://mpra.ub.uni-muenchen.de/72286/ ,

http://ssrn.com/abstract=2802085 .

**114.** Ledenyov V O, Ledenyov D O 2017 Investment in capital markets *Lambert Academic Publishing* Saarbrücken Germany ISBN 978-3-330-05708-1; *MPRA Paper no 77414* Munich University Munich Germany; *SSRN Paper no SSRN-id2930848 Social Sciences Research Network* New York USA pp 1 – 696

www.lap-publishing.com ,

www.investmentcapitalmarkets.com ,

http://mpra.ub.uni-muenchen.de/77414/ ,

http://ssrn.com/abstract=2930848 .

***Juglar economic cycle theory in economics (selected papers):***

**115.** Juglar C 1862 Des crises commerciales et de leur retour périodique en France en Angleterre et aux États-Unis *Guillaumin* Paris France.

**116.** Schumpeter J A 1939 Business cycle *McGraw-Hill* New York USA.

*117.* Grinin L E, Korotayev A V, Malkov S Y 2010 A mathematical model of Juglar cycles and the current global crisis in *History & Mathematics* Grinin L, Korotayev A, Tausch A (editors) *URSS* Moscow Russian Federation.

***Kondratiev economic cycle theory in economics (selected papers):***

*118.* Kondratieff N D 1922 The world economy and its trends during and after war *Regional branch of state publishing house* Vologda Russian Federation.

*119.* Kondratieff N D 1925 The big cycles of conjuncture *The problems of conjuncture* **1** (1) pp 28 – 79.

*120.* Kondratieff N D 1926 Die langen wellen der konjunktur *Archiv fuer Sozialwissenschaft und Sozialpolitik* **56** (3) pp 573 – 609.

*121.* Kondratieff N D 1928 The big cycles of conjuncture *Institute of Economics RANION* Moscow Russian Federation.

*122.* Kondratieff N D, Stolper W F 1935 The long waves in economic life *Review of Economics and Statistics The MIT Press* **17** (6) pp 105 – 115 doi:10.2307/1928486 JSTOR 1928486.

*123.* Kondratieff N D 1984 The Long wave cycle *Richardson & Snyder* New York USA.

*124.* Kondratieff N D 2002 The big cycles of conjuncture and theory of forecast *Economics* Moscow Russian Federation.

*125.* Garvy G 1943 Kondratieff's theory of long cycles *Review of Economic Statistics* **25** (4) pp 203 – 220.

*126.* Silberling N J 1943 The dynamics of business: An analysis of trends, cycles, and time relationships in American economic activity since 1700 and their bearing upon governmental and business policy *McGraw-Hill* New York USA.

*127.* Rostow W W 1975 Kondratieff, Schumpeter and Kuznets: Trend periods revisited *Journal of Economic History* **25** (4) pp 719 – 753.

*128.* Forrester J W 1978 Innovation and the economic long wave *MIT System Dynamics Group Working Paper* Massachusetts Institute of Technology Cambridge USA.

*129.* Forrester J W 1981 The Kondratieff cycle and changing economic conditions *MIT System Dynamics Group Working Paper* Massachusetts Institute of Technology Cambridge USA.

*130.* Forrester J W 1985 Economic conditions ahead: Understanding the Kondratieff wave *Futurist* **19** (3) pp 16 – 20.

*131.* Kuczynski Th 1978 Spectral analysis and cluster analysis as mathematical methods for the periodization of historical processes: Kondratieff cycles – Appearance or reality? *Proceedings of the Seventh International Economic History Congress* vol **2** International Economic History Congress Edinburgh UK pp 79–86.

*132.* Kuczynski Th 1982 Leads and lags in an escalation model of capitalist development: Kondratieff cycles reconsidered *Proceedings of the Eighth International Economic History Congress* vol **B3** International Economic History Congress Budapest Hungary pp 27.

*133.* Barr K 1979 Long waves: A selective annotated bibliography *Review* **2** (4) pp 675 – 718.

*134.* Van Duijn J J 1979 The long wave in economic life *De Economist* **125** (4) pp 544 – 576.

**135.** Van Duijn J J 1981 Fluctuations in innovations over time *Futures* **13**(4) pp 264 – 275.

**136.** Van Duijn J J 1983 The long wave in economic life *Allen and Unwin* Boston MA USA.

**137.** Eklund K 1980 Long waves in the development of capitalism? *Kyklos* **33** (3) pp 383 – 419.

**138.** Mandel E 1980 Long waves of capitalist development *Cambridge University Press* Cambridge UK.

**139.** Van der Zwan A 1980 On the assessment of the Kondratieff cycle and related issues *in Prospects of Economic Growth* Kuipers S K, Lanjouw G J (editors) North-Holland Oxford UK pp 183 – 222.

**140.** Tinbergen J 1981 Kondratiev cycles and so-called long waves: The early research *Futures* **13** (4) pp 258 – 263.

**141.** Van Ewijk C 1982 A spectral analysis of the Kondratieff cycle *Kyklos* **35** (3) pp 468 – 499.

**142.** Cleary M N, Hobbs G D 1983 The fifty year cycle: A look at the empirical evidence *in Long Waves in the World Economy* Freeman Chr (editor) *Butterworth* London UK pp 164 – 182.

**143.** Glismann H H, Rodemer H, Wolter W 1983 Long waves in economic development: Causes and empirical evidence *in* Long Waves in the World Economy Freeman Chr (editor) *Butterworth* London UK pp 135 – 163.

**144.** Bieshaar H, Kleinknecht A 1984 Kondratieff long waves in aggregate output? An econometric test *Konjunkturpolitik* **30** (5) pp 279 – 303.

**145.** Wallerstein I 1984 Economic cycles and socialist policies *Futures* **16** (6) pp 579 – 585.

**146.** Zarnowitz V 1985 Recent work on business cycles in historical perspective: Review of theories and evidence *Journal of Economic Literature* **23** (2) pp 523 – 580.

**147.** Summers L H 1986 Some skeptical observations on real business cycle theory *Federal Reserve Bank of Minneapolis Quarterly Review* **10** pp 23 – 27.

**148.** Freeman C 1987 Technical innovation, diffusion, and long cycles of economic development *in* The long-wave debate Vasko T (editor) *Springer* Berlin Germany pp 295–309.

**149.** Freeman C, Louçã F 2001 As time goes by: From the industrial revolutions to the information revolution *Oxford University Press* Oxford UK.

**150.** Goldstein J 1988 Long cycles: Prosperity and war in the modern age *Yale University Press* New Haven CT USA.

**151.** Solomou S 1989 Phases of economic growth, 1850–1973: Kondratieff waves and Kuznets swings *Cambridge University Press* Cambridge UK.

**152.** Berry B J L 1991 Long wave rhythms in economic development and political behavior *Johns Hopkins University Press* Baltimore MD USA.

**153.** Metz R 1992 Re-examination of long waves in aggregate production series *New Findings in Long Wave Research* Kleinknecht A, Mandel E, Wallerstein I (editors) *St. Martin's* New York USA pp 80 – 119.

*154.* Metz R 1998 Langfristige wachstumsschwankungen – Trends, zyklen, strukturbrüche oder zufall Kondratieffs *Zyklen der Wirtschaft. An der Schwelle neuer Vollbeschäftigung?* Thomas H, Nefiodow L A, Herford (editors) pp 283 – 307.

*155.* Metz R 2006 Empirical evidence and causation of Kondratieff cycles *Kondratieff Waves, Warfare and World Security* Devezas T C (editor) *IOS Press* Amsterdam The Netherlands pp 91 – 99.

*156.* Tylecote A 1992 The long wave in the world economy *Routledge* London UK.

*157.* Cooley Th (editor) 1995 Frontiers of business cycle research *Princeton University Press* USA ISBN 0-691-04323-X.

*158.* Modelski G, Thompson W R 1996 Leading sectors and world politics: The co-evolution of global politics and economics *University of South Carolina Press* Columbia SC USA.

*159.* Modelski G 2001 What causes K-waves? *Technological Forecasting and Social Change* **68** pp 75 – 80.

*160.* Modelski G 2006 Global political evolution, long cycles, and K-waves *Kondratieff Waves, Warfare and World Security* Devezas T C (editor) *IOS Press* Amsterdam The Netherlands pp 293 – 302.

*161.* Perez C 2002 Technological revolutions and financial capital – The dynamics of bubbles and golden ages *Edward Elgar* Cheltenhem UK.

*162.* Rennstich J K 2002 The new economy, the leadership long cycle and the nineteenth K-wave *Review of International Political Economy* **9** pp 150 – 182.

*163.* Rumyantseva S Yu 2003 Long waves in economics: Multifactor analysis *St. Petersburg University Publishing House* St. Petersburg Russian Federation.

*164.* Diebolt C, Doliger C 2006 Economic cycles under test: A spectral analysis *in Kondratieff Waves, Warfare and World Security* Devezas T C (editor) *IOS Press* Amsterdam The Netherlands pp 39 – 47.

*165.* Linstone H A 2006 The information and molecular ages: Will K-waves persist? *Kondratieff Waves, Warfare and World Security* edited by Devezas T C *IOS Press* Amsterdam The Netherlands pp 260 – 269.

*166.* Thompson W 2007 The Kondratieff wave as global social process *in World System History, Encyclopedia of Life Support Systems* Modelski G (editor) *EOLSS Publishers* Oxford UK

http://www.eolss.net.

*167.* Papenhausen Ch 2008 Causal mechanisms of long waves *Futures* **40** pp 788 – 794.

*168.* Korotayev A V, Tsirel S V 2010 A spectral analysis of world GDP dynamics: Kondratieff waves, Kuznets swings, Juglar and Kitchin cycles in global economic development, and the 2008–2009 economic crisis *Structure and Dynamics* vol **4** issue 1 pp 1 – 55

http://www.escholarship.org/uc/item/9jv108xp .

*169.* Grinin L, Korotayev A, Tausch A 2016 Economic cycles, crises, and the global periphery *Springer International Publishing* ISBN 978-3-319-41260-3 eBook ISBN978-3-319-41262-7 DOI10.1007/978-3-319-41262-7 pp 1 – 265.

*170.* Wikipedia 2015a Kondratieff *Wikipedia* USA

www.wikipedia.org .

### Kitchin economic cycle theory in economics (selected papers):

**171.** Kitchin J 1923 Cycles and trends in economic factors *Review of Economics and Statistics The MIT Press* **5** (1) pp 10 – 16 doi:10.2307/1927031 JSTOR 1927031.

### Kuznets economic cycle theory in economics (selected papers):

**172.** Kuznets S 1924 Economic system of Dr. Schumpeter *M. Sc. Thesis under Prof. Wesley Clair Mitchell* Columbia University NY USA.

**173.** Kuznets S 1930a Secular movements in production and prices *Ph. D. Thesis under Prof. Wesley Clair Mitchell* Columbia University NY USA.

**174.** Kuznets S 1930b Secular movements in production and prices. Their nature and their bearing upon cyclical fluctuations *Houghton Mifflin* Boston USA.

**175.** Kuznets S 1934 National income, 1929 - 1932 73[rd] US Congress 2[nd] Session Senate document no 124 pp 5 – 7 https://fraser.stlouisfed.org/scribd/?title_id=971&filepath=/docs/publications/natincome_1934/19340104_nationalinc.pdf .

**176.** Kuznets S 1937 National income and capital formation, 1919 – 1935.

**177.** Kuznets S 1941 National income and its composition, 1919 – 1938.

**178.** Kuznets S March 1955 Economic growth and income inequality *American Economic Review* **45** pp 1 – 28.

**179.** Kuznets S October 1962 How to judge quality *The New Republic*.

**180.** Kuznets S 1963 Quantitative aspects of the economic growth of nations, VIII: The distribution of income by size *Economic Development and Cultural Change* **11** pp 1 – 92.

**181.** Kuznets S 1966 Modern economic growth: Rate, structure, and spread.

**182.** Kuznets S 1968 Toward a theory of economic growth, with reflections on the economic growth of modern nations.

**183.** Kuznets S 1971 Economic growth of nations: Total output and production structure.

**184.** Kuznets S 1973a Population, capital and growth.

**185.** Kuznets S 1973b Modern economic growth: Findings and reflections *American Economic Review* **63** pp 247 – 58.

**186.** Abramovitz M 1961 The nature and significance of Kuznets cycles *Economic Development and Cultural Change* **9** (3) pp 225 – 248.

**187.** Abramovitz M March 1986 Simon Kuznets (1901 – 1985) *The Journal of Economic History* vol **46** no 1 pp 241 – 246.

**188.** Lundberg E 1971 Simon Kuznets contributions to economics *The Swedish Journal of Economics* **73** (4) pp 444 – 459 DOI:10.2307/3439225, JSTOR 3439225.

**189.** Hozelitz B F January 1983 Bibliography of Simon Kuznets *Economic Development and Cultural Change* vol **31** no 2 pp 433 – 454.

**190.** Ben-Porath Y April 1988 Simon Kuznets in person and in writing *Economic Development and Cultural Change* vol **36** no 3 pp 435 – 447.

**191.** Street J H June 1988 The contribution of Simon S. Kuznets to institutionalist development theory *Journal Economic Issues* vol **22** no 2 pp 499 – 509.

**192.** Kapuria-Foreman V, Perlman M November 1995 An economic historian's economist: Remembering Simon Kuznets *The Economic Journal* **105** pp 1524 – 1547.

**193.** Fogel R W 2000 Simon S. Kuznets: April 30, 1901 – July 9, 1985 *NBER Working Paper no W7787* NBER USA.

**194.** Fogel R W, Fogel E M, Guglielmo M, Grotte N 2013 Political arithmetic: Simon Kuznets and the empirical tradition in economics *University of Chicago Press* Chicago USA ISBN 0-226-25661-8.

**195.** Syed M K, Mohammad M J 2004 Revisiting Kuznets hypothesis: An analysis with time series and panel data *Bangladesh Development Studies* **30** (3-4) pp 89 – 112.

**196.** Kleinknecht A, Van der Panne G 2006 Who was right? Kuznets in 1930 or Schumpeter in 1939? in *Kondratieff Waves, Warfare and World Security* Devezas T C (editor) *IOS Press* Amsterdam The Netherlands pp 118 – 127.

**197.** Diebolt C, Doliger C 2008 New international evidence on the cyclical behaviour of output: Kuznets swings reconsidered. Quality & quantity. *International Journal of Methodology* **42** (6) pp 719 – 737.

**198.** Wikipedia 2015b Simon Kuznets Economist *Wikipedia* USA

www.wikipedia.org .

**199.** Wikipedia 2017 Gross domestic product *Wikipedia* USA

www.wikipedia.org .

**200.** Coyle D 2014 GDP: A brief but affectionate history *Princeton University Press* Princeton NJ USA ISBN 978-0-691-15679-8 .

***Akamatsu economic cycle theory in economics (selected papers):***

**201.** Akamatsu K March-August 1962 A historical pattern of economic growth in developing countries *Journal of Developing Economies* **1** (1) pp 3 – 25.

**202.** Ozawa T 2005 Institutions, industrial upgrading, and economic performance in Japan - The "Flying-Geese" paradigm of catch-up growth *Edward Elgar Publishing* Northampton Massachusetts USA.

**203.** Grinin L, Korotayev A, Tausch A 2016 Economic cycles, crises, and the global periphery *Springer International Publishing* Heidelberg Germany ISBN 978-3-31917780-9 .

***Ledenyov economic cycle theory in economics and econodynamics (selected papers):***

**204.** Ledenyov D O, Ledenyov V O 2015d Information money fields of cyclic oscillations in nonlinear dynamic economic system *MPRA Paper no 63565* Munich University Munich Germany, *SSRN Paper no SSRN-id2592975 Social Sciences Research Network* New York USA pp 1 – 40

http://mpra.ub.uni-muenchen.de/63565/ ,

http://ssrn.com/abstract=2592975 .

**205.** Ledenyov D O, Ledenyov V O 2015e On the spectrum of oscillations in economics *MPRA Paper no 64368* Munich University Munich Germany, *SSRN Paper no SSRN-id2606209 Social Sciences Research Network* New York USA pp 1 – 48

http://mpra.ub.uni-muenchen.de/64368/ ,

http://ssrn.com/abstract=2606209 .

**206.** Ledenyov D O, Ledenyov V O 2015f Digital waves in economics *MPRA Paper no 64755* Munich University Munich Germany, *SSRN Paper no SSRN-id2613434 Social Sciences Research Network* New York USA pp 1 – 55

http://mpra.ub.uni-muenchen.de/64755/ ,

http://ssrn.com/abstract=2613434 .

**207.** Ledenyov D O, Ledenyov V O 2016r Precise measurement of macroeconomic variables in time domain using three dimensional wave diagrams *MPRA Paper no 69609* Munich University Munich Germany, *SSRN Paper no SSRN-id2733607 Social Sciences Research Network* New York USA pp 1 – 52

http://mpra.ub.uni-muenchen.de/69609/ ,

http://ssrn.com/abstract=2733607 .

**208.** Ledenyov V O, Ledenyov D O 2016s Forecast in capital markets *Lambert Academic Publishing* Saarbrücken Germany ISBN 978-3-659-91698-4; *MPRA Paper no 72286* Munich University Munich Germany; *SSRN Paper no SSRN-id2802085 Social Sciences Research Network* New York USA pp 1 – 260

www.lap-publishing.com ,

www.forecastcapitalmarkets.com ,

http://mpra.ub.uni-muenchen.de/72286/ ,

http://ssrn.com/abstract=2802085 .

**209.** Ledenyov V O, Ledenyov D O 2017 Investment in capital markets *Lambert Academic Publishing* Saarbrücken Germany ISBN 978-3-330-05708-1; *MPRA Paper no 77414* Munich University Munich Germany; *SSRN Paper no SSRN-id2930848 Social Sciences Research Network* New York USA pp 1 – 696

www.lap-publishing.com ,

www.investmentcapitalmarkets.com ,

http://mpra.ub.uni-muenchen.de/77414/ ,

http://ssrn.com/abstract=2930848 .

***Economic output waves / business cycles nature, properties, parameters spectral analysis in macroeconomics/econodynamics:***

**210.** Ricardo D 1817, 1821 On the principles of political economy and taxation 3[rd] edition *John Murray* Albemarle Street London UK.

**211.** Juglar C 1862 Des crises commerciales et de leur retour périodique en France en Angleterre et aux États-Unis *Guillaumin* Paris France.

**212.** Benner S 1875 Benner's prophecies of future ups and downs in prices: What years to make money on pig iron, hogs, corn and provisions *Robert Clark and Company* Cincinnati USA.

**213.** George H 1881, 2009 Progress and poverty *Kegan Paul* USA; *Cambridge University Press* Cambridge UK ISBN 978-1-108-00361-2.

**214.** Marshall A 1890 Principles of economics *Macmillan* London UK.

**215.** Pareto V 1890 Cours d'économie politique Paris France.

**216.** Walras L 1898 Etudes de'économie politique appliqué Paris France.

**217.** Wicksell 1898 Geldzins und Güterpreise Germany.

**218.** Parvus (pseudonym Helphand A) 1901, 1999 The sturm und drang – Period of capital *Die Handelskrisis und die Gewerkschaften M Ernst* Munich Germany, *in* Louçã F, Reijnders J (editors) pp 21 – 24.

**219.** Tugan-Baranovsky M I 1901 Studien für Geschichte der Handelskrisen in England *Kharkiv* Ukraine.

**220.** Tugan-Baranovsky M I 1923 Periodic industrial crises *"Book" Publishing House* Petrograd-Moscow Russian Federation pp 1 – 386.

**221.** Spiethoff 1902 Vorbemerkungen zu einer Theorie der Überproduction *Jahrbuch für Gestzgebung, Verwaltung und Volkswirtschaft* Germany.

**222.** Schumpeter J A 1908 Das Wesen und der Hauptinhalt der Theorischen Nationalokonomie *Duncker und Humblot* Leipzig Germany.

**223.** Schumpeter J A 1911, 1934, 1955 Theory of economic development *Harvard University Press* Cambridge USA.

**224.** Schumpeter J A 1917, 1956 Das Sozial Produkt und die Rechenpfennige : Glossen und Beitrage zur Geldtheorie von Heute *Archiv für Sozialwissenschaft* **44** pp 627 – 715, Money and the social product Marget A W (translator) *International Economic Papers* **6**.

**225.** Schumpeter J A 1934 The theory of economic development *Harvard University Press* Cambridge Massachusetts USA.

**226.** Schumpeter J A 1936 Review of Keynes' general theory of employment, interest and money *Journal of the American Statistical Association* pp 791 – 795.

**227.** Schumpeter J A 1939, 1989 Business cycles: A theoretical and historical and statistical analysis of the capitalist process vols **1-2** *McGraw-Hill Book Company Inc* New York USA, *Porcupine Press* Philadelphia USA.

**228.** Schumpeter J 1945 Capitalism, socialism and democracy *Harper* New York USA.

**229.** Schumpeter J A 1952 Ten great economists. From Marx to Keynes *George Allen and Unwin* London UK.

**230.** Schumpeter J A 1954 History of economic analysis *George Allen and Unwin*, *Oxford University Press* London UK.

**231.** Schumpeter J A 1961 Konjunkturzyklen Bd 1 *Vandenhoeck & Ruprecht* Göttingen.

**232.** Aftalion A 1909 Essai d'une Théorie des crises générales et périodiques *Revue d'Economie Politique* Paris France.

**233.** Aftalion A 1913 Les crises périodiques de surproduction *M Rivière et C$^{ie}$* Paris France.

**234.** Mitchell W C 1913 Business cycles *University of California Press* Berkeley California USA.

**235.** Mitchell W C 1923, 1951 Business cycles *in* Business cycles and unemployment *National Bureau of Economic Research* New York USA ISBN: 0-87014-003-5, *in* Readings in business cycle theory *Blakiston* Philadelphia USA pp 43 – 60.

**236.** Mitchell W C 1923 Competitive illusion as a cause of business cycles *Quarterly Journal of Economics* **38** pp 631 – 652.

**237.** Mitchell W C, King W I 1923 The economic losses caused by business cycles *in* Business cycles and unemployment *National Bureau of Economic Research* New York USA ISBN: 0-87014-003-5 pp 34 – 42
http://www.nber.org/chapters/c4660 .

**238.** Mitchell W C 1927 Economic organization and business cycles *in* Business cycles: The problem and its setting Mitchell W C (editor) *National Bureau of Economic Research* New York USA ISBN: 0-870-14084-1 pp 61 – 188
http://www.nber.org/chapters/c0681 .

**239.** Mitchell W C 1927, 1928 Business cycles: The problem and its setting *National Bureau of Economic Research* New York USA.

**240.** Mitchell W 1941, 1971 Business cycles and their causes *Reprint Series Edition University of California Press* Berkeley California USA.

**241.** Burns A F, Mitchell W C 1946 Measuring business cycles *National Bureau of Economic Research* New York USA pp 1 – 36.

**242.** Mitchell W C 1951 What happens during business cycles: A progress report *NBER* USA pp 6 – 26.

**243.** Van Gelderen J (alias Fedder J) 1913, 1996 Springvloed: Beschouwingen over industriele ontwikteling en prijsbeweging / Springtide: reflections on industrial development and price movements *De Nieuwe Tijd* **18** pp 253 – 464, *in* Freeman Ch (editor) pp 3 – 55.

**244.** Moore H L 1914, 1967 Economic cycles: Their law and cause, *in* Reprints of economic classics *Augustus M Kelley Publishers* New York USA.

**245.** Persons W M 1914 Books on business cycles: Mitchell, Aftalion, Ilgram *The Quarterly Journal of Economics* **XXVIII** 4 pp 795 – 810.

**246.** Persons W M 1919a An index of general conditions *Review of Economic Statistics* **I** pp 11 – 205.

**247.** Persons W M 1919b General considerations and assumptions *The Review of Economics and Statistics* **I** 1 pp 5 – 8.

**248.** Persons W M 1919c Measurement of secular trend *The Review of Economics and Statistics* **I** 1 pp 8 – 18.

**249.** Persons W M 1919d Measurement of seasonal fluctuations *The Review of Economics and Statistics* **I** 1 pp 18 – 31.

**250.** Persons W M 1919e Measurement of cyclical and irregular fluctuations *The Review of Economics and Statistics* **I** 1 pp 31 – 36.

**251.** Persons W M 1919f Remaining problems of method and interpretation *The Review of Economics and Statistics* **I** 1 pp 36 – 37.

**252.** Persons W M 1919g General considerations relating to secular trend, 1903-1908 *The Review of Economics and Statistics* **I** 1 pp 38 – 39.

**253.** Persons W M 1919h General considerations relating to secular trend, 1903-1908 *The Review of Economics and Statistics* **I** 1 pp 38 – 39.

**254.** Persons W M 1919i Summary of results for each series *The Review of Economics and Statistics* **I** 1 pp 39 – 48.

**255.** Persons W M 1919j I The index: A statement of results *The Review of Economics Statistics* **I** 2 pp 111 – 117.

**256.** Persons W M 1919k IV Application of the method to the data (B) The groups of series *The Review of Economics Statistics* **I** 2 pp 182 – 205.

**257.** Persons W M 1919l The role of prices in the business cycle *The Review of Economics Statistics* **I** 2 pp 206 – 210.

**258.** Persons W M 1921 Bank loans and business cycle *The Review of Economics Statistics* **III** 2 pp 30 – 33.

**259.** Persons W M, Coyle E S 1921 A commodity price index of business cycles *The Review of Economics  Statistics* **III** 11 pp 353 – 369.

260. Persons W M 1922a An index chart based prices and money rates *The Review of Economics Statistics* **IV** 1 pp 7 – 11.

261. Persons W M 1922b The crisis of 1920 in the United States: A quantitative survey *The American Economic Review* **XI** 1 pp 5 – 19.

262. Persons W M 1922c An index of British economic conditions: 1903-1914 *The Review of Economics Statistics* **IV** 2 pp 157 – 175.

263. Persons W M 1924a Cyclical fluctuations of the ratio of bank loans to deposits, 1827-1924 *The Review of Economics Statistics* **VI** 4 pp 260 – 283.

264. Persons W M 1924b Some fundamental concepts of statistics *Journal of the American Statistical Association* **XIX** 145 pp 1 – 8.

265. Persons W M 1926 Theories of business fluctuations *The Quarterly Journal of Economics* **XLI** 1 pp 94 – 128.

266. Persons W M 1927 An index of general business conditions *The Review of Economics Statistics* **IX** 1 pp 20 – 29.

267. Persons W M 1928 Pigou, industrial fluctuations *The Quarterly Journal of Economics* **XLII** 4 pp 669 – 677.

268. Kondratieff N D 1922 The world economy and its trends during and after war *Regional branch of state publishing house* Vologda Russian Federation.

269. Kondratieff N D 1925, 1984 The big cycles of conjuncture *The problems of conjuncture* **1** (1) pp 28 – 79, The long wave cycle *Richardson & Snyder* New York USA.

270. Kondratieff N D December 1926, Spring 1979 Die langen Wellen der Konjunktur *Archiv fuer Sozialwissenschaft und Sozialpolitik* **56** (3) pp 573 – 609, Die langen Wellen der Konjunktur *Review* **2** pp 519 – 562.

271. Kondratieff N D 1928 The big cycles of conjuncture *Institute of Economics RANION* Moscow Russian Federation.

272. Kondratieff N D, Stolper W F 1935, 1996 The long waves in economic life *Review of Economics and Statistics The MIT Press* **17** (6) pp 105 – 115 DOI:10.2307/1928486 JSTOR 1928486, *in* Long wave theory Freeman Ch (editor) *Edward Elgar Publishing* Cheltenham UK.

273. Kondratriev N D 1979 The long waves in economic life *Review* **4** pp 519 – 562.

274. Kondratieff N D 1984 The long wave cycle *Richardson & Snyder* New York USA.

275. Kondratiev N D 1998a Long cycles of economic conjuncture *in* The works of Nikolai D Kondratiev Makasheva N, Samuels W J (editors) vol **I** *Pickering & Chatto* London UK.

276. Kondratyev N D 1998b Main problems of economic statics and dynamics *Economica* Moscow Russian Federation pp 1 – 134.

277. Kondratieff N D 2002 The big cycles of conjuncture and theory of forecast. Selected works Yakovets Yu V, Abalkin L M *Economics Publishing House* Moscow Russian Federation pp 1 – 767.

278. Polanyi K 1922 Sozialistische Rechnungslegung *Archiv für Sozialwissenschaft und Sozialpolitik* vol **49** pp 377 – 420.

279. Kitchin J 1923 Cycles and trends in economic factors *Review of Economics and Statistics The MIT Press* **5** (1) pp 10 – 16 DOI:10.2307/1927031 JSTOR 1927031.

280. De Wolff S 1924 Prosperitats- und Depressionsperioden *in* Der Lebendige Marxismus Jenssen O (editor) *Thuringer Verlagsanstalt* Jena Germany pp 13 – 43.

281. De Wolff S 1924, 1999 Phases of prosperity and depression, *in* Louçã F, Reijnders J (editors) (editors) pp 25 – 44.

282. Kuznets S 1924 Economic system of Dr. Schumpeter *M. Sc. Thesis under Prof. Wesley Clair Mitchell* Columbia University NY USA.

283. Kuznets S 1929 Random events and cyclical oscillations *Journal of the American Statistical Association* **XXIV** 167 pp 258 – 275.

284. Kuznets S 1930a Secular movements in production and prices *Ph. D. Thesis under Prof. Wesley Clair Mitchell* Columbia University NY USA.

285. Kuznets S 1930b Secular movements in production and prices. Their nature and their bearing upon cyclical fluctuations *Houghton Mifflin* Boston USA.

286. Kuznets S 1930c Monetary business cycle theory in Germany *The Journal of Political Economy* **XXXVIII** 2 pp 125 – 163.

287. Kuznets S 1930d Equilibrium economics and business cycle theory *The Quarterly Journal of Economics* **XLIV** 3 pp 381 – 415.

288. Kuznets S 1934 National income, 1929 - 1932 73[rd] US Congress 2[nd] Session Senate document no 124 pp 5 – 7
https://fraser.stlouisfed.org/scribd/?title_id=971&filepath=/docs/publications/natincome_1934/19340104_nationalinc.pdf .

289. Kuznets S 1937 National income and capital formation, 1919 – 1935.

290. Kuznets S 1940 Schumpeter's business cycles *The American Economic Review* vol **30** no 2 Part 1 pp 257 – 271.

291. Kuznets S 1941 National income and its composition, 1919 – 1938.

292. Kuznets S 1954 Economic change *Heineman* London UK.

293. Kuznets S March 1955 Economic growth and income inequality *American Economic Review* **45** (1) pp 1 – 28.

294. Kuznets S October 1962 How to judge quality *The New Republic*.

295. Kuznets S 1963 Quantitative aspects of the economic growth of nations, VIII: The distribution of income by size *Economic Development and Cultural Change* **11** pp 1 – 92.

296. Kuznets S 1965 Economic growth and structure *Norton* New York USA.

297. Kuznets S 1966 Modern economic growth: Rate, structure, and spread *Yale University Press* New Haven Connecticut USA.

298. Kuznets S 1968 Toward a theory of economic growth, with reflections on the economic growth of modern nations.

299. Kuznets S 1971 Economic growth of nations: Total output and production structure.

300. Kuznets S 1973a Population, capital and growth.

301. Kuznets S 1973b Modern economic growth: Findings and reflections *American Economic Review* **63** pp 247 – 258.

302. Yule G U 1926 Why do we sometimes get nonsense correlations between time series? - A study in sampling and the nature of time-series *Journal of the Royal Statistical Society* **89** (1) pp 1 – 63 JSTOR 2341482.

303. Ashby F B 1927 Individual cycles in stock prices *The Journal of Political Economy* **XXXV** 6 pp 835 – 851.

304. Slutzky E 1927 The summation of random causes as the source of cyclic processes *Problems of Economic Conditions* vol **3** no 1 pp 105 – 146.

305. Copeland M A 1929 The national income and its distribution *in* Recent economic changes in the United States, volumes 1 and 2 *National Bureau of Economic Research* Cambridge MA USA pp 761 – 844 ISBN: 0-87014-012-4.

306. Dopsch A 1930 Naturalwirtschaft und Geldwirtschaft in der Weltgeschichte *Seidel* Wien Austria.

307. Heckscher E 1930 Natural- und Geldwirtschaft in der Geschichte *Vierteljahreszeitschrift für Sozial- und Wirtschaftsgeschichte* vol **4** pp 454 – 467.

308. Souter R W 1930 Equilibrium economics and business cycle theory *The Quarterly Journal of Economics* **XLV** 1 pp 40 – 93.

309. Wagemann E 1930 Economic rhythm: A theory of business cycles Blelloch D H (translator) *McGraw-Hill* New York USA.

310. Wilson R May 1930, 22 October 2007 Economic cycles in Australia and New Zealand *Economic Record* vol **6** issue 1 pp 68 – 88, *John Wiley and Sons Inc* DOI: 10.1111/j.1475-4932.1930.tb01083.x .

311. Hayek F A 1931, 1935, 2008 Prices and production 1ˢᵗ edition *Routledge and Sons* London UK, 2ⁿᵈ edition *Routledge and Kegan Paul* London UK, 2008 edition *Ludwig von Mises Institute* Auburn Alabama USA.

312. Hayek F A von 1933 Monetary theory and the trade cycle *Jonathan Cape* London UK.

313. Hayek F A von 1948, 1980 Individualism and economic order *London School of Economics and Political Science* London UK, *University of Chicago Press* Chicago USA.

314. Hayek F A von 1966 Monetary theory and the trade cycle New York USA.

315. Hayek F A von 1974 The pretence of knowledge *Nobel Memorial Prize Lecture* Stockholm Sweden
http://www.nobelprize.org/nobel_prizes/economics/laureates/1974/hayek-lecture.html .

316. Hayek F A von 2012 The collected works of F A Hayek: Business cycles: Part 1 and 2 *The University of Chicago Press* Chicago USA ISBN 9780226320441 pp 1 – 304.

317. Funk J M 1932 The 56-year cycle *in* American Business Activity Ottawa IL USA.

318. Hansen A H 1932 Stabilization in an unbalanced world *Harcourt, Brace* New York USA.

319. Hansen A H 1941 Fiscal policy and business cycles *W W Norton* New York USA.

320. Hansen A H 1951a Business cycles and national income *W W Norton* New York USA.

321. Hansen A H 1951b, 1991 Schumpeter's contribution to business cycle theory *Review of Economic Statistics* **33** pp 129 – 132, *in* J A Schumpeter: critical assessments Cunningham Wood J (editor) *Routledge* London UK vol **I** pp 208 – 213.

322. Hansen A H 1997 Economic cycles and national income *Economica* vol **2** Moscow Russian Federation pp 1 – 298.

323. Mortara G (editor) 1932 Cicli economici *Utet* Torino Italy.

**324.** Fisher I 1933 The debt-deflation theory of great depressions *Econometrica* **1** pp 337 – 357.

**325.** Frisch R 1933 Propagation problems and impulse problems in dynamic economics *in* Readings in business cycles Gordon R, Klein L (editors) *Richard D Irwin* Homewood Illinois USA 1965 pp 155 – 185.

**326.** Working H 1934 A random difference series for use in the analysis of time series *Journal of the American Statistical Association* **XXIX** 185 pp 11 – 24.

**327.** Keynes J M 1919 The economic consequences of the peace *Macmillan* London UK.

**328.** Keynes J M 1930 The applied theory of money: A treatise on money vol **2** *Macmillan* London UK.

**329.** Keynes J M 1934 A treatise on money *Macmillan* London UK.

**330.** Keynes J M 1936 The general theory of employment, interest and money *Macmillan Cambridge University Press* Cambridge UK.

**331.** Keynes J M 1939 Professor Tinbergen´s method *Economic Journal* **XLXIX** pp 558 – 568.

**332.** Keynes J M 1998 The collected writings of John Maynard Keynes *Cambridge University Press* Cambridge UK ISBN 978-0-521-30766-6.

**333.** Leontief W 1936 *Rev Economics and Statistics* **18** p 105.

**334.** Leontief W 1941 The structure of American economy 1919–1939 *Oxford University Press* New York USA.

**335.** Leontief W 1973 Structure of the World economy. Outline of a simple input-output formulation *Nobel Memorial Lecture* Harvard University Cambridge Massachusetts USA.

**336.** Leontief W 1977 The future of the world economy *Oxford University Press* New York USA.

**337.** Haberler G 1937 Prosperity and depression: A theoretical analysis of cyclical movements *League of Nations* Geneva Switzerland.

**338.** Slutsky E E 1937 The summation of random causes as the source of cyclic processes *Econometrica* **5** (April) pp 105 – 146.

**339.** Varga E S 1937 Les crises économiques mondiales: 1848-1935 *OGIZ* Moscow Russian Federation.

**340.** Trakhtenberg I A 1939 Les crises monétaires *Gosfinizdat* Moscow Russian Federation.

**341.** Kolmogorov A N 1941 Interpolation and extrapolation *Bulletin de l'academie des sciences de USSR* Ser Math **5** pp 3 – 14.

**342.** Metzler L 1941 The nature and stability of inventory cycles *Review of Economic Statistics* **23** pp 113 – 129.

**343.** Northrop F S C 1941 The impossibility of a theoretical science of economic dynamics *The Quarterly Journal of Economics* **LVI** 1 1 pp 1 – 17.

**344.** Rose A March 1941 Wars, innovations and long cycles: A brief comment *American Econ Rev* **31** pp 105 – 107.

**345.** Fromm E 1942 The fear of freedom *Routledge / Kegan Paul* London UK.

**346.** Garvy G November 1943 Kondratieff's theory of long cycles *Review of Economic Statistics* **25** (4) pp 203 – 220.

*347.* Samuelson P A 1943 Dynamics, statics, and the stationary state *Review of Economic Statistics* **25** pp 58 – 68.

*348.* Samuelson P A 1947 Foundations of economic analysis *Harvard University Press* Cambridge MA USA.

*349.* Silberling N J 1943 The dynamics of business: An analysis of trends, cycles, and time relationships in American economic activity since 1700 and their bearing upon governmental and business policy *McGraw-Hill* New York USA.

*350.* Ayres C E 1944, 1962 The theory of economic progress 1st edition *University of North Carolina Press* Chapel Hill USA, 2nd edition *Schocken Books* New York USA.

*351.* Åkerman J 1947 Political economic cycles *Kyklos* **1** pp 107 – 117.

*352.* Dupriez L H 1947 Des mouvements economiques generaux vol **2** pt 3 *Institut de Recherches Economiques et Sociales de 1'Universite de Louvain* Belgium.

*353.* Dupriez L H September 1978 1974, a downturn of the long wave? *Banca Nazionale del Lavoro Q Rev* **126** pp 199 – 210.

*354.* Williams D Apr 16 1947 Rhythmic cycles in American business *Henry George School of Social Sciences* New York USA.

*355.* Duesenberry J 1949 Income, saving, and the theory of consumer behavior *Harvard University Press* USA.

*356.* Von Mises L 1949 Human action: A treatise on economics *Yale University Press* New Haven USA.

*357.* Von Mises L 1962 The ultimate foundation of economic science *Nostrand* Princeton USA.

*358.* Dahmén E 1950 Svensk industriell företagarverksamhet. Kausalanalys av den industriella utvecklingen Del I-II *IUI* Stockholm Sweden.

*359.* Dahmén E 1988 'Development blocks' in industrial economics *Scandinavian Economic History Review* vol **XXXVI** (1) pp 3 – 14.

*360.* Hicks J R 1950 A contribution to the theory of the trade cycle *Oxford University Press* Oxford UK.

*361.* Goodwin R M 1951 The nonlinear accelerator and persistence of business cycles *Econometrica* **19** no 1 pp 1 – 17.

*362.* Goodwin R M 1991 Economic evolution, chaotic dynamics and the Marx-Keynes-Schumpeter system *in* Rethinking economics: markets, technology and economic evolution Hodgson G M, Screpanti E (editors) Ch 9 *Edward Elgar Publishing* Cheltenham UK pp 138 – 152.

*363.* Tintner G 1953 Econometrics *John Wiley and Sons Inc* New York USA.

*364.* Wold H O A 1954 Causality and econometrics *Econometrica* **XXII** 2 pp 162 – 177.

*365.* Wold H O 1955 Causality and econometrics: Reply *Econometrica* **XXIII** 2 pp 196 – 197.

*366.* Wold H O 1960 A generalization of causal chain models *Econometrica* **XXVIII** 2 pp 443 – 463.

*367.* Levi-Strauss C 1955, 1961 Tristes tropiques *Plon* Paris France, A world in wane *Criterion* London UK.

*368.* Solow R M 1956 A contribution to the theory of economic growth *Quarterly Journal of Economics* **70** pp 65 – 94.

**369.** Solow R M 1970 Growth theory *Oxford University Press* New York USA.

**370.** Wolfson R J 1958, 1991 The economic dynamics of Joseph Schumpeter *Economic Development and Cultural Change* **7** pp 31 – 54, *in* J A Schumpeter: Critical assessments Cunningham Wood J (editor) vol **II** pp 191 – 215 *Routledge* London UK.

**371.** Mendelson L A 1959-1964 Les théories et l'histoire des crises économiques et des cycles vol **1-3** *Socekgiz* Moscow Russian Federation.

**372.** Oldak P 1959 Modern capitalist theories of the economic cycle *Problems of Economic Transition* vol **2** issue 2 pp 58 – 63.

**373.** Maddison A June 1960 The postwar business cycle in Western Europe and the role of government policy *Banca Nazionale del Lavoro Quarterly Review* pp 100 – 148.

**374.** Maddison A 1981 Les phases du développement capitaliste *Economica* Paris France.

**375.** Maddison A 1995 Monitoring the world economy, 1820-1990 *OECD* Paris France.

**376.** Maddison A 2003 The world economy: Historical statistics *OECD* Paris France.

**377.** Olivera J H G 1960 Cyclical growth under collectivism *Kyklos* **13** pp 229 – 255.

**378.** Abramovitz M 1961 The nature and significance of Kuznets cycles *Economic Development and Cultural Change* **9** (3) pp 225 – 248.

**379.** Abramovitz M 1986 Catching up, forging ahead and falling behind *The Journal of Economic History* vol **46**.

**380.** Abramovitz M March 1986 Simon Kuznets (1901 – 1985) *The Journal of Economic History* vol **46** no 1 pp 241 – 246.

**381.** Date K 1961, 1991 The relation of cycles and trends in Schumpeter's model *Waseda Economic Papers* **5** pp 22 – 34, *in* J A Schumpeter: Critical assessments Cunningham Wood J (editor) *Routledge* London UK vol **II** pp 349 – 359.

**382.** Akamatsu K March-August 1962 A historical pattern of economic growth in developing countries *Journal of Developing Economies* **1** (1) pp 3 – 25.

**383.** Friedman M, Schwartz J A 1963 A monetary history of the United States, 1867–1960 *Princeton University Press* Princeton USA.

**384.** Friedman M 1993 The plucking model of business fluctuations revisited *Economic Inquiry* **31** pp 171 – 177.

**385.** Hickman B 1963 Postwar growth in the United states in light of the long-swing hypothesis *American Economic Review: Papers and Proceedings* **53** pp 490 – 507.

**386.** Korenjak F 1963 Vertical models for economic development and cycles *International Journal for Theoretical and Applied Statistics* vol **7** issue 1 pp 121 – 142.

**387.** Mandel E 1964 The heyday of neo-capitalism and its aftermath *Socialist Register* **14** pp 56 – 67.

**388.** Mandel E 1975 Late capitalism *New Left Books* New York USA.

**389.** Mandel E 1980, 1995 Long waves of capitalist development. The Marxist interpretation *Cambridge University Press* Cambridge UK and *Editions de la Maison des Sciences de l'Homme* Paris France, 2[nd] edition *Verso* London UK.

**390.** Mandel E 1981 Explaining long waves of capitalist development *Futures* **13** pp 332 – 338.

**391.** Adelman I June 1965 Long cycles—fact or artifact? *Amer Econ Rev* **60** pp 444 – 463.

**392.** Perroux F 1965 La pensée économique de Joseph Schumpeter *Editions Droz* Geneva Switzerland.

**393.** Granger C W J 1966 The typical spectral shape of an economic variable *Econometrica* **34** (1) pp 150 – 161.

**394.** Granger C W J 1969 Investigating causal relations by econometric models and cross-spectral methods *Econometrica* **37** (3) pp 424 -438 DOI: 10.2307/1912791 .

**395.** Granger C W J, Newbold P 1978 Spurious regressions in econometrics *Journal of Econometrics* **2** (2) pp 111 – 120 JSTOR 2231972.

**396.** Granger C W J 1980 Testing for causality: A personal viewpoint *Journal of Economic Dynamics and Control* **2** pp 329 – 352 DOI:10.1016/0165-1889(80)90069-X .

**397.** Granger C W J, Teräsvirta T 1993 Modelling nonlinear economic relationships *Oxford University Press* Oxford UK.

**398.** Granger C W J, Inoue T, Morin N 1997 Nonlinear stochastic trends *Journal of Econometrics* **81** pp 65 – 92.

**399.** Granger C W J 2004 Time series analysis, cointegration, and applications *American Economic Review* **94** (3) pp 421 – 425 DOI:10.1257/0002828041464669 .

**400.** Schmookler J 1966 Invention and economic growth *Harvard University Press* Cambridge Massachusetts USA.

**401.** Easterlin R A 1968 Population, labor force and long swings in economic growth *NBER* New York USA.

**402.** Eckstein A 1968 Economic fluctuations in Communist China's domestic development *in* China in crisis Ho P, Tsou T (editors) *University of Chicago Press* Chicago USA pp 702 – 730.

**403.** Frey B S 1968 Economic growth in a democracy: A model *Public Choice* **4** pp 19 – 33.

**404.** Frey B S 1974 The politico-economic system: A simulation model *Kyklos* **27** pp 227 – 254.

**405.** Frey B S 1976a Politico-economic models and cycle *Diskussionsbeiträge des Fachbereichs Wirtschaftswissenschaften der Universität Konstanz no 83* pp 1 – 45

**406.** http://hdl.handle.net/10419/78195 .

**407.** Frey B S 1976b Theorie und Empirie Politischer Konjunktur-zyklen Zeitschrift für Nationalökonomie.

**408.** Frey B S, Ramser H J 1976 The political business cycle: Comment *Review of Economic Studies*.

**409.** Frey B 1978 Politico-economic models and cycles *Journal of Public Economics* vol **9** issue 2 pp 203 – 220.

**410.** Jenkins G M, Watts D G 1968 Spectral analysis and its applications *Holden Day* San Francisco USA.

**411.** Mansfield E 1968 Industrial research and technological innovation *Norton* New York USA.

**412.** Mansfield E 1983 Long waves and technological innovation *Papers and Proceedings American Economic Association* **73** (2) pp 141 – 145.

**413.** Rezneck S 1968 Business depression and financial panic *Essays in American Business and Economic History Greenwood Publishing Co* New York USA.

**414.** De Cecco M 1969 The Italian payment crisis of 1963 – 1964 *in* Problems of the international economy Swoboda A, Mundell R (editors) *Chicago University Press* Chicago USA.

**415.** Hutchings R 1969 Periodic fluctuation in Soviet industrial growth rates *Soviet Studies* **20** (3) pp 331 – 352.

**416.** Matthews R C O 1969 Postwar Business Cycles in the United Kingdom *in* Is the business cycle obsolete? Bronfenbrenner M (editor) John Wiley New York USA pp 103 – 130.

**417.** Palmer R, Colton J 1969 A history of the modern world 3rd edition *Alfred A. Knopf Publishers* New York USA.

**418.** Goodhart C A E, Bhansali R J 1970 Political economy *Political Studies* **18** pp 43 – 106.

**419.** Goodhart C A E 2003 The historical pattern of economic cycles and their interaction with asset prices and financial regulation *in* Asset prices bubbles Hunter W et al (editors) *MIT Press* Cambridge Massachusetts USA.

**420.** Link W 1970 Die Amerikanische Stabilisierungspolitik in Deutschland 1921-32 *Droste* Düsseldorf Germany.

**421.** Bajt A 1971 Investment cycles in European socialist economies: A review article *Journal of Economic Literature* **9** pp 53 – 63.

**422.** Bry G, Boschan C 1971 Cyclical analysis of time series: Selected procedures and computer programs *National Bureau of Economic Research* New York USA.

**423.** Harley C N 1971 The shift from sailing ships to steamships, 1850–1890: A study of technological change and its diffusion *in* Essays on a mature economy McCloskey D M (editor) *Princeton University Press* London UK.

**424.** Luce R D, Krantz D H, Suppes P, Tversky A 1971 Foundations of measurement vol I *Academic Press* New York USA.

**425.** Luce R D, Krantz D H, Suppes P, Tversky A 1989 Foundations of measurement, vol II *Academic Press* New York USA.

**426.** Luce R D, Krantz D H, Suppes P, Tversky A 1990 Foundations of measurement, vol III *Academic Press* New York USA.

**427.** Lundberg E 1971 Simon Kuznets contributions to economics *The Swedish Journal of Economics* **73** (4) pp 444 – 459 DOI:10.2307/3439225, JSTOR 3439225.

**428.** Arrow K J 1972 General economic equilibrium: Purpose, analytic techniques, collective choice *Nobel Memorial Lecture* Stockholm Sweden.

**429.** Attali J 1972 Analyse economique de la vie Politique *Presses Universitaires* Paris France.

**430.** Bernholz P 1972 Grundlagen der Politischen Ökonomie vol **I** *J C B Mohr* Tübingen Germany.

**431.** Inada K, Uzawa H 1972 Economical development and fluctuations *Iwanami* Tokyo Japan.

*432.* Meadows D et al 1972 Limits to growth *Universe Books* New York USA.

*433.* Shuman J B, Rosenau D 1972 The Kondratieff wave: The future of America until 1984 and beyond *Dell* New York USA.

*434.* Kindleberger Ch P 1973 The world in depression, 1929-1939 *University of California Press* Berkeley California USA.

*435.* Kindleberger Ch P 1978 Manias, panics and crashes *Basic Books* New York USA.

*436.* Kindleberger Ch P, Laffargue J P 1979 Financial crises *Cambridge University Press* UK.

*437.* Kindleberger Ch P 1989 Long waves in economics and politics *Economics and Politics* vol **1** issue 2 pp 201 – 206.

*438.* Kindleberger Ch P 1996 Manias, panics and crashes, a history of financial crises 3[rd] edition *MacMillan* London UK.

*439.* Namenwirth Z 1973 The wheels of time and the interdependence of value change *Journal of Interdisciplinary History* **III** pp 649 – 683.

*440.* Ryder H, Heal G 1973 Optimal growth with inter-temporally dependent preferences *Review of Economic Studies* **40** pp 1 – 31.

*441.* Feiwel G R 1974 Reflections on Kalecki's theory of political business cycle *Kyklos* **27** pp 21 – 48.

*442.* Tufte E R 1974 The political manipulation of the economy: Influence of the electoral cycle on macroeconomic performance and policy *mimeo* Princeton University NJ USA.

*443.* Ben-Porath Y 1975 The years of plenty and the years of famine: A political business cycle *Kyklos* **28** pp 400 – 403.

*444.* Bloom H S, Price H D 1975 Voter response to short-run economic conditions: The asymmetric effect of prosperity and recession *American Political Science Review* **69** pp 1240 – 1254.

*445.* Boddy R, Crotty J 1975 Class conflict and macro-policy: The political business cycle *Review of Radical Political Economics* **7** pp 1 – 19.

*446.* Buchanan J M 1975 A contractarian paradigm for applying economic theory *American Economic Review Papers and Proceedings* pp 225 – 230.

*447.* Lindbeck A 1975 Business, cycles, politics and international economic dependence *Skandinaviska Enskilda Banken Quarterly Review* **2** pp 53 – 68.

*448.* MacRae D C 1975 A political model of the business cycle *The Journal of Political Economy*.

*449.* Mass N 1975 Economic cycles: An analysis of underlying causes *MIT Press* Cambridge USA.

*450.* Mass N 1980 Stock and flow variables and the dynamics of supply and demand *in* Elements of the system dynamics method Randers J (editor) *MIT Press* Cambridge pp 95 – 114.

*451.* Mass N, Senge P 1981 Reindustrialization: Aiming for the right targets *Technology Review* Aug/Sept pp 56 – 65.

*452.* McClelland P D 1975 Causal explanation and model building in history, economics and the new economic history *Cornell University Press* Ithaca NY USA/London UK.

*453.* Mensch G 1975a Das Technologische Patt *Umschau* Frankfurt Germany.

**454.** Mensch G 1975b, 1979 Stalemate in technology. Innovations overcome the depression *Ballinger Publishing Company* New York, Cambridge USA.

**455.** Mensch G 1981 Changing capital values and the propensity to innovate *Futures* **13** pp 276–292.

**456.** Mensch G et al 1987 Outline of a formal theory of long-term economic cycles *in* The long wave debate Vasko T (editor) *Springer* Berlin Germany.

**457.** Nordhaus W D 1975 The political business cycle *Review of Economic Studies* pp 169 – 190.

**458.** Pruden H O 1975 The Kondratiev wave *Journal of Marketing* **39** pp 63 – 70.

**459.** Rostow W 1975 Kondratieff, Schumpeter, and Kuznets: Trend periods revisited *Journal of Economic History* **25** (4) pp 719 – 753.

**460.** Rostow W 1978a The world economy: History and prospect *University of Texas Press* Austin Texas USA, *Macmillan* NY USA.

**461.** Rostow W 1978b Getting from here to there *McGraw Hill* New York USA.

**462.** Rostow W, Kennedy M 1979 A simple model of the Kondratief cycle *Research in Economics History*.

**463.** Rostow W W 1980 Why the poor get richer and the rich slow down: Essays in the Marshallian long period *University of Texas Press* Austin TX USA.

**464.** Day R B 1976 The theory of long waves: Kondratieff, Trotsky, Mandel *New Left Review* **99** pp 67 – 82.

**465.** Forrester J W 1976 Business structure, economic cycles, and national policy *Futures* **8** pp 195 – 214.

**466.** Forrester J W, Mass N J, Ryan Ch J July 1976 The system dynamics national model: Understanding socio-economic behavior and policy alternatives *Technological Forecasting and Social Change* **9** pp 51 – 68.

**467.** Forrester J W 1977, 1999 Growth cycles *De Economist* **125** (4) pp 525 – 543, *in* Louçã F, Reijnders J (editors) pp 491 – 509.

**468.** Forrester J W 1978 Innovation and the economic long wave *MIT System Dynamics Group Working Paper* Massachusetts Institute of Technology Cambridge USA.

**469.** Forrester J W 1979 An alternative approach to economic policy: Macrobehavior from Microstructure *in* Economic issues of the eighties Kamrany N, Day R (editors) *Johns Hopkins University Press* Baltimore USA.

**470.** Forrester J W 1981a Innovation and economic change *Futures* **13** pp 323 – 331.

**471.** Forrester J W 1981b The Kondratieff cycle and changing economic conditions *MIT System Dynamics Group Working Paper* Massachusetts Institute of Technology Cambridge USA.

**472.** Forrester J W 1982 A dynamic synthesis of basic macroeconomic theory: Implications for stabilization policy analysis *Ph D Dissertation* MIT Cambridge Massachusetts USA.

**473.** Forrester J W, Graham A, Senge P, Sterman J 1983 An integrated approach to the economic long wave *Working Paper D-3447-1* System Dynamics Group MIT Cambridge Massachusetts USA.

**474.** Forrester J W 1984 The system dynamics national model - objectives philosophy, and status *Proceedings of the 1984 International System Dynamics Conference* Oslo Norway.

**475.** Forrester J W 1985 Economic conditions ahead: Understanding the Kondratieff wave *Futurist* **19** (3) pp 16 – 20.

**476.** Gottlieb M 1976 Long swings in urban development *Columbia University Press* NY USA.

**477.** Keran M W 4 June 1976 The world and the cycle *Economic Letter* Federal Reserve Bank of San Francisco California USA.

**478.** Minsky H 1976, 2008 John Maynard Keynes *McGraw Hill* New York USA.

**479.** Minsky H 1986, 2008 Stabilizing an unstable economy *McGraw-Hill* New York USA.

**480.** Georgescu-Roegen N 1977 Inequality, limits and growth from a bioeconomic point of view *Review of Social Economy* **XXXV** (3) p 361.

**481.** Lucas R E Jr 1977 Understanding business cycles *Journal of Monetary Economics* vol **5**, *in* Stabilization of the domestic and international economy K Brunner, A H Meltzer (editors) *Carnegie-Rochester Conference Series on Public Policy* **5** pp 7 – 29 *North-Holland* Amsterdam The Netherlands.

**482.** Lucas R E Jr November 1980, 1981 Methods and problems in business cycle theory *Journal of Money, Credit and Banking* **12** pp 696 – 715, *Studies in Business-Cycle Theory* pp 271 – 296 *MIT Press* Cambridge Massachusetts USA.

**483.** Lucas Jr R E 1987 Models of business cycles *Basil Blackwell Inc*, *MIT Press* Cambridge USA.

**484.** Lucas Jr R E 1988 On the mechanics of economic development *Journal of Monetary Economics* **22** pp 3 – 42.

**485.** Lucas Jr R E 1990 Supply-side economics: An analytical review *Oxford Economic Papers* **42** pp 293 – 316.

**486.** Lucas Jr R E 1993 Making a miracle *Econometrica* **61** (2) pp 251.

**487.** Lucas Jr R E 2000 Some macroeconomics for the 21st century *Journal of Economic Perspectives* **14** (Winter) pp 159 – 168.

**488.** Lucas Jr R E 2002 The industrial revolution: Past and future. The Kuznets lectures *University of Chicago* USA, *in* Lectures on economic growth *Harvard University Press* Cambridge Massachusetts USA pp 109 – 190.

**489.** Anikine A V, Entov R M (editors) 1978 Mécanisme du cycle économique aux USA *Nauka* Moscow Russian Federation.

**490.** Glissmann H H, Rodemer H, Wolter Fr Juni 1978 Zur Natur der Wachstumsschwache in der Bundesrepublik Deutschland. Eine empirische Analyse langer Zyklen der wirtschaftlichen Entwicklung. *Kieler Diskussionsbeitrage no 55* Institut für Weltwirtschaft Germany.

**491.** Gordon D M 1978 Up and down the long roller coaster in US capitalism in crisis *Union for Radical Political Economists* New York USA.

**492.** Gordon D M 1980 Stages of accumulation and long economic cycles *in* Processes of the world system Hopkins T, Wallerstein I (editors) *Sage* New York USA.

*493.* Gordon G October 1982 Banking panics and business cycles *Federal Reserve Bank of Philadelphia* USA.

*494.* Gordon D M, Weisskopf T E, Bowles D 1983 Long swings and the non-reproductive cycle *American Economic Review* **73** pp 152 – 157.

*495.* Gordon D M, Edwards R, Reich M 1994 Long waves and stages of capitalism *in* Social structures of accumulation Kotz et al (editors) *Cambridge University Press* Cambridge UK.

*496.* Gordon D M, Weisskopf Th, Bowles S 1996 Power, accumulation, and crisis: The rise and demise of the postwar social structure of accumulation *in* Lippit (editor) pp 226 – 246.

*497.* Kuczynski Th 1978 Spectral analysis and cluster analysis as mathematical methods for the periodization of historical processes: Kondratieff cycles – Appearance or reality? *Proceedings of the Seventh International Economic History Congress* vol **2** International Economic History Congress Edinburgh UK pp 79–86.

*498.* Kuczynski Th 1982 Leads and lags in an escalation model of capitalist development: Kondratieff cycles reconsidered *Proceedings of the Eighth International Economic History Congress* vol **B3** International Economic History Congress Budapest Hungary pp 27.

*499.* Lewis W A 1978 Growth and fluctuations 1870-1913 *Allen and Unwin* London UK.

*500.* Modelski G April 1978 The long cycle of global politics and the nation-state *Comparative Studies in Society and History* **20** pp 214 – 235.

*501.* Modelski G, Johnston R, Wu F W March 1979 The long cycle and wars, 1770-1975: A preliminary test of theory *Joint Annual Meetings of International Studies Association/West and Western Political Science Association* Portland Oregon USA.

*502.* Modelski G 1981 Long cycles, Kondratieff's, alternating innovations: Implications for U.S. foreign policy *in* The political economy of foreign policy behavior Kegley C W Jr, McGowan P (editors) *Sage* Beverly Hills California USA.

*503.* Modelski G 1987 Long cycles in world politics *Macmillan* London UK.

*504.* Modelski G, Thompson W R 1988 Sea power in global politics, 1494-1993 *Macmillan* London UK.

*505.* Modelski G, Thompson W R 1996 Leading sectors and world politics: The co-evolution of global politics and economics *University of South Carolina Press* Columbia SC USA.

*506.* Modelski G 2001 What causes K-waves? *Technological Forecasting and Social Change* **68** pp 75 – 80.

*507.* Modelski G 2006 Global political evolution, long cycles, and K-waves *Kondratieff Waves, Warfare and World Security* Devezas T C (editor) *IOS Press* Amsterdam The Netherlands pp 293 – 302.

*508.* Modelski G 2008 Globalization as evolutionary process: Modeling global change *Routledge* New York USA.

*509.* Senge P 1978 The system dynamics national model investment function: A comparison to the neoclassical investment function *Ph D Dissertation* MIT Cambridge Massachusetts USA.

*510.* Senge P 1980 A system dynamics approach to investment function formulation and testing *Socio-Economic Planning Sciences* **14** pp 269 – 280.

*511.* Senge P 1982 The economic long wave: A survey of evidence *Working Paper D-3262-1* System Dynamics Group MIT Cambridge MA USA.

*512.* Shaikh A 1978 Political economy and capitalism: Notes on Dobb's theory of crisis *Cambridge Journal of Economics* **2** pp 233 – 251.

*513.* Shaikh A 1992 The falling rate of profit as the cause of long waves: Theory and empirical evidence *in* New findings in long wave research Kleinknecht A, Mandel E, Wallerstein I (editors) *St Martin's Press* New York USA.

*514.* Aglietta M 1979 A theory of capitalist regulation new left books New York USA.

*515.* Aglietta M 1997 Régulation et crises du capitalisme *Odile Jacob* Paris France.

*516.* Aglietta M, Rebérioux A 27 Apr 2005 Financial crises and the economic cycle Monograph Chapter 7 pp 1 – 41 *in* Corporate governance adrift: A critique of shareholder value *Edward Elgar Publishing* ISBN 9781845425470 pp 1 – 320 http://dx.doi.org/10.4337/9781845425470.00015 .

*517.* Barr K Spring 1979 Long waves: A selective annotated bibliography *Review* **2** (4) pp 675 – 718.

*518.* Bernanke B S 1979 Long-term commitments, dynamic optimization, and the business cycle *Ph. D. Thesis* Department of Economics Massachusetts Institute of Technology USA.

*519.* Bernanke B S, Lown C 1991 The credit crunch *Brookings Papers on Economic Activity* **2** pp 205 – 247.

*520.* Bernanke B S, Gartler M 1989 Agency costs, net worth, and business fluctuations *American Economic Review* **79** (1) pp 14 – 31.

*521.* Bernanke B S 1995 The macroeconomics of the great depression: A comparative approach *Journal of Money, Credit and Banking* **27** pp 1 – 28.

*522.* Bernanke B S 2000 Essays on the great depression *Princeton University Press* Princeton USA.

*523.* Dickey D A, Fuller W A 1979 Distribution of the estimators for autoregressive time series with a unit root *Journal of the American Statistical Association* **74** (366a) pp 427 – 431 DOI:10.1080/01621459.1979.10482531 .

*524.* Freeman Ch 1979 The Kondratiev long waves, technical change, and unemployment *in* Structural determinants of employment and unemployment vol **2** *OECD* Paris France pp 181 – 196.

*525.* Clark J, Freeman Ch, Soete L 1981 Long waves, inventions, and innovations *Futures* **13** pp 308 – 322.

*526.* Freeman Ch, Clark J, Soete L 1982 Unemployment and technical innovation: A study of long waves and economic development *Greenwood Press* Westport Connecticut USA, *Frances Pinter Editors* London UK.

*527.* Freeman Ch 1982 Innovation and long cycles of economic development *Internacional seminar on innovation and development at industrial sector* University of Campinas.

*528.* Freeman Ch 1983 The long wave and the world economy *Butterworths* Boston USA.

*529.* Freeman Ch (editor) 1984 Long waves in the world economy *Pinter* USA.

*530.* Freeman Ch 1987 Technical innovation, diffusion, and long cycles of economic development *in* The long-wave debate Vasko T (editor) *Springer* Berlin Germany pp 295–309.

*531.* Freeman Ch, Pérez C 1988, 1996 Structural crises of adjustment, business cycles and investment behaviour *in* Freeman Ch (editor) pp 242 – 270, *in* Technical change and economic theory Dosi G et al (editor) *Pinter Publishers* London UK, New York USA pp 38 – 66.

*532.* Freeman Ch 1993 Schumpeter's business cycles revisited *in* Evolutionary economics Witt U (editor) *Edward Elgar Publishing* Aldershot pp 17 – 38.

*533.* Freeman Ch (editor) 1996 Long wave theory *Edward Elgar Publishing* Cheltenham UK, *The International library of critical writings in economics* Cambridge UK.

*534.* Freeman Ch 1998 Lange Wellen und Arbeitslosigkeit. Kondratieffs Zyklen der Wirtschaft? An der Schwelle neuer Vollbeschäftigung? Thomas H, Nefiodow L (editors) *BusseSeewald* Herford.

*535.* Freeman Ch 2001 A hard landing for the 'New Economy'? Information technology and the United States national system of innovation *Structural Change and Economic Dynamics* vol **12** no 2 pp 115 – 139.

*536.* Freeman Ch, Louca F 2001 As time goes by: From the industrial revolutions to the information revolution *Oxford University Press* Oxford UK.

*537.* Freeman Ch 2007 The political economy of the long wave *in* The evolution of economic institutions Ch 5 *Edward Elgar Publishing* UK.

*538.* Nerlove M, Grether D M, Carvalho J L 1979 Analysis of economic time series. A Synthesis *Academic Press* New York USA.

*539.* Van Duijn J J 1979 The long wave in economic life *De Economist* **125** (4) pp 544 – 576.

*540.* Van Duijn J J 1981 Fluctuations in innovations over time *Futures* **13** (4) pp 264 – 275.

*541.* Van Duijn J J 1983a Fluctuations in innovations over time *in* Long waves in the world economy Freedman C (editor) *Butterworth* London UK.

*542.* Van Duijn J 1983b The long wave in economic life *George Allen and Unwin* Boston MA USA / London UK.

*543.* Wallerstein I 1979 Kondratiev up or Kondratiev down? *Review* **2** pp 663 – 673.

*544.* Wallerstein I 1980 The capitalist world economy *Cambridge University Press* Cambridge UK.

*545.* Wallerstein I 1982 Crisis as transition in dynamics of global crisis Amin S, Arrighi G, Frank A G, Wallerstein I (editors) *Macmillan* London UK.

*546.* Wallerstein I 1984a Economic cycles and socialist policies *Futures* **16** (6) pp 579 – 585.

*547.* Wallerstein I 1984b Long waves as capitalist process *Review* **7** pp 559 – 575.

*548.* Bousquet N 1980 From hegemony to competition: Cycles of the core? pp 46 – 83 *in* Processes of the world-system Hopkins T K, Wallerstein I (editors) *Sage* Beverly Hills California USA.

**549.** Choudri E, Kochin L 1980 The exchange rate and the international transmission of business cycle disturbances: Some evidence from the Great Depression *Journal of Money, Credit, and Banking* **12** pp 565 – 574.

**550.** Doran C F, Parsons W 1980 War and the cycle of relative power *American Political Science Review* **74** pp 947 – 965.

**551.** Doran C F 2003 Economics, philosophy of history, and the "single dynamic" of power cycle theory: Expectations, competition, and statecraft *International Political Science Review* **24** (1) pp 13 – 49.

**552.** Eklund K 1980 Long waves in the development of capitalism? *Kyklos* **33** (3) pp 383 – 419.

**553.** Graham A, Senge P 1980 A long-wave hypothesis of innovation *Technological Forecasting and Social Change* **17** pp 283 – 311.

**554.** Homer J 1980 The role of consumer demand in business cycle entrainment *Working Paper D-3227-1* System Dynamics Group MIT Cambridge USA.

**555.** Low G 1980 The multiplier-accelerator model of business cycles interpreted from a system dynamics perspective *in* Elements of the system dynamics method Randers J (editor) *MIT Press* Cambridge USA pp 76 – 94.

**556.** Marchetti C 1980 Society as a learning system: Discovery, invention, and innovations cycles revisited *Technological Forecast and Social Change* **18** pp 257 – 282.

**557.** Marchetti C 1988 Kondratiev revisited – After one Kondratiev cycle *International Institute for Applied Systems Analysis* p 7.

**558.** Marchetti C 1998 Lange Wellen durchdringen alles, Ist die menschliche Gesellschaft zyklotym veranlagt? Kondratieffs Zyklen der Wirtschaft? An der Schwelle neuer Vollbeschäftigung? Thomas H, Nefiodow L (editors) *BusseSeewald* Herford.

**559.** Frank A G 1980 Crisis in the world economy *Heinemann* London UK.

**560.** Hodrick R J, Prescott E C 1980, 1997 Postwar U.S. business cycles: An empirical investigation *manuscript*, *Journal of Money, Credit, and Banking* vol **29** no 1 pp 1 – 16.

**561.** Kydland F E, Prescott E C November 1982 Time to build and aggregate fluctuations *Econometrica* **50** (1) pp 345 – 370.

**562.** Prescott E C 1986 Theory ahead of business-cycle measurement *Federal Reserve Bank of Minneapolis Quarterly Review* **10** pp 9 – 22.

**563.** Kydland F E, Prescott E C Spring 1990 Business cycles: Real facts and a monetary myth *Federal Reserve Bank of Minneapolis Quarterly Review* vol **14** no 2 ISSN 0271-5287 pp 3 – 18.

**564.** Parente St L, Prescott E C 1993 Changes in the wealth of nations *Federal Reserve Bank of Minneapolis Quarterly Review* **17** (Spring) pp 3 – 16.

**565.** Parente St L, Prescott E C 1994 Barriers to technology adoption and development *Journal of Political Economy* **102** (April) pp 298 – 321.

**566.** Prescott E C 1998a Business cycles research: Methods and problems *Working Paper 590* Federal Reserve Bank of Minneapolis USA.

**567.** Prescott E C 1998b Needed: A theory of total factor productivity *International Economic Review* **39** (August) pp 525 – 551.

**568.** Parente St L, Prescott E C 2000 Barriers to riches *MIT Press* Cambridge Massachusetts USA.

**569.** Sims Ch 1980 Macroeconomics and reality *Econometrica* **48** (1) pp 1 – 48 DOI:10.2307/1912017 .

**570.** Sterman J D 1980 The use of aggregate production functions in disequilibrium models of energy-economy interactions *Working Paper D-3234* System Dynamics Group MIT Cambridge MA USA.

**571.** Sterman J D 1981 The energy transition and the economy: A system dynamics approach 2 vols *Ph D Dissertation* MIT Cambridge MA USA.

**572.** Sterman J D 1982a Economic vulnerability and the energy transition *Energy Systems and Policy*, *Working Paper D-3356-1* System Dynamics Group MIT Cambridge MA USA.

**573.** Sterman J D 1982b Amplification and self-ordering: Causes of capital overexpansion in the economic long wave *Working Paper D-3366* System Dynamics Group MIT Cambridge Massachusetts USA.

**574.** Sterman J D 18 March 1983a The long wave (letter) *Science* **219** p 1276.

**575.** Sterman J D March 1983b A simple model of the economic long wave *Working Paper WP 1422-83* System Dynamics Group MIT Cambridge Massachusetts USA pp 1 – 72.

**576.** Sterman J D May 1984 An integrated theory of the economic long wave *WP-1563-84* Sloan School of Management Massachusetts Institute of Technology USA pp 1 – 49.

**577.** Sterman J D 1985 A behavioral model of the economic long wave *Journal of Economic Behavior and Organization* vol **6** issue 1 pp 17 – 53.

**578.** Sterman J D June 17 2012 The economic long wave: Theory and evidence *Forgotten Books* pp 1 – 62.

**579.** Sterman J D October 2 2013 Deterministic chaos in an experimental economic system *Nabu Press* ISBN-13: 978-1289800178 ISBN-10: 1289800170 pp 1 – 54.

**580.** Sterman J D, Mosekilde E February 28 2014, February 14 2015, August 25 2017 Business cycles and long waves: A behavioral disequilibrium perspective *Nabu Press* ISBN-13: 978-1294769903 ISBN-10: 1294769901, *Scholar's Choice* ISBN-13: 978-1296021337 ISBN-10: 1296021335, *Andesite Press* ISBN-13: 978-1376331745 ISBN-10: 1376331748 pp 1 – 64.

**581.** Van der Zwan A 1980 On the assessment of the Kondratieff cycle and related issues *in Prospects of Economic Growth* Kuipers S K, Lanjouw G J (editors) *North-Holland* Oxford UK pp 183 – 222.

**582.** Weber R 1980 Society and economy in the Western world system *Social Forces* **59** (4) pp 1130 – 1148.

**583.** Beveridge S, Nelson C R 1981 A new approach to the decomposition of economic time series into permanent and transitory components with particular attention to measurement of the business cycle *Journal of Monetary Economics* **7** pp 151 – 172.

**584.** Brunner K (editor) 1981 The great depression revisited *Kluwer* Boston USA.

*585.* Delbeke J 1981 Recent long-wave theories: A critical survey *Futures* **13** pp 246 – 257.

*586.* Hall P 26 March 1981 The geography of the fifth Kondratiev cycle *New Society* pp 535 – 537.

*587.* Hall P 1997 Macroeconomic fluctuations and the allocation of time *Journal of Labor Economics* **15** pp 223 – 250.

*588.* Kleinknecht A 1981a Observations on the Schumpeterian swarming of innovations *Futures* **13** (4) pp 246 – 257.

*589.* Kleinknecht A 1981b Innovation, accumulation, and crisis: Waves in economic development? *Review* **4** (4) pp 683 – 711.

*590.* Kleinknecht A 1982 Innovation patterns in crisis and prosperity: Schumpeterian long cycles reconsidered *MacMillan* London UK.

*591.* Bieshaar H, Kleinknecht A 1984 Kondratieff long waves in aggregate output? An econometric test *Konjunkturpolitik* vol **30** no 5 pp 279 – 303.

*592.* Kleinknecht A 1986 Long waves, depression and innovation *De Economist* **134** pp 84 – 108.

*593.* Kleinknecht A 1987 Innovation patterns in crisis and prosperity: Schumpeter's long cycle reconsidered *Macmillan* London UK.

*594.* Kleinknecht A 1990 Are there Schumpeterian waves of innovations? *Cambridge Journal of Economics* **14** no 1 pp 81 – 92.

*595.* Kleinknecht A, Mandel E, Wallerstein I (editors) 1992 New findings in long wave research *St Martin's Press* New York USA.

*596.* Kleinknecht A 1993 Recent research on Kondratieff long waves *Les cahiers de l'association Charles Gilde pour l'tude de la pense conomique* **5** pp 171 – 193.

*597.* Kleinknecht A, Van der Panne G 2006 Who was right? Kuznets in 1930 or Schumpeter in 1939? in *Kondratieff Waves, Warfare and World Security* Devezas T C (editor) *IOS Press* Amsterdam The Netherlands pp 118 – 127.

*598.* Kleinknecht A, Van der Panne G 30 July 2008 Technology and long waves in economic growth Chapter 33 pp 1 – 10 *in* The Elgar companion to social economics Davis J B, Dolfsma W (editors) pp 1 – 704 ISBN: 9781845422806 http://dx.doi.org/10.4337/9781848442771.00050 .

*599.* Pasinetti L L 1981 Structural change and economic growth. A theoretical essay on the dynamics of the wealth of nations *Cambridge University Press* Cambridge UK.

*600.* Pasinetti L L 1995 Problemi irrisolti di teoria keynesiana. Riflessioni su domanda effettiva, debito pubblico, disoccupazione e inflazione *Rendiconti dell'Accademia Nazionale dei Lincei* Scienze Morali serie IX vol **VI** pp 213 – 235.

*601.* Pasinetti L L 1998 The myth (or folly) of the 3% deficit-GDP Maastricht 'parameter' *Cambridge Journal of Economics* vol **22** pp 103 – 116.

*602.* Pasinetti L L December 2000 Critique of the neoclassical theory of growth and distribution *Banca Nazionale del Lavoro Quarterly Review* no 215 pp 383 – 431.

*603.* Tinbergen J 1981 Kondratiev cycles and so-called long waves: The early research *Futures* **13** (4) pp 258 – 263.

*604.* Tinbergen J 1984 Kondratiev cycles and so-called long waves. Long waves in the world economy *Frances Pinter* London UK.

**605.** Priestley M B 1981 Spectral analysis and time series *Academic Press* New York USA.

**606.** Warren J P 1981 The fifty year boom-bust cycle *Warren-Cameron* Guildford.

**607.** Boltho A 1982 The European economy: Growth and crisis *Oxford University Press* Oxford UK.

**608.** Chitre V S 1982 Growth cycles in the Indian economy *Artha Vijnana* **24** pp 293 – 450.

**609.** Haustein H D, Neuwirth E 1982 Long waves in world industrial production, energy consumption, innovations, inventions, and patents and their identification by spectral analysis *Technological Forecasting And Social Change* **22** pp 53 – 89.

**610.** Moore G H 1982 Business cycles *in* Encyclopedia of economics Greenwald D (editor-in-Chief) *McGraw Hill Book Company* New York USA.

**611.** Moore G H, Klein P A 1989 Business cycles and inflation cycles in the market-oriented world *Working Paper* Columbia Graduate School of Business Columbia University NY USA.

**612.** Nelson C R, Winter S 1982 An evolutionary theory of economic change *Belknap Press of Harvard University Press* Cambridge USA.

**613.** Nelson C R, Plosser C I 1982 Trends and random walks in macroeconomic time series: Some evidence and implications *Journal of Monetary Economics* **10** pp 139 – 162.

**614.** Senge P 1982 The economic long wave: A survey of evidence *Working Paper D-3262-1* System Dynamics Group MIT Cambridge Massachusetts USA.

**615.** Senge P 1983 A long wave theory of real interest rate behavior *Working Paper D-3470* System Dynamics Group MIT Cambridge Massachusetts USA.

**616.** Stewart H B 1982 Technology innovation and business growth *Nutevco* San Diego California USA.

**617.** Thompson W R, Zuk G L 1 December 1982 War, inflation, and the Kondratieff long wave *Journal of Conflict Resolution* vol **26** issue 4 pp 621 – 644.

**618.** Thompson W R 2007 The Kondratieff wave as global social process *in* World system history, encyclopedia of life support systems Modelski G (editor) *EOLSS Publishers* Oxford UK

http://www.eolss.net.

**619.** Van Ewijk C 1982 A spectral analysis of the Kondratieff cycle *Kyklos* **35** (3) pp 468 – 499.

**620.** Van Ewijk C 1994 Hysteresis in a model of cycles and growth *Discussion Paper TI 94-116* Tinbergen Institute.

**621.** Beckman R C 1983 The downwave *Pan Books* London UK.

**622.** Blatt J M 1983 Economic policy and endogenous cycles *Journal of Post Keynesian Economics* vol **5** issue 4 pp 635 – 647.

**623.** Cass D, Shell K April 1983 Do sunspots matter? *Journal of Political Economy* **91** pp 193 – 227.

**624.** Cleary M N, Hobbs G D 1983 The fifty year cycle: A look at the empirical evidence *in* Long waves in the world economy Freeman Chr (editor) *Butterworth* London UK pp 164 – 182.

**625.** Dickson D 25 February 1983 Technology and cycles of boom and bust *Science* **219** (4587) pp 933 – 936.

**626.** Glismann H H, Rodemer H, Wolter W 1983 Long waves in economic development: Causes and empirical evidence *in* Long waves in the world economy Freeman Chr (editor) *Butterworth* London UK pp 135 – 163.

**627.** Goldstein J S 1983 Long waves and war cycles *M Sc thesis* MIT Cambridge USA.

**628.** Goldstein J S 1988 Long cycles: Prosperity and war in the modern age *Yale University Press* New Haven CT USA.

**629.** Goldstein J S 1991 The possibility of cycles in international relations *International Studies Quarterly* **35** pp 477 – 480.

**630.** Hill C W L 1983 Conglomerate performance over the economic cycle *Journal of Industrial Economics* vol **32** issue 2 pp 197 – 211.

**631.** Hozelitz B F January 1983 Bibliography of Simon Kuznets *Economic Development and Cultural Change* vol **31** no 2 pp 433 – 454.

**632.** Long J, Plosser C 1983 Real business cycles *Journal of Political Economy* **91** pp 1345 – 1370.

**633.** Pérez C 1983, 1996 Structural change and assimilation of new technologies in the economic and social systems *Futures* **15** (5) pp 357 – 375, *in* Freeman C (editor) pp 373 – 391.

**634.** Pérez C 1985 Microelectronics, long waves and world structural change: New perspectives for developing countries *World Development* **13** (3) pp 441 – 463.

**635.** Pérez C 2002 Technological revolutions and financial capital: The dynamics of bubbles and golden ages *Edward Elgar Publishing* Cheltenham UK.

**636.** Pérez C 2015 From long waves to great surges *European Journal of Economic and Social Systems* vol **27** issue 1 – 2 pp 70 – 80.

**637.** Rosenberg N, Frischtak C 1983 Long waves and economic growth: A critical appraisal *Papers and Proceedings American Economic Association* **73** (2) pp 146 – 151.

**638.** Rosier B, Dockes P 1983 Rythmes économiques. Crises et changement social, une perspective historique *La Découverte/Maspero* Paris France.

**639.** Tobin J 1983 Okun, Arthur M *in* The new Palgrave dictionary of economics vol **3** pp 700 – 701 *Macmillan* London UK.

**640.** Bell W R 1984 Signal extraction for nonstationary time series *Annals of Statistics* **12** pp 646 – 664.

**641.** Bergstrom A R 1984 Continuous time stochastic models and issues of aggregation over time *in* Handbook of econometrics Griliches Z, Intriligator M D (editors) **2** pp 1146 – 1212 *North-Holland* Amsterdam The Netherlands.

**642.** Bergstrom A R 1988 The history of continuous-time econometric models *Econometric Theory* **4** pp 350 – 373.

**643.** Bergstrom A R, Nowman K B 2007 A continuous time econometric model of the United Kingdom with stochastic trends *Cambridge University Press* Cambridge UK.

**644.** King R G, Plosser Ch I 1984 Money, credit, and prices in a real business cycle *American Economic Review* **74** (June) pp 363 – 380.

**645.** King R G, Rebelo S T 1988 Business cycles with endogenous growth *Unpublished Manuscript* University of Rochester USA.

**646.** King R G, Plosser Ch I, Rebelo S T 1988a Production, growth and business cycles: I The basic neoclassical model *Journal of Monetary Economics* **21** (2-3) pp 196 – 232.

**647.** King R G, Plosser Ch I, Rebelo S T 1988b Production, growth, and business cycles: II. New directions *Journal of Monetary Economics* **21** (2-3) pp 309 – 341.

**648.** King R G, Plosser Ch I, Stock J, Watson M 1991 Stochastic trends and economic fluctuations *Working Paper Series no 91-4* Macroeconomic Issues Federal Reserve Bank of Chicago, *The American Economic Review* **81** no 4 pp 819 – 840.

**649.** King R G, Rebelo S T 1993 Low frequency filtering and real business cycles *Journal of Economic Dynamics and Control* **17** pp 207 – 231.

**650.** King R G, Rebelo S T 1999 Resuscitating real business cycles *in* Handbook of macroeconomics Taylor J, Woodford M North-Holland The Netherlands.

**651.** Miller D, Friesen P H 1984 A longitudinal study of the corporate life cycle *Management Science* vol **30** no 10 pp 1161 – 1183.

**652.** Screpanti E 1984, 1999 Long economic cycles and recurring proletarian insurgencies *Review, A Journal of the Fernand Braudel Center for the Study of Economies, Historical Systems, and Civilizations* vol **7** no 2 pp 509 – 548, *in* Louçã F, Reijnders J (editors) pp 139 – 178.

**653.** Rothbard M 1984 The Kondratieff cycle: Real or fabricated? *Investment Insights* August and September.

**654.** Foders F, Glismann H H 1985 Long waves in Argentine economic development *Kiel Working Paper no 236* Institut für Weltwirtschaft (IfW) Kiel Germany pp 1 – 30 http://hdl.handle.net/10419/722 .

**655.** Grandmont J M 1985 On endogenous business cycles *Econometrica* **53** pp 995 – 1045.

**656.** Hansen G D 1985 Indivisible labor and the business cycle *Journal of Monetary Economics* **16** pp 309 – 327.

**657.** Hansen G D, Wright R 1992 The labor market in real business cycle theory *Federal Reserve Bank of Minneapolis Quarterly Review* **16** (2) pp 2 – 12.

**658.** Harvey A C 1985 Trends and cycles in macroeconomic time series *Journal of Business and Economic Statistics* **3** pp 216 – 228.

**659.** Harvey A C 1989 Forecasting, structural time series models and the Kalman filter *Cambridge University Press* Cambridge UK.

**660.** Harvey A C, Jaeger A 1993 Detrending, stylized facts and the business cycle *Journal of Applied Econometrics* **8** pp 231 – 247.

**661.** Harvey A C, Koopman S J 1997 Multivariate structural time series models *in* Systematic dynamics in economic and financial models Heij C, Schumacher H, Hanzon B, Praagman C (editors) *John Wiley and Sons Inc* pp 231 – 247.

**662.** Harvey A C, Trimbur T M 2003 General model-based filters for extracting cycles and trends in economic time series *The Review of Economics and Statistics* **85** (2) pp 244 – 255.

*663.* Harold J 1985 The Reichsbank and public finance in Germany, 1924-1933: A Study of the politics of economics during the great depression *Knapp* Frankfurt am Main Germany.

*664.* Harold J 1986 The German slump: Politics and economics, 1924-1936 *Clarendon Press* Oxford UK.

*665.* Elliot J E 1985, 1991 Schumpeter's theory of economic development and social change: Exposition and assessment *International Journal of Social Economics* **12** pp 6 – 33, J A Schumpeter: Critical assessments Cunningham Wood J (editor) *Routledge* London UK vol **IV** pp 324 – 357.

*666.* Zarnowitz V 1985 Recent work on business cycles in historical perspective: Review of theories and evidence *Journal of Economic Literature* **23** (2) pp 523 – 580.

*667.* Zarnowitz V, Moore G 1986 Major changes in cyclical behaviour *in* The American business cycle *University of Chicago Press* Chicago USA.

*668.* Zarnowitz V 1989 Facts and factors in the recent evolution of business cycles in the United States *Working Paper 2865* National Bureau of Economic Research USA.

*669.* Zarnowitz V June 1992 Business cycles theory, history, indicators, and forecasting *University of Chicago Press* Chicago IL USA pp 1 – 614 ISBN: 9780226978901.

*670.* Zarnowitz V, Stock J H 1992 What is a business cycle? *Proceedings* Federal Reserve Bank of St Louis pp 3 – 83.

*671.* Bhargava A 1986 On the theory of testing for unit roots in observed time series *The Review of Economic Studies* **53** (3) pp 369 – 384 DOI:10.2307/2297634 .

*672.* Farmer R E 1986 Deficits and cycles *Journal of Economic Theory* **40** pp 77 – 86.

*673.* Kitwood T 1 January 1986 Long waves in economic life: An image without a method *Journal of Interdisciplinary Economics* vol **1** issue 2 pp 107 – 125 ISSN: 0260-1079
http://journals.sagepub.com/doi/pdf/10.1177/02601079X8600100206 .

*674.* Mosekilde E, Rasmussen St April 1986 Technical economic succession and the economic long wave *European Journal of Operational Research* vol **25** issue 1 pp 27 – 38
https://doi.org/10.1016/0377-2217(86)90111-6 .

*675.* Rasmussen St, Mosekilde E, Holst J 16 October 1989 Empirical indication of economic long waves in aggregate production *European Journal of Operational Research* vol **42** issue 3 pp 279 – 293.

*676.* Mosekilde E 1996-1997 Private communications on economic waves *Technical University of Denmark* Lyngby Denmark.

*677.* Reichlin P 1986 Equilibrium cycles in an overlapping generations economy with production *Journal of Economic Theory* **40** pp 89 – 102.

*678.* Romer P 1986 Increasing returns and long-run growth *Journal of Political Economy* **94** pp 1002 – 1037.

*679.* Romer P 1990 Endogenous technological change *Journal of Political Economy* **98** pp S71 – S102.

*680.* Romer P February 1991 The cyclical behaviour of industrial production, 1889-1984 *Quarterly Journal of Economics* vol **106** pp 1 – 31.

**681.** Romer P September 1994 Re-measuring business cycles *Journal of Economic History* vol **54** pp 573 – 609.

**682.** Romer P Spring 1999 Changes in business cycles: Evidence and explanations *Journal of Economic Perspectives* vol **13** pp 23 – 44.

**683.** Semmler W 1986 On nonlinear theories of economic cycles and the persistence of business cycles *Mathematical Social Sciences* vol **12** issue 1 pp 47 – 76.

**684.** Solomou S 1986a Innovation clusters and Kondratieff long waves in economic growth *Cambridge Journal of Economics* vol **10** issue 2 pp 101 – 112.

**685.** Solomou S 1986b Non-balanced growth and Kondratieff waves in the world economy, 1850-1913 *Journal of Economic History* **46** pp 165 – 169.

**686.** Solomou S 1987, 1989 Phases of economic growth, 1850-1973. Kondratieff waves and Kuznets swings *Cambridge University Press* Cambridge UK.

**687.** Solomou S 1998 Economic cycles: Long cycles and business cycles since 1870 *Manchester University Press* Manchester UK.

**688.** Staley C E 1986 Schumpeter's business cycles *New York Economic Review* **XVI** pp 300 – 313.

**689.** Summers L H 1986 Some skeptical observations on real business cycle theory *Federal Reserve Bank of Minneapolis Quarterly Review* **10** (Fall) pp 23 – 27.

**690.** Summers L H 2005 What caused the great moderation? Some cross country evidence *Federal Reserve Bank of Kansas City Economic Review* Third Quarter pp 5 – 30.

**691.** Summers L H 2014 Reflections on the new secular stagnation hypothesis *in* Secular stagnation: Facts, causes and cures Teulings C, Baldwin R (editors) *CERP Press* www.VoxEU.org .

**692.** Engle R F, Granger C W J 1987 Co-integration, and error correction: Representation, estimation and testing *Econometrica* **55** (2) pp 251 – 276 JSTOR 1913236.

**693.** Kennedy P 1987 The rise and fall of the great powers. Economic change and military conflict from 1500–2000 *Random House* New York USA.

**694.** Kirk R 1987 Are business services immune to the business cycle? Growth and change vol **18** issue 2 pp 15 – 23.

**695.** Klimenko L, Menshikov S 1987 Catastrophe theory applied to the analysis of the long wave *in* The long wave debate Vasko T (editor) Ch 25 *Springer-Verlag* Berlin Germany pp 345 – 358.

**696.** Mager N 1987 The Kondratieff waves *Praeger* New York USA.

**697.** Nakicenovic N 1987 Technological substitution and long waves in the USA *in* The long-wave debate Vasko T (editor) *Springer* Berlin Germany.

**698.** Puu T 1987 Complex dynamics in continuous models of the business cycles *Lecture Notes in Economics and Mathematical Systems Springer-Verlag* Berlin Germany.

**699.** Puu T 2001 Non-linear economic dynamics *Udmurt University* Izhevsk Russian Federation.

**700.** Ritzen J M M 1987 Human capital and economic cycles *Economics of Education Review* vol **6** issue 2 pp 151 – 160.

701. Tremblay R December 28 1987 The discipline of economics and economic cycles *Département de sciences économiques* Université de Montréal Canada, *Annual Meeting of North American Economics and Finance Association* Chicago Illinois USA ISSN 0709-9231 pp 1 – 26.

702. Volland C S A 1987 A comprehensive theory of long wave cycles *Technological Forecasting and Social Change* vol **32** no 2 pp 123 – 145.

703. Alesina A 1988 Macroeconomics and politics *NBER Macroeconomics Annual* pp 13 – 62.

704. Alesina A 1989 Politics and business cycles in industrial democracies *Economic Policy* **4** (8) pp 57 – 98.

705. Alesina A, Tabellini G 1990 A positive theory of fiscal deficits and government debt *The Review of Economic Studies* **57** (3) pp 403 – 414.

706. Alesina A, Ozler S, Roubini N, Swagel P 1992 Political Instability and Economic Growth *NBER Working Paper no 4173* NBER USA.

707. Alesina A, Perotti R 1994 The political economy of budget deficits *Discussion Paper* National Bureau of Economic Research USA.

708. Alesina A, Rodrik D 1994 Distributive politics and economic growth *Quarterly Journal of Economics* **109** pp 465 – 490.

709. Alesina A, Ozler S, Roubini N, Swagel Ph June 1996 Political instability and economic growth *Journal of Economic Growth* **1** (2) pp 149 – 187.

710. Alesina A, Roubini N, Cohen G 1997 Political cycles and the macroeconomy *MIT Press* USA.

711. Ben-Porath Y April 1988 Simon Kuznets in person and in writing *Economic Development and Cultural Change* vol **36** no 3 pp 435 – 447.

712. Cochrane J H 1988 How big is the random walk in GNP? *The Journal of Political Economy* vol **96** issue 5 pp 893 – 920.

713. Cuadrado-Roura J R, Rubalcaba-Bermejo L 1998 The growth of business services and the economic cycle *Department of Applied Economy* University of Alcalá Madrid Spain *38$^{th}$ Congress of the European Regional Science Association* Vienna Austria pp 1 – 36.

714. Greenwald B 1988 Examining alternative macroeconomic theories *Brookings Papers on Economic Activity* **1** pp 207 – 270.

715. Greenwald B 1993 Financial market imperfections and business cycles *Quarterly Journal of Economics* **108** (1) pp 77 – 114.

716. Greenwood J, Hercowitz Z, Huffman G W June 1988 Investment, capacity utilization, and the real business cycle *American Economic Review* **78** (3) pp 402 – 417.

717. Greenwood J, Hercowitz Z, Krusell P 2000 The role of investment-specific technological change in the business cycle *European Economic Review* **44** pp 91 – 115.

718. Schön L 1988 Historiska nationalräkenskaper för Sverige: Industri och hantverk 1800 - 1980 *Ekonomisk-Historiska Institutionen* Lunds Universitet Sweden.

719. Schön L 1989 From war economy to state debt policy *The National Debt Office* Stockholm Sweden.

**720.** Schön L 1991 Development blocks and transformation pressure in a macro-economic perspective - a model of long-term cyclical change *Skandinaviska Enskilda Banklen Quarterly Review* vol **XX** (3-4) pp 67 – 76.

**721.** Schön L 1994 Omvandling och obalans. Mönster i svensk ekonomisk utveckling. Bilaga 4 till Långtidsutredningen 1995 *Finansdepartementet* Stockholm Sweden.

**722.** Schön L 1998 Industrial crises in a model of long cycles; Sweden in an international perspective *in* Myllyntaus T (editor) Economic crises and restructuring in history *Scripta Mercaturae Verlag* Katharinen Sweden.

**723.** Schön L 2000 Electricity, technological change and productivity in Swedish industry 1890-1990 *European Review of Economic History* vol **V** (2) pp 175 – 194.

**724.** Schön L 2004 Total factor productivity in Swedish manufacturing in the period 1870-2000 *in* Exploring economic growth: Essays in measurement and analysis: A festschrift for Riitta Hjerppe on her sixtieth birthday Heikkinen S, Van Zanden J L (editors) *Aksant* Amsterdam The Netherlands.

**725.** Schön L 2009 Technological waves and economic growth - Sweden in an international perspective 1850-2005 *Paper no 2009/06* Centre for Innovation, Research and Competence in the Learning Economy (CIRCLE) Lund University Lund Sweden ISSN 1654-3149 pp 1 – 43

http://www.circle.lu.se/publications .

**726.** Stock J H, Watson W 1988a Testing for common trends *Journal of the Royal Statistical Association* **83** no 404 pp 1097 – 1107.

**727.** Stock J H, Watson W 1988b Variable trends in economic time series *Journal of Economic Perspectives* **2** pp 147 – 174.

**728.** Stock J H, Watson M 1989 New indexes of coincident and leading indicators *in* NBER macroeconomic annual **4** Blanchard O, Fischer S (editors) pp 351 – 394.

**729.** Stock J H, Watson M 1993 A procedure for predicting recessions with leading indicators: Econometric issues and recent performance *in* Business cycles, indicators, and forecasting Stock J, Watson M (editors) *University of Chicago Press* Chicago IL USA.

**730.** Stock J H, Watson M September 2002 Has the business cycle changed and why? *NBER Working Paper no 9127*, *NBER Macroeconomics Annual* NBER USA.

**731.** Stock J H, Watson M W 2004 Understanding changes in international business cycle dynamics *NBER Working Paper no 9859* NBER USA.

**732.** Street J H June 1988 The contribution of Simon S Kuznets to institutionalist development theory *Journal Economic Issues* vol **22** no 2 pp 499 – 509.

**733.** Blanchard O, Fischer S 1989 Lectures on macroeconomics *MIT* Cambridge USA.

**734.** Hamilton J D 1989 A new approach to the economic analysis of nonstationary time series and the business cycles *Econometrica* **57** pp 357 – 384.

**735.** Hamilton J D 1994 Time series analysis *Princeton University Press* Princeton USA.

**736.** İmrohoroglu A 1989 Cost of business cycles with indivisibilities and liquidity constraints *Journal of Political Economy* **97** (6) pp 1364 – 1383.

**737.** Mankiw N G March 1989 Real business cycles: A new Keynesian perspective *Working Paper no 2882* National Bureau of Economic Research USA, *Journal of Economic Perspectives* **3** (Summer) pp 79 – 90.

**738.** Ploser Ch 1989a Money and business cycles: A real business cycle interpretation *RCER Working Paper* Center for Economic Research (RCER) University of Rochester USA.

**739.** Ploser Ch 1989b Understanding real business cycles *RCER Working Paper* Center for Economic Research (RCER) University of Rochester USA.

**740.** Shiller R 1989 Market volatility *MIT Press* Cambridge MA USA.

**741.** Stewart H 1989 Recollecting the future: A view of business, technology and innovation in the next 30 years *Dow Jones-Irwin* USA.

**742.** Tarascio V J 1989 Economic and war cycles *History of Political Economy* vol **21** issue 1 pp 91 – 101.

**743.** West M, Harrison J S 1989 Bayesian forecasting and dynamic models *Springer-Verlag* New York USA.

**744.** Harrison J S, Berman Sh L 2016 Corporate social performance and economic cycles *Journal of Business Ethics* vol **138** issue 2 pp 279 – 294.

**745.** Ayres R U 1990a Technological transformations and long waves, Part I *Technological Forecasting and Social Change* vol **37** no 1 pp 1 – 37.

**746.** Ayres R U 1990b Technological Transformations and Long Waves, Part II *Technological Forecasting and Social Change* vol **37** no 2 pp 111 – 137.

**747.** Ayres R U 2006 Did the fifth K-Wave begin in 1990-92? Has it been aborted by globalization? *in* Kondratieff waves, warfare and world security Devezas T C (editor) *IOS Press* Amsterdam The Netherlands pp 57 – 71.

**748.** Boldrin M, Woodford M 1990 Equilibrium models displaying endogenous fluctuations and chaos *Journal of Monetary Economics* **25** pp 189 – 222.

**749.** Boldrin M, Christiano L, Fisher J 1995 Asset pricing lessons for modeling business cycles *Working Paper* NBER USA.

**750.** Boldrin M September 2000 Growth cycles and market crashes *Research Department Staff Report 279* Federal Reserve Bank of Minneapolis pp 1 – 34.

**751.** Boschan C, Banerji A 1990 A reassessment of composite indexes *in* Analyzing modern business cycles Klein P A, Sharpe M E (editor) New York USA.

**752.** Dua P, Banerji A 1999 An index of coincident economic indicators for the Indian economy *Journal of Quantitative Economics* **15** pp 177 – 201.

**753.** Dua P, Banerji A 2004a Monitoring and predicting business and growth rate cycles in the Indian economy *in* Business cycles and economic growth: An analysis using leading indicators Dua P (editor) *Oxford University Press* Oxford UK.

**754.** Dua P, Banerji A 2004b Economic indicator approach and sectoral analysis: Predicting cycles in growth of Indian exports *in* Business cycles and economic growth: An analysis using leading indicators Dua P (editor) *Oxford University Press* Oxford UK.

**755.** Layton A P, Banerji A 2004 Dating business cycles: Why output alone is not enough business cycles and economic growth: An analysis using leading indicators Dua P (editor) *Oxford University Press* Oxford UK.

**756.** Dua P, Banerji A August 2006 Business cycles in India *Working Paper no 146* Centre for Development Economics Department of Economics Delhi School of Economics Delhi India pp 1 – 15.

**757.** Danthine J-P, Donaldson J 1990 Efficiency wages and the business cycle puzzle *European Economic Review* **34** pp 1275 – 1301.

**758.** Danthine J-P, Neftci S 1990 Business cycles as nonlinear phenomena: Characterizing Swiss and German cycles 1965-1988 *Cahiers de Recherches Economiques du Département d'Econométrie et d'Economie politique (DEEP)* Faculté des HEC DEEP Université de Lausanne Switzerland.

**759.** Danthine J-P, Donaldson J B 1991 Risk sharing in the business cycle *Cahiers de Recherches Economiques du Département d'Econométrie et d'Economie politique (DEEP)* Faculté des HEC DEEP Université de Lausanne Switzerland.

**760.** Danthine J P, Donaldson J B 1993 Methodological and empirical issues in real business cycle theory *European Economic Review* **37** pp l – 35.

**761.** Danthine J-P, Donaldson J 1995 Non-walrasian economies *in* Frontiers of business cycle research Cooley T (editor) *Princeton University Press* USA.

**762.** Escudier J L 1990 Long–term movement of the economy: Terminology and theoretical options *in* Life cycles and long waves Barnard S (translator) Vasco T, Ayres R, Fontvieille L (editors) Chapter 14 *Springer-Verlag* Berlin Germany pp 239 – 260; *in* The foundations of long wave theory Louçã F, Reijnders J (editors) vol **II** *Edward Elgar Publishing* Cheltenham UK.

**763.** Grubler A, Nowotny H 1990 Towards the fifth Kondratiev upswing: Elements of an emerging new growth phase and possible development trajectories *International Journal of Technological Management* **5** (4) pp 431 – 471.

**764.** Hedtke U 1990 Stalin oder Kondratieff *Dietz* Berlin Germany.

**765.** Jovanovic B, Rob R 1990 Long waves and short waves: Growth through intensive and extensive search *Econometrica* **58** (6) pp 1391 – 1409.

**766.** Kontorovich V 1990 Utilization of fixed capital and Soviet industrial growth *Economics of Planning* **23** pp 37 – 50.

**767.** Reijnders J P G 1990 Long waves in economic development *Edward Elgar Publishing* Cheltenham UK.

**768.** Reijnders J P G 2009 Trend movements and inverted Kondratieff waves in the Dutch economy, 1800 - 1913 *Structural Change and Economic Dynamics* vol **20** issue 2 pp 90 – 113.

**769.** Rogoff K 1990 Equilibrium political budget cycles *The American Economic Review* pp 21 – 36.

**770.** Baxter M, King R G 1991 Productive externalities and business cycles *Discussion Paper 53* Institute for Empirical Macroeconomics Federal Reserve Bank of Minneapolis.

**771.** Baxter M, King R G 1999 Measuring business cycles: Approximate band-pass filters for economic time series *Review of Economics and Statistics* **81** (4) pp 575 – 593.

**772.** Berry B J L 1991 Long-wave rhythms in economic development and political behaviour *Johns Hopkins University Press* Baltimore MD USA.

**773.** Berry B J L, Kim H 1993 Are long waves driven by techno-economic transformations? Evidence for the US and the UK *Technological Forecasting and Social Change* vol **44** no 2 pp 111 – 135.

**774.** Berry B J L, Kim H 1994 Leadership generations: A long-wave macro-history *Technological Forecasting and Social Change* vol **46** no 1 pp 1 – 9.

**775.** Berry B J L 2000 A pacemaker for the long wave *Technological Forecasting and Social Change* vol **63** no 1 pp 1 – 23.

**776.** Diebold F X, Rudebusch G 1991 Have postwar economic fluctuations been stabilized? *Working Paper no 116* Economic Activity Section Board of Governors of the Federal Reserve System NY USA.

**777.** Bangia A, Diebold F X, Kronimus A, Schlagen C, Schuermann T 2002 Rating migrations and business cycle with application to credit portfolio stress testing *Journal of Banking and Finance* **26** pp 445 – 474.

**778.** Diebold F X 2004 Elements of forecasting *Thompson* Louseville Canada.

**779.** Aruoba S, Diebold F X, Kose M, Terrones M 2011 Globalization, the business cycle, and macroeconomic monitoring *in* Clarida R, Giavazzi F (editors) NBER international seminar on macroeconomics *University of Chicago Press* Chicago USA pp 245 – 302.

**780.** French M W, Sichel D 1991, 1993 Cyclical patterns in the variance of economic activity *Finance and Economics Discussion Series Paper no 161* Board of Governors of the Federal Reserve System NY USA.

**781.** Grossman G M, Helpman E 1991 Innovation and growth in the global economy *MIT Press* Cambridge USA.

**782.** Jang-Ok Cho, Cooley Th 1991, 1995 The business cycle with nominal contracts *RCER Working Paper* Center for Economic Research (RCER) William E Simon Graduate School of Business Administration University of Rochester USA.

**783.** Jang-Ok Cho, Cooley Th, Hyung Seok Kim 2015 Business cycle uncertainty and economic welfare *Review of Economic Dynamics* vol **18** issue 2 pp 185 – 200.

**784.** Levy J S 1991 Long cycles, hegemonic transitions, and the long peace *in* The long postwar peace Kegley C W Jr (editor) *HarperCollins* New York USA.

**785.** Madaràsz A 1991 Schumpeter's theory of economic development *in* J A Schumpeter: Critical assessments Cunningham Wood J (editor) vol **III** pp 218 – 240 *Routledge* London.

**786.** Schubert A 1991 The credit-anstalt crisis of 1931 *Cambridge University Press* New York USA.

**787.** Scott M F 1991 A new view of economic growth *Oxford University Press* Oxford UK.

**788.** Sterken E 1991 Models of the business cycle; A survey *Working Paper* Groningen State Institute of Economic Research Groningen The Netherlands.

**789.** Thio K B T 1991 On simultaneous explanation of long and medium-term employment cycles *De Economist* **139** pp 331 – 357.

**790.** Pope D 1991 Long waves and crash of 87' *The Economic Record* vol **67** issue 197 pp 158 – 167.

**791.** Aghion Ph, Howitt P A 1992 Model of growth through creative destruction *Econometrica* **60** (2) pp 323 – 351.

**792.** Aghion Ph, Howitt P 1998 Endogenous growth theory *MIT Press* Cambridge Massachusetts USA.

*793.* Aghion Ph, Askenazy P, Berman N, Cette G, Eymard L 2012 Credit constraints and the cyclicality of R&D investment: Evidence from France *Journal of the European Economic Association* vol **10** no 5 pp 1001 – 1024.

*794.* Aghion Ph, Akcigit U, Howitt P 2014 What do we learn from Schumpeterian growth theory? *Handbook of Economic Growth* vol **2** pp 515 – 563.

*795.* Backus D K, Kehoe P, Kydland F 1992 International real business cycles *Journal of Political Economy* **100** pp 745 – 775.

*796.* Backus D K, Kehoe P 1992 International evidence on the historical properties of business cycles *American Economic Review* **82** pp 864 – 888.

*797.* Backus D K, Kehce P J, Kydland F E 1993, 1995 International business cycles: Theory and evidence *NBER Working Papers 4493* National Bureau of Economic Research Inc, *in* Frontiers of business cycle research Cooley T F (editor) *Princeton University Press* Princeton New Jersey USA.

*798.* Christiano L J, Eichenbaum M 1992 Current real-business-cycle theories and aggregate labor-market fluctuations *The American Economic Review* **82** (3) pp 430 – 450.

*799.* Christiano L J, Fitzgerald T J 1998 The business cycle is still a puzzle *Economic Perspectives* Federal Reserve Bank of Chicago vol **22** no 4 pp 56 – 83.

*800.* Christiano L J, Eichenbaum M, Evans C 2002 Nominal rigidities and the dynamic effects of a shock to monetary policy *mimeo*.

*801.* Christiano L J, Fitzgerald T J 2003 The band-pass filter *International Economic Review* **44** pp 435 – 465.

*802.* Christiano L J, Eichenbaum M, Vigfusson R 2004 The response of hours to a technology shock: Evidence based on direct measures of technology *Journal of the European Economic Association* **2** (2-3) pp 381 – 395.

*803.* Christiano L J, Ilut C, Motto R, Rostagno M 2007 Monetary policy and stock market boom-bust cycles *Manuscript* Northwestern University USA, European Central Bank EU.

*804.* Christiano L J, Motto R, Rostagno M 2010 Financial factors in economic fluctuations *Working Paper* European Central Bank.

*805.* Gilles S-P October 1992 Productivity growth and the structure of the business cycle *Centre for Economic Policy Research*

*806.* http://cepr.org/active/publications/discussion_papers/dp.php?dpno=709 .

*807.* Kenwood A, Lougheed A 1992 The growth of the international economy, 1820-1990 *Routledge* London UK.

*808.* Laidler D 1992 Hayek on neutral money and the cycle *Working Paper* Department of Economics University of Western Ontario Canada.

*809.* Mellander E, Vredin A, Warne A 1992 Stochastic trends and economic fluctuations in a small open economy *Journal of Applied Econometrics* **7** pp 369 – 394.

*810.* Merrill B, Szidarovszky F 1992 Limit cycles in dynamic economic models *Pure Mathematics and Applications* vol **3** issue 1 – 4 pp 25 – 33.

*811.* Metz R 1992 Re-examination of long waves in aggregate production series *New Findings in Long Wave Research* Kleinknecht A, Mandel E, Wallerstein I (editors) *St. Martin's* New York USA pp 80 – 119.

*812.* Metz R 1998 Langfristige wachstumsschwankungen – Trends, zyklen, strukturbrüche oder zufall Kondratieffs *Zyklen der Wirtschaft. An der Schwelle neuer Vollbeschäftigung?* Thomas H, Nefiodow L A (editors) Herford pp 283 – 307.

*813.* Metz R 2006 Empirical evidence and causation of Kondratieff cycles *Kondratieff Waves, Warfare and World Security* Devezas T C (editor) *IOS Press* Amsterdam The Netherlands pp 91 – 99.

*814.* Phelan Ch 1992 Incentive, inequality and business cycle *Working Paper* University of Wisconsin-Madison USA.

*815.* Reati A 1992 Are we at the eve of a new long-term expansion induced by technological change? *International Review of Applied Economics* vol **6** no 3 pp 249 – 285.

*816.* Reati A 1998 A long-wave pattern for output and employment in Pasinetti's model of structural change *Economie Appliquée* tome **LI** no 2 pp 27 – 75.

*817.* Reati A, Toporowski J December 2004 An economic policy for the fifth long wave *BNL Quarterly Review* no 231 pp 395 – 437.

*818.* Sowell F 1992 Modeling long run behavior with the fractional ARIMA model *Journal of Monetary Economics* **29** pp 277 – 302.

*819.* Tylecote A 1992 The long wave in the world economy: The present crisis in historical perspective *Routledge* London UK.

*820.* Tylecote A 1994 Long waves, long cycles, and long swings *Journal of Economic Issues* vol **XXVIII** no 2 pp 477 – 488.

*821.* Atkeson A, Kehoe P 1993 Evolution and transition: The role of informational capital *Federal Reserve Bank of Minneapolis Staff Report #162*.

*822.* Atkeson A, Phelan C 1994 Reconsidering the costs of business cycles with incomplete markets *NBER Macroeconomics Annual 1994* Fischer St, Rotemberg J J (editors) *MIT Press* Cambridge Massachusetts USA / London England.

*823.* Bowles S, Edwards R 1993 Understanding capitalism: Competition, command, and change in the U.S. economy 2nd edition *Harper Collins* New York USA.

*824.* Bruno M 1993 Crisis, stabilization, and economic reform: Therapy by consensus *Clarendon Press* Oxford UK.

*825.* Bruno M, Easterly W 1998 Inflation crises and long-run growth *Journal of Monetary Economics* **41** (1) pp 3 – 26.

*826.* Chari V V, Christiano L J, Kehce P J 1993 Optimal fiscal policy in a business cycle model *Federal Reserve Bank of Minneapolis Staff Report no 160*.

*827.* Dolado J, Sebastián M, Vallés J Septiembre 1993 Cyclical patterns of Spanish economy *Investigaciones Económicas* vol **XVII** (3).

*828.* Fischer St 1993 The role of macroeconomic factors in growth *Journal of Monetary Economics* **32** pp 485 – 512.

*829.* Goodwin Th 1993 A business-cycle analysis with a Markov-switching model *Journal of Business and Economic Statistics* vol **11** (3) pp 331 – 339.

*830.* Labini P S 1993 Economic growth and business cycles prices and the process of cyclical development *Edward Elgar Publishing* ISBN: 978 1 85278 833 9 pp 1 – 256.

*831.* Saint-Paul G 1993 Productivity growth and the structure of the business cycle *European Economic Review* **37** pp 861 – 890.

*832.* Silverberg G, Lehnert D 1993 Long waves and 'evolutionary chaos' in a simple Schumpeterian model of embodied technical change *Structural Change and Economic Dynamics* **4** pp 9 – 37.

*833.* Silverberg G, Lehnert D 1996 Evolutionary chaos: Growth fluctuations in a Schumpeterian model of creative destruction *in* Nonlinear dynamics in economics Barnett W, Kirman A, Salmon M (editors) *Cambridge University Press* Cambridge UK.

*834.* Silverberg G, Verspagen B 1996 From the artificial to the endogenous: Modelling evolutionary adaptation and economic growth *in* Behavioral norms, technological progress and economic dynamics: Studies in Schumpeterian economics Helmstädter E, Perlman M (editors) *University of Michigan Press* Ann Arbor MI USA.

*835.* Silverberg G 2003 Long waves: Conceptual, empirical and modelling issues *MERIT Research Memorandum RM2003-15* Maastricht The Netherlands, *in* The Elgar companion to neo-Schumpeterian economics Hanusch H, Pyka A (editors) *Edward Elgar Publishing* Aldershot http://www.merit.unimaas.nl/publications/rmpdf/2003/rm2003-015.pdf .

*836.* Silverberg G, Verspagen B 2003a Breaking the waves: A Poisson regression approach to Schumpeterian clustering of basic innovations *Cambridge Journal of Economics* **27** pp 671 – 693.

*837.* Silverberg G, Verspagen B 2003b Long memory in the world economy since the 19[th] century *in* Long memory and its applications Rangarajan G, Ding M (editors) *Springer* Berlin Germany.

*838.* Silverberg G 2005 When is a wave a wave? Long waves as empirical and theoretical constructs from a complex systems perspective *MERIT-Infonomics Research Memorandum Series Paper 2005-016* MERIT – Maastricht Economic Research Institute on Innovation and Technology Maastricht The Netherlands pp 1 – 12.

*839.* Silverberg G 2006 Long waves: Conceptual, empirical and modelling issues *in* The Elgar companion to neo-Schumpeterian economics Hanusch H, Pyka A (editors) *Edward Elgar Publishing* Cheltenham UK.

*840.* Stiglitz J E 1993 Endogenous growth and cycles *NBER Working Paper no 4286* NBER USA.

*841.* Stiglitz J E 1994a Economic growth revisited *Industrial and Corporate Change* **3** (1).

*842.* Stiglitz J E 1994b Endogenous growth and cycles *in* Innovation in technology, industries, and institutions Shionoya Y, Perlman M (editors) *The University of Michigan Press* Ann Arbor USA pp 121 – 156.

*843.* Stiglitz J E 1999 Economic consequences of income inequality *1998 Symposium Proceedings: Income Inequality: Issues and Policy Options* pp 221 – 263.

*844.* Stiglitz J E 2002 Globalization and its discontents *W W Norton* New York USA.

*845.* Stiglitz J E June 28 2011 Re-thinking macroeconomics: What failed, and how to repair it? *Journal of the European Economic Association* vol **9** issue 4 pp 591 – 645.

*846.* Warne A 1993 A common trend model: Identification, estimation and inference mimeo Institute for International Economic Studies Stockholm University Sweden.

**847.** Bils M, Cho J O 1994 Cyclical factor utilization *Journal of Monetary Economics* **33** pp 319 – 354.

**848.** Bils M, Klenow P 1998 Using consumer theory to test competing business cycle models *Journal of Political Economy* **106** pp 233 – 261.

**849.** Crafts N 1994 The industrial revolution *in* The economic history of Britain since 1700 Flood R, McCloskey D (editors) *Cambridge University Press* New York USA.

**850.** Kashyap A K, Stein J C, Lamont O A August 1994 Credit conditions and the cyclical behavior of inventories *The Quarterly Journal of Economics* vol **109** pp 565 – 592.

**851.** Kim Ch-J 1994 Dynamic linear models with Markov-switching *Journal of Econometrics* **60** pp 1 – 22.

**852.** Kim Ch-J, Yoo J S 1995 New index of coincident indicators: A multivariate Markov switching factor model approach *Journal of Monetary Economics* **36** pp 607 – 630.

**853.** Kim Ch-J, Nelson C 1998 Business cycle turning points, a new coincident index, and tests of duration dependence based on a dynamic factor model with regime switching *Review of Economics and Statistics* **80** pp 188 – 201.

**854.** Kim Ch-J, Nelson Ch 1999a Has the U.S. economy become more stable? A Bayesian approach based on a Markov-switching model of the business cycle *Review of Economics and Statistics*.

**855.** Kim Ch-J, Nelson Ch 1999b Friedman's plucking model of business fluctuations: Tests and estimates of permanent and transitory components *Journal of Money, Credit, and Banking* **31** pp 317 – 334.

**856.** Kim Ch-J, Piger J 2002 Common stochastic trends, common cycles, and asymmetry in economic fluctuations *Journal of Monetary Economics* **49** pp 1189 – 1211.

**857.** Kim Ch-J, Morley J, Piger J 2002 Nonlinearity and the permanent effects of recessions *Federal Reserve Bank of St Louis Working Paper 2002-014*.

**858.** McGrattan E 1994a A progress report on business cycle models *Quarterly Review Federal Reserve Bank of Minneapolis* Fall 1994 pp 2 – 15.

**859.** McGrattan E 1994b The macroeconomic effects of distortionary taxation *Journal of Monetary Economics* **33** pp 573 – 601.

**860.** Barro R J, Sala-I-Martin X 1995 Economic growth *McGraw-Hill* New York USA.

**861.** Barro R J 1996 Democracy and growth *Journal of Economic Growth* **1** (1) pp 1 – 28.

**862.** Barro R J 1997 The determinants of economic growth *MIT Press* Cambridge USA.

**863.** Burnside C, Eichenbaum M, Rebelo S 1995 Capital utilization and returns to scale *NBER Macroeconomics Annual 1995 MIT Press* Cambridge MA USA pp 67 – 110.

**864.** Burnside C, Eichenbaum M, Rebelo S 1996 Sectoral Solow residuals *European Economic Review* **40** pp 861 – 869.

**865.** Burnside C, Eichenbaum M 1996 Factor hoarding and the propagation of business cycle shocks *American Economic Review* **86** pp 1154 – 1174.

**866.** Calomiris C, Himmelberg C, Wachtel P 1995 Commercial paper, corporate finance, and the business cycle: A microeconomic perspective *Carnegie-Rochester Conference Series on Public Policy* **42** pp 203 – 250.

867. Cogley Th F, Nason J 1995 Effects of the Hodrick-Prescott filter on trend and difference stationary time series: Implications of business-cycle research *Journal of Economic Dynamics and Control* **146** pp 155 – 178.

868. Cooley Th F (editor) 1995 Frontiers of business cycle research *Princeton University Press* Princeton New Jersey USA ISBN 0-691-04323-X.

869. Cooley Th F, Prescott E C 1995 Economic growth and business cycles *in* Frontiers of business cycle research Cooley Th F (editor) *Princeton University Press* Princeton New Jersey USA.

870. Kapuria-Foreman V, Perlman M November 1995 An economic historian's economist: Remembering Simon Kuznets *The Economic Journal* **105** pp 1524 – 1547.

871. Pedregal D J 1995 Comparación Teórica, Estructural y Predictiva de Modelos de Componentes no Observables y Extensiones del Modelo de Young *PhD Thesis* Universidad Autónoma de Madrid Madrid Spain.

872. Pedregal D J, Young P C 1996 Modulated cycles, a new approach to modelling seasonal/cyclical behaviour in unobserved component models *Centre for Research on Environmental Systems and Statistics (CRES) Technical Note no 145*.

873. Pedregal D J 2001 Analysis of economic cycles using unobserved components models *Review on Economic Cycles* **II** pp 77 – 92.

874. Pedregal D J, Young P C 2001 Some comments on the use and abuse of the Hodrick-Prescott filter *Review on Economic Cycles* **II** pp 93 – 104.

875. Uhlig H, Xu Y 1995 Effort and the cycle: Cyclical implications of efficiency wages *Working Paper* CentER Tilburg The Netherlands.

876. Witt U 1995 Schumpeter vs Hayek: Two approaches to evolutionary economics *in* New perspectives on Austrian economics Meijer G (editor) *Routledge* London UK pp 81 – 101.

877. Akerlof G, Dickens W, Perry G 1996 The macroeconomics of low inflation *Brookings Papers on Economic Activity* **1** pp 1 – 76.

878. Arena R, Festré A 1996 Banks, credit and the financial system in Schumpeter: An interpretation *in* Joseph Schumpeter, historian of economics Moss L (editor) *Routledge* London UK.

879. Bernard H, Gerlach St 1996 Does the term structure predict recessions? The international evidence *Working Paper no 37* Bank for International Settlements Switzerland.

880. Cheng L K, Dinopoulos E 1996 A multisector general equilibrium model of Schumpeterian growth and fluctuations *Journal of Economic Dynamics and Control* **20** pp 905 – 923.

881. Crafts N, Mills T 1996 Europe's golden age: An econometric investigation *in* Quantitative aspects of postwar economic growth van Ark B, Crafts N (editors) *Cambridge University Press* Cambridge UK.

882. Crafts N, Toniolo G 1996 Postwar growth: An overview *in* Economic growth in Europe since 1945 Crafts N, Toniolo G (editors) *Cambridge University Press* Cambridge UK pp 1 – 37.

883. Davis S J, Haltiwanger J C, Schuh S 1996 Job creation and destruction *MIT Press* Cambridge USA.

**884.** De la Croix D, Deneulin S 1996 Relativit´e de la satisfaction et croissancé economique *Actes du 12` emecongrés des économistes belges de langue Francaise* **1** pp 153 – 169.

**885.** De la Croix D February 29 2000 Standard-of-living aspirations and economic cycles *National Fund for Scientific Research IRES* Universit´e Catholique de Louvain Louvain-la-Neuve Belgium pp 1 – 25.

**886.** Galor O 1996 Convergence? Inferences from theoretical models *The Economic Journal* **106** pp 1056 – 1069.

**887.** Haken H 1996 Synergetics as a bridge between the natural and social sciences: Evolution, order, and complexity *Routledge* London UK.

**888.** Justman M 1996 Swarming mechanics *Economics of Innovation and New Technologies* **4** pp 235 – 244.

**889.** Justman M 1997 Schumpeterian waves of disequilibrium growth: A general equilibrium analysis *Monaster Center for Economic Research Discussion Paper #97-14*.

**890.** Koop G, Pesaran M, Potter S 1996 Impulse response analysis in nonlinear multivariate models *Journal of Econometrics* **74** pp 119 – 147.

**891.** Lee J 1996 Do service temper business cycle?: Implications of the rising service sector *Department of Economics* University of California Irvine California USA.

**892.** McDonald I A 1996 Kaldorian model of the trade cycle *Working Paper no 529* Department of Economics The University of Melbourne Australia.

**893.** Quah D 1996 Empirics for economic growth and convergence *European Economic Review* **40** pp 1353 – 1375.

**894.** Silverberg G, Lehnert D 1996 Evolutionary chaos: Growth fluctuations in a Schumpeterian model of creative destruction *in* Nonlinear dynamics and economics Barnett W A, Kirman A P, Salmon M (editors) Ch 2 *Cambridge University Press* Cambridge UK pp 45 – 74.

**895.** Silverberg G, Verspagen B 2000 Breaking the waves: A Poisson regression approach to Schumpeterian clustering of basic innovations *Cambridge Journal of Economics* **27** pp 671 - 693.

**896.** Silverberg G 2002 The discrete charm of the bourgeoisie: Quantum and continuous perspectives on innovation and growth *Research Policy* **31** pp 1275 – 1289.

**897.** Zimmermann Ch 1996 A real business cycle bibliography *CREFE Working Paper no 43* Université du Québec Montréal Canada.

**898.** Argandoña A, Gamez C, Mochón F 1997 Macroeconomía avanzada II *McGraw-Hill* USA.

**899.** Artis M J, Zhang W 1997 International business cycles and the ERM: Is there a European business cycle? *International Journal of Finance and Economics* **38** pp 1471 – 1487.

**900.** Artis M J, Krolzig H-M, Toro J 2004 The European business cycle *Oxford Economic Papers* **56** pp 1 – 44.

**901.** Artis M J, Marcellino M, Proietti T 2004 Characterizing the business cycle for accession countries *IGIER Working Paper no 261*.

**902.** Balke N, Fomby T 1997 Threshold cointegration *International Economic Review* **38** no 3 pp 627 – 645.

**903.** Berman J, Pfleeger J 1997 Which industries are sensitive to business cycles? *Monthly Labor Review* vol **120** no 2 pp 19 – 25.

**904.** Bierens H 1997 Nonparametric cointegration analysis *Journal of Econometrics* **77** pp 379 – 404.

**905.** Budnevich C, Le Fort Varela G 1997 Fiscal policy and the economic cycle in Chile *Revista CEPAL.*

**906.** Canton E J F 1997 Economic growth and business cycles *Center for Economic Research* Tilburg University *Thesis Publishers* Amsterdam The Netherlands ISBN 90-5668-025-0 pp 1 – 205.

**907.** Gandolfo G 1997 Economic dynamics *Springer* Berlin Germany.

**908.** Harrod R 1997 Theory of economic dynamics *Economica* Moscow Russian Federation pp 48 – 49.

**909.** Kiyotaki N, Moore J 1997 Credit cycles *Journal of Political Economy* **105** pp 211 – 248.

**910.** Gertler M, Kiyotaki N 2009 Financial intermediation and credit policy in business cycle analysis *in* Handbook of monetary economics vol **3** Friedman B, Woodford M (editors) *North-Holland* The Netherlands.

**911.** Klenow P J, Rodriguez-Clare A 1997 The neoclassical revival in growth economics: Has it gone too far? *in* NBER Macroeconomics Annual 1997 vol **12** Bernanke B S, Rotemberg J (editors) *MIT Press* Cambridge Massachusetts USA pp 73 – 114.

**912.** Klepper S 1997 Industry lifecycles *Industrial and Corporate Change* vol **6** pp 145 – 181.

**913.** Krolzig H-M 1997 Markov-switching vector autoregression *Springer* Berlin Germany.

**914.** Krolzig H-M 1999 Statistical analysis of co-integrated VAR processes with Markovian regime shifts *University of Oxford* mimeo.

**915.** Krolzig H-M, Toro J 1999 A new approach to the analysis of shocks and the cycle in a model of output and employment *European University Institute Working Paper* no 99/30.

**916.** Krolzig H-M, Marcellino M, Mizon G 2002 A Markov-switching vector equilibrium correction model of the UK labor market *Empirical Economics* **27** pp 233 – 254.

**917.** Louçã F 1997 Turbulence in economics. An evolutionary appraisal of cycles and complexity in historical processes *Edward Elgar Publishing* Aldershot USA.

**918.** Louçã F 1999 Nikolai Kondratiev and the early consensus and dissensions about history and statistics *History of Political Economy* **31** (1) pp 169 – 206

**919.** http://www.users.qwest.net/~drakete/LoucaKondrat_.PDF.

**920.** Louçã F, Reijnders J 1999 The foundations of long wave theory: Models and methodology *Edward Elgar Publishing* Cheltenham UK pp 1 – 1104 ISBN: 9781858988429.

**921.** Neumann M 1997 The rise and fall of the wealth of nations – long waves in economics and international politics *Edward Elgar Publishing* Cheltenham UK.

*922.* Pagan A 1997 Towards an understanding of some business cycle characteristics *The Australian Economic Review* **30** pp 1 – 15.

*923.* Stein J 1997 Waves of creative destruction: Learning-by-doing and the dynamics of innovation *Review of Economic Studies* **64** pp 265 – 288.

*924.* Barnett V 1998a Kondratiev and the dynamics of economic development *Palgrave Macmillan* New York USA.

*925.* Barnett V 1998b Dating the long cycle turning points: Kondratiev and after *in* The works of Nikolai D Kondratiev Makasheva N, Samuels W J (editors) vol **I** *Pickering & Chatto* London UK.

*926.* Brenner R May-June 1998 Uneven development and the long downturn: The advanced capitalist economies from boom to stagnation, 1950-1998 *New Left Review* **229**.

*927.* Canova F 1998 Detrending and business cycles facts *Journal of Monetary Economics* **41** pp 475 – 512.

*928.* Chauvet M 1998 An econometric characterization of business cycle dynamics with factor structure and regime switches *International Economic Review* **39** pp 969 – 996.

*929.* Chauvet M 1999 Stock market fluctuations and the business cycle *Journal of Economic and Social Measurement* **25** pp 235 – 257.

*930.* Edison H J et al March 1998 Asset bubbles, domino effects and lifeboats: Elements of the East Asian crisis *Board of Governors of the Federal Reserve System IFDP #606* NY USA.

*931.* Estrella A, Mishkin F 1998 A predicting US recessions: Financial variables as leading indicators *The Review of Economics and Statistics* **80** pp 45 – 61.

*932.* Gowdy J, Mesner S Summer 1998 The evolution of Georgescu-Roegen's bioeconomics *Review of Social Economy* vol **LVI** no 2.

*933.* Helpman E, Trajtenberg M 1998 A time to sow and a time to reap: Growth based on general purpose technologies *in* General purpose technologies and economic growth Helpman E (editor) *MIT Press* Cambridge MA USA pp 55 – 83.

*934.* Israel D C November 1998 The environmental impacts of economic cycles *Discussion Paper Series no 98-43* Philippine Institute for Development Studies Makati City Philippine pp 1 – 35.

*935.* Kouparitsas M A 1998 Are business cycles different under fixed and flexible exchange rate regimes? *Economic Perspectives* Federal Reserve Bank of Chicago vol **22** no 1 pp 46 – 64.

*936.* Kouparitsas M A 1999 Is the EMU a viable common currency area? A VAR analysis of regional business cycles *Economic Perspectives* Federal Reserve Bank of Chicago vol **23** no 4 pp 2 – 20.

*937.* Kouparitsas M A 2000 Evidence of the North–South business cycle *Economic Perspectives* Federal Reserve Bank of Chicago vol **25** no 1 pp 46 – 59.

*938.* Kouparitsas M A December 2001 Is there a world business cycle? *Chicago Fed Letter* no 172 The Federal Reserve Bank of Chicago USA ISSN 0895-0164 pp 1 – 4.

*939.* Lee I H 1998 Market crashes and informational avalanches *Review of Economic Studies* **65** pp 741 – 760.

*940.* Peel D, Davidson J 1998 A non-linear error correction mechanism based on the bilinear model *Economics Letters* **58** pp 165 – 170.

*941.* Thomas H, Nefiodow L A (Hrsg) 1998 Kondratieffs Zyklen der Wirtschaft, An der Schwelle neuer Vollbeschäftigung? *Busse Seewald* Herford.

*942.* Volkmann H 1998 Projektionen zur Entwicklung der Informationsgesellschaft im fünften Kondratieff-Zyklus. Kondratieffs Zyklen der Wirtschaft? An der Schwelle neuer Vollbeschäftigung? Thomas H, Nefiodow L (editors) *Busse Seewald* Herford.

*943.* Anderson H M, Ramsey J B 1999 *Economic Research Reports PR # 99-01* New York University NY USA.

*944.* Basu S, Taylor A M 1999 Business cycles in international historical perspectives *Journal of Economic Perspective* vol **13** no 2 pp 45 – 68.

*945.* Gali J 1999 Technology, employment and the business cycle: Do technology shocks explain aggregate fluctuations? *American Economic Review* **89** (1) pp 249 – 271.

*946.* Gordon R J 1999 US economic growth since 1870: One big wave? *American Economic Review* **89** (2) pp 123 – 128 DOI: 10.1257/aer.89.2.123.

*947.* Jones Ch I 1999 Was an industrial revolution inevitable: Economic growth over the very long-run Stanford University California USA.

*948.* Jones Ch I, Klenow P J 2011 Beyond GDP? Welfare across countries and time *Manuscript* Stanford University California USA.

*949.* Koopman S J, Shephard N, Doornik J A 1999 Statistical algorithms for models in state space using SsfPack 2.2 *Econometrics Journal* **2** pp 113 – 166.

*950.* Koopman S J, Valle e Azevedo J 2004 Measuring synchronization and convergence of business cycles *Working Paper* Department of Econometrics *Free University* Amsterdam The Netherlands.

*951.* Lettau M, Uhlig H 1999 Can habit formation be reconciled with business cycle facts? *Review of Economic Dynamics*.

*952.* Ljungqvist L, Uhlig H 1999 Tax policy and aggregate demand management under catching up with the joneses *American Economic Review*.

*953.* Pollock D S G 1999 A handbook of time-series analysis, Signal processing and dynamics *Academic Press* London UK.

*954.* Pollock D S G April 2008 Investigating economic trends and cycles *Working Paper no 07/17* University of Leicester UK.

*955.* Quigley J M 1999 Real estate prices and economic cycles *International Real Estate Review* vol **2** no 1 pp 1 – 20.

*956.* Xiangkang Yin, Zuscovitch E October 1999 Interaction of drastic and incremental innovations: Economic development through Schumpeterian waves *La Trobe University* Bundoora Victoria Australia / BETA-CNRS Ben Gurion University of the Negev Israel pp 1 – 31.

*957.* Zeira J 1999 Informational overshooting, booms and crashes *Journal of Monetary Economics* **43** pp 237 – 258.

*958.* Alt J E, Lassen D 2000 Transparency, political polarization, and political budget cycles in OECD countries *American Journal of Political Science* **50** (3) pp 530 – 550.

*959.* Breschi S, Malerba F, Orsenigo L 2000 Technological regimes and Schumpeterian patterns of innovation *Economic Journal*.

**960.** Chin D, Geweke J, Miller P 2000 Predicting turning points *Federal Reserve Bank of Minneapolis Research Department Staff Report #267* USA.

**961.** Collard F, de la Croix D 2000 Gift exchange and the business cycle: The fair wage strikes back *Review of Economic Dynamics* **3** pp 166 – 193.

**962.** Den Haan W J 2000 The co-movement between output and prices *Journal of Monetary Economics* **46** pp 3 – 30.

**963.** Devezas T C 2000 The long wave phenomenon: Open questions and new insights http://www.unizar.es/sociocybernetics/absg11.html.

**964.** Devezas T C, Corredine J T 2001 The biological determinants of long-wave behavior in socioeconomic growth and development *Technological Forecasting & Social Change* **68** pp 1 – 57.

**965.** Devezas T C, Corredine J T 2002 The nonlinear dynamics of technoeconomic systems. An informational interpretation *Technological Forecasting & Social Change* **69** pp 317 – 357.

**966.** Devezas T C, Modelski G 2003 Power law behavior and world system evolution: A millennial learning process *Technological Forecasting and Social Change* **70** pp 819 – 859.

**967.** Devezas T C, Linstone H A, Santos H J S 2005 The growth dynamics of the internet and the long wave theory *Technological Forecasting and Social Change* vol **72** no 8 pp 913 – 935.

**968.** Devezas T C (editor) 2006 Kondratieff waves, warfare and world security *IOS Press* Amsterdam The Netherlands.

**969.** Devezas T C, Modelski G 2008 The Portugese as system-builders – Technological innovation in early globalization *in* Globalization as evolutionary process: Modelling global change *Taylor-Francis* pp 30 – 57.

**970.** Devezas T, Grinin L, Korotayev A (editors) 2012 Kondratieff waves *Uchitel* Volgograd Russian Federation.

**971.** Drazen A 2000 The political business cycle after 25 Years *NBER Macroeconomics Annual* pp 75 – 117.

**972.** Fogel R W 2000 Simon S. Kuznets: April 30, 1901 – July 9, 1985 *NBER Working Paper no W7787* NBER USA.

**973.** Fogel R W, Fogel E M, Guglielmo M, Grotte N 2013 Political arithmetic: Simon Kuznets and the empirical tradition in economics *University of Chicago Press* Chicago USA ISBN 0-226-25661-8.

**974.** Hornstein A 2000 The business cycle and industry co-movement *Federal Reserve Bank of Richmond* vol **86** no 1 pp 27 – 48.

**975.** Hossein-Zadeh I, Gabb A 1 September 2000 Making sense of the current expansion of the U.S. economy: A long wave approach and a critique *Review of Radical Political Economics* vol **32** issue 3 pp 388 – 397 ISSN: 0486-6134 Online ISSN: 1552-8502 .

**976.** Imbs J 2000 Sectors and the OECD business cycle *CEPR Discussion Paper no 2473* Center for Economic Policy research (CEPR) London UK.

**977.** McConnell M, Pérez-Quirós G 2000 Output fluctuations in the United States: What has changed since the early 1980s? *American Economic Review*.

**978.** Persson T, Tabellini G 2000 Political economics *MIT Press* USA.

**979.** Blanchard O J, Simon J 2001 The long and large decline in US output volatility *Brookings Papers on Economic Activity* **1** pp 135 – 164.

**980.** Brillet J L 2001 The efficiency of the Taylor rule: A stochastic analysis using the Macsim model *INSEE French National Institute for Statistics and Economic Studies* no 160.

**981.** Camacho M 2001 Three essays in nonlinear macro-econometrics *Ph D dissertation* Universitat Autònoma de Barcelona Spain.

**982.** Camacho M 2003 Markov-switching stochastic trends and economic fluctuations *Departamento de Metodos Cuantitativos* Universidad de Murcia Spain pp 1 – 43.

**983.** Chistilin D September 2001 The problems of self-organization in economic theory. Case of transition economy *Conference Proceedings* University of Salerno Italy.

**984.** Chistilin D 2008 To the wave nature of economic cycles *Civilization Problems Study Department* Institute of the World Economy and International Relations National Academy of Science of Ukraine pp 1 – 11.

**985.** Chol-Won Li 2001 Science, diminishing returns and long wave *Manchester School* vol **69** issue 5 pp 553 – 573.

**986.** Ehrmann M, Elison M, Valla N 2001 Regime-dependent impulse response functions in a Markov-switching vector autoregression model *Bank of Finland Working Paper no 11* Helsinki Finland.

**987.** Ehrmann M, Elison M, Valla N 2003 Regime-dependent impulse response functions in a Markov-switching vector autoregression model *Economics Letters* pp 295 – 299.

**988.** Gómez V 2001 The use of Butterworth filters for trend and cycle estimation in economic time series *Journal of Business and Economic Statistics* **19** (3) pp 365 – 373.

**989.** Gonzalo J, Ng S 2001 A systematic framework for analyzing the dynamics effects of permanent and transitory shocks *Journal of Economic Dynamics and Control* **10** no 25 pp 1527 – 1546.

**990.** Inklaar R, de Haan J 2001 Is there really a European business cycle? *Oxford Economic Papers* **53** pp 215 – 220.

**991.** Kongsamut P, Rebelo S, Danyang Xie 2001 Beyond balanced growth Review of Economic Studies **68** (October) pp 869 – 882.

**992.** Kouparitsas M 2001 Is the United States an optimum currency area? An empirical analysis of regional business cycles *mimeo* Federal Reserve Bank of Chicago USA.

**993.** Psaradakis Z, Sola M, Spagnolo F 2001 On Markov error-correction models *mimeo* School of Economics, Mathematics and Statistics Brikbeck College London UK.

**994.** Racorean O 2001 Considerations on gravitational wave in economics pp 1 – 11.

**995.** Rothman P, van Dijk D, Franses P 2001 A multivariate analysis of the relationship between money and output *Macroeconomic Dynamics* **5** pp 506 – 532.

**996.** Trimbur T M 2001 Properties of a general class of stochastic cycles *mimeograph* University of Cambridge UK.

**997.** Vošvrda M January 2001 On economic model of cycles *CeNDEF Workshop Paper no PO3* Center for Nonlinear Dynamics in Economics and Finance Universiteit van Amsterdam The Netherlands.

**998.** Weder M 2001 The great demand depression *CEPR Discussion Papers 3067* Center for Economic Policy research (CEPR) London UK.

**999.** Wen Y 2001 Demand-driven business cycles: Explaining domestic and international co-movements *Working Paper 01-18* Cornell University NY USA.

**1000.** Wen Y 2002 Fickle consumers versus random technology: Explaining domestic and international co-movements *Working Paper* Cornell University NY USA.

**1001.** Wen Y 2004 What does it take to explain pro-cyclical productivity? *Contributions to Macroeconomics* **4** no 1 article 5.

**1002.** Wen Y March 2006 Demand shocks and economic fluctuations *Working Paper 2006-011A* Research Division Federal Reserve Bank of St Louis USA pp 1 – 9 http://research.stlouisfed.org/wp/2006/2006-011.pdf .

**1003.** Yao J 2001 On square-integrability of an AR process with Markov switching *Statistics and Probability Letters* **52** pp 265 – 270.

**1004.** Agénor P-R 2002 Business cycles, economic crises, and the poor *Journal of Economic Policy Reform* vol **5** issue 3 pp 145 – 160.

**1005.** Arnord L 2002 Business cycle theory *Oxford University Press* Oxford UK.

**1006.** Beaudry P, Portier F 2002 The French depression in the 1930s *Review of Economic Dynamics* **5** pp 73 – 99.

**1007.** Beaudry P, Portier F September 2004 An exploration into Pigou's theory of cycles *Journal of Monetary Economics* **51** (1) pp 183 – 216.

**1008.** Beaudry P, Portier F September 2006 Stock prices, news, and economic fluctuations *American Economic Review* **96** (1) pp 293 – 307.

**1009.** Beaudry P, Portier F July 2007 When can changes in expectations cause business cycle fluctuations in neo-classical settings? *Journal of Economic Theory* **135** pp 458–477.

**1010.** Beaudry P, Dupaigne M, Portier F 2011 Modeling news-driven international business cycles *Review of Economic Dynamics* **14** (1) pp 72 – 91.

**1011.** Beaudry P, Fve P, Guay A, Portier F 2015 When is nonfundamentalness in VARs a real problem? An application to news shocks *CEPR Discussion Papers 10763* CEPR Discussion Papers.

**1012.** Festré A 2002 Innovation and business cycles *in* The contribution of Joseph Schumpeter to economics: Economic development and institutional change Arena R, Dangel-Hagnauer C (editors) *Routledge* London UK pp 127 – 145 HAL Id: halshs-00271362 https://halshs.archives-ouvertes.fr/halshs-00271362 .

**1013.** Fisher J, Hornstein A 2002 The role of real wages, productivity, and fiscal policy in Germany's great depression 1928-1937 *Review of Economic Dynamics* **5** (1) pp 100 – 127.

**1014.** Harding D, Pagan A 2002 Dissecting the cycle: A methodological investigation *Journal of Monetary Economics* **49** pp 365 – 381.

**1015.** Harding D, Pagan A 2003 A comparison of two business cycle dating methods *Journal of Economic Dynamics and Control* **27** pp 1681 – 1690.

**1016.** Heathcote J, Perri F 2002 Financial autarky and international business cycles *Journal of Monetary Economics* **49** (3) pp 601 – 627.

1017. Heathcote J, Perri F 2003 Why has the U.S. economy become less correlated with the rest of the World? *American Economic Review* **93** (2) pp 63 – 69.

1018. Jordan J S, Rosengren E S 2002 Economic cycles and bank health *Working Paper 4/10/02* Federal Reserve Bank of Boston USA pp 1 – 28.

1019. Kim M K, Burnie D A 2002 The firm size effect and the economic cycle *Journal of Financial Research* vol **25** issue 1 pp 111 – 124.

1020. Livio M 2002 The Golden ratio: The story of the World's most astonishing number *Broadway Books* New York USA.

1021. O' Hara P A 2002 A new financial social structure of accumulation in the United States for long wave upswing? *Review of Radical Political Economics* vol **34** no 3 pp 295 – 301.

1022. O' Hara P A 2003 Deep recession and financial instability or a new long wave of economic growth for US capitalism? A regulation school approach *Review of Radical Political Economics* vol **35** no 1 pp 18 – 43.

1023. Pedregal D J 2002 Trend models for the prediction of economic cycles *Escuela Técnica Superior de Ingenieros Industriales* Universidad de Castilla-La Mancha Spain.

1024. Ravn M, Uhlig H 2002 On adjusting the Hodrick-Prescott filter for the frequency of observations *Review of Economics and Statistics* **84** (2) pp 371 – 376.

1025. Rennstich J K 2002 The new economy, the leadership long cycle and the nineteenth K-wave *Review of International Political Economy* **9** pp 150 – 182.

1026. Ritschl A 2002a Deutschlands krise und konjunktur, 1924-1934. Binnenkonjunktur, auslandsverschuldung und reparationsproblem zwischen Dawes-plan und transfersperre *Akademie-Verlag* Berlin Germany.

1027. Ritschl A 2002b International capital movements and the onset of the great depression: Some international evidence *in* The interwar depression in an international context James H (editor) *Oldenbourg* Munich Germany.

1028. Ritschl A 2003 Dancing on a volcano: The economic recovery and collapse of the Weimar Republic *in* World economy and national economies in the interwar slump Balderston Th (editor) *Macmillan* London UK.

1029. Ritschl A 2005 The pity of peace: Germany's war economy, 1914-1918 and beyond *in* Broadberry S N, Harrison M (editors) The economics of World War I *Cambridge University Press* Cambridge UK.

1030. Ritschl A, Straumann T January 2009 Business cycles and economic policy, 1914-1945: A survey *Working Papers no 115/09* Department of Economic History London School of Economics Houghton Street London UK pp 1 – 50.

1031. Albu L, Nicolae-Balan M, Iordan M, Caraiani P 2003 Modelling the economic cycles. a theoretical approach *Journal for Economic Forecasting* issue 5 pp 5 – 16.

1032. Álvarez Vázquez N J 2003 The quantitative analysis of economic cycles *Catedrático de Economía Aplicada* Dto de Economía Aplicada Cuantitativa Facultad de Ciencias Económicas y Empresariales UNED Spain pp 1 – 48.

1033. Brian M 2003 European historical statistics *Stockton* New York USA.

1034. Duecker M, Wesche K 2003 European business cycles: New indices and their synchronicity *Economic Inquiry* **41** (1) pp 116 – 131.

1035. Helenius A 2003 Die ökonomische Entwicklung des Luftverkehrs im Lichte der Theorie der Kondratieff Wellen *Cuvillier Verlag* Göttingen Germany.

1036. Helenius A 2009 The global financial crisis and the long waves *Proceedings to ICGFC 2009 Conference* University of Nevada Las Vegas NV USA.

1037. Helenius A 2010 The global financial crisis and its aftermath and the long waves *Journal for Multidisciplinary Thought* **1**.

1038. Helenius A, Pagni J 2011 Allure of the Seas *Ships Monthly* **46** (3) pp 33 – 36.

1039. Helenius A 2012 Waves on waves – Long waves on the seven seas *in* Kondratieff waves: Dimensions and prospects *Volgograd* Russian Federation pp 85 – 106.

1040. Helfat C E, Peteraf M A 2003 The dynamic resource-based view: Capability life cycles *Strategic Management Journal* **24** (10) pp 997 – 1010.

1041. Hirooka M 2003 Nonlinear dynamism of innovations and business cycles *Journal of Evolutionary Economics* **13** pp 549 – 576.

1042. Hirooka M 2006 Innovation dynamism and economic growth: A nonlinear perspective *Edward Elgar Publishing* Cheltenham UK Northampton MA USA.

1043. Mariano R, Murasawa Y 2003 A new coincident index os business cycles based on monthly and quarterly series *Journal of Applied Econometrics* **18** pp 427 – 443.

1044. Mills T C 2003 Modelling trends and cycles in economic time series *Palgrave Macmillan* Basingstoke.

1045. Nakajima T 2003 A business cycle model with variable capacity utilization and demand disturbances *European Economic Review*.

1046. Ogawa K 2003 Daifukyo no Keizai Bunseki (Economic Analysis of the Great Depression in Japan) Nihon-keizai Shimbun Sha Tokyo Japan.

1047. Rumyantseva S Yu 2003 Long waves in economics: Multifactor analysis *St Petersburg University Publishing House* St. Petersburg Russian Federation.

1048. Selover D D, Jensen R V, Kroll J 2003 *Studies in Nonlinear Dynamics & Econometrics* **7** p 1.

1049. Sussmuth B 2003 Business cycles in the contemporary World *Springer* Berlin Heidelberg Germany.

1050. Tsoulfidis L 2003 Economic history of Greece (in Greek) *University of Macedonia Press* Thessaloniki Greece.

1051. Tsoulfidis L October 2006 Falling rate of profit and over-accumulation in Marx and Keynes *Political Economy Quarterly* vol **43** pp 65 – 75.

1052. Zhang J, Zhang J, Lee R 2003 Rising longevity, education, savings, and growth *Journal of Development Economics* **70** (1) pp 83 – 101.

1053. Benhabib J, Wen Y 2004 Indeterminacy, aggregate demand, and the real business cycle, *Journal of Monetary Economics* **51** pp 503 – 530.

1054. Caraiani P 2004 Nominal and real stylized facts of the business cycles in the Romanian economy *Romanian Journal of Economic Forecasting* **5** (4) pp 121 – 132.

1055. Caraiani P 2007a An analysis of economic fluctuations in Romanian economy using the real business cycle approach *Romanian Journal of Economic Forecasting* **8** (2) pp 76 – 86.

1056. Caraiani P 2007b An estimated new Keynesian model for Romania Romanian *Journal of Economic Forecasting* **8** (4) pp 114 – 123.

1057. Jaeger A, Schucknecht L 2004 Boom-bust phases in asset prices and fiscal policy behavior *IMF Working Paper WP/04/54* IMF USA.

1058. Kaminsky G, Reinhart C, Vegh C 2004 When it rains, it pours: Procyclical capital flows and macroeconomic policies *NBER macroeconomics annual 2004* Gertler M, Rogoff (editors) Cambridge UK pp 11 – 82.

1059. Kobayashi K, Inaba M March 20 2004 Monetary cycles *RIETI Discussion Paper Series 04-E-020* The Research Institute of Economy, Trade and Industry Japan pp 1 – 7.

1060. Manfredi P, Fanti L 2004 Cycles in dynamic economic modelling *Economic Modelling* vol **21** issue 3 pp 573 – 594.

1061. McCauley J L 2004 Dynamics of markets: Econophysics and finance *Cambridge University Press* Cambridge UK.

1062. Ngai L R 2004 Barriers and the transition to modern growth *Journal of Monetary Economics* **51** (October) 1 pp 353 – 383.

1063. Ngai L R, Pissarides Ch A 2007 Structural change in a multi-sector model of growth *American Economic Review* **97** (March) pp 429 – 443.

1064. Ohn J, Taylor L W, Pagan A 2004 Testing for duration dependence in economic cycles *Econometrics Journal* vol **7** issue 2 pp 528 – 549.

1065. Pelagatti M M 2004 Business cycle and sector cycles *Department of Statistics* Università degli Studi di Milano-Bicocca Italy pp 1 – 18.

1066. Rampini A 2004 Entrepreneurial activity, risk and the business cycle *Journal of Monetary Economics* vol **5** no 3 pp 555 – 573.

1067. Schnabel I 2004 The German twin crisis of 1931 *Journal of Economic History* **64** pp 822 – 871.

1068. Sergienko J 2004 On the financial mechanism of long-wave technological and economic changes *Voprosy Ekonomiki* vol **1**.

1069. Syed M K, Mohammad M J 2004 Revisiting Kuznets hypothesis: An analysis with time series and panel data *Bangladesh Development Studies* **30** (3-4) pp 89 – 112.

1070. Verspagen B 2004 Innovation and economic growth *in* The Oxford handbook of innovation Fagerberg J, Mowery D, Nelson R (editors) *Oxford University Press* Oxford UK.

1071. Valle e Azevedo J, Koopman S J, Rua A 2004 Tracking the business cycle of the Euro area: A multivariate model-based band-pass filter *Working Paper* Department of Econometrics Free University Amsterdam The Netherlands.

1072. Andergassen R, Nardini F 2005 Endogenous innovation waves and economic growth *Structural Change and Economic Dynamics* vol **16** issue 4 pp 522 – 539.

1073. Banerjee A V, Duflo E 2005 Growth theory through the lens of development economics *in* Handbook of economic growth Aghion Ph, Durlauf St (editors) vol **1A** *North Holland* New York USA pp 473 – 552.

1074. Bilbiie F O, Ghironi F, Melitz M J 2005 Business cycle and firm dynamics *Oxford University* UK *Harvard University* USA.

1075. Cover J P, Pecorino P 2005 The length of US business expansions: When did the break in the data occur? *Journal of Macroeconomics* vol **27** no 3 pp 452 – 471.

1076. Darvas Z, Rose A K, Szapary G 2005 Fiscal divergence and business cycle synchronization: Irresponsibility is idiosyncratic *NBER Working Paper no 11580* NBER USA.

1077. De Groot B, Franses Ph H 2005 Cycles in basic innovations *Econometric Institute Report 2005-35* Econometric Institute Erasmus University Rotterdam The Netherlands pp 1 – 27.

1078. De Groot B, Franses P H 27 March 2006 Stability through cycles *Econometric Institute Report 2006-07* Econometric Institute Erasmus University Rotterdam The Netherlands pp 1 – 30.

1079. De Groot B, Franses P H 2008 Stability through cycles *Technological Forecasting And Social Change* vol **75** no 3 pp 301 – 311.

1080. De Groot B, Franses P H 2012 Common socio-economic cycle periods *Technological Forecasting and Social Change* vol **79** issue 1 pp 59 – 68.

1081. Francis N, Ramey V A 2005 Is the technology-driven real business cycle hypothesis dead? Shocks and aggregate fluctuations revisited *Journal of Monetary Economics* **52** (8) pp 1379 – 1399.

1082. Jonung L, Schucknecht L, Tujula M 2005 The boom-bust cycle in Finland and Sweden 1984 – 1995 in an international perspective *European Economy Economic Papers no 237.*

1083. Miles D, Scott A 2005 Macroeconomics: Understanding the wealth of nations *John Wiley & Sons* Chichester England.

1084. Neumeyer P, Perri F 2005 Business cycles in emerging economies: The role of interest rates *Journal of Monetary Economics* **52** (2) pp 345 – 380.

1085. Ozawa T 2005 Institutions, industrial upgrading, and economic performance in Japan - The "Flying-Geese" paradigm of catch-up growth *Edward Elgar Publishing* Northampton Massachusetts USA.

1086. Peaucelle I 2005 Dynamic analysis of bankruptcy and economic waves *Working Paper no 2005 – 09* Centre National De La Recherche Scientifique – École Des Hautes Études En Sciences Sociales École Nationale Des Ponts Et Chaussées – École Normale Supérieure Paris France pp 1 – 15.

1087. Rebelo S 2005 Business cycles *Annals of Economics and Finance* **6** pp 229 – 250.

1088. Rebelo S 2010 Models of real business cycles: Past, present and future *Problems of Economics* no 10 pp 56 – 67.

1089. Shimer R March 2005 The cyclical behavior of equilibrium unemployment and vacancies *American Economic Review* **95** pp 25 – 49.

1090. Steehouwer H 2005 Macroeconomic scenarios and reality *Ph D Thesis* Free University Amsterdam The Netherlands.

1091. Woo J 2005 Social polarization, fiscal instability, and growth *European Economic Review* **49** (6) pp 1451 – 1477.

1092. Andersen E S 2006 The limits of Schumpeter's business cycles *Industry and Innovation* vol **13** issue 1 pp 107 – 116.

1093. Chian A C-L, Rempel E L, Borotto F A, Rogers C 2006 An example of intermittency in nonlinear economic cycles *Applied Economics Letters* vol **13** issue 4 pp 257 – 263.

1094. Comin D, Gertler M 2006 Medium-term business cycles *American Economic Review* **96** (3) pp 523 – 551.

1095. Congcong Dong 27 February 2006 An introduction to World economic long wave-crises and depressions: From study to anticipation *MPRA Paper no 2106* Munich Germany pp 1 – 8

http://mpra.ub.uni-muenchen.de/2106/ .

1096. Diebolt C, Doliger C 2006 Economic cycles under test: A spectral analysis *in* Kondratieff Waves, Warfare and World Security Devezas T C (editor) *IOS Press* Amsterdam The Netherlands pp 39 – 47.

1097. Diebolt C, Doliger C 2008 New international evidence on the cyclical behaviour of output: Kuznets swings reconsidered. Quality & quantity. *International Journal of Methodology* **42** (6) pp 719 – 737.

1098. Lee R, Sang-Hyop Lee, Mason A July 2006 Charting the economic life cycle *Working Paper 12379* National Bureau Of Economic Research Cambridge USA pp 1 – 32

http://www.nber.org/papers/w12379.

1099. Linstone H A 2006 The information and molecular ages: Will K-waves persist? *Kondratieff Waves, Warfare and World Security* edited by Devezas T C *IOS Press* Amsterdam The Netherlands pp 260 – 269.

1100. McMinn D 2006 Market timing by the moon & the sun *Twin Palms Publishing* USA.

1101. McMinn D 2007 Market timing by the number 56 *Twin Palms Publishing* USA.

1102. Monteiro R 2006 Ppps, the business cycle, and the political cycle *in* Book of abstracts of 13th annual European real estate society conference *ERES Conference* Weimar Germany.

1103. Pantin V I, Lapkin V V 2006 Philosophy of history forecasting: Rhythms of history and prospects of world development in the first half of the 21st century *Feniks* Dubna Russian Federation.

1104. Perotin V 2006 Entry, exit, and the business cycle: Are cooperatives different? *Journal of Comparative Economics* vol **34** no 2 pp 295 – 316.

1105. Yamashiro G, Uesugi I 2006 Economic conditions and Japanese firm financing *Economics Bulletin* vol **5** no 11 pp 1 – 17

http://economicsbulletin.vanderbilt.edu/2006/volume5/EB-06E00001A.pdf .

1106. Aguiar M, Gopinath G 2007 Emerging market business cycles: The cycle is the trend *Journal of Political Economy* **115** pp 69 – 102.

1107. Fernández-Villaverde J, Rubio-Ramirez J F, Sargent Th J, Watson M W June 2007 ABCs (and Ds) for understanding VARs *American Economic Review* **97** (1) pp 21– 26.

1108. Fernández-Villaverde J, Rubio-Ramirez J 2010 Macroeconomics and volatility: Data, models, and estimation *mimeo*.

1109. Fernández-Villaverde J, Guerrn P, Kuester K, Rubio-Ramrez J 2012 Fiscal volatility shocks and economic activity *mimeo*.

1110. Fernández-Villaverde J, Rubio-Ramírez J F, Schorfheide F 2016 *in* Handbook of macroeconomics Taylor J B, Uhlig H (editors) vol **2** pp 527 – 724

http://dx.doi.org/10.1016/S1574-0048(16)30050-7 .

1111. Flodén M 2007 Vintage capital and expectations driven business cycles CEPR *Discussion Paper 6113* Center for Economic Policy research (CEPR) London UK.

1112. Honjo K 1 August 2007 The golden rule and the economic cycles *Working Paper no 07/199* International Monetary Fund (IMF) pp 1 – 23 ISBN/ISSN: 9781451867633/1018-5941.

1113. Knotek E 2007 How useful is Okun's law? *Economic Review Kansas City Federal Reserve* issue **Q IV** pp 73 – 103.

1114. Krantz O, Schön L 2007 Swedish historical national accounts 1800-2000 *Lund Studies in Economic History* **41** Almqvist & Wiksell International.

1115. Rodriguez Mora J V, Schulstad P November 2007 The effect of GNP announcements on fluctuations of GNP growth *European Economic Review* **51** (1) pp 922 – 940.

1116. Adrian T, Estrella A 2008 Monetary tightening cycles and the predictability of economic activity *Economics Letters* **99** (2) pp 260 – 264.

1117. Adrian T, Estrella A, Hyun Song Shin January 2010 Monetary cycles, financial cycles, and the business cycle *Staff Report no 421* Federal Reserve Bank of New York USA pp 1 – 20.

1118. Battilossi St, Foreman-Peck J, Kling G September 2008 European business cycles and economic policy, 1945-2007 *Working Papers in Economic History WP 08-13* Departamento de Historia Economica Instituciones Instituto Figuerola de Historia Economica Universidad Carlos III De Madrid Spain pp 1 – 57

http://hdl.handle.net/10016/2990 .

1119. Kling G, Foreman-Peck J, Battilossi St 2008 European business cycles and economic policy, 1945-2007 *IFCS - Working Papers in Economic History* Instituto Figuerola Universidad Carlos III de Madrid Spain.

1120. Hagedorn M, Manovskii I September 2008 The cyclical behavior of equilibrium unemployment and vacancies revisited *American Economic Review* **98** (1) pp 692 – 706.

1121. Haltmaier J April 2008 Predicting cycles in economic activity *International Finance Discussion Paper no 926* Board of Governors of Federal Reserve System USA pp 1 – 51.

1122. Hori K 2008 Economic growth, unemployment, and business cycles *Keio/Kyoto Global Coe Discussion Paper DP2008-039* Graduate School of Economics and Graduate School of Business and Commerce Keio University, Institute of Economic Research Kyoto University Japan pp 1 – 15.

1123. Ilzetzki E, Vegh C 2008 Procyclical fiscal policy in developing countries: Truth of fiction *mimeo*.

1124. Ilzetzki E 2011 Rent-seeking distortions and fiscal procyclicality *Journal of Development Economics* **96** (1) pp 30 – 46.

1125. Iyetomi H, Aoyama H, Ikeda Y, Souma W, Fujiwara Y 2008 Econophysics *Kyoritsu Shuppan* Tokyo Japan.

**1126.** Iyetomi H, Nakayama Y, Yoshikawa H, Aoyama H, Fujiwara Y, Ikeda Y, Souma W 2011 What causes business cycles? Analysis of the Japanese industrial production data *Journal of the Japanese and International Economies* **25** (3) pp 246 – 272.

**1127.** Iyetomi H, Aoyama H, Fujiwara Y, Sato A-H (editors) 2012 Econophysics 2011 - The Hitchhiker's guide to the economy *Proceedings of the YITP Workshop on Econophysics Japan Progress of Theoretical Physics Supplement* no 194.

**1128.** Jaimovich N, Rebelo S 2008 News and business cycles in open economies *Journal of Money, Credit and Banking* **40** (8) pp 1699 – 1711.

**1129.** Jaimovich N, Rebelo S September 2009 Can news about the future drive the business cycle? *American Economic Review* **99** (4) pp 1097 – 1118.

**1130.** Jourdon Ph 2008 La monnaie unique Europeenne et son lien au developpement economique et social coordonne: une analyse cliometrique *Thèse Universite Montpellier* France.

**1131.** Papenhausen Ch 2008 Causal mechanisms of long waves *Futures* **40** (9) pp 788 – 794.

**1132.** Roa Maria J, Vazquez Francisco Jose, Saura Dulce 2008 Unemployment and economic growth cycles *Studies in Nonlinear Dynamics and Econometrics* vol **12** issue 2 pp 1 – 21.

**1133.** Taniguchi M, Bando M, Nakayama A 2008 Business cycle and conserved quantity in economics *Journal of the Physical Society of Japan* vol **77** no 11.

**1134.** Angeletos G-M, La'O J May 2009 Noisy business cycles *Working Paper 14982* National Bureau of Economic Research Cambridge Massachusetts USA.

**1135.** Boone P, Johnson S 2009 The doomsday cycle *CentrePiece* Winter 2009/10 pp 1 – 6.

**1136.** Boretos G P 2009 The future of the global economy *Technological Forecasting and Social Change* vol **76** n 3 pp 316 – 326.

**1137.** Coccia M 2009a Forecast horizon of 5th – 6th – 7th long wave and short-period of contraction in economic cycles *Working Paper CERIS-CNR no 4/2009* National Research Council CERIS-CNR Italy ISSN (print) 1591-0709 ISSN (on line) 2036-8216 pp 1 – 19.

**1138.** Coccia M 2009b Business cycles and the scale of economic shock *Working Paper CERIS-CNR no 6/2009* National Research Council CERIS-CNR Italy ISSN (print) 1591-0709 ISSN (on line) 2036-8216 pp 1 – 24.

**1139.** Coccia M 2010 The asymmetric path of economic long waves *Technological Forecasting and Social Change* vol **77** pp 730 – 738.

**1140.** Den Haan W J, Kaltenbrunner G April 2009 Anticipated growth and business cycles in matching models *Journal of Monetary Economics* **56** pp 309 – 327.

**1141.** McCauley J 2009 Dynamics of markets: The new financial economics 2[nd] edition *Cambridge University Press*.

**1142.** OECD 2009 International migration and the economic crisis: Understanding the links and shaping policy responses *in* International Migration Outlook 2009 *OECD* Paris France pp 11 – 76.

**1143.** Purica I, Caraiani P 2009 Second order dynamics of economic cycles *Romanian Journal of Economic Forecasting* **1** pp 36 – 47.

1144. Schmitt-Grohé St, Uribe M 2009 What's news in business cycles *manuscript* Duke University NC USA, *Econometrica* **80** (6) pp 2733 – 2764.

1145. Sims E R 2009 Expectations driven business cycles: An empirical evaluation *mimeo* University of Michigan USA.

1146. Walentin K April 2009 Expectation driven business cycles with limited enforcement *Sveriges Riksbank Working Paper Series 229* Riksbank Stockholm Sweden.

1147. Akaev A A, Sadovnichiy V A, Korotaev A V December 2010 On the possibilities to forecast the current crisis and its second wave *Ekonomicheskaya Politika* issue **6** pp 39 – 46.

1148. Akaev A A, Sadovnichiy V A, Korotaev A V 15 May 2011 Huge rise in gold and oil prices as a precursor of a global financial and economic crisis *Doklady Mathematics* vol **83** issue 2 pp 243 – 246
https://doi.org/10.1134/S1064562411020372 .

1149. Akaev A A, Sadovnichiy V A, Korotaev A V 2012 On the dynamics of the world demographic transition and financial-economic crises forecasts *The European Physical Journal* no 205 pp 355 – 373
https://doi.org/10.1140/epjst/e2012-01578-2.

1150. Sadovnichiy V A, Akaev A A Korotaev A V Malkov S Yu 2012 Modeling and forecast of world dynamics *ISPI RAN* Moscow Russian Federation ISBN 978-5-7556-0456-7.

1151. Crafts N, Fearon P 2010 Lessons from the 1930s great depression *Oxford Review of Economic Policy* **26** pp 285 – 317.

1152. Fanti L, Gori L 15 January 2010 PAYG pensions and economic cycles *University of Pisa* Italy *MPRA Paper no 19984* Munich Germany pp 1 – 24
https://mpra.ub.uni-muenchen.de/19984/ .

1153. Grigor'ev L 2010 Theory of cycle under hit of crisis *Problems of Economics* no 10 pp 31 – 55.

1154. Grinin L E, Korotayev A V, Malkov S Y 2010 A mathematical model of Juglar cycles and the current global crisis *in* History and mathematics Grinin L, Korotayev A, Tausch A (editors) *URSS* Moscow Russian Federation.

1155. Korotayev A V, Tsirel S V 2010 A spectral analysis of world GDP dynamics: Kondratieff waves, Kuznets swings, Juglar and Kitchin cycles in global economic development, and the 2008–2009 economic crisis *Journal of Structure and Dynamics, Social Dynamics and Complexity* Institute for Mathematical Behavioral Sciences University of California at Irvine vol **4** issue 1 pp 1 – 55
http://www.escholarship.org/uc/item/9jv108xp .

1156. Grinin L, Korotayev A, Tausch A 2016 Economic cycles, crises, and the global periphery *Springer International Publishing* ISBN 978-3-319-41260-3 eBook ISBN978-3-319-41262-7 DOI10.1007/978-3-319-41262-7 pp 1 – 265.

1157. Ritschl A, Straumann T 2010 *Business cycles and economic policy, 1914-1945 in* Cambridge economic history of modern Europe *Cambridge University Press* Cambridge UK pp 156 – 180 ISBN 9780521882033.

1158. Krusell P, McKay A Fourth Quarter 2010 News shocks and business cycles *Economic Quarterly* vol **96** no 4 pp 373 – 397.

1159. Navarro P, Bromiley P, Sottile P 2010 Business cycle management and firm performance *Journal of Strategy and Management* vol **3** no 1 pp 50 – 71.

1160. Rossiter J 2010 Now casting the global economy *Bank of Canada Discussion Paper no 12*.

1161. Xu Feiqiong 2010 Economic analysis on agricultural disasters: Cyclical fluctuation and comprehensive solutions to the problems *Economic Theory and Economic Management* **8**.

1162. Asteriou D, Hall St G 2011 Vector autoregressive models and causality tests in Applied Econometrics 2$^{nd}$ edition *Palgrave MacMillan* London UK pp 319 – 333.

1163. Barseghyan L, Battaglini M, Coate S 2011 Fiscal policy over the real business cycle: A positive theory *mimeo*.

1164. Barsky R B, Sims E R 2011 News shocks and business cycles *Journal of Monetary Economics* **58** (3) pp 273 – 289.

1165. Bazzi S, Blattman C 2011 Economic shocks and conflict: The (Absence of?) evidence from commodity prices *Working Paper*.

1166. Buera F J, Monge-Naranjo A, Primiceri G E 2011 Learning the wealth of nations Econometrica **79** (1) pp 1 – 45.

1167. Canes-Wrone C, Park J K 2011 Electoral business cycles in OECD countries *American Political Science Review* **106** (1) pp 103 – 122.

1168. Claessens S, Kose A, Terrones M E 2011 Financial cycles: What? How? When? *Working Paper 11/76* IMF USA.

1169. Drehmann M, Borio C, Tsatsaronis K 2011 Anchoring countercyclical capital buffers: The role of credit aggregates *International Journal of Central Banking* vol **7** no 4 pp 189 – 240.

1170. Gokcekus O, Suzuki Yu 2011 Business cycle and corruption *Economics Letters* vol **111** issue 2 pp 138 – 140.

1171. Jacks D, Novy D, Meissner Ch 2011 Trade booms, trade busts, and trade costs *Journal International Econ* **83** (2) pp 185 – 201.

1172. Jenkins S P, Brandolini A, Micklewright J, Nolan B 2011 Introduction: Scope, review of analytical approaches and evidence from the past *in* The great recession and the distribution of household income Jenkins S P, Brandolini A, Micklewright J, Nolan B (editors) *Fondazione Rodolfo Debenedetti* Milan Italy

http://www.frdb.org/upload/file/report_1_palermo.pdf .

1173. Jenkins S P, Taylor M P 2012 Non-employment, age, and the economic cycle *Longitudinal and Life Course Studies* **3** (1) pp 18 – 40 ISSN1757-9597 DOI:10.14301/llcs.v3i1.161

http://eprints.lse.ac.uk/57589/ .

1174. Jin Guantao, Liu Jinfeng 2011 Prosperity and crisis: The discussion of the super-stable structure in Chinese society *Law Press*.

1175. Jordá O, Schularick M, Taylor A M 2011 When credit bites back: Leverage, business cycles and crises *Working Paper Series no 17621* NBER USA.

1176. Lopes M P 2011 A psychosocial explanation of economic cycles *Journal of Behavioral and Experimental Economics* (*The Journal of Socio-Economics*) vol **40** issue 5 pp 652 – 659.

1177. Lostun A M 2011 An entropic perspective on economic crises *MPRA* Munich Germany.

1178. Lucchese M, Pianta M February 2011 Cycles and innovation *Working Papers Series in Economics, Mathematics and Statistics WP-EMS # 2011/03* ISSN 1974-4110 Faculty of Economics Università di Urbino Italy pp 1 – 21.

1179. Lucchese M, Pianta M June 2012 Innovation and employment in economic cycles *Comparative Economic Studies* vol **54** Issue 2 pp 341 – 359
https://doi.org/10.1057/ces.2012.19 .

1180. Patterson K 2011, 2012 Unit root tests in time series 1, 2 *Palgrave Macmillan* London UK.

1181. Qin Duo 2011 Rise of VAR modeling approach *Journal of Economic Surveys* **25** (1) pp 156 – 174 DOI:10.1111/j.1467-6419.2010.00637.x

1182. Restuccia D 3$^{rd}$ Quarter 2011 Recent developments in economic growth *Economic Quarterly* vol **97** (3) pp 329 – 357.

1183. Roper D 2011 Methodology: Do K-waves exist?
http://www.colorado.edu/peacestudies/sustainable-economics/defl-waves/methodology.html .

1184. Shiyan D V 2011 Impact by expectations on mechanism of action of economic cycle pp 40 – 52.

1185. Aimar Th, Bismans Fr, Diebolt Cl 2012 Economic cycles: A synthesis *Working Paper no 12-11* Association Francaise de Cliometrie (AFC) France.

1186. Albers S, Albers A L 31 March 2012 On the mathematic prediction of economic and social crises: Toward a harmonic interpretation of the Kondratiev wave *MPRA Paper no 37771* Munich Germany pp 1 – 86
http://mpra.ub.uni-muenchen.de/37771/ .

1187. Ales L, Maziero P, Yared P 2012, 2013, 2014 A theory of political and economic cycles *Working Paper 18354* National Bureau of Economic Research Cambridge USA pp 1 – 66, *2013 Society for Economic Dynamics Meeting Paper no 1261*, *Journal of Economic Theory* vol **153** issue C pp 224 – 251
http://www.nber.org/papers/w18354.

1188. Bandura O 2012 Relationship for economic and financial cycles *Ukrainian Journal Ekonomist* issue 1 pages 13 – 15.

1189. Borio C December 2012 The financial cycle and macroeconomics: What have we learned and what are the policy implications *BIS Working Papers* no 395 Switzerland.

1190. Borio C, Karroubi E, Upper C, Zampoli F 14 April 2016 Financial cycles, labor misallocation, and economic stagnation
www.VoxEU.org .

1191. Branca A S, Pina J, Catalão-Lopes M 2012 Corporate giving, competition and the economic cycle *WP 15/2012/DE/UTL/UNIV NOVA* Department of Economics School of Economics and Management Technical University Of Lisbon Portugal ISSN 0874-4548 pp 1 – 38.

1192. Camacho M, Perez Quiros G, Poncela P 2012 Markov-switching dynamic factor models in real time *CEPR Working Paper no 8866* Center for Economic Policy research (CEPR) London UK.

1193. Camacho M, Martinez-Martin J 2014 Real-time forecasting US GDP from small-scale factor models *Empirical Economics* **47** pp 347 – 364.

1194. Camacho M, Martínez-Martín J February 2015 Monitoring the world business cycle *Working Paper no 15/06* BBVA Research Spain pp 1 – 24.

1195. Camacho M, Perez Quiros G, Poncela P 2015 Extracting nonlinear signals from several economic indicators *Journal of Applied Econometrics*.

1196. Dementiev V Ye 2012 Long waves in the economy: Investment aspect *Working Paper # WP/2012/297* CEMI RAS Moscow Russian Federation.

1197. Dementiev V Ye 2013 Structural factors of technological development *Economics and Mathematical Methods* vol **49** no 4 pp 33 – 46.

1198. Dementiev V Ye December 2014 The long waves in the post-industrial economy *Montenegrin Journal of Economics* vol **10** no 2 pp 63 – 70.

1199. Forni M, Gambetti L, Sala L 2014 No news in business cycles *Economic Journal* **124** (581) pp 1168 – 1191.

1200. Ikeda Y, Aoyama H, Fujiwara Y, Iyetomi H, Ogimoto K, Souma W, Yoshikawa H 2012 Coupled oscillator model of the business cycle with fluctuating goods markets *Proceedings of the YITP Workshop on Econophysics Japan Progress of Theoretical Physics Supplement* no 194 pp 111 – 121
arXiv:1110.6679v1 .

1201. Ikeda Y, Aoyama H, Yoshikawa H 2013a Synchronization and the coupled oscillator model in international business cycles *RIETI Discussion Paper October 13-E-089* The Research Institute of Economy, Trade and Industry Japan
http://www.rieti.go.jp/en/ .

1202. Ikeda Y, Aoyama H, Yoshikawa H 2013b Direct evidence for synchronization in international business cycles *Financial Networks and Systemic Risk*.

1203. Ikeda Y 2013 Direct evidence for synchronization in Japanese business cycles *Evolutionary and Institutional Economic Review* **10** (2) pp 1 – 13
arXiv:1305.2263v1 .

1204. Lee Y, Mukoyama T 2012 Entry, exit and plant-level dynamics over the business cycle *Sogang University University of Virginia* USA.

1205. Papageorgiou A, Tsoulfidis L 20 July 2012 Kondratiev, Marx and the long cycle *MPRA Paper no 31355* Munich Germany pp 1 – 20
https://mpra.ub.uni-muenchen.de/31355/ .

1206. Pérez Quirós G December 2012 The role of credit as a predictor of the economic cycle *Directorate General Economics, Statistics and Research* María Dolores Gadea Rivas University of Zaragoza Spain, *Economic Bulletin* Banco de España pp 11 – 16.

1207. Podlesnaya V G 2012a Dialectics of development of social-economic cycles of economic production *Businessinform* no 6 pp 21 – 25.

1208. Podlesnaya V G 2012b Role of innovations in development of social-economic cycles *Businessinform* no 9 pp 26 – 32.

1209. Pustovoit O V 2012 Effect of imposition of waves of external- and internal-economic conjuncture in economy of Ukraine pp 99 – 117.

1210. Swiss National Bank 2012 Swiss National Bank financial stability report 2012 http://www.snb.ch/en/mmr/reference/stabrep_2012/source/stabrep_2012.en.pdf .

1211. Swiss National Bank 2013 Countercyclical capital buffer: Proposal of the Swiss National Bank and decision of the Federal Council http://www.snb.ch/en/mmr/reference/pre_20130213/source/pre_20130213.en.pdf .

1212. Uechi L, Akutsu T 2012 Conservation laws and symmetries in competitive systems *Progress of Theoretical Physics Supplement* no 194 pp 210 – 222.

1213. Azzimonti M, Talbert M November 2013 Polarized business cycles *Working Paper no 13-44* Research Department Federal Reserve Bank of Philadelphia USA pp 1 – 44.

1214. Bachmann R, Bai J 2013 Public consumption over the business cycle *Quantitative Economics.*

1215. Beber A, Brandt M, Luisi M 2013 Economic cycles and expected stock returns *CEPR Discussion Paper no 9528* Center for Economic Policy research (CEPR) London UK.

1216. Central Banking Newsdesk 2013 Swiss board member supports counter-cyclical capital buffer

http://www.centralbanking.com/central-banking/speech/2203857/swiss-board-member-supportscountercyclical-capital-buffer .

1217. Duran H E 2013 Convergence of regional economic cycles in Turkey *Review of Urban and Regional Development Studies* vol **25** issue 3 pp 152 – 175.

1218. Klapper L, Love I, Randall 2013 New firm registration and the business cycle *World Bank.*

1219. Klepach A, Kuranov G 2013 Cyclical waves in the economic development of the US and Russia (Issues of Methodology and Analysis) *Voprosy Economiki* (*Issues of Economics*) no 11 pp 4 – 33.

1220. Krolzig H M 2013 Markov-switching vector autoregressions: Modelling, statistical inference, and application to business cycle analysis *Springer Science & Business Media* 454.

1221. Ledenyov D O, Ledenyov V O 2013c On the accurate characterization of business cycles in nonlinear dynamic financial and economic systems *Cornell University* NY USA pp 1 – 26

www.arxiv.org 1304.4807.pdf .

1222. Ledenyov D O, Ledenyov V O 2015d Information money fields of cyclic oscillations in nonlinear dynamic economic system *MPRA Paper no 63565* Munich University Munich Germany, *SSRN Paper no SSRN-id2592975 Social Sciences Research Network* New York USA pp 1 – 40

http://mpra.ub.uni-muenchen.de/63565/ ,

http://ssrn.com/abstract=2592975 .

1223. Ledenyov D O, Ledenyov V O 2015e On the spectrum of oscillations in economics *MPRA Paper no 64368* Munich University Munich Germany, *SSRN Paper no SSRN-id2606209 Social Sciences Research Network* New York USA pp 1 – 48

http://mpra.ub.uni-muenchen.de/64368/ ,

http://ssrn.com/abstract=2606209 .

1224. Ledenyov D O, Ledenyov V O 2015f Digital waves in economics *MPRA Paper no 64755* Munich University Munich Germany, *SSRN Paper no SSRN-id2613434 Social Sciences Research Network* New York USA pp 1 – 55

http://mpra.ub.uni-muenchen.de/64755/ ,

http://ssrn.com/abstract=2613434 .

*1225.* Ledenyov D O, Ledenyov V O 2015g General information product theory in economics science *MPRA Paper no 64991* Munich University Munich Germany, *SSRN Paper no SSRN-id2617310 Social Sciences Research Network* New York USA pp 1 – 54

http://mpra.ub.uni-muenchen.de/64991/ ,

http://ssrn.com/abstract=2617310 .

*1226.* Ledenyov D O, Ledenyov V O 2015h Quantum macroeconomics theory *MPRA Paper no 65566* Munich University Munich Germany, *SSRN Paper no SSRN-id2627086 Social Sciences Research Network* New York USA pp 1 – 55

http://mpra.ub.uni-muenchen.de/65566/ ,

http://ssrn.com/abstract=2627086 .

*1227.* Ledenyov D O, Ledenyov V O 2016r Precise measurement of macroeconomic variables in time domain using three dimensional wave diagrams *MPRA Paper no 69609* Munich University Munich Germany, *SSRN Paper no SSRN-id2733607 Social Sciences Research Network* New York USA pp 1 – 52

http://mpra.ub.uni-muenchen.de/69609/ ,

http://ssrn.com/abstract=2733607 .

*1228.* Ledenyov V O, Ledenyov D O 2016s Forecast in capital markets *Lambert Academic Publishing* Saarbrücken Germany ISBN 978-3-659-91698-4; *MPRA Paper no 72286* Munich University Munich Germany; *SSRN Paper no SSRN-id2802085 Social Sciences Research Network* New York USA pp 1 – 260

www.lap-publishing.com ,

http://mpra.ub.uni-muenchen.de/72286/ ,

http://ssrn.com/abstract=2802085 .

*1229.* Ledenyov V O, Ledenyov D O 2017 Investment in capital markets *Lambert Academic Publishing* Saarbrücken Germany ISBN 978-3-330-05708-1; *MPRA Paper no 77414* Munich University Munich Germany; *SSRN Paper no SSRN-id2930848 Social Sciences Research Network* New York USA pp 1 – 696

www.lap-publishing.com ,

www.investmentcapitalmarkets.com ,

http://mpra.ub.uni-muenchen.de/77414/ ,

http://ssrn.com/abstract=2930848 .

*1230.* Maziero P, Yared P, Ales L 2013 A theory of economic and political cycles *Society For Economic Dynamics Meeting Paper no 1261.*

*1231.* Raicu G, Stanca C, Raicu A 2013 Business cycles and economic distortions *Constanta Maritime University Annals* Year **XIII** Vol **17** pp 295 – 298.

*1232.* Sabol A, Šander M, Fučkan D 2013 The concept of industry life cycle and development in business strategies *Management, Knowledge and Learning International Conference* Croatia.

*1233.* Sakane K M 2013 News-driven international business cycles *The B.E. Journal of Macroeconomics* **13** (1) p 43.

1234. Ternyik S I January 1 2013a Global wave compression *MPRA Paper no 43831* Techno-Logos Inc pp 1 – 19

https://mpra.ub.uni-muenchen.de/43831/ .

1235. Ternyik S I 2013b Monetary wave theory and quantum economics *Vook* New York USA.

1236. Union Bank of Switzerland 2013 UBS outlook Switzerland http://www.ubs.com/global/en/wealth_management/wealth_management_research/ubs_outlook_ch.html .

1237. Coyle D 2014 GDP: A brief but affectionate history *Princeton University Press* Princeton NJ USA ISBN 978-0-691-15679-8 .

1238. Davis S J 2014 Financial integration and international business cycle co-movement *Journal of Monetary Economics* **64** (C) pp 99 – 111.

1239. Estrada M R 2014 Economic waves: The effect of the U.S. economy on the world economy *Contemporary Economics* vol **8** issue 3.

1240. Ferrara L, Marsilli C 2014 Now casting global economic growth: A factor-augmented mixed-frequency approach *Bank of France Working Papers No 515* Paris France.

1241. Golinelli R, Parigi G 2014 Tracking world trade and GDP in real time *International Journal of Forecasting* **30** pp 847 – 862.

1242. Grossman V, Mack A, Martinez-Garcia E 2014 A contribution to the chronology of turning points in global economic activity (1980-2012) Federal Reserve Bank of Dallas *Working Paper no 169* Dallas Texas USA.

1243. Lucas B, Gevorkyan A, Palley Th 2014 Time scales and mechanism of economic cycles: A review of theories of long waves *Review of Keynesian Economics* vol **2** issue 1 pp 87 – 107.

1244. Peters B, Dachs B, Dünser M, Hud M, Köhler Ch, Rammer Ch 2014 Firm growth, innovation and the business cycle *Austrian Institute of Technology Background Report for the 2014 Competitiveness Report* Austria.

1245. Quast T, Gonzalez F 2014 Economic cycles and heart disease in Mexico *Social Science & Medicine* vol **109** issue pp 19 – 25.

1246. Sedlacek P, Sterk V 2014 The growth potential of startups over the business cycle *University of Bonn* Germany *University College London* UK.

1247. Da Costa 2015 Weak first-quarter growth due to seasonal issues after all, SF Fed says *The Wall Street Journal* New York USA.

1248. Desai M, King St, Goodhart Ch 2015 Hubris: Why economists failed to predict the crisis and how to avoid the next one *Public Lecture on 27.05.2015* London School of Economics and Political Science London UK

http://media.rawvoice.com/lse_publiclecturesandevents/richmedia.lse.ac.uk/publiclecturesandevents/20150527_1830_hubris.mp4 .

1249. Federal Reserve Bank of St Louis 2015 US Federal Reserve Economic Data (FRED) *Federal Reserve Bank of St Louis* USA

http://research.stlouisfed.org/fred .

1250. Galli J 2015 Monetary policy, inflation, and the business cycle *Princeton University Press* Princeton USA.

1251. Gomis R M, Khatiwada S 2015 Firm dynamics and business cycle: Better understanding the effects of regressions *International Labour Organization Research Department*.

1252. Levchenko A A, Pandalai-Nayar N 2015 TFP, news, and sentiments: The international transmission of business cycles *NBER Working Papers 21010* National Bureau of Economic Research Inc USA.

1253. Lucarelli St, Paulré B 2015 Introduction *EJESS* vol **27** no 1-2/2015 pp 1 – 4 www.ejess.revuesonline.com .

1254. Maloney J, Pickering A 2015 Voting and the economic cycle *Public Choice* vol **162** issue 1 pp 119 – 133.

1255. Moreira S 2015 Firm dynamics, persistent effects of entry conditions, and business cycles *The University of Chicago* USA.

1256. Motoki F Y S, Gutierrez C E C 2015 Firm performance and business cycles: Implications for managerial accountability *Applied Finance and Accounting* vol **1** no 1 pp 47 – 59.

1257. Yaguang Zhang, Guo Fan, Whalley J October 2015 Economic cycles in ancient China *Working Paper 21672* National Bureau Of Economic Research Cambridge USA pp 1 – 33 http://www.nber.org/papers/w21672 .

1258. Wikipedia 2015a Kondratieff *Wikipedia* USA www.wikipedia.org .

1259. Wikipedia 2015b Simon Kuznets Economist *Wikipedia* USA www.wikipedia.org .

1260. Wikipedia 2017a Gross domestic product *Wikipedia* USA www.wikipedia.org .

1261. Wikipedia 2017c Business cycle *Wikipedia* California USA www.wikipedia.org .

1262. Achilleas T March 2016 Financial versus real economic variables in explaining growth and cycles *Thesis ID:8/15* University of Macedonia pp 1 – 51.

1263. Bai C Y 2016 Linear Bayesian inference theory applied in complex analysis of economic forecasting and management *Chemical Engineering Transactions* **51** pp 937 – 942 DOI:10.3303/CET1651157 ISBN 978-88-95608-43-3 ISSN 2283-9216.

1264. Bazillier R, Magris F, Mirza D February 2016 Out-migration and economic cycles Document de Recherche du Laboratoire d'Économie d'Orléans DR LEO 2016-03 Laboratoire d'Économie d'Orléans Collegium DEG France pp 1 – 39 www.leo-univ-orleans.fr/ .

1265. Begenau J, Salomao J 2016 Firm financing over the business cycle *Harvard Business School* Boston MA USA.

1266. Chunyang Bai 2016 Linear Bayesian inference theory applied in complex analysis of economic forecasting and management *Chemical Engineering Transactions* vol **51** pp 937 – 942.

1267. D'Agostino A, Giannone D, Lenza M, Modugno M 2016 Nowcasting business cycles: A Bayesian approach to dynamic heterogeneous factor models *Advances in Econometrics* Emerald Group Publishing Limited vol **35** pp 569 – 594.

1268. Gîlcă V-F 2016 The economic cycle and sustainable development pp 1 – 12.

1269. Kronick J July 2016 Taking the economic pulse: An improved tool to help track economic cycles in Canada *Research Paper no 243* C D Howe Institute Canada.

1270. Legrand M D-P, Assous M, Hagemann H 2016 Business cycles and economic growth *in* Handbook of the history of economic analysis Faccarello G, Kurz H (editors) *Edward Elgar Publishing* Cheltenham UK.

1271. Nopphawan Photphisutthiphong, Weder M 2016 Observations on the Australian business cycle *Journal of Business Cycle Research* vol **12** issue 2 pp 141 – 164.

1272. Ojamäe K 2016 Ettevõtte alustamise ja lõpetamise õjutegurid majandustsükli eri faasides (Euroopa riikide andmetel) Factors affecting entry and exit of companies under different phases of the economic cycle (Based upon European data) *Tallinna Tehnikaülikool* Tallinn Estonia pp 1 – 92.

1273. Olkhov V October 2016 On economic space notion *International Review of Financial Analysis* vol **47** issue C pp 372 – 381 https://doi.org/10.1016/j.irfa.2016.01.001 .

1274. Olkhov V 2017 Econophysics of macroeconomics: "Action-at-a-distance" and waves *TVEL* Moscow Russian Federation pp 1 – 24.

1275. Yaguang Zhang, Guo Fan, Whalley J January 2016 Economic cycles in ancient China *Working Paper no 262* Centre for Competitive Advantage in Global Economy Department of Economics University of Warwick UK pp 1 – 33.

1276. Xu Y, Ni Y, van Leeuwen B 2016 Calculating China's historical economic aggregate: A GDP-centered overview *Soc Sci China* **37** (2) pp 56 – 75.

1277. Bazillier R, Magris Fr, Mirza D 2017 Out-migration and economic cycles *Review of World Economic* vol **153** issue 1 pp 39 – 69.

1278. Daianu D 2017 When policies fuel economic cycles *Romanian Journal of Economic Forecasting* **XX** (1) pp 167 – 190.

1279. Falter J-M, Bossi F, Scheurer R, Hanimann D, Schönholzer U, Chabloz A, Näf W, Kobel R December 2017 Business cycle signals: Results of the SNB company talks *Quarterly Bulletin 4/2017* Swiss National Bank Economic Affairs Börsenstrasse 15 CH-8022 Zurich Switzerland ISSN 1662-257X pp 29 – 37 www.snb.ch .

1280. McGregor Th, Wills S March 2017 Surfing a wave of economic growth *Economics Working Paper Series 2017 – 05* University of Sydney Australia pp 1 – 59.

1281. Siena D 15 February 2017 What's news in international business cycles *Banque de France* Paris France pp 1 – 32.

1282. Villarreal C C, Bielma L H 2017 Economic integration, economic crises and economic cycles in Mexico *Contaduría Y Administración* vol **62** no 62-1 ISSN: 0186-1042 (Print) 2448-8410 (Online) www.cya.unam.mx/index.php/cya .

1283. National Bureau of Economic Research 2018 US business cycle expansions and contractions *National Bureau of Economic Research* Cambridge MA USA http://www.nber.org/cycles/cyclesmain.html .

1284. US Department of Commerce 2018 National economic accounts *US Department of Commerce* USA www.bea.gov/national .

### *Disruptive technological and social innovations in economics and finances:*

**1285.** Schumpeter J A 1911; 1939, 1961 Theorie der wirtschaftlichen entwicklung; The theory of economic development: An inquiry into profits, capital, credit, interest and the business cycle Redvers Opie (translator) *OUP* New York USA.

**1286.** Schumpeter J A 1939 Business cycle *McGraw-Hill* New York USA.

**1287.** Schumpeter J A 1947 The creative response in economic history *Journal of Economic History* vol **7** pp 149 – 159.

**1288.** Solow R H August 1957 Technical change and the aggregate production function *Review of Economics and Statistics* **39** pp 214 – 231.

**1289.** Christensen C M June 16, 1977 Fatal attraction: The dangers of too much technology *Computerworld Leadership Series* pp 3 – 11.

**1290.** Christensen C M Fall 1992a Exploring the limits of the technology S-curve, Part 1: Component Technologies *Production and Operations Management* **1** pp 334 – 357.

**1291.** Christensen C M Fall 1992b Exploring the limits of the technology S-curve, Part 2: Architectural technologies *Production and Operations Management* **1** pp 358 – 366.

**1292.** Bower J L, Christensen C M January February 1995 Disruptive technologies: Catching the wave *Harvard Business Review* **73** no 1 pp 43 – 53.

**1293.** Bower J L, Christensen C M 1997 Disruptive technologies: Catching the wave *in* Seeing differently: Insights on innovation Brown J S (editor) *Harvard Business School Press* Boston MA USA.

**1294.** Christensen C M 1997 The innovator's dilemma: When new technologies cause great firms to fail *Harvard Business School Press* Boston MA USA.

**1295.** Christensen C M, Armstrong E G Spring 1998 Disruptive technologies: A credible threat to leading programs in continuing medical education? *Journal of Continuing Education in the Health Professions* **69** no 80 pp 69 – 80.

**1296.** Christensen C M 1998 The evolution of innovation *in* Technology management handbook Dorf R (editor) *CRC Press* Boca Raton FL USA.

**1297.** Christensen C M December 1998 Disruptive technologies: Catching the wave TN *Harvard Business School Teaching Note 699 - 125.*

**1298.** Christensen C M, Cape E G December 1998 Disruptive technology a heartbeat away: Ecton, Inc *Harvard Business School Case 699 - 018.*

**1299.** Christensen C M April 1999a Value networks and the impetus to change: Managing innovation: Overview teaching note for module 1 *Harvard Business School Teaching Note 699 - 163.*

**1300.** Christensen C M April 1999b Finding new markets for new and disruptive technologies: Managing innovation, overview teaching note for module 2 *Harvard Business School Teaching Note 699 - 164.*

**1301.** Christensen C M April 1999c Teradyne: The Aurora project & Teradyne: Corporate management of disruptive change, TN *Harvard Business School Teaching Note 399 - 087.*

**1302.** Christensen C M, Dann J June 1999 Processes of strategy definition and implementation, The *Harvard Business School Background Note 399 - 179.*

1303. Bower J L, Christensen C M 1999 Disruptive technologies: Catching the wave Ch 29 *in* The entrepreneurial venture 2nd edition Sahlman W A, Stevenson H H, Roberts M J, Bhide A V pp 506 – 520 *Harvard Business School Press* Boston MA USA.

1304. Christensen C M 1999a Innovation and the general manager *Irwin McGraw-Hill* Homewood IL USA.

1305. Christensen C M 1999b Impact of disruptive technologies in telecommunications in Bringing PC economies to the telecommunications industry *PulsePoint Communications*.

1306. Christensen C M, Tedlow R S January February 2000 Patterns of disruption in retailing *Harvard Business Review* **78** no 1 pp 42 – 45.

1307. Christensen C M, Donovan T March 2000 Disruptive technology a heartbeat away: Ecton, Inc TN *Harvard Business School Teaching Note 600 - 129*.

1308. Christensen C M, Overdorf M March April 2000 Meeting the challenge of disruptive change *Harvard Business Review* **78** no 2 pp 66 – 76.

1309. Christensen C M, Bohmer R M J, Kenagy J September October 2000 Will disruptive innovations cure health care? *Harvard Business Review* **78** no 5 pp 102 – 117.

1310. Christensen C M, Craig Th, Hart S March April 2001 The great disruption *Foreign Affairs* **80** no 2.

1311. Christensen C M Summer 2001 Assessing your organization's innovation capabilities *Leader to Leader* no 21 pp 27 – 37.

1312. Christensen C M, Milunovich S March 2002 Technology strategy: The theory and application of the Christensen model *Merrill Lynch Report Series*.

1313. Bass M J, Christensen C M April 2002 The future of the microprocessor business *IEEE Spectrum* **39** no 4.

1314. Anthony S D, Roth E A, Christensen C M April 2002 The policymaker's dilemma: The impact of government intervention on innovation in the telecommunications industry *Harvard Business School Working Paper no 02 - 075*.

1315. Kenagy J, Christensen C M May 2002 Disruptive innovation: A new diagnosis for health care's 'Financial flu' *Healthcare Financial Management* pp 62 – 66.

1316. Christensen C M, Johnson M W, Rigby D K Spring 2002 Foundations for growth: How to identify and build disruptive new businesses *MIT Sloan Management Review* **43** no 3.

1317. Kenagy J W, Christensen C M 2002 Disruptive innovation - New diagnosis and treatment for the systemic maladies of healthcare *World Markets Series Business Briefing Global Healthcare 2002* pp 14 – 17.

1318. Christensen C M June 2002 The rules of innovation *Technology Review*.

1319. Hart S L, Christensen C M Fall 2002 The great leap: Driving innovation from the base of the global pyramid *MIT Sloan Management Review* **44** no 1 pp 51 – 56.

1320. Christensen C M, Verlinden M, Westerman G November 2002 Disruption, disintegration, and the dissipation of differentiability *Industrial and Corporate Change* **11** no 5 pp 955 – 993.

1321. Christensen C M 2003 The opportunity and threat of disruptive technologies *Harvard Business School Publishing Class Lecture* HBSP Product Number 1482C Boston MA USA.

1322. Shah Ch D, Brennan T A, Christensen C M April 2003 Interventional radiology: Disrupting invasive medicine.

1323. Christensen C M March April 2003 Beyond the innovator's dilemma *Strategy & Innovation* **1** no 1.

1324. Christensen C M, Raynor M E 2003 The innovator's solution: Creating and sustaining successful growth *Harvard Business School Press* Boston MA USA.

1325. Burgelman R A, Christensen C M, Wheelwright S C 2003 Strategic management of technology and innovation 4th edition *McGraw-Hill Irwin* USA.

1326. Christensen C M, Anthony S D January February 2004 Cheaper, faster, easier: Disruption in the service sector *Strategy & Innovation* **2** no 1.

1327. Christensen C M, Anthony S D, Roth E A 2004 Seeing what's next: Using the theories of innovation to predict industry change *Harvard Business School Press* Boston MA USA.

1328. Christensen C M January 2006 The ongoing process of building a theory of disruption *Journal of Product Innovation Management* **23** pp 39 – 55.

1329. Christensen C M, Baumann H, Ruggles R, Sadtler Th M December 2006 Disruptive innovation for social change *Harvard Business Review* **84** no 12.

1330. Christensen C M, Horn M B, Johnson C W 2008 Disrupting class: How disruptive innovation will change the way the World learns *McGraw-Hill* USA.

1331. Christensen C M, Grossman J H, Hwang J 2009 The innovator's prescription: A disruptive solution for health care *McGraw-Hill* USA.

1332. Dyer J H, Gregersen H B, Christensen C M December 2009 The innovator's DNA *Harvard Business Review* **87** no 12.

1333. Christensen C M, Donovan T May 2010 Disruptive IPOs? WR Hambrecht & Co *Harvard Business School Case 610-065*.

1334. Dyer J H, Gregersen H B, Christensen C M 2011 The innovator's DNA: Mastering the five skills of disruptive innovators *Harvard Business Press* Boston MA USA.

1335. Christensen C M, Talukdar Sh, Alton R, Horn M B Spring 2011 Picking green tech's winners and losers *Stanford Social Innovation Review* USA.

1336. Christensen C M, Wang D, van Bever D October 2013 Consulting on the cusp of disruption *Harvard Business Review* **91** no 10 pp 106 – 114.

1337. Christensen C M, Raynor M E, McDonald R December 2015 What is disruptive innovation? *Harvard Business Review* Cambridge MA USA pp 44 – 53 https://hbr.org/2015/12/what-is-disruptive-innovation .

1338. Christensen C M, Denning St December 2015 Disruptive innovation *Forbes* New York USA http://www.forbes.com/sites#/sites/stevedenning/2015/12/02/fresh-insights-from-clayton-christensen-on-disruptive-innovation/ .

1339. Christensen C M 2015 Disruptive strategy *Course for Senior Executives* Harvard Business School Harvard University Cambridge USA.

1340. Bhattacharya S, Ritter J R 1983 Innovation and communication: Signaling with partial disclosure *Review of Economic Studies* **50** pp 331 – 346.

1341. Scherer F M 1984 Innovation and growth: Schumpeterian perspectives *MIT Press* Cambridge MA USA.

1342. Rogers E M August 16 2003 Diffusion of innovations 5th edition *Free Press* ISBN-13: 978-0743222099 pp 1 – 576.

1343. Rodin J 2015 Managing disruption, avoiding disaster and growing stronger in an unpredictable World *Public Lecture on 19.01.2015* London School of Economics and Political Science London UK

http://media.rawvoice.com/lse_publiclecturesandevents/richmedia.lse.ac.uk/publiclectures andevents/20150119_1830_managingDisruption.mp4 .

1344. Dobbs R, Woetzel J, Flanders St 2015 No ordinary disruption: The four global forces breaking all the trends *Public Lecture on 08.06.2015* London School of Economics and Political Science London UK

http://media.rawvoice.com/lse_publiclecturesandevents/richmedia.lse.ac.uk/publiclectures andevents/20150608_1830_noOrdinaryDisruption.mp4 .

1345. Barber L 2015 Making news for the new World *Public Lecture on 12.11.2015* London School of Economics and Political Science London UK http://media.rawvoice.com/lse_publiclecturesandevents/richmedia.lse.ac.uk/publiclect uresandevents/20151112_1830_makingNewsForTheNewWorld.mp4 .

1346. Isaacson W 6 October 2015 The innovators: How a group of hackers, geniuses, and geeks created the digital revolution *Simon & Schuster* ISBN-13 978-1476708706 pp 1 – 560.

***Metal coins, paper money, electronic money, network money, electronic cash, digital cash, bit coin, electronic payments, debit cards, credit cards, stored value cards, smart cards (electronic purses) in finances:***

1347. Smith A 1776, 1991 An inquiry into the nature and causes of the wealth of nations London UK, *Alfred A Knopf Inc* New York USA.

1348. Ricardo D 1816, 1951 Proposals for an economical and secure currency *in* The works and correspondence of David Ricardo vol **IV**: Pamphlets and Papers, 1815-1823 Piero Sraffa (editor), *2nd* edition *Cambridge University Press* London Cambridge UK.

1349. Del Mar A 1894 History of money in ancient countries New-York USA.

1350. Fisher I 1933 Stamped scrip *Adelphi & Co* New York USA

http://userpage.fu-berlin.de/~roehrigw/fisher/

1351. Keynes J M 1936 The general theory of employment, interest, and money *Harcourt Brace Jovanovich* New York USA.

1352. Redlich F 1951 The molding of American banking: Men and ideas *Hafner Publishing Company Inc* New York USA.

1353. Baumol W 1952 The transactions demand for cash – An inventory theoretic approach *Quarterly Journal of Economics* **66** pp 545 – 556.

1354. Butlin S J 1953 Foundations of the Australian monetary system 1788-1851 *Sydney University Press* Sydney Australia.

*1355.* Tobin J 1956 The interest rate elasticity of transactions demand for money *Review of Economics and Statistics* **38** (3) pp 241 – 247.

*1356.* Tobin J May 1961 Money, capital, and other stores of value *American Economic Review* **LI** pp 26 – 37.

*1357.* Tobin J 1963 Commercial banks as creators of money *in* Banking and monetary studies Carson D (editor) *Irwin* Homewood IL USA pp 408 – 419.

*1358.* Tobin J 1965 Money and economic growth *Econometrica* vol **33** no 3 pp 671 – 684.

*1359.* Cook R M 1958 Speculations on the origins of coinage *Historia* 7 pp.257 – 262.

*1360.* Carson R A G 1962 Coins of the World New-York USA.

*1361.* Friedman M, Jacobson Schwartz A 1963 A monetary history of the United States, 1867-1960 *Princeton University Press* Princeton NJ USA pp 1 – 442.

*1362.* Friedman M, Jacobson Schwartz A 1986 Has government any role in money? *Journal of Monetary Economics* vol **17** (1) pp 37 – 62.

*1363.* Black F August 1970 Banking and interest rates in a World without money *Journal of Banking Research* **1** pp 8 – 20.

*1364.* Crawford M 1970 Money and exchange in the Roman world *Journal of Roman Studies* **60** pp 40 – 48.

*1365.* Balmuth M S 1971 Remarks on the appearance of the earliest coins *in* Studies Presented to George M.A. Hanfmann *Harvard University Press* Cambridge USA pp 1 – 7.

*1366.* Rousseas St W 1972 Monetary theory *Enopf* New York USA.

*1367.* Thompson M, Kraay C M, Morkholm O (editors) 1973 An inventory of Greek coin hoards New York USA.

*1368.* Hayek F A 1974, 1976a Choice in currency: A way to stop inflation *Occasional Paper 48* The Institute of Economic Affairs London UK.

*1369.* Hayek F A 1976b Denationalization of money: An analysis of the theory and practice of concurrent currencies *Hobart Paper Special 70* The Institute of Economic Affairs London UK.

*1370.* Hayek F A 1978 Denationalization of money: The argument refined *The Institute of Economic Affairs* London UK.

*1371.* Checkland S G 1975 Adam Smith and the bankers *in* Essays on Adam Smith Skinner A S, Wilson T (editors) *Clarendon Press* London UK.

*1372.* Galbraith J K 1976 Money: Whence it came, where it went *Houghton Mifflin Company* Boston USA.

*1373.* McKinnon R I 1979 Money in international exchange: The convertible currency system *Oxford University Press* UK.

*1374.* Fama E F 1980 Banking in a theory of finance *Journal of Monetary Economics* vol **6** pp 39 – 57.

*1375.* Kagan D 1982 The dates of the earliest coins *American Journal of Archaeology* **86** (2) pp 343 – 360.

*1376.* Price M J 1983 Thoughts on the beginnings of coinage *in* Studies in numismatic method presented to Philip Grierson Brooke C N L et al (editors) *Cambridge University Press* Cambridge UK pp 1 – 10.

1377. White L H September 1984 Competitive payments systems and the unit of account *American Economic Review* vol **74** no 4 pp 699 – 712.

1378. White L H 1989 Competition and currencies *New York University Press* NY USA.

1379. White L H (editor) 1993 Free banking vols **1**, **2**, **3** *E Elgar Publishing* Aldershot Hants UK.

1380. White L H 1999 The theory of monetary institutions *Blackwell Publishers* Oxford UK.

1381. Hellwig M F 1985 What do we know about currency competition? *Zeitschrift Wirtschafts-und Sozialwissenschaften* **5** pp 565 – 588.

1382. Lawrence C, Shay R P (editors) Technological innovation, regulation, and the monetary economy *Ballinger Publishing Company* Cambridge USA.

1383. Wallace N 1986 The impact of new payment technologies: A macro view *in* Technological innovation, regulation, and the monetary economy Lawrence C, Shay R P (editors) *Ballinger Publishing Company* Cambridge pp 201 – 206.

1384. Prescott E 1987 Multiple means-of-payment model *in* New approaches to money economics Barnett W, Singleton K (editors) *Cambridge University Press* Cambridge, New York USA.

1385. Suhr D 1989 The capitalistic cost-benefit structure of money: An analysis of money's structural non-neutrality and its effects on the economy *Springer* Berlin Heidelberg New York USA.

1386. Wallace R B 1987 The origin of the electrum coinage *American Journal of Archaeology* **91** pp 385 – 397.

1387. Wallace R B 1989 On the production and exchange of early Anatolian electrum coinages *Revue des Etudes Anciennes* **91** pp 87 – 94.

1388. Goodhart Ch 1989 Money, information, and uncertainty *Macmillan* London UK.

1389. Goodhart Ch 2000 Can central banks survive the IT revolution? *International Finance* vol **3** no 2 pp 189 – 209.

1390. Kennedy M 1989 Interest and inflation free money: How to create an exchange medium that works for everyone *Permakultur Instiute e.V.*

1391. Whitesell W 1989 The demand for currency versus debitable accounts *Journal of Money, Credit and Banking* **21** (2) pp 246 – 251.

1392. Whitesell W 1992 Deposit banks and the market for payment media *Journal of Money, Credit and Banking* **24** (4) pp 484 – 496.

1393. Howgego Ch J 1990 Why did ancient states strike coins? *The Numismatic Chronicle* **150** pp 1 – 25.

1394. Karwiese St 1991 The Artemisian hoard and the first coins of Ephesus *Revue Belge de Numismatique* **13** (7 ) pp 1 – 28.

1395. Selgin G A, White L W December 1994 How would the invisible hand handle money? *Journal of Economic Literature* pp 1718 – 1749.

1396. Bauer P W October 1 1995 Making payments in cyberspace *Economic Commentary* Federal Reserve Bank of Cleveland USA.

1397. Crede A 1995 Electronic commerce and the banking industry: The requirement and opportunities for new payment systems using the Internet *Journal of Computer Mediated Communication* vol **1/3**

http://www.ascusc.org/jcmc/vol1/issue3/vol1no3.html .

1398. Duca J V, Whitesell W C 1995 Credit cards and money demand: A cross-sectional study *Journal of Money, Credit and Banking* **27** (2) pp 604 – 623.

1399. Humphrey D B, Pulley L B, Vesala J M 1996 Cash, paper, and electronic payments: A cross-country analysis *Journal of Money, Credit and Banking* vol **28** no 4 part 2 pp 914 – 939.

1400. Humphrey D B 2004 Replacement of cash by cards in US consumer payments *Journal of Economics and Business* **56** pp 211 – 225.

1401. Kezar M Winter 1995/1996 Logging on to electronic means of payment *Cross Sections* Federal Reserve Bank of Richmond USA pp 10 – 18.

1402. Matonis J W 1995 Digital cash and monetary freedom *INET 95 June 26 - 30* Honolulu Hawaii USA.

1403. Thiveaud J-M, Sylvain P 1995 De la monnaie électronique à l'invention de la monnaie d'électron : en Lydie au VIIe sièclleavant Jésus-Christ *Revue d'économie financière* N°32 1995 Les technologies bancaires et financières pp 271 – 293.

doi : 10.3406/ecofi.1995.2178

http://www.persee.fr/web/revues/home/prescript/article/ecofi_0987-3368_1995_num_32_1_2178

1404. Wenninger J, Laster D April 1995 The electronic purse *Federal Reserve Bank of New York Current Issues in Economics and Finance* **1** (1) Federal Reserve Bank of New York US   pp 1 – 6.

1405. Bank for International Settlements (BIS) 1996a Security of digital money *Committee on Payment and Settlement Systems Bank for International Settlements* Basel Switzerland

www.bis.org/publ/index.htm .

1406. Bank for International Settlements (BIS) 1996b Implications for central banks of the development of digital money *Committee on Payment and Settlement Systems Bank for International Settlements* Basel Switzerland

www.bis.org/publ/index.htm .

1407. Bank for International Settlements (BIS) December 1998 Statistics on payment systems in the group of ten countries *Committee on Payment and Settlement Systems Bank for International Settlements* Basel Switzerland

www.bis.org/publ/index.htm .

1408. Bank for International Settlements (BIS) September 1999 Retail payments in selected countries: A comparative study *Committee on Payment and Settlement Systems Bank for International Settlements* Basel Switzerland

www.bis.org/publ/index.htm .

1409. Bank for International Settlements (BIS) 2000 Survey of electronic money developments *Committee on Payment and Settlement Systems Bank for International Settlements* Basel Switzerland

www.bis.org/publ/index.htm .

1410. Bank for International Settlements (BIS) 2001a Survey of electronic money developments *Committee on Payment and Settlement Systems Bank for International Settlements* Basel Switzerland

www.bis.org/publ/index.htm .

1411. Bank for International Settlements (BIS) 2001b Electronic finance: A new perspective and challenges *BIS Papers no 7 Monetary and Economic Department Bank for International Settlements* Basel Switzerland

www.bis.org/publ/index.htm .

1412. Bank for International Settlements (BIS) 2004 Survey of developments in electronic money and Internet and mobile payments *Committee on Payment and Settlement Systems Bank for International Settlements* Basel Switzerland

www.bis.org/publ/index.htm .

1413. Bernkopf M 1996 Electronic cash and monetary policy *First Monday Munksgaard International Publishers* Copenhagen Denmark

http://www.firstmonday.dk .

1414. Browne F X, Cronin D 1996 Payment technologies, financial innovation, and Laissez-Faire banking: A further discussion of the issues *in* The future of money in the information age Dorn J A (editor) *Cato Institute* Washington D C USA

http://www.cato.org/pubs/books/money/money18.htm .

1415. Dorn J A (editor) 1996 The future of money in the information age *Cato Institute* Washington DC USA.

1416. Jordan J L, Stevens E J 1996 Money in the 21$^{st}$ Century *in* The future of money in the information age Dorn J A (editor) *Cato Institute* Washington DC USA http://www.cato.org/pubs/books/money/money15.htm .

1417. Lynch D, Lundquist L 1996 Digital money: The new era of Internet commerce *John Willey and Sons Inc* New York USA.

1418. Mitchell R December 1996 Lots of calls for phone cards *Credit Card Management* vol **9** no 9 pp 14 – 18.

1419. Santomero A, Seater J 1996 Alternative monies and the demand for media of exchange *Journal of Money, Credit and Banking* **28** (4) pp 942 – 960.

1420. US Treasury September 1996 An introduction to electronic money issues *US Treasury* USA.

1421. Berentsen A 1997a Digital money, liquidity, and monetary policy *First Monday Munksgaard International Publishers* Copenhagen Denmark

http://www.firstmonday.dk/issues/issue2_7/berentsen/index.html .

1422. Berentsen A 1997b Supervision and regulation of network banks *First Monday Munksgaard International Publishers* Copenhagen Denmark

http://www.firstmonday.dk/issues/issue2_8/berentsen/index.html .

1423. Berentsen A 1997c March 2012 Monetary policy implications of digital money *MPRA Paper no 37392* University of Munich Germany pp 1 – 29

http://mpra.ub.uni-muenchen.de/37392/ .

1424. Choi S-Y, Stahl D, Whinston A 1997 The economics of electronic commerce *Macmillan Technical Publishing* Indianapolis USA.

1425. Cronin M J (editor) Banking and finance on the Internet *Van Nostrand Reinhold Press* New York USA.

1426. Frei F, Kalakota R 1997 Frontiers of online financial services *in* Banking and finance on the Internet Cronin M J (editor) *Van Nostrand Reinhold Press* New York USA.

1427. Hitachi Research Institute 1997 Electronic money: Its impact on retail banking and electronic commerce *Hitachi Research Institute* Japan, *FIA Financial Publishing Company* USA.

1428. Kennickell A B, Kwast M L July 1997 Who uses electronic banking? Results from the 1995 survey of consumer finance *Finance and Economics Discussion Series Paper 1997-35* Board of Governors of the Federal Reserve System USA.

1429. Kobrin S J 1997 Electronic cash and the end of national markets *Foreign Policy* pp 65 – 77.

1430. Marimon R, Nicolini J P, Teles P August 1997 Electronic money: The end of inflation? *Discussion Paper 122* Institute for Empirical Macroeconomics Federal Reserve Bank of Minneapolis USA pp 1 – 36.

1431. McAndrews J J January/February 1997 Making payments on the Internet *Federal Reserve Bank of Philadelphia Business Review* USA pp 3 – 14.

1432. McAndrews J J November/December 1997 Network issues and payment systems *Federal Reserve Bank of Philadelphia Business Review* USA pp 15 – 25.

1433. McAndrews J J 1997 Banking and payment system stability in an electronic money World *Working Paper 97-9* Federal Reserve Bank of Philadelphia USA.

1434. McAndrews J J July 1999 E-money and payment system risk *Contemporary Economic Policy* **17** pp 348 – 357.

1435. McKnight L, Bailey J (editors) 1997 Internet economics *MIT Press* Cambridge MA USA.

1436. Neuman C, Medvinsky G 1997 Internet payment services *in* Internet economics McKnight L, Bailey J (editors) *MIT Press* Cambridge MA USA.

1437. Schreft S L 1997 Looking forward: The role for government in regulating electronic cash *Economic Review - 4th Quarter 1997* Federal Reserve Bank of Kansas City Kansas USA pp 59 – 84.

1438. Woodford M September 1997 Doing without money: Controlling inflation in a post-monetary world *NBER Working Paper no 6188* National Bureau of Economic Research Cambridge Massachusetts USA.

1439. Woodford M 2000 Monetary policy in a world without money *International Finance* vol **3** no 2 pp 229 – 260.

1440. Woodford M 2003 Interest and prices: Foundations of a theory of monetary policy *Princeton University Press* Princeton NJ USA.

1441. European Central Bank August 1998 Report on electronic money *European Central Bank* Frankfurt am Main Germany www.ecb.int/pub/pdf/emoney.pdf .

1442. Furst K, Lang W W, Nolle D E September 1998 Technological innovation in banking and payments: Industry trends and implications for banks *Office of the Comptroller of the Currency Quarterly Journal* **17** pp 23 – 31.

1443. Hatakka T 1998 Payment methods in Finland and selected EU countries: Electronic banking and developments *Bank of Finland Bulletin* **72** (2) pp 3 – 7.

1444. Phillips A L Winter 1998 Migration of corporate payments from check to electronic format: A report on the current status of payments *Financial Management* **27** pp 92 – 105.

1445. Shy O, Tarkka J 1998 The market for electronic cash cards *Suomen Pankin Monistuskeskus* Helsinki Finland ISBN 951-686-591-7 pp 1 – 35.

1446. Stalder F, Clement A July 1998 Exploring policy issues of electronic cash: The Mondex case *Working Paper 8* Information Policy Research Program Faculty of Information Studies University of Toronto Canada www.fis.utoronto.ca/research/iprp/dipcii/workpap8.htm.

1447. US General Accounting Office July 1998 Experience with electronic check presentment *GAO/GGD-98-145* US General Accounting Office USA.

1448. Gowrisankaran G, Stavins J May 1999 Are there network externalities in electronic payments? *Federal Reserve Bank of Chicago Conference on Bank Structure and Competition* Federal Reserve Bank of Boston USA.

1449. Hankel T, Ize A, Kovanen A 1999 Central banking without a central bank *IMF Working Paper no 99/92* International Monetary Fund Washington DC USA.

1450. Hitt L M, Frei F X April 1999 Do better customers utilize electronic distribution channels? The case of PC banking *Working Paper 99-21* Wharton Financial Institutions Center USA.

1451. Hogarth J M, O'Donnell K H July 1999 Banking relationships of lower-income families and the governmental trend toward electronic payment *Federal Reserve Bulletin* **85** pp 459 – 473.

1452. King M August 27 1999, November 1999 Challenges for monetary policy: New and old *New Challenges for Monetary Policy Symposium* Federal Reserve Bank of Kansas City Jackson Hole Wyoming USA, *Bank of England Quarterly Bulletin* vol **39** pp 397 – 415.

1453. Orr B July 1999a At last Internet banking takes off *ABA Banking Journal* p 36.

1454. Orr B July 1999b E-banks or E-branches? Both are in play as early adopters make them work *ABA Banking Journal* pp 32 – 34.

1455. Prinz A 1999 Money in the real and virtual world: E-money, C-money and the demand for Cb-money *Netnomics* vol **1** pp 11 – 35.

1456. Schulz K August 1999 The future of digital cash *Banking Policy Report* **18** pp. 8 – 13.

1457. Van Hove L 1999 Electronic money and the network externalities theory: Lessons for real life *Netnomics* vol **1** pp 137 – 171.

1458. Freedman Ch 2000 Monetary policy implementation: Past, present and future - Will electronic money lead to the eventual demise of central banking? *International Finance* vol **3** no 2 pp 211 – 227.

1459. Friedman B M 2000 Decoupling at the margins: The threat to monetary policy from the electronic revolution in banking *NBER Working Paper no 7955* National Bureau of Economic Research Cambridge Massachusetts USA.

1460. Huber J, Robertson J 2000 Creating new money: A monetary reform for the information age New Economics Foundation London UK.

1461. Mester L J March/April 2000 The changing nature of the payments system: Should new players mean new rules? *Business Review* Federal Reserve Bank of Philadelphia USA pp 3 – 26.

1462. Rahn R W 2000 The impact of digital money on central banks *Cato Journal Institute's 18th Annual Monetary Conference*
www.cato.org .

1463. Workshop October 20 – 21 2000 The analysis of new electronic payments systems based on Carl Menger's institutional theory of the origin of money *Workshop* Vienna Austria.

1464. Arnone M February 26 2001 E-money and the transmission of monetary policy *2nd Monetary Conference: Electronic Banking: Challenges and Opportunities for Central Banks* Central Bank of Philippines Manila Philippines.

1465. Arnone M, Bandiera L 2003 E-money: Chances of success and consequences for monetary policy *Working Paper no 401* Kiel Institute for World Economics Kiel Germany.

1466. Arnone M, Bandiera L July 2004 Monetary policy, monetary areas, and financial development with electronic money *IMF Working Paper WP/04/122* Monetary and Financial Systems Department International Monetary Fund USA pp 1 – 43.

1467. Beck H 2001 Banking is essential, banks are not. The future of financial intermediation in the age of the Internet *Netnomics* vol 3 pp 7 – 22.

1468. Bootle R 2001 The future of electronic money - Why the Nok will not replace the Dollar *The Business Economist* vol **32** no 1 pp 7 – 15.

1469. Cohen B J 2001 Electronic money. New day or false dawn? *Review of International Political Economy* vol **8** (2) pp 197 – 225.

1470. Costa Storti C, De Grauwe P February 2001 Monetary policy in a cashless society *CEPR Discussion Paper Series no 2696* Centre for Economic Policy Research London UK.

1471. Costa Storti C, De Grauwe P May 2002 Electronic money and optimal size of monetary union *CEPR Discussion Paper Series no 3391* Centre for Economic Policy Research London UK.

1472. Hawkins J 2001 Electronic finance and monetary policy *in* Electronic finance: A new perspective and challenges *BIS Papers no 7* Bank for International Settlements Basel Switzerland.

1473. Mesonnier J-S July 2001 Monnaie electronique et politique monétaire *Bulletin de la Banque de France* no 91 Paris France.

1474. Sato S, Hawkins J November 2001 Electronic finance: An overview of the issues *in* Electronic finance: A new perspective and challenges *BIS Paper no 7* Bank for International Settlements Basel Switzerland.

1475. Spencer P January–March 2001 E-money: Will it take off? *World Economics* vol **2** no 1 pp 121 – 136.

1476. Berk J M September 2002 Central banking and financial innovation. A survey of the modern literature *Banca Nazionale del Lavoro Quarterly Review no 222*.

1477. Davies G 2002 A history of money from ancient times to the present day 3rd edition *University of Wales Press* Cardiff UK.

1478. Drehmann M, Goodhart Ch, Krüger M 2002 The challenges facing currency usage: Will the traditional transaction medium be able to resist competition from the new technologies? *Economic Policy* vol **17** pp 193 – 227.

1479. Organization for Economic Cooperation and Development (OECD) 2002 The future of money *Organization for Economic Cooperation and Development* Paris France.

1480. Palley Th 2002 The e-money revolution: Challenges and implications for monetary policy *Journal of Post Keynesian Economics* vol **24** no 2 pp 217 – 233.

1481. Shy O, Tarkka J May 2002 The market for electronic cash cards *Journal of Money, Credit and Banking* vol **34** no 2.

1482. Stevens E March 2002 Electronic money and the future of central banks *Federal Reserve Bank of Cleveland* USA pp 1 – 4.

1483. Gormez Y, Budd C H 2003 Electronic money free banking and some implications for central banking *Working Paper* Research Department The Central Bank of the Republic of Turkey.

1484. Markose Sh M, Yiing Jia Loke 2003 Network effects on cash-card substitution in transactions and low interest rate regimes *Economic Journal* **113** (487) pp 456 – 476.

1485. Rysman M 2004 An empirical analysis of payment card usage *Journal of Industrial Economics* **55** (1) pp 1 – 36.

1486. Stix H 2004 How do debit cards affect cash demand? Survey data evidence *Empirica* **31** (2-3) pp 93 – 115.

1487. Amromin G, Chakravorti S 2007 Debit card and cash usage: A cross-country analysis *Working Paper Series: WP-07-04* Federal Reserve Bank of Chicago USA.

1488. Nakata M 2007 Effect of electronic money to existing currency demand (in Japanese) *PRI Discussion Paper Series No 07A-19* Policy Research Institute Ministry of Finance Government of Japan.

1489. Williams M M, Anderson R G March 2007 Handicapping currency design: Counterfeit deterrence and visual accessibility in the United States and abroad *Working Paper 2007-011B* Federal Reserve Bank of St. Louis MO USA pp 1 – 74 http://research.stlouisfed.org/wp/2007/2007-011.pdf .

1490. Bank of Japan 2008 Recent developments in electronic money in Japan *BOJ Report and Research Papers* Payment and Settlement Systems Department Bank of Japan 2-1-1 Nihonbashi-Hongokucho Chuo-Ku Tokyo 103-8660 Japan.

1491. Bank of Japan 2009 Recent developments in electronic money in Japan (in Japanese) BOJ Report and Research Papers Payment and Settlement Systems Department Bank of Japan 2-1-1 Nihonbashi-Hongokucho Chuo-Ku Tokyo 103-8660 Japan.

1492. Boaden A March 2008 Recent trends and developments in currency *Reserve Bank of New Zealand: Bulletin* vol **71** no 1 New Zealand pp 16 – 24.

1493. Godschalk H July 28 2008 Electronic money: "Op or a new run-up?" Everyday *digital monies: Innovation in money cultures and technologies conference* Paper Session 1.

1494. Fujiki H, Tanaka M 2009 Demand for currency, new technology and the adoption of electronic money: Evidence using individual household data *Discussion Paper No*

*2009-E-27* Institute for Monetary and Economic Studies Bank of Japan 2-1-1 Nihonbashi-Hongokucho Chuo-Ku Tokyo 103-8660 Japan pp 1 – 39

http://www.imes.boj.or.jp .

**1495.** Turnbull Sh 2010 How might cell phone money change the financial system? *The Capco Institute Journal of Financial Transformation* pp 33 – 42

www.capco.com ,

http://ssrn.com/abstract=1602323 .

**1496.** Moroz V S, Moroz V S September 2014 Roman coin on the Podillia and South-Eastern Volyn' *III International Scientific Conference "Bar's Land Podill'ya: European Heritage and Innovative Development Perspectives* Bar Vinnytsia Region Ukraine ISBN 978-617-7171-10-1 pp 162 – 169.

**1497.** Yeoman R S 2014 A guide book of United States Coins 2015 68th edition *Whitman Publishing LLC* Atlanta Georgia USA ISBN 978-0-7948-4180-5.

***Central banks, federal reserve banks, federal reserve system in finances:***

**1498.** Owen R L 1919 The Federal Reserve Act: Its origin and principles *Century Company* New York USA.

**1499.** Willis H P 1923 The Federal Reserve System: Legislation, Organization, and Operation *Ronald Press Company* New York USA.

**1455.** Bernanke B S 1979 Long-term commitments, dynamic optimization, and the business cycle *Ph. D. Thesis* Department of Economics Massachusetts Institute of Technology USA.

**1456.** Bernanke B S, Blinder A S 1992 The federal funds rate and the channels of monetary transmission *American Economic Review* **82** (4) pp 901 – 921.

**1457.** Bernanke B S, Gertler M 1995 Inside the black box: The credit channel of monetary policy transmission *Journal of Economic Perspectives* **9** (4) pp 27 – 48.

**1458.** Bernanke B S 2002 Deflation: "Making sure it doesn't happen here." *Speech before the National Economists Club* Washington DC

http://www.federalreserve.gov.

**1459.** Bernanke B S 2004 The great moderation

www.federalreserve.gov.

**1460.** Bernanke B S, Reinhart V R 2004 Conducting monetary policy at very low short-term interest rates *The American Economic Review* vol **94** no 2 pp 85 – 90.

**1461.** Bernanke B S, Reinhart V R, Sack B P 2004 Monetary policy alternatives at the zero bound: an empirical assessment *Brookings Papers on Economic Activity* Issue 2 pp 1 – 78.

**1462.** Bernanke B S 2007 The financial accelerator and the credit channel *Speech at The Credit Channel of Monetary Policy in the Twenty-first Century Conference* Federal Reserve Bank of Atlanta Georgia USA.

**1463.** Bernanke B S 2009a The crisis and the policy response *Federal Reserve* USA.

**1464.** Bernanke B S 2009b On the outlook for the economy and policy *Bank for International Settlements* Basel Switzerland

http://www.bis.org/review/r091119a.pdf .

**1465.** Bernanke B S 2009c The Federal Reserve's balance sheet – an update *Bank for International Settlements* Basel Switzerland

http://www.bis.org/review/r091013a.pdf .

**1466.** Bernanke B S 2009d Regulatory reform *Bank for International Settlements* Basel Switzerland

http://www.bis.org/review/r091006a.pdf .

**1467.** Bernanke B S 2009e Policy responses to the financial crisis *Public Lecture on 13.01.2009* London School of Economics and Political Science London UK. http://richmedia.lse.ac.uk/publicLecturesAndEvents/20090113_1300_policyResponses ToTheFinancialCrisis.mp3 .

**1468.** Bernanke B S 2010a Monetary policy and the housing bubble *Annual Meeting of the American Economic Association* Atlanta Georgia USA.

**1469.** Bernanke B S 2010b Causes of the recent financial and economic crisis *testimony before the Financial Crisis Inquiry Commission* Washington USA www.federalreserve.gov/newsevents/testimony/bernanke20100902a.htm .

**1470.** Bernanke B S 2012a Some reflections on the crisis and the policy response *Rethinking Finance: Perspectives on the Crisis conference* sponsored by the Russell Sage Foundation and The Century Foundation New York USA www.federalreserve.gov/newsevents/speech/bernanke20120413a.htm .

**1471.** Bernanke B S 2012b Monetary policy since the onset of the crisis *The Changing Policy Landscape symposium* sponsored by the Federal Reserve Bank of Kansas City Jackson Hole Wyoming USA

www.federalreserve.gov/newsevents/speech/bernanke20120831a.htm .

**1472.** Bernanke B S 2013a Financial and economic education *13th Annual Redefining Investment Strategy Education (RISE) Forum* Dayton Ohio USA.

**1473.** Bernanke B S 2013b Stress testing banks: What have we learned? *Maintaining Financial Stability: Holding a Tiger by the Tail conference* sponsored by the Federal Reserve Bank of Atlanta Stone Mountain Ga USA www.federalreserve.gov/newsevents/speech/bernanke20130408a.htm .

**1474.** Bernanke B S 2013c Monitoring the financial system *49th Annual Conference on Bank Structure and Competition sponsored by the Federal Reserve Bank of Chicago* Chicago Illinois USA pp 1 – 16.

**1475.** Bernanke B S 2013d A century of U.S. central banking: Goals, frameworks, accountability *The first 100 years of the Federal Reserve: The policy record, lessons learned, and prospects for the future conference* sponsored by the *National Bureau of Economic Research* Cambridge Massachusetts USA.

**1476.** Bernanke B S, Blanchard O, Summers L H, Weber A A 2013 What should economists and policymakers learn from the financial crisis? *Public Lecture on 25.03.2013* London School of Economics and Political Science London UK

http://media.rawvoice.com/lse_publiclecturesandevents/richmedia.lse.ac.uk/publiclectures andevents/20130325_1715_whatShouldEconomistsAndPolicymakersLearn.mp4 ; http://www.federalreserve.gov/newsevents/speech/bernanke20130325a.htm.

**1477.** Bernanke 2013a Teaching and learning about the Federal Reserve *A Teacher Town Hall Meeting: 100 Years of the Federal Reserve* Washington DC USA pp 1 – 5.

**1478.** Bernanke 2013c The crisis as a classic financial panic *Fourteenth Jacques Polak Annual Research Conference* Washington DC USA pp 1 – 9.

**1500.** Bernanke B S 2013b A century of U.S. central banking: Goals, frameworks, accountability *The first 100 years of the Federal Reserve: The policy record, lessons learned, and prospects for the future conference* sponsored by the *National Bureau of Economic Research* Cambridge Massachusetts USA.

**1501.** Alesina A, Summers L H 1993 Central Bank independence and macroeconomic performance: Some comparative evidence *Journal of Money, Credit and Banking* vol **25** pp 151 – 162.

**1502.** Capie F, Fischer St, Goodhart Ch, Schnadt N 1994 The development of central banking The future of central banking the tercentenary symposium of the Bank of England *Cambridge University Press* Cambridge UK ISBN 9780521496346

http://eprints.lse.ac.uk/39606/ .

**1503.** Fischer St 1993 The role of macroeconomic factors in growth *Journal of Monetary Economics* **32** pp 485 – 512.

**1504.** Taylor J B 1999 Monetary policy rules *University of Chicago Press* Chicago USA.

**1505.** Ferguson R W Jr 2003 Rules and flexibility in monetary policy *Remarks at the University of Georgia* Athens Georgia USA

http://www.federalreserve.gov/boarddocs/speeches/2003/20030212/default.htm .

**1506.** Meltzer A H 2003 A history of the Federal Reserve vol **1**: 1913-1951 *University of Chicago Press* Chicago Illinois USA.

**1507.** Meltzer A H 2009a A history of the Federal Reserve vol **2** Book 1: 1951-1969 *University of Chicago Press* Chicago Illinois USA.

**1508.** Meltzer A H 2009b A history of the Federal Reserve vol **2** Book 2: 1970-1986 *University of Chicago Press* Chicago Illinois USA.

**1509.** Fox L S, Alvarez S G, Braunstein S, Emerson M M, Johnson J J, Johnson K H, Malphrus S R, Reinhart V R, Roseman L L, Spillenkothen R, and Stockton D J 2005 The Federal Reserve System: Purposes and Functions *Board of Governors of Federal Reserve System* Washington DC 20551 USA *9th edition* Library of Congress Control Number 39026719 pp 1 – 146.

**1510.** Quinn St, Roberts W 2006 An economic explanation of the early Bank of Amsterdam, debasement, bills of exchange, and the emergence of the first central bank Working Paper 2006-13 Federal Reserve Bank of Atlanta USA

http://papers.ssrn.com/sol3/papers.cfm?abstract_id=934871 .

**1511.** Baltensperger E, Hildebrand P M, Jordan T J 2007 The Swiss National Bank's monetary policy concept – an example of a 'principles-based' policy framework *Swiss National Bank Economic Studies* no 3 Swiss National Bank Switzerland ISSN 1661-142X pp 1 - 28.

**1512.** Ledenyov D O, Ledenyov V O 2013g On the Stratonovich - Kalman - Bucy filtering algorithm application for accurate characterization of financial time series with use of state-space model by central banks *MPRA Paper no 50235* Munich University Munich Germany pp 1 – 52, *SSRN Paper no SSRN-id2594333 Social Sciences Research Network* New York USA

http://mpra.ub.uni-muenchen.de/50235/ ,

http://ssrn.com/abstract=2594333 .

1513. Ledenyov D O, Ledenyov V O December 11 - 12 2015 On the Stratonovich - Kalman - Bucy filtering algorithm application for accurate characterization of financial time series with use of state-space model by central banks *23rd Conference on the Theories and Practices of Securities and Financial Markets* National Sun Yat-Sen University Kaohsiung Taipei Taiwan 52 p http://sfm.finance.nsysu.edu.tw/php/Papers/CompletePaper/014-1856280412.pdf .

***Stock exchange history, stock exchange operation principles, company valuation, company stock emission, company stock valuation by market, company stock valuation by rating agencies in economics and finances:***

1514. Mortimer Th 1765 Every man his own broker *4th edition* London UK.

1515. Courtois A 1855 Des opérations de bourse Paris France.

1516. Maddison E 1875 The Paris Bourse and the London Stock Exchange. A comparison of the course of business on each exchange London UK.

1517. Bachelier L 1900 Theorie de la speculation *Annales de l'Ecole Normale Superieure* vol **17** pp 21-86.

1518. Lowenfeld H 1907 The investment of trust funds in the safest and most productive manner *Effingham Wilson* London UK.

1519. Lowenfeld H 1910 All about investment *2nd Edition The Financial Review of Reviews* London UK.

1520. Fisher I 1922 The making of index numbers. A study of their varieties, tests, and reliability New York USA.

1521. Morgan E V, Thomas W A 1961 The stock exchange, its history and functions *Elek Books Ltd* London UK.

1522. Michie R 1988 Different in name only? The London Stock Exchange and foreign bourses c. 1850 - 1914 *Business History* **30** pp 46 – 68.

1523. Michie R 1999, 2001 The London Stock Exchange: A history *Oxford University Press* UK 978-0-19-924255-9 pp 1 – 688.

1524. Arbulu P 1998a Le marché parisien des actions françaises au XIXe siècle: Performance et efficience d'un marché émergent Thèse de doctorat Université d'Orléans France.

1525. Arbulu P 1998b La Bourse de Paris au xixème siecle: l'exemple d'un marche emergent devenu efficient Revue d'Économie Financière pp 213 – 249.

1526. Neal L 2005 The London Stock Exchange in the 19th century: Ownership structures, growth and performance *Working Paper 115 Oesterreichische Nationalbank* Wien Austria pp 1 – 38.

1527. Petit M 2006 Inventaire des séries de données historiques du projet Old Paris stock exchange 1919-1939 *Economies et Sociétés* **40** no 8 pp 1089 – 1119.

1528. P C Hautcoeur 2006 Why and how to measure stock market fluctuations? The early history of stock market indices, with special reference to the French case *Working paper no 10 Paris School of Economics* Paris France.

1529. Hautcoeur P-C, Riva A 2007 The Paris financial market in the 19th century: An efficient multi-polar organization ? *Working Paper no.31 Paris School of Economics* Paris France.

*1530.* Le Bris D, Hautcoeur P-C 2011 A challenge to triumphant optimists? *halshs-00586765 version 1 Paris School of Economics – EHESS* Paris France www.pse.ens.fr .

*1531.* Gallais-Hamonno, Georges (ed.) 2007 Le marché financier français au 19e siècle, Aspects antitatifs des acteurs et des instruments à la Bourse de Paris *Paris* France.

*1532.* Hamao Y, Hoshi T, Okazaki T 2005 The genesis and development of the capital market in pre-war Japan CARF-F-023 Tokyo Japan http://www.carf.e.u-tokyo.ac.jp/workingpaper/index.cgi .

*1533.* Hart O, Moore J 1996 The governance of exchanges: Members' cooperatives versus outside ownership *Oxford Review of Economic Policy* **12** pp 53 – 69.

*1534.* Goetzmann W, Ibbotson R, Peng C 2000 A new historical database for the NYSE 1815 to 1925: Performance and predictability *The Journal of Financial Markets* **4** no 1 pp. 1 – 32.

*1535.* Davis L E, Neal L 1998 Micro rules and macro outcomes: The impact of the structure of organizational rules on the efficiency of security exchanges, London, New York, and Paris, 1800-1914 *American Economic Review* **88** (2) pp 40 – 45.

*1536.* Davis L E, Neal L, White E N 2003 How it all began: The rise of listing requirements on the London, Berlin, Paris, and New York stock exchanges *The International Journal of Accounting* **38** (2) pp 117 – 143.

*1537.* Pirrong C 1999 The organization of financial exchange markets: Theory and evidence *Journal of Financial Markets* **2** (4) pp 329 – 357.

*1538.* Pirrong C 2000 A theory of financial exchange organization *Journal of Law & Economics* **43** pp 437 – 471.

*1539.* Di Noia C 2001 Competition and integration among stock exchanges in Europe: Network effects, implicit mergers and remote access *European Financial Management* **7** (1) pp 39 – 72.

*1540.* Gregory A, Harris D F, Michou M 2001 An analysis of contrarian investment strategies in the UK *Journal of Business Finance and Accounting* **28** (9-10) pp 1192 – 1228.

*1541.* Munro J H 2003 The medieval origins of the 'Financial Revolution': usury, rentes, and negotiability *The International History Review* **XXV 3** ISSN 0707-5332 pp 505 – 756 *Munich University* Munich Germany http://mpra.ub.uni-muenchen.de/10925/ .

*1542.* Landes D 1998, 1999 The wealth and poverty of nations *W W Norton & Company Inc USA*; *Little Brown & Company UK*; *Abacus UK* ISBN 0 349 11166 9 pp 1 – 650.

*1543.* Shiryaev A N 1998a Foundations of stochastic financial mathematics vol **1** *Fazis Scientific and Publishing House* Moscow Russian Federation ISBN 5-7036-0044-8 pp 1 – 492.

*1544.* Shiryaev A N 1998b Foundations of stochastic financial mathematics vol **2** *Fazis Scientific and Publishing House* Moscow Russian Federation ISBN 5-7036-0044-8 pp 493 – 1017.

*1545.* Schnoor I 2006 Comparable analysis and data manipulation tools *The Marquee Group* Toronto Canada pp 1 – 66.

1546. Schnoor I 2005-2006 Private communications on business valuation methodologies *Rotman School of Management University of Toronto* Canada.

1547. Łuniewska M A 2007 A statistical analysis of dividend in chosen companies listed on the Warsaw Stock Exchange *Folia Oeconomica Stetinensia* DOI: 10.2478/v10031-007-0013-4 pp 76 – 85.

1548. Bartram S M, Bodnar G M 2009 Crossing the lines: The conditional relation between exchange rate exposure and stock returns in emerging and developed markets *MPRA Paper No. 14018 Munich University Munich Germany* pp 1 – 43 http://mpra.ub.uni-muenchen.de/14018/ .

1549. Granger C W J 1969 Investigating causal relations by econometrics models and cross spectral methods *Econometrica* **37** pp 424 – 438.

1550. Granger C W J, Huang B, Yang C W, 2000 Bivariate causality between stock prices and exchange rates: Evidence from recent Asian flu *The Quarterly Review of Economics and Finance* **40** pp 337 – 354.

1551. Aggarwal R 1981 Exchange rates and stock prices: A study of U.S. capital market under floating exchange rates *Akron Business and Economic Review* **12** pp 7 – 12.

1552. Soenen L A, Hennigar E S 1988 An analysis of exchange rates and stock prices - the U.S. experience between 1980 and 1986 *Akron Business and Economic Review* **19** pp 7 – 16.

1553. Giovannini A, Jorion P 1989 The time-variation of risk and return in the foreign exchange and stock markets *Journal of Finance* **44** pp 307 – 325.

1554. Jorion P 1991 The Pricing of exchange rate risk in the stock market *Journal of Financial and Quantitative Analysis* **26** (3) pp 363 – 376.

1555. Loudon G F 1993 The foreign exchange operating exposure of Australian stocks *Accounting and Finance* **33** (1) pp 19 – 32.

1556. Bailey W, Chung Y P 1995 Exchange rate fluctuations, political risk, and stock returns: Some evidence from an emerging market *Journal of Financial and Quantitative Analysis* **30** (4) pp 541 – 560.

1557. Prasad A M, Rajan M 1995 The role of exchange and interest risk in equity valuation: A comparative study of international stock markets *Journal of Economics and Business* **47** (5) pp 457 – 472.

1558. Ajayi R A, Mougoue M 1996 On the dynamic relation between stock prices and exchange rates *Journal of Financial Research* **19** pp 193 – 207.

1559. Ajayi R A, Friedman J, Mehdian S M 1998 On the relationship between stock returns and exchange rates: Tests of Granger causality *Global Finance Journal* **9** pp 241 – 251.

1560. Issam S, Abdalla A, Murinde V 1997 Exchange rate and stock price interactions in emerging financial markets: Evidence on India, Korea, Pakistan and the Philippines *Applied Financial Economics* **7** pp 25 – 35.

1561. Yu Q 1997 Stock prices and exchange rates: Experience in leading East Asian financial countries: Tokyo, Hong Kong and Singapore *Singapore Economic Review* **41** pp 47 – 56.

1562. Chow E H, Lee W Y, Solt M E 1997 The exchange rate risk exposure of asset returns *Journal of Business* **70** (1) pp 107 – 123.

**1563.** Choi J J, Hiraki T, Takezawa N 1998 Is foreign exchange rate risk priced in the Japanese stock market? *Journal of Financial and Quantitative Analysis* **33** (3) pp 361 – 382.

**1564.** Stavarek D 2004 Stock prices and exchange rates in the EU and the USA: Evidence of their mutual interactions *MPRA Paper No 7297* pp 1 – 23

http://mpra.ub.uni-muenchen.de/7297/ .

**1565.** Kolari J W, Moorman T C, Sorescu S M 2005 Foreign exchange risk and the cross-section of stock returns *Texas A&M Working Paper* USA.

**1566.** Tabak B M 2006 The dynamic relationship between stock prices and exchange rates: Evidence for Brasil *The Banco Central do Brasil Working Paper Series no 124* ISSN 1518-3548 pp 1 – 37.

**1567.** Hartmann D, Pierdzioch Ch 2006 Nonlinear links between stock returns and exchange rate movements *MPRA Paper No. 2918 Munich University* Munich Germany pp 1 – 15 http://mpra.ub.uni-muenchen.de/2918/ .

**1568.** Hyde S J 2007 The response of industry stock returns to market, exchange rate and interest rate risks *Manchester Business School Working Paper No 2007-491 MPRA Paper No. 9679 Munich University* Munich Germany

http://mpra.ub.uni-muenchen.de/9679/ .

**1569.** Alagidede P, Panagiotidis T, Xu Zhang 2010 Causal relationship between stock prices and exchange rates *Stirling Economics Discussion Paper 2010-05 Stirling Management School* Stirling University UK

http://www.economics.stir.ac.uk .

**1570.** Thai-Ha Le, Youngho Chang 2011a The impact of oil price fluctuations on stock markets in developed and emerging economies *MPRA Paper No. 31936 Munich University Germany* http://mpra.ub.uni-muenchen.de/31936/ .

**1571.** Thai-Ha Le, Youngho Chang 2011b Dynamic relationships between the price of oil, gold and financial variables in Japan: A bounds testing approach *MPRA Paper No. 33030 Munich University Germany* pp 1 – 30

http://mpra.ub.uni-muenchen.de/33030/ .

**1572.** Hamilton J D 1983 Oil and the macroeconomy since World War II *Journal of Political Economy* **91** pp 228 – 248.

**1573.** Gisser M, Goodwin T H 1986 Crude oil and the macroeconomy: Tests of some popular notions *Journal of Money, Credit, and Banking* **18** pp 95 – 103.

**1574.** Fortune N J 1987 The inflation rate of the price of gold, expected prices and interest rates *Journal of Macroeconomics* **9** (1) pp 71 – 82.

**1575.** Mork K A 1989 Oil and the macroeconomy. When prices go up and down: An Extension of Hamilton's results *Journal of Political Economy* **91** pp 740 – 744.

**1576.** Sjaastad L A, Scacciallani F 1996 The price of gold and the exchange rate *Journal of Money and Finance* **15** pp 879 – 897.

**1577.** Chan H W H, Faff R 1998 The sensitivity of Australian industry equity returns to a gold price factor *Accounting and Finance* **38** pp 223 – 244.

**1578.** Cai J, Cheung Y-L, Wong M C S 2001 What moves the gold market? *Journal of Futures Markets* **21** pp 257 – 278.

1579. Basher S A, Sadorsky P 2006 Oil price risk and emerging stock markets *Global Finance Journal* **17** pp 224 – 251.

1580. Park J, Ratti R 2008 Oil prices shocks and stock markets in the U.S. and 13 European countries *Energy Economics* **30** (5) pp 2587 – 2608.

1581. Kilian L, Park C 2009 The impact of oil price shocks on the US stock market International *Economic Review* **50** (4) pp 1267 – 1287.

1582. Wang M L Wang C P, Huang T Y 2010 Relationships among oil price, gold price, exchange rate and international stock markets *International Research Journal of Finance and Economics* **47** pp 82 – 91.

1583. Busch Th, Christensen B J Nielsen M Ø 2009 The role of implied volatility in forecasting future realized volatility and jumps in foreign exchange, stock, and bond markets *CREATES Research Paper 2007-9* Danske Bank Copenhagen Denmark; Center for Research in Econometric Analysis of Time Series Aarhus Denmark; School of Economics and Management University of Aarhus Denmark; Queen's University Kingston Ontario Canada pp 1 – 20.

1584. Schwert G W 1989 Why does stock market volatility change over time? *Journal of Finance* **44** pp 1115 – 1153.

1585. Égert B, Koubaa Y 2004 Modelling stock returns in the G-7 and in selected CEE economies: a non-linear GARCH approach *William Davidson Institute Working Paper no 663*.

1586. Kavkler A, Festić M 2011 Modeling stock exchange index returns in different GDP growth regimes *Prague Economic Papers* **1** pp 3 – 22.

1587. Elton E J, Gruber M J 1995 Modern portfolio theory and investment analysis *J Wiley and Sons Inc* New York USA.

1588. Statman M 1987 How many stocks make a diversified portfolio *Journal of Financial and Quantitative Analysis* vol **22** no 3 pp 353 – 363.

1589. Anghelache G 2000 The stock exchange and the OTC *Economic Publishing House* Bucharest Romania.

1590. Stoica V, Ionescu E 2000 Capital markets and stock exchange *Economic Publishing House* Bucharest Romania.

1591. Utsugi A, Ino K, Oshikawa M 2003 Random matrix theory analysis of cross correlations in financial markets *Cornell University* NY USA
http://arxiv.org/abs/cond-mat/0312643v1 .

1592. Abhyankar A H, Copeland L S, Wong W W 1997 Uncovering nonlinear structure in real-time stock-market indexes: the S&P 500, the DAX, the Nikkei 225, and the FTSE-100 *Journal of Business and Economic Statistics* **15** (1) pp 1 – 14.

1593. Chappell D, Panagiotidis T 2005 Using the correlation dimension to detect nonlinear dynamics: Evidence from the Athens Stock exchange *Finance Letters* **3** (4) pp 29 – 32.

### *Investment capital, investment portfolio, investment portfolio risk management in finances:*

1594. Markowitz H M 1952 Portfolio selection *The Journal of Finance* vol **7** (1) pp 77 – 91.

1595. Markowitz H M 1956 The optimization of a quadratic function subject to linear constraints *Naval Research Logistics Quarterly* vol **3**.

1596. Markowitz H M 1959 Portfolio selection: Efficient diversification of investments *John Wiley & Sons Inc* NY USA.

1597. Markowitz H M 1987 Mean-variance analysis in portfolio choice and capital markets *Basil Blackwell* USA.

1598. Lintner J 1956 Distribution of incomes of corporations among dividends, retained earnings, and taxes *American Economic Review* **46** (2) pp 97 – 113.

1599. Lintner J 1965 The valuation of risk assets and the selection of risky investments in stock portfolios and capital budgets *Review of Economics and Statistics* **47** pp 13 – 37.

1600. Tobin J 1958 Liquidity preference as behavior towards risk *Review of Economic Studies* vol **25** pp 65 – 86.

1601. Osborne M F M 1959 Brownian motion in the stock market *Operations Research* **7** pp 145 – 173.

1602. Alexander S S 1961 Price movements in speculative markets: Trends or random walks *Industrial Management Review* **2** pp 7 – 26.

1603. Shiryaev A N 1961 The problem of the most rapid detection of a disturbance in a stationary process *Soviet Mathematical Doklady* **2** pp 795 – 799.

1604. Shiryaev A N 1963 On optimal methods in quickest detection problems *Theory of Probability and its Applications* **8** (1) pp 22 – 46.

1605. Shiryaev A N 1964 On Markov sufficient statistics in non-additive Bayes problems of sequential analysis *Theory of Probability and its Applications* **9** (4) pp 670 – 686.

1606. Shiryaev A N 1965 Some exact formulas in a 'disorder' problem *Theory of Probability and its Applications* **10** pp 348 – 354.

1607. Grigelionis B I, Shiryaev A N 1966 On Stefan's problem and optimal stopping rules for Markov processes *Theory of Probability and its Applications* **11** pp 541 – 558.

1608. Shiryaev A N 1967 Two problems of sequential analysis *Cybernetics* **3** pp 63 – 69.

1609. Shiryaev A N 1978 Optimal stopping rules *Springer* Berlin Germany.

1610. Shiryaev A N 1998a Foundations of stochastic financial mathematics vol **1** *Fazis Scientific and Publishing House* Moscow Russian Federation ISBN 5-7036-0044-8 pp 1 – 492.

1611. Shiryaev A N 1998b Foundations of stochastic financial mathematics vol **2** *Fazis Scientific and Publishing House* Moscow Russian Federation ISBN 5-7036-0044-8 pp 493 – 1017.

1612. Graversen S E, Peskir G, Shiryaev A N 2001 Stopping Brownian motion without anticipation as close as possible to its ultimate maximum *Theory of Probability and its Applications* **45** pp 125–136 MR1810977
http://www.ams.org/mathscinet-getitem?mr=1810977 .

1613. Kallsen J, Shiryaev A N 2001 Time change representation of stochastic integrals *Theory of Probability and its Applications* **46** pp 579–585 MR1978671 http://www.ams.org/mathscinet-getitem?mr=1978671 .

1614. Kallsen J, Shiryaev A N 2002 The cumulant process and Esscher's change of measure *Finance Stoch* **6** pp 397–428 MR1932378
http://www.ams.org/mathscinet-getitem?mr=1932378 .

1615. Shiryaev A N 2002 Quickest detection problems in the technical analysis of the financial data *Proceedings Mathematical Finance Bachelier Congress* Paris France (2000) *Springer* Germany pp 487–521 MR1960576

http://www.ams.org/mathscinet-getitem?mr=1960576 .

1616. Jacod J, Shiryaev A N 2003 Limit theorems for stochastic processes 2nd edition Grundlehren der Mathematischen Wissenschaften [Fundamental Principles of Mathematical Sciences] **288** *Springer* Berlin Germany MR1943877

http://www.ams.org/mathscinet-getitem?mr=1943877

1617. Peskir G, Shiryaev A N 2006 Optimal stopping and free-boundary problems *Lectures in Mathematics* ETH Zürich *Birkhäuser* Switzerland MR2256030

http://www.ams.org/mathscinet-getitem?mr=2256030 .

1618. Feinberg E A, Shiryaev A N 2006 Quickest detection of drift change for Brownian motion in generalized Bayesian and mini-max settings *Statistics & Decisions* **24** (4) pp 445 – 470.

1619. du Toit J, Peskir G, Shiryaev A N 2007 Predicting the last zero of Brownian motion with drift Cornell University NY USA pp 1- 17

http://arxiv.org/abs/0712.3415v1 .

1620. Shiryaev A N 2008a Generalized Bayesian nonlinear quickest detection problems: on Markov family of sucient statistics Mathematical Control Theory and Finance Proceedings of the Workshop of April 10–14 2007 Lisbon Portugal Sarychev A et al editors *Springer* Berlin Germany pp 377 – 386.

1621. Shiryaev A N 2008b Optimal stopping rules 3rd edition *Springer* Germany.

1622. Eberlein E, Papapantoleon A, Shiryaev A N 2008 On the duality principle in option pricing: Semimartingale setting *Finance Stoch* **12** pp 265 – 292

http://www.ams.org/mathscinet-getitem?mr=2390191 .

1623. Shiryaev A N, Novikov A A 2009 On a stochastic version of the trading rule "Buy and Hold" *Statistics & Decisions* **26** (4) pp 289 – 302.

1624. Eberlein E, Papapantoleon A, Shiryaev A N 2009 Esscher transform and the duality principle for multidimensional semimartingales *The Annals of Applied Probability* vol **19** no 5 pp 1944 – 1971

http://dx.doi.org/10.1214/09-AAP600 http://arxiv.org/abs/0809.0301v5 .

1625. Shiryaev A N, Zryumov P Y 2009 On the linear and nonlinear generalized Bayesian disorder problem (discrete time case) optimality and risk – modern trends in mathematical finance The Kabanov Festschrift Delbaen F et al editors *Springer* Berlin Germany pp 227 – 235.

1626. Gapeev P V, Shiryaev A N 2010 Bayesian quickest detection problems for some diffusion processes Cornell University NY USA pp 1 – 25

http://arxiv.org/abs/1010.3430v2 .

1627. Karatzas I, Shiryaev A N, Shkolnikov M 2011 The one-sided Tanaka equation with drift Cornell University NY USA

http://arxiv.org/abs/1108.4069v1 .

1628. Shiryaev A N, Zhitlukhin M V 2012 Optimal stopping problems for a Brownian motion with a disorder on a finite interval Cornell University NY USA pp 1 – 10 http://arxiv.org/abs/1212.3709v1 .

1629. Zhitlukhin M V, Shiryaev A N 2012 Bayesian disorder detection problems on filtered probability spaces *Theory of Probability and Its Applications* **57** (3) pp 453 – 470.

1630. Feinberg E A, Mandava M, Shiryaev A N 2013 On solutions of Kolmogorov's equations for nonhomogeneous jump Markov processes Cornell University NY USA pp 1 – 15 http://arxiv.org/abs/1301.6998v3 .

1631. Rothbard M N 1962, 2004 Man, economy, and state *Ludwig von Mises Institute* Auburn Alabama USA

http://www.mises.org/rothbard/mes.asp .

1632. Cootner P 1962 Stock prices: Random vs. systematic changes *Industrial Management Review* **3** pp 24 – 45.

1633. Cootner P 1964 The random character of stock prices *MIT Press* Cambridge USA.

1634. Mandelbrot B 1963 The variation of certain speculative prices *Journal of Business* **36** pp 394 – 419.

1635. Fama E F 1963 Mandelbrot and the stable Paretian hypothesis *Journal of Business* **36** (4) pp 420 – 429.

1636. Fama E F 1965 The behavior of stock market prices *Journal of Business* **38** pp 34 - 105.

1637. Fama E F 1970 Efficient capital markets: a review of theory and empirical work *Journal of Finance* **25** pp 383 - 417.

1638. Fama E F 1976 Foundations of finance: Portfolio decisions and securities prices *Basic Books* New York USA.

1639. Fama E F 1984 Forward and spot exchange rates *Journal of Monetary Economics* **14** pp 319 – 338.

1640. Fama E F 1991 Efficient capital markets II *Journal of Finance* **46** pp 1575 – 1618.

1641. Fama E F, French K R 1993 Common risk factors in the returns on stocks and bonds *Journal of Financial Economics* **33** (1) pp 3 – 56.

1642. Fama E 1998 Market efficiency, long-term returns, and behavioral finance *Journal of Financial Economics* **49** pp 283 – 306.

1643. Fama E F, Blume M E 1966 Filter rules and stock market trading *Journal of Business* **39** (II) pp 226 – 241.

1644. Fama E F, Fisher L, Jensen M, Roll R 1969 The adjustment of stock prices to new information *International Economic Review* **10** pp 1 – 21.

1645. Fama E F 1970 Efficient capital markets: A review of theory and empirical work *Journal of Finance* **25** (2) pp 383 – 417.

1646. Fama E F, MacBeth J D 1973 Risk, return and equilibrium: empirical tests *Journal of Political Economy* **81** pp 607 – 636.

1647. Fama E F, Schwert G W 1977 Asset returns and inflation *Journal of Financial Economics* **5** pp 115 – 146.

1648. Fama E F, Bliss R R 1987 The information in long-maturity forward rates *American Economic Review* **77** pp 680 – 692.

1649. Fama E F, French K R 1988a Dividend yields and expected stock returns *Journal of Financial Economics* **22** pp 3 – 26.

1650. Fama E F, French K R 1988b Permanent and temporary components of stock prices *Journal of Political Economy* **96** pp 246 – 273.

1651. Fama E F and French K R 1989 Business conditions and expected returns on stocks and bonds *Journal of Financial Economics* **25** pp 23 - 50.

1652. Fama E F 1991 Efficient capital markets: II *Journal of Finance* **46** (5) pp 1575 – 1617.

1653. Fama, E.F. and French K R 1992 The cross-section of expected stock returns *Journal of Finance* **47** pp 427 – 466.

1654. Fama E F, French K R 1993 Common risk factors in the returns on stocks and bonds *Journal of Financial Economics* **33** pp 3 – 56.

1655. Fama E F, French K R 1995 Size and book-to-market factors in earnings and returns *Journal of Finance* **50** pp 131 – 156.

1656. Fama E F, French K R 1996 Multifactor explanations for asset pricing anomalies *Journal of Finance* **51** (1) pp 55 – 84.

1657. Fama E F, French K R 1998 Value versus growth: The international evidence Journal of Finance **53** (6) pp 1975 – 1999.

1658. Fama E F, French K R 2004 The capital asset pricing model: Theory and evidence *Journal of Economic Perspectives* **18** (3) pp 25 – 46.

1659. Fama E F, French K R 2010 Luck versus skill in the cross-section of mutual fund returns *Journal of Finance* **65** (5) pp 1915 – 1947.

1660. Davis J L, Fama E F, French K R 2000 Characteristics, covariances, and expected returns: 1928-1997 *Journal of Finance* **55** pp 389 – 406.

1661. Fama E F, Litterman R 2012 An experienced view on markets and investing *Financial Analysts Journal* **68** (6) pp 1 – 5.

1662. Sharkovsky A N 1964 Co-existence of cycles of a continuous map of a line in itself *Ukrainian Mathematical Journal* vol **16** pp 61 – 71.

1663. Sharkovsky A N 1965 On the cycles and structure of continuous mapping *Ukrainian Mathematical Journal* vol **17** p 104.

1664. Sharkovsky A N, Maistrenko Yu L, Romanenko E Yu 1986 Differential equations and their applications *Naukova Dumka* Kiev Ukraine pp 1 – 280.

1665. Sharpe W F 1964 Capital asset prices: A theory of market equilibrium under conditions of risk *Journal of Finance* vol **19** pp 425 – 442.

1666. Sharpe W 1965 Risk aversion in the stock market *The Journal of Finance* vol **20** Issue 3 pp 416-422.

1667. Sharpe W F 1966 Mutual fund performance *Journal of Business* vol **39** no 1 part 2: Supplement on security prices pp 119 – 138.

1668. Sharpe W F 1968 Mutual fund performance and the theory of capital asset pricing: Reply *The Journal of Business* vol **41** no 2 pp 235 – 236.

1669. Sharpe W 1992 Asset allocation: Management style and performance measurement *Journal of Portfolio Management* **18** no 2 pp 7 – 19.

1670. Sharpe W F 1994 The Sharpe ratio *Journal of Portfolio Management* Fall pp 49 – 59.

1671. Sharpe W F, Alexander G J, Bailey J V 1999 Investments 6th edition *Prentice-Hall Inc* USA.

1672. Samuelson P A 1965 Proof that properly anticipated prices fluctuate randomly *Industrial Management Review* **6** pp 41 – 49.

1673. Treynor J 1965 How to rate management of investment funds *Harvard Business Review* **43** no 1 pp 63 – 75.

1674. Mossin J 1966 Equilibrium in a capital asset market *Econometrica* **34** (4) pp 768 – 783.

1675. Jensen M C 1968 The performance of mutual funds in the period 1945-1964 *Journal of Finance* **23** (2) pp 389 – 416.

1676. Blumethal R M, Getoor R K 1968 Markov processes and potential theory *Academic Press* USA MR0264757

http://www.ams.org/mathscinet-getitem?mr=0264757 .

1677. Ball R, Brown P 1968 An empirical evaluation of accounting income numbers *Journal of Accounting Research* **6** pp 159 – 178.

1678. Merton R C 1969 Lifetime portfolio selection under uncertainty: The continuous time case *Review of Economics and Statistics* vol **51** pp 247 – 257.

1679. Merton R C 1970 A dynamic general equilibrium model of the asset market and its application to the pricing of the capital structure of the firm *Working Paper No. 497-70 Sloan School of Management* Massachusetts Institute of Technology Cambridge MA [Reprinted in Merton R C 1992 Ch 11)].

1680. Merton R C 1971 Optimum consumption and portfolio rules in a continuous-time model *Journal of Economic Theory* vol **3** pp 373 – 413.

1681. Merton R C 1972 Appendix: Continuous-time speculative processes *in* Day R H, Robinson S M editors *Mathematical Topics in Economic Theory and Computation Society for Industrial and Applied Mathematics* Philadelphia USA [Reprinted in Merton R C 1973 *SIAM Review* vol **15** pp 34 – 38].

1682. Merton R C 1973a Theory of rational option pricing *Bell Journal of Economics and Management Science* vol **4** pp 141 – 183.

1683. Merton R C 1973b An inter-temporal Capital Asset Pricing Model *Econometrica* vol **41** pp 867 – 887.

1684. Merton R C 1977a An analytic derivation of the cost of deposit insurance and loan guarantees: An application of modern option pricing theory *Journal of Banking and Finance* vol **1** pp 3 – 11.

1685. Merton R C 1977b A re-examination of the Capital Asset Pricing Model Studies in risk and return Bicksler J, Friend I editors *Ballinger* vols **I** & **II** pp 141 – 159.

1686. Merton R C 1982 On the microeconomic theory of investment under uncertainty in Handbook of Mathematical Economics vol **II** Arrow K, Intriligator M editors *North-Holland Publishing Company* Amsterdam The Netherlands.

1687. Merton R C 1983a On consumption-indexed public pension plans *in* Financial Aspects of the US Pension System in Bodie Z, Shoven J editors *University of Chicago Press* Chicago USA.

1688. Merton R C 1983b Financial economics *in* Brown E C, Solow R M editors Paul Samuelson and Modern Economic Theory *McGraw-Hill* NY USA.

1689. Merton R C 1990 The financial system and economic performance *Journal of Financial Services Research* vol **4** pp 263 – 300.

1690. Merton R C 1992 Continuous-time finance Revised Edition *Basil Blackwell* Cambridge MA USA.

1691. Merton R C 1993a Operation and regulation in financial intermediation: A functional perspective *in* Englund P editor Operation and regulation of financial markets *The Economic Council* Stockholm Sweden.

1692. Merton R C 1993b Optimal investment strategies for university endowment funds in Clotfelter C, Rothschild M editors Studies of supply and demand in higher education *University of Chicago Press* Chicago USA.

1693. Merton R C 1994 Influence of mathematical models in finance on practice: Past, present and future *Philosophical Transactions of the Royal Society of London* vol **347** pp 451 – 463.

1694. Merton R C 1995a A functional perspective of financial intermediation *Financial Management* vol **24** no 2 pp 23 – 41.

1695. Merton R C 1995b Financial innovation and the management and regulation of financial institutions *Journal of Banking and Finance* vol **19** pp 461 – 481.

1696. Merton R C 1997 On the role of the Wiener process in finance theory and practice: The case of replicating portfolios *in* Jerison D, Singer I M, and Stroock D W editors The Legacy of Norbert Wiener: A Centennial Symposium PSPM Series vol **60** *American Mathematical Society* Providence RI USA.

1697. Merton R C 1998 Applications of option-pricing theory: Twenty-five years later Les Prix Nobel 1997 Nobel Foundation Stockholm Sweden [Reprinted in American Economic Review pp 323 – 349].

1698. Merton R C 1999 Commentary: Finance theory and future trends: The shift to integration risk pp 48 – 50.

1699. Merton R C, Bodie Z 1992 On the management of financial guarantees *Financial Management* vol **21** pp 87 – 109.

1700. Merton R C, Bodie Z 1993 Deposit insurance reform: A functional approach in Meltzer A, Plosser C editors *Carnegie-Rochester Conference Series on Public Policy* vol **38** pp 1 – 34

1701. Merton R C, Scholes M S 1995 Fischer Black *Journal of Finance* vol **50**.

1702. Merton R C 2001 Future possibilities in finance: Theory and finance practice *Working Paper 01-030 Harvard Business School* Harvard University Boston USA pp 1 – 43.

1703. Black F, Jensen M C, Scholes M 1972 The capital asset pricing model: Some empirical tests *in* Jensen M editor Studies in the theory of capital markets *Praeger*.

1704. Black F, Scholes M 1973 The pricing of options and corporate liabilities *Journal of Political Economy* **81** pp 637 – 654.

1705. Fischer St 1977a Long-term contracts, rational expectations, and the optimal money supply rule *Journal of Political Economy* vol **85** no 1 pp 191 – 205.

1706. Fischer St 1977b Long-term contracting, sticky prices, and monetary policy *Journal of Monetary Economics* vol **3** pp 317 – 323.

1707. Shiller R J 1979 The volatility of long term interest rates and expectations models of the term structure *Journal of Political Economy* **87** pp 1190 – 1219.

1708. Shiller R J 1981a Do stock prices move too much to be justified by subsequent changes in dividends? *American Economic Review* **71** pp 421 – 436.

1709. Shiller R J 1981b The use of volatility measures in assessing market efficiency *Journal of Finance* **36** (2) pp 291 – 304.

1710. Shiller R J 1982 Consumption, asset markets and macroeconomic fluctuations *Carnegie Rochester Conference Series on Public Policy* **17** pp 203 – 238.

1711. Shiller R J 1984 Stock prices and social dynamics *Carnegie Rochester Conference Series on Public Policy* pp 457 – 510.

1712. Shiller R J 1987 Investor behavior in the 1987 stock market crash: Survey evidence *NBER working paper no 2446*.

1713. Shiller R J 1988 Portfolio insurance and other investor fashions as factors in the 1987 stock market crash *NBER Macroeconomics Annual* **3** pp 287 – 297.

1714. Shiller R J 1989 Market volatility *MIT Press* Cambridge USA.

1715. Shiller R J 2000 Irrational exuberance *Princeton University Press* USA.

1716. Shiller R J 2008 The subprime solution: How today's global financial crisis happened, and what to do about I *Princeton University Press* USA.

1717. Shiller R J, Campbell J Y, Schoenholtz K L 1983 Forward rates and future policy: interpreting the term structure of interest rates *Brookings Papers on Economic Activity* **1** pp 173 – 223.

1718. Shiller R J, Perron P 1985 Testing the random walk hypothesis: Power vs. frequency of observation *Economics Letters* **18** pp 381 – 386.

1719. Shiller R J, Pound J 1989 Survey evidence on the diffusion of interest and information among investors *Journal of Economic Behavior and Organization* **12** pp 47 – 66.

1720. Bernanke B S 1979 Long-term commitments, dynamic optimization, and the business cycle *Ph. D. Thesis Department of Economics* Massachusetts Institute of Technology USA.

1721. Bernanke B S, Blinder A S 1992 The federal funds rate and the channels of monetary transmission *American Economic Review* **82** (4) pp 901 – 921.

1722. Bernanke B S, Gertler M 1995 Inside the black box: The credit channel of monetary policy transmission *Journal of Economic Perspectives* **9** (4) pp 27 – 48.

1723. Bernanke B S 2002 Deflation: "Making sure it doesn't happen here." Speech before the *National Economists Club* Washington DC http://www.federalreserve.gov .

1724. Bernanke B S 2004 The great moderation www.federalreserve.gov .

1725. Bernanke B S, Reinhart V R 2004 Conducting monetary policy at very low short-term interest rates *The American Economic Review* vol **94** no 2 pp 85 – 90.

1726. Bernanke B S, Reinhart V R, Sack B P 2004 Monetary policy alternatives at the zero bound: an empirical assessment *Brookings Papers on Economic Activity* Issue 2 pp 1 – 78.

1727. Bernanke B S 2007 The financial accelerator and the credit channel *Speech at The Credit Channel of Monetary Policy in the Twenty-first Century Conference* Federal Reserve Bank of Atlanta Georgia USA.

**1728.** Bernanke B S 2009a The crisis and the policy response *Federal Reserve* USA.

**1729.** Bernanke B S 2009b On the outlook for the economy and policy *Bank for International Settlements* Basel Switzerland

http://www.bis.org/review/r091119a.pdf .

**1730.** Bernanke B S 2009c The Federal Reserve's balance sheet – an update *Bank for International Settlements* Basel Switzerland

http://www.bis.org/review/r091013a.pdf .

**1731.** Bernanke B S 2009d Regulatory reform *Bank for International Settlements* Basel Switzerland

http://www.bis.org/review/r091006a.pdf .

**1732.** Bernanke B S 2009e Policy responses to the financial crisis Public Lecture on 13.01.2009 *London School of Economics and Political Science* London UK. http://richmedia.lse.ac.uk/publicLecturesAndEvents/20090113_1300_policyResponses ToTheFinancialCrisis.mp3 .

**1733.** Bernanke B S 2010a Monetary policy and the housing bubble *Annual Meeting of the American Economic Association* Atlanta Georgia USA.

**1734.** Bernanke B S 2010b Causes of the recent financial and economic crisis testimony before the *Financial Crisis Inquiry Commission* Washington USA www.federalreserve.gov/newsevents/testimony/bernanke20100902a.htm .

**1735.** Bernanke B S 2012a Some reflections on the crisis and the policy response Rethinking Finance: Perspectives on the Crisis conference sponsored by the *Russell Sage Foundation* and *The Century Foundation* New York USA www.federalreserve.gov/newsevents/speech/bernanke20120413a.htm .

**1736.** Bernanke B S 2012b Monetary policy since the onset of the crisis *The Changing Policy Landscape symposium* sponsored by the *Federal Reserve Bank of Kansas City* Jackson Hole Wyoming USA www.federalreserve.gov/newsevents/speech/bernanke20120831a.htm .

**1737.** Bernanke B S 2013a Financial and economic education *13th Annual Redefining Investment Strategy Education (RISE) Forum* Dayton Ohio USA.

**1738.** Bernanke B S 2013b Stress testing banks: What have we learned? *Maintaining Financial Stability: Holding a Tiger by the Tail conference* sponsored by the *Federal Reserve Bank of Atlanta* Stone Mountain Ga USA www.federalreserve.gov/newsevents/speech/bernanke20130408a.htm .

**1739.** Bernanke B S 2013c Monitoring the financial system *49th Annual Conference on Bank Structure and Competition* sponsored by the *Federal Reserve Bank of Chicago* Chicago Illinois USA pp 1 – 16.

**1740.** Bernanke B S 2013d A century of U.S. central banking: Goals, frameworks, accountability *The first 100 years of the Federal Reserve: The policy record, lessons learned, and prospects for the future conference* sponsored by the *National Bureau of Economic Research* Cambridge Massachusetts USA.

**1741.** Bernanke B S, Blanchard O, Summers L H, Weber A A 2013 What should economists and policymakers learn from the financial crisis? Public Lecture on 25.03.2013 *London School of Economics and Political Science* London UK

http://media.rawvoice.com/lse_publiclecturesandevents/richmedia.lse.ac.uk/publiclectures andevents/20130325_1715_whatShouldEconomistsAndPolicymakersLearn.mp4 ; http://www.federalreserve.gov/newsevents/speech/bernanke20130325a.htm.

1742. Hansen L P, Sargent T J 1980 Formulating and estimating dynamic linear rational expectation models *Journal of Economic Dynamics and Control* **2** pp 7 – 46.

1743. Hansen L P, Hodrick R J 1980 Forward exchange rates as optimal predictors of future spot rates: An econometric analysis *Journal of Political Economy* **88** (5) pp 829 – 853.

1744. Hansen L P 1982 Large sample properties of generalized method of moments estimators *Econometrica* **50** pp 1029 – 1054.

1745. Hansen L P, Singleton K J 1982 Generalized instrumental variable estimation of nonlinear rational expectations models *Econometrica* **50** (5) pp 1269 – 1286.

1746. Hansen L P, Singleton K J 1983 Stochastic consumption, risk aversion, and the temporal behavior of asset prices *Journal of Political Economy* **91** (2) pp 249 – 265.

1747. Hansen L P, Singleton K J 1984 Erratum: Generalized instrumental variable estimation of nonlinear rational expectations models *Econometrica* **52** (1) pp 267 – 268.

1748. Hansen L P 1985 A method for calculating bounds on the asymptotic covariance matrices of generalized method of moments estimators *Journal of Econometrics* **30** pp 203 – 238.

1749. Hansen L P, Richard S F 1987 The role of conditioning information in deducing testable restrictions implied by dynamic asset pricing models *Econometrica* **55** pp 587 – 613.

1750. Hansen L P, Heaton J C, Ogaki M 1988 Efficiency bounds implied by multiperiod conditional moment restrictions *Journal of the American Statistical Association* **83**.

1751. Hansen L P, Jagannathan R 1991 Implications of security market data for models of dynamic economies *Journal of Political Economy* **99** pp 225 – 262.

1752. Hansen L P, Scheinkman J A 1995 Back to the future: Generating moment implications for continuous-time Markov processes *Econometrica* **63** pp 767 – 804.

1753. Hansen L P, Heaton J C, Yaron A 1996 Finite-sample properties of some alternative GMM estimators *Journal of Business & Economic Statistics* **14**.

1754. Hansen L P Jagannathan R 1997 Assessing specification errors in stochastic discount factor models Journal of Finance **52** (2) pp 557 – 590.

1755. Hansen L P, Sargent T J 2001 Robust control and model uncertainty *American Economic Review* **91** (2) pp 60 – 66.

1756. Hansen L P, West K D 2002 Generalized method of moments and macroeconomics *Journal of Business and Economic Statistics* **20** (4) pp 460 – 469.

1757. Hansen L P, Heaton J C, Li N 2008 Consumption strikes back? Measuring long-run risk *Journal of Political Economy* **116** pp 260 – 302.

1758. Hansen L P, Sargent T J 2008 Robustness *Princeton University Press* USA**.**

1759. Engle R F, Ta-Chung Liu 1972 Effects of aggregation over time on dynamic characteristics of an econometric model *National Bureau of Economic Research Chapters* in *Econometric Models of Cyclical Behavior* vols **1** and **2** pp 673 – 738.

1760. Engle R F 1974 Band spectrum regression *International Economic Review* Department of Economics University of Pennsylvania and Osaka University Institute of Social and Economic Research Association vol **15** (1) pp 1 – 11.

1761. Engle R F, Foley D K 1975 An asset price model of aggregate investment *International Economic Review* Department of Economics University of Pennsylvania and Osaka University Institute of Social and Economic Research Association vol **16** (3) pp 625 – 647.

1762. Engle R F, Gardner R 1976 Some finite sample properties of spectral estimators of a linear regression *Econometrica* Econometric Society vol **44** (1) pp 149 – 165.

1763. Engle R F 1976 Interpreting spectral analyses in terms of time-domain models *National Bureau of Economic Research Chapters* in *Annals of Economic and Social Measurement* vol **5** no 1 pp 89 – 109.

1764. Engle R F 1978 Testing price equations for stability across spectral frequency bands *Econometrica* Econometric Society vol **46** (4) pp 869 – 881.

1765. Engle R F 1980 Exact maximum likelihood methods for dynamic regressions and band spectrum regressions *International Economic Review* Department of Economics University of Pennsylvania and Osaka University Institute of Social and Economic Research Association vol **21** (2) pp 391 – 407.

1766. Engle R F 1982a Autoregressive conditional heteroskedasticity with estimates of the variance of UK inflation *Econometrica* 50 pp 987 – 1008.

1767. Engle R F 1982b A general approach to Lagrange multiplier model diagnostics *Journal of Econometrics* Elsevier vol **20** (1) pp 83 – 104.

1768. Engle R F 1983 Estimates of the variance of US inflation based upon the ARCH model *Journal of Money, Credit, and Banking* **15** pp 286 – 301.

1769. Engle R F, Watson M 1983 Alternative algorithms for the estimation of dynamic factor, MIMIC and varying coefficient regression models *Journal of Econometrics* vol **23** pp 385 – 400.

1770. Engle R F, Granger C W J, Kraft D 1984 Combining competing forecasts of inflation using a bivariate arch model *Journal of Economic Dynamics and Control* Elsevier vol **8** (2) pp 151 – 165.

1771. Engle R F, Hendry D F, Trumble D 1985 Small-sample properties of ARCH estimators and tests Canadian *Journal of Economics* Canadian Economics Association vol **18** (1) pp 66 – 93.

1772. Engle R F, Lilien D M, Watson M 1985 A Dynamic model of housing price determination *Journal of Econometrics* vol **28** pp 307 – 326.

1773. Watson M W, Engle R F 1985 Testing for regression coefficient stability with a stationary AR(1) alternative The Review of Economics and Statistics *MIT Press* vol **67** (2)    pp 341 – 346.

1774. Bollerslev T, Engle R F, Nelson D B 1986 Arch models *in* Engle R F, McFadden D editors Handbook of Econometrics *Elsevier* edition 1 vol **4** chapter 49 pp 2959 – 3038.

1775. Engle R F, Yoo B S 1987 Forecasting and testing in co-integrated systems Journal of Econometrics *Elsevier* vol **35** (1) pp 143 – 159.

1776. Engle R F, Granger C W J 1987 Co-integration and error correction: Representation, estimation, and testing *Econometrica* Econometric Society vol **55** (2) pp 251 – 276.

1777. Engle R F, Lilien D M, Robins R P 1987 Estimating time varying risk premia in the term structure: The ARCH-M model *Econometrica* **55** pp 391 – 407.

1778. Bollerslev T, Engle R F, Wooldridge J M 1988 A capital asset pricing model with time-varying covariances Journal of Political Economy *University of Chicago Press* vol **96** (1) pp 116 – 131.

1779. Engle R F 1988 Estimates of the variance of US inflation based upon the ARCH model: Reply Journal of Money, Credit and Banking *Blackwell Publishing* vol **20** (3) pp 422 – 423.

1780. Engle R F 1990 Stock volatility and the crash of '87: Discussion *Review of Financial Studies Society for Financial Studies* vol **3** (1) pp 103 – 106.

1781. Engle R F, Ito T, Lin W L 1990 Meteor-showers or heat waves – Heteroskedastic intradaily volatility in the foreign-exchange market *Econometrica* **58** pp 525 – 542.

1782. Engle R F, Ng V K, Rothschild M 1990 Asset pricing with a factor-ARCH covariance structure: Empirical estimates for treasury bills *Journal of Econometrics* **45** pp 213 – 237.

1783. Engle R F, Granger C W J editors 1991 Long-run economic relationships: Readings in Cointegration *Oxford University Press* UK OUP catalogue number 9780198283393.

1784. Engle R F, Gonzalez-Rivera G 1991 Semiparametric ARCH models *Journal of Business & Economic Statistics* American Statistical Association vol **9** (4) pp 345 – 359.

1785. Chou R, Engle R F, Kane A 1992 Measuring risk-aversion from excess returns on a stock index *Journal of Econometrics* **52** pp 201 – 224.

1786. Ng V, Engle R F, Rothschild M 1992 A multi-dynamic-factor model for stock returns *Journal of Econometrics Elsevier* vol **52** (1-2) pp 245 – 266.

1787. Engle R F, Navarro P, Carson R 1992 On the theory of growth controls *Journal of Urban Economics Elsevier* vol **32** (3) pp 269 – 283.

1788. Engle R F, Mustafa Ch 1992 Implied ARCH models from options prices *Journal of Econometrics Elsevier* vol **52** (1-2) pp 289 – 311.

1789. Ding Zh, Granger C W J, Engle R F 1993 A long memory property of stock market returns and a new model *Journal of Empirical Finance Elsevier* vol **1** (1) pp 83 – 106.

1790. Engle R F, Ng V K 1993 Measuring and testing the impact of news on volatility *Journal of Finance* American Finance Association vol **48** (5) pp 1749 – 1778.

1791. Engle R F, Hendry D F 1993 Testing superexogeneity and invariance in regression models *Journal of Econometrics Elsevier* vol **56** (1-2) pp 119 – 139.

1792. Engle R F, Kozicki Sh 1993 Testing for common features *Journal of Business & Economic Statistics* American Statistical Association vol **11** (4) pp 369 – 380.

1793. Vahid F, Engle R F 1993 Common trends and common cycles *Journal of Applied Econometrics John Wiley & Sons Ltd* vol **8** (4) pp 341 – 360.

1794. Engle R F, Susmel R 1993 Common volatility in international equity markets *Journal of Business & Economic Statistics American Statistical Association* vol **11** (2) pp 167 – 176.

1795. Engle R F, Ng V K 1993 Time-varying volatility and the dynamic behavior of the term structure Journal of Money, Credit and Banking *Blackwell Publishing* vol **25** (3) pp 336 – 349.

1796. Susmel R, Engle R F 1994 Hourly volatility spillovers between international equity markets *Journal of International Money and Finance Elsevier* vol **13** (1) pp 3 – 25.

1797. Lin W-L, Engle R F, Ito T 1994 Do bulls and bears move across borders? International transmission of stock returns and volatility *Review of Financial Studies* Society for Financial Studies vol **7** (3) pp 507 – 538.

1798. Engle R F 1994 Bayesian analysis of stochastic volatility models: Comment *Journal of Business & Economic Statistics* American Statistical Association vol **12** (4) pp 395 – 396.

1799. Engle R F, Kroner K F 1995 Multivariate simultaneous generalized ARCH Econometric Theory *Cambridge University Press* UK vol **11** (01) pp 122 – 150.

1800. Engle R F, Issler J V 1995 Estimating common sectoral cycles *Journal of Monetary Economics Elsevier* vol **35** (1) pp 83 – 113.

1801. Engle R F editor 1995 ARCH: Selected readings *Oxford University Press* UK OUP Catalogue number 9780198774327.

1802. Engle R F, Russell J R 1997 Forecasting the frequency of changes in quoted foreign exchange prices with the autoregressive conditional duration model *Journal of Empirical Finance Elsevier* vol **4** (2-3) pp 187 – 212.

1803. Vahid F, Engle R F 1997 Codependent cycles *Journal of Econometrics Elsevier* vol **80** (2) pp 199 – 221.

1804. Engle R F, Russell J R 1998 Autoregressive conditional duration: A new model for irregularly spaced transaction data *Econometrica* **66** pp 1127 – 1162.

1805. Burns P, Engle R F, Mezrich J 1998 Correlations and volatilities of asynchronous data *Journal of Derivatives* pp 1 – 12.

1806. Engle R F, White H editors 1999 Cointegration, causality, and forecasting: Festschrift in honour of Clive W J Granger *Oxford University Press* UK OUP Catalogue number 9780198296836.

1807. Engle R F, Smith A D 1999 Stochastic permanent breaks *The Review of Economics and Statistics MIT Press* Cambridge USA vol **81** (4) pp 553 – 574.

1808. Engle R F 2000 The econometrics of ultra-high-frequency data *Econometrica* **68** (1) pp 1 – 22.

1809. Alfonso D, Engle R F 2000 Time and the price impact of a trade *Journal of Finance* American Finance Association vol **55** (6) pp 2467 – 2498.

1810. Engle R F 2001a GARCH 101: The use of ARCH/GARCH models in applied econometrics *Journal of Economic Perspectives* American Economic Association vol **15** (4) pp 157 – 168.

1811. Engle R F 2001b Financial econometrics - A new discipline with new methods *Journal of Econometrics Elsevier* vol **100** (1) pp 53 – 56.

1812. Engle R F, Lange J 2001 Predicting VNET: A model of the dynamics of market depth *Journal of Financial Markets Elsevier* vol **4** (2) pp 113 – 142.

1813. Engle R F, Patton A J 2001 What good is a volatility model? *Quantitative Finance Taylor and Francis Journals* vol **1** (2) pp 237 – 245.

1814. Manganelli S, Engle R F 2001 Value at risk models in finance *European Central Bank Working Paper no 75* ISSN 1561-0810 European Central Bank Kaiserstrasse 29 D-60311 Frankfurt am Main Germany pp 1 – 41.

1815. Engle R 2002a Dynamic conditional correlation: A simple class of multivariate generalized autoregressive conditional heteroskedasticity models *Journal of Business & Economic Statistics* **20** (3) pp 339 – 350.

1816. Engle R 2002b New frontiers for ARCH *Journal of Applied Econometrics* **17** (5) pp 425 – 446.

1817. Engle R, Ishida I 2002 Forecasting variance of variance: The square-root, the affine, and the CEV Garch models *Department of Finance Working Papers* New York University NY USA.

1818. Rosenberg J V, Engle R F 2002 Empirical pricing kernels *Journal of Financial Economics* **64** (3) pp 341 – 372.

1819. Engle R F 2003 Risk and volatility: Econometrics models and financial practice Nobel Lecture Stockholm Sweden pp 326 – 349 www.nobel.se .

1820. Engle R F, Lunde A 2003 Trades and quotes: A bivariate point process *Journal of Financial Econometrics Society for Financial Econometrics* vol **1** (2) pp 159 – 188.

1821. Engle R F, Manganelli S 2004 CAViaR: Conditional autoregressive value at risk by regression quantiles *Journal of Business & Economic Statistics* American Statistical Association vol **22** pp 367 – 381.

1822. Engle R F 2004a Robert F Engle: Understanding volatility as a process *Quantitative Finance Taylor and Francis Journals* vol **4** (2) pp 19 – 20.

1823. Engle R F 2004b Risk and volatility: Econometric models and financial practice *American Economic Review* American Economic Association vol **94** (3) pp 405 – 420.

1824. Engle R F, Patton A J 2004 Impacts of trades in an error-correction model of quote prices *Journal of Financial Markets Elsevier* vol **7** (1) pp 1 – 25.

1825. Russell J R, Engle R F 2005 A discrete-state continuous-time model of financial transactions prices and times: The autoregressive conditional multinomial-autoregressive conditional duration model *Journal of Business & Economic Statistics* American Statistical Association vol **23** pp 166 - 180.

1826. Cappiello L, Engle R F, Sheppard K 2006 Asymmetric dynamics in the correlations of global equity and bond returns *Journal of Financial Econometrics* Society for Financial Econometrics vol **4** (4) pp 537 – 572.

1827. Engle R F, Gallo G M 2006 A multiple indicators model for volatility using intra-daily data *Journal of Econometrics Elsevier* vol **131** (1-2) pp 3 – 27.

1828. Diebold F X, Engle R F, Favero C, Gallo G M, Schorfheide F 2006 The econometrics of macroeconomics, finance, and the interface *Journal of Econometrics Elsevier* vol **131** (1-2) pp 1 – 2.

1829. Engle R F, Marcucci J 2006 A long-run pure variance common features model for the common volatilities of the Dow Jones *Journal of Econometrics Elsevier* vol **132** (1) pp 7 – 42.

1830. Engle R F, Colacito R 2006 Testing and valuing dynamic correlations for asset allocation *Journal of Business & Economic Statistics* American Statistical Association vol **24** pp 238 – 253.

1831. Engle R F 2006a Private communications on the modern portfolio, risk management and nonlinear dynamic chaos theories in finances *Rotman School of Management* University of Toronto Ontario Canada.

1832. Engle R F 2006b Private communications on the Stratonovich – Kalman – Bucy filtering algorithm *Rotman School of Management* University of Toronto Canada.

1833. Barone-Adesi G, Engle R F, Mancini L 2007 A GARCH option pricing model in incomplete markets *Swiss Finance Institute Research Paper Series 07-03* Swiss Finance Institute Switzerland.

1834. Engle R F, Rangel J G 2008 The Spline-GARCH model for low-frequency volatility and its global macroeconomic causes *Review of Financial Studies Society for Financial Studies* vol **21** (3) pp 1187 – 1222.

1835. Easley D, Engle R F, O'Hara M, Wu L 2008 Time-varying arrival rates of informed and uninformed trades *Journal of Financial Econometrics Society for Financial Econometrics* vol **6** (2) pp 171 – 207.

1836. Giovanni B-A, Engle R F, Mancini L 2008 A GARCH option pricing model with filtered historical simulation *Review of Financial Studies* Society for Financial Studies vol **21** (3) pp 1223 – 1258.

1837. Bali T G, Engle R F 2010 The intertemporal capital asset pricing model with dynamic conditional correlations *Journal of Monetary Economics Elsevier* vol **57** (4) pp 377 – 390.

1838. Engle R F 2011 Long-term skewness and systemic risk *Journal of Financial Econometrics* Society for Financial Econometrics vol **9** (3) pp 437 – 468.

1839. Colacito R, Engle R F, Ghysels E 2011 A component model for dynamic correlations *Journal of Econometrics Elsevier* vol **164** (1) pp 45 – 59.

1840. Engle R F, Kelly B 2012 Dynamic equicorrelation *Journal of Business & Economic Statistics* American Statistical Association vol **30** (2) pp 212 – 228.

1841. Engle R F, Gallo G M, Velucchi M 2012 Volatility spillovers in East Asian financial markets: A mem-based approach *The Review of Economics and Statistics MIT Press* vol **94** (1) pp 222 – 223.

1842. Engle R F 2012 Forecasting intraday volatility in the US equity market. Multiplicative component GARCH *Journal of Financial Econometrics* Society for Financial Econometrics vol **10** (1) pp 54 – 83.

1843. Acharya V, Engle R F, Richardson M 2012 Capital shortfall: A new approach to ranking and regulating systemic risks *American Economic Review* American Economic Association vol **102** (3) pp 59 – 64.

1844. Rangel J G, Engle R F 2012 The Factor--Spline--GARCH model for high and low frequency correlations *Journal of Business & Economic Statistics* American Statistical Association vol **30** (1) pp 109 – 124.

1845. Engle R F, Ghysels E, Sohn B 2013 Stock market volatility and macroeconomic fundamentals *The Review of Economics and Statistics MIT Press* vol **95** (3) pp 776 – 797.

1846. Bollerslev T 1986 Generalized Autoregressive Conditional Heteroskedasticity *Journal of Econometrics* vol **31** pp 307 – 327.

1847. Bollerslev T, Russell J, Watson M 2010 Volatility and time series econometrics *Oxford University Press* Oxford UK ISBN13: 9780199549498 ISBN10: 0199549494 pp 1 – 384.

1848. Barone-Adesi G, Giannopoulos K, Vosper L 1999 VaR without correlations for non-linear portfolios *Journal of Futures Markets* vol **19** pp 583 – 602.

1849. McNeil A, Frey R 2000 Estimation of tail related risk measure for heteroscedastic financial time series: An extreme value approach *Journal of Empirical Finance* vol **7** pp 271 – 300

1850. Campbell J Y 1987 Stock returns and the term structure *Journal of Financial Economics* **18** (2) pp 373 – 399.

1851. Campbell J Y 1993 Intertemporal asset pricing without consumption data *American Economic Review* **83** (3) pp 487 – 512.

1852. Campbell J Y, Cochrane J H 1999 By force of habit: A consumption-based explanation of aggregate stock market behavior *Journal of Political Economy* **107** pp 205 – 251.

1853. Campbell J Y, Polk C, Vuolteenaho T 2009 Growth or glamour: Fundamentals and systematic risk in stock returns *Review of Financial Studies* **23** (1) pp 305 – 344.

1854. Campbell J Y , Giglio S, Polk C, Turley R 2012 An intertemporal CAPM with stochastic volatility *NBER Working Paper 18411*.

1855. Campbell J Y, Shiller R J 1987 Cointegration and tests of present value models *Journal of Political Economy* **95** pp 1062 – 1088.

1856. Campbell J Y, Shiller R J 1988a The dividend-price ratio and expectations of future dividends and discount factors *Review of Financial Studies* **1** pp 195 – 227.

1857. Campbell J Y, Shiller R J 1988b Stock prices, earnings, and expected dividends *Journal of Finance* **43** pp 661 – 676.

1858. Campbell J Y, Shiller R J 1991 Yield spreads and interest rate movements: A bird's eye view *Review of Economic Studies* **58** pp 495 – 514.

1859. Campbell J Y, Vuolteenaho T 2004 Bad beta, good beta *American Economic Review* **94** (5) pp 1249 – 1275.

1860. Jegadeesh N, Titman S 1993 Returns to buying winners and selling losers: Implications for stock market efficiency *The Journal of Finance* **48** (1) pp 65 – 91.

1861. Higuchi K 2016 Private communication on the interest rates setting by the Bank of Japan V. N. Karazin Kharkiv National University Kharkiv Ukraine.

***Land investment, land valuation, land ownership, land exchange, financial capital investment product, financial capital investment medium in finances:***

1862. Ricardo D 1817, 1821 On the principles of political economy and taxation 3[rd] edition *John Murray* Albemarle Street London UK.

1863. Denman D R 1956 The paradox of rural land investment in Britain *Land Economics* **32** (2) pp 109 – 117.

1864. Silberberg R 1975 Rates of return on Melbourne land investment, 1880-92 *The Economic Record* **5**1 (134) pp 203 – 217.

1865. Veenman H, Zonen 1961 An assessment of investments in land reclamation *Indian Journal of Agricultural Economics* **16** (2).

1866. Elvin C A, Ervin D E 1982 Factors affecting the use of soil conservation practices: hypotheses, evidence and policy implications *Land Economics* **58** (3) pp 277 – 292.

1867. Mills W L Jr, Hoover W L 1982 Investment in forest land: Aspects of risk and diversification *Land Economics* **58** (1) pp 33 – 51.

1868. McFarlane L A 1983 British investment and the land: Nebraska, 1877–1946 *Business History Review* **57** (02) pp 258 – 272.

1869. Eaton J December 1984 Foreign-owned land *NBER Working Paper no 1512* National Bureau of Economic Research Inc Cambridge USA pp 1 – 37.

1870. Phipps T T 1984 Land prices and farm - based returns *American Journal of Agricultural Economics* **66** (4) pp 422 – 429.

1871. Kaplan H M Winter 1985 Farm land as a portfolio investment *The Journal Portfolio Management* pp 73 – 78.

1872. Fiske J R October 7-8 1986 The interaction of investment and financing decisions and the demand for farmland *Paper no 127214* Proceedings: 1986 Regional Committee NC-161 St Paul Minnesota USA pp 1 – 21.

1873. Wiltshaw G D 1986 Land investment, planning permission, and uncertainty: A state-preference analysis *Environment and Planning A* **18** (2) pp 207 – 215.

1874. Guzhvin P 1987 The yield of land and the structure of investment *Problems of Economic Transition* **29** (10) pp 38 – 54.

1875. Eaton J 1988 Foreign-owned land *American Economic Review* **78** pp 76 – 88.

1876. Mercier J E 1988 Farmland as an asset in the context of portfolio investment *M Sc Thesis* Department of Rural Economy University of Alberta Canada.

1877. Reeve I J 1988 A squandered land: 200 Years of land degradation in Australia *Rural Development Unit* University of New England Armidale New South Wales Australia.

1878. Asako K, Kuninori M, Inoue T, Murase H 1989 Tochi Hyouka to Tobin no q/Multiple q no Keisoku (Land valuation and Tobin's q: Estimation by multiple q) *Japan Development Bank* Keizai Keiei Kenkyu 10-3 (in Japanese).

1879. Asako K, Kuninori M, Inoue T, Murase H 1997 Setsubi Toushi to Tochi Toushi: 1977-1994 (Capital investment and land investment: 1977-1994) *in* Gendai Makuro Keizai Dougaku (Contemporary Dynamic Macroeconomics) Asako K, Otaki M (editors) pp 323 – 349 *University of Tokyo Press* Tokyo Japan (in Japanese).

1880. Feder G, Onchan T 1989 Land ownership security and farm investment: Reply *American Journal of Agricultural Economics* **71** (1) pp 215 – 216.

1881. Phillips W E, Bauer L, Mercier J E, Mumey G A 1989 Alberta farmland asset returns in the context of portfolio investment *Farming for the Future Project Report no 89-12* Department of Rural Economy University of Alberta Canada.

1882. Phillips W E, Bauer L, Akabua K 1993 Returns to farmland investment in Alberta, 1964-89 *Paper no 232373* Project Report Series Department of Resource Economics and Environmental Sociology University of Alberta Edmonton Canada pp 1 – 49 http://ageconsearch.umn.edu//handle/232373 .

1883. Roth M, Barrows R, Carter M, Kanel D 1989 Land ownership security and farm investment: Comment *American Journal of Agricultural Economics* **71** (1) pp 211 – 214.

1884. Schnitkey G D, Taylor C R, Barry P J 1989 Evaluating farmland investments considering dynamic stochastic returns and farmland prices *Western Journal of Agricultural Economics* **14** (01).

1885. Barrett S 1991 Optimal soil conservation and the reform of agricultural pricing policies *Journal of Development Economics* **36** pp 167 – 187.

1886. Kanemoto Y 1991 Land tax and urban land supply *Japanese Economy* **20** (2) pp 53 – 93.

1887. Lins D A, Hoffman C, Kowalski A September 23-24, 1991 Institutional investment diversification: Foreign stocks vs US farmland *Paper no 130948* Proceedings: 1991 Regional Committee NC-161 St Louis Missouri pp 1 – 27.

1888. Yoshida T 1991 Usr cost of land investment *ISER Discussion Paper* Institute of Social and Economic Research Osaka University Japan.

1889. Butler G S, DeBraal J P 1993 Foreign ownership of U.S. agricultural land through December 31, 1992 *Statistical Bulletin no 154796* Economic Research Service United States Department of Agriculture USA.

1890. Miller E N, Andrews G February 9 - 11 1993 Factors influencing landholders' investments in soil conservation activities *Research Paper no 147759* Australian Agricultural and Resource Economics Society Conference (37[th]) Sydney Australia pp 1 – 14

http://ageconsearch.umn.edu/handle/147759 .

1891. Capozza D, Yuming Li 1994 The intensity and timing of investment: The case of land *American Economic Review* **84** (4) pp 889 – 904.

1892. Werner R A 1994 Japanese foreign investment and the "Land bubble" *Review of International Economics* **2** (2) pp 166 – 178.

1893. Nieuwoudt W L 1995 The impact of a land tax on future investments: A note *South African Journal of Economics* **63** (1) pp 47 – 50.

1894. Ogawa K, Suzuki K 1995 Land value and corporate investment: Evidence from Japanese panel data *ISER Discussion Paper* Institute of Social and Economic Research Osaka University Japan.

1895. Ogawa Kazuo, Shin-Ichi Kitasaka, Hiroshi Yamaoka, Yasuharu Iwata 1996 Borrowing constraints and role of land asset in Japanese corporate investment decision *Journal of the Japanese and International Economies* pp 122 – 149.

1896. Grepperud S February 1996 Soil conservation as an investment in land *Discussion Paper no 163* Research Department Statistics Norway Oslo Norway pp 1 – 21

http://www.ssb.no/a/publikasjoner/pdf/DP/dp_163.pdf .

1897. Echevarria E 1997 Non-homothetic preferences, growth, trade and land *Working Paper* Instituto Valenciano de Investigaciones Económicas SA *Ivie*.

1898. Deininger K, Feder G 2001 Land institutions and land markets *Chapter 06* Handbook of Agricultural Economics *Elsevier* **1** (1) pp 288 – 331.

1899. Kiyotaki N, West K 2004 Land prices and business fixed investments in Japan *NBER Working Paper no 10909* National Bureau of Economic Research Inc USA

http://www.nber.org/papers/w10909.pdf .

*1900.* Sekine T, Tachibana T March 2004 Land investment by Japanese firms during and after the bubble period *Working Paper no 04-E-2* Bank of Japan Tokyo Japan pp 1 – 48

http://www.boj.or.jp/en/research/wps_rev/wps_2004/data/wp04e02.pdf .

*1901.* Feinerman E, Peerlings J 2005 Uncertain land availability and investment decisions: The case of Dutch dairy farms *Journal of Agricultural Economics* **56** (1) pp 59 – 80.

*1902.* Hennings E, Sherrick B J, Barry P J 2005 Portfolio diversification using farmland investments *Paper no 19273* Annual meeting of Agricultural and Applied Economics Association July 24-27 Providence RI USA.

*1903.* Ray T 2005 Sharecropping, land exploitation and land-improving investments *The Japanese Economic Review* **56** (2) pp 127 – 143.

*1904.* Turnbull G 2005 The investment incentive effects of land use regulations *The Journal of Real Estate Finance and Economics* **31** (4) pp 357 – 395.

*1905.* Kiyotaki N, West K 2006 Land prices and business fixed investment in Japan Chapter 12 *in* Long-run Growth and Short-run Stabilization *Edward Elgar Publishing* London UK.

*1906.* Holden S, Deininger K, Ghebru H 2007 Impacts of low-cost land certification on investment and productivity *American Journal of Agricultural Economics* **91** (2) pp 359 – 373.

*1907.* Sekine T, Tachibana T 2007 Land as production input and collateral: Land investment by Japanese firms *The Journal of Real Estate Finance and Economics* **35** (4) pp 497 – 526.

*1908.* Głuszak M 2008 Land acquisition in development projects: Investment value and risk *SRE-Discussion 2008/06* Department of Socioeconomics Institute for Multilevel Governance and Development Vienna University of Economics and Business Vienna Austria pp 1 – 16.

*1909.* Von Braun J, Meinzen-Dick R April 2009 Land grabbing by foreign investors in developing countries: Risks and opportunities *IFPRI Policy Brief* **13**.

*1910.* Cotula L, Vermeulen S, Leonard R, Keeley J 2009 Land grab or development opportunity? Agricultural investment and international land deals in Africa *International Fund for Agricultural Development*, Food and Agriculture Organization London, Rome UK, Italy.

*1911.* Cotula L 2012 The international political economy of the global land rush: A critical appraisal of trends, scale, geography and drivers *Journal of Peasant Studies* **39** (3-4) pp 649 – 680.

*1912.* Fakton B, van der Post W 2010 The practice of land pricing across Europe *European Real Estate Society (ERES)*.

*1913.* Fischer G, Shah M 2010 Farmland investments and food security, statistical annex Laxenburg Austria (World Bank IIASA contract: Lessons for the large-scale acquisition of land from a agricultural land use analysis).

*1914.* Zoomers A 2010 Globalisation and the foreignisation of space: Seven processes driving the current global land grab *Journal of Peasant Studies* **37** (2) pp 429 – 447.

**1915.** Abdulai A, Owusu V, Goetz R 2011 Land tenure differences and investment in land improvement measures: Theoretical and empirical analyses *Journal of Development Economics* **96** (1) pp 66 – 78.

**1916.** Arezki R, Deininger K, Selod H December 2011 What drives the global land rush? *CESifo Working Paper no 3666* CESifo Group Munich Germany pp 1 – 35 www.CESifo-group.org/wp .

**1917.** De Schutter O 2011 The green rush: The global race for farmland and the rights of land users *Harvard International Law Journal* **52** (2) pp 503 – 559.

**1918.** Hui-Wen Koo 2011 Property rights, land prices, and investment: A study of the Taiwanese land registration system *Journal of Institutional and Theoretical Economics (JITE)* **167** (3) pp 515 – 535.

**1919.** Palyanychko N 2011 Investment in the sustainable use of agricultural land *Economics of Nature and the Environment* pp 170 – 179.

**1920.** Zagema B September 22 2011 Land and power: The growing scandal surrounding the new wave of investments in land *Oxfam Briefing Paper 151* Oxfam International UK http://www.oxfam.org/sites/www.oxfam.org/files/bp151-land-power-rights-acquisitions-220911-en.pdf .

**1921.** Di Corato L, Hess S September 26-28 2012 Foreign investment in farmland under uncertainty *Paper no 133017* 52[nd] Annual Conference of the German Association of Agricultural Economists (GEWISOLA) Stuttgart Germany pp 1 – 3 http://ageconsearch.umn.edu/handle/133017 .

**1922.** Domeher D, Abdulai R 2012 Land registration, credit and agricultural investment in Africa *Agricultural Finance Review* **72** (1) pp 87 – 103.

**1923.** Van der Kr Pasmans E, Van der Heijden R 2012 Explanations for the private investment decision in the industrial land market *19th Annual European Real Estate Society Conference* Edinburgh Scotland DOI: 10.15396/eres2012_264 .

**1924.** Todorovic S Z, Vasiljevic Z R, Rajic Z N 2012 Economic efficiency of investments in agricultural land *International Journal of Sustainable Economies Management (IJSEM)* **1** (1) pp 61 – 74.

**1925.** Oya C 2013 Methodological reflections on "land grab" databases and the "land grab" literature "rush" *Journal of Peasant Studies* **40** (3) pp 503 – 520.

**1926.** Xianlei Ma, Heerink N, van Ierland E, van den Berg M, Xiaoping Shi 2013 Land tenure security and land investments in Northwest China *China Agricultural Economic Review* **5** (2) pp 281 – 307.

**1927.** Zolin M B, Luzi L 2013 Unexpected and growing interest in land investments? The Asian case *Working Paper no 2013:26* Department of Economics University of Venice "Ca' Foscari" Venice Italy pp 1 – 20.

**1928.** Boehlje M, Baker T G, Langemeier M January 3-5 2014 Farmland: Is it currently priced as an attractive investment? *Paper no 161650* Allied Social Science Association (ASSA) Annual Meeting Philadelphia PA USA.

**1929.** Malashchuk O 2014 The land use ranking by the degree of investment attractiveness *Balanced Nature Using* **5** (3) pp 116 – 120.

***1930.*** Palenychak O 2014 Formation of investment attractiveness of the ecological-safety land use *Balanced Nature Using* **5** (3) pp 120 – 124.

***1931.*** Arezki R, Deininger K, Selod H 2015 What drives the global "Land rush"? *World Bank Economic Review* **29** (2) pp 207 – 233.

***1932.*** Busha D 2015 Land market infrastructure development as a basic prerequisite for improving the investment attractiveness of lands *Ukrainian Journal Ekonomist* issue **3** pp 21 – 25.

***1933.*** Szreder J 2015 Success factors in land development investments *World of Real Estate Journal (Swiat Nieruchomosci)* **93** pages 55 – 60.

***1934.*** Bochco O 2016 Increasing the level of investment attractiveness of land resources in Lviv region *Balanced Nature Using* **4** (2) pp 43 – 47.

***1935.*** Borghesi S, Giovannetti G, Iannucci G, Russu P 2016 The dynamics of foreign direct investments in land and pollution accumulation *SEEDS Working Paper no 1116* Sustainability Environmental Economics and Dynamics Studies pp 1 – 22 http://www.sustainability-seeds.org/ .

***Commodity investment, commodity valuation, commodity derivatives, commodity futures, commodities exchange, financial capital investment product, financial capital investment medium in finances:***

***1936.*** Keynes J M 1923 Some aspects of commodity markets *Manchester Guardian Commercial European Reconstruction Series Section* **13** pp 784 – 786.

***1937.*** Keynes J M September 1938 the policy of government storage of food-stuffs and raw materials *Economic Journal* **XLVIII** pp 449 – 460.

***1938.*** Keynes J M 1942, 1980 On commodities control *in* Collected writings of John Maynard Keynes Moggridge D (editor) **XXVII**: Activities 1940–46 Shaping the post-war world: Employment and commodities *Macmillan* London UK.

***1939.*** Keynes J M 1943 The objective of price stability *Economic Journal* **53** no 210 pp 185–187.

***1940.*** Hotelling H 1931 The economics of exhaustible resources *Journal of Political Economy* **39** pp 137 – 175.

***1941.*** Graham B 1937 Storage and stability: A modern ever-normal granary *McGraw* Hill NY USA.

***1942.*** Kaldor N 1939 Speculation and economic stability *Review of Economic Studies* **7** pp 1 – 27.

***1943.*** Working H February 1948 Theory of the inverse carrying charge in futures markets *Journal of Farm Economics* **30** pp 1 – 28.

***1944.*** Working H 1949 The theory of the price of storage *American Economic Review* **39** pp 1254 – 1262.

***1945.*** Working H 1960 Speculation on hedging markets *Food Research Institute Series* **1** pp 185 – 220.

***1946.*** Brennan M 1958 The supply of storage *American Economic Review* **48** pp 50 – 72.

***1947.*** Brennan M, Schwartz E S 1985 Evaluating natural resource investments *Journal of Business* **58** (2) pp 135 – 157.

*1948.* Brennan M 1991 The price of convenience and the valuation of commodity contingent claims *in* Stochastic Models and Option Values Lund D, Oksendal B (editors) *North Holland* The Netherlands.

*1949.* Sharpe W F 1964 Capital asset prices: A theory of market equilibrium under conditions of risk *Journal of Finance* **19** (3) pp 425 – 442.

*1950.* Sharpe W F, Alexander G J, Bailey J V 1995 Investments 5[th] edition New York USA.

*1951.* Telser L 1968 Futures trading and the storage of cotton and wheat *Journal of Political Economy* **66** pp 233 – 255.

*1952.* Dusak K 1973 Futures trading and investor returns: An investigation of commodity market risk premiums *Journal of Political Economy* **81** (6) pp 1387 – 1406.

*1953.* Lovell M C, Vogel R C 1973 A CPI-futures market *Journal of Political Economy* vol **81** no 4 pp 1009 – 1012.

*1954.* Merton R C 1973 An intertemporal capital asset pricing model *Econometrica* **41** pp 867 – 888.

*1955.* Stoll H R 1979 Commodity futures and spot price determination and hedging in capital market equilibrium *Journal of Financial and Quantitative Analysis* **14** (4) pp 873 – 894.

*1956.* Stoll H R, Whaley R E 2009, 2010 Commodity index investing and commodity futures prices *Working Paper Owen Graduate School of Management* Vanderbilt University USA, *Journal of Applied Finance* **1** pp 1 – 40.

*1957.* Bodie Z, Rosansky V I May June 1980 Risk and return in commodity futures *Financial Analysts Journal* vol **36** pp 3 – 14.

*1958.* Breeden D T 1980 Consumption risk in futures markets *Journal of Finance* **35** (2) pp 503 – 520.

*1959.* Spraos J March 1980 The statistical debate on the net barter terms of trade between primary commodities and manufacturers *Economic Journal* **90** pp 107 – 28.

*1960.* Newbery D M, Stiglitz J E 1981 The theory of commodity price stabilization. A study of the economics of risk *Claredon Press Oxford University Press* Oxford UK.

*1961.* Wright B, Williams J C 1982 The economic role of commodity storage *Economic Journal* **92** (367) pp 596 – 614.

*1962.* Abel A B 1983 Optimal investment under uncertainty *American Economic Review* **73** (1) pp 228 – 233.

*1963.* Abel A B, Blanchard O J 1986 The present value of profits and cyclical movements in investment *Econometrica* **54** (2) pp 249 – 273.

*1964.* Abel A B, Mankiw N G, Summers L H, Zeckhauser R J 1989 Assessing dynamic efficiency: Theory and evidence *Review of Economic Studies* **56** pp 1 – 20.

*1965.* Abel A B, Eberly J C 1994 A unified model of investment under uncertainty *American Economic Review* **84** (5) pp 1369 – 1384.

*1966.* Bernanke B S 1983 Irreversibility, uncertainty, and cyclical investment *The Quarterly Journal of Economics* **98** pp 85 – 106.

*1967.* Carter C A, Rausser G C, Schmitz A April 1983 Efficient asset portfolios and the theory of normal backwardation *The Journal of Political Economy* vol **91** no 2 pp 319 – 333.

1968. Garbade K, Silber W L 1983 Price movements and price discovery in futures and cash markets *Review of Economics and Statistics* **65** pp 289 – 297.

1969. Edwards G W, Freebairn J W 1984 The gains from research into tradable commodities *American Journal Agriculture Economics* **66** pp 41 – 49.

1970. Kanbur S M R 1984 How to analyse commodity price stabilization? A review article *Oxford Economic Papers* **36** pp 336 – 358.

1971. Cox J C, Rubinstein M 1985 Option markets *Prentice Hall Inc* New Jersey USA.

1972. Gilbert C L September 1985 Futures trading and the welfare evaluation of commodity price stabilization *Economic Journal* **95** pp 637 – 661.

1973. Gilbert C L 1996 International commodity agreements: An obituary notice *World Development* **24** (1) pp 1 – 19.

1974. Jagannathan R 1985 An investigation of commodity futures prices using the consumption-based intertemporal capital asset pricing model *The Journal of Finance* vol **40** no 1 pp 175 – 191.

1975. Taylor S J 1985 The behavior of futures prices over time *Applied Economics* **17** no 4 pp 713 – 734.

1976. Fama E F, French K R January 1987 Commodity future prices: Some evidence on forecast power, premiums, and the theory of storage *Journal of Business* **60** (1) pp 55 – 73.

1977. Fama E F, French K H December 1988 Business cycles and the behavior of metals prices *The Journal of Finance* **43** (5) pp 1075 – 1093.

1978. Hartzmark M 1987 Returns to individual traders of futures: Aggregate results *Journal of Political Economy* **95** (6) pp 1292 – 1306.

1979. Chinloy P 1988 Real estate—investment and financial strategy *Kluwer Academic Publishers* Boston USA.

1980. Gyourko J, Linneman P 1988 Owner-occupied homes, income-producing real estate and REIT as inflation hedges *Journal of Real Estate Finance and Economics* vol **1** no 4 pp 347 – 372.

1981. Hirshleifer D 1988 Residual risk, trading costs and commodity futures risk premia *Review of Financial Studies* **1** (2) pp 173 – 193.

1982. Hirshleifer D 1989 Determinants of hedging and risk premia in commodity futures markets *Journal of Financial and Quantitative Analysis* **24** (3) pp 313 – 331.

1983. Hirshleifer D 1990 Hedging pressure and futures price movements in a general equilibrium model *Econometrica* **58** pp 411 – 428.

1984. Lichtenberg A J, Ujihara A 1988 Application of non-linear mapping theory to commodity price fluctuations *Journal Econ Dyn Control* **13** pp 225 – 246.

1985. Paddock J L, Siegel D R, Smith J L 1988 Option valuation of claims on real assets: The case of offshore petroleum leases *Quarterly Journal of Economics* **103** pp 479 – 508.

1986. Baumol W J, Benhabib J 1989 Chaos: Significance, mechanism, and economic applications *Journal Econ Perspect* **3** pp 77 – 105.

1987. Frank M, Stengos T 1989 Measuring the strangeness of gold and silver rates of return *Rev Econ Stud* **456** pp 553 – 567.

**1988.** Jaffee J F 1989 Gold and gold stocks as investments for institutional portfolios *Financial Analysts Journal* **45** pp 53 – 60.

**1989.** Rubens J H, Bond M T, Webb J R 1989 The inflation-hedging effectiveness of real estate *Journal of Real Estate Research* vol **4** no 2 pp 45 – 55.

**1990.** De Gorter H, Zilberman D 1990 On the political economy of public good inputs in agriculture *American Journal Agriculture Economics* **72** pp 131 – 137.

**1991.** De Gorter H, Tsur Y 1991 Explaining price policy bias in agriculture: The calculus of support-maximizing politicians *American Journal Agriculture Economics* **73** pp 1244 – 1254.

**1992.** De Gorter H, Nielson D J, Rausser G C 1992 Productive and predatory public policies *American Journal Agriculture Economics* **74** pp 27 – 37.

**1993.** De Gorter H, Swinnen J 1998 The impact of economic development on public research and commodity policies in agriculture *Review Develop Econ* **2** (1) pp 41 – 60.

**1994.** Gibson R, Schwartz E S 1990 Stochastic convenience yield and the pricing of oil contingent claims *Journal of Finance* **45** pp 959 – 976.

**1995.** Kaminsky G, Kumar M S 1990 Efficiency in commodity futures markets *IMF Staff Papers* vol **37** no 3 pp 670 – 699.

**1996.** Pindyck R, Rotemberg J 1990 The excess co-movement of commodity prices *Economic Journal* **100** pp 1173 – 1189.

**1997.** Shleifer A, Summers L H 1990 The noise trader approach to finance *The Journal of Economic Perspectives* **4** (2) pp 19 – 33.

**1998.** Siegel D R, Siegel D F 1990 The futures markets *McGraw Hill Book Company Ltd* Maidenhead UK.

**1999.** Blank S C 1991 "Chaos" in futures markets? A non-linear dynamical analysis *Journal Futures Markets* **11** pp 711 – 728.

**2000.** Brennan M 1991 The price of convenience and the valuation of commodity contingent claims *in* Stochastic Models and Option Models Lund D, Oksendal B (editors) *North Holland* Amsterdam The Netherlands.

**2001.** Chan K, Chan K C, Karolyi G A 1991 Intraday volatility in the stock index and stock index futures markets *Review of Financial Studies* **4** pp 657 – 684.

**2002.** Subrahmanyam A 1991 A theory of trading in stock index futures *Review of Financial Studies* **4** pp 17 – 51.

**2003.** Williams J C, Wright B D 1991 Storage and commodity markets *Cambridge University Press* Cambridge UK.

**2004.** Bessembinder H 1992 Systemic risk, hedging pressure, and risk premiums in futures markets *Review of Financial Studies* **5** (4) pp 637 – 667.

**2005.** Bessembinder H, Chan K 1992 Time-varying risk premia and forecastable returns in futures markets *Journal of Financial Economics* **32** pp 169 – 193.

**2006.** Bessembinder H, Seguin P J 1992 Futures-trading activity and stock price volatility *Journal of Finance* **47** pp 2015 – 2034.

**2007.** Bessembinder H, Seguin P J 1993 Price volatility, trading volume, and market depth: Evidence from futures markets *Journal of Financial and Quantitative Analysis* **28** pp 21 – 39.

**2008.** Bessembinder H, Coughenour J F, Seguin P J, Smoller M M 1995 Mean-reversion in equilibrium asset prices: Evidence from the futures term structure *Journal of Finance* **50** (1) pp 361 – 375.

**2009.** Bessembinder H, Coughenour J P, Seguin P J, Smoller M M Winter 1996 Is there a term structure of futures volatilities? Reevaluating the Samuelson hypothesis *Journal Derivatives* pp 45 – 58.

**2010.** Deaton A, Laroque G January 1992 On the behaviour of commodity prices *The Review of Economic Studies* **59** (1) pp 1 – 23.

**2011.** Deaton A, Laroque G 1996 Competitive storage and commodity price dynamics Journal *of Political Economy* **104** pp 896 – 923.

**2012.** DeCoster G P, Labys W C, Mitchell D W 1992 Evidence of chaos in commodity futures prices *Journal Futures Markets* **12** pp 291 – 305.

**2013.** Edwards F M, Ma C W 1992 Futures and options *McGraw-Hill* New York USA.

**2014.** Kolb R W 1992 Is normal backwardation normal? *Journal of Futures Markets* **12** pp 75 – 91.

**2015.** Kolb R 1996 The systematic risk of futures contracts *Journal of Futures Markets* **16** no 6 pp 631 – 654.

**2016.** Kolb R 1997 Understanding futures markets 5th edition *Blackwell Publishers* Oxford UK.

**2017.** Maizels A 1992 The commodity price collapse of the 1980s, The impact on the developing countries *in* Commodities in crisis: The commodity crisis of the 1980s and the political economy of international commodity policies *Oxford University Press* Oxford UK.

**2018.** Ankrim E M, Hensel C R May/June 1993 Commodities in asset allocation: A real-asset alternative to real estate? *Financial Analysts Journal* **49** (3) pp 20 – 29.

**2019.** Bleaney M, Geenaway D July 1993 Long-run trends in the relative price of primary commodities and in the terms of trade of developing countries *Oxford Economic Papers* **45** (3), pp 349 – 363.

**2020.** Brueggeman B W, Fisher D J 1993 Real estate finance and investments 9th edition *Irwin* BurrRidge USA pp 1 – 908 ISBN: 0-256-08290-1 .

**2021.** Lummer S L, Siegel L B Summer 1993 GSCI collateralized futures: A hedging and diversification tool for institutional investors *Journal of Investing* pp75 – 82.

**2022.** Yang S R, Brorsen B W 1993 Nonlinear dynamics of daily futures prices: Conditional heteroskedasticity or chaos? *The Journal of Futures Markets* **13** (2) pp 175 – 191.

**2023.** Dixit A, Pindyck R 1994 Investment under uncertainty *Princeton University Press* Princeton New Jersey USA.

**2024.** Satyanarayan S, Varangis P March 1994 An efficient frontier for international portfolios with commodity assets *Policy Research Working Paper 1266* The World Bank

http://www-wds.worldbank.org/.../WDSP/IB/1994/03/01/000009265_3961006020009/Rendered/PDF/multi_page.pdf .

**2025.** Litzenberger R H, Rabinowitz N December 1995 Backwardation in oil futures markets: Theory and empirical evidence *The Journal of Finance* vol **50** issue 5 pp 1517 – 1545.

**2026.** O'Hara M 1995 Market microstructure theory *Blackwell* Oxford UK.

**2027.** Sachs J D, Warner A M 1995a Natural resource abundance and economic growth *NBER Working Paper no 5398* Cambridge MA USA.

**2028.** Sachs J D, Warner A M 1995b Economic convergence and economic policy *NBER Working Paper no 5039* Cambridge MA USA.

**2029.** Sachs J D, Warner A M 1997 Sources of slow growth in African economies *Journal of African Economies* **6** pp 335 – 376.

**2030.** Sachs J D, Warner A M 1999 The big push, natural resource booms and growth *Journal of Development Economics* **59** pp 43 – 76.

**2031.** Sachs J D, Warner A M 2001 Natural resources and economic development. The curse of natural resources *European Economic Review* **45** pp 827 – 838.

**2032.** Chambers M, Bailey R 1996 A theory of commodity price fluctuations *Journal of Political Economy* **104** (5) pp 924 – 957.

**2033.** Duffie D 1996 Dynamic asset pricing theory *Princeton University Press* Princeton USA.

**2034.** Hamilton J D 1996 This is what happened to the oil-price macroeconomy relationship *Journal of Monetary Economics* **38** pp 215 – 220.

**2035.** Hamilton J D 2005 Oil and the macroeconomy *in* The New Palgrave Dictionary of Economics Durlauf S, Blume L (editors) 2nd edition *Palgrave MacMillan Ltd*.

**2036.** Hamilton J D Spring 2009a Causes and consequences of the oil shock of 2007-08 *Brookings Papers on Economic Activity* **40** (1) pp 215 – 293.

**2037.** Hamilton J 2009b Understanding crude oil prices *The Energy Journal* **30** (2) pp 179 – 206.

**2038.** Hamilton J, Jing Wu 2013a Effects of index-fund investing on commodity futures prices *Working Paper* University of California San Diego USA.

**2039.** Hamilton J, Jing Wu 2013b Risk premia in crude oil futures prices *Working Paper* University of California San Diego USA.

**2040.** Hamilton J D, Cynthia J Wu 2014 Effects of index-fund investing on commodity futures prices *NBER Working Paper 19892* National Bureau of Economic Research 1050 Massachusetts Avenue Cambridge MA 02138 USA pp 1 – 43 http://www.nber.org/papers/w19892 .

**2041.** Sanders D R, Irwin S H, Leuthold R M 1996 Noise trader demand in futures markets *Working Paper no 96-02* Department of Agricultural and Consumer Economics University of Illinois at Urbana-Champaign.

**2042.** Sanders D R, Irwin S H 2010a A speculative bubble in commodity futures prices? Cross-sectional evidence *Agricultural Economics* **41** (1) pp 25 – 32.

**2043.** Sanders D R, Irwin S H 2010b The impact of index and swap funds on commodity futures markets *OECD Food, Agriculture and Fisheries Working Papers no 27 OECD Publishing*.

2044. Sanders D R, Irwin S H, Merrin R 2010 The adequacy of speculation in agricultural futures markets: Too much of a good thing? *Applied Economic Perspectives and Policy* **32** (1) pp 77 – 94.

2045. Sanders D R, Irwin S H 2011a The impact of index funds in commodity futures markets: A systems approach *Journal of Alternative Investments* **14** pp 40 – 49.

2046. Sanders D R, Irwin S H 2011b New evidence on the impact of index funds in U.S. grain futures markets *Canadian Journal of Agricultural Economics* **59** pp 519 – 532.

2047. Sanders D W, S H Irwin 2013 Measuring index investment in commodity futures markets *The Energy Journal* **34** (3) pp 105 – 127.

2048. Hoesli M, MacGregor B D, Matysiak G, Nanthakumaran N 1997 The short-term inflation-hedging characteristics of UK real estate *Journal of Real Estate Finance and Economics* vol **15** no 1 pp 59 – 76.

2049. Kocagil A E 1997 Does futures speculation stabilize spot prices? Evidence from metals markets *Applied Financial Economics* **7** pp 115 – 125.

2050. Schneeweis T, Spurgin R 1997 Energy based investment products and investor asset allocation *Center for International Securities and Derivatives Markets (CISDM) Isenberg School of Management* University of Massachusetts USA.

2051. Schwartz E S July 1997 The stochastic behavior of commodity prices: Implications for valuation and hedging *Journal of Finance* **52** (3) pp 923 – 973.

2052. Schwartz E S, Smith J E July 2000 Short-term variations and long-term dynamics in commodity prices *Management Science* **47** (2) pp 893 – 911.

2053. Mabro R 1998 The oil price crisis of 1998 *Oxford Institute for Energy Studies* Oxford UK.

2054. Pilipovic D 1998 Energy risk – valuing and managing energy derivatives *McGraw-Hill* NY USA.

2055. Cashin P, McDermott J C, Scott A 1999 The myth of co-moving commodity prices *IMF Working Paper no 99/169* IMF USA.

2056. Greenaway D, Morgan C W 1999 Introduction *The Economics of Commodity Markets* The International Library of Critical Writings in Economics An Elgar Reference Collection *Edward Elgar Publishing Ltd* Cheltenham UK.

2057. Hielscher U 1999 Investmentanalyse *3rd edition* München Germany.

2058. Irwin S H, Yoshimaru S 1999 Managed futures, positive feedback trading, and futures price volatility *Journal of Futures Markets* **19** pp 759 – 776.

2059. Irwin S H, Holt B R 2004 The effect of large hedge fund and CTA trading on futures market volatility *in* Commodity Trading Advisors: Risk, Performance Analysis and Selection Gregoriou G N, Karavas V N, Lhabitant F-S and Rouah F (editors) *Wiley* Hoboken NJ USA.

2060. Irwin S H, Sanders D R, Merrin R P April 20-21 2009a A speculative bubble in commodity futures prices? *Cross-sectional evidence NCCC-134 Conference on Applied Commodity Price Analysis, Forecasting, and Market Risk Management* St Louis Missouri USA.

2061. Irwin S H, Sanders D R, Merrin R P August 2009b Devil or angel? The role of speculation in the recent commodity price boom (and bust) *Journal of Agricultural and Applied Economics* **41** (2) pp 377 – 391.

2062. Irwin S H, Sanders D R February 2011 Index funds, financialization, and commodity futures markets *Applied Economic Perspectives and Policy* **33** pp 1 – 31.

2063. Irwin S H, Sanders D R 2012a Testing the masters hypothesis in commodity futures markets *Energy Economics* **34** pp 256 – 269.

2064. Irwin S H, Sanders D R 2012b A reappraisal of investing in commodity futures markets *Applied Economic Perspectives and Policy* **34** pp 515 – 530.

2065. Labys W C, Achouch A, Terraza M 1999 Metal prices and the business cycle *Resources Policy* no 25 pp 229 – 238.

2066. Becker K, Finnerty J 2000 Indexed commodity futures and the risk and return of institutional portfolios *Office of Futures and Options Research Working Paper*.

2067. Borenstein S, Bushnell J, Stoft S 2000 The competitive effects of transmission capacity in a deregulated electricity industry *RAND Journal of Economics* **31** (2) pp 294 – 325.

2068. Cashin P, Liang H, McDermott C J 2000 How persistent are hocks to world commodity prices *IMF Staff Papers* **47** (2) pp 177 – 217.

2069. Cashin P, McDermott C 2002 The long-run behaviour of commodity prices: Small trends and big variability *IMF Staff Papers* **49** pp 175 – 199.

2070. Christie-David R, Chaudry M, Koch T W 2000 Do macroeconomics news releases affect gold and silver prices? *Journal of Economics and Business* **52** pp 405 – 421.

2071. De Roon F, Nijman T, Veld Ch 2000 Hedging pressure effects in futures markets *Journal of Finance* **55** (3) pp 1437 – 1456.

2072. De Roon F, Szymanowska M 2010 The cross-section of commodity futures returns *SSRN Working Paper #891073* Social Science Research Network USA.

2073. Greer R J 2000 The nature of commodity index returns *Journal of Alternative Investments* vol **3** no 1 pp 45 – 52.

2074. Jensen G R, Johnson R R, Mercer J M 2000 Efficient use of commodity futures in diversified portfolios *The Journal of Futures Markets* vol **20** issue 5 pp 489 – 506.

2075. Jensen G R, Johnson R R, Mercer J M Summer 2002 Tactical asset allocation and commodity futures *Journal of Portfolio Management* vol **28** issue 4 pp 100 – 111.

2076. Routledge B R, Seppi D J, Spatt C S 2000 Equilibrium forward curves for commodities *The Journal of Finance* **55** (3) pp 1297 – 1338.

2077. Swinnena J F M, De Gorterc H, Rausserd G C, Banerjeea A N 2000 The political economy of public research investment and commodity policies in agriculture: An empirical study *Agricultural Economics* **22** pp 111 – 122 www.elsevier.com/locate/agecon .

2078. Till H Fall 2000a Passive strategies in the commodity futures markets *Derivatives Quarterly* pp 49 – 54.

2079. Till H September 2000b Systematic returns in commodity futures *Commodities Now* pp 75 – 79.

2080. Till H April 2003 Actively timing an investment in the Goldman Sachs commodity index http://www.premiacap.com/publications.php .

2081. Till H 2006 Portfolio risk measurement in commodity futures investments http://www.premiacap.com/publications.php .

2082. Till H, Eagleeye J May 2003 The risks of commodity investing

http://www.premiacap.com/publications.php .

*2083.* Till H, Eagleeye J Fall 2005 Commodities: Active strategies for enhanced return *Journal of Wealth Management* vol **8** no 2 pp 42 – 61.

*2084.* Cochrane J H 2001 Asset pricing *Princeton University Press* Princeton NJ USA.

*2085.* Fabozzi F J 2001 The handbook of fixed income securities 6th Edition *McGraw-Hill* New York USA.

*2086.* Kogan L November 2001 An equilibrium model of irreversible investment *Journal of Financial Economics* **62** (2) pp 201 – 245.

*2087.* Kogan L 2004 Asset prices and real investment *Journal of Financial Economics* **73** pp 411 – 431.

*2088.* Kogan L, Livdan D, Yaron A 2005 Futures prices in a production economy with investment constraints *NBER Working Paper* National Bureau of Economic Research USA.

*2089.* Richards T J, Padilla L August 5 - 8 2001 Commodity R&D, patenting and promotion *Selected Paper Presented at AAEA Annual Meetings* pp 1 – 21.

*2090.* Barsky R B, Kilian L May 2002 Do we really know that oil caused the great stagflation? A monetary alternative *in NBER Macroeconomics Annual 2001* Bernanke B, Rogoff K (editors) pp 137 – 183.

*2091.* Barsky R B, Kilian L 2004 Oil and the macroeconomy since the 1970s *Journal of Economic Perspectives* **18** (4) pp 115 – 134.

*2092.* Chatrath A, Adrangi B, Dhanda K K 2002 Are commodity prices chaotic? *Agricultural Economics* **27** pp 123 – 137

www.elsevier.com/locate/agecon .

*2093.* Cremer H, Laffont J J 2002 Competition in gas markets *European Economic Review* **46** pp 928 – 935.

*2094.* Cremer H, Gasmi F, Laffont J J 2003 Access to pipelines for competitive gas markets *Journal of Regulatory Economics* **24** (1) pp 5 – 33.

*2095.* Ederington L, Lee J H 2002 Who trades futures and how: Evidence from the heating oil futures market *Journal of Business* **75** (2) pp 353 – 373.

*2096.* Weiner R J 2002 Sheep in wolves' clothing? Speculators and price volatility in petroleum futures *Quarterly Review of Economics and Finance* **42** pp 391 – 400.

*2097.* Bower J, Kamel N June 2003 Commodity price insurance: A Keynesian idea revisited *Oxford Institute for Energy Studies* Oxford UK ISBN 1 901 795 26 8 pp 1 – 50.

*2098.* Sorenson V L 2003 Trade and international commodity programs *Michigan State University* USA pp 1 – 12.

*2099.* Vrugt E B 2003 Tactical commodity strategies in reality: An economic theory approach *ABP Working Paper*

http://www.fdewb.unimaas.nl/finance/workingpapers .

*2100.* Wang C 2003 The behavior and performance of major types of futures traders *Journal of Futures Markets* **23** pp 1 – 31.

*2101.* Weiser S 2003 The strategic case for commodities in portfolio diversification *Commodities Now* pp 7 – 11.

2102. Greer R J Summer 2004 The nature of commodity index returns *The Journal of Alternative Investments* pp 45 – 53.

2103. Greer R J April 2005 Commodity indexes for real return and efficient diversification *An Investor Guide to Commodities Deutsche Bank* pp 24 – 34.

2104. Gorton G, Rouwenhorst K 2004 Facts and fantasies about commodity futures *NBER Working Papers no 10595* USA.

2105. Gorton G, Rouwenhorst K G 2006 Facts and fantasies about commodity futures *Yale ICF Working Paper no 04-20* Yale School of Management Yale University USA, *Financial Analysts Journal* **62** (2) pp 47 – 68.

2106. Gorton G B, Hayashi F, Rouwenhorst K G 2013 The fundamentals of commodity futures returns *Review of Finance* **17** (1) pp 35 – 105.

2107. Longstaff F, Wang A W 2004 Electricity forward prices: A high-frequency empirical analysis *The Journal of Finance* **59** (4) pp 1877 – 1900.

2108. Beenen J 2005 Commodities as a strategic investment for PGGM *in* An investor guide to commodities Michael Lewis (editor) *Deutsche Bank* Germany.

2109. Casassus J, Collin-Dufresne P 2005 Stochastic convenience yield implied from commodity futures and interest rates *Journal of Finance* **60** pp 2283 – 2331.

2110. Casassus J, Collin-Dufresne P, Routledge B R December 2005 Equilibrium commodity prices with irreversible investment and non-linear technology *NBER Working Paper 11864* National Bureau of Economic Research 1050 Massachusetts Avenue Cambridge MA 02138 USA pp 1 – 74
http://www.nber.org/papers/w11864 .

2111. Erb C, Harvey C May 2005 The tactical and strategic value of commodity futures *NBER Working Papers no 11222* USA.

2112. Erb C, Harvey C 2006 The strategic and tactical value of commodity futures *Financial Analysts Journal* **62** (2) pp 69 – 97.

2113. Haigh M, Hranaiova J, Oswald J 2005 Price dynamics, price discovery and large futures trader interactions in the energy complex *US Commodity Futures Trading Commission Working Paper* USA.

2114. Lautier D Summer 2005 Term structure models of commodity prices: A review *Journal of Alternative Investments* vol **8** issue 1 pp 42 – 64.

2115. Lewis M April 2005 Convenience yields, term structures and volatility across commodity markets *An Investor Guide to Commodities Deutsche Bank* pp 18 – 23.

2116. Micu M December 2005 Declining risk premia in the crude oil futures market *BIS Quarterly Review* pp 50 – 1.

2117. O'Connell R December 2005 What sets the precious metals apart from other commodities? *World Gold Council*.

2118. Pulvermacher K March 2005a What are commodities?
http://www.gold.org/value/stats/research/index.html .

2119. Pulvermacher K 2005b Commodity returns and the economic cycle
http://www.gold.org/value/stats/research/index.html .

2120. Taylor S J 2005 Asset price dynamics: Volatility and prediction *Princeton University Press* Princeton USA.

2121. Brown S P January February 2006 The commodity question: Can adding commodities to a portfolio improve performance? *CFA Magazine* pp 44 – 45.

2122. Bryant H L, Bessler D A, Haigh. S 2006 Causality in futures markets *Journal of Futures Markets* **26** (11) pp 1039 – 1057.

2123. Campbell P, Orskaug B-E, Williams R Spring 2006 The forward market for oil *The Bank of England Quarterly Bulletin* pp 66 – 74.

2124. Feldman, B Till H 2006 Separating the wheat from the chaff: Backwardation as the long-term driver of commodity futures performance; evidence from soy, corn, and wheat futures from 1950 to 2004 *EDHEC Risk and Asset Management Research Centre* Nice France.

2125. Frankel J 2006 The effect of monetary policy on real commodity prices *NBER Working Paper #12713* National Bureau of Economic Research USA, *in* Asset Prices and Monetary Policy John Campbell (editor) *University of Chicago Press* Chicago USA.

2126. Fusaro, P, Vasey G September 2006 Energy & environmental funds continuing to offer superior opportunities? *Commodities Now* pp 1 – 3.

2127. Holmes D 2006 A financial feast: A-la-carte commodity investing *Alchemy* issue **43** *The London Bullion Market Association* pp 10 – 12.

2128. McNee A July 2006 Investors slake commodities thirst with structured products *The Banker* pp 40 – 42.

2129. Upperman F 2006 Positions of traders *Wiley* Hoboken NJ USA.

2130. Blanchard O J, Gali J 2007 The macroeconomic effects of oil shocks: Why are the 2000s so different from the 1970s? *NBER Working Paper 13368* National Bureau of Economic Research USA.

2131. Demidova-Menzel N; Heidorn Th August 2007 Commodities in asset management Working *Paper Series Frankfurt School of Finance and Management no 81* Frankfurt Germany pp 1 – 67

http://nbn-resolving.de/urn:nbn:de:101:1-20080827282 ,

http://hdl.handle.net/10419/27848 ,

www.frankfurt-school.de .

2132. Domanski D, Heath A March 2007 Financial investors and commodity markets *BIS Quarterly Review* pp 53 – 67.

2133. Kat H M, Oomen R C A 2007 What every investor should know about commodities Part II: Univariate return analysis *Journal of Investment Management* **5** (1) pp 1 – 25.

2134. Kat H M, Oomen R C A 2007 What every investor should know about commodities Part II: Multivariate return analysis *Journal of Investment Management* **5** (3) pp 16 – 40.

2135. Miffre J, Rallis G 2007 Momentum strategies in commodity futures markets *Journal of Banking and Finance* **31** (6) pp 1863 – 1886.

2136. Chong J, Miffre J 2010 Conditional correlation and volatility in commodity futures and traditional asset markets *Journal of Alternative Investments* **12** (3) pp 61 – 75.

2137. Basu D, Miffre J 2013 Capturing the risk premium of commodity futures: The role of hedging pressure *Journal of Banking and Finance* **37** pp 2652 – 2664.

2138. Miffre J, Brooks Ch 2013 Do long-short speculators destabilize commodity futures markets? *EDHEC Business School* 393 Promenade des Anglais 06 202 Nice France pp 1 – 30.

2139. Röthig A, Chiarella C 2007 Investigating nonlinear speculation in cattle, corn, and hog futures markets using logistic smooth transition regression models *Journal of Futures Markets* **27** pp 719 – 737.

2140. Worthington A C, Pahlavani M 2007 Gold investment as an inflationary hedge: Cointegration evidence with allowance for endogenous structural breaks *Applied Financial Economics Letters* vol **3** no 4 pp 259 – 262.

2141. Adams Z, Füss R, Kaiser D 2008 Macroeconomic determinants of commodity futures returns *in* Handbook of Commodity Investment Fabozzi F J, Füss R, Kaiser D (editors) *John Wiley & Sons Inc* Hoboken USA.

2142. Bachmeier L, Qi Li, Liu D 2008 Should oil prices receive so much attention? An evaluation of the predictive power of oil prices for the U.S. economy *Economic Inquiry* **46** pp 528 – 539.

2143. Bhardwaj G, Gorton G, Rouwenhorst K G 2008 Fooling some of the people all of the time: The inefficient performance of commodity trading advisors *Yale ICF Working paper 08-21* Yale University USA.

2144. Büyükşahin B, Haigh M, Harris J, Overdahl J, Robe M 2008 Fundamentals, trader activity and derivative pricing *EFA 2009 Bergen Meetings Paper*, Social Sciences Research Network NY USA
http://papers.ssrn.com/sol3/papers.cfm?abstract_id=966692.

2145. Büyükşahin B, Haigh M, Robe M 2010 Commodities and equities: Ever a "market of one"? *Journal of Alternative Investments* **12** (3) pp 76 – 95.

2146. Büyükşahin B, Robe M A 2010 Speculators, commodities and cross-market linkages Social Sciences Research Network NY USA
http://ssrn.com/abstract=1707103.

2147. Büyükşahin B, Brunetti C, Harris J 2010 Is speculation destabilizing? *Commodity Futures Trading Commission Working Paper*.

2148. Büyükşahin B, Robe M A 2011 Does 'paper oil' matter? Energy markets' financialization and equity-commodity co-movements Social Sciences Research Network NY USA
http://ssrn.com/abstract=1855264.

2149. Büyükşahin B, Harris J H 2011 Do speculators drive crude oil futures prices? *Energy Journal* **32** (2) pp 167 – 202.

2150. Büyükşahin B, Robe M 2014 Speculation, commodities and cross-market linkages *Journal of International Money and Finance* **42** pp 38 – 70.

2151. Caballero R J, Farhi E, Gourinchas P O 2008 Financial crash, commodity prices, and global imbalances *Brookings Papers on Economic Activity Fall* pp 1 – 55.

2152. Carney M 2008 Capitalizing on the commodity boom – The role of monetary policy *Haskayne School of Business* Calgary Alberta
http://www.bis.org/review/r080620d.pdf .

2153. Fabozzi F J, Füss R, Kaiser D G 2008 A primer on commodity investing *in* The Handbook of Commodity Investing Fabozzi F J, Füss R, Kaiser DG (editors) *Wiley* Hoboken NJ USA pp 3 – 37.

2154. Johnson Matthey Plc 2008 Platinum 2008 *Johnson Matthey* Royston, UK.

2155. Khan S A, Khoker Z I, Simin T T 2008 Expected commodity futures returns Social Sciences Research Network NY USA

http://ssrn.com/abstract=1107377

2156. Kilian L 2008a The economic effects of energy price shocks *Journal of Economic Literature* **46** (4) pp 871 – 909.

2157. Kilian L 2008b Exogenous oil supply shocks: How big are they and how much do they matter for the U.S. economy? *Review of Economics and Statistics* **90** pp 216 – 240.

2158. Kilian L 2009 Not all oil price shocks are alike: Disentangling demand and supply shocks in the crude oil market *American Economic Review* **99** pp 1053 – 1069.

2159. Kilian L, Park Ch 2009 The impact of oil price shocks on the U.S. stock market *International Economic Review* **50** (4) pp 1267 – 1287.

2160. Kilian L 2010 Explaining fluctuations in gasoline prices: A joint model of the global crude oil market and the U.S. retail gasoline market *The Energy Journal* **31** (2) pp 87 – 104.

2161. Kilian L, Vega C 2011 Do energy prices respond to U.S. macroeconomic news? A test of the hypothesis of predetermined energy prices *Review of Economics and Statistics* **93** (2) pp 660 – 671.

2162. Kilian L, Vigfusson R J 2011 Nonlinearities in the oil price-output relationship *Macroeconomic Dynamics* **15** (3) pp 337 – 363.

2163. Kilian L, Murphy D 2013 The role of inventories and speculative trading in the global market for crude oil *Journal of Applied Econometrics*.

2164. Kilian L, Hicks B 2013 Did unexpectedly strong economic growth cause the oil price shock of 2003-2008? *Journal of Forecasting* **32** (5) pp 385 – 394.

2165. Kilian L 2014 Oil price shocks: Causes and consequences *Annual Review of Resource Economics* 6 pp 133 – 154.

2166. Kilian L, Murphy D P 2014 The role of inventories and speculative trading in the global market for crude oil *Journal of Applied Econometrics* **29** pp 454 – 478.

2167. Kolos S P, Ronn E I March 2008 Estimating the commodity market price of risk for energy prices *Energy Economics* vol **30** issue 2 pp 621 – 641.

2168. Reisen H July 18 2008 How to spend it: Commodity and non-commodity sovereign wealth funds *Policy Brief no 38* OECD Development Centre pp 1 – 22, *Research Notes Working Paper Series Deutsche Bank Research no 28* Frankfurt am Main Germany

www.dbresearch.com .

2169. Roache S K September 2008 Commodities and the market price of risk *IMF Working Paper WP/08/221* International Monetary Fund NY USA pp 1 – 25.

2170. Roache S K, Rossi M July 2009 The effects of economic news on commodity prices: Is gold just another commodity? *IMF Working Paper 140* International Monetary Fund USA.

2171. Scherer B, He L 2008 The diversification benefits of commodity futures indexes: A mean-variance spanning test *in* The Handbook of Commodity Investing Fabozzi F J, Füss R, Kaiser D G (editors) *Wiley* Hoboken NJ USA pp 241 – 265.

2172. Woodward J D 2008 Commodity futures investments: A review of strategic motivations and tactical opportunities *in* The Handbook of Commodity Investing Fabozzi F J, Füss R, Kaiser D G (editors) *Wiley* Hoboken NJ USA pp 56 – 86.

2173. Attié A P, Roache1 S K April 2009 Inflation hedging for long-term investors *IMF Working Paper WP/09/90* International Monetary Fund USA pp 1 – 39.

2174. Bose S 2009 The role of futures market in aggravating commodity price inflation and the future of commodity futures in India *Money and Finance ICRA Bulletin* http://www.icra.in/Files/MoneyFinance/4.%20Suchismita%20Bose.pdf .

2175. Du X, Yu C L, Hayes D J 2009 Speculation and volatility spillover in the crude oil and agricultural commodity markets: A Bayesian analysis *Working Paper no 09* Iowa State University USA.

2176. Frankel J A, Rose A K 2009 Determinants of agricultural and mineral commodity prices *in* Inflation in an era of relative price shocks Fry R, Jones C, Kent C (editors) *Proceedings of Conference* Reserve Bank of Australia Sydney Australia pp 9 – 43.

2177. Kaufmann R K, Ullman B 2009 Oil prices, speculation, and fundamentals: Interpreting causal relations among spot and futures prices *Energy Economics* **31** pp 550 – 558.

2178. Korniotis G 2009 Does speculation affect spot price levels? The case of metals with and without futures markets *Working Paper Federal Reserve Board* Washington DC USA.

2179. Mayer J 2009 The growing interdependence between financial and commodity markets UNCTAD Discussion Paper 195 Geneva Switzerland.

2180. Mayer J 2012 The growing financialization of commodity markets: Divergences between index investors and money managers *Journal of Development Studies* **48** (6) pp 751 – 767.

2181. Reitz S, Slopek U 2009 Non-linear oil price dynamics: A tale of heterogeneous speculators? *German Economic Review* **10** pp 270 – 283.

2182. Smith J 2009 World oil: Market or mayhem? *Journal of Economic Perspectives* **23** pp 145 – 164.

2183. Smith J, Thompson R, Lee Th 2013 The informational role of spot prices and inventories *Working Paper* Southern Methodist University USA.

2184. Yamori N September 7 2009 Characteristics of Japan's commodities index and its correlation with stock index *Nagoya University* Japan, *MPRA Paper no 17160* Munich University Germany, *Journal of Applied Research in Finance* **I** (2) pp 187 – 192. http://mpra.ub.uni-muenchen.de/17160/ .

2185. Yamori N May 20 2011 Commodity ETFs in the Japanese stock exchanges *Nagoya University* Japan, *MPRA Paper no 31003* Munich University Munich Germany https://mpra.ub.uni-muenchen.de/31003/ .

2186. Yamori N 2011 Co-movement between commodity market and equity market: Does commodity market change? *Modern Economy*.

2187. Yung K, Liu Y C 2009 Implications of futures trading volume: Hedgers versus speculators *Journal of Asset Management* **10** pp 318 – 337.

2188. Alquist R, Kilian L 2010 What do we learn from the price of crude oil futures? *Journal of Applied Econometrics* **25** pp 539 – 573.

2189. Alquist R, Kilian L, Vigfusson R J 2013 Forecasting the price of oil *in* Handbook of Economic Forecasting Elliott G, Timmermann A (editors) **2** pp 427 – 507.

2190. Anzuini A, Lombardi M J, Pagano P 2010 The impact of monetary policy shocks on commodity prices *European Central Bank Working Paper no 1232.*

2191. Aulerich N M, Irwin S H, Garcia P 2010 The price impact of index funds in commodity futures markets: Evidence from the CFTC's daily large trader reporting system *Working Paper Department of Agricultural and Consumer Economics* University of Illinois at Urbana- Champaign USA.

2192. Baffes J, Haniotis T July 2010 Placing the 2006/08 commodity price boom into perspective *Policy Research Working Paper 5371* The World Bank Development Prospects Group World Bank USA.

2193. Baker St 2012 The financialization of storable commodities *Working Paper* Carnegie Mellon University NY USA.

2194. Baker St, Routledge B 2012 The price of oil risk *Working Paper* Carnegie Mellon University NY USA.

2195. Basak S, Pavlova A 2012 A model of financialization of commodities *Working Paper* London Business School London UK.

2196. Blanchard O, Gali J 2010 The macroeconomic effects of oil shocks: Why are the 2000s so different from the 1970s? *in* International Dimensions of Monetary Policy Gali J, Gertler M (editors) University of Chicago Press Chicago USA.

2197. Calvo-Gonzalez O, Shankar R, Trezzi R 2010 Are commodity prices more volatile now: a long-run perspective *World Bank Policy Research Working Paper 5460* World Bank NY USA.

2198. Cifarelli G, Paladino G 2010 Oil price dynamics and speculation: A multivariate financial approach *Energy Economics* **32** pp 363 – 372.

2199. Fuertes A-M, Miffre J, Rallis G 2010 Tactical allocation in commodity futures markets: Combining momentum and term structure signals *Journal of Banking and Finance* **34** (10) pp 2530 – 2548.

2200. Elder J, Serletis A 2010 Oil price uncertainty *Journal of Money, Credit and Banking* **42** pp 1137 – 1159.

2201. Etula E 2010 Broker-dealer risk appetite and commodity returns *Working Paper* Federal Reserve Bank of New York NY USA.

2202. Gilbert C L March 2010 Speculative influences on commodity futures prices 2006-08 *UNCTAD Discussion Paper no 197* United Nations Conference on Trade and Development NY USA pp 1 – 40.

2203. Gilbert C L 2010 How to understand high food prices *Journal of Agricultural Economics* **61** (2) pp 398 – 425.

2204. Hailu G, Weersink A 2010 Commodity price volatility: The impact of commodity index traders *CATPRN Commissioned Paper 2010-02.*

2205. Hernandez M, Torero M 2010 Examining the dynamic relationship between spot and future prices of agricultural commodities *IFPRI Discussion Paper no 988*.

2206. Nissanke M December 2010 Commodity markets and excess volatility: Sources and strategies to reduce adverse development impacts *CFC Conference* University of London Brussels Belgium.

2207. Silvennoinen A, Thorp S 2010 Financialization, crisis, and commodity correlation dynamics *Working Paper no 267* Quantitative Finance Research Centre Sydney University of Technology Sydney Australia.

2208. Tang K, Xiong W 2010 Index investment and financialization of commodities *NBER Working Paper no 16325* National Bureau of Economic Research 1050 Massachusetts Avenue Cambridge MA 02138 USA, *Financial Analysts Journal* **68** pp 54 – 74.

2209. Tang K, Wei Xiong 2012 Index investment and financialization of commodities *Financial Analysts Journal* **68** (6) pp 54 – 74.

2210. Tang K, Zhu H 2015 Commodities as collateral *MIT Sloan Working Paper* MIT USA.

2211. Wong T, Smith A July 25 - 27 2010 Commodity markets: Rational expectations in markets with irrational investors *Selected Paper prepared for presentation at the Agricultural and Applied Economics Associations 2010 AAEA, CAES and WAEA Joint Annual Meeting* Denver Colorado USA pp 1 – 12.

2212. Basu P, Gavin W T 2011 What explains the growth in commodity derivatives? *Federal Reserve Bank of St Louis Review* vol **93** no 1 pp 37 – 48.

2213. Cárdenas M, Ramírez S, Tuzemen D December 2011 Commodity dependence and fiscal capacity *RWP 11-08* Federal Reserve Bank of Kansas City USA.

2214. Connolly E, Orsmond D 2011 The mining industry: From bust to boom *RBA Research Discussion Paper no 2011-08*.

2215. Daskalaki C, Skiadopoulos G 2011 Should investors include commodities in their portfolios after all? *Journal of Banking and Finance* **35** (10) pp 2606 – 2626.

2216. Dwyer A, Gardner G, Williams T 2011 Global commodity markets – price volatility and financialization *RBA Bulletin June* pp 49 – 57.

2217. Dwyer A, Holloway J, Wright M 2012 Commodity market financialization: A closer look at the evidence *Reserve Bank of Australia Bulletin March Quarter 2012* Reserve Bank of Australia Canberra Australia pp 65 – 78.

2218. Fattouh B 2011 An anatomy of the crude oil pricing system *The Oxford Institute for Energy Studies no 40* Oxford UK.

2219. Ghosh J 2011 Implications of regulating commodity derivatives markets in the USA and EU *PSL Quarterly Review* vol **64** no 258 pp 287 – 304.

2220. G20 Study Group on Commodities November 2011 Report of the G20 study group on commodities under the chairmanship of Mr Hiroshi Nakaso *Banque de France* France
http://www.banque-france.fr/fileadmin/user_upload/banque_de_france/Economie_et_Statistiques/Tendances_Regionales__ne_pas_ecraser_/mois_impairs/Study_group_report_final.pdf .

2221. Hong H, Yogo M 2011 What does futures market interest tell us about the macroeconomy and asset prices? *Journal of Financial Economics* **105** (3) pp 473 – 490.

2222. Inamura Y, Kimata T, Kimura T, Muto T March 2011 Recent surge in global commodity prices – Impact of financialization of commodities and globally accommodative monetary conditions *Bank of Japan Review* Tokyo Japan.

2223. Lombardi M J, Robays I V 2011 Do financial investors destabilize the oil price? *Working Paper European Central Bank* Belgium.

2224. Mou Y 2011 Limits to arbitrage and commodity index investment: Front-running the Goldman roll *Working Paper* Columbia University NY USA.

2225. Peñaranda F, Micola A 2011 On the drivers of commodity co-movement: Evidence from biofuels *Working Paper* Universitat Pompeu Fabra.

2226. Reichsfeld D A, Roache S K 2011, Do commodity futures help forecast spot prices? *IMF Working Paper WP/11/254* IMF USA.

2227. Tilton J E, Humphreys D, Radetzki M 2011 Investor demand and spot commodity prices *Resources Policy* **36** pp 187 – 195.

2228. Yiuman Tse, Williams M December 20 2011 Does index speculation impact commodity prices? An intraday futures analysis *WP # 007FIN-257-2011* Department of Finance University of Texas San Antonio Texas USA pp 1 – 44.

2229. UNCTAD 2011 Price formation in financialized commodity markets http://www.unctad.org/en/docs/gds20111_en.pdf .

2230. Arbatli E, Vasishtha G 2012 Growth in emerging market economies and the commodity boom of 2003–2008: Evidence from growth forecast revisions *Bank of Canada Working Paper no 2012-8* Ottawa Canada.

2231. Cespedes L F, Velasco A 2012 Macroeconomic performance during commodity price booms and busts *NBER Working Paper 18569* National Bureau of Economic Research Cambridge MA USA

2232. Fattouh B, Kilian L, Mahadeva L 2012 The role of speculation in oil markets: What have we learned so far? *Working Paper* University of Michigan USA.

2233. Gasmi F, Oviedo J D October 2012 Investment in transport capacity and regulation of regional monopolies in natural gas commodity markets *Serie Documentos De Trabajo no 125* Universidad del Rosario Columbia pp 1 – 54.

2234. Girardi D January 2 2012 Do financial investors affect commodity prices? The case of Hard Red Winter wheat *University of Siena* Italy, *MPRA Paper no 35670* Munich University Germany pp 1 – 39

https://mpra.ub.uni-muenchen.de/35670/ .

2235. Girardi D October 9 2012 Do financial investors affect the price of wheat? *CRESME* University of Siena Italy, *MPRA Paper no 40285* Munich University Munich Germany pp 1 – 32

https://mpra.ub.uni-muenchen.de/40285/ .

2236. Gruber J, Vigfusson R 2012 Interest rates and the volatility and correlation of commodity prices *Working Paper* Federal Reserve Board USA.

2237. Henderson B, Pearson N, Li Wang 2012 New evidence on the financialization of commodity markets *Working Paper* University of Illinois USA.

2238. Hong H, Yogo M 2012 What does futures market interest tell us about the macroeconomy and asset prices? *Journal of Financial Economics* **105** pp 473 – 490.

2239. Juvenal L, Petrella I 2012 Speculation in oil market *Working Paper* Federal Reserve Bank of Saint Loius USA.

2240. Nissanke M 2012 Commodity markets linkages in the global financial crisis: Excess volatility and development impacts *Journal of Development Studies* **48** (6) pp 732 – 750.

2241. Rouwenhorst G K, Ke Tang 2012 Commodity investing *Annual Review of Financial Economics* **4** pp 447 – 467.

2242. Sockin M, Wei Xiong 2012 Informational frictions and commodity markets *Working Paper* Princeton University Princeton USA.

2243. Sockin M, Wei Xiong 2013 Feedback effects of commodity futures prices *NBER Working Paper 18906* National Bureau of Economic Research 1050 Massachusetts Avenue Cambridge MA 02138 USA.

2244. Varadi V K March 2012 An evidence of speculation in Indian commodity markets *MPRA Paper no 38337* Munich University Munich Germany pp 1 – 24 https://mpra.ub.uni-muenchen.de/38337/ .

2245. Acharya V, Lochstoer L, Ramadorai T 2013 Limits to arbitrage and hedging: Evidence from commodity markets *Journal of Financial Economics* **109** pp 441 – 465.

2246. Anghelache G V, Dincă Z, Sacală C, Lixandru D 2013 The main indicators used in the analysis of commodity exchanges *Academy of Economic Studies* Bucharest Romania pp 1 – 6.

2247. Aulerich N M, Irwin S H, Garcia Ph 2013 Bubbles, food prices, and speculation: Evidence from the CFTC's daily large trader data files *NBER Working Paper 19065* National Bureau of Economic Research 1050 Massachusetts Avenue Cambridge MA 02138 USA.

2248. Baumeister Ch, Peersman G 2013 The role of time-varying price elasticities in accounting for volatility changes in the crude oil market *Journal of Applied Econometrics* **28** (7) pp 1087 – 1109.

2249. Baumeister Ch, Kilian L 2014 Do oil price increases cause higher food prices? *Economic Policy* **80** pp 691 – 747.

2250. Baumeister Ch, Kilian L 2016 Understanding the decline in the price of oil since June 2014 *Journal of the Association of Environmental and Resource Economists* (forthcoming).

2251. Chevallier J, Ielpo F, Ling-Ni Boon 2013 Common risk factors in commodities *Economics Bulletin* **33** (4) pp 2801 – 2916.

2252. Creti A, Joêts M, Mignon V 2013 On the links between stock and commodity markets volatility *Energy Economics* **37** pp 16 – 28.

2253. Fattouh B, Kilian L, Mahadeva L 2013 The role of speculation in oil markets: What have we learned so far? *The Energy Journal* **34** (3) pp 7 – 33.

2254. Heumesser Ch, Staritz C October 2013 Financialization and the microstructure of commodity markets - a qualitative investigation of trading strategies of financial investors and commercial traders *Working Paper Austrian Foundation for Development Research (ÖFSE) no 44* Vienna Sensengasse 3 Austria pp 1 – 59

http://hdl.handle.net/10419/98809 ,

http://www.centrum3.at ,

http://www.eza.at .

*2255.* Hurduzeu G, Hurduzeu R 2013 International diversification of the asset portfolio by investing in agricultural commodities. why not use the CAPM futures markets? *135 EAAE Seminar Challenges for the Global Agricultural Trade Regime after Doha* pp 183 – 190.

*2256.* Ing-Haw Cheng, Wei Xiong 2013 Why do hedgers trade so much? *Working Paper* Dartmouth College USA.

*2257.* Ing-Haw Cheng, Wei Xiong November 2013, 2014 The financialization of commodity markets *Working Paper* Princeton University Princeton USA, *Working Paper 19642* National Bureau of Economic Research 1050 Massachusetts Avenue Cambridge MA 02138 USA pp 1 – 37, *Annual Review of Financial Economics* **6** (1) pp 419 – 441.

http://www.nber.org/papers/w19642 .

*2258.* Ing-Haw Cheng, Kirilenko A, Wei Xiong 2014 Convective risk flows in commodity futures markets *Working Paper* Dartmouth College USA, *Review of Finance* **1** pp 1 – 49.

*2259.* Jovanovic B 2013 Bubbles in prices of exhaustible resources *International Economic Review* **54** pp 1 – 34.

*2260.* Kang W, Rouwenhorst G K, Ke Tang 2013 The role of hedgers and speculators in liquidity provision to commodity futures markets *Working Paper* Yale University USA.

*2261.* Le Pen Y, Sévi B 2013 Futures trading and the excess co-movement of commodity prices *SSRN Working Paper #2191659* Social Sciences Research Network NY USA.

*2262.* Marshall B, Nguyen N, Visaltanachoti N 2013 Liquidity commonality in commodities *Journal of Banking and Finance* **37** (1) pp 11 – 20.

*2263.* Grimes A, Hyland S Passing the buck: Impacts of commodity price shocks on local outcomes *Motu Working Paper 13-10* Motu Economic and Public Policy Research Wellington New Zealand ISSN 1176-2667 pp 1 – 46

www.motu.org.nz .

*2264.* Santos J M 2013 Trading grain now and then: the relative performance of early grain futures markets *Applied Economics* **45** (3) pp 287 – 298.

*2265.* Van Robays I 2013 Macroeconomic uncertainty and the impacts of oil shocks *ECB Working Paper 1479* European Central Bank.

*2266.* Venditti F 2013 From oil to consumer energy prices: how much asymmetry along the way? *Energy Economics* **40** pp 468 – 473.

*2267.* Adams Z, Glück T June 2014, August 2015 Financialization in commodity markets: A passing trend or the new normal? *Working Papers on Finance no 2014/13* Swiss Institute of Banking and Finance (S/BF – HSG) University of St Gallen Rosenbergstrasse 52 9000 St Gallen Switzerland pp 1 – 52, *Journal of Banking and Finance* **60** pp 93 – 111.

*2268.* Basak S, Pavlova A 2014 A model of financialization of commodities *Working Paper London Business School* London UK.

2269. Daskalakis C, Kostakis A, Skiadopoulos G 2014 Are there Common Factors in Commodity Futures Returns? *Journal of Banking and Finance* **40** pp 346 – 363.

2270. Fornero J, Kirchner M, Yany A 2014 Terms of trade shocks and investment in commodity-exporting economies *Central Bank of Chile Working Paper* Central Bank of Chile Santiago Chile.

2271. Gruss B 2014 After the boom–commodity prices and economic growth in Latin America and the Caribbean *IMF Working Paper 14/154* International Monetary Fund Washington USA.

2272. Jo S 2014 The effect of oil price uncertainty on global real economic activity *Journal of Money, Credit, and Banking* **46** (6) pp 1113 – 1135.

2273. Mykhailovska O 2014 Fractal characteristics of world market commodity derivatives pp 36 – 48.

2274. Sadorsky P 2014 Modeling volatility and correlations between emerging market stock prices and the prices of copper, oil and wheat *Energy Economics* **43** pp 72 – 81.

2275. Bonato M, Taschini L November 2015 Comovement and the financialization of commodities *Centre for Climate Change Economics and Policy Working Paper no 240, Grantham Research Institute on Climate Change and the Environment Working Paper no 215* UK pp 1 – 31.

2276. Henderson B J, N D Pearson, Wang L 2015 New evidence on the financialization of commodity markets *Review of Financial Studies* **28** (5) pp 1287 – 1311.

2277. Novotný J 2015 Commodity investment model and its significance for investors *International Scientific Forum (ISF)* European Scientific Institute University of Oxford UK pp 68 – 78 ISBN: 978-608-4642-42-8.

2278. Novotný J, Polach J 2016 Real silver and its investment and business options *International Journal of Entrepreneurial Knowledge* vol **4** issue 1/2016 pp 40 – 51 DOI: 10.1515/ijek-2016-0003.

2279. Joëts M, Mignon V, Razafindrabe T November 2016 Does the volatility of commodity prices reflect macroeconomic uncertainty? *Banque de France* Paris France pp 1 – 61 www.banque-france.fr .

2280. Babbage Ch 1832 On the economy of machinery and manufacturers *Charles Knight* 13 Pall Mall East London UK.

***Precious metal investment, precious metal valuation, precious metals exchange, financial capital investment product, financial capital investment medium in finances:***

2281. Hourwich I A 1902 The production and consumption of the precious metals *Journal of Political Economy* **10** (4) pp 575 – 610

http://dx.doi.org/10.1086/250881 ,

http://www.journals.uchicago.edu/doi/10.1086/250881 .

2282. Hourwich I A 1903 The production and consumption of the precious metals: II *Journal of Political Economy* **11** (4) pp 503 – 539

http://dx.doi.org/10.1086/250987 ,

http://www.journals.uchicago.edu/doi/10.1086/250987 .

2283. Goodman B 1956 The price of gold and international liquidity *Journal of Finance* **11** pp 15 – 28.

2284. Tschoegl A E 1980 Efficiency in the gold market *Journal of Banking and Finance* **4** pp 371 – 379.

2285. Solt M E, Swanson P J 1981 On the efficiency of the markets for gold and silver *Journal of Business* **54** pp 453 – 478.

2286. Burke W June 4 1982 Not-so-precious metal *FRBSF Economic Letter* issue **jun4** pp 1 – 4

https://fraser.stlouisfed.org/scribd/?item_id=517473&filepath=/files/docs/historical/frbsf/frbsf_let/frbsf_let_19820604.pdf&start_page=1 .

2287. Mate M E 1984 Precious metals in the later medieval and early modern worlds *Business History Review* **58** (3) pp 457 – 458.

2288. Ho Y-K 1985 Test of the incrementally efficient market hypothesis for the London gold market *Economics Letters* **19** pp 67 – 70.

2289. Aggarwal R, Sundararaghavan P 1987 Efficiency of the silver futures market: An empirical study using daily data *Journal of Banking and Finance* **11** (1) pp 49 – 64.

2290. Aggarwal R, Soenen L 1988 Nature and efficiency of the gold market *Journal of Portfolio Management* **14** (3) pp 18 – 21.

2291. Aggarwal R, Mohanty S, Song F 1995 Are survey forecasts of macroeconomic variables rational? *Journal of Business* **68** (1) pp 99 – 119.

2292. Aggarwal R 2004 Persistent puzzles in international finance and economics *The Economic and Social Review* **35** (3) pp 241 – 250.

2293. Aggarwal R, Lucey B 2007 Psychological barriers in gold prices? *Review of Financial Economics* **16** pp 217 – 230.

2294. Aggarwal R, Zong S 2008 Behavioral biases in forward rates as forecasts of future exchange rates: Evidence of systematic pessimism and under-reaction *Multinational Finance Journal* **12** (3) pp 241 – 277.

2295. Aggarwal R, Lucey B, O'Connor F November 2014 Rationality in precious metals forward markets: Evidence of behavioural deviations in the gold markets *Discussion Paper no 462* The Institute for International Integration Studies pp 1 – 32

http://www.tcd.ie/iiis/documents/discussion/pdfs/iiisdp462.pdf .

2296. Aggarwal R, Lucey B, O'Connor F 2015 Chapter 10 *in* The World Scientific Handbook of Futures Markets *World Scientific Publishing Co Pte Ltd* ISBN: 978-981-4566-91-9 pp 325 – 347

http://dx.doi.org/10.1142/9789814566926_0010 .

2297. Crowson P C F 1987 Metal prices and exchange rates *Resources Policy* **13** (3) pp 250 – 251.

2298. Anikin A V 1988 Gold: International economical aspect 2[nd] edition *International Relationships* Moscow Russian Federation.

2299. Fama E F, French K R 1988 Business cycles and the behaviour of metal prices *Journal of Finance* **43** pp 1075 – 1093.

2300. Luke Chan M W, Mountain D 1988 The interactive and causal relationships involving precious metal *Journal of Business & Economic Statistics* **6** (1) pp 69 – 77.

2301. Frank M, Stengos Th 1989 Nearest neighbor forecasts of precious metal rates of return *Working Paper* Department of Economics and Finance University of Guelph Canada.

*1.* Jaffee J F 1989 Gold and gold stocks as investments for institutional portfolios *Financial Analysts Journal* **45** pp 53 – 60.

*2302.* Kaufmann T D, Winters R A 1989 The price of gold: A simple model *Resources Policy* **15** pp 309 – 313.

*2303.* Radetzki M 1989 Precious metals: The fundamental determinants of their price behaviour *Resources Policy* **15** (3) pp 194 – 208.

*2304.* Sephton P, Cochrane D K 1990 A note on the efficiency of the London metal exchange *Economics Letters* **33** (4) pp 341 – 345.

*2305.* Akgiray V, Booth G, Hatem J C M August 1991 Conditional dependence in precious metal prices *The Financial Review* **26** (3) pp 367 – 386.

*2306.* Agbeyegbe T D 1992 Common stochastic trends: Evidence from the London Metal Exchange *Bulletin of Economic Research* **44** (2) pp 141 – 151.

*2307.* Cheung Y-W, Lai K S May 1993 Do gold market returns have long memory? *The Financial Review* **28** (2) pp 181 – 202.

*2308.* Chaudhuri K N 1994 Precious metals and mining in the New World: 1500–1800 *European Review* **2** (4) pp 261 – 270.

*2309.* Brunetti C, Gilbert C L 1995 Metals price volatility 1972–95 *Resources Policy* **21** pp 237 – 254.

*2310.* Moore M, Cullen U 1995 Speculative efficiency on the London metal exchange *The Manchester School of Economic and Social Studies* **63** (3) pp 235 – 256.

*2311.* Qiang Y, Weber E J 1995 Forecasting world metal prices *Discussion Paper 95.17* Department of Economics University of Western Australia.

*2312.* Qiang Y 1998 World metal prices: A database *Discussion Paper 98.03* Department of Economics University of Western Australia.

*2313.* Wahab M 1995 Conditional dynamics and optimal spreading in the precious metals futures markets *Journal of Futures Markets* **15** (2) pp 131 – 166.

*2314.* Sjaastad L A, Scacciavillani F December 1996 The price of gold and the exchange rate *Journal of International Money and Finance* **15** (6) pp 879 – 897.

*2315.* Sjaastad L A June 2008 The price of gold and the exchange rates: Once again *Resources Policy* **33** (2) pp 118 – 124.

*2316.* Escribano A, Granger C 1998 Investigating the relationship between gold and silver prices *Journal of Forecasting* 17 pp 81 – 107.

*2317.* Taylor N 1998 Precious metals and inflation *Applied Financial Economics* **8** (2) pp 201 – 210.

*2318.* Labys W C, Achouch A, Terraza M 1999 Metal prices and the business cycle *Resources Policy* **25** (4) pp 229 – 238.

*2319.* Rockerbie D W 1999 Gold prices and gold production: Evidence for South Africa *Resources Policy* **25** pp 69 – 76.

*2320.* Christie-David R, Chaudhry M, Koch T W 2000 Do macroeconomics news releases affect gold and silver prices? *Journal of Economics and Business* **52** (5) pp 405 – 421.

*2321.* Cai J, Cheung Y-L, Wong M C S 2001 What moves the gold market? *Journal of Futures Markets* **21** (3) pp 257 – 278.

*2322.* Ciner C 2001 On the long-run relationship between gold and silver: A note *Global Finance Journal* **12** (2) pp 299 – 303.

2323. Ciner C, Gurdgiev C, Lucey B M 2013 Hedges and safe havens: An examination of stocks, bonds, gold, oil and exchange rates *International Review of Financial Analysis* **29** pp 202 – 211.

2324. Hammoudeh Sh, Malik F, McAleer M 2001 Risk management of precious metals The *Quarterly Review of Economics and Finance* **51** pp 435 – 441.

2325. Hammoudeh Sh, Yuan Y 2008 Metal volatility in presence of oil and interest rate shocks *Energy Economics* **30** (2) pp 606 – 620.

2326. Hammoudeh Sh, Yuan Y, McAleer M, Thompson M 2009, 2010 Precious metals-exchange rate volatility transmissions and hedging strategies *Econometric Institute Research Paper no EI 2009-38* Erasmus School of Economics (ESE) Econometric Institute Erasmus University Rotterdam The Netherlands pp 1 – 43, *International Review of Economics & Finance* **19** (4) pp 633 – 647

http://repub.eur.nl/pub/17308/EI2009-38.pdf .

2327. Hammoudeh Sh, Malik F, McAleer M March 2011 Risk management of precious metals *KIER Working Paper no 765* Institute of Economic Research Kyoto University Japan pp 1 – 29, *The Quarterly Review of Economics and Finance* **51** (4) pp 435 – 441

http://www.kier.kyoto-u.ac.jp/DP/DP765.pdf .

2328. Hammoudeh Sh, Santos P A, Al-Hassan A 2013 Downside risk management and VaR-based optimal portfolios for precious metals, oil and stocks *The North American Journal of Economics and Finance* **25** (C) pp 318 – 334.

2329. Mackenzie M, Mitchell H, Brooks R, Faff R 2001 Power ARCH modeling of commodity futures data on the London's metal market *European Journal of Finance* **7** pp 22 – 38.

2330. Adrangi B, Chatrath A April 2002 The dynamics of palladium and platinum prices *Computational Economics* **17** pp 179 – 197.

2331. Smith G 2002 Tests of the random walk hypothesis for London gold prices *Applied Economics Letters* **9** pp 671 – 674.

2332. Cavaletti C, Factor T, All D 2004 Will gold hold its luster? *Futures* **33** pp 24 – 31.

2333. Baron M 2005 The price paid by the Romanian National Bank for the precious metals purchased from the local producers between 1920–1948 *Annals of the University of Petrosani: Economics* **5** pp 21 – 38.

2334. Capie F, Mills T C, Wood G 2005 Gold as a hedge against the dollar *Journal of International Financial Markets, Institution and Money* **15** pp 343 – 352.

2335. Conover C M, Jensen G R, Johnson R R, Mercer J M 2005 Can precious metals make your portfolio shine? *Financial Analysts Journal* **46** pp 76 – 79.

2336. Conover C M, Jensen G R, Johnson R R, Mercer J M 2009 Can precious metals make your portfolio shine? *Journal of Investing* **18** (1) pp 75 – 86.

2337. Drelichman M 2005 All that glitters: Precious metals, rent seeking and the decline of Spain *European Review of Economic History* **9** (3) pp 313 – 336.

2338. Papyrakis E, Gerlagh R 2005 Institutional explanations of economic development: The role of precious metals *Working Paper no 2005.131* Fondazione Eni Enrico Mattei Italy pp 1 – 35

http://www.feem.it/userfiles/attach/Publication/NDL2005/NDL2005-131.pdf .

2339. Pulvermacher K March 2005a What are commodities?

http://www.gold.org/value/stats/research/index.html .

*2340.* Pulvermacher K May 2005b Commodity returns and the economic cycle
http://www.gold.org/value/stats/research/index.html .

*2341.* Xiaoqing Eleanor Xu, Hung-Gay Fung 2005 Cross-market linkages between US and Japanese precious metals futures trading *Journal of International Financial Markets Institutions and Money* **15** (2) pp 107 – 124.

*2342.* Banken R 2006 National socialist plundering of precious metals, 1933-1945: The role of Degussa *Working Paper* Institute of European Studies UC Berkeley USA pp 1 – 22

http://escholarship.org/uc/item/5615g5pm .

*2343.* Draper P, Faff R W, Hillier D 2006 Do precious metals shine? An investment perspective *Financial Analysts Journal* **62** (2) pp 98 – 106.

*2344.* Hillier, Draper D, Robert F 2006 Do precious metals shine? An investment perspective *Financial Analysts Journal* **62** pp 98 – 106.

*2345.* Batten J M, Lucey B M 2007, 2010 Volatility in the gold futures market *Discussion Paper no 225* Institute for International Integration Studies, *Applied Economics Letter* **17** (2) pp 187 – 190.

*2346.* Batten J M, Ciner C, Lucey B M June 2008, 2010 The macroeconomic determinants of volatility in precious metals markets *Discussion Paper no 255* The Institute for International Integration Studies IIIS pp 1 – 26, *Resources Policy* **35** (2) pp 65 – 71
http://www.tcd.ie/iiis/documents/discussion/pdfs/iiisdp255.pdf .

*2347.* Batten J M, Ciner C, Lucey B M 2013 On the economic determinants of the gold-inflation relation *Working Paper*.

*2348.* Batten J, Ciner C, Lucey B M 2015 Which precious metals spill over on which, when and why? Some evidence *Applied Economics Letters* **22** (6) pp 466 – 473.

*2349.* Demidova-Menzel N; Heidorn Th August 2007a Commodities in asset management *Working Paper no 81* Frankfurt School of Finance and Management Frankfurt Germany pp 1 – 67
http://nbn-resolving.de/urn:nbn:de:101:1-20080827282 ,
http://hdl.handle.net/10419/27848 ,
www.frankfurt-school.de .

*2350.* Demidova-Menzel N; Heidorn Th August 2007b Gold in the investment portfolio *Working Paper no 87* Frankfurt School of Finance and Management Frankfurt Germany pp 1 – 50

*2351.* Kyrtsou, Labys 2007 Detecting feedback in multivariate time series: The case of metal prices and US inflation *Physica A* **377** pp 227 – 229.

*2352.* Sari R, Hammoudeh S, Ewing B T 2007 Dynamic relationships between oil and metal commodity futures prices *Geopolitics of Energy* **29** (7) pp 2 – 13.

*2353.* Sari R, Hammoudeh Sh, Soytas U 2010 Dynamics of oil price, precious metal prices, and exchange rate *Energy Economics* **32** (2) pp 351 – 362.

*2354.* Tully E, Lucey B 2007 A power GARCH examination of the gold market *Research in International Business and Finance* **21** (2) pp 316 – 325.

2355. Worthington A C, Pahlavani M 2007 Gold investment as an inflationary hedge: Cointegration evidence with allowance for endogenous structural breaks *Applied Financial Economics Letters* vol **3** no 4 pp 259 – 262.

2356. Jerrett D, Cuddington J T 2008 Broadening the statistical search for metal price super cycles to steel and related metals *Resources Policy* **33** (4) pp 188 – 195.

2357. LBMA (London Bullion Market Association) 2008 A guide to the London precious metals markets *The London Bullion Market Association*, *The London Platinum and Palladium Market*.

2358. Watkins C, McAleer M 2008 How has the volatility in metals markets changed? *Mathematics and Computers in Simulation* **78** pp 237 – 249.

2359. Roberts M C 2009 Duration and characteristics of metal price cycles *Resources Policy* **34** (3) pp 87 – 102.

2360. Soytas U, Sari R, Hammoudeh Sh, Hacihasanoglu E 2009 World oil prices, precious metal prices and macroeconomy in Turkey *Energy Policy* **37** (12) pp 5557 – 5566.

2361. Baur D G, Lucey B M 2010 Is gold a hedge or a safe heaven? An analysis of stocks, bonds and gold *The Financial Review* **45** (2) pp 217 – 229.

2362. Baur D G, Mcdermott T K 2010 Is gold a safe haven? International evidence *Journal of Banking and Finance* **34** pp 1886 – 1898.

2363. Baur D G 2012 Asymmetric volatility in the gold market *The Journal of Alternative Investments* **14** (4) pp 26 – 38.

2364. Baur D G, Tran D T 2014 The long-run relationship of gold and silver and the influence of bubbles and financial crisis *Empirical Economics* DOI: 10.1007/s00181-013-0787-1.

2365. Chen M-H September 2010 Understanding world metals prices–returns, volatility and diversification *Resources Policy* **35** (3) pp 127 – 140.

2366. Humphreys D 2010 The great metals boom: A retrospective *Resources Policy* **35** (1) pp 1 – 13.

2367. Kovalenko S A 2010 Development operation with precious metal and their role in stabilizations of the economy *Вісник Економіки Транспорту і Промисловості* **31** pp 102 – 104
http://cyberleninka.ru/article/n/development-operation-with-precious-metal-and-their-role-in-stabilizations-of-the-economy .

2368. Lucey B M 2010 Lunar seasonality in precious metal returns? *Applied Economics Letters* **17** (9) pp 835 – 838.

2369. Lucey B M, Larkin C, O'Connor F 2013 New York or London: Where and when does the gold price originate? *Applied Economics Letters* **20** pp 813 – 817.

2370. Lucey B M, O'Connor F 2014 Do bubbles occur in gold prices? Evidence from gold lease rates and Markov switching models *Borsa Istanbul Review*.

2371. Lucey B M, Sile Li 2015 What precious metals act as safe havens, and when? Some US evidence *Applied Economics Letters* **22** (1) pp 35 – 45.

2372. Riley C Summer 2010 A new gold rush: Investing in precious metals *Journal of Investing* pp 95 – 100.

2373. Roache S K, Rossi M 2010 The effects of economic news on commodity prices: Is gold just another commodity *The Quarterly Review of Economics and Finance* **50** pp 377 – 385.

2374. Shafiee S, Topal E 2010 An overview of global gold market and gold price forecasting *Resources Policy* **35** pp 178 – 189.

2375. Tsuchiya Y 2010 Linkages among precious metals commodity futures prices: Evidence from Tokyo *Economics Bulletin* **30** (3) pp 1772 – 1777

http://www.accessecon.com/Pubs/EB/2010/Volume30/EB-10-V30-I3-P162.pdf .

2376. Zhang Y-J, Wei Y-M September 2010 The crude oil market and the gold market: Evidence for cointegration, causality and price discovery *Resources Policy* **35** (3) pp 168 – 177.

2377. Khalifa A A A, Miao H, Ramchander S 2011 Return distribution and volatility forecasting in metal futures markets: Evidence from gold, silver and copper *Journal of Future Markets* **31** (1) pp 55 – 80.

2378. Morales L, Andreosso-O'Callaghan B 2011 Comparative analysis on the effects of the Asian and global financial crises on precious metal markets *Research in International Business and Finance* **25** (2) pp 203 – 227.

2379. Morales L, Andreosso-O'Callaghan B 2014 Volatility analysis of precious metals returns and oil returns: An ICSS approach *Journal of Economics and Finance* **38** (3) pp 492 – 517.

2380. Pukthuanthong K, Roll R 2011 Gold and the dollar (and the euro, pound, and yen) *Journal of Banking and Finance* **35** pp 2070 – 2083.

2381. Arouri M E H, Hammoudeh Sh, Lahiani A, Nguyen D K 2012 Long memory and structural breaks in modeling the return and volatility dynamics of precious metals *The Quarterly Review of Economics and Finance* **52** (2) pp 207 – 218.

2382. Arouri M E H, Hammoudeh Sh, Nguyen D K, Lahiani A 2013 On the short- and long-run efficiency of energy and precious metal markets *Energy Economics* **40** (C) pp 832 – 844

https://hal.archives-ouvertes.fr/hal-00798036/document .

2383. Cochran S J, Mansur I, Odusami B 2012 Volatility persistence in metal returns: A FIGARCH approach *Journal of Economics and Business* **64** (4) pp 287 – 305.

2384. Elder J, Miao H, Ramchander S 2012 Impact of macroeconomic news on metal futures *Journal of Banking & Finance* **36** (1) pp 51 – 65.

2385. Krezolek D 2012 Non-classical measures of investment risk on the market of precious non-ferrous metals using the methodology of stable distributions *Dynamic Econometric Models* **12** pp 89 – 104.

2386. Mutafoglu T H, Tokat E, Tokat H A 2012 Forecasting precious metal price movements using trader positions *Resources Policy* **37** (3) pp 273 – 280.

2387. Papież M, Śmiech S 2012 Causality in mean and variance between returns of crude oil and metal prices, agricultural prices and financial market prices *in* Proceedings of 30[th] International Conference Mathematical Methods in Economics Ramík J, Stavárek D (editors) *School of Business Administration Karviná Silesian University* pp 675 – 680.

2388. Śmiech S, Papież M 2012 A dynamic analysis of causality between prices on the metals market *in* Proceedings of the International Conference Quantitative Methods in Economics (Multiple Criteria Decision Making XVI) Reiff M (editor) Slovakia Bratislava pp 221 – 225.

2389. Yermilova M 2012 Possible scenarios for the world monetary system development in conditions of returning to using of precious metals *Ukrainian Journal Ekonomist* **10** pp 4 – 7.

2390. Caporin M, Ranaldo A, Velo G G 2013 Stylized facts and dynamic modeling of high-frequency data on precious metals *Working Paper on Finance no 1318* School of Finance  University of St Gallen Switzerland pp 1 – 40

http://ux-tauri.unisg.ch/RePEc/usg/sfwpfi/WPF-1318.pdf .

2391. Caporin M, Ranaldo A, Velo G G 2015 Precious metals under the microscope: A high-frequency analysis *Quantitative Finance* **15** (5) pp 743 – 759

http://ux-tauri.unisg.ch/RePEc/usg/sfwpfi/WPF-1409.pdf .

2392. Emmirich O, McGroarty F J 2013 Should gold be included in institutional investment portfolios? *Applied Financial Economics* **23** (19) pp 1553 – 1565.

2393. Ewing B, Malik F 2013 Volatility transmission between gold and oil futures under structural breaks *International Review of Economics and Finance* **25** pp 113 – 121.

2394. Hood M, Malik M 2013 Is gold the best hedge and a safe haven under changing stock market volatility? *Review of Financial Economics* **22** pp 47 – 52.

2395. Jain A, Ghosh S 2013 Dynamics of global oil prices, exchange rate and precious metal prices in India *Resources Policy* **38** (1) pp 88 – 93.

2396. Öztek M F, Ocal N 2013 Financial crises, financialization of commodity markets and correlation of agricultural commodity index with precious metal index and S&P500 *ERC Working Paper no 1302* ERC - Economic Research Center Middle East Technical University Turkey

http://www.erc.metu.edu.tr/menu/series13/1302.pdf .

2397. Reboredo J C 2013a Is gold a hedge or safe haven against oil price movements? *Resources Policy* **38** (2) pp 130 – 137.

2398. Reboredo J C 2013b Is gold a safe haven or a hedge for the US dollar? Implications for risk management *Journal of Banking and Finance* **37** (8) pp 2665 – 2676.

2399. Reboredo J C, Ugolini A 2015 A downside/upside price spillovers between precious metals: A vine copula approach *The North American Journal of Economics and Finance* **34** (C) pp 84 – 102.

2400. Revenda Z 2013 Peníze a zlato (2[nd] edition) *Management Press* Prague Czech ISBN 978-80-7261-260-4.

2401. Revenda Z 2016 Investment in precious metals and stocks *Acta Oeconomica Pragensia* **2016** (4) pp 25 – 36

http://www.vse.cz/polek/download.php?jnl=aop&pdf=542.pdf .

2402. Rizea Raluca D, Sârbu R and Condrea E 2013 The effects of the global economical crisis on the global precious metals market *Ovidius University Annals Economic Sciences Series* **13** (2) pp 60 – 64.

2403. Sensoy A 2013 Dynamic relationship between precious metals *Resources Policy* **38** (4) pp 504 – 511.

2404. Smales L A 2013 News sentiment in the gold futures market *Social Science Research Network* NY USA

http://ssrn.com/abstract=2309868 .

2405. Westerlund J 2013 Simple unit root testing in generally trending data with an application to precious metal prices in Asia *Journal of Asian Economics* **28** (C) pp 12 – 27.

2406. Agyei-Ampomah S, Gounopoulos D, Mazouz K 2014 Does gold offer a better protection against losses in sovereign debt bonds than other metals? *Journal of Banking and Finance* **40** (C) pp 507 – 521.

2407. Apergis N, Christou Ch, Payne J 2014 Precious metal markets, stock markets and the macroeconomic environment: a FAVAR model approach *Applied Financial Economics* **24** (10) pp 691 – 703.

2408. Charles A, Darné O, Kim J June 19 2014 Precious metals shine? A market efficiency perspective *Working Paper* HAL France

https://hal.archives-ouvertes.fr/hal-01010516/document .

2409. Charlot Ph, Vêlayoudom Marimoutou 2014 On the relationship between the prices of oil and the precious metals: Revisiting with a multivariate regime-switching decision tree *Energy Economics* **44** (C) pp 456 – 467.

2410. Demiralay S, Ulusoy V January 27 2014a Value-at-risk predictions of precious metals with long memory volatility models *MPRA Paper no 53229* University Library of Munich Germany pp 1 – 25

https://mpra.ub.uni-muenchen.de/53229/1/MPRA_paper_53229.pdf .

2411. Demiralay S, Ulusoy V 2014b Non-linear volatility dynamics and risk management of precious metals The North American *Journal of Economics and Finance* **30** (C) pp 183 – 202.

2412. Giles D, Qinlu Chen 2014 Risk analysis for three precious metals: An application of extreme value theory *Econometrics Working Paper no EWP1402* Department of Economics University of Victoria Canada pp 1 – 28

http://www.uvic.ca/socialsciences/economics/assets/docs/econometrics/ewp1402.pdf .

2413. Golosnoy V, Rossen A 2014 Modeling dynamics of metal price series via state space approach with two common factors *HWWI Research Paper no 156* Hamburg Institute of International Economics (HWWI) pp 1 – 30

https://www.econstor.eu/bitstream/10419/104543/1/806861886.pdf .

2414. Issler J V, Rodrigues C, Burjack R 2014 Using common features to understand the behavior of metal-commodity prices and forecast them at different horizons *Journal of International Money and Finance* **42** pp 310 – 335.

2415. Papadamou St, Markopoulos Th 2014 Investigating intraday interdependence between gold, silver and three major currencies: the Euro, British Pound and Japanese Yen *International Advances in Economic Research* **20** (4) pp 399 – 410.

2416. Tsolas I E 2014 Precious metal mutual fund performance appraisal using DEA modeling *Resources Policy* **39** (C) pp 54 – 60.

2417. Walczak M 2014 Precious metals roles in investment portfolio - comparative analysis of Czech, Hungary and Poland *Economy & Business Journal* **8** (1) pp 1220 – 1230.

**2418.** Wanat S, Papież M, Śmiech S June 15 2014 The conditional dependence structure between precious metals: A copula-GARCH approach *MPRA Paper no 56664* University Library of Munich Germany pp 1 – 15

https://mpra.ub.uni-muenchen.de/56664/1/MPRA_paper_56664.pdf .

**2419.** Antonakakis N, Kizys R 2015 Dynamic spillovers between commodity and currency markets *International Review of Financial Analysis* **41** (C) pp 303 – 319.

**2420.** Auer B R 2015 Superstitious seasonality in precious metals markets? Evidence from GARCH models with time-varying skewness and kurtosis *Applied Economics* **47** (27) pp 2844 – 2859.

**2421.** Balcilar M, Katzke N, Gupta R 2015 Do precious metal prices help in forecasting South African inflation *Working Paper no 15-05* Department of Economics Eastern Mediterranean University Turkey pp 1 – 22

http://repec.economics.emu.edu.tr/RePEc/emu/wpaper/15-05.pdf ,

http://www.ekon.sun.ac.za/wpapers/2015/wp032015/wp-03-2015.pdf .

**2422.** Balcilar M, Hammoudeh Sh, Nwin-Anefo Fru Asaba 2015 A regime-dependent assessment of the information transmission dynamics between oil prices, precious metal prices and exchange rates *International Review of Economics & Finance* **40** (C) pp 72 – 89.

**2423.** Bildirici M, Türkmen N 2015 Nonlinear causality between oil and precious metals *Resources Policy* **46** (P2) pp 202 – 211.

**2424.** Bosch D, Pradkhan E 2015 The impact of speculation on precious metals futures markets *Resources Policy* **44** (C) pp 118 – 134.

**2425.** Gil-Alana L, Chang Sh, Balcilar M, Aye G C, Gupta R 2015 Persistence of precious metal prices: A fractional integration approach with structural breaks *Resources Policy* **44** (C) pp 57 – 64

http://ncid.unav.edu/download/file/fid/590 .

**2426.** Mensi W, Hammoudeh Sh, Sang Hoon Kang 2015 Precious metals, cereal, oil and stock market linkages and portfolio risk management: Evidence from Saudi Arabia *Economic Modelling* **51** (C) pp 340 – 358.

**2427.** Figuerola-Ferretti I, McCrorie J R 2016 The shine of precious metals around the global financial crisis *Journal of Empirical Finance* **38** (PB) pp 717 – 738.

**2428.** Novotný J, Polach J 2016 Real silver and its investment and business options *International Journal of Entrepreneurial Knowledge* vol **4** issue 1/2016 pp 40 – 51 DOI: 10.1515/ijek-2016-0003.

**2429.** Pierdzioch Ch, Risse M, Rohloff S 2016 Are precious metals a hedge against exchange-rate movements? An empirical exploration using Bayesian additive regression trees *The North American Journal of Economics and Finance* **38** (C) pp 27 – 38.

**2430.** Pradkhan E 2016 Information content of trading activity in precious metals futures markets *Journal of Futures Markets* **36** (5) pp 421 – 456.

**2431.** Rand Kwong Yew Low, Yiran Yao, Faff R 2016 Diamonds vs precious metals: What shines brightest in your investment portfolio? *International Review of Financial Analysis* **43** (C) pp 1 – 14.

***Real estate investment, real estate valuation, real estate exchange financial capital investment product, financial capital investment medium in finances:***

*2432.* Grebler L 1954 Real estate investment experience in New York City *Journal of Finance* **9** (2) pp 216 – 218.

*2433.* Taylor G S, Bailey E N 1963 Real estate investment trusts *Business Horizons* **6** (2) pp 71 – 80.

*2434.* Wendt P F, Wong S N December 1965 Investment performance: Common stocks versus apartment houses *Journal of Finance* pp 633 – 646.

*2435.* Friedman H C December 1970 Real estate investment and portfolio theory *Journal of Financial and Quantitative Analysis* pp 861 – 874.

*2436.* Friedman H C March 1971 Real estate investment and portfolio theory *Journal of Financial and Quantitative Analysis* **6** (2) pp 861 – 874 DOI: https://doi.org/10.2307/2329720 .

*2437.* Roulac S E 1976 Can real estate returns outperform common stocks? *Journal of Portfolio Management* **2** pp 26 – 43.

*2438.* Roulac S E 1996 Real estate market cycles, transformation forces and structural change *Journal of Real Estate Portfolio Management* **2** (1) pp 1 – 17.

*2439.* Smith K V, Shulman D September–October 1976 The performance of equity real estate investment trusts *Financial Analysts Journal* pp 61 – 66.

*2440.* Lusht K M 1978 Inflation and real estate investment value *Real Estate Economics* **6** (1) pp 37 – 49.

*2441.* Pellatt P G K 1972 The analysis of real estate investments under uncertainty *Journal of Finance* **27** (2) pp 459 – 471.

*2442.* Chapman Findlay III M, Hamilton C W, Messner S D, Yormark J S September 1979 Optimal real estate portfolios *Real Estate Economics* DOI: 10.1111/1540-6229.t01-11-00201

http://onlinelibrary.wiley.com/doi/10.1111/1540-6229.t01-11-00201/abstract;jsessionid=0F84C68B32F80FF1F91F04EEC5EA6ADB.f01t02 .

*2443.* Ibbotson R G, Fall C L 1979 The United States market wealth portfolio *Journal of Portfolio Management* **6** pp 82 – 92.

*2444.* Ibbotson R G, Sinquefield R A 1982 Stocks, bonds, bills and inflation: The past and the future *Financial analysts Research Foundation* Charlottesville VA USA.

*2445.* Ibbotson R G, Siegel L B 1983 The world market wealth portfolios *Journal of Portfolio Management* **9** pp 5 – 17.

*2446.* Ibbotson R G, Siegel L B 1984 Real estate returns: A comparison with other investments *Real Estate Economics* **12** (3) pp 219 – 242.

*2447.* Ibbotson R G, Siegel L, Love K Fall 1985 World wealth: Market values and returns *Journal of Portfolio Management* pp 4 – 23.

*2448.* Penny P E 1980 Modern investment theory and real estate analysis South African *Journal of Economics* **48** (1) pp 39 – 52.

*2449.* Rystrom D 1980 Inflation and real estate investment value: A comment *Real Estate Economics* **8** (4) pp 395 – 401.

*2450.* Burns W L, Epley D R Winter 1982 The performance of portfolio of REITs + stocks *Journal of Portfolio Management* pp 37 – 41.

**2451.** Miles M E, Estey A S Winter 1982 How well do real estate funds perform? *Journal of Portfolio Management* pp 62 – 68.

**2452.** Miles M E, McCue T Summer 1982 Historic returns and institutional real estate portfolios *AREUEA Journal* pp 184 – 199.

**2453.** Miles M, McCue T 1984 Commercial real estate returns *AREUEA Journal* **12** (3) pp 355 – 377.

**2454.** Brueggeman W B H, Chen A H, Thibodeau T G 1984 Real estate investment funds: Performance and portfolio considerations *AREUEA Journal* **12** (3) pp 333 – 354.

**2455.** Fogler H R 1984 20 percent in real estate: Can theory justify it? *Journal of Portfolio Management* **10** pp 6 – 13.

**2456.** Webb J R 1984 Real estate investment acquisition rules for life insurance companies and pension funds *Journal of the American Real Estate and Urban Economics Association* **12** (4) pp 495 – 520.

**2457.** Webb J R, McIntosh W 1986 Real estate investment acquisition rules for REITs: A survey *Journal of Real Estate Research* **1** (1) pp 77 – 98.

**2458.** Gau G W 1985 Public information and abnormal returns in real estate investment *Real Estate Economics* **13** (1) pp 15 – 31.

**2459.** Gau G W, Wang K 1990 Capital structure decisions in real estate investment *Real Estate Economics* **18** (4) pp 501 – 521.

**2460.** Hartzell D J, Mengden A August 27 1986 Real estate investment trusts - Are they stocks or real estate? *Salomon Brothers Real Estate Research* USA.

**2461.** Hartzell D J, Hekman J S, Miles M E Spring 1987 Real estate returns and inflation *AREUEA Journal* pp 617 – 637.

**2462.** Hartzell D J, Webb J 1988 Real estate risk and return expectations: Recent survey results *Journal of Real Estate Research* **3** (3) pp 31 – 38.

**2463.** Hartzell J C, Liu C H, Kallberg J G 2004 The role of underlying real asset market in REIT IPOs *Social Science Research Network* NY USA
http://ssrn.com/abstract=516662 .

**2464.** Kuhle J L, Walther C H, Wurtzebach C H 1986 The financial performance of real estate investment trusts *Journal of Real Estate Research* **1** (1) pp 67 – 75.

**2465.** Kuhle J L 1987 Portfolio diversification and return benefits - Common stocks vs. real estate investment trusts *Journal of Real Estate Research* **2** (2) pp 1 – 9.

**2466.** Titman S, Warga A Fall 1986 Risks and the performance of real estate investment trusts: A multiple index approach *AREUEA Journal* **14** pp 414 – 431.

**2467.** Haight G T, Fort D A 1987 REITs, new opportunities in real estate investment trust securities *Probus Publishing* Chicago Illinois USA.

**2468.** Chen K C, Tzang D D 1988 Interest-rate sensitivity of real estate investment trusts *Journal of Real Estate Research* **3** (3) pp 13 – 22.

**2469.** Chinloy P 1988 Real estate - investment and financial strategy *Kluwer Academic Publishers* Boston USA.

**2470.** Firstenberg P M, Ross S A, Zisler R C 1988 Real estate: The whole story *The Journal of Portfolio Management* **14** pp 22 – 34.

*2471.* Gyourko J, Linneman P 1988 Owner - occupied homes, income producing properties, and REITs as inflation hedges: Empirical findings *Journal of Real Estate Finance and Economics* **2** (3) pp 350 – 360.

*2472.* Gyourko J, Keim D B 1992 What does the stock market tell us about real estate returns? *AREUEA Journal* **20** pp 457 – 485.

*2473.* Hines M March 1988 International dimensions of real estate *Appraisal Journal* pp 492 – 501.

*2474.* Lusht K M 1988 The real estate pricing puzzle *AREUEA Journal* 16 (2) pp 95 – 104.

*2475.* Miller N, Sklarz M, Ordway N 1988 Japanese purchases, exchange rates and speculation in residential real estate markets *Journal of Real Estate Research* **3** (3) pp 39 – 49.

*2476.* Mooney S, Mooney K 1988 Considerations for the foreign investor in the U.S. real estate market *Journal of Valuation* 7 pp 22 – 34.

*2477.* Wittner D A Fall 1988 Japanese investment in US real estate: The Oriental express *Real Estate Law Journal* **17.**

*2478.* Geltner D 1989 Estimating real estate's systematic risk from aggregated level appraisal-based returns *AREUEA Journal* **17** pp 463 – 481.

*2479.* Goebel P R, Kim K S 1989 Performance evaluation of finite-life real estate investment trusts *Journal of Real Estate Research* **4** (2) pp 57 – 69.

*2480.* Rubens J H, Bond M T, Webb J R 1989 The inflation-hedging effectiveness of real estate *Journal of Real Estate Research* vol **4** no 2 pp 45 – 55.

*2481.* Sweeney F 1989 International real estate diversification: A viable strategy *Journal of Property Management* **5** pp 317 – 326.

*2482.* Sweeney F 1993 Mapping a European property investment strategy *Journal of Property Valuation and Investment* **11** pp 259 – 267.

*2483.* Chan K C, Hendershott P H, Sanders A B Winter 1990 Risk and return on real estate: Evidence from equity REITs *AREUEA Journal* **18** (4) pp 431 – 452.

*2484.* Colwell P F, Park H Y September 1990 Seasonality and size effects: The case of real-estate-related investment *Journal of Real Estate Finance and Economics* **3** pp 251 – 259.

*2485.* Giliberto S M Summer 1990 Equity real estate investment trusts and real estate returns *Journal of Real Estate Research* **5** pp 259 – 263.

*2486.* Goetzmann W N, Ibbotson R G 1990 The performance of real estate as an asset class *Journal of Applied Corporate Finance* pp 450 – 460.

*2487.* Howe J S, Shilling J D 1990 REIT advisor performance *AREUEA Journal* **18** (4) pp 479 – 500.

*2488.* Liu C H, Hartzell D J, Grissom T V, Wylie G 1990 The composition of the market portfolio and real estate investment performance *AREUEA Journal* pp 18 (1) pp 49 – 75.

*2489.* McMahan J 1990 Foreign investment in U.S. real estate *Real Estate Issues* **15** pp 35 – 37.

*2490.* Sagalyn L B 1990 Real estate risk and the business cycle: Evidence from security markets *Journal of Real Estate Research* **5** (2) pp 203 – 219.

**2491.** Wheaton W C, Torto R G 1990 An investment model of the demand and supply for industrial real estate *Real Estate Economics* **18** (4) pp 530 – 547.

**2492.** Asabere P, Kleiman R, McGowan C 1991 The risk-return attributes of international real estate equities *Journal of Real Estate Research* **6** (2) pp 143 – 151.

**2493.** Martin J D, Cook D O 1991 A comparison of the recent performance of publicly traded real property portfolios and common stocks *AREUEA Journal* **19** (2) pp 184 – 212.

**2494.** McIntosh W, Liang Y, Tompkins D L Spring 1991 An examination of the small firm effect within the REIT industry *Journal of Real Estate Research* **6** pp 9 – 17.

**2495.** Ross S, Zisler R 1991 Risk and return in real estate *Journal of Real Estate Finance and Economics* **4** pp 129 – 146.

**2496.** Ziobrowski A J, Boyd J W 1991 Leverage and foreign investment in U. S. real estate *Journal of Real Estate Research* **7** (1) pp 33 – 57.

**2497.** Ziobrowski A J, Curcio R J 1991 Diversification benefits of U.S. real estate to foreign investors *Journal of Real Estate Research* **6** (2) pp 119 – 142.

**2498.** Ziobrowski A J, Ziobrowski B J 1993 Hedging foreign investments in U.S. real estate with currency options *Journal of Real Estate Research* **8** (1) pp 27 – 54.

**2499.** Ziobrowski B J, Ziobrowski A J 1995 Using forward contracts to hedge foreign investment in U.S. real estate *Journal of Property Valuation & Investment* **13** pp 22 – 43.

**2500.** Ziobrowski A J, McAlum H, Ziobrowski B J 1996 Taxes and foreign real estate investment *The Journal of Real Estate Research* **11** (2) pp 197 – 213.

**2501.** Ambrose B W, Ancel E, Griffiths M D Spring 1992 The fractal structure of real estate investment returns: The search for evidence of market segmentation and nonlinear dependency *Journal of the American Real Estate and Urban Economics Association* **20** pp 23 – 54.

**2502.** DiPascuale D, Wheaton W C 1992 The markets for real estate and space: A conceptual framework *Journal of the American Real Estate and Urban Economics Association* **20** (1) pp 181 – 197.

**2503.** Kleiman R, Farragher E 1992 Investment characteristics of international real estate equities *Real Estate Review* **22** pp 33 – 39.

**2504.** Liu C H, Mei J 1992 The predictability of returns on equity REITs and their co-movement with other assets *Journal of Real Estate Finance and Economics* **5** (4) pp 401 – 418.

**2505.** Brueggeman B W, Fisher D J 1993 Real estate finance and investments 9[th] edition *Irwin BurrRidge* USA pp 1 – 908 ISBN: 0-256-08290-1 .

**2506.** Newell G, MacFarlane J March 1994 Property: More volatile than you thought *Journal of Australian Securities Institute* pp 25 – 37.

**2507.** Worzala E 1994 Overseas property investments: How are they perceived by the institutional investor? *Journal of Property Valuation and Investment* **12** pp 31 – 47.

**2508.** Worzala E, Newel G 1997 International real estate: A review of strategic investment issues *Journal of Real Estate Portfolio Management* **3** (2) pp 87 – 96.

2509. Zumpano L V, Elder H W 1994 Economies of scope and density in the market for real estate brokerage services *Journal of the American Real Estate and Urban Economics Association* **22** (3) pp 497 – 513.

2510. Baum A 1995 Can foreign investment be successful? *Real Estate Finance* **12** (1) pp 81 – 89.

2511. Baum A 2009 Commercial real estate investment: A strategic approach 2nd edition *EG Books*.

2512. Corgel J B, McIntosh W, Ott S H 1995 Real estate investment trusts: A Review of the literature *Journal of Real Estate Literature* **3** (1) pp 13 – 46.

2513. Jun Han, Youguo Liang 1995 The historical performance of real estate investment trusts *The Journal of Real Estate Research* **10** (3) pp 235 – 262.

2514. Mueller G R Spring 1995 Understanding real estate's physical and financial cycles *Journal of Real Estate Finance* pp 47 – 52.

2515. Barkham R J, Ward C W R, Henry O T 1996 The inflation-hedging characteristics of U.K. property *Journal of Property Finance* **7** (1) pp 16 – 26.

2516. Geurts T G, Jaffe A J 1996 Risk and real estate investment: An international perspective *The Journal of Real Estate Research* **11** (2) pp 117 – 130.

2517. Han J 1996 Targeting markets is popular: A survey of pension real estate investment advisors *Real Estate Finance* **13** (1) pp 66 – 75.

2518. Newell G 1996 The inflation-hedging characteristics of Australian commercial property: 1984-1995 *Journal of Property Finance* **7** (1) pp 1 – 15.

2519. Newell G, Webb J 1996 Assessing risk for international real estate investments *The Journal of Real Estate Research* **11** (2) pp 103 – 115.

2520. Seck D 1996 The substitutability of real estate assets *Real Estate Economics* **24** (1) pp 75 – 95.

2521. Solnik B 1996 International investments 3rd edition *Addison-Wesley* Massachusetts USA.

2522. Bers M, Springer Th M 1997 Economies-of-scale for real estate investment trusts *Journal Of Real Estate Research* **14** (3) pp 275 – 291.

2523. Brueggeman W B, Fisher J D 1997 Real estate finance and investments *Irwin McGraw-Hill* Boston MA USA.

2524. Farrell R 1997 Japanese foreign direct investment in real estate 1985–1994 *Pacific Economic Papers no 272* Australia–Japan Research Centre Research School of Pacific and Asian Studies The Australian National University Canberra Australia ISBN 0 86413 216 6 pp 1 – 53.

2525. Hoesli M, MacGregor B D, Matysiak G, Nanthakumaran N 1997 The short-term inflation-hedging characteristics of UK real estate *Journal of Real Estate Finance and Economics* vol **15** no 1 pp 59 – 76.

2526. Mei J, Saunders A 1997 Have U.S. financial institutions' real estate investments exhibited "trend-chasing" behavior? *Review of Economics and Statistics* **79** (2) pp 248 – 258.

2527. Redman A L, Manakyan H, Liano K 1997 Real estate investment trusts and calendar anomalies *Journal Of Real Estate Research* **14** (1/2) pp 19 – 28.

2528. Ball M, Lizieri C, MacGregor B N 1998 The economics of commercial property markets *Routledge* London UK.

2529. Chun G H, Shilling J D 1998 Real estate asset allocations and international real estate markets *Journal of the Asian Real Estate Society* **1** (1) pp 17 – 44.

2530. D'Arcy E, Keogh G 1998 Territorial competition and property market process: An explanatory analysis *Journal of Urban Studies* **35** (8) pp 1215 – 1230.

2531. Martens C-P 1998 Real estate investment in Germany *5th European Real Estate Society Conference* Maastricht The Netherlands.

2532. Svensson J 1998 Investment, property rights and political instability: Theory and evidence *European Economic Review* **42** (7) pp 1317 – 1341.

2533. Adair A, Berry J, McGreal S, Syacutekora L, Ghanbari Parsa A, Redding B Spring 1999 Globalization of real estate markets in Central Europe *European Planning Studies* **7** (3)        pp 295 – 305.

2534. Keogh G, D'Arcy E 1999 Property market efficiency: An institutional economics perspective *Journal of Urban Studies* **36** (13) pp 2401 – 2414.

2535. Liao H-H, Mei J 1999 Institutional factors and real estate returns – A cross country study *International Real Estate Review* **2** (1) pp 21 – 34.

2536. Moshirian F, Pham T 2000 Determinants of US investment in real estate abroad *Journal of Multinational Financial Management* **10** (1) pp 63 – 72.

2537. Brounen D, Cools T, Schweizer M Summer 2001 Information transparency pays: Evidence from European property shares *Real Estate Finance* pp 39 – 49.

2538. Lee St 2001 The risk of investing in the real estate markets of the Asian region *Working Paper of Department of Land Management no 6* University of Reading UK.

2539. Lee St 2005 Gauging the investment potential of international real estate markets *Working Paper of Department of Real Estate and Planning no 19* University of Reading UK.

2540. Lee St 2006a The impact of country risk on international real estate returns *Working Paper of Department of Real Estate and Planning no 10* University of Reading UK.

2541. Lee St 2006b The impact of exchanges rates on international real estate portfolio allocation *Working Paper of Department of Real Estate and Planning no 4* University of Reading UK.

2542. McGreal S, Parsa A, Keivani R 2001 Perceptions of real estate markets in central Europe: A survey of European investor *Journal of Real Estate Literature* **9** (2) pp 147 – 160.

2543. Thrall G I 2002 Business geography and new real estate market analysis *Oxford University Press* New York and Oxford USA and UK.

2544. Blazenko G W, Pavlov A D 2004 The economics of maintenance for real estate investments *Real Estate Economics* **32** (1) pp 55 – 84.

2545. Deqing Diane Li, Yung K 2004 Short interests in real estate investment trusts *International Real Estate Review* **7** (1) pp 56 – 70.

2546. Engelberts R, Suarez J L 2004 The real estate industry in The Netherlands *Working Paper WP 558* IESE Business School University of Navarra Barcelona Spain pp 1 – 62.

2547. Hoskin N, Higgins D, Cardew R Spring 2004 Macroeconomic variables and real estate returns: An international comparison *The Appraisal Journal* pp 163 – 170.

2548. Loo Lee Sim, Xing Quan Zhang, Jieming Zhu 2004 Globalization and real estate investment in Beijing *11th European Real Estate Society Conference* Milano Italy.

2549. Pi-Ying, Lai Peddy 2004 The competitiveness of real estate industry in Taiwan *11th European Real Estate Society Conference* Milano Italy.

2550. Hardin W G, Liano K, Gow-Cheng Huang 2005 Real estate investment trusts and calendar anomalies: Revisited *International Real Estate Review* **8** (1) pp 83 – 94.

2551. Laposa S, Lizieri C April 13 – 16 2005 Real estate capital flows and transitional economies *Conference Paper* ARES Meeting Santa Fe NM USA.

2552. Holsapple E J, Ozawa T, Olienyk J 2006 Foreign "direct" and "portfolio" investment in real estate *Journal of Real Estate Portfolio Management* 12 (1) pp 37 – 47.

2553. Johnson D T 2006 Real estate investing *Managerial Finance* **32** (12) pp 953 – 954.

2554. Lim L Ch, McGreal S, Webb J R 2006 Perception of real estate investment opportunities in Central/South America and Africa *Journal of Real Estate Portfolio Management* **12** (3) pp 261 – 276.

2555. Adlington G, Grover R, Heywood M, Keith S, Munro-Faure P, Perrotta L December 6-8 2008 Developing real estate markets in transition economies UN Intergovernmental Conference Paper RICS Research Foundation.

2556. Chan K C, Hardin III W G, Liano K, Zheng Yu 2008 The internationalization of real estate research *Journal Real Estate Research* **30** (1) pp 91 – 124.

2557. Lambie-Hanson L Fall 2008 Addressing the prevalence of real estate investments in the new markets tax credit program *Working Paper 2008-04* Federal Reserve Bank of San Francisco California USA pp 1 – 43

http://frbsf.org/cdinvestments/ .

2558. Minye Zhang, Yongheng Deng July 9 2008 The relationship between stock returns and the past performance of hotel real estate industry in the U.S.: Is hotel real estate prone to overinvestment? *School of Policy, Planning, and Development* USC California USA pp 1 – 52.

2559. Falkenbach H 2009 Market selection for international real estate investments *International Journal of Strategic Property Management* **13** (4) pp 299 – 308.

2560. Kurzrock B M, Rottke N, Schiereck D 2009 Influence factors on the performance of direct property investments *Journal of Real Estate Portfolio Management* **15** (1) pp 59 – 73.

2561. Edelstein R, Qian W, Tsang D 2010 How do institutional factors affect international real estate returns? *Journal of Real Estate Economics and Finance* DOI: 10.1007/s11146-010-9245-4.

2562. Lieser K, Groh A P 2010 The attractiveness of 66 countries for institutional real estate investment: A composite index approach *Working Paper WP 868* IESE Business School University of Navarra Barcelona Spain pp 1 – 62.

2563. Lieser K, Groh A P July 2011 The determinants of international commercial real estate investments *Working Paper WP 935* IESE Business School University of Navarra Barcelona Spain pp 1 – 55.

2564. Peralta-Alva A 2011 Real estate bubbles and weak recoveries *Economic Synopses* no 39 Federal Reserve Bank of St. Louis USA pp 1 – 2.

2565. Chaney Th, Sraer D, Thesmar D 2012 The collateral channel: How real estate shocks affect corporate investment *American Economic Review* **102** (6) pp 2381 – 2409.

2566. Baker H K, Chinloy P (editors) September 4 2014 Private real estate markets and investments *Oxford University Press* ISBN: 9780199388752 pp 1 – 312.

2567. Baker H K, Chinloy P (editors) September 8 2014 Public real estate markets and investments *Oxford University Press* ISBN: 9780199993277pp 1 – 336.

2568. Cunat V, Cvijanovic D, Yuan K 2014 Within-bank transmission of real estate shocks *SSRN Working Paper*
http://ssrn.com/abstract=2332177 ,
http://dx.doi.org/10.2139/ssrn.2332177 .

2569. Gauder M, Houssard C, Orsmond D 2014 Foreign investment in residential real estate *Reserve Bank of Australia Bulletin June Quarter 2014* Australia pp 11 – 18.

2570. Lieser K, Groh A 2014 The determinants of international commercial real estate investment *The Journal of Real Estate Finance and Economics* **48** (4) pp 611 – 659.

2571. Anderson R I, Rottke N B, Krautz S 2015 Is real estate private equity real estate? *22nd Annual European Real Estate Society Conference* Istanbul Turkey.

2572. Hazama M, Uesugi L October 2015 Heterogeneous impact of real estate prices on firm investment *Working Paper Series no 30* Institute of Economic Research, Hitotsubashi University Tokyo Japan
http://hdl.handle.net/10086/27541 .

2573. Maksimov S, Bachurinskaya I 2015 Real estate markets of Russia and St. Petersburg in the context of globalization: Challenges and prospects *Izvestia Journal of University of Economics* Varna Bulgaria pp 21 – 35.

2574. Papastamos D, Matysiak G, Stevenson S 2015 Assessing the accuracy and dispersion of real estate investment forecasts *International Review of Financial Analysis* **42** (C) pp 141 – 152.

***Private and public firms theories in economics and finances:***

2575. Ueda 1904 Shogyo Dai Jiten (The dictionary of commerce) Japan.

2576. Ueda 1937 Keieikeizaigaku Saran (The science of business administration, *Allgemeine Betriebswirtschaftslehre*) Japan.

2577. Marshall A 1923 Money, credit, and commerce *Prometheus Books* Amherst New York USA ISBN 13: 9781591020363.

2578. Berle A 1932 For whom corporate managers are trustees: A note *Harvard Law Review* vol **45** pp 1365 – 1072.

2579. Berle A A, Means G 1932 The modern corporation and private property *Harcourt Brace and Word Inc, The Commerce Clearing House*, *The MacMillan Company* New York USA pp 1 – 396.

2580. Dodd M 1932 For whom are corporate managers trustees? *Harvard Law Review* vol **45** pp 1145 – 1163.

2581. Ohlin B 1933 Interregional and international trade *Harvard University Press* Cambridge Massachusetts USA.

2582. Coase R H 1937 The nature of the firm *Economica* vol **4** (16) pp 386 – 405 doi:10.1111/j.1468-0335.1937.tb00002.x year=1937.

2583. Coase R H October 1960 The problem of social cost *Journal of Law and Economics* **3** pp 1 – 44.

2584. Coase R H 1988 The nature of the firm: Influence *Journal of Law, Economics, and Organization* **4** (1) pp 33 – 47.

2585. Barnard C I 1938 The functions of the executive *Harvard University Press* Cambridge MA USA pp 241 – 244.

2586. Barnard C I 1948 Organization and management: Selected papers *Harvard University Press* Cambridge MA USA.

2587. Barnard C I 1949 The entrepreneur and formal organization *Change and the Entrepreneur. Harvard University Press* Cambridge MA USA.

2588. Barnard C I 1958 Elementary conditions of business morale *California Management Review* vol **1** no 1.

2589. Alchian A A June 1950 Uncertainty, evolution and economic theory *Journal of Political Economy* **58** no 3 pp 211 – 221.

2590. Alchian A A, Demsetz H December 1972 Production, information costs, and economic organisation *American Economic Review* **62** no 5 pp 777 – 795.

2591. Solow R M August 1957 Technical changes and the aggregate production function *Review of Economics and Statistics* **39** pp 214 – 231.

2592. March J, Simon H 1958 Organizations *John Wiley and Sons Inc* New York USA.

2593. Modigliani F, Miller M June 1958 The cost of capital, corporation finance, and the theory of investment *American Economic Review* pp 261 – 297.

2594. Ward B 1958 The firm in Illyria: Market syndicalism *American Economic Review* **48** (4)    pp 566 – 589.

2595. Baumol W 1959 Business behaviour, value and growth.

2596. Baumol W 1962 The theory of expansion of the firm *American Economic Review.*

2597. Penrose E T 1959 The theory of the growth of the firm *Oxford University Press* Oxford UK, *John Wiley and Sons Inc* New York USA.

2598. March J G 1962 The business firm as a political coalition *Journal of Politics* **24** pp 662 – 678.

2599. Cyert R, March J G 1963, 1992 A behavioral theory of the firm *Wiley-Blackwell* 2nd edition ISBN 0-631-17451-6.

2600. Marris R May 1963 A model of the managerial enterprise *Quarterly Journal of Economics* **77** pp 185 – 209.

2601. Telser L 1963 Cutthroat competition and the long purse *Journal of Law and Economics* **9**    pp 259 – 277.

2602. Fogel R 1964 Railroads and American economic growth: Essays in econometric history *Johns Hopkins Press* Baltimore USA.

2603. Williamson O E 1964 The economics of discretionary behavior: Managerial objectives in a theory of the firm *Prentice-Hall* Englewood Cliffs NJ USA.

2604. Williamson O E 1975 Markets and hierarchies: Analysis and antitrust implications *Free Press* New York USA.

2605. Williamson O E 1984 Corporate governance *Yale Law Journal* vol **88** pp 1183 – 1200.

2606. Williamson O E December 1985 The modern corporation: Origins, evolution, attributes *Journal of Economic Literature* pp 537 – 568.

2607. Williamson O E 1988 Corporate finance and corporate governance *Journal of Finance* **43** (3) pp 28 – 30.

2608. Williamson O E 1996 The mechanisms of governance *Oxford University Press* New York USA.

2609. Williamson O E 2002 The theory of the firm as governance structure: From choice to contract *Journal of Economic Perspectives* vol **16** (3) pp 171 – 195.

2610. Williamson O E 2007 Corporate boards of directors: In principle and in practice *Journal of Law, Economics, and Organization*.

2611. Manne H G 1965 Mergers and the market for corporate control *Journal of Political Economy* **73-74** pp 110 – 120.

2612. Galbraith J K 1967, 1978 The new industrial state 1$^{st}$ edition, 2$^{nd}$ edition *Penguin Books* USA.

2613. Stigler G 1968 The organization of industry *Richard Irwin Inc* Homewood USA.

2614. Mano O 1968-1969 On the science of business administration in Japan *Hokudai Economic Papers* vol **1** pp 77 – 93 Hokkaido University Japan.

2615. Mano O 1970-1971 The development of the science of business administration in Japan since 1955 *Hokudai Economic Papers* vol **2** pp 30 – 42 Hokkaido University Japan.

2616. Mano O 1972-1973 An approach to the organization economy - The development of Barnard's theory *Hokudai Economic Papers* vol **3** Hokkaido University Japan.

2617. Mano O 1975-1976 Social responsibility of the firm-one development of Barnard's theory *Hokudai Economic Papers* vol **5** Hokkaido University Japan.

2618. Mano O 1978 Soshiki Keizai no Kaimei (Organization economy - A study of the management theory of C. L. Barnard) *Bunshindo* Tokyo Japan.

2619. Mano O 1980-1981 Barnard's theory of education for executives *Hokudai Economic Papers* vol **10** Hokkaido University Japan.

2620. Mano O 1987 Barnard no Keiei Riron (Management Theory of C. I. Barnard) *Bunshindo* Tokyo Japan.

2621. Mano O 1994 The differences between Barnard's and Simon's concept of organization equilibrium-Simon's misunderstanding about Barnard's intention *Economic Journal of Hokkaido University* vol **23** Hokkaido University Japan.

2622. Mano O 1995 On the significance of lateral organization *in* C I Barnard's theory *Economic Journal of Hokkaido University* vol **24** pp 1 – 13 Hokkaido University Japan.

2623. Lewellen W 1969 Management and ownership in the large firms *Journal of Finance* **24** pp 299 – 329.

2624. Hirschman A O 1970 Exit, voice, and loyalty: Responses to decline in firms, organizations, and states *Harvard University Press* Cambridge MA USA.

2625. Meade J March 1972 The theory of labour-managed firms and profit sharing *Economic Journal* **82** pp 402 – 428.

2626. Meade J 1986 Alternative systems of business organisation and of workers' remuneration *Allen & Unwin* London UK.

2627. Meade J 1989 Agathotopia: The economics of partnerships *Cambridge University Press* Cambridge UK.

2628. Black F, Scholes M 1973 The pricing of options and corporate liabilities *Journal of Political Economy* vol **81** pp 637 – 654.

2629. Black F, Cox J C 1976 Valuing corporate securities: Some effects of bond indenture provisions *Journal of Finance* vol **31** pp 351 – 367.

2630. Merton R C 1973 Theory of rational option pricing *Bell Journal of Economics and Management Science* vol **4** pp 141 – 183.

2631. Mintzberg H 1973 The nature of managerial work *Harper & Row* New York USA.

2632. Arrow K J 1974 The limits of organization pp 69 – 70.

2633. Merton R C 1974 On the pricing of corporate debt: The risk structure of interest rates *Journal of Finance* vol **29** pp 449 – 470.

2634. Crew M A 1975 Theory of the firm *Longman* New York USA ISBN 0-582-44042-4 pp 1 – 182.

2635. Lee, Jung Hwan 1975 An essay on the theory of the firm 慶應経営研究 vol **2** pp 133 – 153.

2636. Jensen M C, Meckling W H 1976 Theory of the firm: Managerial behavior, agency costs and ownership structure *Journal of Financial Economics* **3** pp 305 – 360.

2637. Jensen M C, Meckling W H October 1979 Rights and production functions: An application to labor managed firms and codetermination *Journal of Business* **52** no 4 pp 469 – 506.

2638. Jensen M C, Ruback R S 1983 The market for corporate control: The scientific evidence *Journal of Financial Economics* **11** (1-4) pp 5 – 50.

2639. Jensen M C 1986 Agency costs of free cash flow, corporate finance and takeovers *American Economic Review* Papers and Proceedings of the 98[th] Annual Meeting of the American Economic Association **76** (2) pp 323 – 329.

2640. Jensen M C September-October 1989 The eclipse of the public corporation *Harvard Business Review* **67** (5) pp 61 – 74
http://papers.ssrn.com/abstract=146149 .

2641. Jensen M C, Zimmermann J L April 1985 Management compensation and the managerial labor market *Journal of Accounting and Economics* **7** no 1-3 pp 3 – 9.

2642. Jensen M C, Murphy K J 1990 Performance pay and top management incentives *Journal of Political Economy* **98** (2) pp 225 – 264.

2643. Jensen M C 1993 The modern industrial revolution: Exit and the failure of internal control systems *Journal of Finance* **48** (3) pp 831 – 880.

2644. Jensen M C 2007 The economic case for private equity *Unpublished Harvard NOM Research Paper no 07-02*.

2645. Simon Y, Tezenas Du Montcel H Mai 1977 Théorie de la firme et réforme de l'entreprise *Revue Économique* **1** pp 321 – 351.

2646. Pfeffer J, Salancik G R 1978 The external control of organizations: A resource-dependency perspective *Harper & Row* New York USA.

2647. Pfeffer J 1981 Power in organizations *Harper Business*, *Pitman* Marshfield MA USA.

2648. Pfeffer J 1982 Organizations and organization theory *Ballinger Publishing Company* USA.

2649. Pfeffer J 1983 Organizational demography *in* Research in organizational behavior Cummings L L, Staw B M (editors) *JAI Press* **5** Greenwich pp 449 – 461.

2650. Pfeffer J 1991 Organizational theory and structural perspectives on management *Journal of Management* **17** pp 789 – 803.

2651. Pfeffer J, Salancik G R 2003 The external control of organizations: A resource dependency perspective *Stanford Business Classics* Stanford USA.

2652. Simon Y, Tezenas Du Montcel H 1978 Économie des ressources humaines dans l'entreprise *Masson*.

2653. Fama E April 1980 Agency problems and the theory of the firm *Journal of Political Economy* **88** (2) pp 288 – 307.

2654. Fama E, Jensen M 1983a Agency problems and residual claims Journal of Law ans Economics vol 26.

2655. Fama E, Jensen M 1983b Separation of ownership and control *Journal of Law and Economics* **26** pp 301 – 325.

2656. Fama E, Jensen M 1985 Organizational forms and investment decisions *Journal of Financial Economics* vol **14** pp 101 – 119.

2657. Mintzberg H 1982 Structure et dynamique des organisations *Les Editions d'Organisation*.

2658. Demsetz H June 1983 The structure of ownership and the theory of the firm *Journal of Law and Economics* vol **26** pp 375 – 390.

2659. Demsetz D, Lehn K 1985 The structure of corporate ownership: Causes and consequences *Journal of Political Economy* **93-6** pp 1155 – 1177.

2660. Demsetz H 1997 Theories of the firm: Their contrasts and implications *Chung-Hua Series of Lectures* no 25 *The Institute of Economics Academia Sinica* Nankang Taipei Taiwan Republic of China.

2661. Demsetz D, Villalonga B 2001 Ownership structure and corporate performance *Journal of Corporate Finance* vol **7** pp 209 – 233.

2662. Wernerfelt B 1984 A resource-based view of the firm *Strategic Management Journal* **5** pp 171 – 180.

2663. Wernerfelt B 1995 The resource-based view of the firm: Ten years after *Strategic Management Journal* **16** pp 171 – 174.

2664. Lode Li 1986 A stochastic theory of the firm *Working Paper no 1844-86* Sloan School of Management MIT USA.

2665. Perrow C 1986 Complex organizations *Random House* New York USA.

2666. Tirole J 1986 Hierarchies and bureaucracies: On the role of collusion in organizations *Journal of Law Economics, and Organization* **2** (2) pp 181 – 214.

2667. Tirole J 1988 The theory of the firm *in* The theory of industrial organization *MIT Press* pp 15 – 60.

2668. Aghion P, Tirole J 1997 Formal and real authority in organisations *Journal of Political Economy* **105** (1) pp 1 – 29.

**2669.** Tirole J 2001 Corporate governance *Econometrica* vol **69** no 1 pp 1 – 35.

**2670.** Tirole J 2006 The theory of corporate finance *Princeton University Press* Princeton USA.

**2671.** Donaldson L 1990 The ethereal hand: Organizational economics and management theory *Academy of Management Review* **15** (3) pp 369 – 381.

**2672.** Hart O, Moore J 1990 Property rights and the nature of the firm *Journal of Political Economy* vol **98** no 6 pp 1119 – 1158.

**2673.** Hart O 2011 Thinking about the firm: A review of Daniel Spulber's the theory of the firm *Journal of Economic Literature* **49** (1) pp 101 – 113

http:www.aeaweb.org/articles.php?doi=10.1257/jel.49.1.101 .

**2674.** Kogut B, Zander U 1992 Knowledge of the firm, combinative capabilities, and the replication of technology *Organizational Science* **3** (3) pp 383 – 397.

**2675.** Kogut B, Zander U 2000 The network as knowledge: Generative rules and the emergence of structure *Strategic Management Journal* **21** pp 405 – 425.

**2676.** Prowse S 1992 The structure of corporate ownership in Japan *Journal of Finance* vol **47** pp 1121 – 1140.

**2677.** Barnhart S, Marr M W, Rosenstein S 1994 Firm performance and board composition some new evidence *Managerial and Decision Economics* **15** (4) pp 329 – 340.

**2678.** Short H 1994 Ownership, control, financial structure and the performance of the firms *Journal of Economic Surveys* vol **8** no 3 pp 204 – 247.

**2679.** Nonaka I, Takeuchi H 1995 The knowledge creating company: How Japanese companies create the dynamics of innovation *Oxford University Press* New York USA.

**2680.** Aharony J, Falk H, Chan-Jane Lin 1996 Changes in ownership structure and the value of the firm: The case of mutual-to-stock converting thrift institutions *Journal of Corporate Finance* vol **2** no 3 pp 301 – 316.

**2681.** Agrawal A, Knoeber C R 1996 Firm performance and mechanisms to control agency problems between managers an shareholders *Journal of Financial and Quantitative Analysis* vol **31** pp 377 – 397.

**2682.** Foss N J 1996 More critical comments on knowledge-based theories of the firm *Organizational Science* **7** (5) pp 519 – 523.

**2683.** Grant R M 1996a Toward a knowledge-based theory of the firm *Strategic Management Journal* **17** Winter Special Issue pp 109 – 122.

**2684.** Grant R M 1996b Prospering in dynamically-competitive environments: Organizational capability as knowledge integration *Organizational Science* **7** (4) pp 375 – 387.

**2685.** Grant R M 1996c The knowledge-based view of the firm *in* The strategic management of intellectual capital and organizational knowledge Choo C W, Bontis N (editors) *Oxford University Press* Oxford UK pp 133 – 148.

**2686.** Hansmann H 1996 Ownership of the enterprise *Oxford University Press* Oxford UK.

**2687.** Spender J C 1996 Making knowledge the basis of a dynamic theory of firm *Strategic Management Journal* **17** Special Issues pp 45 – 62.

2688. Rajan R, Zingales L May 1998 Power in a theory of the firm *Quarterly Journal of Economics*.

2689. Blair M 1999 Firm-specific human capital and theories of the firm *in* Employees and corporate governance Blair M, Roe M (editors) *Brookings Institution Press* Brookings institution Washington DC USA.

2690. Sterman 2000 Business dynamics *McGraw Hill* USA.

2691. Alavi M, Leidner D E 2001 Review: Knowledge management and knowledge management systems *MIS Quarterly* **25** (1) pp 1107 – 136.

2692. Donaldson L 2001 The contingency theory of organizations *Sage* London UK.

2693. Gompers P A, Metrick A, Ishii J L 2003 Corporate governance and equity prices *Quarterly Journal of Economics* **118** pp 107 – 156.

2694. Bhagat S, Black B, Blair M 2004 Relational investing and firm performance *Journal of Financial Research* **27** (1) pp 1 – 30.

2695. Nickerson J, Zenger T 2004 A knowledge –based theory of the firm: The problem-solving perspective *Organizational Science* **15** (6) pp 617 – 632.

2696. Perez-Gonzalez F 2006 Inherited control and firm performance *American Economic Review* 96 pp 1559 – 1588.

2697. Biondi Y, Canziani A, Kirat T (editors) 2007 The firm as an entity: Implications for economics, accounting and law *Routledge*.

2698. Kantarelis D 2007, 2017 Theories of the firm 5[th] edition *Inderscience* Genève Switzerland ISBN 0-907776-34-5 ISBN: 0-907776-35-3, ISBN 0-907776-62-0.

2699. Jacobides M G 2008 Competitive environments and redefining firm and industry boundaries *Public Lecture* London School of Economics and Political Science London UK.

2700. Jacobides M G 2008 Can firms shape their environments to gain an architectural advantage? *Public Lecture* London School of Economics and Political Science London UK.

2701. Spulber D F 2009 The theory of the firm: Microeconomics with endogenous entrepreneurs, firms, markets, and organizations *Cambridge University Press* UK http://www.cambridge.org/9780521736602 .

2702. Ledenyov D O, Ledenyov V O 2013b On the theory of firm in nonlinear dynamic financial and economic systems *Cornell University* NY USA pp 1 – 27 www.arxiv.org 1206.4426v2.pdf .

2703. Ledenyov D O, Ledenyov V O 2015c Information theory of firm *MPRA Paper no 63380* Munich University Munich Germany, *SSRN Paper no SSRN-id2587716 Social Sciences Research Network* New York USA pp 1 – 185 http://mpra.ub.uni-muenchen.de/63380/ , http://ssrn.com/abstract=2587716 .

2704. Wikipedia 2015k Microeconomics *Wikipedia* USA www.wikipedia.org .

***Public company investment, public company initial public offering valuation by rating agency/open market, stock exchange, financial capital investment product, financial capital investment medium in finances:***

2705. Berney R E 1964 The auction of long-term government securities *Journal Finance* **19** pp 470 – 479.

2706. Smith V L 1967 Experimental studies of discrimination versus competition in sealed-bid auction markets *Journal Business* **40** pp 56 – 60.

2707. Fama E F, Fisher L, Jensen M C, Roll R 1969 The adjustment of stock prices to new information *International Economic Review* **10** pp 1 – 21.

2708. Fama E F 1970 Efficient capital markets: A review of theory and empirical work *Journal of Finance* vol **25** pp 383 – 417.

2709. Fama E F, French K R 1992 The cross-section of expected stock returns *Journal of Finance* **47** (2) pp 427 – 465.

2710. Fama E F, French K R 1993 Common risk factors in the returns on stocks and bonds *Journal of Financial Economics* **33** (1) pp 3 – 56.

2711. Fama E F, French K 1996 Multifactor explanations of asset pricing anomalies *Journal of Finance* vol **51** no 1 pp 55 – 84.

2712. Fama E F 1998 Market efficiency, long-term returns, and behavioral finance *Journal of Financial Economics* **49** pp 283 – 306.

2713. Fama E, Hansen L P, Shiller R 2013 Lectures: 2013 Nobel Prize in economic sciences http://www.youtube.com/watch?v=WzxZGvrpFu4 , www.nobelprize.org .

2714. Akerlof G A 1970 The market for 'lemons': Quality uncertainty and the market mechanism *Quarterly Journal of Economics* **84** pp 488 – 500.

2715. Stoll H R, Curley A J 1970 Small business and the new issues market for equities *Journal of Financial and Quantitative Analysis* **5** pp 309 – 322.

2716. Logue D E 1973 On the pricing of unseasoned equity issues: 1965-1969 *Journal of Financial and Quantitative Analysis* **8** pp 91 – 103.

2717. Logue D E, Rogalski R, Seward J, Foster-Johnson L 2001 What's special about the role of underwriter reputation and market activities in IPOs? *Journal of Business*.

2718. Reilly F K 1973 Further evidence on short-run results for new issues investors *Journal of Financial and Quantitative Analysis* **8** pp 83 – 90.

2719. McDonald J G, Jacquillat B C 1974 Pricing of initial equity issues: The French sealed-bid auction *Journal Business* **47** pp 37 – 39.

2720. Ibbotson R G, Jaffe J 1975 Hot issue markets *Journal of Finance* vol **30** (5) pp 1027 – 1042.

2721. Ibbotson R G 1975 Price performance of common stocks new issues *Journal of Financial Economics* **2** (3) pp 235 – 272.

2722. Ibbotson R G, Sindeler J L, Ritter J R 1988 Initial public offering *Journal of Applied Corporate Finance* **1** (2) pp 37 – 45.

2723. Ibbotson R G, Sindelar J, Ritter J 1994 The market's problems with the pricing of initial public offerings *Journal of Applied Corporate Finance* **7** pp 66 – 74.

2724. Ibbotson R G, Ritter 1995 Initial public offerings *in* Jarrow R A, Maksimovic V, Ziemba W T (editors) Handbooks in operations research and management science: Finance pp 993 – 1016 *North-Holland Publishing* Amsterdam The Netherlands.

2725. Benston G, Smith C 1976 A transactions cost approach to the theory of financial intermediation *Journal of Finance* **31** pp 218 – 231.

2726. Jensen M C, Meckling H 1976 Theory of the firm: Managerial behavior, agency costs and ownership structure *Journal of Financial Economics* vol **3** (4) pp 305 – 360.

2727. Jensen M C 1986 Agency costs of free cash flow, corporate finance, and takeovers *The American Economic Review* **76** pp 323 – 329.

2728. Miller E M 1977 Risk, uncertainty, and divergence of opinion *Journal of Finance* pp 1151 – 1168.

2729. Leland H, Pyle D 1977 Informational asymmetries, financial structure, and financial intermediation *The Journal of Finance* **32** (2) pp 371 – 387.

2730. Weinstein M I 1978 The seasoning process of new corporate bond issue *Journal of Finance* **33** (5) pp 1343 – 1355.

2731. Kahneman D, Tversky A 1979 Prospect theory: An analysis of decision under risk *Econometrica* **47** pp 263 – 291.

2732. Wilson R 1979 Auctions of shares *Quarterly Journal of Economics* **93** pp 675 – 689.

2733. Brown S, Warner J B 1980 Measuring security price performance *Journal of Financial Economics* **8** pp 205 – 258.

2734. Buckland R, Herbert P J, Yeomans K A 1981 Price discount on new equity issues in the UK and their relationship to investor subscription in the period 1965 - 75 *Journal of Business Finance and Accounting* **8** pp 79 – 95.

2735. Milgrom P 1981 Rational expectations, information acquisition, and competitive bidding *Econometrica* **49** pp 921 – 943.

2736. Milgrom P, Weber R J 1982 A theory of auctions and competitive bidding *Econometrica* **50** pp 1089 – 1122.

2737. Myerson R 1981 Optimal auction design *Mathematics of Operation Research* **6** pp 58 – 73.

2738. Baron D P 1982 A model of the demand for investment banking advising and distribution services for new issues *Journal of Finance* **37** (4) pp 955 – 976.

2739. Dretske F 1983 The flow of information *MIT Press* Cambridge Massachusetts USA.

2740. Bhattacharya S, Ritter J R 1983 Innovation and communication: Signaling with partial disclosure *Review of Economic Studies* **50** pp 331 – 346.

2741. Ritter J R 1984 The "hot issue" market of 1980 *Journal of Business* **57** (2) pp 215 – 240.

2742. Ritter J R 1984 Signaling and the valuation of unseasoned new issues: A comment *Journal of Finance* **39** (4) pp 1231 – 1237.

2743. Beatty R P, Ritter J R 1986 Investment banking, reputation, and the underpricing of initial public offerings *Journal of Financial Economics* **15** pp 213 – 232.

2744. Ritter J R 1987 The costs of going public *Journal of Financial Economics* **19** pp 269 – 282.

2745. Ritter J R 1988 The buying and selling behavior of individual investors at the turn of the year *Journal of Finance* **43** (3) pp 701 – 717.

2746. Ritter J R 1991 The long-run underperformance of initial public offerings *Journal of Finance* vol **46** (1) pp 3 – 27.

2747. Chopra N, Lakonishok J, Ritter J R 1992 Measuring abnormal performance: Do stocks overreact? *Journal of Financial Economics* **31** (2) pp 235 – 268.

**2748.** Ritter J R 1992 The turn-of-the-year effect in The new Palgrave dictionary of money and finance Eatwell J, Milgate M, Newman P (editors) *MacMillan* London UK pp 705 – 706.

**2749.** Hanley K W, Ritter J R 1992 Going public pp 248 – 255 in The new Palgrave dictionary of money and finance Eatwell J, Milgate M, Newman P (editors) *MacMillan* London UK.

**2750.** Ibbotson R G, Sindelar J L, Ritter J R 1994 The market's problems with the pricing of initial public offerings *Journal of Applied Corporate Finance* **7** (1) pp 66 – 74.

**2751.** Loughran T, Ritter J R, Rydqvist K 1994 Initial public offerings: International insights *Pacific-Basin Finance Journal* **2** (2-3) pp 165 – 199.

**2752.** Ibbotson R G, Ritter J R 1995 Initial public offerings *Chapter 30* pp 993 – 1016 *in* Handbooks of operations research and management science: Finance Jarrow R, Maksimovic V, Ziemba W *North-Holland Publishing* Amsterdam The Netherlands.

**2753.** Loughran T, Ritter J R 1995 The new issue puzzle *Journal of Finance* **50** (1) pp 23 – 51.

**2754.** Lee I, Lochhead S, Ritter J R, Zhao Q 1996 The costs of raising capital *Journal of Financial Research* **19** (1) pp 59 – 74.

**2755.** Loughran T, Ritter J R 1996 Long-term market overreaction: The effect of low-priced stocks *Journal of Finance* **51** (5) pp 1959 – 1970.

**2756.** Loughran T, Ritter J R 1997 The operating performance of firms conducting seasoned equity offerings *Journal of Finance* 52 5 pp 1823 - 1850.

**2757.** Ritter J R 1998a Initial public offerings *in* Warren, Gorham, and Lamont Handbook of modern finance Logue D, Seward J *WGL/RIA* Boston and New York USA; *Contemporary Financial Digest* **2** (1) pp 5 – 30.

**2758.** Ritter J R 1998b Initial public offerings *in* Warren, Gorham, and Lamont *Handbook of modern finance* Logue D, Seward J (editors) Boston and New York USA.

**2759.** Hamao Y, Packer F, Ritter J R 1998 Institutional affiliation and the role of venture capital: Evidence from initial public offerings in Japan *Working Paper* Columbia University New York USA.

**2760.** Kim M, Ritter J R 1999 Valuing IPOs *Journal of Financial Economics* **53** (3) pp 409 – 437.

**2761.** Chen H C, Ritter J R 2000 The seven percent solution *Journal of Finance* **55** (3) pp 1105 – 1131.

**2762.** Hamao Y, Packer F, Ritter J R 2000 Institutional affiliation and the role of venture capital: Evidence from initial public offerings in Japan *Pacific-Basin Finance Journal* **8** (5) pp 529 – 558.

**2763.** Ritter J R 2002 The future of the new issues market *Brookings-Wharton Papers on Financial Services* pp 293 – 308.

**2764.** Ritter J R, Welch I 2002 A review of IPO activity, pricing and allocations *Journal of Finance* **57** (4) pp 1795 – 1828.

**2765.** Ritter J R, Warr R S 2002 The decline of inflation and the bull market of 1982 – 1999 *Journal of Financial and Quantitative Analysis* **37** (1) pp 29 – 61.

**2766.** Loughran T, Ritter J R 2002 Why don't issuers get upset about leaving money on the table in IPOs? *Review of Financial Studies* **15** (2) pp 413 – 443.

2767. Ritter J R 2003a Behavioral finance *Pacific-Basin Finance Journal* **11** (4) pp 429 – 437.

2768. Ritter J R 2003b Differences between European and American IPO markets *European Financial Management* **9** (4) pp 421 – 434.

2769. Ritter J R 2003c Investment banking and securities issuance *in* Handbook of the economics of finance Constantinides G M, Harris M, Stulz R (editors) *North-Holland Publishing* Amsterdam The Netherland pp 255 – 306.

2770. Ritter J R, Bradley D, Jordan B D 2003 The quiet period goes out with a bang *Journal of Finance* **58** (1) pp 1 – 36.

2771. Ritter J R, Bradley D J, Jordan B D, Wolf J G 2004 The IPO quiet period revisited *Journal of Investment Management* **2** (3) pp 3 – 13.

2772. Ritter J R, Loughran T 2004 Why has IPO underpricing changed over time? *Financial Management* **33** (3) pp 5 – 37.

2773. Ritter J R (editor) 2005 Introduction to recent developments in corporate finance *in* Recent developments in corporate finance vols **1**, **2** *Edward Elgar Publishing* Northampton MA.

2774. Ritter J R 2005 Economic growth and equity returns *Pacific-Basin Finance Journal* **13** (5) pp 489 – 503.

2775. Ritter J R 2005 Some facts about the 2004 IPO market *Working Paper* University of Florida USA http://bear.cba.ufl.edu/ritter .

2776. Ritter J R, Asquith P, Pathak P A 2005 Short interest, institutional ownership, and stock returns *Journal of Financial Economics* **78** (2) pp 243 – 276.

2777. Ritter J R, Nimalendran M, Donghang Zhang 2007 Do today's trades affect tomorrow's IPO allocation? *Journal of Financial Economics* **84** (1) pp 87 – 109.

2778. Ritter J R, Donghang Zhang 2007 Affiliated mutual funds and the allocation of initial public offerings *Journal of Financial Economics* **86** (2) pp 337 – 368.

2779. Bradley D J, Jordan B D, Ritter J R 2008 Analyst behavior following IPOs: The "bubble period" evidence *Review of Financial Studies* **21** (1) pp 101 – 133.

2780. Ritter J R 2008 Forensic finance *Journal of Economic Perspectives* **22** (3) pp 127 – 147.

2781. Ritter J R, Huang R 2009 Testing theories of capital structure and estimating the speed of adjustment *Journal of Financial and Quantitative Analysis* **44** (2) 237 – 271.

2782. Ritter J R, Xiaoding Liu 2010 The economic consequences of IPO spinning *Review of Financial Studies* **23** (5) pp 2024 – 2059.

2783. Ritter J R, Xiaohui Gao 2010 The marketing of seasoned equity offerings *Journal of Financial Economics* **97** (1) pp 33 – 52.

2784. Ritter J R, Xiaoding Liu 2011 Local underwriter oligopolies and IPO underpricing *Journal of Financial Economics* **102** (3) pp 579 – 601.

2785. Ritter J R 2011 Equilibrium in the initial public offerings market *Annual Review of Financial Economics* **3** pp 347 – 374.

2786. Ritter J R, Kenney M, Patton D 2012 Post-IPO employment and revenue growth for US IPOs, June 1996-2010 *Kauffman Foundation Report Ewing Marion Kauffman Foundation*.

**2787.** Ritter J R, Vismara S, Paleari S 2012 Europe's second markets for small companies *European Financial Management* **18** (3) pp 352 – 388.

**2788.** Ritter J R 2013 Re-energizing the IPO market *Chapter 4* pp 123 – 145 *in* Restructuring to speed economic recovery Bailey M, Herring R, Seki Y (editors) *Brookings Press* USA.

**2789.** Ritter J R, Xiaohui Gao, Zhongyan Zhu 2013 Where have all the IPOs gone? **48** (6) *Journal of Financial and Quantitative Analysis*.

**2790.** Ritter J R, Signori A, Vismara S 2013 Economies of scope and IPO activity in Europe *Chapter 1* pp 11 - 34 *in* Handbook of research on IPOs Levis M, Vismara S (editors) *Edward Elgar Publishing* Cheltenham UK.

**2791.** Myers S C, Majluf N S 1984 Corporate financing and investment decisions when firms have information that investors do not have *Journal of Financial Economics* **13** pp 187 – 22.

**2792.** Kyle A S 1985 Continuous auction and insider trading *Econometrica* **53** (6) pp 1315 – 1336.

**2793.** Amihud Y, Mendelson H 1986 Asset pricing and the bid-ask spread *Journal of Financial Economics* **17** (2) pp 223 – 249.

**2794.** Amihud Y, Mendelson H, Uno J 1999 Number of shareholders and stock prices: Evidence from Japan *Journal of Finance* **54** pp 1169 – 1184.

**2795.** Amihud Y, Hauser S, Kirsh A 2001 Allocations, adverse selection and cascades in IPOs: Evidence from Israel *Working Paper*.

**2796.** Amihud Y, Hauser S, Kirsh A 2003 Allocations, adverse selection and cascades in IPOs: Evidence from the Tel Aviv exchange *Journal of Financial Economics* **68** (1) pp 137 – 158.

**2797.** Beatty R P, Ritter J R 1986 Investment banking, reputation, and the underpricing of initial public offerings *Journal of Financial Economics* **15** pp 213 – 232.

**2798.** Beatty R P, Zajac E J 1994 Managerial incentives, monitoring, and risk bearing: A study of executive compensation, ownership, and board structure in initial public offerings *Administrative Science Quarterly* **39** pp 313 – 335.

**2799.** Beatty R P, Welch I 1996 Issuer expenses and legal liability in initial public offerings *Journal of Law and Economics* **39** pp 545 – 602.

**2800.** Booth J, Smith R 1986 Capital raising, underwriting and the certification hypothesis *Journal of Finance* **15** pp 261 – 281.

**2801.** Ridder A 1986 Access to the stock market - An empirical study on the efficiency of the British and the Swedish primary markets *Casion Press* Stockholm Sweden.

**2802.** Rock K 1986 Why new issues are underpriced *Journal of Financial Economics* **15** (1-2) pp 187 – 212.

**2803.** Shleifer A, Vishny R 1986 Large shareholders and corporate control *Journal of Political Economy* **94** (3) pp 461 – 488.

**2804.** Shleifer A, Wolfenzon D 2002 Investor protection and equity markets *Journal of Financial Economics* **66** (1) pp 3 – 27.

**2805.** Titman Sh, Trueman B 1986 Information quality and the valuation of new issues *Journal of Accounting and Economics* **8** pp 159 – 172.

2806. Bernheim D, Peleg B, Whinston M D 1987 Coalition-proof Nash equilibria I. Concepts *Journal of Economic Theory* **42** pp 1 – 12.

2807. McAfee R P, McMillan J 1987 Auctions and bidding *Journal Econ Lit* **25** pp 699 – 702.

2808. Miller R E, Reilly F K 1987 An examination of mispricing, returns and uncertainty for initial public offerings *Financial Management* **16** (2) pp 33 – 38.

2809. Balvers R J, McDonald B D, Miller R E 1988 Underpricing of new issues and the choice of auditor as a signal of investment banker reputation *The Accounting Review* **63** pp 605 – 622.

2810. Johnson J M, Miller R E 1988 Investment banker prestige and the underpricing of initial public offerings *Financial Management* **17** pp 19 – 29.

2811. Tinic S M 1988 Anatomy of initial public offerings of common stocks *Journal of Finance* **43** (4) pp 789 – 822.

2812. Allen F, Faulhaber G A 1989 Signaling by under-pricing in the IPO market *Journal of Financial Economics* **23** (2) pp 303 – 323.

2813. Barry C B 1989 Initial public offerings underpricing: The issuer's view — A comment *Journal of Finance* **44** pp 1099 – 1103.

2814. Barry C B, Muscarella C J, Peavy J W, Vetsuypens M R 1990 The role of venture capital in the creation of public companies: Evidence from the going-public process *Journal of Financial Economics* **27** pp 447 – 471.

2815. Benveniste L M, Spindt P A 1989 How investment bankers determine the offer price and allocation of new issues *Journal of Financial Economics* **24** (2) pp 343 – 362.

2816. Benveniste L M, Wilhelm W 1990 A comparative analysis of IPO proceeds under alternative regulatory environments *Journal of Financial Economics* **28** (1-2) pp 173 – 208.

2817. Benveniste L M, Busaba W Y, Wilhelm W J Jr 1996 Price stabilization as a bonding mechanism in new equity issues *Journal of Financial Economics* **42** pp 223 – 256.

2818. Benveniste S, Busaba W Y 1997 Book-building vs fixed price: An analysis of competing strategies for marketing IPOs *Journal of Financial and Quantitative Analysis* **32** (4) pp 383 – 403.

2819. Benveniste L M, Wilhelm W J 1997 Initial public offerings: Going by the book *Journal of Applied Corporate Finance* **10** pp 98 – 108.

2820. Benveniste L M, Erdal S M, Wilhelm W J Jr 1998 Who benefits from secondary market price stabilization of IPOs *Journal of Banking and Finance* **22** pp 741 – 767.

2821. Benveniste L M, Ljungqvist A P, Wilhelm W J, Yu X 2003 Evidence of information spillovers in the production of investment banking services *Journal of Finance* **58** pp 577 – 608.

2822. Grinblatt M, Hwang C 1989 Signaling and the pricing of new issues *Journal of Finance* **44** (2) pp 393 – 420.

2823. Koh F, Walter T 1989 A direct test of Rock's model of the pricing of unseasoned issues *Journal of Financial Economics* **23** pp 251 – 272.

2824. Maskin E S, Riley J G 1989 Optimal multi unit auctions *in* Franck Hahn (editor) The economics of missing markets, information, and game *Oxford University Press, Clarendon Press* Oxford UK pp 312 – 335.

2825. Muscarella C J, Vetsuypens M R 1989a A simple test of Baron's model of IPO underpricing *Journal of Financial Economics* **24** pp 125 – 136.

2826. Muscarella C J, Vetsuypens M R 1989b The under-pricing of second initial public offering *Journal of Financial Research* **12** (3) pp 183 – 192.

2827. Muscarella C, Vetsuypens M 1990 Firm age, uncertainty, and IPO underpricing: Some new empirical evidence *SMU Working Paper* USA.

2828. Uhlir H 1989 Der gang an die börse und das underpricing - phänomen: Eine empirische untersuchung Deutscher emissionen (1977-1987) *Zeitschrift für Bankrecht und Betriebswirtschaft* **1** pp 2 - 16.

2829. Welch I 1989 Seasoned offerings and the pricing of new issues *Journal of Finance* **44** (2) pp 421 – 450.

2830. Welch I 1992 Sequential sales, learning and cascades *Journal of Finance* **47** (2) pp 695 – 732.

2831. Welch I 1996 Equity offerings following the IPO: Theory and evidence *Journal of Corporate Finance* **2** (3) pp 227 – 259.

2832. Welch I, Ritter J 2002 A review of IPO activity, pricing and allocations *Yale ICF Working Paper no 02-01* Yale School of Management Yale University, University of Florida USA pp 1 - 45 http://papers.ssrn.com/abstract=296393 .

2833. Carter R B, Manaster S 1990 Initial public offerings and underwriter reputation *Journal of Finance* **45** (4) pp 1045 – 1068.

2834. Carter R B, Dark F A, Singh A K 1998 Underwriter reputation, initial returns, and the long run performance of IPO stocks *The Journal of Finance* **53** (1) pp 285 – 311.

2835. Clarkson P M, Thompson R 1990 Empirical estimates of beta when investors face estimation risk *Journal of Finance* **45** (2) pp 431 – 453.

2836. Husson B, Jacquillat B 1990 Sous-évaluation des titres et méthodes d'introduction au marché *Marché Finance* **11** (1) pp 123 – 134.

2837. Levis M 1990 The winner's curse problem, interest costs and the under-pricing of initial public offerings *Economic Journal* **100** (399) pp 76 – 89.

2838. Lucas D, McDonald R 1990 Equity issues and stock price dynamics *Journal of Finance* **45** pp 1019 – 1043.

2839. Sahlman W A 1990 The structure and governance of venture capital organizations *Journal of Financial Economics* **27** pp 473 – 524.

2840. Allen S N 1991 A Lawyer's guide to the operation of underwriting syndicates *New Eng Law Review* **26** (94) pp 350 – 351.

2841. Hasbrouck J 1991 Measuring the information content of stock trades *Journal of Finance* **46** (1) pp 179 – 208.

2842. Lee Ch M C, Shleifer A, Thaler R H 1991 Investor sentiment and the closed-end fund puzzle *Journal of Finance* **46** pp 75 – 109.

2843. Megginson W L, Weiss K A 1991 Venture capitalists certification in initial public offerings *Journal of Finance* **46** (3) pp 879 – 903.

2844. Megginson W L, Smart S B 2009 Introduction to corporate finance 2[nd] edition *Cengage Learning Inc* Florence Kentucky USA.

2845. Menon K, Williams D D 1991 Auditor credibility and initial public offerings *The Accounting Review* **66** pp 313 – 332.

2846. Spatt C, Srivastava S 1991 Preplay communication, participation restrictions and efficiency in initial public offerings *Review of Financial Studies* **4** (4) pp 709 – 726.

2847. Cotter J F 1992 The long run efficiency of IPO pricing *Working Paper* University of North Carolina at Chapel Hill NC USA.

2848. Hughes P J, Thakor A V 1992 Litigation risk, intermediation, and the underpricing of initial public offerings *Review of Financial Studies* **5** (4) pp 709 – 742.

2849. Mauer D C, Senbet W 1992 The effect of the secondary market on the pricing of initial public offerings: Theory and evidence *Journal of Financial and Quantitative Analysis* **27** (1) pp 55 – 79.

2850. Aggarwal R, Leal R, Hernandez F 1993 The aftermarket performance of initial public offerings in Latin America *Financial Management* **22** pp 42 – 53.

2851. Aggarwal R 2000 Stabilization activities by underwriters after initial public offerings *Journal of Finance* **55** (3) pp 1075 – 1103.

2852. Aggarwal R, Conway P 2000 Price discovery in initial public offerings and the role of the lead underwriter *Journal of Finance* **55** pp 2093 – 2922.

2853. Aggarwal R, Prabhala N, Puri M 2002 Institutional allocation in initial public offerings: Empirical evidence *Journal of Finance* **57** pp 1421 – 1442.

2854. Aggarwal R, Krigman L, Womack K 2002 Strategic IPO underpricing, information momentum, and lockup expiration selling *Journal of financial Economics* **66** (1) pp 106 – 137.

2855. Aggarwal R 2003 Allocation of initial public offerings and flipping activity *Financial Economics* **68** (1) pp 111 – 112.

2856. Affleck-Graves J, Hegde S, Miller R E, Reilly F K 1993 The effect of the trading system on the underpricing of initial public offerings *Financial Management* **22** (1) pp 99 – 108.

2857. Choe H, Masulis R, Nanda V 1993 Common stock offerings across the business cycle: Theory and evidence *Journal of Empirical Finance* **1** pp 3 – 31.

2858. Back K, Zender J 1993 Auctions of divisible goods: On the rationale for the treasury experiment *Review of Financial Studies* **6** pp 733 – 764.

2859. Back K, Zender J 2001 Auctions of divisible goods with endogenous supply *Economics Letters* **73** pp 29 – 34.

2860. Chemmanur Th J 1993 The pricing of initial public offerings: A dynamic model with information production *Journal of Finance* **48** (1) pp 285 – 304.

2861. Chemmanur Th J, Fulghieri P 1997 Why include warrants in new equity issues? A theory of unit IPOs *Journal of Financial and Quantitative Analysis* **32** (1).

2862. Chemmanur Th J, Fulghieri P 1999 A theory of the going-public decision *Review of Financial Studies* **12** pp 249 – 279.

2863. Chemmanur Th J, Liu H 2003 How should a firm go public? A dynamic model of the choice between fixed-price offerings and auctions in IPOs and privatizations *Working Paper* Boston College USA.

2864. Chemmanur T J, Hu G 2007 Institutional trading, allocation sales, and private information in IPOs *Western Finance Association 2007 Meetings Paper*.

2865. Chemmanur T J, Yan A 2009 Product market advertising and new equity issues *Journal of Financial Economics* **92** pp 40 – 65.

2866. Chemmanur T J, He S, Hu G 2009 The role of institutional investors in seasoned equity offerings *Journal of Financial Economics* **94** pp 384 – 411.

2867. Chemmanur Th J, He Sh, Nandy D 2010 The going public decision and the product market *Review of Financial Studies* **23** pp 1855 – 1908.

2868. Chemmanur T J, Krishnan K 2012 Heterogeneous beliefs, short sale constraints, and the economic role of the underwriter in IPOs *Financial Management* **41** pp 769 – 811.

2869. Chemmanur T J, He J 2012 IPO waves, product market competition, and the going public decision: theory and evidence *Discussion Paper CES 12-07* Center for Economic Studies US Census Bureau Washington DC USA pp 1 – 72.

2870. Conrad J, Kaul G 1993 Long-term market overreaction or biases in computed returns? *Journal of Finance* **48** pp 39 – 63.

2871. Dhatt M, Kim Y, Lim U 1993 The short-run and long-run performance of Korean IPOs: 1980-1990 *Working Paper* University of Cincinnati USA, Yonsei University South Korea.

2872. Drake P D, Vetsuypens M R 1993 IPO underpricing and insurance against legal liability *Financial Management* **22** pp 64 – 73.

2873. Figlewski S, Webb G P 1993 Options, short sales, and market completeness *Journal of Finance* **48** pp 761 – 777.

2874. Hanley K W 1993 The underpricing of initial public offerings and the partial adjustment Phenomenon *Journal of Financial Economics* **34** (2) pp 231 – 250.

2875. Hanley K W, Wilhelm W J 1995 Evidence on the strategic allocation of initial public offerings *Journal of Financial Economics* **37** (2) pp 239 – 257.

2876. Hebner K J, Hiraki T 1993 Japanese initial public offerings. Restructuring Japan's financial markets *in* Walter I, Hiraki T (editors) *Business One / Irwin* USA pp 79 – 113.

2877. Jegadeesh N, Weinstein M, Titman S 1993a Returns to buying winners and selling losers: Implications for stock market efficiency *Journal of Finance* **48** (1) pp 65 – 91.

2878. Jegadeesh N, Weinstein M, Welch I 1993b An empirical investigation of IPO returns and subsequent equity offerings *Journal of Financial Economics* **34** (2) pp 153 – 175.

2879. Keloharju M 1993 The winner's curse, legal liability, and the long-run price performance of initial public offerings in Finland *Journal of Financial Economics* **34** (2) pp 251 – 277.

2880. Keloharju M, Kulp K 1996 Market-to-book ratios, equity retention, and management ownership in Finnish initial public offerings *Journal of Banking and Finance* **20** pp 1583 – 1599.

2881. Keloharju M, Nyborg K, Rydqvist K 2004 Strategic behavior and underpricing in uniform price auctions: Evidence from Finnish treasury auctions *Journal of Finance*.

2882. Leleux B F 1993 Post-IPO performance: A French appraisal *Finance* vol **14** (2) pp 79 – 106.

2883. Leleux B F, Muzyka D F 1997 European IPO markets: The post-issue performance imperative *Entrepreneurship Theory and Practice* **21** (4) pp 111 – 118.

2884. Levis M 1993 The long-run performance of initial public offerings: The UK experience 1980 - 1988 *Financial Management* **22** (1) pp 28 – 41.

2885. Levis M 2004 The UK IPO market 2000 *Working Paper* Cass Business School UK.

2886. Loughran T 1993 NYSE vs NASDAQ: Market microstructure or the poor performance of initial public offerings *Journal of Financial Economics* **33** (2) pp 241 – 260.

2887. Loughran T, Ritter J R, Rydqvist K 1994 Initial public offerings: International insights *Pacific-Basin Finance Journal* **2** (2-3) pp 165 – 199.

2888. Loughran T, Ritter J R 1995 The new issue puzzle *Journal of Finance* **50** (1) pp 23 – 51.

2889. Loughran T, Ritter J R 1997 The operating performance of firms conducting seasoned equity offering *Journal of Finance* vol **52** (5) pp 1823 – 1850.

2890. Loughran T, Ritter J R 2000 Uniformly least powerful tests of market efficiency *Journal of Financial Economics* **55** (3) pp 361 – 389.

2891. Loughran T, Ritter J R 2002 Why don't issuers get upset about leaving money on the table in IPOs? *Review of Financial Studies* **15** (2) pp 413 – 443.

2892. Loughran T, Ritter J R 2003 Why has IPO underpricing changed over time? *Working Paper* University of Florida USA.

2893. Loughran T, Ritter J R 2004 Why has IPO underpricing changed over time? *Financial Management* **33** (3) pp 5 – 37.

2894. Ruud J S 1993 Underwriter price support and the IPO underpricing puzzle *Journal of Financial Economics* **34** (2) pp 135 – 151.

2895. Rydqvist K 1993 Compensation, participation, restrictions, and the underpricing of initial public offerings: Evidence from Sweden *Working Paper* Stockholm School of Economics Stockholm Sweden.

2896. Rydqvist K, Högholm K 1995 Going public in the 1980s: Evidence from Sweden *European Financial Management* **1** (3) pp 287 – 315.

2897. Rydqvist K 1997 IPO underpricing as tax-efficient compensation *Journal of Banking and Finance* **21** (3) pp 295 – 313.

2898. Vos E A, Cheung J 1993 New Zealand IPO underpricing: A reputation based model *Small Enterprise Research* **1** pp 13 – 22.

2899. Friedlan J 1994 Accounting choices by issuers of initial public offerings *Contemporary Accounting Research* **11** pp 1 – 32.

2900. Jain B A, Kini O 1994 The post-issue operating performance of IPO firms *Journal of Finance* vol **49** (5) pp 1699 – 1726.

2901. Jog V M, Srivastava A 1994 Underpricing of Canadian initial public offerings 1971 - 1992: An update *FINECO* **4** (1) pp 81 – 89.

2902. Jog V M, McConomy B J 1999 Voluntary disclosure of management earnings forecasts in IPOs and the impact on underpricing and post-issue return performance *Working Paper* School of Business Carlton University Ottawa Canada.

2903. Kunz R, Aggarwal R 1994 Why initial public offerings are underpriced: Evidence from Switzerland *Journal of Banking and Finance* **18** (4) pp 705 – 723.

*2904.* Lerner J 1994 Venture capitalists and the decision to go public *Journal of Financial Economics* **35** pp 293 – 316.

*2905.* Michaely R, Shaw W H 1994 The pricing of initial public offerings: Tests of the adverse-selection and signaling theories *Review of Financial Studies* **7** (2) pp 279 – 319.

*2906.* Michaely R, Womack K 1999 Conflict of interest and the credibility of underwriter analyst recommendations *Review of Financial Studies* **12** (4) pp 653 – 686.

*2907.* Schultz P H, Zaman M A 1994 Aftermarket support and underpricing of initial public offerings *Journal of Financial Economics* **35** (2) pp 199 – 219.

*2908.* Schultz P H 2003 Pseudo market timing and the long-run underperformance of IPOs *Journal of Finance* **58** (2) pp 483 – 518.

*2909.* Degeorge F 1995 Underwriter price support and the IPO underpricing puzzle: A comment *Working Paper* HEC Paris France.

*2910.* Degeorge F, Derrien F, Womack K L 2005 Quid pro quo in IPOs: Why book-building is dominating auctions? *Working Paper* University of Lugano Switzerland, University of Toronto Canada, Dartmouth College USA.

*2911.* Gerstein F S 1995 Private communications on the IPO underpricing, long term performance and emerging issues markets Copenhagen Denmark www.scornik-gerstein.com, www.scornik.com .

*2912.* Gerstein F S 1996 Private communications on the IPO underpricing, long term performance and emerging issues markets Brighton, London UK www.scornik-gerstein.com, www.scornik.com .

*2913.* Gompers P A 1995 Optimal investment, monitoring, and the staging of venture capital *Journal of Finance* **50** (5) pp 1461 – 1489.

*2914.* Gompers P, Lerner J 1997 Venture capital and the creation of public companies: Do venture capitalists really bring more than money? *Journal of Private Equity* **1** (1) pp 15 – 30.

*2915.* Gompers P, Lerner J 2001 The really long-term performance of initial public offerings: The pre-Nasdaq evidence *Unpublished Harvard Business School Working Paper* Harvard University USA.

*2916.* Gompers P A, Lerner J 2003a The really long-run performance of initial public offerings: The pre-Nasdaq evidence *Harvard Business School Working Paper* Harvard University USA.

*2917.* Gompers P A, Lerner J 2003b The really long-run performance of initial public offerings: The pre-Nasdaq evidence *Journal of Finance* **63** pp 1355 – 1392.

*2918.* Kim J-B, Krinsky I, Lee J 1995 The aftermarket performance of initial public offerings in Korea *Pacific-Basin Finance Journal* vol **3** no 4 pp 429 – 448.

*2919.* Spiess D K, Affleck-Graves J 1995 Underperformance in long-run stock returns following seasoned equity offerings *Journal of Financial Economics* **38** pp 243 – 267.

*2920.* Spiess D K, Pettway R H 1997 The IPO and the first seasoned equity sale: Issue proceeds, owner/manager's wealth, and the under-pricing signal *Journal of Banking and Finance* **21** (7) pp 967 – 988.

*2921.* Spiess D K, Affleck-Graves J 1999 The long-run performance of stock returns following debt offerings *Journal of Financial Economics* **54** pp 45 – 73.

2922. Zingales L 1995 Insider ownership and the decision to go public *Review of Economic Studies* **62** pp 425 – 448.

2923. Barber B M, Lyon J D 1996a Detecting long-run abnormal stock returns: The empirical power and specification of test statistics *Working Paper* University California - Davis California USA.

2924. Barber B M, Lyon J D 1996b How can long-run abnormal stock returns be both positively and negatively biased? *Working Paper* University California - Davis California USA.

2925. Barber B M, Lyon J D 1997 Detecting long-run abnormal stock returns: The empirical power and specification of test statistics *Journal of Financial Economics* **43** pp 341 – 372.

2926. Barber B M, Odean T 2008 All that glitters: The effect of attention on the buying behaviour of individual and institutional investors *Review of Financial Studies* **21** pp 785 – 818.

2927. Booth J R, Chua L 1996 Ownership dispersion, costly information, and IPO underpricing *Journal of Financial Economics* **41** (2) pp 291 – 310.

2928. Booth J R, Booth L 2003 Technology shocks, regulation, and the IPO market *Working Paper* Arizona State University Arizona USA.

2929. Borggreve H, Dobrikat J 1996 Telekom-börsengang beeinflusst mmarkt – pläne anderer unternehmen. Aktienemissionen werden weiter eine wichtige rolle spielen *Handelsblatt 1996/90* B6.

2930. Brealey R A, Myers S C 1996 Principles of corporate finance *McGraw-Hill* New York USA pp 1 – 998.

2931. Brennan M J, Subrahmanyam A 1996 Market microstructure and asset pricing: On the compensation for illiquidity in stock returns *Journal of Financial Economics* **41** (3) pp 441 – 464.

2932. Chowdhry B, Sherman A 1996 The winner's curse and international methods of allocating initial public offerings *Pacific-Basin Finance Journal* **4** pp 15 – 30.

2933. Chowdhry B, Nanda V 1996 Stabilization, syndication, and pricing of IPOs *Journal of Financial and Quantitative Analysis* **31** pp 25 – 42.

2934. Easley D, Kiefer N M, O'Hara M, Paperman J B 1996 Liquidity, information, and infrequently traded stocks *Journal of Finance* **51** (4) pp 1405 – 1436.

2935. Hazen Th L 1996 Treatise on the law of securities regulation *3rd edition*.

2936. Houston J, James C 1996 Bank information monopolies and the mix of private and public debt claims *Journal of Finance* **51** pp 1863 – 1899.

2937. Houston J, James C, Karceski J 2004 What a difference a month makes: Stock analyst valuations following initial public offerings *University of Florida Working Paper* USA.

2938. Kogut B, Zander U 1996 What firms do? Coordination, identity and learning *Organizational Science* **7** pp 502 – 518.

2939. Kothari S P, Warner J B 1996 Measuring long-horizon security price performance *Working Paper* University of Rochester USA.

2940. Kothari S, Warner J 1997 Measuring long-horizon security price performance *Journal of Financial Economics* **43** (3) pp 301 – 339.

*2941.* Kothari S P 2001 Capital markets research in accounting *Journal of Accounting and Economics* **31** pp 105 – 231.

*2942.* Lee P J, Taylor S, Walter T 1996a Australian IPO underpricing in the short and long run *Journal of Banking and Finance* **20** pp 1189 – 1210.

*2943.* Lee P J, Taylor S L, Walter T S 1996b Expected and realized returns for Singaporean IPOs: Initial and long-run analysis *Pacific-Basin Finance Journal* vol **4** no 2-3 pp 153 – 180.

*2944.* Lee P J, Taylor S L, Walter T S 1999 IPO underpricing explanations: Implications from investor application and allocation schedules *Journal of Financial and Quantitative Analysis* **34** pp 425 – 444.

*2945.* Nyborg K, Sundaresan S 1996 Discriminatory versus uniform treasury auctions: Evidence from when-issued transactions *Journal of Financial Economics* **42** pp 63 – 104.

*2946.* Pettway R H, Kaneko T 1996 The effect of removing price limits and introducing auctions upon short-term IPO returns: The case of Japanese IPOs *Pacific-Basin Finance Journal* **4** (2-3) pp 241 – 258.

*2947.* Périer S 1996 Gestion des résultats comptables et introduction en bourse *Ph D Thesis* University of Grenoble France.

*2948.* Aussenegg W 1997 Short and long-run performance of initial public offerings in the Austrian stock market *Working Paper no 24* Austrian Working Group on Banking and Finance Austria.

*2949.* Brav A, Gompers P 1997 Myth or reality? The long-run performance of initial public offerings: Evidence from venture and non-venture capital-backed companies *Journal of Finance* **52** (5) pp 1791 – 1821.

*2950.* Brav A 2000 Inference in long-horizon event studies: A Bayesian approach with applications to initial public offerings *Journal of Finance* **55** pp 1979 – 2016.

*2951.* Brav A, Geczy C, Gompers P A 2000 Is the abnormal return following equity issuances anomalous? *Journal of Financial Economics* **56** (2) pp 209 – 249.

*2952.* Brav A, Gompers P A 2002 Insider trading subsequent to initial public offerings: Evidence from expirations of lockup provisions *Review of Financial Studies*.

*2953.* Brav A, Gompers P A 2003 The role of lockups in initial public offerings *Review of Financial Studies* **16** p 1 - 29.

*2954.* Brennan M J, Franks J 1997 Underpricing, ownership and control in initial public offerings of equity securities in the UK *Journal of Financial Economics* **45** (3) pp 391 – 413.

*2955.* Cai J, Wei J 1997 The investment and operating performance of Japanese initial offerings *Pacific-Basin Finance Journal* vol **5** pp 389 – 417.

*2956.* Carhart M M 1997 On persistence in mutual funds performance *Journal of Finance* **52** (1) pp 57 – 82.

*2957.* Datta S, Iskandar-Datta M, Patel A 1997 The pricing of initial public offers of corporate straight debt *Journal of Finance* **52** (1) pp 379 – 396.

*2958.* Dechow P M, Sloan R G 1997 Returns to contrarian investment: Tests of the naïve expectations hypotheses *Journal of Financial Economics* **43** pp 3 – 28.

2959. Dechow P, Hutton A, Sloan R 1999 The relation between analysts' forecasts of long-term earnings growth and stock price performance following equity offerings *Working Paper* University of Michigan USA.

2960. Dechow P M, Hutton A, Sloan R G 2000 The relation between analysts' forecasts of long-term earnings growth and stock price performance following equity offerings *Contemporary Accounting Research* **17** (1) pp 1 – 32.

2961. Ehrhardt O 1997 Börseneinführungen von aktien am deutschen kapitalmarkt *Wiesbaden* Germany.

2962. Firth M 1997 An analysis of the stock market performance of new issues in New Zealand *Pacific-Basin Finance Journal* **5** pp 63 – 85.

2963. Gande A, Puri M, Saunders A, Walter I 1997 Bank underwriting of debt securities: Modern evidence *Review of Financial Studies* **10** pp 1175 – 1202.

2964. Gande A, Puri M, Saunders A 1999 Bank entry, competition, and the market for corporate securities underwriting *Review of Financial Economics* **54** pp 165 – 195.

2965. Gregg S 1997 Regulation "A" initial public offerings on the Internet: A new opportunity for small businesses? *Journal of Small and Emerging Business* **50** pp 417 – 433.

2966. Huang R D, Stoll H R 1997 The components of the bid-ask spread: A general approach *Review of Financial Studies* **10** (4) pp 995 – 1034.

2967. Kooli M 2000 La sous-évaluation des émissions initiales: Le cas du Canada *Gestion* **25** (3) pp 78 – 91.

2968. La Porta R, Lopez-de-Silanes F, Shleifer A, Vishny R W 1997 Legal determinants of external finance *Journal of Finance* **52** pp 1131 – 1150.

2969. La Porta R, Lopez-de-Silanes F, Shleifer A, Vishny R 1998 Law and finance *Journal of Political Economy* **106** (6) pp 1113 – 1156.

2970. La Porta R, Lopez-de-Silanes F, Shleifer A 2002 Investor protection and corporate valuation *Journal of Finance* **57** (3) pp 1147 – 1170.

2971. La Porta R, Lopez-de-Silanes F, Shleifer A 2006 What works in securities laws? *The Journal of Finance* **61** (1) pp 1 – 32.

2972. Lee I 1997 Do firms knowingly sell overvalued equity? *Journal of Finance* **52** (4) pp 1439 – 1466.

2973. Ljungqvist A P 1997 Pricing initial public offerings: Further evidence from Germany *European Economic Review* **41** (7) pp 1309 – 1320.

2974. Ljungqvist A, Nanda V, Singh R 2001 Hot markets, investor sentiment, and IPO pricing, *Unpublished New York University Working Paper* New York USA.

2975. Ljungqvist A P, Wilhelm W J 2002 IPO allocations: Discriminatory or discretionary? *Journal of Financial Economics* **65** (2) pp 167 – 201.

2976. Ljungqvist A P, Wilhelm W J 2003 IPO pricing in the dot-com bubble *Journal of Finance* **58** (2) pp 723 – 752.

2977. Ljungqvist A, Jenkinson T, Wilhelm W 2003 Global integration in primary equity markets: The role of US banks and US investors *Review of Financial Studies* **16** (1) pp 63 – 99.

*2978.* Ljungqvist A P, Marston F, Wilhelm W J 2003 Competing for securities underwriting mandates: Banking relationships and analyst recommendations *NYU Working Paper* New York USA.

*2979.* Ljungqvist A P, Nanda V, Singh R 2003 Hot markets, investor sentiment, and IPO pricing *Stern School of Business Working Paper* New York University New York USA.

*2980.* Ljungqvist A P, Nanda V, Singh R 2006 Hot markets, investor sentiment, and IPO pricing *The Journal of Business* **79** (4) pp 1667 – 1702.

*2981.* Ljungqvist A 2006 IPO underpricing *in* Handbook of empirical corporate finance Eckbo B E (editor) chapter 3 *North-Holland* Amsterdam The Netherlands.

*2982.* Mikkelson W H, Partch M M, Shah K 1997 Ownership and operating performance of companies that go public *Journal of Financial Economics* vol **44** pp 281 – 307.

*2983.* Nanda V, Youngkeol Yun 1997 Reputation and financial intermediation: An empirical investigation of the impact of IPO mispricing on underwriter market value *Journal of Financial Intermediation* **6** pp 39 – 63.

*2984.* Page M J, Reyneke I 1997 The timing and subsequent performance of initial public offerings (IPOs) on the Johannesburg stock exchange *Journal of Business Finance & Accounting* **24** pp 1401 – 1420.

*2985.* Rajan R G, Servaes H 1997a Analyst following of initial public offerings *Journal of Finance* **52** (2) pp 507 – 529.

*2986.* Rajan R G, Servaes H 1997b The effect of market conditions on initial public offerings *Journal of Finance* vol **52** (2) pp 507 – 529.

*2987.* Stehle R 1997 Der size-effekt am Deutschen kapitalmarkt *Zeitschrift für Bankrecht und Bankwirtschaft* **9** pp 237 – 260.

*2988.* Stehle R, Ehrhardt O 1999 Renditen bei börseneinführungen am Deutschen kapitalmarkt *Zeitschrift für Betriebswirtschaft* **69** pp 1395 – 1422.

*2989.* Stehle R, Ehrhardt O, Przyborowsky R 2000 Long-run stock performance of German initial public offerings and seasoned equity issues *European Financial Management* **6** pp 173 – 196.

*2990.* Steib S, Mohan N 1997 The German re-unification, changing capital market conditions, and the performance of German initial public offerings *Quarterly Review of Economics & Finance* **37** (1) pp 115 – 137.

*2991.* Su D, Fleisher B M 1997 An empirical investigation of underpricing in Chinese IPOs *American Economic Association* New Orleans USA.

*2992.* Arkebauer J 1998 Going public *Dearborn*.

*2993.* Asquith D, Jones J, Kieschnick R 1998 Evidence on price stabilization and underpricing in early IPO returns *Journal of Finance* **53** (5) pp 1759 – 1773.

*2994.* Ausubel L, Cramton P 1998a Demand reduction and inefficiency in multi - unit auctions *Working Paper* University of Maryland USA.

*2995.* Ausubel L, Cramton P 1998b Auctioning securities *Working Paper* University of Maryland USA.

*2996.* Black B S, Gilson R J 1998 Venture capital and the structure of capital markets: Banks versus stock markets *Journal of Financial Economics* **47** pp 243 – 277.

**2997.** Daniel K, Hirshleifer D, Subrahmanyam A 1998 Investor psychology and security market under- and over- reactions *Journal of Finance* **53** pp 1839 – 1886.

**2998.** Goergen M 1998 Corporate governance and financial performance *Edward Elgar Publishing Ltd* Cheltenham UK.

**2999.** Goergen M, Renneboog L 2002 Prediction of control concentration in German and UK IPOs *in* Renneboog L, McCahery J, Moerland P, Raaijmakers T (editors) Convergence and diversity of corporate governance regimes and capital markets *Oxford University Press* Oxford UK.

**3000.** Helwege J, Kleiman P 1998 The pricing of high-yield debt IPOs *Journal of Fixed Income* **8** pp 61 – 68.

**3001.** Helwege J, Liang N 2001 Initial public offerings in hot and cold markets *Working Paper* Ohio State University USA.

**3002.** Helwege J, Liang N 2004 Initial public offerings in hot and cold markets *Journal of Financial and Quantitative Analysis* vol **39** pp 541 – 569.

**3003.** Kahn C, Winton A 1998 Ownership structure, speculation, and shareholder intervention *Journal of Finance* **53** (1) pp 99 – 129.

**3004.** Malvey P, Archibald C 1998 Uniform price auctions: Update of the treasury experience *Report of the US Department of the Treasury* US Government USA.

**3005.** Mello A, Parsons J 1998 Going public and the ownership structure of the firm *Journal of Financial Economics* **49** pp 79 – 109.

**3006.** Mok H M K, Hui Y V 1998 Underpricing and aftermarket performance of IPOs in Shanghai, China *Pacific-Basin Finance Journal* **6** (5) pp 453 – 474.

**3007.** Pagano M, Panetta F, Zingales L 1998 Why do companies go public: An empirical analysis *Journal of Finance* vol **53** (1) pp 27 – 64.

**3008.** Pagano M, Röell A 1998 The choice of stock ownership structure: Agency costs, monitoring and the decision to go public *The Quarterly Journal of Economics (February)* pp 187 – 225.

**3009.** Paudyal K, Saadouni B, Briston R J 1998 Privatisation initial public offerings in Malaysia: Initial premium and long-term performance *Pacific-Basin Finance Journal* **6** (5) pp 427 – 451.

**3010.** Poon W P H, Firth M, Fung H G 1998 Asset pricing in segmented capital markets: Preliminary evidence from China-domiciled companies *Pacific-Basin Finance Journal* **6** pp 307 – 319.

**3011.** Rangan S 1998 Earnings management and the performances of seasoned equity offerings *Journal of Financial Economics* vol **13** pp 1987 – 2221.

**3012.** Reese W A 1998 IPO underpricing, trading volume, and investor interest Working Paper Tulane University.

**3013.** Sapusek A 1998 Empirical evidence on the long-run performance of initial public offerings in Germany *Banque & Marchés* **34** pp 38 – 45.

**3014.** Stoughton N, Zechner J 1998 IPO-mechanisms, monitoring and ownership structure *Journal of Financial Economics* **49** (1) pp 45 – 77.

**3015.** Taylor S, Whittred G 1998 Security design and the allocation of voting rights: Evidence from the Australian IPO market *Journal of Corporate Finance* vol **4** pp 107 – 131.

3016. Theoh S H, Welch I, Wong T J 1998a Earnings management and the underperformance of seasoned equity offerings *Journal of Financial Economics* vol **50** pp 63 – 99.

3017. Theoh S H, Welch I, Wong T 1998b Earnings management and the long-run market performance of initial public offerings *Journal of Finance* **53** (6) pp 1935 – 1974.

3018. Ansotegui O C, Fabregat J 1999 Initial public offerings on the Spanish stock exchange *Working Paper* Escuela Superior de Administración y Dirección de Empresas (ESADE) Spain.

3019. Arcas M, Ruiz F 1999 Las ofertas públicas de venta (OPVs) de acciones en el mercado bursatil Español: Privatizationes frente a no privatizaciones *Cuadernos de Economia y Direccion de la Empresa* **4** pp 325 – 347.

3020. Baker M, Gompers P A 1999 An analysis of executive compensation, ownership, and control in closely held firms *Working Paper* Harvard Business School Harvard University.

3021. Baker M, Wurgler J 2000 The equity share in new issues and aggregate stock returns *The Journal of Finance* **55** pp 2219 – 2257.

3022. Baker M, Gompers P A 2001 The determinants of board structure at the initial public offering *Working Paper* Harvard Business School Harvard University USA.

3023. Baker H K, Nofsinger J R, Weaver D G 2002 International cross-listing and visibility *Journal of Financial and Quantitative Analysis* **37** pp 495 – 521.

3024. Brown E 1999 Long-run performance analysis of a new sample of UK IPOs *Working Paper* University of Edinburgh UK.

3025. Cornelli F, Goldreich D 1999 Book building and strategic allocation *IFA Working Paper no 286* http://ssrn.com/abstract=157352 .

3026. Cornelli F, Goldreich D 2001 Book-building and strategic allocations *Journal of Finance* **56** (6) pp 2337 – 2370.

3027. Cornelli F, Goldreich D 2002 Book building: How informative is the order book? *Unpublished London Business School Working Paper* London UK.

3028. Cornelli F, Goldreich D 2003 Book building: How informative is the order book? *Journal of Finance* **58** pp 1415 – 1444.

3029. Cornelli F, Goldreich G, Ljungqvist A 2006 Investor sentiment and pre-IPO markets *Journal of Finance* vol **61** issue 3 pp 1187 – 1216.

3030. Field L C 1999 Control considerations of newly public firms: The implementation of anti-takeover provisions and dual class shares before the IPO *Working Paper* Pennsylvania State University USA.

3031. Field L C, Hanka G 2001 The expiration of IPO share lockups *Journal of Finance* **56** pp 471 – 500.

3032. Field L C, Sheehan D P 2001 Underpricing in IPOs: Control, monitoring, or liquidity? *Unpublished Penn State Working Paper* USA.

3033. Field L C, Karpoff J 2002 Takeover defenses of IPO firms *Journal of Finance* **57** pp 1857 – 1890.

3034. Field L C, Sheehan D P 2002 IPO underpricing and outside block holdings *Working Paper* Penn State University USA www.ssrn.com .

3035. Field L C, Lowry M 2009 Institutional versus individual investment in IPOs: The importance of firm fundamentals *Journal of Financial and Quantitative Analysis* **44** pp 489 – 516.

3036. Kandel S, Sarig O, Wohl A 1999 The demand for stocks: An analysis of IPO auctions *Review of Financial Studies* **12** pp 227 – 248.

3037. Khurshed A, Mudambi R, Goergen M 1999 On the long-run performance of IPOs *Working Paper* ISMA Centre University of Reading UK.

3038. Khurshed A, Mudambi R 2002 The short-run price performance of investment trust IPOs on the UK main market *Applied Financial Economics* **12** pp 697 – 706.

3039. Krigman L, Shaw W H, Womack K L 1999 The persistence of IPO mispricing and the predictive power of flipping *Journal of Finance* **54** (3) pp 1015 – 1044.

3040. Krigman L, Shaw W, Womack K 2001 Why do firms switch underwriters? *Journal of Financial Economics* **60** (2-3) pp 245 – 284.

3041. Lyon J D, Barber B M, Tsai C - L 1999 Improved methods for tests of long-run abnormal stock returns *Journal of Finance* **54** (1) pp 165 – 201.

3042. Olson J F, Nelson D W 1999 What makes a company a good candidate for going public? Criteria, advantages, and disadvantages related to going public *American Law Institute-American Bar Association Continuing Legal Education* 7/1999 pp 591 – 593.

3043. Short H, Keasey K 1999 Managerial ownership and the performance of firms: Evidence from the UK *Journal of Corporate Finance* vol **5** pp 79 – 101.

3044. Stulz R M 1999 Globalization, corporate finance, and the cost of capital *Journal of Applied Corporate Finance* **12** pp 8 – 25.

3045. Stulz R M 2005 The limits of financial globalization *Journal of Finance* **60** pp 1595 – 1638.

3046. Stulz R M 2009 Securities laws, disclosure, and national capital markets in the age of financial globalization *Journal of Accounting Research* **47** pp 349 – 390.

3047. Subramanyam A, Titman Sh 1999 The going public decision and the development of financial markets *Journal of Finance* **54** pp 1045 – 1082.

3048. Thomas J, Zhang X 1999 Identifying unexpected accruals: A comparison of current approaches *Working Paper* Columbia University New York USA www.ssrn.com .

3049. Arosio R, Giudici G, Paleari S 2000 What drives the initial market performance of Italian IPOs? An empirical investigation on underpricing and price support *Working Paper* University of Bergamo Italy.

3050. Arosio R, Giudici G, Paleari S 2001 Why do (or did?) Internet-stock IPOs leave so much money on the table? *Politecnico di Milano Working Paper* Milan Italy.

3051. Aussenegg W 2000 Privatization versus private sector initial public offerings in Poland *Multinational Finance Journal* **4** (1-2) pp 69 – 99.

3052. Berkman H, Bradbury M E, Ferguson J 2000 The accuracy of price-earnings and discounted cash flow methods of IPO equity valuation *Journal of International Financial Management and Accounting* **11** (2) pp 71 – 83.

3053. Binmore K, Swierzbinski J 2000 Treasury auctions: Uniform or discriminatory? *Review of Economic Design* **5** pp 387 – 410.

3054. Boehmer E, Fishe R P H 2000 Do underwriters encourage stock flipping? A new explanation for the under-pricing of IPOs *Working Paper* University of Miami USA.

3055. Boehmer E, Fishe P R 2001 Equilibrium rationing in initial public offerings of equity *Unpublished University of Miami Working Paper* University of Miami USA.

3056. Boehmer E, Fishe R P H 2005 Who receives IPO allocations? An analysis of 'regular' investors *Working Paper* American Finance Association (AFA) 2005 Meeting Philadelphia.

3057. Brailsford T, Heaney R, Powell J G, Shi J 2000 Hot and cold IPO markets: Identification using a regime switching model *Multinational Finance Journal* **4** (1/2) pp 35 – 68.

3058. Brailsford T, Heaney R, Shi J 2004 Modeling the behavior of the new issue market *International Review of Financial Analysis* **13** (2) pp 119 – 132.

3059. Chen G M, Firth M, Kim J 2000 The post-issue market performance of IPOs in China's new stock markets *Review of Quantitative Finance and Accounting* **14** pp 319 – 339.

3060. D'Mello R, Ferris S P 2000 The information effects of analyst activity at the announcement of new equity issues *Financial Management* **29** (1) pp 78 – 95.

3061. Draho J 2000 The timing of initial public offerings: A real option approach SSRN: 271351.

3062. Dunbar C 2000 Factors affecting investment bank initial public offering market share *Journal of Financial Economics* **55** pp 3 – 41.

3063. Dunbar C G, Foerster S 2008 Second time lucky? Withdrawn IPOs that return to the market *Journal of Financial Economics* **87** pp 610 – 635.

3064. Duque J, Almeida M 2000 Ownership structure and initial public offerings in small economies – The case of Portugal *Working Paper* Technical University of Lisbon Portugal.

3065. Eckbo B E, Masulis R W, Norli O 2000 Seasoned public offerings: Resolution of the new issues puzzle *Journal of Financial Economics* **56** (2) pp 251 – 291.

3066. Eckbo E, Norli O 2001 Leverage, liquidity, and long-run IPO returns *Unpublished Dartmouth Working Paper*.

3067. Eckbo B E, Norli Ø 2002 Liquidity risk, leverage, and long-run IPO returns *Dartmouth Tuck School of Business Working Paper* USA.

3068. Eckbo B, Norli Ø 2005 Liquidity risk, leverage and long-run IPO returns *Journal of Corporate Finance* **11** pp 1 – 35.

3069. Eckbo B E 2008 Handbook of corporate finance vol **1** Handbook of Finance Series *Elsevier / North-Holland* The Netherlands.

3070. Ellis K, Michaely R, O'Hara M 2000 When the underwriter is the market maker: An examination of trading in the IPO aftermarket *Journal of Finance* **55** no 3 pp 1039 – 1074.

3071. Ellis K, Michaely R, O'Hara M 2002 The making of a dealer market: From entry to equilibrium in the trading of Nasdaq stocks *Journal of Finance* **57** no 5 pp 2289 – 2316.

3072. Fabrizio S 2000 Asymmetric information and underpricing of IPOs: The role of the underwriter, the Prospectus and the analysts – An empirical examination of the Italian situation *Economic Research Department Italian Securities Exchange Commission* Rome Italy.

3073. Foerster B 2000 Securities industry association's capital markets handbook *2nd edition Aspen Publishers* Aspen Colorado USA.

3074. Gilbert R, Klemperer P 2000 An equilibrium theory of rationing *Rand Journal of Economics* **31** pp 1 – 21.

3075. Jain B, Kini O 2000 Does the presence of venture capitalists improve the survival profile of IPO firms? *Journal of Business Finance and Accounting* **27** (9-10) pp 1139 – 1176.

3076. Kiymaz H 2000 The initial and after market performance of IPOs in an emerging market: Evidence from Istanbul stock exchange *Journal of Multinational Financial Management* **10** (2) pp 213 – 227 UK.

3077. Koskie J L, Michaely R 2000 Prices, liquidity, and the information content of trades *Review of Financial Studies* **13** pp 659 – 696.

3078. Lewis C, Seward J, Foster-Johnson L 2000 Busted IPOs and windows of misopportunity *Working Paper* Dartmouth College.

3079. Löffler T 2000 Investorenunsicherheit bei börsenerstemissionen *Finanzmarkt und Portfolio Management* **14** pp 57 – 75.

3080. Reuschenbach H 2000 Der börsengang der Deutschen Telekom AG *in* Volk G Going public – der gang an die börse *Beratung, Erfahrung, Begleitung* 3rd edition Stuttgart Germany pp 159 – 187.

3081. Sapusek A 2000 Benchmark - sensitivity of IPO-long-run performance: An empirical study for Germany *Schmalenbach Business Review* **52** pp 374 – 405.

3082. Schultz P 2000 The timing of initial public offerings *Unpublished University of Notre Dame Working Paper.*

3083. Schultz P 2001 Pseudo market timing and the long-run underperformance of IPOs *Unpublished University of Notre Dame Working Paper.*

3084. Schultz P, Zaman M 2001 Do the individuals closest to internet firms believe they are overvalued? *Journal of Financial Economics* **59** (3) pp 347 – 381.

3085. Sinclair G 2000 Internet direct public offerings: New opportunities for small business capital finance *Manitoba Law Journal* vol **27** no 3.

3086. Sherman A E 2000 IPOs and long-term relationships: An advantage of book building *Review Financial Studies* **13** (3) pp 697 – 714.

3087. Sherman A 2001 Global trends in IPO methods: Book building versus auctions *Unpublished University of Notre Dame Working Paper.*

3088. Sherman A E, Titman S 2002 Building the IPO order book: Under-pricing and participation limits with costly information *Journal of Financial Economics* **65** (1) pp 3 – 29.

3089. Sherman A E 2003 Global trends in IPO methods: Book-building vs auctions *Working Paper* University of Notre-Dame.

3090. Smart S B, Zutter C J 2000 Control as a motivation for underpricing: A comparison of dual- and single- class IPOs *Working Paper* Indiana University USA.

3091. Stehle R, Ehrhardt O, Przyborowsky R 2000 Long-run stock performance of German initial public offerings and seasoned equity issues *European Financial Management* **6** (2) pp 173 – 196.

*3092.* Westerholm J 2000 Essays on initial public offerings – Empirical findings from the Helsinki stock exchange *Research Reports 48* Hanken Swedish School of Economics and Business Administration Stockholm Sweden.

*3093.* Von Eije J H, de Witte M C, van der Zwaan A H 2000 IPO-related organizational change and long-term performance *Faculty of Management and Organization* University of Groningen The Netherlands pp 1 - 18 http://som.rug.nl/ .

*3094.* Bernardo A E, Welch I 2001 On the evolution of overconfidence and entrepreneurs *Journal of Economics and Management Strategy* **10** pp 301 – 330.

*3095.* Bradley D J, Jordan B D, Ha-Chin Yi, Roten I C 2001 Venture capital and IPO lockup expiration: An empirical analysis *Journal of financial Research* **24** pp 465 – 494.

*3096.* Bradley D J, Jordan B 2002 Partial adjustment to public information and IPO underpricing *Journal of Financial and Quantitative Analysis* **37** pp 595 – 616.

*3097.* Bradley D J, Jordan B, Ritter J R 2003 The quiet period goes out with a bang *Journal of Finance* **58** pp 1 – 36.

*3098.* Bradley D J, Jordan B, Ritter J 2008a Analyst behavior following IPOs: The 'bubble period' evidence *Review of Financial Studies* **21** (1) pp 101 – 133.

*3099.* Bradley D J, Chan K, Kim J, Singh A 2008b Are there long-run implications of analysts' coverage for IPOs? *Journal of Banking and Finance* **32** (6) pp 1120 – 1132.

*3100.* Busaba W Y, Benveniste L M, Guo R J 2001 The option to withdraw IPOs during the premarket: Empirical analysis *Journal of Financial Economics* **60** pp 73 – 102.

*3101.* Certo S T, Covin J G, Daily C M, Dalton D R 2001 Wealth and the effects of founder management among IPO-stage new ventures *Strategic Management Journal* **22** pp 641 – 658.

*3102.* Chan K, Wang J, Wei K C 2001 Underpricing and long-term performance of IPOs in China *Journal of Corporate Finance* **183** pp 1 – 22.

*3103.* Cooney J W, Singh A K, Carter R B, Dark F H 2001 IPO initial returns and underwriter reputation: Has the inverse relationship flipped in the 1990s? *University of Kentucky Case Western Reserve University, Iowa State University Working Paper* USA.

*3104.* Daines R, Klausner M 2001 Do IPO charters maximize firm value? Antitakeover protection in IPOs *Journal of Law, Economics and Organization* vol **17** pp 83 – 120.

*3105.* Danielsen B R, Sorescu S M 2001 Why do option introductions depress stock prices? An empirical study of diminishing short-sale constraints *Journal of Financial and Quantitative Analysis* **36** pp 451 – 484.

*3106.* Degeorge F, Derrien F 2001a IPO Performance and earnings expectations: Some French evidence *Working Paper* HEC Paris France.

*3107.* Degeorge F, Derrien F 2001b Les déterminants de la performance à long terme des introductions en bourse : Le cas français *Banque & Marchés* **55** pp 8 – 18.

*3108.* Derrien F, Womack K L 2002 Auctions vs book building and the control of underpricing in hot IPO markets *Review of Financial Studies*.

*3109.* Derrien F 2005 IPO pricing in hot market conditions: Who leaves money on the table? *Journal of Finance* **60** (1) pp 487 – 521.

3110. Derrien F, Kecskés A 2006 The initial public offerings of listed firms *Journal of Finance*.

3111. Derrien F 2007 Currying favor to win IPO mandates *Working Paper AFA Conference* Chicago USA.

3112. DuCharme L L, Rajgopal S, Sefcik S E 2001 Why was Internet IPO underpricing so severe? *University of Washington Working Paper* Washington DC USA.

3113. Francis B B, Hasan I 2001 The underpricing of venture and non-venture capital IPOs: An empirical investigation *Journal of Financial Services Research* **19** pp 93 – 113.

3114. Gerke W, Fleischer J 2001 Die performance der börsengänge am neuen markt *Zeitschrift für betriebswirtschaftliche Forschung* **53** pp 827 – 839.

3115. Habib M, Ljungqvist A 2001 Underpricing and entrepreneurial wealth losses: Theory and evidence *Review of Financial Studies* **14** pp 433 – 458.

3116. Hahn T, Ligon J A 2004 Liquidity and initial public offering underpricing *College of Business and Economics* University of Idaho USA.

3117. Hansen R S 2001 Do investment banks compete in IPOs?: The advent of the "7% plus contract" *Journal of Financial Economics* **59** pp 313 – 346.

3118. Heaton J B 2001 Managerial optimism and corporate finance *Unpublished Bartlet Beck Working Paper* Chicago Illinois USA.

3119. Hoffmann-Burchardi U 2001 Clustering of initial public offerings, information revelation and underpricing *European Economic Review* **45** (2) pp 353 – 383.

3120. Holmén M, Högfeldt P 2001 A law and finance analysis of initial public offerings *Working Paper* Stockholm University Stockholm Sweden.

3121. Houge T, Loughran T, Suchanek G, Yan X 2001 Divergence of opinions, uncertainty, and the quality of initial public offerings *Financial Management* **30** (4) pp 5 – 23.

3122. Jakobsen J B, Sørensen O 2001 Decomposing and testing long-term returns: An application on Danish IPOs *European Financial Management* **7** pp 393 – 417.

3123. Jenkinson T, Ljungqvist A 2001 Going public, the theory and evidence on how companies raise equity finance *2nd edition University of Oxford Press* New York USA.

3124. Jenkinson T, Jones H 2004 Bids and allocation in European IPO book-building *Journal of Finance* **59** (5) pp 2309 – 2339.

3125. Jenkinson T, Morrison A, Wilhelm W 2006 Why are European IPOs so rarely prices outside the indicative price range? *Journal of Financial Economics* **80** pp 185 – 209.

3126. Jenkinson T, Jones H 2007 The economics of IPO stabilization, syndicates and naked shorts *European Financial Management* **13** pp 616 – 642.

3127. Killian L, Smith K, Smith W 2001 IPOs for everyone: The 12 secrets of investing in IPOs *John Wiley & Sons* Indiana USA.

3128. Lowry M, Schwert G W 2001 Biases in the IPO pricing process *Working Paper* Penn State University USA.

3129. Lowry M, Schwert G W 2002 IPO market cycles: Bubbles or sequential learning? *Journal of Finance* **57** (3) pp 1171 – 1200.

3130. Lowry M, Shu S 2002 Litigation risk and IPO under-pricing *Journal of Financial Economics* **65** (3) pp 309 – 335.

3131. Lowry M 2003 Why does IPO volume fluctuate so much? *Journal of Financial Economics* **67** pp 3 – 40.

3132. Mager F B 2001 Die performance von unternehmen vor und nach dem börsengang *Wiesbaden* Germany.

3133. Maksimovic V, Pichler P 2001 Technological innovation and initial public offerings *Review of Financial Studies* **14** pp 459 – 494.

3134. Purnanandam A K, Swaminathan B 2001 Are IPOs underpriced? *Unpublished Cornell University Working Paper*.

3135. Rehkugler H, Schenek A 2001 Underpricing oder overpricing? IPOs am Deutschen kapitalmarkt *in* Wirtz B W, Salzer E IPO - management – strukturen und erfolgsfaktoren Wiesbaden Germany pp 277 – 308.

3136. Schatt A, Roy T 2001 Modification de la structure de propriété et valorisation des sociétés introduites en bourse *Banque et Marchés* vol **55** pp 19 – 30.

3137. Schatt A, Broye G 2003 Pourquoi certains actionnaires cèdent plus d'actions que d'autres lors de l'introduction en bourse *Banque et Marchés* vol **65** pp 29 – 35.

3138. Sentis P 2001 Performances opérationnelles et boursières des sociétés introduites en bourse: Le cas français 1991-1995 *Finance* vol **22** (1) pp 87 – 118 France.

3139. Sentis P 2002 Initial public offerings: the good, the bad and the liars *Working Paper* University of Montpellier France.

3140. Sentis P 2004 Introduction en bourse, une approche internationale *Economica* Paris France.

3141. Severin E 2001 Ownership structure and the performance of firms: Evidence from France *European Journal of Economic and Social Systems* vol **15** (2) pp 85 – 107.

3142. Stoughton N M, Wong K P, Zechner J 2001 IPOs and product quality *Journal of Business* **74** pp 375 – 408.

3143. Torstila S 2001 What determines IPO gross spreads in Europe? *European Financial Management* **7** (4) pp 523 – 541.

3144. Torstila S 2003 The clustering of IPO gross spreads: International evidence *Journal of Financial and Quantitative Analysis* **38** (3) pp 673 – 694.

3145. Van Bommel J, Vermaelen Th 2001 Post-IPO capital expenditures and market feedback *Journal of Banking and Finance*.

3146. Van Frederikslust R A I, van der Geest R A 2001 Initial returns and long-run performance of private equity-backed initial public offerings on the Amsterdam stock exchange *Working Paper* Rotterdam School of Management The Netherlands.

3147. Vayanos D 2001 Strategic trading in a dynamic noisy market *Journal of Finance* **56** pp 131 – 171.

3148. Zhang D 2004 Why do IPO underwriters allocate extra shares when they expect to buy them back? *Journal of Financial and Quantitative Analysis* **39** (3) pp 571 – 594.

3149. Biais B, Bossaerts P, Rochet J C 2002 An optimal IPO mechanism *Review of Economic Studies* **69** pp 117 – 146.

3150. Biais B, Faugeron-Crouzet A M 2002 IPO auctions: English, Dutch, ... French and Internet *Journal of Financial Intermediation* **11** (1) pp 9 – 36.

3151. Blondell D, Hoang P, Powell J G, Shi J 2002 Detection of financial time series turning points: A new CUSUM approach applied to IPO cycles *Review of Quantitative Finance and Accounting* **18** (3) pp 293 – 315.

3152. Brau J C, Francis B, Kohers N 2002 The choice of IPO versus takeover: Empirical evidence *Journal of Business*.

3153. Brounen D, Eichholtz P M A 2002 Initial public offerings: Evidence from the British, Swedish and French property share markets *Journal of Real Estate Finance and Economics* **24** (1) pp 103 – 117.

3154. Bulow J, Klemperer P 2002 Prices and the winners curse *RAND Journal of Economics* **33**　pp 1 – 21.

3155. Cheng L T W, Mak B S C, Chan K C 2002 Strategic share allocation, information content of pre-listing characteristics, listing-day trading activities, and under-pricings of IPOs *Working Paper Hong Kong Polytechnic University* Hong Kong P R China.

3156. Deloof M, de Maeseneire W, Inghelbrecht K 2002 The valuation of IPOs by investment banks and the stock market: Empirical evidence *Working Paper* Faculty of Applied Economics University of Antwerp Ghent University Belgium pp 1 – 29.

3157. Easton P, Taylor G, Shroff P, Sougiannis T 2002 Using forecasts of earnings to simultaneously estimate growth and the rate of return on equity investments *Journal of Accounting Research* **40** pp 657 – 676.

3158. Easton P 2004 PE rations, PEG rations, and estimating the implied expected rate of return on equity capital *Accounting Review* **79** pp 73 – 95.

3159. Easton P 2006 Use of forecasts of earnings to estimate and compare cost of capital across regimes *Journal of Business, Finance, and Accounting* **33** pp 374 – 394.

3160. Easton P, Sommers G 2007 Effect of analysts' optimism on estimates of the expected rate of return implied by earnings forecasts *Journal of Accounting Research* **45** (5) pp 983 – 1015.

3161. Faugeron-Crouzet A M, Ginglinger E 2002 Introductions en bourse, signal et émissions d'actions nouvelles sur le second marché français *Finance* **22** (2) pp 51 – 74.

3162. Filatotchev I, Bishop K 2002 Board composition, share ownership, and 'underpricing' of UK IPO firms *Strategic Management Journal* **23** pp 941 – 955.

3163. Fishe R P H 2002 How stock flippers affect IPO pricing and stabilization *Journal Financial and Quantitative Analysis* **37** pp 319 – 320.

3164. Gao Y, Mao C, Zhong R 2002 Divergence of opinion and IPO long-term performance *Working Paper* City University of New-York NY USA.

3165. Giudici G, Roosenboom P 2002 Pricing initial public offerings in Europe: What has Changed? *Working Paper* Erasmus University Germany.

3166. Giudici G, Roosenboom P 2005 Pricing initial public offerings on 'new' European stock markets *Working Paper* Università degli Studi di Bergamo and Tilburg University Germany.

3167. Houge T, Loughran T, Suchanek G, Xuemin Yan 2002 Divergence of opinion, uncertainty, and the quality of initial public offerings *Financial Management* **30** pp 5 – 23.

3168. Kim K, Kitsabunnarat P, Nofsinger J 2002 Ownership control and operating performance in an emerging market: Evidence from Thai IPO firms *Journal of Corporate Finance* vol **156** pp 1 – 27.

3169. Kiss I, Stehle R 2002 Underpricing and long-term performance of initial public offerings at Germany's neuer markt, 1997 – 2001 *Working Paper* Humboldt-University Berlin Germany.

3170. Kutsuna K, Okamura H, Cowling M 2002 Ownership structure pre and post-IPOs and the operating performance of JASDAQ companies *Pacific-Basin Finance Journal* vol **10** pp 163 – 181.

3171. Logue D E, Rogalski R J, Seward J K, Foster-Johnson L 2002 What is special about the roles of underwriter reputation and market activities in initial public offerings? *Journal of Business* **75** pp 213 – 243.

3172. Martimort D 2002 Multi unit auctions: A survey of recent literature *Revue d'Economie Politique* **112** pp 303 – 348.

3173. Moerland P W 2002 Corporate supervision in the Netherlands *in* Renneboog L, McCahery J, Moerland P, Raaijmakers T (editors) Convergence and diversity of corporate governance regimes and capital markets *Oxford University Press* Oxford UK.

3174. NASD Regulation 2002 NASD regulation charges Credit Suisse First Boston with siphoning tens of millions of dollars of customers' profits in exchange for "hot" IPO shares http://www.nasdr.com/news .

3175. Schiereck D, Wagner Ch 2002 Kursentwicklung von börsenneulingen am neuen markt und die reputation des konsortialführers *Zeitschrift für Betriebswirtschaft* **72** pp 823 – 845.

3176. Schuster J A 2002 The cross-section of European IPO returns *Working Paper* London School of Economics and Political Science London UK.

3177. Schuster J A 2003 IPOs: Insights from seven European countries *Working Paper* London School of Economics and Political Science London UK.

3178. Wang J, Zender J 2002 Auctioning divisible goods *Economic Theory* **19** pp 673 – 705.

3179. Xie F 2002 The endogeneity of IPOs being underwritten by prestigious underwriters SSRN: 352160.

3180. Baginski S, Wahlen J 2003 Residual income risk, intrinsic values and share prices *Accounting Review* **78** pp 327 – 351.

3181. Barondes R D R, Nyce C, Sanger G C 2003 Law firm prestige and performance in IPOs: Underwriters' counsel as gatekeeper or turnstile *Contracting and Organizations Research Institute* University of Missouri Columbia Missouri USA.

3182. Bartlett J, Shulman Z 2003 IPO reform: Some immodest proposals *Journal of Private Equity* Summer pp 1 – 11.

3183. Binay M, Pirinsky C 2003 Institutional participation in IPOs *Working Paper* Koc University Istanbul Turkey.

3184. Bourjade S 2003, 2008 Strategic price discounting and rationing in uniform price auctions *Toulouse Business School* Ecole Superieure de Commerce de Toulouse France *MPRA Paper no 7260* Munich University Munich Germany pp 1 – 32

http://mpra.ub.uni-muenchen.de/7260/ .

3185. Clarke J E, Dunbar C, Kahle K M 2003 All-star analyst turnover, investment bank market share, and the performance of initial public offerings *Georgia Tech Working Paper* USA.

3186. Derrien F, Womack K 2003 Auctions versus book building and the control of underpricing in hot IPO markets *Review of Financial Studies* **16** (1) pp 31 – 61.

3187. Doeswijk R Q, Hemmes H S K, Venekamp R 2005 25 years of Dutch IPOs – An examination of frequently cited IPO anomalies within main sectors and during hot and cold issue periods *Working Paper* IRIS Fortis and Erasmus University The Netherlands.

3188. Ellul A, Pagano M 2003 IPO underpricing and after-market liquidity *CSEF Working Paper no 99* University of Salerno Centre for Studies in Economics and Finance Italy.

3189. Goergen M, Khurshed A, McCahery J, Renneboog L 2003 The rise and fall of European new markets: On the short and long-run performance of high-tech initial public offerings *in* McCahery J, Renneboog L (editors) Venture capital contracting and the valuation of high technology firms *Oxford University Press* UK.

3190. Gounopoulos D 2003 The initial and aftermarket performance of IPOs: Evidence from Athens stock exchange *Working Paper* Manchester School of Management Manchester University UK.

3191. Gulati R, Higgins M C 2003 Which ties matter when? The contingent effects of inter-organizational partnerships on IPO success *Strategic Management Journal* **24** pp 127 – 144.

3192. Higgins M C, Gulati R 2003 Getting off to a good start: The effects of upper echelon affiliations on underwriter prestige *Organization Science* **14** pp 244 – 263.

3193. Hoberg G 2003 Strategic underwriting in initial public offerings *Yale University Working Paper* USA.

3194. Hong H, Kubik J D 2003 Analyzing the analysts: Career concerns and biased earnings forecasts *Journal of Finance* **58** pp 313 – 351.

3195. Huyghebaert N, Van Hulle C 2003 Structuring the IPO: Empirical evidence on the primary and secondary portion.

3196. Jelic R, Briston R 2003 Privatization initial public offerings: The Polish experience *European Financial Management* **9** (4) pp 457 – 484.

3197. Kaneko T, Pettway R H 2003 Auctions versus book-building of Japanese IPOs *Pacific-Basin Finance Journal* **11** (4) pp 439 – 462.

3198. Karolyi G A, Stulz R M 2003 Are financial assets priced locally or globally? *in* Handbook of the economics of finance vol **1B** Asset pricing Constantinides G M, Harris M, Stulz R M (editors) *Elsevier North Holland* Amsterdam The Netherlands.

3199. Kraus T, Burghof H P 2003 Post-IPO performance and the exit of venture capitalists *Working Paper Universität München* Germany.

3200. Lemmens G 2003, 2007 The choice of the investment bank when going public *Catholic University of Leuven* Belgium *MPRA Paper no 4692* Munich University Munich Germany pp 1 – 95

http://mpra.ub.uni-muenchen.de/4692/ .

3201. Lemmens G 2004, 2007 Internet & capital raising: The perfect match? *Catholic University of Leuven* Belgium *MPRA Paper no 4691* Munich University Munich Germany pp 1 – 41 http://mpra.ub.uni-muenchen.de/4691/ .

3202. Manigart S, de Maeseneire W 2003 Initial returns: Underpricing or overvaluation? Evidence from EASDAQ and Euro N M *Working Paper* Vierick Leuven Gent Belgium.

3203. Neuhaus S, Schremper R 2003 Langfristige performance von initial public offerings am Deutschen kapitalmarkt *Zeitschrift für Betriebswirtschaft* **73** pp 445 – 472.

3204. Nounis C 2003 Greek initial public offerings, 1994 – 2002 *Working Paper* National and Kapodistrian University of Athens Greece.

3205. Ofek E, Richardson M 2003 Dotcom mania: The rise and fall of Internet stock prices *Journal of Finance* **58** pp 1113 – 1138.

3206. Peristiani S 2003 Evaluating the riskiness of initial public offerings: 1980-2000 *Working Paper Federal Reserve Bank of New* York USA .

3207. Pham P K, Kalev P S, Steen A B 2003 Underpricing, stock allocation, ownership structure and post-listing liquidity of newly listed firms *Journal of Banking and Finance* **27** (5) pp 919 – 947.

3208. Roosenboom P, van der Goot T 2003 Takeover defenses and IPO firm value in the Netherlands *ERIM Report Series Research in Management no ERS-2003-049-F&A* Erasmus Research Institute of Management Rotterdam School of Management Rotterdam School of Economics Erasmus Universiteit Rotterdam University of Amsterdam The Netherlands      pp 1 – 51.

3209. Roosenboom P G, van der Goot T, Mertens G 2003 Earnings management and initial public offerings: Evidence from the Netherlands *International Journal of Accounting* **38** (3) pp 243 – 266.

3210. Roosenboom P G, van der Goot T 2005 The effect of ownership and control on market valuation: Evidence from initial public offerings in the Netherlands *International Review of Financial Analysis* **14** (1) pp 43 – 59.

3211. Roosenboom P G J 2007 How do underwriters value IPOs? An empirical analysis of the French IPO market *Contemporary Accounting Research* **24** (4) pp 1217 – 1243.

3212. Smart S, Zutter C J 2003 Control as a motivation for underpricing: A comparison of dual- and single-class IPOs *Journal of Financial Economics* **69** pp 85 – 110.

3213. Van Bommel J, Vermaelen T 2003 Post-IPO capital expenditures and market feedback *Journal of Banking and Finance* **27** (2) pp 275 – 315.

3214. Van der Goot T 2003 Risk, the quality of intermediaries and legal liability in the Netherlands IPO market *International Review of Law & Economics* **23** (2) pp 121 – 140.

3215. Weber J, Willenborg M 2003 Do expert informational intermediaries add value? Evidence from auditors in microcap IPOs *Journal of Accounting Research* **41** pp 681 – 720.

3216. Arugaslan O, Cook D O, Kieschnick R 2004 Monitoring as a motivation for IPO underpricing *The Journal of Finance* **59** pp 2403 – 2420.

3217. Bodnaruk A, Kandel E, Massa M, Simonov A 2004 Shareholder diversification and IPOs *Working Paper* CEPR Discussion Papers Sweden.

3218. Burrowes A, Jones K 2004 Initial public offerings: Evidence from the UK *Managerial Finance* **30** pp 46 – 62.

3219. Cassia L, Giudici G, Paleari S, Redondi R 2004 IPO underpricing in Italy *Applied Financial Economics* **14** (3) pp 179 – 194.

3220. Cassia L, Paleari S, Vismara S 2004 The valuation of firms listed on the Nuovo Mercato: The peer comparables approach *Advances in Financial Economics* **10** pp 113 – 129.

3221. Cassia L, Vismara S 2009 Valuation accuracy and infinity horizon forecast: Empirical evidence from Europe *Journal of International Financial Management and Accounting* **20** (2).

3222. Chahine S 2004a Long-run abnormal return after IPOs and optimistic analysts' forecasts *International Review of Financial Analysis* **13** (1) pp 83 – 103.

3223. Chahine S 2004b Underpricing versus gross spreads: New evidence on the effects of sold shares at the time of IPOs *Working Paper* 2004 European Financial Management Association (EFMA) Meeting Basel Switzerland.

3224. Chiang K C H, Harikumar T 2004 Offering price clusters and underpricing in the US primary market *Applied Financial Economics* **14** pp 809 – 811.

3225. Cliff M, Denis D 2004 Do initial public offering firms purchase analysts' coverage with underpricing? *Journal of Finance* **59** (6) pp 2871 – 2901.

3226. Corwin S A, Harris J H, Lipson M L 2004 The development of secondary market liquidity for NYSE-listed IPOs *Journal of Finance* **59** (5) 2339 – 2373.

3227. Durnev A, Morck R, Yeung B 2004 Value enhancing capital budgeting and firm-specific stock return variation *Journal of Finance* **59** pp 65 – 105.

3228. Fernando C S, Gatchev V A, Spindt P A 2004 Wanna dance? How firms and underwriters choose each other *Tulane University Working Paper*.

3229. Foerster B 2004 Securities industry association's capital markets handbook *5th edition Aspen Publishers* Aspen Colorado USA.

3230. Griffith S J 2004 Spinning and underpricing: A legal and economic analysis of the preferential allocation of shares in initial public offerings *Brook Law Review* **69** pp 583 – 659.

3231. Ganor M 2004 A proposal to restrict manipulative strategy in auction IPOs *unpublished manuscript*
http://ssrn.com/abstract=572243 .

3232. Hahn T, Ligon J A 2004 Liquidity and initial public offering underpricing *Working Paper* College of Business and Economics University of Idaho University of Alabama USA.

3233. Hao Q 2004 Laddering in initial public offerings *Working Paper* University of Florida Miami USA.

3234. Hoberg G 2004 Strategic underwriting in initial public offers *Yale ICF Working Paper no 04-07* Yale University USA.

3235. Kooli M, Suret J M 2004 The aftermarket performance of initial public offerings in Canada *Journal of Multinational Financial Management* **14** pp 47 – 66.

3236. Kremer I, Nyborg K 2004a Divisible good auctions – The role of allocation rules *Rand Journal of Economics* **35** pp 147 – 159.

**3237.** Kremer I, Nyborg K 2004b Underpricing and market power in uniform price auctions *Review of Financial Studies* **17** pp 849 – 877.

**3238.** Kutsuma K, Smith R 2004 Why does book-building drive out auction methods of IPO issuance? Evidence from *Japan Review of Financial Studies* **17** (4) pp 1129 – 1166.

**3239.** Lamont O 2004 Short sale constraints and overpricing *in* Fabozzi F J (editors) The theory and practice of short selling: Risks, rewards, strategies *John Wiley & Sons* USA.

**3240.** Lee M, Wahal S 2004 Grandstanding, certification and the underpricing of venture capital backed IPOs *Journal of Financial Economics* **73** pp 375 – 407.

**3241.** Levy E R 2004 The law and economics of IPO favoritism and regulatory spin *Sw U L Review* **33** pp 185 – 203.

**3242.** Lubig D 2004 Underpricing und langfristige performance der IPOs am neuen markt *Frankfurt am Main* Germany.

**3243.** Mayhew S, Mihov V 2004 How do exchanges select stocks for option listing *Journal of Finance* **59** pp 447 – 471.

**3244.** Mayhew S, Mihov V 2005 Short sale constraints, overvaluation, and the introduction of options *Working Paper Texas Christian University Texas* USA.

**3245.** Mira G 2004 A proposal to restrict manipulative strategy in auction IPO's *Working Paper Series* Berkeley Program in Law and Economics University of California Berkeley USA pp 1 – 23 http://www.escholarship.org/uc/item/4sg5191h .

**3246.** Peggy M L, Wahal S 2004 Grandstanding, certification and the underpricing of venture capital backed IPOs *Journal of Financial Economics* **73** pp 375 – 407.

**3247.** Pollock T G, Porac J F, Wade J B 2004 Constructing deal networks: Brokers as network "architects" in the US IPO market and other examples *The Academy of Management Review* **29** pp 50 – 72.

**3248.** Pollock T G, Chen G, Jackson E M, Hambrick D 2005 Symbolic certification or substantive resources? Over tallying the signalling value of IPOs' prestigious affiliates *in Academy of Management Annual Meeting* Honolulu HI USA.

**3249.** Pritsker M 2004 Large investors: Implications for equilibrium asset returns, shock absorption and liquidity *The Federal Reserve Board* USA.

**3250.** Pritsker M 2004, 2005 A fully-rational liquidity-based theory of IPO underpricing and underperformance *Board of Governors Federal Reserve System* Washington USA pp 1 – 55.

**3251.** Pritsker M 2006 A fully-rational liquidity-based theory of IPO underpricing and underperformance *Staff Working Paper 2006-12* Finance and Economics Discussion Series Divisions of Research & Statistics and Monetary Affairs Federal Reserve Board Washington DC USA pp 1 – 69.

**3252.** Purnanandam A K, Swaminathan B 2004 Are IPOs really underpriced *Review of Financial Studies* **17** (3) pp 811 – 848.

**3253.** Rath N, Tebroke H-J, Tietze Ch 2004 Marktstimmung" zur erklärung der langfristigen kursentwicklung von aktien nach dem börsengang *Zeitschrift für Bankrecht und Bankwirtschaft* **16** pp 269 – 352.

**3254.** Reuter J 2004 Are IPO allocations for sale? Evidence from the mutual fund industry *University of Oregon Working Paper* USA.

**3255.** Rice D T 2004 The nanotech IPO *Nanotechnology L and Bus* **1** pp 315 – 322.

**3256.** Rice D T 2006 When the nanotech company goes public: Using the electronic Dutch auction *Nanotechnology L and Bus* **3** pp 185 – 188.

**3257.** Rindermann G 2004 The performance of venture-backed IPOs on Europe's new stock markets: Evidence from France, Germany and the UK *Advances in Financial Economics* **10** pp 231 – 294.

**3258.** Sanders W G, Boivie S 2004 Sorting things out: Valuation of new firms in uncertain markets *Strategic Management Journal* **25** pp 167 – 186.

**3259.** Schenone C 2004 The effect of bank relationships on the firm's cost of equity in its IPO *Journal of Finance* **59** (6) pp 2903 – 2958.

**3260.** Serve S 2004 The operating performance of French IPO firms *THEMA Universite de Cergy-Pontoise* Cedex France.

**3261.** Alti A 2005 IPO market timing *Review of Financial Studies* **18** (3) pp 1105 – 1138.

**3262.** Alti A 2006 How persistent is the impact of market timing on capital structure *Journal of Finance* **61** pp 1681 – 1710.

**3263.** Alvarez S, Gonzalez V 2005 Long-run performance of initial public offerings in the Spanish capital market *Journal of Business Finance and Accounting* **32** (1-2) pp 325 – 350.

**3264.** Benninga S, Helmantel M, Sarig O 2005 The timing of initial public offerings *Journal of Financial Economics* **75** (1) pp 115 – 132.

**3265.** Berg J E, Neumann G R, Rietz T A 2005 Searching for Google's value: Using prediction markets to forecast market capitalization prior to an initial public offering *Working Paper* University of Iowa USA.

**3266.** Butler A W, Grullon G, Weston J P 2005 Stock market liquidity and the cost of issuing equity *Journal of Financial and Quantitative Analysis* **40** (2) pp 331 – 348.

**3267.** Choo E 2005 Note, Going Dutch: The Google IPO *Berkeley Technology L Journal* **20** p 405.

**3268.** Corwin Sh A, Schultz P 2005 The role of IPO underwriting syndicates: Pricing, information production, and underwriter competition *The Journal of Finance* **60** (1) pp 443 – 486.

**3269.** Dolvin S 2005 Venture capitalist certification of IPOs *Venture Capital: An International Journal of Entrepreneurial Finance* **7** pp 131 – 148.

**3270.** Drobetz W, Kammermann M, Wälchli U 2005 Long-run performance of initial public offerings: The evidence for Switzerland *Schmalenbach Business Review* **57** (3) pp 253 – 275.

**3271.** Forestieri G 2005 Corporate and investment banking *Egea* Milan Italy.

**3272.** Hess B M 2005 Google, Inc.: The Dutch auction approach as an alternative to firm commitment underwriting *Duque Bus L Journal* **7** pp 90 – 91.

**3273.** Hurt Ch 2005 Moral hazard and the initial public offering *Cardozo Law Review* **26** pp 711 – 769.

**3274.** Hurt Ch 2006 What Google can't tell us about Internet auctions (And what it can) *U Tol Law Review* **37** pp 423 – 424.

*3275.* Jagannathan R, Gao Y R 2005 Are IPOs underpriced? A closer examination *Working Paper* EFMA Conference Milan Italy.

*3276.* Jain B A, Kini O 2005 Industry clustering of initial public offerings *Managerial and Decision Economics* **27** (1) pp 1 – 20.

*3277.* Jaskiewicz P, Gonzàlez V M, Menéndez S, Schiereck D 2005 Long-run IPO performance analysis of German and Spanish family-owned businesses *Family Business Review* **18** (3)  pp 179 – 202.

*3278.* Khanna N, Noe T H, Sonti R 2005 Good IPOs draw in bad: Inelastic banking capacity in the primary issue market *Michigan St University Working Paper* USA.

*3279.* Khanna N, Noe T H, Sonti R 2008 Good IPOs draw in bad: Inelastic banking capacity and hot markets *Review of Financial Studies* **21** pp 1873 – 1906.

*3280.* Li M, McInish T H, Wongchoti U 2005a Asymmetric information in the IPO aftermarket *The Financial Review* **40** (2) pp 131 – 153.

*3281.* Li M, Zheng S X, Melancon M V 2005b Underpricing, share retention, and the IPO aftermarket liquidity *International Journal of Managerial Finance* **1** (2) pp 76 – 94.

*3282.* LiCalzi M, Pavan A 2005 Tilting the supply schedule to enhance competition in uniform price auctions *European Economic Review* **49** pp 227 – 250.

*3283.* Malloy C J 2005 The geography of equity analysis *The Journal of Finance* **60** pp 719 – 755.

*3284.* Nounis Ch 2005 The Greek IPO initial returns and the price cap constraints: evidence from the Athens stock exchange (1994-2003) *Working Paper* National and Kapodistrian University of Athens Greece.

*3285.* Pandey A 2005 Initial returns, long run performance, and characteristics of issuers: Differences in Indian IPOs following fixed price and book building processes *Working Paper Indian Institute of Management* India.

*3286.* Parlour C, Rajan U 2005 Rationing in IPOs *Review of Finance* **9** pp 33 – 63.

*3287.* Pastor L, Veronesi P 2005 Rational IPO waves *Journal of Finance* **60** (4) pp 1713 – 1757.

*3288.* Pastor L, Taylor L, Veronesi P 2009 Entrepreneurial learning, the IPO decision, and the post-IPO drop in firm profitability *Review of Financial Studies* **22** pp 3005 – 3046.

*3289.* Pons-Sanz V 2005 Who benefits from IPO underpricing? Evidence from hybrid book building offerings *Working Paper* European Central Bank (ECB).

*3290.* Sherman A 2005 Global trends in IPO methods: Book building versus auctions with endogenous entry *Journal of Financial Economics* **78** (3) pp 615 – 649.

*3291.* Yan H 2005 Natural selection in financial markets: Does it work? *Thesis* London School of Economics and Political Science London UK.

*3292.* Anand A I 2006 Is the Dutch auction IPO a good idea? *Stanford Journal Law Business and Finance* **11** pp 233 – 238.

*3293.* Aussenegg W 2006 Underpricing and the aftermarket performance of initial public offerings: the case of Austria *in* Gregoriou G N (editor) Initial public offerings: An international perspective *Elsevier* Quantitative Finance Series Amsterdam The Netherlands pp 187 – 213.

*3294.* Aussenegg W, Pichler P, Stomper A 2006 IPO pricing with book building and a when-issued market *Journal of Financial and Quantitative Analysis*.

3295. Boot A W A, Gopalan R, Thakor A V 2006 The entrepreneur's choice between private and public ownership *Journal of Finance* **61** pp 803 – 836.

3296. Das S, Guo R, Zhang H 2006 Analysts' selective coverage and subsequent performance of newly public firms *Journal of Finance* **61** (3) pp 1159 – 1185.

3297. Damodaran A 2006 Damodaran on valuation *John Wiley and Sons* New York USA.

3298. Ellul A, Pagano M 2006 IPO underpricing and after-market liquidity *Review of Financial Studies* **19** (2) pp 381 – 421.

3299. Gajewski J - F, Gresse C 2006 A survey of the European IPO market *Université Paris, Université Paris X Nanterre* France *European Capital Markets Institute Paper no 2* ISBN 92-9079-658-8 pp 1 – 94 Brussels Belgium www.ceps.be .

3300. Goergen M, Renneboog L, Khurshed A 2006 Explaining the diversity in shareholder lockup agreements *Journal of Financial Intermediation* **15** (2) pp 254 – 280.

3301. Hong L 2006 The new development of the Chinese capital market *China Securities Journal* **17**.

3302. Jagannathan R, Sherman A 2006 Why do IPO auctions fail? *Working Paper* University of Notre-Dame.

3303. James C, Karceski J 2006 Strength of analyst coverage following IPOs *Journal of Financial Economics* **82** (1) pp 1 – 34.

3304. Pastor-Llorca M J, Poveda-Fuentes F 2006 Earnings management and the long-run performance of Spanish initial public offerings *in* Gregoriou G N (editor) Initial public offerings: An international perspective *Elsevier Butterworth-Heinemann* chapter 7 pp 81 – 112.

3305. Tirole J 2006 The theory of corporate finance *Princeton University Press* New Jersey USA.

3306. Trauten A, Schulz R C 2006 IPO investment strategies and pseudo market timing *Internetökonomie und Hybridität no 36, ZEW Discussion Papers no 36* Leibniz Information Centre for Economics Germany pp 1 – 44 http://hdl.handle.net/10419/46583 .

3307. Yung C, Colak G, Wang W 2006 Cycles in the IPO Market *Working paper*.

3308. Zhang X F 2006 Information uncertainty and analyst forecast behavior *Contemporary Accounting Research* **23** (2) pp 565 – 590.

3309. Arnold T, Fishe R, North D 2007 The effects of "risk-factor" disclosure on the pricing of IPOs and long run returns *Working Paper* University of Richmond VA USA.

3310. Berkeley A J 2007 Some background and simple FAQs about Dutch auctions and the Google IPO *in* Fundamentals of securities law *American Law Institute American Bar Association* USA pp 275 – 279.

3311. Doran J, Jiang D, Peterson D 2007, 2009 Short-sale constraints and the idiosyncratic volatility puzzle: An event study approach *Florida State University MPRA Paper no 8261* Munich University Munich Germany pp 1 – 39 http://mpra.ub.uni-muenchen.de/8261/ .

**3312.** Hopp Ch, Dreherdo A 2007 Differences in institutional and legal environments explain cross-country variations in IPO underpricing? *CESIFO Working Paper no 2082* pp 1 – 48.

**3313.** Jog V M, Sun Ch 2007 Blank check IPOs: A home run for management http://ssrn.com/abstract=1018242 ,

http://dx.doi.org/10.2139/ssrn.1018242 .

**3314.** Kerins F, Kutsuna K, Smith R 2007 Why are IPOs underpriced? Evidence from Japan's hybrid auction-method offerings *Journal of Financial Economics* **85** pp 637 – 666.

**3315.** Leite T 2007 Adverse selection, public information and underpricing in IPOs *Journal of Corporate Finance*.

**3316.** Paleari S, Vismara S 2007 Over-optimism when pricing IPOs *Managerial Finance* **33** (6) pp 352 – 367.

**3317.** Paleari S, Pellizzoni E, Vismara S 2008 The going public decision: Evidence from the IPOs in Italy and in the UK International *Journal of Applied Decision Sciences* **1** pp 131 – 152.

**3318.** Paleari S, Ritter J R, Vismara S 2010 Explaining the simultaneous consolidation and fragmentation of Europe's stock markets *Working Paper University of Florida* USA.

**3319.** Penman S H 2007 Financial statement analysis and security valuation *3rd edition McGraw-Hill* New York USA.

**3320.** Roosenboom P G J 2007 How do underwriters value IPOs? An empirical analysis of the French IPO market *Contemporary Accounting Research* **24** (4) pp 1217 – 1243.

**3321.** Thomas J K 2007 Discussion of how do underwriters value IPOs? An empirical analysis of the French IPO market *Contemporary Accounting Research* **24** (4) pp 1245 – 1254.

**3322.** Toniato J B A 2007 Hot issue" IPO markets and its consequences for issuing firms and investors: The UK market of 2000 *Brazilian Business Review* vol **4** no 1 pp 1 – 26 www.bbronline.com.br .

**3323.** Zheng S X, Stangeland D A 2007 IPO underpricing, firm quality, and analyst forecasts *Financial Management* **36** (2) pp 45 – 64.

**3324.** An H H, Chan K C 2008 Credit ratings and IPO pricing *Journal of Corporate Finance* 14 (5) pp 584 – 595.

**3325.** Casotti F P, Motta L F J 2008 IPO abordagem de múltiplos e custo de capital próprio *Revista Brasileira de Finanças* vol **6** no 2.

**3326.** Farina V 2008 Network embeddedness, specialization choices and performance in investment banking industry *University of Rome Tor Vergata* Italy *MPRA Paper no 11701* Munich University Munich Germany pp 1 – 26 http://mpra.ub.uni-muenchen.de/11701/ .

**3327.** Hale G, Santos J A C 2008 Do banks price their informational monopoly? *Working Paper 2008-14* Federal Reserve Bank of San Francisco USA pp 1 – 49 http://ssrn.com/abstract=935225

http://www.frbsf.org/publications/economics/papers/2008/wp08-14bk.pdf .

**3328.** Hild M 2008 The Google IPO *Journal of Business and Technology Law* vol **3** issue 1 pp 41 – 59

http://digitalcommons.law.umaryland.edu/jbtl/vol3/iss1/4 .

*3329.* Kaustia M, Knupfer S 2008 Do investors overweight personal experience? Evidence from IPO subscriptions *Journal of Finance* **63** pp 2679 – 2702.

*3330.* Khurshed A, Pande A, Singh A 2008 Subscription patterns, offer prices and the underpricing of IPOs

http://papers.ssrn.com/sol3/papers.cfm?abstract_id=1343024 .

*3331.* Khurshed A, Paleari S, Pande A, Vismara S 2011 Grading, transparent books and initial public offerings

http://www.unibg.it/dati/persone/1823/4211-Grading%20paper.pdf .

*3332.* Rossetto S 2008 The price of rapid exit in venture capital-backed IPOs *Annals of Finance* **4** pp 29 – 53.

*3333.* Kim W, Weisbach M S 2008 Motivations for public equity offers: An international perspective *Journal of Financial Economics* **87** pp 281 – 307.

*3334.* Poudyal S 2008 Grading initial public offerings (IPOs) in India's capital markets: A globally unique concept *Working Paper no 2008-12-08* Indian Institute of Management Ahmedabad India pp 1 – 30.

*3335.* Yongyuan Qiao 2008 Analysis into IPO underpricing and clustering in Hong Kong equity market *School of Economics and Finance* University of St Andrews UK *MPRA Paper no 7876* Munich University Munich Germany pp 1 – 28

http://mpra.ub.uni-muenchen.de/7876/ .

*3336.* Yung C, Colak G, Wang W 2008 Cycles in the IPO market *Journal of Financial Economics* **89** (1) pp 192 – 208.

*3337.* Colak G, Gunay H 2011 Strategic waiting in the IPO markets *Journal of Corporate Finance* **17** (3) pp 555 – 583.

*3338.* Bouis R 2009 The short-term timing of initial public offerings *Journal of Corporate Finance* **15** (5) pp 587 – 601.

*3339.* Coakley J, Hadass L, Wood A 2009 UK IPO underpricing and venture capitalists *The European Journal of Finance* **15** pp 421 – 435.

*3340.* Deloof M, De Maeseneire W, Inghelbrecht K 2009 How do investment banks value IPOs? *Journal of Business Finance and Accounting*.

*3341.* Jiang F, Leger L A 2009 The impact on IPO performance of reforming IPO allocation regulations: An event study of Shanghai stock exchange A-shares *Working Paper 2009 - 04* Department of Economics Loughborough University Loughborough UK ISSN 1750-4171 pp 1 – 24.

*3342.* Zhang J X 2009 Shareholding by venture capitalists and patent applications of Japanese firms in the pre- and post- IPO periods *IIR Working Paper WP#09-02* Institute of Innovation Research Hitotsubashi University Tokyo Japan pp 1 – 32

http://hdl.handle.net/10086/17349 ,

http://www.iir.hit-u.ac.jp .

*3343.* Arikawa Y, Imad'eddine G 2010 Venture capital affiliation with underwriters and the underpricing of initial public offerings in Japan *Journal of Economics and Business* **62** pp 502 – 516.

3344. Bonardo D, Paleari S, Vismara S 2010 When academia comes to market: Does university affiliation reduce the uncertainty of IPOs? *International Journal of Entrepreneurship and Innovation*.

3345. Bonardo D, Paleari S, Vismara S 2010 Valuing university-based firms: The effects of academic affiliation on IPO performance *Entrepreneurship Theory and Practice*.

3346. Caglio C, Weiss-Hanley K, Marietta-Westberg J 2010 Going public abroad: The role of international markets for IPOs *US Securities and Exchange Commission Working Paper* http://ssrn.com/abstract=1572949 .

3347. Cogliati G, Paleari S, Vismara S 2010 IPO pricing: Growth rates implied in offer prices *Working Paper no 02 – 2010* Department of Economics and Technology Management University of Bergamo Italy.

3348. Chod J, Lyandres E 2010 Strategic IPOs and product market competition *Journal of Financial Economics*.

3349. Deb S S, Marisetty V B 2010 Information content of IPO grading *Journal of Banking & Finance* **34** (9) pp 2294 – 2305.

3350. Elston J A, Yang J J 2010 Venture capital, ownership structure, accounting standards and IPO underpricing: Evidence from Germany *Journal of Economics and Business* **62** pp 517 – 536.

3351. Guo H, Brooks R, Shami R 2010 Detecting hot and cold cycles using a Markov regime switching model - Evidence from the Chinese A-share IPO market International *Review of Economics & Finance* **19** (2) pp 196 – 210.

3352. Hsu H-Ch, Reed A V, Rocholl J 2010 The new game in town: Competitive effects of IPOs *Journal of Finance* **65** pp 495 – 528.

3353. Hussinger K 2010 Absorptive capacity and post-acquisition inventor productivity *Katholieke Universiteit Leuven Belgium Centre for European Economic Research (ZEW) Mannheim Germany ZEW Discussion Papers no 10-066* Leibniz Information Centre for Economics Germany pp 1 – 32

http://hdl.handle.net/10419/41435 ,

ftp://ftp.zew.de/pub/zew-docs/dp/dp10066.pdf.

3354. Hussinger K 2012 Absorptive capacity and post-acquisition inventor productivity *Journal Technology Transfer* **37** pp 490 – 507 DOI 10.1007/s10961-010-9199-y .

3355. Jagannathan R, Jirnyi A, Sherman A 2010 Why don't issuers choose IPO auctions? The complexity of indirect mechanisms *Technical Report*

http://www.nber.org/papers/w16214 .

3356. Pennacchio L, Del Monte A, Acconcia A 2010 Underpricing and distance: An empirical analysis *MPRA Paper no 21737* Munich University Munich Germany pp 1 – 11 http://mpra.ub.uni-muenchen.de/21737/ .

3357. Acconcia A, Del Monte A, Pennacchio L 2011 Underpricing and firms' distance from financial centre: Evidence from three European countries *Working Paper no 295* CSEF University of Naples Italy.

3358. Pennacchio L 2013 The causal effect of venture capital backing on the underpricing of Italian IPOs *University of Napoli "Federico II" Italy MPRA Paper no 48695* Munich University Munich Germany pp 1 – 44

http://mpra.ub.uni-muenchen.de/48695/ .

3359. Sahoo S, Rajib P 2010 After market performance of initial public offerings (IPOs): Indian IPO market 2002 - 2006 *The Journal for Decision Makers* **35** (4) pp 27 – 43.

3360. Shao X, Wu H, Qin J Wang D 2010 The IPO market cycles in China: The analysis based on the investor sentiment and government's market timing *Journal of Financial Research* **11** pp 123 – 143.

3361. Yao-Min Chiang, Hirshleifer D, Yiming Qian, Sherman A 2010 Learning to fail? Evidence from frequent IPO investors *Department of Finance* National Chengchi University Taipei Taiwan, *The Paul Merage School of Business* University of California Irvine California USA, *Department of Finance* University of Iowa USA, *Department of Finance* DePaul University Chicago Illinois USA *MPRA Paper no 25231* Munich University Munich Germany pp 1 – 40

http://mpra.ub.uni-muenchen.de/25231/ .

3362. Boeh K K, Southam C 2011 Impact of initial public offering coalition on deal completion *Venture Capital: An International Journal of Entrepreneurial Finance* **13** (4) pp 313 – 336 *Routledge: Taylor and Francis Group* DOI: 10.1080/13691066.2011.642148 .

3363. Bubna A, Prabhala N R 2011 IPOs with and without allocation discretion: Empirical evidence *Journal of Financial Intermediation* **20** (4) pp 530 – 561.

3364. Doidge C, Karolyi G A, Stulz R M 2011 The US left behind: The rise of IPO activity around the World *Fisher College of Business WP 2011-03-008, NBER Working Paper 16916* Rotman School of Management University of Toronto Canada, Johnson Graduate School of Management Cornell University USA, Charles A Dice Center for Research in Financial Economics Fisher College of Business Department of Finance The Ohio State University USA pp 1 – 58

http://www.ssrn.com/abstract=1795423,

http://www.nber.org/papers/w16916.

3365. Ferretti R, Meles A 2011 Underpricing, wealth loss for pre-existing shareholders and the cost of going public: The role of private equity backing in Italian IPOs *Venture Capital* **13** pp 23 – 47.

3366. Adesoye A B, Atanda A A-M 2012 Monetary policy and share pricing business in Nigeria *Department of Economics* Olabisi Onabanjo University Nigeria *MPRA Paper no 35846* Munich University Munich Germany pp 1 – 20

http://mpra.ub.uni-muenchen.de/35846/ .

3367. Boissin R 2012 Orphan versus non-orphan IPOs: The difference analyst coverage makes *MPRA Paper no 41584* Munich University Munich Germany pp 1 – 31

http://mpra.ub.uni-muenchen.de/41584/ .

3368. Cumming D J, Hass L H, Schweizer D 2012 The fast track IPO – Success factors for taking firms public with SPACs

http://ssrn.com/abstract=2144892 ,

http://dx.doi.org/10.2139/ssrn.2144892 .

3369. Datar V, Emm E, Ince U 2012 Going public through the back door: A comparative analysis of SPACs and IPOs *Banking and Finance Review* **4** (1).

3370. Jacob J 2012 Mandatory IPO grading: Does it help pricing efficiency? *Working Paper no 2012-12-07* Indian Institute of Management Ahmedabad India pp 1 – 39.

*3371.* Rodrigues U, Stegemoller M A 2012 What all-cash companies tell us about IPOs and acquisitions SSRN

http://ssrn.com/abstract=2101830 ,

http://dx.doi.org/10.2139/ssrn.2101830 .

*3372.* Saturnino O, Saturnino V, Lucena P, Caetano M, dos Santos J F 2012 Initial Public Offer of stocks in Brazil: An analysis of returns from stocks with low price/earnings ratio *Universidade Federal de Pernambuco* Brasil *MPRA Paper no 48106* Munich University Munich Germany pp 1 – 13

http://mpra.ub.uni-muenchen.de/48106/ .

*3373.* Chang-Yi Hsu, Jean Yu, Shiow-Ying Wen 2013 The analysts' forecast of IPO firms during the global financial crisis International *Journal of Economics and Financial Issues* vol **3** no 3 pp 673 – 682 ISSN: 2146-4138

www.econjournals.com .

*3374.* Zhiqiang Hu, Yizhu Wang 2013 The IPO cycles in China's A-share IPO market: Detection based on a three regimes Markov switching model *Romanian Journal of Economic Forecasting* **16** (3) pp 115 – 131 rjef3_2013p115-131.pdf .

*3375.* Lakicevic M, Shachmurove Y, Vulanovic M 2013 Institutional changes of SPACs *University of Montenegro, City College City University of New York, Western New England University MPRA Paper no 44181* Munich University Munich Germany pp 1 – 35 http://mpra.ub.uni-muenchen.de/44181/ .

***Private company investment, private company valuation, venture capital investment, venture capital fund, angel capital investment, financial capital investment product, financial capital investment medium in finances:***

*3376.* Akerlof G, Stiglitz J E 1966 Investment, income and wages *Econometrica* **34** (5) p 118.

*3377.* Stiglitz J E 1969 Theory of innovation: Discussion *American Economic Review* **59** (2) pp 46 – 49.

*3378.* Rothschild M, Stiglitz J 1976 Equilibrium in competitive insurance markets: An essay on the economics of imperfect information *Quarterly Journal of Economics* **90** pp 629 – 649.

*3379.* Stiglitz J, Weiss A 1981 Credit rationing in markets with incomplete information *American Economic Review* **71** pp 393 – 409.

*3380.* Stiglitz J E 1988 Randomization with asymmetric information *RAND Journal of Economics* **19** (3) pp 344 – 362.

*3381.* Greenwald B, Stiglitz J E 1990 Asymmetric information and the new theory of the firm: Financial constraints and risk behavior *American Economic Review* **80** (2) pp 160 – 165.

*3382.* J E Stiglitz 2001, 2002 Joseph E Stiglitz - Biographical *The Sveriges Riksbank Prize in Economic Sciences in Memory of Alfred Nobel 2001* p 1 http://www.nobelprize.org/nobel_prizes/economic-sciences/laureates/2001/stiglitz-bio.html .

*3383.* Richiardi M, Gallegati M, Greenwald B, Stiglitz J 2007 The asymmetric effect of diffusion processes: Risk sharing and contagion *Global Economy Journal* vol **8** issue 3 pp 1 - 10 ISSN 1524-5861 DOI: 10.2202/1524-5861.1365 .

3384. Stiglitz J E 2013 Five years in limbo *Columbia University* 3022 Broadway NY USA http://www8.gsb.columbia.edu/chazen/globalinsights/node/218/Five%20Years%20in%20Limbo# .

3385. Kirzner I 1973 Competition and entrepreneurship *University of Chicago Press* Chicago USA.

3386. Jensen M C, Meckling W H 1976 Theory of the firm: Managerial behavior, agency costs and ownership structure *Journal of Financial Economics* **3** pp 305 – 360.

3387. Lucas R E 1978 On the size distribution of business firm *Bell Journal of Economics* **9** (2) pp 508 – 523.

3388. Rind K W 1981 The role of venture capital in corporate development *Strategic Management Journal* **2** pp 169 – 180.

3389. Tyebjee T T, Bruno A V 1981 Venture capital decision making: Preliminary results from three empirical studies *Frontiers of Entrepreneurship Research* pp 316 – 334.

3390. Bruno A V, Tyebjee T T 1983 The one that got away: A study of ventures rejected by venture capitalists *in* Hornaday J A, Timmons J A, Vesper K H (editors) Frontiers of entrepreneurship research *Babson College* Wellesley MA pp 289 – 306.

3391. Tyebjee T, Bruno A 1984 A model of venture capitalist investment activity *Management Science* **30** pp 1051 – 1066.

3392. Bruno A V, Tyebjee T T 1986 The destinies of rejected venture capital deals *MIT Sloan Management Review* **27** pp 43 – 53.

3393. Chan Y S 1983 On the positive role of financial intermediation in allocation of venture capital in a market with imperfect information *The Journal of Finance* **38** (5) pp 1543 – 1568.

3394. Chan Y-S, Siegel D, Thakor A 1990 Learning, corporate control and performance requirements in venture capital contracts *International Economic Review* **31** pp 365 – 381.

3395. Felda H G, DeNino M J, Salter M S 1983 When corporate venture capital doesn't work *Harvard Business Review* **61** May - June pp 114 – 120.

3396. Wilson J W 1983 The new ventures – inside the high stakes world of venture capital *Addison Wesley Publishing Company* Boston MA USA.

3397. Merkle E 1984 Venture capital als instrument des technologiemanagements *Betriebs-Berater* **39** pp 1060 – 1064.

3398. Tyebjee T T, Bruno A V 1984 A model of venture capitalist investment activity *Management Science* **30** (9) pp 1051 – 1066.

3399. Hutt R W, Thomas B 1985 Venture capital in Arizona *Frontiers of Entrepreneurship Research* pp 155 – 169.

3400. MacMillan I, Siegel R, Narasimha P 1985 Criteria used by venture capitalists to evaluate new venture proposals *Journal of Business Venturing* **1** (1) pp 119 – 128.

3401. MacMillan I C, Zemann L, Narasimha P N 1987 Criteria distinguishing successful from unsuccessful ventures in the venture screening process *Journal of Business Venturing* **2** (2) pp 123 – 137.

3402. Beatty R P, Ritter J 1986 Investment banking, reputation, and the underpricing of initial public offerings *Journal of Financial Economics* **15** pp 213 – 232.

3403. Nevermann H, Falk D 1986 Venture capital - Ein betriebswirtschaftlicher und steuerlicher vergleich zwischen den USA und der Bundesrepublik Deutschland *NOMOS Verlagsgesellschaft* Baden-Baden Germany.

3404. Timmons J A, Bygrave W D 1986 Venture capital's role in financing innovation for economic growth *Journal of Business Venturing* **1** pp 161 – 176.

3405. Block Z, Ornati O A 1987 Compensating corporate venture managers *Journal of Business Venturing* **2** pp 41 – 52.

3406. Bygrave W D 1987 Syndicated investments by venture capital firms: A networking perspective *Journal of Business Venturing* **2** (2) pp 139 – 154.

3407. Bygrave W D, Timmons J A 1992 Venture capital at the crossroads *Harvard Business School Press* Boston MA pp 1 – 79.

3408. Robinson R B 1987 Emerging strategies in venture capital industry *Journal of Business Venturing* vol **2** pp 53 – 77.

3409. Ruhnka J C, Young J E 1987 A venture capital model of the development process for new ventures *Journal of Business Venturing* vol **2** pp 167 – 184.

3410. Ruhnka J C, Young J E 1991 Some hypotheses about risk in venture capital investing *Journal of Business Venturing* vol **6** (2) pp 115 – 133.

3411. Ruhnka J C, Felman H D, Dean T J, 1992 The "living dead" phenomenon in venture capital investments *Journal of Business Venturing* vol **7** issue 2 pp 137 – 155.

3412. Sandberg W R, Hofer C W 1987 Improving new venture performance: The role of strategy, industry structure, and the entrepreneur *Journal of Business Venturing* **2** (1) pp 5 – 28.

3413. Stedler H 1987 Venture capital und geregelter freiverkehr: Eine empirische studie *Frankfurt am Main* Germany.

3414. Brophy D J, Guthner M W 1988 Publicly traded venture capital funds: Implications for institutional 'Fund of Funds' investors *Journal of Business Venturing* **3** (3) pp 187 – 206.

3415. Clark R 1988 Venture capital in Britain, America, and Japan *Croom Helm* London UK.

3416. Eisinger P 1988 The rise of the entrepreneurial state: State and local economic development policy in the United States *University of Wisconsin Press* Madison USA.

3417. Eisinger P 1993 State venture capitalism, state politics, and world of high risk investment *Economic Development Quarterly* **7** (2) pp 131 – 140.

3418. Florida R L, Kenney M 1988 Venture capital-financed innovation and technological change in the USA *Research Policy* **17** (3) pp 119 – 137.

3419. Florida R L, Smith D F 1993 Keep the government out of venture capital *Issues in Science and Technology* **9** pp 61 – 68.

3420. Gladstone D 1988 Venture capital handbook. New and revised. An entrepreneur's guide to obtaining capital to start a business, buy a business, or existing business *Prentice Hall* USA pp 1 – 350.

3421. Harris M, Raviv A 1988 Corporate governance: Voting rights and majority rules *Journal of Financial Economics* **20** pp 203 – 235.

3422. Sandberg W R, Schweiger D M, Hofer Ch W 1988 Use of verbal protocols in determining venture capitalists' decision process *Entrepreneurship Theory and Practice* **13** pp 182 – 196.

3423. Schmidt R H 1988 Venture capital in Deutschland: Ein problem der qualität *Die Bank* **4** pp 184 – 187.

3424. Siegel R, Siegel E, and MacMillan I 1988 Corporate venture capitalists: Autonomy, obstacles and performance *Journal of Business Venturing* **3** pp 233 – 247.

3425. Tirole J 1988 The theory of industrial organization Cambridge *MIT Press* Cambridge Massachusetts USA.

3426. Benveniste L M, Spindt P A 1989 How investment bankers determine the offer price and allocation of new issues *Journal of Financial Economics* **24** pp 343 – 361.

3427. Gorman M, Sahlman W 1989 What do venture capitalists do? *Journal of Business Venturing* **4** pp 231 – 248.

3428. Holmstrom B, Tirole J 1989 The theory of the firm *in* Schmalensee R, Willig R (editors) Handbook of industrial economics Chapter 2 Part 1 *Elsevier Publishing BV* Amsterdam The Netherlands.

3429. Poterba J 1989a Venture capital and capital gains taxation *Tax Policy and the Economy* **3** p 47.

3430. Poterba J 1989b Capital gains tax policy toward entrepreneurship *National Tax Journal* **42** pp 375 – 390.

3431. Shiller R J, Pound J 1989 Survey evidence on the diffusion of interest and information among investors *Journal of Economic Behavior and Organization* **12** pp 47 – 66.

3432. Amit R, Glosten L, Muller E 1990a Entrepreneurial ability, venture investments, and risk sharing *Management Science* **36** pp 1232 – 1245.

3433. Amit R, Glosten L, Muller E 1990b Does venture capital foster the most promising entrepreneurial firms? *California Management Review* pp 102 – 111.

3434. Amit R, Brander J, Zott C 1998 Why do venture capital firms exist? Theory and Canadian evidence *Journal of Business Venturing* **13** pp 441 – 466.

3435. Barry C, Muscarella C, Peavy J, Vetsuypens M 1990 The role of venture capital in the creation of public companies: evidence from the going public process *Journal of Financial Economics* **27** pp 447 – 471.

3436. Barry C 1994 New directions in research on venture capital finance *Financial Management* **23** (3) pp 3 – 15.

3437. Hisrich R D, Jankowitz A D 1990 Intuition in venture capital decisions: An exploratory study using a new technique *Journal of Business Venturing* **5** pp 49 – 62.

3438. Sahlman W 1990 The structure and governance of venture-capital organizations *Journal of Financial Economics* **27** (2) pp 473 – 521.

3439. Sykes H B 1990 Corporate venture capital: Strategies for success *Journal of Business Venturing* **5** pp 37 – 47.

3440. Dixon R 1991 Venture capitalists and the appraisal of investments OMEGA The *International Journal of Management Science* **19** (5) pp 333 – 344.

3441. Megginson W, Weiss K 1991 Venture capitalist certification in initial public offerings *Journal of Finance* **46** pp 879 – 903.

3442. Megginson W 2004 Towards a global model of venture capital? *Journal of Applied Corporate Finance* **16** (1) pp 8 – 26.

3443. Riquelme H, Rickards T 1992 Hybrid conjoint analysis: An estimation probe in new venture decisions *Journal of Business Venturing* **7** (6) pp 505 – 518.

3444. Sapienza H 1992 When do venture capitalists add value? *Journal of Business Venturing* **7** pp 9 – 27.

3445. Sapienza H, Gupta A 1994 Impact of agency risks and task uncertainty on venture capitalist-CEO interaction *Academy of Management Journal* **37** pp 1618 – 1632.

3446. Hall J, Hofer C 1993 Venture capitalists' decision criteria in new venture evaluation *Journal of Business Venturing* **8** (1) pp 25 – 42.

3447. Norton E, Tenenbaum B H 1993 Specialization versus diversification as a venture capital investment strategy *Journal of Business Venturing* **8** (5) pp 431 – 442.

3448. Rosenstein J A, Bruno V, Bygrave W D, Taylor N T 1993 The CEO, venture capitalists, and the board *Journal of Business Venturing* **8** (2) pp 99 – 113.

3449. Sahlman W A 1993 Aspects of financial contracting in venture capital *in* Chew D H (editor) The new corporate finance. Where theory meets practice *McGraw-Hill* pp 229 – 242.

3450. Admati A, Pfleiderer P 1994 Robust financial contracting and the role of venture capitalists *Journal of Finance* **49** pp 371 – 402.

3451. Aghion P, Tirole J 1994 The management of innovation *Quarterly Journal of Economics* pp 109 1185 – 1209.

3452. Anton J, Yao D 1994 Expropriation and inventions *American Economic Review* **84** pp 190 – 209.

3453. Barry C B 1994 New directions in research on venture capital finance *Financial management* pp 3 – 15.

3454. Berglöf E 1994 A control theory of venture capital finance *Journal of Law, Economics and Organization* **10** pp 247 -267.

3455. Bhidé A 1994 The origin and evolution of new businesses *Oxford University Press* Oxford UK.

3456. Fried V, Hisrich R 1994 Toward a model of venture capital investment decision making *Financial Management* **23** (3) pp 28 – 37.

3457. Gompers P 1994 The rise and fall of venture capital *Business and Economic History* **23** pp 1 – 26.

3458. Gompers P 1995 Optimal investment, monitoring, and the staging of venture capital *Journal of Finance* **50** pp 1461 – 1489.

3459. Gompers P 1996 Grandstanding in the venture capital industry *Journal of Financial Economics* **42** pp 133 – 156.

3460. Gompers P A 1998 Venture capital growing pains: Should the market diet? *Journal of Banking and Finance* **22** (6-8) pp 1089 – 1104.

3461. Gompers P A 2002 Gompers P A 2002 Corporations and the financing of innovation: The corporate venturing experience *Federal Reserve Bank of Atlanta Economic Review Fourth Quarter* pp 1 – 18.

3462. Gompers P, Lerner J 1996 The use of covenants: An empirical analysis of venture partnership agreements *Journal of Law and Economics* **39** pp 463 – 498.

3463. Gompers P, Lerner J 1997 Risk and reward in private equity investments: The challenge of performance assessment *Journal of Private Equity* **1** pp 5 – 12.

3464. Gompers P, Lerner J 1998a Venture capital distributions: Short-run and long-run reactions *Journal of Finance* **53** pp 2161 – 2183.

3465. Gompers P, Lerner J 1998b What drives venture capital fundraising*? Brookings Papers on Economic Activity Microeconomics* pp 149 – 192.

3466. Gompers P, Lerner J 1998c The determinants of corporate venture capital success: organizational structure, incentives, and complementarities *National Bureau of Economic Research Conference* volume on Concentrated Ownership *NBER* Cambridge Massachusetts USA.

3467. Gompers P, Lerner J 1999a Conflict of interest in the issuance of public securities: Evidence from venture capital *Journal of Law and Economics* **42** pp 1 – 28.

3468. Gompers P, Lerner J 1999b An analysis of compensation in the U.S. venture capital partnership *Journal of Financial Economics* **51** pp 3 – 44.

3469. Gompers P, Lerner J 1999c Capital market imperfections in venture markets: A report to the advanced technology program *US Department of Commerce US Government* Washington USA.

3470. Gompers P, Lerner J 1999d The venture capital cycle *MIT Press* Cambridge Massachusetts USA.

3471. Gompers P, Lerner J 2000a The determinants of corporate venture capital success in Morck R (editor) Concentrated corporate ownership *University of Chicago Press* Chicago USA

3472. Gompers P, Lerner J 2000b Money chasing deals? The impact of fund inflows on private equity valuation *Journal of Financial Economics* **55** pp 281 – 325.

3473. Gompers P, Lerner J 2001a The venture capital revolution *Journal of Economic Perspectives* **15** pp 145 – 168.

3474. Gompers P, Lerner J 2001b The money of invention *Harvard Business School Press* Boston USA.

3475. Baker M, Gompers P 2003 The determinants of board structure at the initial public offering *Journal of Law and Economics* **46** pp 569 – 598.

3476. Gompers P, Lerner J, Scharfstein D 2005 Entrepreneurial spawning: Public corporations and the genesis of new ventures, 1986 to 1999 *Journal of Finance* **60** pp 577 – 614.

3477. Gompers P A, Kovner A, Lerner J, Scharfstein D 2006 Skill vs luck in entrepreneurship and venture capital: Evidence from serial entrepreneurs *NBER Working Paper no 12592*.

3478. Gompers P 2007 Venture capital in Eckbo E (editor) Handbook of corporate finance vol **1** *North- Holland* Amsterdam The Netherlands.

3479. Gompers P, Kovner A, Lerner J, Scharfstein D 2008 Venture capital investment cycles: The impact of public markets *Journal of Financial Economics* **87** pp 1 – 23.

3480. Gompers P, Kovner A, Lerner J 2009 Specialization and success: evidence from venture capital *Journal of Economics and Management Strategy* **18** pp 817 – 844.

3481. Gompers P, Lerner J, Scharfstein D, Kovner A 2010 Performance persistence in entrepreneurship *Journal of Financial Economics* **96** pp 18 – 32.

3482. Knight R 1994 Criteria used by venture capitalists: A cross cultural analysis *International Small Business Journal* vol **13** no 1 pp 26 – 37.

3483. Kroszner R, Rajan R 1994 Is the Glass-Steagall act justified? A study of the US experience with universal banking before 1933 *American Economic Review* **84** pp 810 – 832.

3484. Lerner J 1994a Venture capitalists and the decision to go public *Journal of Financial Economics* **35** pp 293 – 316.

3485. Lerner J 1994b The syndication of venture capital investments *Financial Management* **23** pp 16 – 27.

3486. Lerner J 1995a Venture capitalists and the oversight of private firms *Journal of Finance* **50** pp 301 – 318.

3487. Lerner J 1995b Patenting in the shadow of competitors *Journal of Law and Economics* vol **38** pp 463 – 488.

3488. Lerner J 1998 Angel financing and public policy: An overview *Journal of Banking and Finance* **22** pp 773 – 783.

3489. Lerner J 1999 The government as venture capitalist: The long-run impact of the SBIR program *Journal of Business* **72** pp 285 – 318.

3490. Lerner J 2002 Boom and bust in the venture capital industry and the impact on innovation *Federal Reserve Bank of Atlanta Economic Review Fourth Quarter* USA pp 1 – 15.

3491. Kortum S, Lerner J 1998 Does venture capital spur innovation? *NBER Working Paper no W6846.*

3492. Kortum S, Lerner J 2000 Assessing the contribution of venture capital to innovation *RAND Journal of Economics* **31** (4) pp 674 – 692.

3493. Lerner J, Shane H, Tsai A 2003 Do equity financing cycles matter? Evidence from biotechnology alliances *Journal of Financial Economics* **67** pp 411 – 446.

3494. Lerner J, Schoar A 2004 The illiquidity puzzle: theory and evidence from private equity *Journal of Financial Economics* **72** pp 3 – 40.

3495. Lerner J, Schoar A 2005 Does legal enforcement affect financial transactions? The contractual channel in private equity *Quarterly Journal of Economics* **120** pp 223 – 246.

3496. Lerner J, Moore D, Shepherd S 2005 A study of New Zealand's venture capital market and implications for public policy *A report to the Ministry of Research, Science and Technology* New Zealand http://papers.ssrn.com/sol3/papers.cfm?abstract_id=1459175 .

3497. Lerner J, Schoar A, Wongsunwai W 2007 Smart institutions, foolish choices? The limited partner performance puzzle *Journal of Finance* **62** pp 731 – 764.

3498. Lerner J 2008, 2009 Boulevard of broken dreams: Why public efforts to boost entrepreneurship and venture capital have failed *Princeton University Press* Princeton USA.

3499. Lerner J, Sorensen M, Strömberg P 2009 What drives private equity activity and success globally *in* Gurung A, Lerner J (editors) Globalization of alternative investments *Working Papers* vol **2** The global economic impact of private equity report 2009 *World Economic Forum* Geneva Switzerland  pp 63 – 98.

3500. Chen H, Gompers P, Kovner A, Lerner J 2009 Buy local? The geography of successful and unsuccessful venture capital expansion *Working Paper 15102* National Bureau of Economic Research Cambridge MA USA

http://www.nber.org/papers/w15102 .

3501. Lerner J, Tåg J 2012, 2013 Institutions and venture capital. Working Paper 2012 : 17 *Swedish Entrepreneurship Forum* pp 1 – 31, *Industrial and Corporate Change* **22** (1) pp 153 – 182.

3502. Puri M 1994 The long-term default performance of bank underwritten security issues *Journal of Banking and Finance* **18** pp 397 – 418.

3503. Puri M 1996 Commercial banks in investment banking: Conflict of interest or certification role? *Journal of Financial Economics* **40** pp 373 – 401.

3504. Puri M, Robinson D 2011 Optimism and economic choice *Journal of Financial Economics* **86** pp 71 – 99.

3505. Timmons J A, Spinelli S 1994 New venture creation: Entrepreneurship for the 21st century *Irwin* Burr Ridge IL USA.

3506. Anton J, Yao D 1995 Start-ups, spin-offs, and internal projects *Journal of Law, Economics and Organization* **11** pp 362 – 378.

3507. Elango B, Fried V H, Hisrich R D, Polonchek A 1995 How venture capital firms differ *Journal of Business Venturing* vol **10** no 2 pp 157 – 179.

3508. Fiet J O 1995 Risk avoidance strategies in venture capital markets *Journal of Management Studies* **32** (4) pp 551 – 574.

3509. Hart O 1995 Firms, contracts, and financial structure *Oxford University Press* UK.

3510. Jain B, Kini O 1995 Venture capitalist participation and the post-issue operating performance of IPO firms *Managerial and Decision Economics* **5** pp 593 – 606.

3511. Loughran T, Ritter J 1995 The new issues puzzle *Journal of Finance* **50** pp 23 – 51.

3512. Willner R 1995 Valuing start-up venture growth options Real options in capital investment Trigeorgies L (editor) *Praeger* pp 221 – 239.

3513. Mason C M, Harrison R T 1996 Informal venture capital: A study of the investment process, the post-investment experience and investment performance *Entrepreneurship & Regional Development* **8** (2) pp 105 – 126.

3514. Mason C, Stark M 2004 What do investors look for in a business plan? A comparison of the investment criteria of bankers, venture capitalists and business angels *International Small Business Journal* 22 (3) pp 227 – 248.

3515. Sapienza H, Manigart S, Vermeir 1996 Venture capital governance and value-added in four countries *Journal of Business Venturing* **11** pp 439 – 469.

3516. Manigart S, Wright M, Robbie K, Desbrieres P, De Waele K 1997 Venture capitalists appraisal of investment projects: An empirical *European study Entrepreneurship Theory and Practice* vol **21** (4) pp 29 – 43.

3517. Manigart S, De Waele K, Wright M, Robbie K, Desbrieres P, Sapienza H, Beekman A 2000 Venture capitalists, investment appraisal and accounting information: A comparative study of the USA, UK, France, Belgium and Holland *European Financial Management* **6** (3) pp 389 – 403.

3518. Manigart S, De Waele K, Wright M, Robbie K, Desbrieres P, Sapienza H J, Beekman A 2002 The determinants of the required returns in venture capital investments: a five-country study *Journal of Business Venturing* **17** (4) pp 291 – 312.

3519. Manigart S, Baeyens K, Hyfte W V 2002 The survival of venture capital backed companies *Venture Capital* vol **4** no 2 pp 103 – 124.

3520. Muzyka D, Birley S, Leleux B 1996 Trade-offs in the investment decisions of European VCs *Journal of Business Venturing* **11** (4) pp 273 – 288.

3521. Packer F 1996 Venture capital, bank shareholding, and IPO underpricing in Japan in Levis M (editor) Empirical issues in raising equity capital *North-Holland* Amsterdam The Netherlands.

3522. Pettway R, Kaneko T 1996 The effects of removing price limits and introducing auctions upon short-term IPO returns: The case of Japanese IPOs *Pacific-Basin Finance Journal* **4** pp 241 – 158.

3523. Amit A R, Brander J, Zott C 1997 Venture capital financing of entrepreneurship in Canada *in* Halpern P (editor) Financing innovative enterprise in Canada *University of Calgary Press* Calgary Alberta Canada pp 237 – 277.

3524. Brav A, Gompers P 1997 Myth or reality? The long-run underperformance of initial public offerings: Evidence from venture- and non venture-capital-backed companies *Journal of Finance* **52** pp 1791 – 1821.

3525. Cai J, Wei K C J 1997 The investment and operating performance of Japanese Initial Public Offerings *Pacific - Basin Finance Journal* **5** pp 389 – 417.

3526. Chevalier J, Ellison G 1997 Risk taking by mutual funds as a response to incentives *Journal of Political Economy* **105** pp 1167 – 1200.

3527. Fredriksen O 1997 Venture capital firms' relationship and cooperations with entrepreneurial companies *Thesis no 625* Linköping Studies in Science and Technology Linköpings Universitet SE-581 83 Linköping Sverige.

3528. Gilford S 1997 Limited attention and the role of the venture capitalist *Journal of Business Venturing* vol **12** pp 459 – 482.

3529. Karsai J, Wright M, Filatotchev I 1997 Venture capital in transition economies: The case of Hungary *Entrepreneurship Theory and Practice* **21** (4) pp 93 – 110.

3530. Shepherd D A, Zacharakis A 1997 Conjoint analysis: A window of opportunity for entrepreneurship research *Advances in entrepreneurship, firm emergence and growth* **3** pp 203 – 248.

3531. Shepherd D A 1999 Venture capitalists' assessment of new venture survival *Management Science* **45** (5) pp 621 – 632.

3532. Shepherd D A, Zacharakis A 1999 Conjoint analysis: A new methodology for researching the decision policies of venture capitalists *Venture Capital* **1** pp 197 – 217.

3533. Shepherd D A, Ettenson R, Crouch A 2000 New venture strategy and profitability: A venture capitalist's assessment *Journal of Business Venturing* **15** (5) pp 449 – 467.

3534. Shepherd D A, Zacharakis A 2002 Venture capitalists' expertise: A call for research into decision aids and cognitive feedback *Journal of Business Venturing* **17** (1) pp 1 – 20.

3535. Wright M, Robbie K, Ennew Ch 1997 Venture capitalists and serial entrepreneurs *Journal of Business Venturing* **12** pp 227 – 249.

3536. Amit R, Brander J, Zott C 1998 Why do venture capital firms exist? Theory and Canadian evidence *Journal of business Venturing* **13** (6) pp 441 – 466.

3537. Bergemann D, Hege U 1998 Venture capital financing, moral hazard, and learning *Journal of Banking and Finance* **22** pp 703 – 735.

3538. Berger A, Udell G 1998 The economics of small business finance: the roles of private equity and debt markets in the financial growth cycle *Journal of Banking and Finance* **22** pp 613 – 673.

3539. Berger A, Schaek K 2011 Small and medium-sized enterprises, bank relationship strength, and the use of venture capital *Journal of Money, Credit and Banking* **43** pp 461 – 490.

3540. Black B, Gilson R 1998 Venture capital and the structure of capital markets: Banks versus stock markets *Journal of Financial Economics* **47** pp 243 – 277.

3541. Cornelius B, Isaksson A 1998 Venture capital incentives: A two country comparison *10th Nordic Conference on Small Business Research* Växjö University Växjö.

3542. Fried V H, Bruton G D, Hisrich R D 1998 Strategy and the board of directors in venture capital - backed firms *Journal of Business Venturing* vol **13** pp 493 – 503.

3543. Gerke W 1998 Market failure in venture capital markets for new medium and small enterprises *in* Hopt K J et al (editors) Comparative corporate governance – The state of the art and emerging research *Clarendon Press* Oxford UK pp 607 – 635.

3544. Hellmann T 1998 The allocation of control rights in venture capital contracts *RAND Journal of Economics* **29** pp 57 – 76.

3545. Hellmann Th F 2000 Venture capitalists: The coaches of Silicon Valley *in* Lee Ch-M, Miller W F Hancock M G, Rowen H S (editors) The Silicon Valley edge: A habitat for innovation and entrepreneurship *Stanford University Press* Stanford USA pp 276 – 294.

3546. Hellmann T, Puri M 2000 The interaction between product market and financing strategy: The role of venture capital *Review of Financial Studies* **13** pp 959 – 984.

3547. Hellmann T 2002 A theory of strategic venture investing *Journal of Financial Economics* **64** pp 285 – 314.

3548. Hellmann T, Puri M 2002 Venture capital and the professionalization of start-up firms: Empirical evidence *Journal of Finance* **57** (1) pp 169 – 197.

3549. Hellmann Th, Lindsey L, Puri M 2004 Building relationships early: Banks in venture capital *Working Paper 10535* National Bureau of Economic Research USA http://www.nber.org/papers/w10535 .

3550. Hellmann T 2006 IPOs, acquisitions, and the use of convertible securities in venture capital *Journal of Financial Economics* **81** pp 649 – 679.

3551. Hellmann T 2007 Entrepreneurs and the process of obtaining resources *Journal of Economics and Management Strategy* **16** pp 81 – 109.

3552. Hellmann T, Lindsey L, Puri M 2008 Building relationships early: Banks in venture capital *Review of Financial Studies* **21** pp 513 – 541.

3553. Hellmann T, Thiele V 2015 Friends or foes? The interrelationship between angel and venture capital markets *Journal of Financial Economics* **115** (3) pp 639 – 653.

3554. Hyde A 1998 Silicon valley's high-velocity labor market *Journal of Applied Corporate Finance* **11** pp 28 – 37.

3555. Jacobs O, Scheffler W 1998 Unternehmensbesteuerung und rechtsform, handbuch zur besteuerung deutscher unternehmen *2nd edition Beck* München Germany.

3556. Karsai J 1998 The appearance of venture capital in the Hungarian market and its participation in innovative processes *HFEP Research Booklet no 9* Hungarian Foundation for Enterprise Promotion and its Network Budapest Hungary.

3557. Karsai J, Wright M, Dudzinski Z, Morovic J 1999 Venture capital in transition economies: The cases of Hungary, Poland and Slovakia *in* Wright M, Robbie K (editors) Management buy-outs and venture capital *Edward Elgar* Cheltenham pp 81 – 114.

3558. Karsai J 2003 What has the state got to do with the venture capital market? Public financing of venture capital in Hungary *Acta Oeconomica* vol **53** no 3 pp 271 – 291.

3559. Karsai J 2004 Can the state replace private capital investors? Public financing of venture capital in Hungary *KTK/IE Discussion Papers 2004/9* Institute of Economics Hungarian Academy of Sciences Budapest Hungary HU ISSN 1785-377X ISBN 963 9588 09 1 pp 1 – 24.

3560. Lin T, Smith R 1998 Insider reputation and selling decisions: The unwinding of venture capital investments during equity IPOs *Journal of Corporate Finance* **4** pp 241 – 263.

3561. Marx L 1998 Efficient venture capital financing combining debt and equity *Review of Economic Design* **3** pp 371 – 387.

3562. Marx L, Strumsky D, Fleming L 2009 Mobility, skills, and the Michigan non-compete experiment *Management Science* **55** pp 875 – 889.

3563. Murray G C, Marriott R 1998 Why has the investment performance of technology-specialist European venture capital funds been so poor? *Research Policy* **27** pp 947 – 976.

3564. Prowse S 1998 Angel investors and the market for angel investments *Journal of Banking and Finance* **22** pp 785 – 792.

3565. Rajan R, Zingales L 1998 Financial dependence and growth *American Economic Review* **88** pp 59 – 86.

3566. Trester J 1998 Venture capital contracting under asymmetric information *Journal of Banking and Finance* **22** pp 675 – 699.

3567. Wright M, Robbie K 1998 Venture capital and private equity: A review and synthesis *Journal of Business Finance and Accounting* **25** (5) pp 521 – 570.

3568. Zider B 1998 How venture capital works *Harvard Business Review* pp 131 - 139.

3569. Aernoudt J 1999 European policy towards venture capital: Myth or reality? *Venture Capital* **1** (1) pp 47 – 57.

3570. Bliss R 1999 A venture capital model for transitioning economies: The case of Poland venture capital *International Journal of Entrepreneurial Finance* **1** (3) pp 241 – 257.

3571. Bygrave W D, Hay M, Peeters J B 1999 The venture capital handbook *Financial Times Prentice Hall* London UK.

3572. Gilson R 1999 The legal infrastructure of high technology industrial districts: Silicon Valley Route 128, and covenants not to compete *York University Law Review* **74** pp 575 – 629.

3573. Gilson R, Schizer D 2003 Understanding venture capital structure: A tax explanation for convertible preferred stock *Harvard Law Review* **116** pp 874 – 916.

3574. Gulati R, Gargiulo M 1999 Where do inter-organizational networks come from? *American Journal of Sociology* vol **104** (5) pp 1439 – 1493 *University of Chicago Press* Chicago USA.

3575. Hamao Y, Packer F, Ritter J R 1999 Institutional affiliation and the role of venture capital: Evidence from initial public offerings in Japan *Marshall School of Business University of Southern California Capital Markets Department Federal Reserve Bank of New York Warrington College of Business Administration University of Florida* USA pp 1 – 50.

3576. Leopold G 1999 Venture capital - das eigenkapitalgeschaft mit kleinen und mittleren unternehmen *Deutsches Steuerrecht* **37** pp 470 – 476.

3577. Neher D 1999 Staged financing: An agency perspective *Review of Economic Studies* **66** pp 255 – 274.

3578. Stillman R, Sunderland J, Heyl L, Swart H 1999 A venture capital programme for South Africa: Study and recommendations *Nathan Associates Working Paper* Arlington Virginia USA.

3579. Baygan G, Freudenberg M 2000 The internationalization of venture capital activity in OECD countries: Implications for measurement and policy *OECD Science, Technology and Industry Working Papers 2000/07 OECD Publishing* pp 1 – 55 http://dx.doi.org/10.1787/084236411045 .

3580. Baygan G 2003 Venture capital policies in Korea *OECD Science, Technology and Industry Working Papers 2003/02 OECD Publishing* Directorate for Science, Technology and Industry Organization for Economic Co-operation and Development pp 1 – 21 http://www.oecd.org/sti/working-papers http://dx.doi.org/10.1787/248000716362 .

3581. Bharat A, Galetovic A 2000 Weak property rights and holdup in R&D *Journal of Economics and Management Strategy* **9** pp 615 – 642.

3582. Cumming D J 2000 The convertible preferred equity puzzle in Canadian venture capital finance *Working Paper* University of Alberta
www.ssrn.com .

3583. Cumming D J 2001 The determinants of venture capital portfolio size: Empirical evidence *Working Paper* University of Alberta
www.ssrn.com .

3584. Cumming D J, MacIntosh J G 2000 Venture capital exits in Canada and the United States *University of Toronto Law Journal 2003*
www.ssrn.com .

3585. Cumming D J, MacIntosh J G 2001 Venture capital investment duration in Canada and the United States *Journal of Multinational Financial Management* **11** pp 445 – 463.

3586. Cumming D J, MacIntosh J G 2002a Crowding out private equity: Canadian evidence *Working Paper* University of Alberta and University of Toronto
www.ssrn.com .

3587. Cumming D J, MacIntosh J G 2002b A cross-country comparison of full and partial venture capital exits *Journal of Banking and Finance*

www.ssrn.com .

3588. Cumming D J, MacIntosh J G 2002c The extent of venture capital exits: Evidence from Canada and the United States in McCahery J, Renneboog L D R (editors) Venture capital contracting and the valuation of high-tech firms *Oxford University Press*

www.ssrn.com .

3589. Cumming D J, MacIntosh J G 2002d Economic and institutional determinants of venture capital investment duration in Libecap G (editor) Advances in the study of entrepreneurship innovation and economic growth *JAI Press*.

3590. Cumming D, Fleming G 2002 A law and finance analysis of venture capital exits in emerging markets *Working Paper Series in Finance 02-03* The Australian National University Canberra Australia.

3591. Cumming D J, MacIntosh J G 2003a A cross-country comparison of full and partial venture capital exits *Journal of Banking and Finance* **27** pp 511 – 548.

3592. Cumming D J, MacIntosh J G 2003b Venture-capital exits in Canada and the United States *The University of Toronto Law Journal* **53** (2) pp 101 – 199.

3593. Cumming D J, MacIntosh J F 2003c Comparative venture capital governance: private versus labour sponsored venture capital funds *CESifo Working Paper no 853* 2002 CESifo Area Conference on Venture Capital, Entrepreneurship and Public Policy pp 1 – 41.

3594. Cumming D, Fleming G, Schwienbacher A, 2005 Liquidity risk and venture capital finance *Financial Management* **34** (4) pp 77 – 105.

3595. Cumming D, Fleming G, Suchard J 2005 Venture capitalist value-added activities, fundraising and drawdowns *Journal of Banking and Finance* **29** pp 295 – 331.

3596. Cumming D, Fleming G, Schwienbacher A 2006 Legality and venture capital exits *Journal of Corporate Finance* **12** pp 214 – 245.

3597. Cumming D, MacIntosh J 2006 Crowding out private equity: Canadian evidence *Journal of Business Venturing* **21** pp 569 – 609.

3598. Cumming D 2008 Contracts and exits in venture capital finance *Review of Financial Studies* **21** (5) pp 1947 – 1982.

3599. Cumming D, Johan 2008 Preplanned exit strategies in venture capital *European Economic Review* **52** pp 1209 – 1241.

3600. Cumming D, Fleming G, Schwienbacher A 2009 Style drift in private equity *Journal of Business Finance and Accounting* **36** pp 645 – 678.

3601. Cumming D, Walz U 2010 Private equity returns and disclosure around the world *Journal of International Business Studies* **41** pp 727 – 754.

3602. Gans J, Stern S 2000 Incumbency and R&D incentives: Licensing the gale of creative destruction *Journal of Economics and Management Strategy* **9** pp 485 – 511.

3603. Gans J, Hsu D, Stern S 2002 When does start-up innovation spur the gale of creative destruction? *RAND Journal of Economics* **33** pp 571 – 586.

3604. Gans J, Stern S 2003 When does funding research by smaller firms bear fruit? Evidence from the SBIR program *Economics of Innovation and New Technology* **16** pp 361 – 384.

3605. Jain A, Kini O 2000 Does the presence of venture capitalists improve the survival profile of IPO firms? *Journal of Business Finance and Accounting* vol **27** issues 9, 10 pp 1139 – 1176.

3606. Jeng L, Wells P 2000 The determinants of venture capital funding: Evidence across countries *Journal of Corporate Finance* **6** pp 241 – 289.

3607. Kaplan S, Strömberg P 2000 Financial contracting theory meets the real world: An empirical analysis of venture capital contracts *NBER Working Paper no 7660*.

3608. Kaplan S, Strömberg P 2001 Venture capitalists as principals: Contracting, screening, and monitoring *American Economic Review* **91** pp 426 – 430.

3609. Kaplan S, Strömberg P 2002 Financial contracting theory meets the real world: An empirical analysis of venture capital contracts *Review of Economic Studies* vol **70** no 2 pp 281 – 315.

3610. Kaplan S, Strömberg P 2003 Financial contracting theory meets the real world: An empirical analysis of venture capital contracts *Review of Economic Studies* **70** pp 281 – 315.

3611. Kaplan S, Strömberg P 2004 Characteristics, contracts, and actions: Evidence from venture capitalist analyses *Journal of Finance* **59** pp 2177 – 2210.

3612. Kaplan S, Schoar A 2005 Private equity performance: returns, persistence, and capital flows *Journal of Finance* **60** pp 1791 – 1823.

3613. Kaplan S, Martel F, Strömberg P 2007 How do legal differences and experience affect financial contracts? *Journal of Financial Intermediation* **16** pp 273 – 311.

3614. Kaplan S, Strömberg P 2009 Leveraged buyouts and private equity *Journal of Economic Perspectives* **23** pp 121 – 146.

3615. Kaplan S, Sensoy B, Strömberg P 2009 Should investors bet on the jockey or the horse? Evidence from the evolution of firms from early business plans to public companies *Journal of Finance* **64** (1) pp 75 – 115.

3616. Kaplan S, Lerner J 2010 It ain't broke: The past, present, and future of venture capital *Journal of Applied Corporate Finance* **22** pp 36 – 47.

3617. Karaömerlioğlu D Ç, Jacobsson S 2000 The Swedish venture capital industry – an infant, adolescent or grown-up? *Venture Capital* **2** pp 61 – 68.

3618. Koski T 2000 Success potential of a venture capital investment. A model to simulate venture capitalists' perception using fuzzy set theory *PhD Dissertation A-177 HeSE Print* Helsinki School of Economics and Business Administration Helsinki Finland.

3619. Lee J 2000 Challenges of Korean technology-based ventures and governmental policies in the emergent-technology sector *Technovation* **20**.

3620. Lehtonen S 2000 Venture capitalist's exit vehicles and their effects on perceived utility. Allocation of rewards and contract structure *PhD Dissertation no 89* Swedish School of Economics and Business Administration Helsingfors Sweden.

3621. Quindlen R 2000 Confessions of a venture capitalist: Inside the high-stakes world of new venture financing *Warner Books* New York USA.

*3622.* Schefczyk M 2000 Erfolgsstrategien deutscher venture capital - gesellschaften *2nd edition Schaeffer-Poeschel* Stuttgart Germany.

*3623.* Schertler A 2000 The impact of public subsidies on venture capital investments in start-up enterprises *Kiel Working Paper no 1018* p 23.

*3624.* Van Osnabrugge M 2000 A comparison of business angel and venture capitalist investment procedures: An agency theory-based analysis *Venture Capital: An International Journal of Entrepreneurial Finance* **2** (2) pp 91 – 109.

*3625.* Zacharakis A L, Meyer G D 2000 The potential of actuarial decision models: Can they improve the venture capital investment decision? *Journal of Business Venturing* **15** (4) pp 323 – 346.

*3626.* Zacharakis A L, McMullen, J S, Shepherd D A 2007 Venture capitalists' decision policies across three countries: An institutional theory perspective *Journal of International Business Studies* **38** (5) pp 691 – 708.

*3627.* Bascha A, Walz U 2001 Convertible securities and optimal exit decisions in venture capital finance *Journal of Corporate Finance* **7** pp 285 – 306.

*3628.* Engel D 2001a Höheres beschäftigungswachstum durch venture capital? *ZEW Discussion Paper no 01-34* Mannheim Germany.

*3629.* Engel D 2001b Die identifizierung VC-finanzierter unternehmen in den *ZEW Gründungspanels Unveröffentlichtes Manuskript* Mannheim Germany.

*3630.* Engel D 2002 Welche regionen profitieren von venture capital - aktivitäten? *ZEW Discussion Papers no 02-37*
http://hdl.handle.net/10419/24770 ,
ftp://ftp.zew.de/pub/zew-docs/dp/dp0237.pdf .

*3631.* Engel D, Keilbach M 2007 Firm-level implications of early stage venture capital investment—An empirical investigation *Journal of Empirical Finance* **14** (2) pp 150 – 167.

*3632.* Francis B B, Hasan I 2001 The underpricing of venture and non-venture capital IPOs: An empirical investigation *Journal of Financial Services Research* **19** pp 93 – 113.

*3633.* Fredriksen Ø, Klofsten M 2001 Venture capitalists' governance of their portfolio companies *Journal of Enterprising Culture* **9** pp 201 – 219.

*3634.* Hyytinen A, Pajarinen M 2001 Financial systems and venture capital in Nordic Countries: A comparative study *Discussion Paper 774* The Research Institute of the Finnish Economy Helsinki Finland.

*3635.* Keuschnigg Ch, Nielsen S B 2001 Public policy for venture capital *International Tax and Public Finance* **8** (4) pp 557 – 572.

*3636.* Keuschnigg Ch 2003 Optimal public policy for venture capital backed innovation *CEPR DP 3850.*

*3637.* Keuschnigg Ch, Nielsen S B 2003a Tax policy, venture capital, and entrepreneurship *Journal of Public Economics* **87** (1) pp 175 – 203.

*3638.* Keuschnigg C, Nielsen S B 2003b Taxes and venture capital support *European Finance Review* **7** pp 515 – 539.

*3639.* Keuschnigg, Ch 2004a Taxation of a venture capitalist with a portfolio of firms *Oxford Economic Papers* **56** (1) pp 285 – 306.

*3640.* Keuschnigg Ch 2004b Venture capital backed growth *Journal of Economic Growth* **9** (2) pp 239 – 261.

*3641.* Kanniainen V, Keuschnigg Ch 2004 Start-up investment with scarce venture capital support *Journal of Banking and Finance* **28** pp 1935 – 1959.

*3642.* Keuschnigg Ch, Nielsen S 2004a Start-ups, venture capitalists, and the capital gains tax *Journal of Public Economics* **88** pp 1011 – 1042.

*3643.* Keuschnigg C, Nielsen S B 2004b Taxation and venture capital backed entrepreneurship *International Tax and Public Finance* **11** pp 369 – 390.

*3644.* Kirilenko A 2001 Valuation and control in venture finance *Journal of Finance* **56** pp 565 – 587.

*3645.* Lockett A, Wright M 2001 The syndication of venture capital investments Omega The International *Journal of Management Science* **29** pp 375 – 390.

*3646.* Maula M, Murray G 2001 Corporate venture capital and the creation of US public companies *in* Hitt A, Amit R, Lucier C, Nixon D (editors) Creating value: Winners in the new business environment *Blackwell* New York USA.

*3647.* Peng L 2001 Building a venture capital index *Social Science Research Network Electronic Paper Collection* Economics Department Yale University pp 1 – 54 http://paper.ssrn.com/abstract=281804 .

*3648.* Seppä T J, Laamanen T 2001 Valuation of venture capital investments: Empirical evidence *R&D Management* vol **31** (2) pp 215 – 230.

*3649.* Seppä T J 2003 Essays on the valuation and syndication of venture capital investments *PhD Dissertation 2003/1* Espoo Finland.

*3650.* Shachmurove Y 2001 Annualized returns of venture-backed public companies categorized by stage of financing *Journal of Entrepreneurial Finance* vol **6** no 1 pp 44 – 58.

*3651.* Shachmurove Y 2007a Innovation and trade: Introduction and comments in Sheshinski E, Strom R J, Baumol W J (editors) Entrepreneurship, innovation, and the growth mechanism of the free-enterprise economies *Princeton University Press* pp 247 – 260.

*3652.* Shachmurove Y 2007b Geography and industry meets venture capital *Departments of Economics The City College of the City University of New York University of Pennsylvania* USA pp 1 – 34

http://ssrn.com/abstract=977989 .

*3653.* Shachmurove A, Shachmurove Y 2004 Annualized and cumulative returns on venture-backed public companies categorized by industry *The Journal of Entrepreneurial Finance and Business Ventures* vol **9** issue 3 pp 41 – 60.

*3654.* Shachmurove E, Shachmurove Y 2004 Annualized returns of ventured-backed public companies stratified by decades and by stage of financing *The Journal of Entrepreneurial Finance and Business Ventures* vol **9** issue 2 pp 109 – 123.

*3655.* Sorenson O, Stuart T 2001 Syndication networks and the spatial distribution of venture capital investments *American Journal of Sociology* **106** pp 1546 – 1588.

*3656.* Allen F, Song W 2002 Venture capital and corporate governance http://fic.wharton.upenn.edu/fic/papers/03/0305.pdf .

3657. Audretsch D, Lehmann E 2002 Debt or equity? The role of venture capital in financing the new economy Germany *CEPR Working Paper no 3656*.

3658. Bottazzi L, Da Rin M 2002a Venture capital in Europe and the financing of innovative companies *Economic Policy* **34** pp 229 – 269.

3659. Bottazzi L, Da Rin M 2002b Europe's "New" stock markets *CEPR Discussion Paper no 3521*.

3660. Bottazzi L, Da Rin M, Giavazzi F 2003 Research, patents, and the financing of ideas: Why is the EU growth potential so low?' *in* Sapir A, Nava M (editors) Economic policy-making in the European Union *European Commission* Brussels Belgium.

3661. Bottazzi L, Da Rin M 2004 Financing European entrepreneurial firms: Facts, issues and research agenda *in* Keuschnigg Ch, Kanniainen V (editors) Venture capital, entrepreneurship and public policy Cambridge MA *MIT Press* USA.

3662. Bottazzi L, Da Rin M, Hellmann T 2004a The changing face of the European venture capital industry: Facts and analysis *Journal of Private Equity* **7** pp 26 – 53.

3663. Bottazzi L, Da Rin M, Hellmann Th 2004b Active financial intermediation: Evidence on the role of organizational specialization and human capital *RICAFE Working Paper no 12*.

3664. Bottazzi L, Da Rin M, Hellmann T 2008 Who are the active investors? Evidence from venture capital *Journal of Financial Economics* **89** pp 488 – 512.

3665. Bottazzi L, Da Rin M, Hellmann T 2009 What is the role of legal systems in financial intermediation? Theory and evidence *Journal of Financial Intermediation* **18** pp 559 – 598.

3666. Brander J, Amit R, Antweiler W 2002 Venture capital syndication: Improved venture selection versus value-added hypothesis *Journal of Economics and Management Strategy* **11** (3) pp 423 – 452.

3667. Brander J, Egan E, Hellmann T 2008 Government sponsored versus private venture capital: Canadian evidence *in* Lerner J and Schoar A (editors) *NBER Conference Volume on International Differences in Entrepreneurship.*

3668. Brander J, De Bettignies J 2009 Venture capital investment: the role of predator-prey synamics with learning by doing *Economics of Innovation and New Technology* **18** pp 1 – 19.

3669. Brander J, Du Q, Hellmann T 2010 Governments as venture capitalists: Striking the right balance in *Globalization of alternative investments* vol **3** The global economic impact of private equity report 2010 *World Economic Forum* Davos Switzerland.

3670. Chesbrough H W 2002 Making sense of corporate venture capital *Harvard Business Review* **80** (3) pp 4 – 11.

3671. Cestone G 2002 Venture capital meets contract theory: Risky claims or formal control? *CEPR Discussion Paper 3462*.

3672. Davis M, Schachermayer W, Tompkins R 2002 The evaluation of venture capital as an instalment option: Valuing real options using real options pp 1 – 31.

3673. Dossani R, Kenney M 2002 Creating an environment for venture capital in India *World Development* **30** (2) pp 227 – 253.

3674. Eisele F, Habermann M, Oesterle R 2002 Die beteiligungskriterien fur eine venture capital finanzierung - eine empirische analyse der phasenbezogenen bedeutung *Tubinger Diskussionsbeitrag no 238* Germany pp 1 – 35.

3675. Everts M 2002 Cash dilution in illiquid funds *MPRA Paper no 4655* Munich University Munich Germany pp 1 – 15

http://mpra.ub.uni-muenchen.de/4655/ .

3676. Gilson R, Schizer D 2002 Understanding venture capital structure: A tax explanation for convertible preferred stock *Columbia Law School Working Paper no 230* Columbia University New York USA.

3677. Kenney M, Han K, Tanaka S 2002 Scattering geese: The venture capital industries of East Asia *A Report to the World Bank BRIE Working Paper 146* Berkeley California USA http://brie.berkeley.edu/publications/wp146.pdf .

3678. Koh F C C, Koh W T H 2002 Venture capital and economic growth: An industry overview and Singapore's experience *Working Paper no 21 - 2002* http://www.research.smu.edu.sg/faculty/edge/entrep_fin/papers/SERVCandGrowth_WP_dec_2002.pdf .

3679. McGlue D 2002 The funding of venture capital in Europe: Issues for public policy *Venture Capital* **4** (1) pp 45 – 58.

3680. Moskowitz T, Vissing-Jørgensen A 2002 The returns to entrepreneurial investment: A private equity premium puzzle? *American Economic Review* **92** pp 745 – 778.

3681. Shane S, Cable D 2002 Network ties, reputation, and the financing of new ventures *Management Science* **48** (3) pp 364 – 381.

3682. Shane S, Stuart T 2002 Organizational endowments and the performance of university start-ups *Management Science* **48** pp 154 – 170.

3683. Zook M 2002 Grounded capital: Venture financing and the geography of the Internet industry, 1994-2000 *Journal of Economic Geography* **2** pp 151 – 177.

3684. Brav A, Gompers P 2003 The role of lockups in Initial Public Offerings *Review of Financial Studies* **16** pp 1 – 29.

3685. Becker R, Hellman T 2003 The genesis of venture capital: Lessons from the German experience CESifo *Working Paper no 883* Sauder School of Business.

3686. Bergemann D, Hege U 2003 The value of benchmarking *in* McCahery J A, Renneboog L (editors) Venture capital contracting and the valuation of high tech firms *Oxford University Press* Oxford UK pp 83 – 107.

3687. Casamatta C 2003 Financing and advising: optimal financial contracts with venture capitalists *Journal of Finance* **58** pp 2059 - 2086.

3688. Casamatta C, Haritchabalet C 2007 Experience, screening and syndication in venture capital investments *Journal of Financial Intermediation* **16** pp 368 – 398.

3689. Cornelli F, Yosha O 2003 Stage financing and the role of convertible securities *Review of Economic Studies* **70** pp 1 – 32.

3690. Das S R, Jagannathan M, Sarin A 2003 Private equity returns: An empirical examination of the exit of venture-backed companies (Digest summary) *Journal of Investment Management* 1 (1152-177).

3691. Davila A, Foster G, Gupta M 2003 Venture capital financing and the growth of start-up firms *Journal of Business Venturing* **18** pp 689 – 708.

3692. Franzke S A, Grohs S, Laux Ch 2003 Initial public offerings and venture capital in Germany *CFS Working Paper no 2003/26* Center for Financial Studies an der Johann Wolfgang Goethe-*Universität Frankfurt am Main* Leibniz Information Centre for Economics Germany pp 1 – 39

http://hdl.handle.net/10419/25395 ,

www.ifk-cfs.de www.econstor.eu.

3693. Gawlik R, Teczke J 2003 Spin of enterprises as a financing solution for the innovative clients of science and technology parks *MPRA Paper no 45224* Munich University Munich Germany pp 1 – 7

http://mpra.ub.uni-muenchen.de/45224/ .

3694. Gilson R, Schizer D 2003 Venture capital structure: A tax explanation for convertible preferred stock *Harvard Law Review* **116** pp 875 – 916.

3695. Hirukawa M, Ueda M 2003 Venture capital and productivity *University of Wisconsin-Madison* USA.

3696. Inderst R, Muller H 2003 The effect of capital market characteristics on the value of start-up firms *Journal of Financial Economics* **72** pp 319 – 356.

3697. Jones C M, Rhodes-Kropf M 2003 The price of diversifiable risk in venture capital and private equity *unpublished working paper* Columbia University NY USA.

3698. Keilbach M, Engel D 2003 Firm level implications of early stage venture capital investment: An empirical investigation *ZEW Discussion Papers no 02-82* Leibniz Information Centre for Economics Germany p 1 – 25

http://hdl.handle.net/10419/23962 ,

ftp://ftp.zew.de/pub/zew-docs/dp/dp0282.pdf ,

www.econstor.eu.

3699. Nielsen S B, Keuschnigg Ch 2006 Public policy, start-up entrepreneurship, and the market for venture capital *Working Paper 15-06* Department of Economics Copenhagen Business School Copenhagen Denmark pp 1 – 43.

3700. Leleux B, Surlemont B 2003 Public versus private venture capital: seeding or crowding out? A pan-European analysis *Journal of Business Venturing* **18** pp 81 – 104.

3701. Rindermann G 2003 Venture capitalist participation and the performance of IPO firms: Empirical evidence from France, Germany and the UK *EFMA 2003 Helsinki Meeting* Finland.

3702. Schmidt K 2003 Convertible securities and venture capital finance *Journal of Finance* **58** pp 1139 – 1166.

3703. Schmidt D, Wahrenburg M 2003 Contractual relations between European - VC funds and investors: The impact of reputation and bargaining power on contractual design *RICAFE Working Paper 12*

http://ww.lse.ac.uk/ricafe.

3704. Schertler A 2003 Driving forces of venture capital investments in Europe: A dynamic panel data analysis. European Integration. Financial Systems and Corporate Performance *EFIC Working Paper no 03-27* United Nations University.

3705. Stuart T, Sorenson O 2003 Liquidity events and the geographic distribution of entrepreneurial activity *Administrative Science Quarterly* **48** pp 175 – 201.

3706. Wang C K, Wang K, Lu Q 2003 Effects of venture capitalists' participation in listed companies *Journal Banking Finance* **27** pp 2015 – 2034.

3707. Wasserman N 2003 Founder-CEO succession and the paradox of entrepreneurial success *Organization Science* **14** pp 149 – 172.

3708. Wasserman N 2006 Stewards, agents, and the founder discount: Executive compensation in new ventures *Academy of Management Journal* **49** pp 960 – 976.

3709. Woodward S, Hall R 2003 Benchmarking the returns to venture *NBER Working Paper 10202*

http://www.nber.org/papers/w10202.

3710. Aghion P, Bolton P, Tirole J 2004 Exit options in corporate finance: Liquidity versus incentives *Review of Finance* **8** pp 327 – 353.

3711. Avnimelech G, Kenney M, Teubal M 2004 Building venture capital industries: understanding the US and Israeli experiences *Berkeley Roundtable on the International Economy* UC Berkeley USA.

3712. Avnimelech G, Teubal M 2004 Venture capital start-up co-evolution and the emergence & development of Israel's new high tech cluster *Economics of Innovation & New Technology* **13** (1) pp 33 – 60.

3713. Audretsch D B, Keilbach M 2004 Entrepreneurship capital and economic performance *Regional Studies* **38** (8) pp 949 – 959.

3714. Baum J, Silverman B 2004 Picking winners or building them? Alliances, patents, and human capital as selection criteria in venture financing of biotechnology startups *Journal of Business Venturing* **19** pp 411 – 436.

3715. Berk J, Green R, Naik V 2004 Valuation and return dynamics of new ventures *Review of Financial Studies* **17** pp 1 – 35.

3716. Da Rin M, Nicodano G, Sembenelli A 2004 Public policy and the creation of active venture capital markets *Working Paper no 270* IGIER – Università Bocconi Italy pp 1 – 38 http://www.igier.uni-bocconi.it .

3717. Da Rin M, Nicodano G, Sembenelli A 2005, 2006 Public policy and the creation of active venture capital markets *Journal of Public Economics* **90** pp 1699 – 1723

http://www.ecb.int ,

http://ssrn.com/abstract_id=634025 .

3718. Da Rin M, Hege M U, Llobet G, Walz U 2005 The law and finance of venture capital financing in Europe: Findings from the RICAFE research project *MPRA Paper no 39552* Munich University Munich Germany pp 1 – 30

http://mpra.ub.uni-muenchen.de/39552/ .

3719. Da Rin M, Hellmann Th, Puri M 2011 A survey of venture capital research *TILEC Discussion Paper no 2011-044 CentER Discussion Paper no 2011-111* Tilburg University Tilburg Netherlands University of British Columbia Vancouver British Columbia Canada Duke University Durham North Carolina USA ISSN 0924-7815 pp 1 – 136 http://ssrn.com/abstract=1942821.

3720. De Clercq D, Dimov D 2004 Explaining venture capital firms' syndication behavior: A longitudinal study. *Venture Capital: An International Journal of Entrepreneurial Finance* **6** (4) pp 243 – 256.

3721. De Clercq D, Dimov D 2010 Doing it not alone. Antecedents, dynamics, and outcomes of venture capital syndication *Venture capital* pp 221 – 242 *John Wiley & Sons Inc* USA.

3722. Dittmann I, Maug E, Kemper J 2004 How fundamental are fundamental values? Valuation methods and their impact on the performance of German venture capitalists *European Financial Management* vol **10** (4) pp 609 – 638.

3723. Hsu D 2004 What do entrepreneurs pay for venture capital affiliation? *Journal of Finance* **59** pp 1805 – 1844.

3724. Inderst R, Müller H 2004 The effect of capital market characteristics on the value of startup firms *Journal of Financial Economics* **72** pp 319 – 356.

3725. Inderst R, Müller H, Muennich F 2007 Financing a portfolio of projects *Review of Financial Studies* **20** pp 1289 – 1325.

3726. Inderst R, Müller H 2009 Early-stage financing and firm growth in new industries *Journal of Financial Economics* **93** pp 276 – 291.

3727. Jones C M, Rhodes-Kropf M 2004 The price of diversifiable risk in venture capital and private equity pp 1 – 53.

3728. Lee P, Wahal S 2004 Grandstanding, certification and the underpricing of venture capital backed IPOs *Journal of Financial Economics* **73** pp 375 – 407.

3729. Michelacci C, Suarez J 2004 Business creation and the stock market *Review of Economic Studies* **71** pp 459 – 481.

3730. Mishra A 2004 Indian venture capitalists (VCs): Investment evaluation criteria *ICFAI Journal of Applied Finance* **10** (7) pp 71 – 93.

3731. Peggy M L, Wahal S 2004 Grandstanding, certification and the underpricing of venture capital backed IPOs *Journal of Financial Economics* **73** pp 375 – 407.

3732. Repullo R, Suarez J 2004 Venture capital finance: A security design approach *Review of Finance* **8** pp 75 – 108.

3733. Roman A, van Pottelsberghe de la Potterie B 2004 The determinants of venture capital: A panel analysis of 16 OECD countries *Working paper WF CEB 04/015* Universite Libre de Bruxelles Brussels Belgium.

3734. Sternberg R J 2004 Successful intelligence as a basis for entrepreneurship *Journal Business Venture* **19** pp 189 – 201.

3735. Ueda M 2004 Banks versus venture capital: Project evaluation, screening, and expropriation *Journal of Finance* **59** pp 601 – 621.

3736. Becker R, Hellmann T 2005 The genesis of venture capital: Lessons from the German experience *in* Keuschnigg C, Kanniainen V (editors) Venture capital, entrepreneurship, and public policy Chapter 2 *MIT Press* Massachusetts Institute of Technology USA pp 33 – 67.

3737. Bergemann D, Hege U 2005 The financing of innovation: learning and stopping *RAND Journal of Economics* **36** pp 719 – 752.

3738. Cochrane J 2005 The risk and return of venture capital *Journal of Financial Economics* **75** (1) pp 3 – 52.

3739. De Carvalho A G, Calomiris C W, De Matos J A 2005 Venture capital as human resource management *NBER Working Paper no 11350*.

3740. Dessein W 2005 Information and control in ventures and alliances *Journal of Finance* **60** pp 2513 – 2549.

3741. Dessí R 2005 Start-up finance, monitoring and collusion *RAND Journal of Economics* **36** pp 255 – 274.

3742. Dimov D, Shepherd D 2005 Human capital theory and venture capital firms: Exploring "home runs" and "strike outs" *Journal of Business Venturing* **20** pp 1 – 21.

3743. Dushnitsky G, Lenox M 2005a When do firms undertake R&D by investing in new ventures? *Strategic Management Journal* **26** pp 947 – 965.

3744. Dushnitsky G, Lenox M J 2005b When do incumbents learn from entrepreneurial ventures? Corporate venture capital and investing firm innovation rates *Research Policy* **34** pp 615 – 639.

3745. Dushnitsky G, Lenox M J 2006 When does corporate venture capital create firm value? *Journal of Business Venturing* **21** pp 753 – 772.

3746. Dushnitsky G, Lavie D 2008 Interdependence in the evolution of inter-firm networks: How alliances shape corporate venture capital in the software industry *2007 West Coast Research Symp on Technology Entrepreneurship* University of Washington Seattle USA.

3747. Ernst H, Witt P, Brachtendorf G 2005 Corporate venture capital as a strategy for external innovation: An exploratory empirical study *R&D Management* **25** pp 233 – 242.

3748. Ge D, Mahoney J M, Mahoney J T 2005 New venture valuation by venture capitalists: An integrative approach *Federal Reserve Bank of New York* NY USA.

3749. Hsu D, Kenney M 2005 Organizing venture capital: The rise and demise of American Research & Development Corporation, 1946-1973 *Industrial and Corporate Change* **14** pp 579 – 616.

3750. Hsu D 2006 Venture capitalists and cooperative start-up commercialization strategy *Management Science* **52** pp 204 – 219.

3751. Hsu D 2007 Experienced entrepreneurial founders, organizational capital, and venture capital funding *Research Policy* **36** pp 722 – 741.

3752. Klepper S, Sleeper S 2005 Entry by spinoffs *Management Science* **51** pp 1291 – 1306.

3753. Klepper S, Thompson P 2010 Disagreements and intra-industry spinoffs *International Journal of Industrial Organization* **28** pp 526 – 538.

3754. Kõomägi M 2005a Riskikapitaliinvesteeringud eestis. Haridus ja majandus *Rahvusvahelise teaduskonverentsi materjalid* Tallinn Estonia pp 53 – 56.

3755. Kõomägi M 2005b Venture capital financing and deal structuring problems in Estonia. Business development possibilities in the new European area *Scientific Proceedings* Part 1 Vilnius Lithuania pp 125 – 130.

3756. Kõomägi M 2005c Riskikapitali hinnakujunemise mehhanism. Eesti ettevõtluse perspektiivid Euroopa liidus *Pärnu Mattimar OÜ* Tallinn-Pärnu Estonia pp 94 – 110.

3757. Kõomägi M, Sander P 2006 Venture capital investments and financing in Estonia: A case study approach *Tartu University Press* Faculty of Economics and Business Administration University of Tartu Estonia ISSN 1406–5967 ISBN 9949–11–264–8 pp 1 – 77

www.tyk.ee .

3758. Lai R 2005 Does competition kill ties? *MPRA Paper no 4759* Munich University Munich Germany pp 1 – 37

http://mpra.ub.uni-muenchen.de/4759/ .

3759. Mäkelä M M, Maula M M 2005 Cross-border venture capital and new venture internationalization: An isomorphism perspective *Venture Capital* vol **7** no 3 pp 227 – 257.

3760. Mayer C, Schoors K, Yafeh Y 2005 Sources of funds and investment activities of venture capital funds: Evidence from Germany, Israel, Japan and the United Kingdom *Journal of Corporate Finance* **11** pp 586 – 608.

3761. Neus W, Walz U 2005 Exit timing of venture capitalists in the course of an initial public offering *Journal of Financial Intermediation* **14** pp 253 – 277.

3762. Wong L H 2005 Venture capital fund management: A comprehensive approach to investment practices and the operations of a VC firm *Aspatore Books* Singapore pp 1 – 485.

3763. Zook M A 2005 The geography of the Internet industry: Venture capital, dotcoms, and local knowledge *Blackwell Publishing* Malden MA USA.

3764. Antonelli C, Teubal M 2006 Venture capitalism as a mechanism for knowledge governance *Working paper series no 4* Department of Economics Università di Torino Italy.

3765. Cassiman B, Ueda M 2006 Optimal project rejection and new firm start-ups *Management Science* **52** pp 262 – 275.

3766. Colombo M, Grilli L, Piva E 2006 In search of complementary assets: the determinants of alliance formation of high-tech start-ups *Research Policy* **35** pp 1166 – 1199.

3767. Dimov D, De Clercq D 2006 Venture capital investment strategy and portfolio failure rate: A longitudinal study *Entrepreneurship Theory and Practice* vol **30** no 2 pp 207 – 223.

3768. Dimov D, Shepherd D A, Sutcliffe K M 2007 Requisite expertise, firm reputation, and status in venture capital investment allocation decisions *Journal of Business Venturing* **22** (4) pp 481 – 502.

3769. Eckhardt J, Shane S, Delmar F 2006 Multistage selection and the financing of new ventures. *Management Science* **52** pp 220 – 232.

3770. Ellul A, Pagano M 2006 IPO underpricing and after-market liquidity *Review of Financial Studies* **19** pp 381 – 421.

3771. Gebhardt G, Schmidt K 2006 Conditional allocation of control rights in venture capital firms *CEPR Discussion Paper 5758.*

3772. Fallick B, Fleischman C, Rebitzer J 2006 Job-hopping in Silicon Valley: Some evidence concerning the microfoundations of a high-technology cluster *Review of Economics and Statistics* **88** pp 472 – 481.

3773. Franke N, Gruber M, Harhoff D, Henkel J 2006 What you are is what you like—similarity biases in venture capitalists evaluations of start-up teams *Journal of Business Venturing* **21** (6) pp 802 – 826.

3774. Franke N, Gruber M, Harhoff D, Henkel J 2008 Venture capitalists' evaluations of start-up teams: Trade-offs, knock-out criteria, and the impact of VC experience *Entrepreneurship Theory and Practice* **32** (3) pp 459 – 483.

3775. Franco A, Filson D 2006 Spinouts: Knowledge diffusion through employee mobility *RAND Journal of Economics* **37** pp 841 – 860.

3776. Franco A, Mitchell M 2008 Covenants not to compete, labor mobility, and industry dynamics *Journal of Economics and Management Strategy* **17** pp 581 – 606.

3777. Isaksson A 2006 Studies on the venture capital process *PhD Thesis* Umeå University Umeå Sweden.

3778. Isaksson A 2010 Staten som venture-capitalist: En sammanställning av internationell empirisk forskning *in* Tillväxtanalys, Staten och riskkapitalet - Delrapport 1: Metodbeskrivning och kunskapsöversikt Rapport 2010:01 Östersund: Tillväxtanalys pp 33 – 84.

3779. Mathews R 2006 Strategic alliances, equity stakes, and entry deterrence *Journal of Financial Economics* **80** pp 35 – 79.

3780. Motohashi K 2006 R&D networks of small and medium size companies *in* Goto and Kodama editors Innovation system in Japan *University of Tokyo Press* (in Japanese) Tokyo Japan pp 137 – 167.

3781. Motohashi K 2010 Comparative analysis of biotechnology startups between Japan and the US *MPRA Paper no 50095* Munich University Munich Germany pp 1 – 25 http://mpra.ub.uni-muenchen.de/50095/ .

3782. Proimos A, Murray W 2006 Entrepreneuring into venture capital *The Journal of Private Equity* **9** (3) pp 23 – 34 doi: 10.3905/jpe.2006.635426 .

3783. Riyanto Y, Schwienbacher A 2006 On the strategic use of corporate venture financing for securing demand *Journal of Banking and Finance* **31** pp 679 – 702.

3784. Tirole J 2006 The theory of corporate finance *Princeton University Press* New Jersey USA.

3785. Wadhwa A, Kotha S 2006 Knowledge creation through external venturing: Evidence from the telecommunications equipment manufacturing industry *Academy of Management Journal* **49** (4) pp 819 – 835.

3786. Zhang J 2006 A study of academic entrepreneurs using venture capital data *Working Paper no 2006.01* Public Policy Institute of California San Francisco California USA.

3787. Zhang J 2007a Access to venture capital and the performance of venture capital-backed star-ups in Silicon Valley *Economics Developments Quarterly* **21** pp 124 – 147.

3788. Zhang J 2007b The advantage of experienced start-up founders in venture capital acquisition: Evidence from serial entrepreneurs *IZA Discussion Paper no 2964* Department of Economics Clark University USA pp 1 – 42.

3789. Bernile G, Cumming D, Lyandres E 2007 The size of venture capital and private equity fund portfolios *Journal of Corporate Finance* **13** pp 564 – 590.

3790. Campbell R A, Kraeussl R 2007 A survey of the venture capital industry in Central and Eastern Europe *in* Gregoriou G N, Kooli M, Kraeussl R (editors) Venture capital in Europe Chapter 4 *Elsevier – North Holland* Quantitative Finance Series New York USA.

*3791.* Casamatta C, Haritchabalet C 2007 Experience, screening and syndication in venture capital investments *Journal of Financial Intermediation* 16 (3) pp 368 – 398.

*3792.* Colombo L, Dawid H, Kabus K 2007 When do Thick Venture Capital Markets Foster Innovation? An Evolutionary Analysis *IEF0074 - marzo - 2007* Istituto di Economia e Finanza *Università Cattolica del Sacro Cuore* Milano Italy pp 1 – 31.

*3793.* de Bettignies J-E, Brander J 2007 Financing entrepreneurship: Bank finance versus venture capital *Journal of Business Venturing* **22** pp 808 – 832.

*3794.* de Bettignies J-E, Chemla G 2008 Corporate venturing, allocation of talent, and competition for star managers *Management Science* **54** pp 505 – 521.

*3795.* de Bettignies J-E 2008 Financing the entrepreneurial venture *Management Science* **54** pp 151 – 166.

*3796.* Hochberg Y, Ljungqvist A, Lu Y 2007 Whom you know matters: Venture capital networks and investment performance *Journal of Finance* **62** pp 251 – 301.

*3797.* Hochberg Y, Ljungqvist A, Lu Y 2010 Networking as a barrier to entry and the competitive supply of venture capital *Journal of Finance* **65** pp 829 – 859.

*3798.* Giot P, Schwienbacher A 2007 IPOs, trade sales and liquidations: Modelling venture capital exits using survival analysis *Journal of Banking & Finance* **31** (3) pp 679 – 702.

*3799.* Hsu D H 2007 Experienced entrepreneurial founders and venture capital funding *Research Policy* **36** pp 722 – 741.

*3800.* Jovanovic B, Szentes B 2007 On the return to venture capital *NBER Working Paper no 12874*.

*3801.* Lai R 2007 Why funds of funds? *MPRA Paper no 4762* Munich University Munich Germany pp 1 – 53

http://mpra.ub.uni-muenchen.de/4762/ .

*3802.* Li K, Prabhala N 2007 Self-selection models in corporate finance *in* Gecko E (editor) Handbook of corporate finance: Empirical corporate finance vol **1** *North Holland* Amsterdam The Netherlands.

*3803.* Luukkonen T 2007 Understanding the strategies of venture capital investors in helping their portfolio firms to become international *Discussion Paper no 1099* The Research Institute of the Finnish Economy Helsinki Finland ISSN 0781-6847 pp 1 – 28.

*3804.* Luukkonen T 2008 Different types of venture capital investors and value-added to high-tech portfolio firms *Discussion Paper no 1149* The Research Institute of the Finnish Economy Helsinki Finland ISSN 0781-6847 pp 1 – 28.

*3805.* Mann R, Sager T 2007 Patents, venture capital, and software start-ups *Research Policy* **36** pp 193 – 208.

*3806.* Pintado T R, De Lema D G P, Van Auken H 2007 Venture capital in Spain by stage of development *Journal of Small Business Management* **45** (1) pp 68 – 88.

*3807.* Robinson D, Stuart T 2007 Financial contracting in biotech strategic alliances *Journal of Law and Economics* **50** pp 559 - 596.

*3808.* Sau L 2007 New pecking order financing for innovative firms: An overview *MPRA Paper no 3659* Munich University Munich Germany pp 1 – 28

http://mpra.ub.uni-muenchen.de/3659/ .

3809. Schwienbacher A 2007 A theoretical analysis of optimal financing strategies for different types of capital-constrained entrepreneurs *Journal of Business Venturing* **22** pp 753 – 781.

3810. Schwienbacher A 2008 Innovation and venture capital exits *Economic Journal* **118** pp 1888 – 1916.

3811. Sørensen M 2007 How smart is smart money? A two-sided matching model of venture capital *Journal of Finance* **62** pp 2725 – 2762.

3812. Tykvová T 2007 Who chooses whom? Syndication, skills, and reputation *Review of Financial Economics* **16** pp 5 – 28.

3813. Aizenman J, Kendall J 2008 The internationalization of venture capital and private equity *NBER Working Paper no 14344*.

3814. Broughman B 2008 Independent directors and board control in venture finance *American Law and Economics Association Papers* 41 pp 1 – 24.

3815. Broughman B, Fried J 2010 Renegotiation of cash flow rights in the sale of VC-backed firms *Journal of Financial Economics* **95** pp 384 – 399.

3816. Cumming D 2008 Contracts and exits in venture capital finance *Review of Financial Studies* **21** pp 1947 – 1982.

3817. Davidsson P, Steffens P R, Gordon S R, Senyard J M 2008 Characteristics of high-potential start-ups: Some early observations from the CAUSEE project http://eprints.qut.edu.au/19649/1/c19649.pdf.

3818. Geronikolaou G, Papachristou G 2008 Venture capital and innovation in Europe *MPRA Paper no 36706* Munich University Munich Germany pp 1 – 16

http://mpra.ub.uni-muenchen.de/36706/ ,

http://ssrn.com/abstract=1309186 .

3819. Hand J 2008 Give everyone a prize? Employee stock options in private venture-backed firms *Journal of Business Venturing* **23** pp 385 – 404.

3820. Hirukawa M, Ueda M 2008a Venture capital and industrial innovation *CEPR Discussion Paper 7089*.

3821. Hirukawa M, Ueda M 2008b Venture capital and innovation: Which is first? *CEPR Discussion Paper 7090*.

3822. Katila R, Rosenberger J, Eisenhardt K 2008 Swimming with sharks: Technology ventures and corporate relationships *Administrative Science Quarterly* **53** pp 295 – 332.

3823. Lindsey L 2008 Blurring firm boundaries: The role of venture capital in strategic alliance *Journal of Finance* **63** pp 1137 – 1168.

3824. McMillan I, Roberts E, Livada V, Wang A 2008 Corporate venture capital: Seeking innovation and corporate growth *National Institute for Standards and Technology* US Department of Commerce US Government USA.

3825. Nahata R 2008 Venture capital reputation and investment performance *Journal of Financial Economics* **90** pp 127 – 151.

3826. Orman C 2008 Organization of innovation and capital markets *MPRA Paper no 23499* Munich University Munich Germany pp 1 – 39

http://mpra.ub.uni-muenchen.de/23499/ .

3827. Rossetto S 2008 The price of rapid exit in venture capital-backed IPOs *Annals of Finance* **4** pp 29 – 53.

3828. Schwienbacher A 2008 Innovation and venture capital exits *Economic Journal* **118** pp 1888 – 1916.

3829. Schwienbacher A 2009 Venture capital exits *in* Cumming D (editor) Companion to venture capital *Robert W Kolb Companion to finance Series Wiley/Blackwell* USA.

3830. Sorenson O, Stuart T 2008 Bringing the context back in: Settings and the search for syndicate partners in venture capital investment networks *Administrative Science Quarterly* **53** pp 266 – 294.

3831. Phalippou L 2008 The hazards of using IRR to measure performance: The case of private equity *Journal of Performance Measurement* **4** pp 55 – 67.

3832. Phalippou L, Gottschalg O 2009 The performance of private equity funds *Review of Financial Studies* **22** pp 1747 – 1776.

3833. Puri M, Zarutskie R 2008 On the lifecycle dynamics of venture-capital- and non-venture capital- financed firms *EFA 2007 Ljubljana Meetings Paper US Census Bureau Center for Economic Studies Paper no CES-WP-08-13* http://ssrn.com/abstract=967841 .

3834. Van Deventer B, Mlambo Ch 2008, 2009 Factors influencing venture capitalists' project financing decisions in South Africa *MPRA Paper no 24970* Munich University Munich Germany pp 1 – 10 http://mpra.ub.uni-muenchen.de/24970/ ; *South Africa Journal Business Management* **40** (1) pp 33 – 41.

3835. Winton A, Yerrmilli V 2008 Entrepreneurial finance: Banks versus venture capital *Journal of Financial Economics* **88** pp 51 – 79.

3836. Aberman J 2009 The decline of the United States venture capital industry: What the federal government should do about it *Amplifier Ventures* Washington DC USA.

3837. Block J, Sandner Ph 2009 What is the effect of the current financial crisis on venture capital financing? Empirical evidence from US Internet start-ups *MPRA Paper no 14727* Munich University Munich Germany pp 1 – 23 http://mpra.ub.uni-muenchen.de/14727/ .

3838. Casamatta C 2003 Financing and advising: Optimal financial contracts with venture capitalists *Journal of Finance* **58** pp 2059 – 2086.

3839. Clarysse B, Knockaert M, Wright M 2009 Benchmarking UK venture capital to the US and Israel: What lessons can be learned? *Report prepared for the British Private Equity and Venture Capital Association (BVCA)*.

3840. Cockburn I, MacGarvie M 2009 Patents, thickets and the financing of early-stage firms: Evidence from the software industry *Journal of Economics and Management Strategy* **18** pp 729 – 773.

3841. Fitza M, Matusik S, Mosakowski E 2009 Do VCs matter? The importance of owners on performance variance in start-up firms *Strategic Management Journal* **30** pp 387 – 404.

3842. Fulghieri P, Sevilir M 2009a Organization and financing of innovation, and the choice between corporate and independent venture capital *Journal of Financial and Quantitative Analysis* **44** pp 1291 – 1321.

3843. Fulghieri P, Sevilir M 2009b Size and focus of a venture capitalist's portfolio *Review of Financial Studies* **22** pp 4643 – 4680.

3844. Duffner S, Schmid M, Zimmermann H 2009 Trust and success in venture capital financing—an empirical analysis with German survey data *Kyklos* **62** pp 15 – 43.

3845. Hege U, Palomino F, Schwienbacher A 2009 Venture capital performance: The disparity between Europe and the United States *MPRA Paper no 39551* Munich University Munich Germany pp 1 – 42

http://mpra.ub.uni-muenchen.de/39551/ .

3846. Jones M, Mlambo Ch 2009 Early-stage venture capital in South Africa: Challenges and prospects *MPRA Paper no 42890* Munich University Munich Germany pp 1 – 29 http://mpra.ub.uni-muenchen.de/42890/ .

3847. Krohmer P, Lauterbach R, Calanog V 2009 The bright and dark side of staging: Investment performance and the varying motivations of private equity *Journal of Banking and Finance* **33** pp 1597 – 1609.

3848. Lingelbach D, Murray G, Gilbert E 2009 The rise and fall of South African venture capital: A co-production perspective http://papers.ssrn.com/sol3/papers.cfm?abstract_id=1459175 .

3849. Litvak K 2009a Venture capital partnership agreements: Understanding compensation arrangements *University of Chicago Law Review* **76** pp 161 – 218.

3850. Litvak K 2009b Governing by exit: Default penalties and walkway options in venture capital partnership agreements *Willamette Law Review* **40** pp 771 – 812.

3851. Masulis R, Nahata R 2009 Financial contracting with strategic investors: Evidence from corporate venture capital backed IPOs *Journal of Financial Intermediation* **18** pp 599 – 631.

3852. Masulis R, Nahata R 2011 Venture capital conflicts of interest: Evidence from acquisitions of venture backed firms *Journal of Financial and Quantitative Analysis* **46** pp 395 – 430.

3853. Norbäck P, Persson L 2009 The organization of the innovation industry: entrepreneurs, venture capitalists, and oligopolists *Journal of the European Economic Association* **7** pp 1261 – 1290.

3854. Samuelsson M, Davidsson P 2009 Does venture opportunity variation matter? Investigating systematic differences between innovative and imitative new ventures *Small Business Economics* **33** (2) pp 229 – 255.

3855. Van de Vrande V, Vanhaverbeke W, Duysters G 2009 Additivity and complementarity in external technology sourcing: The added value of corporate venture capital investments *MPRA Paper no 26419* Munich University Munich Germany pp 1 – 45

http://mpra.ub.uni-muenchen.de/26419/ .

3856. Arikawa Y, Imad'eddine G 2010 Venture capital affiliation with underwriters and the underpricing of initial public offerings in Japan *Journal of Economics and Business* **62** pp 502 – 516.

3857. Benson D, Ziedonis R 2010 Corporate venture capital and the returns to acquiring portfolio companies *Journal of Financial Economics* **98** pp 478 – 499.

3858. Bienz C, Walz U 2010 Venture capital exit rights *Journal of Economics and Management Strategy* **19** pp 1071 – 1116.

3859. Cantner U, Stützer M 2010 The use and effect of social capital in new venture creation: Solo entrepreneurs vs new venture teams *Jena Economics Research Papers* **12**.

3860. Cowling M, Murray G, Liu W 2010 An independent econometric analysis of the "Innovation Investment Fund" programme (IIF) of the Australian commonwealth government: Findings and implications *Department of Innovation, Industry, Science and Research* Canberra Australia.

3861. Dushnitsky G, Shapira Z 2010 Entrepreneurial finance meets organizational reality: Comparing investment practices by corporate and independent venture capitalists *Strategic Management Journal* **31** pp 990 – 1017.

3862. Elston J A, Yang J J 2010 Venture capital, ownership structure, accounting standards and IPO underpricing: Evidence from Germany *Journal of Economics and Business* **62** pp 517 – 536.

3863. Groh A, Liechtenstein H 2010 The European venture capital and private equity attractiveness indices *Journal of Corporate Finance* **16**.

3864. Inci E, Barlo M 2010 Banks versus venture capital when the venture capitalist values private benefits of control *MPRA Paper no 25566* Munich University Munich Germany pp 1 – 43

http://mpra.ub.uni-muenchen.de/25566/ .

3865. Ivanov V, Xie F 2010 Do corporate venture capitalists add value to start-up firms? Evidence from IPOs and acquisitions of VC-backed companies *Financial Management* **35** pp 129 – 152.

3866. Jegadeesh N, Kräussl R, Pollet J 2010 Risk and expected returns of private equity investments: evidence based on market prices *NBER Working Paper 15335* USA.

3867. Hall R, Woodward S 2010 The burden of the non-diversifiable risk of entrepreneurship *American Economic Review* **100** pp 1163 – 1194.

3868. Korteweg A, Sørensen M 2010 Risk and return characteristics of venture capital-backed entrepreneurial companies *Review of Financial Studies* **23** pp 3738 – 3772.

3869. Metrick A, Yasuda A 2010a The economics of private equity funds *Review of Economic Studies* **23** pp 2303 – 2341.

3870. Metrick A, Yasuda A 2010b Venture capital and the finance of innovation *John Wiley and Sons Inc* NY USA.

3871. Metrick A, Yasuda A 2011 Venture capital and other private equity: A survey *European Financial Management* **17** pp 619 – 654.

3872. Obschonka M, Silbereisen R K, Schmitt-Rodermund E, StuetzerNascent M 2010 Entrepreneurship and the developing individual: Early entrepreneurial competence in adolescence and venture creation success during the career *MPRA Paper no 32021* Munich University Munich Germany pp 1 – 36 http://mpra.ub.uni-muenchen.de/32021/

http://www.sciencedirect.com/science/article/pii/S000187911000206X .

*3873.* Stuetzer M, Obschonka M, Davidsson P, Schmitt-Rodermund E 2013 Where do entrepreneurial skills come from? *MPRA Paper no 48274* Munich University Munich Germany pp 1 – 10

http://mpra.ub.uni-muenchen.de/48274/ .

*3874.* Samila S, Sorenson O 2010 Venture capital as a catalyst to innovation *Research Policy* **39** pp 1348 – 1360.

*3875.* Samila S, Sorenson O 2011 Venture capital, entrepreneurship, and economic growth *Review of Economics and Statistics* **93** pp 338 – 349.

*3876.* Schertler A, Tykvová T 2010 What lures cross-border venture capital inflows? *ZEW Discussion Papers no 10-001* pp 1 – 36

http://hdl.handle.net/10419/30016 .

*3877.* Sevilir M 2010 Human capital investment, new firm creation and venture capital *Journal of Financial Intermediation* **19** pp 483 – 508.

*3878.* Zacharakis A, Erikson T, George B 2010 Conflict between the VC and entrepreneur: The entrepreneur's perspective *Venture Capital* **12** (2) pp 109 – 126 doi:10.1080/13691061003771663

*3879.* Zarutskie R 2010 The role of top management team human capital in venture capital markets: Evidence from first-time funds *Journal of Business Venturing* **25** pp 155 – 172.

*3880.* Ball E, Chiu H, Smith R 2011 Can VCs time the market? An analysis of exit choice for venture-backed firms *Review of Financial Studies* **24** pp 3105 – 3138.

*3881.* Bengtsson O, Hand J 2011 CEO compensation in venture capital markets *Journal of Business Venturing* **26** pp 391 – 411.

*3882.* Bengtsson O, Sensoy B 2011 Investor abilities and financial contracting: Evidence from venture capital *Journal of Financial Intermediation* **20** pp 477 – 502.

*3883.* Cherif M, Gazdar K 2011 What drives venture capital in Europe? New results from a panel data analysis *Journal of Applied Business and Economics* **12** (3) pp 122 – 139.

*3884.* Das S, Jo H, Kim Y 2011 Polishing diamonds in the rough: The sources of syndicated venture performance *Journal of Financial Intermediation* **20** pp 199 – 230.

*3885.* Ferretti R, Meles A 2011 Underpricing, wealth loss for pre-existing shareholders and the cost of going public: The role of private equity backing in Italian IPOs *Venture Capital* **13** pp 23 – 47.

*3886.* Kandel E, Leshchinskii D, Yuklea H 2011 VC funds: Aging brings myopia *Journal of Financial and Quantitative Analysis* **46** pp 431 – 457.

*3887.* Kerr W, Nanda R 2011 Financing constraints and entrepreneurship *in* Audretsch D, Falck O, Heblich S (editors) Handbook on research on innovation and entrepreneurship *Edward Elgar* Cheltenham UK.

*3888.* Kraeussl R, Krause S 2011 Has Europe been catching up? An industry level analysis of venture capital success over 1985-2009 *Document de Travail no 327* Direction Générale des Etudes et des Relations Internationales Banque de France pp 1 – 42

http://www.banque-france.fr/ .

*3889.* Li J, Abrahamsson J T 2011 New money, new problems: A qualitative study of the conflicts between venture capitalists and entrepreneurs in Sweden *Ph.D. dissertation* Umeå University Sweden

http://umu.divaportal.org/smash/record.jsf?pid=diva2:426612 .

*3890.* Samila S, Sorenson O 2011 Venture capital, entrepreneurship, and economic growth *The Review of Economics and Statistics* **93** (1) pp 338 – 349 doi:10.1162/REST_a_00066 .

*3891.* Tian X 2011 The causes and consequences of venture capital stage financing *Journal of Financial Economics* **101** pp 132 – 159.

*3892.* Diaconu M 2012 Characteristics and drivers of venture capital investment activity in Romania *Theoretical and Applied Economics* vol **XIX** no 7 (572) pp 111 – 132.

*3893.* Gvazdaitytė A 2012 The analysis of the conditions needed for building venture capital industry in Lithuania *Journal of Business Management and Economics* vol **3** (3) pp 96 – 105 ISSN 2141-7482.

*3894.* Hochberg Y V 2012 Venture capital and corporate governance in the newly public firm *Review of Finance* **16** (2) pp 429 – 480.

*3895.* Lazarevski D, Mrsik J, Smokvarski E 2012 Evolution of the venture capital financing for growing small and medium enterprises in Central and Eastern Europe countries: The case of Macedonia *MPRA Paper No. 41997* Munich University Munich Germany pp 1 – 12 http://mpra.ub.uni-muenchen.de/41997/ .

*3896.* Lim K, Cu B 2012 The effects of social networks and contractual characteristics on the relationship between venture capitalists and entrepreneurs *Asia Pacific Journal of Management* **29** (3) pp 573 – 596 doi:10.1007/s10490-010-9212-x .

*3897.* Nyman M, Lundgren J, Rösiö C C 2012 Riskkapital – Private equity - och venture capital - investeringar *Norstedts Juridik* Stockholm Sweden.

*3898.* Rosenbusch N, Brinckmann J, Müller V 2012 Does acquiring venture capital pay off for the funded firms? A meta-analysis on the relationship between venture capital investment and funded firm financial performance *Journal of Business Venturing* http://dx.doi.org/10.1016/j.jbusvent.2012.04.002 .

*3899.* Pommet S 2012 The survival of venture capital backed companies: An analysis of the French case *GREDEG Working Paper no2012-14* University of Nice-Sophia Antipolis and GREDEG CNRS France pp 1 – 24 http://www.gredeg.cnrs.fr/working-papers.html .

*3900.* Yitshaki R 2012 Relational norms and entrepreneurs' confidence in venture capitalists' cooperation: The mediating role of venture capitalists' strategic and managerial involvement *Venture Capital* **14** (1) pp 3 – 59 doi:10.1080/13691066.2012.662839 .

*3901.* Alqatawni T 2013 The relationship conflict between venture capital and entrepreneur *MPRA Paper no 48006* Munich University Munich Germany pp 1 – 12 http://mpra.ub.uni-muenchen.de/48006/ .

*3902.* Brettel M, Mauer R, Appelhoff D 2013 The entrepreneur's perception in the entrepreneur–VCF relationship: the impact of conflict types on investor value *Venture Capital* pp 1 – 25 doi:10.1080/13691066.2013.782625 .

*3903.* Pennacchio L 2013 The causal effect of venture capital backing on the underpricing of Italian IPOs *MPRA Paper no 48695* Munich University Munich Germany pp 1 – 44 http://mpra.ub.uni-muenchen.de/48695/ .

3904. Stuetzer M, Obschonka M, Schmitt-Rodermund E 2013 Balanced skills among nascent entrepreneurs *MPRA Paper no 48641* Munich University Munich Germany pp 1 – 39

http://mpra.ub.uni-muenchen.de/48641/                                        ;
http://www.springerlink.com/content/e73m7t6j82733411/ .

3905. Hsu D K, Haynie J M, Simmons S A, McKelvie A 2014 What matters, matters differently: A conjoint analysis of the decision policies of angel and venture capital investors *Venture Capital* **16** (1) pp 1 – 25.

3906. Lahr H, Mina A 2014 Liquidity, technological opportunities, and the stage distribution of venture capital investments *Financial Management* **43** (2) pp 291 – 325.

3907. Woike J K, Hoffrage U, Petty J S 2015 Picking profitable investments: The success of equal weighting in simulated venture capitalist decision making *Journal of Business Research* **68** (8) pp 1705 – 1716.

3908. Kvist Ph, Rosengren J May 29 2016 Venture capitalisters Investeringskriterier: Vilka finansiella investeringskriterier är mest betydelsefulla vid en investeringsmöjlighet? *MSc Thesis (Karl Wennberg supervisor)* Linköpings Universitet SE-581 83 Linköping Sverige pp 1 – 99 www.liu.se

***Firm's stock option investment, traded stock options, employee/executive stock options, equity options in finances:***

3909. Weinberg S J, Patton A 1963 Rx for stock option reform *Business Horizons* **6** (3) pp 45 – 52

http://www.sciencedirect.com/science/article/pii/000768136390048X .

3910. James Boness A April 1964 Elements of a theory of stock-option value *Journal of Political Economy* **72** (2) pp 163 – 175

http://www.journals.uchicago.edu/doi/10.1086/258885 .

3911. Bierman H September 1967 The valuation of stock options *Journal of Financial and Quantitative Analysis* **2** (3) pp 327 – 335

https://www.cambridge.org/core/journals/journal-of-financial-and-quantitative-analysis/article/div-classtitlethe-valuation-of-stock-optionsa-hreffn01-ref-typefnadiv/4D5E56090E9AE7227B94765F060319FD .

3912. Lewellen W G 1968 Stock options Chapter *in* Executive Compensation in Large Industrial Corporations *National Bureau of Economic Research Inc* USA pp 46 – 69

http://www.nber.org/chapters/c9345.pdf .

3913. Hirshleifer J 1970 Investment, interest, and capital *Prentice-Hall* London UK.

3914. Black F, Scholes M 1973 The pricing of options and corporate liabilities *Journal of Political Economy* **81** pp 637 – 659.

3915. Black F 1975 Fact and fantasy in the use of options *Financial Analysts Journal* **31** pp 36 – 41, pp 61 – 72.

3916. Merton R C 1973 Theory of rational option pricing *Bell Journal of Economics and Management Science* **4** pp 141 – 183.

3917. Merton R C 1997 Continuous-time finance *Blackwell Publishers* USA.

3918. Fisher S March 1978 Call option pricing when the exercise price is uncertain and the valuation of index bonds *Journal of Finance* pp 169 – 176.

3919. Klemkosky R C 1978 The impact of option expirations on stock prices *Journal of Financial and Quantitative Analysis* **13** (3) pp 507 – 518.

3920. Margrabe W March 1978 The value of an option to exchange one asset for another *Journal of Finance* pp 177 – 186.

3921. Cox J C, Ross S R, Rubinstein M 1979 Option pricing: A simplified approach *Journal of Financial Economics* **7** pp 229 – 263.

3922. MacBeth J D, Merville L J 1979 An empirical examination of the Black-Scholes call option pricing model *Journal of Finance* **34** pp 1173 – 1186.

3923. Bookstaber R M 1981 Observed option mispricing and the nonsimultaneity of stock and option quotations *The Journal of Business* **54** (1) pp 141 – 155.

3924. Brown R L, Rainbow K A 1981 Exercising of options in the Australian options market *Australian Journal of Management* **6** (1) pp 1 – 22.

3925. Hite G L, Long M S 1982 Taxes and executive stock options *Journal of Accounting and Economics* **4** (1) pp 3 – 14.

3926. Whaley R E 1982 Valuation of American call options on dividend paying stocks: Empirical tests *Journal of Financial Economics* **10** pp 29 – 57.

3927. Geske R, Roll R 1984 On valuing American call options with the Black-Scholes formula *Journal of Finance* **39** pp 443 – 455.

3928. Leland H E 1985 Option pricing and replication with transactions costs *Journal of Finance* **40** pp 1283 – 1301.

3929. Rubinstein M 1985 Nonparametric tests for alternative option pricing models using all reported trades and quotes on the 30 most active CBOE option classes from August 23, 1976 through August 31, 1978 *Journal of Finance* **40** pp 455 – 480.

3930. Barone-Adesi G, Whaley R E 1987 Efficient analytic approximation of American option values *Journal of Finance* **42** pp 301 – 320.

3931. Hull J, White A 1987 The pricing of options on assets with stochastic volatilities *Journal of Finance* **42** pp 281 – 300.

3932. Hull J 1997 Options, futures, and other derivative securities 3[rd] edition *Prentice Hall* New Jersey USA.

3933. Hull J, White A January-February 2004 How to value employee stock options *Financial Analysts Journal* **60** (1).

3934. Hull J 2005 – 2006 Private communications on the options pricing *Rotman School of Management* University of Toronto Canada.

3935. Blomeyer E C, Johnson H E 1988 An empirical examination of the pricing of American put options *Journal of Financial and Quantitative Analysis* **23** pp 13 – 22.

3936. Detemple, Jorion 1989 Option listing and stock returns *Working Paper* Columbia - Graduate School of Business Columbia University NY USA.

3937. Jorion Ph, Stoughton N M 1989 An empirical investigation of the early exercise premium of foreign currency options *Journal of Futures Markets* **9** pp 365 – 376.

3938. Lambert R A, Lanen W N, Larker D F 1989 Executive stock option plans and corporate dividend policy *Journal of Financial and Quantitative Analysis* **24** pp 409 – 424.

3939. Defusco R A, Johnson R R, Zorn T S 1990 The effect of executive stock option plans on stockholders and bondholders *Journal of Finance* **45** (2) pp 617 – 627.

3940. Vijh A M 1990 Liquidity of the CBOE equity options *Journal of Finance* **45** pp 1157 – 1179.

3941. Zivney T L 1991 The value of early exercise in option prices: An empirical investigation *Journal of Financial and Quantitative Analysis* **26** pp 129 – 138.

3942. Boyle Ph P, Vorst T 1992 Option replication in discrete time with transaction costs *Journal of Finance* **47** pp 271 – 293.

3943. French D W, Maberly E D 1992 Early exercise of American index options *Journal of Financial Research* **15** pp 127 – 138.

3944. Harvey C, Whaley R E 1992 Dividends and S&P 100 index option valuation *Journal of Futures Markets* **12** pp 123 – 137.

3945. Bizjak J M, Brickley J A, Coles J L 1993 Stock-based incentive compensation and investment behavior *Journal of Accounting and Economics* **16** pp 349 – 372.

3946. George T, Longstaff F 1993 Bid-ask spreads and trading activity in the S&P 100 index options market *Journal of Financial and Quantitative Analysis* **28** pp 381 – 397.

3947. Heston S L 1993 A closed form solution for options with stochastic volatility and application to bond and currency options *The Review of Financial Studies* **6** (2) pp 326 – 343.

3948. Dawson P 1994 Comparative pricing of American and European index options: An empirical analysis *Journal of Futures Markets* **14** pp 363 – 378.

3949. Dawson P 2000 The intraday distribution of volatility and the value of wildcard options *Journal of Futures Markets* **20** pp 307 – 320.

3950. Diz F, Finucane Th J 1994 The rationality of early exercise decisions: Evidence from the S&P 100 index options market *Review of Financial Studies* **6** pp 765 – 797.

3951. Fleming J, Whaley R E March 1994 The value of wildcard options *Journal of Finance* **49** pp 215 – 236.

3952. Huddart St 1994 Employee stock options *Journal of Accounting and Economics* **18** (2) pp 207 – 231.

3953. Huddart S, Lang M February 1996 Employee stock option exercises: An empirical analysis *Journal of Accounting and Economics* **21** (1) pp 5 – 43.

3954. Saly J 1994 Re-pricing executive stock options in a down market *Journal of Accounting and Economics* **18** (3) pp 325 – 356.

3955. Sheikh A M, Ronn E I 1994 A characterization of the daily and intra-day behavior of returns on options *Journal of Finance* **49** pp 557 – 579.

3956. Kamara A, Miller Th W Jr 1995 Daily and intradaily tests of European put-call parity *Journal of Financial and Quantitative Analysis* **30** pp 519 – 539.

3957. Mas-Colell A, Whinston M D, Green J R 1995 Microeconomic theory *Oxford University Press* New York USA, Oxford UK.

3958. Rubinstien M Fall 1995 On the accounting valuation of employee stock options *Journal of Derivatives*.

3959. Sung H M 1995 The early exercise premia of American put options on stocks *Review of Quantitative Finance and Accounting* **5** pp 365 – 373.

3960. Abken P A, Madan D B, Buddhavarapu Sailesh Ramamurtie 1996 Pricing S&P 500 index options using a Hilbert space basis *FRB Atlanta Working Paper no 96-21* Federal Reserve Bank of Atlanta USA.

**3961.** Aboody D 1996 Market valuation of employee stock options *Journal of Accounting and Economics* **22** (1-3) pp 357 – 391.

**3962.** Bergman Y, Grundy B, Weiner Z 1996 General properties of option prices *The Journal of Finance* **51** pp 1573 – 1610.

**3963.** Chicago Board Options Exchange 1996 CBOE constitution and rules: The official constitution and rules of the Chicago Board Options Exchange Inc *Commerce Clearing House* Chicago IL USA.

**3964.** Corrado Ch J, Miller Th W Jr 1996 Efficient option-implied volatility estimators *Journal of Futures Markets* **16** pp 247 – 242.

**3965.** Ikenberry D L, Vermaelen Th 1996 The option to repurchase stock *Financial Management* **25** (4).

**3966.** Scott L O 1997 Pricing stock options in a jump-diffusion model with stochastic volatility and interest rates: Applications of Fourier inversion methods *Mathematical Finance* **7** (4) pp 413 – 426.

**3967.** Easley D, O'Hara M, Subrahmanya Srinivas P 1998 Option volume and stock prices: Evidence on where informed traders trade *Journal of Finance* **53** pp 431 – 465.

**3968.** Alexander D, Veronesi P 1999 Option prices with uncertain fundamentals theory and evidence on the dynamics of implied volatilities *Finance and Economics Discussion Paper no 1999-47* Board of Governors of the Federal Reserve System USA pp 1 – 66

http://www.federalreserve.gov/pubs/feds/1999/199947/199947pap.pdf.

**3969.** Bates D S 1999 Post-'87 crash fears in the S&P 500 futures option market *Journal of Econometrics*.

**3970.** Campbell S D, Canlin Li 1999 Option prices with unobserved and regime-switching volatility *Technical Report* Department of Economics University of Pennsylvania USA.

**3971.** Heath Ch, Huddart St, Lang M May 1999 Psychological factors and stock options exercises *Quarterly Journal of Economics* **114** pp 601 – 628.

**3972.** Chance D M, Kumar R, Todd R B 2000 The 'repricing' of executive stock options *Journal of Financial Economics* **57** (1) pp 129 – 154.

**3973.** Etling Ch, Miller Th W Jr November 2000 The relationship between index option moneyness and relative liquidity *Journal of Futures Markets* **20** pp 971 – 987.

**3974.** Hall B J, Murphy K J May 2000a Optimal exercise prices for executive stock options *American Economic Review* **90** pp 209 – 214.

**3975.** Hall B J, Murphy K 2000b, 2002 Stock options for undiversified executives *NBER Working Paper no 8052* National Bureau of Economic Research Inc USA, *Journal of Accounting and Economics* **33** (1) pp 3 – 42.

**3976.** Hall B J, Murphy K J 2003 The trouble with stock options *Journal of Economic Perspectives* **17** (3) pp 49 – 70.

**3977.** Johnson Sh, Tian Y S 2000 Indexed executive stock options *Journal of Financial Economics* **57** (1) pp 35 – 64.

**3978.** Lee J H, Nayar N 2000 Can the American option sell for less than the matched European option? *Working Paper* University of Oklahoma Norman USA.

3979. McMurray L, Yadav P 2000 The early exercise premium in American option prices: Direct empirical evidence *Derivatives Use, Trading & Regulation* **6** (1) pp 411 – 435.

3980. Core J E, Guay W R August 2001 Stock option plans for non-executive employees *Journal of Financial Economics* **61** (2) pp 253 – 287.

3981. Coval J D, Shumway T 2001 Expected option returns *Journal of Finance* **56** pp 983 – 1009.

3982. Dueker M, Miller Th W 2002 Directly measuring early exercise premiums using American and European S&P 500 index options *Working Paper no 2002-016* Federal Reserve Bank of St Louis USA pp 1 – 38

http://research.stlouisfed.org/wp/2002/2002-016.pdf .

3983. Kalok Chan, Peter Chung Y, Wai-Ming Fong 2002 The informational role of stock and option volume *Review of Financial Studies* **15** (4) pp 1049 – 1075.

3984. Poteshman A, Serbin V 2002 Clearly irrational financial market behavior: Evidence from the early exercise of exchange traded stock options *Journal of Finance*.

3985. Sahlman W A December 2002 Expensing options solves nothing *Harvard Business Review* pp 91 – 96.

3986. Taylor St L 2002 Executive stock options: An accounting dilemma *Australian Accounting Review* **12** (26) pp 2-2.

3987. Utsunomiya K 2002 An analysis of employee stock options in Japan *Discussion Paper no 101* Center for Intergenerational Studies Institute of Economic Research Hitotsubashi University Japan

http://hermes-ir.lib.hit-u.ac.jp/rs/bitstream/10086/14471/1/pie_dp101.pdf .

3988. Bliss R April 2003 Common sense about executive stock options *Chicago Fed Letter no 188* Federal Reserve Bank of Chicago USA.

3989. Bodie Z, Kaplan R S, Merton R C March 2003 For the last time: Stock options are an expense *Harvard Business Review*.

3990. Campbell B A June 2003 Firm volatility and stock option incidence *Working Paper* Institute for Research on Labor and Employment, Institute of Industrial Relations UC Berkeley California USA pp 1 – 37

http://escholarship.org/uc/item/7gt1r0pn .

3991. Cassano M 2003 Executive stock options: Back to basics *Australian Economic Review* **36** (3) pp 306 – 315.

3992. Craig B, Glatzer E, Keller J G, Scheicher M 2003 The forecasting performance of German stock option densities *Working Paper no 312* Federal Reserve Bank of Cleveland USA.

3993. Guay W, Kothari S P, Sloan R May 2003 Accounting for employee stock options *American Economic Review* **93** (2), pp 405 – 409.

3994. Haubrich J November 2003 Expensing stock options *Economic Commentary* Federal Reserve Bank of Cleveland pp 1 – 4.

3995. Raupach P 2003 The cost of employee stock options *Working Paper no 123* Department of Finance Goethe University Frankfurt am Main Germany.

3996. Barenbaum L, Schubert W, O'Rourke B December 16-20 2004 Valuing employee stock options using a lattice model *The CPA Journal*.

**3997.** Battalio R, Hatch B, Jennings R 2004 Toward a national market system for US exchange–listed equity options *Journal of Finance* **59** pp 933 – 962.

**3998.** Calomiris Ch, Glenn Hubbard R January 2004 Options pricing and accounting practice *Working Paper no 103* American Enterprise Institute USA.

**3999.** Arnold M C, Gillenkirch R M September 2005 Stock options and dividend protection *Journal of Institutional and Theoretical Economics JITE* **161** (3) pp 453 – 472

http://www.ingentaconnect.com/content/mohr/jite/2005/00000161/00000003/art00006 .

**4000.** Bulow J, Shoven J B Fall 2005 Accounting for stock options *Journal of Economic Perspectives* **19** (4) pp 115 – 134.

**4001.** Ciccotello C, Terry Grant C, Mark Wilder W September 2005 Finance, politics, and the accounting for stock options *Journal of Applied Corporate Finance* **17** (4) pp 125 – 133.

**4002.** Dahlgren M, Korn R 2005 The swing option on the stock market *International Journal of Theoretical and Applied Finance (IJTAF)* **8** (1) pp 123 – 139.

**4003.** Eberhart A C 2005 Employee stock options as warrants *Journal of Banking and Finance* **29** (10) pp 2409 – 2433.

**4004.** Lie E May 2005 On the timing of CEO stock option awards *Management Science* **51** (5) pp 802 – 812.

**4005.** Manzon G 2005 Good options, bad stock *Journal of Business Research* **58** (7) pp 1006 – 1008.

**4006.** Oyer P, Schaefer S April 2005 Why do some firms give stock options to all employees?: An empirical examination of alternative theories *Journal of Financial Economics* **76** pp 99 – 133.

**4007.** Ross S, Westerfield R, Jaffe J 2005 Corporate finance *McGraw-Hill* Irwin Boston USA.

**4008.** Sundaram R K 2005 On rescissions in executive stock options *The Journal of Business* **78** (5) pp 1809 – 1836.

**4009.** Sautner Z, Weber M 2005 Stock options and employee behavior *Working Paper no 05-26* Sonderforschungsbereich 504 Universität Mannheim Germany.

**4010.** Cetin U, Jarrow R, Protter Ph, Warachka M 2006 Pricing options in an extended Black Scholes economy with illiquidity: Theory and empirical evidence *Review of Financial Studies* **19** pp 493 – 529.

**4011.** Chongwoo Choe, Xiangkang Yin 2006 Should executive stock options be abandoned? *Australian Journal of Management* **31** (2) pp 163 – 179.

**4012.** Deshmukh S, Howe K M, Luft C Spring 2006 Executive stock options: To expense or not? *Financial Management* **35** (1) pp 87 – 106.

**4013.** Macminn R D, Page F January 2006 Stock options and capital structure *Annals of Finance* **2** (1) pp 39 – 50

http://link.springer.com/article/10.1007%2Fs10436-005-0029-4 .

**4014.** Pan J, Poteshman A M 2006 The information in option volume for future stock prices *Review of Financial Studies* **19** pp 871 – 908.

**4015.** Rhee Th A 2006 On valuing employee stock option plans with the requisite service period requirement *Journal of Entrepreneurial Finance* **11** (1) pp 39 – 50.

4016. Sharma K 2006 Understanding employee stock options *Global Business Review* **7** (1) pp 77 – 93.

4017. Tirole J 2006 The theory of corporate finance *Princeton University Press* Princeton USA, Oxford UK.

4018. Uchida K 2006 Determinants of stock option use by Japanese companies *Review of Financial Economics* **15** (3) pp 251 – 269.

4019. Campbell D Winter 2007 Options on the outs *Econ Focus* pages 16 – 22 https://www.richmondfed.org/~/media/richmondfedorg/publications/research/region_focus/2007/winter/pdf/cover_story.pdf .

4020. Lakonishok J, Inmoo Lee, Pearson N D, Poteshman A M 2007 Option market activity *Review of Financial Studies* **20** pp 813 – 857.

4021. Taylor St J December 2007 Valuing options Chapter 9 *in* Modelling Financial Time Series *World Scientific Publishing Co Pte Ltd* ISBN: 978-981-277-084-4 pp 225 – 237 http://dx.doi.org/10.1142/9789812770851_0009 .

4022. Hurtt D N 2008 Stock buybacks and their association with stock options exercised in the IT industry *American Journal of Business* **23** (1) pp 13 – 22.

4023. Ni S X, Pan J, Poteshman A M 2008 Volatility information trading in the option market *Journal of Finance* **63** pp 1059 – 1091.

4024. Ye G L 2008 Asian options versus vanilla options: A boundary analysis *Journal of Risk Finance* **9** (2) pp 188 – 199.

4025. Cuny Ch J, Martin G S, Puthenpurackal J J 2009 Stock options and total payout *Journal of Financial and Quantitative Analysis* **44** (2) pp 391 – 410.

4026. Gârleanu N, Pedersen L H, Poteshman A M 2009 Demand-based option pricing *Review of Financial Studies* **22** pp 4259 – 4299.

4027. Chen Y-R, Bong Soo Lee 2010 A dynamic analysis of executive stock options: Determinants and consequences *Journal of Corporate Finance* **16** (1) pp 88 – 103.

4028. Engle R, Neri B 2010 The impact of hedging costs on the bid and ask spread in the options market *Working Paper* New York University NY USA.

4029. Hallock K F, Olson C A 2010 New data for answering old questions regarding employee stock options *in* Labor in the New Economy Abraham K G, Spletzer J R, Harper M J (editors) *The University of Chicago Press* Chicago USA, London UK pp 149 – 180.

4030. Babenko I, Lemmon M, Tserlukevich Y 2011 Employee stock options and investment *Journal of Finance* **66** (3) pp 981 – 1009.

4031. Børsum Ø 2011 Employee stock options *Memorandum no 11/2010* Department of Economics Oslo University Oslo Norway pp 1 – 37 https://www.sv.uio.no/econ/english/research/unpublished-works/working-papers/pdf-files/2010/Memo-11-2010.pdf .

4032. Schürhoff N, Ziegler A 2011 Variance risk, financial intermediation, and the cross-section of expected option returns *Working Paper* University of Lausanne, University of Zurich Switzerland.

4033. Yang J, Carleton W 2011 Repricing of executive stock options *Review of Quantitative Finance and Accounting* **36** (3) pp 459 – 490.

*4034.* Zhongjin Yang, Cassidy Yang 2011 A model of stock option prices *International Journal of Financial Markets and Derivatives* **2** (4) pp 288 – 297.

*4035.* Korn O, Paschke C, Uhrig-Homburg M 2012 Robust stock option plans *Review of Quantitative Finance and Accounting* **39** (1) pp 77 – 103.

*4036.* Morikawa M 2012 Stock options and productivity *Discussion Paper 12-J-002* Research Institute of Economy Trade and Industry (RIETI) Japan pp 1 – 31 http://www.rieti.go.jp/jp/publications/dp/12j002.pdf .

*4037.* Perobelli F F C, de Souza Lopes B, da Silveira A D M 2012 Employee stock options plans and the value of Brazilian companies *Brazilian Review of Finance* **10** (1) pp 105 – 147.

*4038.* Timraz G A, Al-Shubiri F N 2012 The impact of stock options trading on the market value of companies listed in Kuwait stock exchange *Business Excellence and Management* **2** (3) pp 63 – 76.

*4039.* Baule R, Tallau Ch 2013 Stock price dynamics of listed growth companies: Evidence from the options market *The IUP Journal of Applied Finance* **19** (1) pp 5 – 26.

*4040.* Cao J, Bing Han 2013 Cross section of option returns and idiosyncratic stock volatility *Journal of Financial Economics* **108** pp 231 – 249.

*4041.* Aldatmaz S, Ouimet P, Van Wesep E D 2014 The option to quit: The effect of employee stock options on turnover *Working Paper* Center for Economic Studies US Census Bureau USA.

*4042.* An B-J, Ang A, Bali T G, Cakici N 2014 The joint cross section of stocks and options *Journal of Finance* **69** pp 2279 – 2337.

*4043.* Boyer B H, Vorkink K 2014 Stock options as lotteries *Journal of Finance* **69** (4) pp 1485 – 1527.

*4044.* Câmara A, Popova I, Simkins B 2014 Options on troubled stock *Journal of Futures Markets* **34** (7) pp 637 – 657.

*4045.* Karakaya M 2014 Characteristics and expected returns in individual equity options *Working Paper* University of Chicago USA.

*4046.* Sonika R, Carline N F, Shackleton M 2014 The option and decision to repurchase stock *Financial Management* **43** (4) pp 833 – 855.

*4047.* Christoffersen P, Goyenko R, Jacobs K, Karoui M 2015 Illiquidity premia in the equity options market *Working Paper* McGill University, University of Houston, University of Toronto Canada, USA.

*4048.* Goyenko R Y, Ornthanalai Ch, Shengzhe Tang 2015 Options illiquidity: Determinants and implications for stock returns *Working Paper* McGill University, University of Toronto Canada.

*4049.* Choy S-K, Wei J 2016 Liquidity risk and expected option returns *Working Paper* Shanghai University, University of Toronto P R China, Canada.

*4050.* Investopedia 2016 Options basics: How options work *Investopedia* http://www.investopedia.com/university/options/ .

*4051.* Kanne St, Korn O, Uhrig-Homburg M 2016 Stock illiquidity, option prices, and option returns *CFR Working Paper no 16-08* Centre for Financial Research (CFR) University of Cologne Germany pp 1 – 52

***Bond investment, bond valuation, financial securities investment, financial securities exchange, financial capital investment product, financial capital investment medium in finances:***

*4052.* Macaulay F R 1938 Some theoretical problems suggested by the movement of interest rates, bond yields, and stock prices in the U.S. since 1856 *National Bureau of Economic Research* New York USA.

*4053.* Dougall H E 1953 Savings bonds in personal investment programs *Journal of Finance* **8** (2) pp 224 – 225.

*4054.* Burrell K O 1953 Savings bonds in personal investment programs *Journal of Finance* **8** (2) pp 212 – 223.

*4055.* Hickman W B 1957 Corporate bonds: Quality and investment performance *National Bureau of Economic Research Inc* USA pp 1 – 529

http://www.nber.org/books/hick57-1 .

*4056.* Fisher L June 1959 Determinants of risk premiums on corporate bonds *Journal of Political Economy* **LXVII** 217 – 237.

*4057.* Francis J C, Archer S H 1971 Portfolio analysis *Prentice-Hall* Englewood Cliffs USA.

*4058.* Jensen M C 1972 Capital markets: Theory and evidence *Bell Journal of Economics and Management Science* **III** pp 357 – 398.

*4059.* Barro B l974 Are government bonds net wealth? *Journal of Political Economy*.

*4060.* Merton R C 1974 On the pricing of corporate debt: The risk structure of interest rates *Journal of Finance* **29** pp 449 – 470.

*4061.* Ederlngton L H December 1974 The yield spread on new issues of corporate bonds *Journal of Finance* **XXIX** pp 1531 – 1543.

*4062.* Fischer St June 1975 The demand for index bonds *Journal of Political Economy*.

*4063.*

*4064.* Black F, Cox J C 1976 Valuing corporate securities: Some effects of bond indenture provisions *Journal of Finance* **31** (2) pp 351 – 367.

*4065.* Black F 1980 The tax advantages of pension fund investments in bonds *Working Paper no 533* NBER Working Paper Series NBER USA pp 1 – 21.

*4066.* Jackson W D 1976 Determinants of long-term bond risk *Working Paper no 76-03* Federal Reserve Bank of Richmond NC USA pp 1 – 37.

*4067.* Khang Ch December 1979 Bond immunization when short-term rates fluctuate more than long-term rates *Journal of Financial and Quantitative Analysis* pp 1085 – 1095.

*4068.* Ibbotson R G, Sinquefield R A 1982 Stocks, bonds, bills and inflation.

*4069.* Bodie Z, Kane A, McDonald R March 1983 Inflation and the role of bonds in investor portfolios *Working Paper no 1091* NBER Working Paper Series NBER USA pp 1 – 43.

*4070.* Brennan M J, Schwartz E S 1983 Duration, bond pricing and portfolio management *in* Innovations in Bond Portfolio Management: Duration Analysis and Immunization Kaufman G G, Bierwag G O, Toevs A (editors) Greenwich CT *JAI Press* pp 61 – 101.

*4071.* Kaufman G G, Bierwag G O, Toevs A (editors) 1983 Innovations in bond portfolio management: Duration analysis and immunization JAI Press Greenwich CT USA.

*4072.* Blume M E 1987 Risk and return characteristics of lower grade bonds *Financial Analysts Journal* vol **43** July/August pp 26 – 33.

*4073.* Blume M E, Keim D B, Patel S A March 1991 Returns and volatility of low-grade bonds *Journal of Finance* **46** (1) pp. 49 – 74.

*4074.* Fons J S March 1987 The default premium and corporate bond experience *Journal of Finance*.

*4075.* Fons J S 1990 Default risk and duration analysis *in* The High Yield Debt Market Altman E I (editor) *Dow Jones Irwin* New York USA.

*4076.* Levy H, Lerman Z 1988 The benefits of international diversification in bonds *Financial Analysts Journal* **44** (5) pp 56 – 64.

*4077.* Altman E September 1989 Measuring corporate bond mortality and performance *Journal of Finance* vol **44** pp 909 – 921.

*4078.* Altman E I Summer 1993 Revisiting the high-yield bond market *Financial Management* pp 78 – 92.

*4079.* Asquith P, Mullins Jr D, Wolff E September 1989 Original issue high yield bonds: Aging analyses of defaults, exchanges and calls *Journal of Finance* vol **44** pp 923 – 954.

*4080.* Tucker J F Fall 1989 Junk bonds for individual investors: Key characteristics and comparisons with other investments *Cross Sections* pp 1 – 4.

*4081.* Varma J 1989 Equilibrium pricing of special bearer bonds *IIMA Working Paper no WP1989-08-01_00893* Indian Institute of Management Ahmedabad India.

*4082.* Ambarish R, Subrahmanyam M G 1990 Default risk and the valuation of high-yield bonds: A methodological critique *in* The High Yield Debt Market Altman E I (editor) *Dow Jones Irwin* New York USA pp 79 – 95.

*4083.* Chance D June 1990 Default risk and the duration of zero coupon bonds *Journal of Finance* **45** (2) pp 265 – 274.

*4084.* Rosengren E May 1990 The case for junk bonds *New England Economic Review* pp 40 – 49.

*4085.* Litterman R, Iben Th March-April 1991 Corporate bond valuation and the term structure of credit spreads *Financial Analysts Journal* vol **47** pp 52 – 64.

*4086.* Shreve S E, Soner H M, Xu G-L 1991 Optimal investment and consumption with two bonds and transaction costs *Mathematical Finance* **1** (3) pp 53 – 84.

*4087.* Leland H November 1994 Bond prices, yield spreads, and optimal capital structure with default risk *Institute of Business and Economic Research Working Paper* University of California Berkeley USA.

*4088.* Merrill C January 1994 The valuation of corporate bonds: Theory and evidence *Ph D Dissertation* University of Pennsylvania USA.

*4089.* Morgan D 1994 Will the shift to stocks and bonds by households be destabilizing? *Economic Review* issue Q II pp 31 – 44.

*4090.* Babbel D F, Merrill C, Panning W September 1995 Default risk and the effective duration of bonds *Policy Research Working Paper 1511* Financial Sector Development Department The World Bank USA pp 1 – 22.

**4091.** Whittingham M Winter 1997 The Canadian market for zero-coupon bonds *Bank of Canada Review* **1996-1997** pp 47 – 62.

**4092.** Zatti F 1998 Le obbligazioni come mezzo di provvista delle banche *Studi e note di economia* vol **3** pp 49 – 81.

**4093.** Narayan D August 1999 Bonds and bridge: Social capital and poverty *Policy Research Working Paper 2167* Poverty Reduction and Economic Management Network Poverty Division The World Bank pp 1 – 60.

**4094.** Bolton P, Freixas X 2000 Equity, bonds, and bank debt: Capital structure and financial market equilibrium under asymmetric information *Journal of Political Economy* **108** (2) pp 324 – 351.

**4095.** Brennan M J, Xia Y 2000 Stochastic interest rates and the bond-stock mix *European Finance Review* **4** (2) pp 197 – 210.

**4096.** Canterbery E R 2000 The sacred college of bonds and money Chapter 4 *in* Wall Street capitalism: The theory of the bondholding class *World Scientific Publishing Co Pte Ltd* pp 47 – 63.

**4097.** Hong G, Warga A 2000 An empirical study of bond market transactions *Financial Analyst Journal* **56** pp 32 – 46.

**4098.** Landen C 2000 Bond pricing in a hidden Markov model of the short rate *Finance and Stochastics* **4** (4) pp 371 – 389.

**4099.** Campbell J Y, Viceira L M 2001 Who should buy long-term bonds? *The American Economic Review* **91** (1) pp 99 – 127.

**4100.** Campbell J Y, Taksler G B 2003 equity volatility and corporate bond yields *Journal of Finance* **58** pp 2321 – 2349.

**4101.** Elliott R J, Van Der Hoek J 2001 Stochastic flows and forward measure *Finance and Stochastics* **5** (4) pp 511 – 525.

**4102.** Elliott R J, Nishide K 2013 Pricing of discount bonds with a Markov switching regime *KIER Working Paper no 859* Institute of Economic Research Kyoto University Kyoto Japan pp 1 – 18.

**4103.** Schultz P 2001 Corporate bond trading costs: A peek behind the curtain *Journal of Finance* **56** pp 677 – 698.

**4104.** Livingston M, Zhou L 2002 The impact of rule 144A debt offerings upon bond yields and underwriter fees *Financial Management* **31** pp 5 – 27.

**4105.** Livingston M, Zhou L 2010 Split bond ratings and information opacity premiums *Financial Management* **39** pp 515 – 532.

**4106.** Jacoby G A 2003 Duration model for defaultable bonds *Journal of Financial Research* **26** (1) pp 129 – 146.

**4107.** Hunter D, Simon D 2004 Benefits of international bond diversification *Journal of Fixed Income* **13** (4) pp 57 – 72.

**4108.** Cochrane J H, Piazzesi M 2005 Bond risk premia *The American Economic Review* **95** (1) pp 138 – 160.

**4109.** Koijen R S J, Nijman Th, Werker B 2005 Labor income and the demand for long-term bonds *Discussion Paper no 2005-95* Center for Economic Research Tilburg University The Netherlands pp 1 – 67.

*4110.* Pu Shen 2005 How long is a long-term investment? *Economic Review First Quarter 2005 Federal Reserve Bank of Kansas City* USA pp 1 – 28.

*4111.* Alderson M J, Betker B L, Stock D R 2006 Investment and financing activity following calls of convertible bonds *Journal of Banking & Finance* **30** (3) pp 895 – 914.

*4112.* Anghelache G 2006 Bonds evaluation - A pillar for investment decision *Theoretical and Applied Economics* **3** (498).

*4113.* Harris L, Piwowar M 2006 Secondary trading costs in the municipal bond market *Journal of Finance* **61** pp 1361 – 1397.

*4114.* Pilotte E A, Sterbenz F P 2006 Sharpe and Treynor ratios on treasury bonds *The Journal of Business* **79** (1) pp 149 – 180.

*4115.* Bierman H 2007 Stocks versus bonds Chapter 7 *in* The bare essentials of investing: Teaching the horse to talk *World Scientific Publishing Co Pte Ltd* pp 111 – 118.

*4116.* Bierman H 2010 Convertible bonds Chapter 16 *in* An introduction to accounting and managerial finance: A merger of equals *World Scientific Publishing Co Pte Ltd* pp 323 – 342.

*4117.* Chen L, Lesmond D A, Wei J 2007 Corporate yield spreads and bond liquidity *Journal of Finance* **62** pp 117 – 149.

*4118.* Connolly M 2007 Measuring the effect of corruption on sovereign bond ratings *Journal of Economic Policy Reform* **10** (4) pp 309 – 323.

*4119.* Edwards A, Harris L, Piwowar M 2007 Corporate bond market transaction costs and transparency *Journal of Finance* **62** pp 1421 – 1451.

*4120.* Goldstein M, Hotchkiss E 2007 Dealer behavior and the trading of newly issued corporate bonds *Working Paper* Boston College Boston USA.

*4121.* Green R C, Hollifield B, Schurhoff N 2007a Dealer intermediation and price behavior in the aftermarket for new bond issues *Journal of Financial Economics* **86** pp 643 – 682.

*4122.* Green R C, Hollifield B, Schurhoff N 2007b Financial intermediation and the costs of trading in an opaque market *Review of Financial Studies* **20** (2) pp 275 – 314.

*4123.* Bessembinder H, Maxwell W 2008 Transparency and the corporate bond market *Journal of Economic Perspectives* **22** (2) pp 217 – 234.

*4124.* Hanson M, Liljeblom E, Löflund A 2008 International bond diversification strategies: The impact of currency, country, and credit risk *Social Science Research Network* NY USA

http://ssrn.com/abstract=1082789 .

*4125.* Rosengren E S The case for junk bonds *Federal Reserve Bank of Boston* USA pp 1 – 10.

*4126.* Santos J A C, Winton A 2008 Bank loans, bonds, and information monopolies across the business cycle *Journal of Finance* **63** pp 1315 – 1359.

*4127.* Beber A, Brandt M, Kavajecz K 2009 Flight-to-quality or flight-to-liquidity? Evidence from the Euro-area bond market *Review of Financial Studies* **22** pp 925 – 957.

*4128.* Dejun Xie 2009 Theoretical and numerical valuation of callable bonds *The International Journal of Business and Finance Research* **3** (2) pp 71 – 82.

**4129.** Drut B March 2009 Sovereign bonds and socially responsible investment *Working Paper 2009-17* Université Paris X Nanterre Paris France pp 1 – 23

http://economix.u-paris10.fr/ .

**4130.** Webb A 2009 The case for investing in bonds during retirement issues in brief *Center for Retirement Research* Boston College Boston USA

http://crr.bc.edu/briefs/the-case-for-investing-in-bonds-during-retirement/ .

**4131.** Banko J C, Lei Zhou 2010 Callable bonds revisited *Financial Management* **39** (2) pp 613 – 641.

**4132.** Chulho Jung, Shambora W, Kyongwook Choi 2010 Are stocks really riskier than bonds? *Applied Economics* **42** (4) pp 403 – 412.

**4133.** Grasso R, Linciano N, Pierantoni L, Siciliano G 2010 Le obbligazioni emesse dalle banche italiane *Consob Quaderni di Finanza no 67.*

**4134.** Güntay L, Hackbarth D 2010 Corporate bond spreads and forecast dispersion *Journal of Banking & Finance* **34** pp 2328 – 2345.

**4135.** Răscolean I, Szabo R 2010 Investments in bonds on Romania's capital market *Annals of the University of Petrosani: Economics* **10** (4) pp 281 – 288.

**4136.** Schneider R, Ciobanu G 2010 Capital-protected structured bonds *Romanian Economic Journal* **13** (37) pp 69 – 93.

**4137.** Chang E C, Krueger Th M 2011 Does index investing work in bonds? *Managerial Finance* **37** (5) pp 451 – 464.

**4138.** Ibarra-Ramirez R June 2011 Stocks, bonds and the investment horizon: A spatial dominance approach *Banco de Mexico Working Paper no 2011-03* Direccion General de Investigacion Economica *Banco de Mexico* pp 1 – 32.

**4139.** Puzyrewicz T 2011 Obligacje na rynku Catayst. Przewodnik dla inwestorów (Bonds on Catalyst. The guide for investors) *Wydawnictwo GPW* Warszawa Poland.

**4140.** Fabozzi F J January 16, 2012 Bond markets, analysis and strategies 8th edition *Prentice Hall* ISBN-13: 978-0132743549 pp 1 – 744.

**4141.** Kovner A, Chenyang Wei March 2012 The private premium in public bonds *Staff Report no 553* Federal Reserve Bank of New York NY USA pp 1 – 57.

**4142.** Prewysz-Kwinto P 2012 Catalyst – Rynek obligacji GPW w 2,5 roku od otwarcia (Catalyst – WSE bond market 2,5 year after opening) *Annales Universitatis Mariae Curie-Skłodowska Sectio H Oeconomia* **56** pp 699 – 710.

**4143.** Prewysz-Kwinto P, Voss G 2014 Development of public market of corporate bonds in Poland *Eurasian Journal of Business and Management* **2** (3) pp 17 – 25

http://www.eurasianpublications.com .

**4144.** Zhiguo He, Milbradt K January 13 2012 Endogenous liquidity and defaultable bonds *Booth School of Business University of Chicago*, *MIT Sloan School of Management* Cambridge USA pp 1 – 58.

**4145.** Compton W, Kunkel R A, Kuhlemeyer G 2013 Calendar anomalies in Russian stocks and bonds *Managerial Finance* **39** (12) pp 1138 – 1154.

**4146.** Hung-Gay Fung, Derrick Tzau, Jot Yau 2013 Offshore Renminbi-denominated bonds *Chinese Economy* **46** (2) pp 6 – 28.

**4147.** Janska A 2013 Risk appraisal in the award of contract bonds *Copernican Journal of Finance & Accounting* **2** (1) pp 91 – 105.

4148. Mazurek J 2013 Obligacje korporacyjne na Catalyst – Przewodnik dla potencjalnych emitentów (Corporate bonds on Catalyst – The guide for potential issuers) *Wydawnictwo GPW* Warszawa Poland.

4149. Booth L, Gounopoulos D, Skinner F 2014 The choice between callable and non-callable bonds *Journal of Financial Research* **37** (4) pp 435 – 460.

4150. Morellec E, Valta Ph, Zhdanov A 2015 Financing investment: The choice between bonds and bank loans *Management Science* **61** (11) pp 2580 – 2602.

4151. Coletta M, Santioni R October 2016 Le obbligazioni bancarie nel portafoglio delle famiglie italiane *Questioni di Economia e Finanza no 359* Dipartimento di Economia e Statistica Banca d'Italia ISSN 1972-6627 pp 1 – 25.

4152. Ranosz R 2016 The raw materials convertible into bonds *Gospodarka Surowcami Mineralnymi / Mineral Resources Management* **32** (2) pp 79 – 94.

4153. Yu-Sheng Lai 2016 Modeling and forecasting the conditional covariance matrix between stock and bond returns using a multivariate high-frequency-based volatility (HEAVY) model Department of Banking and Finance National Chi Nan University Taiwan, SFM 24 NSYSU Taiwan

http://sfm.finance.nsysu.edu.tw .

***Credit derivative investment, credit derivative pricing, credit derivatives exchange, financial capital investment product, financial capital investment medium in finances:***

4154. Knight F 1921 Risk, uncertainty and profit *Houghton Mifflin* New York USA.

4155. Fisher R, Tippett L 1928 Limiting forms of the frequency distribution of the largest or smallest member of a sample *Proceedings of Cambridge Philosophical Society* **24** pp 180 – 190.

4156. Working H 1948 Theory of the inverse carrying charge in futures markets *Journal of Farm Economics* **30** no 1 pp 1 – 28.

4157. Modigliani F, Miller M 1958 The cost of capital, corporation finance and the theory of investment *American Economic Review* vol **48**.

4158. Borch K 1960 Reciprocal reinsurance treaties *The Astin Bulletin* **1** pp 170 – 191.

4159. Borch K 1961 Some elements of a theory of reinsurance *Journal of Risk and Insurance* **28** pp 35 – 43.

4160. Borch K 1962 Equilibrium in a reinsurance market *Econometrica* **30** pp 424 – 444.

4161. Sharpe W 1964 Capital asset prices: A theory of market equilibrium under conditions of risk *Journal of Finance* pp 425 – 441.

4162. Akerlof G A 1970 The market for "lemons": Quality uncertainty and the market mechanism *Quarterly Journal of Economics* vol **84** pp 488 – 500.

4163. Cox D R 1972 Regression models and life tables *Journal of the Royal Statistical Society* Series **B 34/2** pp 187 – 220.

4164. Black F, Scholes M 1973 The pricing of options and corporate liabilities *The Journal of Political Economy* **81** (3) pp 637 – 654.

4165. Black F, Cox J C 1976 Valuing corporate securities: Some effects of bond indenture provisions *Journal of Finance* vol **31** pp 351 – 367.

4166. Fama E, MacBeth J 1973 Risk, return and equilibrium: Empirical tests *Journal of Political Economy* **81** pp 607 – 636.

4167. Merton R C 1974a On the pricing of corporate debt: The risk structure of interest rates *Journal of Finance* **29** pp 449 – 470.

4168. Merton R C 1974b Theory of rational option pricing *Bell Journal of Econom*ics **4** (1) pp 141 – 183.

4169. Jaffee D, Russel T 1976 Imperfect information, uncertainty, and credit rationing *Quarterly Journal of Economics* **90** pp 651 – 666.

4170. Leland H E, Pyle D H 1977 Informational asymmetries, financial structure and financial intermediation *Journal Finance* **32** pp 371 – 387.

4171. Leland H E 1994 Risky debt, bond covenants and optimal capital structure *Journal of Finance* **49** pp 1213 – 1252.

4172. Stiglitz J E, Weiss A 1981 Credit rationing in markets with imperfect information *American Economic Review* **71** pp 393 – 410.

4173. Stiglitz J E, Weiss A 1983 Incentive effects of terminations: Applications to credit and labor markets *American Economic Review* **73** pp 912 – 927.

4174. Diamond D W 1984 Financial intermediation and delegated monitoring *Review of Economic Studies* **51** pp 393 – 414.

4175. Diamond D W 1991 Monitoring and reputation: The Choice between bank loans and directly placed debt *Journal of Political Economy* **99** pp 393 – 414.

4176. Diamond D W, Rajan R G 2000 A theory of bank capital *Journal of Finance* **55** pp 2431 – 2465.

4177. Fama E F 1985 What's different about banks? *Journal of Monetary Economics* **15** pp 29 – 40.

4178. Blazenko G 1986 The economics of reinsurance *Journal of Risk and Insurance* **53** pp 258 – 277.

4179. James C 1987 Some evidence on the uniqueness of bank loans *Journal of Financial Economics* **17** pp 113 – 134.

4180. Rudolph B 1987 Innovationen zur steuerung und begrenzung bankbetrieblicher risiken *in* Bankmanagement für neue märkte Krümmel H J, Rudolph B (editors) Frankfurt Germany pp 19 – 45.

4181. Rudolph B 1994 Ansätze zur kalkulation von risikokosten für kreditgeschäfte *in* Handbuch Bankcontrolling Schierenbeck H, Moser H Wiesbaden Germany.

4182. Berger A, Udell G 1990 Collateral, loan quality, and bank risk *Journal of Monetary Economics* **25** (1) pp 21 – 42.

4183. Cronon W 1991 Nature's metropolis: Chicago and the Great West *W W Norton* New York USA.

4184. Hellwig M 1991 Banking, financial  intermediation and corporate finance *in* European Financial Integration Giovannini A, Mayer C et al (editors) pp 35 – 68.

4185. Berlin M 1992 Securitization *in* The New Palgrave Dictionary of Money and Finance vol **3** Newman P, Murray M, Eatwell J (editors) London UK pp 433 – 435.

4186. Heath D, Jarrow R, Morton A 1992 Bond pricing and the term structure of interest rates *Econometrica* **60** (1) pp 77 – 106.

4187. Dewatripont M, Tirole J 1993 The prudential regulation of banks *MIT Press* USA.

4188. Froot K, Scharfstein D, Stein J 1993 Risk management: Coordinating corporate investment and financing policies *Journal of Finance* **48** pp 1629 – 1658.

*4189.* Froot K, Stein J 1998 Risk management, capital budgeting and capital structure policy for financial institutions: An integrated approach *Journal of Financial Economics* **47** pp 55 – 82.

*4190.* Cocheo S 1994 Business - loan securitization: Trickle or flood? *ABA Banking Journal* **86** (12) pp 35 – 38.

*4191.* Figlewski St Summer 1994 The birth of the AAA derivatives subsidiary *Journal of Derivatives* pp 80 – 84.

*4192.* Hull J, White A 1994a Numerical procedures for implementing term structure models i:Single-factor models *Journal of Derivatives* **2** (1) pp 7 – 16.

*4193.* Hull J, White A 1994b Numerical procedures for implementing term structure models i: Two-factor models *Journal of Derivatives* **2** (2) pp 37 – 48.

*4194.* Hull J, White A 1996 Using Hull-White interest rate trees *Journal of Derivatives* **4** (1) pp 26 – 36.

*4195.* Hull J, White A 2000 Valuing credit default swaps I: No counterparty default risk *Journal of Derivatives* **8** (1) pp 29 – 40.

*4196.* Hull J, White A 2001 Valuing credit default swaps II: Modeling default correlations *Journal of Derivatives* **8** (3) pp 12 – 22.

*4197.* Hull J, White A 2004 Valuation of CDO and n[th] to default CDS without Monte Carlo simulation *Journal of Derivatives* vol **12** (2) pp 8 – 23.

*4198.* Hull J, Predescu M, White A 2004 The relationship between credit default swap spreads, bond yields, and credit rating announcements *Journal of Banking & Finance* **28** (11) pp 2789 – 2811.

*4199.* Hull J, White A 2005a The valuation of correlation-dependent credit derivatives using a structural model *Joseph L Rotman School of Management* University of Toronto Canada.

*4200.* Hull J, White A 2005b The perfect copula *Working Paper* University of Toronto Canada.

*4201.* Hull J, White A 2008 Dynamic models of portfolio credit risk: A simplified approach *Journal of Derivatives* vol **15**.

*4202.* Hull J C 2013 Fundamentals of futures and options markets 8th edition *Pearson* ISBN-13: 978-0132993340 pp 1 – 624.

*4203.* Hull J January 25 2014 Options, futures and other derivatives 9[th] edition USA ISBN-13: 978-0133456318 pp 1 – 896.

*4204.* Jarrow R, Lando D, Turnbull S 1994 A Markov model of the term structure of credit spreads *Working Paper* Cornell University NY USA.

*4205.* Jarrow R, Turnbull S 1995 Pricing derivatives on financial securities subject to credit risk *Journal of Finance* **50** (1) pp 53 – 85.

*4206.* Jarrow R, Lando D, Turnbull S M 1997 A Markov model for the term structure of credit risk spread *Review of Financial Studies* vol **10** pp 481 – 523.

*4207.* Jarrow R, Turnbull S 2000a The intersection of market and credit risk *Journal of Banking and Finance* **24**.

*4208.* Jarrow R, Turnbull S 2000b Derivatives securities 2[nd] edition *South Western College Publishing* USA.

**4209.** Jarrow R 2001 Default parameter estimation using market prices *Financial Analysts Journal* vol **5**.

**4210.** Jarrow R, Yu F 2001 Counterparty risk and the pricing of defaultable securities *The Journal of Finance* **56** pp 1765 – 1799.

**4211.** Jarrow R, Yildirim Y 2002 Valuing default swaps under market and credit risk correlation *The Journal of Fixed Income* **11** pp 7 – 19.

**4212.** McAllister P H, Mingo J J May 1994 Commercial loan risk management, credit scoring, and pricing: The need for a new shared database *Journal of Commercial Lending* pp 6 – 20.

**4213.** Neuberger D 1994 Kreditvergabe durch banken *Tübingen* Germany.

**4214.** Das S Spring 1995 Credit risk derivatives *Journal of Derivatives* pp 7 – 23.

**4215.** Das S 1998a Credit derivatives – instruments *in* Credit derivatives: Trading and management of credit and default risk Das S (editor) *John Willey and Sons Inc* Singapore pp 7 – 97.

**4216.** Das S 1998b Credit-linked structured notes *in* Credit derivatives: Trading and management of credit and default risk Das S (editor) *John Willey and Sons Inc* pp 99 – 121.

**4217.** Das S R 2000 Credit derivatives: Trading and management of credit and default risk *John Wiley & Sons Inc* Singapore.

**4218.** Das S R, Hanouna P, Sarin A 2009 Accounting-based versus market-based cross-sectional models of CDS spreads *Journal of Banking and Finance* **33** pp 719 – 730.

**4219.** Edwards B August 1995 Let's shuffle those loans *Euromoney* no 316 pp 22 – 26.

**4220.** Gorton G B, Pennacchi G 1995 Banks and loan sales: Marketing nonmarketable assets *Journal of Monetary Economics* **35** pp 389 – 411.

**4221.** Gorton G, Winton A 1998 Liquidity provision, the cost of bank capital and the macroeconomic *NBER Working Paper* NBER USA.

**4222.** Gorton G, Souleles N 2005 Special purpose vehicles and securitization *in* The Risks of Financial Institutions Carey M, Stulz R (editors) *University of Chicago Press* Chicago USA.

**4223.** Hasbrouck J 1995 One security, many markets: Determining the contributions to price discovery *Journal of Finance* **50** pp 1175 – 1199.

**4224.** Longstaff F A, Schwartz E June 1995a Valuing credit derivatives *Journal of Fixed Income* pp 6 – 14.

**4225.** Longstaff F A, Schwartz E 1995b A simple approach to valuing risky fixed and floating rate debt *Journal of Finance* **50** pp 789 – 819.

**4226.** Longstaff F A, Mithal S, Neis E August 2003 The credit-default swap market: Is credit protection priced correctly? *NBER Working Paper* NBER USA.

**4227.** Longstaff F A, Mithal S, Neis E 2005 Corporate yield spreads: Default risk or liquidity? New evidence from the credit default swap market *Journal of Finance* **60** (5) pp 2213 – 2253.

**4228.** Longstaff F A, Rajan A 2008 An empirical analysis of the pricing of collateralized debt obligations *Journal of Finance* **63** pp 529 – 563.

**4229.** Whittaker G, Kumar S 1995 Credit derivatives: A primer *Working Paper* Chemical Bank USA *in* The Handbook of Fixed Income Derivatives Dattateya R (editor) *Probus Publishing Company* Chicago USA.

**4230.** Whittaker G, Frost J May 1997 An introduction to credit derivatives *Journal of Lending and Credit Risk Management* vol **79** no 9 pp 15 – 25.

**4231.** Altman E I September 1996 Corporate bond and commercial loan portfolio analysis *Working Paper.*

**4232.** Duffee G R 1996 Rethinking risk management for banks: Lessons from credit derivatives *Proceedings of the 32^{nd} Annual Conference on Bank Structure and Competition* Federal Reserve Bank of Chicago USA pp 381 – 400.

**4233.** Duffee G R, Zhou Ch February 1997 Credit derivatives in banking: Useful tools for loan risk management? *Federal Reserve Board Working Paper* NY USA.

**4234.** Duffee G R, Zhou Ch November 1999 Credit derivatives in banking: Useful tools for managing risk? *Research Program in Finance Working Paper RPF-289* University of California – Berkeley California USA pp 1 – 38

http://www.escholarship.org/uc/item/7g67n911 .

**4235.** Duffee G R, Zhou C 2001 Credit derivatives in banking: Useful tools for managing risk? *Journal of Monetary Economics* **48** (1) pp 25 – 54.

**4236.** Duffie D 1996a Dynamic asset pricing theory *Princeton University Press* Princeton USA.

**4237.** Duffie D 1996b Recursive valuation of defaultable securities and the timing of resolution of uncertainty *Annals of Applied Probability* **6** (4).

**4238.** Duffie D 1999 Credit swap valuation *Financial Analysts Journal* **55** pp 73 – 87.

**4239.** Duffie D, Singleton K 1999 Simulating correlated defaults *Graduate School of Business* Stanford University California USA.

**4240.** Duffie D, Pan J, Singleton J November 2000 Transform analysis and asset pricing for affine jump diffusions *Econometrica* **68** (6) pp 1343 – 1376.

**4241.** Duffie D, Gârleanu N G 2001 Risk and valuation of collateralized debt obligations *Working Paper* Graduate School of Business Stanford University, *Financial Analysts Journal* **57** (1) pp 41 – 59.

**4242.** Duffie D, Lando D 2001 Term structures of credit spreads with incomplete accounting information *Econometrica* **69** (3) pp 633 – 664.

**4243.** Duffie D, Singleton K 2002 Credit risk: Pricing, measurement, and management *Princeton University Press* Princeton NJ USA.

**4244.** Duffie D, Gârleanu N 2003 Risk and valuation of collateralized debt obligations *Graduate School of Business* Stanford University Stanford USA.

**4245.** Duffie D, Filipovic D, Schachermayer W 2003 Affine processes and applications in finances *Annals of Applied Probability* **13** (3) pp 985 – 1053.

**4246.** Duffie D July 2008 Innovations in credit risk transfer: Implications for financial stability *BIS Working Paper no 255* Bank for International Settlements Switzerland.

**4247.** Hattori P K 1996 The Chase guide to credit derivatives in Europe *London* UK.

**4248.** Hyder I, Bolger R M, Leung C July 1996 How S&P evaluates commercial loan-backed securitizations *Journal of Lending and Credit Risk Management* vol **78** no 11 pp 14 – 34.

4249. Neal R S May 1996 Credit derivatives: New financial instruments for controlling credit risk *Economic Review* Second Quarter Federal Reserve Bank of Kansas City Kansas USA pp 15 – 27.

4250. Neal R S, Rolph D S 1999 An introduction to credit derivatives *in* Handbook of Credit Derivatives Francis J C, Frost J A, Whittaker J G (editors) *McGraw-Hill* New York USA pp 3 – 24.

4251. Reoch R, Masters B March 1996 Credit derivatives: Structures and applications *Financial Derivatives and Risk Management* no 5 pp 4 – 10.

4252. Schönbucher P J August 1996 The term structure of defaultable bond prices *Discussion Paper B-384* University of Bonn SFB 303.

4253. Schönbucher P J Term structure modelling of defaultable bonds *The Review of Derivatives Studies* Special Issue: Credit Risk **2** (2/3) pp 161 – 192.

4254. Schönbucher P J Fall 1999 Credit risk modelling and credit derivatives *Ph D Thesis* Faculty of Economics University of Bonn Germany.

4255. Schönbucher P J 2000a Pricing of credit risk and credit risk derivatives *Bonn Graduate School of Economics* Department of Economics University of Bonn Germany.

4256. Schönbucher P J 2000b A tree implementation of a credit spread model for credit *Discussion Paper 17/2001* Bonn Graduate School of Economics Department of Economics University of Bonn Germany pp 1 – 37.

4257. Schönbucher P J 2003 Credit derivatives pricing models: Models, pricing, implementation *Wiley Finance John Wiley and Sons Inc* Chichester UK.

4258. Schönbucher P J 2006 Portfolio losses and the term structure of loss transition rates: A new methodology for the pricing of portfolio credit derivatives *Working Paper* ETHZ Switzerland.

4259. Smithson Ch, Holappa H, Rai Sh June 1996 Credit derivatives (2) *Risk* **9** (6) pp 47 – 48.

4260. Drzik J P, Kuritzkes A July 1997 Credit derivatives: The tip of the iceberg *Credit Risk A Risk Special Supplement* pp 12 – 14.

4261. Freixas X, Rochet J-C 1997 Microeconomics of banking *MIT Press* Cambridge MA USA.

4262. Géczy C, Minton B, Schrand C 1997 Why firms use currency derivatives *Journal of Finance* **52** pp 1323 – 1354.

4263. Hill C A 1997 Securitization: A low-cost Sweetener for Lemons *Journal of Applied Corporate Finance* vol **10** pp 64 – 71.

4264. Hüttemann P 1997 Derivative instrumente für den transfer von kreditrisiken *München* Germany.

4265. Hüttemann P 1998 Derivate instrumente für den transfer von kreditrisiken *in* Oehler A Credit Risk und Value-at-Risk Alternativen – Herausforderungen für das Risk Management *Schäffer-Poeschel Verlag* Stuttgart Germany.

4266. Joe H 1997 Multivariate models and dependence concepts *Chapman & Hall* London UK.

4267. Kaufman F 1997 Besonderheiten der finanzierung kleiner und mittlerer unternehmen *Kredit und Kapital* vol **30** pp 140 – 155.

**4268.** Mahtani A August 23 1997 Credit derivatives go structured *International Financial Review* no 1197.

**4269.** Ogden J 1997 Credit derivatives can cut companies' financing costs *Global Finance* **11** (4) pp 6 – 8.

**4270.** Wilson Th C September 1997a Portfolio credit risk (I) *Risk* **10** (9) pp 111 – 117.

**4271.** Wilson Th C October 1997b Portfolio credit risk (II) *Risk* **10** (10) pp 56 – 61.

**4272.** Burghof H-P, Henschel C 1998 Credit and Information in Universal Banking - A Clinical Study *CFS Working Paper*.

**4273.** Burghof H-P, Henke S, Rudolph B, Schönbucher Ph J, Sommer D 2005 Kreditderivate – Handbuch für die bank und anlagepraxis *Schäffer-Poeschel Verlag* Stuttgart Germany.

**4274.** Henke S, Burghof H-P, Rudolph B 1998 Credit securitization and credit derivatives: Financial instruments and the credit risk management of middle market commercial loan portfolios *CFS Working Paper no 1998/07* pp 1 – 30

http://nbn-resolving.de/urn:nbn:de:hebis:30-9462 ,

http://hdl.handle.net/10419/78086 .

**4275.** Fabozzi F J January 1998 Valuation of fixed income securities and derivatives 3rd edition *John Wiley and Sons Inc* ISBN-13: 978-1883249250 pp 1 – 288.

**4276.** Fabozzi F J, Davis H A, Choudhry M October 20 2006 Introduction to structured finance *John Wiley and Sons Inc* ISBN-13: 978-0470045350 pp 1 – 400.

**4277.** Fabozzi F J, Kothari V 2007 Securitization: The tool of financial transformation *Journal of Financial Transformation*.

**4278.** Lancaster B P, Schultz G M, Fabozzi F J April 25, 2008 Structured products and related credit derivatives: A comprehensive guide for investors *John Wiley and Sons Inc* ISBN-13: 978-0470129852 pp 1 – 544.

**4279.** Fabozzi F J January 16, 2012 Bond markets, analysis and strategies 8th edition *Prentice Hall* ISBN-13: 978-0132743549 pp 1 – 744.

**4280.** Lando D 1998 On Cox processes and credit risky securities *Review of Derivatives Research* **2** (2/3) pp 99 – 120.

**4281.** Lando D 2004 Credit risk modeling – Theory and applications *Princeton University Press* Princeton New Jersey USA.

**4282.** Scott–Quinn B, Walmsley J 1998 The impact of credit derivatives on securities markets *International Securities Market Association* Zurich Switzerland.

**4283.** Steinherr A 1998 Derivatives: The wild beast of finance *John Wiley and Sons Inc* Hoboken NJ USA.

**4284.** Tavakoli J 1998 Credit derivatives: A guide to instruments and applications *John Wiley and Sons Inc* New York USA.

**4285.** Tavakoli J 2001 Credit derivatives and synthetic structures: A guide to instruments and applications *John Wiley and Sons Inc* New York USA.

**4286.** Tavakoli J M 2003 Collateralized debt obligations and structured finance *John Wiley and Sons Inc* Hoboken New Jersey USA.

**4287.** Arvanitis A, Gregory J, Laurent J Spring 1999 Building models for credit spreads *Journal of Derivatives* **1** pp 27 – 43.

4288. DeMarzo P, Duffie D 1999 A liquidity-based model of security design *Econometrica* **67** pp 65 – 99.

4289. DeMarzo P 2005 The pooling and tranching of securities: A model of informed intermediation *Review of Financial Studies* **18** pp 1 – 35.

4290. Gauvin A September 8 1999 Nature et régime juridique des dérivés de crédit *Ph D Thesis* Sorbonne Paris 1 France.

4291. Heidorn T 1999 Kreditderivate *Hochschul* Frankfurt am Main Germany.

4292. Embrechts P, McNeil A, Straumann D 1999 Correlation and dependence in risk management: Properties and pitfalls *Working Paper* Department Mathematics ETH Zurich Switzerland.

4293. Embrechts P, Lindskog F, McNeil A 2001 Modelling dependence with copulas and applications to risk management *Working Paper* Department Mathematics ETH Zurich Switzerland.

4294. Master B, Bryson K 1999 Credit derivatives and loan portfolio management *in* Handbook of Credit Derivatives Francis J C, Frost J A, Whittaker J G (editors) *McGraw-Hill* New York USA pp 43 – 85.

4295. Nelken I 1999 Implementing credit derivatives: Strategies and techniques for using credit derivatives in risk management *McGraw-Hill* New York USA.

4296. Rappoport P 1999 Valuing credit fundamentals: Rock bottom spreads *JP Morgan Research* USA.

4297. Staehle D, Cumming C 1999 The supervision of credit derivative activities of banking organizations *in* Handbook of Credit Derivatives Francis J C, Frost J A, Whittaker J G (editors) *McGraw-Hill* New York pp 293 – 326.

4298. Allen F, Gale D 2000 Financial contagion *Journal of Political Economy* **108** (1) pp 1 – 33.

4299. Allen F, Carletti E 2006 Credit risk transfer and contagion *Journal of Monetary Economics* **53** (1) pp 80 – 111.

4300. British Bankers' Association 2000 Credit derivatives survey *BBA* London UK.

4301. Brewer III E, Minton B A, Moser J T 2000 Interest-rate derivatives and bank lending *Journal of Banking and Finance* **24** pp 353 – 379.

4302. Kiff J, Morrow R Autumn 2000 Credit derivatives *Bank of Canada Review* Autumn 2000 pp 3 – 11.

4303. Kiff J, Michaud F-L, Mitchell J 2002 Instruments of credit risk transfer: Effects on financial contracting and financial stability *Working Paper* Bank of Canada Ottawa Canada.

4304. Kiff J, Michaud F-L, Mitchell J Juin 2003a Une revue analytique des instruments de transfert du risque de crédit *Revue de la Stabilité Financière Banque de France, Banque de France Financial Stability Review no 2* pp 106 – 131.

4305. Kiff J, Michaudand F-L, Mitchell J 2003b Instruments of credit risk transfer: Effects on financial contracting and financial stability *NBB Working Paper* NBB.

4306. Li D 2000 On default correlation: a copula function approach *Journal of Fixed Income* **9** pp 43 – 54.

4307. Müller F 2000 Kreditderivate und risikomanagement *Bankakademie-Verlag GmbH* Frankfurt am Main Germany.

*4308.* Saunders A 2000 Financial institutions management: A modern perspective *Irwin McGraw-Hill* International Edition Homewood IL USA.

*4309.* Wahl J E, Broll U 2000 Financial hedging and bank's assets and liabilities management *in* Risk Management Frenkel M, Hommel U, Rudolf M (editors) *Springer* Heidelberg Germany.

*4310.* Arvantis A, Gregory Jon 2001 Credit: The complete guide to pricing, hedging and risk management *Risk Waters Group Ltd* London UK.

*4311.* Bomfim A N July 11 2001 Understanding credit derivatives and their potential to synthesize riskless assets *Monetary and Financial Market Analysis* Division of Monetary Affairs Federal Reserve Board Washington DC USA pp 1 – 31.

*4312.* Cossin D, Hricko May 2001 Exploring of credit risk in credit default swap transaction data *Working Paper* HEC Lausanne Switzerland.

*4313.* Collin-Dufresne P, Goldstein R, Martin S 2001 The determinants of credit spread changes *Journal of Finance* **56** (6) pp 2177 – 2207.

*4314.* Craddock M, Platen E June 28 2001 Benchmark pricing of credit derivatives under a standard market model *Department of Mathematical Sciences* School of Finance and Economics University of Technology Sydney Australia pp 1 – 27.

*4315.* Delianedis G, Geske R 2001 The components of corporate credit spreads: Default, recovery, tax, jumps, liquidity, and market factors *Working Paper* University of California USA.

*4316.* Elton E, Gruber D, Agrawal D, Mann C 2001 Explaining the rate spread on corporate bonds *Journal of Finance* **56** (1) pp 247 – 277.

*4317.* Giesecke K 2001 Default compensator, incomplete information, and the term structure of credit spreads *Department of Economics der Humboldt-Universität Berlin* Germany.

*4318.* Giesecke K 2002 Credit risk modeling and valuation: An introduction *Humboldt-Universität* Berlin Germany.

*4319.* Giesecke K 2003 A simple exponential model for dependent defaults *Journal of Fixed Income* vol **13**.

*4320.* Giesecke K, Weber St 2003 Cyclical correlations, credit contagion, and portfolio losses Cornell University, Technische Universität Berlin Ithaca, Berlin USA, Germany.

*4321.* Giesecke K 2004 Credit risk modeling and valuation: An introduction Cornell University Ithaca USA.

*4322.* Giesecke K, Goldber L R 2004 Forecasting default in the face of uncertainty *The Journal of Derivatives* vol **Fall/04**.

*4323.* Giesecke K 2006 Default and information *Journal of Economic Dynamics & Control* vol **30**.

*4324.* Goodman L S 2001 Synthetic CDOs *in* Investing collateralized debt obligations *John Wiley & Sons Inc* Fabozzi F J, Goodman L S (editors) Hoboken NJ USA.

*4325.* Lopez J A 2001 Financial instruments of mitigating credit risk *Federal Reserve Bank of San Francisco Economic Letter* pp 2001 – 2034.

4326. Lucas A, Klaassen P, Spreij P, Staetmans S 2001 An analytic approach to credit risk of large corporate bond and loan portfolios *Journal of Banking and Finance* **9** pp 1635 – 1664.

4327. Mashal R, Naldi M 2001 Pricing multi-name credit derivatives: Heavy tailed approach, *Working Paper* Columbia University NY USA.

4328. Ranciere R G 2001 Credit derivatives in emerging markets *IMF Policy Discussion Paper* IMF USA.

4329. Rehm F Ch 2001 Kreditrisikomodelle – Bewertung von Kreditderivaten und Portfoliomodelle zur Kreditrisikomessung *Dissertation an der Wissenschaftlichen Hochschule für Unternehmensführung – Otto-Beisheim-Hochschule – Vallendar* Koblenz Germany.

4330. Rule D June 2001 The credit derivatives market: Its development and possible implications for financial stability *Financial Stability Review* pp 117 – 140.

4331. Schmidt W März 2001 Credit default swaps: Analyse und bewertung Deutsche *Bank Global Markets Research & Analytics* Frankfurt Germany.

4332. Amihud Y 2002 Illiquidity and stock returns: Cross-section and time-series effects *Journal of Financial Markets* **5** pp 31 – 56.

4333. Angelini E 2002 Il profilo di rischio di un portafoglio con derivati creditizi: L'elaborazione di un modello di simulazione Montecarlo *Banca Impresa Società* **XXI** no 2 pp 259 – 282.

4334. Arping S May 2002 Playing hardball: Relationship banking in the age of credit derivatives *Research Paper no 49* FAME - International Center for Financial Asset Management and Engineering Geneva Switzerland
http://www.fame.ch .

4335. Aunon-Nerin D, Cossin D, Hricko T, Huang Z 2002 Exploring for the determinants of credit risk in credit default swap transaction data: Is fixed-income markets information sufficient to evaluate credit risk? *Working Paper* HEC-University of Lausanne and FAME Switzerland.

4336. Batten J, Hogan W 2002 A perspective on credit derivaties *International Review of Financial Analysis* **11**.

4337. Bielecki T R, Rutkowski M 2002 Credit risk: Modeling, valuation and hedging *Springer Verlag* Germany.

4338. Bielecki T R, Rutkowski M 2004 Credit risk: Modeling, valuation and hedging *Springer Verlag* Berlin Heidelberg Germany 2004.

4339. Broll U, Welzel P 2002 Bankrisiko und risikosteuerung mit derivaten *Volkswirtschaftliche Diskussionsreihe Beitrag nr 227* Institut für Volkswirtschaftslehre Universität Augsburg Germany.

4340. Broll U, Pausch Th, Welzel P July 2002 Credit risk and credit derivatives in banking *Volkswirtschaftliche Diskussionsreihe Beitrag nr 228* Institut für Volkswirtschaftslehre Universität Augsburg Germany pp 1 – 9.

4341. Broll U, Schweimayer G, Welzel P 2003 Managing credit risk with credit and macro derivatives *Volkswirschaftliche Diskussionsreihe Discussion Paper no 252* Institut für Volkswirtschaftslehre Universität Augsburg Germany.

*4342.* Di Graziano G, Rogers C 2002 A dynamic approach to the modelling of correlation credit derivatives using Markov chains *Cambridge University Press* Cambridge UK.

*4343.* Dodd R 2002 The structure of OTC derivatives markets *The Financier* vol **9** (1–4) http://www.financialpolicy.org/dscotcstructure.pdf .

*4344.* Dodd R March 2012 Markets: Exchange or over-the-counter *Finance & Development*
http://www.imf.org/external/pubs/ft/fandd/basics/markets.htm .

*4345.* Dunne P G, Moore M J, Portes R 2002 Defining benchmark status: An application using Euro-area bonds *CEPR Discussion Paper no 3490*.

*4346.* Grill H, Perczynski H, Int-Veen Th, Muthig J, Platz S 2002 Wirtschaftslehre des kreditwesens *Bildungsverlag* EINS Troisdorf Germany.

*4347.* Jobst A 2002 The pricing puzzle: The default term structure of collateralized loan obligation *CFS Working Paper no 2002/14*.

*4348.* Läger V 2002 Bewertung von kreditrisiken und kreditderivaten *Uhlenbruch Verlag* Bad Soden/Ts Germany.

*4349.* Prato O November 2002 Credit derivatives: A new source of financial instability? *FSR* pp 69 – 83.

*4350.* Ranciere R G April 2002 Credit derivatives in emerging markets *IMF Policy Discussion Paper* IMF USA pp 1 – 24.

*4351.* Skinner F S, Townend T G 2002 An empirical analysis of credit default swaps *International Review of Financial Analysis* **11**.

*4352.* Albanese C, Campolieti J, Chen O, Zavidonov A 2003 Credit barrier models *Risk* **16** (6) pp 109 – 113.

*4353.* Albanese C, Chen O 2005 Discrete credit barrier models *Quantitative Finance* 5 pp 247 – 256.

*4354.* Albanese C, Vidler A October 9 2007 A structural model for credit-equity derivatives and bespoke CDOs *MPRA Paper no 5227* pp 1 – 28
http://mpra.ub.uni-muenchen.de/5227/ .

*4355.* Amato J D, Remolona E December 2003 The credit spread puzzle *BIS Quarterly Review* pp 51 – 63.

*4356.* Amato J D December 2005 Risk aversion and risk premia in the CDS market *BIS Quarterly Review* pp 55 – 68.

*4357.* BIS 2003 Credit risk transfer *Bank for International Settlements* ISBN 92-9131-641-5 pp 1 – 57.

*4358.* Blanco R, Brennan S, Marsch I W 2003 An empirical analysis of the dynamic relationship between investment-grade bonds and credit default swaps *Bank of England Working Paper no 211* London UK.

*4359.* Blanco R, Brennan S, March W 2004 An empirical  analysis of the dynamic relationship between investment-grade bonds and credit default swaps *Bank of Spain Working Paper no 0401* Madrid Spain.

*4360.* Codogno L, Favero C, Missale A 2003 Government bond spreads *Economic Policy* **18** pp 504 – 532.

*4361.* Fage P November 24 2003 Hedging with credit default swaps *Credit Suisse First Boston Emerging Markets Sovereign Strategy Focus*.

4362. Francis C, Kakodkar A, Martin B April 2003 Credit derivative handbook 2003. A guide to products, valuation, strategies and risks *Merrill Lynch Global Securities Research and Economics Group*.

4363. Galiani S 2003 Copula functions and their application in pricing and risk managing multi-name credit derivative products *M Sc Thesis* King's College London UK.

4364. Grundke P 2003 Modellierung und bewertung von kreditrisiken *Ph D Dissertation* Universität zu Köln *Der Deutsche Universitätsverlag* Wiesbaden Germany.

4365. Haas F June 2003 Towards a 'market continuum'? Structural models and interaction between credit and equity markets *Banque de France Financial Stability Review* no 2.

4366. Li Chen, Filipović D July 4 2003 Credit derivatives in an affine framework *Information Sciences and Systems* Princeton University Princeton USA pp 1 – 21.

4367. MacKenzie D, Millo Y 2003 Constructing a market, performing theory: The historical sociology of a financial derivatives exchange *American Journal of Sociology* **109** pp 107 – 145.

4368. Packer F, Suthiphongchai C December 2003 Sovereign credit default swaps *BIS Quarterly Review* USA.

4369. Schmidt Th 2003 Credit risk modeling with random fields *Inaugural-Dissertation zur Erlangung des Doktorgrades an den Naturwissenschaftlichen Fachbereichen (Mathematik) der Justus-Liebig-Universität Gießen* Germany.

4370. Xu D, Wilder C May 2003 Emerging markets credit derivatives: Market overview, products, analyses and applications *Deutsche Bank Global Markets Research* Germany.

4371. Blanco R, Brennan S, March W 2004 An empirical analysis of the dynamic relationship between investment-grade bonds and credit default swaps *Working Paper no 0401* Bank of Spain Madrid Spain, *Bank of England Working Paper no 211* London UK.

4372. Cebenoyan A S, Strahan P E 2004 Risk management, capital structure and lending at banks *Journal of Banking and Finance* **28** pp 19 – 43.

4373. Chan-Lau J A, Kim Y S 2004 Equity prices, credit default swaps, and bond spreads in emerging markets *IMF Working Paper* IMF USA.

4374. Edwards F R, Morrison E R 2004 Derivatives and the Bankruptcy Code: Why the Special Treatment? *Working Paper no 258* The Center for Law and Economic Studies Columbia Law School New York USA.

4375. Eller R, Markus H 2004 Kreditderivate in der praktischen anwendung *Deutscher Sparkassen Verlag* Stuttgart Germany.

4376. Eller R 2005 Handbuch derivativer Instrumente – Produkte, Strategien und Risikomanagement *Schäffer-Poeschel-Verlag* Stuttgart Germany.

4377. Ericsson J, Jacobs K, Oviedo R A 2004 The determinants of credit default swap premia *CIRANO 2004s-55* Montréal Canada.

4378. Ericsson J, Reneby J, Wang H 2006 Can structural models price default risk? Evidence from bond and derivative markets *Working Paper McGill University*, *Stockholm School of Economics* Montreal Canada, Stockholm Sweden.

4379. Favero C, Von Thadden E-L 2004 The European bond markets under EMU *Oxford Review of Economic Policy* **4** pp 531 – 554.

**4380.** Favero C, Pagano M, Von Thadden E-L 2009 How does liquidity affect government bond yields? *Journal of Financial and Quantitative Analysis*.

**4381.** Felsenheimer J June 2 2004 CDS: Mechanism, pricing and application *HVB Global Markets Research*.

**4382.** Felsenheimer J September 2004 Kreditderivate spezial – CDS: Funktionsweise, bewertung, anwendung *Global Markets Research* HVB Corporates & Markets.

**4383.** Felsenheimer J, Gisdakis Ph, Zaiser M 2005 Kreditderivate spezial – Das kreditmarkt *Global Markets Research* HVB Corporates & Markets.

**4384.** Garcia T, Maghakian A, Sharma S 2004 Implications of stochastic recovery rates in evaluating CDO tranches *The Journal of Fixed Income* pp 64 – 71.

**4385.** Gibson M S 2004 Understanding the risk of synthetic CDOs finance and economics *Discussion Series 2004-36 Board of Governors of the Federal Reserve System* NY USA

www.federalreserve.gov/pubs/feds/2004/200436/200436pap.pdf .

**4386.** Gibson M May 22 2007 Credit derivatives and risk management *Board of Governors of Federal Reserve System Finance and Economics Discussion Series Paper 2007-47* NY USA pp 1 – 23.

**4387.** Johnson, McBride Ph, Lee Hazen Th 2004 Derivatives regulation vol **1** *Aspen* New York USA.

**4388.** Madan D B, Konikov M, Marinescu M 2004 Credit and basket default swaps *Working Paper Robert H Smith School of Business* and *Bloomberg LP*.

**4389.** Meneguzzo D, Vecchiato W 2004 Copula sensitivity in collateralized debt obligations and basket default swaps *The Journal of Futures Markets* vol **24** pp 37 – 70.

**4390.** Norden L, Weber M 2004a Informational efficiency of credit default swap and stock markets: The impact of credit rating announcements *Working Paper* University of Mannheim Germany, *Journal of Banking and Finance* **28** (11) pp 2813 – 2843.

**4391.** Norden L, Weber M 2004b The co-movement of credit default swap, bond and stock markets: An empirical analysis *CEPR Discussion Paper Series no 4674, Center for Financial Studies Working Paper no 2004/20* London UK.

**4392.** Norden L, Wagner W March 26 2007 Credit derivatives and loan pricing *TILEC Discussion Paper DP 2007-015* University of Mannheim, Tilburg University Germany ISSN 1572-4042 pp 1 – 37.

**4393.** Olléon-Assouan E June 2004 Techniques used on the credit derivatives market: Credit default swaps *Banque de France Financial Stability Review* no 4 Paris France pp 94 – 107.

**4394.** Pagano M, Von Thadden E-L 2004 The European bond markets under the EMU *Oxford Review of Economic Policy* **20** pp 531 – 554.

**4395.** Pausch Th, Schweimayer G March 2004 Hedging with credit derivatives and its strategic role in banking competition *Beitrag nr 260* Faculty of Business Administration and Economics University of Augsburg Germany pp 1 – 39.

**4396.** Puffer M 2004 Einsatzmöglichkeiten von Kreditderivaten *in* Aktuelle Herausforderungen des Finanzmanagements Gerke W(editor) Stuttgart Germany pp 51 – 60.

**4397.** Shelton D August 2004 Back to normal, proxy integration: A fast accurate method for CDO and CDO-squared pricing Citigroup Structured Credit Research USA.

**4398.** Shimko D 2004 Credit risk: Models and management *Risk Books* Division of Incisive Financial Publishing Ltd London UK.

**4399.** Tavares P A C, Nguyen T U, Chapovsky A, Vaysburd I 2004 Composite basket model *Working Paper* Merrill Lynch Credit Derivatives NY USA.

**4400.** Verdier P-H 2004 Credit derivatives and the sovereign debt restructuring process *Harvard Law School* Harvard University USA.

**4401.** Zhu H 2004 An empirical comparison of credit spreads between the bond market and the credit default swap market *BIS Working Papers no 160* BIS Switzerland.

**4402.** Abid F, Naifar N 2005 The impact of stock returns volatility on credit default swap rates: A copula study *International Journal of Theoretical and Applied Finance* vol **8** no 8 pp 1135 - 1155.

**4403.** Abid F, Naifar N 2006a The determinants of credit default swap rates: An explanatory study *International Journal of Theoretical and Applied Finance* **9** (1) pp 23 – 42.

**4404.** Abid F, Naifar N August 2006b Credit default swap rates and equity volatility: a nonlinear relationship *Journal of Risk Finance* **7** (4) p 348 – 371.

**4405.** Acharya V, Pedersen L 2005 Asset pricing with liquidity risk *Journal of Financial Economics* **77** pp 375 – 410.

**4406.** Acharya V, Schaefer S 2006 Liquidity risk and correlation risk: Implications for risk management *Working Paper* London Business School London UK.

**4407.** Acharya V, Johnson T 2007 Insider trading in credit derivatives *Journal of Financial Economics* **84** pp 110 – 141.

**4408.** Acharya V, Schaefer S, Zhang Y May 2007 Liquidity risk and correlation risk: A clinical study of the General Motors and Ford *Working Paper* SSRN NY USA.

**4409.** Albrecht P 2005 Kreditrisiken – Modellierung und management: Ein überblick *Risk and Insurance Review*.

**4410.** Bielecki T R, Crepey S, Jeanblanc M, Rutkowski M March 2005 Valuation of basket credit derivatives in the credit migrations environment *Working Paper*.

**4411.** Bielecki T R, Vidozzi A, Vidozzi L May 2006 An efficient approach to valuation of credit basket products and rating triggered step-up bonds *Working Paper*.

**4412.** Blanco R, Brennan S, Marsh I W 2005 An empirical analysis of the dynamic relationship between investment grade bonds and credit default Swaps *Journal of Finance* **60** (5) pp 2255 – 2281.

**4413.** Bomfim A N 2005 Understanding credit derivatives and related instruments *Elsevier Academic Press* San Diego USA.

**4414.** Brigo D January 2005 Market models for CDS options and callable floaters *Risk*, *and in* Derivatives Trading and Options Pricing Dunbar N (editor) *Risk Books*.

**4415.** Brigo D, Alfonsi A 2005 Credit default swap calibration and derivatives pricing with SSRD stochastic intensity module *Finance and Stochastic* **9** (1) http://www.damianobrigo.it/cirppcredit.pdf .

**4416.** Brigo D, Cousot L 2006 A comparison between SSRD module and the market module for CDS options pricing *International Journal of Theoretical and Applied Finance* **9** (3).

**4417.** Brigo D, El-Bachir N 2006 Credit derivatives pricing with a smile extended jump stochastic intensity model *ICMA Discussion Paper in Finance DP 2006 – 13* University of Reading UK pp 1 – 22.

**4418.** Brigo D, Pallavicini A, Torresetti R June 2007 CDO calibration with the dynamical generalized Poisson loss model *Risk Magazine*.

**4419.** Burtschell X, Gregory J, Laurent J-P 2005, 2008 A comparative analysis of CDO pricing models *Working Paper* BNP Paribas, Université de Lyon Paris, Lyon France.

**4420.** Chiesa G 2005 Risk transfer, lending capacity and real investment activity *Working Paper* Department of Economics University of Bologna Italy.

**4421.** Cossin D, Lu H June 2005 Are European corporate bonds and default swap markets segmented? *Working Paper 153* FAME.

**4422.** Cousseran O, Rahmouni I June 2005 The CDO market functioning and implications in terms of financial stability *Banque de France Financial Stability Review* no 6 pp 43 – 59.

**4423.** Franke G, Krahnen J 2005 Default risk sharing between banks and markets: The contribution of collateralized debt obligations *NBER Working Paper 11741* National Bureau of Economic Research USA, *in* The Risks of Financial Institutions Carey M, Stulz R *University of Chicago Press* Chicago USA.

**4424.** Franke G 2005 Risikomanagement mit kreditderivaten *in* Burghof H-P et al Kreditderivate – Handbuch für die bank und anlagepraxis *Schäffer-Poeschel Verlag* Stuttgart.

**4425.** Greenspan A May 5 2005 Risk transfer and financial stability *Remarks at the Federal Reserve Bank of Chicago's Forty-first Annual Conference on Bank Structure* Chicago IL USA.

**4426.** Heinrich M 2005 Kreditderivate *in* Eller R Handbuch derivativer instrumente – Produkte, strategien und risikomanagement *Schäffer-Poeschel-Verlag* Stuttgart Germany.

**4427.** Houweling P, Vorst T 2005 Pricing default swaps: Empirical evidence *Journal of International Money and Finance* **24** pp 1200 – 1225.

**4428.** Instefjord N 2005 Risk and hedging: Do credit derivatives increase bank risk? *Journal of Banking and Finance* **29** pp 333 – 345.

**4429.** Jorion Ph 2005 Financial risk manager handbook 3[rd] edition *Wiley–Finance John Wiley and Sons Inc* Hoboken NJ USA.

**4430.** Jortzik S 2005 Semi-analytische und simulative Kreditrisikomessung synthetischer Collateralized Debt Obligations bei heterogenen Referenzportfolios *Ph D Dissertation* Universität Göttingen Göttingen.

**4431.** Joshi M S, Stacey A 2005 Intensity gamma: A new approach to pricing portfolio credit derivatives *Working Paper* Royal Bank of Scotland UK.

**4432.** Kalemanova A, Schmid B, Werner R 2005 The normal inverse Gaussian distribution for synthethic CDO pricing *Risklab*, *Algorithmics Inc*, *Allianz* München, London Germany, UK.

**4433.** Kim S J, Lucey B M, Wu E 2005 Dynamics of bond market integration between established and new European Union countries *Journal of International Financial Markets, Institutions and Money* vol **16** no 1 pp 41 – 46.

**4434.** Kim S J, Moshirian F, Wu E 2006 Evolution of international stock and bond market integration: Influence of the European monetary union *Journal of Banking and Finance* vol **30** pp 1507 – 1534.

**4435.** Lucas D J, Goodman L S, Fabozzi F J 2006 Collateralized debt obligations: Structures and analysis 2$^{nd}$ edition *John Wiley & Sons Inc* Hoboken NJ USA ISBN-13: 978-0471718871 pp 1 – 528.

**4436.** Lucas D J, Goodman L S, Fabozzi F J, Manning R 2007 Developments in the collateralized debt obligations markets: New products and insights *John Wiley and Sons Inc* New Jersey USA.

**4437.** Lucas D J, Goodman L S, Fabozzi F J 2007 Collateralized debt obligations and credit risk transfer *Yale ICF Working Paper no 07-06* Yale University USA pp 1 – 14 http://ssrn.com/abstract=997276 .

**4438.** Minton B, Stulz R, Williamson R August 2005, June 2006 How much do banks use credit derivatives to reduce risk? *NBER Working Paper* 11579 National Bureau of Economic Research Massachusetts Avenue Cambridge USA pp 1 – 40, *Fisher College of Business Working Paper 2006-03-001* http://www.nber.org/papers/w11579 .

**4439.** Morrison A 2005 Credit derivatives, disintermediation and investment decisions *Journal of Business* **78** (2) pp 621 – 647.

**4440.** Nési C March 2005 Reforming the legal framework for securisation in France *Banque de France Bulletin Digest no 135* Paris France.

**4441.** Neske Ch 2005 Grundformen von kreditderivaten *in* Burghof H-P et al Kreditderivate – Handbuch für die Bank und Anlagepraxis *Schäffer-Poeschel Verlag* Stuttgart Germany.

**4442.** Nicolò A, Pelizzon L October 17 2005 Credit derivatives: Capital requirements and strategic contracting Department of Economics University of Padova Italy pp 1 – 33.

**4443.** Parlour C, Plantin G 2005 Credit Risk Transfer *Working Paper* Tepper School of Business Carnegie Mellon University NY USA.

**4444.** Posthaus A 2005 Exotische kreditderivate *in* Burghof H-P et al Kreditderivate – Handbuch für die Bank und Anlagepraxis *Schäffer-Poeschel Verlag* Stuttgart Germany.

**4445.** Pelizzon L, Schaefer S 2005 Pillar 1 vs Pillar 2 under risk management *in* Risks of Financial Institutions and of the Financial Sector Carey M, Stulz R (editors) *Oxford University Press* Oxford UK.

**4446.** Skinner F 2005 Pricing and hedging interest and credit risk sensitive instruments *Elsevier Butterworth-Heinemann* Burlington Massachusetts USA.

**4447.** Stamicar R, Finger C C 2005 Incorporating equity derivatives into the credit grades model *RiskMetrics Group* New York USA.

**4448.** Allen F, Carletti E 2006 Credit risk transfer and contagion *Journal of Monetary Economics* **53** pp 89 – 111.

**4449.** Beitel P, Dürr J, Pritsch G, Stegemann U 2006 Actively managing the credit portfolio *in* Banking in a changing world McKinsey & Company (editors) pp 173 – 178.

**4450.** Beck A, Lesko M, Schlottmann F, Wimmer K Ausgabe 14 2006 Copulas im risikomanagement *Zeitschrift für das gesamte Kreditwesen*.

**4451.** Chilcote E 2006 Credit derivatives and financial fragility *Policy Note 2006 / 1* The Levy Economics Institute of Bard College Annandale-on-Hudson NY USA pp 1 – 6.

**4452.** Cremers H 2006 Bankcontrolling und risiko *in* Vorlesungsskript des Sommersemesters 2006 im Bachelor-Studiengang der HfB *Business School of Finance and Management* Frankfurt am Main Germany.

**4453.** Cremers H, Walzner J Juni 2007 Risikosteuerung mit kreditderivaten unter besonderer berücksichtigung von credit default swaps *Working Paper Frankfurt School of Finance & Management no 80* ISSN: 14369753 pp 1 – 66

http://nbn-resolving.de/urn:nbn:de:101:1-20080827273 ,

http://hdl.handle.net/10419/27847 .

**4454.** Cremers H, Walzner J 2009a Modellierung des kreditrisikos im einwertpapierfall *Working Paper Frankfurt School of Finance & Management no 126* ISSN: 14369753 pp 1 – 90

http://nbn-resolving.de/urn:nbn:de:101:1-20090826287 ,

http://hdl.handle.net/10419/27932 .

**4455.** Cremers H, Walzner J 2009b Modellierung des kreditrisikos im portfoliofall *Working Paper Frankfurt School of Finance & Management no 127* ISSN: 14369753 pp 1 – 72

http://nbn-resolving.de/urn:nbn:de:101:1-20090826290 ,

http://hdl.handle.net/10419/27933 .

**4456.** De Wit J 2006 Exploring the CDS-bond basis *Working Paper Research no 104 Research Series 200611-16* National Bank of Belgium Brussels Belgium.

**4457.** Felsenheimer J, Gisdakis Ph, Zaiser M 2006a Active portfolio management *John Wiley and Sons Inc* pp 1 – 581.

**4458.** Felsenheimer J, Gisdakis Ph, Zaiser M 2006b Active credit portfolio management – A practical guide to credit risk management strategies *Wiley Verlag* Weinheim Germany.

**4459.** Geithner T F May 16 2006 Implications of growth in credit derivatives for financial stability *Remarks at New York University Stern School of Business Third Credit Risk Conference* New York USA.

**4460.** Gikhman I July 2006 Some critical comments on credit risk modeling *MPRA Paper no 1451* Munich University Germany pp 1 – 7

http://mpra.ub.uni-muenchen.de/1451/ .

**4461.** Gikhman I February 4 2008, July 2008 Risky swaps *MPRA Paper no 7078* Munich University Germany pp 1 – 39, *ICFAI Journal of Derivative Markets*.

http://mpra.ub.uni-muenchen.de/7078/ .

**4462.** Gikhman I 2008, November 2011 Multiple risky securities valuation I *MPRA Paper no 34511* Munich University Germany pp 1 – 25

http://mpra.ub.uni-muenchen.de/34511/ .

*4463.* Goderis B, Marsh I W, Castello J V, Wagner W 2006 Bank behavior with access to credit risk transfer markets SSRN USA

http://ssrn.com/abstract=937287 .

*4464.* Gruber J, Gruber W, Braun H 2005 Praktike Gruber r-handbuch asset-backed-securities und kreditderivate – Strukturen, preisbildung, anwendungsmöglichkeiten, aufsichtliche behandlung *Schäffer-Poeschel-Verlag* Stuttgart Germany.

*4465.* Huault I, Rainelli-Le Montagner H 2006 Constructing a globalized over-the-counter financial market. Limits of performativity and problems of cognitive framing. The case of credit derivatives pp 1 – 29.

*4466.* Huault I, Rainelli-Le Montagner H 2008 Market shaping as an answer to ambiguities: The case of credit derivatives *University Paris Dauphine, University Paris I Pantheon Sorbonne* Paris France pp 1 – 30.

*4467.* Joshi M, Stacey A July 2006 Intensity gamma: A new approach to pricing portfolio credit derivatives *RISK* **19** pp 78 – 83.

*4468.* Juselius K 2006 The co-integrated VAR model: Methodology and applications *Oxford University Press* Oxford UK.

*4469.* Ludovici A November December 2006 The application of neural networks to the pricing of credit derivatives *Rivista Di Politica Economica* pp 187 – 221.

*4470.* Marsh I 2006 The effect of lenders' credit risk transfer activities on borrowing firms' equity returns *Cass Business School*.

*4471.* Martin M R W, Reitz S, Wehn C S 2006 Kreditderivate und kreditrisikomodelle – Eine mathematische einführung *Vieweg Verlag* Wiesbaden Germany.

*4472.* McDonald R L 2006 Derivatives markets 2$^{nd}$ edition *Addison Wesley* Boston USA.

*4473.* Minton B, Stulz R, Williamson R June 2006 How much do banks use credit derivatives to reduce risk? *Working Paper no 2006-03-001* Fisher College of Business.

*4474.* Mortensen A 2006 Semi-analytical valuation of basket credit derivatives in intensity-based models *Journal of Derivatives* **13** no 4 pp 8 – 26.

*4475.* Nicolò A, Pelizzon L November 2006 Credit derivatives, capital requirements and opaque OTC markets *Working Paper no 58/WP/2006* Department of Economics Ca' Foscari University of Venice Italy ISSN 1827-336X pp 1 – 38.

*4476.* Papenbrock J 2006 Price calibration and hedging of correlation dependent credit derivatives using a structural model with alpha-stable distributions *Diploma Thesis Institute for Statistics and Mathematical Economic Theory*.

*4477.* Partnoy F, Skeel D 2006 The promises and perils of credit derivatives *School of Law Legal Studies Research Paper no 07-74* University of San Diego San Diego USA.

*4478.* Predescu M 2006 The performance of structural models of default for firms with liquid CDS spreads *Working Paper* Rotman School of Management University of Toronto Canada.

*4479.* Pugachevsky D 2006 Pricing counterparty risk in unfunded synthetic CDO tranches *in* Counterparty Credit Risk Modeling Pykhtin M (editor) *Risk Books* London UK pp 371 – 394.

*4480.* Sircar R, Zariphopoulou T 2006 Utility valuation of credit derivatives and application to CDOs *Working Paper* Princeton University Princeton USA.

**4481.** Zhu H 2006 An empirical comparison of credit spreads between the bond market and the credit default swap market *Journal of Financial Services Research* **29** (3) pp 211 – 235.

**4482.** Wagner W, Marsh I 2006 Credit risk transfer and financial sector stability *Journal of Financial Stability* **2** (2) pp 173 – 193.

**4483.** Ammer J, Cai F 2007 Sovereign CDS and bond pricing dynamics in emerging markets: Does the cheapest-to-deliver option matter? *Board of Governors of Federal Reserve System International Finance Discussion Papers no 912* Federal Reserve System NY USA.

**4484.** Ashcraft A B, Santos J A C July 2007 Has the credit default swap market lowered the cost of corporate debt? *Staff Report no 290* Federal Reserve Bank of New York NY USA pp 1 – 44.

**4485.** Baba N, Inada M 2007 Price discovery of credit spreads for Japanese mega-banks: Subordinated bond and CDS *Discussion Paper 2007-E-6* Institute for Monetary and Economic Studies Bank of Japan Tokyo Japan.

**4486.** Boulier J F, Brière M, Viala J R 2007 Do leveraged credit derivatives modify credit allocation? *CEB Working Paper no 08/014* Solvay Business School Université Libre de Bruxelles Belgium pp 1 – 18.

**4487.** Byström H 2007 Back to the future: Futures margins in a future credit default swap index futures market *The Journal of Futures Markets* **27** (1) pp 85 – 104.

**4488.** Byström H 2008 Credit default swaps and equity prices: The iTraxx CDS index market *in* Credit Risk – Models, Derivatives, and Management *Financial Mathematics Series* vol **6** *Chapman & Hall*, Wagner N (editor) *CRC Press* pp 69 – 84.

**4489.** Byström H 2010 The age of turbulence, credit derivatives style *Department of Economics* Lund University Lund Sweden pp 1 – 20.

**4490.** Chen L, Lesmond D A, Wei J 2007 Corporate yield spreads and bond liquidity *The Journal of Finance* **62** pp 119 – 149.

**4491.** Davydenko S A, Strebulaev I A 2007 Strategic actions and credit spreads: An empirical investigation *Journal of Finance* **62** pp 2633 – 2671.

**4492.** Dötz N 2007 Time-varying contributions by the corporate bond and CDS markets to credit risk price discovery *Discussion Paper Series 2 Banking and Financial Studies no 08* Deutsche Bundesbank Germany.

**4493.** Drucker S, Puri M 2007 On loan sales, loan contracting, and lending relationships FDIC *Center for Financial Research Working Paper no WP 2007-04*, *Social Science Research Network* NY USA

http://ssrn.com/abstract=920877 .

**4494.** Fathi A, Nader N 2007 Copula based simulation procedures for pricing basket credit derivatives *Faculty of Business and Economics* University of Sfax Tunisia, *MPRA Paper no 6014* Munich University Germany pp 1 – 31

http://mpra.ub.uni-muenchen.de/6014/ .

**4495.** Fulop A, Lescourret L 2007 An analysis of intra-daily patterns on the CDS market *Working Paper* ESSEC Business School.

**4496.** Henrard M 2007 The irony in the derivative discounting *Working Paper* Social Science Research Network NY USA

http://ssrn.com/abstract=970509 .

*4497.* Henrard M 2009 The irony in the derivative discounting part II: The Crisis *Working Paper* Social Science Research Network NY USA

http://ssrn.com/abstract=1433022 .

*4498.* Hirtle B February 2007, March 2008 Credit derivatives and bank credit supply *Staff Report 276* Federal Reserve Bank of New York NY USA pp 1 – 44.

*4499.* In F, Kang B U, Kim S T 2007 Sovereign credit default swaps, sovereign debt and volatility transmission across emerging markets *Korea Advanced Institute of Science and Technology Working Paper Series* Seoul South Korea.

*4500.* Mengle D 2007 Credit derivatives: An overview *Economic Review Federal Reserve Bank of Atlanta* issue Q4 pp 1 – 24.

*4501.* Nashikkar A J, Subrahmanyam M G 2007 Latent liquidity and corporate bond yield spreads *NYU Working Paper no FIN-07-013* New York University NY USA.

*4502.* Nashikkar A J, Subrahmanyam M G, Mahanti S 2009 Limited arbitrage and liquidity in the market for credit risk *NYU Working Paper no FIN-08-011* New York University NY USA.

*4503.* Pan J, Singleton K J 2007 Default and recovery implicit in the term structure of sovereign CDS spreads *Review of Financial Studies*.

*4504.* Papageorgiou E, Sircar R 2007 Multiscale intensity models and name grouping for valuation of multi-name credit derivatives *Applied Mathematical Finance* **15** (1) pp 73 – 105.

*4505.* Pausch Th 2007 Endogenous credit derivatives and bank behavior *Discussion Paper Series 2: Banking and Financial Studies no 16/2007* Deutsche Bundesbank Frankfurt am Main Germany ISBN 978-3–86558–361–1 pp 1 – 48.

*4506.* Pykhtin M, Zhu S A July August 2007 Guide to modeling counterparty credit risk *GARP Risk Review*.

*4507.* Rajan A, McDermott G 2007 The structured credit handbook *Wiley Finance* New Jersey USA.

*4508.* Schulz A, Wolf G 2007 Sovereign bond market integration: The Euro, trading platforms and globalization *Deutsche Bundesbank Discussion Paper (Series 1) 12* Frankfurt Germany.

*4509.* Summer Ch 2007 Credit risk – Advanced models *Skriptum des Fachbereichs Bankbetriebslehre der Wirtschaftsuniversität* Wien Austria.

*4510.* Tang D Y, Yan H 2007 Liquidity and credit default swap spreads *Working Paper* Kennesaw State University.

*4511.* Thompson J R 2007 Credit risk transfer: To sell or to insure? *Working Paper 1131* Queen's University Kingston Ontario Canada.

*4512.* Truslow D March 22 2007 Effectiveness of credit risk management in allocating risk *Remarks at Federal Reserve Bank of Richmond 2007 Credit Market Symposium* Charlotte NC USA.

http://www.richmondfed.org/news_and_speeches/conferences/pdf/cms_risk%20allocation_truslow.pdf

*4513.* Weithers T Fourth Quarter 2007 Credit derivatives, macro risks, and systemic risks *Federal Reserve Bank of Atlanta Economic Review* Atlanta USA pp 43 – 69.

*4514.* Yan B, Zivot E 2007 The dynamics of price discovery *Working Paper* Series University of Washington USA.

*4515.* Abid F, Naifar N 2008 The application of copulas in pricing dependent credit derivatives instruments *Journal of Applied Economic Sciences* vol **III** issue2 (4) pp 65 – 74.

*4516.* Alexander C, Kaeck A 2008 Regime dependent determinants of credit default swap spreads *Journal of Banking and Finance* **32** pp 1008 – 1021.

*4517.* Arnsdorf M, Halperin I 2008 Bslp: Markovian bivariate spread-loss model for portfolio credit derivatives *Journal of Computational Finance* **12** pp 77 – 100.

*4518.* Chen R-R, Cheng X, Liu B 2008 Estimation and evaluation of the term structure of credit default swaps: An empirical study *Insurance: Mathematics and Economics* **43** pp 339 – 349.

*4519.* Claes A, De Ceuster M 2008 Single name credit default swap valuation: A review *in* Credit Risk Models, Derivatives, and Management Wagner N (editor) *CRC Press* pp 69 – 84.

*4520.* Cont R, Minca A 2008 Extracting portfolio default rates from CDO spreads *Working Paper* Columbia University NY USA.

*4521.* Coudert V, Gex M 2008 Co-movements in the CDS market and relationship to other financial markets: The case of the GM and Ford crisis in 2005 *Working Paper*.

*4522.* Cserna B, Imbierowicz B 2008 How efficient are credit default swap markets? An empirical study of capital structure arbitrage based on structural pricing models *Working Paper* Goethe University Frankfurt Germany.

*4523.* Curto J D, Nunes J P, Oliveira L 2008 The determinants of sovereign credit spread changes in the Euro-zone *Working Paper* ISCTE Business School.

*4524.* Jankowitsch R, Pullirsch R, Veža T 2008 The delivery option in credit default swaps *Journal of Banking and Finance* **32** pp 1269 – 1285.

*4525.* Le Roux S J October 2008 Measuring counterparty credit risk: An overview of the theory and practice *Ph D Dissertation* University of Pretoria South Africa.

*4526.* Li X M, Zou L P 2008 How do policy and information shocks impact co-movements of China's T-bond and stock markets? *Journal of Banking and Finance* vol **32** pp 347 – 359.

*4527.* Morgan G May 2008 Market formation and governance in international financial markets: The case of OTC derivatives *Human Relations* **61** (5) pp 637 – 660.

*4528.* O'Kane D 2008 Modelling single-name and multi-name credit derivatives *Wiley Finance Series John Wiley and Sons Inc* NY USA.

*4529.* Peng X, Kou S S G 2008 Default clustering and valuation of collateralized debt obligations *Working Paper* Columbia University NY USA.

*4530.* Sougné D, Heuchenne C, Hübner G 2008 The determinants of credit default swap prices: An industry-based investigation *in* Credit Risk Models Derivatives, and Management Wagner N (editor) *CRC Press* pp 85 – 96.

*4531.* Ametrano F M, Bianchetti M 2009 Bootstrapping the illiquidity: Multiple curve construction for coherent forward rates estimation *in* Modelling Interest Rates: Latest Advances for Derivatives Pricing Mercurio F (editor) *Risk Books* London UK.

4532. Ametrano F M January 18 2011 Rates curves for forward euribor estimation and CSA-Discounting *QuantLib Forum* London UK.

4533. Bayraktar E, Yang B 2009 Multi-scale time-changed birth processes for pricing multi-name credit derivatives *Applied Mathematical Finance* **16** (5) pp 429 – 449.

4534. Deutsche Börse AG September 2009 The global derivatives market - A blueprint for market safety and integrity White paper *Deutsche Börse AG* Germany

http://deutsche-boerse.com/dbg/dispatch/en/binary/gdb_content_pool/imported_files/public_files/10_downloads/80_misc/whitepaper_derivatives2.pdf.

4535. Elizalde A, Doctor S 2009 The bond-CDS funding basis *J P Morgan European Credit Derivatives Research*.

4536. Elizalde A, Doctor S, Saltuk Y 2009 Bond-CDS basis handbook *J P Morgan European Credit Derivatives Research.*

4537. Forte S, Peña J I 2009 Credit spreads: Theory and evidence about the information content of stocks, bonds and CDSs *Journal of Banking and Finance* **33** (11) pp 2013 – 2025.

4538. Fujii M, Shimada Y, Takahashi A 2009a A survey on modelling and analysis of basis spread *CARF Working Paper CARF-F-195*

http://ssrn.com/abstract=1520619 .

4539. Fujii M, Shimada Y, Takahashi A 2009b A market model of interest rates with dynamic basis spreads in the presence of collateral and multiple currencies *CARF Working Paper CARF-F-196*

http://ssrn.com/abstract=1520618

4540. Fujii M, Shimada Y, Takahashi A 2010a A note on construction of multiple swap curves with and without collateral *CARF Working Paper CARF-F-154* http://ssrn.com/abstract=1440633 .

4541. Fujii M, Shimada Y, Takahashi A 2010b Collateral posting and choice of collateral currency – implications for derivative pricing and risk management *CARF Working Paper CARF-F-216*

http://ssrn.com/abstract=1601866 .

4542. Fujii M, Takahashi A 2011 Choice of collateral currency *Risk Magazine* **24** (1) pp 120 – 125.

4543. Goderis B, Wagner W March 19 2009 Credit derivatives and sovereign debt crises *Department of Economics* University of Oxford UK, *MPRA Paper no 17314* Munich University Germany pp 1 – 43

http://mpra.ub.uni-muenchen.de/17314/ .

4544. Mayordomo S, Peña J I, Romo J September 2009 Are there arbitrage opportunities in credit derivatives markets? A new test and an application to the case of CDS and ASPs *Working Paper 09-63* Business Economic Series 06 Departamento de Economía de la Empresa Business Economic Series 06 Universidad Carlos III de Madrid Spain pp 1 – 62.

4545. Mayordomo S, Peña J I, Romo J 2012 The effect of liquidity on the price discovery process in credit derivatives markets in times of financial distress *Department of*

*Research and Statistics* Comisión Nacional del Mercado de Valores Madrid Spain pp 1 – 31.

4546. Rey N 2009 Credit derivatives: Instruments of hedging and factors of instability. The example of "Credit Default Swaps" on French reference entities *CEPN University of Paris 13* France pp 1 – 27.

4547. Panchenko V, Wu E 2009 Time-varying market integration and stock and bond return concordance in emerging markets *Journal of Banking and Financ*e vol **33** pp 1014 – 1021.

4548. Rey N November 20 2009 Credit derivatives: instruments of hedging and factors of instability. The example of "Credit Default Swaps" on French reference entities *CEPN University of Paris 13* France hal-00433883 version 1 pp 1 – 27.

4549. Varga L May 2009 The information content of Hungarian sovereign CDS spreads *MNB Occasional Papers 78* Magyar Nemzeti Bank Budapest Hungary ISSN 1585-5678 pp 1 – 30 www.mnb.hu .

4550. Buraschi A, Porchia P, Porchia F 2010 Correlation risk and optimal portfolio choice *Journal of Finance* vol **65** (1) pp 393 – 420.

4551. Ciolpan M, Adam C 2010 Developments of credit default swap contracts under the influence of global crisis *Finance – Challenges of the Future* **IX** (11) pp 254 – 259.

4552. Coudert V, Gex M 2010 Contagion inside the credit default swaps market: The case of the GM and Ford crisis in 2005 *Journal of International Financial Markets, Institutions and Money* **20** (2) pp 109 – 134.

4553. Courtney S May 19 2010 2008 SEC short selling ban: Impacts on the credit default swap market *Stanford University* California USA, *MPRA Paper no 35390* Munich University Germany pp 1 – 38 http://mpra.ub.uni-muenchen.de/35390/ .

4554. De Wit J 2006 Exploring the CDS-bond basis *Working Paper* National Bank of Belgium Brussels Belgium.

4555. Gómez S M 2010 Four essays on the interaction between credit derivatives and fixed income markets *Ph D Thesis* Departamento De Economía De La Empresa Universidad Carlos Iii De Madrid Spain pp 1 – 267.

4556. Gregory J 2010 Counterparty credit risk: The new challenge for global financial markets *John Wiley and Sons Inc* USA.

4557. Fries C P 2010 Discounting revisited: Valuation under funding, counterparty risk and collateralization Working Paper Social Science Research Network NY USA http://ssrn.com/abstract=1609587

4558. Tsui L K 2010 Exact numerical algorithm for n-th order derivative of a single variable function *Working Paper* University of Pittsburgh USA.

4559. Tsui L K May 18 2010 Multi-factor bottom-up model for pricing credit derivatives *Department of Mathematics* University of Pittsburgh USA, *MPRA Paper no 23090* Munich University Germany pp 1 – 43 http://mpra.ub.uni-muenchen.de/23090/ .

4560. Piterbarg V 2010a Funding beyond discounting: Collateral agreements and derivative pricing *Risk Magazine* **23** (2) pp 97 – 102.

**4561.** Piterbarg V May 2010b Effects of funding and collateral *Global Derivatives and Risk Management Conference* Paris France.

**4562.** Bianchetti M October 31 2011 The Zeeman effect in finance: Libor spectroscopy and basis risk management *Market Risk Management* Intesa Sanpaolo Milan Italy, *MPRA Paper no 42247* Munich University Germany pp 1 – 23

http://mpra.ub.uni-muenchen.de/42247/ .

**4563.** Bianchetti M, Mattia C March 28 2012 Interest rates after the credit crunch: Multiple-curve vanilla derivatives and SABR *Market Risk Management* Intesa Sanpaolo Milan Italy, *MPRA Paper no 42248* Munich University Germany pp 1 – 27

http://mpra.ub.uni-muenchen.de/42248/ .

**4564.** Calice G, Ioannidis Ch, Williams J 2011 Credit derivatives and the default risk of large complex financial institutions *CESIFO Working Paper no 3583* Category 7: Monetary Policy and International Finance pp 1 – 46

www.CESifo-group.org/wp .

**4565.** Coşkun Y 2011 The limitations of transparency policy in OTC markets and derivatives trading *Journal of Securities Operations and Custody* **4** (2) pp 122 – 133.

**4566.** Lahusen R, Speyer B 2011 Dérivés de crédit: La transformation d'un métier traditionnel de la banque *Allianz AG*, *Deutsche Bank Research* Frankfurt Germany *Revue D'économie Financière* **78** (1) pp 39 – 53.

**4567.** Lijun Bo, Ying Jiao, Xuewei Yang December 13 2011 Credit derivatives pricing with default density term structure modeled by Lévy random fields *Department of Mathematics* Xidian University Xi'an P R China *HAL – CCSd* hal-00651397 version 1, arXiv:1112.2952v1 [q-fin.PR] 13 Dec 2011 pp 1 – 21

www.arxiv.org 1112.2952v1.pdf .

**4568.** Ojo M May 24 2011 Capital, liquidity standards and macro prudential policy tools in financial supervision: Addressing sovereign debt problems *MPRA Paper no 31096* Munich University Germany pp 1 – 21

http://mpra.ub.uni-muenchen.de/31096/ .

**4569.** Ojo M April 30 2012 Bailouts and longer term refinancing operations (LTROs): When temporary cures generate longer term economic concerns *Covenant University*, *MPRA Paper no 38483* Munich University Germany pp 1 – 32

http://mpra.ub.uni-muenchen.de/38483/ .

**4570.** Rennie A, Lipton A 2011 Technical introduction *Chapter 2* ISBN 9780199546787 pp 17 – 36.

**4571.** Avino D, Lazar E 2012 Rethinking capital structure arbitrage *ICMA Centre* University of Reading Henley School of Business UK, *MPRA Paper no 42850* Munich University Germany pp 1 – 28

http://mpra.ub.uni-muenchen.de/42850/ .

**4572.** Heller D, Vause N 2012 Collateral requirements for mandatory central clearing of over-the-counter derivatives *BIS Working Papers no 373* Switzerland.

**4573.** Roman A, Şargu A C 2012 Credit risk transfer mechanisms in the EU banking sector *Revista Economică Supplement no 4/2012 Journal of Economic-Financial Theory and Practice* Romania ISSN: 1582-6260 pp 552 – 560

http://economice.ulbsibiu.ro/revista.economica .

4574. Tim Leung, Peng Liu October 2 2012 Risk premia and optimal liquidation of credit derivatives *Columbia University*, *Johns Hopkins University* arXiv:1110.0220v2 [q-fin.PR] 2 Oct 2012 pp 1 – 30
www.arxiv.org 1110.0220v2.pdf .

4575. Yildirim B D, Coskun Y, Caglar O, Yildirak K 2012 How dangerous is the counterparty risk of OTC derivatives in Turkey? *Capital Markets Board of Turkey* Istanbul Turkey, *MPRA Paper no 40600* Munich University Germany pp 1 – 11
http://mpra.ub.uni-muenchen.de/40600/ .

4576. Heckinger R, Mengle D 2013 Understanding derivatives - Markets and infrastructure: Derivatives overview *Federal Reserve Bank of Chicago* USA pp 1 – 11.

4577. Heckinger R, Ruffini I, Wells K 2014 Understanding derivatives - Markets and infrastructure: Over-the-Counter (OTC) derivatives *Federal Reserve Bank of Chicago* USA pp 27 – 38.

4578. Steigerwald R S 2013 Understanding derivatives - Markets and infrastructure: Central Counterparty *Federal Reserve Bank of Chicago* USA pp 12 – 26.

***Foreign currencies investment, foreign currencies exchange rates valuation, ultra high frequency electronic trading, foreign currencies exchange, financial capital investment product, financial capital investment medium in finances:***

4579. Ellis H, Metzler L (editors) 1949 Readings in the theory of international trade *Blakiston* Philadelphia USA.

4580. Machlup F 1949 The theory of foreign exchanges *in* Readings in the theory of international trade Ellis H, Metzler L (editors) *Blakiston* Philadelphia USA.

4581. Robinson J 1949 The foreign exchanges *in* Readings in the theory of international trade Ellis H, Metzler L (editors*) Blakiston* Philadelphia USA.

4582. Friedman M 1953 The case for flexible exchange rates *in* Essay in positive economics *University of Chicago Press* Chicago USA.

4583. Friedman M (editor) 1953 Essays in positive economics *Chicago University Press* Chicago USA.

4584. Baumol W 1957 Speculation, profitability, and stability *Review of Economics and Statistics* **39** pp 263 – 271.

4585. Debreu G 1959 Theory of value *Cowles Foundation Monograph* vol **17** *John Wiley & Sons Inc* New York USA.

4586. Shiryaev A N 1961 The problem of the most rapid detection of a disturbance in a stationary process *Soviet Mathematical Doklady* **2** pp 795 – 799.

4587. Shiryaev A N 1963 On optimal methods in quickest detection problems *Theory of Probability and its Applications* **8** (1) pp 22 – 46.

4588. Shiryaev A N 1964 On Markov sufficient statistics in non-additive Bayes problems of sequential analysis *Theory of Probability and its Applications* **9** (4) pp 670 – 686.

4589. Shiryaev A N 1965 Some exact formulas in a 'disorder' problem *Theory of Probability and its Applications* **10** pp 348 – 354.

4590. Grigelionis B I, Shiryaev A N 1966 On Stefan's problem and optimal stopping rules for Markov processes *Theory of Probability and its Applications* **11** pp 541 – 558.

4591. Shiryaev A N 1967 Two problems of sequential analysis *Cybernetics* **3** pp 63 – 69.

4592. Liptser R S, Shiryaev A N 1977 Statistics of random processes *Springer-Verlag* New York USA.

4593. Shiryaev A N 1972 Random processes *Moscow State University Press* Russian Federation.

4594. A. N. Shiryaev A N 1973, 1974 Probability, statistics, random processes *Moscow State University Press* vols **1**, **2** Russian Federation.

4595. Shiryaev A N 1978, 2008b Optimal stopping rules $1^{st}$ edition, $3^{rd}$ edition *Springer ISSN 0172-4568 Library of Congress Control Number: 2007934268* Berlin Germany pp 1 – 217.

4596. Shiryaev A N 1988 Probability *Springer-Verlag* Berlin Heidelberg Germany.

4597. Shiryaev A N 1995 Probability $2^{nd}$ edition *Springer - Verlag* ISBN 0-387-94549-0 New York USA pp 1 – 621.

4598. Shiryaev A N 1998a Foundations of stochastic financial mathematics vol **1** *Fazis Scientific and Publishing House* Moscow Russian Federation ISBN 5-7036-0044-8 pp 1 – 492.

4599. Shiryaev A N 1998b Foundations of stochastic financial mathematics vol **2** *Fazis Scientific and Publishing House* Moscow Russian Federation ISBN 5-7036-0044-8 pp 493 – 1017.

4600. Shiryaev A N 1999 Essentials of stochastic finance: Facts, models, theory *Advanced Series on Statistical Science & Applied Probability* vol **3** *World Scientific Publishing Co Pte Ltd* Kruzhilin N (translator) ISBN 981-02-3605-0 Singapore pp 1 – 834.

4601. Shiryaev A N, Spokoiny V G 2000 Statistical experiments and decisions: Asymptotic theory *World Scientific Publishing Co Pte Ltd* ISBN 9810241011 Singapore pp 1 – 283.

4602. Graversen S E, Peskir G, Shiryaev A N 2001 Stopping Brownian motion without anticipation as close as possible to its ultimate maximum *Theory of Probability and its Applications* **45** pp 125 – 136 MR1810977 http://www.ams.org/mathscinetgetitem?mr=1810977 .

4603. Kallsen J, Shiryaev A N 2001 Time change representation of stochastic integrals *Theory of Probability and its Applications* **46** pp 579 - 585 MR1978671 http://www.ams.org/mathscinet-getitem?mr=1978671 .

4604. Kallsen J, Shiryaev A N 2002 The cumulant process and Esscher's change of measure *Finance Stoch* **6** pp 397 – 428 MR1932378 http://www.ams.org/mathscinetgetitem?mr=1932378 .

4605. Shiryaev A N 2002 Quickest detection problems in the technical analysis of the financial data *Proceedings Mathematical Finance Bachelier Congress* Paris France (2000) *Springer* Germany pp 487 – 521 MR1960576 http://www.ams.org/mathscinet-getitem?mr=1960576 .

4606. Jacod J, Shiryaev A N 2003 Limit theorems for stochastic processes *2nd edition* Grundlehren der Mathematischen Wissenschaften [Fundamental Principles of Mathematical Sciences] **288** *Springer* Berlin Germany *MR1943877* http://www.ams.org/mathscinetgetitem?mr=1943877 .

**4607.** Shiryaev A N 2004 Kolmogorov and modern mathematics *International Conference at Mathematical Institute named after V A Steklov June 16-21, 2003* Russian Academy of Sciences Moscow Russian Federation ISBN 5-98419-003-6 pp 1 – 195.

**4608.** Shiryaev A N, Grossinho M R, Oliveira P E, Esquível M L (editors) 2006 Stochastic finance *Springer* Germany ISBN-10:0-387-28262-9 pp 1 – 364.

**4609.** Peskir G, Shiryaev A N 2006 Optimal stopping and free-boundary problems *Lectures in Mathematics* ETH Zürich *Birkhäuser* Switzerland MR2256030 http://www.ams.org/mathscinet-getitem?mr=2256030 .

**4610.** Feinberg E A, Shiryaev A N 2006 Quickest detection of drift change for Brownian motion in generalized Bayesian and mini-max settings *Statistics & Decisions* **24** (4) pp 445 – 470.

**4611.** Kabanov Yu, Lipster R, Stoyanov J 2006 The Shiryaev festschrift: From stochastic calculus to mathematical finance *Springer* Germany pp 1 – 668.

**4612.** du Toit J, Peskir G, Shiryaev A N 2007 Predicting the last zero of Brownian motion with drift *Cornell University* NY USA pp 1- 17 http://arxiv.org/abs/0712.3415v1 .

**4613.** Shiryaev A N 2008a Generalized Bayesian nonlinear quickest detection problems: on Markov family of sucient statistics *Mathematical Control Theory and Finance Proceedings of the Workshop of April 10–14 2007* Lisbon Portugal Sarychev A et al (editors) *Springer* Berlin Germany pp 377 – 386.

**4614.** Eberlein E, Papapantoleon A, Shiryaev A N 2008 On the duality principle in option pricing: Semimartingale setting *Finance Stoch* **12** pp 265 – 292 http://www.ams.org/mathscinet-getitem?mr=2390191 .

**4615.** Shiryaev A N, Novikov A A 2009 On a stochastic version of the trading rule "Buy and hold" *Statistics & Decisions* **26** (4) pp 289 – 302.

**4616.** Eberlein E, Papapantoleon A, Shiryaev A N 2009 Esscher transform and the duality principle for multidimensional semimartingales *The Annals of Applied Probability* vol **19** no 5 pp 1944 – 1971 http://dx.doi.org/10.1214/09-AAP600 http://arxiv.org/abs/0809.0301v5 .

**4617.** Shiryaev A N, Zryumov P Y 2009 On the linear and nonlinear generalized Bayesian disorder problem (discrete time case) optimality and risk – modern trends in mathematical finance *The Kabanov Festschrift* Delbaen F et al (editors) *Springer* Berlin Germany pp 227 – 235.

**4618.** Gapeev P V, Shiryaev A N 2010 Bayesian quickest detection problems for some diffusion processes *Cornell University* NY USA pp 1 – 25 http://arxiv.org/abs/1010.3430v2 .

**4619.** Karatzas I, Shiryaev A N, Shkolnikov M 2011 The one-sided Tanaka equation with drift *Cornell University NY USA* http://arxiv.org/abs/1108.4069v1 .

**4620.** Shiryaev A N, Zhitlukhin M V 2012 Optimal stopping problems for a Brownian motion with a disorder on a finite interval *Cornell University NY USA* pp 1 – 10 http://arxiv.org/abs/1212.3709v1 .

4621. Zhitlukhin M V, Shiryaev A N 2012 Bayesian disorder detection problems on filtered probability spaces *Theory of Probability and Its Applications* **57** (3) pp 453 – 470.

4622. Feinberg E A, Mandava M, Shiryaev A N 2013 On solutions of Kolmogorov's equations for nonhomogeneous jump Markov processes *Cornell University* NY USA pp 1 – 15 http://arxiv.org/abs/1301.6998v3 .

4623. Fama E F 1965 The behavior of stock market prices *Journal of Business* **38** pp 34 – 105.

4624. Fama E F, Blume M 1966 Filter rules and stock market trading profits *Journal of Business* **39** pp 226 – 241.

4625. Fama E F 1970 Efficient capital markets: A review of theory and empirical work *Journal of Finance* **25** (2) pp 383 – 417.

4626. Fama E 1984 Forward and spot exchange rates *Journal of Monetary Economics* **14** pp 319 – 338.

4627. Fama E, French K 1988 Permanent and temporary components of stock prices *Journal of Political Economy* **96** pp 246 – 273.

4628. Fama E F, French K R 1996 Multifactor explanations of asset pricing anomalies *Journal of Finance* **51** (1) pp 55 – 84.

4629. Fama E F 1998 Market efficiency, long-term returns, and behavioral finance *Journal of Financial Economics* **49** pp 283 – 306.

4630. Fama E, Hansen L P, Shiller R 2013 Lectures: 2013 Nobel prize in economic sciences http://www.youtube.com/watch?v=WzxZGvrpFu4 , www.nobelprize.org .

4631. Demsetz H 1968 The cost of transacting *Quarterly Journal of Economics* **82** pp 33 – 53.

4632. Radner R 1968 Competitive equilibrium under uncertainty *Econometrica* **36** pp 31 – 58.

4633. Bates J M, Granger C W J 1969 The combination of forecasts *Operations Research Quarterly* **20** pp 451 – 468.

4634. Akerlof G A 1970 The market for lemons: Qualitative uncertainty and the market mechanism *Quarterly Journal of Economics* **84** (3) pp 488 – 500.

4635. Akerlof G A (29 August) 2014 Writing the "The Market for 'Lemons'": A Personal Interpretive Essay". Nobelprize.org. Nobel Media AB 2014. Web. 29 Aug 2014. http://www.nobelprize.org/nobel_prizes/economic-sciences/laureates/2001/akerlof-article.html?utm_source=facebook&utm_medium=social&utm_campaign=facebook_page

4636. Arrow K 1970 Essays in the theory of risk bearing *Markham* Chicago USA.

4637. Black F 1971 Toward a fully automated exchange *Financial Analysts Journal* **27** pp 29 – 35 and pp 86 – 87.

4638. Black F, Scholes M 1973 The pricing of options and corporate liability *Journal of Political Economics* **81** pp 637 – 654.

4639. Black F 1986 Noise *Journal of Finance* **41** (3) pp 529 – 543.

4640. Merton R C 1973 Theory of rational option pricing *Bell Journal of Economics and Management Science* **4** pp 141 – 183.

**4641.** Newbold P, Granger C W J 1974 Experience with forecasting univariate time series and the combination of forecasts *Journal of the Royal Statistical Society* **137** pp 131 – 165.

**4642.** Fleming J M 1975 Floating exchange rates, asymmetrical intervention and the management of international liquidity *IMF* Washington USA http://www.imf.org .

**4643.** Shapiro A C 1975 Exchange rate changes, inflation, and the value of the multinational corporation

**4644.** Dooley M P, Shafer J R 1976 Analysis of short-run exchange rate behaviour: March 1973 to September 1975 *Federal Reserve Board International Finance Discussion Paper no 123* Federal Reserve Board USA.

**4645.** Dornbusch R 1976 Expectations and exchange rate dynamics *Journal of Political Economy* **84** (6) pp 1161 – 1176.

**4646.** Dornbusch R 1987 Exchange rates and prices *American Economic Review* **77** (1) pp 93 – 106.

**4647.** Frankel J A 1976 A monetary approach to the exchange rate: Doctrinal aspects and empirical evidence *Scandinavian Journal of Economics* **78** pp 200 – 224.

**4648.** Frankel J A 1979 On the mark: A theory of floating exchange rates based on real interest differentials *American Economic Review* **69** pp 610 – 622.

**4649.** Frankel J A 1982a In search of the exchange risk premium: A six currency test assuming mean-variance optimization *Journal of International Money and Finance* **1** pp 255 – 274.

**4650.** Frankel J A 1982b A test of perfect substitutability in the foreign exchange market *Southern Economic Journal* **49** pp 406 – 416.

**4651.** Frankel J A (editor) 1983 Exchange rate and international macroeconomics *University of Chicago Press* Chicago USA.

**4652.** Frankel J A, Froot K 1987 Using survey data to test standard propositions regarding exchange rate expectations *American Economic Review* **77** (1) pp 133 – 153.

**4653.** Frankel J A, Froot K 1990a Chartists, fundamentalists, and trading in the foreign exchange market *American Economic Review* **80** pp 181 – 185.

**4654.** Frankel J A, Froot K 1990b Chartists, fundamentalists, and the demand for dollars *in* Private behavior and government policy in interdependent economies Courakis A, Taylor M P *Clarendon* Oxford UK.

**4655.** Frankel J A, Froot K 1990c Exchange rate forecasting techniques, survey data, and implications for the foreign exchange market *Working Paper no 3470* National Bureau of Economic Research Cambridge Massachusetts USA.

**4656.** Frankel J A, Goldstein M, Mason P 1991 Characteristics of a successful exchange rate system *IMF* Washington USA http://www.imf.org .

**4657.** Frankel J A 1992 In search of the exchange rate premium: A six-currency test assuming mean-variance optimization *Journal of International Money and Finance* **1**.

**4658.** Frankel J A (editor) 1993 On exchange rates *MIT Press* Cambridge MA USA.

**4659.** Frankel J A, Rose A K 1994 A survey of empirical research on nominal exchange rates *NBER Working Paper no 4865* NBER USA.

4660. Frankel J A, Rose A 1995 Empirical research on nominal exchange rates *in* Handbook of international economics Grossman G, Rogoff K (editors) *Elsevier Science* vol **3** pp 1689 – 1729.

4661. Frankel J A, Galli G, Giovannini A (editors) 1996 Introduction *in* The microstructure of foreign exchange markets *University of Chicago Press* Chicago USA ISBN: 0-226-26000-3 pp 1 – 15 http://www.nber.org/books/fran96-1 , http://www.nber.org/chapters/c11360 .

4662. Frankel J A, Galli G, Giovannini A (editors) 1996 The microstructure of foreign exchange markets *University of Chicago Press* Chicago USA.

4663. Frankel J A, Poonawala J 2004 The forward market in emerging currencies: Less biased than in major currencies *Working Paper* Harvard University USA.

4664. Garman M 1976 Market microstructure *Journal of Financial Economics* **3** pp 257 – 275.

4665. Grossman S 1976 On the efficiency of competitive stock markets when agents have diverse information *Journal of Finance* **31** pp 573 – 585.

4666. Grossman S, Stiglitz J 1980 On the impossibility of informationally efficient markets *American Economic Review* **70** pp 393 – 408.

4667. Grossman S, Miller M 1988 Liquidity and market structure *Journal of Finance* **43** pp 617 – 633.

4668. Kouri P J K 1976 The exchange rate and the balance of payments in the short run and in the long run: A monetary approach *The Scandinavian Journal of Economics* **78** (2) pp 280 – 304.

4669. McKinnon R 1976 Floating exchange rates, 1973-74: The emperor's new clothes *Carnegie-Rochester Conference Series on Public Policy* **3** pp 79 – 114.

4670. Mussa M 1976 The exchange rate, the balance of payments, and monetary and fiscal policy under a regime of controlled floating *Scandinavian Journal of Economics* **78** pp 229 – 248.

4671. Mussa M 1979 Empirical regularities in the behaviour of exchange rates and theories of the foreign exchange market *in* Brunner K, Meltzer A H (editors) Policies for employment, prices and exchange rates *Carnegie-Rochester Conference Series on Public Policy* **11** *North-Holland Publishing Company Elsevier* Amsterdam The Netherlands pp 9 – 57.

4672. Mussa M 1981 The role of official intervention *Group of Thirty* New York NY USA.

4673. Mussa M 1984 The theory of exchange rate determination *in* Exchange rates in theory and practice Bilson J, Marston R (editors) *University of Chicago Press* Chicago USA.

4674. Williamson J 1976 Exchange rate flexibility and reserve use *Scandinavian Journal of Economics* **78** (2) pp 327 – 339.

4675. Branson W 1977 Asset markets and relative prices in exchange rate determination *Sozialwissenschaftliche Annalen* **1** pp 69 – 89.

4676. Branson W, Halttunen H, Masson P 1977 Exchange rates in the short run: The Deutschemark rate *European Economic Review* **10** pp 303 – 324.

4677. Branson W, Henderson D 1985 The specification and influence of asset markets *in* Handbook of international economics vol **2** Jones R, Kenen P (editors) *North-Holland Publishing Company* Amsterdam The Netherlands.

4678. Clark, Logue, Sweeney (editors) 1977 The effects of exchange rate adjustment *Department of the Treasury* Washington DC USA.

4679. Girton L, Henderson D 1977 Central bank operations in foreign and domestic assets under fixed and flexible exchange rates *in* The effects of exchange rate adjustment Clark P, Logue D, Sweeney R (editors) *Department of the Treasury* Washington DC pp 151 – 179.

4680. Cornell W B, Dietrich J K 1978 The efficiency of the market for foreign exchange under floating exchange rates *Review of Economics and Statistics* **60** (1) pp 111 – 120.

4681. Cornell W B 1982 Money supply announcements, interest rates, and foreign exchange *Journal of International Money and Finance* **1** pp 201 – 208.

4682. Stoll H R 1978 The supply of dealer services in securities markets *Journal of Finance* **33** pp 1133 – 1151.

4683. Stoll H R 1985 The stock exchange specialist system: An economic analysis *Monograph Series in Finance and Economics: Monograph 1985-2* New York University NY USA.

4684. Stoll H 1989 Inferring the components of the bid-ask spread: Theory and empirical tests *Journal of Finance* **44** pp 115 – 134.

4685. Stoll H R 1995 The importance of equity trading costs: Evidence from securities firms' revenues *in* Global equity markets: Technological, competitive, and regulatory challenges Schwartz R (editor) *Irwin* Homewood Illinois USA pp 98 – 120.

4686. Huang R, Stoll H 1996 Dealer versus auction markets: A paired comparison of execution costs on NASDAQ and the NYSE *Journal of Financial Economics* **41** pp 313 – 357.

4687. Huang R, Stoll H 1997 The components of the bid-ask spread: A general approach *Review of Financial Studies* **10** pp 995 – 1034.

4688. Stoll H R 1998 Reconsidering the affirmative obligation of market-makers *Financial Analysts Journal* **54** (5) pp 72 – 82.

4689. Stoll H R, Schenzler Ch 2005 Trades outside the quotes: Reporting delay, trading option, or trade size? *Journal of Financial Economics*.

4690. Stoll H R 2006 Electronic trading in stock markets *Journal of Economic Perspectives* **20** (1) pp 153 – 174.

4691. Blanchard O 1979 Speculative bubbles, crashes, and rational expectations *Economics Letters* **14** pp 387 – 389.

4692. Brunner K, Meltzer A H (editors) 1979 Policies for employment, prices and exchange rates *Carnegie-Rochester Conference Series on Public Policy* **11** *North-Holland Publishing Company Elsevier* Amsterdam The Netherlands.

4693. Deardorff A 1979 One way arbitrage and its implications for the foreign exchange markets *Journal of Political Economy* **87** pp 351 - 364.

4694. Goodman S 1979 Foreign exchange rate forecasting techniques: Implications for business and policy *The Journal of Finance* **34** pp 415 – 424.

**4695.** Aliber R (October) 1980 The integration of the offshore and domestic banking system *Journal of Monetary Economics* vol **6** issue 4 pp 509 – 526.

**4696.** Aliber R 2002 The new international money game 6[th] edition *University of Chicago Press* Chicago USA.

**4697.** Allen P, Kenen P 1980 Asset markets, exchange rates, and economic integration *Cambridge University Press* New York USA.

**4698.** Amihud Y, Mendelson H 1980 Dealership markets: Market making with inventory *Journal of Financial Economics* **8** pp 31 – 53.

**4699.** Amihud Y, Ho T, Schwartz R (editors) 1985 Market making and the changing structure of the securities industry *Lexington* Massachusetts USA.

**4700.** Amihud Y 1994a Evidence on exchange rates and valuation of equity shares *in* Exchange rates and corporate finance Amihud Y, Levich R M (editors) *Business One Irwin* Homewood IL USA.

**4701.** Amihud Y 1994b Exchange rates and the valuation of equity shares *in* Exchange rates and corporate performance Amihud Y, Levich R M (editors) *Irwin* New York USA pp 49 – 59.

**4702.** Amihud Y, Levich R M (editors) 1994 Exchange rates and corporate finance *Business One Irwin* Homewood IL USA.

**4703.** Hansen L P, Hodrick R J 1980 Forward exchange rates as optimal predictors of future spot rates: An econometric analysis *Journal of Political Economy* **88** (5) pp 829 – 853.

**4704.** Hellwig M 1980 On the aggregation of information in complete markets *Journal of Economic Theory* 26 pp 279 – 312.

**4705.** Hellwig M 1982 Rational expectations equilibrium with conditioning on past prices: A mean-variance example *Journal of Economic Theory* 26 pp 279 – 312.

**4706.** Krugman P 1980 Vehicle currencies and the structure of international exchange *Journal of Money, Credit, and Banking* **12** pp 503 – 526.

**4707.** Krugman P 1984 The international role of the dollar: Theory and prospect *in* Exchange rate theory and practice Bilson J, Marston R (editors) *University of Chicago Press* Chicago USA pp 261 – 278.

**4708.** Krugman P 1991 Target zones and exchange rate dynamics *Quarterly Journal of Economics* **106** (3) pp 669 – 682.

**4709.** Krugman P, Miller M 1993 Why have a target zone? *Carnegie-Rochester Conference Series on Public Policy* **38** pp 279 – 314.

**4710.** Krugman P 1999 The return of depression economics *W W Norton & Company* New York USA.

**4711.** Callier P 1981 One way arbitrage, foreign exchange and securities markets: A note *Journal of Finance* **36** pp 1177 – 1186.

**4712.** Cohen K, Maier S, Schwartz R, Whitcomb D 1981 Transaction costs, order placement strategy , and existence of the bid - ask spread *Journal of Political Economy* **89** (2) pp 287 – 305.

**4713.** Cox J C, Ingersoll Jr J E, Ross S A 1981 The relation between forward and futures prices *Journal of Financial Economics* **9** pp 321 – 346.

*4714.* Diamond D, Verrecchia R 1981 Information aggregation in a noisy rational expectations economy *Journal of Financial Economics* **9** pp 221 – 235.

*4715.* Diamond D 1982 Aggregate demand management in search equilibrium *Journal of Political Economy* **90** pp 881 – 894.

*4716.* Fieleke N (February) 1981 Foreign-currency positioning by US firms: Some new evidence *Review of Economics and Statistics* **63** no 1 pp 35 – 43.

*4717.* Ho Th, Stoll H 1981 Optimal dealer pricing under transaction and return uncertainty *Journal of Financial Economics* **9** (1) pp 47 – 73.

*4718.* Ho Th, Stoll H 1983 The dynamics of dealer markets under competition *Journal of Finance* **38** pp 1053 – 1074.

*4719.* Loosignian A M 1981 Foreign exchange futures *Dow Jones - Irwin* Homewood IL USA.

*4720.* Mussa M 1981 The role of official intervention *Group of Thirty* New York NY USA.

*4721.* Stigum M 1981 Money market calculations: Yields, break - evens, and arbitrage *Dow Jones - Irwin* Homewood IL USA.

*4722.* Stigum M 1990 The money market *Dow Jones - Irwin* Homewood IL USA.

*4723.* Dooley M, Isard P 1982 A portfolio balance rational expectations model of the Dollar-Mark exchange rate *Journal of International Economics* **12** pp 257 – 276.

*4724.* Hansen L 1982 Large sample properties of generalized method of moments estimators *Econometrica* **50** pp 1029 – 1054.

*4725.* Hodder J E 1982 Exposure to foreign exchange-rate movements *Journal of International Economics* **13** (11) pp 375 – 386.

*4726.* Milgrom P, Stokey N 1982 Information, trade and common knowledge *Journal of Economic Theory* **26** pp 17 – 27.

*4727.* Taylor D 1982 Official intervention in the foreign exchange market, or, bet against the central bank *Journal of Political Economy* **90** (2) pp 356 – 368.

*4728.* Bigman D, Taya T (editors) 1983 Exchange rate and trade instability *Ballinger* Cambridge Massachusetts USA.

*4729.* Copeland T E, Galai D 1983 Information effects on the bid-ask spread *The Journal of Finance* **38** pp 1457 – 1469.

*4730.* Dooley M P, Shafer J R 1983 Analysis of short-run exchange rate behaviour: March 1973 to November 1981 *in* Exchange rate and trade instability Bigman D, Taya T (editors) *Ballinger* Cambridge Massachusetts USA pp 43 - 69.

*4731.* Edwards S 1983 The demand for international reserves and exchange rate adjustments: The case of LDCs, 1964–1972 *Economica* **50** pp 269 – 280.

*4732.* French K R 1983 A comparison of futures and forward prices *Journal of Financial Economics* **12** pp 311 – 342.

*4733.* Garman M B, Kohlhagen S W 1983 Foreign currency option values *Journal of International Money and Finance* **2** pp 231 – 237.

*4734.* Meese R A, Rogoff K 1983a Empirical exchange rate models of the seventies: Do they fit out of sample? *Journal of International Economics* **14** pp 3 – 24.

4735. Meese R A, Rogoff K 1983b The out-of-sample failure of empirical exchange rate models *in* Exchange rate and international macroeconomics Frankel J (editor) *University of Chicago Press* Chicago USA.

4736. Rogoff K 1984 On the effects of sterilized intervention: An analysis of weekly data *Journal of Monetary Economics* **14** pp 133 – 150.

4737. Rogoff K 1985 Can exchange rate predictability be achieved without monetary convergence? Evidence from the EMS *European Economic Review* **28** pp 93 – 115.

4738. Meese R A 1986 Testing for bubbles in exchange markets *Journal of Political Economy* **94** pp 345 – 373.

4739. Meese R A, Rogoff K 1988 Was it real? The exchange rate-interest differential relation of the modern floating-rate period *Journal of Finance* **43** pp 933 - 948.

4740. Meese R A 1990 Currency fluctuations in the post-Bretton Woods era *Journal of Economic Perspectives* **4** pp 117 – 134.

4741. Obstfeld M, Rogoff K 1995 Exchange rate dynamics redux *Journal of Political Economy* **103** pp 624 – 660.

4742. Rogoff K 1996 The purchasing power parity puzzle *Journal of Economic Literature* **34** pp 647 – 668.

4743. Obstfeld M, Rogoff K (August) 1998 Risk and exchange rates *NBER Working Paper 6694 NBER USA in* Helpman E, Sadka E (editors) Contemporary economic policy: Essays in honor of Assaf Razin *Cambridge University Press* Cambridge U.K.

4744. Robinson P 1983 Nonparametric estimators for time series *Journal of Time Series Analysis* **4** pp 185 – 207.

4745. Adler M, Dumas B 1984 Exposure to currency risk: Definition and measurement *Financial Management* **13** pp 41 – 50.

4746. Backus D 1984 Empirical models of the exchange rate: Separating the wheat from the chaff *Canadian Journal of Economics* **17** pp 826 – 846.

4747. Bilson J, Marston R (editors) 1984 Exchange rate theory and practice *University of Chicago Press* Chicago USA.

4748. Booth L D 1984 Bid-ask spreads in the market for forward exchange *Journal of International Money and Finance* **3** (2) pp 209 – 222.

4749. Engel Ch M, Frankel J A 1984a Why interest rates react to money announcements: An answer from the foreign exchange market *Journal of Monetary Economics* **13** pp 31 – 39.

4750. Engel Ch M, Frankel J A 1984b Do asset demand functions optimize over the mean and variance of the real returns? A six-currency test *Journal of International Economics* **17**.

4751. Engel Ch M, Hamilton J D 1990 Long swings in the dollar: Are they in the data and do markets know it? *American Economic Review* **80** pp 689 – 713.

4752. Engel Ch M 1992 Can the Markov switching model forecast exchange rates? *NBER Working Paper no 4210* NBER USA.

4753. Engel Ch M 1995 The forward discount anomaly and the risk premium: A survey of recent evidence *Technical Report 5312* National Bureau of Economic Research USA.

4754. Engel Ch M 1996 The forward discount anomaly and the risk premium: A survey of recent evidence *Journal of Empirical Finance* **3** (2) pp 123 – 191.

4755. Engel Ch M 1999 On the foreign exchange risk premium in sticky-price general equilibrium models *in* International finance and financial crises: Essays in honor of Robert P. Flood Isard P, Razin A, Rose A (editors) *IMF* and *Kluwer* The Netherlands.

4756. Devereux M, Engel Ch M 1999 The optimal choice of exchange-rate regime: Price setting rules and internationalized production *NBER Working Paper 6992* NBER USA.

4757. Devereux M B, Engel Ch M 2002 Exchange rate pass-through, exchange rate volatility, and exchange rate disconnect *Journal of Monetary Economics* **49** pp 913 – 940.

4758. Devereux M B, Shi S 2005 Vehicle currency *Working Paper* University of British Columbia Vancouver Canada http://www.econ.ubc.ca/devereux/vc2.pdf .

4759. Engel Ch M, West K (May) 2004a Accounting for exchange rate variability in present value models when the discount factor is near one *American Economic Review* **94** pp 118 – 125.

4760. Engel Ch M, West K (August) 2004b, 2005 Exchange rates and fundamentals *Working Paper 10723* NBER USA*, Journal of Political Economy* **113** pp 485 – 517.

4761. Engel Ch M, West K D 2006 Taylor rules and the Deutschmark - Dollar real exchange rate *Journal of Money, Credit, and Banking* **38** (5) pp 1175 – 1194.

4762. Engel Ch M, Mark N, West K D 2007 Exchange rate models are not as bad as you think *NBER Working Paper* NBER USA.

4763. Garner C K, Shapiro A C 1984 A practical method of assessing foreign exchange risk *Midland Corporate Finance Journal* pp 6 – 17.

4764. Loopesko B 1984 Relationships among exchange rates, intervention, and interest rates: An empirical investigation *Journal of International Money and Finance* **3** pp 257 – 277.

4765. Roll R 1984 A simple implicit measure of the effective bid - ask spread in an efficient market *The Journal of Finance* **39** pp 1127 – 1139.

4766. French K, Roll R 1986 Stock return variances: The arrival of information and the reaction of traders *Journal of Financial Economics* **17** pp 5 – 26.

4767. Roll R 1988 $R^2$ *Journal of Finance* **43** pp 541 – 566.

4768. Urich T, Watchel P 1984 The effects of inflation and money supply announcements on interest rates *Journal of Finance* **39** pp 1177 – 1188.

4769. White H, Domowitz I 1984 Nonlinear regression with dependent observations *Econometrica* **52** (1) pp 143 – 161.

4770. Bahmani-Oskooee M, Das S 1985 Transaction costs and the interest parity theorem *Journal of Political Economy* **93** pp 793 – 799.

4771. Cohen K, Conroy R, Maier S 1985 Order flow and the quality of the market *in* Market making and the changing structure of the securities industry Amihud Y, Ho T, Schwartz R (editors) *Lexington* Massachusetts USA.

4772. Glosten L R, Milgrom P (March) 1985 Bid, ask, and transaction prices in a specialist market with heterogeneously informed agents *Journal of Financial Economics* **14** pp 71 – 100.

4773. Glosten L R, Harris L 1988 Estimating the components of the bid - ask spread *Journal of Financial Economics* **21** pp 123 – 142.

4774. Glosten L R 1989 Insider trading, liquidity, and the role of the monopolist specialist *Journal of Business* **62** (2) pp 211 – 235.

4775. Glosten L R 1994 Is the electronic open limit order book inevitable? *Journal of Finance* **49** pp 1127 – 1162.

4776. Hakkio C, Pearce D 1985 The reaction of exchange rates to economic news *Economic Inquiry* **23** pp 621 – 635.

4777. Hardouvelis G A 1985 Exchange rates, interest rates, and money-stock announcements: A Theoretical exposition *Journal of International Money and Finance* **4** pp 443 – 454.

4778. Jones R, Kenen P (editors) 1985 Handbook of international economics *North-Holland Publishing Company* Amsterdam The Netherlands.

4779. Kearney C, Macdonald R 1985 Intervention and sterilization under floating exchange rates: The UK 1973-1983 *European Economic Review* **30**.

4780. Kyle A 1985 Continuous auctions and insider trading *Econometrica* **53** pp 1315 – 1335.

4781. Kyle A 1989 Informed speculation with imperfect competition *Review of Economic Studies* **56** pp 317 – 356.

4782. Kyle A, Xiong W 2001 Contagion as a wealth effect *Typescript* Duke University North Carolina USA.

4783. Levich R M 1985 Empirical studies of exchange rates: Price behaviour, rate determination and market efficiency *in* Handbook of international economics Jones R W, Kenen P B (editors) vol **2** *North-Holland Publishing Company* Amsterdam The Netherlands.

4784. McInish T H, Wood R A 1985 An analysis of transactions data for the Toronto Stock Exchange *The Journal of Banking and Finance* **14** pp 441 – 458.

4785. Dominguez K M 1986 Are foreign exchange forecasts rational? New evidence from survey data *Economic Letters* **21** pp 277 – 281.

4786. Dominguez K M 1990 Market responses to coordinated central bank intervention *Carnegie-Rochester Series on Public Policy* **32** pp 121 – 163.

4787. Dominguez K M 1992 Exchange rate efficiency and the behavior of international asset markets *Garland* New York USA.

4788. Dominguez K M 1993 Does central bank intervention increase the volatility of foreign exchange rates? *Technical Report 4532* National Bureau of Economic Research Cambridge MA USA.

4789. Dominguez K M, Frankel J 1993a Does foreign-exchange intervention matter? The portfolio effect *American Economic Review* **83** (5) pp 1356 – 1369.

4790. Dominguez K M, Frankel J 1993b Does foreign-exchange intervention work? *Institute for International Economics* Washington DC USA.

4791. Dominguez K M, Frankel J A 1993c Foreign exchange intervention: An empirical assessment *in* On exchange rates Frankel J A (editor) *MIT Press* Cambridge MA USA.

4792. Dominguez K M 1998 Central bank intervention and exchange rate volatility *Journal of International Money and Finance* **18** pp 161 – 190.

4793. Dominguez K M 2003a The market microstructure of central bank intervention *Journal of International Economics* **59** pp 25 – 45.

**4794.** Dominguez K M 2003b Foreign exchange intervention: Did it work in the 1990s? *in* Dollar overvaluation and the World economy Bergsten C F, Williamson J (editors) *Institute for International Economics* Washington DC USA.

**4795.** Bollerslev T 1986 Generalized autoregressive conditional heteroskedasticity *Journal of Econometrics* **21** pp 307 – 328.

**4796.** Baillie R, Bollerslev T 1989 The daily message in exchange rates: A conditional variance tale *Journal of Business and Economic Statistics* **7** pp 297 – 305.

**4797.** Baillie R, Bollerslev T 1990 Intra-day and inter market volatility in foreign exchange rates *Review of Economic Studies* **58** pp 565 – 585.

**4798.** Bollerslev T 1990 Modeling the coherence in short-run nominal exchange rates: A multivariate generalized ARCH model *Review of Economics and Statistics* **72** pp 498 – 595.

**4799.** Bollerslev T, Chou R Y, Jayaraman N, Kroner K F 1990 ARCH modeling in finance: A review of the theory and empirical evidence *Journal of Econometrics* **52** (1) pp 5 – 60.

**4800.** Bollerslev T, Domowitz I 1991 Price volatility, spread variability and the role of alternative market mechanisms *Review of Futures Markets* **10** pp 78 – 102.

**4801.** Baillie R T, Bollerslev T 1991 Intra - day and inter - market volatility in foreign exchange rates *Review of Economic Studies* **58** pp 565 – 585.

**4802.** Bollerslev T, Domowitz I (September) 1993 Trading patterns and prices in the interbank foreign exchange market *Journal of Finance* **48** (4) pp 1421 – 1443.

**4803.** Bollerslev T, Melvin M 1994 Bid - ask spreads and volatility in the foreign exchange market: An empirical analysis *Journal of International Economics* **36** pp 355 – 372.

**4804.** Andersen T, Bollerslev T 1994 Intraday seasonality and volatility persistence in foreign exchange and equity markets *Working Paper no 186* Department of Finance Northwestern University USA.

**4805.** Bollerslev T, Engle R F, Nelson D B 1995 ARCH models *in* Handbook of econometrics vol **4** *North-Holland Publishing Company* New York USA.

**4806.** Andersen T, Bollerslev T 1998 Deutsche mark-dollar volatility: Intraday activity patterns, macroeconomic announcements, and longer run dependencies *Journal of Finance* **53** pp 219 – 266.

**4807.** Bollerslev T, Cai J, Song F 2000 Intraday periodicity, long-memory volatility, and macroeconomic announcement effects in the US treasury bond market *Journal of Empirical Finance* **7** pp 37 – 55.

**4808.** Andersen T G, Bollerslev T, Diebold F X, Labys P 2000 Exchange rate returns standardized by realized volatility are (nearly) Gaussian *Multinational Finance Journal* **4** pp 159 – 179.

**4809.** Andersen T, Bollerslev T, Diebold F, Vega C (September) 2001, 2003 Micro effects of macro announcements: Real-time price discovery in foreign exchange *Typescript* Northwestern University USA; *American Economic Review* **93** pp 38 – 62.

**4810.** Andersen T, Bollerslev T, Diebold F X, Labys P 2001 The distribution of realized exchange rate volatility *Journal of the American Statistical Association* **96** (453) pp 42 – 55.

4811. Andersen T G, Bollerslev T, Diebold F X, Labys P 2003 Modeling and forecasting realized volatility *Econometrica* **71** pp 579 - 625.

4812. Andersen T G, Bollerslev T, Diebold F X 2007 Roughing it up: Including jump components *in* The measurement, modeling and forecasting of return volatility Review *of Economics and Statistics* **89** pp 701 – 720.

4813. Engle R F 1982 Autoregressive conditional heteroskedasticity with estimates of the variance of United Kingdom inflation *Econometrica* **50** pp 987 – 1007.

4814. Engle R F, Bollerslev T 1986 Modeling the persistence of conditional variance *Econometrics Reviews* **5** pp 1 – 50.

4815. Engle R F, Granger C 1987 Cointegration and error correction: Representation, estimation and testing *Econometrica* **55** pp 251 – 276.

4816. Engle R F, Rodriguez A P 1989 Tests of international CAPM with time varying co-variances *Journal Of Applied Econometrics* **4**.

4817. Engle R F, Ito T, Lin Wen-Ling 1990 Meteor showers or heat waves? Heteroskedastic intra-daily volatility in the foreign exchange market *Econometrica* **58** pp 525 – 542.

4818. Engle R F, Russell J R 1995 Forecasting transaction rates: The autoregressive conditional duration model *Proceedings of the First International Conference on High Frequency Data in Finance* (*HFDF-1*) vol **4** *Research Institute for Applied Economics Olsen & Associates* Zürich Switzerland.

4819. Engle R F, Gallo G M 2006 A multiple indicators model for volatility using intra-daily data *Journal of Econometrics* **131** pp 3 – 27.

4820. Evans G 1986 A test for speculative bubbles in the sterling-dollar exchange rate *American Economic Review* **76** pp 621 – 636.

4821. Flood E Jr, Lessard D R 1986 On the measurement of operating exposure to exchange rates: A conceptual approach *Financial Management* **15** pp 25 – 37.

4822. Grammatikos T, Saunders A, Swary I 1986 Returns and risks of US Bank foreign currency activities *Journal of Finance* **41** (3) pp 671 – 682.

4823. Harris L 1986 A transaction data survey of weekly and intraday patterns in stock returns *Journal of Financial Economics* **16** pp 99 – 117.

4824. Harris L 1990 Statistical properties of the roll serial covariance bid/ask spread estimator *Journal of Finance* **45** pp 579 – 590.

4825. Hart O D, Kreps D M 1986 Price destabilizing speculation *Journal of Political Economy* **94** pp 927 – 952.

4826. Lyons R K 1986 Tests of the foreign exchange risk premium using the expected second moments implied by option pricing *International Finance Discussion Papers 290* Board of Governors of the Federal Reserve System USA.

4827. Lyons R K (March) 1988 Tests of the foreign exchange risk premium using the expected second moments implied by option pricing *Journal of International Money and Finance Elsevier* **7** (1) pp 91 – 108.

4828. Lyons R K (November) 1990 Whence exchange rate overshooting: Money stock or flow? *Journal of International Economics Elsevier* **29** (3 - 4) pp 369 – 384.

**4829.** Lyons R K 1991 Private beliefs and information externalities in the foreign exchange market *NBER Working Papers 3889* National Bureau of Economic Research Inc.

**4830.** Lyons R K (January) 1992 Floating exchange rates in Peru, 1950-1954 *Journal of Development Economics Elsevier* **38** (1) pp 99 – 118.

**4831.** Lyons R 1993a Information intermediation in the microstructure of the foreign exchange market *NBER Working Paper #3889* Berkeley Business School USA.

**4832.** Lyons R 1993b Tests of microstructural hypothesis in the foreign exchange market *NBER Working Paper #4471* Berkeley Business School USA.

**4833.** Lyons R K 1993c Optimal transparency in a dealership market with an application to foreign exchange *NBER Working Papers 4467* National Bureau of Economic Research Inc.

**4834.** Baldwin R E, Lyons R K (January) 1994 Exchange rate hysteresis? Large versus small policy misalignments *European Economic Review Elsevier* **38** (1) pp 1 – 22.

**4835.** Lyons R 1994 Foreign exchange volume: Sound and fury signifying nothing? *Berkeley Business School* USA.

**4836.** Lyons R K 1995 Tests of microstructural hypotheses in the foreign exchange market *Journal of Financial Economics Elsevier* **39** (2 - 3) pp 321 – 351.

**4837.** Lyons R K, Rose A K (September) 1995 Explaining forward exchange bias . . . intraday *Journal of Finance American Finance Association* **50** (4) pp 1321 – 1329.

**4838.** Lyons R K 1996a Foreign exchange volume: Sound and fury signifying nothing? *in* The microstructure of foreign exchange markets *National Bureau of Economic Research Inc* pp 183 – 208.

**4839.** Lyons R K (July) 1996b Optimal transparency in a dealer market with an application to foreign exchange *Journal of Financial Intermediation Elsevier* **5** (3) pp 225 – 254.

**4840.** Lyons R K 1997a Explaining trading volume in foreign exchange: Lessons from Tokyo *FRBSF Economic Letter* Federal Reserve Bank of San Francisco issue December 26.

**4841.** Lyons R K (May) 1997b A simultaneous trade model of the foreign exchange hot potato *Journal of International Economics Elsevier* **42** (3 - 4) pp 275 – 298.

**4842.** Lyons R K 1997c Profits and position control: A week of FX dealing *Research Program in Finance Working Papers RPF-273* University of California at Berkeley.

**4843.** Lyons R K (February) 1998a Profits and position control: A week of FX dealing *Journal of International Money and Finance Elsevier* **17** (1) pp 97 – 115.

**4844.** Lyons R K (December) 1998b Introduction to the international market microstructure issue *Journal of International Financial Markets, Institutions and Money Elsevier* **8** (3 - 4) pp 219 – 223.

**4845.** Lyons R K (Summer) 2001 New perspective on FX markets: Order-flow analysis *International Finance Wiley Blackwell* **4** (2) pp 303 – 320.

**4846.** Fan M, Lyons R (July) 2001 Customer-dealer trading in the foreign exchange market *Typescript* UC Berkeley USA.

**4847.** Killeen W, Lyons R, Moore M (September) 2001 Fixed versus flexible: Lessons from EMS order flow *NBER Working Paper 8491* NBER USA.

**4848.** Killeen W, Hau H, Moore M 2001 The euro as an international currency: Explaining puzzling first evidence from the foreign exchange markets *Journal of International Money and Finance*.

**4849.** Lyons R K (October) 2002 Theoretical perspective on euro liquidity *Economic Policy* CEPR & CES & MSH **17** (35) pp 571 – 597.

**4850.** Lyons R K 2002 Foreign exchange: Macro puzzles, micro tools *Economic Review* Federal Reserve Bank of San Francisco pp 51 – 69.

**4851.** Lyons R K 2003 Explaining and forecasting exchange rates with order flows *Economie Internationale* CEPII research center issue 96 pp 107 – 127.

**4852.** Fan M, Lyons R K, 2003, Customer trades and extreme events in foreign exchange *in* Central banking, monetary theory and practice: Essays in honor of Charles Goodhart Mizen P (editor) vol **2** pp 160 – 179 *Edward Elgar* Cheltenham UK.

**4853.** Killeen W P, Lyons R K, Moore M J (June) 2006 Fixed versus flexible: Lessons from EMS order flow *Journal of International Money and Finance Elsevier* **25** (4) pp 551 – 579.

**4854.** Lyons R K (January) 2006 The microstructure approach to exchange rates *MIT Press* edition 1 vol **1** ISBN 026262205x Cambridge MA USA.

**4855.** O'Hara M, Oldfield G 1986 The microeconomics of market making *Journal of Financial and Quantitative Analysis* **21** pp 361 - 376.

**4856.** Burdett K, O'Hara M 1987 Building blocks: An introduction to block trading *Journal of Banking and Finance* **13** pp 397 – 419.

**4857.** O'Hara M 1995, 1998 Market microstructure theory *Blackwell Business* Cambridge MA USA, *John Wiley and Sons Inc* USA.

**4858.** Shleifer A 1986 Do demand for stock slope down? *Journal of Finance* **41** (3) pp 579 – 590.

**4859.** Shleifer A, Summers L 1990 The noise trader approach to finance *Journal of Economic Perspectives* **4** (2) pp 19 – 33.

**4860.** Sweeney R 1986 Beating the foreign exchange market *The Journal of Finance* **41** pp 163 – 182.

**4861.** DeLong B, Shleifer A, Summers L, Waldmann R 1990 Positive feedback investment strategies, and destabilizing rational speculation *Journal of Finance* **45** pp 379 – 396.

**4862.** Bilson J F, Hsieh D 1987 The profitability of currency speculation *International Journal of Forecasting* **3** pp 115 – 130.

**4863.** Glassman D 1987 Exchange rate risk and transactions costs: Evidence from bid-ask spreads *Journal of International Money and Finance* **6** (4) pp 479 – 490.

**4864.** Gerlach S 1987 Exchange rates: A review essay *Journal of Monetary Economics* **19** pp 137 –142.

**4865.** Hasbrouck J, Ho T S H 1987 Order arrival, quote behaviour and the return - generating process *The Journal of Finance* **42** (4) pp 1035 – 1048.

**4866.** Hasbrouck J 1988 Trades, quotes, inventories, and information *Journal of Financial Economics* **22** pp 229 – 252.

**4867.** Hasbrouck J 1991 Measuring the information content of stock trades *Journal of Finance* **46** pp 179 – 207.

*4868.* Hasbrouck J, Sofianos G 1993 The trades of market makers: An empirical analysis of NYSE specialists *Journal of Finance* **48** pp 1565 – 1593.

*4869.* Hasbrouck J, Seppi D 2001 Common factors in prices, order flows, and liquidity *Journal of Financial Economics* **59** pp 383 – 411.

*4870.* Hodrick R 1987 The empirical evidence on the efficiency of forward and futures foreign exchange markets *in* Fundamentals of pure and applied economics vol **24** *Harwood Academic Publishers* New York USA.

*4871.* Ito T, Roley V 1987 News from the US and Japan: Which moves the Yen/Dollar exchange rate? *Journal of Monetary Economics* **19** pp 255 – 277.

*4872.* Ito T, Roley V V 1990 Intraday Yen/Dollar exchange rate movements: News or noise? *Journal of International Financial Markets, Institutions and Money* vol **1** no 1.

*4873.* Canova F, Ito T 1991 The time series properties of the risk premium in the Yen/Dollar exchange market *Journal of Applied Econometrics* **6** pp 125 – 142.

*4874.* Ito T, Engle R F, Lin W-L 1992 Where does the meteor shower come from? The role of stochastic policy coordination *Journal of International Economics* **32** pp 221 – 240.

*4875.* Ito T, Lin W 1992 Lunch break and intraday volatility of stock returns: An hourly data analysis of Tokyo and New York stock markets *Economics Letters* **39** pp 85 - 90.

*4876.* Ito T, Isard P, Symansky St, Bayoumi T 1996 Exchange rate movements and their impact on trade and investment in the APEC region *IMF Occasional Paper no 145* International Monetary Fund.

*4877.* Ito T, Lyons R K, Melvin M T 1998 Is there private information in the FX market? The Tokyo experiment *Journal of Finance American Finance Association* **53** (3) pp 1111 – 1130.

*4878.* Ito T (April) 2002 Is foreign exchange intervention effective? The Japanese experience in the 1990s *NBER Working Paper no 8914* NBER MA USA.

*4879.* Ito T 2005a The exchange rate in the Japanese economy: The past, puzzles, and prospects *Japanese Economic Review* **56** no 1 pp 1 – 38.

*4880.* Ito T 2005b The Yen and the Japanese economy: 2004 *in* Dollar adjustment: How far? Against what? Bergsten F, Williamson J (editors) *Institute of International Economics* Washington DC pp 171 – 196.

*4881.* Ito T, Hashimoto Y 2006 Intraday seasonality in activities of the foreign exchange markets: Evidence from the electronic broking system *Journal of Japanese and International Economics* **20** pp 637 – 664.

*4882.* Mendelson H 1987 Consolidation, fragmentation, and market performance *Journal of Financial and Quantitative Analysis* **22** pp 189 – 208.

*4883.* Newey W, West K 1987 A simple positive semi-definite, heteroskedasticity and autocorrelation consistent covariance matrix *Econometrica* **55** pp 703 – 708.

*4884.* Rubinstein A, Wolinsky A 1987 Middlemen *Quarterly Journal of Economics* **102** pp 581 – 593.

*4885.* Taylor M P 1987 Covered interest parity: A high-frequency, high-quality data study *Economica* **54** (216) pp 429 – 438.

*4886.* Taylor M P 1989 Covered interest arbitrage and market turbulence *Economic Journal* **99** pp 376 – 391.

4887. Allen H, Taylor M P 1989 Chartists, noise and fundamentals: A study of the London foreign exchange market *Working Paper no 341* Centre for Economic Policy Research London UK.

4888. Taylor M P, Allen H 1992 The use of technical analysis in the foreign exchange market *Journal of International Money and Finance* **11** (3) pp 304 – 314.

4889. Taylor M P 1995 The economics of exchange rates *Journal of Economic Literature* **33** pp 13 – 47.

4890. Sarno L, Taylor M P 2000 Official intervention in the foreign exchange market *University of Oxford* UK.

4891. Sarno L, Taylor M P 2001a Official intervention in the foreign exchange market: Is it effective and if so how does it work? *Journal of Economic Literature* **39** pp 839 – 868.

4892. Sarno L, Taylor M P 2001b The microstructure of the foreign exchange market. A selective survey of the literature *Princeton Studies in International Economics Series no 89 Princeton University Press* Princeton NJ USA.

4893. Taylor M P 2005 Official foreign exchange intervention as a coordinating signal to the Dollar-Yen market *Pacific Economic Review* **10** pp 73 – 82.

4894. Sager M, Taylor M P 2005 Order flow and exchange rate movements *Typescript* University of Warwick UK.

4895. Reitz S, Taylor M P 2006 The coordination channel of foreign exchange intervention: A non-linear microstructural analysis *Deutsche Bundesbank Discussion Paper no 08/2006* Germany.

4896. Sager M, Taylor M P 2006 Under the microscope: The structure of the foreign exchange market *International Journal of Finance and Economics* **11** pp 81 – 95.

4897. Sager M J, Taylor M P 2008 Commercially available order flow data and exchange rate movements: Caveat emptor *Journal of Money, Credit and Banking* **40** (4) pp 583 – 625.

4898. Schulmeister St 1987 An essay on exchange rate dynamics *Research Unit Labor Market and Employment Discussion Paper no 87-8* Wissenschaftzentrum Berlin fur Sozialforschung Berlin Germany.

4899. Melvin M, Taylor M P 2009 The crisis in the foreign exchange market *Journal of International Money and Finance* **28** (8) pp 1317 – 1330.

4900. Newey W, West K A 1987 Simple, positive semidefinite, heteroskedasticity and autocorrelation consistent covariance matrix *Econometrica* **55** pp 703 – 708.

4901. Wolff Ch C P 1987 Forward foreign exchange rates, expected spot rates, and premia: A signal-extraction approach *Journal of Finance* **42** (2) pp 395 – 406.

4902. Admati A, Pfleiderer P 1988 A theory of intraday patterns: Volume and price variability *The Review of Financial Studies* **1** pp 3 – 40.

4903. Admati A, Pfleiderer P 1989 Divide and conquer: A theory of intraday and day - of - the - week mean effects *The Review of Financial Studies* **2** (2) pp 189 – 223.

4904. Boothe P 1988 Exchange rate risk and the bid-ask spread *Economic Inquiry* ***XXVI*** pp 485 – 492.

4905. Choi J Y, Salandro D, Shastri K 1988 On the estimation of bid - ask spreads: Theory and evidence *Journal of Financial Analysis* **23** pp 219 – 230.

4906. Clinton K 1988 Transactions costs and covered interest arbitrage: Theory and evidence *Journal of Political Economy* **96** pp 358 - 370.

4907. Goodhart Ch A E 1988 The foreign exchange market: A random walk with a dragging anchor *Economica* **55** (220) pp 437 – 460.

4908. Goodhart Ch A E (October) 1989 "News" and the foreign exchange market *Proceedings of Manchester Statistical Society* Manchester UK pp 1 – 79.

4909. Goodhart Ch A E, Demos A (Winter) 1990 Reuters screen images of the foreign exchange market: The Deutschemark / Dollar spot rate *Journal of International Securities Markets* **4** pp 333 – 348.

4910. Goodhart Ch A E, Curcio R (January) 1991 The clustering of bid / ask pries and spreads in the foreign exchange market *Discussion Paper 110* Financial Markets Group London School of Economics and Political Science London UK.

4911. Goodhart Ch A E, Demos A 1991a Reuters screen images of the foreign exchange market: The Yen / Dollar and Sterling / Dollar spot market *Journal of International Securities Markets* **5** Spring pp 35 – 64.

4912. Goodhart Ch A E, Demos A (September 2nd) 1991b The Asian surprise in the forex markets *Financial Times* p 13.

4913. Goodhart Ch A E, Figliuoli L 1991 Every minute counts in financial markets *Journal of International Money and Finance* **10** (1) pp 23 – 52.

4914. Goodhart Ch A E 1992 News effects in a high-frequency model of the Sterling-Dollar exchange rate *Journal of Applied Econometrics* **7**.

4915. Goodhart Ch A E, Hall S, Henry S, Pesaran B 1993 News effects in a high frequency model of the Sterling-Dollar exchange rate *Journal of Applied Econometrics* **8** pp 1 – 13.

4916. Goodhart Ch A E, Hesse T 1993 Central bank forex intervention assessed in continuous time *Journal of International Money and Finance* **12** pp 368 – 389.

4917. Goodhart Ch A E, Ito T, Payne R 1995, 1996 One day in June, 1993: A study of the working of Reuters 2000-2 electronic foreign exchange trading system *National Bureau of Economic Research* Cambridge MA USA pp 1 – 133, *in* The microstructure of foreign exchange markets Frankel J, Galli G, Giovannini A (editors) *University of Chicago Press* Chicago IL USA pp 107 – 179.

4918. Goodhart Ch A E, O'Hara M 1995 High frequency data in financial markets: Issues and applications *Introductory Lecture Proceedings of the First International Conference on High Frequency Data in Finance (HFDF-1) Research Institute for Applied Economics Olsen & Associates* Zürich Switzerland.

4919. Goodhart Ch A E, Payne R G 1996 Microstructural dynamics in a foreign exchange electronic broking system *Journal of International Money and Finance* **15** (6) pp 829 – 852.

4920. Goodhart Ch A E, O'Hara M 1997 High frequency data in financial markets: Issues and applications' *Journal of Empirical Finance* **4** pp 73 – 114.

4921. Goodhart Ch A E, Love R, Payne R, Rime D 2002 Analysis of spreads in the Dollar/Euro and Deutschemark/Dollar foreign exchange markets *Economic Policy* **17** (35) pp 537 – 552.

4922. Hardouvelis G 1988 Economic news, exchange rates, and interest rates *Journal of International Money and Finance* **7** pp 23 – 25.

4923. Lewis K 1988 Testing the portfolio balance model: A multilateral approach *Journal of International Economics* **7** pp 273 – 288.

4924. Lewis K 1995 Puzzles in international financial markets *in* Handbook of international economics Grossman G, Rogoff K (editors) vol **3** *North Holland Publishing Company* Amsterdam The Netherlands.

4925. Baldwin R, Krugman P 1989 Persistent trade effects of large exchange rate shocks *The Quarterly Journal of Economics* **104** (4) pp 635 – 654.

4926. Baxter M, Stockman A 1989 Business cycles and the exchange rate regime: Some international evidence *Journal of Monetary Economics* **23** pp 377 – 400.

4927. Dooley M, Lizondo S, Mathieson D 1989 The currency composition of foreign exchange reserves *IMF Staff Papers* **36** no 2 pp 385 – 434.

4928. Giovannini A 1989 How do fixed exchange rate regimes work? Evidence from the gold standard, Bretton Woods and the EMS *in* Blueprints for exchange rate management Miller M, Eichengreen B, Portes R (editors) *Academic* New York USA.

4929. Golub S 1989 Foreign currency government debt, asset markets, and the balance of payments *Journal of International Money and Finance* **8** pp 285 – 294.

4930. Humpage O 1989 On the effectiveness of exchange market intervention *Federal Reserve Bank of Cleveland* USA.

4931. Leach C, Madhavan 1989 Price experimentation and market structure *Working Paper* Wharton School University of Pennsylvania USA.

4932. Leahy M 1989 The profitability of US intervention *Technical Report 343* Board of Governors Federal Reserve Bank Washington DC USA.

4933. Miller M, Eichengreen B, Portes R (editors) 1989 Blueprints for exchange rate management *Academic* New York USA.

4934. Van Hagen J 1989 Monetary targeting with exchange rate constraints: The Bundesbank in the 1980s *Federal Reserve Bank of St Louis* USA.

4935. Allen H L, Taylor M P 1990 Charts, noise and fundamentals in the London foreign exchange market *Economic Journal* **100** (Supplement) pp 49 – 59.

4936. Allen H L, Karjalainen R 1999 Using genetic algorithms to find technical trading rules *Journal of Financial Economics* **51** pp 245 – 271.

4937. Courakis A, Taylor M P (editors) 1990 Private behavior and government policy in interdependent economies *Clarendon* Oxford UK.

4938. Diebold F X, Nason J 1990 Nonparametric exchange rate prediction? *Journal of International Economics* **28** pp 315 – 332.

4939. Flood R, Hodrick R 1990 On testing for speculative bubbles *Journal of Economic Perspectives* **4** pp 85 – 101.

4940. Flood R, Rose A 1995 Fixing exchange rates: A virtual quest for fundamentals *Journal of Monetary Economics* **36** pp 3 – 37.

4941. Flood R, Taylor M 1996 Exchange rate economics: What's wrong with the conventional macro approach? *in* The microstructure of foreign exchange markets Frankel J, Galli G, Giovannini A (editors) *The University of Chicago Press* Chicago USA pp 261 – 294.

4942. Flood R, Marion N 2001 Holding international reserves in an era of high capital mobility in Brookings Trade Forum 2001 Collins S, Rodrik D (editors) *Brookings Institution Press* Washington DC USA.

4943. Foster D, Viswanathan S 1990 A theory of inter-day variations in volumes variances, and trading costs in securities markets *Review of Financial Studies* **3** pp 593 – 624.

4944. Foster D, Viswanathan S 1993 Variations in trading volume, return volatility, and trading costs: Evidence on recent price formation models *Journal of Finance* **48** 187 – 211.

4945. Holthausen R W, Leftwich R W, Mayers D 1990 Large-block transactions, the speed of response, and temporary and permanent stock-price effects *Journal of Financial Economics* **26** (1) pp 71 – 95.

4946. De Long J B, Shleifer A, Summers L H, Waldmann R J 1990 Noise trader risk in financial markets *Journal of Political Economy* **98** (4) pp 703 – 738.

4947. Domowitz I (June) 1990 The mechanics of automated trade execution systems *Journal of Financial Intermediation* **1** pp 167 – 194.

4948. Domowitz I 1993 A taxonomy of automated trade execution systems *Journal of International Money and Finance* **12** (6) pp 607 – 631.

4949. Domowitz I, Steil B (September) 1999 Automation, trading costs and the structure of the securities trading industry *Brookings-Wharton Papers on Financial Services* pp 33 – 92.

4950. Johansen S, Juselius K 1990 Maximum likelihood estimation and inference on cointegration with applications to the demand for money *Oxford Bulletin of Economics and Statistics* **52** (2) pp 169 – 210.

4951. Johansen S 1991 Estimation and hypothesis testing of cointegration vectors in Gaussian vector autoregressive models *Econometrica* **59** (6) pp 1551 – 1580.

4952. Johansen S 1992 Cointegration in partial systems and the efficiency of single equation analysis *Journal of Econometrics* **52** pp 389 – 402.

4953. Jorion P 1990 The exchange-rate exposure of United-States multinationals *Journal of Business* **63** pp 331 – 345.

4954. Jorion P 1991 The pricing of exchange rate risk in the stock market *Journal of Financial and Quantitative Analysis* **26** (3) pp 363 – 376.

4955. Jorion P 1996 Risk and turnover in the foreign exchange market *in* The microstructure of foreign exchange markets Frankel J A, Galli G, Giovannini A (editors) *University of Chicago Press* Chicago USA pp 19 – 37.

4956. Lo A W, MacKinley A C 1990 An econometric analysis of nonsynchronous trading *Journal of Econometrics* **45** pp 181 – 211.

4957. Melino A, Turnbull S M 1990 Pricing foreign currency options with stochastic volatility *Journal of Econometrics* **45** pp 239 – 265.

4958. Melino A, Turnbull S M 1995 Misspecification and the pricing and hedging of long-term foreign currency options *Journal of International Money and Finance* **14** pp 373 – 393.

4959. Mishkin F 1990 What does the term structure tell us about future inflation? *Journal of Monetary Economics* **25** (1) pp 77 – 95.

4960. Müller U A, Dacorogna M M, Olsen R B, Pictet O, Schwarz M, Morgenegg C 1990 Statistical study of foreign exchange rates, empirical evidence of a price scaling law, and intraday analysis *Journal of Banking and Finance* **14** pp 1189 – 1208.

4961. Müller U A, Dacorogna M M, Dave R D, Pictet O V, Olsen R B, Ward J R 1993 Fractals and intrinsic time - A challenge to econometricians *Technical Report UAM 1993-08-16 Research Institute for Applied Economics Olsen & Associates* Zurich Switzerland.

4962. Müller U A, Dacorogna M M, Dave R D, Olsen R B, Pictet O V, von Weizsäcker J E 1995 Volatilities of different time resolutions - Analyzing the dynamics of market components *Preprint UAM 1995-01-12 Research institute for Applied economics Olsen & Associates* Zurich Switzerland.

4963. Roell A 1990 Dual capacity trading and the quality of the market *Journal of Financial Intermediation* **1** pp 105 – 124.

4964. Seppi D 1990 Equilibrium block trading and asymmetric information *Journal of Finance* **45** pp 73 – 94.

4965. Bali T 1991 An empirical comparison of continuous time models of the short term interest rate *Journal of Futures Markets* **19** (7) pp 777 – 797.

4966. Bhattacharya U, Spiegel M 1991 Insiders, outsiders, and market breakdowns *Review of Financial Studies* **4** pp 255 – 282.

4967. Black S 1991 Transaction costs and vehicle currencies *Journal of International Money and Finance* **10** pp 512 - 527.

4968. Bossaerts P, Hillion P 1991 Market microstructure effects of government intervention in the foreign exchange market *Review of Financial Studies* **4** pp 513 – 541.

4969. Burnham J B 1991 Current structure and recent developments in foreign exchange markets *in* Recent developments in international banking and finance Khonry S J (editor) *Elsevier Science Publishing North Holland Publishing Company* pp 123 – 163.

4970. Campbell J, LaMaster S, Smith V, Van Boening M 1991 Off-floor trading, disintegration, and the bid-ask spread in experimental markets *Journal of Business* **64** pp 495 – 522.

4971. Campbell J, Lo A, MacKinlay A 1997 The econometrics of financial markets *Princeton University Press* USA.

4972. Chinn M D 1991 Some linear and non-linear thoughts on exchange rates *Journal of International Money and Finance* **10** pp 214 – 230.

4973. Chinn M D, Meese R A 1995 Banking on currency forecasts: How predictable is change in money *Journal of International Economics* **38** pp 161 – 178.

4974. Chowdhry B, Nanda V 1991 Multimarket trading and market liquidity *Review of Financial Studies* **4** pp 483 – 511.

4975. Edwards S 1991 Real exchange rates, devaluation, and adjustment – Exchange rate policy in developing countries *MIT Press* USA.

4976. Froot K A, Obstfeld M 1991 Exchange rate dynamics under stochastic regime shifts: A unified approach *Journal of International Economics* **31** pp 203 – 229.

**4977.** Froot K A, Rogoff K 1995 Perspectives on PPP and long-run real exchange rates *in* Handbook of international economics Grossman G, Rogoff K (editors) *Elsevier Science* Amsterdam pp 1647 – 1688.

**4978.** Froot K A, Ramadorai T (August) 2002 Currency returns, institutional investor flows, and exchange rate fundamentals *NBER Working Paper 9101* NBER USA.

**4979.** Froot K A, Donohue 2004 Decomposing the persistence of international equity flows *Finance Research Letters* pp 154 – 170.

**4980.** Froot K A, Ramadorai T 2005 Currency returns, intrinsic value, and institutional-investor flows *Journal of Finance* **60** pp 1535 – 1566.

**4981.** Georg Th, Kaul G, Nimalendran M 1991 Estimation of the bid-ask spread and its components: A new approach *Review of Financial Studies* **4** pp 623 – 656.

**4982.** Grabbe J O 1991 International financial markets 2nd edition *Elsevier Science Publishing Co Inc* New York USA.

**4983.** Harvey C R, Huang R D Volatility in the foreign currency futures market *Review of Financial Studies* **4** pp 543 – 569.

**4984.** Khonry S J (editor) 1991 Recent developments in international banking and finance *Elsevier Science Publishing North Holland Publishing Company* The Netherlands.

**4985.** Kim O, Verrecchia R 1991 Trading volume and price reactions to public announcements *Journal of Accounting Research* **29** pp 302 – 321.

**4986.** Kim O, Verrecchia R 1994 Market liquidity and volume around earnings announcements *Journal of Accounting and Economics* **17** pp 41 – 67.

**4987.** Kim O, Verrecchia R 1997 Pre-announcement and event-period information *Journal of Accounting and Economics* **24** pp 395 – 419.

**4988.** Klein M 1991 Managing the dollar: Has the Plaza agreement mattered? *Journal of Money, Credit, and Banking* **23** pp 742 – 751.

**4989.** Klein M, Rosengren E 1991 Foreign exchange intervention as a signal of monetary policy New England *Economic Review* pp 39 – 50.

**4990.** Lease R, Masulis R, Page J 1991 An investigation of market microstructure impacts on event study returns *The Journal of Finance* **46** pp 1523 – 1536.

**4991.** LeBaron B 1991 Technical trading rules and regime shifts in foreign exchange *Technical Report* University of Wisconsin - Madison WI USA.

**4992.** Lee Ch M C, Ready M J 1991 Inferring trade direction from intraday data *Journal of Finance* **46** pp 733 – 746.

**4993.** Messe R A, Rose A K 1991 An empirical assessment of nonlinearities in models of exchange rate determination *Review of Economic Studies* **58** 603-2-19.

**4994.** Subrahmanyam A 1991 Risk aversion, market liquidity, and price efficiency *Review of Financial Studies* **4** pp 417 – 442.

**4995.** Spiegel M, Subrahmanyam A 1992 Informed speculation and hedging in a non-competitive securities market *Review of Financial Studies* **5** (2).

**4996.** Spiegel M, Subrahmanyam A 1995 On intraday risk premia *Journal of Finance* **50** pp 319 - 339.

**4997.** Williamson J (May) 1991 Advice on the choice of an exchange rate policy *Working Paper no 3* ICEG.

**4998.** Bekaert G, Hodrick R J 1992 Characterizing predictable components in excess returns on equity and foreign exchange markets *The Journal of Finance* **47** pp 467 – 511.

**4999.** Choi J J, Elyasiani E, Kopecky K J 1992 The sensitivity of bank stock returns to market, interest and exchange rate risk *Journal of Banking and Finance* **16** (5) pp 983 – 1005.

**5000.** Choi J J, Elyasiani E 1997 Derivatives exposure and the interest rate and exchange rate risks of US banks *Journal of Financial Services Research* **12** (2/3) pp 267 – 286.

**5001.** Curcio R, Goodhart Ch 1992 When support / resistance levels are broken, can profits be made? Evidence from the foreign exchange market *Discussion Paper no 142* Financial Markets Group London School of Economics London UK.

**5002.** Curcio R, Goodhart Ch, Guillaume D, Payne R 1997 Do technical trading rules generate profits? Conclusions from the intra-day foreign exchange market *International Journal of Finance and Economics* **2** (4) pp 267 – 280.

**5003.** De Grauwe P, Decupere D 1992 Psychological barriers in the foreign exchange market *Journal of International and Comparative Economics* **1** (2) pp 87 – 101.

**5004.** De Grauwe P, Grimaldi M 2006a The exchange rate in a behavioural framework *Princeton University Press* Princeton USA.

**5005.** De Grauwe P, Grimaldi G 2006b Exchange rate puzzles: A tale of switching attractors *European Economic Review* **50** pp 1 – 33.

**5006.** Edison H J 1992, 1993 The effectiveness of central bank intervention: A survey of the post - 1982 literature *Working Paper* Federal Reserve Board of Governors Washington DC USA, *Princeton Studies in International Economics* **18** Princeton University Princeton NJ USA.

**5007.** Edison H J, Liang H 1999 Foreign exchange intervention and the Australian dollar: Has it mattered? *IMF Working Paper WP/03/99*.

**5008.** Edison H J (September) 2003 Are foreign exchange reserves in Asia too high? *in* World economics outlook (September 2003) *International Monetary Fund* Washington DC USA.

**5009.** Flood M D 1994 Market structure and inefficiency in the foreign exchange market *Journal of International Money and Finance* **13** (2) pp 131 – 158.

**5010.** Flood M D, Rose A K 1995 Fixing exchange rates: A virtual quest for fundamentals *Journal of Monetary Economics* **36** pp 3 – 37.

**5011.** Flood M D, Huisman R, Koedijk K, Mahieu R (December) 1996 Price discovery in multiple-dealer financial markets: The effect of pre-trade transparency *Typescript* Concordia University.

**5012.** Flood M D, Huisman R, Koedijk K, Mahieu R 1998 Quote disclosure and price discovery in multiple dealer financial markets *Review of Financial Studies*.

**5013.** Gosh A 1992 Is it signaling? Exchange intervention and the Dollar Deutsche-mark rate *Journal Of International Economics* **32**.

**5014.** Guillaume D M, Dacorogna M M, Dave R R, Müller U A, Olsen R B, Hamon O V, Jacquillat B 1992 Le marche Francais des actions *Presses Universitaires de France* Paris France.

5015. Guillaume D M, Pictet O V, Dacorogna M M 1995 On the intra-day performance of GARCH processes *Proceedings of the First International Conference on High Frequency Data in Finance (HFDF-1)* vol **3** *Research Institute for Applied Economics Olsen & Associates* Zürich Switzerland.

5016. Guillaume D M, Dacorogna M M, Dave R R, Müller U A, Olsen R B, Pictet O V 1997 From the bird's eye to the microscope: A survey of new stylized facts of the intra - daily foreign exchange markets *Finance and Stochastics* **1** (2) pp 95 – 129.

5017. Hansen B 1992 Tests for parameter instability in regressions with I (1) processes *Journal of Business and Economic Statistics* **10** pp 321 – 335.

5018. Holden C, Subrahmanyam A 1992 Long-lived private information and imperfect competition *Journal of Finance* **47** pp 247 – 270.

5019. Neal R 1992 A comparison of transaction costs between competitive market maker and specialist market structures *Journal of Business* **65** pp 317 – 334.

5020. Pesaran M H, Samiei H 1992 Estimating limited - dependent rational expectations models with an application to exchange rate determination in a target zone *Journal of Econometrics* **53** pp 141 – 163.

5021. Rhee S G, Chang R P 1992 Intra–day arbitrage opportunities in foreign exchange and Euro-currency markets *Journal of Finance* **47** pp 363 – 379.

5022. Svensson L E O 1992 An interpretation of recent research on exchange rate target zones *Journal of Economic Perspectives* **6** (1) pp 19 – 44.

5023. Svensson L E O 1993 Assessing target zone credibility *European Economic Review* **37** pp 763 – 802.

5024. Bertola G, Svensson L E O 1993 Stochastic devaluation risk and the empirical fit of target zone models *Review of Economic Studies* **60** pp 689 – 712.

5025. Rose A K, Svensson L E O 1994 European exchange rate credibility before the fall *European Economics Review* **38** pp 1185 – 1216.

5026. Taylor S J 1992 Rewards available to currency futures speculators: Compensation for risk or evidence of inefficient pricing? *Economic Record* **68** pp 105 – 116.

5027. Zhou B 1992a High frequency data and volatility in foreign exchange rates *Manuscript* Department of Finance Sloan School of Management MIT Cambridge MA USA.

5028. Zhou B 1992b Forecasting foreign exchange rates subject to de-volatilization *Working Paper no 3510* Sloan School of Management Massachusetts Institute of Technology Cambridge MA USA.

5029. Zhou B 1997 Currency exchange in a random search model *Review of Economic Studies* **64** pp 289 – 310.

5030. Bank for International Settlements 1993 Survey of foreign exchange market activity *Monetary and Economic Department Bank for International Settlements* Basel Switzerland www.bis.org .

5031. Bank for International Settlements (May) 1999 Central bank survey of foreign exchange market activity in April 1998 *Monetary and Economics Department Bank for International Settlements* Basel Switzerland
www.bis.org .

**5032.** Bank for International Settlements (October) 1999 A review of financial market events in autumn 1998 *Committee on the Global Financial System Bank for International Settlements* Basel Switzerland

www.bis.org .

**5033.** Bank for International Settlements (June) 2001 70[th] *annual report Bank for International Settlements* Basel Switzerland

www.bis.org .

**5034.** Bank for International Settlements 2002 Central bank survey of foreign exchange market activity in April 2001 *Monetary and Economics Department Bank for International Settlements* Basel Switzerland

www.bis.org .

**5035.** Bank for International Settlements 2004 Triennial central bank survey: Foreign exchange and derivatives market activity in 2004 *Monetary and Economics Department Bank for International Settlements* Basel Switzerland

www.bis.org .

**5036.** Bank for International Settlements (March) 2005 Triennial central bank survey of foreign exchange and derivatives market activity in 2004 *Bank for International Settlements* Basel Switzerland ISSN 1814-7356

www.bis.org .

**5037.** Bank for International Settlements (May) 2005 Foreign exchange market intervention in emerging markets: motives, techniques and implications *BIS Paper no 24* ISSN 1609-0381 pp 1 – 307 Monetary and Economic Department *Bank for International Settlements* Basel Switzerland

www.bis.org .

**5038.** Bank for International Settlements (December) 2007 Triennial central bank survey of foreign exchange and derivatives market activity in 2007 *Bank for International Settlements* Basel Switzerland ISSN 1814-7356

http://www.bis.org/publ/rpfxf07t.pdf?noframes=1 .

**5039.** Bank for International Settlements 2010 Triennial central bank survey. Foreign exchange and derivatives market activity in 2010 *Bank for International Settlements* Basel Switzerland ISSN 1814-7356

www.bis.org .

**5040.** Bertola G, Svensson L E O 1993 Stochastic devaluation risk and the empirical fit of target zone models *Review of Economic Studies* **60** pp 689 – 712.

**5041.** Biais B 1993 Price formation and equilibrium liquidity in fragmented and centralized markets *Journal of Finance* **48** pp 157 – 184.

**5042.** Chan Y-S, Weinstein M 1993 Reputation, bid-ask spread and market structure *Financial Analysts Journal* July/August pp 57 – 62.

**5043.** Cheung Y - W 1993 Exchange rate risk premiums *Journal of International Money and Finance* **12** pp 182 – 194.

**5044.** Cheung Y-W, Ng L 1996 A causality-in-variance test and its application to financial market prices *Journal of Econometrics* **72** pp 33 – 48.

5045. Cheung Y - W, Chinn M 1998 Integration, cointegration, and the forecast consistency of structural exchange rate models *Journal of International Money and Finance* **17** pp 813 – 830.

5046. Cheung Y-W, Wong C Y-P 1999 Foreign exchange traders in Hong Kong, Tokyo, and Singapore *in* Advances in pacific basin financial markets vol **5** Bos Th, Fetherstone Th A (editors) *JAI Press* Greenwich Connecticut pp 111 – 134.

5047. Cheung Y - W, Wong C 2000 A survey of market practitioners' views on exchange rate dynamics *Journal of International Economics* **51** pp 401 – 423.

5048. Cheung Y - W, Chinn M D 2001 Currency traders and exchange rate dynamics: A survey of the US market *Journal of International Money and Finance* **20** (4) pp 439 – 471.

5049. Cheung Y-W, Chinn M D, Marsh I W 2004 How do UK-based foreign exchange dealers think their market operates? *International Journal of Finance and Economics* **9** pp 289 – 306.

5050. Cheung Y - W, Chinn M, Pascual A G 2004, 2005 Empirical exchange rate models of the nineties: Are any fit to survive? *IMF Working Paper WP/04/73* IMF Washington USA, *Journal of International Money and Finance* **24** pp 1150 – 1175.

5051. Dacorogna M M, Müller U A, Nagrel R J, Olsen R B, Pictet O V 1993 A geographical model for the daily and weekly seasonal volatility in the FX market *Journal of International Money and Finance* **12** (4) pp 413 – 438.

5052. Dacorogna M M, Müller U A, Pictet O V. de Vries C G (March 17) 1995 The distribution of external foreign exchange rate returns in extremely large data sets *Preprint UAM 1992-10-22* Research Institute for Applied Economics *Olsen and Associates* Zurich Switzerland.

5053. Dominguez K M E, Frankel J A 1993 Does foreign exchange intervention work? Institute for International Economics Washington DC USA.

5054. Dominguez K M E 1998 Central bank intervention and exchange rate volatility Journal of International Money and Finance **17** pp 161 – 190.

5055. Dominguez K M E 2006 When do central bank interventions influence intra-daily and longer-term exchange rate movements? *Journal of International Money and Finance* **25** (7) pp 1051 – 1071.

5056. Dominguez K M E, Panthaki F 2006 What defines 'news' in foreign exchange markets? *Journal of International Money and Finance* **25** (1) pp 168 – 198.

5057. Ederington L, Lee J 1993 How markets process information: News releases and volatility *Journal of Finance* **48** pp 1161 – 1191.

5058. Edin P A, Vredin A 1993 Devaluation risk in target zones: Evidence from the Nordic countries *Economic Journal* **103** pp 161 – 175.

5059. Goldstein M, Folkerts-Landau D, Garber P, Rojas-Suarez L, Spencer M 1993 International capital markets Parts I and II *The International Monetary Fund* Washington DC USA.

5060. Griffiths M D, White R W (June) 1993 Tax - induced trading and the turn - of - the - year anomaly: An intraday study *The Journal of Finance* **48** (2) pp 575 – 598.

5061. Grimes A 1993 International reserves under floating exchange rates: Two paradoxes explained *The Economic Record* **69** pp 411 – 415.

5062. Harris M, Raviv A 1993 Differences of opinion make a horse race *Review of Financial Studies* **6** pp 473 – 506.

5063. Klein M W 1993 The accuracy of reports of foreign exchange intervention *The Journal of International Money and Finance* **12** (6) pp 644 – 653.

5064. Levich R M, Thomas L R 1993 The significance of technical trading-rule profits in the foreign exchange market: A bootstrap approach *Journal of International Money and Finance* **12** pp 451 – 474.

5065. Matsuyama K, Kiyotaki N, Matsui A 1993 Toward a theory of international currency *Review of Economic Studies* **60** pp 283 – 320.

5066. Romer D 1993 Rational asset-price movements without news *American Economic Review* **83** pp 1112 – 1130.

5067. Schmidt H, Iversen P, Treske K 1993 Parkett oder computer? *Zeitschrift für Bankrecht und Bankwirtschaft* **5** pp 209 – 221.

5068. Schmidt H, Iversen P 1993 Automating German equity trading: Bid-ask spreads on competing systems *Journal of Financial Services Research* **6** pp 373 – 397.

5069. Schmidt H, Oesterhelweg O, Treske K 1996 Deutsche Börsen im leistungsvergleich: IBIS und BOSS-CUBE *Kredit und Kapital* **29** pp 90 – 122.

5070. Wolinsky A 1990 Information revelation in a market with pair wise meetings *Econometrica* **58** pp 1 – 23.

5071. Ammer J, Brunner A 1994 Are banks market timers or market makers? Explaining foreign exchange trading profits *International Finance Discussion Paper #484* Board of Governors of the Federal Reserve System USA.

5072. Andrew R, Broadbent J 1994 Reserve bank operations in the foreign exchange market: Effectiveness and profitability *Research Discussion Paper 9406* Reserve Bank of Australia Sydney Australia.

5073. Backus D, Kehoe P, Kydland F 1994 Dynamics of the trade balance and the terms of trade: The J-curve? *American Economic Review* **84** pp 84 – 103.

5074. Bakker A, Boot H, Sleijpen O, Vanthoor W (editors) 1994 Monetary stability through international cooperation *Kluwer Academic Publishers* Dordrecht The Netherlands.

5075. Bartov E, Bodnar G M 1994 Firm valuation, earnings expectations, and the exchange rate exposure effect *Journal of Finance* **44** (5) pp 1755 – 1785.

5076. Bartov E, Bodnar G M 1995 Foreign currency translation reporting and the exchange-rate exposure effect *Journal of International Financial Management & Accounting* **6** (2) pp 93 – 115.

5077. Berry T, Howe K 1994 Public information arrival *Journal of Finance* **49** pp 1331 – 1346.

5078. Bessembinder H (June) 1994 Bid - ask spreads in the inter-bank foreign exchange markets *Journal of Financial Economics* **35** (3) pp 317 – 348.

5079. Ball C, Roma A 1994 Target zone modelling and estimation for European monetary system exchange rates *Journal of Empirical Finance* **1** pp 385 – 420.

5080. Brousseau V, Czarnecki M O (August 16) 1994 Modelling exchange rates: The stable model *Preprint* Ecole Polytechnique Paris France.

*5081.* De Jong F 1994 A univariate analysis of EMS exchange rates using a target zone model *Journal of Applied Econometrics* **9** pp 35 – 45.

*5082.* De Jong F, Nijman Th, Röell A 1995 A comparison of the cost of trading French shares on the Paris bourse and on SEAQ *International European Economic Review* **39** pp 1277 – 1301.

*5083.* De Jong F, Nijman Th, Röell A 1996 Price effects of trading and components of the bid-ask spread on the Paris bourse *Journal of Empirical Finance* **3** pp 193 – 213.

*5084.* De Jong F, Mahieu R, Schotman P 1998 Price discovery in the foreign exchange market: An empirical analysis of the Yen / Dmark rate *Journal of International Money and Finance* **17** pp 5 – 27.

*5085.* De Jong F, Ligterink J, Macrae V 2006 A firm specific analysis of the exchange rate exposure of Dutch firms *Journal of International Financial Management and Accounting* **17** (1) pp 1 – 28.

*5086.* De Jong F, Verschoor W F C, Zwinkels R C J 2010 Heterogeneity of agents and exchange rate dynamics: Evidence from the EMS *Journal of International Money and Finance* **29** (8) pp 1652 – 1669.

*5087.* Degryse H, de Jong F, van Kervel V 2011 The impact of dark trading and visible fragmentation on market quality *Discussion Paper 8630 CEPR*.

*5088.* Dini L 1994 Turbulence in the foreign exchange markets: Old and new lessons *in* Monetary stability through international cooperation Bakker A, Boot H, Sleijpen O, Vanthoor W (editors) *Kluwer Academic Publishers* Dordrecht The Netherlands.

*5089.* Fialkowski D, Petersen 1994 Posted versus effective spreads: Good prices or bad quotes? *Journal of Financial Economics* **35** pp 269 – 292.

*5090.* Glass G R 1994 Multinet *Panel Discussion*: Clearing house arrangements in the foreign exchange markets *Federal Reserve Board's International Symposium*: Banking and payment services Washington DC USA.

*5091.* Grünbichler A, Longstaff F, Schwartz E 1994 Electronic screen trading and the transmission of information: An empirical examination *Journal of Financial Intermediation* **3** pp 166 – 187.

*5092.* Hansch O, Naik N, Viswanathan S (November) 1994 Trading profits, inventory control and market share in a competitive dealer market *Typescript* Duke University USA.

*5093.* Hirschleifer D, Subrahmanyam A, Titman S 1994 Security analysis and trading patterns when some investors receive information before others *Journal of Finance* **49** pp 1665 – 1698.

*5094.* Hogan K C Jr, Melvin M 1994 Sources of meteor showers and heat waves in the foreign exchange market *Journal of International Economics* **37** pp 239 – 247.

*5095.* Jones C, Kaul G, Lipson M 1994 Transactions, volume, and volatility *Review of Financial Studies* **7** pp 631 – 651.

*5096.* Jones C, Lipson M 1999 Execution costs of institutional equity orders *Journal of Financial Intermediation* **8** pp 123 – 140.

*5097.* Kraus A, Smith M (September) 1994 Beliefs about beliefs *Working Paper* University of British Columbia Vancouver Canada.

**5098.** Massib M, Phelps B 1994 Electronic trading, market structure and liquidity *Financial Analysts Journal* pp 39 – 50.

**5099.** Mendelson M, Peake J (September) 1994 Equity markets in economies in transition *Journal of Banking and Finance* **15** (5) pp 913 – 929.

**5100.** Naidu G N, Rozeff M 1994 Volume, volatility, liquidity and efficiency of the Singapore stock exchange before and after automation *Pacific-Basin Finance Journal* **2** pp 23 – 42.

**5101.** Nieuwland F G M C, Verschoor W F C, Wolff C C P 1994 Stochastic jumps in EMS exchange rates *Journal of International Money and Finance* **13** pp 699 – 727.

**5102.** Pictet O V, Dacorogna M M, Muller U A, De Vries C G 1994 The distribution of external foreign exchange rate returns and extremely large data sets *Olsen and Associates Research Group* Zurich Switzerland.

**5103.** Sharpe W A (Fall) 1994 The Sharpe ratio *Journal of Portfolio Management* pp 49 – 58.

**5104.** Silber W L 1994 Technical trading: When it works and when it doesn't *The Journal of Derivatives* **1** (3).

**5105.** Slezak S 1994 A theory of the dynamics of security returns around market closures *Journal of Finance* **49** pp 1163 – 1211.

**5106.** Szpiro G G 1994 Exchange rate speculation and chaos inducing intervention *Journal of Economic Behavior and Organization* **24** pp 363 – 368.

**5107.** Yadav P, Pope P, Paudyal K 1994 Threshold autoregressive modeling in finance: The rice differences of equivalent assets *Mathematical Finance* **4** (2) pp 205 – 221.

**5108.** Walsh E J 1994 Operating income, exchange rate changes, and the value of the firm: An empirical analysis *Journal of Accounting, Auditing and Finance* **9** (4) pp 703 – 724.

**5109.** Wei, Sh-J (May) 1994 Anticipations of foreign exchange volatility and bid-ask spreads *Working Paper no 4737* National Bureau of Economic Research Cambridge Massachusetts USA.

**5110.** Watanabe T 1992 The signaling effect of foreign exchange intervention: The case of Japan *Bank of Japan* Tokyo Japan.

**5111.** Watanabe T, Harada K 2004 Effects of the Bank of Japan's intervention on yen/dollar exchange rate volatility *Journal of the Japanese and International Economies*.

**5112.** Watanabe T, Yabu T (June) 2007 The great intervention and massive money injection: The Japanese experience 2003-2004 *Working Paper* Institute of Economic Research Hitotsubashi University Japan.

**5113.** Almekinders G J 1995 Foreign exchange intervention: Theory and evidence *E Elgar* Brookfield VT USA.

**5114.** Chiang T, Jiang C 1995 Foreign exchange returns over short and long horizons *International Review of Economics & Finance* **4** pp 267 – 282.

**5115.** Dumas B, Solnik B 1995 The world price of foreign exchange risk *Journal of Finance* **50** pp 445 – 479.

**5116.** Ederington L, Lee J 1995 The short-run dynamics of price adjustment to new information *Journal of Financial and Quantitative Analysis* **30** pp 117 – 134.

*5117.* Evertsz C J G 1995 Self - similarity of high - frequency USD/DEM exchange rates *Proceedings of the First International Conference on High Frequency Data in Finance* (*HFDF-1*) vol **3** *Research Institute for Applied Economics Olsen & Associates Zürich* Switzerland.

*5118.* Faruqee H 1995 Long-run determinants of the real exchange rate: A stock-flow perspective *IMF Staff Papers* **42** (1) pp 80 – 107.

*5119.* Frino A, McCorry M 1995 Why are spreads tighter on the Australian Stock Exchange than the NYSE? An electronic open limit order book versus the specialist structure *Working Paper* University of Sydney Australia.

*5120.* Frino A, McInish Th, Toner M 1998 The liquidity of automated exchanges: New evidence from German bund futures *Journal of International Financial Markets, Institutions and Money* **8** pp 225 – 241.

*5121.* Ghysels E, Jasiak J 1995 Trading patterns: Time deformation and stochastic volatility in foreign exchange markets *Proceedings of the First International Conference on High Frequency Data in Finance* (*HFDF-1*) vol **1** *Research Institute for Applied Economics Olsen & Associates Zürich* Switzerland.

*5122.* Grossman G, Rogoff K (editors) 1995 Handbook of international economics *Elsevier Science* The Netherlands.

*5123.* Havrilesky T 1995 The pressures on American monetary policy 2[nd] edition *Kluwer Academic Publishing* Boston USA.

*5124.* Hong H, Wang J (July) 1995 Trading and returns under periodic market closures *Working Paper* Massachusetts Institute of Technology USA.

*5125.* Isard P 1995 Exchange rate economics *Cambridge University Press* Cambridge UK.

*5126.* Kandel E, Pearson N 1995 Differential interpretation of public signals and trade in speculative markets *Journal of Political Economy* **103** pp 831 – 872.

*5127.* Lewis K K 1995 Are foreign exchange intervention and monetary policy related and does it really matter? *Journal of Business* **68** pp 185 – 214.

*5128.* Lin J-C, Sanger G, Booth G 1995 Trade size and components of the bid-ask spread *Review of Financial Studies* **8** pp 1153 – 1183.

*5129.* Mantegna R N, Stanley H E 1995 Scaling behaviour in economic index *Nature* vol **376** pp 46 – 49.

*5130.* Mark N 1995 Exchange rates and fundamentals: Evidence on long-horizon predictability *American Economic Review* **85** pp 201 – 218.

*5131.* Mark N, Wu Y 1998 Rethinking deviations from uncovered interest parity: The role of covariance risk and noise *Economic Journal* **108** pp 1686 – 1706.

*5132.* Mark N 2001 International macroeconomics and finance *Blackwell Publishers* Oxford UK.

*5133.* Mark N 2009 Changing monetary policy rules, learning, and real exchange rate dynamics *Journal of Money, Credit and Banking*.

*5134.* Obstfeld M, Rogoff K 1995 Exchange rate dynamics redux *Journal of Political Economy* **103** pp 624-660.

*5135.* Obstfeld M, Rogoff K (August) 1998 Risk and exchange rates *NBER Working Paper 6694 in* Contemporary economic policy: Essays in honor of Assaf Razin; Helpman E, Sadka E (editors) *Cambridge University Press* Cambridge UK.

**5136.** Osler C L 1995 Exchange rate dynamics and speculator horizon *Journal of International Money and Finance* **14** (5) pp 695 – 720.

**5137.** Osler C L 1998 Short-term speculators and the puzzling behavior of exchange rates *Journal of International Economics* **45** pp 37 – 57.

**5138.** Carlson J A, Osler C L (March) 1999 Determinants of currency risk premiums *Federal Reserve Bank of New York Staff Reports Series* no 70.

**5139.** Kevin C P H, Osler C L 1999 Methodical madness: Technical analysis and the irrationality of exchange-rate forecasts *Economic Journal* **109** (458) pp 636 – 661.

**5140.** Osler C L 2000 Support for resistance: Technical analysis and intraday exchange rates *Federal Reserve Bank of New York Economic Policy Review* **6** (2) pp 53 – 68.

**5141.** Osler C L 2003 Currency orders and exchange-rate dynamics: Explaining the success of technical analysis *Journal of Finance* **58** (5) pp 1791 – 1819.

**5142.** Osler C L 2005 Stop-loss orders and price cascades in currency markets *Journal of International Money and Finance* **24** (2) pp 219 – 241.

**5143.** Carlson A J, Osler C L 2005 Short-run exchange rate dynamics: Theory and evidence.

**5144.** Osler C L 2006 Macro lessons from microstructure *International Journal of Finance and Economics* **11** (1) pp 55 – 80.

**5145.** Osler C L 2008 Foreign exchange microstructure: A survey *in* Springer encyclopedia of complexity and system science *Springer* Germany.

**5146.** Osler C L 2009 Market microstructure, foreign exchange *in* Encyclopedia of complexity and system science Meyers R A (ed.) *Springer* pp 5404 – 5438.

**5147.** Osler C L, Vandrovych V 2009 Hedge funds and the origins of private information in currency markets *Typescript* Brandeis University.

**5148.** Osler C L, Yusim R 2009 Intraday dynamics of foreign-exchange spreads *Typescript* Brandeis University.

**5149.** Osler C L, Mende A, Menkhoff L 2011 Price discovery in currency markets *Journal of International Money and Finance* **30** (8) pp 1696 – 1718.

**5150.** Osler C L, Savaser T 2011 Extreme returns: The case of currencies *Journal of Banking and Finance* **35** (11) pp 2868 – 2880.

**5151.** Dahl Ch, Carlson J A, Osler C L 2011 Short-run exchange-rate dynamics: Theory and evidence *Research Paper* Brandeis University.

**5152.** Osler C L 2012 Currency market microstructure, the carry trade, and technical trading *Annual Review of Financial Economics* **4** (1).

**5153.** Peiers B (October) 1995 Informed traders, intervention, and price leadership: A deeper view of the microstructure of the foreign exchange market *University California Los Angeles* California USA.

**5154.** Prasad A M, Rajan M 1995 The role of exchange and interest risk in equity valuation: A comparative study of international stock markets *Journal of Economics and Business* **47** (5) pp 457 – 472.

**5155.** Schnidrig R, Würtz D (March) 1995 Investigation of the volatility and autocorrelation function of the USD/DEM exchange rate on operational time scales *Proceedings of the First International Conference on High Frequency Data in*

Finance (HFDF-1) vol **3** *Research Institute for Applied Economics Olsen & Associates* Zürich Switzerland.

5156. Schwartz R (editor) 1995 Global equity markets: Technological, competitive, and regulatory challenges *Irwin* Homewood Illinois USA.

5157. Shyy G, Lee J 1995 Price transmission and information asymmetry in bund futures markets: LIFFE vs DTB *Journal of Futures Markets* **15** pp 87 – 99.

5158. Shyy G, Vijayraghavan V, Scott-Quinn B 1996 A further investigation of the lead-lag relationship between the cash market and stock index futures market with the use of bid/ask quotes: The case of France *Journal of Futures Markets* **16** pp 405 – 420.

5159. Vivex X 1995 The speed of information revelation in a financial market mechanism *Journal of Economic Theory* **67** pp 178 – 204.

5160. Zaheer A, Zaheer S (December) 1995 Catching the wave: Alertness, responsiveness, and market influence in global electronic networks *Typescript* University of Minnesota USA.

5161. Bonser – Neal C, Tanner G 1996 Central bank intervention and volatility of foreign exchange rates: Evidence from options market *Journal of International Money and Finance* vol **15** pp 853 – 878.

5162. Claassen E M 1996 Global monetary economics *Oxford University Press* Oxford UK.

5163. Danker D, Haas R A, Henderson D W, Symanski S, Tryon R W 1996 Small empirical models of exchange market intervention: Applications to Germany, Japan, and Canada *Journal of Policy Modeling* **9**.

5164. Dukas S P, Fatemi A M, Tavakkol A 1996 Foreign exchange rate exposure and the pricing of exchange rate risk *Global Finance Journal* **7** (2) pp 169 – 189.

5165. Dwyer G, Locke P, Yu W 1996 Index arbitrage and nonlinear dynamics between the S&P 500 futures and cash *Review of Financial Studies* **9** (1) pp 301 – 332.

5166. Easley D, Kiefer N, O'Hara M, Paperman J 1996 Liquidity, information and infrequently traded stocks *Journal of Finance* **51** pp 1405 – 1436.

5167. Easley D, Kiefer N, O'Hara M 1997a The information content of the trading process *Journal of Empirical Finance* **4** pp 159 – 185.

5168. Easley D, Kiefer N, O'Hara M 1997b One day in the life of a very common stock *Review of Financial Studies*.

5169. Easley D, O'Hara M, Srinivas P S 1998 Option volume and stock prices: Evidence on where informed traders trade *Journal of Finance* **53** pp 431 – 465.

5170. Flemming J, Ostdiek B, Whaley R 1996 Trading costs and the relative rates of price discovery in stocks, futures and options markets *Journal of Futures Markets* **16** pp 353 – 387.

5171. Gagnon J E 1996 Net foreign assets and equilibrium exchange rates: Panel evidence *Board of Governors of the Federal Reserve System International Finance Discussion Papers 574* USA.

5172. Ghashghaie S, Breymann W, Peinke J, Talkner P, Dodge Y (June) 1996 Turbulent cascades in foreign exchange markets *Nature* **381** (27) pp 767 – 770.

5173. Hsieh D, Kleidon A 1996 Bid-ask spreads in foreign exchange markets: Implications for models of asymmetric information *in* The Microstructure of foreign exchange markets Frankel J A, Galli G, Giovannini A (editors) *University of Chicago Press* Chicago pp 41 – 65.

5174. Ingersoll J E Jr 1996 Valuing foreign exchange rate derivatives with a bounded exchange process *Review of Derivatives Research* **1** pp 159 – 181.

5175. Kaminsky G, Lewis K 1996 Does foreign exchange intervention signal future monetary policy? *Journal Of Monetary Economics* **37** pp 285 – 312.

5176. LeBaron B (March) 1996 Technical trading rule, profitability and foreign exchange intervention *Working Paper 5505* NBER USA pp 1 – 18.

5177. MacDonald R, Marsh I W 1996 Currency forecasters are heterogeneous: Confirmation and consequences *Journal of International Money and Finance* **15** (5) pp 665 – 685.

5178. Madrigal V 1996 Non-fundamental speculation *Journal of Finance* **51** pp 553 – 578.

5179. Pirrong C 1996 Market liquidity and depth on computerized and open outcry trading systems: A comparison of DTB and LIFFE bund contracts *Journal of Futures Markets* **16** pp 519 – 543.

5180. Rosenberg M 1996 Currency forecasting: A guide to fundamental and technical models of exchange rate determination *Irwin Professional Publishing* Chicago USA.

5181. Tsang Sh-K 1996 A study of the linked exchange rate system and policy options for Hong Kong *Hong Kong Policy Research Institute* Hong Kong P R China.

5182. Tsang Sh-K 1998 The case for adopting the convertible reserves system in Hong Kong *Pacific Economic Review* **3** pp 265 – 275.

5183. Tsang Sh-K, Sin Ch-Y, Cheng Y-Sh 1999 The robustness of Hong Kong's linked exchange rate system as a currency board arrangement *The 54th European Meeting of the Econometric Society* Hong Kong P R China.

5184. Tsang Sh-K 1999a A study of the linked exchange rate system and policy options for Hong Kong *Hong Kong Policy Research Institute Ltd* Hong Kong P R China.

5185. Tsang Sh-K 1999b Fixing the exchange rate through a currency board arrangement: Efficiency risk, systemic risk and exit cost *Asian Economic Journal* **13** pp 239 – 266.

5186. Tsang Sh-K, Yue Ma 2002 Currency substitution and speculative attacks on a currency board system *Journal of International Money and Finance* **21** (1) pp 53 – 78.

5187. Vermeiren D, Ková T Z 1996 Foreign exchange risk management and auditing *Basic Documentation Version 1.0a* Prague CNB – Banking Supervision 1996.

5188. Balke N, Fomby T 1997 Threshold cointegration *International Economic Review* **38** pp 627 – 646.

5189. Balke N, Wohar M 1998 Nonlinear dynamics and covered interest rate parity *Empirical Economics* **23** pp 535 – 559.

5190. Bhattacharya U, Weller P 1997 The advantage to hiding one's hand: Speculation and central bank intervention in the foreign exchange market *Journal of Monetary Economics* **39** pp 251 – 277.

5191. Campbell J Y, Lo A W, MacKinlay A C 1997 The econometrics of financial markets *Princeton University Press* Princeton New Jersey USA.

5192. Campbell J Y, Viceira L 2002 Strategic asset allocation: Portfolio choice for long term investors *Clarendon Lectures in Economics Oxford University Press* Oxford UK.

**5193.** Chamberlain S, Howe J S, Popper H 1997 The exchange rate exposure of US and Japanese banking institution *Journal of Banking and Finance* **21** (6) pp 871 – 892.

**5194.** Clarida R H, Taylor M P 1997 The term structure of forward exchange premiums and the forecastability of spot exchange rates: Correcting the errors *Review of Economics and Statistics* **79** pp 353 – 361.

**5195.** Clarida R H, Sarno L, Taylor M P, Valente G 2003 The out-of-sample success of term structure models as exchange rate predictors: A step beyond *Journal of International Economics* **60** pp 61 – 83.

**5196.** Copejans M, Domowitz I 1997 The performance of an automated trading market in an illiquid environment *Working Paper* Duke University, Northwestern University USA.

**5197.** DeGennaro R, Shrieves R 1997 Public information releases, private information arrival, and volatility in the foreign exchange market *Journal of Empirical Finance* **4** pp 295 – 315.

**5198.** Dewachter H 1997 Sign predictions of exchange rate changes: Charts as proxies for Bayesian inferences *Review of World Economics* **133** (1) pp 39 – 55.

**5199.** Dewachter H 2001 Can Markov switching models replicate chartist profits in the foreign exchange market? *Journal of International Money and Finance* **20** pp 25 – 41.

**5200.** Dewachter H, Lyrio M 2005 The economic value of technical trading rules: A nonparametric utility-based approach *International Journal of Finance and Economics* **10** pp 41 – 62.

**5201.** Embrechts P, Klueppelberg C, Mikosch T 1997 Modeling external events for insurance and finance *Springer - Verlag* Berlin Germany.

**5202.** Evans M D D (November) 1997 The microstructure of foreign exchange dynamics *Typescript* Georgetown University USA.

**5203.** Evans M D D, Lyons R K 1999, (February) 2002a Order flow and exchange rate dynamics *Typescript* UC Berkeley, *Journal of Political Economy* **110** (1) pp 170 – 180.

**5204.** Evans M D D, Lyons R K 2001a Why order flow explains exchange rates *Unpublished Manuscript* University California Berkeley USA.

**5205.** Evans M D D, Lyons R K (July) 2001b Portfolio balance, price impact, and secret intervention *NBER Working Paper 8356* NBER USA.

**5206.** Evans M D D (February) 2001c, 2002b FX trading and exchange rate dynamics *NBER Working Paper 8116; Journal of Finance* **57** (6) pp 2405 – 2447.

**5207.** Evans M D D, Lyons R K (July) 2002c Time-varying liquidity in foreign exchange *Journal of Monetary Economics Elsevier* **49** (5) pp 1025 – 1051.

**5208.** Evans M D D, Lyons R K (November) 2002d Informational integration and FX Trading *Journal of International Money and Finance* **21** (6) pp 807 – 831.

**5209.** Evans M D D, Lyons R K (January) 2003 How is macro news transmitted to exchange rates? *NBER Working Paper 9433* National Bureau of Economic Research Inc USA.

**5210.** Cao H, Evans M, Lyons R (August) 2003 Inventory information *NBER Working Paper 9893* NBER USA http://www.nber.org/papers/w9893, *Journal of Business*, USA.

**5211.** Evans M D D, Lyons R K (March) 2004a A new micro model of exchange rate dynamics *NBER Working Papers 10379* National Bureau of Economic Research Inc USA http://www.nber.org/papers/w10379 .

**5212.** Evans M D D, Lyons R K 2004b, 2007 Exchange rate fundamentals and order flow *Working Papers gueconwpa~05-05-03* Department of Economics Georgetown University; *Working Paper 13151* National Bureau of Economic Research Cambridge MA USA http://www.nber.org/papers/w13151 .

**5213.** Evans M D D, Lyons R K 2005a Meese-Rogoff redux: Micro-based exchange-rate forecasting *NBER Working Paper 11042* National Bureau of Economic Research http://www.nber.org/papers/w11042, *American Economic Review Papers and Proceedings* **95** (2) pp 405 – 414.

**5214.** Evans M D D, Lyons R K 2005b Do currency markets absorb news quickly? *Working Paper 11041* NBE USA pp 1 – 25, *Journal of International Money and Finance* **24** (2) pp 197 – 217.

**5215.** Evans M D D, Lyons R 2005c Are different-currency assets imperfect substitutes? *in* Exchange rate economics: Where do we stand? DeGrauwe P (editor) *MIT Press* Cambridge USA.

**5216.** Evans M D D, Lyon R K 2005d, 2006 Understanding order flow NBER Working Paper 11748 NBER USA http://www.nber.org/papers/w11748, *International Journal of Finance & Economics John Wiley & Sons Inc* **11** (1) pp 3 – 23.

**5217.** Evans M D D 2005 Where are we know? Real-time estimates of the macroeconomy *International Journal of Central Banking* **1** (6) pp 127 – 175.

**5218.** Evans M D D, Hnatkovska V 2005 International capital flows, returns and world financial integration *NBER Working Paper* NBER USA.

**5219.** Evans M D D, Lyons R K 2008 How is macro news transmitted to exchange rates? *Journal of Financial Economics* **88** (1) pp 26 – 50.

**5220.** Evans M D D, Lyons R K 2009 Forecasting exchange rate fundamentals with order flow *Working Paper* Georgetown University USA.

**5221.** Evans M D D 2010 Order flows and the exchange rate disconnect puzzle *Journal of International Economics* **80** (1) pp 58 – 71.

**5222.** Evans M D D 2011 Exchange-rate dynamics *Princeton University Press* USA.

**5223.** Fleming M, Remolona E 1997 What moves the bond market? *Federal Reserve Bank of New York Economic Policy Review* **3** pp 31 – 50.

**5224.** Fleming M, Remolona E 1999 Price formation and liquidity in the US treasury market *Journal of Finance* **54** pp 1901 – 1915.

**5225.** Fleming M (September) 2002 Price discovery in the US treasury market: The impact of order flow and liquidity on the yield curve *Typescript* New York Federal Reserve Bank New York USA.

**5226.** Fleming M 2003 Measuring treasury market liquidity *Federal Reserve Bank of New York Economic Policy Review* **9** pp 83 – 108.

**5227.** Franke G, Hess D 1997, 2000 Information diffusion in electronic and floor trading *Working Paper* Universität Konstanz Germany, *Journal of Empirical Finance* **7** (5) pp 455 – 478.

5228. Goldberg L, Tenorio R 1997 Strategic trading in a two-sided foreign exchange auction *Journal of International Economics* **42** pp 299 – 326.

5229. Gosh A R, Ostry J D, Gulde A M, Wolf H C 1997 Does the exchange rate regime matter for inflation and growth? *IMF* Washington USA http://www.imf.org .

5230. Harris J, Schultz P 1997 The importance of firm quotes and rapid executions: Evidence from the January 1994 SOES rule change *Journal of Financial Economics* **45** pp 135 – 166.

5231. Hartmann P 1997 The currency denomination of international trade after European Monetary Union *Typescript* European Central Bank.

5232. Hartmann Ph 1998 Do Reuters spreads reflect currencies' differences in global trading activity? *Journal of International Money and Finance* **17** (5) pp 757 – 784.

5233. Hartmann P 1998 Currency competition and foreign exchange markets: The dollar, the yen, and the euro *Cambridge University Press* Cambridge UK.

5234. Hartmann P 1999 Trading volumes and transaction costs in the foreign exchange market: Evidence from daily dollar-yen spot data *Journal of Banking and Finance* **23** pp 801 – 824.

5235. Hung J 1997 Intervention strategies and exchange rate volatility: A noise trading perspective *Journal of International Money and Finance* **16** (5) pp 779 – 793.

5236. Kirilenko A 1997 Endogenous trading arrangements in emerging foreign exchange markets *Typescript* International Monetary Fund USA.

5237. Lamoureux C, Schnitzlein C 1997 When it's not the only game in town: The effect of bilateral search on the quality of a dealer market *Journal of Finance* **52** pp 683 – 712.

5238. Madhavan A, Smidt S 1991 A Bayesian model of intraday specialist pricing *Journal of Financial Economics* **30** pp 99 – 134.

5239. Madhavan A, Smidt S 1993 An analysis of changes in specialist inventories and quotations *Journal of Finance* **48** (5) pp 1595 – 1628.

5240. Leach J, Madhavan A 1993 Price experimentation and security market structure *Review of Financial Studies* **6** pp 375 – 404.

5241. Keim D, Madhavan A 1996 The upstairs market for large-block transactions: Analysis and measurement of price effects *Review of Financial Studies* **9** pp 1 – 36.

5242. Madhavan A, Cheng M 1997 In search of liquidity: Block trades in the upstairs and downstairs market *Review of Financial Studies* **10** pp 175 – 203.

5243. Madhavan A, Richardson M, Roomans M 1997 Why do security prices change? A transaction-level analysis of NYSE stocks *Review of Financial Studies* **10** pp 1035 – 1064.

5244. Madhavan A, Sofianos G 1997 An empirical analysis of NYSE specialist trading *Journal of Financial Economics* **48** pp 189 – 210.

5245. Madhavan A (March) 2000 Market microstructure: A survey *University of Southern California* USA.

5246. Madhavan A 2000 Market microstructure: A survey *Journal of Financial Markets* **3** pp 205 – 258.

5247. Madhavan A (October) 2000 In search of liquidity in the internet era *9th Annual Financial Markets Conference of the Federal Reserve Bank of Atlanta* USA.

5248. Martens M 1997 Interaction between financial markets *Tinbergen Institute Research Series no 139* Rotterdam The Netherlands.

5249. Montiel P J 1997 Exchange rate policy and macroeconomic management tin ASEAN countries in macroeconomic issues facing ASEAN countries *International Monetary Fund* Washington USA.

5250. Pagano M, Roell A 1997 Front running: Market professionals as quasi-insiders *Typescript* Tilburg University.

5251. Peiers B 1997 Informed traders, intervention, and price leadership: A deeper view of the microstructure of the foreign exchange market *Journal of Finance* **52** pp 1589 – 1614.

5252. Reiss P, Werner I (February) 1997 Interdealer trading: Evidence from London *Stanford Graduate School of Business Research Paper no 1430* University of Stanford California USA.

5253. Sweeney R J 1997 Do central banks lose on foreign exchange intervention? A review article *Journal of Banking & Finance* **21** pp 1667 – 1684.

5254. Sweeney R J 2000 Does the Fed beat the foreign exchange market? *Journal of Banking & Finance* **24** pp 665 – 694.

5255. Szakmary A, Mathur I 1997 Central bank intervention and trading rule profits in foreign exchange markets *Journal of International Money and Finance* **16** pp 513 – 535.

5256. Vogler K 1997 Risk allocation and interdealer trading *European Economic Review* **41** pp 417 – 441.

5257. Wei S, Kim J (November) 1997 The big players in the foreign exchange market: Do they trade on information or noise? *NBER Working Paper 6256* NBER USA.

5258. Werner I (September) 1997 A double auction model of interdealer trading *Research Paper no 1454* Stanford University California USA.

5259. Wren-Lewis S (July) 1997 The choice of exchange rate regime *Economic Journal*.

5260. Abhyankar A H 1998 Linear and nonlinear Granger causality: Evidence from the UK stock index futures market *Journal of Futures Markets* **18** (5) pp 519 – 540.

5261. Abrams R K, Beato P 1998 The prudential regulation and management of foreign exchange risk *International Monetary Fund* Washington DC USA.

5262. Anthony M, MacDonald R 1998 On the mean reverting properties of target zone exchange rates: Some evidence from the ERM *European Economic Review* **42** pp 1493 – 1523.

5263. Anthony M, MacDonald R 1999 The width of the band and exchange rate mean reversion: Some further ERM-Based Results *Journal of International Money and Finance* **18** pp 411 – 428.

5264. Bjønnes G H, Rime D (December) 1998 FX trading ... live: Impact of new trading environments *Typescript* Norwegian School of Management University of Oslo Norway.

5265. Bjønnes G H, Rime D (March) 2001 FX trading live! Dealer behavior and trading systems in foreign exchange markets *Typescript* Norwegian School of Management University of Oslo Norway
www.uio.no/~dagfinri .

5266. Bjønnes G H, Rime D 2005 Dealer behavior and trading systems in foreign exchange markets *Journal of Financial Economics* **75** (3) pp 571 – 605.

5267. Bjønnes G H, Rime D, Solheim H O A 2005 Liquidity provision in the overnight foreign exchange market *Journal of International Money and Finance* **24** (2) pp 177 – 198.

5268. Bjønnes G H, Osler C, Rime D (September 15) 2007 Asymmetric information in the interbank foreign exchange market *3rd Annual Conference on Market Microstructure* Budapest Hungary.

5269. Bjønnes G H, Osler C L, Rime D 2011 Sources of information advantage in the foreign exchange market *Working Paper* Norges Bank Norway.

5270. Blennerhasset M, Bowman R G 1998 A change in market microstructure: The switch to electronic screen trading on the New Zealand stock exchange *Journal of International Financial Markets, Institutions and Money* **8** pp 261 – 276.

5271. Bodnar G, Hayt G, Marston R 1998 Wharton survey of financial risk management by US non-financial firms *Financial Management* **27** (4) pp 70 – 91.

5272. Caramazza F, Aziz J 1998 Fixed or flexible? Getting the exchange rate right in the 1990s *IMF* Washington USA

http://www.imf.org .

5273. Chang Y, Taylor S 1998 Intraday effects of foreign exchange intervention by the Bank of Japan *Journal of International Money and Finance* **18** pp 191 – 210.

5274. Choi J J, Hiraki T, Takezawa N 1998 Is foreign exchange risk priced in the Japanese stock market *Journal of Financial and Quantitative Analysis* **33** pp 361 – 382.

5275. Chow E H, Chen H L 1998 The determinants of foreign exchange rate exposure: Evidence on Japanese firms *Pacific-Basin Finance Journal* **6** (1/2) pp 153 – 174.

5276. Clark P B, Macdonald R 1998 Exchange rates and economic fundamentals: A methodological comparison of BEERS and FEERS *IMF Working Paper WP/98/67* IMF USA.

5277. Covrig V, Melvin M 1998, 2002 Asymmetric information and price discovery in the FX market: Does Tokyo know more about the yen? *Typescript* Arizona State University, *Journal of Empirical Finance* **9** pp 271 – 285.

5278. Eddelbuttel D, McCurdy T 1998 The impact of news on foreign exchange rates: Evidence from high frequency data *Typescript* University of Toronto Canada.

5279. Edison H (February) 1998 On foreign exchange intervention: An assessment of the US experience *Typescript* Board of Governors of the Federal Reserve System USA.

5280. Fleming J, Kirby C, Ostdiek B 1998 Information and volatility linkages in the stock, bond, and money markets *Journal of Financial Economics* **49** pp 111 – 137.

5281. Garfinkel J, Nimalendran M 1998 Market structure and trader anonymity: An analysis of insider trading *Working Paper* Loyola University of Chicago University of Florida USA.

5282. George E 1998 Exchange rates: An intractable aspect of monetary policy *Bank of England Quarterly Bulletin* (May 1998) **38** no 2.

5283. Hansch O, Naik N, Viswanathan S 1998 Do inventories matter in dealership markets? Evidence from the London stock exchange *Journal of Finance* **53** pp 1623 – 1656.

*5284.* Hau H 1998 Competitive entry and endogenous risk in the foreign exchange market *Review of Financial Studies* **11** pp 757 – 788.

*5285.* Hau H, Killeen W, Moore M (July) 2000, 2002a The Euro as an international currency: Explaining puzzling first evidence from the foreign exchange markets *CEPR Discussion Paper no 2510* CEPR, *Journal of International Money and Finance* **21** (3) pp 351 – 383.

*5286.* Hau H, Killeen W, Moore M (April) 2002b Euro's forex role: How has the Euro changed the foreign exchange market? *Economic Policy* pp 151 – 191.

*5287.* Hau H, Rey H (December) 2002 Exchange rates, equity prices, and capital flows *NBER Working Paper 9398* NBER USA.

*5288.* Hau H, Rey H 2003 Can portfolio rebalancing explain the dynamics of equity returns, equity flows, and exchange rates? *American Economic Review*.

*5289.* He J, Ng L K 1998 The foreign exchange exposure of Japanese multinational corporations *Journal of Finance* **53** (2) pp 733 – 753.

*5290.* Helpman E, Sadka E (editors) 1998 Contemporary economic policy: Essays in honor of Assaf Razin *Cambridge University Press* Cambridge U.K.

*5291.* Hong Kong Monetary Authority 1998 Strengthening of currency board arrangements in Hong Kong *Quarterly Bulletin November* pp 1 – 5.

*5292.* Isard P, Faruqee H 1998 Exchange rate assessment: Extensions of the macroeconomic balance approach *IMF Occasional Paper no 167* IMF Washington USA.

*5293.* Isard P, Faruqee H, Kincaid G R, Fetherston M 2001 Methodology for current account and exchange rate assessments *IMF Occasional Paper no 209* International Monetary Fund Washington USA.

*5294.* Kanas A 1998 Testing for a unit root in ERM exchange rates in the presence of structural breaks: Evidence from the boot-strap *Applied Economics Letters* **5** pp 407 – 410.

*5295.* Lee R 1998 What is an exchange? The automation, management and regulation of financial markets *Oxford University Press* Oxford UK.

*5296.* Litterman R, Winkelmann K (January) 1998 Estimating covariance matrices *Goldman Sachs Risk Management Series*.

*5297.* Lui Y - H, Mole D 1998 The use of fundamental and technical analyses by foreign exchange dealers: Hong Kong evidence *Journal of International Money and Finance* **17** pp 535 – 545.

*5298.* Menkhoff L 1998 The noise trading approach — Questionnaire evidence from foreign exchange *Journal of International Money and Finance* **17** pp 547 – 564.

*5299.* Gehrig Th, Menkhoff L 2000, 2004 The use of flow analysis in foreign exchange: Exploratory evidence *Typescript* Department of Economics University of Freiburg Germany, *Journal of International Money and Finance* **23** (4) pp 573 – 594.

*5300.* Mende A, Menkhoff L (March) 2003 Different counterparties, different foreign exchange trading? The perspective of a median bank.

*5301.* Mende A, Menkhoff L 2006 Profits and speculation in intra-day foreign exchange trading *Journal of Financial Markets* **9** (3) pp 223 – 245.

5302. Menkhoff L, Taylor M P 2007 The obstinate passion of foreign exchange professionals: Technical analysis, *Warwick Economic Research Papers no 769* Department Of Economics The University of Warwick UK pp 1 – 61, *Journal of Economic Literature* **45** (4) pp 936 – 972.

5303. Frömmel M, Mende A, Menkhoff L 2008 Order flows, news, and exchange rate volatility *Journal of International Money and Finance* **27** (6) pp 994 – 1012.

5304. Menkhoff L, Schmeling M 2008 Local information in foreign exchange markets *Journal of International Money and Finance* **27** (8) pp 1383 – 1406.

5305. Menkhoff L, Osler C L, Schmeling M (May 11) 2010 Limit-order submission strategies under asymmetric information *CESIFO Working Paper no 3054* Category 7: Monetary Policy and International Finance Department of Economics Leibniz University Hannover Germany pp 1 – 42.

5306. Menkhoff L 2010 High-frequency analysis of foreign exchange interventions: What do we learn? *Journal of Economic Surveys* **24** (1) pp 85 – 112.

5307. Menkhoff L, Schmeling M 2010 Trader see, trader do: How do (small) FX traders react to large counterparties' trades? *Journal of International Money and Finance*.

5308. Miller K D, Reuer J J 1998 Asymmetric corporate exposures to foreign exchange rate changes *Strategic Management Journal* **19** (12) pp 1183 – 1191.

5309. Miville M, DiMillo J 1998 Survey of the Canadian foreign exchange and derivatives markets *Bank of Canada Review Winter 1995-1996* Financial Markets Department Bank of Canada Ottawa Canada.

5310. Nagayasu J 1998 Japanese effective exchange rates and determinants: Prices, real interest rates, and actual and optimal current accounts *IMF Working Paper WP/98/86* IMF USA.

5311. Neely Ch 1998 Technical analysis and the profitability of US foreign exchange intervention *Federal Reserve Bank of St Louis Review* **80** pp 3 – 17.

5312. Neely Ch J 2000a The practice of central bank intervention: Looking under the hood *Central Banking* **XI** pp 24 – 37.

5313. Neely Ch J 2000b Are changes in foreign exchange reserves well correlated with official intervention? *Economic Review of the Federal Reserve Bank of St Louis September/October* pp 17 – 30.

5314. Neely C J 2004 Forecasting foreign exchange volatility: Why is implied volatility biased and inefficient? And does it matter? *Working Paper 2002-017D* Federal Reserve Bank of St Louis MO USA.

5315. Neely Ch J 2005 An analysis of recent studies of the effect of foreign exchange intervention *Federal Reserve Bank of St Louis Review November/December* **87** (6) pp 685 – 717.

5316. Pesaran M, Hasem P M, Smith R P 1998 Structural analysis of co-integrating VARs *Journal of Economic Survey* **12** (5) pp 471 – 505.

5317. Portes R, Rey H 1998 The emergence of the Euro as an international currency *Economic Policy* **26** pp 307 – 332.

5318. Rey H 2001 International trade and currency exchange *Review of Economic Studies* **68** pp 443 – 464.

5319. Reiss P, Werner I 1998 Does risk sharing motivate interdealer trading? *Journal of Finance* **53** pp 1657 – 1704.

5320. Sarkar A, Tozzi M (January) 1998 Electronic trading on futures exchanges *Current Issues in Economics and Finance* Federal Reserve Bank of New York **4** (1).

5321. Viswanathan S, Wang J 1998 Why is interdealer trade so pervasive in financial markets? *Working Paper* Duke University North Carolina USA.

5322. Viswanathan S, Wang J D 2000 Inter-dealer trading in financial markets *Working Paper* Duke University Durham North Carolina USA.

5323. Vitale P 1998 Two months in the life of several gilt-edged market makers on the London Stock Exchange *Journal of International Financial Markets, Institutions, & Money* **8** pp 301 – 326.

5324. Vitale P 1999 Sterilized central bank intervention in the foreign exchange market *Journal of International Economics* **49** pp 245 – 267.

5325. Vitale P 2000 Speculative noise trading and manipulation in the foreign exchange market *Journal of International Money and Finance* **19** pp 689 – 712.

5326. Vitale P 2003 Foreign exchange intervention: How to signal policy objectives and stabilize the economy *Journal of Monetary Economics* **50** pp 841 – 870.

5327. Vitale P 2004 A guided tour of the market microstructure approach to exchange rate determination *CEPR Working Paper 4530*.

5328. Vitale P 2006 A market microstructure analysis of foreign exchange intervention *Working Paper series no 629 / MAY 2006* European Central Bank Frankfurt am Main Germany ISSN 1561-0810 (print) ISSN 1725-2806 (online) pp 1 - 59 http://ssrn.com/abstract_id=902528 http://www.ecb.int .

5329. Yao J 1998 Market making in the interbank foreign exchange market *Salomon Center Working Paper #S-98-4* New York University NY USA.

5330. Alberola E, Cervero S G, Lopez H, Ubide A 1999 Global equilibrium exchange rates: Euro, Dollar, "ins", "outs", and other major currencies in a panel cointegration framework *IMF Working Paper WP/99/175* IMF USA.

5331. Bos Th, Fetherstone Th A (editors) 1999 Advances in pacific basin financial markets *JAI Press* Greenwich Connecticut USA.

5332. Carrera J 1999 Speculative attacks to currency target zones: A market microstructure approach *Journal of Empirical Finance* **6** pp 555 – 582.

5333. Chaboud A P, LeBaron B (July)1999 Foreign exchange market trading volume and Federal Reserve intervention *Typescript* Brandeis University, *Journal of Banking and Finance*.

5334. Chaboud A P, LeBaron B 2001 Foreign exchange market trading volume and Federal Reserve intervention *Journal of Futures Markets* **21** pp 851 – 860.

5335. Chaboud A P, Humpage O (January) 2005 An assessment of the impact of Japanese foreign exchange intervention: 1991-2004 *International Finance Discussion Paper 824* Board of Governors of the Federal Reserve System USA.

5336. Chaboud A P, Chernenko S, Wright J 2008 Trading activity and macroeconomic announcements in high-frequency exchange rate data *Journal of the European Economic Association* **6** pp 589 – 596.

5337. Chaboud A P, Chiquoine B, Hjalmarsson E, Vega C 2009 Rise of the machines: Algorithmic trading in the foreign exchange market *International Finance Discussion Papers 980* Federal Reserve Board of Governors USA.

5338. Chaboud A P, Chiquoine B, Hjalmarsson E, Loretan M 2009 Frequency of observation and the estimation of integrated volatility in deep and liquid financial markets *Federal Reserve Board of Governors* USA.

5339. Fiess N, MacDonald R 1999 Technical analysis in the foreign exchange market: A cointegration-based approach *Multinational Finance Journal* **3** (3) pp 147 – 172.

5340. Fiess N, MacDonald R 2002 Towards the fundamentals of technical analysis: Analyzing the information content of high, low and close prices *Economic Modelling* **19** (3) pp 353 – 374.

5341. Fleming M, Lopez J 1999 Heat waves, meteor showers, and trading volume: An analysis of volatility spillovers in the US Treasury market *Federal Reserve Bank of New York Staff Reports #82* NY USA.

5342. Freihube Th, Kehr C-H, J. Krahnen J, Theissen E 1999 Was leisten die kursmakler? Eine empirische untersuchung am beispiel der Frankfurter wertpapierbörse *Kredit und Kapital* **32** pp 426 – 460.

5343. Grammig J, Schiereck D, Theissen E 1999 Informationsbasierter aktienhandel über *IBIS Working Paper* Johann Wolfgang Goethe-Universität Frankfurt *Zeitschrift für betriebswirtschaftliche Forschung* Germany.

5344. Isard P, Razin A, Rose A (editors) 1999 *IMF* and *Kluwer* The Netherlands.

5345. Jeanne O, Rose A (April) 1999 Noise trading and exchange rate regimes *NBER Working Paper #7104* NBER USA, *Quarterly Journal of Economics*.

5346. Kandel E, Marx L 1999 Payments for order flow on Nasdaq *Journal of Finance* **54** pp 35 – 66.

5347. LeBaron B 1999 Technical trading rule profitability and foreign exchange intervention *Journal of International Economics* **49** pp 125 – 214.

5348. Marks J 1999 The impact of the internet on users and suppliers of financial services *Brookings-Wharton Papers on Financial Services* pp 147 – 185.

5349. Macey J, O'Hara M 1999 Globalization, exchange governance and the future of exchanges *Brookings-Wharton Papers on Financial Services* pp 1 – 32.

5350. Naik N Y, Neuberger A, Viswanathan S 1999 Trade disclosure regulation in markets with negotiated trades *Review of Financial Studies* **12** (4) pp 873 – 900.

5351. Naik N Y, Yadav P 1999 The effects of market reform on trading costs of public investors: Evidence from the London stock exchange *Working Paper* London Business School and University of Strathclyde UK.

5352. Payne R (January) 1999, 2003 Informed trade in spot foreign exchange markets: An empirical investigation *Typescript* London School of Economics and Political Science London UK, *Journal of International Economics* **61** (2) pp 307 – 329.

5353. Payne R, Vitale P 2003, A transaction level study of the effects of central bank intervention on exchange rates *Journal of International Economics* **61** pp 331 – 352.

5354. Moore M J, Payne R 2011 On the sources of private information in FX markets *Journal of Banking and Finance* **35** (5) pp 1250 – 1262.

5355. Love R, Payne R 2004, 2008 Macroeconomic news, order flows, and exchange rates *Typescript* London School of Economics and Political Science London UK, *Journal of Financial and Quantitative Analysis* **43** pp 467 – 488.

5356. Rigobon R (September) 1999 On the measurement of the international propagation of shocks *NBER Working Paper 7354* NBER USA.

5357. Saar G (July) 1999 Demand uncertainty and the information content of order flow *Typescript Johnson School* Cornell University NY USA.

5358. Scalia A, Vacca V (October) 1999 Does market transparency matter? A case study *Banca d'Italia Discussion Paper 359* Bank of Italy Rome Italy.

5359. Scalia A (August) 2004 Is foreign exchange intervention effective? Some micro-analytical evidence from Central Europe *Typescript* Bank of Italy Rome Italy.

5360. Scalia A 2008 Is foreign exchange intervention effective? Some micro-analytical evidence from the Czech Republic *Journal of International Money and Finance* **27** (4) pp 529 – 546.

5361. Shapiro C, Varian H 1999 Information rules *Harvard Business School Press* Harvard University USA.

5362. Theissen E 1999 Floor versus screen trading: Evidence from the German stock market *Département Finance et Economie Groupe HEC* France, *Johann Wolfgang Goethe-Universität* Frankfurt/Main Frankfurt Germany pp 1 – 37.

5363. Vayanos D 1999 Strategic trading and welfare in a dynamic market *Review of Economic Studies* **66** pp 219 – 234.

5364. Vayanos D 2001 Strategic trading in a dynamic noisy market *Journal of Finance* **56** pp 131 – 171.

5365. Wang J 1999 Asymmetric information and the bid-ask spread: An empirical comparison between automated order execution and open outcry auction *Journal of International Financial Markets, Institutions and Money* **9** pp 115 – 128.

5366. Aliber R Z, Chowdhry Bh, Yan Sh 2000 Transactions costs in the foreign exchange market *University Of Chicago, The Anderson Graduate School of Management UCLA, University of Arizona* USA http://www.escholarship.org/uc/item/4qw3p6rp .

5367. Ausloos M 2000 Statistical physics in foreign exchange currency and stock markets *Physica A* **285** pp 48 – 65.

5368. Baillie R, Humpage O, Osterberg W 2000 Intervention from an information perspective *Journal of International Financial Markets, Institutions and Money* **10** pp 407 – 421.

5369. Carlson J, Osler C 2000 Rational speculators and exchange rate volatility *European Economic Review* **44** pp 231 – 253.

5370. Carlson J (August) 2002 One minute in the life of the DM/$: Public information in an electronic market *Typescript* Purdue University USA.

5371. Ebrahim S K 2000 Volatility transmission between foreign exchange and money markets *Working Paper 2000-16* Bank of Canada Ottawa Canada.

5372. Eichengreen B, Mathieson D J 2000 The currency composition of foreign exchange reserves: Retrospect and prospect *IMF Working Paper 00/131* International Monetary Fund Washington DC USA.

*5373.* Greenspan A (October) 2000 Remarks [on e-finance] *9th Annual Financial Markets Conference of the Federal Reserve Bank of Atlanta* USA.

*5374.* Hüfner F P 2000 The British foreign exchange reserves puzzle *ZEW Discussion Papers no 00-55* ZEW - Zentrum für Europäische Wirtschaftsforschung / Center for European Economic Research http://hdl.handle.net/10419/24403 .

*5375.* Fujiwara I (June) 2000 Liquidity and leverage risk in the Dollar/Yen market *Typescript* Nuffield College Oxford UK.

*5376.* Kanas A 2000 Volatility spillovers between stock returns and exchange rate changes: International evidence *Journal of Business Finance and Accounting* **27** pp 447 – 467.

*5377.* Kaul A, Mehrotra V, Morck R 2000 Demand curves for stock do slope down: New evidence from an index weights adjustment *Journal of Finance* **55** pp 893 – 912.

*5378.* Kim S-J, Kortian T, Sheen J 2000 Identifying central bank intervention and exchange rate volatility – Australian evidence *Journal of International Financial Markets Institutions and Money* vol **10** pp 381 – 405.

*5379.* Kim S-J, Sheen J 2002 The determinants of foreign exchange intervention by central banks: Evidence from *Australia Journal of International Money and Finance* vol **21** pp 619 – 649.

*5380.* Kim S-J 2003 Monetary policy, foreign exchange intervention, and the exchange rate in a unifying framework *Journal of International Economics* **60** pp 355 – 386.

*5381.* Lane Ph R, and Milesi-Ferretti G M 2000 The transfer problem revisited: Net foreign assets and real exchange rates *Hong Kong Institute for Monetary Research Working Paper 6/2000*.

*5382.* Lo Ch-K (editor) 2000 Financial markets in Hong Kong *Springer* Singapore.

*5383.* Lee C, Swaminathan B 2000 Price momentum and trading volume *The Journal of Finance* **55** pp 2017 – 2069.

*5384.* Ma Y, Kanas A 2000 Testing nonlinear relationship among fundamentals and exchange rates in the ERM *Journal of International Money and Finance* **19** (1) pp 135 – 152.

*5385.* Ma Y, Tsang Sh-K, Yiu M S, Wai-Yip Alex Ho 2010 A target-zone model with two types of assets *Working Paper* Hong Kong Institute for Monetary Research Hong Kong P R China.

*5386.* Martin A D 2000 Exchange rate exposure of the key financial institutions in the foreign exchange market *International Review of Economics and Finance* **9** (3) pp 267 – 286.

*5387.* Martin A D, Mauer L J 2003 Exchange rate exposures of US banks: A cash flow-based methodology *Journal of Banking and Finance* **27** (5) pp 851 – 865.

*5388.* Martin A D, Mauer L J 2005 A note on common methods used to estimate foreign exchange exposure *Journal of International Financial Markets, Institutions & Money* **15** (2) pp 125 – 140.

*5389.* McCallum B (April) 2000 Theoretical analysis regarding a zero lower bound on nominal interest rates *NBER Working Paper no 7677* NBER USA.

*5390.* Melvin M, Yin X 2000 Public information arrival, exchange rate volatility, and quote frequency *Economic Journal* **110** pp 644 – 661.

5391. Melvin M, Melvin B P 2003 The global transmission of volatility in the foreign exchange market *The Review of Economics and Statistics* **85** pp 670 – 679.

5392. Melvin M, Taylor M P 2009 The crisis in the foreign exchange market *Journal of International Money and Finance* **28** (8) pp 1317 – 1330.

5393. Naranjo A, Nimalendran M 2000 Government intervention and adverse selection costs in foreign exchange markets *Review of Financial Studies* **13** pp 453 – 477.

5394. Ng A 2000 Volatility spillover effects from Japan and the US to the Pacific-Basin *Journal of International Money and Finance* **19** pp 207 – 233.

5395. Ramaswamy R, Samiei H (June) 2000 The Yen-Dollar rate: Have interventions mattered? *IMF Working Paper no 00/95* IMF USA.

5396. Rime D (March) 2000 Private or public information in foreign exchange markets? An empirical analysis *Typescript* University of Oslo Norway www.uio.no/~dagfinri .

5397. Rime D 2001 Private or public information in foreign exchange markets? An empirical analysis *Typescript Central Bank of Norway* Oslo Norway.

5398. Rime D 2003 New electronic trading systems in the foreign exchange markets *in* New economy handbook Jones D C (editor) chap 21 *Academic Press* San Diego USA pp 471 – 504.

5399. Akram Q, Rime D, Sarno L (February) 2005 Arbitrage in the foreign exchange market: Turning on the microscope *Working Paper no 2005/12* Norges Bank Norway.

5400. Rime D, Sarno L, Sojli E 2006 Exchange rate dynamics and order flow: A step beyond *Typescript* Warwick University UK.

5401. Rime D, Sarno L, Sojli E 2007 Exchange rate forecasting, order flow and macroeconomic information *Working Paper no 2* Norges Bank Norway.

5402. Rime D, Sarno L, Sojli E 2010 Exchange rate forecasting, order flow and macroeconomic information *Journal of International Economics* **80** (1) pp 72 – 88.

5403. Schwartz A J 2000 The rise and fall of foreign exchange market intervention *NBER Working Paper W7751* National Bureau of Economic Research Cambridge MA USA.

5404. US General Accounting Office (May) 2000 On-line trading: Better investor protection information needed on brokers' web sites *US General Accounting Office* USA.

5405. Allayannis G, Ofek E 2001 Exchange rate exposure, hedging, and the use of foreign currency derivatives *Journal of International Money and Finance* **20** pp 273 – 296.

5406. Anderson H, Vahid F 2001 Market architecture and nonlinear dynamics of Australian stock and futures indices *Australian Economic Papers* **40** (4) pp 541 – 566.

5407. Brandt M W, Edelen R, Kavajecz K A 2001 Liquidity in the US treasury market: Asymmetric information and inventory effects *Manuscript* Department of Finance The Wharton School University of Pennsylvania USA.

5408. Brown G W 2001 Managing foreign exchange risk with derivatives *Journal of Financial Economics* **60** pp 401 – 448.

5409. Cai J, Cheung Y - L, Lee R S K, Melvin M 2001 Once-in-a-generation Yen volatility in 1998: Fundamentals, intervention, and order flow *Journal of International Money and Finance* **20** (3) pp 327 – 347.

5410. Claessens S, Forbes K 2001 International and financial contagion *Springer*.

*5411.* Clark T, McCracken M 2001 Evaluating long-horizon forecasts *Working Paper no 01-14* Federal Reserve Bank of Kansas City USA.

*5412.* Collins S, Rodrik D (editors) 2001 Brookings Trade Forum 2001 *Brookings Institution Press* Washington DC USA.

*5413.* Corsetti G, Pesenti P, Roubini N (May) 2001 Does one Soros make a difference? The role of a large trader in currency crises *NBER Working Paper 8303* NBER USA, *Review of Economic Studies*.

*5414.* Coval J D, Shumway T 2001 Is sound just noise? *The Journal of Finance* **56** pp 1887 – 1910.

*5415.* Croushore D, Stark T 2001 A real-time data set for macroeconomists *Journal of Econometrics* **105** pp 111 – 130.

*5416.* Dacorogna M M, Gencay R, Mueller U A, Olsen R B, Pictet O V 2001 An introduction to high-frequency finance *Academic Press* San Diego CA USA.

*5417.* D'Souza C (March) 2001 A market-microstructure analysis of FX intervention in Canada *Working Paper* Financial Markets Division Bank of Canada Ottawa Canada.

*5418.* Duarte M, Stockman A (July) 2001 Rational speculation and exchange rates *NBER Working Paper 8362* NBER USA.

*5419.* Galati G 2001 Why has global FX turnover declined? Explaining the 2001 triennial survey *BIS Quarterly Review* (December) pp 39 – 47.

*5420.* Griffin J M, Stulz R M 2001 International competition and exchange rate shocks: A cross-country industry analysis of stock returns *Review of Financial Studies* **14** pp 215 – 241.

*5421.* Guembel A, Sussman O 2001 Optimal exchange rates: A market-microstructure approach *Typescript* Said Business School Oxford University Oxford UK.

*5422.* Hong Y 2001 A test for volatility spillover with application to exchange rates *Journal of Econometrics* **103** pp 183 – 224.

*5423.* Lane P 2001 The new open-economy macroeconomics: A survey *Journal of International Economics* **54** pp 235 – 266.

*5424.* Montgomery J D, Popper H A 2001 Information sharing and central bank intervention in the foreign exchange market *Journal of International Economics* **55** pp 295 – 316.

*5425.* Moore M, Roche M (May) 2001 Liquidity in the forward exchange market *Journal of Empirical Finance* **8** pp 157 – 170.

*5426.* Moore M, Roche M 2002 Less of a puzzle: A new look at the forward forex market *Journal of International Economics* **58** pp 387 – 411.

*5427.* Rey H 2001 International trade and currency exchange *Review of Economic Studies* **68** pp 443 – 464.

*5428.* Sato S, Hawkins J 2001 Electronic finance: An overview of the issues *BIS Paper no 7* Switzerland.

*5429.* Sinn H, Westermann F (July) 2001 Why has the euro been falling? An investigation into the determinants of the exchange rate *NBER Working Paper 8352* NBER USA.

*5430.* Tse Y, Zabotina T 2001 Transaction costs and market quality: Open outcry versus electronic trading *Journal of Futures Markets* **21** (8) pp 713 – 735.

**5431.** Williamson R 2001 Exchange rate exposure and competition: Evidence from the automotive industry *Journal of Financial Economics* **59** pp 441 – 475.

**5432.** Yamaguchi Y 2001 The implications of electronic trading in financial markets *Committee on the Global Financial System* Bank for International Settlements ISBN 92-9131-613-X pp 1 – 37.

**5433.** Aguiar M (March) 2002 Informed speculation and the choice of exchange rate regime *Typescript* University of Chicago USA.

**5434.** Beine M et al 2002 Central bank intervention and foreign exchange rates: New evidence from FIGARCH estimations *Journal of International Money and Finance* **21**.

**5435.** Cavallo M, Perri F, Roubini N, Kisselev K (March) 2002 Exchange rate overshooting and the costs of floating *Typescript* New York University.

**5436.** Chari V, Kehoe P, McGrattan E 2002 Can sticky price models generate volatile and persistent real exchange rates? *Review of Economic Studies* **69** pp 533 – 564.

**5437.** Chari A 2006 Heterogeneous market making in foreign exchange markets: Evidence from individual bank responses to central bank interventions *Journal of Money, Credit, and Banking*.

**5438.** Chordia T, Roll R, Subrahmanyam A 2002 Order imbalance, liquidity, and market returns *Journal of Financial Economics* **65** (1) pp 111 – 130.

**5439.** Daníelsson J, Payne R 2002 Real trading patterns and prices in spot foreign exchange markets *Journal of International Money and Finance* **21** (2) pp 203 – 222.

**5440.** Danielsson J, Payne R, Luo J (July) 2002 Exchange rate determination and inter-market order flow effects *Typescript* Financial Markets Group London School of Economics and Political Science London UK.

**5441.** Daníelsson J, Love R 2006 Feedback trading *International Journal of Finance and Economics* **11** (1) pp 35 – 53.

**5442.** Daníelsson J, Payne R 2011 Liquidity determination in an order-driven market *European Journal of Finance*.

**5443.** Deutsche Bundesbank (January) 2002 Capital flows and the exchange rate *Deutsche Bundesbank Monthly Report* pp 15 – 26.

**5444.** Doyne F J, Joshi Sh 2002 The price dynamics of common trading strategies *Journal of Economic Behavior and Organization* **49** pp 149 – 171.

**5445.** Fatum R, Hutchison M 2002 Is foreign exchange intervention an alternative to monetary policy? Evidence from Japan *Working Paper 02-11* Department of Economics University of Copenhagen Denmark.

**5446.** Fatum R, King M R 2005 Rules versus discretion in foreign exchange intervention: Evidence from official Bank of Canada high-frequency data *Working Paper 05-06* Department of Economics University of Copenhagen Denmark.

**5447.** King M R, Sarno L, Sojli E 2010 Timing exchange rates using order flow: The case of the Loonie *Journal of Banking and Finance* **34** (12) pp 2917 – 2928.

**5448.** King M R, Rime D 2010 The $4 trillion question: What explains FX growth since the 2007 survey? *BIS Quarterly Review* **4** pp 27 – 42.

**5449.** King M R, Mallo C 2010 A user's guide to the Triennial Central Bank Survey of foreign exchange market activity *BIS Quarterly Review* **4** pp 71 – 83.

*5450.* King M R, Osler C, Rime D 2011 Instruments, players and the foreign exchange trading environment' *in* The handbook of exchange rates James J, Marsh I W, Sarno L (editors) *John Wiley & Sons Inc* USA.

*5451.* King M R, Osler C, Rime D 2011 Foreign exchange market structure, players and evolution *Working Paper no 2011 / 10* Norges Bank Oslo Norway ISSN 1502-8143 ISBN 978-82-7553-616-5 pp 1 – 47.

*5452.* King M R, Osler C, Rime D 2012 The market microstructure approach to foreign exchange: Looking back and looking forward *Working Paper 2012 / 54* Ivey Business School University of Western Ontario, Brandeis International Business School Brandeis University, Norges Bank Oslo Norway pp 1 – 40.

*5453.* Kantelhardt J, Zschiegner St, Koscielny-Bunde E, Havlin Sh, Bunde A, Stanley E 2002 Multifractal de-trended fluctuation analysis of nonstationary time series *Physica A* **316** pp 87 – 114.

*5454.* Galati G 2002 Settlement risk in foreign exchange markets and CLS Bank *BIS Quarterly Review* **4** pp 55 – 65.

*5455.* Girardin E, Horsewood N 2002 New transmission mechanisms and instruments of monetary policy at near zero interest rates: The case of Japan in the 1990s *in* Essays in honour of Maxwell J. Fry Dickinson D, Mullineux A (editors) *Edward Elgar* UK.

*5456.* Huang R, Cai J, Wang X 2002 Information-based trading in the interdealer market *Journal of Financial Intermediation* **11** pp 269 – 296.

*5457.* Jeanne O, Rose A 2002 Noise trading and exchange rate regimes *Quarterly Journal of Economics* **117** pp 537 – 569.

*5458.* Kaul A, Mehrotra V (June) 2002 Ticker or trade? How prices adjust in international markets *Typescript* University of Alberta Edmonton Alberta Canada.

*5459.* Obadan M I 2002 Towards exchange rate stability in Nigeria *The Year 2002 One-Day Seminar of the Nigerian Economic Society (NES)* Federal Palace Hotel Lagos Nigeria.

*5460.* Ryan S, Worthington A C 2002 Time-varying market, interest rate and exchange rate risk in Australian Bank portfolio stock returns: A GARCH-M approach *Discussion Papers and Working Paper Series no 112* School of Economics and Finance Queensland University of Technology Brisbane Queensland Australia.

*5461.* Abreu D, Brunnermeier M K 2003 Bubbles and crashes *Econometrica* **71** (1) pp 173 – 204.

*5462.* Aliber R Z, Chowdry B, Yan S 2003 Some evidence that a Tobin tax on foreign exchange transactions may increase volatility *European Finance Review* **7** pp 481 – 510.

*5463.* Bacchetta P, van Wincoop E (February) 2003 Can information dispersion explain the exchange rate disconnect puzzle? *NBER Working Paper 9498* NBER USA.

*5464.* Bergsten C F, Williamson J (editors) 2003 Dollar overvaluation and the World economy *Institute for International Economics* Washington DC USA.

*5465.* Bergsten C F, Williamson J (editors) 2005 Dollar adjustment: How far? Against what? *Institute of International Economics* Washington DC USA.

*5466.* Bodnar G M, Wong M H F 2003 Estimating exchange rate exposures: Issues in model structure *Financial Management* **32** pp 35 – 67.

5467. Burstein A T, Neves J, Rebelo S (September) 2003 Distribution costs and real exchange rate dynamics during exchange rate-based stabilizations *Journal of Monetary Economics* vol **50** (6) pp 1189 – 1214.

5468. Carpenter A, Wang J (January) 2003 Sources of private information in FX trading *Typescript* University of New South Wales Sydney Australia.

5469. Choi C, Baek S-G 2004 Exchange rate regimes and international reserves *Working Paper*.

5470. Derviz A 2003 Asset return dynamics and the FX risk premium in a decentralized dealer market *Typescript* Czech National Bank, *European Economic Review*.

5471. Dominguez K M E 2003 The market microstructure of central bank intervention *Journal of International Economics* **59** pp 25 – 45.

5472. Dominguez K M E, Panthaki F 2006 What defines "News" in foreign exchange markets? *Journal of International Money and Finance* **25** pp 168 – 198.

5473. Doukas J A, Hall P H, Lang L H P 2003 Exchange rate exposure at the firm and industry level *Financial Markets, Institutions & Instruments* **12** (5) pp 291 – 347.

5474. Fatum R, Hutchison M M 2003 Is sterilized foreign exchange intervention effective after all? An event study approach *Economic Journal* **113** pp 390 – 411.

5475. Fatum R, Hutchison M M 2006 Effectiveness of official daily foreign exchange market operations in Japan *Journal of International Money and Finance* **25** pp 199 – 219.

5476. Faust J, Rogers J H, Wright J H 2003 Exchange rate forecasting: The errors we've really made *Journal of International Economics* **60** (1) pp 35 – 59.

5477. Gordon M 2003 Estimates of time-varying term premia for New Zealand and Australia *Discussion Paper Series DP2003/06* Reserve Bank of New Zealand New Zealand.

5478. Humpage O F 2003 Government intervention in the foreign exchange market *Federal Reserve Bank of Cleveland Working Paper 0315* USA.

5479. Koutmos G, Martin A D 2003 Asymmetric exchange rate exposure: theory and evidence *Journal of International Money and Finance* **22** (3) pp 365 – 384.

5480. Laurenceson J, Chai J C H 2003 Financial reform and economic development in China *Edward Elgar* Cheltenham UK.

5481. Mathisen J 2003 Estimation of the equilibrium real exchange rate for Malawi *IMF Working papers 03/104* IMF USA.

5482. Okunev J, White D 2003 Do momentum-based strategies still work in foreign currency markets? *Journal of Financial and Quantitative Analysis* **38** pp 425 – 447.

5483. Peng W, Shu Ch, Chow K (May) 2003 The Yen exchange rate and net foreign assets *Research Department* Economic Research Division Hong Kong Monetary Authority pp 1 – 19.

5484. Rogers J M, Siklos P L 2003 Foreign exchange market intervention in two small open economies: The Canadian and Australian experience *Journal of International Money and Finance*.

5485. Spiegel M 2003 Japanese foreign exchange intervention *Federal Reserve Bank of San Francisco Newsletter* 2003-36 San Francisco USA.

*5486.* Westerhoff F H 2003 Market-maker, inventory control and foreign exchange dynamics *Quantitative Finance* **83** (3) pp 363 – 369.

*5487.* Wright J H 2003 Bayesian model averaging and exchange rate forecasts *International Finance Discussion Papers no 779* Board of Governors of the Federal Reserve System USA.

*5488.* Aitken M, Frino A, Hill A, Jarnecic E 2004 The impact of electronic trading on bid - ask spreads: Evidence from futures markets in Hong Kong, London, and Sydney *Journal of Futures Markets* **24** (7) pp 675 – 696.

*5489.* Anwar T 2004 Recent macroeconomic developments and implications for poverty and employment in Pakistan: The cost of foreign exchange reserve holdings in South Asia Australia South Asia *ASARC Working Paper 2004-14 2* Research Centre The Research School of Pacific & Asian Studies The National Institute of Economics and Business Australian National University Canberra Australia pp 1 – 23.

*5490.* Bacchetta P, van Wincoop E (January) 2004 A scapegoat model of exchange rate fluctuation *NBER Working Paper 10245* NBER USA.

*5491.* Bacchetta P, van Wincoop E 2006 Can information dispersion explain the exchange rate disconnect puzzle? *American Economic Review* **93** pp 552 – 576.

*5492.* Bartram S M 2004 Linear and nonlinear foreign exchange rate exposures of German nonfinancial corporations *Journal of International Money and Finance* **23** (4) pp 673 – 699.

*5493.* Bartram S M, Bodnar G M 2004 The foreign exchange exposure puzzle *Johns Hopkins University Working Paper* USA.

*5494.* Bartram S M, Brown G, Minton B 2005 Resolving the exposure puzzle: The many facets of exchange rate exposure *Working Paper* Lancaster University UK, University of North Carolina at Chapel Hill, Ohio State University USA.

*5495.* Bartram S M, Karolyi G A 2006 The impact of the introduction of the Euro on foreign exchange rate risk exposures *Journal of Empirical Finance* **13** (4-5) pp 519 – 549.

*5496.* Bhanumurthy N R 2004 Microstructures in the Indian foreign exchange market *Working Paper* Delhi University India.

*5497.* Brandt M W, Kavajecz K A 2004 Price discovery in the US treasury market: The impact of order flow and liquidity on the yield curve *Journal of Finance* **59** (6) pp 2623 – 2654.

*5498.* Breedon F, Vitale P 2004 An empirical study of information and liquidity effects of order flow on exchange rates *CEPR Working Paper 4586*.

*5499.* Cashin P, Cespedes L, Sahay R 2004 Commodity currencies and the real exchange rate *Journal of Development Economies* vol **75** pp 239 – 268.

*5500.* De Wet W A, Gebreselasie T G 2004 The exchange rate exposure of major commercial banks in South Africa *The African Finance Journal* **6** (2) pp 21 – 35.

*5501.* Dunne P, Hau H, Moore M (November) 2004 Macroeconomic order flows: Explaining equity and exchange rate returns *Typescript*.

*5502.* Fratzscher M 2004 Exchange rate policy strategies and foreign exchange interventions in the group of three economies in Dollar adjustment: How far? Against what? Bergsten C F, Williamson J (editors) *Institute for International Economics* Washington DC USA.

5503. Hahm J H 2004 Interest rate and exchange rate exposures of banking institutions in pre-crisis Korea *Applied Economics* **36** (13) pp 1409 – 1419.

5504. Hui Ch-H, Neely Ch J, Higbee J (November) 2004, (June) 2007 Foreign exchange volatility is priced in equities *Working Paper 2004-029F* Research Division Federal Reserve Bank of St Louis MO USA http://research.stlouisfed.org/wp/2004/2004-029.pdf pp 1 – 39.

5505. Hui Ch-H, Yeung V, Fung L, Lo Ch-F 2007 Valuing foreign currency options with a mean-reverting process: a study of Hong Kong dollar *Working Paper 08/2007* Research Department Market Research Division Hong Kong Monetary Authority Hong Kong P R China, Physics Department The Chinese University of Hong Kong P R China pp 1 – 28.

5506. Hui Ch-H, Fong T 2007 Is the Hong Kong dollar exchange rate "bounded" in the convertibility zone? *Working Paper 13/2007* Research Department Market Research Division Hong Kong Monetary Authority Hong Kong P R China pp 1 - 14.

5507. Hui Ch-H, Genberg H, Chung T-K 2009 Liquidity, risk appetite and exchange rate movements during the financial crisis of 2007-2009 *Working Paper 11/2009* Research Department Market Research Division Hong Kong Monetary Authority Hong Kong P R China pp 1 – 23.

5508. Kim K, Yoon S-M 2004 Multifractal features of financial markets *Physica A* **344** pp 272 – 278.

5509. Nagayasu J 2004 The effectiveness of Japanese foreign exchange interventions during 1991-2001 *Economics Letters* **84** pp 377 – 381.

5510. National Bank of Poland 2004 Turnover in the Polish foreign exchange and OTC derivatives markets in April 2004 Result Summary *National Bank of Poland* Warsaw Poland.

5511. National Bank of Poland 2007 Turnover in the Polish foreign exchange and OTC derivatives markets in April 2007 Result Summary *National Bank of Poland* Warsaw Poland.

5512. Reinhart C M, Rogoff K S 2004 The modern history of exchange rate arrangements: A reinterpretation *Quarterly Journal of Economics* **119** (1) pp 1 – 48.

5513. Rigobon R, Sack B 2004 The impact of monetary policy on asset prices *Journal of Monetary Economics* **51** pp 1553 – 1575.

5514. Simatele M 2004 Foreign exchange intervention and the exchange rate in Zambia *Economics Studies Goteborg University* Sweden.

5515. Akram Q F, Rime D, Sarno L 2005 Arbitrage in the foreign exchange market: Turning on the microscope *Working Paper ANO 2005/12* ISSN 0801-2504 (printed) 1502-8143 (online) Norges Bank Oslo Norway.

5516. Ates A, Wang G H 2005 Information transmission in electronic versus open-outcry systems: An analysis of US equity index futures markets *Journal of Futures Markets* **25** (7) pp 679 – 715.

**5517.** Bauwens L, Omrane W B, Giot P 2005 News announcements, market activity and volatility in the euro/dollar foreign exchange market *Journal of International Money and Finance* **24** (7) pp 1108 – 1125.

**5518.** Campa J M, Goldberg L S 2005 Exchange rate pass through into import prices *Review of Economics and Statistics November* **87** (4) pp 679 – 690.

**5519.** Campa J M, Goldberg L S 2006a Distribution margins, imported inputs, and the sensitivity of the CPI to exchange rates *NBER Working Paper no 12121* NBER USA.

**5520.** Campa J M, Goldberg L S 2006b Pass-through of exchange rates to consumption prices: What has changed and why? *Working Paper* Federal Reserve Bank of New York USA.

**5521.** Chui M, Gerlach S, Yu I W 2005 The recent appreciation of the Hong Kong dollar *BIS Papers* **23** pp 150 – 155.

**5522.** DeGrauwe P (editor) 2005 Exchange rate economics: Where do we stand? *MIT Press* Cambridge USA.

**5523.** Dueker M, Neely Ch J 2005 Can Markov switching models predict excess foreign exchange returns? *Journal of Banking and Finance*.

**5524.** Eichengreen B 2005 Sterling's past, Dollar's future: Historical perspectives on reserve currency competition *NBER Working Paper no 11336* National Bureau of Economic Research Cambridge MA USA.

**5525.** Fung J K W, Lien D, Tse Y M, Tse Y K 2005 Effects of electronic trading on the Hang Seng index futures market international *Review of Economics and Finance* **14** (4) pp 415 – 425.

**5526.** Hau H, Rey H 2005 Exchange rates, equity prices and capital flows *Review of Financial Studies*.

**5527.** Hull J C (2005-2006) Private communications on the electronic trading strategies in the foreign currencies exchange markets *Electronic Trade Laboratory Rotman School of Management* University of Toronto Canada.

**5528.** Hull J C 2012a Options, futures, and other derivatives *Prentice Hall* 8th edition ISBN: 0-13-216484-9 USA pp 1 – 816.

**5529.** Hull J C 2012b Risk management and financial institutions *John Wiley and Sons Inc* 3$^{rd}$ edition ISBN: 978-1-1182-6903-9 USA pp 1 – 672.

**5530.** Hull J C 2010 Fundamentals of futures and options markets *Prentice Hall* 7th edition ISBN-10: 0136103227 ISBN-13: 978-0136103226 USA pp 1 – 624.

**5531.** Inoue A, Kilian L 2005 In-sample or out-of-sample tests of predictability: Which one should we use? *Econometric Reviews* **23** (4) pp 371 – 402.

**5532.** Marsh I W, O'Rourke C 2005 Customer order flow and exchange rate movements: Is there really information content? *Working Paper* Cass Business School UK.

**5533.** Newsome J 2006 La criée ou le tout électronique: Le marché décide *Revue D'économie Financière* pp 1 – 5.

**5534.** Vaubel R 2005 Foreign exchange accumulation by emerging and transition economies: An explanation and critique *in* Aspekte der internationalen ökonomie El-Shagi M, Rübel G (editors) *Wiesbaden* Gabler Germany pp 77 – 84.

**5535.** Yu I-W, Fung L, Hongyi Ch (November) 2005 Exchange rate risk premiums in Hong Kong dollar: A signal-extraction approach *Research Department* Economic Research Division Hong Kong Monetary Authority pp 1 – 16.

**5536.** Alexander K, Barbosa A 2006 The impact of electronic trading and exchange traded funds on the effectiveness of minimum variance hedging *ICMA Centre Discussion Paper in Finance DP2006-04* University of Reading UK pp 1 – 23.

**5537.** Bacchetta P, van Wincoop E 2006 Can information heterogeneity explain the exchange rate determination puzzle? *American Economic Review* **96** (3) pp 552 – 576.

**5538.** Bayoumi T, Lee J, Jayanthi S 2006 New rates from new weights IMF Staff Papers vol **53** no 2 *IMF* Washington USA.

**5539.** Boyen M M, Van Norden S 2006 Exchange rates and order flow in the long run *Finance Research Letters* **3** (4).

**5540.** Cai F, Howorka E, Wongswan J 2006 Transmission of volatility and trading activity in the global interdealer foreign exchange market: Evidence from electronic broking services (EBS) data *International Finance Discussion Papers* Board of Governors of the Federal Reserve System USA.

**5541.** Cai F, Howorka E, Wongswan J 2008 Informational linkages across trading regions: Evidence from foreign exchange markets *Journal of International Money and Finance* **27** (8) pp 1215 – 1243.

**5542.** Cao H H, Evans M D D, Lyons R K 2006 Inventory information *Journal of Business* **79** (1) pp 325 – 364.

**5543.** Carlson J A, Lo M 2006 One minute in the life of the DM/US$: Public news in an electronic market *Journal of International Money and Finance* **25** (7) pp 1090 – 1102.

**5544.** Charlebois M, Sapp St 2006 Temporal patterns in foreign exchange returns and options *Richard Ivey School of Business* University of Western Ontario Canada.

**5545.** Chu C, Mo Y K, Wong G, Lim P 2006 Financial integration in Asia *Hong Kong Monetary Authority Quarterly Bulletin* **49**.

**5546.** Gilbert C, Rijken H 2006 How is futures trading affected by the move to a computerized trading system? Lessons from the LIFFE FTSE 100 contract *Journal of Business Finance and Accounting* pp 1 – 31.

**5547.** Jeon J, Oh Y, Yang D Y 2006 Financial market integration in East Asia: Regional or global *Asian Economic Papers* **5** (1) pp 73 – 89.

**5548.** Escribano A, Pascual R 2006 Asymmetries in bid and ask responses to innovations in the trading process *Empirical Economics* **30** pp 913 – 946.

**5549.** Kaul A, Sapp S 2006 Y2K fears and safe haven trading of the US dollar *Journal of International Money and Finance* **25** (5) pp 760 – 779.

**5550.** Killeen W P, Lyons R K, Moore M J 2006 Fixed versus flexible: Lessons from EMS order flow *Journal of International Money and Finance* **25** (4) pp 551 – 579.

**5551.** Kim S, Lee J W, Shin K 2006 Regional and global financial integration in East Asia *Working Paper Series 0602* Institute of Economic Research Korea University South Korea.

**5552.** Kočenda E, Valachy J 2006 Exchange rate volatility and regime change: Visegrad comparison *Journal of Comparative Economics* **34** (4) pp 727 – 753.

*5553.* Kočenda E, Kutan A M, Yigit T 2008 Fiscal convergence in the European Union *North-American Journal of Economics and Finance* **19** (3) pp 319 – 330.

*5554.* Kočenda E, Poghosyan T 2009 Macroeconomic sources of foreign exchange risk in new EU members *Journal of Banking and Finance* **33** (11) pp 2164 – 2173.

*5555.* LeBaron B 2006 Agent-based computational finance vol **2** Tesfatsion L, Judd K L (editors) *North-Holland Publishing Company / Elsevier* Amsterdam The Netherlands.

*5556.* Mende A 2006 09/11 on the USD/EUR foreign exchange market *Applied Financial Economics* **16** (3) pp 213 – 222.

*5557.* Mende A, Menkhoff L 2006 Profits and speculation in intra-day foreign exchange trading *Journal of Financial Markets* 9 (3) pp 223 – 245.

*5558.* Muller A, Verschoor W F C 2006 Foreign exchange risk exposure: Survey and suggestions *Journal of Multinational Financial Management* **16** (4) pp 385 – 410.

*5559.* Norouzzadeh P, Rahmani B A 2006 Multifractal de-trended fluctuation description of Iranian-US dollar exchange rate *Physica A* **367** pp 328 – 336.

*5560.* Pelham M 2006 Automation creates level playing field for FX traders *The Banker* (August) pp 30 – 33.

*5561.* Rodrik D 2006 The social cost of foreign exchange reserves *NBER Working Paper no 11952* National Bureau of Economic Research Cambridge MA USA.

*5562.* Sager M J, Taylor M P 2006 Under the microscope: The structure of the foreign exchange market *International Journal of Finance and Economics* **11** (1) pp 81 – 95.

*5563.* Starks L T, Wei K D 2006 Foreign exchange rate exposure and short-term cash flow sensitivity *Working Paper* University of Texas TX USA.

*5564.* Tabak B, Cajueiro D 2006 Assessing inefficiency in euro bilateral exchange rates *Physica A* **367** pp 319 – 327.

*5565.* Taylor A, Farstrup A 2006 Active currency management: Arguments, considerations, and performance for institutional investors *CRA Rogers Casey International Equity Research* Darien Connecticut USA.

*5566.* Taylor J B (September 14) 2006 Lessons from the recovery from the 'lost decade' in Japan: The case of the great intervention and money injection *ESRI international Conference* Cabinet Office Government of Japan Tokyo Japan.

*5567.* Tesfatsion L, Judd K L (editors) 2006 *North-Holland Publishing Company / Elsevier* Amsterdam The Netherlands.

*5568.* Wong A 2006 Analyzing foreign exchange reserve diversification *Institute for International Economics Working Paper* Institute for International Economics Washington DC USA.

*5569.* Adebiyi M A (July) 2007 An evaluation of foreign exchange intervention and monetary aggregates in Nigeria (1986 - 2003) *Department of Economics* University of Lagos Nigeria *MPRA Paper no 3817* pp 1 – 21 http://mpra.ub.uni-muenchen.de/3817/ .

*5570.* Barker W 2007 The global foreign exchange market: Growth and transformation *Bank of Canada Review* (Autumn) pp 3 – 12.

*5571.* Bhansali V 2007 Volatility and the carry trade *Journal of Fixed Income* **17** (3) pp 72 - 84.

5572. Broz J L, Frieden J, Weymouth S 2007 Exchange rate policy attitude: Direct evidence from survey data *IMF* Washington USA http://www.imf.org .

5573. Burnside C, Eichenbaum M S, Rebelo S 2007 The returns to currency speculation in emerging markets *American Economic Review Papers and Proceedings* **97** (2) pp 333 – 338.

5574. Burnside C, Eichenbaum M S, Rebelo S 2009 Understanding the forward premium puzzle: A microstructure approach *American Economic Journal: Macroeconomics* **1** (2) pp 127 – 154.

5575. Burnside C 2012 Carry trades and risk *in* The handbook of exchange rates James J, Marsh I W, Sarno L (editors) *John Wiley & Sons Inc.* USA.

5576. Canto B, Kräussl R 2007 Electronic trading systems and intraday non-linear dynamics: An examination of the FTSE 100 cash and futures returns *CFS Working Paper no 2007/20* Center for Financial Studies an der Johann Wolfgang Goethe-Universität Frankfurt am Main Leibniz Information Centre for Economics Germany pp 1 – 45 http://hdl.handle.net/10419/25521 www.ifk-cfs.de www.econstor.eu .

5577. Chi J, Tripe D, Young M (24-25 September) 2007 Do exchange rate affect the stock performance of Australian Banks? 12[th] Finsia-Melbourne Centre for Financial Studies Banking and Finance Conference Melbourne Australia.

5578. Christodoulou G, O'Connor P 2007 The foreign exchange and over-the-counter derivatives markets in the United Kingdom *Bank of England Quarterly Bulletin* (Q4) pp 548 – 563.

5579. Dreher A, Vaubel R 2007 Foreign exchange intervention and the political business cycle: A panel data analysis *KOF Working Paper 159* KOF Swiss Economic Institute Swiss Federal Institute of Technology Zurich Switzerland pp 1 – 28 www.kof.ethz.ch .

5580. DuCharme M 2007 First steps in foreign exchange transaction cost analysis *Journal of Performance Measurement* pp 19 – 27.

5581. Egstrup R, Fischer B D (4th Quarter) 2007 Foreign exchange and derivatives markets in 2007 Monetary review *Danmarks Nationalbank* Copenhagen Denmark http:// www. natioanlbanken.dk/DNUK/Publications.nsf .

5582. Fleming M J, Mizrach B 2007 The microstructure of a US treasury ECN: The BrokerTec platform pp 1 – 40 http://ssrn.com/abstract=1433488 .

5583. Fung L, Yu I W 2007 Assessing the credibility of the convertibility zone of the Hong Kong dollar *Working Paper 19/2007* Hong Kong Monetary Authority Hong Kong P R China.

5584. Genberg H, He D, Leung F 2007 Recent performance of the Hong Kong dollar linked exchange rate system *Research Note 02/2007* Hong Kong Monetary Authority Hong Kong  P R China.

5585. Genberg H, He D, Leung F 2007 The 'Three refinements' of the Hong Kong dollar linked exchange rate system two years on *Hong Kong Monetary Authority Quarterly Bulletin* **51** pp 5 – 11.

5586. Genberg H, Hui C H 2009 The credibility of the LINK from the perspective of modern financial theory *Working Paper 02/2009* Hong Kong Monetary Authority Hong Kong P R China.

5587. Hong Kong Monetary Authority (December) 2007 The foreign exchange and derivatives markets in Hong Kong *Hong Kong Monetary Authority Quarterly Bulletin* Hong Kong P R China.

5588. Jiang J, Ma K, Cai X 2007 Scaling and correlations in foreign exchange market *Physica A* **375** pp 274 – 280.

5589. Leung F, Ng P 2007 Is the Hong Kong dollar real exchange rate misaligned? *Working Paper 21/2007* Research Department Market Research Division Hong Kong Monetary Authority Hong Kong P R China pp 1 – 31.

5590. Leung F, Ng P 2008 Impact of IPO activities on the Hong Kong dollar interbank market *Working Paper 2008-11* Hong Kong Monetary Authority Hong Kong P R China.

5591. Mitchell M, Pedersen L H, Pulvino T 2007 Slow moving capital *American Economic Review* 97 (2) pp 215 – 220.

5592. Pasquariello P 2007 Informative trading or just costly noise? An analysis of central bank interventions *Journal of Financial Markets* **10** pp 107 – 143.

5593. Sahminan S 2007 Effects of exchange rate depreciation on commercial bank failures in Indonesia *Journal of Financial Stability* **3** (2) pp 175 – 193.

5594. Scarlat E, Stan Cr, Cristescu C 2007 Self-similar characteristics of the currency exchange rate in an economy in transition *Physica A* **370** pp 188 – 198.

5595. Van Wincoop E, Tille C 2007 International capital flows *NBER Working Paper 33* NBER USA.

5596. Wong J, Wong E, Fong T, Choi K-F 2007 Testing for collusion in the Hong Kong banking sector *Working Paper 01/2007* Research Department Market Research Division Hong Kong Monetary Authority Hong Kong P R China pp 1 – 20.

5597. Wong E, Wong J, Leung Ph 2008 The foreign exchange exposure of Chinese banks *Working Paper 07/2008* Research Department Market Research Division Hong Kong Monetary Authority Hong Kong P R China pp 1 – 25.

5598. Yu I-W, Fung L, Tam Ch-S 2007 Assessing financial market integration in Asia equity markets *Working Paper 04/2007* Research Department Market Research Division Hong Kong Monetary Authority Hong Kong P R China pp 1 – 37.

5599. Acemoglu D, Rogoff K, Woodford M (editors) 2008 NBER macroeconomics annual 2008 *University of Chicago Press* Cambridge MA USA.

5600. Baglioni A, Monticini A 2008 The intraday price of money: Evidence from the e-MID market *Journal of Money, Credit and Banking* **40** (7).

5601. Barndorff-Nielsen O E, Hansen P R, Lunde A, Shephard N 2008 Realized kernels in practice: Trades and quotes *Econometrics Journal* **4** pp 1 – 33.

5602. Bartram S M (January) 2008 What lies beneath: Foreign exchange rate exposure, hedging and cash flows *Department of Accounting and Finance* Management School Lancaster University UK MPRA Paper no 6661 pp 1 – 31 http://mpra.ub.uni-muenchen.de/6661/ .

5603. Beaupain R, Durré A 2008 The inter-day and intra-day patterns of the overnight market: Evidence from an electronic platform *ECB Working Paper no 988*.

5604. Berger D W, Chaboud A P, Chernenko S V, Howorka E, Wright J H 2008 Order flow and exchange rate dynamics in electronic brokerage system data *Journal of International Economics* **75** (1) pp 93 – 109.

5605. Brunnermeier M K, Nagel S, Pedersen L H 2008 Carry trades and currency crashes *NBER Macroeconomics Annual 2008* NBER USA.

5606. Burnside A C 2008 Comment on "Carry trades and currency crashes" *in* NBER chapters *NBER Macroeconomics Annual 2008* Acemoglu D, Rogoff K, Woodford M (editors) National Bureau of Economic Research Inc USA.

5607. Burnside A, Eichenbaum M, Kleshchelski I, Rebelo S, Hall L, Hall H 2008 Do Peso problems explain the returns to the carry trade? *NBER Working Papers 14054* NBER USA.

5608. Chinn M D, Moore M J 2008 Private information and the monetary model of exchange rates: Evidence from a novel data set http://www.imf.org/External/NP/seminars/eng/2007/macrofin/index.htm .

5609. Chinn M D, Moore M J 2011 Order flow and the monetary model of exchange rates: Evidence from a novel data set *Journal of Money, Credit and Banking* **43** (8) pp 1599 – 1624.

5610. Gagnon J E, Chaboud A 2008 What Can the Data Tell Us About Carry Trades in Japanese Yen? *FRB International Finance Discussion Paper 899*.

5611. Lam L, Fung L, Yu I-W 2008 Comparing forecast performance of exchange rate models *Working Paper 08/2008* Research Department Market Research Division Hong Kong Monetary Authority Hong Kong P R China pp 1 – 23.

5612. Lien K 2008 Day trading and swing trading the currency market: Technical and fundamental strategies to profit from market moves *John Wiley and Sons* New York USA.

5613. Lindley R 2008 Reducing foreign exchange settlement risk *BIS Quarterly Review* **3** pp 53 – 65.

5614. Liu L-G, Tsang A 2008 Exchange rate pass-through to domestic inflation in Hong Kong *Working Paper 02/2008* Research Department Market Research Division Hong Kong Monetary Authority Hong Kong P R China pp 1 – 23.

5615. Liu Q, Fung H-G, Tse Y 2008 An analysis of price linkages among DJIA index, futures, and exchange-traded fund markets *Review of Futures Markets*.

5616. Lo I, Sapp S G 2008 The submission of limit orders or market orders: The role of timing and information in the Reuters D2000-2 system *Journal of International Money and Finance* **27** (7) pp 1056 – 1073.

5617. Lo I, Sapp S G 2010 Order aggressiveness and quantity: How are they determined in a limit order market? *Journal of International Financial Markets, Institutions and Money* **20** (3)    pp 213 – 237.

5618. Ramadorai T 2008 What determines transaction costs in foreign exchange markets? International *Journal of Finance and Economics* **13** (1) pp 14 – 25.

5619. Sebastião H M C V 2008 The partial adjustment factors of FTSE 100 stock index and stock index futures: The informational impact of electronic trading systems *Working Paper no 7* Faculdade de Economia da Universidade de Coimbra Portugal pp 1 – 48.

5620. Terada T, Higashio N, Iwasaki J 2008 Recent trends in Japanese foreign-exchange margin trading *Bank of Japan Review no 2008-E-3* Tokyo Japan.

5621. Adrian T, Etula E, Shin H S 2009 Risk appetite and exchange rates *Staff Report no 361* Reserve Bank of New York NY USA.

5622. Bacchetta Ph, Mertens E, Van Wincoop E 2009 Predictability in financial markets: What do survey expectations tell us? *Journal of International Money and Finance* **28** (3) pp 406 – 426.

5623. Baba N, Packer F 2009 From turmoil to crisis: Dislocations in the FX swap market before and after the failure of Lehman Brothers *BIS Working Papers 285* Bank for International Settlements Basel Switzerland.

5624. Brunnermeier M K, Nagel S, Pedersen L H 2009 Carry trades and currency crashes *in* NBER macroeconomics annual 2008 vol **23** Acemoglu D, Rogoff K, Woodford M (editors) *University of Chicago Press* Cambridge MA USA pp 313 – 347.

5625. Brunnermeier M, Crockett A, Goodhart C, Persaud A D, Shin H 2009 The fundamental principles of financial regulation *Geneva Reports on the World Economy 11* Preliminary Conference Draft www.voxeu.org/reports/Geneva11.pdf .

5626. Bubák V, Zikes F 2009 Distribution and dynamics of Central-European exchange rates: Evidence from intraday data *Czech Journal of Economics and Finance* **4** pp 334 – 359.

5627. Bubák V, Kočenda E, Žikeš F 2010 Volatility transmission in emerging European foreign exchange markets *CESIFO Working Paper no 3063* pp 1 – 36.

5628. De Zwart G, Markwat T, Swinkels L, van Dijk D 2009 The economic value of fundamental and technical information in emerging currency markets *Journal of International Money and Finance* **28** (4) pp 581 – 604.

5629. Ding L 2009 Bid-ask spread and order size in the foreign exchange market: An empirical investigation *International Journal of Finance and Economics* **14** (1) pp 98 – 105.

5630. Gallardo P, Heath A (March) 2009 Execution methods in foreign exchange markets *BIS Quarterly Review* pp 83 – 91.

5631. Gençay R, Gradojevic N 2009 Informed trading in an electronic foreign exchange market *The Rimini Centre for Economic Analysis* Bologna University Italy pp 1 – 12.

5632. Jiang Zh-Q, Zhou W-X 2009 De-trended fluctuation analysis of inter-trade durations *Physica A* **388** pp 433 – 440.

5633. Hattori M, Shin H S 2009 Yen carry trade and the subprime crisis *IMF Staff Papers* IMF USA.

5634. He D, Zhang Z, Wang H 2009 Hong Kong's financial market interactions with the US and mainland China in crisis and tranquil times *Working Paper 10/2009* Research Department Market Research Division Hong Kong Monetary Authority Hong Kong P R China pp 1 – 27.

5635. Heath A, Whitelaw J June 2011 Electronic trading and the Australian foreign exchange market *Bulletin* Reserve Bank of Australia Canberra Australia pp 41 – 48.

5636. McGuire P, von Peter G 2009 The US dollar shortage in global banking *BIS Quarterly Review (March 2009)* pp 47 – 63.

5637. Meyers R A (editor) 2009 Encyclopedia of complexity and system science *Springer.*

5638. Muller A, Verschoor W F 2009 The effect of exchange rate variability on US shareholder wealth *Journal of Banking & Finance* pp 1963 – 1972.

5639. Nolte I, Nolte S 2009 Customer trading in the foreign exchange market. Empirical evidence from an Internet trading platform *Working Paper 09-01* FERC.

5640. Nolte I, Nolte S 2011 How do individual investors trade? *European Journal of Finance*  pp 1 – 27.

5641. Serban A F (November) 2009 Combining mean reversion and momentum trading strategies in foreign exchange markets *Department of Economics* West Virginia University USA pp 1 – 30.

5642. Simwaka K, Mkandawire L 2009 The efficacy of foreign exchange market intervention in Malawi *Reserve Bank of Malawi MPRA Paper no 15946* pp 1 – 39
http://mpra.ub.uni-muenchen.de/15946/ .

5643. Breedon F, Vitale P 2010 An empirical study of portfolio-balance and information effects of order flow on exchange rates *Journal of International Money and Finance* **29** (3) pp 504 – 524.

5644. Breedon F, Rime D, Vitale P 2011 Carry trades, order flow and the forward bias puzzle *Working Paper* Norges Bank Oslo Norway.

5645. Dunne P, Hau H, Moore M 2010 International order flows: Explaining equity and exchange rate returns *Journal of International Money and Finance* **29** (2) pp 358 – 386.

5646. Fukuda Sh-I, Kon Y (February) 2010 Macroeconomic impacts of foreign exchange reserve accumulation: Theory and international evidence ADBI *Working Paper Series no 197* Asian Development Bank Institute Tokyo Japan
http://www.adbi.org/working-
paper/2010/02/19/3515.macroeconomic.impact.forex.reserve.accumulation/ .

5647. Liu L-Zh, Qian X-Y, Lu H-Y 2010 Cross-sample entropy of foreign exchange time series *Physica A* **389** pp 4785 – 4792.

5648. Maurer K-O, Schäfer C 2010 Analysis of binary trading patterns in Xetra *CFS Working Paper no 2010/12* Center for Financial Studies an der Johann Wolfgang Goethe-Universität Frankfurt am Main Leibniz Information Centre for Economics Germany pp 1 – 22.

5649. Nightingale S, Ossolinski C, Zurawski A December 2010 Activity in global foreign exchange markets *RBA Bulletin* pp 45 – 51.

5650. Pasquariello P 2010 Central bank intervention and the intraday process of price formation in the currency markets *Journal of International Money and Finance* **29** (6) pp 1045 – 1061.

5651. Yiu M S, Ho W-Y A, Ma Y, Tsang Sh-K 2010 An analytical framework for the Hong Kong dollar exchange rate dynamics under strong capital inflows *Working Paper 05/2010* Hong Kong Monetary Authority Hong Kong P R China.

5652. Diamond R (April 4) 2011 Banks' profits could take hit in fight over forex fees *Pensions and Investments*.

5653. Durčáková J 2011 Foreign exchange rate regimes and foreign exchange markets in transitive economies *Prague Economic Papers* **4** pp Prague Check Republic pp 309 – 328.

5654. Heimer R Z, Simon D 2011 Facebook finance: How social interaction propagates active investing *Working Paper* Brandeis University.

5655. Marzo M, Zagaglia P P 2011 Trading directions and the pricing of euro interbank deposits in the long run *Working Paper 11-20* The Rimini Centre for Economic Analysis University of Bologna Italy.

5656. Moore M J, Payne R 2011 On the sources of private information in FX markets *Journal of Banking and Finance* **35** (5) pp 1250 – 1262.

5657. Plantin G, Shin H H 2011 Carry trades, monetary policy and speculative dynamics *Princeton University* USA.

5658. Rafferty B 2011 Currency returns, skewness and crash risk *Working Paper* Duke University North Carolina Durham USA.

5659. Wang Y, Wu Ch, Pan Zh 2011 Multifractal de-trending moving average analysis on the US dollar exchange rates *Physica A* **390** pp 3512 – 3523.

5660. Banti Ch, Phylaktis K, Sarno L 2012 Global liquidity risk in the foreign exchange market *Journal of International Money and Finance* **31** (2) pp 267 – 291.

5661. James J, Marsh I W, Sarno L (editors) 2012 The handbook of exchange rates *John Wiley & Sons Inc* USA.

5662. Mancini L, Ranaldo A, Wrampelmeyer J 2012 Liquidity in the foreign exchange market: Measurement, commonality, and risk premiums *Journal of Finance*.

5663. Sheng A (February) 2012a Hong Kong's global challenge - How to build on success pp 1 – 3
http://www.fungglobalinstitute.org/en/hong-kong's-global-challenge-how-build-success ,
http://www.fungglobalinstitute.org/en/experts/andrew-sheng .

5664. Sheng A (August) 2012b The future of central banking *Fung Global Institute* Hong Kong P R China, *Central Banking Publications* London UK
http://riskbooks.com/the-future-of-central-banking ,
http://www.fungglobalinstitute.org/en/future-central-banking ,
http://www.fungglobalinstitute.org/en/experts/andrew-sheng .

5665. Sheng A (April) 2014 Invited speech *6th London School of Economics Asia Forum 2014* Kuala Lumpur Malaysia
http://media.rawvoice.com/lse_publiclecturesandevents/richmedia.lse.ac.uk/publiclectures andevents/20140403_1540_plenary4.mp4 .

5666. Trenca I, Plesoianu A, Căpusan R 2012 Multifractal structure of Central and Eastern European foreign exchange markets *Faculty of Economics and Business Administration Babes-Bolyai University, Faculty of Finance, Insurance, Banking and Stock Exchange Academy of Economic Studies* pp 784 – 790.

5667. Wang D-H, Yu X-W, Suo Y-Y 2012 Statistical properties of the Yuan exchange rate index *Physica A* **391** pp 3503 – 3512.

5668. Lassmann A 2013 Exchange rate transmission and export activity at the firm level *Working Paper 331* KOF Swiss Economic Institute Swiss Federal Institute of Technology Zurich Switzerland pp 1 – 39 www.kof.ethz.ch .

5669. Ingves St, Danielsson J, Goodhart Ch (July 7) 2014 Towards a safer and more stable financial system: Stefan Ingves *Public Lecture London School of Economics and Political Science* London

UKhttp://media.rawvoice.com/lse_publiclecturesandevents/richmedia.lse.ac.uk/publicl
ecturesandevents/20140707_1830_saferStableFinancial.mp4 .

***Intellectual property investment, intellectual property exchange in finances:***

*5670.* Plant A 1934a The economic theory concerning patents for inventions *Economica* **1** pp 30 – 51.

*5671.* Plant A 1934b The economic aspects of copyright in books *Economica* **1** pp 167 – 195.

*5672.* Callmann R 1947 Unfair competition without competition?: The importance of the property concept in the law of trade-marks *University of Pennsylvania Law Review* **95** 443 ff.

*5673.* Penrose E T 1951 The economics of the international patent system *John Hopkins University Press* Baltimore MD USA.

*5674.* Prager F D 1952 The early growth and influence of intellectual property *Journal of the Patent Office Society* pp 106 – 110.

*5675.* Arrow K J 1962 Economic welfare and the allocation of resources for invention *in* The Rate and Direction of Inventive Activity: Economic and Social Factors Nelson R R (editor) *Princeton University Press* New York USA.

*5676.* Scherer F M 1965 Firm size, market structure, opportunity and the output of patented invention *American Economic Review* **55** pp 1097 – 1125.

*5677.* Scherer F M 1984 Innovation and growth: Schumpeterian perspectives *MIT Press* Cambridge MA USA.

*5678.* Baxter W F 1966 Legal restrictions on exploitation of the patent monopoly: An economic analysis *Yale Law Journal* **76** pp 267 – 370.

*5679.* Hurt, Schuchman 1966 The economic rationale for copyright *American Economic Review* **56** pp 421 – 432.

*5680.* Barzel Y 1968 Optimal timing of innovations *Review of Economics and Statistics* **50** pp 348 – 355.

*5681.* Nordhaus W 1969 Invention, growth and welfare: A theoretical treatment of technological change *MIT Press* Cambridge MA USA.

*5682.* Bowman W S Jr 1973 Patent and antitrust law: A legal and economic appraisal *University of Chicago Press* Chicago USA.

*5683.* Bowman W S Jr 1977 The incentive to invent in competitive as contrasted to monopolistic industries *Journal of Law and Economics* **20** pp 227 – 228.

*5684.* Taylor C, Silberston Z A 1973 The economic impact of the patent system *Cambridge University Press* Cambridge UK.

*5685.* Roffe P 1974 Abuses of patent monopoly: A legal appraisal *World Development* **2** (9) pp 15 – 26.

*5686.* Adelman M 1977 Property rights theory and patent-antitrust: The role of compulsory licensing *New York University Law Review* **52** pp 977 – 1013.

*5687.* Loury G C 1979 Market structure and innovation *Quarterly Journal of Economics* **93** (3) pp 395 – 410.

*5688.* Cheung St N S 1982 Property rights in trade secrets *Economic Inquiry* **20** pp 40 – 53.

*5689.* Gilbert R J, Newbery D M G 1982 Preemptive patenting and the persistence of monopoly *American Economic Review* **72** (3) pp 514 – 526.

*5690.* Gilbert R J, Newbery D M G 1984 Preemptive patenting and the persistence of monopoly: Reply *American Economic Review* **74** pp 251 – 253.

*5691.* Gilbert R J, Shapiro C 1990 Optimal patent protection and breadth *Rand Journal of Economics* **21** pp 106 – 112.

*5692.* Mackaay E 1982 Economics of information and law *Kluwer-Nijhoff* Boston USA pp 1 – 293.

*5693.* Mackaay E 1985a De hersenschim als rustig bezit - Moet Alle informatie voorwerp van eigendom zijn? (The chimera as peaceful possession - Must all information be the object of a property right?) *Computerrecht* pp 12 – 16.

*5694.* Mackaay E 1985b Informational goods - Property of a mirage *Computer Law and Practice* pp 193 – 197.

*5695.* Mackaay E 1986 Les Biens informationnels ou le droit de suite dans les idées (Informational goods or the droit de suite in ideas) *in* l'Appropriation de l'Information Chamoux J-P (editor) *Librairies Techniques* pp 26 – 48.

*5696.* Mackaay E 1989 Les droits intellectuels - Entre propriété et monopole (Intellectual property rights - Between property and monopoly *Journal des Économistes et des Études Humaines* **1**.

*5697.* Mackaay E 1990a Les droits intellectuels - Entre propriété et monopole (Intellectual property rights - Between property and monopoly) *Revue des Économistes et des Études Humaines* **1** pp 61 – 358.

*5698.* Mackaay E 1990b Economic incentives in markets for information and innovation *Harvard Journal of Law and Public Policy* **13** pp 867 – 909.

*5699.* Mackaay E 1991a La propriété est-elle en voie d'extinction? (Is property about to become extinct?) *in* Nouvelles Technologies et Propriété Mackaay E (editor) *Éd Thémis* Montréal Canada, *LITEC* Paris France pp 217 – 247.

*5700.* Mackaay E 1991b Economisch-filosofische aspecten van de intellectuele rechten (The economic and philosophical aspects of intellectual property rights) *in* De Sociaal Economische Rol van Intellectuele Rechten Van Hoecke M (editor) *Story-Scientia* Brussels Belgium pp 1 – 30.

*5701.* Mackaay E (editor) 1991c Nouvelles technologies et propriété (New technologies and property rights) *Éd. Thémis* Montréal Canada, *LITEC* Paris France.

*5702.* Mackaay E 1992a An economic view of information law *in* Information Law Towards the 21st Century Korthals Altes W F, Dommering E J, Hugenholtz P B, Kabel J J C (editors) *Kluwer* Deventer pp 43 – 65.

*5703.* Mackaay E 1992b Los derechos intelectuales: Entre propriedad y monopolio (Intellectual property rights: Between property and monopoly) *DAT Derecho de la Alta Tecnologia* pp 1 – 15.

*5704.* Mackaay E 1994 Legal hybrids: Beyond property and monopoly? *Columbia Law Review* **94** pp 2630 – 2643.

*5705.* Sieghart P 1982 Information technology and intellectual property *European Intellectual Property Review* pp 187 – 188.

5706. Ashford N A, Heaton G E 1983 Regulation and technological innovation in the chemical industry *Law and Contemporary Problems* **46** pp 109 – 157.

5707. Ashford N A et al 1985 Using regulation to change the market for innovation *Harvard Environmental Law Review* **9** pp 419 – 466.

5708. Baird D G 1983 Common law intellectual property and the legacy of International News Service v Associated Press *University of Chicago Law Review* **50** pp 411 – 429.

5709. Beck R L 1983 The prospect theory of the patent system and unproductive competition *Research in Law and Economics* **5** pp 193 – 209.

5710. Fudenberg D, Gilbert R J, Stiglitz J E, Tirole J 1983 Preemption, leapfrogging and competition in patent races *European Economic Review* **77** pp 176 – 183.

5711. Wright B D 1983 The economics of invention incentives: Patents prizes and research contracts *American Economic Review* **73** pp 691 – 707.

5712. Mossinghoff G J 1984 The importance of intellectual property protection in international trade *Boston College International and Comparative Law Review* **7** pp 235 – 249.

5713. Adelstein R P, Peretz St I 1985 The competition of technologies in markets for ideas: Copyright and fair use in evolutionary perspective *International Review of Law and Economics* **5** pp 209 – 238.

5714. Cave J 1985 A further comment on preemptive patenting and the persistence of monopoly *American Economic Review* **75** pp 256 – 258.

5715. David P A 1985 New technology, diffusion, public policy and industrial competitiveness center for economic policy research Stanford University California USA.

5716. David P A 1993 Intellectual property and the panda's thumb: patents, copyrights and trade secrets in economic theory and history *in* Global Dimensions of Intellectual Property Rights in Science and Technology *National Academy Press* Washington USA.

5717. Farrell J, Saloner G 1985 Standardization, compatibility and innovation *Rand Journal of Economics* **16** pp 70 – 83.

5718. Farrell J 1989 Standardization and intellectual property *Jurimetrics Journal* **30** pp 35 - 50.

5719. Gallini N, Winter R A 1985 Licensing in the theory of innovation *Rand Journal of Economics* **16** pp 237 – 252.

5720. Gallini N 1992 Patent policy and costly imitation *Rand Journal of Economics* **23** pp 52 – 63.

5721. Judd K 1985 On the performance of patents *Econometrica* **53** pp 567 – 585.

5722. Lehmann M 1985 The theory of property rights and the protection of intellectual and industrial property *IIC* **16** pp 525 – 540.

5723. Lehmann M 1989 Property and intellectual property - Property rights as restrictions on competition in furtherance of competition *IIC* **20** pp 1 – 15.

5724. Lehmann M 1990 La teoría de los 'property rights' y la protección de la propriedad intelectual e industrial (The property rights theory and the protection of intellectual and industrial property) Revista General de Derecho pp 544 – 545.

5725. Pendleton M 1985 Intellectual property, information- based society and a new international economic order - the policy options? *European Intellectual Property Review* pp 31 – 34.

5726. Samuelson P 1985 Creating a new kind of intellectual property: Applying lessons of the chip law to computer programs *Minnesota Law Review* **70** pp 471 – 531.

5727. Department of Trade and Industry 1986 Intellectual property and innovation *Her Majesty's Stationary Office* London UK pp 1 – 78.

5728. Hay F 1986 Canada's role in international negotiations concerning intellectual property *Laws Research in Law and Economics* **8** pp 239 – 263.

5729. Mansfield E 1986 Patents and innovation: An empirical study *Management Science* **32** pp 173 – 181.

5730. Priest G L 1986 What economists can tell lawyers about intellectual property: Comment on Cheung *Research in Law and Economics* **8** pp 19 – 24.

5731. Evenson R E, Putnam J D 1987 Institutional change in intellectual property rights *American Journal of Agricultural Economics* **69** pp 403 – 409.

5732. Menell P S 1987 Tailoring legal protection for computer software *Stanford Law Review* 39 pp 1329 – 1372.

5733. Menell P S 1989 An analysis of the scope of copyright protection for application programs *Stanford Law Review* **41** pp 1045 – 1104.

5734. Menell P S 1994 The challenges of reforming intellectual property protection for computer software *Columbia Law Review* **94** pp 2644 – 2654.

5735. Menell P S 1998 An epitaph for traditional copyright protection of network features of computer software *The Antitrust Bulletin* **43** pp 651 – 713.

5736. Rozek R P 1987 Protection of intellectual property rights: Research and development decisions and economic growth *Contemporary Policy Issues* **5** (3) pp 54 – 65.

5737. Sirilli G 1987 Patents and inventors: An empirical study *Research Policy* **16** pp 157 – 174.

5738. Tullock G 1987 Intellectual property *in* Direct protection of innovation Kingston W (editor) *Kluwer Academic Publishers* Dordrecht pp 171 – 199.

5739. Feinberg R M 1988 Intellectual property, injury and international trade **22** (2) *Journal of World Trade Law* pp 45 – 56.

5740. Feinberg R M, Rousslang D J 1990 The economic effects of intellectual property right infringements *Journal of Business* **63** pp 79 – 90.

5741. Hughes J 1988a The philosophy of intellectual property *Georgetown Law Journal* **77** 287 ff.

5742. Hughes J 1988b The Personality Interest of Artists and Inventors in Intellectual Property *Cardozo Arts and Entertainment Law Journal* pp 81 – 181.

5743. Merges R P 1988 Commercial success and patent standards: Economic perspectives on innovation *California Law Review* **76** pp 803 – 876.

5744. Merges R P 1992 Rent control in the patent district: Observations on the Grady-Alexander thesis *Virginia Law Review* **78** pp 359 – 381.

5745. Merges R P 1994a Intellectual property rights and bargaining breakdown: The case of blocking patents *Tennessee Law Review* **62** pp 75 – 106.

5746. Merges R P 1994b Of property rules, coase and intellectual property *Columbia Law Review* **94** pp 2655 – 2673.

5747. Merges R P 1995a Expanding boundaries of the law: Intellectual property and the costs of commercial exchange: A review essay *Michigan Law Review* **93** pp 1570 – 1615.

5748. Merges R P 1995b The economic impact of intellectual property rights: An overview and guide *Journal of Cultural Economics* **19** pp 103 – 117.

5749. Merges R P 1996a Contracting into liability rules: Intellectual property rights and collective rights organizations *California Law Review* **84** pp 1293 – 1393.

5750. Merges R P 1996b Property rights theory and the commons: The case of scientific research *Social Philosophy and Policy* **13** pp 144 – 167.

5751. Merges R P, Menell P S, Lemley M A, Jorde Th M 1997 Intellectual property in the new technological age *Aspen Publishers Inc* New York USA.

5752. Von Hippel E 1988 The sources of innovation *Oxford University Press* New York USA.

5753. Walker Ch E, Bloomfield M A (editors) 1988 Intellectual property rights and capital formation *University Press of America* Lanham MD USA pp 1 – 189.

5754. Beier F-K, Shricker G 1989 GATT or WIPO? New ways in the international protection of intellectual property *IIC Studies Weinheim* **11** VCH-Verlagsgesellschaft.

5755. Besen St M, Kirby Sh N 1989a Private copying, appropriability, and optimal copying royalties *Journal of Law and Economics* **32** pp 255 – 280.

5756. Besen St M, Kirby Sh N 1989b Compensating creators of intellectual property: Collectives that collect *Rand Corporation no R-3751-MF*.

5757. Besen St M, Raskind L J 1991 An Introduction to the Law and Economics of Intellectual Property *Journal of Economic Perspectives* **5** pp 3 – 27.

5758. Besen St M, Kirby Sh N, Salop St 1992 An economic analysis of copyright collectives *Virginia Law Review* **78** pp 383 – 411.

5759. Braga C A 1989 The economics of intellectual property rights and the GATT *Vanderbilt Journal of Transnational Law* **22** pp 243 – 264.

5760. Centner T J 1989 Invent in America: Pitfalls for foreigners under US intellectual property rights legislation *Canadian Journal of Agricultural Economics* **37** pp 1307 – 1313.

5761. Centner T J, White F C 1989 Protecting inventors' intellectual property rights in biotechnology *Western Journal of Agricultural Economics* **14** pp 189 – 199.

5762. Davis L N 1989 Skydd for innovationer (Protecting innovations) *in* Foretaget - et Kontraksekonomisk Analys Bjurggren P-O, Skogh G (editors) *SNS forlag* Stockholm Sweden pp 151 – 158.

5763. Krauss M I 1989 Property, monopoly and intellectual rights *Hamline Law Review* **12** pp 305 – 320.

5764. Palmer T G 1989 Intellectual property: A non-Posnerian law and economics approach *Hamline Law Review* **2** pp 261 – 304.

5765. Palmer T G 1990 Are patents and copyrights morally justified? The philosophy of property rights and ideal objects *Harvard Journal of Law and Public Policy* **13** pp 817 – 865.

*5766.* Brenner R 1990 Inventions et innovations dans le monde des affaires et des sciences (Inventions and innovations in business and science) *Études Françaises* **26** pp 51 – 78.

*5767.* Chin J, Grossman G M 1990 Intellectual property rights and north-south trade *in* The Political Economy of International Trade Jones R W, Krueger A O (editors) *Basil Blackwell Publishers* Cambridge MA USA.

*5768.* Easterbrook F H 1990 Intellectual property is still property *Harvard Journal of Law and Public Policy* **13** pp 108 – 118.

*5769.* Gilbert R, Shapiro C 1990 Optimal patent length and breadth *The RAND Journal of Economics* **21** pp 106 – 112.

*5770.* Klemperer P 1990 How Broad Should the Scope of Patent Protection Be? *RAND Journal of Economics* **21** pp 113 – 130.

*5771.* Rushing F W, Brown C G (editors) 1990 Intellectual property rights in science, technology and economic performance *Westview Press* Boulder CO USA pp 1 – 354.

*5772.* Caves R E, Whinston M D, Hurwitz M A 1991 Patent expiration, entry, and competition in the U.S. pharmaceutical industry *Brookings Papers on Economic Activity: Microeconomics* pp 1 – 48.

*5773.* Coombe R J 1991 Objects of property and subjects of politics: Intellectual property laws and democratic dialogue *Texas Law Review* **69** pp 1853 – 1880.

*5774.* Heald P J 1991 Federal intellectual property law and the economics of preemption *Iowa Law Review* **76** pp 959 – 1010.

*5775.* Scotchmer S 1991 Standing on the shoulders of giants: Cumulative research and the patent Law *Journal of Economic Perspectives* **5** pp 29 – 41.

*5776.* Segerstrom P 1991 Innovation, imitation and economic growth *Journal of Political Economy* **99** (4) pp 807 – 827.

*5777.* Teijl R, Holzhauer R W 1991 De toenemende complexiteit van het intellectuele eigendomsrecht. Een economische analyse (The growing complexity of intellectual property law. An economic analysis) *in* Rechtseconomische Verkenningen Deel **1** *Kluwer* Deventer.

*5778.* Deardorff A V 1992 Welfare effects of global patent protection *Economica* **59** pp 35 – 51.

*5779.* Gallini N 1992 Patent policy and costly imitation *RAND Journal of Economics* **23** pp 52 – 63.

*5780.* Gordon W J 1992a On owning information: Intellectual property and the restitutionary impulse *Virginia Law Review* **78** pp 149 – 281.

*5781.* Gordon W J 1992b Of harms and benefits: Torts, restitution and intellectual property *Journal of Legal Studies* **21** pp 449 – 482.

*5782.* Gordon W J 1993 A property right in self-expression: Equality and individualism in the natural law of intellectual property *Yale Law Journal* **102** pp 1533 – 1609.

*5783.* Grady M F, Alexander J I 1992 Patent law and rent dissipation *Virginia Law Review* **78** pp 305 – 350.

*5784.* Quaedvlieg A A 1992 The Economic Analysis of Intellectual Property Law *in* Information Law Towards the 21st Century Korthals Altes, W F, Dommering E J, Hugenholtz P B, Kabel J J C (editors) *Kluwer* Deventer pp 379 – 393.

5785. Aoki K 1993-1994 Authors, inventors and trademark owners: Private intellectual property and the public domain (parts 1 and 2) *Columbia - VLA Journal of Law and the Arts* **18** pp 197 – 267.

5786. Aoki K 1996a (Intellectual) property and sovereignty: Notes toward a cultural geography of authorship *Stanford Law Review* **48** pp 1293 – 1355.

5787. Aoki K 1996b Foreword: innovation and the information environment: Interrogating the entrepreneur *Oregon Law Review* **75** pp 1 – 18.

5788. Becker L C 1993 Deserving to own intellectual property *Chicago-Kent Law Review* **68** pp 609 – 629.

5789. Brennan T J 1993 Copyright, property and the right to deny *Chicago-Kent Law Review* **68** pp 675 – 714.

5790. Carter St L 1993 Does it matter whether intellectual property is property? *Chicago-Kent Law Review* **68** pp 715 – 723.

5791. Chou Chien-Fu, Shy O 1993 The crowding-out effects of long duration of patents *Rand Journal of Economics* **24** pp 304 – 312.

5792. Helpman E 1993 Innovation, imitation and intellectual property rights *Econometrica* **61** (6) pp 1247 – 1280.

5793. Kay J 1993 The economics of intellectual property rights *International Review of Law and Economics* **13** (4) pp 337 – 348.

5794. Lanjouw J 1993 Patent protection: Of what value and how long? *NBER Working Paper 4475* National Bureau of Economic Research USA.

5795. Lanjouw J, Pakes A, Putnam J 1998 How to count patents and value intellectual property: The uses of patent renewal and application data *Journal of Industrial Economics* **46** pp 405 – 432.

5796. Lanjouw J O, Schankerman M 2001a Characteristics of patent litigation: A window on competition *RAND Journal of Economics* 32 (1) pp 129 – 151.

5797. Lanjouw J O, Schankerman M December 2001b Enforcing intellectual property rights *STICERD - Economics of Industry Paper no EI/30* Suntory and Toyota International Centres for Economics and Related Disciplines LSE London UK pp 1 – 44

http://sticerd.lse.ac.uk/dps/ei/ei30.pdf .

5798. Nelson R R 1993 National innovation systems: A comparative analysis *Oxford University Press* New York USA.

5799. Nelson R R 1994 Intellectual property protection for cumulative systems technology *Columbia Law Review* **94** pp 2674 – 2677.

5800. Barlow J P 1994 The framework for economy of ideas: Rethinking patents and copyrights in the digital age *1994 WIRED* pp 83 – 97.

5801. Dam K W 1994 Economic underpinnings of patent law *Journal of Legal Studies* **23** pp 247 – 271.

5802. Dam K W 1995 Some economic considerations in the intellectual property protection of software *Journal of Legal Studies* **24** pp 321 – 377.

5803. Japan Institute of Intellectual Property 1994 Report on the basic issues concerning economic effects of intellectual property *Japan Institute of Intellectual Property* Tokyo Japan.

*5804.* Karjala D S 1994 Misappropriation as a third intellectual property paradigm *Columbia Law Review* **94** pp 2594 – 2609.

*5805.* Lerner J 1994 The importance of patent scope: An empirical analysis *RAND Journal of Economics* **25** pp 319 – 333.

*5806.* Lerner J 1995 Patenting in the shadow of competitors *Journal of Law and Economics* **38** pp 463 – 496.

*5807.* Lerner J 2002 150 years of patent protection *American Economic Review* **92** (2) pp 221 – 225.

*5808.* Chang H F 1995 Patent scope, antitrust policy and cumulative innovation *Rand Journal of Economics* **26** pp 34 – 57.

*5809.* Lemley M A 1995 Intellectual property and shrink-wrap licenses *Southern California Law Review* **68** 1239 ff.

*5810.* Gould D M, Gruben W C 1996 The role of intellectual property rights in economic growth *Journal of Development Economics* **48** (2) pp 323 – 350.

*5811.* Matutes C, Regibeau P, Rockett K 1996 Optimal patent design and the diffusion of innovations *Rand Journal of Economics* **27** pp 60 – 83.

*5812.* Brousseau E, Bessy Ch September 19 - 21 1997 The governance of intellectual property rights: Patents and copyrights in France and in the US *Inaugural Conference for the International Society for New Institutional Economics The Present and Future of the New Institutional Economics* Washington University St Louis Missouri USA.

*5813.* Ginarte J C, Park W G 1997 Determinants of patent rights: A cross-national study *Research Policy* **26** (3) pp 283 – 301.

*5814.* Grindley P, Teece D 1997 Managing intellectual capital: Licensing and cross-licensing in semiconductors and electronics *California Management Review* **39** (2) pp 8 – 41.

*5815.* Park W G, Ginarte J C 1997 Intellectual property rights and economic growth *Contemporary Economic Policy* **15** (1) pp 51 – 61.

*5816.* Besen St M 1998 Intellectual property *in* The New Palgrave Dictionary of Economics and the Law Newman P (editor) London Macmillan UK pp 348 – 351.

*5817.* Maskus K 1998 The international regulation of intellectual property *Review of World Economics (Weltwirtschaftliches Archiv)* **134** (2) pp 186 – 208.

*5818.* Schankerman M 1998 How valuable is patent protection: Estimates by technology field *RAND Journal of Economics* **29** (1) pp 77 – 107.

*5819.* Templeman S 1998 Intellectual property *Journal of International Economic Law* **1** (4) pp 603 – 606.

*5820.* Reilly R F, Schweihs R P 1999 Valuing intangible assets *Mc Graw Hill* USA.

*5821.* Reilly R F 2013 The intellectual property valuation process *License Journal* **33** (2).

*5822.* Gallini N, Scotchmer S 2001 Intellectual property: When is it the best incentive system? *in* Innovation Policy and the Economy vol **2** Jaffe A, Lerner J, Stern S (editors) *MIT Press* Cambridge MA USA.

*5823.* Hall B, Zeidonis R 2001 The patent paradox revisited: An empirical study of patenting in the semiconductor industry, 1979-1999 *RAND Journal of Economics* **32** (1) pp 101 – 128.

5824. McCalman Ph 2001 Reaping what you sow: An empirical analysis of international patent harmonization *Journal of International Economics* **55** pp 161 – 185.

5825. Sakakibara M, Branstetter L 2001 Do stronger patents induce more innovation? Evidence from the 1988 Japanese patent law reforms *RAND Journal of Economics* **32** (1) pp 77 – 100.

5826. Scotchmer S 2001 The political economy of intellectual property treaties *Working Paper no E01-305* Institute of Business and Economic Research University of California at Berkeley California USA.

5827. Shapiro C 2001 Navigating the patent thicket: Cross licenses, patent pools and standard setting *in* Innovation Policy and the Economy Jaffe A, Lerner J, Stern S (editors) *MIT Press* vol **1** pp119 – 150.

5828. Boldrin M, Levine D K 2002 The case against intellectual property *American Economic Review Papers and Proceedings* **92** pp 209 – 212.

5829. Boldrin M, Levine D K 2004 IER Lawrence Klein lecture: The case against intellectual monopoly *International Economic Review* **45** pp 327 – 350.

5830. Boldrin M, Levine D K 2004a Rent Seeking and Innovation *Journal of Monetary Economics* **51** pp 127 – 160.

5831. Boldrin M, Levine D K 2004b Intellectual property and the scale of the market" http://www.dklevine.com/papers/scale22.pdf .

5832. Boldrin M, Levine D K 2005 The economics of ideas and intellectual property *Proceedings of the National Academy of Sciences* **102** pp 1252 – 1256 http://www.dklevine.com/papers/pnas18.pdf .

5833. Boldrin M, Levine D K 2006 Growth and intellectual property *NBER Working Paper no 12769* National Bureau of Economic Research Inc USA pp 1 – 29.

5834. Boldrin M, Levine D K February 6 2007 Appropriation and intellectual property Levine's *Working Paper Archive from David K. Levine* pp 1 – 21 http://www.dklevine.com/papers/marginalip.pdf .

5835. Grossman G, Lai E 2002, 2004 International protection of intellectual property *CESifo Working Paper no 790* CESifo Group Munich Germany pp 1 – 43, *American Economic Review* **94** (5) pp 1635 – 1653 DOI: 10.1257/0002828043052312 http://www.cesifo-group.de/portal/page/portal/DocB ... -10/cesifo_wp790.pdf , https://www.aeaweb.org/articles?id=10.1257/0002828043052312 .

5836. Lasinski M J 2002 Valuation of intellectual property assets in mergers and acquisitions *in* Intellectual Property Assets in Mergers and Acquisitions Bryer L, Simensky M (editors) *John Wiley and Sons Inc* USA.

5837. Maskus K E 2000a Intellectual property rights in the global economy *The Institute for International Economics* Washington DC USA.

5838. Maskus K E 2000b Parallel imports *The World Economy* **23** pp 1269 – 1284.

5839. Deli Yang 2003 The development of intellectual property in China *World Patent Information* **25** (2) pp 131 – 142.

5840. Menell P S 2003 Intellectual property: General theories *Levine's Working Paper Archive from David K. Levine* pp 1 – 60 http://www.dklevine.com/archive/ittheory.pdf .

*5841.* Anson W, Suchy D P, Ahya C 2005 Fundamentals of intellectual property valuation. A Primer for identifying and determining value *The American Bar Association* Chicago USA.

*5842.* Anson W, Noble D, Samala J 2014 IP valuation: What methods are used to value intellectual property and intangible assets? *Licensing Journal* **34** (2).

*5843.* Blair R D, Cotter T F June 2005 Intellectual property *Cambridge University Press* Cambridge UK ISBN: 9780521540674 pp 1 – 316.

*5844.* Ramello G 2005 Intellectual property and the markets of ideas *Review of Network Economics* **4** (2) pp 1 – 20.

*5845.* Smith G, Parr R 2005 Intellectual property, valuation, exploitation and infringement damages *John Willey and Sons Inc* NY USA.

*5846.* Andersen B (editor) January 1 2006 Intellectual property rights *Edward Elgar Publishing* ISBN:9781845422691 pp 1 – 384

DOI: http://dx.doi.org/10.4337/9781847201522 ,

https://www.elgaronline.com/view/9781845422691.xml .

*5847.* Hisamitsu Arai 2006 Japan's intellectual property strategy *World Patent Information* **28** (4) pp 323 – 326.

*5848.* Kamiyama Sh, Sheehan J, Martínez C 2006 Valuation and exploitation of intellectual property OECD Science, Technology and Industry *Working Paper no 2006/5 OECD Publishing* Paris France.

*5849.* Kanwar S 2006 Innovation and intellectual property rights *Working Paper no 142* Centre for Development Economics Delhi School of Economics India pp 1 – 16

http://www.cdedse.org/pdf/work142.pdf .

*5850.* Kumar J 2006 Intellectual property securitization: How far possible and effective *Journal of Intellectual Property Rights* **11**.

*5851.* Lakdawalla D, Philipson T, Wang Y R October 2006 Intellectual property and marketing *NBER Working Paper no 12577* National Bureau of Economic Research Inc pp 1 – 53

http://www.nber.org/papers/w12577.pdf .

*5852.* Moerman L, Laan S V 2006 Accounting for intellectual property: Inconsistencies and challenges *Journal of Intellectual Property* **11** (4).

*5853.* Aoki R, Schiff A 2007 Intellectual property access systems *Discussion Paper no a491* Institute of Economic Research Hitotsubashi University Japan.

*5854.* Bittelmeyer C 2007 Patente und finanzierung am kapitalmarkt *Deutsche Universitätsverlag* Wiesbaden Germany.

*5855.* Holland C J, III V A, Reed D M, Lee S H, Kimmel A I, Peterson W K 2007 Intellectual property, patents, trademarks, copyrights and trade secrets *Entrepreneur Media* Canada.

*5856.* Holland C J, Benedikt M B 2014 Intellectual property valuation *in* Intellectual Property in Business Transactions Weakley S L (editor) *CEB* California USA.

*5857.* Malackowski J E, Cardoza K, Gray C, Conroy R 2007 The intellectual property marketplace: Emerging transaction and investment vehicles *Licensing Journal* **27** (2).

*5858.* Menell P S, Scotchmer S 2007 Intellectual property law Chapter 19 *in* Handbook of Law and Economics **2** pp 1473 – 1570 *Elsevier.*

5859. Parr R 2007 Royalty rates for licensing intellectual property *John Willey and Sons Inc* NY USA.

5860. Siegel D, Wright M 2007 Intellectual property: The assessment *Oxford Review of Economic Policy* **23** (4) pp 529 – 540.

5861. Van Caenegem W May 2007 Intellectual property law and innovation *Cambridge University Press* Cambridge UK ISBN: 9780521837576 pp 1 – 240.

5862. Ruder D S 2008 Strategies for investing in intellectual property *Beard Books* Washington USA.

5863. Kite S 2009 Intellectual property may soon be traded as its own asset class' *Securities Industry News* **8** (3).

5864. Blakeney M, Ullrich H, Stauder D, Llewelyn D, MacQueen H, Justice Jacob, Justice Laddie, Chisum D, Benyamini A, Straus J, Llewellyn M, McCarthy T, Dworkin G, Soltysinski S, Lahore J, Dufty A, Ricketson S, Ginsburg J, Christie A, Goldstein P, Tapper C B, Kamina P December 2010 Intellectual property in the new millennium Vaver D, Bently L (editors) *Cambridge University Press* Cambridge UK ISBN: 9780521173414 pp 1 – 322.

5865. Flanagan A, Montagnani M L (editors) January 2010 Intellectual property law *Edward Elgar Publishing* ISBN:9781848446274 pp 1 – 232

DOI: http://dx.doi.org/10.4337/9781849806701 ,

https://www.elgaronline.com/view/9781848446274.xml .

5866. Baker S, Pak Yee Lee, Mezzetti C 2011 Intellectual property disclosure as threat *International Journal of Economic Theory* **7** (1) pp 21 – 38.

5867. Bryer L G, Lebson S J, Asbell M D 2011 Intellectual property strategies for the 21st -century corporation. A shift in strategic and financial management *John Wiley & Sons Inc* New Jersey USA.

5868. Cottier T, Veron P 2011 Concise international and European IP law *Kluwer Law International* The Netherlands.

5869. McCoy M D, Barton R, McDermott R 2011 Royalty rate trends in patent and technology licensing *Licensing Journal* **31** (3).

5870. Palfrey J October 2011 Intellectual property strategy, vol 1 *The MIT Press* Cambridge MA USA ISBN: 9780262516792 pp 1 – 192

https://mitpress.mit.edu/books/intellectual-property-strategy .

5871. Bouchoux D E 2012 Intellectual property. The law of trademarks, copyrights, patents and trade secrets 4[th] edition *Recording for Blind & Dyslexic* Princeton New Jersey USA.

5872. George A 2012 Constructing intellectual property *Cambridge University Press* New York USA.

5873. Rüther F 2012 Patent aggregating companies. Their strategies, activities and options for producing companies *Springer Gabler* Wiesbaden Germany.

5874. Boldrin M, Levine D 2013 What's intellectual property good for? *Revue Économique* **64** (1) pp 29 – 53.

5875. Frey C B 2013 Intellectual property rights and the financing of technological innovation *Edward Elgar Publishing* Cheltenham Massachusetts USA.

5876. Howe H, Griffiths J, Sherman B, Pottage A, Gangjee D, Bently L, Hudson A, Dreier Th, Breakey H, Balganesh Sh, Carrier M, Burrell R, Hudson E, Lametti D, Dussollier S September 2013 Concepts of property in intellectual property law *Cambridge University Press* Cambridge UK ISBN: 9781107041820 pp 1 – 330.

5877. Buchanan J, Wilson B 2014 An experiment on protecting intellectual property *Experimental Economics* **17** (4) pp 691 – 716.

5878. Fawcett D 2014 Valuation of intellectual property *Licensing Journal* **34** (5).

5879. Gervais D (editor) 2014 Intellectual property, trade and development 2nd edition *Oxford University Press* Oxford UK ISBN: 9780199684700 pp 1 – 416

https://global.oup.com/academic/product/intellectual-property-trade-and-development-9780199684700?cc=us&lang=en& .

5880. Guellec D, Ménière Y 2014 Markets for patents: Actors, workings and recent trends *in* Patent Markets in the Global Knowledge Economy Madiès T, Guellec D, Prager J-C *Cambridge University Press* New York USA.

5881. Sople V D 2014 Managing intellectual property. The strategic imperative 4th edition *PHI* Delhi India.

5882. Schmitt Fr 2016 Intellectual property and investment funds, vol 1 *EIKV-Schriftenreihe zum Wissens- und Wertemanagement* European Institute for Knowledge and Value Management (EIKV) Luxembourg pp 1 – 177

https://www.econstor.eu/bitstream/10419/126181/1/846558815.pdf ,

http://hdl.handle.net/10419/126181 .

5883. Searle N, Brassell M August 2 2016 Economic approaches to intellectual property *Oxford University Press* Oxford UK ISBN: 9780198736264 pp 1 – 280

https://global.oup.com/academic/product/economics-for-intellectual-property-lawyers-9780198736264?cc=us&lang=en&

***Investment bank, financial capital investment vehicle in finances:***

5884. Howell P L 1953 Competition in investment banking *Journal of Finance* **8** (2) pp 278 – 282.

5885. O'Donnell L J 1957 The financial operations of a regional investment bank *Journal of Finance* **12** (1) pp 84 – 85.

5886. Pontecorvo G 1958 Investment banking and security speculation in the late 1920's *Business History Review* **32** (02) pp 166 – 191.

5887. Mandelker G, Raviv A 1977 Investment banking: An economic analysis of optimal underwriting contracts *Journal of Finance* **32** (3) pp 683 – 694.

5888. Beatty R P, Ritter J 1986 Investment banking, reputation, and the underpricing of initial public offerings *Journal of Financial Economics* **15** (1-2) p 213 – 232.

5889. Smith C 1986 Investment banking and the capital acquisition process *Journal of Financial Economics* **15** (1-2) pp 3 – 29.

5890. Keeley M C, Pozdena R June 19 1987 Uniting investment and commercial banking *FRBSF Economic Letter.*

5891. McDonough W J 1987 Financial innovations and the merging of commercial and investment banking activities *Research Paper no 139* Proceedings Federal Reserve Bank of Chicago USA.

5892. Walter I, Smith R C 1989 Investment banking in Europe after 1992 *Research Paper no 243 Proceedings* Federal Reserve Bank of Chicago USA.

5893. Carter R B, Dark F H 1992 An empirical examination of investment banking reputation measures *The Financial Review* **27** (3) pp 355 – 374.

5894. Chemmanur Th, Fulghieri P 1994 Investment bank reputation, information production, and financial intermediation *Journal of Finance* **49** (1) pp 57 – 79.

5895. Sussman O 1994 Investment and banking: Some international comparisons *Oxford Review of Economic Policy* **10** (4) pp 79 – 93.

5896. Clark M A April 1995 Commercial & investment banking: should this divorce be saved? *The Regional Economist* pp 5 – 9.

5897. Brockman P 1996 The role of reputation capital in the investment banking industry *Applied Economics Letters* **3** (7) p 455 – 458.

5898. Ferguson R 1996 An investment banking perspective on the future of the financial system Chapter *in* The Future of the Financial System *Reserve Bank of Australia* ISBN: 0 642 25621 7

http://www.rba.gov.au/publications/confs/1996/pdf/ferguson-r.pdf .

5899. Grant K C 1999 Internet investment banking and corporate debt issuance *Journal of Applied Corporate Finance* **12** (2) pp 4-4.

5900. Parkan C, Ming-Lu Wu 1999 Measurement of the performance of an investment bank using the operational competitiveness rating procedure *Omega* **27** (2) pp 201 – 217.

5901. Fleuriet M 2000 The location of investment banking business *Revue d'Économie Financière* **57** (2) pp 119 – 126.

5902. Anand B N, Galetovic A P July 01 2001 Investment banking and security market development: Does finance follow industry? *IMF Working Paper no 01/90* International Monetary Fund USA pp 1 – 26

http://www.imf.org/external/pubs/cat/longres.aspx?sk=15169 .

5903. Smith R C 2001 Strategic directions in investment banking-a retrospective analysis *Journal of Applied Corporate Finance* **14** (1) pp 111 – 124.

5904. Perez C 2002 Technological revolutions and financial capital *Elgar* Cheltenham UK.

5905. Benveniste L M, Ljungqvist A, Wilhelm W J, Xiaoyun Yu 2003 Evidence of Information spillovers in the production of investment banking services *Journal of Finance* **58** (2) pp 577 – 608.

5906. Ritter J 2003 Investment banking and securities issuance Chapter 05 *in* Handbook of the Economics of Finance 1 part 1 *Elsevier* The Netherlands pp 255 – 306.

5907. Sirri E 2004 Investment banks, scope, and unavoidable conflicts of interest Economic Review issue Q 4 pp 23 – 35.

5908. Abor J Summer 2005 The role of investment banking in raising capital in Ghana *Journal of International Financial Analyst* **5**.

5909. Morrison A D, Wilhelm Jr W J 2007 Investment of banking: Past, present, and future *Journal of Applied Corporate Finance* **19** (1) pp 42 – 54.

5910. Morrison A D, Wilhelm Jr W J November 15 2008a Investment banking: Institutions, politics, and law *Oxford University Press* ISBN: 9780199544189 pp 1 – 360

https://global.oup.com/academic/product/investment-banking-9780199544189?cc=us&lang=en& .

5911. Morrison A D, Wilhelm W J 2008b The demise of investment banking partnerships: Theory and evidence *Journal of Finance* **63** (1) pp 311 – 350.

5912. Brambilla C, Piluso G 2008 Italian investment and merchant banking up to 1914: Hybridising international models and practices *Discussion Paper no 69* Dipartimento di Economia e Management (DEM) University of Pisa Italy pp 1 – 42

http://www.ec.unipi.it/documents/Ricerca/papers/2008-69.pdf .

5913. Chindris-Vasioiu O 2008 Investment banking *Revista Economica* **42-43** (5-6) pp 42 – 54.

5914. Jaffee Dwight M, Perlow M 2008 Investment banking regulation after Bear Stearns *The Economists' Voice* **5** (5) pp 1 – 5.

5915. Bodnaruk A, Massa M, Simonov A 2009 Investment banks as insiders and the market for corporate control *Review of Financial Studies* **22** (12) pp 4989 – 5026.

5916. Bao J, Edmans A 2011 Do investment banks matter for M&A returns? *Review of Financial Studies* **24** (7) pp 2286 – 2315.

5917. Blundell-Wignall A 2011 On the necessity of separating investment and commercial banking *Intereconomics: Review of European Economic Policy* **46** (6) pp 298 – 299.

5918. Rubinton B J June 17 2011 Crowdfunding: Disintermediated investment banking *MPRA Paper no 31649* University Library of Munich Germany pp 1 – 23

http://mpra.ub.uni-muenchen.de/31649/ .

5919. Ziman I 2011 RAD applied in the context of investment banking *Informatica Economica* **15** (4) pp 134 – 146.

5920. Berzins J, Liu C H, Trzcinka Ch 2013 Asset management and investment banking *Journal of Financial Economics* **110** (1) pp 215 – 231.

5921. Clare A, Gulamhussen M A, Pinheiro C 2013 What factors cause foreign banks to stay in London? *Journal of International Money and Finance* **32** issue C pp 739 – 761.

5922. Erkan Celik I, Hacioglu U, Dincer H 2013 Regional development and effects of investment banks *International Journal of Finance & Banking Studies* **2** (1) pp 48 – 57.

5923. Bonin H 2014 From history to present: Investment banking at stake *HAL*.

5924. Grullon G, Underwood Sh, Weston J P 2014 Comovement and investment banking networks *Journal of Financial Economics* **113** (1) pp 73 – 89.

5925. Balluck K 2015 Investment banking: Linkages to the real economy and the financial system *Bank of England Quarterly Bulletin* **55** (1) pp 4 – 22.

5926. Corovei E-A 2015 Investment banks and their role in the financial crisis *Annals - Economy Series* **2** pages 118 – 123.

5927. Mishra R 2016 Asian infrastructure investment bank: An assessment *India Quarterly: A Journal of International Affairs* **72** (2) pp 163 – 176.

***Hedge fund, financial capital investment vehicle in finances:***

**5928.** Brown S J, Harlow V, Starks L 1996 Of tournaments and temptations: An analysis of managerial incentives in the mutual fund industry *Journal of Finance* **51** (1) pp 85 – 110.

**5929.** Brown S J, Goetzmann W N, Park J 1997 Conditions for survival: Changing risk and the performance of hedge fund managers and CTAs SSRN-id58477.pdf.

**5930.** Brown S J, Goetzmann W N, Ibbotson R G 1998 Offshore hedge funds: survival & performance 1989-1995 *Journal of Business* vol **72** pp 91 – 117.

**5931.** Brown S J, Goetzmann W N, Ibbotson R G 1999 Offshore hedge funds: Survival and performance 1989-1995 *The Journal of Business* vol **72** no 1 pp 91 – 117.

**5932.** Brown S J, Goetzmann W, Park J 2000 Hedge funds and the Asian currency crisis *Journal of Portfolio Management* **26** (4) pp 95 – 101.

**5933.** Brown S J 2001 Hedge funds: Omniscient or just plain wrong *Pacific-Basin Finance Journal* **9** (4) pp 301 – 311.

**5934.** Brown S J, Goetzmann W N, Park J 2001 Careers and survival: Competition and risk in the hedge fund and CTA industry *Journal of Finance* **61** pp 1869 – 1886.

**5935.** Brown S J, Goetzmann W N 2001 Hedge funds with style Yale International Center for Finance Working Paper No 00-29.

**5936.** Brown S J, Goetzmann W N 2003 Hedge funds with style *The Journal of Portfolio Management* vol **29** (2) pp 101 – 112.

**5937.** Brown S J, Fraser Th L, Liang B 2008 Hedge fund due diligence: A source of alpha in a hedge fund portfolio strategy
http://ssrn.com/paper=1016904.

**5938.** Brown S J, Goetzmann W N, Liang B, Schwarz Ch 2008 Mandatory disclosure and operational risk: Evidence from hedge fund registration *Journal of Finance* **63** (6) pp 2785 – 2815.

**5939.** Brown S J, Goetzmann W N, Liang B, Schwarz C 2010 Trust and Delegation *NBER Working Papers no 15529* National Bureau of Economic Research New York USA.

**5940.** Fung W, Hsieh D A 1997a Empirical characteristics of dynamic trading strategies: The case of hedge funds *Review of Financial Studies* **10** (2) pp 275 – 307.

**5941.** Fung W, Hsieh D A 1997b Investment style and survivorship bias in the returns of CTAs: The information content of track records *Journal of Portfolio Management* **24** pp 30 – 41.

**5942.** Fung W, Hsieh D A 1999a A primer on hedge funds *Journal of Empirical Finance* **6** (3) pp 309 – 331.

**5943.** Fung W, Hsieh D A 1999b Is mean-variance analysis applicable to hedge funds? *Economic Letters* **62** (1) pp 53 – 58.

**5944.** Fung W, Hsieh D A 2000a Measuring the market impact of hedge funds *Journal of Empirical Finance* **7** pp 1 – 36.

**5945.** Fung W, Hsieh D A 2000b Performance characteristics of hedge funds and commodity funds: Natural vs. spurious biases *The Journal of Financial and Quantitative Analysis* vol **35** no 3 pp 291 -307.

**5946.** Fung W, Hsieh D A 2001 The risk in hedge fund strategies: Theory and evidence from trend followers *Review of Financial Studies* **14** (2) pp 313 – 341.

*5947.* Fung W, Hsieh D A 2002a Risk in fixed-income hedge fund styles *Journal of Fixed Income* **12** no 2 pp 6 – 27.

*5948.* Fung W, Hsieh D A 2002b Asset-based style factors for hedge funds *Financial Analyst Journal* **58** no 5 pp 16 – 27.

*5949.* Fung W, Hsieh D A 2002c Benchmarks of hedge funds performance: information content and measurement bias *Financial Analyst Journal* **58** no 1 pp 22 – 34.

*5950.* Fung W, Hsieh D A 2003 The risks in hedge fund strategies: alternative alphas and alternative betas in The new generation of risk management for hedge funds and private equity funds Jaeger L editor pp 72 – 87 *Euromoney Institutional Investor PLC* London UK.

*5951.* Fung W, Hsieh D A 2004a Extracting portable alphas from equity long-short hedge funds *Journal of Investment Management* **2** no 4 pp 57 – 75.

*5952.* Fung W, Hsieh D A 2004b Hedge fund benchmarks: A risk-based approach *Financial Analysts Journal* **60** no 5 pp 65 – 80.

*5953.* Fung W, Hsieh D A 2006a The risk in hedge fund strategies: theory and evidence from long/short equity hedge funds *Duke University Working Paper* Raleigh North Carolina USA.

*5954.* Fung W, Hsieh D A 2006b Hedge funds: An industry in its adolescence *Federal Reserve Bank of Atlanta Economic Review* **91** Fourth Quarter pp 1 – 33.

*5955.* Fung W, Hsieh D A 2007 Hedge fund replication strategies: Implications for investors and regulators *Financial Stability Review* **10** pp 55-66.

*5956.* Fung W, Hsieh D A, Naik N, Ramadorai T 2006 Hedge funds: Performance, risk and capital formation *American Finance Association 2007 Chicago Meetings Paper* (July 19) http://ssrn.com/abstract=778124 (November 17, 2006).

*5957.* Fung W, Hsieh D A, Naik N Y, Ramadorai T 2008 Hedge funds: Performance, risk, and capital formation *Journal of Finance* **63** no 4 pp 1777 – 1803.

*5958.* Ackermann C, Ravenscraft D 1998 The impact of regulatory restrictions on fund performance: A comparative study of hedge funds and mutual funds *University of North Carolina Dissertation* Raleigh North Carolina USA.

*5959.* Ackermann C, McEnally R, Ravenscrat D 1999 The performance of hedge funds: Risk, returns and incentives *Journal of Finance* vol **53** pp 833 – 874.

*5960.* Eichengreen B, Mathieson D, Chadha B, Jansen A, Kodres L, Sharma S 1998 Hedge funds and financial market dynamics *International Monetary Fund Occasional Paper 166* Washington DC USA.

*5961.* Mathieson D, Chadha B, Jansen A, Kodres L, Eichengreen B, Sharma S 1998 Hedge fund and financial market dynamics *International Monetary Fund.*

*5962.* Edwards F R 1999 Hedge funds and the collapse of long-term capital management *Journal of Economic Perspectives* **13** (2) pp 189-210.

*5963.* Edwards F 2000a Measuring the market impact of hedge funds *Journal of Empirical Finance* **7** no 1 pp 1 – 36.

*5964.* Edwards F 2000b Performance characteristics of hedge funds and commodity funds: Natural vs. spurious biases *Journal of Financial and Quantitative Analysis* **35** no 3 pp 291 – 307.

5965. Edwards F, Caglayan M O 2001 Hedge fund and commodity fund investment styles in bull and bear markets *The Journal of Portfolio Management* **27** (4) pp 97 – 108.

5966. Edwards F 2003 The risks in hedge fund strategies: Alternative alphas and alternative betas in The new generation of risk management for hedge funds and private equity funds Jaeger L editor pp 72 – 87 London: *Euromoney Institutional Investor PLC*.

5967. Edwards F R, Gaon S 2003 Hedge funds: What do we know? *Journal of Applied Corporate Finance* **15** no 4 pp 58 – 71.

5968. Edwards F 2004a Extracting portable alphas from equity long-short hedge funds *Journal of Investment Management* **2** no 4 pp 57 – 75.

5969. Edwards F 2004b Hedge fund benchmarks: A risk-based approach *Financial Analysts Journal* **60** no 5 pp 65 – 80.

5970. Edwards F R 2006 Hedge funds and investor protection regulation *Economic Review Fourth Quarter* pp 35 – 48.

5971. Liang B 1999 On the performance of hedge funds *Financial Analysts Journal* **55** (4).

5972. Liang B 2000 Hedge funds: The living and the dead *The Journal of Financial and Quantitative Analysis* vol **35** no 3 pp 309 – 326.

5973. Liang B 2003 Hedge fund returns: Auditing and accuracy *Journal of Portfolio Management* vol **29** pp 111 – 122.

5974. Liang B 2004 Alternative investments: CTAs, hedge funds, and funds-of-funds *Journal of Investment Management* **3** (4) pp 76 – 93.

5975. President's Working Group on Financial Markets 1999 Hedge funds, leverage, and the lessons of long-term capital management *Report of the President's Working Group on Financial Markets*.

5976. Stonham P 1999 Too close to the hedge: The case of long term capital management LP Part one: Hedge fund analytics *European Management Journal* vol **17** pp 282 – 289.

5977. Stonham P 1999 Too close to the hedge: The case of long term capital management Part two: Near-collapse and rescue *European Management Journal* vol **17** issue 4 pp 382 – 390.

5978. Tatsaronis K 2000 Hedge funds *BIS Quarterly Review* vol **61** pp 61 – 71.

5979. Agarwal V, Naik N Y 2000 Performance evaluation of hedge funds with option based and buy-and-hold strategies *Working Paper London Business School* London UK.

5980. Agarwal V, Naik N Y 2004 Risks and portfolio decisions involving hedge funds *Review of Financial Studies* **17** (1) pp 63 – 98.

5981. Asness C, Krail R, Liew J 2001 Do hedge funds hedge? *Journal of Portfolio Management* **28** (1) pp 6 – 19.

5982. Braga M D 2001 Problematiche di performance measurement nell'hedge fund industry *Lettera Newfin* vol **14** no 2.

5983. Brealy R, Kaplanis E 2001 Hedge funds and financial stability: An analysis of factor exposures *International Finance* **4** (2) pp 161 – 187.

5984. Cochrane J H 2001 Asset pricing *Princeton University Press* USA.

*5985.* Brooks C, Kat H M 2001 The statistical properties of hedge fund index returns and their implications for investors *Journal of Alternative Investments* vol **5** no 3 pp 26 – 44.

*5986.* Amin G S, Kat H M 2001 Hedge fund performance 1990-2000: Do the "money machines" really add value? *ISMA Centre Discussion Papers in Finance 2001 - 05* University of Reading UK pp 1 – 33.

*5987.* Amin G S, Kat H M 2003a Hedge fund performance 1990-2000: Do the "money machines" really add value? *Journal of Financial and Quantitative Analysis* **38** (2) pp 251 – 274.

*5988.* Amin G, Kat H 2003b Stocks, bonds, and hedge funds: Not a free lunch! *The Journal of Portfolio Management* **29** (4) pp 113 - 120.

*5989.* Kat H M 2003 10 things that investors should know about hedge funds *Institutional Investor* pp 72 – 81.

*5990.* Kat H M, Menexe F 2003 Persistence in hedge fund performance: the true value of track record *Journal of Alternative Investments* vol **5** pp 66 – 72.

*5991.* Kat H M, Palaro H P 2005 Hedge fund returns – You can make them yourself! *Journal of Wealth Management* vol **8** no 2 pp 62 – 68.

*5992.* Kat H M, Palaro H P 2006 Replication and evaluation of funds of hedge funds returns *in* Fund of hedge funds: Performance, assessment, diversification and statistical properties editor Gregoriou G Chapter 3 *Elsevier Press* The Netherlands.

*5993.* Kat H M 2007 Alternative routes to hedge fund return replication *Journal of Wealth Management* **10** (3) pp 25 – 39.

*5994.* Kat H M 2010 Things that investors should know about hedge funds *Institutional Investor* pp 72 – 81.

*5995.* Capocci D, Hübner G 2001 L'Univers des hedge funds, une perspective empirique *Revue Bancaire et Financiere September* no 6 pp 361 – 369.

*5996.* Capocci D, Corhay A, Hübner G 2003 Hedge fund performance and persistence in bull and bear markets *Department of Management* Universite de Liege Belgium 0402018.pdf       pp 1 – 40.

*5997.* Capocci D, Hübner G 2004 An analysis of hedge fund performance *Journal of Empirical Finance* **11** (1) pp 55 – 89.

*5998.* Kramer D 2001 Hedge fund disasters: Avoiding the next catastrophe *Alternative Investment Quarterly* **1**.

*5999.* Goetzmann W N, Ingersoll J Jr, Ross S A 2001 High-water marks and hedge fund management contracts *Yale International Center for Finance Social Science Research Network* pp 1 – 41
http://papers.ssrn.com/paper.taf?abstract_id=270290 .

*6000.* Anson M J P 2002 Hedge funds Chapter 29 *in* Fabozzi F and Markowitz H editors The Theory and Practice of Investment Management *John Wiley & Sons Inc* New York USA.

*6001.* Favre L, Galeano J-A 2002 Mean –modified Value-at-Risk optimization with hedge funds *Journal of Alternative Investments* Fall vol **5**.

*6002.* Gimbel Th, Gupta F, Pines D 2002 Entry and exit: The lifecycle of a hedge fund 0407002.pdf pp 1 – 12 *Credit Suisse Asset Management* New York USA.

6003. Ineichen A 2002 Absolute returns: Risks and opportunities of hedge fund investing *John Wiley & Sons Inc* Hoboken New Jersey USA pp 1 – 514.

6004. Kao D 2002 Battle for alphas: Hedge funds versus long-only portfolios *Financial Analysts Journal* **58** pp 16 – 36.

6005. Locho R 2002 Hedge funds and hope *The Journal of Portfolio Management* **28** pp 92 – 99.

6006. Schneeweis, Kazemi, Martin 2002 Understanding hedge fund performance: Research issues revisited - Part I *Journal of Alternative Investments* Winter pp 6 – 22.

6007. Weismann A 2002 Informationless investing and hedge fund measurement bias *Journal of Portfolio Management* Summer pp 80 – 91.

6008. Amenc N, El Bied S, Martellini L 2003 Evidence of predictability in hedge fund returns *Financial Analysts Journal* **59** pp 32 – 46.

6009. Amenc N, Géhin W, Martellini L, Meyfredi J - Ch 2007 The myths and limits of passive hedge fund replication *EDHEC Risk and Asset Management Research Centre Working Paper*.

6010. Amenc N, Géhin W, Martellini L, Meyfredi J - Ch, Ziemann V 2008 Passive hedge fund replication - Beyond the linear case *EDHEC Risk and Asset Management Research Centre Working Paper*.

6011. Bacmann J F, Scholz S 2003 Alternative performance measures for hedge funds *AIMA Journal* vol **1** pp 1 – 9.

6012. Bares P-A, Gibson R, Gyger S 2003 Performance in the hedge fund industry: An analysis of short and long-term persistence *Journal of Alternative Investments* **6** pp 25 – 41.

6013. Geman H, Kharoubi C 2003 Hedge funds revisited: Distributional characteristics dependence structure, and diversification *Journal of Risk* **5** (4) pp 55 – 74.

6014. Gregoriou G N 2003 Performance appraisal of funds of hedge funds using data envelopment analysis *Journal of Wealth Management* **5** pp 88 – 95.

6015. Gregoriou G N, Gueyie J P 2003 Risk adjusted performance of funds of hedge funds using a modified Sharpe ratio *Journal of Wealth Management* vol **6** pp 77 – 83.

6016. Gregoriou G N, Sedzro K, Zhu J 2005 Hedge fund performance appraisal using data envelopment analysis *European Journal of Operational Research* **164** (2) p 555.

6017. Gregoriou G N, Kooli M, Rouah F 2008 Survival of strategic, market defensive, diversified and conservative fund of hedge funds: 1994-2005 *Journal of Derivatives and Hedge Funds* vol **13** no 4 pp 273 – 286.

6018. Goetzmann W N, Ingersoll J E Jr, Ross S A 2003 High-water marks and hedge fund management contracts *Journal of Finance* **58** pp 1685 – 1717.

6019. Gulko L 2003 Performance metrics for hedge funds *Journal of Alternative Investments* vol **5** pp 88 – 95.

6020. Ennis, Sebastian 2003 A critical look at the case for hedge funds Journal of Portfolio Management **29** (4) pp 103 – 123.

6021. Schneeweis, Kazemi, Martin 2003 Understanding hedge fund performance: Research issues revisited - Part II *Journal of Alternative Investments* Spring pp 8 – 30.

6022. Popova I, Morton D P, Popova E 2003 Optimal hedge fund allocation with asymmetric preferences and distributions *Technical Report* University of Texas at Austin Texas USA.

6023. Popova I, Morton D P, Popova E 2006, Efficient fund of hedge fund construction under downside risk measures *Journal of Banking and Finance* **30** pp 503 – 518.

6024. Morton D P, Popova E, Popova I 2006 Efficient fund of hedge funds construction under downside risk measures *Journal of Banking & Finance* vol **30** (2).

6025. Agarwal V, Naik N J 2004 Risk and portfolio decisions involving hedge funds The Review of Financial Studies pp 63 – 98.

6026. Aggarwal R K, Jorion Ph 2010 Hidden survivorship in hedge fund returns *Financial Analysts Journal* vol **66** no 2.

6027. Bacmann J F, Gawron G 2004 Fat tail risk in portfolios of hedge funds and traditional investments *Working Paper RMF Investment Management*.

6028. Baquero G, ter Horst J, Verbeek M 2004 Survival, look-ahead bias, and persistence in hedge fund performance *Journal of Financial and Quantitative Analysis* **40** pp 493 – 517.

6029. ter Horst J, Verbeek M 2004 Fund liquidation, self-selection and look-ahead bias in the hedge fund industry *ERIM Rep Ser Research in Management* ERS-2004-104-F&A pp 1 – 32.

6030. ter Horst J, Verbeek M 2007 Fund liquidation, self-selection and look-ahead bias in the hedge fund industry *Review of Finance* **11** no 4 pp 605 – 632.

6031. Boido C., Riente E 2004 Hedge fund: dal mito alla realtà *Banche e Banchieri* vol **5** pp 406 – 420.

6032. Abreu D, Brunnermeier M 2002 Synchronization risk and delayed arbitrage *Journal of Financial Economics* **66** (2-3) pp 341 – 360.

6033. Brunnermeier M, Nagel S 2004 Hedge funds and the technology bubble *Journal of Finance* **59** pp 2013 – 2040.

6034. Brunnermeier M 2009 Deciphering the liquidity and credit crunch of 2007-2008 *Journal of Economic Perspectives* **23** (1) pp 77 – 100.

6035. Brunnermeier M, Pedersen L H 2009 Market liquidity and funding liquidity *Review of Financial Studies* **22** (6) pp 2201 – 2238.

6036. Feiger G, Botteron P 2004 Should you, would you, could you invest in hedge funds? *Journal of Financial Transformation* vol **10** pp 57 – 65.

6037. Getmansky M, Lo A W, Mei S X 2004 Sifting through the wreckage: Lessons from recent hedge-fund liquidations *Journal of Investment Management* **2** pp 6 – 38.

6038. Getmansky M, Lo A W, Makarov I 2004 An econometric model of serial correlation and illiquidity in the hedge fund returns *Journal of Financial Economics* vol **74** pp 529 – 609.

6039. Hedges J R 2004 Size vs performance in the hedge fund industry *Journal of Financial Transformation* vol **10** April pp 14 – 17.

6040. Huber C, Kaiser H 2004 Hedge fund factors with option-like structures: Examples and explanations *Journal of Wealth Management* **7** pp 49 – 60.

**6041.** Nguyen-Thi-Thanh Huyen 2004 Hedge fund behavior: An ex-post analysis *Working Paper LEO* Université d'Orléans Rue de Blois BP 6739 45067 Orléans Cedex 2 France.

**6042.** Nguyen-Thi-Thanh Huyen 2006 On the use data envelopment analysis in hedge fund performance appraisal *Working Paper LEO* Université d'Orléans Rue de Blois BP 6739 45067 Orléans Cedex 2 France pp 1 – 34

http://halshs.archives-ouvertes.fr/halshs-00120292/fr/ .

**6043.** Lhabitant F S 2004 Hedge funds with quantitative insights *John Wiley & Sons Inc* USA.

**6044.** Posthuma N, van der Sluis P J 2004 A critical examination of historical hedge fund returns Chapter 13 *in* Intelligent hedge fund investing: Successfully avoiding pitfalls through better risk evaluation Schachter B editor *Risk Books*.

**6045.** Al-Sharkas A A 2005 The return in hedge fund strategies *International Journal of Business* vol **10** no 3.

**6046.** Alexander C, Dimitriu A 2005 Detecting switching strategies in equity hedge funds *Journal of Alternative Investments* **8** pp 7 – 13.

**6047.** Carretta A, Mattarocci G 2005 The performance evaluation of hedge funds: a comparison of different approaches using European data *MPRA Paper No. 4294* Munich University Munich Germany pp 1 – 18

http://mpra.ub.uni-muenchen.de/4294/ .

**6048.** Chan N T, Getmansky M, Haas Sh M, Lo A W 2005 Systemic risk and hedge funds *The Risks of Financial Institutions* (NBER Book Chapter).

**6049.** Chan N T, Getmansky M, Haas Sh M, Lo A W 2006 Do hedge funds increase systemic risk? *Economic Review - Federal Reserve Bank of Atlanta* **91** (4) pp 49 – 80.

**6050.** Chan N T, Getmansky M, Lo A W, Haas Sh M 2007 Systemic risk and hedge funds *in* Carey M, Stulz R editors The Risks of Financial Institutions and the Financial Sector *University of Chicago Press* Chicago IL USA.

**6051.** Cremers J, Kritzman M, Page S 2005 Optimal hedge fund allocation *Journal of Alternative Investments* pp 70 – 81.

**6052.** Danielsson J, Taylor A, Zigrand J P 2005 Highwaymen or heroes: Should hedge funds be regulated?: A survey *Journal of Financial Stability* **1** (4) pp 522 – 543.

**6053.** Do V, Faff R, Wickramanayake J 2005 An empirical analysis of hedge fund performance: The case of Australian hedge funds industry *Journal of Multinational Financial Management* **15** (4-5) pp 377 – 393.

**6054.** Eling M, Schuhmacher F 2005 Performance-Maße für Hedgefonds-Indizes – wie geeignetist die Sharpe-Ratio? *Absolut Report no 29* December pp 36 – 43.

**6055.** Jaeger L, Wagner C 2005 Factor modelling and benchmarking of hedge funds: Can passive investments in hedge funds deliver? *Journal of Alternative Investments* **8** pp 9 – 36.

**6056.** Hodder J E, Jackwerth J C 2005 Incentive contracts and hedge fund management *Finance Department* School of Business University of Wisconsin-Madison USA; Department of Economics University of Konstanz Germany pp 1 – 34.

**6057.** Malkiel B G, Saha A 2005 Hedge funds: Risk and return *Financial Analysts Journal* **61** pp 80 – 88.

**6058.** Kaiser D G, Kisling K 2005 Der einfluss von kapitalbindungsfristen auf die Sharpe ratio aktienbasierter hedgefonds-strategien *Absolut Report no 28 10/2005* pp 26 – 33.

**6059.** Garbaravičius T 2005 Hedge funds and their implications for financial stability *ECB Occasional Paper Series 34.*

**6060.** Garbaravičius T, Dierick F 2005 Hedge funds and their implications for financial stability *ECB Occasional Paper no 34* August.

**6061.** Gilroy B M, Lukas E 2005 Economic theory in everyday life: Hedge funds *MPRA Paper No. 22047* Munich University Munich Germany pp 1 – 6

http://mpra.ub.uni-muenchen.de/22047/.

**6062.** Gupta A, Lang B 2005 Do hedge funds have enough capital? A value-at-risk approach *Journal of Financial Economics* **77** pp 219-253.

**6063.** Azman-Saini W N W 2006 Hedge funds, exchange rates and causality: Evidence from Thailand and Malaysia *MPRA Paper No. 716* Munich University Munich Germany http://mpra.ub.uni-muenchen.de/716/ .

**6064.** Baba N, Goko H 2006 Survival analysis of hedge funds *Bank of Japan Working Paper Series no 06-E-05.*

**6065.** Heidorn T, Hoppe C, Kaiser D G 2006a Hedgefondszertifikate in Deutschland - marktanalyse, strukturierungsvarianten und eignung für privatinvestoren *BankArchiv Ausgabe 2/2006* Germany pp 87 – 97.

**6066.** Heidorn T, Hoppe C, Kaiser D G 2006b Konstruktion und verzerrungen von hedgefonds-indizes in Busack M, Kaiser D G Editors Handbuch Alternative Investments Band 1 *Gabler Verlag* Wiesbaden Germany.

**6067.** Jagannathan R, Malakhov A, Novikov D 2006 Do hot hands exist among hedge fund managers? An empirical evaluation *Working Paper 12015 National Bureau of Economic Research* Cambridge Massachusetts USA

http://www.nber.org/papers/w12015 .

**6068.** Sadka R 2006 Momentum and post-earnings-announcement drift anomalies: The role of liquidity risk *Journal of Financial Economics* **80** pp 309 – 349.

**6069.** Boyson N M, Stahel Ch W, Stulz R M 2006 Is there hedge fund contagion? *Working Paper 12090 National Bureau of Economic Research* Cambridge Massachusetts USA http://www.nber.org/papers/w12090 .

**6070.** Boyson N M, Stahel Ch W, Stulz R M 2008 Hedge fund contagion and liquidity *Working Paper 14068 National Bureau of Economic Research* Cambridge Massachusetts USA pp 1 – 46

http://www.nber.org/papers/w14068 .

**6071.** Izzo P 2006 Moving the market: Economists see hedge-fund risks — Survey indicates concerns about a lack of oversight, use of borrowed money *Wall Street Journal* Oct. 13 p C3.

**6072.** Jackwerth J C, Hodder J E 2006 Incentive contracts and hedge fund management *MPRA Paper No. 11632* Munich University Munich Germany

http://mpra.ub.uni-muenchen.de/11632/ .

**6073.** Ding B, Shawky H A 2006 The performance of hedge fund strategies and the asymmetry of return distributions *Center for Institutional Investment Management Working Paper* Department of Finance School of Business University at Albany USA.

6074. Heidorn Th, Hoppe Ch, Kaiser D G 2006 Heterogenität von Hedgefondsindizes *Working paper series HfB - Business School of Finance & Management* no 71 pp 1 – 39

http://nbn-resolving.de/urn:nbn:de:101:1-2008082777 ,

http://hdl.handle.net/10419/27839 .

6075. Adrian T 2007 Measuring risk in the hedge fund sector *Federal Reserve Bank of New York Current Issues in Economics and Finance* **13** (3).

6076. Becker Ch, Clifton K 2007 Hedge fund activity and carry trades in Research on Global Financial Stability: The use of BIS International Financial Statistics *CGFS Publications no 29* pp 156 – 175.

6077. Bowler B 2007 The emergence of synthetic hedge funds http://www.pacificprospect.com/jsp_2007/downloads/a/2.pdf .

6078. Billio M, Getmansky M, Pelizzon L 2007 Dynamic risk exposure in hedge funds working paper *University of Massachusetts* Amherst MA USA.

6079. Kambhu J, Schuermann T, Stiroh K 2007 Hedge funds, financial intermediation, and systemic risk *FRBNY Economic Policy Review* **13** pp 1 – 18.

6080. Kosowski R, Naik N Y, Teo M 2007 Do hedge funds deliver Alpha? A Bayesian and Bootstrap analysis *Journal of Financial Economics* **84** pp 229 – 264.

6081. Li Sh, Linton O 2007 Evaluating hedge fund performance: a stochastic dominance approach *Discussion Paper no 591 London School of Economics* London UK ISSN 0956-8549-591 pp 1 – 20.

6082. Smedts K, Smedts J 2007 Dynamic investment strategies of hedge funds **AFI 0622** *Department of Accountancy, Finance and Insurance (AFI)* Faculty of Economics and Applied Economics Catholic University of Leuven Naamsestraat 69 3000 Leuven Belgium pp 1 – 27.

6083. Stulz R M 2007 Hedge funds: Past, present and future *Journal of Economic Perspectives* **21** (2) pp 175 – 194.

6084. Goltz F, Martellini L, Vaissié M 2007 Hedge fund indices: Reconciling investability and representativity *European Financial Management* vol **13** no 2 pp 257 – 286.

6085. King M R, Maier P 2007 Hedge funds and financial stability: The state of the debate *Bank of Canada*

http://www.bankofcanada.ca/wp-content/uploads/2010/01/dp07-9.pdf .

6086. Hakamada T, Takahashi A, Yamamoto K 2007 Selection and performance analysis of Asia-Pacific hedge funds *Journal of Alternative Investments* vol **10** (3) pp 7 – 29.

6087. Hasanhodzic J, Lo A 2007 Can hedge-fund returns be replicated?: The linear case *Journal of Investment Management* **5** (2) pp 5 – 45.

6088. Papademos L D 2007 Monitoring hedge funds: A financial stability perspective *Banque de France Financial Stability Review* – Special Issue on Hedge Funds no 10 pp 113 – 125.

6089. Weber A A 2007 Hedge funds: A central bank perspective (Deutsche Bundesbank) *Banque de France Financial Stability Review* – special issue on hedge funds no 10 April 2007 pp 1 – 8.

**6090.** Billio M, Getmansky M, Pelizzon L 2008 Non-parametric analysis of hedge fund returns: New insight from high frequency date *Working Paper no 1 1 /WP/2008* ISSN 1827-336X Ca'Foscari University of Venice Italy pp 1 – 41.

**6091.** Carlson M, Steinman J 2008 Market conditions and hedge fund survival *Finance and Economics Discussion Series Divisions of Research & Statistics and Monetary Affairs* Federal Reserve Board Washington DC USA.

**6092.** Lo A W 2008 Hedge funds, systemic risk, and the financial crisis of 2007-2008 *Written testimony prepared for the U.S. House of Representatives Committee on Oversight and Government Reform* November 13, 2008 Hearing on Hedge Funds.

**6093.** McGuire P, Tsatsaronis K 2008 Estimating hedge fund leverage *BIS Quarterly Review no 260*.

**6094.** Kazemi H B, Tu F, Li Y 2008 Replication and benchmarking of hedge funds *Journal of Alternative Investments* vol **11** no 2 pp 40 – 59.

**6095.** Nahum R, Aldrich D 2008 Hedge fund operational risk: meeting the demand for higher transparency and best practices *Journal of Financial Transformation* **22** pp 104 – 107.

**6096.** Gray W 2008 Information exchange and the limits of arbitrage *MPRA Paper No 11918* Munich University Munich Germany pp 1 – 30

http://mpra.ub.uni-muenchen.de/11918/ .

**6097.** Gray W, Kern A 2008 Fundamental value investors: Characteristics and performance *MPRA Paper No 12620* Munich University Munich Germany pp 1 – 30

http://mpra.ub.uni-muenchen.de/12620/ .

**6098.** Gupta B, Szado E, Spurgin W 2008 Performance characteristics of hedge fund replication programs *Journal of Alternative Investments* vol **11** no 2 pp 61 – 68.

**6099.** Roncalli Th, Teiletche J 2008 An alternative approach to alternative beta *Journal of Financial Transformation* Cass-Capco Institute Paper Series on Risk

http://www.thierry-roncalli.com/#gauss_l17 .

**6100.** Roncalli Th, Weisang G 2008 Tracking problems, hedge fund replication and alternative beta *MPRA Paper No. 37358* Munich University Munich Germany pp 1 – 67 http://mpra.ub.unimuenchen.de/37358/ .

**6101.** de los Rios A D, Garcia R 2008 Assessing and valuing the non-linear structure of hedge fund returns *Bank of Canada Working Paper* Ottawa Canada.

**6102.** Jackwerth J C, Kolokolova O, Hodder J E 2008 Recovering delisting returns of hedge funds *MPRA Paper No. 11641* Munich University Munich Germany

http://mpra.ub.uni-muenchen.de/11641/ .

**6103.** Takahashi A, Yamamoto K 2008 Hedge fund replication *CIRJE-F-592* Graduate School of Economics University of Tokyo Japan pp 1 – 32

http://www.e.u-tokyo.ac.jp/cirje/research/03research02dp.html .

**6104.** Hedge Fund Working Group & Hedge Fund Standards Board 2008 Hedge fund standards: Final report *Hedge Fund Working Group & Hedge Fund Standards Board* http://www.hfsb.org/files/final_report.pdf .

**6105.** Bollen N P B, Pool V K 2009 Do hedge fund managers misreport returns? Evidence from the pooled distribution *Journal of Finance* **64** (5) pp 2257 – 2288.

6106. Brophy D J, Ouimet P P, Sialm C 2009 Hedge funds as investors of last resort? *Review of Financial Studies* **22** pp 541 – 574.

6107. Füss R, Kaiser D G, Strittmatter A 2009 Measuring funds of hedge funds performance using quantile regressions: Do experience and size matter? *Journal of Alternative Investments* vol **12** no 2 pp 41 – 53.

6108. Heidorn T, Kaiser D G, Roder C 2009 The risk of funds of hedge funds: An empirical analysis of the maximum drawdown *Journal of Wealth Management* vol **12** no 2 pp 89 – 100.

6109. Jaeger L 2009 Alternative beta strategies and hedge fund replication 1st edition *John Wiley & Sons Inc* New York USA.

6110. Khanniche S 2009 Evaluation of hedge fund returns value at risk using GARCH models *Working Paper 2009 - 46* Groupama Asset Management 58 bis rue de la Boétie 75008 Paris; Université de Paris Ouest Nanterre La Défense 200 Avenue de la République 92001 Nanterre CEDEX France pp 1 – 39

http://economix.u-paris10.fr/ .

6111. Minsky B, Obradovic M, Tang Q, Thapar R 2009 Applying a global optimization algorithm to fund of hedge funds portfolio optimization *MPRA Paper No 17099* Munich University Munich Germany pp 1 – 24

http://mpra.ub.uni-muenchen.de/17099/ .

6112. Mitra S 2009 An introduction to hedge funds *Cornell University* NY USA pp 1 – 25 http://arxiv.org/abs/0904.2731v2 .

6113. Xiong J, Idzorek T M, Chen P, Ibbotson R 2009 Impact of size and flows on performance for funds of hedge funds *Journal of Portfolio Management* **35** (4) pp 118 – 130.

6114. Gibson R, Wang S 2010 Hedge fund alphas: do they reflect managerial skills or mere compensation for liquidity risk bearing? *Research Paper Series N°08 – 37* Swiss Finance Institute University of Genève Switzerland pp 1 – 66

http://ssrn.com/abstract=1304541 .

6115. Heidorn Th, Kaiser D G, Voinea A 2010 The value-added of investable hedge fund indices *Working paper series Frankfurt School of Finance & Management no 141* ISSN: 14369753 pp 1 – 58

http://hdl.handle.net/10419/36695 .

6116. Maillard S, Roncalli Th, Teiletche J 2010 The properties of equally weighted risk contributions portfolios *Journal of Portfolio Management* **36** (4) pp 60 – 70.

6117. Ramadorai T 2010 Investor interest and hedge fund returns http://www.cepr.org/pubs/dps/DP8092 .

6118. Sadka R 2010 Liquidity risk and the cross section of hedge fund returns *Journal of Financial Economics* **98** (1) pp 54 – 71.

6119. Titman S 2010 The leverage of hedge funds *Finance Research Letters* **7** pp 2 – 7.

6120. Wallerstein E, Tuchschmid N S, Zaker S 2010 How do hedge fund clones manage the real world? *Journal of Alternative Investments* vol **12** no 3 pp. 51 – 60.

6121. Ang A, Gorovyy S, van Inwegen G B 2011 Hedge fund leverage *Working Paper 16801 National Bureau of Economic Research* Cambridge Massachusetts USA http://www.nber.org/papers/w16801 .

6122. Cao Y, Ogden J P, Tiu C I 2011 Who benefits from funds of hedge funds? A critique of alternative organizational structures in the hedge fund industry (i) *Business Excellence and Management* vol **1** issue 1 pp 19 – 36.

6123. Cao Y, Ogden J P, Tiu C I 2012 Who benefits from funds of hedge funds? A critique of alternative organizational structures in the hedge fund industry (i) *Business Excellence and Management* vol **2** issue 1 pp 5 – 20.

6124. Freed M F, McMillan B 2011 Investible benchmarks & hedge fund liquidity *MPRA Paper No 32226* Munich University Munich Germany pp 1 – 16

http://mpra.ub.uni-muenchen.de/32226/ .

6125. Piluso F, Amerise I L 2011 The asset allocation of hedge funds during the financial crisis: An empirical investigation *MPRA Paper No 28178* Munich University Munich Germany pp 1 – 26

http://mpra.ub.uni-muenchen.de/28178/ .

6126. Eychenne K, Martinetti S, Roncalli Th 2011 Strategic asset allocation Lyxor *White Paper Series 6*

www.lyxor.com .

6127. Ben Dor B A, Eisenthal-Berkovitz Y, Xu J 2012 A quantitative framework for analyzing the performance of an individual hedge fund vs its peers *Barclays Research* UK.

6128. Bruder B, Roncalli Th 2012 Managing risk exposures using the risk budgeting approach *Working Paper*

www.ssrn.com/abstract=2009778 .

6129. Chakravarty S, Deb S S 2012 Capacity constraints and the opening of new hedge funds pp 1 – 59.

6130. Chen J, Tindall M L 2012 Hedge fund dynamic market sensitivity *Occasional Paper 12-01* Financial Industry Studies Department Federal Reserve Bank of Dallas Texas USA.

6131. Chen J, Tindall M L 2013 Understanding hedge fund alpha using improved replication methodologies *Occasional Paper 13-02* Financial Industry Studies Department Federal Reserve Bank of Dallas USA pp 1 – 21.

6132. Roncalli Th, Weisang G 2012 Risk parity portfolios with risk factors *MPRA Paper No 44017* Munich University Munich Germany pp 1 – 32

http://mpra.ub.uni-muenchen.de/44017/ .

6133. Hassine M, Roncalli Th 2013 Measuring performance of exchange traded funds *MPRA Paper No. 44298* Munich University Munich Germany pp 1 – 32

http://mpra.ub.uni-muenchen.de/44298/ .

6134. Agarwal V, Vikram N, Sugata R 2013 Institutional investment and intermediation in the hedge fund industry *CFR Working Paper no 13-03* Centre for Financial Research (CFR) Leibniz Information Centre for Economics University of Cologne Germany pp 1 – 56 http://hdl.handle.net/10419/76876 .

6135. European Commission Working Document of the Commission Services (DG Internal Market): Consultation paper on hedge funds http://ec.europa.eu/internal_market/consultations/docs/hedgefunds/consultation_paper_en.pdf .

6136. Stock J H, Watson M W 2002 Macroeconomic forecasting using diffusion indexes *Journal of Business and Economic Statistics* **20** no 2 pp 147 – 162.

6137. Jorion P 2003 Portfolio optimization with tracking-error constraints *Financial Analysts Journal* **59** (5) pp 70 – 82.

6138. Gikhman I I, Skorokhod A V 2004 The theory of stochastic processes II *Springer-Verlag* Berlin Germany.

6139. Avramov D 2004 Stock return predictability and asset pricing models *Review of Financial Studies* **17** 699 – 738.

6140. Avramov D, Wermers R 2006 Investing in mutual funds when returns are predictable *Journal of Financial Economics* **81** pp 339 – 377.

6141. Avramov D, Chordia T 2006 Predicting stock returns *Journal of Financial Economics* **82** pp 387 – 415.

6142. Hull J C 2005-2006 Private communications on investment portfolio allocation *Rotman School of Management* University of Toronto Ontario Canada.

6143. Hull J C 2010 Fundamentals of futures and options markets *Prentice Hall* 7th Edition ISBN-10: 0136103227 ISBN-13: 978-0136103226 USA pp 1 – 624.

6144. Hull J C 2012a Options, futures, and other derivatives *Prentice Hall* 8th Edition ISBN: 0-13-216484-9 USA pp 1 – 816.

6145. Hull J C 2012b Risk management and financial institutions *John Wiley and Sons Inc* 3rd Edition ISBN: 978-1-1182-6903-9 USA pp 1 – 672.

6146. Schnoor I 2005-2006 Private communications on risk management *Rotman School of Management* University of Toronto Canada.

6147. Schnoor I 2006 Comparable analysis and data manipulation tools *The Marquee Group* Toronto Canada pp 1 – 66.

6148. Basel Committee on Banking Supervision 2006 International convergence of capital measurement and capital standards: A revised framework *Bank for International Settlements (BIS)* Switzerland

http://www.bis.org/publ/bcbsca.htm .

6149. Basel Committee on Banking Supervision 2009 Principles for sound stress testing practices and supervision - final paper *Bank for International Settlements (BIS)* Switzerland

http://www.bis.org/publ/bcbs155.htm .

6150. Scherer B 2007 Portfolio construction & risk budgeting 3rd edition *Risk Books*.

6151. Xiaohong Chen, Hansen L P, Carrasco M 2009 Nonlinearities and temporal dependence *CIRANO ISSN 1198-8177* pp 1 – 33.

6152. Caporin M, Ranaldo A, Santucci de Magistris P 2011 On the predictability of stock prices: A case for high and low prices *Swiss National Bank Working Paper 2011 - 11* Swiss National Bank Börsenstrasse 15 Zurich Switzerland ISSN 1660-7716 (printed version) ISSN 1660-7724 (online version) pp 1 – 34.

6153. Derman E September 2016 My life as a quant: Reflections on physics and finance pp 1 – 308 ISBN-13: 978-0470192733.

6154. Whitehead Ch September 23 2016 Private communication on capital investment, hedge funds, and corporate law *VIII International Economic Forum: Innovations Investments Kharkiv Initiatives* Kharkiv Ukraine.

***Pension fund investment, financial capital investment vehicle in finances:***

**6155.** MacIntosh M R 1976 The great pension fund robbery *Canadian Public Policy* **2** (2) pp 256 – 261.

**6156.** Sharpe W F June 1976 Corporate pension funding policy *Journal of Financial Economics* pp 183 – 193.

**6157.** Treynor J May 1977 The principles of corporate pension finance *Journal of Finance* pp 627 – 638.

**6158.** Winklevass H E 1977 Pension mathematics: With numerical fluctuations *Richard D Irwin Inc* USA.

**6159.** Bulow J I 1979 Analysis of pension funding under Erisa *NBER Working Paper no 402* National Bureau of Economic Research Inc USA.

**6160.** Bulow J I August 1982 What are corporate pension liabilities *Quarterly Journal of Economics* **97**.

**6161.** Bulow J I, Morck R, Summers L 1987 How does the market value unfunded pension liabilities? *in* Issues in Pension Economics Bodie Z, Shoven J, Wise D (editors) *University of Chicago Press* Chicago USA.

**6162.** Bodie Z Fall 1980 An innovation for stable real retirement income *The Journal of Portfolio Management* pp 5 – 13.

**6163.** Bodie Z, Light J O, Morck R, Taggart Jr R H 1987 Corporate pension policy: An empirical investigation *in* Issues in Pension Economics Bodie, Shoven, Wise (editors) *University of Chicago Press* Chicago USA.

**6164.** Bodie Z, Shoven J, Wise D (editors) 1987 Issues in pension economics *University of Chicago Press* Chicago USA.

**6165.** Bodie Z October 1988 Pension fund investment policy *NBER Working Paper no 2752* National Bureau of Economic Research Inc USA

http://www.nber.org/papers/w2752.pdf .

**6166.** Bodie Z, Kane A, Marcus A J 1989 Investments *Irwin* Homewood Illinois.

**6167.** Black F September—October 1980 The tax consequences of long run pension policy *Financial Analysts Journal* pp 17 — 23.

**6168.** Black F, Dewhurst M P Summer 1981 A new investment strategy for pension funds *Journal of Portfolio Management.*

**6169.** Feldstein M S, Seligman S 1981 Pension funding, share prices and national savings *Journal of Finance* **36** pp 801 – 824.

**6170.** Feldstein M S 1982 Private pensions as corporate debt *in* Changing Roles of Debt and Equity in Financing U.S. Capital Formation Friedman B (editor) *University of Chicago Press* Chicago USA.

**6171.** Feldstein M S, Morck R I 1983 Pension funding decisions, interest rate assumptions, and share prices *in* Financial Aspects of the U.S. Pension System Bodie Z, Shoven S (editors) *University of Chicago Press* Chicago USA pp 177 – 207.

**6172.** Frankfurter G M, Hill J M 1981 A normative approach to pension fund management *Journal of Financial and Quantitative Analysis* **16** (4) pp 533 – 555.

**6173.** Tepper I March 1981 Taxation and corporate pension policy *Journal of Finance* pp 1 – 13.

6174. Harrison M J, Sharpe W F 1983 Optimal funding and asset allocation rules for defined benefit pension plans Chapter 4 *in* Financial Aspects of U.S. Pension Systems Bodie, Shoven (editors) *University of Chicago Press* Chicago USA.

6175. Friedman B M 1983 Pension funding, pension asset allocation and corporate finance: Evidence from individual company data *in* Financial Aspects of U.S. Pension Systems Bodie, Shoven (editors) *University of Chicago Press* Chicago USA.

6176. Kotlikoff L S, Smith D E 1983 Pensions in the American economy *NBER - University of Chicago Press* USA.

6177. Warshawsky M J 1987 The funding of private pension plans *Staff Studies Paper no 155* Board of Governors of the Federal Reserve System USA.

6178. Warshawsky M J November 1988 Pension plans: Funding, assets, and regulatory environment *Federal Reserve Bulletin* issue November pp 717 – 730.

6179. Warshawsky M J 1989 The adequacy of funding of private defined benefit pension plans *Finance and Economics Discussion Paper no 58* Board of Governors of the Federal Reserve System USA.

6180. Warshawsky M J 1990 Financial accounting for pensions: Measures of funding status *Finance and Economics Discussion Series Paper no 145* Board of Governors of the Federal Reserve System USA.

6181. Bernheim B D, Shoven J B 1985 Pension funding and saving *NBER Working Paper no 1622* National Bureau of Economic Research Inc USA pp 1 – 42.

6182. Ang J, Tsong-Yue Lai 1988 On optimal pension funding policy *Journal of Economics and Business* **40** (3) pp 229 – 238.

6183. Bernheim B D, Shoven J B 1988 Pension funding and saving Chapter *in* Pensions in the US Economy *National Bureau of Economic Research Inc* pp 85 – 114 http://www.nber.org/chapters/c6045 .

6184. Coggin T D, Fabozzi F J, Rahman S 1993 The investment performance of US equity pension fund managers: An empirical investigation *Journal of Finance* **48** pp 1039 – 1055.

6185. Haberman S, Joo-Ho Sung 1994 Dynamic approaches to pension funding *Insurance: Mathematics and Economics* **15** (2-3) pp 151 – 162.

6186. Mitchell O, Smith R S 1994 Pension funding in the public sector *The Review of Economics and Statistics* **76** (2) pp 278 – 290.

6187. Davis E P 1995 Pension funds. Retirement-income security and capital markets. An international perspective *Clarendon Press* London UK.

6188. Dyson A C L, Exley C J 1995 Pension fund asset valuation and investment *British Actuarial Journal* **1** (3) pp 471 – 557.

6189. Wahal S 1996 Pension fund activism and firm performance *Journal of Financial and Quantitative Analysis* **31** (1) pp 1 – 23.

6190. Brown G, Draper P, McKenzie E March 1997 Consistency of UK pension fund performance *Journal of Business Finance and Accounting* **24** pp 155 – 178.

6191. Cairns A J G, Parker G 1997 Stochastic pension fund modelling *Insurance: Mathematics and Economics* **21** (1) pp 43 – 79.

**6192.** Blake D 1998 Pension schemes as options on pension fund assets: Implications for pension fund management *Insurance: Mathematics and Economics* **23** (3) pp 263 – 286.

**6193.** Blake D, Lehmann B, Timmermann A 1999 Asset allocation dynamics and pension fund performance *Journal of Business* **72** pp 429 – 461.

**6194.** Cangiano M, Cottarelli C, Cubeddu L 1998 Pension development and reforms in transitional economies *IMF Working Paper no 151* IMF USA.

**6195.** Hemming R 1998 Should public pensions be funded? *IMF Working Paper no 98/35* International Monetary Fund USA.

**6196.** Mitchell O S 1998 Administrative costs in public and private pension plans *in* Privatizing Social Security Feldstein M (editor) *University of Chicago Press* Chicago USA pp 403 – 456.

**6197.** Clark G July 13 2000 Pension fund capitalism *Oxford University Press* Oxford UK ISBN: 9780199240487 pp 1 – 368.

**6198.** Clark G March 27 2003 European pensions and global finance *Oxford University Press* Oxford UK ISBN: 9780199253630 pp 1 – 272.

**6199.** Clark G L 2008 Pension fund governance: Expertise and organizational form Chapter 2 *in* Pension Fund Governance Evans J, Orszag M, Piggott J (editors) *Edward Elgar Publishing* UK ISBN: 9781847204851 pp 1 – 21

https://www.elgaronline.com/view/9781847204851.00010.xml .

**6200.** Sinn H-W 2000 Why a funded pension system is useful and why it is not useful *International Tax and Public Finance* **7** pp 389 – 410.

**6201.** Srinivas P S, Whitehouse E, Yermo J 2000 Regulating private pension funds' structure, performance and investments: Cross-country evidence *The World Bank - Social Protection Discussion Paper no 113*.

**6202.** Whitehouse E 2000 Paying for pensions, an international comparison of administrative charges in funded retirement-income systems *FSA Occasional Paper Series no 13*.

**6203.** Chapman R J, Gordon T J, Speed C A 2001 Pensions, funding and risk *British Actuarial Journal* **7** (04) pp 605 – 662.

**6204.** Head S J, Adkins D R, Cairns A J G, Corvesor A J, Cule D O, Exley C J, Johnson I S, Spain J G, Wise A J 2001 Pension fund valuations and market values *British Actuarial Journal* **7** (1) pp 103 – 122.

**6205.** Thomas A, Tonks I 2001 Equity performance of segregated pension funds in the UK *Journal of Asset Management* **1** (4) pp 321 – 343.

**6206.** Tonks I 2002 Performance persistence of pension fund managers *London School of Economics and Political Science London* UK pp 1 – 26

http://eprints.lse.ac.uk/24942/ .

**6207.** Besley T, Prat A 2003 Pension funds governance and the choice between defined Benefit and defined contribution plans *CEPR Discussion Paper no 3955*.

**6208.** Kakabadse N, Kakabadse A, Kouzmin A 2003 Pension fund trustees: Role and contribution *European Management Journal* **21** (3) pp 376 – 386.

**6209.** Bateman H, Mitchell O S 2004 New evidence on pension plan design and administrative expenses: The Australian experience *Journal of Pension Economics and Finance* **3** (1) pp 63 – 76.

**6210.** Bergstresser D, Desai M A, Rauh J 2004 Earnings manipulation and managerial investment decisions: Evidence from sponsored pension plans *NBER Working Paper no 10543* NBER USA.

**6211.** Eaton T V, Nofsinger J R 2004 The effect of financial constraints and political pressure on the management of public pension plans *Journal of Accounting and Public Policy* **23** pp 161 – 189.

**6212.** Munnell A H, Sundén A 2004 Coming up short: The challenge of 401(k) plans *The Brookings Institution Press* USA.

**6213.** Owadally M I, Haberman S 2004 The treatment of assets in pension funding *ASTIN Bulletin: The Journal of the International Actuarial Association* **34** (02) pp 425 – 433.

**6214.** Sundén A 2004 How do individual accounts work in the Swedish pension system? *Issue in Brief Center for Retirement Research no 22.*

**6215.** Cowling C A, Gordon T J, Speed C A 2005 Funding defined benefit pension schemes *British Actuarial Journal* **11** (1) pp 63 – 97.

**6216.** Diamond P 2005 Reforming public pensions in the US and the UK *UBS PRP Discussion Paper no 38.*

**6217.** Dobronogov A, Murthi M 2005 Administrative fees and costs of mandatory private pensions in transition economies *Journal of Pension Economics and Finance* **4** (1) pp 31 – 55.

**6218.** Franzoni F, Marin J January 25 2005 Pension plan funding and market efficiency *Working Paper no 31* Barcelona Graduate School of Economics Barcelona University Barcelona Spain pp 1 – 50
http://www.barcelonagse.eu/sites/default/files/working_paper_pdfs/31.pdf .

**6219.** McCarthy D, Neuberger A 2005 The pension protection fund *Fiscal Studies* **26** (2) pp 139 – 167.

**6220.** Greco L G 2006 The optimal design of funded pensions *Financial Market Group* London School of Economics and Political Science London UK pp 1 – 47
http://eprints.lse.ac.uk/24519/.

**6221.** Miao Jerry C Y, Wang J L 2006 Intertemporal stable pension funding *Asia-Pacific Journal of Risk and Insurance* **1** (2) pp 1 – 15.

**6222.** Sweeting P 2006 Correlation and the pension protection fund *Fiscal Studies* **27** (2) pp 157 – 182.

**6223.** Perotti E, Schwienbacher A 2007 The political origin of pension funding *CEPR Discussion Paper no 6100* CEPR
www.cepr.org/active/publications/discussion_papers/dp.php?dpno=6100 .

**6224.** Antolin P 2008 Pension fund performance *OECD Working Paper no 20 on Insurance and Private Pensions* OECD Publishing DOI: 10.1787/240401404057 pp 1 – 20
http://www.oecd-ilibrary.org/finance-and-investment/pension-fund-performance_240401404057.

**6225.** Evans J, Orszag M, Piggott J (editors) January 1 2008 Pension fund governance *Edward Elgar Publishing* UK ISBN: 9781847204851 pp 1 – 288

https://www.elgaronline.com/view/9781847204851.xml.

**6226.** Impavido G 2008 Governance of public pension plans: The importance of residual claimants Chapter 6 *in* Pension Fund Governance Evans J, Orszag M, Piggott J (editors) *Edward Elgar Publishing* UK ISBN: 9781847204851 pp 1 – 19

https://www.elgaronline.com/view/9781847204851.00015.xml .

**6227.** Adam A, Moutos Th 2009 Pension funding in a unionized economy *Scottish Journal of Political Economy* **56** (2) pp 213 – 231.

**6228.** Inderst G January 1 2009 Pension fund investment in infrastructure *OECD Working Paper no 32 on Insurance and Private Pensions* OECD Publishing DOI: 10.1787/227416754242 pp 1 – 45

http://www.oecd-ilibrary.org/finance-and-investment/pension-fund-investment-in-infrastructure_227416754242 .

**6229.** Kleinow T 2011 Pension fund management and conditional indexation *ASTIN Bulletin: The Journal of the International Actuarial Association* **41** (1) pp 61 – 86.

**6230.** Munnell A, Aubry J-P, Quinby L 2011 Public pension funding in practice *Journal of Pension Economics and Finance* **10** (2) pp 247 – 268.

**6231.** Bovenberg L, Mehlkopf R 2014 Optimal design of funded pension schemes *Annual Review of Economics* **6** (1) pp 445 – 474.

**6232.** Rossi A G, Blake D, Timmermann A, Tonks I, Wermers R 2015 Network centrality and pension fund performance *CFR Working Paper no 15-16* pp 1 – 62

https://www.econstor.eu/bitstream/10419/123714/1/841338175.pdf .

**6233.** Garon J-D 2016 The commitment value of funding pensions *Economics Letters* **145** issue **C** pp 11 – 14.

**6234.** Dahlquist M, Setty O, Vestman R 2016 On the asset allocation of a default pension fund *CEPR Discussion Paper no 11052*.

**6235.** Mitsel A, Rekundal O 2016 Pension capital investment in the context of a private pension fund *Economic Studies Journal* issue **1** pp 112 – 125.

***Mutual fund, financial capital investment vehicle in finances:***

**6236.** Sharpe W January 1966 Mutual fund performance *The Journal of Business* **39** (1) pp 119 – 138

http://dx.doi.org/10.1086/294846 ,

http://www.jstor.org/stable/2351741?seq=1#page_scan_tab_contents .

**6237.** Treynor J L, Mazuy K K 1966 Can mutual funds outguess the market *Harvard Business Review* **44** pp 66 – 86.

**6238.** Treynor J L, Mazuy K K July-August 1996 Can mutual funds outguess the market? *Harvard Business Review* pp 131 – 136.

**6239.** Allerdice F B, Farrar D E 1967 Factors that affect mutual fund growth *Journal of Financial and Quantitative Analysis* **2** (4) pp 365 – 382.

**6240.** Jensen M June 1968 The performance of mutual funds in the period 1945-1964 *Journal of Finance* **23** pp 389 – 416.

**6241.** Arditti F D 1971 Another look at mutual fund performance *Journal of Financial and Quantitative Analysis* **6** (3) pp 909 – 912.

6242. Fama E June 1972 Components of investment performance *The Journal of Finance* **27** (3) pp 551 – 567.

6243. Maurice Joy O, Burr Porter R 1974 Stochastic dominance and mutual fund performance *Journal of Financial and Quantitative Analysis* **9** (1) pp 25 – 31.

6244. Scott D, Klemkosky R C 1975 Mutual fund performance and unrealized expectations *Journal of Business Research* **3** (1) pp 25 – 32.

6245. Fabozzi F, Francis J C March 1978 Beta as a random coefficient *Journal of Financial and Quantitative Analysis* **13** (1) pp 101 – 116.

6246. Fabozzi F, Francis J C December 1979 Mutual fund systematic risk for bull and bear markets: An empirical examination *The Journal of Finance* **34** (5) pp 1243 – 1250.

6247. Fabozzi F, Francis J C 1980 Stability of mutual fund systematic risk statistic *Journal of Business Research* **8** (2) pp 263 – 275.

6248. Kon S J, Jen F C May 1978 Estimation of the time-varying systematic risk and performance for mutual fund portfolios: An application of switching regression *The Journal of Finance* **33** (2) pp 457 – 475.

6249. Kon S J, Jen F C April 1979 The investment performance of mutual funds: An empirical investigation of timing, selectivity and market efficiency *Journal of Business* **52** (2) pp 263 – 289.

6250. Kon S J 1983 The market –timing performance of mutual fund managers *Journal of Business* **54** (3) pp 323 – 347.

6251. Kim T 1978 An assessment of the performance of mutual fund management: 1969–1975 *Journal of Financial and Quantitative Analysis* **13** (3) pp 385 – 406.

6252. Gatto M A, Geske R, Litzenberger R, Sosin H 1980 Mutual fund insurance *Journal of Financial Economics* **8** (3) pp 283 – 317.

6253. Miller T W, Gressis N 1980 Nonstationarity and evaluation of mutual fund performance *Journal of Financial and Quantitative Analysis* **15** (3) pp 639 – 654.

6254. Eckardt W L, Bagamery B D 1983 Short selling: the mutual fund alternative *Journal of Financial Research* **6** (3) pp 231 – 238.

6255. Cowen T, Kroszner R 1990 Mutual fund banking: A market approach *Cato Journal* **10** (1) pp 223 – 237.

6256. Cook W D, Hebner K J 1992 A multicriteria approach to mutual fund selection *Financial Services Review* **2** (1) pp 1 – 20.

6257. Grinblatt M, Titman S 1992 The persistence of mutual fund performance *Journal of Finance* **47** (5) pp 1977 – 1984.

6258. Grinblatt M, Titman S 1993 Performance measurement without benchmarks: An examination of mutual fund returns *Journal of Business* **66** (1) pp 47 – 68.

6259. Grinblatt M, Titman S, Wermers R 1995 Momentum investment strategies, portfolio performance and herding: A study of mutual fund behavior *American Economic Review* **85** pp 1088 – 1105.

6260. Hendricks D, Patel J, Zeckhauser R 1993 Hot hands in mutual funds: Short run persistence of performance, 1974-88 *Journal of Finance* **48** pp 93 – 130.

6261. Mack Ph R November 1993 Recent trends in the mutual fund industry *Federal Reserve Bulletin* pp 1001 – 1012.

*6262.* Wohlever J E 1993 Excess capacity in the mutual fund industry *Research Paper no 9318* Federal Reserve Bank of New York NY USA.

*6263.* Dickson J, Shoven J 1995 Taxation and mutual funds: An investor perspective *Tax Policy and the Economy* **9** pp 151 – 180.

*6264.* Dickson J, Shoven J, Sialm C 2000 Tax externalities of equity mutual funds *National Tax Journal* **53** pp 607 – 628.

*6265.* Malkiel B G 1995 Returns from investing in equity mutual funds *Journal of Finance* **50** pp 549 – 572.

*6266.* Neely M C 1995 Will the mutual fund boom be a bust for banks? *The Regional Economist* issue October pp 12 – 13.

https://www.stlouisfed.org/Publications/Regional-Economist/October-1995/Will-the-Mutual-Fund-Boom-Be-a-Bust-for-Banks .

*6267.* Chordia T 1996 The structure of mutual fund charges *Journal of Financial Economics* **41** (1) pp 3 – 39.

*6268.* Droms W G, Walker D A 1996 Mutual fund investment performance *The Quarterly Review of Economics and Finance* **36** (3) pp 347 – 363.

*6269.* Elton E J, Gruber M J, Blake Ch R 1996 Survivorship bias and mutual fund performance *Review of Financial Studies* **9** (4) pp 1097 – 1120.

*6270.* Livingston M, O'Neal E S 1996 Mutual fund brokerage commissions *Journal of Financial Research* **19** (2) pp 273 – 292.

*6271.* Livingston M, O'Neal E S 1998 The cost of mutual fund distribution fees *Journal of Financial Research* **21** (2) pp 205 – 218.

*6272.* Livingston M, Lei Zhou 2015 Brokerage commissions and mutual fund performance *Journal of Financial Research* **38** (3) pp 283 – 303.

*6273.* Lockwood L J 1996 Macroeconomic forces and mutual fund betas *The Financial Review* **31** (4) pp 747 – 763.

*6274.* Brown St, Goetzmann W 1997 Mutual fund styles *Journal of Financial Economics* **43** (3) pp 373 – 399.

*6275.* Carhart M 1997 On persistence in mutual fund performance *Journal of Finance* **52** (1) pp 57 – 82.

*6276.* Carhart M, Carpenter J N, Lynch A W, Musto D K 2002 Mutual fund survivorship *Review of Financial Studies* **15** (5) pp 1439 – 1463.

*6277.* Cortez M D C R, Armada M J D R 1997 On mutual fund performance evaluation *Portuguese Journal of Management Studies* **III** (3) pp 145 – 163.

*6278.* Cortez M D C R, Paxson D, Armada M J D R 1999 Persistence in Portuguese mutual fund performance *The European Journal of Finance* **5** (4) pp 342 – 365.

*6279.* Daniel K M, Grinblatt M, Titman S, Wermers R 1997 Measuring mutual fund performance with characteristic based benchmarks *Journal of Finance* **52** pp 1035 – 1058.

*6280.* Goetzmann W, Peles N 1997 Cognitive dissonance and mutual fund investors *Journal of Financial Research* **20** (2) pp 145 – 158.

*6281.* Goetzmann W, Brown St 1998 Mutual fund styles *Working Paper F-46* Yale School of Management USA

https://papers.ssrn.com/sol3/papers.cfm?abstract_id=6482 .

**6282.** Goetzmann W, Ivkovic Z, Rouwenhorst G 2001 Day trading international mutual funds: Evidence and policy solutions *Journal of Financial and Quantitative Analysis* **36** (3) pp 287 – 310.

**6283.** Malhotra K D, McLeod R W 1997 An empirical analysis of mutual fund expenses *Journal of Financial Research* **20** (2) pp 175 – 190.

**6284.** Tufano P, Sevick M 1997 Board structure and fee-setting in the US mutual fund industry *Journal of Financial Economics* **46** pp 321 – 355.

**6285.** Barclay M, Pearson N, Weisbach M 1998 Open end mutual funds and capital gains taxes *Journal of Financial Economics* **49** pp 3 – 43.

**6286.** Blake D, Timmermann A 1998 Mutual fund performance: Evidence from the UK *European Finance Review* **2** pp 57 – 77.

**6287.** Blake D, Timmermann A 2003 Performance persistence in mutual funds : An independent assessment of the studies prepared by Charles River Associates for the Investment Management Association *Financial Services Authority* www.pensions-institue.org/reports .

**6288.** Fortune P 1998 Mutual fund myths *Regional Review* **Q 2** pp 5 – 7.

**6289.** Sirri E, Tufano P 1998 Costly search and mutual fund Flows *Journal of Finance* **53** (5) pp 1589 – 1622.

**6290.** Bogle J C 1999 Common sense on mutual funds: New imperatives for the intelligent investor *John Wiley & Sons Inc* USA.

**6291.** Chevalier J, Ellison G 1999 Are some mutual fund managers better than others? Cross-sectional patterns in behavior and performance *Journal of Finance* **54** (3) pp 875 – 899.

**6292.** James E, Ferrier G, Smalhout J, Vittas D 1999 Mutual funds and institutional investments: What is the most efficient way to set up individual accounts in a social security system? *NBER Working Paper no 7049* NBER USA.

**6293.** Khorana A, Servaes H 1999 The determinants of mutual fund starts *Review of Financial Studies* **12** (5) pp 1043 – 1074.

**6294.** Kulldorff M, Khanna A 1999 A generalization of the mutual fund theorem *Finance and Stochastics* **3** (2) pp 167 – 185.

**6295.** Latzko D A 1999 Economies of scale in mutual fund administration *Journal of Financial Research* **22** (3) pp 331 – 339.

**6296.** Blake C A, Morey M 2000 Morningstar ratings and mutual fund performance *Journal of Financial and Quantitative Analysis* **35** (3) pp 451 – 483.

**6297.** Chen H L, Jegadeesh N, Wermers R 2000 The value of active mutual fund management: An examination of the stockholdings and trades of fund managers *Journal of Financial and Quantitative Analysis* **35** (3) pp 343 – 368.

**6298.** Kim M, Shukla R, Tomas M 2000 Mutual fund objective misclassification *Journal of Economics and Business* **52** (4) pp 309 – 323.

**6299.** Wermers R 2000 Mutual fund performance: An empirical decomposition into stock picking talent, style, transactions costs, and expenses *Journal of Finance* **55** (4) pp 1655 – 1703.

**6300.** Wermers R 2003 Is money really "smart"? New evidence on the relation between mutual fund flows and performance persistence *Department of Finance* University of Maryland USA.

**6301.** Bollen N P B 2001 On the timing ability of mutual fund managers *Journal of Finance* **56** (3) pp 1075 – 1094.

**6302.** Kothari S P 2001 Evaluating mutual fund performance *Journal of Finance* **56** (5) pp 1985 – 2010.

**6303.** Teo M, Woo Sung-Jun 2001 Persistence in style-adjusted mutual fund returns *Manuscript* Harvard University Cambridge USA.

**6304.** Bergstresser D, Poterba J 2002 Do after-tax returns affect mutual fund inflows? *Journal of Financial Economics* **63** pp 381 – 414.

**6305.** Chan L K C, Hsiu-Lang Chen, Lakonishok J 2002 On mutual fund investment styles *Review of Financial Studies* **15** (5) pp 1407 – 1437.

**6306.** Otten R, Bams D 2002 European mutual fund performance *European Financial Management* **8** (1) pp 75 – 101.

**6307.** Pastor L, Stamburgh R 2002 Mutual fund performance and seemingly unrelated assets *Journal of Financial Economics* **63** (3) pp 315 – 350.

**6308.** Sengupta J 2003 Efficiency tests for mutual fund portfolios *Applied Financial Economics* **13** (12) pp 869 – 876.

**6309.** Berk J B, Green R C 2004 Mutual fund flows and performance in rational markets *Journal of Political Economy* **112** (6) pp 1269 – 1295.

**6310.** Kosowski R, Timmermann A, White H, Wermers R April 2004 Can mutual fund "stars" really pick stocks? New evidence from a bootstrap analysis *Discussion Paper* INSEAD.

**6311.** Mahoney P 2004 Manager-investor conflicts in mutual funds *Journal of Economic Perspectives* **18** (2) pp 161 – 182.

**6312.** Mamaysky H, Spiegel M, Zhang H 2004 Improved forecasting of mutual fund alphas and betas' *ICF Working Paper 04-23* Yale School of Management Yale University USA.

**6313.** Nanda V, Wang J, Zheng L 2004 Family values and the star phenomenon: Strategies of mutual fund families *Review of Financial Studies* **17** (3) pp 667 – 698.

**6314.** Prather L, Bertin W J, Henker Th 2004 Mutual fund characteristics, managerial attributes, and fund performance *Review of Financial Economics* **13** (4) pp 305 – 326.

**6315.** Anderson S C, Ahmed P 2005 Mutual funds: Fifty years of research findings Chapter 3 *in* Mutual Fund Fees and Expenses *Springer* pp 57 – 69.

**6316.** Barber B, Odean T, Zheng L 2005 Out of sight, out of mind: The effects of expenses on mutual fund flows *Journal of Business* 78 pp 2095 – 2120.

**6317.** Bollen N P B, Busse J A 2005 Short-term persistence in mutual fund performance *Review of Financial Studies*.

**6318.** Cuthbertson K, Nitzsche D, O'Sullivan N February 2005 Mutual fund performance: Skill or luck? *Cass Business School* City University London UK pp 1 – 50.

**6319.** Kacperczyk M, Sialm C, Zheng L 2005 On the industry concentration of actively managed equity mutual funds *Journal of Finance* **60** pp 1983 – 2011.

*6320.* Kacperczyk M, Sialm C, Zheng L 2008 Unobserved actions of mutual funds *Review of Financial Studies* **21** pp 2379 – 2416.

*6321.* Nitzsche D, Cuthbertson K, O'Sullivan N 2005 Mutual fund performance: Skill or luck? *Money Macro and Finance (MMF) Research Group Conference* pp 1 – 50 http://repec.org/mmfc05/paper4.pdf .

*6322.* Raychaudhuri A 2005 Persistence in the Indian mutual fund market *The IUP Journal of Financial Economics* **III** (1) pp 6 – 25.

*6323.* Busse J A, Irvine P J 2006 Bayesian alphas and mutual fund persistence *Journal of Finance* **61** (5) pp 2251 – 2288.

*6324.* Gaspar J, Massa M, Matos P 2006 Favoritism in mutual fund families? Evidence on strategic cross-fund subsidization *Journal of Finance* **61** (1) pp 73 – 104.

*6325.* Morey M, O'Neal E 2006 Window dressing in bond mutual funds *Journal of Financial Research* **29** pp 325 – 347.

*6326.* Alves C, Mendes V 2007 Are mutual fund investors in jail? *Applied Financial Economics* **17** (16) pp 1301 – 1312.

*6327.* Bollen N P B 2007 Mutual fund attributes and investor behavior *Journal of Financial and Quantitative Analysis* **42** (3) pp 683 – 708.

*6328.* Cohen L, Frazzini A, Malloy C 2007 The small world of investing: Board connections and mutual fund returns *Journal of Political Economy* **116** (5) pp 951 – 979.

*6329.* Chen Q, Goldstein I, Jiang W 2008 Directors' ownership in the US mutual fund industry *Journal of Finance* **63** (6) pp 2629 – 2677.

*6330.* Chen Q, Goldstein I, Jiang W 2009 Payoff complementarities and financial fragility: Evidence from mutual fund outflows *Working Paper* Duke University, University of Pennsylvania, Columbia University USA.

*6331.* Ferris S, Yan X 2007 Do independent directors and chairmen matter? The role of boards of directors in mutual fund governance *Journal of Corporate Finance* **13** (2-3) pp 392 – 420.

*6332.* Khorana A, Tufano P, Wedge L 2007 Board structure, mergers, and shareholder wealth: A study of the mutual fund industry *Journal of Financial Economics* **85** (2) pp 571 – 598.

*6333.* Khorana A, Servaes H, Tufano P 2008 Mutual fund fees around the world *Review of Financial Studies* **22** (3) pp 1279 – 1310.

*6334.* Meschke F 2007 An empirical examination of mutual fund boards *SSRN Paper 676901* SSRN USA.

*6335.* Haslem J A, Baker H K, Smith D M 2008 Performance and characteristics of actively managed retail equity mutual funds with diverse expense ratios *Financial Services Review* **17** pp 49 – 68.

*6336.* Jans R, Otten R 2008 Tournaments in the UK mutual fund industry *Managerial Finance* **34** (11) pp 786 – 798.

*6337.* Jin-Li Hu, Tzu-Pu Chang 2008 Decomposition of mutual fund underperformance *Applied Financial Economics Letters* **4** (5) pp 363 – 367.

*6338.* Kacperczyk M, Sialm C, Zheng L 2008 Unobserved actions of mutual funds *Review of Financial Studies* **21** (6) pp 2379 – 2416.

**6339.** Kacperczyk M, van Nieuwerburgh S, Veldkamp L 2014 Time-varying fund manager skill *Journal of Finance* **69** (4) pp 1455 – 1484.

**6340.** Kempf A, Ruenzi St 2008 Tournaments in mutual-fund families *Review of Financial Studies* **21** (2) pp 1013 – 1036.

**6341.** Kong S, Tang D 2008 Unitary boards and mutual fund governance *Journal of Financial Research* **31** (3) pp 193 – 224.

**6342.** Palmiter A, Taha A 2008 Mutual fund investors: Sharp enough? *Journal of Financial Transformation* **24** pp 113 – 121.

**6343.** Sophie Xiaofei Kong, Dragon Yongjun Tang 2008 Unitary boards and mutual fund governance *Journal of Financial Research* **31** (3) pp 193 – 224.

**6344.** Bergstresser D, Chalmers J M, Tufano P 2009 Assessing the costs and benefits of brokers in the mutual fund industry *Review of Financial Studies* **22** (10) pp 4129 – 4156.

**6345.** Coggins F, Beaulieu M-C, Gendron M 2009 Mutual fund daily conditional performance *Journal of Financial Research* **32** (2) pp 95 – 122.

**6346.** Cremers M, Driessen J, Maenhout P, Weinbaum D 2009 Does skin in the game matter? Director incentives and governance in the mutual fund industry *Journal of Financial and Quantitative Analysis* **44** (6) pp 1345 – 1373.

**6347.** Karoui A, Meier I 2009 Performance and characteristics of mutual fund starts *The European Journal of Finance* **15** (5-6) pp 487 – 509.

**6348.** Khorana A, Servaes H, Tufano P 2009 Mutual fund fees around the world *Review of Financial Studies* **22** (3) pp 1279 – 1310.

**6349.** Koehler J, Mercer M 2009 Selection neglect in mutual fund advertisements *Management Science* **55** (7) pp 1107 – 1121.

**6350.** Kuhnen C 2009 Business networks, corporate governance, and contracting in the mutual fund industry *Journal of Finance* **64** (5) pp 2185 – 2220.

**6351.** Lilian Ng, Qinghai Wang, Zaiats N 2009 Firm performance and mutual fund voting *Journal of Banking & Finance* **33** (12) pp 2207 – 2217.

**6352.** Qiang Bu, Lacey N 2009 On understanding mutual fund terminations *Journal of Economics and Finance* **33** (1) pp 80 – 99.

**6353.** Sialm C, Starks L 2009 Mutual fund tax clienteles *NBER Working Paper no 15327* National Bureau of Economic Research Inc USA pp 1 – 54 http://www.nber.org/papers/w15327.pdf .

**6354.** Evans R B 2010 Mutual fund incubation *Journal of Finance* **65** (4) pp 1581 – 1611.

**6355.** Hsuan-Chi Chen, Lai Ch W 2010 Reputation stretching in mutual fund starts *Journal of Banking & Finance* **34** (1) pp 193 – 207.

**6356.** Jin Zhang, Maringer D 2010 Index mutual fund replication *Working Paper no 35* COMISEF pp 1 – 27 www.comisef.eu .

**6357.** Soongswang A, Sanohdontree Y 2011 Equity mutual fund: Performances, persistence and fund rankings *Journal of Knowledge Management, Economics and Information Technology* **1** (6) p 27.

**6358.** English Ph C, Demiralp I, Dukes W P 2011 Mutual fund exit and mutual fund fees *Journal of Law and Economics* **54** (3) pp 723 – 749.

**6359.** Adams J C, Mansi S A, Nishikawa T 2012 Are mutual fund fees excessive? *Journal of Banking & Finance* **36** (8) pp 2245 – 2259.

**6360.** Bhojraj S, Young Jun Cho, Yehuda N 2012 Mutual fund family size and mutual fund performance: The role of regulatory changes *Journal of Accounting Research* **50** (3) pp 647 – 684.

**6361.** Cashman G D 2012 Convenience in the mutual fund industry *Journal of Corporate Finance* **18** (5) pp 1326 – 1336.

**6362.** Gottesman A, Morey M 2012 Mutual fund corporate culture and performance *Review of Financial Economics* **21** (2) pp 69 – 81.

**6363.** Sialm C, Starks L 2012 Mutual fund tax clienteles *Journal of Finance* **67** (4) pp 1397 – 1422.

**6364.** Christensen M 2013 Danish mutual fund performance *Applied Economics Letters* **20** (8) pp 818 – 820.

**6365.** Eisele A, Nefedova T, Parise G 2013 Predation versus Cooperation in Mutual Fund Families *Swiss Finance Institute Research Paper 13-19* Switzerland.

**6366.** Vidal-García J 2013 The persistence of European mutual fund performance *Research in International Business and Finance* **28** (C) pp 45 – 67.

**6367.** Agarwal V, Gay G, Ling L 2014 Window dressing in mutual funds *Review of Financial Studies* **27** (11) pp 3133 – 3170.

**6368.** Jingjing Yang, Jing Chi, Martin Young 2014 Mutual fund investment strategies and preferences *Chinese Economy* **47** (1) pp 5 – 37.

**6369.** Simutin M 2014 Cash holdings and mutual fund performance *Review of Finance* **18** (4) pp 1425 – 1464.

**6370.** Berk J B, van Binsbergen J H 2015 Measuring skill in the mutual fund industry *Journal of Financial Economics* **118** (1) pp 1 – 20.

**6371.** Chuprinin O, Massa M, Schumacher D 2015 Outsourcing in the international mutual fund industry: An equilibrium view *Journal of Finance* **70** pp 2275 – 2308.

**6372.** Gallaher S, Kaniel R, Starks L 2015 Advertising and mutual funds: From families to individual funds *CEPR Discussion Paper no DP10329*.

**6373.** Kopsch F, Han-Suck Song, Wilhelmsson M 2015 Determinants of mutual fund flows *Managerial Finance* pp 10 – 25.

**6374.** Meifen Qian, Bin Yu 2015 Do mutual fund managers manipulate? *Applied Economics Letters* **22** (12) pp 967 – 971.

**6375.** Ortiz C, Ramírez G, Vicente L 2015 Mutual fund trading and portfolio disclosures *Journal of Financial Services Research* **48** (1) pp 83 – 102.

**6376.** Panda B, Mahapatra R P, Moharana S 2015 Myth of equity mutual fund performance *Vision: The Journal of Business Perspective* **19** (3) pp 200 – 209.

**6377.** Sisli-Ciamarra E, Hornstein A 2015 Board overlaps in mutual fund families *Working Paper no 92* Department of Economics, International Business School Brandeis University Waltham MA USA http://www.brandeis.edu/departments/economics/RePEc/brd/doc/Brandeis_WP92.pdf .

**6378.** Grinblatt M, Ikäheimo S, Keloharju M, Knüpfer S 2016 IQ and mutual fund choice *Management Science* **62** (4) pp 924 – 944.

6379. Matallín-Sáez J C, Soler-Domínguez A, Tortosa-Ausina E 2016 On the robustness of persistence in mutual fund performance The North American *Journal of Economics and Finance* **36** (C) pp 192 – 231.

*Angel investor, business angel, financial investment vehicle in finances:*

6380. Rubenstein A H 1958 Problems of financing and managing new research-based enterprises in New England *Federal Reserve Bank of Boston* MA USA.

6381. Wetzel Jr W E 1981 Informal risk capital in New England *in* Frontiers of Entrepreneurship Research Vesper K H (editor) *Babson College* Wellesley MA USA pp 217 – 245.

6382. Wetzel Jr W E 1983 Angels and informal venture capital *Sloan Management Review* **24** (4) pp 23 – 34.

6383. Wetzel Jr W E 1986 Informal risk capital: Knowns and unknowns *in* The Art and Science of Entrepreneurship Sexton D, Smilor R (editors) Cambridge MA USA.

6384. Wetzel Jr W E 1987 The informal venture capital market: Aspects of scale and market efficiency *Journal of Business Venturing* **2** (4) pp 299 – 313.

6385. Landström H 1992 The relationship between private investors and small firms: An agency theory approach *Entrepreneurship and Regional Development* **4** pp 199 – 223.

6386. Landström H 1993 Informal risk capital in Sweden and some international comparisons *Journal of Business Venturing* **8** pp 525 – 540.

6387. Landström H 1995 A pilot study on the investment decision-making behavior of informal investors in Sweden *Journal of Small Business Management* **33** (3) pp 67 – 76.

6388. Landström H 1998 Informal investors as entrepreneurs *Technovation* **18** pp 321 – 333.

6389. Landström H, Manigart S, Mason C, Sapienza H 1998 Contracts between entrepreneurs and investors: Terms and negotiation processes *in* Frontiers of Entrepreneurship Research Reynolds P D, Bygrave W D, Carter N M, Manigart S, Mason C M, Meyer G D, Shaver K G (editors) Babson College Babson Park MA USA pp 571 – 585.

6390. Landström H 2007 Handbook of research on venture capital *Edward Elgar Publishing* Cheltenham UK Northampton MA USA.

6391. Landström H, Mason C (editors) 2012 Handbook of research on venture capital: Volume 2 *Edward Elgar Publishing* Cheltenham UK Northampton MA USA.

6392. Landström H, Mason C 2016a Business angels as a research field Chapter 1 *in* Handbook of Research on Business Angels *Edward Elgar Publishing* UK ISBN: 9781783471713 pp 1 – 22
http://dx.doi.org/10.4337/9781783471720.00005 ,
https://www.elgaronline.com/view/9781783471713.00005.xml .

6393. Landström H, Mason C (editors) October 29 2016b Handbook of research on business angels *Edward Elgar Publishing* UK ISBN: 9781783471713 pp 1 – 432
http://dx.doi.org/10.4337/9781783471720 ,
http://www.e-elgar.com/shop/isbn/9781783471713 ,
http://www.e-elgar.com/shop/handbook-of-research-on-business-angels .

**6394.** Mason C M, Harrison R T 1994 The informal venture capital market in the UK *in* Financing small firms Hughes A, Storey D J (editors) *Routledge* London UK pp 64 – 111.

**6395.** Mason C M, Harrison R T 1995 Closing the regional equity gap – The role of informal venture capital *Small Business Economics* **7** pp 153 – 172.

**6396.** Mason C M, Rogers A 1997 Understanding the business angel's investment decision *Working Paper no 14* Venture Finance Research Project Southampton University UK.

**6397.** Mason C M, Harrison R T 2000a The size of the informal venture capital market in the United Kingdom *Small Business Economics* **15** pp 137 – 148.

**6398.** Mason C M, Harrison R T 2000b Informal venture capital and the financing of emergent growth businesses Handbook of Entrepreneurship Sexton D, Landström H (editors) *Blackwell* Oxford UK pp 221 – 229.

**6399.** Mason C M, Harrison R T 2002 Is it worth it? The rates of return from informal venture capital investments *Journal of Business Venturing* **17** pp 211 – 236.

**6400.** Mason C M, Harrison R T 2003 Closing the regional equity gap? A critique of the Department of Trade and Industry's regional venture capital funds initiative *Regional Studies* **37** (8) pp 855 – 868.

**6401.** Mason C M, Stark D 2004 What do investors look for in a business plan?: A comparison of the investment criteria for bankers, venture capitalists and business angel *International Small Business Journal* **22** (3) pp 227 – 248.

**6402.** Mason C M 2006 Informal sources of venture finance *in* The life cycle of entrepreneurial ventures Parker S (editor) *Springer* New York USA pp 259 – 299.

**6403.** Mason C M 2007 Informal sources of venture finance *in* The Life Cycle of Entrepreneurial Ventures, Volume 3 Parker S (editor) *Springer* New York USA pp 258 – 300.

**6404.** Mason C M, Harrison R T 2008 Measuring business angel investment activity in the United Kingdom: A review of potential data sources *Venture Capital* **10** (4) pp 309 – 330.

**6405.** Mason C M 2009 Public policy support for the informal venture capital market: A critical review *International Small Business Journal* **27** (5) pp 536 – 556.

**6406.** Mason C M 2011a Business angels *in* World Encyclopedia of Entrepreneurship Dana L-P (editor) *Edward Elgar Publishing* Cheltenham UK.

**6407.** Mason C M 2011b Investment activity by Canadian angel group *The National Angel Capital Organization*.

**6408.** Mason C M, Harrison R T 2015 Business angel investment activity in the financial crisis: UK evidence and policy implications *Environment and Planning C: Government and Policy* **33** (1) pp 43 – 60.

**6409.** Mason C M 2016 Researching business angels: Definitional and data challenges Chapter 2 *in* Handbook of Research on Business Angels *Edward Elgar Publishing* UK ISBN: 9781783471713 pp 25 – 52
http://dx.doi.org/10.4337/9781783471720.00007 ,
https://www.elgaronline.com/view/9781783471713.00007.xml .

6410. Mason C M, Botelho T 2016 The role of the exit in the initial screening of the investment opportunities: The case of business angel syndicate gatekeepers *International Small Business Journal* **34** pp 157 – 175.

6411. Mason C M, Harrison R T, Botelho T 2016 Business angels and beyond: The changing nature of the early risk capital market Chapter 7 *in* International Research Handbook on Entrepreneurial Finance Hussain J G, Scott J S (editors) *Edward Elgar Publishing* Cheltenham UK, Northampton MA USA.

6412. Riding A, Duxbury L, Haines G 1994 Financing enterprise development: Decision-making by Canadian angels *Carleton University* Ottawa Canada.

6413. Coveney P, Moore K 1998 Business angels: Securing start-up finance *John Wiley and Sons Inc* Chichester UK.

6414. Lerner J 1998 Angel financing and public policy: An overview *Journal of Banking and Finance* **22** (6-8) pp 773 – 783.

6415. Lerner J 2009 Boulevard of broken dreams *Princeton University Press* Princeton NJ USA.

6416. Lerner J, Schoar A, Sokolinski St, Wilson K E August 2015 The globalization of angel investments *Working Paper 2015/09* Bruegel pp 1 – 45

http://bruegel.org/wp-content/uploads/2015/09/WP-2015-09-SG-210915.pdf .

6417. Prowse St 1998 Angel investors and the market for angel investments *Journal of Banking & Finance* **22** (6-8) pp 785 – 792.

6418. Aernoudt R 1999 Business Angels: Should they fly on their own wings? *Venture Capital* **1** (2) pp 187 – 195.

6419. Tashiro Y 1999 Business angels in Japan *Venture Capital* **1** (3) pp 259 – 273.

6420. Farrell H 2000 A literature review and industry analysis of informal investment in Canada: A research agenda on angels *Industry Canada* Ottawa Canada pp 1 – 57.

6421. Freear J, Sohl J E, Wetzel W 2000 Angels and non-angels: Are there differences? *in* Advances in Entrepreneurship vol **1** *Edward Elgar Publishing* Cheltenham UK, Northampton MA USA pp 520 – 534.

6422. Freear J, Sohl J E, Wetzel W 2002 Angles on angels: Financing technology-based ventures – a historical perspective *Venture Capital* **4** pp 275 – 287.

6423. Freear J, Sohl J 2012 Angels on angels and venture capital *in* Financing Economic Development in the 21st Century 2nd edition Kotval Z, White S (editors) *M E Sharpe Inc* pp 224 – 244.

6424. Kelly P, Hay M 2000 Deal-makers: Reputation attracts quality *Venture Capital* **2** (3) pp 183 – 202.

6425. Kelly P, Hay M 2003 Business angel contracts: The influence of context *Venture Capital* **5** (4) pp 287 – 312.

6426. Kelly P 2007 Business angels research: The road traveled and the journey ahead *in* Handbook of Research on Venture Capital Landström H (editor) *Edward Elgar* Cheltenham UK, Northampton MA USA pp 315 – 331.

6427. Prasad D, Bruton G, Vozikis G 2000 Signaling value to business angels: The proportion of the entrepreneur's net worth invested in a venture as a decision signal *Venture Capital* **2** (3) pp 167 – 182.

6428. Van Osnabrugge M 2000 A comparison of business angels and venture capital investment procedures: An agency theory-based analysis *Venture Capital* **2** (2) pp 91 – 109.

6429. Amis D, Stevenson H 2001 Winning angels: The seven fundamentals of early-stage investing *Financial Times Prentice Hall* London, New York UK, USA.

6430. Amis D, Stevenson H, Liechtenstein H 2003 Winning angels: Mentoren im netzwerk des erfolges *Signum Wirtschaftsverlag* Wien Austria.

6431. Benjamin G A, Margulis J 2001 The angel investor's handbook *Bloomberg Press* Princeton NJ USA.

6432. Sørheim R, Landström H 2001 Informal investors: A categorization with policy implications *Entrepreneurship and Regional Development* **13** (4) pp 351 – 370.

6433. Sørheim R 2003 The pre-investment behavior of business angels: a social capital approach *Venture Capital* **5** (4) pp 337 – 364.

6434. Brettel M 2002 German business angels in international comparison *Journal of Private Equity* **5** (2) p 53.

6435. Brettel M 2003 Business angels in Germany: A research note *Venture Capital: An International Journal of Entrepreneurial Finance* **5** (3) pp 251 – 268.

6436. Jensen M 2002 Angel investors: Opportunity amidst chaos *Venture Capital* **4** (4) pp 295 – 304.

6437. Payne W H, Macarty M J 2002 The anatomy of an angel investing network: Tech coast angels *Venture Capital* **4** (4) pp 331 – 336.

6438. Stadler H, Peters H H 2003 Business angels in Germany: An empirical study *Venture Capital* **5** (3) pp 269 – 276.

6439. Sohl J E 2003 The US angel and venture capital market: Recent trends and developments *The Journal of Private Equity* **6** pp 7 – 17.

6440. Sohl J E 2006 Angel investing: Changing strategies during volatile times *Journal of Entrepreneurial Finance* **11** (2) pp 27 – 48.

6441. Sohl J E, Hill L 2007 Women business angels: Insights from angel groups *Venture Capital* **9** (3) pp 207 – 222.

6442. Sohl J E 2012 The changing nature of the angel market *in* Handbook of Research on Venture Capital *Edward Elgar Publishing* UK.

6443. San Jose A, Roure J, Aernoudt R 2004 Business angel academies: Unleashing the potential for business angel investment *Venture Capital* **7** (2) pp 149 – 165.

6444. Jossi F July 2005 Angels we have heard on high *Fedgazette* pp 11 – 13.

6445. Månsson N, Landström H 2005 Business angels in a changing economy: The case of Sweden *Venture Capital* **8** (4) pp 281 – 301.

6446. Maula M, Autio E, Arenius P 2005 What drives micro-angel investments? *Small Business Economics* **25** (5) pp 459 – 475.

6447. EBAN (European Business Angel Network) (2005) European directory of business angel networks in Europe *EBAN* Brussels Belgium.

6448. Shane S A October 1 2005 Angel investing: A report prepared for the Federal Reserve Banks of Atlanta Cleveland Kansas City Philadelphia, Richmond North Carolina
SSRN: http://ssrn.com/abstract=1142687 .

**6449.** Shane S A January 1 2008 Angel groups: An examination of the angel capital association survey SSRN NY USA

http://ssrn.com/abstract=1142645.

**6450.** Shane S A 2009 Fool's gold: The truth behind angel investing in America *Oxford University Press* Oxford UK.

**6451.** Vance D E 2005 Raising capital *Springer* New York USA.

**6452.** Wiltbank R October 2005 Investment practices and outcomes of information angel investors *Venture Capital* **7** pp 343 – 357.

**6453.** Wiltbank R, Boeker W November 2007a Returns to angel investors in groups *Ewing Marion Kauffman Foundation*, *Angel Capital Education Foundation* pp 1 – 16.

**6454.** Wiltbank R, Boeker W 2007b Angel investor performance project: Data overview *Kauffman Symposium on Entrepreneurship and Innovation Data*.

**6455.** Wiltbank R, Read St, Dew N, Sarasvathy S D 2009 Prediction and control under uncertainty: Outcomes in angel investing *Journal of Business Venturing* **24** (2) pp 116 – 133.

**6456.** Wiltbank R 2009 Siding with angels: Business angel investing – promising outcomes and effective strategies *Research Report* National Endowment for Science Technology, and the Arts, British Business Angels Association London UK.

**6457.** Amatucci F, Sohl J 2006 Business angels: Investment processes, outcomes and current trends *in* Entrepreneurship: The Engine of Growth, Vol 2: The Process *Greenwood Publishing Group* pp 87 – 104.

**6458.** Heukamp F, Liechtenstein H, Wakeling N October 2006 Do business angels alter the risk-return equation in early stage investments? Business angels as seen by venture capitalists in the German speaking countries *IESE Research Paper no D/655* IESE Business School University of Navarra Barcelona Spain pp 1 – 33

http://www.iese.edu/research/pdfs/DI-0655-E.pdf .

**6459.** Mercil St 2006 Organizing angel investment to benefit angels, companies, and communities *Community Development Investment Review* Federal Reserve Bank of san Francisco **2** (3) pp 43 – 49

www.frbsf.org/cdinvestments ,

http://www.frbsf.org/publications/community/review/122006/mercil.pdf .

**6460.** Sudek R 2006 Angel investment criteria *Journal of Small Business Strategy* **17** (2) pp 89 – 103.

**6461.** Chahine S, Filatochev I, Wright M 2007 Venture capitalist, business angels, and performance of entrepreneurial IPOs in the UK and France *Journal of Business Finance and Accounting* **34** pp 505 – 528.

**6462.** Harrison R T, Mason C M 2007 Does gender matter? Women business angels and the supply of entrepreneurial finance *Entrepreneurship Theory and Practice* **31** (3) pp 445 – 472.

**6463.** Harrison R T, Mason C, Robson P 2010 Determinants of long-distance investing by business angels in the UK *Entrepreneurship & Regional Development* **22** (2) pp 113 – 137.

**6464.** Knyphausen-Aufseß D Z, Westphal R 2007 Do business angel networks deliver value to business angels? *Venture Capital* **10** (2) pp 149 – 169.

6465. Preston S L 2007 Angel financing for entrepreneurs: Early-stage funding for long-term success *Jossey-Bass*.

6466. Clark C 2008 The impact of entrepreneurs' oral "pitch" presentation skills on business angels' initial screening investment decision *Venture Capital* **10** (3) pp 257 – 279.

6467. Goldfarb B, Hoberg G, Kirsch D, Triantis A 2008 Does angel participation matter? An analysis of early venture financing *Working Paper 06-072* Robert H Smith School of Business

http://papers.ssrn.com/sol3/papers.cfm?abstract_id=1024186 .

6468. Ibrahim D 2008 The (not so) puzzling behavior of angel investors *Vanderbilt Law Review* **61** pp 1405 – 1452.

6469. Ibrahim D M 2011 Should angel-backed start-ups reject venture capital? *Legal Studies Research Paper no 1170* University of Wisconsin USA.

6470. Riding A L 2008 Business angels and love money investors: Segments of the informal market for risk capital *Venture Capital* **10** (4) pp 355 – 369.

6471. Collewaert V, Manigart S 2009 First-round valuation of angel-backed companies: The role of investor human capital *Working Paper no 2009/624* Faculty of Economics and Business Administration Ghent University Belgium

http://wps-feb.ugent.be/Papers/wp_09_624.pdf .

6472. Collewaert V, Manigart S, Aernoudt R 2010 Assessment of government funding of business angel networks in Flanders *Regional Studies* **44** (1) pp 119 – 130.

6473. Collewaert V 2016 Angel–entrepreneur relationships: Demystifying their conflicts Chapter 8 *in* Handbook of Research on Business Angels *Edward Elgar Publishing* UK ISBN: 9781783471713 pp 176 – 198

http://dx.doi.org/10.4337/9781783471720.00014 ,

https://www.elgaronline.com/view/9781783471713.00014.xml .

6474. DeGennaro R P, Dwyer G P 2009 Expected returns to angel investors *Working Paper* Federal Reserve Bank of Atlanta

http://www.clevelandfed.org/research/Conferences/2009/3-12-2009/DeGennaro_Dwyer-Revised.pdf.

6475. DeGennaro R, Dwyer G 2010 Expected returns to stock investments by angel investors in groups *FRB Atlanta Working Paper no 2010-14* Federal Reserve Bank of Atlanta USA

http://www.frbatlanta.org/documents/pubs/wp/wp1014.pdf .

6476. DeGennaro R P 2010, 2012 Angel investors and their investments *Working Paper, in* The Oxford Handbook of Entrepreneurial Finance *Oxford University Press 13* Oxford UK.

6477. Goldfarb B, Hoberg G, Kirsch D, Triantis A 2009 Does angel participation matter? An analysis of early venture financing *Working Paper* University of Maryland USA.

6478. Goldfarb B D, Hoberg G, Kirsch D, Triantis A J 2012 Does angel participation matter? An analysis of early venture financing *R H Smith Research Paper no 06-072*.

6479. Mehrotra V 2009 Fool's gold: The truth behind angel investing in America *The Economic Record* **85** (271) pp 495 – 496.

*6480.* Roach G 2009 Is angel investing worth the effort? A study of Keiretsu forum *Venture Capital* **12** (2) pp 153 – 166

http://dx.doi.org/10.1080/13691061003643276 .

*6481.* Wallisch M 2009 Unternehmensfinanzierung durch business angels *Zeitschrift für Wirtschaftsgeographie* **53** (1-2) pp 47 – 68.

*6482.* Wong A, Bhatia M, Freeman Z 2009 Angel finance: The other venture capital *Strategic Change* **18** (7-8) pp 221 – 230.

*6483.* Markova S, Petkovska-Mirčevska T 2010 Entrepreneurial finance: Angel investing as a source of funding high-growth start-up firms *Annals of the University of Petrosani: Economics* **10** (3) pp 217 – 224.

*6484.* Lahti T 2011a Angel investing: An examination of the evolution of the Finnish market *Venture Capital* **13** (2) pp 147 – 173.

*6485.* Lahti T 2011b Categorization of angel investments and explorative analysis of risk reduction strategies in Finland *Venture Capital* **13** (1) pp 49 – 74.

*6486.* Lahti T, Keinonen H 2016 Business angel networks: a review and assessment of their value to entrepreneurship Chapter 14 *in* Handbook of Research on Business Angels *Edward Elgar Publishing* UK ISBN: 9781783471713 pp 354 – 378

http://dx.doi.org/10.4337/9781783471720.00022 ,

https://www.elgaronline.com/view/9781783471713.00022.xml .

*6487.* Maxwell A L, Jeffrey S A, Lévesque M 2011 Business angel early stage decision making *Journal of Business Venturing* **26** (2) pp 212 – 225.

*6488.* Maxwell A L, Levesque M Jeffrey S 2014 The non-compensatory relationship between risk and return in business angel investment decisions *Academy of Management Proceedings* **2014** (1) 10463.

*6489.* Maxwell A L 2016 Investment decision-making by business angels Chapter 6 *in* Handbook of Research on Business Angels *Edward Elgar Publishing* UK ISBN: 9781783471713 pp 115 – 146

http://dx.doi.org/10.4337/9781783471720.00012 ,

https://www.elgaronline.com/view/9781783471713.00012.xml .

*6490.* OECD 2011 Financing high growth firms. The role of angel investors *OECD Publishing* Paris France.

*6491.* Johnson W C, Sohl J E 2012a Initial public offerings and pre-IPO shareholders: Angels versus venture capitalists *Journal of Developmental Entrepreneurship* (JDE) **17** (4) pp 1250022-1 – 1250022-23.

*6492.* Johnson W C, Sohl J E 2012b Angels and venture capitalists in the initial public offering market *Venture Capital* **14** (1) pp 27 – 42.

*6493.* Mitteness C, Sudek R, Cardon M S 2012 Angel investor characteristics that determine whether perceived passion leads to higher evaluations of funding potential *Journal of Business Venturing* **27** (5) pp 592 – 606.

*6494.* Scheela W, Jittrapanun T 2012 Do institutions matter for business angel investing in emerging Asian markets? *Venture Capital* **14** (4) pp 289 – 308.

*6495.* Wu Zhenyu, Yuan Wenlong, Wei Xueqi 2012 The effects of new ventures' resource strategies on angels' investing outcomes: Big gains and big losses in angel investments *Entrepreneurship Research Journal* **2** (3) pp 1 – 27.

**6496.** Festel G W, De Cleyn S H 2013 Founding angels as an emerging subtype of angel investment model in high-tech businesses *Venture Capital* **15** (3) pp 261 – 282.

**6497.** Florin J, Dino R, Huvaj M N 2013 Research on angel investing: A multilevel framework for an emerging domain of inquiry *Venture Capital* **15** (1) pp 1 – 27.

**6498.** Gregson G, Mann S, Harrison R 2013 Business angel syndication and the evolution of risk capital in a small market economy: Evidence from Scotland *Managerial and Decision Economics* **34** (2) pp 95 – 107.

**6499.** Hoyos J, Santos M S 2013 The informal investment context: Specific issues concerned with business angels *Investigaciones Regionales* (*Journal of Regional Research*) issue **26** pp 179 – 198.

**6500.** Zachary R, Mishra C S 2013 Research on angel investments: The intersection of equity investments and entrepreneurship *Entrepreneurship Research Journal* **3** (2) pp 160 – 170.

**6501.** Carpentier C, Suret J-M 2014 Business angels' perspectives on exit by IPO *CIRANO Working Paper no 2014s-21* CIRANO Montreal Canada pp 1 – 35.

**6502.** Carpentier C, Suret J-M 2016 The effectiveness of tax incentives for business angels Chapter 13 *in* Handbook of Research on Business Angels *Edward Elgar Publishing* UK ISBN: 9781783471713 pp 327 – 353

http://dx.doi.org/10.4337/9781783471720.00021 ,

https://www.elgaronline.com/view/9781783471713.00021.xml .

**6503.** Parhankangas A-L, Ehrlich M 2014 How entrepreneurs seduce business angels: An impression management approach *Journal of Business Venturing* **29** (4) pp 543 – 564.

**6504.** Zhujun Ding, Sunny Sun, Kevin Au 2014 Angel investors' selection criteria: A comparative institutional perspective *Asia Pacific Journal of Management* **31** (3) pp 705 – 731.

**6505.** Altuntas B 2015 Turkey Chapter 14 *in* Angels without Borders: Trends and Policies Shaping Angel Investment Worldwide *World Scientific Publishing Co Pte Ltd* ISBN: 978-981-4733-05-2 pp 151 – 155

http://dx.doi.org/10.1142/9789814733816_0014 ,

http://www.worldscientific.com/doi/abs/10.1142/9789814733816_0014 .

**6506.** Anselmo E P, Amati L 2015 Italy Chapter 15 in Angels without Borders: Trends and Policies Shaping Angel Investment Worldwide World Scientific Publishing Co Pte Ltd ISBN: 978-981-4733-05-2 pp 157 – 162

http://dx.doi.org/10.1142/9789814733816_0015 ,

http://www.worldscientific.com/doi/abs/10.1142/9789814733816_0015 .

**6507.** Ashish D, Agarwal M 2015 India Chapter 21 *in* Angels without Borders: Trends and Policies Shaping Angel Investment Worldwide *World Scientific Publishing Co Pte Ltd* ISBN: 978-981-4733-05-2 pp 215 – 224

http://dx.doi.org/10.1142/9789814733816_0021 ,

http://www.worldscientific.com/doi/abs/10.1142/9789814733816_0021 .

**6508.** Banga F, Lewis D 2015 New Zealand Chapter 26 *in* Angels without Borders: Trends and Policies Shaping Angel Investment Worldwide *World Scientific Publishing Co Pte Ltd* ISBN: 978-981-4733-05-2 pp 255 – 262

http://dx.doi.org/10.1142/9789814733816_0026 ,

http://www.worldscientific.com/doi/abs/10.1142/9789814733816_0026 .

**6509.** Baumann B 2015 Switzerland Chapter 31 *in* Angels without Borders: Trends and Policies Shaping Angel Investment Worldwide *World Scientific Publishing Co Pte Ltd* ISBN: 978-981-4733-05-2 pp 285 – 291

http://dx.doi.org/10.1142/9789814733816_0031 ,

http://www.worldscientific.com/doi/abs/10.1142/9789814733816_0031 .

**6510.** Dollinger D T, Rhodes S 2015 Israel Chapter 6 *in* Angels without Borders: Trends and Policies Shaping Angel Investment Worldwide *World Scientific Publishing Co Pte Ltd* ISBN: 978-981-4733-05-2 pp 61 – 73

http://dx.doi.org/10.1142/9789814733816_0006 ,

http://www.worldscientific.com/doi/abs/10.1142/9789814733816_0006 .

**6511.** Finlay R, Witkin B 2015 Canada Chapter 24 *in* Angels without Borders: Trends and Policies Shaping Angel Investment Worldwide *World Scientific Publishing Co Pte Ltd* ISBN: 978-981-4733-05-2 pp 243 – 248

http://dx.doi.org/10.1142/9789814733816_0024 ,

http://www.worldscientific.com/doi/abs/10.1142/9789814733816_0024 .

**6512.** Gluntz Ph 2015 France Chapter 7 *in* Angels without Borders Trends and Policies Shaping Angel Investment Worldwide *World Scientific Publishing Co Pte Ltd* ISBN: 978-981-4733-05-2 pp 75 – 82

http://dx.doi.org/10.1142/9789814733816_0007 ,

http://www.worldscientific.com/doi/abs/10.1142/9789814733816_0007 .

**6513.** Gray N 2015 Scotland Chapter 12 *in* Angels without Borders: Trends and Policies Shaping Angel Investment Worldwide *World Scientific Publishing Co Pte Ltd* ISBN: 978-981-4733-05-2 pp 123 – 137

http://dx.doi.org/10.1142/9789814733816_0012 ,

http://www.worldscientific.com/doi/abs/10.1142/9789814733816_0012 .

**6514.** Green J 2015 Australia Chapter 16 *in* Angels without Borders: Trends and Policies Shaping Angel Investment Worldwide *World Scientific Publishing Co Pte Ltd* ISBN: 978-981-4733-05-2 pp 163 – 175

http://dx.doi.org/10.1142/9789814733816_0016 ,

http://www.worldscientific.com/doi/abs/10.1142/9789814733816_0016 .

**6515.** Gunther U 2015 Germany Chapter 11 *in* Angels without Borders: Trends and Policies Shaping Angel Investment Worldwide *World Scientific Publishing Co Pte Ltd* ISBN: 978-981-4733-05-2 pp 115 – 122

http://dx.doi.org/10.1142/9789814733816_0011 ,

http://www.worldscientific.com/doi/abs/10.1142/9789814733816_0011 .

**6516.** Hellmann Th, Thiele V 2015 Friends or foes? The interrelationship between angel and venture capital markets *Journal of Financial Economics* **115** (3) pp 639 – 653.

**6517.** Henyon H 2015 United Arab Emirates Chapter 19 *in* Angels without Borders: Trends and Policies Shaping Angel Investment Worldwide *World Scientific Publishing Co Pte Ltd* ISBN: 978-981-4733-05-2 pp 195 – 201

http://dx.doi.org/10.1142/9789814733816_0019 ,

http://www.worldscientific.com/doi/abs/10.1142/9789814733816_0019 .

**6518.** Hudson M 2015 United States Chapter 10 *in* Angels without Borders: Trends and Policies Shaping Angel Investment Worldwide *World Scientific Publishing Co Pte Ltd* ISBN: 978-981-4733-05-2 pp 99 – 114

http://dx.doi.org/10.1142/9789814733816_0010 ,

http://www.worldscientific.com/doi/abs/10.1142/9789814733816_0010 .

**6519.** Litzka B 2015 Austria Chapter 29 *in* Angels without Borders: Trends and Policies Shaping Angel Investment Worldwide *World Scientific Publishing Co Pte Ltd* ISBN: 978-981-4733-05-2 pp 277 – 279

http://dx.doi.org/10.1142/9789814733816_0029 ,

http://www.worldscientific.com/doi/abs/10.1142/9789814733816_0029 .

**6520.** May J, Liu M M (editors) 2015 Front Matter *in* Angels without borders: Trends and policies shaping angel investment worldwide *World Scientific Publishing Co Pte Ltd* ISBN: 978-981-4733-05-2 pp 1 – 308

DOI: 10.1142/9789814733816_fmatter ,

http://www.worldscientific.com/worldscibooks/10.1142/9881 .

**6521.** May J, Manhong Mannie Liu 2015 Over the Horizon Chapter 32 *in* Angels without Borders: Trends and Policies Shaping Angel Investment Worldwide *World Scientific Publishing Co Pte Ltd* ISBN: 978-981-4733-05-2 pp 293 – 295

http://dx.doi.org/10.1142/9789814733816_0032 ,

http://www.worldscientific.com/doi/abs/10.1142/9789814733816_0032 .

**6522.** Mullett C 2015 South Africa Chapter 8 *in* Angels without Borders: Trends and Policies Shaping Angel Investment Worldwide *World Scientific Publishing Co Pte Ltd* ISBN: 978-981-4733-05-2 pp 83 – 90

http://dx.doi.org/10.1142/9789814733816_0008 ,

http://www.worldscientific.com/doi/abs/10.1142/9789814733816_0008 .

**6523.** Neira J P R 2015 Colombia Chapter 28 *in* Angels without Borders: Trends and Policies Shaping Angel Investment Worldwide *World Scientific Publishing Co Pte Ltd* ISBN: 978-981-4733-05-2 pp 271 – 276

http://dx.doi.org/10.1142/9789814733816_0028 ,

http://www.worldscientific.com/doi/abs/10.1142/9789814733816_0028 .

**6524.** Oker-Blom J D 2015 Finland Chapter 30 *in* Angels without Borders: Trends and Policies Shaping Angel Investment Worldwide *World Scientific Publishing Co Pte Ltd* ISBN: 978-981-4733-05-2 pp 281 – 283

http://dx.doi.org/10.1142/9789814733816_0030 ,

http://www.worldscientific.com/doi/abs/10.1142/9789814733816_0030 .

**6525.** Payne B 2015 Sources of capital for start-ups Chapter 2 *in* Angels without Borders Trends and Policies Shaping Angel Investment Worldwide *World Scientific Publishing Co Pte Ltd* pp 7 – 24

http://dx.doi.org/10.1142/9789814733816_0002 ,

http://www.worldscientific.com/doi/abs/10.1142/9789814733816_0002 .

**6526.** Poh-Kam Wong 2015 Singapore Chapter 23 *in* Angels without Borders: Trends and Policies Shaping Angel Investment Worldwide *World Scientific Publishing Co Pte Ltd* ISBN: 978-981-4733-05-2 pp 233 – 242

http://dx.doi.org/10.1142/9789814733816_0023 ,

http://www.worldscientific.com/doi/abs/10.1142/9789814733816_0023 .

**6527.** Protopopov I, Fokin K 2015 Russia Chapter 27 *in* Angels without Borders: Trends and Policies Shaping Angel Investment Worldwide *World Scientific Publishing Co Pte Ltd* ISBN: 978-981-4733-05-2 pp 263 – 270

http://dx.doi.org/10.1142/9789814733816_0027 ,

http://www.worldscientific.com/doi/abs/10.1142/9789814733816_0027 .

**6528.** Reijtenbagh R A G 2015 The Netherlands Chapter 22 *in* Angels without Borders: Trends and Policies Shaping Angel Investment Worldwide *World Scientific Publishing Co Pte Ltd* ISBN: 978-981-4733-05-2 pp 225 – 231

http://dx.doi.org/10.1142/9789814733816_0022 ,

http://www.worldscientific.com/doi/abs/10.1142/9789814733816_0022 .

**6529.** Roure J, San Jose A D 2015 Spain Chapter 18 *in* Angels without Borders: Trends and Policies Shaping Angel Investment Worldwide *World Scientific Publishing Co Pte Ltd* ISBN: 978-981-4733-05-2 pp 187 – 194

http://dx.doi.org/10.1142/9789814733816_0018 ,

http://www.worldscientific.com/doi/abs/10.1142/9789814733816_0018 .

**6530.** Roza J T D, Banha F 2015 Portugal Chapter 25 *in* Angels without Borders: Trends and Policies Shaping Angel Investment Worldwide *World Scientific Publishing Co Pte Ltd* ISBN: 978-981-4733-05-2 pp 249 – 254

http://dx.doi.org/10.1142/9789814733816_0025 ,

http://www.worldscientific.com/doi/abs/10.1142/9789814733816_0025 .

**6531.** Sidman Ch 2015 Crowdfunding and angel investing Chapter 5 *in* Angels without Borders: Trends and Policies Shaping Angel Investment Worldwide *World Scientific Publishing Co Pte Ltd* ISBN: 978-981-4733-05-2 pp 43 – 57

http://dx.doi.org/10.1142/9789814733816_0005 ,

http://www.worldscientific.com/doi/abs/10.1142/9789814733816_0005 .

**6532.** Silby W, Nicholas J 2015 Angel impact investing Chapter 4 *in* Angels without Borders: Trends and Policies Shaping Angel Investment Worldwide *World Scientific Publishing Co Pte Ltd* ISBN: 978-981-4733-05-2 pp 33 – 41

http://dx.doi.org/10.1142/9789814733816_0004 ,

http://www.worldscientific.com/doi/abs/10.1142/9789814733816_0004 .

**6533.** Tooth J 2015 United Kingdom Chapter 20 *in* Angels without borders: Trends and policies shaping angel investment worldwide *World Scientific Publishing Co Pte Ltd* ISBN: 978-981-4733-05-2 pp 203 – 213

http://dx.doi.org/10.1142/9789814733816_0020 ,

http://www.worldscientific.com/doi/abs/10.1142/9789814733816_0020 .

**6534.** Vossen R, Munck C 2015 Belgium Chapter 17 *in* Angels without borders: Trends and policies Shaping Angel Investment Worldwide *World Scientific Publishing Co Pte Ltd* ISBN: 978-981-4733-05-2 pp 177 – 186

http://dx.doi.org/10.1142/9789814733816_0017 ,

http://www.worldscientific.com/doi/abs/10.1142/9789814733816_0017 .

**6535.** Wallace P, Conti R 2015 Women angel investors Chapter 3 *in* Angels without borders: Trends and policies shaping angel investment worldwide *World Scientific Publishing Co Pte Ltd* ISBN: 978-981-4733-05-2 pp 25 – 31

http://dx.doi.org/10.1142/9789814733816_0003 ,

http://www.worldscientific.com/doi/abs/10.1142/9789814733816_0003 .

*6536.* Wang Jiani, Chen Su 2015 China Chapter 13 *in* Angels without Borders: Trends and Policies Shaping Angel Investment Worldwide *World Scientific Publishing Co Pte Ltd* ISBN: 978-981-4733-05-2 pp 139 – 150

http://dx.doi.org/10.1142/9789814733816_0013 ,

http://www.worldscientific.com/doi/abs/10.1142/9789814733816_0013 .

*6537.* Wang Jiani, Yi Tan, Manhong Liu 2016 Business angels in China: Characteristics, policies and international comparison Chapter 9 in Handbook of Research on Business Angels *Edward Elgar Publishing* UK ISBN: 9781783471713 pp 201 – 232

http://dx.doi.org/10.4337/9781783471720.00016 ,

https://www.elgaronline.com/view/9781783471713.00016.xml .

*6538.* Yeung A 2015 Hong Kong Chapter 9 *in* Angels without Borders: Trends and Policies Shaping Angel Investment Worldwide *World Scientific Publishing Co Pte Ltd* ISBN: 978-981-4733-05-2 pp 91 – 98

http://dx.doi.org/10.1142/9789814733816_0009 ,

http://www.worldscientific.com/doi/abs/10.1142/9789814733816_0009 .

*6539.* Zhujun Ding, Au K, Chiang F 2015 Social trust and angel investors' decisions: A multilevel analysis across nations *Journal of Business Venturing* **30** (2) pp 307 – 321.

*6540.* Avdeitchikova S, Landström H 2016 The economic significance of business angels: Toward comparable indicators Chapter 3 *in* Handbook of Research on Business Angels *Edward Elgar Publishing* UK ISBN: 9781783471713 pp 53 – 75

http://dx.doi.org/10.4337/9781783471720.00008 ,

https://www.elgaronline.com/view/9781783471713.00008.xml .

*6541.* Amatucci F M 2016 Women business angels: Theory and practice Chapter 5 in Handbook of Research on Business Angels *Edward Elgar Publishing* UK ISBN: 9781783471713 pp 92 – 112

http://dx.doi.org/10.4337/9781783471720.00010 ,

https://www.elgaronline.com/view/9781783471713.00010.xml .

*6542.* Estapé-Dubreuil G, Ashta A, Hédou J-P 2016 Micro-equity for sustainable development: Selection, monitoring and exit strategies of micro-angels *Ecological Economics* **130** (C) pp 117 – 129.

*6543.* Hornuf L, Schwienbacher A October 29 2016 Crowdinvesting: Angel investing for the masses? Chapter 15 *in* Handbook of Research on Business Angels *Edward Elgar Publishing* UK ISBN: 9781783471713 pp 381 – 398

http://dx.doi.org/10.4337/9781783471720.00024 ,

https://www.elgaronline.com/view/9781783471713.00024.xml .

*6544.* Lingelbach D 2016 Business angels in sub-Saharan Africa Chapter 11 in Handbook of Research on Business Angels *Edward Elgar Publishing* UK ISBN: 9781783471713 pp 256 – 281

http://dx.doi.org/10.4337/9781783471720.00018 ,

https://www.elgaronline.com/view/9781783471713.00018.xml .

6545. Murnieks Ch Y, Cardon M S, Sudek R, White T D, Brooks W T 2016 Drawn to the fire: The role of passion, tenacity and inspirational leadership in angel investing *Journal of Business Venturing* **31** (4) pp 468 – 484.

6546. Politis D 2016 Business angels as smart investors: A systematic review of the evidence Chapter 7 *in* Handbook of Research on Business Angels *Edward Elgar Publishing* UK ISBN: 9781783471713 pp 147 – 175

http://dx.doi.org/10.4337/9781783471720.00013 ,

https://www.elgaronline.com/view/9781783471713.00013.xml .

6547. Qoqiauri L 2016 Principles of venture and business-angel investments *Asian Economic and Financial Review* **6** (10) pp 583 – 601.

6548. Romaní G, Atienza M 2016 Business angels in developing economies: the experience of Latin America Chapter 12 *in* Handbook of Research on Business Angels *Edward Elgar Publishing* UK ISBN: 9781783471713 pp 282 – 324

http://dx.doi.org/10.4337/9781783471720.00019 ,

https://www.elgaronline.com/view/9781783471713.00019.xml .

6549. Scheela W 2016 Business angels in emerging economies: Southeast Asia Chapter 10 *in* Handbook of Research on Business Angels *Edward Elgar Publishing* UK ISBN: 9781783471713 pp 233 – 255

http://dx.doi.org/10.4337/9781783471720.00017 ,

https://www.elgaronline.com/view/9781783471713.00017.xml .

6550. Sørheim R, Botelho T 2016 Categorizations of business angels: An overview Chapter 4 *in* Handbook of Research on Business Angels *Edward Elgar Publishing* UK ISBN: 9781783471713 pp 76 – 91

http://dx.doi.org/10.4337/9781783471720.00009 ,

https://www.elgaronline.com/view/9781783471713.00009.xml .

*Investment boutique firm, investment boutique bank, financial capital investment vehicle in finances:*

6551. Thrift N 1994 On the social and cultural determinants of international financial centres: The case study of the City of London *in* Money, Power and Space Corbridge S, Thrift N, Martin R (editors) *Blackwell* Oxford UK pp 327 – 355.

6552. Luenberger D G 1997 Investment science *Oxford University Press* Oxford USA ISBN: 9780195108095.

6553. Hall S 2007 "Relational marketplaces" and the rise of boutiques in London's corporate finance industry *Environment and Planning A* **39** (8) pp 1838 – 1854

http://journals.sagepub.com/doi/abs/10.1068/a38233?id=a38233& .

6554. Morrison A D, Wilhelm W J Winter 2007 Investment banking: Past, present and future *Journal of Applied Corporate Finance* **19** (1) pp 8 – 20.

6555. Office of Career Services August 2012 Investment banking boutique firms Columbia University NY USA pp 1 – 8

http://www.sipa.columbia.edu/resources_services/career_services/current_students/career_resources/documents/InvestmentBankingBoutiqueFirms.pdf .

6556. Weihong Song Wei Jie, Lei Zhou 2013 The value of "boutique" financial advisors in mergers and acquisitions *Journal of Corporate Finance* **20** (C) pp 94 – 114.

6557. Goodhart Ch, Schoenmaker D March 2016 The United States dominates global investment banking: Does it matter for Europe? *Policy Contributions Bruegel* issue **2016/06** pp 1 – 14

http://bruegel.org/wp-content/uploads/2016/03/pc_2016_06-1.pdf .

6558. Thomson Reuters 2016 Global investment banking review full year 2015 *Thomson Reuters Deals Business Intelligence* New York USA.

6559. Wikipedia 2016 Boutique investment bank Wikipedia Inc

https://en.wikipedia.org/wiki/Boutique_investment_bank .

***Probability theory, statistics theory, Brownian movement theory, diffusion theory and chaos theory in econometrics, econophysics and physics:***

6560. Huygens 1657 De ratiociniis in aleae ludo (On calculations in games of chance).

6561. Newton I 5 July 1687, 1999 The Principia: Mathematical principles of natural philosophy (Newton's laws of motion) *Jussu Societatis Regiæ ac Typis Joseph Streater* Londini UK, University of California Press California USA.

6562. Bernoulli J 1713 Ars conjectandi (The art of guessing).

6563. Bernoulli D 1738, 1954 Specimen theoria novae de mensura sortis *Commentarii Academiae Scientiarium Imperialis Petropolitanae* Petropoli vol **5** pp 175 – 192; Exposition of a new theory on the measurements of risk Sommer L (translator) *Econometrica* vol **22** pp 23 – 36.

6564. De Moivre 1730 Miscellanea analytica supplementum (The analytic method).

6565. De Laplace 1812 Théorie analytique des probabilités *Paris* France.

6566. Bunyakovsky V Ya 1825 Heat propagation in solids *Ph D Thesis* under Prof. Augustin -Louis Cauchy supervision *École Polytechnique* Paris France.

6567. Bunyakovsky V Ya 1846 Foundations of the mathematical theory of probability *St. Petersburg* Russian Federation.

6568. Connor J J, Robertson E F July 2000 Viktor Yakovlevich Bunyakovsky (December 16, 1804 - December 12, 1889) *School of Mathematics and Statistics* University of St Andrews Scotland UK

http://www-history.mcs.st-andrews.ac.uk/Biographies/Bunyakovsky.html .

6569. V Ya Bunyakovsky International Conference (August 20 - 21) 2004 Private communications with conference participants on V Ya Bunyakovsky's mathematical theory of probability and its applications in econophysics and econometrics during tour to Town of Bar Vinnytsia Region Ukraine *V Ya Bunyakovsky International Conference Institute of Mathematics of National Academy of Sciences of Ukraine (NASU)* Kyiv Ukraine

www.imath.kiev.ua/~syta/bunyak .

6570. Brown R 1828 *Philosophical Magazine* N S **4** pp 161 – 173.

6571. Brown R 1829 *Philosophical Magazine* N S **6** pp 161 – 166.

6572. Chebyshev P L 1846 An experience in the elementary analysis of the probability theory *Crelle's Journal fur die Reine und Angewandte Mathematik*.

6573. Chebyshev P L 1867 Des valuers moyennes *Journal de Math'ematics Pures et Appliqu'ees* vol **12** pp 177 – 184.

6574. Chebyshev P L 1891 Sur deux theoremes relatifs aux probabilities *Acta Mathematica* vol **14**.

6575. Chebyshev P L 1936 Theory of probability: Lectures given in 1879 and 1880 Lyapunov A N (lecture notes writer) Krylov A N (editor) *Moscow - St Petersburg* Russian Federation.

6576. Markov A A 1890 On one problem by D I Mendeleev *Zapiski Imperatorskoi Akademii Nauk SPb* **62** pp 1 – 24.

6577. Markov A A 1899 Application des functions continues au calcul des probabilit´es *Kazan Bulletin* **9** (2) pp 29 – 34 Russian Federation.

6578. Markov A A 1900, 1912, 1913 Calculation of probabilities *St Petersburg* Russian Federation; Wahrscheinlichkeits-Rechnung *Teubner* Leipzig-Berlin Germany; 3<sup>rd</sup> edition *St Petersburg* Russian Federation.

6579. Markov A A 1906 Extension of law of big numbers on variables, depending from each other *Izvestiya Fiziko-Matematicheskogo Obschestva pri Kazanskom Universitete* 2<sup>nd</sup> series vol **15** (94) pp 135 – 156 Russian Federation.

6580. Markov A A 1907, 1910 Research on fine case of depending trials *Izvestiya Akademii Nauk SPb* 6<sup>th</sup> series vol **1** (93) pp 61 – 80; Recherches sur un cas remarquable d'epreuves dependantes *Acta Mathematica* **33** pp 87 – 104 Stockholm Sweden.

6581. Markov A A 1908, 1912, 1971 Extension of limit theorems of calculation of probabilities to sum of variables, connected in chain *Zapiski Akademii Nauk po Fiziko-Matematicheskomu Otdeleniyu* 8<sup>th</sup> series vol **25** (3); Ausdehnung der Satze uber die Grenzwerte in der Wahrscheinlichkeitsrechnung auf eine Summe verketteter Grossen Liebmann H (translator) *in* Wahrscheinlichkeitsrechnung Markov A A (author) pp 272 – 298 *Teubner B G* Leipzig Germany; Extension of the limit theorems of probability theory to a sum of variables connected in a chain Petelin S (translator) *in* Dynamic probabilities systems Howard R A (editor) vol **1** pp 552 – 576 *John Wiley and Sons Inc* New York USA.

6582. Markov A A 1910 Research on common case of trials, connected in chain *Zapiski Akademii Nauk po Fiziko-Matematicheskomu Otdeleniyu* 8<sup>th</sup> series vol **25** (93) Russian Federation.

6583. Markov A A 1911 On one case of trials, connected in complex chain *Izvestiya Akademii Nauk SPb* 6<sup>th</sup> series vol **5** (93) pp 171 – 186 Russian Federation.

6584. Markov A A 1912 On trials of connected in chain unobserved events *Izvestiya Akademii Nauk SPb* 6<sup>th</sup> series vol **6** (98) pp 551 – 572 Russian Federation.

6585. Markov A A 1913 Example of statistical research on text of "Eugene Onegin", illustrating interconnection of trials in chain *Izvestiya Akademii Nauk SPb* 6<sup>th</sup> series vol **7** (93) pp 153 – 162 Russian Federation.

6586. Fisher I 1892 Mathematical investigations in the theory of value and prices *Transactions of the Connecticut Academy* **9** pp 1 – 124.

6587. Einstein A 1905 On the movement of small particles suspended in a stationary liquid demanded by the molecular-kinetic theory of heat *Annalen der Physik* **17** pp 549 – 560.

6588. Einstein A 1956 Investigation on the theory of the Brownian motion Furth R (editor) *Dover* New York USA.

6589. Einstein A, Smolukhovsky M 1936 Brownian movement: Collection of research papers *ONTI* Moscow Russian Federation.

6590. Bowley A L 1924 The mathematical groundwork of economic *Clarendon Press* Oxford UK.

6591. Lévy P 1925 Calcul des probabilités *Gauthier-Villars* Paris France.

6592. Kolmogorov A N 1937 Markov chains with countable many states *Bulletin Moscow University* **1**.

6593. Kolmogorov A N 1938 On analytic methods in probability theory *in* Selected works of Kolmogorov A N vol **2** Probability theory and mathematical statistics Shiryaev A N (editor) *Springer* Germany.

6594. Kolmogorov A N 1941 Interpolation and extrapolation *Bulletin de l'academie des sciences de USSR* Ser Math **5** pp 3 – 14.

6595. Kolmogorov A N 1947 The contribution of Russian science to the development of probability theory *Uchenye Zapiski Moskovskogo Universiteta* no 91.

6596. Kolmogorov A N 1956 Probability theory in mathematics: Its contents, methods, and meaning *Academy of Sciences USSR* vol **2**.

6597. Kolmogorov A N 1956 Foundations of the theory of probability *Chelsea* New York USA.

6598. Kolmogorov A N 1985 Mathematics and mechanics Selected works vol **1** *Nauka Publishing House* Moscow Russian Federation.

6599. Kolmogorov A N 1986 Probability theory and mathematical statistics Selected works vol **2** *Nauka Publishing House* Moscow Russian Federation.

6600. Allen R G D 1938 Mathematical analysis for economists *Macmillan* London UK.

6601. Cramer H 1940 On the theory of stationary random processes *Ann Math* vol **41** pp 215 – 230.

6602. Cramer H 1946 Mathematical methods of statistics *Princeton University Press* USA.

6603. Cramer H, Leadbetter M 1967 Stationary and related stochastic processes. Sample function properties and their applications *John Wiley and Sons Inc* NY USA.

6604. Tintner G 1940 The variate difference method *John Wiley and Sons Inc* Bloomington Indiana USA.

6605. Tintner G 1953 Econometrics *John Wiley and Sons Inc* New York USA.

6606. Bemshtein S N 1946 Theory of probability 4[th] edition *Gostehizdat* Moscow Russian Federation.

6607. Hannan E J 1960 Time series analysis *Methuen* London.

6608. Hannan E J 1970 Multiple time series *John Wiley and Sons Inc* New York USA.

6609. Mandelbrot B B 1960 The Pareto-Levy law and the distribution of income *International Economic Review* no 1.

6610. Mandelbrot B B 1963a The stable Paretian income distribution when the apparent exponent is near two *International Economic Review* no 4.

6611. Mandelbrot B B 1963b The variation of certain speculative prices *Journal of Business* vol **36** pp 394 – 419.

**6612.** Mandelbrot B B 1965 Une classe de processus stochastiques homothetiques a soi: Application a la loi climatologique de H. E. Hurst Comptes Rendus de l'Academie des Sciences vol **240** pp 3274 – 3277 Paris France.

**6613.** Mandelbrot B B 1967a The variation of some other speculative prices Joural of Business vol **40** pp 393 – 413.

**6614.** Mandelbrot B B (April) 1967b Some noises with 1/f spectrum: A bridge between direct current and white noise *IEEE Transactions on Information* Theory USA.

**6615.** Mandelbrot B B, Taylor H M 1967 On the distribution of stock price difference *Operations Research* vol **15** no 6 pp 1057 – 1062.

**6616.** Mandelbrot B B, van Ness J W 1968 Fractional Brownian motions, fractional noises and applications *SIAM Review* vol **10** no 4 pp 422 – 437.

**6617.** Mandelbrot B B 1969 Robustness of the rescaled range R/S in the measurement of non-cyclic long-run statistical dependence *Water Resources Research* vol **5** no 5 pp 967 – 988.

**6618.** Mandelbrot B B, Wallis J R 1969 Computer experiments with fractional Gaussian noises I, II, III *Water Resources Research* vol **5** pp 228 – 267.

**6619.** Mandelbrot B B 1971 When can price be arbitrated efficiently? A limit of the validity of the random walk and martingale models *Review of Economics and Statistic* vol **53** pp 225 – 236.

**6620.** Mandelbrot B B 1972 Statistical methodology for non-periodic cycles: From the covariance to R/S analysis Annals of Economic and Social Measurement vol **1** no 3 pp 259 – 290.

**6621.** Mandelbrot B B 1975a Les objects fractals *Flammarion* Paris France.

**6622.** Mandelbrot B B 1975b Limit theorems on the self-normalized range for weakly and strongly dependent process *Zeitschrift Wahrscheinlichkeitsttheorie und Verwandte Gebiete* vol **31** pp 271 – 285.

**6623.** Mandelbrot B B 1977 Fractals: Form, chance and dimension *W H Freeman* San Francisco USA.

**6624.** Mandelbrot B B 1982 The fractal geometry of nature *W H Freeman* San Francisco USA.

**6625.** Mandelbrot B B 1997 Fractals and scaling in finance *Springer* New York USA.

**6626.** Mandelbrot B B, Hudson L March 7 2006 The misbehavior of markets: A fractal view of financial turbulence *Basic Books* pp 1 – 368 ISBN-13: 978-0465043576.

**6627.** Gnedenko B V, Khinchin A Ya 1961 An elementary introduction to the theory of probability *Freeman* San Francisco USA.

**6628.** Gnedenko B V 1988 The theory of probability *Mir* Moscow Russian Federation.

**6629.** Abramowitz M, Stegun I A (editors) 1964 Handbook of mathematical functions *National Bureau of Standards Applied Mathematics Series* vol **55** USA.

**6630.** Kubilius J 1964 Probabilistic methods in the theory of numbers American Mathematical Society Providence USA.

**6631.** Akhiezer N I, Glazman I M 1966 Theory of linear operators in Hilbert space *Nauka* Moscow Russian Federation.

**6632.** Lamperti J 1966 Probability *Benjamin* New York USA.

**6633.** Kai-Lai Chung 1967 Markov chains with stationary transition probabilities *Springer-Verlag* New York USA.

**6634.** Skorohod A V 1967 Random processes with independent increments *Nauka* Moscow Russian Federation.

**6635.** Gikhman I I, Skorohod A V 1968 Stochastic differential equations *Naukova Dumka* Kiev Ukraine.

**6636.** Gikhman I I, Skorohod A V 1969 Introduction to the theory of random processes 1[st] edition *Saunders* Philadelphia USA.

**6637.** Gikhman I I, Skorohod A V 1974-1979 Theory of stochastic processes vols **1**, **2**, **3** *Springer-Verlag* New York-Berlin USA-Germany.

**6638.** Breiman L 1968 Probability *Addison-Wesley* Reading MA USA.

**6639.** Feller W 1968 An introduction to probability theory and its applications vols **1**, **2** 3[rd] edition *John Wiley and Sons Inc* New York USA.

**6640.** Brush S G 1968, 1977 A history of random processes: 1. Brownian movement *in* Study history statistics and probability Kendall M G, Plackett R L (editors) **2** pp 347 – 382 London UK.

**6641.** Glesjer H 1969 A new test for heteroskedasticity *Journal of the American Statistical Association* **64** pp 316 – 323.

**6642.** Ash R B 1970 Basic probability theory *John Wiley and Sons Inc* New York USA.

**6643.** Ash R B 1972 Real analysis and probability *Academic Press* New York USA.

**6644.** Ash R B, Gardner M F 1975 Topics in stochastic processes *Academic Press* New York USA.

**6645.** Box G E P, Jenkins G M 1970 Time series analysis: Forecasting and control *Holden Day* San Francisco California USA.

**6646.** Renyi A 1970 Probability theory *North-Holland Publishing Company* Amsterdam The Netherlands.

**6647.** Isihara A 1971 Statistical physics *Academic Press* New York USA.

**6648.** Andreev A F February 1976 Diffusion in quantum crystals *Uspekhi Fizicheskih Nauk* (*UFN*) vol **118** (2) pp 251 – 271.

**6649.** Borovkov A A 1976 Wahrscheinlichkeitstheorie: Eine EinjUhrung 1[st] edition *Birkhiuser* Basel-Stuttgart Switzerland-Germany.

**6650.** Grangel C W J, Newbold P 1977 Forecasting economic time series *Academic Press* New York USA.

**6651.** Pugachev V S 1979b Theory of probability and mathematical statistics 1[st] edition *Nauka* Moscow Russian Federation, 2[nd] edition *Fizmatlit* Moscow Russian Federation ISBN 5–92210254–0 pp 1 – 496.

**6652.** Grangel C W J, Teräsvirta T 1993 Modeling nonlinear economic relationships *Oxford University Press* Oxford New York UK USA.

**6653.** Karlin S, Taylor H M 1981 A second course in stochastic processes *Academic Press* New York USA.

**6654.** Venttsel A D 1981 A course in the theory of stochastic processes *McGraw-Hill* New York USA.

**6655.** Yaglom A M, Yaglom I M 1983 Probability and information *Reidel Dordrecht*.

6656. Pagan A 1984 Econometric issues in the analysis of regressions with generated regressors *International Economic Review* **25** pp 221 – 247.

6657. Van Horne J C 1984 Financial market rates and flows *Prentice Hall* Englewood Cliffs NJ USA.

6658. Taylor S 1986 Modeling financial time series *John Willey and Sons Inc* New York USA.

6659. Tong H 1986 Nonlinear time series *Oxford University Press* Oxford UK.

6660. Sharkovsky A N, Maistrenko Yu L, Romanenko E Yu 1986 Differential equations and their applications *Naukova Dumka* Kiev Ukraine pp 1 – 280.

6661. Newey W, West K 1987 A simple positive semi-definite, heteroskedasticity and autocorrelation consistent covariance matrix *Econometrica* **55** pp 703 – 708.

6662. Luukkonen R, Saikkonen P, Terasvirta T 1988 Testing linearity against smooth transition autoregressive models *Biometrika* **75** pp 491 – 499.

6663. Judge G, Hill C, Griffiths W, Lee T, Lutkepol H 1988 An introduction to the theory and practice of econometrics *John Wiley and Sons Inc* New York USA.

6664. Priestley M B 1989 Spectral analysis and time series *Academic Press* London UK.

6665. Hardle W 1990 Applied nonparametric regression *Econometric Society Monograph Cambridge University Press* Cambridge UK.

6666. Tong H 1990 Nonlinear time series: A dynamical system approach *Clarendon Press* Oxford UK.

6667. Johansen S 1992 Cointegration in partial systems and the efficiency of single equation analysis *Journal of Econometrics* **52** pp 389 – 402.

6668. Pesaran M H, Potter S M (editors) 1993 Nonlinear dynamics, chaos and econometrics *John Willey and Sons Inc* New York USA.

6669. Banerjee A, Dolado J J, Galbraith J W, Hendry D F 1993 Cointegration, error correction, and the econometric analysis of nonstationary data *Oxford University Press* Oxford UK.

6670. Hamilton J D 1994 Time series analysis *Princeton University Press* Princeton, NJ USA.

6671. Peters E E 1994 Fractal market analysis: Applying chaos theory to investment and economics *John Wiley and Sons Inc* New York USA.

6672. Enders W 1995 Applied econometric time series *John Wiley and Sons Inc* New York USA.

6673. Johansen S 1995 Likelihood based inference in co-integrated vector autoregressive models *Oxford University Press* Oxford UK.

6674. Karatzas I, Shreve S 1995 Methods of mathematical finance *Columbia University Press* New York USA.

6675. Moore G E 1995 Lithography and the future of Moore's law *Proceedings SPIE Symposium Optical Microlithography Conference VIII* **2440** 2.

6676. Shiryaev A N 1995 Probability $2^{nd}$ edition *Springer - Verlag* ISBN 0-387-94549-0 New York USA pp 1 – 621.

6677. Moore G E 2003 No exponential is forever – but we can delay forever *ISSCC*.

6678. Campbell J Y, Lo A W, MacKinlay A C 1996 The econometrics of financial markets *Princeton University Press* Princeton USA.

**6679.** Mosekilde E 1996 Topics in nonlinear dynamics: Applications to physics, biology and economic systems *World Scientific Publishing Pte Ltd* Singapore.

**6680.** Rogers L C G, Talay D (editors) 1997 Numerical methods in finance *Cambridge University Press* Cambridge UK.

**6681.** Campbell J, Lo A, MacKinlay C 1997 The econometrics of financial markets *Princeton University Press* Princeton NJ USA.

**6682.** Mosekilde E (1996-1997) Private communications on quantum chaos *Department of Physics* Technical University of Denmark Lyngby Denmark.

**6683.** Greene W H 1997, 2003 Econometric analysis 1$^{st}$ edition, 5$^{th}$ edition *Prentice Hall* Upper Saddle River USA.

**6684.** Hasem P M, Pesaran B 1997 Working with Microfit 4.0: Interactive econometric analysis *Oxford University Press* Oxford UK.

**6685.** Lo A W, MacKinlay A C 1997 The econometrics of financial markets *Princeton University Press* Princeton New Jersey USA.

**6686.** Anderson H M, Vahid F 1998 Testing multiple equation systems for common nonlinear factors *Journal of Econometrics* **84** pp 1 – 37.

**6687.** Escribano, Jorda 1999 Improved testing and specification of smooth transition regression models *in* Nonlinear time series analysis of economic and financial data Rothman (editor) *Kluwer Academic Press* Amsterdam The Netherlands.

**6688.** Hasem P M, Shin Y 1999 An autoregressive distributed lag modelling approach to cointegration analysis *in* Econometrics and economic theory in the 20th century: The Ranger Frisch centennial symposium Strom S, Holly A, Diamond P (editors) *Cambridge University Press* Cambridge UK www.econ.cam.ac.uk/faculty/pesaran/ADL.pdf .

**6689.** Hasem P M, Shin Y, Smith R J 2001 Bounds testing approaches to the analysis of level relationships *Journal of Applied Econometrics* **16** (3) pp 289 – 326.

**6690.** Potter S 1999 Non-linear time series modelling: An introduction *Typescript* Federal Reserve Bank of New York NY USA.

**6691.** Rothman (editor) 1999 Nonlinear time series analysis of economic and financial data *Kluwer Academic Press* Amsterdam The Netherlands.

**6692.** Hayashi F 2000 Econometrics *Princeton University Press* Princeton NJ USA.

**6693.** Durbin J, Koopman S J 2000 Time series analysis of non-Gaussian observations based on state-space models from both classical and Bayesian perspectives *Journal of Royal Statistical Society Series B* **62** pp 3 – 56.

**6694.** Durbin J, Koopman S J 2002 A simple and efficient simulation smoother for state space time series analysis *Biometrika* **89** pp 603 – 615.

**6695.** Durbin J, Koopman S J 2012 Time series analysis by state space methods 2nd edition *Oxford University Press* Oxford UK.

**6696.** Ilinski K 2001 Physics of finance: Gauge modelling in non-equilibrium pricing *John Wiley and Sons Inc* New York USA ISBN-10: 0471877387 pp 1 – 300.

**6697.** Nicolau J 2002 Stationary processes that look like random walks – The bounded random walk process in discrete and continuous time *Econometric Theory* **18** pp 99 – 118.

**6698.** Koop G 2003 Bayesian econometrics *John Wiley and Sons Inc* New York USA.

**6699.** Ledenyov V O, Ledenyov O P, Ledenyov D O 2002 A quantum random number generator on magnetic flux qubits *Proceedings of the 2[nd] Institute of Electrical and Electronics Engineers Conference IEEE-NANO 2002* Chicago Washington DC USA IEEE Catalog no 02TH86302002 Library of Congress number: 2002106799 ISBN: 0-7803-7538-6.

**6700.** Davidson R, MacKinnon J 2004 Econometric theory and methods *Oxford University Press* Oxford UK.

**6701.** Protter P E 2005 Stochastic integration and differential equations *Springer* Germany.

**6702.** Ledenyov V O, Ledenyov D O, Ledenyov O P 2006, 2012 Features of oxygen and its vacancies diffusion in superconducting composition $YBa_2Cu_3O_{7-\delta}$ near to magnetic quantum lines *Problems of Atomic Science and Technology* vol **15** no **1** pp 76 – 82 ISSN 1562-6016, Cornell University NY USA www.arxiv.org 1206.5635v1.pdf .

**6703.** Demenok S L 2011 Fractal: Between the myth and craft *Rinvol, Academy of Culture Research Publishing Houses* St Petersburg Russian Federation pp 1 – 296 ISBN 978-5-903931-62-0.

**6704.** Jakimowicz A 2016 Econophysics as a new school of economic thought: Twenty years of research *Proceedings of the 8[th] Polish Symposium of Physics in Economy and Social Sciences* FENS Rzeszów Poland November 4–6 2015 *Acta Physica Polonica A* no 5 vol **129** 2016 pp 897 – 907 DOI: 10.12693/APhysPolA.129.897 .

**6705.** Wilson 2016 Non Western mathematics Public Lecture *London School of Economics and Political Science* London UK http://media.rawvoice.com/lse_publiclecturesandevents/richmedia.lse.ac.uk/publiclectures andevents/20160118_1830_nonWesternMathematics.mp4 .

*Quantum diffusion in physics, econophysics and finances:*

**6706.** Feynman R P, Hibbs A R 1965 Quantum mechanics and path integrals *McGraw-Hill* ISBN 0-07-020650-3.

**6707.** Andreev A F February 1976 Diffusion in quantum crystals *Uspekhi Fizicheskih Nauk* (*UFN*) vol **118** (2) pp 251 – 271.

**6708.** Kleinert H 2004 Path integrals in quantum mechanics, statistics, polymer physics, and financial markets 4[th] edition *World Scientific* Singapore ISBN 981-238-107-4.

*Stability of investment portfolio in nonlinear dynamics, mathematics and finances:*

**6709.** Lyapunov A M 1892, 1966, 1992 The general problem of the stability of motion *Ph D dissertation Imperial University of Kharkov* Kharkov Ukraine, Stability of motion *Academic Press* New-York & London USA & UK, The general problem of the stability of motion (Fuller A T translator) *Taylor & Francis* London UK.

**6710.** Kolmogorov A N, Petrovsky I G, Piskunov N S 1937 Research on the equation of diffusion, connected with increase of quantity of matter and its application to one biological problem *Moscow State University Bulletin: Mathematics and Mechanics* vol **1** p 1.

**6711.** Kolmogorov A N 1940 Wienersche spiralen und einige andere interessante kurven im Hilbertschen raum *DAN USSR* vol **26** no 2 pp 115 – 118.

**6712.** Kolmogorov A N 1941 Local structure of turbulence in non-compressed viscous liquid at very big Reynolds numbers *DAN USSR* vol **30** pp 299 – 303.

**6713.** Kolmogorov A N 1959 On the entropy on unit of time as metric invariant of automorphism *DAN USSR* vol **124** pp 754 – 755.

**6714.** Kolmogorov A N 1985 Mathematics and mechanics. Selected works vol **1** *Nauka Publishing House* Moscow Russian Federation.

**6715.** Kolmogorov A N 1986 Probability theory and mathematical statistics. Selected works vol **2** *Nauka Publishing House* Moscow Russian Federation.

**6716.** Alexandrov P S, Khinchin A Ya 1953 Andrei Nikolaevich Kolmogorov (To the 50[th] Birthday Anniversary) *Uspekhi Matematicheskih Nauk* (*UMN*) vol **8** no 3 pp 178 – 200.

**6717.** Gleick J 1988 Chaos: Making a new science *Penguin Books Ltd* USA ISBN 0 14 00.92501 pp 1 – 354.

**6718.** Ulam S M, von Neumann J 1947 *Bulletin American Mathematical Society* vol **53** no 11 p 1120.

**6719.** Kubo R 1957 Statistical-mechanical theory of irreversible processes *Journal Physics Society Japan* vol **12** p 570.

**6720.** Kubo R, Toda M, Hashitsume N 1983 Statistical physics *Springer* Heidelberg Germany.

**6721.** Kalman R E, Bertram J F 1960 Control system analysis and design via the second method of Lyapunov *Journal of Basic Engineering* **88** 371 – 394.

**6722.** Kalman R E 1963 Lyapunov functions for the problem of Lur'e in automatic control *Proc National Academy Science USA* **49** (2) pp 201 – 205.

**6723.** Sharkovsky A N 1964 Co-existence of cycles of a continuous map of a line in itself *Ukrainian Mathematical Journal* vol **16** pp 61 – 71.

**6724.** Sharkovsky A N 1965 On the cycles and structure of continuous mapping *Ukrainian Mathematical Journal* vol **17** p 104.

**6725.** Sharkovsky A N, Maistrenko Yu L, Romanenko E Yu 1986 Differential equations and their applications *Naukova Dumka* Kiev Ukraine pp 1 – 280.

**6726.** May R 1974 Biological populations with Non-overlapping generations: Stable points, stable cycles, and chaos *Science* vol **186** pp 645 – 647.

**6727.** May R 1976 Simple mathematical models with very complicated dynamics *Nature* vol **261** pp 459 – 467.

**6728.** May R, Oster G F 1976 Bifurcations and dynamic complexity in simple ecological models *The American Naturalist* vol **110** pp 573 – 579.

**6729.** Li T-Y, Yorke J A 1975 Period three implies chaos *American Mathematics Monthly* vol **82** pp 982 – 985.

**6730.** Feigenbaum M J 1978 Quantitative universality for a class of nonlinear transformations *Journal of Statistical Physics* vol **12** no 1 pp 25 – 52.

**6731.** Feigenbaum M J 1979a The universal metric properties of nonlinear transformations *Journal of Statistical Physics* vol **21** no 6 pp 669 – 706.

**6732.** Feigenbaum M J 1979b The onset spectrum of turbulence *Physics Letters* vol **A74** no 6 pp 375 – 378.

**6733.** Feigenbaum M J 1980 The transition to aperiodic behaviour in turbulent systems *Communications Mathematical Physics* vol **77** no 1 pp 65 – 86.

**6734.** Feigenbaum M J, Kadanoff L P, Shenker S J 1982 Quasiperiodicity in dissipative systems: A renormalization group analysis *Physica* vol **D5** p 370.

**6735.** Lanford O A 1982 Computer-assisted proof of the Feigenbaum conjectures *American Mathematical Society Bulletin* vol **6** no 3 pp 427 – 434.

**6736.** Grebogi C, Ott E, Yorke J A 1982 Chaotic attractor in crisis *Physical Review Letters* vol **48** pp 1507 – 1510.

**6737.** Klimontovich Yu L 1982 Statistical physics *Nauka Publishers* Moscow Russian Federation.

**6738.** Klimontovich Yu L 1990 Turbulent movement and structure of chaos *Nauka Publishers* Moscow Russian Federation ISBN 5-02-014038-4 pp 320.

**6739.** Anishenko V S 1990 Complex oscillations in simple systems: Mechanisms of origination, structure and properties of dynamic chaos in radiophysics systems *Nauka Publishers* Moscow Russian Federation pp 1 – 312.

**6740.** Anishenko V S 2000 Introduction to Nonlinear Dynamics: Lectures by Soros Professor: Training Manual *College* Saratov Russian Federation.

**6741.** Anishenko V S, Vadivasova T E, Astakhov V V 1999 Nonlinear dynamics of chaotic and stochastic systems *Saratov University Publishing House* Saratov Russian Federation.

**6742.** Hsieh D 1991 Chaos and nonlinear dynamics *The Journal of Finance* vol **16** p 5.

**6743.** Strogatz S 1994 Nonlinear dynamics and chaos *Addison Wesley* NY USA.

**6744.** Gandolfo G 1996 Economic dynamics 3$^{rd}$ edition *Springer* Berlin Germany pp 407 – 428 ISBN 3-540-60988-1 .

**6745.** Kuznetsov S P 1996-1997 Private communications on dynamic chaos *Technical University of Denmark* Lyngby Copenhagen Denmark.

**6746.** Kuznetsov S P 2001 Dynamic chaos *Izdatel'stvo Fiziko-Matematicheskoi Literatury* Moscow Russian Federation pp 1 – 296.

**6747.** Mosekilde E 1996-1997 Private communications on quantum chaos *Department of Physics* Technical University of Denmark Lyngby Copenhagen Denmark.

**6748.** Mosekilde E 1996 Topics in nonlinear dynamics: Applications to physics, biology and economic systems *World Scientific Publishing Pte Ltd* Singapore.

**6749.** Peters E E 1996 Chaos and order in the capital markets: A new view of cycles, prices, and market volatility *John Wiley & Sons, Inc* 2$^{nd}$ edition NY USA ISBN 0-471-13938-6.

**6750.** Sauer T et al 1996 Chaos: An introduction to dynamic systems *Springer Verlag* NY USA.

**6751.** Shiryaev A N 1998a Foundations of Stochastic Financial Mathematics vol **1** *Fazis Scientific and Publishing House* Moscow Russian Federation ISBN 5-7036-0044-8 pp 1 – 492.

**6752.** Shiryaev A N 1998b Foundations of Stochastic Financial Mathematics vol **2** *Fazis Scientific and Publishing House* Moscow Russian Federation ISBN 5-7036-0044-8 pp 493 – 1017.

**6753.** Medvedeva N V 2000 Dynamics of the logistic function *Soros Educational Journal* vol **6** no 8 pp 121 – 127.

**6754.** Kuznetsov N V, Leonov G A 2005 On stability by the first approximation for discrete systems *2005 International Conference on Physics and Control* PhysCon 2005 Proceedings Volume 2005 pp 596 – 599 doi:10.1109/PHYCON.2005.1514053 .

**6755.** Nikulchev E V 2007 Geometric approach to modeling of nonlinear systems from experimental data ISBN 978-5-8122-0926-1 pp 1 – 162.

**6756.** Nikulchev E V 2011 Private communications on modeling of nonlinear systems *Int. Conference on Design of Engineering and Scientific Applications in Matlab* KhPI Kharkov Ukraine.

**6757.** Bunde A, Havlin S 2009 Fractal geometry: A brief introduction to «Encyclopedia of Complexity and Systems Science» *Springer* Berlin Heidelberg Germany.

**6758.** Smirnov A D 2010 Macrofinances II: Model of bulbs and crises *Economic Journal of High School of Economics* no 4 pp 401 – 439.

**6759.** Saratov group of theoretical nonlinear dynamics 2012 Information on the logistic equation Saratov Russian Federation.

**6760.** Ledenyov D O, Ledenyov V O 2012 3D Bifurcation diagram for accurate characterization of dynamic properties of combining risky investments in investment portfolio in nonlinear dynamic financial system *Software in Matlab R2012* Townsville Australia, Kharkov Ukraine.

***Continuous-time signals, analog signals spectrum, analog signals processing, analog signals filtering in physics, electrical and computer engineering:***

**6761.** Bessel F W 1832 Versuche über die Kraft mit welcher die Erde Körper von verschiedener Beschaffenheit anzieht (Experiments on the force with which the earth attracts things of different matter) Berlin Germany.

**6762.** Chebyshev P L 1854 Théorie des mécanismes connus sous le nom de parallélogrammes *Mémoires des Savants étrangers présentés à l'Académie de Saint-Pétersbourg* Russia vol 7 pp 539 – 586.

**6763.** Maxwell J C 1890 Introductory lecture on experimental physics *in* Scientific papers of J C Maxwell Niven W D (editor) vols **1**, **2** Cambridge UK.

**6764.** Cauer W 1927 Über die Variablen eines passiven Vierpols (On the variables of some passive quadripoles) *Sitzungsberichte Preuß Akademie Wissenschaften phys-math Klasse* pp 268 – 274.

**6765.** Butterworth S 1930 On the theory of filter amplifiers *Wireless Engineer* vol **7** pp 536 – 541.

**6766.** Wanhammar L June 2 2009 Analog filters using MATLAB *Springer* Germany ASIN B008BBH22JM ISBN 13: 978-0387927664 pp 1 – 331.

**6767.** Ledenyov D O, Ledenyov V O 2015a Nonlinearities in microwave superconductivity 8[th] edition *Cornell University* NY USA pp 1 – 923 www.arxiv.org 1206.4426v8.pdf .

***Discrete-time digital signals spectrum, discrete-time digital signals processing, discrete-time digital signals filtering, Wiener filtering theory, Pugachev filtering theory, Stratonovich optimal nonlinear filtering theory, Stratonovich-Kalman-Bucy filtering***

***algorithm, Stratonovich-Kalman-Bucy filter, particle filter in econometrics, econophysics, electrical and computer engineering:***

*6768.* Edgeworth F I 1905 The law of error *Proceedings Cambridge Philosophical Society* vol **20** pp 36 – 65.

*6769.* Whittaker E T 1915 On the functions which are represented by the expansions of the interpolation theory *Proc Royal Soc Edinburg* **35** pp 181 – 194 DOI:10.1017/s0370164600017806 .

*6770.* Whittaker E T 1923 On a new method of graduation *Proceedings of the Royal Society of Edinburgh* **44** pp 77 – 83.

*6771.* Whittaker E T 1935 Interpolation function theory *Cambridge University Press* Cambridge England.

*6772.* Walsh J L 1923a A closed set of normal orthogonal functions *American J Math* **45** pp 5 – 24.

*6773.* Walsh J L 1923b A property of Haar's system of orthogonal functions *Math Ann* **90** p 3845.

*6774.* Wikipedia 2015d Joseph L Walsh *Wikipedia* USA www.wikipedia.org .

*6775.* Whittaker E T 1923 On a new method of graduation Proceedings *of the Royal Society of Edinburgh* **44** pp 77 – 83.

*6776.* Wiener N 1923 Differential space *Journal of Mathematical Physics Math Inst Tech* vol **2** pp 131 – 174.

*6777.* Wiener N 1930 Generalized harmonic analysis *Acta Math* vol **55** no 2 - 3 pp 117 – 258.

*6778.* Wiener N 1941, 1949 The extrapolation, interpolation and smoothing of stationary time series *Report on the Services Research Project DIC-6037, MIT Technology Press/John Wiley & Sons Inc* New York USA.

*6779.* Nyquist H 1928 Certain topics in telegraph transmission theory *Trans AIEE* **47** pp 617 – 644 DOI: 10.1109/t-aiee.1928.5055024 .

*6780.* Küpfmüller K 1928 Uber die Dynamik der selbsttatigen Verstarkungsregler *Elektrische Nachrichtentechnik* **5** (11) pp 459 – 467.

*6781.* Andronov A A, Vitt A A, Pontryagin L S 1933 On statistical consideration of dynamic systems *Soviet Journal of Experimental and Theoretical Physics* vol **3** no 3 pp 165 – 180.

*6782.* Kotelnikov V A 1933 On the carrying capacity of the ether and wire in telecommunications *Materials for the first all-Union conference on questions of communication* Izd Red Upr Svyazi RKKA.

*6783.* Tintner G 1940 The variate difference method *John Wiley and Sons* Bloomington Indiana.

*6784.* Ito K 1944 Stochastic integral *Proceedings Imperial Academy Tokyo* vol **20** pp 519 – 524.

*6785.* Ito K 1951a On a formula concerning stochastic differentials *Nagoya Mathematics Journal* vol **3** pp 55 – 65.

*6786.* Ito K 1951b On stochastic differential equations *Mem American Mathematical Society* vol **4** pp 1 – 51.

6787. Ito K, Xiong K 2000 Gaussian filters for nonlinear filtering problems *IEEE Transactions on Automatic Control* vol **45** no 5.

6788. Pugachev V S 1944 Random functions, defined by common differential equations *Works by Air Forces Military Academy named after Zhukovsky N E* vol **118** pp 3 – 36.

6789. Pugachev V S 1956a The use of canonical expansions of random functions in determining an optimum linear system *Automatics and Remote Control (USSR)* vol **17** pp 489 – 499.

6790. Pugachev V S 1956b On a possible general solution of the problem of determining optimum dynamic systems *Automatics and Remote Control (USSR)* vol **17** pp 585 – 589.

6791. Pugachev V S 1960 Theory of random functions and its application in problems of automatic control *State Publishing House of Physical Mathematical Literature (Fizmatlit)* Moscow Russian Federation pp 1 – 883.

6792. Pugachev V S 1961 Application of theory of Markov processes in analysis of accuracy of automatic systems *News Academy of Sciences USSR Energetics and Automatics* no 3 pp 46 – 57.

6793. Pugachev V S 1962 Theory of random functions and its application to problems of automatic control *Fizmatgiz* Moscow Russian Federation.

6794. Pugachev V S 1971 On distribution of computing number of random process *Works of 1st All Union Symposium on Statistics Problems in Technical Cybernetics* Moscow USSR February 14 - 18, 1967 *in* Nonlinear and optimal systems *Nauka* Moscow Russian Federation pp 374 – 381.

6795. Pugachev V S 1973, 1974, 1975 Stochastic systems *Nauka* Moscow Russian Federation issues 7-9, 11-12, 10.

6796. Pugachev V S (editor) 1974 Foundations of automatic control *Nauka* Moscow Russian Federation.

6797. Pugachev V S 1978 Estimation of variables and parameters in stochastic systems, described by differential equations *DAN USSR* vol **241** no 5 pp 1031 – 1034.

6798. Pugachev V S 1979a Estimation of state and parameters of continuous nonlinear systems *Automatics and Tele-mechanics* no 6 pp 63 – 79.

6799. Pugachev V S 1979b Theory of probability and mathematical statistics 1st edition *Nauka* Moscow Russian Federation, 2nd edition *Fizmatlit* Moscow Russian Federation ISBN 5–92210254–0 pp 1 – 496.

6800. Pugachev V S 1980a Estimation of Markov processes. Time series *Proceedings International Conference Nottingham March, 1979; North Holland Publishing House* Amsterdam New York London pp 389 – 400.

6801. Pugachev V S 1980b Finite distributions of processes, defined by stochastic differential equations, and extrapolation of these processes *DAN USSR* vol **251** no 1 pp 40 – 43.

6802. Pugachev V S 1981 The finite-dimensional distributions of a random process determined by a stochastic differential equation and their application to control problems *Problems of Control and Theory of Information* vol **10** no 2 pp 95 – 114.

6803. Pugachev V S 1982a Generalization of theory of conditional estimation and extrapolation *DAN USSR* vol **262** no 3 pp 535 - 538.

*6804.* Pugachev V S 1982b Conditionally optimal estimation in stochastic differential systems *Automatics* vol **118** no 6 pp 685 – 696.

*6805.* Pugachev V S 1984 Conditionally optimal filtering and extrapolation of continuous processes *Automatics and Tele-mechanics* no 2 pp 82 – 89.

*6806.* Pugachev V S 1985 Conditionally optimal estimation in systems with randomly varying structure *Proceedings of the 9$^{th}$ World Congress of IFAC Budapest Hungary July 2-6, 1984* vol **2** pp 773 – 777 *Pergamon Press* Oxford UK.

*6807.* Pugachev V S 1986 Approximate methods for findings finite-dimensional distributions of random sequences determined by difference equations *Problems of Control and Theory of Information* vol **15** no 2 pp 101 – 109.

*6808.* Pugachev V S, Sinitsyn I N 1986 Directions of development of mathematical support for stochastic systems research in Modern informatics techniques *Nauka* Moscow Russian Federation pp 166 – 174.

*6809.* Pugachev V S, Sinitsyn I N, Shin V I 1986a Conditionally optimal discrete filtering of processes in continuous - discrete systems *DAN USSR* vol **289** no 2 pp 297 – 301.

*6810.* Pugachev V S, Sinitsyn I N, Shin V I 1986b Problems of analysis and on-line conditionally optimal filtering of processes in nonlinear stochastic systems *Preprints of the 2$^{nd}$ IFAC Symposium on Stochastic Control Vilnius USSR* May 19-23, 1986 no 1 pp 4 – 18.

*6811.* Pugachev V S, Sinitsyn I N, Shin V I 1987a On one program realization of method of normal approximation in problems of analysis of nonlinear stochastic differential systems *in* Computers for numerous applications *Nauka* Moscow Russian Federation pp 55 – 60.

*6812.* Pugachev V S, Sinitsyn I N, Shin V I 1987b Program realization of method of normal approximation in problems of analysis of nonlinear stochastic systems *Automatics and Tele-mechanics* no 2 pp 62 – 68.

*6813.* Pugachev V S, Sinitsyn I N, Shin V I 1987c Problems of analysis and conditionally optimal filtering in real time scale processes in nonlinear stochastic systems (review) *Automatics and Tele-mechanics* no 12 pp 3 – 24.

*6814.* Pugachev V S, Sinitsyn I N (editors) 1989 Principles of development of dialogue packets of applied programs for research of linear and nonlinear stochastic differential systems. Software pack "StS Analysis" version **1** *Pre-print Institute of Informatics Problems Academy of Sciences USSR* Moscow Russian Federation.

*6815.* Pugachev V S, Sinitsyn I N 1990, 2004 Stochastic differential systems: Analysis and filtering *Nauka* Moscow Russian Federation pp 1 – 642, *Logos* Moscow Russian Federation ISBN 5-94010-199-2 pp 1 – 1000.

*6816.* Pugachev V S, Sinitsyn I N 1999 Lectures on functional analysis and applications *World Scientific* Singapore ISBN 9810237227 ISBN 9810237235 pp 1 – 730.

*6817.* Shannon C E July 1948 A mathematical theory of communication *Bell System Technical Journal* **27** (3) pp 379 – 423 DOI: 10.1002/j.1538-7305.1948.tb01338.x , **27** (4) pp 623 – 666 DOI: 10.1002/j.1538-7305.1948.tb00917x .

*6818.* Shannon C E January 1949 Communication in presence of noise *Proceedings of the Institute of Radio Engineers* **37** (1) pp 10 – 21 DOI:10.1109/jrproc.1949.232969 .

6819. Bode H W, Shannon C E 1950 A simplified derivation of linear least-squares smoothing and prediction theory *Proceedings IRE* vol **38** pp 417 – 425.

6820. Zadeh L A, Ragazzini J R 1950 An extension of Wiener's theory of prediction *Journal of Applied Physics* vol **21** pp 645 – 655.

6821. Booton R C 1952 An optimization theory for time-varying linear systems with nonstationary statistical inputs *Proceedings IRE* vol **40** pp 977 – 981.

6822. Davis R C 1952 On the theory of prediction of nonstationary stochastic processes *Journal of Applied Physics* vol **23** pp 1047 – 1053.

6823. Bartlett M S 1954 Problemes de l'analyse spectral des series temporelles stationnaires *Publ Inst Statist University Paris III–3* pp 119 – 134.

6824. Doob J L 1955 Stochastic processes *John Wiley & Sons Inc* New York N Y USA.

6825. Franklin G 1955 The optimum synthesis of sampled-data systems *Ph D Thesis* Department of Electrical Engineering Columbia University New York USA.

6826. Laning J H, Battin R H 1956 Random processes in automatic control *McGraw-Hill Book Company Inc* New York NY, USA.

6827. Lees A B 1956 Interpolation and extrapolation of sampled data *Trans IRE Prof Group on Information Theory* **IT-2** 1956 pp 173 – 175.

6828. Solodovnikov V V, Batkov A M 1956 On the theory of self-optimizing systems *Proc Heidelberg Conference on Automatic Control* pp 308 – 323.

6829. Newton G C Jr, Gould L A, Kaiser J F 1957 Analytical design of linear feedback controls *John Wiley & Sons Inc* New York USA.

6830. Tukey J W 1957 On the comparative anatomy of transformations *Annals of Mathematical Statistics* **28** pp 602 – 632.

6831. Rytov S M 1957 Development of theory of nonlinear oscillations in the USSR *Radio-Technique and Electronics* no 11 pp 1435 – 1450.

6832. Cramer H 1957 Mathematical methods of statistics *Princeton University Press* Princeton NJ USA.

6833. Bellman R E, Glicksberg I, Gross O A 1958 Some aspects of the mathematical theory of control processes *RAND Report R-313* pp 1 – 244.

6834. Blum M 1958 Recursion formulas for growing memory digital filters *Trans IRE Prof Group on Information Theory* **IT-4** pp 24 – 30.

6835. Darlington S 1958 Linear least-squares smoothing and prediction with applications *Bell System Tech Journal* vol **37** pp 1221 – 1294.

6836. Davenport W B Jr, Root W L 1958 An introduction to the theory of random signals and noise *McGraw-Hill Book Company Inc* New York NY USA.

6837. Sherman S 1958 Non-mean-square error criteria *Trans IRE Prof Group on Information Theory* **IT-4** pp 125 – 126.

6838. Shinbrot M 1958 Optimization of time-varying linear systems with nonstationary inputs *Trans ASME* vol **80** pp 457 – 462.

6839. Smith O J M 1958 Feedback control systems *McGraw-Hill Book Company Inc* New York USA.

6840. Kalman R E, Koepcke R W 1958 Optimal synthesis of linear sampling control systems using generalized performance indexes *Transactions of the ASME* vol **80** pp 1820 – 1826.

*6841.* Kalman R E, Koepcke R W 1959 The role of digital computers in the dynamic optimization of chemical reactors *Proceedings of the Western Joint Computer Conference* pp 107 – 116.

*6842.* Kalman R E, Bertram J E 1958 General synthesis procedure for computer control of single and multi-loop linear systems *Transactions of the AIEE* vol **77** II pp 602 – 609.

*6843.* Kalman R E, Bertram J E 1959 A unified approach to the theory of sampling systems *Journal of the Franklin Institute* vol **267** pp 405 – 436.

*6844.* Kalman R E 1960a On the general theory of control systems *Proceedings of the First International Conference on Automatic Control* Moscow USSR.

*6845.* Kalman R E 1960b A new approach to linear filtering and prediction problems *Journal of Basic Engineering Transactions ASME Series D* **82** pp 35 – 45; **59** pp 1551 – 1580.

*6846.* Kalman R E, Bucy R S 1961 New results in linear filtering and prediction theory *Journal of Basic Engineering Transactions ASME Series D* **83** pp 95 – 108.

*6847.* Kalman R E 1963 New methods in Wiener filtering theory *in* Proceedings of the First Symposium of Engineering Applications of Random Function Theory and Probability Bogdanoff J L, Kozin F (editors) *John Wiley and Sons Inc* New York USA pp 270 – 388.

*6848.* Merriam C W III 1959 A class of optimum control systems *Journal of the Franklin Institute* vol **267** pp 267 – 281.

*6849.* Stratonovich R L 1958 On a method of calculating quantum distribution functions *Soviet Physics Doklady* Band 2 pp 41.

*6850.* Stratonovich R L 1959a Optimum nonlinear systems which bring about a separation of a signal with constant parameters from noise *Radiofizika* **2** (6) pp 892 – 901.

*6851.* Stratonovich R L 1959b On the theory of optimal non-linear filtering of random functions *Theory of Probability and its Applications* **4** pp 223 – 225.

*6852.* Stratonovich R L 1960a Application of the Markov processes theory to optimal filtering *Radio Engineering and Electronic Physics* **5** (11) pp 1 – 19.

*6853.* Stratonovich R L 1960b Conditional Markov processes *Theory of Probability and its Applications* **5** pp 156 – 178.

*6854.* Stratonovich R L 1961 Selected problems in theory of oscillations in radio-technique *Soviet Radio* Moscow Russian Federation pp 1 – 558.

*6855.* Stratonovich R L 1964 New form of formulation of stochastic integrals and equations *Moscow State University Bulletin* Series 1 Mathematics and Mechanics Moscow Russian Federation.

*6856.* Stratonovich R L 1965 On value of information *Izvestiya of USSR Academy of Sciences: Technical Cybernetics* **5** pp 3 – 12.

*6857.* Stratonovich R L, Kuznetsov P I, Tikhonov V I 1965 Nonlinear transformation of stochastic processes *Pergamon Press* USA pp 1 – 500.

*6858.* Stratonovich R L 1966, 1968 Conditional Markov processes and their application in theory of optimal control *Moscow State University Publishing House* Moscow Russian Federation pp 1 – 319, Elsevier, The Netherlands.

*6859.* Stratonovich R L 1967a Topics in the theory of random noise, Vol **1** *Gordon and Breach* ISBN 0677007906, ISBN 9780677007908 pp 1 – 297.

**6860.** Stratonovich R L 1967b Topics in the theory of random noise, Vol **2** *CRC Press* ISBN 0677007906 ISBN 9780677007908 pp 1 – 330.

**6861.** Stratonovich R L 1975 Theory of information *Soviet Radio* Moscow Russian Federation pp 1 – 424.

**6862.** Bunkin F V et al 1997 In memory of Ruslan Leont'evich Stratonovich *Uspekhi Fizicheskih Nauk (UFN)* **40** pp 751 – 752.

**6863.** Romanovski Yu M 2007 Professor R L Stratonovich: Reminiscences of relatives, colleagues and friends *Publishing House of Computer Research Institute* Moscow-Izhevsk ISBN 978-5-93972-606-1 pp 1 – 174.

**6864.** Volterra V 1959 Theory of functionals and integral and integro-differential equations *Dover Publications Inc* New York USA.

**6865.** Middleton D 1960 An introduction to statistical communication theory *McGraw - Hill* New York USA.

**6866.** US Air Forces Office of Scientific Research 1960 – 2016 Full digital collection of technical research reports completed under US AFOSR contracts in 1960 – 2016 *US Air Forces Office of Scientific Research (US AFOSR)* Arlington DC USA.

**6867.** Friedman M 1962 The interpolation of time series by related series *Journal of the American Statistical Association* **57** pp 729 – 757.

**6868.** Kushner H J 1964a On the differential equations satisfied by the conditional densities of Markov processes with applications *Journal SIAM Control* Ser A vol **2** pp 106 – 119.

**6869.** Kushner H J 1964b On the dynamical equations of conditional probability density functions with applications to optimal stochastic control theory *Journal Math Anal Appl* vol **8** pp 332 – 334.

**6870.** Kushner H J 1967a Dynamical equations for optimal nonlinear filtering *Journal Differential Equations* vol **3** pp 179 – 190.

**6871.** Kushner H J 1967b Approximations to optimal nonlinear filters *IEEE Transactions on Automatic Control* vol **12**.

**6872.** Kushner H J, Budhiraja A S 2000 A nonlinear filtering algorithm based on an approximation of the conditional distribution *IEEE Transactions on Automatic Control* vol **45** no 3.

**6873.** Busy R S 1967 Optimal filtering for correlated noise *Journal of Mathematical Analysis and Applications* vol **20** no 1 pp 1 – 8.

**6874.** Fisher I R 1967 Optimal nonlinear filtering *in* Advances in control systems. Theory and application Leondes C T (editor) vol **5** pp 199 – 300 *Academic Press* New York London USA UK.

**6875.** Liptser R Sh, Shiryaev A N 1968 Nonlinear filtering of diffusive Markov processes *Steklov Institute Research Works Academy of Sciences USSR* vol **104** pp 135 – 180.

**6876.** Liptser R Sh, Shiryaev A N 1974 Statistics of random processes *Nauka* Moscow Russian Federation.

**6877.** Bryson A E, Ho Y C 1969 Applied optimal control: Optimization, estimation, and control *Blaisdell Publishing* Waltham Massachusetts USA.

**6878.** Jazwinski A H 1970 Stochastic processes and filtering theory *Academic Press* New York USA.

*6879.* Sorenson H W 1970 Least-squares estimation: From Gauss to Kalman *IEEE Spectrum* vol **7** pp 63 – 68.

*6880.* Box G E P, Jenkins G M 1970, 1976 Time series analysis, forecasting and control *Holden-Day* San Francisco California USA.

*6881.* Box G E P, Jenkins G M, Reinsel G C 1994 Time series analysis, forecasting and control *Prentice Hall International* Englewood Cliffs New Jersey USA.

*6882.* Box G E P, Hillmer S C, Tiao G C 1978 Analysis and modelling of seasonal time series *in* Seasonal analysis of economic time series Zellner A (editor) *US Dept of Commerce-Bureau of the Census* Washington DC USA pp 309 – 334.

*6883.* Bucy R S, Joseph P D 1970 Filtering for stochastic processes with applications to guidance *John Wiley & Sons Inc* New York USA.

*6884.* Wright-Patterson Air Forces Base (AFB) 1970 – 2018 Full extended complemented digital collection of technical research reports and research seminars minutes *Wright-Patterson Air Forces Base (AFB)* Ohio USA.

*6885.* Chow G C, Lin A 1971 Best linear unbiased interpolation, distribution, and extrapolation of time series by related series *Review of Economics and Statistics* **53** pp 372 – 375.

*6886.* Chow G C, Lin A 1976 Best linear unbiased estimation of missing observations in an economic time series *Journal of the American Statistical Association* **71** pp 719 – 721.

*6887.* Chow Y S, Teicher H 1978 Probability theory: Independence, interchangeability, Martingales *Springer-Verlag* New York USA.

*6888.* Maybeck P S 1972 The Kalman filter—An introduction for potential users ***TM-72-3 Air Force Flight Dynamics Laboratory*** Wright-Patterson Air Forces Base (AFB) Ohio USA.

*6889.* Maybeck P S 1974 Applied optimal estimation - Kalman filter design and implementation *Air Force Institute of Technology* Wright-Patterson Air Forces Base (AFB) Ohio USA.

*6890.* Maybeck P S 1990 The Kalman filter: An introduction to concepts *Autonomous Robot Vehicles* editors I J Cox and G T Wilfong *Springer-Verlag* New York USA pp 194 – 204.

*6891.* Willner D 1973 Observation and control of partially unknown systems *Ph D Thesis Department of Electrical Engineering Massachusetts Institute of Technology* USA.

*6892.* Leondes C T, Pearson J O 1973 Kalman filtering of systems with parameter uncertainties: A survey *International Journal of Control* vol **17** no 4 pp 785 – 801.

*6893.* Akaike H 1974 A New look at the statistical model identification *IEEE Transactions on Automatic Control* **AC-19** pp 716 – 723.

*6894.* Athans M 1974 The importance of Kalman filtering methods for economics *Annals of Economic and Social Measurement* vol **3** no 1 pp 49 – 64.

*6895.* Harrison P J, Stevens C F 1976 Bayesian forecasting *Journal Royal Stat Soc Series B* **38** pp 205 – 247.

*6896.* Dempster A P, Laird N M, Rubin D B 1977 Maximum likelihood estimation from incomplete data *Journal of the Royal Statistical Society* **14** pp 1 – 38.

**6897.** Griffiths L J 1977 A continuously adaptive filter implemented as a lattice structure *Proceedings of IEEE International Conference on Acoustics, Speech, and Signal Processing* Hartford CT USA pp 683 – 686.

**6898.** Schwarz G 1978 Estimating the dimension of a model *Annals of Statistics* **6** pp 147 – 164.

**6899.** Falconer D D, Ljung L 1978 Application of fast Kalman estimation to adaptive equalization *IEEE Transactions Comm* vol **COM-26** pp 1439 – 1446.

**6900.** Anderson B D O, Moore J B 1979 Optimal filtering *Prentice-Hall* Englewood Cliffs NJ USA.

**6901.** Bozic S M 1979 Digital and Kalman filtering *Edward Arnold* London USA.

**6902.** Julier S J, Uhlmann J K 1997 A new extension of the Kalman filter to nonlinear systems *Proceedings of Aero-Sense: The 11th International Symposium on Aerospace/Defense Sensing, Simulation and Controls.*

**6903.** Priestley M B 1981 Spectral Analysis and Time Series *John Wiley and Sons Inc* USA.

**6904.** Geweke J F, Singleton K J 1981 Maximum likelihood confirmatory factor analysis of economic time series *International Economic Review* **22** p 1980.

**6905.** Fernandez R B 1981 A methodological note on the estimation of time series *Review of Economics and Statistics* **63** pp 471 – 476.

**6906.** Whittle P 1983 Prediction and regulation *English Universities Press* London UK.

**6907.** Gersch W, Kitagawa G 1983 The prediction of time series with trends and seasonalities *Journal of Buss and Econ Statist* vol **1** no3 pp 253 – 264.

**6908.** Gersch W, Kitagawa G 1984 A smoothness priors state space modelling of time series with trend and seasonality *Journal of the American Statist Assoc* **79** pp 378 – 389.

**6909.** Litterman R B 1983 A random walk, Markov model for the distribution of time series *Journal of Business and Economic Statistics* **1** pp 169 – 173.

**6910.** Meinhold R J, Singpurwalla N D 1983 Understanding the Kalman filter *The American Statistician* **37** (2) pp 123 – 127.

**6911.** Ahlbehrendt N, Kempe V 1984 Analyse stochastischer systeme *Academie-Verlag* Berlin Germany.

**6912.** Bell W R 1984 Signal extraction for nonstationary time series *Annals of Statistics* **12** pp 646 – 664.

**6913.** Harvey A C, Pierse R G 1984 Estimating missing observations in economic time series *Journal of the American Statistical Association* **79** pp 125 – 131.

**6914.** Harvey A C 1987, 1994 Applications of the Kalman filter in econometrics *in* Advances in econometrics: Fifth World congress Bewley T F (editor) vol **1** *Econometric Society Monograph no 13 Cambridge University Press* Cambridge UK, Truman, Bewley (editor) *Cambridge University Press* Cambridge UK pp 1 – 285 ISBN 0-521-46726-8.

**6915.** Harvey A C 1989 Forecasting, structural time series models and the Kalman filter *Cambridge University Press* Cambridge UK.

**6916.** Lewis F 1986 Optimal estimation *John Wiley & Sons Inc* USA.

*6917.* Watson M W 1986 Univariate de-trending methods with stochastic trends *Journal of Monetary Economics* **18** pp 49 – 75.

*6918.* Lanning S G 1986 Missing observations: A simultaneous approach versus interpolation by related series *Journal of Economic and Social Measurement* **14** pp 155 – 163.

*6919.* Wolff Ch C P 1987 Forward foreign exchange rates, expected spot rates, and premia: A signal-extraction approach *Journal of Finance* **42** (2) pp 395 – 406.

*6920.* Burridge P, Wallis K F 1988 Prediction theory for autoregressive-moving average processes *Econometric Reviews* **7** pp 65 – 69.

*6921.* Proakis J G, Manolakis D G 1988 Introduction to digital signal processing *Macmillan* New York USA.

*6922.* Caines P E 1988 Linear stochastic systems *Wiley Series in Probability and Mathematical Statistics John Wiley & Sons* New York USA.

*6923.* De Jong P 1988 The likelihood for a state space model *Biometrika* **75** pp 165 – 169.

*6924.* De Jong P 1989 Smoothing and interpolation with the state space model *Journal of the American Statistical Association* **84** pp 1085 – 1088.

*6925.* De Jong P 1991 The diffuse Kalman filter *Annals of Statistics* **19** pp 1073 – 1083.

*6926.* De Jong P, Chu-Chun-Lin S 1994 Fast likelihood evaluation and prediction for nonstationary state space models *Biometrika* **81** pp 133 – 142.

*6927.* De Jong P, Penzer J 2004 The ARMA model in state space form *Statistics and Probability Letters* **70** pp 119 – 125.

*6928.* Young P C 1988 Recursive extrapolation, interpolation and smoothing of nonstationary time series *in* Identification and system parameter estimation Chen C F (editor) *Pergamon Press* Oxford UK pp 33 – 44.

*6929.* Young P C 1993 Time variable and state dependent modelling of nonstationary and nonlinear time series *in* Developments in time series, volume in honour of Maurice Priestley Subba Rao T (editor) *Chapman and Hall* London UK.

*6930.* Young P C 1994 Time-variable parameter and trend estimation in non-stationary economic time series *Journal of Forecasting* **13** pp 179 – 210.

*6931.* Young P C 1998 Data-based mechanistic modelling of engineering systems *Journal of Vibration and Control* **4** pp 5 – 28.

*6932.* Young P C 1999 Data-based mechanistic modelling, generalised sensitivity and dominant mode analysis *Computer Physics Communications* **115** pp 1 – 17.

*6933.* Young P C, Pedregal D J 1996 Recursive and En-Block Approaches to Signal Extraction *Journal of Applied Statistics* **26** pp 103 – 128.

*6934.* Young P C, Pedregal D J 1997 Data-based mechanistic modelling in System dynamics in economic and financial models Heij C et al (editors) *John Wiley and Sons Inc* Chichester UK.

*6935.* Young P C, Pedregal D J 1999 Macro-economic relativity: Government Spending, private investment and unemployment in the USA *Structural Change and Economic Dynamics* 10 pp 359 – 380.

*6936.* Young P C, Ng C N, Armitage P 1989 A systems approach to economic forecasting and seasonal adjustment *International Journal on Computers and Mathematics with*

*Applications* issue System Theoretic Methods in Economic Modelling **18** pp 481 – 501.

**6937.** Young P C, Pedregal D J, Tych W 1999 Dynamic harmonic regression *Journal of Forecasting* **18** pp 369 – 394.

**6938.** West M, Harrison J 1989 Bayesian forecasting and dynamic models *Springer-Verlag* New York USA.

**6939.** Franklin G F, Powell J D, Workman M L 1990 Digital control of dynamic systems *2nd edition Addison-Wesley* USA.

**6940.** Brockwell P J, Davis R A 1991 Time series: Theory and methods *Springer* Germany.

**6941.** Jang J-S R 1991 Fuzzy modeling using generalized neural networks and Kalman filter algorithm *Proceedings of the 9th National Conference on Artificial Intelligence (AAAI-91)* pp 762 – 767.

**6942.** Doran E 1992 Constraining Kalman filter and smoothing estimates to satisfy time varying restrictions *Review of Economics and Statistics* **74** pp 568 – 572.

**6943.** Brown R G, Hwang P Y C 1992, 1997 Introduction to random signals and applied Kalman filtering $3^{rd}$ *edition John Wiley and Sons Inc* New York USA.

**6944.** Gordon N J, Salmond D J, Smith A F M 1993 A novel approach to non-linear and non-Gaussian Bayesian state estimation *IEE-Proceedings* **F 140** pp 107 – 113.

**6945.** Tanizaki H 1993 Non-linear filters: Estimation and applications *Lecture Notes in economics and mathematical systems Springer Verlag* Germany.

**6946.** Pinheiro M, Coimbra C 1993 Distribution and extrapolation of a time series by related series using logarithms and smoothing penalties *Economica* **12** pp 359 – 374.

**6947.** Bar-Shalom, Xiao-Rong Li 1993 Estimation and tracking: Principles, techniques and software *Artech House* Boston USA.

**6948.** Farhmeir L, Tutz G 1994 Multivariate statistical modeling based generalized linear models *Springer-Verlag* New-York USA.

**6949.** Grimble M J 1994 Robust industrial control: Optimal design approach for polynomial systems *Prentice Hall* USA.

**6950.** Bomhoff E 1994 Financial forecasting for business and economics *Dryden* London UK.

**6951.** Lee J H, Ricker N L 1994 Extended Kalman filter based nonlinear model predictive control *Ind Eng Chem Res* vol **33** no 6 pp 1530 – 1541.

**6952.** Ricker N L, Lee J H 1995 Nonlinear model predictive control of the Tennessee Eastman challenge process *Computers & Chemical Engineering* vol **19** no 9 pp 961 – 981.

**6953.** Kleeman L 1995 Understanding and applying Kalman filtering *Department of Electrical and Computer Systems Engineering Monash University* Clayton Australia pp 1 – 37.

**6954.** Venegas F, de Alba E, Ordorica M 1995 An economist's guide to the Kalman filter *Estudious Economicos* **10** (2) pp 123 – 145.

**6955.** Golub G H, van Loan C F 1996 Matrix computations $3^{rd}$ *edition The John Hopkins University Press* USA.

**6956.** Hayes M H 1996 Statistical digital signal processing and modeling *John Wiley and Sons Inc* USA.

**6957.** Haykin S 1996 Adaptive filter theory *3rd edition Prentice-Hall* Inc Upper Saddle River New Jersey USA.

**6958.** Haykin S (editor) 2001 Kalman filtering and neural networks *Wiley Inter-Science* USA.

**6959.** Fuller W A 1996 Introduction to statistical time series *John Wiley & Sons Inc* USA.

**6960.** Roncalli Th 1996 TSM - Time series and wavelets for finance *Global Design* Paris France.

**6961.** Wells C 1996 The Kalman filter in finance *Advanced Studies in Theoretical and Applied Econometrics Kluwer Academic Publishers* vol **32** The Netherlands.

**6962.** Hodrick R, Prescott E C 1997 Postwar U.S. business cycle: An empirical investigation, *Journal of Money, Credit and Banking* **29** (1) pp 1 – 16.

**6963.** Krelle W 1997 How to deal with unobservable variables in economics *Discussion Paper no B 414 Bonn University* Germany.

**6964.** Babbs S H, Nowman K B 1999 Kalman filtering of generalized Vasicek term structure models *Journal of Financial and Quantitative Analysis* vol **34** no 1.

**6965.** Kim C J, Nelson C 1999 State-space models with regime-switching *MIT Press* Cambridge MA USA.

**6966.** Mantegna R N, Stanley H E 1999 Introduction to Econophysics *Cambridge University Press* Cambridge UK.

**6967.** Pitt M K, Shephard N 1999 Filtering via simulation: Auxiliary particle filters *Journal of the American Statistical Association* **94** (446) pp 590 – 599.

**6968.** Pollock D S G 1999 A handbook of time-series analysis signal processing and dynamics *Academic Press* London UK.

**6969.** Pollock D S G 2000 Trend estimation and de-trending via rational square wave filters *Journal of Econometrics* 99 pp 317 – 334.

**6970.** Pollock D S G 2003a Sharp Filters for Short Sequences *Journal of Statistical Inference and Planning* **113** pp 663 – 683.

**6971.** Pollock D S G 2003b Recursive estimation in econometrics *Journal of Computational Statistics and Data Analysis* **44** pp 37 – 75.

**6972.** Pollock D S G 2006 Wiener–Kolmogorov filtering, frequency-selective filtering and polynomial regression *Econometric Theory* **23** pp 71 – 83.

**6973.** Shiryaev A N 1999 Essentials of stochastic finance: Facts, models, theory Advanced Series on Statistical Science & Applied Probability vol **3** *World Scientific Publishing Co Pte* Ltd Kruzhilin N (translator) ISBN 981-02-3605-0 Singapore pp 1 – 834.

**6974.** Wanhammar L February 24 1999 DSP integrated circuits *Academic Press* San Diego USA ISBN 13: 978-0-12-734530-2 pp 1– 577.

**6975.** Durbin J, Koopman, S J 2000 Time series analysis of non-Gaussian observations based on state-space models from both classical and Bayesian perspectives *Journal of Royal Statistical Society* Series **B 62** pp 3 – 56.

**6976.** Durbin J, Koopman S J 2001 Time series analysis by state space methods *Oxford University Press* New York USA.

6977. Durbin J, Koopman S J 2002 A simple and efficient simulation smoother for state space time series analysis *Biometrika* **89** pp 603 – 615.

6978. Durbin J, Koopman S J 2012 Time series analysis by state space methods *2<sup>nd</sup> edition Oxford University Press* Oxford UK.

6979. Cuche N A, Hess M K 2000 Estimating monthly GDP in a general Kalman filter framework: Evidence from Switzerland *Economic & Financial Modelling Winter 2000* pp 153 – 193.

6980. Doucet A, de Freitas J F G, Gordon N J 2001 Sequential Monte Carlo methods in practice *Springer-Verlag* New York USA.

6981. Welch G, Bishop G 2001 An introduction to the Kalman filter *Department of Computer Science University of North Carolina at Chapel Hill* Chapel Hill USA.

6982. Arulampalam S, Maskell S, Gordon N J, Clapp T 2002 A tutorial on particle filters for online nonlinear/non-Gaussian Bayesian tracking *IEEE Transaction on Signal Processing* **50** (2) pp 174 – 188.

6983. Javaheri A, Lautier D, Galli A 2002 Filtering in finance *RBC Capital Markets Universit´e Paris IX Ecole Nationale Sup´erieure des Mines de Paris Ecole Nationale Sup´erieure des Mines de Paris* France Filteringinfinance.pdf pp 1 – 26.

6984. Doucet A, Tadic V B 2003 Parameter estimation in general state-space models using particle methods *Annals of the Institute of Statistical Mathematics* **55** (2) pp 409 – 422.

6985. Bahmani O, Brown F 2004 Kalman filter approach to estimate the demand for international reserves *Applied Economics* **36** (15) pp 1655 – 1668.

6986. Broto C, Ruiz E 2004 Estimation methods for stochastic volatility models: A survey, *Journal of Economic Surveys* **18** (5) pp 613 – 637.

6987. Ristic B, Arulampalam S, Gordon N J 2004 Beyond the Kalman Filter: Particle filters for tracking applications *1st edition Artech House* Boston USA.

6988. Cappé O, Moulines E 2005 On the use of particle filtering for maximum likelihood parameter estimation in *European Signal Processing Conference* Antalya Turkey.

6989. Ozbek L, Ozale U 2005 Employing the extended Kalman filter in measuring the output gap *Journal of Economic Dynamics and Control* **29** pp 1611 – 1622.

6990. Poyiadjis G, Doucet A, Singh S S 2005a Particle methods for optimal filter derivative: application to parameter estimation in *Proceedings IEEE International Conference on Acoustics, Speech, and Signal Processing*.

6991. Poyiadjis G, Doucet A, Singh S S 2005b Maximum likelihood parameter estimation in general state-space models using particle methods in *Proceedings of the American Statistical Association JSM 05*.

6992. Proietti T 2006 Trend–cycle decompositions with correlated components *Econometric Reviews* **25** pp 61 – 84.

6993. Litvin A, Konrad J Karl W C 2003 Probabilistic video stabilization using Kalman filtering and mosaicking *IS&T/SPIE Symposium on Electronic Imaging, Image and Video Communications and Proc*.

6994. Van Willigenburg L G, De Koning W L 2004 UDU factored discrete-time Lyapunov recursions solve optimal reduced-order LQG problems *European Journal of Control* **10** pp 588 – 601.

**6995.** Voss H U, Timmer J, Kurths J 2004 Nonlinear dynamical system identification from uncertain and indirect measurements *International Journal Bifurcation and Chaos* **14** pp 1905 – 1933.

**6996.** Capp´e O, Moulines E, Ryd´en T 2005 Inference in hidden Markov models *Springer Series in Statistics Springer* New York USA.

**6997.** Fernàndez-Villaverde J, Rubio-Ramirez J F 2005 Estimating dynamic equilibrium economies: Linear versus non-linear likelihood *Journal of Applied Econometrics* 20 891910.

**6998.** Fernàndez-Villaverde J, Rubio-Ramrez J F 2007 Estimating macroeconomic models: A likelihood approach *Review of Economic Studies* **74** pp 1059 – 1087.

**6999.** Fernàndez-Villaverde J 2010 The econometrics of DSGE models *Journal of the Spanish Economic Association* **1** pp 3 – 49.

**7000.** Frühwirth-Schnatter S 2006 Finite mixture and Markov switching models *Springer Series in Statistics Springer* New York USA.

**7001.** Pasricha G K 2006 Kalman filter and its economic applications *MPRA Paper no 22734 Munich University Munich Germany* pp 1 – 14 http://mpra.ub.uni-muenchen.de/22734/ .

**7002.** Misra P, Enge P 2006 Global Positioning System signals, measurements, and performance *2^{nd} edition* USA.

**7003.** Gamerman D, Lopes H F 2006 Markov chain Monte Carlo. Stochastic simulation for Bayesian inference *2^{nd} edition Chapman & Hall* London UK.

**7004.** Pasricha G K 2006 Kalman filter and its economic applications *MPRA Paper no 22734 Munich University Munich Germany* pp 1 – 14

http://mpra.ub.uni-muenchen.de/22734/ .

**7005.** Rajamani M R 2007 Data-based techniques to improve state estimation in model predictive control *PhD Thesis* University of Wisconsin-Madison USA.

**7006.** Bignasca F, Rossi E 2007 Applying the Hirose-Kamada filter to Swiss data: Output gap and exchange rate pass-through estimates *Swiss National Bank working Papers 2007 – 10* Swiss National Bank Switzerland ISSN 1660-7716 pp 1 – 27.

**7007.** Andreasen M M 2008 Non-linear DSGE models, the central difference Kalman filter, and the mean shifted particle filter *CREATES Research Paper 2008-33* School of Economics and Management University of Aarhus Denmark pp 1 – 46.

**7008.** Olsson J, Cappé O, Douc R, Moulines E 2008 Sequential Monte Carlo smoothing with application to parameter estimation in nonlinear state space models *Bernoulli* **14** (1) pp 155 – 179.

**7009.** Roncalli T, Weisang G 2008 Tracking problems, hedge fund replication and alternative beta *MPRA Paper no 37358 Munich University Munich Germany* http://mpra.ub.unimuenchen.de/37358/ .

**7010.** Rajamani M R, Rawlings J B 2009 Estimation of the disturbance structure from data using semidefinite programming and optimal weighting *Automatica* **45** pp 142 – 148.

**7011.** Bationo R, Hounkpodote H 2009 Estimated changes in prices of coffee and cocoa: Kalman filter, Hodrick-Prescott filter and modeling from Markov switching *MPRA Paper no 26980 Munich University Munich Germany* pp 1 – 22

http://mpra.ub.unimuenchen.de/26980/ .

7012. Chang Y, Miller J I, Park J Y 2009 Extracting a common stochastic trend: Theory with some applications *Journal of Econometrics* **15** pp 231 – 247.

7013. Mapa D S, Sandoval M F B, Yap J E B 2009 Investigating the presence of regional economic growth convergence in the Philippines using Kalman filter MPRA *Paper no 20681 Munich University Munich Germany*
http://mpra.ub.uni-muenchen.de/20681/ .

7014. Yakovenko V M, Rosser J B 2009 Colloquium: Statistical mechanics of money, wealth, and income *Cornell University* NY USA arXiv:0905.1518v1 [q-fin.ST].

7015. Chakrabarti B K, Chakrabarti A 2010 Fifteen years of econophysics research *Science and Culture* vol **76** nos 9-10 pp 293-295; *Cornell University* NY USA arXiv:1010.3401 [q-fin.ST].

7016. Francke M K , Koopman S J, de Vos A 2010 Likelihood functions for state space models with diffuse initial conditions *Journal of Time Series Analysis* **31** pp 407 – 414.

7017. Luati A, Proietti T 2010 Hyper-spherical and elliptical stochastic cycles *Journal of Time Series Analysis* **31** pp 169 – 181.

7018. Sinha S, Chatterje A, Chakraborti A, Chakrabarti B K 2010 Econophysics: An Introduction *Wiley-VCH* Berlin.

7019. Theoret R, and Racicot F - E 2010 Forecasting stochastic volatility using the Kalman filter: an application to Canadian interest rates and price-earnings ratio *MPRA Paper no 35911 Munich University Munich Germany*
http://mpra.ub.uni-muenchen.de/35911/ .

7020. Winschel W, Kratzig M 2010 Solving, estimating, and selecting nonlinear dynamic models without the curse of dimensionality *Econometrica* **39** (1) pp 3 – 33.

7021. Xia Y, Tong H 2011 Feature matching in time series modeling *Statistical Science* **26** (1) pp 21 – 46.

7022. Jungbacker B, Koopman S J, van der Wel M 2011 Maximum likelihood estimation for dynamic factor models with missing data *Journal of Economic Dynamics and Control* **35** (8) pp 1358 – 1368.

7023. Moghaddam B A, Haleh H, Ebrahimijam S 2011 Forecasting trend and stock price with adaptive extended Kalman filter data *2011 International Conference on Economics and Finance Research* IPEDR vol **4** *IACSIT Press* Singapore.

7024. Aoyama H, Fujiwara Y, Iyetomi H, Sato A-H 2012 Preface *Progress of Theoretical Physics* Supplement No 194 pp i-ii.

7025. Darvas Z, Varga B 2012 Uncovering time-varying parameters with the Kalman-filter and the flexible least squares: A Monte Carlo study *Working Paper 2012 / 4 Department of Mathematical Economics and Economic Analysis Corvinus University of Budapest Hungary* pp 1 – 19.

7026. Hang Qian 2012 A flexible state space model and its applications *MPRA Paper No 38455 Munich University Munich Germany* pp 1 - 27 http://mpra.ub.uni-muenchen.de/38455/ .

7027. Proietti T, Luati A 2012a A maximum likelihood estimation of time series models: the Kalman filter and beyond *MPRA Paper no 41981 Munich University Munich German* pp 1 – 30

http://mpra.ub.uni-muenchen.de/41981/ .

7028. Proietti T, Luati A 2012b The generalised autocovariance function *MPRA Paper no 43711 Munich University Munich Germany* pp 1 – 30

http://mpra.ub.unimuenchen.de/43711/.

7029. Creal D 2012 A survey of sequential Monte Carlo methods for economics and finance *Econometric Reviews* vol **31** 3 pp 245 – 296.

7030. Matisko P, Havlena V 2012 Optimality tests and adaptive Kalman filter *Proceedings of 16th IFAC System Identification Symposium* Brussels Belgium.

7031. Durbin J, Koopman S J 2012 Time series analysis by state space methods *2nd edition Oxford University Press* Oxford UK.

7032. Wikipedia 2016 Kalman filter *Wikipedia Foundation, Inc.*

***Information communication with discrete-time digital signals, discrete-time digital signals spectrum, discrete-time digital signal processing in physics and digital electronics engineering:***

7033. Gabor D 1946 Theory of communication Part 1 The analysis of information *J Inst Elect Eng* **93** pp 429 – 441.

7034. Shannon C E 1948 A mathematical theory of communication *Bell System Technical Journal* vol **27** pp 379 – 423, 623 – 656

http://cm.bell-labs.com/cm/ms/what/shannonday/paper.html .

7035. Bose R C, Shrikhande S S 1959 A note on a result in the theory of code construction *Information and Control* **2** (2) pp 183 – 194 doi:10.1016/S0019-9958(59)90376-6 CiteSeerX: 10.1.1.154.2879

http://dx.doi.org/10.1016%2FS0019-9958%2859%2990376-6

http://citeseerx.ist.psu.edu/viewdoc/summary?doi=10.1.1.154.2879 .

7036. Granger C W J, Hatanaka M 1964 Spectral analysis of economic time series *Princeton University Press* Princeton USA.

7037. Yuen C-K 1972 Remarks on the ordering of Walsh functions *IEEE Transactions on Computers* **21** (12) p 1452 doi:10.1109/T-C.1972.223524

http://dx.doi.org/10.1109%2FT-C.1972.223524 .

7038. Hwang K, Briggs F A 1984 Computer architecture and parallel processing *McGraw-Hill* New York USA.

7039. Orfanidis S J 1985 Optimum signal processing: An introduction 2nd edition *Macmillan* New York USA.

7040. Orfanidis S J 1995 Introduction to signal processing *Prentice-Hall* Englewood Cliffs NJ USA.

7041. Anceau F 1986 The architectures of microprocessors *Addison-Wesley* Wokingham England.

7042. Fountain T 1987 Processor arrays, architecture and applications *Academic Press* London UK.

7043. Chen C H (editor) 1988 Signal processing handbook *Marcel Dekker* New York USA.

7044. Kay S M 1988 Modern spectral estimation: Theory and application *Prentice-Hall* Englewood Cliffs NJ USA.

**7045.** Oppenheim A V, Schafer R W 1989 Discrete-time signal processing *Prentice-Hall* Englewood Cliffs NJ USA.

**7046.** Van de Goor A J 1989 Computer architecture and design *Addison-Wesley* Wokingham England.

**7047.** Priemer R 1991 Introductory signal processing *World Scientific* Singapore ISBN 9971509199.

**7048.** Jeruchim M C, Balaban Ph, Shanmugan K S 1992 Simulation of communication systems *Plenum Press* New York USA.

**7049.** Witte R A 1993, 2001 Spectrum and network measurements $1^{st}$ edition *Prentice Hall Inc* Upper Saddle River NJ USA, $2^{nd}$ edition *Noble Pub Corp* Atlanta GA USA ISBN 10 1884932169 LC TK7879.4.W58 2001 pp 1 – 297.

**7050.** Hsu P H 1995 Schaum's theory and problems: Signals and systems *McGraw-Hill* ISBN 0-07-030641-9.

**7051.** Simon M K, Hinedi S M, Lindsey W C 1995 Digital communication techniques – Signal design and detection *Prentice-Hall* Englewood Cliffs NJ USA.

**7052.** Simon M K, Alouini M S 2000 Digital communication over fading channels – A unified approach to performance analysis $1^{st}$ edition *John Wiley and Sons Inc* USA.

**7053.** Proakis J G, Manolakis D G 1996 Digital signal processing $3^{rd}$ edition *Prentice Hall* Upper Saddle River NJ USA.

**7054.** Proakis J G 2001 Digital communications $4^{th}$ edition *McGraw-Hill* USA.

**7055.** Lathi B P 1998 Signal processing and linear systems *Berkeley-Cambridge Press* ISBN 0-941413-35-7.

**7056.** Prisch P 1998 Architectures for digital signal processing *John Wiley and Sons Inc* Chichester UK.

**7057.** Gershenfeld N A 1999 The nature of mathematical modeling *Cambridge University Press* UK ISBN 0-521-57095-6.

**7058.** Wanhammar L February 24 1999 DSP integrated circuits *Academic Press* San Diego USA ISBN 13: 978-0-12-734530-2 pp 1 – 577.

**7059.** Sklar B 2001 Digital communications $2^{nd}$ edition *Prentice-Hall* Englewood Cliffs NJ USA.

**7060.** McMahon D 2007 Signals and systems demystified *McGraw Hill* New York USA ISBN 978-0-07-147578-5.

**7061.** Rice M 2008 Digital communications - A discrete-time approach *Prentice Hall* Englewood Cliffs NJ USA.

**7062.** Wikipedia 2015e Signal (electrical engineering) *Wikipedia Inc* USA
www.wikipedia.org .

**7063.** Wikipedia 2015f Continuous wave *Wikipedia Inc* USA
www.wikipedia.org .

**7064.** Wikipedia 2015g Discrete-time signal *Wikipedia Inc* USA
www.wikipedia.org .

**7065.** Wikipedia 2015h Hadamard code *Wikipedia* USA
www.wikipedia.org .

**7066.** Wikipedia 2016i Polarization *Wikipedia* USA
https://en.wikipedia.org/wiki/Polarization_(waves) .

**7067.** Wikipedia 2016j Circular polarization *Wikipedia* USA
https://en.wikipedia.org/wiki/Circular_polarization .

**7068.** Wikipedia 2016k In phase and quadrature components Wikipedia USA
https://en.wikipedia.org/wiki/In_phase_and_quadrature_components .

**7069.** Wikipedia 2016l Constellation diagram Wikipedia USA
https://en.wikipedia.org/wiki/Constellation_diagram .

**7070.** Matlab 2014 IQ diagram *MathSoft* California USA.

**7071.** Ledenyov D O, Ledenyov V O 2015a Nonlinearities in microwave
superconductivity $8^{th}$ edition *Cornell University* NY USA pp 1 – 923
www.arxiv.org 1206.4426v8.pdf .

***Time Division Duplex/Frequency Division Duplex spread spectrum burst
communications, UMTS/WCDMA/LTE wireless communications, WCDMA optical
communications, digital signal processing in telecommunications, electronics
engineering, and physics:***

**7072.** Dixon R C 1976 Spread spectrum systems *Wiley-Blackwell* NY USA pp 1 – 318
ISBN-13: 978-0471216292.

**7073.** Viterbi A J May 1979 Spread spectrum communications - Myths and realities *IEEE
Communications Magazine* pp 11 – 18.

**7074.** Pickholtz R L, Schilling D L, Milstein L B May 1982 Theory of spread spectrum
communication - A tutorial *IEEE Transactions on Communication* vol **COM-30** no 5
pp 855 – 884.

**7075.** Simon M K, Omura J K, Scholtz R A, Levitt B K 1985 Spread spectrum
communications vols **1-3** *Computer Science Press Inc* USA.

**7076.** Simon M K, Omura J K, Scholtz R A, Levitt B K 1994 Spread spectrum
communications handbook *The McGraw-Hill Companies* Inc NY USA ISBN-13: 978-
0070576292.

**7077.** Simon M K, Alouini M S 2000 Digital communication over fading channels – A
unified approach to performance analysis $1^{st}$ edition *John Wiley and Sons Inc* USA.

**7078.** Simon M K May 2006 On the bit-error probability of differentially encoded QPSK
and offset QPSK in the presence of carrier synchronization *IEEE Trans Commun* vol
**54** pp 806 – 812.

**7079.** Simon H, Moher M 2008 Communication systems $5^{th}$ edition *John Wiley & Sons
Inc* USA pp 1 – 440 ISBN-13: 978-0471697909.

**7080.** Ledenyov V O 1998, 1999 Research consulting on spread spectrum systems
*WaveRider* Calgary Alberta Canada.

**7081.** Ledenyov V O 2000, 2001 Research consulting on WCDMA systems *Nortel
Networks* Calgary Alberta Canada.

**7082.** Rappaport T January 2010 Wireless communications principles and practice $2^{nd}$
edition *Prentice-Hall Inc* pp 1 – 736 ISBN-13: 978-0130422323.

**7083.** Miao G, Zander J, Sung K-W, Slimane B 2016 Fundamentals of mobile data
networks *Cambridge University Press* UK pp 1 – 322 ISBN 9781107143210.

***Microprocesssor design, Low Voltage Very Large Scale Integrated Circuit (LVVLSI)
design, Digital Signal Processor (DSP) design, Field Programmable Gate Array
(FPGA) design, Application Specific Integrated Circuit (ASIC) design:***

7084. Wanhammar L February 24 1999 DSP integrated circuits *Academic Press* San Diego USA ISBN 13: 978-0-12-734530-2 pp 1 – 577.

7085. Ball S R Embedded microprocessor systems: Real world design 2nd edition *Newnes Butterworth-Heinemann* ISBN 0-7506-7234-X pp 1 – 327.

7086. Chandrakasan A, Bowhill W J, Fox F (editors) October 2000 Design of high performance microprocessor circuits *Willey IEEE Press* ISBN 978-0780360013 pp 1 – 584.

7087. Harris D, Harris S August 7 2012 Digital design and computer architecture *Morgan Kaufmann* ISBN 978-0123944245 pp 1 – 712.

7088. Hennessy J L, Patterson D A December 17 2017 Computer architecture, Sixth edition: A quantitative approach *Morgan Kaufmann* ISBN: 978-0128119051 pp 1 – 936.

### Quantum physics, electronics engineering, mechanics engineering:

7089. Thomson J J 1897a, 30 April 1897 Cathode rays *The Electrician* **39** p 104, in *Proceedings of the Royal Institution* UK pp 1 – 14.

7090. Thomson J J 1897b Cathode rays *Philosophical Magazine* **44** pp 293.

7091. Thomson J J 1912 Further experiments on positive rays *Philosophical Magazine* **24** pp 209 – 253.

7092. Thomson J J 1913 Rays of positive electricity *Proceedings of the Royal Society A* **89** pp 1 –20.

7093. Thomson J J 1904 On the structure of the atom: An investigation of the stability and periods of oscillation of a number of corpuscles arranged at equal intervals around the circumference of a circle; with application of the results to the theory of atomic structure *Philosophical Magazine Series* 6 vol **7** no 39 pp 237 – 265.

7094. Thomson J J 1923 The electron in chemistry *Franklin Institute* Philadelphia USA.

7095. Planck M 1900a Über eine Verbesserung der Wienschen Spektralgleichung On an improvement of Wien's equation for the spectrum *Verhandlungen der Deutschen Physikalischen Gesellschaft* **2** pp 202 – 204

http://archive.org/stream/verhandlungende01goog#page/n212/mode/2up .

7096. Planck M 1900b Zur Theorie des Gesetzes der Energieverteilung im Normalspektrum *Verhandlungen der Deutschen Physikalischen Gesellschaft* **2** p 237

http://archive.org/stream/verhandlungende01goog#page/n246/mode/2up .

7097. Planck M 1900c Entropie und Temperatur strahlender Wärme Entropy and temperature of radiant heat *Annalen der Physik* **306** (4) pp 719 – 737

http://adsabs.harvard.edu/abs/1900AnP...306..719P ,

https://dx.doi.org/10.1002%2Fandp.19003060410 .

7098. Planck M 1900d Über irreversible Strahlungsvorgänge On irreversible radiation processes *Annalen der Physik* **306** (1) pp 69 – 122

http://adsabs.harvard.edu/abs/1900AnP...306...69P ,

https://dx.doi.org/10.1002%2Fandp.19003060105 .

7099. Planck M 1901 Über das Gesetz der Energieverteilung im Normalspektrum On the law of distribution of energy in the normal spectrum *Annalen der Physik* **309** (3) pp 553 – 563.

http://adsabs.harvard.edu/abs/1901AnP...309..553P ,

https://dx.doi.org/10.1002%2Fandp.19013090310 ,

http://theochem.kuchem.kyoto-u.ac.jp/Ando/planck1901.pdf .

**7100.** Planck M 1903 Treatise on thermodynamics *Longmans, Green & Co* London UK

http://archive.org/stream/treatiseonthermo00planuoft#page/n7/mode/2up ,

http://openlibrary.org/books/OL7246691M .

**7101.** Planck M 1906 Vorlesungen über die Theorie der Wärmestrahlung *JA Barth* Leipzig Germany

http://lccn.loc.gov/07004527 .

**7102.** Planck M 1914 The theory of heat radiation 2$^{nd}$ edition *P Blakiston's Son & Co* http://openlibrary.org/books/OL7154661M .

**7103.** Planck M 1915 Eight lectures on theoretical physics *Dover Publications* ISBN 0-486-69730-4.

**7104.** Planck M 1943 Zur Geschichte der Auffindung des physikalischen Wirkungsquantums *Naturwissenschaften* **31** (14–15) pp 153 – 159

http://adsabs.harvard.edu/abs/1943NW.....31..153P ,

https://dx.doi.org/10.1007%2FBF01475738 .

**7105.** Einstein A 1905 Zur Elektrodynamik bewegter Körper On the electrodynamics of moving bodies *Annalen der Physik* Berlin Germany (in German) **322** (10) pp 891 – 921

http://onlinelibrary.wiley.com/doi/10.1002/andp.19053221004/pdf ,

http://adsabs.harvard.edu/abs/1905AnP...322..891E) ,

http://dx.doi.org/10.1002%2Fandp.19053221004 .

**7106.** Einstein A 1916 Strahlungs-emission und -absorption nach der Quantentheorie *Verhandlungen der Deutschen Physikalischen Gesellschaft* **18** pp 318 – 323 Bibcode:1916DPhyG..18..318E .

**7107.** Einstein A 1917 Zur Quantentheorie der Strahlung On the quantum mechanics of radiation *Physikalische Zeitschrift* (in German) **18** pp 121 – 128 Bibcode:1917PhyZ...18..121E.

http://adsabs.harvard.edu/abs/1917PhyZ...18..121E .

**7108.** Einstein A 1924 Quantentheorie des einatomigen idealen gases Quantum theory of monatomic ideal gases *Sitzungsberichte der Preussischen Akademie der Wissenschaften Physikalisch-Mathematische Klasse* (in German) pp 261 – 267

http://echo.mpiwg-berlin.mpg.de/MPIWG:DRQK5WYB .

**7109.** Einstein A, Podolsky B, Rosen N 1935 Can quantum-mechanical description of physical reality be considered complete? *Physical Review* American Physical Society **47** (10) pp 777 – 780

http://journals.aps.org/pr/pdf/10.1103/PhysRev.47.777 ,

http://adsabs.harvard.edu/abs/1935PhRv...47..777E ,

https://dx.doi.org/10.1103%2FPhysRev.47.777 .

**7110.** Bohr N 1913a On the constitution of atoms and molecules, part I *Philosophical Magazine* **26** (151) pp 1 – 24 DOI:10.1080/14786441308634955.

**7111.** Bohr N 1913b On the constitution of atoms and molecules, part II Systems containing only a single nucleus *Philosophical Magazine* **26** (153) pp 476 – 502. DOI:10.1080/14786441308634993.

7112. Bohr N 1913c On the constitution of atoms and molecules, part III Systems containing several nuclei *Philosophical Magazine* **26** pp 857–875. DOI:10.1080/14786441308635031.

7113. Bohr N 1921 Atomic structure *Nature* **107** (2682) pp 104 – 107 BIBCODE:1921Natur.107..104B. DOI:10.1038/107104a0.

7114. Bohr N 1922 The structure of the atom Nobel prize lecture *in* Niels Bohr A centenary volume French A P, Kennedy P J (editors) *Harvard University Press* Cambridge Massachusetts pp 91 – 97 ISBN 978-0-674-62415-3.

7115. Bohr N, Kramers H A, Slater J C 1924 The quantum theory of radiation *Philosophical Magazine* **6 76** (287) pp 785 – 802

http://www.cond-mat.physik.uni-mainz.de/~oettel/ws10/bks_PhilMag_47_785_1924.pdf, https://dx.doi.org/10.1080%2F14786442408565262 .

7116. Ledenyov V O 1996-1997 Memories about study on Niels Bohr scientific achievements and heritage *Niels Bohr Institute for Theoretical Physics* Copenhagen Denmark.

7117. Ledenyov D O 2017 Memories about visit to *Niels Bohr Institute for Theoretical Physics* Copenhagen Denmark.

7118. De Broglie L 1924, 1925 Recherches sur la théorie des quanta Researches on the quantum theory *Ph D Thesis* Sorbonne Paris France, *Annales de Physique* **10** (3) pp 22 – 128.

7119. De Broglie L 1926 Ondes et mouvements Waves and motions *Gauthier-Villars* Paris France.

7120. De Broglie L 1927 Rapport au 5e Conseil de Physique Solvay Brussels Belgium.

7121. De Broglie L 1928 La mécanique ondulatoire Wave mechanics *Gauthier-Villars* Paris France.

7122. De Broglie L 1929 The wave nature of electron Nobel lecture Stockholm Sweden.

https://www.nobelprize.org/nobel_prizes/physics/laureates/1929/broglie-lecture.pdf .

7123. Compton A 1926 X-Rays and electrons: An outline of recent X-Ray theory *D Van Nostrand Company Inc* New York USA

https://www.worldcat.org/oclc/1871779 .

7124. Compton A; Allison S K 1935 X-Rays in theory and experiment *D Van Nostrand Company Inc* New York USA

https://www.worldcat.org/oclc/853654 .

7125. Schrödinger E 1926 Quantisierung als Eigenwertproblem *Annalen der Phys* **384** (4) pp 273 – 376

http://onlinelibrary.wiley.com/doi/10.1002/andp.19263840404/pdf ,

http://adsabs.harvard.edu/abs/1926AnP...384..361S ,

https://dx.doi.org/10.1002%2Fandp.19263840404 .

7126. Heisenberg W 1927 Über den anschaulichen Inhalt der quantentheoretischen Kinematik und Mechanik *Zeitschrift für Physik* **43** (3–4) pp 172 – 198 Bibcode:1927ZPhy...43..172H, doi:10.1007/BF01397280..

7127. Kennard E H 1927 Zur Quantenmechanik einfacher Bewegungstypen *Zeitschrift für Physik* **44** (4-5) pp 326 – 352.

***7128.*** Fermi E 1934 Radioattività indotta da bombardamento di neutroni La Ricerca scientifica **1** (5) p 283 (in Italian)

http://www.phys.uniroma1.it/DipWeb/museo/collezione%20Fermi/documento2.htm .

***7129.*** Fermi E, Amaldi E, d'Agostino O, Rasetti F, Segre E 1934 Artificial radioactivity produced by neutron bombardment *Proceedings of the Royal Society A: Mathematical, Physical and Engineering Sciences* **146** (857) p 483

http://adsabs.harvard.edu/abs/1934RSPSA.146..483F ,

https://dx.doi.org/10.1098%2Frspa.1934.0168 .

***7130.*** Townes Ch 1939 Concentration of the heavy isotope of carbon and measurement of its nuclear spin *PhD thesis* Caltech California USA

http://thesis.library.caltech.edu/4202/ .

***7131.*** Townes Ch, Schawlow A 1955 Microwave spectroscopy *McGraw-Hill* USA ISBN 978-0-07-065095-4.

***7132.*** Gordon J P, Zeiger H J, Townes Ch H 15 August 1955 The maser — new type of microwave amplifier, frequency standard, and spectrometer *Physical Review* **99** (4) p 1264

https://journals.aps.org/pr/abstract/10.1103/PhysRev.99.1264

http://adsabs.harvard.edu/abs/1955PhRv...99.1264G ,

https://dx.doi.org/10.1103%2FPhysRev.99.1264 .

***7133.*** Shimoda K, Wang T, Townes Ch 1956 Further aspects of the theory of the maser *Physical Review* **102** (5) p 1308

http://adsabs.harvard.edu/abs/1956PhRv..102.1308S ,

https://dx.doi.org/10.1103%2FPhysRev.102.1308 .

***7134.*** Townes Ch H 1964 Nobel Prize in Physics Stockholm Sweden

http://nobelprize.org/nobel_prizes/physics/laureates/1964/townes-bio.html .

***7135.*** Townes Ch H 1966 Obtaining of coherent radiation with help of atoms and molecules *Uspekhi Fizicheskih Nauk (UFN)* vol **88** no 3.

***7136.*** Townes Ch H 1969 Quantum electronics and technical progress *Uspekhi Fizicheskih Nauk (UFN)* vol **98** no 5.

***7137.*** Townes Ch 1995 Making waves *American Institute of Physics Press* New York USA ISBN 978-1-56396-381-0.

***7138.*** Townes Ch 1999 How the laser happened: Adventures of a scientist *Oxford University Press* ISBN 978-0-19-512268-8.

***7139.*** Schiff L I 1949 Quantum mechanics *McGraw - Hill Book Company Inc* New York USA pp 1 – 404.

***7140.*** Blokhintsev D I 1954 Development of first nuclear reactor for nuclear power plant Moscow Russian Federation.

***7141.*** Blokhintsev D I 2004 Foundations of quantum mechanics 7[th] edition *Lan' Publishing House* St Petersburg Russian Federation ISBN 5-8114-0554-5 pp 1 – 664.

***7142.*** Prokhorov A M, Basov N G 1955 Molecular generator and amplifier *Uspekhi Fizicheskih Nauk (UFN)* vol **57** no 3 pp 485 – 501.

***7143.*** Prokhorov A M, Fedorov V B 1963 *Soviet Journal of Experimental and Theoretical Physics JETP* **16** 1489.

***7144.*** Prokhorov A M 1964 Nobel Prize in Physics Stockholm Sweden

http://nobelprize.org/nobel_prizes/physics/laureates/1964/prokhorov-bio.html .

*7145.* Prokhorov A M Quantum electronics 1965 *Uspekhi Fizicheskih Nauk (UFN)* vol **85** no 4 pp 599 – 604.

*7146.* Karlov N V, Prokhorov A M 1976 Laser's separation of isotopes *Uspekhi Fizicheskih Nauk (UFN)* vol **118** no 4 pp 583 – 609.

*7147.* Prokhorov A M 1979 To 25[th] anniversary of laser *Uspekhi Fizicheskih Nauk (UFN)* vol **128** no 3.

*7148.* Prokhorov A M (Editor in Chief), Buzzi J M, Sprangle P, Wille K 1992 Coherent radiation generation and particle acceleration *Research Trends in Physics Series American Institute of Physics Press* New York USA (Springer, Germany) ISBN 0-88318-926-7

http://www.springer-sbm.de/index.php?id=121&L=0 .

*7149.* Bardeen J 1956 Nobel Prize in Physics Stockholm Sweden

http://nobelprize.org/nobel_prizes/physics/laureates/1956/bardeen-bio.html .

*7150.* Bardeen J 1972 Nobel Prize in Physics Stockholm Sweden

http://nobelprize.org/nobel_prizes/physics/laureates/1972/bardeen-bio.html .

*7151.* Bardeen J 1990 Superconductivity and other macroscopic quantum phenomena *Physics Today* **43** (12) pp 25 – 31 doi: 10.1063/1.881218.

*7152.* Schawlow A, Townes Ch 1958 Infrared and optical masers *Physical Review* **112** (6) p 1940

http://dx.doi.org/10.1103%2FPhysRev.112.1940 ,

http://adsabs.harvard.edu/abs/1958PhRv..112.1940S .

*7153.* Schawlow A 1963 Modern optical quantum generators *Uspekhi Fizicheskih Nauk (UFN)* vol **81** no 12.

*7154.* Schawlow A 1964 Nobel Prize in Physics Stockholm Sweden

http://nobelprize.org/nobel_prizes/physics/laureates/1964/schawlow-bio.html .

*7155.* Gould G R 1959 The LASER, Light amplification by stimulated emission of radiation *in* Franken P A, Sands R H (editors) *The Ann Arbor Conference on Optical Pumping* The University of Michigan 15 June - 18 June 1959 p 128 OCLC 02460155

https://www.worldcat.org/oclc/02460155 .

*7156.* Gordon Gould 2018 Wikipedia Inc

https://en.wikipedia.org/wiki/Gordon_Gould .

*7157.* Merzbacher E 1961 Quantum mechanics *John Willey and Sons Inc* New York USA pp 1 – 621.

*7158.* Josephson B D 1962 Possible new effects in superconductive tunneling *Physical Letters* vol **1** p 251.

*7159.* Josephson B D 1964 Coupled superconductors *Review Modern Physics* vol **36** p 216.

*7160.* Josephson B D 1965 Super currents through barriers *Advances in Physics* vol **14** p 419.

*7161.* Basov N G 1964 Nobel Prize in Physics Stockholm Sweden

http://nobelprize.org/nobel_prizes/physics/laureates/1964/basov-bio.html .

*7162.* Basov N G 1965 Semiconductor quantum generators *Uspekhi Fizicheskih Nauk (UFN)* vol **85** no 4.

*7163.* Landau L D, Lifshits E M 1977 Quantum mechanics 3$^{rd}$ edition *Pergamon Press* Oxford UK.

*7164.* Tesche C D, Clarke J 1977 DC SQUID: Noise and optimization *Journal of Low Temperature Physics* **29** pp 301 – 331.

*7165.* Clarke J 1989 Principles and applications of SQUIDs *Proc IEEE* **77** pp 1208 – 1223.

*7166.* Fulton T A, Dolan G J 1987 Observation of single-electron charging effects in small tunnel junctions *Phys Review Letters* **59** pp 109 – 112.

*7167.* Galindo A, Pascual P 1990, 1991 Quantum mechanics vols **1**, **2** *Springer-Verlag* Berlin Germany pp 1 – 417, 1 – 415.

*7168.* Grabert H, Devoret M H (editors) 1992 Single charge tunneling: Coulomb blockade phenomena in nanostructures *Plenum Press* New York USA.

*7169.* Yokoyama H, Ujihara K 1995 Spontaneous emission and laser oscillation in micro-cavities *CRC Press* Boca Raton USA ISBN 0-8493-3786-0.

*7170.* Alferov Zh I 1996 The history and future of semiconductor heterostructures *in* Proceedings 99$^{th}$ Nobel Symposium Arild June 4-8 1996 *Physica Scripta* **T68** 32.

*7171.* Mygind J 1996-1997 Private communications on the new sources of noise (1/f) in the single electron transistors *Department of Physics* Technical University of Denmark Lyngby Denmark.

*7172.* Milonni P W, Eberly J H 1998 Lasers *John Wiley and Sons Inc* USA ISBN 0-471-62731-3.

*7173.* Muck M 1998 Radio frequency superconducting quantum interference devices *Institute of Applied Physics* University of Giessen Germany.

*7174.* Bimberg D, Grundmann M, Ledentsov N N 1999 Quantum dot heterostructures *John Wiley and Sons Inc* New York USA.

*7175.* Loudon R 2001 The quantum theory of light 3$^{rd}$ edition *Oxford University Press* New York USA Oxford UK.

*7176.* Ledenyov V O, Ledenyov O P, Ledenyov D O 2002 A quantum random number generator on magnetic flux qubits *Proceedings of the 2$^{nd}$ Institute of Electrical and Electronics Engineers Conference IEEE-NANO 2002* Chicago Washington DC USA IEEE Catalog no 02TH86302002 Library of Congress number: 2002106799 ISBN: 0-7803-7538-6.

*7177.* Fortov V E 2016 Extreme states of matter Springer Series in Materials Science *Springer International Publishing* Switzerland ISBN 978-3-319-18952-9 pp 1 – 700.

***Wave function in Schrödinger quantum mechanical wave equation in quantum mechanics:***

*7178.* De Broglie L 1924, 1925 Recherches sur la théorie des quanta Researches on the quantum theory *Ph D Thesis* Sorbonne Paris France, *Annales de Physique* **10** (3) pp 22 – 128.

*7179.* Schrödinger E 1926a Quantisierung als Eigenwertproblem *Annalen der Phys* **384** (4) pp 273 – 376 doi:10.1002/andp.19263840404 Bibcode:1926AnP...384..361S https://dx.doi.org/10.1002%2Fandp.19263840404 , http://adsabs.harvard.edu/abs/1926AnP...384..361S .

7180. Schrödinger E 1926b An undulatory theory of the mechanics of atoms and molecules *Physical Review* **28** (6) pp 1049 – 1070 doi:10.1103/PhysRev.28.1049 Bibcode:1926PhRv...28.1049S

https://dx.doi.org/10.1103%2FPhysRev.28.1049 ,

http://adsabs.harvard.edu/abs/1926PhRv...28.1049S .

7181. Schrödinger E 1982 Collected papers on wave mechanics 3rd edition *American Mathematical Society* ISBN 978-0-8218-3524-1.

7182. Schrödinger E 1984 Collected papers *Friedrich Vieweg und Sohn* ISBN 3-7001-0573-8.

7183. Einstein A 1917 Zur Quantentheorie der Strahlung On the quantum mechanics of radiation *Physikalische Zeitschrift* (in German) 18 pp 121 – 128

http://adsabs.harvard.edu/abs/1917PhyZ...18..121E .

7184. Einstein A, Podolsky B, Rosen N 1935 Can quantum-mechanical description of physical reality be considered complete? *Physical Review* American Physical Society **47** (10) pp 777 – 780

http://journals.aps.org/pr/pdf/10.1103/PhysRev.47.777 ,

http://adsabs.harvard.edu/abs/1935PhRv...47..777E ,

https://dx.doi.org/10.1103%2FPhysRev.47.777 .

7185. Akhiezer A I, Berestetsky V B 1953 Quantum electrodynamics *Gostekhteorizdat* Moscow Russian Federation pp 1 – 428.

7186. Akhiezer A I, Berestetsky V B 1964 Quantum electrodynamics 3rd edition *Nauka* Moscow Russian Federation pp 1 – 624.

7187. Akhiezer A I, Berestetsky V B 1980 Quantum electrodynamics 4th edition *Nauka* Moscow Russian Federation pp 1 – 432.

7188. Berestetsky V B, Lifshits E M, Pitaevsky L P 1980 Quantum electrodynamics *Nauka* Moscow Russian Federation pp 1 – 704.

7189. Dirac P A M 1958 The principles of quantum mechanics 4th edition *Oxford University Press* UK.

7190. Merzbacher E 1961 Quantum mechanics *John Willey and Sons Inc* New York USA pp 1 – 621.

7191. Feynman R, Leighton R B, Sands M 1965 Feynman lectures on physics vol **3** *Addison-Wesley* USA ISBN 0-201-02118-8.

7192. Atkins P W 1974 Quanta: A handbook of concepts *Oxford University Press* UK ISBN 0-19-855493-1.

7193. Atkins P W 1977 Molecular quantum mechanics parts I and II: An introduction to quantum chemistry vol **1** *Oxford University Press* UK ISBN 0-19-855129-0.

7194. Atkins P W 1978 Physical chemistry *Oxford University Press* UK ISBN 0-19-855148-7.

7195. Landau L D, Lifshits E M 1977 Quantum mechanics 3rd edition *Pergamon Press* Oxford UK.

7196. Bransden B H, Joachain C J 1983 Physics of atoms and molecules *Longman* ISBN 0-582-44401-2.

7197. Resnick R, Eisberg R 1985 Quantum physics of atoms, molecules, solids, nuclei and particles 2nd edition *John Wiley & Sons Inc* USA ISBN 978-0-471-87373-0.

**7198.** Galindo A, Pascual P 1990, 1991 Quantum mechanics vols **1**, **2** *Springer-Verlag* Berlin Germany pp 1 – 417, 1 – 415.

**7199.** Shankar R 1994 Principles of quantum mechanics $2^{nd}$ edition *Kluwer Academic/Plenum Publishers* ISBN 978-0-306-44790-7.

**7200.** Ballentine L 1998 Quantum mechanics: A modern development *World Scientific Publishing Co* Singapore ISBN 9810241054.

**7201.** Bransden B H, Joachain C J 2000 Quantum mechanics $2^{nd}$ edition *Prentice Hall* PTR ISBN 0-582-35691-1.

**7202.** Liboff R 2002 Introductory quantum mechanics $4^{th}$ edition *Addison Wesley* ISBN 0-8053-8714-5.

**7203.** Abers E, Pearson Ed 2004 Quantum mechanics *Addison Wesley Prentice Hall Inc* ISBN 978-0-13-146100-0.

**7204.** Blokhintsev D I 2004 Foundations of quantum mechanics $7^{th}$ edition Lan' *Publishing House* St Petersburg Russian Federation ISBN 5-8114-0554-5 pp 1 – 664.

**7205.** Griffiths D J 2004 Introduction to quantum mechanics $2^{nd}$ edition *Prentice Hall* NJ USA ISBN 0-13-111892-7.

**7206.** Vakarchuk I O 2004 Quantum mechanics *L'viv National University Publishing House* L'viv Ukraine.

**7207.** McMahon D 2006 Quantum mechanics demystified *McGraw Hill* USA ISBN (10) 0 07 145546 9.

**7208.** Halliday D 2007 Fundamentals of physics $8^{th}$ edition *John Wiley & Sons Inc* NY USA ISBN 0-471-15950-6.

**7209.** Hand L N, Finch J D 2008 Analytical mechanics *Cambridge University Press* UK ISBN 978-0-521-57572-0.

**7210.** Teschl G 2009 Mathematical methods in quantum mechanics with applications to Schrödinger operators *American Mathematical Society* Providence USA ISBN 978-0-8218-4660-5.

**7211.** Zettili N 2009 Quantum mechanics: Concepts and applications *John Wiley & Sons Inc* NY USA ISBN 978-0-470-02678-6.

**7212.** Laloe F 2012 Do we really understand quantum mechanics *Cambridge University Press* UK ISBN 978-1-107-02501-1.

**7213.** Rylov Y A 2015 What is the wave function and why is it used in quantum mechanics? pp 1 – 18

http://gasdyn-ipm.ipmnet.ru/~rylov/yrylov.htm .

**7214.** Wikipedia 2015i Erwin Schrödinger *Wikipedia* USA

www.wikipedia.org .

**7215.** Wikipedia 2015j Schrödinger equation *Wikipedia* USA

www.wikipedia.org .

***Optics, optical crystals, in-fiber optical devices physics and engineering:***

**7216.** Kerr J 1875a A new relation between electricity and light: Dielectrified media birefringent *Philosophical Magazine* **4** 50 (332) pp 337 – 348.

**7217.** Kerr J 1875b A new relation between electricity and light: Dielectrified media birefringent *Philosophical Magazine* **4** 50 (333) pp 446 – 458.

**7218.** Brillouin L 1922 *Ann Phys Paris* **17** 88.

*7219.* Mandelstam L I 1926 *Zh Russ Fiz-Khim* Ova **58** 381.

*7220.* Landsberg G, Mandelstam L 1928 Eine neue Erscheinung bei der Lichtzerstreuung in Krystallen *Naturwissenschaften* **16** (28) p 557.

*7221.* Raman C V 1928 A new radiation *Indian J Phys* **2** pp 387.

*7222.* Stolen R, Lin C April 1978 Self-phase-modulation in silica optical fibers *Phys Rev A* **17** (4) pp 1448 – 1453 Bibcode:1978PhRvA..17.1448S DOI:10.1103/PhysRevA.17.1448.

*7223.* Dutton H J R 1998 Understanding optical communications *Prentice-Hall Inc* NJ USA ISBN 0-13-020141-3 pp 1 – 760.

*7224.* Goff D R 1999 Fiber optics reference guide: A practical guide to the technology 2[nd] edition *Focal Press* Boston USA ISBN 0-240-80360-4 pp 1 – 220.

*7225.* Ledenyov V O 1999 Research consulting on active/passive in-fiber optics devices design *JDS Uniphase* Ottawa Canada.

*7226.* Taylor N 2000 LASER: The inventor, the Nobel laureate, and the thirty-year patent war *Simon & Schuster* New York USA ISBN 0-684-83515-0. OCLC 122973716 pp 1 – 304.

*7227.* Loudon R 2001 The quantum theory of light 3[rd] edition *Oxford University Press* New York USA Oxford UK.

***Artificial intelligence, digital electronics engineering and computer science:***

*7228.* Turing A October 1950 Computing machinery and intelligence *Mind* **LIX** 236 pp 433 – 460 doi:10.1093/mind/LIX.236.433 ISSN 0026-4423.

*7229.* Rich E 1983 Artificial intelligence *McGraw-Hill* USA ISBN 0-07-052261-8.

*7230.* Winston P H 1984 Artificial intelligence *Addison-Wesley* Reading Massachusetts USA ISBN 0-201-08259-4.

*7231.* Haugeland J 1985 Artificial intelligence: The very idea *MIT Press* Cambridge MA USA ISBN 0-262-08153-9.

*7232.* Edelson E 1991 The nervous system *Chelsea House* New York USA ISBN 978-0-7910-0464-7.

*7233.* Jang J-S R July 1991 Fuzzy modeling using generalized neural networks and Kalman filter algorithm *Proceedings of the 9th National Conference on Artificial Intelligence AAAI-91* pp 762 – 767.

*7234.* Sun R, Bookman L (editors) 1994 Computational architectures: Integrating neural and symbolic processes *Kluwer Academic Publishers* Needham MA USA.

*7235.* John G H, Langley P 1995 Estimating continuous distributions in Bayesian classifiers *The 11th Conference on Uncertainty in Artificial Intelligence*.

*7236.* Dowe D L, Hajek A R 1997 A computational extension to the Turing test *Proceedings of the 4[th] Conference of the Australasian Cognitive Science Society*.

*7237.* Kohavi G J R 1997 Wrappers for feature subset selection *Artificial Intelligence* vol **97** no 1-2 pp 272 – 324.

*7238.* Mitchell T 1997 Machine learning *McGraw Hill* USA.

*7239.* Nilsson N 1998 Artificial intelligence: A new synthesis *Morgan Kaufmann Publishers* ISBN 978-1-55860-467-4.

*7240.* Nilsson N 2010 The quest for artificial intelligence: A history of ideas and achievements *Cambridge University Press* New York USA ISBN 978-0-521-12293-1.

**7241.** Poole D, Mackworth A, Goebel R 1998 Computational intelligence: A logical approach *Oxford University Press* New York USA ISBN 0-19-510270-3.

**7242.** Calmet J, Benhamou B, Caprotti O, Henocque L, Sorge V (editors) 2002 Artificial intelligence, automated reasoning, and symbolic computation *Springer-Verlag* Berlin Germany.

**7243.** Russell S J, Norvig P 2003 Artificial intelligence: A modern approach 2[nd] edition *Prentice Hall* Upper Saddle River New Jersey USA ISBN 0-13-790395-2.

**7244.** Luger G, Stubblefield W 2004 Artificial intelligence: Structures and strategies for complex problem solving 5[th] edition *The Benjamin Cummings Publishing Company Inc* ISBN 0-8053-4780-1.

**7245.** McNelis P D 2005 Neural networks in finance *Elsevier Academic Press* San Diego USA.

**7246.** Bach J 2008 Seven principles of synthetic intelligence *in* Artificial general intelligence 2008 Proceedings of the First AGI Conference Wang P, Goertzel B, Franklin S (editors) *IOS Press* pp 63 – 74 ISBN 978-1-58603-833-5.

**7247.** Neapolitan R, Jiang X 2012 Contemporary artificial intelligence *Chapman & Hall* CRC ISBN 978-1-4398-4469-4.

***Deoxyribonucleic acid (DNA), digital DNA of economy of scale and scope in biology economics and finances:***

**7248.** Miescher Fr 1871 Ueber die chemische Zusammensetzung der Eiterzellen (On the chemical composition of pus cells) *Medicinisch-chemische Untersuchungen* **4** pp 441 – 460.

**7249.** Kol'tsov N K December 12, 1927 The physical-chemical basis of morphology *3rd All-Union Meeting of Zoologist, Anatomists, and Histologists* Leningrad USSR.

**7250.** Watson J D, Crick F H 1953 A structure for deoxyribose nucleic acid *Nature* **171** (4356) pp 737 – 738 Bibcode:1953Natur.171..737W , doi:10.1038/171737a0 , PMID 13054692

http://www.nature.com/nature/dna50/watsoncrick.pdf ,

http://adsabs.harvard.edu/abs/1953Natur.171..737W ,

https://dx.doi.org/10.1038%2F171737a0 ,

https://www.ncbi.nlm.nih.gov/pubmed/13054692 .

**7251.** Watson J D 2002 Genes, girls, and Gamow: After the double helix *Random House* New York USA ISBN 0-375-41283-2 OCLC 47716375 .

**7252.** Watson J D 2004 DNA: The secret of life *Random House* New York USA ISBN 978-0-09-945184-6 .

**7253.** Gamow G July 2 1954a Letter to Martynas Ycas *Library of Congress* Washington USA

www.loc.gov/exhibits/treasures/trr115.html .

**7254.** Gamow G 1954b Possible mathematical relation between deoxyribonucleic acid and proteins *Det Kongelige Danske Videnskabernes Selskab* Copenhagen Denmark pp 1 – 2.

**7255.** Library of Congress 2015 DNA: An "amateur" makes a real contribution *American Treasures of the Library of Congress* Library of Congress Washington USA

www.loc.gov/exhibits/treasures/trr115.html .

**7256.** DeVinne (editor) (1985) DNA *American Heritage Dictionary* USA p 413 ISBN 0-395-32943-4 .

**7257.** Dahm R 2008 Discovering DNA: Friedrich Miescher and the early years of nucleic acid research *Hum Genet* **122** (6) pp 565 – 581 doi:10.1007/s00439-007-0433-0 PMID 17901982 https://dx.doi.org/10.1007%2Fs00439-007-0433-0 ,
https://www.ncbi.nlm.nih.gov/pubmed/17901982 .

**7258.** Wikipedia 2015i DNA Wikipedia California USA https://en.wikipedia.org/wiki/DNA .

**7259.** Ledenyov D O, Ledenyov V O 2016p Digital DNA of economy of scale and scope *MPRA Paper no 68960* Munich University Munich Germany, *SSRN Paper no SSRN-id2718931 Social Sciences Research Network* New York USA pp 1 – 58
http://mpra.ub.uni-muenchen.de/68960/ ,
http://ssrn.com/abstract=2718931 .

***Strategy theory in management, business administration, psychology, and mathematics:***

**7260.** Ueda 1904 Shogyo Dai Jiten (The dictionary of commerce) Japan.

**7261.** Ueda 1937 Keieikeizaigaku Saran (The science of business administration, Allgemeine Betriebswirtschaftslehre) Japan.

**7262.** Mano O 1968-1969 On the science of business administration in Japan *Hokudai Economic Papers* vol **1** pp 77 – 93.

**7263.** Mano O 1970 The development of the science of business administration in Japan since 1955 *Hokudai Economic Papers* vol **2** pp 30 – 42.

**7264.** Chandler A D Jr 1962, 1998 Strategy and structure: Chapters in the history of the American industrial enterprise *Beard Books* USA ISBN-10: 158798198X ISBN-13: 978-1587981982 pp 1 – 480.

**7265.** Chandler A D Jr 1977, 1993 The visible hand: The managerial revolution in American business *Belknap Press* ISBN-10 0674940520 ISBN-13 978-0674940529 pp 1 – 624.

**7266.** Chandler A D Jr, Daems H 1980 Managerial hierarchies: Comparative perspectives on the rise of the modern industrial enterprise *Harvard University Press* ISBN 9780674547414 .

**7267.** Chandler A D Jr 1994 Scale and scope: The dynamics of industrial capitalism *Belknap Press* USA ISBN-10: 0674789954 ISBN-13: 978-0674789951 pp 1 – 780.

**7268.** Chandler A D Jr 2001 Inventing the electronic century: The epic story of the consumer electronics and computer industries *Free Press* USA ISBN-10: 0743215672 ISBN-13: 978-0743215671 pp 1 – 336.

**7269.** Chandler A D Jr 2005 Shaping the industrial century: The remarkable story of chemical and pharmaceutical industries *Harvard University Press* Cambridge Massachusetts USA ISBN 0-674-01720-X pp 1 – 366.

**7270.** Andrews K R 1971a The concept of corporate strategy *Richard D Irwin* Homewood USA.

**7271.** Andrews K R 1971b New horizons in corporate strategy *McKinsey Quarterly* vol **7** no 3 pp 34 – 43.

**7272.** Andrews K R 1980 Directors' responsibility for corporate strategy *Harvard Business Review* vol **58** no 6 pp 30 – 42.

7273. Andrews K R 1981a Corporate strategy as a vital function of the board *Harvard Business Review* vol **59** no 6 pp 174 – 180.

7274. Andrews K R 1981b Replaying the board's role in formulating strategy *Harvard Business Review* vol **59** no 3 pp 18 – 23.

7275. Andrews K R 1984 Corporate strategy: The essential intangibles *McKinsey Quarterly* no 4 pp 43 – 49.

7276. Rumelt R P 1974 Strategy, structure and economic performance *Harvard Business School Press* Boston MA USA.

7277. Rumelt R P 1982 Diversification strategy and profitability *Strategic Management Journal* **3** pp 359 – 369.

7278. Porter M E March-April 1979 How competitive forces shape strategy *Harvard Business Review* **57** (2) pp 137 – 145.

7279. Porter M E 1980, 1998 Competitive strategy: Techniques for analyzing industries and competitors *Free Press* New York USA.

7280. Porter M E, Harrigan K R 1981 A framework for looking at endgame strategies *in* Strategic management and business policy Glueck B (editor) *McGraw-Hill* USA.

7281. Porter M E 1982a Cases in competitive strategy *Free Press* New York USA.

7282. Porter M E 1982b Industrial organization and the evolution of concepts for strategic planning: The new learning *in* Corporate strategy: The integration of corporation planning models and economics Taylor T H (editor) *North-Holland Publishing Company* Amsterdam The Netherlands.

7283. Porter M E, Salter M S March 1982, June 1986 Note on diversification as a strategy *Harvard Business School Background Note* Harvard University pp 382 – 129.

7284. Porter M E 1983 Analyzing competitors: Predicting competitor behavior and formulating offensive and defensive strategy *in* Policy, strategy, and implementation Leontiades M (editor) *Random House* USA.

7285. Porter M E 1985 Defensive strategy *Strategy* **7** (1).

7286. Porter M E, Millar V July 1985 How information gives you competitive advantage *Harvard Business Review*
http://hbr.org/1985/07/how-information-gives-you-competitive-advantage/ar/1 .

7287. Porter M E 1985 Competitive advantage: Creating and sustaining superior performance *Free Press* New York USA

7288. Porter M E May 1987a The state of strategic thinking *Economist* London UK.

7289. Porter M E 1987b From competitive advantage to corporate strategy *Harvard Business Review* pp 43 – 59.

7290. Porter M E April 1991 America's green strategy *Scientific American* **264** (4).

7291. Porter M E 1991 Toward a dynamic theory of strategy *Strategic Management Journal* **12** pp 95 – 117.

7292. Montgomery C A, Porter M E (editors) 1991 Strategy: Seeking and securing competitive advantage *Harvard Business School Press* Boston Massachusetts USA.

7293. Porter M E 1994a Global strategy: Winning in the World-wide marketplace *in* The portable MBA in strategy Fahey L, Randall R M (editors) *John Willey & Sons* NY USA.

7294. Porter M E 1994b Competitive strategy revisited: A view from the 1990s *in* The relevance of a decade: Essays to mark the first ten years of the *Harvard Business School Press* Duffy P B (editor) *Harvard Business School Press* Boston Massachusetts USA.

7295. Porter M E, Van der Linde C 1995 Toward a new conception of the environment-competitiveness relationship *Journal of Economic Perspectives* **9** (4) pp 97 – 118.

7296. Porter M E 1996a What is strategy? *Harvard Business Review* **74** (6) pp 61 – 78.

7297. Porter M E December 1996b Tradeoffs, activity systems, and the theory of competitive strategy *Unpublished Work* Harvard University USA.

7298. Porter M E March February 1997 New strategies for inner-city economic development *Economic Development Quarterly* **11** (1).

7299. Schwab K, Conrnelius P, Porter M E 1999 The global competitiveness report *Oxford University Press* New York USA.

7300. Porter M E, Rivkin J W January 2000, March 2001 Competition & strategy: Course structure TN *Harvard Business School Teaching Note* Harvard University pp 700 – 091.

7301. Porter M E March 2001a Strategy and the Internet *Harvard Business Review* **79** (3) http://hbr.org/2001/03/strategy-and-the-internet/ar/1 .

7302. Porter M E 2001b The technological dimension of competitive strategy *in* Research on technological innovation, management and policy vol **7** Burgelman R A, Chesbrough H (editors) *JAI Press* Greenwich CT USA.

7303. Porter M E, Kramer M R 2002 The competitive advantage of corporate philanthropy *Harvard Business Review* **80** (12) pp 56 – 68.

7304. Porter M E, Sakakibara M 2004 Competition in Japan *Journal of Economic Perspectives* Winter Issue.

7305. Anand B N, Bradley S P, Ghemawat P, Khanna T, Montgomery C A, Porter M E, Rivkin J W, Rukstad M G, Wells J R, Yoffie D B June 2005, September 2008 Strategy: Building and sustaining competitive advantage *Harvard Business School Class Lecture* Harvard University USA pp 705 – 509.

7306. Porter M E, Kramer M R December 2006 Strategy and society: The link between competitive advantage and corporate social responsibility *Harvard Business Review* **84** (12).

7307. Porter M E January 2008 The five competitive forces that shape strategy *Special Issue on HBS Centennial Harvard Business Review* **86** (1) http://hbr.org/2008/01/the-five-competitive-forces-that-shape-strategy/ar/1 .

7308. Porter M E, Kramer M R January-February 2011 Creating shared value *Harvard Business Review* Harvard Business School USA https://hbr.org/2011/01/the-big-idea-creating-shared-value .

7309. Porter M December 2013 Fundamental purpose *Value Investor Insight* pp 8 – 20 www.valueinvestorinsight.com .

7310. Porter M E, Heppelmann J E November 2014 How smart, connected products are transforming competition *Harvard Business Review* November USA http://hbr.org/2014/11/how-smart-connected-products-are-transforming-competition/ar/1 .

7311. Porter M E 2015 Strategy award *Thinkers50* London UK

www.thinkers50.org .

***7312.*** Schendel D E, Hofer Ch W 1979 Strategic management. A new view of business policy and planning *Little Brown* Boston USA p 9.

***7313.*** Yelle L E 1979 The learning curve: Historical review and comprehensive survey *Decision Sciences* **10** (2) pp 302 – 328.

***7314.*** Schelling Th C 1980 The strategy of conflict *Harvard University Press* Cambridge USA.

***7315.*** Dess G G, Davis P S 1984 Porter's (1980) generic strategies as determinants of strategic group membership and organizational performance *Academy of Management Journal* **27** (3) pp 467 – 488.

***7316.*** Schwenk C R 1984 Cognitive simplification processes in strategic decision making *Strategic Management Journal* **5** pp 111 – 128.

***7317.*** Hambrick D C 1985 Turnaround strategies *in* Handbook of strategic management Guth W D (editor) *Warren, Gorham and Lamont* New York USA pp 10-1 to 10-32.

***7318.*** Palepu K G 1985 Diversification strategy, profit performance and the entropy measure *Strategic Management Journal* **6** pp 239 – 255.

***7319.*** Barney J B 1986 Strategic factor markets: Expectations, luck, and business strategy *Management Science* **32** (10) pp 1231 – 1241.

***7320.*** Barney J B 1991 Firm resources and sustained competitive advantage *Journal of Management* **17** (1) pp 99 – 120.

***7321.*** Miller D, Friesen P H 1986a Porter's (1980) generic strategies and performance: An empirical examination with American data, Part I: Testing Porter *Organization Studies* **7** pp 37 – 55.

***7322.*** Miller D, Friesen P H 1986b Porter's (1980) generic strategies and performance: An empirical examination with American data, Part II: Performance implications *Organization Studies* **7** pp 255 – 261.

***7323.*** Miller D 1988 Relating Porter's business strategies to environment and structure: Analysis and performance implications *Academy of Management Journal* **31** pp 280 – 308.

***7324.*** Huff A S, Reger R K 1987 A review of strategic process research *Journal of Management* vol **13** no 2 p 211.

***7325.*** Hill C W L, Snell S A 1988 External control, corporate strategy, and firm performance in research intensive industries *Strategic Management Journal* **9** pp 577 – 590.

***7326.*** Baysinger B D, Hoskisson R E 1989 Diversification strategy and R&D intensity in large multiproduct firms *Academy of Management Journal* **32** pp 310 – 332.

***7327.*** Rue L W, Holland P G 1989 Strategic management: Concepts and experiences 2[nd] edition *McGraw-Hill* Singapore; *Sage* Beverly Hills California USA.

***7328.*** Cohen W M, Levinthal D A 1990 Absorptive capacity: A new perspective on learning and innovation *Administrative Science Quarterly* **35** pp 128 – 152.

***7329.*** Goold M 1991 Strategic control in the decentralized firm *Sloan Management Review* **32** (2) pp 69 – 81.

***7330.*** Goold M, Luchs K 1993 Why diversify? Four decades of managed thinking *Academy of Management Executive* **7** (3) pp 7 – 25.

7331. Goold M et al. 1994 Corporate level strategy: Creating value in the multi-business company *John Willey & Sons Inc* New York USA.

7332. Goold M, Campbell A September, October 1998 Desperately seeking synergy *Harvard Business Review* pp 131 – 143.

7333. Alexander M, Goold M, Collis D J, Campbell A, Lieberthal K, Montgomery C A, Palepu K, Prahalad C K, Stalk G, Khanna T, Hart S L, Shulman L F, Evans Ph 1992, 1995, 1996, 1997, 1998, 1999 Harvard Business Review on corporate strategy *Harvard University Press* Cambridge Massachusetts USA ISBN 1-57851-699-4.

7334. Yip G 1992 Total global strategy: Managing for worldwide competitive advantage *Prentice Hall* NY USA.

7335. Yip G 1998 Asian advantage: Key strategies for winning in the Asia-Pacific region *Addison Wesley/Perseus Books* USA.

7336. Yip G 2000 Strategies for Central and Eastern Europe *Macmillan Business* USA.

7337. Yip G 2007 Managing global customers *Oxford University Press* Oxford UK.

7338. Campbell A et al 1995 Corporate strategy: The quest for parenting advantage *Harvard Business Review* **73** (2) pp 120 – 132.

7339. Johnson G, Scholes K 1997 Exploring corporate strategy *Prentice- Hall* London UK.

7340. Johnson G, Scholes K, Whittington R 1998 Exploring corporate strategy *Simon & Shuster* UK ISBN 0-2736-8734-4.

7341. Johnson G, Scholes K, Whittington R 2002, 2003 Exploring corporate strategy 7th Edition *Prentice Hall* Pearson Education Limited UK ISBN 0-2736-8734-4.

7342. McKiernan P 1997 Strategy past, strategy futures *Long range planning* vol **30** no 5 p 792.

7343. Child J, Faulkner D 1998 Strategies of cooperation: Managing alliances, networks and joint ventures *Oxford University Press* Oxford UK.

7344. Hill C, Jones G 1998 Strategic management 1st edition *Houghton Mifflin Co* Boston USA.

7345. Hill C, Jones G 2004 Cases in strategic management 1st edition *Houghton Mifflin Co* Boston USA.

7346. Martin R L (1998-1999, 2005-2006) Private communications on the theory of strategy *Rotman School of Management* University of Toronto Canada.

7347. Moldoveanu M, Martin R L 2001 Agency theory and the design of efficient governance mechanisms *Joint Committee on Corporate Governance Meeting* Rotman School of Management University of Toronto Ontario Canada pp 1 – 57.

7348. Martin R L 2004 Strategic choice structuring: A set of good choices positions a firm for competitive advantage *Rotman School of Management* University of Toronto Canada pp 1 – 14

www.rotman.utoronto.ca strategicChoiceStructuring.pdf .

7349. Martin R L 2007 Becoming an integrative thinker *Rotman Magazine* Rotman School of Management University of Toronto Ontario Canada pp 4 – 9.

7350. Martin R L 2007 Designing the thinker *Rotman Magazine* Rotman School of Management University of Toronto Ontario Canada pp 4 – 8.

**7351.** Martin R L 2008 The opposable mind *Harvard Business Press* Cambridge Massachusetts USA.

**7352.** Martin R L 2009 The design of business *Harvard Business School Press* ISBN 1422177807 pp 1 – 256.

**7353.** Lafley A G, Martin R L 2013 Playing to win: How strategy really works *Harvard Business Review Press* ISBN-10: 142218739X ISBN-13: 978-1422187395 pp 1 – 272.

**7354.** Martin R L 2013 Strategy award *Thinkers50* London UK www.thinkers50.org .

**7355.** Shiryaev A N 1999 Essentials of stochastic finance: Facts, models, theory *Advanced Series on Statistical Science & Applied Probability* vol **3** *World Scientific Publishing Co Pte Ltd* Kruzhilin N (translator) ISBN 981-02-3605-0 Singapore pp 383 – 395, 633 – 646.

**7356.** Laffont J-J, Tirole J 1999 Competition in telecommunications *MIT Press* USA.

**7357.** Grant R 2001 Corporate strategy: Managing scope and strategy content *in* Handbook of strategy and management Pettigrew A, Thomas H, Whittington R (editors) *Sage* Newbury Park California USA pp 72 – 98.

**7358.** Welch J 2001 Straight from the gut *Business Plus* ISBN-10: 0446528382 pp 1 – 496.

**7359.** Welch J 2001 Winning *Warner Business Books* USA.

**7360.** Choo C, Bontis N 2002 The strategic management of intellectual capital and organizational knowledge $1^{st}$ edition *Oxford University Press* Oxford UK.

**7361.** Drejer A 2002 Strategic management and core competencies $1^{st}$ edition *Quorum Books* Westport Connecticut USA.

**7362.** Sadler P 2003 Strategic management $1^{st}$ edition *Kogan Page* Sterling VA USA.

**7363.** Gavetti G, Levinthal D A 2004 The strategy field from the perspective of management science: Divergent strands and possible integration *Management Science* vol **50** no 10 pp 1309–1318 ISSN 0025-1909 EISSN 1526-5501.

**7364.** Gavetti G, Rivkin J W 2007 On the origin of strategy: Action and cognition over time *Organization Science* vol **18** no 3 pp 420 – 439 ISSN 1047-7039 EISSN 1526-5455.

**7365.** Kim W C, Mauborgne R January–February 2004 Value innovation – The strategic logic of high growth *Harvard Business Review* **75** pp 103 – 112 http://hbr.org/2004/07/value-innovation-the-strategic-logic-of-high-growth/ar/1 .

**7366.** Kim W C, Mauborgne R 2005, 2015 Blue ocean strategy: How to create uncontested market space and make the competition irrelevant *Harvard Business School Press* Boston USA ISBN 978-1591396192, ISBN 978-1-62527-449-6 (expanded edition) pp 1 – 240, pp 1 – 287. www.blueoceanstrategy.com , https://smart.ly/blue-ocean-strategy/ .

**7367.** Kim W C, Mauborgne R 2011 Strategy award *Thinkers50* London UK www.thinkers50.org .

**7368.** Roney C 2004 Strategic management methodology $1^{st}$ edition *Praeger* Westport Connecticut USA.

7369. Ireland R, Hoskisson R, Hitt M 2006 Understanding business strategy 1[st] edition *Thomson Higher Education* Mason OH USA.

7370. Besanko D, Shanley M, Dranove D 2007 Economics of strategy *John Wiley &Sons Inc* USA.

7371. Hitt M, Ireland R, Hoskisson R 2007 Management of strategy 1[st] edition *Thomson/South-Western* Australia.

7372. Kirkbride P S 2007 Developing a leadership and talent architecture *MBS Leader-casts* Melbourne Business School Melbourne Australia.

7373. Murphy T, Galunic Ch 2007 Leading in the age of talent wars *INSEAD Leader-casts* INSEAD France.

7374. Sekhar G 2007 Management information systems 1[st] edition *Excel Books* New Delhi India.

7375. Sull D 2007a Simple rules: Strategy as simple rules Part II *Public Lecture* London School of Economics and Political Science London UK.

7376. Sull D 2007b Closing the gap between strategy and execution: Strategy and its discontents *Public Lecture* London School of Economics and Political Science London UK.

7377. Sull D 2007c Closing the gap between strategy and execution: Making hard choices *Public Lecture* London School of Economics and Political Science London UK.

7378. Sull D 2007d Closing the gap between strategy and execution: The strategy loop in action *Public Lecture* London School of Economics and Political Science London UK.

7379. Sull D 2008 An iterative approach to the strategy *Public Lecture* London School of Economics and Political Science London UK.

7380. Teece D J, Winter S 2007 Dynamic capabilities: Understanding strategic change in organizations *Blackwell* Oxford UK.

7381. Samuels R 2008 Japan's grand strategy *Public Lecture on 13.10.2008* London School of Economics and Political Science London UK

http://www.lse.ac.uk/collections/LSEPublicLecturesAndEvents/events/2008/20080819t13 16z001.htm

http://richmedia.lse.ac.uk/publicLecturesAndEvents/20081013_1830_japansGrandStrateg y.mp3

7382. Chamberlain G P 2010 Understanding strategy *Create Space* Charleston South Carolina USA.

7383. Holt D, Cameron D 2010 Cultural strategy *Oxford University Press* Oxford UK ISBN 978-0-19-958740-7.

7384. Heracleous 2013 Quantum strategy by Apple Inc *Organizational Dynamics* **42** pp 92 – 99

www.elsevier.com/locate/orgdyn .

7385. Ive J, Foulkes N March 6 2015 The man behind the Apple watch *How to Spend It Financial Times* London UK

http://howtospendit.ft.com/articles/77791 .

7386. Ledenyov D O, Ledenyov V O 2015b Winning virtuous strategy creation by interlocking interconnecting directors in boards of directors in firms in information century *MPRA Paper no 61681* Munich University Munich Germany, *SSRN Paper no SSRN-id2553938 Social Sciences Research Network* New York USA pp 1 – 108

http://mpra.ub.uni-muenchen.de/61681/ ,

http://ssrn.com/abstract=2553938 .

**7387.** Ledenyov D O, Ledenyov V O 2015n Quantum strategy creation by interlocking interconnecting directors in boards of directors in modern organizations at time of globalization *MPRA Paper no 68404* Munich University Munich Germany, *SSRN Paper no SSRN-id2704417 Social Sciences Research Network* New York USA pp 1 – 104

http://mpra.ub.uni-muenchen.de/68404/ ,

http://ssrn.com/abstract=2704417 .

**7388.** Ledenyov D O, Ledenyov V O 2015o Multivector strategy vs quantum strategy by Apple Inc *MPRA Paper no 68730* Munich University Munich Germany, *SSRN Paper no SSRN-id2707662 Social Sciences Research Network* New York USA pp 1 – 109

http://mpra.ub.uni-muenchen.de/68730/ ,

http://ssrn.com/abstract=2707662 .

**7389.** Ledenyov D O, Ledenyov V O 2016q Quantum strategy synthesis by Alphabet Inc *MPRA Paper no 69405* Munich University Munich Germany, *SSRN Paper no SSRN-id2729207 Social Sciences Research Network* New York USA pp 1 – 104

http://mpra.ub.uni-muenchen.de/69405/ ,

http://ssrn.com/abstract=2729207 .

**7390.** Ledenyov V O, Ledenyov D O 2016s Forecast in capital markets *Lambert Academic Publishing* Saarbrücken Germany ISBN 978-3-659-91698-4; *MPRA Paper no 72286* Munich University Munich Germany; *SSRN Paper no SSRN-id2802085 Social Sciences Research Network* New York USA pp 1 – 260

www.lap-publishing.com ,

http://mpra.ub.uni-muenchen.de/72286/ ,

http://ssrn.com/abstract=2802085 .

**7391.** Grant R M 2016 Contemporary strategy analysis: Text and cases edition *John Wiley and Sons Inc* ISBN 1119120845 ISBN 9781119120841 pp 1 – 776.

***Information absorption, accumulation, asymmetric flows at knowledge base creation in economics, finances, business administration:***

**7392.** Hayek F A 1945 The use of knowledge in society *American Economic Review* **35** pp 519 – 530.

**7393.** Hayek F A 1973, 1980 Law, legislation and liberty (in English) *Routledge & Kegan*, Droit, législation et liberté (in French) *PUF*.

**7394.** Akerlof G A 1970 The market for lemons: Qualitative uncertainty and the market mechanism *Quarterly Journal of Economics* **84** (3) pp 488 – 500.

**7395.** Akerlof G A August 29 2014 Writing the "The Market for 'Lemons'": A Personal Interpretive Essay Nobelprize.org. Nobel Media AB 2014. Web. 29 Aug 2014. http://www.nobelprize.org/nobel_prizes/economic-sciences/laureates/2001/akerlof-article.html?utm_source=facebook&utm_medium=social&utm_campaign=facebook_page .

7396. Alchian A, Demsetz H 1972 Production, information costs, and economic organization *American Economic Review* vol **62** pp 777 – 795.

7397. Leland H, Pyle D 1977 Informational asymmetries, financial structure and financial intermediation *Journal of Finance* **32** pp 371 – 387.

7398. Crawford V P, Sobel J 1982 Strategic information transmission *Econometrica* **50** pp 1431 – 1451.

7399. Aoki M 1988 Information, incentives and bargaining in the Japanese economy *Cambridge University Press* UK.

7400. Cohen W M, Levinthal D A 1989 Innovation and learning: The two faces of R&D *Economic Journal* **99** pp 569 – 596.

7401. Cohen W M, Levinthal D A 1990 Absorptive capacity: A new perspective on learning and innovation *Administrative Science Quarterly* **35** pp 128 – 152.

7402. Nonaka I 1994 A dynamic theory of organizational knowledge creation *Organization Science* **5** (1) pp 14 – 37.

7403. Kumar R, Nti K O 1998 Differential learning and interaction in alliance dynamics: A process and outcome discrepancy model *Organization Science* **9** pp 356 – 367.

7404. Lane R J, Lubatkin M 1998 Relative absorptive capacity and inter-organizational learning *Strategic Management Journal* **19** pp 461 – 477.

7405. Teece D 1998 Capturing value from knowledge assets *California Management Review* **40** (3) pp 62 – 78.

7406. Garicano L 2000 Hierarchies and the organization of knowledge in production *Journal of Political Economy* **108** pp 874 – 904.

7407. Lev B, Nissim D, Thomas J 2002 On the informational usefulness of R&D capitalization and amortization *Working Paper* Columbia University NY USA.

7408. Foray D 2004 The economics of knowledge *MIT Press* Boston USA.

7409. Petersen M A 2004 Information: Hard and soft *Kellogg School of Management* Northwestern University Evanston Illinois USA.

7410. Angeletos G-M, Pavan A 2007 Use of information and social value of information *Econometrica* **75** pp 1103 – 1114.

7411. Farina V 2008 Network embeddedness, specialization choices and performance in investment banking industry *University of Rome Tor Vergata* Italy *MPRA Paper no 11701* Munich University Munich Germany pp 1 – 26 http://mpra.ub.uni-muenchen.de/11701/ .

## *Game theory in strategy, management, business administration, psychology, and mathematics:*

7412. Cournot A A 1838, 1897 Recherches sur les principles mathematiques de la théorie des richesses Libraire des sciences politiques et sociales *M. Rivière & C.ie* Paris France, Researches into the mathematical principles of the theory of wealth *Macmillan* New York USA.

7413. Edgeworth F Y 1881 Mathematical psychics *Kegan Paul* London UK.

7414. von Neumann J 1928, 1959 Zur Theorie der Gesellschaftsspiele, On the theory of games of strategy *Mathematische Annalen* **100** (1) pp 295 – 320 doi:10.1007/BF01448847 , Contributions to the theory of games vol **4** pp 13 – 42 *Princeton University Press* Princeton USA.

**7415.** von Neumann J, Morgerstern O 1944 Theory of games and economic behavior *Princeton University Press* Princeton NJ USA pp 1 – (625)776 ISBN: 9780691130613
http://press.princeton.edu/titles/7802.html ;
https://en.wikipedia.org/wiki/Theory_of_Games_and_Economic_Behavior .

**7416.** Hayek F A September 1945 The use of scientific knowledge in society *American Economic Review* **35** no 4 .

**7417.** Marschak J 1946 Neumann's and Morgenstern's new approach to static economics *Journal of Political Economy* **54** pp 97 – 115.

**7418.** Nash J F 1950a Equilibrium points in n-person games *Proceedings of the National Academy of Sciences of the United States of America* **36** (1) pp 48 – 49 doi:10.1073/pnas.36.1.48.

**7419.** Nash J F 1950b The bargaining problem *Econometrica* **18** (2) pp 155 – 162 doi:10.2307/1907266 .

**7420.** Nash J May 1950c Non-cooperative games *Ph.D. thesis* Princeton University Princeton NJ USA
https://www.princeton.edu/mudd/news/faq/topics/Non-Cooperative_Games_Nash.pdf .

**7421.** Nash J 1951 Non-cooperative games *Annals of Mathematics* **54** (2) pp 286 – 295 doi:10.2307/1969529.

**7422.** Nash J F 1953 Two-person cooperative games *Econometrica* **21** (1) pp 128 – 140 doi:10.2307/1906951 .

**7423.** Shapley L S 1953 Stochastic games *Proceedings of National Academy of Sciences of the United States of America* vol **39** pp 1095–1100.

**7424.** Williams J D 1954 The compliant strategist: Being a primer on the theory of games of strategy *RAND Corporation* Santa Monica California USA ISBN 978-0-8330-4222-4 .

**7425.** Luce R D, Raiffa H 1957, 1989 Games and decisions: Introduction and critical survey *John Wiley and Sons Inc*, *Dover Publications* ISBN 978-0-486-65943-5 New York USA.

**7426.** Shubik M 1953a Game theory and operations research *Journal of the Research Society of America* **1** p 152.

**7427.** Shubik M March 1953b Non-cooperative games and economic theory *in* Report of Conference on the Theory of N-Person Games pp 20 – 23.

**7428.** Shubik M 1953c The role of game theory in economics *Kyklos* **7** (2) p 21.

**7429.** Shubik M 1954 Readings in game theory and political behavior *Doubleday* New York USA.

**7430.** Shubik M 1955 The uses of game theory in management science *Management Science* **2** (1) pp 40 – 54.

**7431.** Shubik M 1956 A game theorist looks at the antitrust laws and the automobile industry *Stanford Law Review* **8** (4) pp 594 – 630.

**7432.** Shubik M 1958a Economics and operations research: A symposium *The Review of Economics and Statistics* **40** (3) pp 214 – 220.

**7433.** Shubik M 1958b Simulation of the firm *The Journal of Industrial Engineering* vol **IX** no 5 pp 391 – 392.

**7434.** Shubik M 1958c Studies and theories of decision making *Administrative Science Quarterly* pp 289 – 306.

**7435.** Shubik M 1959 Strategy and market structure: Competition, oligopoly, and the theory of games *John Wiley and Sons Inc* NY USA.

**7436.** Shubik M 1975 Games for society, business and war *Elsevier* Amsterdam The Netherlands.

**7437.** Brewer G, Shubik M 1979 The war game *Harvard University Press* Cambridge USA.

**7438.** Shubik M, Levitan R 1980 Market structure and behavior *Harvard University Press* Cambridge USA.

**7439.** Shubik M 1981 Game theory models and methods in political economy *in* Handbook of Mathematical Economics vol **1** pp 285–330 doi:10.1016/S1573-4382(81)01011-4 .

**7440.** Shubik M 1987 A game-theoretic approach to political economy *MIT Press* Cambridge MA USA.

**7441.** Shubik M 1988 What is an application and when is theory a waste of time? *Management Science* **33** 12.

**7442.** Shubik M July 12 2000 Game theory: Some observations *Working Paper #132 Yale School of Management* Yale University New Haven CT06520 USA pp 1 – 7 http://papers.ssrn.com/paper.taf?abstract_id=238964 .

**7443.** Shubik M May 2001 Game theory and operations research: Some musings 50 years later *Working Paper#14 Yale School of Managemen*t Yale University New Haven CT06520 USA    pp 1 – 12 http://papers.ssrn.com/abstract=271029 .

**7444.** Simon H A 1955 A behavioral model of rational choice *Quarterly Journal of Economics* **69** pp 99 – 118.

**7445.** Tucker A W, Luce R D (editors) 1959 Contributions to the theory of games vol **4** *Princeton University Press* Princeton NJ USA.

**7446.** Blackett P M S 1962 Studies of war *Oliver and Boyd* Edinborough Scotland UK.

**7447.** Farquharson R 1969 Theory of voting *Blackwell* ISBN 0-631-12460-8 .

**7448.** Morse P E, Kimball G E 1970 Methods of operations research *Peninsula Publishing* Los Altos California USA.

**7449.** Arrow K J 1971 Control in large organizations: Essays in the theory of risk-bearing *Markham Publishing Co*.

**7450.** Howard N 1971 Paradoxes of rationality: Games, metagames, and political behavior *MIT Press* Cambridge MA USA ISBN 978-0-262-58237-7 .

**7451.** Alchian A A, Demsetz H December 1972 Production, information  costs, and economic organization *American Economic Review* **LXII** no 5 pp 777 – 795.

**7452.** Maynard Smith J, Price G R 1973 The logic of animal conflict *Nature* 246 (5427) pp 15 – 18 doi:10.1038/246015a0 .

**7453.** Maynard Smith J 1974 The theory of games and the evolution of animal conflicts *Journal of Theoretical Biology* **47** (1) pp 209–221 doi:10.1016/0022-5193(74)90110-6.

**7454.** Maynard Smith J 1982 Evolution and the theory of games *Cambridge University Press* Cambridge UK ISBN 978-0-521-28884-2 .

**7455.** Aumann R J, Shapley L S 1974 Values of non-atomic games *Princeton University Press* Princeton NJ USA.

**7456.** Harsanyi J C 1974 An equilibrium-point interpretation of stable sets and a proposed alternative definition *Management Science* **20** (11) p 1472 doi:10.1287/mnsc.20.11.1472 .

**7457.** Selten R 1975 Reexamination of the perfectness concept for equilibrium points in extensive games *International Journal of Game Theory* **4** pp 25 – 55.

**7458.** Selten R, Harsanyi J C 1988 A general theory of equilibrium selection in games *MIT Press* Cambridge MA USA.

**7459.** Selten R 1988 Models of strategic rationality *Kluwer Academic Publishers*, *Springer* Theory and Decision Library Series C: Game Theory, Mathematical Programming and Operations Research Dordrecht-Boston-London The Netherlands, USA pp 1 – 336 ISBN 978-9-027-72663-6.

**7460.** Jensen M C, Meckling W H October 1976 Theory of the firm: Managerial behavior, agency costs, and ownership structure *Journal of Financial Economics* **3** no 4 pp 305 – 360.

**7461.** Jensen M C, Meckling W H Summer 1994, 1998 The nature of man *Journal of Applied Corporate Finance* vol **7** no 2 pp 4 – 19, Foundations of organizational strategy *Harvard University Press* Cambridge USA
http://papers.ssrn.com/abstract=5471 .

**7462.** Jensen M C 1995 Economics, organizations, and non-rational behavior *Economic Inquiry*.

**7463.** Morgenstern O 1976 The collaboration between Oskar Morgenstern and John von Neumann on the theory of games *Journal of Economic Literature* **14** (3) pp 805 – 816.

**7464.** Bicchieri C 1989 Self-refuting theories of strategic interaction: A paradox of common knowledge *Erkenntnis* **30** (1–2) pp 69–85 doi:10.1007/BF00184816 .

**7465.** Bicchieri C 1993 Rationality and coordination *Cambridge University Press* Cambridge UK ISBN 0-521-57444-7 .

**7466.** Bicchieri C, Jeffrey R, Skyrms B (editors) 1999 Knowledge, belief, and counterfactual reasoning in games The logic of strategy *Oxford University Press* New York USA ISBN 0195117158 .

**7467.** Bicchieri C 2006 The grammar of society: The nature and dynamics of social norms *Cambridge University Press* Cambridge UK ISBN 0521573726 .

**7468.** Fudenberg D, Tirole J 1991 Game theory *MIT Press* Cambridge MA USA ISBN 978-0-262-06141-4 .

**7469.** Myerson R B 1991 Game theory: Analysis of conflict *Harvard University Press* Cambridge USA.

**7470.** Gibbons R D 1992 Game theory for applied economists *Princeton University Press* Princeton NJ USA ISBN 978-0-691-00395-5 .

**7471.** Poundstone W 1992 Prisoner's dilemma: John von Neumann, game theory and the puzzle of the bomb *Anchor* ISBN 978-0-385-41580-4 .

7472. Fudenberg D, Tirole J 1993 Game theory *MIT Press* Cambridge Massachusetts USA.

7473. Ben David S, Borodin A, Karp R, Tardos G, Wigderson A 1994 On the power of randomization in on-line algorithms *Algorithmica* **11** (1) pp 2–14 doi:10.1007/BF01294260 .

7474. Osborne M J, Rubinstein A 1994 A course in game theory *MIT Press* Cambridge MA USA ISBN 978-0-262-65040-3 .

7475. Osborne M J 2004 An introduction to game theory *Oxford University Press* Oxford UK ISBN 978-0-19-512895-6 .

7476. Green J R, Mas-Colell A, Whinston M D 1995 Microeconomic theory *Oxford University Press* Oxford UK ISBN 978-0-19-507340-9 .

7477. Owen G 1995 Game theory: Third edition *Emerald Group Publishing* Bingley ISBN 0-12-531151-6 .

7478. Petrosyan L, Zenkevich N 1996 Game theory, optimization (3) *World Scientific Publishers* ISBN 978-981-02-2396-0 .

7479. Fernandez L F, Bierman H S 1998 Game theory with economic applications *Addison-Wesley* ISBN 978-0-201-84758-1 .

7480. Dutta P K 1999 Strategies and games: Theory and practice *MIT Press* Cambridge MA USA ISBN 978-0-262-04169-0 .

7481. Isaacs R 1999 Differential games: A mathematical theory with applications to warfare and pursuit, control and optimization *Dover Publications* NY USA ISBN 978-0-486-40682-4.

7482. Gintis H 2000 Game theory evolving: A problem-centered introduction to modeling strategic behavior *Princeton University Press* Princeton NJ USA ISBN 978-0-691-00943-8 .

7483. Camerer C 2003 Introduction. Behavioral game theory: Experiments in strategic interaction *Russell Sage Foundation* pp 1 – 25 ISBN 978-0-691-09039-9 .

7484. Miller J H 2003 Game theory at work: How to use game theory to outthink and outmaneuver your competition *McGraw-Hill* New York USA ISBN 978-0-07-140020-6 .

7485. Rasmusen E 2006 Games and information: An introduction to game theory 4th edition *Wiley-Blackwell* ISBN 978-1-4051-3666-2 .

7486. Bellhouse D 2007 The problem of Waldegrave *Journal Électronique d'Histoire des Probabilités et de la Statistique* **3** (2).

7487. Hansen P G, Hendricks V F (editors) 2007 Game theory: 5 questions *Automatic Press / VIP* New York, London USA, UK ISBN 978-87-991013-4-4 .

7488. Webb J N 2007 Game theory: Decisions, interaction and evolution Undergraduate mathematics *Springer* ISBN 1-84628-423-6 .

7489. Harrington J E 2008 Games, strategies, and decision making *Worth* ISBN 0-7167-6630-2.

7490. Fudenberg D, Levine D K (editors) 2008 A long-run collaboration on long-run games *World Scientific* Hackensack New Jersey USA pp 1 – 416 ISBN 978-981-281-846-1 .

**7491.** Leyton-Brown K, Shoham Y 2008 Essentials of game theory: A concise, multidisciplinary introduction *Morgan & Claypool Publishers* San Rafael CA USA ISBN 978-1-59829-593-1 .

**7492.** Shoham Y, Leyton-Brown K 2009 Multiagent systems: Algorithmic, game-theoretic, and logical foundations *Cambridge University Press* Cambridge UK ISBN 978-0-521-89943-7 .

**7493.** Leonard R 2010 Von Neumann, Morgenstern, and the creation of game theory *Cambridge University Press* New York USA ISBN 9780521562669 .

**7494.** Papayoanou P 2010 Game theory for business *Probabilistic Publishing* ISBN 978-0-9647938-7-3 .

**7495.** Julmi Ch 2012 Introduction to game theory *BookBooN* Copenhagen Denmark ISBN 978-87-403-0280-6 .

**7496.** Brandenburger A 2014 The language of game theory: Putting epistemics into the mathematics of games World Scientific Series in Economic Theory 5 *World Scientific* Hackensack NJ USA pp 1 – 300 ISBN 978-981-4513-43-2 .

**7497.** McCain R A 2014 Game theory: A nontechnical introduction to the analysis of strategy 3rd edition *World Scientific* Hackensack NJ pp 1 – 600 ISBN 978-981-4578-87-5 .

**7498.** RAND Corporation 2018 Digital archive of analytical research publications *RAND Corporation* USA.

***Selected research papers in macroeconomics, microeconomics and nanoeconomics:***

**7499.** Ledenyov V O, Ledenyov D O 2012a Shaping the international financial system in century of globalization *Cornell University* NY USA pp 1 – 20 www.arxiv.org 1206.2022.pdf .

**7500.** Ledenyov V O, Ledenyov D O 2012b Designing the new architecture of international financial system in era of great changes by globalization *Cornell University* NY USA pp 1 – 18 www.arxiv.org 1206.2778.pdf .

**7501.** Ledenyov D O, Ledenyov V O 2012a On the new central bank strategy toward monetary and financial instabilities management in finances: econophysical analysis of nonlinear dynamical financial systems Cornell University NY USA pp 1 – 8 www.arxiv.org 1211.1897.pdf .

**7502.** Ledenyov D O, Ledenyov V O 2012b On the risk management with application of econophysics analysis in central banks and financial institutions *Cornell University* NY USA pp 1 – 10 www.arxiv.org 1211.4108.pdf .

**7503.** Ledenyov D O, Ledenyov V O 2013a On the optimal allocation of assets in investment portfolio with application of modern portfolio management and nonlinear dynamic chaos theories in investment, commercial and central banks *Cornell University* NY USA pp 1 – 34 www.arxiv.org 1301.4881.pdf .

**7504.** Ledenyov D O, Ledenyov V O 2013b On the theory of firm in nonlinear dynamic financial and economic systems *Cornell University* NY USA pp 1 – 27 www.arxiv.org 1206.4426v2.pdf .

**7505.** Ledenyov D O, Ledenyov V O 2013c On the accurate characterization of business cycles in nonlinear dynamic financial and economic systems *Cornell University* NY USA pp 1 – 26

www.arxiv.org 1304.4807.pdf .

**7506.** Ledenyov D O, Ledenyov V O 2013d To the problem of turbulence in quantitative easing transmission channels and transactions network channels at quantitative easing policy implementation by central banks *Cornell University* NY USA pp 1 – 40

www.arxiv.org 1305.5656.pdf .

**7507.** Ledenyov D O, Ledenyov V O 2013e To the problem of evaluation of market risk of global equity index portfolio in global capital markets *MPRA Paper no 47708* Munich University Munich Germany pp 1 – 25

http://mpra.ub.uni-muenchen.de/47708/ .

**7508.** Ledenyov D O, Ledenyov V O 2013f Some thoughts on accurate characterization of stock market indexes trends in conditions of nonlinear capital flows during electronic trading at stock exchanges in global capital markets *MPRA Paper no 49964* Munich University Munich Germany pp 1 – 52

http://mpra.ub.uni-muenchen.de/49964/ .

**7509.** Ledenyov D O, Ledenyov V O 2013g On the Stratonovich - Kalman - Bucy filtering algorithm application for accurate characterization of financial time series with use of state-space model by central banks *MPRA Paper no 50235* Munich University Munich Germany pp 1 – 52, *SSRN Paper no SSRN-id2594333 Social Sciences Research Network* New York USA

http://sfm.finance.nsysu.edu.tw/php/Papers/CompletePaper/014-1856280412.pdf ,

http://mpra.ub.uni-muenchen.de/50235/ ,

http://ssrn.com/abstract=2594333 .

**7510.** Ledenyov D O, Ledenyov V O 2013h Tracking and replication of hedge fund optimal investment portfolio strategies in global capital markets in presence of nonlinearities *MPRA Paper no 51176* Munich University Munich Germany pp 1 – 92, *SSRN Paper no SSRN-id2588380 Social Sciences Research Network* New York USA

http://mpra.ub.uni-muenchen.de/51176/ ,

http://ssrn.com/abstract=2588380 .

**7511.** Ledenyov D O, Ledenyov V O 2013i Venture capital optimal investment portfolio strategies selection in diffusion - type financial systems in global capital markets with nonlinearities *MPRA Paper no 51903* Munich University Munich Germany pp 1 – 81, *SSRN Paper no SSRN-id2592989 Social Sciences Research Network* New York USA

http://mpra.ub.uni-muenchen.de/51903/ ,

http://ssrn.com/abstract=2592989 .

**7512.** Ledenyov D O, Ledenyov V O 2014a Mergers and acquisitions transactions strategies in diffusion - type financial systems in highly volatile global capital markets with nonlinearities *MPRA Paper no 61946* Munich University Munich Germany, *SSRN Paper no SSRN-id2561300 Social Sciences Research Network* New York USA pp 1 – 160

http://mpra.ub.uni-muenchen.de/61946/ ,

http://ssrn.com/abstract=2561300 .

7513. Ledenyov D O, Ledenyov V O 2014b Strategies on initial public offering of company equity at stock exchanges in imperfect highly volatile global capital markets with induced nonlinearities *MPRA Paper no 53780* Munich University Munich Germany, *SSRN Paper no SSRN-id2577767 Social Sciences Research Network* New York USA pp 1 – 138

http://mpra.ub.uni-muenchen.de/53780/ ,

http://ssrn.com/abstract=2577767 .

7514. Ledenyov D O, Ledenyov V O 2014c On the winning virtuous strategies for ultra high frequency electronic trading in foreign currencies exchange markets *MPRA Paper no 61863* Munich University Munich Germany, *SSRN Paper no SSRN-id2560297 Social Sciences Research Network* New York USA pp 1 – 175

http://mpra.ub.uni-muenchen.de/61863/ ,

http://ssrn.com/abstract=2560297 .

7515. Ledenyov D O, Ledenyov V O 2014d On the fundamentals of winning virtuous strategies creation toward leveraged buyout transactions implementation during private equity investment in conditions of resonant absorption of discrete information in diffusion - type financial system with induced nonlinearities *MPRA Paper no 61805* Munich University Munich Germany pp 1 – 161, *SSRN Paper no SSRN-id2559168 Social Sciences Research Network* New York USA

http://mpra.ub.uni-muenchen.de/61805/ ,

http://ssrn.com/abstract=2559168 .

7516. Ledenyov D O, Ledenyov V O 2014e *MicroFX* foreign currencies ultra high frequencies trading software platform with embedded optimized Stratonovich – Kalman - Bucy filtering algorithm, particle filtering algorithm, macroeconomic analysis algorithm, market microstructure analysis algorithm, order flow analysis algorithm, comparative analysis algorithm, and artificial intelligence algorithm for near-real-time decision making / instant switching on / between optimal trading strategies *ECE James Cook University* Townsville Australia, Kharkov Ukraine.

7517. Ledenyov D O, Ledenyov V O 2014f *MicroLBO* software program with the embedded optimized near-real-time artificial intelligence algorithm to create winning virtuous strategies toward leveraged buyout transactions implementation and to compute direct/reverse leverage buyout transaction default probability number for selected public/private companies during private equity investment in conditions of resonant absorption of discrete information in diffusion - type financial system with induced nonlinearities *ECE James Cook University* Townsville Australia, Kharkov Ukraine.

7518. Ledenyov D O, Ledenyov V O 2015a Nonlinearities in microwave superconductivity 8[th] edition *Cornell University* NY USA pp 1 – 923

www.arxiv.org 1206.4426v8.pdf .

7519. Ledenyov D O, Ledenyov V O 2015b Winning virtuous strategy creation by interlocking interconnecting directors in boards of directors in firms in information century *MPRA Paper no 61681* Munich University Munich Germany, *SSRN Paper no SSRN-id2553938 Social Sciences Research Network* New York USA pp 1 – 108

http://mpra.ub.uni-muenchen.de/61681/ ,

http://ssrn.com/abstract=2553938 .

**7520.** Ledenyov D O, Ledenyov V O 2015c Information theory of firm *MPRA Paper no 63380* Munich University Munich Germany, *SSRN Paper no SSRN-id2587716 Social Sciences Research Network* New York USA pp 1 – 185

http://mpra.ub.uni-muenchen.de/63380/ ,

http://ssrn.com/abstract=2587716 .

**7521.** Ledenyov D O, Ledenyov V O 2015d Information money fields of cyclic oscillations in nonlinear dynamic economic system *MPRA Paper no 63565* Munich University Munich Germany, *SSRN Paper no SSRN-id2592975 Social Sciences Research Network* New York USA pp 1 – 40

http://mpra.ub.uni-muenchen.de/63565/ ,

http://ssrn.com/abstract=2592975 .

**7522.** Ledenyov D O, Ledenyov V O 2015e On the spectrum of oscillations in economics *MPRA Paper no 64368* Munich University Munich Germany, *SSRN Paper no SSRN-id2606209 Social Sciences Research Network* New York USA pp 1 – 48

http://mpra.ub.uni-muenchen.de/64368/ ,

http://ssrn.com/abstract=2606209 .

**7523.** Ledenyov D O, Ledenyov V O 2015f Digital waves in economics *MPRA Paper no 64755* Munich University Munich Germany, *SSRN Paper no SSRN-id2613434 Social Sciences Research Network* New York USA pp 1 – 55

http://mpra.ub.uni-muenchen.de/64755/ ,

http://ssrn.com/abstract=2613434 .

**7524.** Ledenyov D O, Ledenyov V O 2015g General information product theory in economics science *MPRA Paper no 64991* Munich University Munich Germany, *SSRN Paper no SSRN-id2617310 Social Sciences Research Network* New York USA pp 1 – 54

http://mpra.ub.uni-muenchen.de/64991/ ,

http://ssrn.com/abstract=2617310 .

**7525.** Ledenyov D O, Ledenyov V O 2015h Quantum macroeconomics theory *MPRA Paper no 65566* Munich University Munich Germany, *SSRN Paper no SSRN-id2627086 Social Sciences Research Network* New York USA pp 1 – 55

http://mpra.ub.uni-muenchen.de/65566/ ,

http://ssrn.com/abstract=2627086 .

**7526.** Ledenyov D O, Ledenyov V O 2015i Wave function in economics *MPRA Paper no 66577* Munich University Munich Germany, *SSRN Paper no SSRN-id2659054 Social Sciences Research Network* New York USA pp 1 – 71

http://mpra.ub.uni-muenchen.de/66577/ ,

http://ssrn.com/abstract=2659054 .

**7527.** Ledenyov D O, Ledenyov V O 2015j Quantum microeconomics theory *MPRA Paper no 67010* Munich University Munich Germany, *SSRN Paper no SSRN-id2667016 Social Sciences Research Network* New York USA pp 1 – 71

http://mpra.ub.uni-muenchen.de/67010/ ,

http://ssrn.com/abstract=2667016 .

**7528.** Ledenyov D O, Ledenyov V O 2015k Quantum theory of firm *MPRA Paper no 67162* Munich University Munich Germany, *SSRN Paper no SSRN-id2672288 Social Sciences Research Network* New York USA pp 1 – 73

http://mpra.ub.uni-muenchen.de/67162/ ,

http://ssrn.com/abstract=2672288 .

**7529.** Ledenyov D O, Ledenyov V O 2015l Wave function method to forecast foreign currencies exchange rates at ultra high frequency electronic trading in foreign currencies exchange markets*MPRA Paper no 67470* Munich University Munich Germany, *SSRN Paper no SSRN-id2681183 Social Sciences Research Network* New York USA pp 1 – 156

http://mpra.ub.uni-muenchen.de/67470/ ,

http://ssrn.com/abstract=2681183 .

**7530.** Ledenyov D O, Ledenyov V O 2015m Quantum money *MPRA Paper no 67982* Munich University Munich Germany, *SSRN Paper no SSRN-id2693128 Social Sciences Research Network* New York USA pp 1 – 70

http://mpra.ub.uni-muenchen.de/67982/ ,

http://ssrn.com/abstract=2693128 .

**7531.** Ledenyov D O, Ledenyov V O 2015n Quantum strategy creation by interlocking interconnecting directors in boards of directors in modern organizations at time of globalization *MPRA Paper no 68404* Munich University Munich Germany, *SSRN Paper no SSRN-id2704417 Social Sciences Research Network* New York USA pp 1 – 104

http://mpra.ub.uni-muenchen.de/68404/ ,

http://ssrn.com/abstract=2704417 .

**7532.** Ledenyov D O, Ledenyov V O 2015o Multivector strategy vs quantum strategy by Apple Inc *MPRA Paper no 68730* Munich University Munich Germany, *SSRN Paper no SSRN-id2707662 Social Sciences Research Network* New York USA pp 1 – 109

http://mpra.ub.uni-muenchen.de/68730/ ,

http://ssrn.com/abstract=2707662 .

**7533.** Ledenyov D O, Ledenyov V O 2016p Digital DNA of economy of scale and scope *MPRA Paper no 68960* Munich University Munich Germany, *SSRN Paper no SSRN-id2718931 Social Sciences Research Network* New York USA pp 1 – 58

http://mpra.ub.uni-muenchen.de/68960/ ,

http://ssrn.com/abstract=2718931 .

**7534.** Ledenyov D O, Ledenyov V O 2016q Quantum strategy synthesis by Alphabet Inc *MPRA Paper no 69405* Munich University Munich Germany, *SSRN Paper no SSRN-id2729207 Social Sciences Research Network* New York USA pp 1 – 104

http://mpra.ub.uni-muenchen.de/69405/ ,

http://ssrn.com/abstract=2729207 .

**7535.** Ledenyov D O, Ledenyov V O 2016r Precise measurement of macroeconomic variables in time domain using three dimensional wave diagrams *MPRA Paper no 69609* Munich University Munich Germany, *SSRN Paper no SSRN-id2733607 Social Sciences Research Network* New York USA pp 1 – 52

http://mpra.ub.uni-muenchen.de/69609/ ,

http://ssrn.com/abstract=2733607 .

**7536.** Ledenyov V O, Ledenyov D O 2016s Forecast in capital markets *Lambert Academic Publishing* Saarbrücken Germany ISBN 978-3-659-91698-4; *MPRA Paper no 72286* Munich University Munich Germany; *SSRN Paper no SSRN-id2802085 Social Sciences Research Network* New York USA pp 1 – 260

www.lap-publishing.com ,

www.forecastcapitalmarkets.com ,

http://mpra.ub.uni-muenchen.de/72286/ ,

http://ssrn.com/abstract=2802085 .

**7537.** Ledenyov V O, Ledenyov D O 2017 Investment in capital markets *Lambert Academic Publishing* Saarbrücken Germany ISBN 978-3-330-05708-1; *MPRA Paper no 77414* Munich University Munich Germany; *SSRN Paper no SSRN-id2930848 Social Sciences Research Network* New York USA pp 1 – 696

www.lap-publishing.com ,

www.investmentcapitalmarkets.com ,

http://mpra.ub.uni-muenchen.de/77414/ ,

http://ssrn.com/abstract=2930848 .

**7538.** Ledenyov D O, Ledenyov V O 2012 3D Bifurcation diagram for accurate characterization of dynamic properties of combining risky investments in investment portfolio in nonlinear dynamic financial system *Software in Matlab R2012* Townsville Australia, Kharkov Ukraine.

**7539.** Ledenyov D O, Ledenyov V O 2015t *MicroID* software program with the embedded optimized near-real-time artificial intelligence algorithm to create the winning virtuous business strategies and to predict the director's election / appointment in the boards of directors in the firms, taking to the consideration both the director's technical characteristics and the interconnecting interlocking director's network parameters in conditions of the resonant absorption of discrete information in diffusion - type financial economic system with induced nonlinearities *ECE James Cook University* Townsville Australia, Kharkov Ukraine.

**7540.** Ledenyov D O, Ledenyov V O 2015u *MicroITF* operation system and software programs: **1)** the operation system to control the firm operation by means of the information resources near-real-time processing in the modern firms in the case of the diffusion - type financial economic system with the induced nonlinearities; **2)** the software program to accurately characterize the director's performance by means of a) the filtering of the generated/transmitted/received information by the director into the separate virtual channels, depending on the information content, and b) the measurement of the levels of signals in every virtual channel with the generated/transmitted/received information by the director, in the overlapping interconnecting interlocking directors networks in the boards of directors in the firms during the Quality of Service (QofS) measurements process; and **3)** the software program to create the winning virtuous business strategies by the interlocking interconnecting directors in the boards of directors in the modern firms in the case of the diffusion - type financial economic system with the induced nonlinearities, using the patented recursive artificial intelligence algorithm *ECE James Cook University* Townsville Australia, Kharkov Ukraine.

**7541.** Ledenyov D O, Ledenyov V O 2015v *MicroIMF* software program: the *MicroIMF* software program to make the computer modeling of **1)** the interactions between the information money fields of one cyclic oscillation and the information money fields of other cyclic oscillation(s) in the nonlinear dynamic economic system, **2)** the interactions between the information money fields of cyclic oscillation and the nonlinear dynamic economic system itself, and 3) the density distributions of the information money fields by different cyclic oscillations (the economic continuous waves) in the nonlinear dynamic economic system *ECE James Cook University* Townsville Australia, Kharkov Ukraine.

**7542.** Ledenyov D O, Ledenyov V O 2015w *MicroSA* software program **1)** to perform the spectrum analysis of the cyclic oscillations of the economic variables in the nonlinear dynamic economic system, including the discrete-time signals and the continuous-time signals; **2)** to make the computer modeling and to forecast the business cycles for **a)** the central banks with the purpose to make the strategic decisions on the monetary policies, financial stability policies, and **b)** the commercial/investment banks with the aim to make the business decisions on the minimum capital allocation, countercyclical capital buffer creation, and capital investments *ECE James Cook University* Townsville Australia, Kharkov Ukraine.

**7543.** Ledenyov D O, Ledenyov V O 2015x *DNACode* software program *1)* to model the Digital DNA's complex knowledge base structure for the selected economy of the scale and scope in the case of the G20 nations; *2)* to accurately forecast the generation/propagation of the Ledenyov discrete time digital waves of GIP(t)/GDP(t)/GNP(t)/PPP(t) (the discrete-time digital business cycles of GIP(t)/GDP(t)/GNP(t)/PPP(t)) in the G20 economies of the scales and scopes) *ECE James Cook University* Townsville Australia, Kharkov Ukraine.

**7544.** Ledenyov D O, Ledenyov V O 2016y *MacroSoft* software program, which creates the Ledenyov three dimensional (3D) wave diagram to accurately characterize and to finely display the GIP(t), GDP(t), GNP(t), PPP(t) dependences for the G7 economies of the scales and scopes in the time domain for the two possible cases: 1) the continuous-time waves of GIP(t), GDP(t), GNP(t), PPP(t), and 2) the discrete-time waves of GIP(t), GDP(t), GNP(t), PPP(t).

# List of Figures:

## *Chapter 3:*

## *Chapter 5:*

## *Chapter 6:*

## *Chapter 7:*

space. GDP$_s$(t, monetary base) is negative projection of EAV on scale of speculative economic sector in XY coordinates space. EAV changes in time scale similar to continuous-/discrete- time wave with rotating polarization in Ledenyov classic and quantum econodynamics ............................................................................................**150**

**Fig. 72.** Economic activity vector (EAV), defined by phase angle α in XY coordinates space, for dependences of GIP(t, monetary base), GDP(t, monetary base), GNP(t, monetary base), PPP(t, monetary base) in economy of scale and scope. GDP$_r$(t) is negative projection of EAV on scale of real economic sector in XY coordinates space. GDP$_s$(t, monetary base) is positive projection of EAV on scale of speculative economic sector in XY coordinates space. EAV changes in time scale similar to continuous-/discrete- time wave with rotating polarization in Ledenyov classic and quantum econodynamics .......**151**

# List of Tables:

# Subject Index

# Authors Index

Foreman-Peck, **60**
Forrester, **28, 55**
Fortov, **122**
Fountain, **139**
Fox, **74**
Frank, **56**
Francis, **60**
Franses, **59, 60**
Freeman, **28, 56**
French, **57**
Frey, **55**
Friedman, **22, 55, 137**
Frisch, **38, 54**
Frischtak, **56**
Fromm, **54**
Fujiwara, **23, 29, 31, 37, 38, 60, 61, 138, 143**
Fuller, **37, 56**
Funk, **54**
Fve, **59**

**G**abb, **59**
Gabor, **138**
Galbraith, **138**
Gali, **59, 61**
Galor, **58**
Gamez, **58**
Gandolfo, **58**
Gartler, **56**
Garvy, **28, 54**
George, **23, 31, 54, 142**
Georgescu-Roegen, **55**
Gerlach, **58**
Gertler, **58, 60**
Gevorkyan, **61**
Geweke, **59**
Gîlcă, **61**
Gilles, **57**
Glismann, **28, 55, 56**
Goebel, **162**
Goergen, **138**
Gokcekus, **61**
Goldstein, **28, 56**
Golinelli, **61**
Gómez, **37, 59**
Gonzalez, **61**
Gonzalo, **59**
Goodhart, **55, 61, 137**
Goodwin, **55, 58, 142**
Gopinath, **60**
Gordon, **55, 59, 122**

Gori, **60**
Gottlieb, **55**
Gould, **122**
Gowdy, **58**
Graham, **55, 56**
Grandmont, **56**
Grangel, **138**
Granger, **38, 55, 57**
Greene, **138**
Greenwald, **57**
Greenwood, **57**
Gregersen, **88, 141**
Grether, **56**
Griffiths, **138**
Grigor'ev, **60**
Grinin, **28, 59, 60**
Grossman, **57, 61, 88, 141**
Grotte, **59**
Grubler, **57**
Grundmann, **122, 135**
Guay, **59**
Guerrn, **60**
Guglielmo, **59**
Guo Fan, **61**

**H**aberler, **54**
Hagedorn, **60**
Hagemann, **61**
Hajek, **162**
Hall, **37, 56, 60**
Haltmaier, **60**
Haken, **58**
Hamilton, **57, 138**
Hansen, **54, 56**
Harding, **59**
Hardle, **138**
Harl, **17**
Harley, **55**
Harold, **56**
Harris, **17, 74**
Harrison, **37, 57**
Harrod, **58**
Hart, **88, 141**
Harvey, **37, 56**
Haugeland, **162**
Haustein, **56**
Hayashi, **138**
Hayek, **17, 22, 31, 54, 137**
Heal, **55**
Heckscher, **54**
Hedtke, **57**

# About Authors

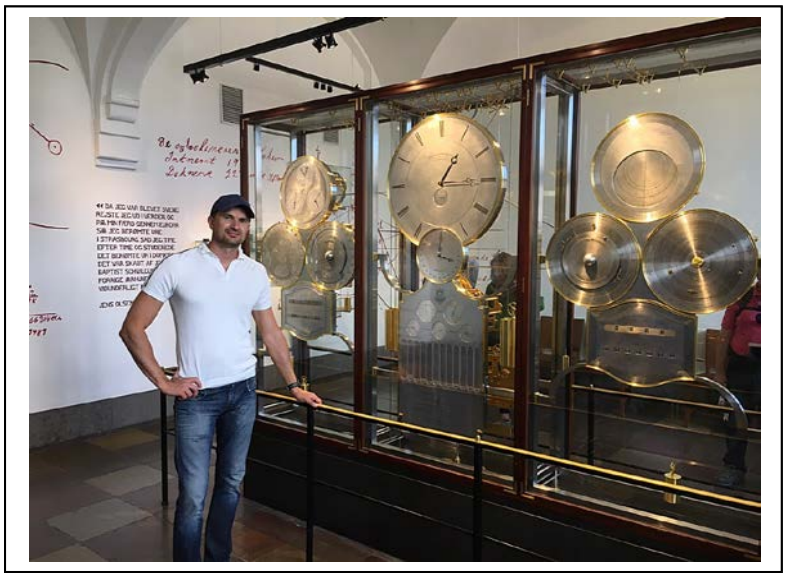

**Dimitri O. Ledenyov** graduated from Department of Radio-Physics and Electronics at V. N. Karazin Kharkiv National University in Kharkiv, Ukraine in 1994 - 1999. Dimitri makes the research on the financial capital investment in the global capital markets in the economics and finances. He also conducts the research in the microwave superconductivity at James Cook University in Townsville, Queensland, Australia in 2000 - 2018. He authored the numerous research papers, reviews and books in the social and natural sciences. He participated in a big number of the international conferences, symposiums, forums, seminars and business meetings around the World.

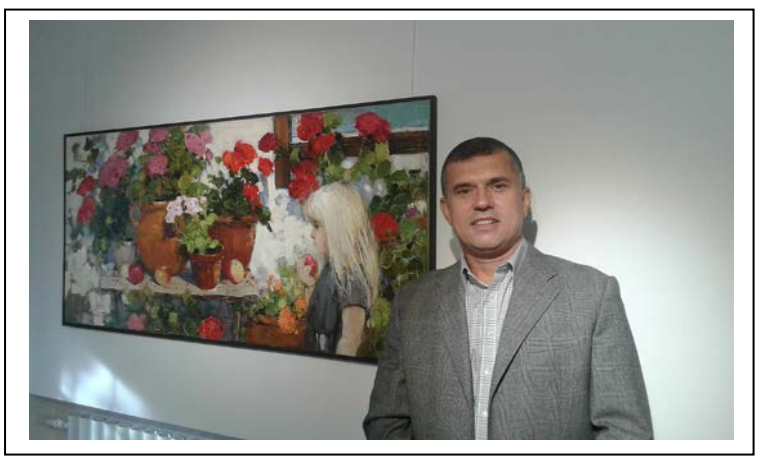

**Viktor O. Ledenyov** graduated from Department of Radio-Physics and Electronics at V. N. Karazin Kharkiv National University in Kharkiv, Ukraine in 1988 - 1993. Viktor makes the research on the financial capital investment in the global capital markets in the economics and finances. He also conducts the research on the microwave superconductivity, the condensed matter physics, the telecommunications, the quantum computing at a number of the research institutions and universities in Europe and in North America in 1993 – 2018. He authored the numerous research papers, reviews and books in the social and natural sciences. He participated in a big number of the international conferences, symposiums, forums, seminars and business meetings around the World.

**Business Cycles in Economics** creates an unified theory of the business cycles in the economics, describing all the possible types of the continuous-time and the discrete-time economic output waves in the economies of the scales and the scopes. The book represents a considerable interest to the prospective and seasoned global investors, because the investors' business philosophy toward the financial capital investment in the real- and speculative- sectors of the economies of the scales and the scopes is mainly based on the forward looking assumptions on the business cycles. The book consists of the eight chapters. Chapter 1 reviews the economies of the scales and the scopes from the ancient days to the modern time. Chapter 2 provides some information on the continuous-time economic output waves in the economy of the scale and the scope in the Ledenyov classic econodynamics. Chapter 3 introduces the Ledenyov discrete-time digital economic output waves in the form of the vector-modulated discrete-time digital direct sequence spread spectrum signal in the economy of the scale and the scope in the Ledenyov classic econodynamics. Chapters 4 describes the Ledenyov discrete-time digital economic output waves in the form of the vector-modulated discrete-time digital direct sequence spread spectrum signals bursts in the economy of the scale and the scope in the Ledenyov classic econodynamics. Chapters 5 writes on the Ledenyov discrete-time digital economic output waves in the form of the vector-modulated discrete-time digital direct sequence spread spectrum signals short/long/ultra long pulses in the economy of the scale and the scope in the Ledenyov classic econodynamics. Chapter 6 discusses the Ledenyov discrete-time digital economic output waves in the form of the vector-modulated discrete-time digital direct sequence spread spectrum signals' short/long/ultra long pulses generated by the quantum leaps in the economy of the scale and the scope in the Ledenyov quantum econodynamics. Chapter 7 considers the precise measurement of the econodynamic variables in the economy of the scale and the scope in the Ledenyov classic and quantum econodynamics. Chapter 8 focuses on some aspects of the accurate forecast of the economic and financial trends with the business cycles oscillation dynamics analysis in the economy of the scale and the scope in the Ledenyov classic and quantum econodynamics sciences.

The private and institutional investors, investment portfolio managers, financiers, economists, financial analysts, financial traders, financial advisers, financial lawmakers, economics policy analysts, economics subject experts, administrative law graduate students, professional lawyers, professors, master of business administration (MBA) students, master of science (M. Sc.) pretenders, philosophy doctor (Ph. D.) candidates and many other interested readers will certainly benefit by reading the book and learning on the new original theoretical propositions on the business cycles in the economies of the scales and the scopes.

**Key Features:**

* Overviews a historical evolution of the economies of the scales and the scopes from the ancient days to the modern time.

* Formulates an unified theory on the business cycles in the economies of the scales and the scopes in the economics.

* Explains the theoretical fundamentals on the business cycles by considering the continuous-time waves and the discrete-time digital waves in the frames of an unified theory on the business cycles in the economies of the scales and the scopes in the economics.

* Introduces the new waveforms to describe the business cycles, including Ledenyov discrete-time digital economic output waves in the form of the vector-modulated discrete-time digital direct sequence spread spectrum signal in the economy of the scale and the scope in the Ledenyov classic econodynamics.

* Discusses the new complex waveforms to characterize the business cycles, including the Ledenyov discrete-time digital economic output waves in the form of the vector-modulated discrete-time digital direct sequence spread spectrum signals short/long/ultra long bursts/pulses in the economy of the scale and the scope in the Ledenyov classic econodynamics.

* Writes about the possible quantum nature of the phase transitions in the new complex waveforms to describe the business cycles, including the Ledenyov discrete-time digital economic output waves in the form of the vector-modulated discrete-time digital direct sequence spread spectrum signals' short/long/ultra long pulses, generated by the quantum leaps, in the economy of the scale and the scope in the Ledenyov quantum econodynamics.

* Presents the scientific thoughts on the precise measurement of the econodynamic variables in the economy of the scale and the scope in the Ledenyov classic and quantum econodynamics.

* Considers some aspects on the accurate forecast of the economic and financial trends with the business cycles oscillation dynamics analysis in the economy of the scale and the scope in the Ledenyov classic and quantum econodynamics.